\PassOptionsToPackage{unicode}{hyperref}
\PassOptionsToPackage{hyphens}{url}
\PassOptionsToPackage{dvipsnames,svgnames,x11names}{xcolor}
\documentclass[
]{article}
\usepackage{amsmath,amssymb}
\usepackage{iftex}
\ifPDFTeX
  \usepackage[T1]{fontenc}
  \usepackage[utf8]{inputenc}
  \usepackage{textcomp} % provide euro and other symbols
\else % if luatex or xetex
  \usepackage{unicode-math} % this also loads fontspec
  \defaultfontfeatures{Scale=MatchLowercase}
  \defaultfontfeatures[\rmfamily]{Ligatures=TeX,Scale=1}
\fi
\usepackage{lmodern}
\ifPDFTeX\else
\fi
\IfFileExists{upquote.sty}{\usepackage{upquote}}{}
\IfFileExists{microtype.sty}{% use microtype if available
  \usepackage[]{microtype}
  \UseMicrotypeSet[protrusion]{basicmath} % disable protrusion for tt fonts
}{}
\makeatletter
\@ifundefined{KOMAClassName}{% if non-KOMA class
  \IfFileExists{parskip.sty}{%
    \usepackage{parskip}
  }{% else
    \setlength{\parindent}{0pt}
    \setlength{\parskip}{6pt plus 2pt minus 1pt}}
}{% if KOMA class
  \KOMAoptions{parskip=half}}
\makeatother
\usepackage{xcolor}
\usepackage[margin=1in]{geometry}
\usepackage{color}
\usepackage{fancyvrb}

\DefineVerbatimEnvironment{Highlighting}{Verbatim}{commandchars=\\\{\}}
\usepackage{framed}
\definecolor{shadecolor}{RGB}{248,248,248}
\newenvironment{Shaded}{\begin{snugshade}}{\end{snugshade}}

\newcommand{\AttributeTok}[1]{\textcolor[rgb]{0.77,0.63,0.00}{#1}}

\newcommand{\CommentTok}[1]{\textcolor[rgb]{0.56,0.35,0.01}{\textit{#1}}}

\newcommand{\ConstantTok}[1]{\textcolor[rgb]{0.00,0.00,0.00}{#1}}
\newcommand{\ControlFlowTok}[1]{\textcolor[rgb]{0.13,0.29,0.53}{\textbf{#1}}}

\newcommand{\DecValTok}[1]{\textcolor[rgb]{0.00,0.00,0.81}{#1}}

\newcommand{\FloatTok}[1]{\textcolor[rgb]{0.00,0.00,0.81}{#1}}
\newcommand{\FunctionTok}[1]{\textcolor[rgb]{0.00,0.00,0.00}{#1}}

\newcommand{\NormalTok}[1]{#1}

\newcommand{\OtherTok}[1]{\textcolor[rgb]{0.56,0.35,0.01}{#1}}

\newcommand{\SpecialCharTok}[1]{\textcolor[rgb]{0.00,0.00,0.00}{#1}}

\newcommand{\StringTok}[1]{\textcolor[rgb]{0.31,0.60,0.02}{#1}}

\usepackage{longtable,booktabs,array}
\usepackage{calc} % for calculating minipage widths
\usepackage{etoolbox}
\makeatletter
\patchcmd\longtable{\par}{\if@noskipsec\mbox{}\fi\par}{}{}
\makeatother
\IfFileExists{footnotehyper.sty}{\usepackage{footnotehyper}}{\usepackage{footnote}}
\makesavenoteenv{longtable}
\usepackage{graphicx}
\makeatletter
\def\maxwidth{\ifdim\Gin@nat@width>\linewidth\linewidth\else\Gin@nat@width\fi}
\def\maxheight{\ifdim\Gin@nat@height>\textheight\textheight\else\Gin@nat@height\fi}
\makeatother
\setkeys{Gin}{width=\maxwidth,height=\maxheight,keepaspectratio}
\makeatletter
\def\fps@figure{htbp}
\makeatother
\providecommand{\tightlist}{%
  \setlength{\itemsep}{0pt}\setlength{\parskip}{0pt}}
\newlength{\cslhangindent}
\newlength{\csllabelwidth}
\newlength{\cslentryspacingunit} % times entry-spacing
\newenvironment{CSLReferences}[2] % #1 hanging-ident, #2 entry spacing
 {% don't indent paragraphs
  \setlength{\parindent}{0pt}
  \ifodd #1
  \let\oldpar\par
  \def\par{\hangindent=\cslhangindent\oldpar}
  \fi
  \setlength{\parskip}{#2\cslentryspacingunit}
 }%
 {}
\usepackage{calc}

\usepackage{bm}
\usepackage{mathtools}
\usepackage{xurl}
\ifLuaTeX
  \usepackage{selnolig}  % disable illegal ligatures
\fi
\IfFileExists{bookmark.sty}{\usepackage{bookmark}}{\usepackage{hyperref}}
\IfFileExists{xurl.sty}{\usepackage{xurl}}{} % add URL line breaks if available
\hypersetup{
  pdftitle={Technical Reports Compilation: Detecting the Fire Drill Anti-pattern Using Source Code and Issue-Tracking Data},
  pdfauthor={Sebastian Hönel},
  colorlinks=true,
  linkcolor={Maroon},
  filecolor={Maroon},
  citecolor={Blue},
  urlcolor={blue},
  pdfcreator={LaTeX via pandoc}}

\title{Technical Reports Compilation: Detecting the Fire Drill Anti-pattern Using Source Code and Issue-Tracking Data}
\author{Sebastian Hönel}
\date{January 30, 2023}

\begin{document}
\maketitle
\begin{abstract}
\noindent Detecting the presence of project management anti-patterns (AP) currently requires experts on the matter and is an expensive endeavor. Worse, experts may introduce their individual subjectivity or bias. Using the Fire Drill AP, we first introduce a novel way to translate descriptions into detectable AP that are comprised of arbitrary metrics and events such as logged time or maintenance activities, which are mined from the underlying source code or issue-tracking data, thus making the description objective as it becomes data-based. Secondly, we demonstrate a novel method to quantify and score the deviations of real-world projects to data-based AP descriptions. Using fifteen real-world projects that exhibit a Fire Drill to some degree, we show how to further enhance the translated AP. The ground truth in these projects was extracted from two individual experts and consensus was found between them. We introduce a novel method called automatic calibration, that optimizes a pattern such that only necessary and important scores remain that suffice to confidently detect the degree to which the AP is present. Without automatic calibration, the proposed patterns show only weak potential for detecting the presence. Enriching the AP with data from real-world projects significantly improves the potential. We also introduce a no-pattern approach that exploits the ground truth for establishing a new, quantitative understanding of the phenomenon, as well as for finding gray-/black-box predictive models. We conclude that the presence detection and severity assessment of the Fire Drill anti-pattern, as well as some of its related and similar patterns, is certainly possible using some of the presented approaches.
\end{abstract}

{
\hypersetup{linkcolor=}
\setcounter{tocdepth}{6}
\tableofcontents
}
\clearpage
\listoffigures
\clearpage
\listoftables

\newcommand*\mean[1]{\overline{#1}}
\newcommand{\abs}[1]{\left\lvert\,#1\,\right\rvert}
\newcommand{\norm}[1]{\left\lVert\,#1\,\right\rVert}
\newcommand{\infdiv}[2]{#1\;\|\;#2}
\newcommand\argmax[1]{\underset{#1}{arg\,max}}
\newcommand\argmin[1]{\underset{#1}{arg\,min}}
\clearpage
\hypertarget{overview}{%
\section{Overview}\label{overview}}

This document is a compilation of five separate technical reports.
The canonical reference to this report is (Hönel 2023).
In all detail, the development of methods for detecting the presence of so-called \emph{"anti-patterns"} in software development projects is presented.

The first technical report is concerned with this concrete problem, and it facilitates two major building blocks: The first is the application of a new method for time warping, called self-regularizing boundary time/amplitude warping (srBTAW). The second building block is a detailed walk-through of creating a classifier for commits, based on source code density. Both these blocks have dedicated technical reports.

The second technical report is concerned with the same problem, but it facilitates issue-tracking data, as well as additional methods for detecting, such as inhomogeneous confidence intervals and vector fields.

The third technical report is an attempt to replace the approach of unsupervised scoring with a robust regression model that has good generalizability using stable learning through adaptive training.
Also, it add a slight adaptation of the first source code pattern.

Notebooks four and five are concerned with an optimization framework for curve warping and -transformation, as well as classifying commits, respectively.

All of the data, source code, and raw materials can be found online. These reports and resources are made available for reproduction purposes. The interested reader is welcome and enabled to re-run all of the computations and to extend upon our ideas. The repository is to be found at \url{https://github.com/MrShoenel/anti-pattern-models}. The data is made available online (Hönel, Pícha, et al. 2023).

\clearpage

\hypertarget{technical-report-detecting-the-fire-drill-anti-pattern-using-source-code}{%
\section{\texorpdfstring{Technical Report: Detecting the Fire Drill anti-pattern using Source Code\label{tr:fire-drill-technical-report}}{Technical Report: Detecting the Fire Drill anti-pattern using Source Code}}\label{technical-report-detecting-the-fire-drill-anti-pattern-using-source-code}}

This is the self-contained technical report for detecting the Fire Drill anti-pattern using source code.

\hypertarget{introduction}{%
\subsection{Introduction}\label{introduction}}

This is the complementary technical report for the paper/article tentatively entitled ``Multivariate Continuous Processes: Modeling, Instantiation, Goodness-of-fit, Forecasting''. Here, we import the ground truth as well as all projects' data, and instantiate our model based on \emph{self-regularizing Boundary Time Warping and Boundary Amplitude Warping}. Given a few patterns that represent the \textbf{Fire Drill} anti-pattern (AP), the goal is evaluate these patterns and their aptitude for detecting the AP in concordance with the ground truth.

All complementary data and results can be found at Zenodo (Hönel, Pícha, et al. 2023). This notebook was written in a way that it can be run without any additional efforts to reproduce the outputs (using the pre-computed results). This notebook has a canonical URL\textsuperscript{\href{https://github.com/MrShoenel/anti-pattern-models/blob/master/notebooks/fire-drill-technical-report.Rmd}{{[}Link{]}}} and can be read online as a rendered markdown\textsuperscript{\href{https://github.com/MrShoenel/anti-pattern-models/blob/master/notebooks/fire-drill-technical-report.md}{{[}Link{]}}} version. All code can be found in this repository, too.

\hypertarget{fire-drill---anti-pattern}{%
\subsection{Fire Drill - anti-pattern}\label{fire-drill---anti-pattern}}

We describe the Fire Drill (FD) anti-pattern for usage in models that are based on the source code (i.e., not from a managerial or project management perspective). The purpose also is to start with a best guess, and then to iteratively improve the description when new evidence is available.

FD is described now both from a managerial and a technical perspective\textsuperscript{\href{https://github.com/ReliSA/Software-process-antipatterns-catalogue/pull/13/commits/78c06c30b1880795e6c1dd0f20f146c548212675?short_path=01589ac\#diff-01589ac85c3fc29739823b5a41ab1bbfba7fbc2579aaf63de0f1ce31713689ab}{{[}Link{]}}}. The technical description is limited to variables we can observe, such as the amount (frequency) of commits and source code, the density of source code, and maintenance activities (a/c/p).

In literature, FD is described in (Silva, Moreno, and Peters 2015) and (Brown et al. 1998), as well as at\textsuperscript{\href{https://web.archive.org/web/20210414150555/https://sourcemaking.com/antipatterns/fire-drill}{{[}Link{]}}}.

Currently, FD is defined to have these symptoms and consequences:

\begin{itemize}
\tightlist
\item
  Rock-edge burndown (especially when viewing implementation tasks only)
\item
  Long period at project start where activities connected to requirements, analysis and planning prevale, and design and implementation activities are rare
\item
  Only analytical or documentational artefacts for a long time
\item
  Relatively short period towards project end with sudden increase in development efforts
\item
  Little testing/QA and project progress tracking activities during development period
\item
  Final product with poor code quality, many open bug reports, poor or patchy documentation
\item
  Stark contrast between inter-level communication in project hierarchy (management - developers) during the first period (close to silence) and after realizing the problem (panic and loud noise
\end{itemize}

From these descriptions, we have attempted to derive the following symptoms and consequences in source code:

\begin{itemize}
\tightlist
\item
  Rock-edge burndown of esp.~implementation tasks mean there are no or just very few adaptive maintenance activities before the burning down starts
\item
  The long period at project start translates to few modifications made to the source code, resulting in fewer commits (lower overall relative frequency)
\item
  Likewise, documentational artifacts have a lower source code density, as less functionality is delivered; this density should increase as soon as adaptive activities are registered
\item
  The short period at project end is characterized by a higher frequency of higher-density implementation tasks, with little to no perfective or corrective work
\item
  At the end of the project, code quality is comparatively lower, while complexity is probably higher, due to pressure exerted on developers in the burst phase
\end{itemize}

\hypertarget{prerequisites}{%
\subsubsection{Prerequisites}\label{prerequisites}}

Through source code, we can observe the following variables (and how they change over time). We have the means to model and match complex behavior for each variable over time. By temporally subdividing the course of a random variable, we can introduce additional measures for a pattern, that are based on comparing the two intervals (e.g., mean, steepest slope, comparisons of the shape etc.).

\begin{itemize}
\tightlist
\item
  Amount of commits over interval of time (frequency) -- We can observe any commit, both at when it was authored first, and when it was added to the repository (author-/committer-date).
\item
  Amount/Frequency of each maintenance activity separately
\item
  Density of the source code (also possibly per activity if required)
  Any other metric, if available (e.g., total Quality or Complexity of project at each commit) -- however, we need to distinguish random variables by whether their relative frequency (when they are changing or simply when they are observed) changes the shape of the function, or whether it only leads to different sample rates. In case of metrics, the latter is the case. In other words, for some variables their occurrence is important, while for others it is the observed value.
\end{itemize}

It is probably the most straightforward way to decompose a complex pattern such as Fire Drill into sub-processes, one for each random variable. That has several advantages:

\begin{itemize}
\tightlist
\item
  We are not bound/limited to only one global aggregated match, which could hide alignment details.
\item
  We can quantify the goodness of match for each variable separately, including details such as the interval in which it matched, and how well it matched in it.
\item
  Matching separately allows us to come up with our own scoring methods; e.g., it could be that the matching-score of one variable needs to be differently computed than the score of another, or simply the weights between variables are different.
\item
  If a larger process was temporally subdivided, we may want to score a variable in one of the intervals differently, or not at all. This is useful for when we cannot make sufficient assumptions.
\end{itemize}

\hypertarget{modeling-the-fire-drill}{%
\subsubsection{Modeling the Fire Drill}\label{modeling-the-fire-drill}}

In this section, we will collect attempts to model the Fire Drill anti-pattern. The first attempt is our initial guess, and subsequent attempts are based on new input (evidence, opinions/discussions etc.).

\hypertarget{initial-best-guess}{%
\paragraph{Initial Best Guess}\label{initial-best-guess}}

\begin{figure}[ht!]

{\centering \includegraphics[width=0.75\linewidth]{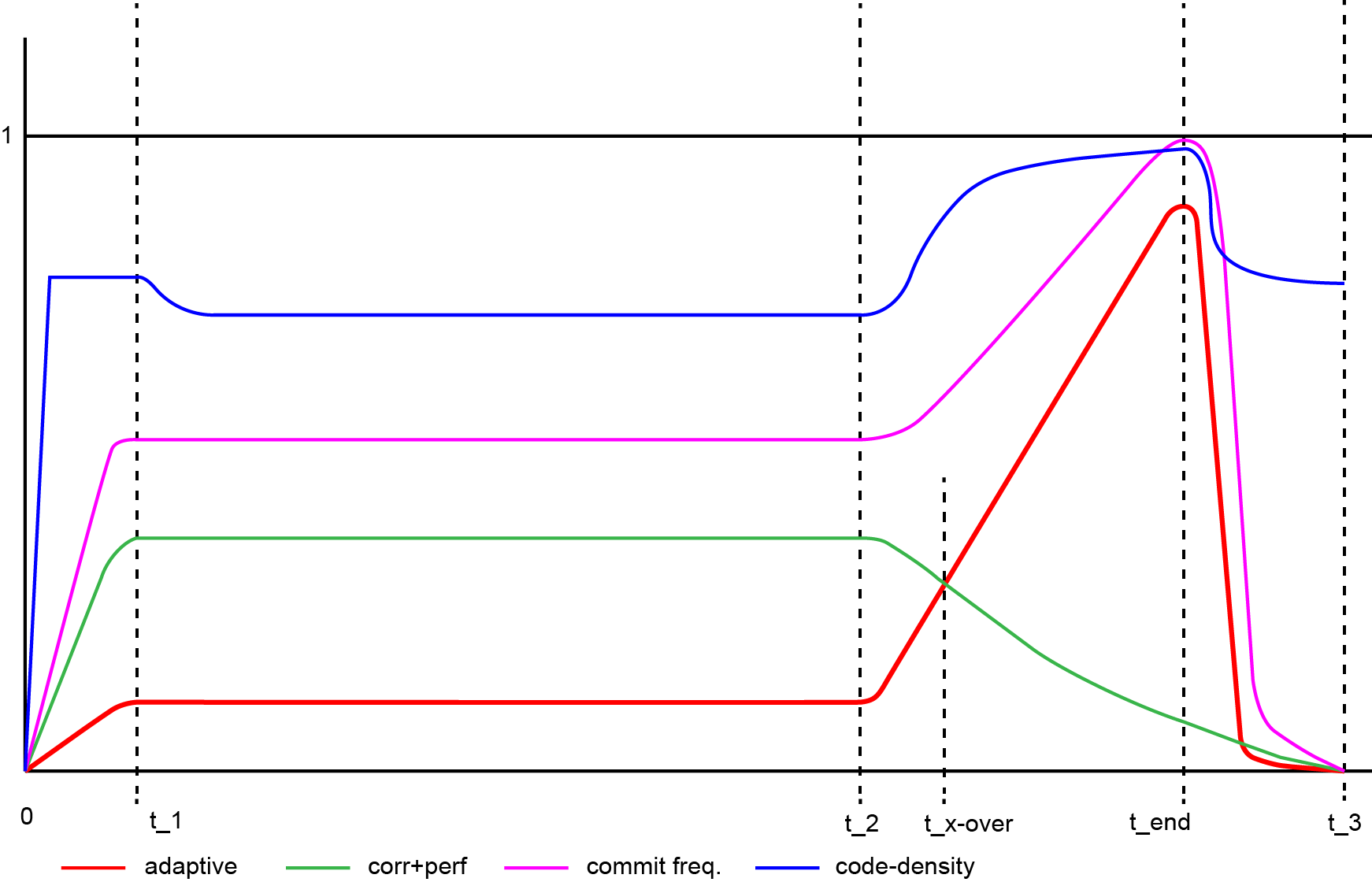} 

}

\caption{Modeling of the Fire Drill anti-pattern over the course of an entire project, according to what we know from literature.}\label{fig:initial-best-guess}
\end{figure}

Our initial best guess is solely based on the literature and descriptions from above, and no dataset was inspected yet. We chose to model our initial best guess using a visual process. Figure \ref{fig:initial-best-guess} must be understood as a simplified and idealized approximation. While we could add a confidence interval for each variable represented, we will later show how align (fit) a project (process) to this pattern (process model), and then measure its deviation from it. The modeled pattern is a \textbf{continuous-time stochastic process model}, and we will demonstrate means to quantify the difference between this \textbf{process model} and a \textbf{process}, which is shown in section \ref{ssec:model-metrics-kde}.

The pattern is divided into four intervals (five if one counts \(t_{\text{x-over}}\) as delimiter, but it is more like an additional point of interest). These intervals are:

\begin{enumerate}
\def\labelenumi{\arabic{enumi}.}
\tightlist
\item
  \([0, t_1)\) -- Begin
\item
  \([t_1, t_2)\) -- Long Stretch
\item
  \([t_2, t_{\text{end}}]\) -- Fire Drill
\item
  \((t_{\text{end}}, t_3]\) -- Aftermath
\end{enumerate}

In each interval and for each of the random variables modeled, we can perform matching. This also means that a) we do not have to attempt matching in each interval, b) we do not have to perform matching for each variable, and c) that we can select a different set of appropriate measures for each variable in each interval (this is useful if, e.g., we do not have much information for some of them).

Each variable is its own sub-pattern. As of now, we track the maintenance activities, and their frequency over time. A higher accumulation results in a higher peak. One additional variable, the source code density (blue), is not measured by its frequency (occurrences), but rather by its value. We may include and model additional metrics, such as complexity or quality.

Whenever we temporally subdivide the pattern into two intervals, we can take these measurements:

\begin{itemize}
\tightlist
\item
  Compute the goodness-of-fit of the curve of a variable, compared to its behavior in the data. As of now, that includes a rich set of metrics, all of which can quantify these differences, and for all of which we have developed scores. While we can compute scores for the actual match, we do also have the means to compute the score of the dynamic time warping. Among others, we have these metrics:

  \begin{itemize}
  \tightlist
  \item
    Reference and Query signal: Start/end (cut-in/-out) for both absolute \& relative
    About the DTW match: (relative) monotonicity (continuity of the warping function), residuals of (relative) warping function (against its LM or optimized own version (fitted using RSS))
  \item
    Between two curves (all of these are normalized as they are computed in the unit square, see R notebook): area (by integration), generic statistics by sampling (mae, rmse, correlation (pearson, kendall, spearman), covariance, standard-deviation, variance, symmetric Kullback-Leibler divergence, symmetric Jenson-Shannon Divergence)
  \end{itemize}
\item
  Sub-division allows for comparisons of properties of the two intervals, e.g.,

  \begin{itemize}
  \tightlist
  \item
    Compare averages of the variable in each interval. This can be easily implemented as a score, as we know the min/max and also do have expectations (lower/higher).
  \item
    Perform a linear regression/create a linear model (LM) over the variable in each interval, so that we can compare slopes, residuals etc.
  \end{itemize}
\item
  Cross-over of two variables: This means that a) the two slopes of their respective LM converge and b) that there is a crossover within the interval.
\end{itemize}

\hypertarget{description-of-figure}{%
\subparagraph{\texorpdfstring{Description of Figure \ref{fig:initial-best-guess}}{Description of Figure }}\label{description-of-figure}}

\begin{itemize}
\tightlist
\item
  The frequency of activities (purple) is the sum of all activities, i.e., this curve is green (corrective + perfective) plus red (adaptive).
\item
  The frequency is the only variable that may be modeled with its actual maximum of 1, as we expect it to reach its maximum at \(t_{\text{end}}\). The frequency also has to be actually 0, before the first commit is made.

  \begin{itemize}
  \tightlist
  \item
    Some of our metrics can measure how well one curve resembles the other, regardless of their ``vertical difference''. Other metrics can describe the distance better. What I want to say is, it is not very important what the value of a variable in our modeled pattern actually is. But it is important however if it touches 0 or 1. A variable should only be modeled as touching 0 or 1 if it is a) monotonically increasing over the course of the project and b) actually assumes its theoretical min/max.
  \end{itemize}
\item
  Corrective and Perfective have been aggregated into one variable, as according to the description of Fire Drill, there is only a distinction between adding features and performing other, such as Q/A related, tasks.
\item
  The source code density (blue) should, with the first commit, jump to its expected value, which is the average over all commits in the project. A steep but short increase is expected in \([t_2, t_{\text{end}}]\) as less focus is spent on documentational artifacts.
\item
  We do not know much about the activities' frequency and relative distribution up till \(t_2\). That leaves us with two options: either, we make only very modest assumptions about the progression of a variable, such as the slope of its LM in that interval or the residuals. Otherwise, we can also choose not to assess the variable in that interval. It is not yet clear how useful the results from the DTW would be, as we can only model a straight line (that's why I am suggesting LMs instead). For Fire Drill, the interval \([t_2, t_3]\) is most characteristic. We can however extract some constraints for our expectations between the intervals \([t_2, t_3]\) and everything that happens before. For example, in {[}BRO'98{]} it is described that \([0, t_2)\) is about 5-7x longer than \([t_2, t_{\text{end}}]\).
\item
  \(t_{\text{end}}\) is not a delimiter we set manually, but it is rather discovered by the sub-patterns' matches. However, it needs to be sufficiently close to the project's actual end (or there needs to be a viable explanation for the difference, such as holidays in between etc.)
\item
  \(t_{\text{x-over}}\) must happen; but other than that, there is not much we can assume about it. We could specify properties as to where approximately we'd expect it to happen (I suppose in the first half of the interval) or how steep the crossover actually is but it is probably hard to rely on.
\end{itemize}

\hypertarget{description-phase-by-phase}{%
\subparagraph{Description Phase-by-Phase}\label{description-phase-by-phase}}

\textbf{Begin}: A short phase of a few days, probably not more than a week. While there will be few commits, \(t_1\) does not really start until the frequency stabilizes. We expect the maintenance activities' relative frequency to decrease towards the end of Begin, before they become rather constant in the next phase. In this phase, the source code density is expected to be close to its average, as initial code as well as documentation are added.

\textbf{Long Stretch}: This is the phase we do not know much about, except for that the amount of adaptive activities is comparatively lower, especially when compared to the aggregated corrective and perfective activities (approx. less than half of these two). While the activities' variables will most likely not be perfectly linear, the LM over these should show rather small residuals. Also the slope of the LM is expected to be rather flat (probably less than +/-10°). The source code density is expected to fall slightly below its average after Begin, as less code is shipped.

\textbf{Fire Drill}: The actual Fire Drill happens in \([t_2, t_{\text{end}}]\), and we detect \(t_{\text{end}}\) by finding the apex of the frequency. However, we choose to extend this interval to include \((t_{\text{end}}, t_3]\), as by doing so, we can craft more characteristics of the anti-pattern and impose more assumptions. These are a) that the steep decline in that last phase has a more extreme slope than its increase before \(t_{\text{end}}\) (because after shipping/project end, probably no activity is performed longer). B) This last sub-phase should be shorter than the phase before (probably just up to a few days; note that the phase \([t_2, t_{\text{end}}]\) is described to be approximately one week long in literature).

With somewhat greater confidence, we can define the following:

\begin{itemize}
\tightlist
\item
  The source code density will rise suddenly and approach its maximum of 1 (however we should not model it with its maximum value to improve matching). It is expected to last until \(t_{\text{end}}\), with a sudden decline back to its average from Begin. We do not have more information for after \(t_{\text{end}}\), so the average is the expected value.
\item
  Perfective and corrective activities will vanish quickly and together become the least frequent activity in the project. The average of these activities is expected to be less than half compared to the Long Stretch. Until \(t_3\) (the very end), the amount of these activities keeps monotonically decreasing.
\item
  At the same time, we will see a steep increase of adaptive activity. The increase is expected to be greater than or equal to the decrease of perfective and corrective activities. In other words, the average of adaptive activities is expected to be more than double, compared to what it was in the Long Stretch. Also, adaptive activities will reach their maximum frequency over the course of the project here.
\item
  The nature of a Fire Drill is a frantic and desperate phase. While adaptive approaches its maximum, the commit frequency also approaches its maximum, even though perfective and corrective activities decline (that is why the purple curve is less steep than the adaptive one but still goes to its global maximum).
\item
  There will be a sharp crossover between perfective+corrective and adaptive activities. It is expected to happen sooner than later in the phase \([t_2, t_{\text{end}}]\).
\end{itemize}

\textbf{Aftermath}: Again, we do not know how this phase looks, but it will help us to more confidently identify the Fire Drill, as the curves of the activities, frequencies and other metrics have very characteristic curves that we can efficiently match. All metrics that are invariant to the frequency are expected to approach their expected value, without much variance (rather constant slope of their resp. LMs). Any of the maintenance activities will continue to fall. In case of adaptive activities we will see an extraordinary steep decline, as after project end/shipping, no one adds functionality. It should probably even fall below all other activities, resulting in another crossover. We do not set any of the activities to be exactly zero however, to allow more efficient matching.

\hypertarget{data}{%
\subsection{Data}\label{data}}

We have \(9\) projects conducted by students, and two raters have \textbf{independently}, i.e., without prior communication, assessed to what degree the AP is present in each project. This was done using a scale from zero to ten, where zero means that the AP was not present, and ten would indicate a strong manifestation.

\hypertarget{the-ground-truth}{%
\subsubsection{The Ground Truth}\label{the-ground-truth}}

\begin{Shaded}
\begin{Highlighting}[]
\NormalTok{ground\_truth }\OtherTok{\textless{}{-}} \FunctionTok{read.csv}\NormalTok{(}\AttributeTok{file =} \StringTok{"../data/ground{-}truth.csv"}\NormalTok{, }\AttributeTok{sep =} \StringTok{";"}\NormalTok{)}
\end{Highlighting}
\end{Shaded}

\begin{table}

\caption{\label{tab:groundtruth}Entire ground truth as of both raters}
\centering
\begin{tabular}[t]{lrrrr}
\toprule
project & rater.a & rater.b & consensus & rater.mean\\
\midrule
project\_1 & 2 & 0 & 1 & 1.0\\
project\_2 & 0 & 0 & 0 & 0.0\\
project\_3 & 8 & 5 & 6 & 6.5\\
project\_4 & 8 & 6 & 8 & 7.0\\
project\_5 & 1 & 1 & 1 & 1.0\\
\addlinespace
project\_6 & 4 & 1 & 2 & 2.5\\
project\_7 & 2 & 3 & 3 & 2.5\\
project\_8 & 0 & 0 & 0 & 0.0\\
project\_9 & 1 & 4 & 5 & 2.5\\
\bottomrule
\end{tabular}
\end{table}

Using the \emph{quadratic weighted Kappa} (Cohen 1968), we can report an unadjusted agreement of \textbf{0.715} for both raters. A Kappa value in the range \([0.6,0.8]\) is considered \emph{substantial}, and values beyond that as \emph{almost perfect} (Landis and Koch 1977). As for the Pearson-correlation, we report a slightly higher value of \textbf{0.771}. The entire ground truth is shown in table \ref{tab:groundtruth}. The final consensus was reached after both raters exchanged their opinions, and it is the consensus that we will use as the actual ground truth from here on and out.

The assessors initially met to agree on the project artifacts to be used in the assessment, the method itself and the scale. The examined artifacts included notes taken by the teams at meetings, both internal and with the customer, iteration retrospectives (required by the process), experience reports compiled by each team at the end of the project, mentor evaluation notes taken after meeting with the teams at the end of each iteration and at the project closure, voluntary comments on teams' performance, and all other artifacts of similar nature -- e.g., describing progress, successes, problems, challenges, practices, and experiences during the projects from a project management and process standpoint.

Expressly excluded were the tasks, time logs and code commits to achieve a higher level of independence between the automatic detection (which is to use these) and the ground truth assessment. The assessors were to examine the materials individually, making notes of what they found, where and whether it indicates presence or absence of a Fire Drill.

After weighing their findings for each project they would evaluate on a zero-to-ten scale their confidence of a present FD, where zero meant no sign of presence or express signs of absence, and ten marked certainty of an FD occurrence.
The initial meeting took approximately an hour and an assessment of each project by a single assessor ranged between one and two hours.

After both assessors processed all the projects, they met for the final time to reach a consensus on the ground truth assessment.
For each individual project, first both assessors presented their case, arguments for and against the presence of Fire Drill, evidence found in the materials and their own assessment value. If the values from both assessors for a single project matched, the value was taken as a consensus outright with no further discussion.
If not, the assessors discussed in depth, going over the materials together whenever needed until reaching an assessment they both agreed on. When a precise agreement could not be reached between two neighboring values (the resulting value needed to be integer, as per previous agreement), more weight was always given to the second assessor, as they were more independent from the projects themselves (having no part in their actual execution).
The individual assessments were seldomly much distant from each other and the consensus value was almost always reached somewhere in the interval between them.
During the discussions, the notes on the consensus were also taken. The consensus meeting took approximately two hours.

\hypertarget{the-student-projects}{%
\subsubsection{The Student Projects}\label{the-student-projects}}

The ground truth was extracted from nine student-conducted projects. Seven of these were implemented simultaneously between March and June 2020, and two the year before in a similar timeframe.

\begin{Shaded}
\begin{Highlighting}[]
\NormalTok{student\_projects }\OtherTok{\textless{}{-}} \FunctionTok{read.csv}\NormalTok{(}\AttributeTok{file =} \StringTok{"../data/student{-}projects.csv"}\NormalTok{, }\AttributeTok{sep =} \StringTok{";"}\NormalTok{)}
\end{Highlighting}
\end{Shaded}

In the first batch, we have a total of:

\begin{itemize}
\tightlist
\item
  Nine projects,
\item
  37 authors that authored 1219 commits total which are of type
\item
  Adaptive / Corrective / Perfective (\texttt{a/c/p}) commits: 392 / 416 / 411
\end{itemize}

\begin{figure}[ht!]
\includegraphics{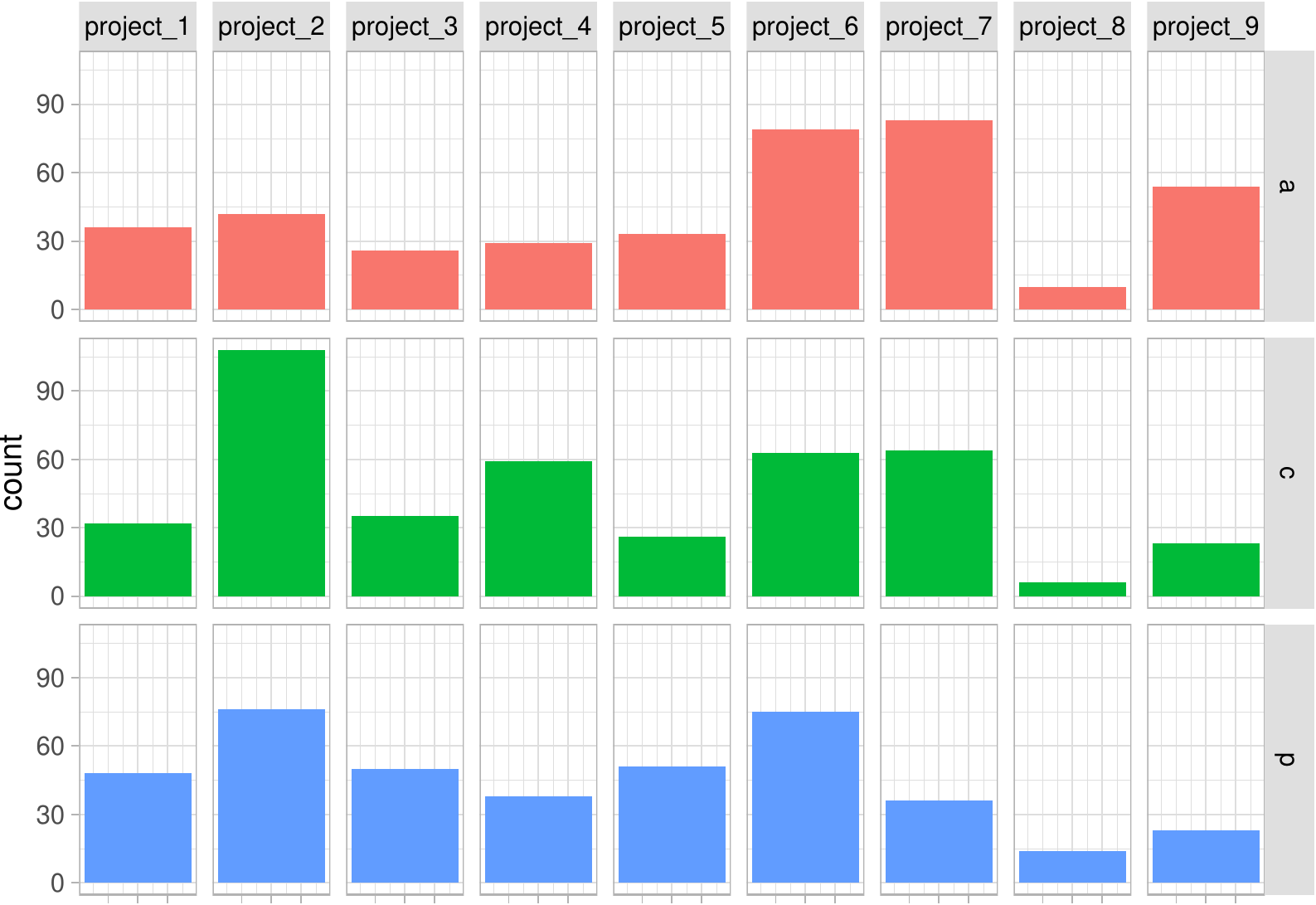} \caption{Commit activities across projects}\label{fig:project-activity}
\end{figure}

We have a complete breakdown of all activities across all projects in figure \ref{fig:project-activity}.

\begin{Shaded}
\begin{Highlighting}[]
\NormalTok{student\_projects\_info }\OtherTok{\textless{}{-}} \ConstantTok{NULL}

\ControlFlowTok{for}\NormalTok{ (pId }\ControlFlowTok{in} \FunctionTok{unique}\NormalTok{(student\_projects}\SpecialCharTok{$}\NormalTok{project)) \{}
\NormalTok{  temp }\OtherTok{\textless{}{-}}\NormalTok{ student\_projects[student\_projects}\SpecialCharTok{$}\NormalTok{project }\SpecialCharTok{==}\NormalTok{ pId, ]}
\NormalTok{  student\_projects\_info }\OtherTok{\textless{}{-}} \FunctionTok{rbind}\NormalTok{(student\_projects\_info, }\FunctionTok{data.frame}\NormalTok{(}\AttributeTok{project =}\NormalTok{ pId,}
    \AttributeTok{authors =} \FunctionTok{length}\NormalTok{(}\FunctionTok{unique}\NormalTok{(temp}\SpecialCharTok{$}\NormalTok{AuthorNominalLabel)), }\AttributeTok{commits =} \FunctionTok{nrow}\NormalTok{(temp),}
    \AttributeTok{a =} \FunctionTok{nrow}\NormalTok{(temp[temp}\SpecialCharTok{$}\NormalTok{label }\SpecialCharTok{==} \StringTok{"a"}\NormalTok{, ]), }\AttributeTok{c =} \FunctionTok{nrow}\NormalTok{(temp[temp}\SpecialCharTok{$}\NormalTok{label }\SpecialCharTok{==} \StringTok{"c"}\NormalTok{, ]),}
    \AttributeTok{p =} \FunctionTok{nrow}\NormalTok{(temp[temp}\SpecialCharTok{$}\NormalTok{label }\SpecialCharTok{==} \StringTok{"p"}\NormalTok{, ]), }\AttributeTok{avgDens =} \FunctionTok{round}\NormalTok{(}\FunctionTok{mean}\NormalTok{(temp}\SpecialCharTok{$}\NormalTok{Density),}
      \DecValTok{3}\NormalTok{)))}
\NormalTok{\}}
\end{Highlighting}
\end{Shaded}

\begin{table}

\caption{\label{tab:studentprojects}Per-project overview of the student projects}
\centering
\begin{tabular}[t]{lrrrrrr}
\toprule
project & authors & commits & a & c & p & avgDens\\
\midrule
project\_1 & 4 & 116 & 36 & 32 & 48 & 0.879\\
project\_2 & 5 & 226 & 42 & 108 & 76 & 0.891\\
project\_3 & 4 & 111 & 26 & 35 & 50 & 0.785\\
project\_4 & 4 & 126 & 29 & 59 & 38 & 0.870\\
project\_5 & 2 & 110 & 33 & 26 & 51 & 0.814\\
\addlinespace
project\_6 & 4 & 217 & 79 & 63 & 75 & 0.784\\
project\_7 & 5 & 183 & 83 & 64 & 36 & 0.813\\
project\_8 & 4 & 30 & 10 & 6 & 14 & 0.687\\
project\_9 & 5 & 100 & 54 & 23 & 23 & 0.743\\
\bottomrule
\end{tabular}
\end{table}

We have slightly different begin- and end-times in each project. However, the data for all projects was previously cropped, so that each project's extent marks the absolute begin and end of it -- it starts with the first commit and ends with the last. As for our methods here, we only need to make sure that we scale the timestamps into a relative \([0,1]\)-range, where \(1\) marks the project's end.

For each project, we model \textbf{four} variables: The activities \textbf{adaptive} (\textbf{\texttt{A}}), \textbf{corrective+perfective} (\textbf{\texttt{CP}}), the frequency of all activities, regardless of their type (\textbf{\texttt{FREQ}}), and the \textbf{source code density} (\textbf{\texttt{SCD}}). While for the first three variables we estimate a Kernel density, the last variable is a metric collected with each commit. The data for it is mined using \texttt{Git-Density} (Hönel 2022), and we use a highly efficient commit classification model\footnote{\url{https://github.com/MrShoenel/anti-pattern-models/blob/master/notebooks/comm-class-models.Rmd}} (\(\approx83.6\%\) accuracy, \(\approx0.745\) Kappa) (Hönel et al. 2020) to attach maintenance activity labels to each commit, based on size- and keyword-data only.

Technically, we will compose each variable into an instance of our \texttt{Signal}-class. Before we start, we will do some normalizations and conversions, like converting the timestamps. This has to be done on a per-project basis.

\begin{Shaded}
\begin{Highlighting}[]
\NormalTok{student\_projects}\SpecialCharTok{$}\NormalTok{label }\OtherTok{\textless{}{-}} \FunctionTok{as.factor}\NormalTok{(student\_projects}\SpecialCharTok{$}\NormalTok{label)}
\NormalTok{student\_projects}\SpecialCharTok{$}\NormalTok{project }\OtherTok{\textless{}{-}} \FunctionTok{as.factor}\NormalTok{(student\_projects}\SpecialCharTok{$}\NormalTok{project)}
\NormalTok{student\_projects}\SpecialCharTok{$}\NormalTok{AuthorTimeNormalized }\OtherTok{\textless{}{-}} \ConstantTok{NA\_real\_}
\end{Highlighting}
\end{Shaded}

\begin{Shaded}
\begin{Highlighting}[]
\ControlFlowTok{for}\NormalTok{ (pId }\ControlFlowTok{in} \FunctionTok{levels}\NormalTok{(student\_projects}\SpecialCharTok{$}\NormalTok{project)) \{}
\NormalTok{  student\_projects[student\_projects}\SpecialCharTok{$}\NormalTok{project }\SpecialCharTok{==}\NormalTok{ pId, ]}\SpecialCharTok{$}\NormalTok{AuthorTimeNormalized }\OtherTok{\textless{}{-}}\NormalTok{ (student\_projects[student\_projects}\SpecialCharTok{$}\NormalTok{project }\SpecialCharTok{==}
\NormalTok{    pId, ]}\SpecialCharTok{$}\NormalTok{AuthorTimeUnixEpochMilliSecs }\SpecialCharTok{{-}} \FunctionTok{min}\NormalTok{(student\_projects[student\_projects}\SpecialCharTok{$}\NormalTok{project }\SpecialCharTok{==}
\NormalTok{    pId, ]}\SpecialCharTok{$}\NormalTok{AuthorTimeUnixEpochMilliSecs))}
\NormalTok{  student\_projects[student\_projects}\SpecialCharTok{$}\NormalTok{project }\SpecialCharTok{==}\NormalTok{ pId, ]}\SpecialCharTok{$}\NormalTok{AuthorTimeNormalized }\OtherTok{\textless{}{-}}\NormalTok{ (student\_projects[student\_projects}\SpecialCharTok{$}\NormalTok{project }\SpecialCharTok{==}
\NormalTok{    pId, ]}\SpecialCharTok{$}\NormalTok{AuthorTimeNormalized}\SpecialCharTok{/}\FunctionTok{max}\NormalTok{(student\_projects[student\_projects}\SpecialCharTok{$}\NormalTok{project }\SpecialCharTok{==}
\NormalTok{    pId, ]}\SpecialCharTok{$}\NormalTok{AuthorTimeNormalized))}
\NormalTok{\}}
\end{Highlighting}
\end{Shaded}

And now for the actual signals: Since the timestamps have been normalized for each project, we model each variable to actually start at \(0\) and end at \(1\) (the support). We will begin with activity-related variables before we model the source code density, as the process is different. When using Kernel density estimation (KDE), we obtain an empirical probability density function (PDF) that integrates to \(1\). This is fine when looking at all activities combined (\textbf{\texttt{FREQ}}). However, when we are interested in a specific fraction of the activities, say \textbf{\texttt{A}}, then we should scale its activities according to its overall ratio. Adding all scaled activities together should again integrate to \(1\). When this is done, we scale one last time such that no empirical PDF has a co-domain larger than \(1\).

\begin{Shaded}
\begin{Highlighting}[]
\NormalTok{project\_signals }\OtherTok{\textless{}{-}} \FunctionTok{list}\NormalTok{()}

\CommentTok{\# passed to stats::density}
\NormalTok{use\_kernel }\OtherTok{\textless{}{-}} \StringTok{"gauss"}  \CommentTok{\# \textquotesingle{}rect\textquotesingle{}}

\ControlFlowTok{for}\NormalTok{ (pId }\ControlFlowTok{in} \FunctionTok{levels}\NormalTok{(student\_projects}\SpecialCharTok{$}\NormalTok{project)) \{}
\NormalTok{  temp }\OtherTok{\textless{}{-}}\NormalTok{ student\_projects[student\_projects}\SpecialCharTok{$}\NormalTok{project }\SpecialCharTok{==}\NormalTok{ pId, ]}

  \CommentTok{\# We\textquotesingle{}ll need these for the densities:}
\NormalTok{  acp\_ratios }\OtherTok{\textless{}{-}} \FunctionTok{table}\NormalTok{(temp}\SpecialCharTok{$}\NormalTok{label)}\SpecialCharTok{/}\FunctionTok{sum}\NormalTok{(}\FunctionTok{table}\NormalTok{(temp}\SpecialCharTok{$}\NormalTok{label))}

\NormalTok{  dens\_a }\OtherTok{\textless{}{-}} \FunctionTok{densitySafe}\NormalTok{(}\AttributeTok{from =} \DecValTok{0}\NormalTok{, }\AttributeTok{to =} \DecValTok{1}\NormalTok{, }\AttributeTok{safeVal =} \ConstantTok{NA\_real\_}\NormalTok{, }\AttributeTok{data =}\NormalTok{ temp[temp}\SpecialCharTok{$}\NormalTok{label }\SpecialCharTok{==}
    \StringTok{"a"}\NormalTok{, ]}\SpecialCharTok{$}\NormalTok{AuthorTimeNormalized, }\AttributeTok{ratio =}\NormalTok{ acp\_ratios[[}\StringTok{"a"}\NormalTok{]], }\AttributeTok{kernel =}\NormalTok{ use\_kernel)}

\NormalTok{  dens\_cp }\OtherTok{\textless{}{-}} \FunctionTok{densitySafe}\NormalTok{(}\AttributeTok{from =} \DecValTok{0}\NormalTok{, }\AttributeTok{to =} \DecValTok{1}\NormalTok{, }\AttributeTok{safeVal =} \ConstantTok{NA\_real\_}\NormalTok{, }\AttributeTok{data =}\NormalTok{ temp[temp}\SpecialCharTok{$}\NormalTok{label }\SpecialCharTok{==}
    \StringTok{"c"} \SpecialCharTok{|}\NormalTok{ temp}\SpecialCharTok{$}\NormalTok{label }\SpecialCharTok{==} \StringTok{"p"}\NormalTok{, ]}\SpecialCharTok{$}\NormalTok{AuthorTimeNormalized, }\AttributeTok{ratio =}\NormalTok{ acp\_ratios[[}\StringTok{"c"}\NormalTok{]] }\SpecialCharTok{+}
\NormalTok{    acp\_ratios[[}\StringTok{"p"}\NormalTok{]], }\AttributeTok{kernel =}\NormalTok{ use\_kernel)}

\NormalTok{  dens\_freq }\OtherTok{\textless{}{-}} \FunctionTok{densitySafe}\NormalTok{(}\AttributeTok{from =} \DecValTok{0}\NormalTok{, }\AttributeTok{to =} \DecValTok{1}\NormalTok{, }\AttributeTok{safeVal =} \ConstantTok{NA\_real\_}\NormalTok{, }\AttributeTok{data =}\NormalTok{ temp}\SpecialCharTok{$}\NormalTok{AuthorTimeNormalized,}
    \AttributeTok{ratio =} \DecValTok{1}\NormalTok{, }\AttributeTok{kernel =}\NormalTok{ use\_kernel)}

  \CommentTok{\# All densities need to be scaled together once more, by dividing for the}
  \CommentTok{\# maximum value of the FREQ{-}variable.}
\NormalTok{  ymax }\OtherTok{\textless{}{-}} \FunctionTok{max}\NormalTok{(}\FunctionTok{c}\NormalTok{(}\FunctionTok{attr}\NormalTok{(dens\_a, }\StringTok{"ymax"}\NormalTok{), }\FunctionTok{attr}\NormalTok{(dens\_cp, }\StringTok{"ymax"}\NormalTok{), }\FunctionTok{attr}\NormalTok{(dens\_freq, }\StringTok{"ymax"}\NormalTok{)))}
\NormalTok{  dens\_a }\OtherTok{\textless{}{-}}\NormalTok{ stats}\SpecialCharTok{::}\FunctionTok{approxfun}\NormalTok{(}\AttributeTok{x =} \FunctionTok{attr}\NormalTok{(dens\_a, }\StringTok{"x"}\NormalTok{), }\AttributeTok{y =} \FunctionTok{sapply}\NormalTok{(}\AttributeTok{X =} \FunctionTok{attr}\NormalTok{(dens\_a,}
    \StringTok{"x"}\NormalTok{), }\AttributeTok{FUN =}\NormalTok{ dens\_a)}\SpecialCharTok{/}\NormalTok{ymax)}
\NormalTok{  dens\_cp }\OtherTok{\textless{}{-}}\NormalTok{ stats}\SpecialCharTok{::}\FunctionTok{approxfun}\NormalTok{(}\AttributeTok{x =} \FunctionTok{attr}\NormalTok{(dens\_cp, }\StringTok{"x"}\NormalTok{), }\AttributeTok{y =} \FunctionTok{sapply}\NormalTok{(}\AttributeTok{X =} \FunctionTok{attr}\NormalTok{(dens\_cp,}
    \StringTok{"x"}\NormalTok{), }\AttributeTok{FUN =}\NormalTok{ dens\_cp)}\SpecialCharTok{/}\NormalTok{ymax)}
\NormalTok{  dens\_freq }\OtherTok{\textless{}{-}}\NormalTok{ stats}\SpecialCharTok{::}\FunctionTok{approxfun}\NormalTok{(}\AttributeTok{x =} \FunctionTok{attr}\NormalTok{(dens\_freq, }\StringTok{"x"}\NormalTok{), }\AttributeTok{y =} \FunctionTok{sapply}\NormalTok{(}\AttributeTok{X =} \FunctionTok{attr}\NormalTok{(dens\_freq,}
    \StringTok{"x"}\NormalTok{), }\AttributeTok{FUN =}\NormalTok{ dens\_freq)}\SpecialCharTok{/}\NormalTok{ymax)}

\NormalTok{  project\_signals[[pId]] }\OtherTok{\textless{}{-}} \FunctionTok{list}\NormalTok{(}\AttributeTok{A =}\NormalTok{ Signal}\SpecialCharTok{$}\FunctionTok{new}\NormalTok{(}\AttributeTok{name =} \FunctionTok{paste}\NormalTok{(pId, }\StringTok{"A"}\NormalTok{, }\AttributeTok{sep =} \StringTok{"\_"}\NormalTok{),}
    \AttributeTok{func =}\NormalTok{ dens\_a, }\AttributeTok{support =} \FunctionTok{c}\NormalTok{(}\DecValTok{0}\NormalTok{, }\DecValTok{1}\NormalTok{), }\AttributeTok{isWp =} \ConstantTok{FALSE}\NormalTok{), }\AttributeTok{CP =}\NormalTok{ Signal}\SpecialCharTok{$}\FunctionTok{new}\NormalTok{(}\AttributeTok{name =} \FunctionTok{paste}\NormalTok{(pId,}
    \StringTok{"CP"}\NormalTok{, }\AttributeTok{sep =} \StringTok{"\_"}\NormalTok{), }\AttributeTok{func =}\NormalTok{ dens\_cp, }\AttributeTok{support =} \FunctionTok{c}\NormalTok{(}\DecValTok{0}\NormalTok{, }\DecValTok{1}\NormalTok{), }\AttributeTok{isWp =} \ConstantTok{FALSE}\NormalTok{), }\AttributeTok{FREQ =}\NormalTok{ Signal}\SpecialCharTok{$}\FunctionTok{new}\NormalTok{(}\AttributeTok{name =} \FunctionTok{paste}\NormalTok{(pId,}
    \StringTok{"FREQ"}\NormalTok{, }\AttributeTok{sep =} \StringTok{"\_"}\NormalTok{), }\AttributeTok{func =}\NormalTok{ dens\_freq, }\AttributeTok{support =} \FunctionTok{c}\NormalTok{(}\DecValTok{0}\NormalTok{, }\DecValTok{1}\NormalTok{), }\AttributeTok{isWp =} \ConstantTok{FALSE}\NormalTok{))}
\NormalTok{\}}
\end{Highlighting}
\end{Shaded}

Now, for each project, we estimate the variable for the source code density as follows:

\begin{Shaded}
\begin{Highlighting}[]
\ControlFlowTok{for}\NormalTok{ (pId }\ControlFlowTok{in} \FunctionTok{levels}\NormalTok{(student\_projects}\SpecialCharTok{$}\NormalTok{project)) \{}
\NormalTok{  temp }\OtherTok{\textless{}{-}} \FunctionTok{data.frame}\NormalTok{(}\AttributeTok{x =}\NormalTok{ student\_projects[student\_projects}\SpecialCharTok{$}\NormalTok{project }\SpecialCharTok{==}\NormalTok{ pId, ]}\SpecialCharTok{$}\NormalTok{AuthorTimeNormalized,}
    \AttributeTok{y =}\NormalTok{ student\_projects[student\_projects}\SpecialCharTok{$}\NormalTok{project }\SpecialCharTok{==}\NormalTok{ pId, ]}\SpecialCharTok{$}\NormalTok{Density)}
\NormalTok{  temp }\OtherTok{\textless{}{-}}\NormalTok{ temp[}\FunctionTok{with}\NormalTok{(temp, }\FunctionTok{order}\NormalTok{(x)), ]}

  \CommentTok{\# Using a polynomial with maximum possible degree, we smooth the SCD{-}data, as}
  \CommentTok{\# it can be quite \textquotesingle{}peaky\textquotesingle{}}
\NormalTok{  temp\_poly }\OtherTok{\textless{}{-}} \FunctionTok{poly\_autofit\_max}\NormalTok{(}\AttributeTok{x =}\NormalTok{ temp}\SpecialCharTok{$}\NormalTok{x, }\AttributeTok{y =}\NormalTok{ temp}\SpecialCharTok{$}\NormalTok{y, }\AttributeTok{startDeg =} \DecValTok{13}\NormalTok{)}

\NormalTok{  dens\_scd }\OtherTok{\textless{}{-}} \FunctionTok{Vectorize}\NormalTok{((}\ControlFlowTok{function}\NormalTok{() \{}
\NormalTok{    rx }\OtherTok{\textless{}{-}} \FunctionTok{range}\NormalTok{(temp}\SpecialCharTok{$}\NormalTok{x)}
\NormalTok{    ry }\OtherTok{\textless{}{-}} \FunctionTok{range}\NormalTok{(temp}\SpecialCharTok{$}\NormalTok{y)}
\NormalTok{    poly\_y }\OtherTok{\textless{}{-}}\NormalTok{ stats}\SpecialCharTok{::}\FunctionTok{predict}\NormalTok{(temp\_poly, }\AttributeTok{x =}\NormalTok{ temp}\SpecialCharTok{$}\NormalTok{x)}
\NormalTok{    tempf }\OtherTok{\textless{}{-}}\NormalTok{ stats}\SpecialCharTok{::}\FunctionTok{approxfun}\NormalTok{(}\AttributeTok{x =}\NormalTok{ temp}\SpecialCharTok{$}\NormalTok{x, }\AttributeTok{y =}\NormalTok{ poly\_y, }\AttributeTok{ties =} \StringTok{"ordered"}\NormalTok{)}
    \ControlFlowTok{function}\NormalTok{(x) \{}
      \ControlFlowTok{if}\NormalTok{ (x }\SpecialCharTok{\textless{}}\NormalTok{ rx[}\DecValTok{1}\NormalTok{] }\SpecialCharTok{||}\NormalTok{ x }\SpecialCharTok{\textgreater{}}\NormalTok{ rx[}\DecValTok{2}\NormalTok{]) \{}
        \FunctionTok{return}\NormalTok{(}\ConstantTok{NA\_real\_}\NormalTok{)}
\NormalTok{      \}}
      \FunctionTok{max}\NormalTok{(ry[}\DecValTok{1}\NormalTok{], }\FunctionTok{min}\NormalTok{(ry[}\DecValTok{2}\NormalTok{], }\FunctionTok{tempf}\NormalTok{(x)))}
\NormalTok{    \}}
\NormalTok{  \})())}

\NormalTok{  project\_signals[[pId]][[}\StringTok{"SCD"}\NormalTok{]] }\OtherTok{\textless{}{-}}\NormalTok{ Signal}\SpecialCharTok{$}\FunctionTok{new}\NormalTok{(}\AttributeTok{name =} \FunctionTok{paste}\NormalTok{(pId, }\StringTok{"SCD"}\NormalTok{, }\AttributeTok{sep =} \StringTok{"\_"}\NormalTok{),}
    \AttributeTok{func =}\NormalTok{ dens\_scd, }\AttributeTok{support =} \FunctionTok{c}\NormalTok{(}\DecValTok{0}\NormalTok{, }\DecValTok{1}\NormalTok{), }\AttributeTok{isWp =} \ConstantTok{FALSE}\NormalTok{)}
\NormalTok{\}}

\FunctionTok{invisible}\NormalTok{(}\FunctionTok{loadResultsOrCompute}\NormalTok{(}\AttributeTok{file =} \StringTok{"../results/project\_signals\_sc.rds"}\NormalTok{, }\AttributeTok{computeExpr =}\NormalTok{ \{}
\NormalTok{  project\_signals}
\NormalTok{\}))}
\end{Highlighting}
\end{Shaded}

Let's plot all the projects:

\begin{figure}[ht!]
\includegraphics{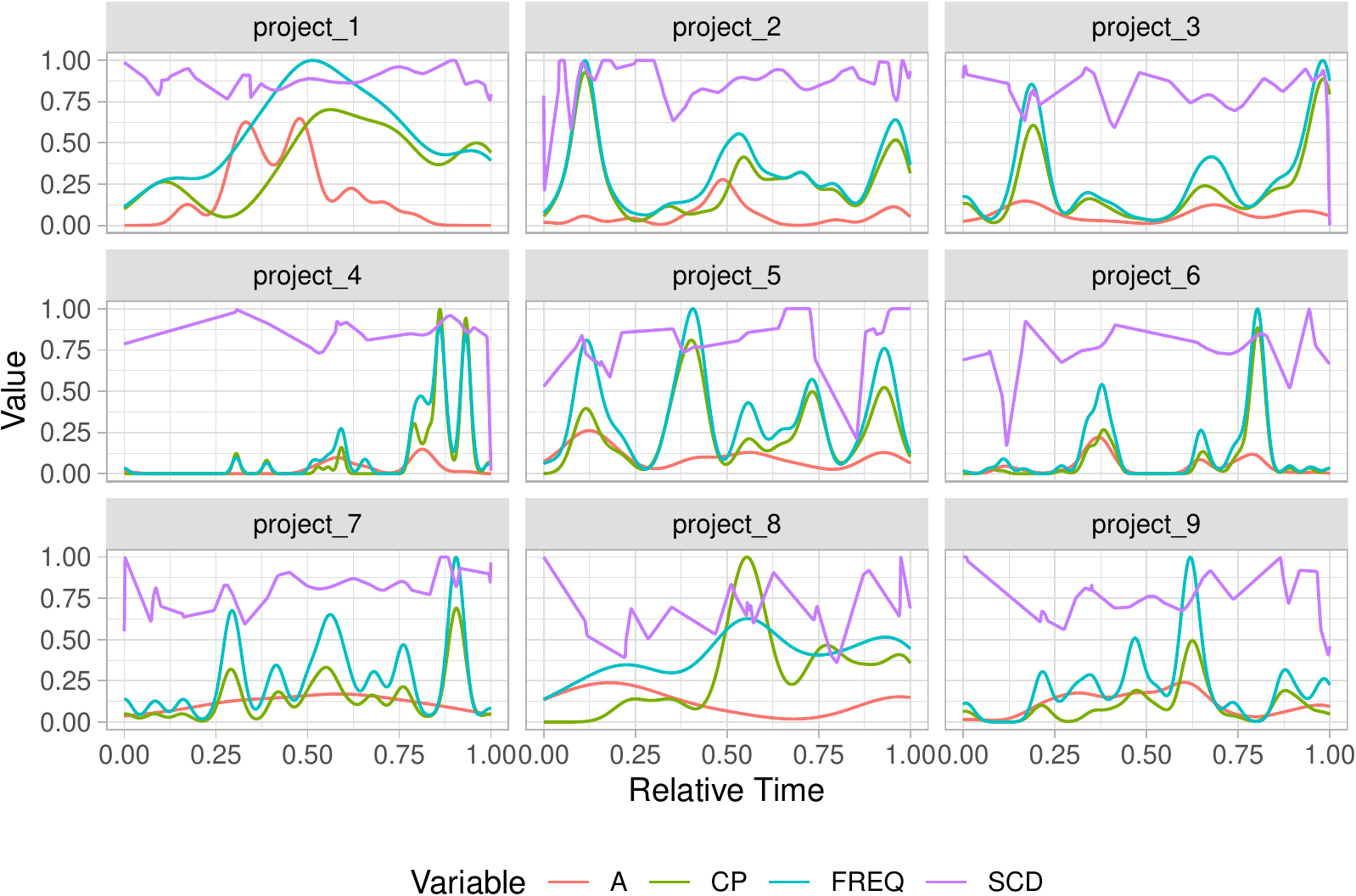} \caption{All variables over each project's time span (first batch of projects).}\label{fig:project-vars}
\end{figure}

\hypertarget{modeling-of-metrics-and-events-using-kde}{%
\subsubsection{\texorpdfstring{Modeling of metrics and events using KDE\label{ssec:model-metrics-kde}}{Modeling of metrics and events using KDE}}\label{modeling-of-metrics-and-events-using-kde}}

We need to make an important distinction between events and metrics. An event does not carry other information, other than that it occurred. One could thus say that such an event is \emph{nulli}-variate. If an event were to carry extra information, such as a measurement that was taken, it would be \emph{uni}-variate. That is the case for many metrics in software: the time of their measurement coincides with an event, such as a commit that was made. On the time-axis we thus know \textbf{when} it occurred and \textbf{what} was its value. Such a metric could be easily understood as a \emph{bivariate x/y} variable and be plotted in a two-dimensional Cartesian coordinate system.

An event however does not have any associated y-value we could plot. Given a time-axis, we could make a mark whenever it occurred. Some of the markers would probably be closer to each other or be more or less accumulated. The y-value could express these accumulations relative to each other. These are called \emph{densities}. This is exactly what KDE does: it expresses the relative accumulations of data on the x-axis as density on the y-axis. For KDE, the actual values on the x-axis have another meaning, and that is to compare the relative likelihoods of the values on it, since the axis is ordered. For our case however, the axis is linear time and carries no such meaning. The project data we analyze is a kind of sampling over the project's events. We subdivide the gathered project data hence into these two types of data series:

\begin{itemize}
\tightlist
\item
  \textbf{Events}: They do not carry any extra information or measurements. As for the projects we analyze, events usually are occurrences of specific types of commits, for example. The time of occurrence is the x-value on the time-axis, and the y-value is obtained through KDE. We model all maintenance activities as such variables.
\item
  \textbf{Metrics}: Extracted from the project at specific times, for example at every commit. We can extract any number or type of metric, but each becomes its own variable, where the x-value is on the time-axis, and the y-value is the metric's value. We model the source code density as such a variable.
\end{itemize}

\hypertarget{patterns-for-scoring-the-projects}{%
\subsection{Patterns for scoring the projects}\label{patterns-for-scoring-the-projects}}

Our overall goal is to propose a single model that is able to detect the presence of the Fire Drill AP, and how strong its manifestation is. In order to do that, we require a pattern that defines how a Fire Drill looks in practice. Any real-world project can never follow such a pattern perfectly, because of, e.g., time dilation and compression. Even after correcting these, some distance between the project and the pattern will remain. The projects from figure \ref{fig:project-vars} indicate that certain phases occur, but that their occurrence happens at different points in time, and lasts for various durations.

Given some pattern, we first attempt to remove any distortions in the data, by using our new model \emph{self-regularizing Boundary Time Warping} (sr-BTW). This model takes a pattern that is subdivided into one or more intervals, and aligns the project data such that the loss in each interval is minimized. After alignment, we calculate a score that quantifies the remaining differences. Ideally, we hope to find a (strong) positive correlation of these scores with the ground truth.

\hypertarget{pattern-i-initial-best-guess}{%
\subsubsection{Pattern I: Initial best guess}\label{pattern-i-initial-best-guess}}

\begin{Shaded}
\begin{Highlighting}[]
\NormalTok{fd\_data\_concat }\OtherTok{\textless{}{-}} \FunctionTok{readRDS}\NormalTok{(}\StringTok{"../data/fd\_data\_concat.rds"}\NormalTok{)}
\end{Highlighting}
\end{Shaded}

This pattern was created based on all available literature, \textbf{without} inspecting any of the projects. It is subdivided into four intervals:

\begin{enumerate}
\def\labelenumi{\arabic{enumi}.}
\tightlist
\item
  Begin -- Short project warm-up phase
\item
  Long Stretch -- The longest phase in the project, about which we do not know much about, except for that there should be a rather constant amount of activities over time.
\item
  Fire Drill -- Characteristic is a sudden and steep increase of adaptive activities. This phase is over once these activities reached their apex.
\item
  Aftermath -- Everything after the apex. We should see even steeper declines.
\end{enumerate}

Brown et al. (1998) describe a typical scenario where about six months are spent on non-developmental activities, and the actual software is then developed in less than four weeks. If we were to include some of the aftermath, the above first guess would describe a project of about eight weeks.

We define the boundaries as follows (there are three boundaries to split the pattern into four intervals):

\begin{Shaded}
\begin{Highlighting}[]
\NormalTok{fd\_data\_boundaries }\OtherTok{\textless{}{-}} \FunctionTok{c}\NormalTok{(}\AttributeTok{b1 =} \FloatTok{0.085}\NormalTok{, }\AttributeTok{b2 =} \FloatTok{0.625}\NormalTok{, }\AttributeTok{b3 =} \FloatTok{0.875}\NormalTok{)}
\end{Highlighting}
\end{Shaded}

The pattern and its boundaries look like this:

\begin{Shaded}
\begin{Highlighting}[]
\FunctionTok{plot\_project\_data}\NormalTok{(}\AttributeTok{data =}\NormalTok{ fd\_data\_concat, }\AttributeTok{boundaries =}\NormalTok{ fd\_data\_boundaries)}
\end{Highlighting}
\end{Shaded}

\begin{figure}[ht!]
\includegraphics{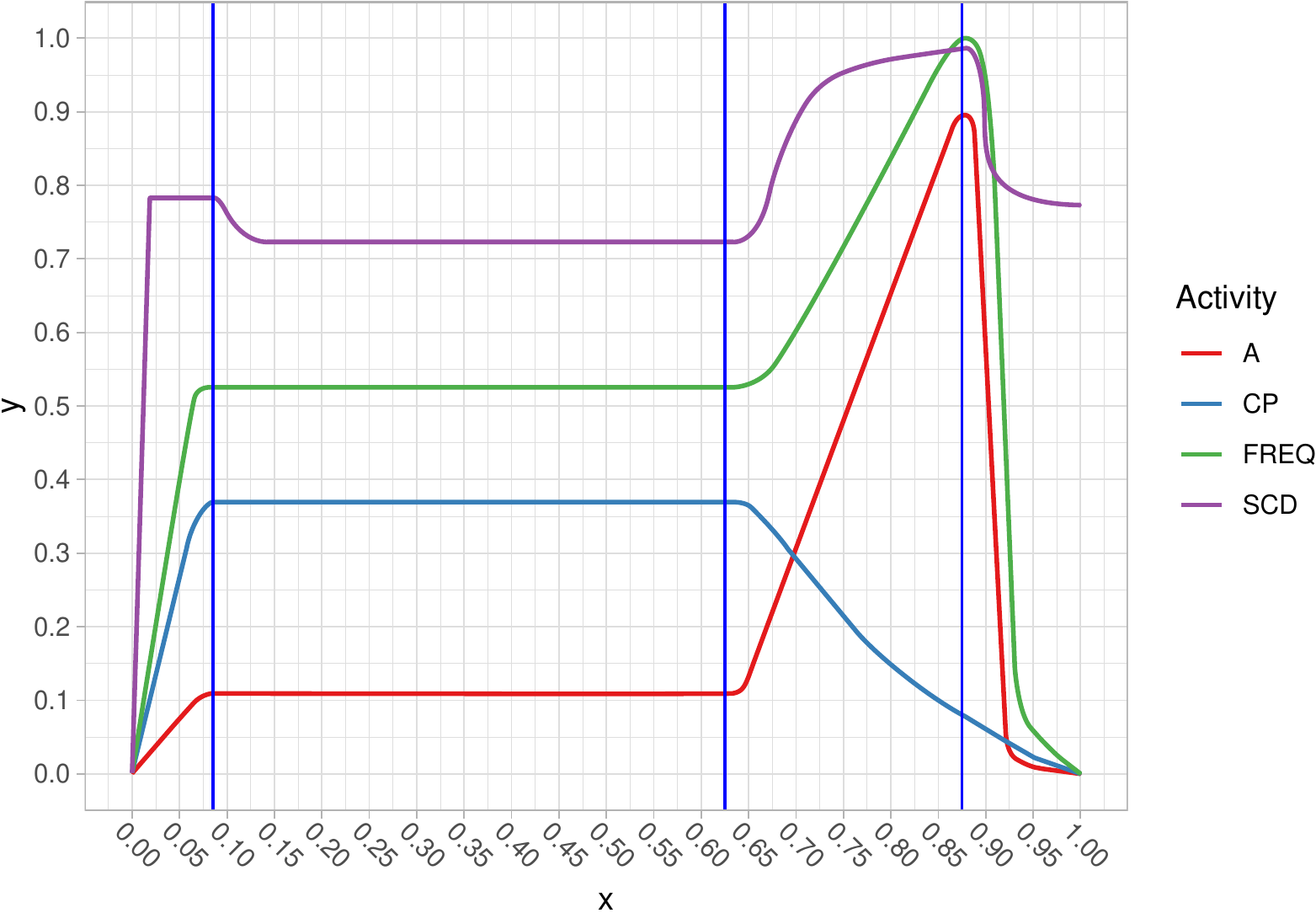} \caption{The pattern that was our initial best guess}\label{fig:pattern-1}
\end{figure}

\hypertarget{initialize-the-pattern}{%
\paragraph{Initialize the pattern}\label{initialize-the-pattern}}

The pattern as shown in \ref{fig:pattern-1} is just a collection of x/y coordinate-data, and for us being able to use it, we need to instantiate it. We do this by storing each variable in an instance of \texttt{Signal}.

\begin{Shaded}
\begin{Highlighting}[]
\NormalTok{p1\_signals }\OtherTok{\textless{}{-}} \FunctionTok{list}\NormalTok{(}\AttributeTok{A =}\NormalTok{ Signal}\SpecialCharTok{$}\FunctionTok{new}\NormalTok{(}\AttributeTok{name =} \StringTok{"p1\_A"}\NormalTok{, }\AttributeTok{support =} \FunctionTok{c}\NormalTok{(}\DecValTok{0}\NormalTok{, }\DecValTok{1}\NormalTok{), }\AttributeTok{isWp =} \ConstantTok{TRUE}\NormalTok{,}
  \AttributeTok{func =}\NormalTok{ stats}\SpecialCharTok{::}\FunctionTok{approxfun}\NormalTok{(}\AttributeTok{x =}\NormalTok{ fd\_data\_concat[fd\_data\_concat}\SpecialCharTok{$}\NormalTok{t }\SpecialCharTok{==} \StringTok{"A"}\NormalTok{, ]}\SpecialCharTok{$}\NormalTok{x, }\AttributeTok{y =}\NormalTok{ fd\_data\_concat[fd\_data\_concat}\SpecialCharTok{$}\NormalTok{t }\SpecialCharTok{==}
    \StringTok{"A"}\NormalTok{, ]}\SpecialCharTok{$}\NormalTok{y)), }\AttributeTok{CP =}\NormalTok{ Signal}\SpecialCharTok{$}\FunctionTok{new}\NormalTok{(}\AttributeTok{name =} \StringTok{"p1\_CP"}\NormalTok{, }\AttributeTok{support =} \FunctionTok{c}\NormalTok{(}\DecValTok{0}\NormalTok{, }\DecValTok{1}\NormalTok{), }\AttributeTok{isWp =} \ConstantTok{TRUE}\NormalTok{,}
  \AttributeTok{func =}\NormalTok{ stats}\SpecialCharTok{::}\FunctionTok{approxfun}\NormalTok{(}\AttributeTok{x =}\NormalTok{ fd\_data\_concat[fd\_data\_concat}\SpecialCharTok{$}\NormalTok{t }\SpecialCharTok{==} \StringTok{"CP"}\NormalTok{, ]}\SpecialCharTok{$}\NormalTok{x, }\AttributeTok{y =}\NormalTok{ fd\_data\_concat[fd\_data\_concat}\SpecialCharTok{$}\NormalTok{t }\SpecialCharTok{==}
    \StringTok{"CP"}\NormalTok{, ]}\SpecialCharTok{$}\NormalTok{y)), }\AttributeTok{FREQ =}\NormalTok{ Signal}\SpecialCharTok{$}\FunctionTok{new}\NormalTok{(}\AttributeTok{name =} \StringTok{"p1\_FREQ"}\NormalTok{, }\AttributeTok{support =} \FunctionTok{c}\NormalTok{(}\DecValTok{0}\NormalTok{, }\DecValTok{1}\NormalTok{), }\AttributeTok{isWp =} \ConstantTok{TRUE}\NormalTok{,}
  \AttributeTok{func =}\NormalTok{ stats}\SpecialCharTok{::}\FunctionTok{approxfun}\NormalTok{(}\AttributeTok{x =}\NormalTok{ fd\_data\_concat[fd\_data\_concat}\SpecialCharTok{$}\NormalTok{t }\SpecialCharTok{==} \StringTok{"FREQ"}\NormalTok{, ]}\SpecialCharTok{$}\NormalTok{x, }\AttributeTok{y =}\NormalTok{ fd\_data\_concat[fd\_data\_concat}\SpecialCharTok{$}\NormalTok{t }\SpecialCharTok{==}
    \StringTok{"FREQ"}\NormalTok{, ]}\SpecialCharTok{$}\NormalTok{y)), }\AttributeTok{SCD =}\NormalTok{ Signal}\SpecialCharTok{$}\FunctionTok{new}\NormalTok{(}\AttributeTok{name =} \StringTok{"p1\_SCD"}\NormalTok{, }\AttributeTok{support =} \FunctionTok{c}\NormalTok{(}\DecValTok{0}\NormalTok{, }\DecValTok{1}\NormalTok{), }\AttributeTok{isWp =} \ConstantTok{TRUE}\NormalTok{,}
  \AttributeTok{func =}\NormalTok{ stats}\SpecialCharTok{::}\FunctionTok{approxfun}\NormalTok{(}\AttributeTok{x =}\NormalTok{ fd\_data\_concat[fd\_data\_concat}\SpecialCharTok{$}\NormalTok{t }\SpecialCharTok{==} \StringTok{"SCD"}\NormalTok{, ]}\SpecialCharTok{$}\NormalTok{x, }\AttributeTok{y =}\NormalTok{ fd\_data\_concat[fd\_data\_concat}\SpecialCharTok{$}\NormalTok{t }\SpecialCharTok{==}
    \StringTok{"SCD"}\NormalTok{, ]}\SpecialCharTok{$}\NormalTok{y)))}
\end{Highlighting}
\end{Shaded}

\begin{figure}[ht!]
\includegraphics{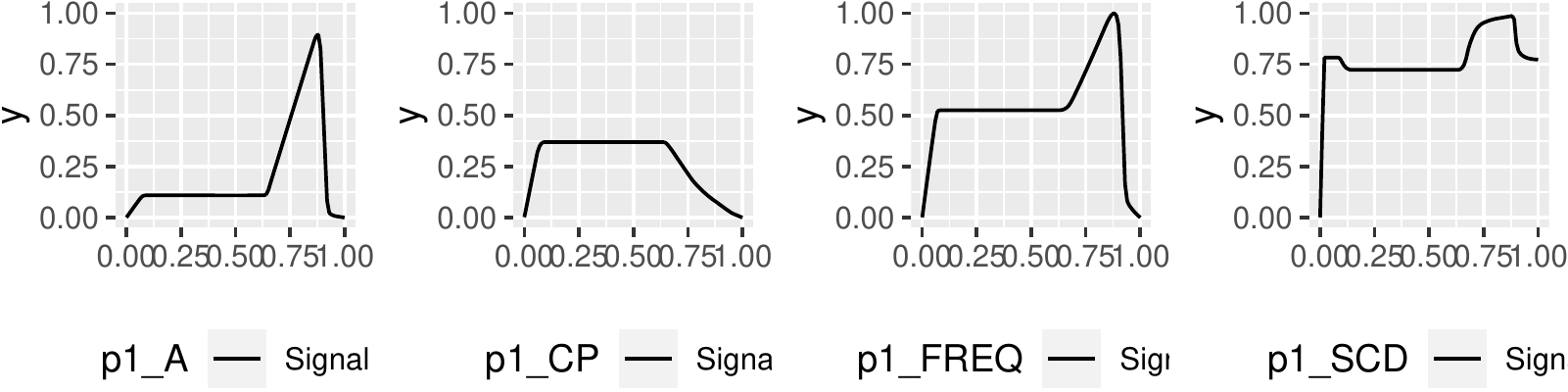} \caption{The separate signals of pattern I.}\label{fig:unnamed-chunk-18}
\end{figure}

\hypertarget{pattern-ii-adaptation-of-best-guess}{%
\subsubsection{Pattern II: Adaptation of best guess}\label{pattern-ii-adaptation-of-best-guess}}

The second pattern is a compromise between the first and the third: While we want to keep as much of the initial best guess, we also want to adjust the pattern based on the projects and the ground truth. Adjusting means, that we will keep what is in each interval, but we allow each interval to stretch and compress, and we allow each interval to impose a vertical translation both at then begin and end (a somewhat trapezoidal translation). In any case, each such alteration is a linear affine transformation. Additionally to sr-BTW, we will also apply \textbf{sr-BAW} (self-regularizing Boundary Amplitude Warping) to accomplish this. This model is called \textbf{\texttt{srBTAW}} and the process is the following:

\begin{itemize}
\tightlist
\item
  The pattern is decomposed into its four variables first, as we can adapt these (almost) independently from each other.
\item
  Then, for each type of variable, an instance of \texttt{srBTAW} is created. As \textbf{Warping Candidates} (WC) we add all of the projects' corresponding variables. The \textbf{Warping Pattern} (WP) is the single variable from the pattern in this case -- again, we warp the project data, however, eventually the learned warping gets inversed and applied to the WC.
\item
  All four \texttt{srBTAW} instances are then fitted simultaneously: While we allow the y-translations to adapt independently for each type of variable, all instances share the same intervals, as eventually we have to assemble the variables back into a common pattern.
\end{itemize}

\hypertarget{preparation}{%
\paragraph{Preparation}\label{preparation}}

We already have the \texttt{srBTAW} \textbf{Multilevel model}, which can keep track of arbitrary many variables and losses. The intention behind this however was, to track variables of the \textbf{same type}, i.e., signals that are logically of the same type. In our case this means that any single instance should only track variables that are either \texttt{A}, \texttt{CP}, \texttt{FREQ} or \texttt{SCD}. For this pattern, the WP is a single signal per variable, and the WC is the corresponding signal from each of the nine projects. This is furthermore important to give different weights to different variables. In our case, we want to give a lower weight to the \texttt{SCD}-variable.

As for the loss, we will first test a combined loss that measures \textbf{\texttt{3}} properties: The area between curves (or alternatively the residual sum of squares), the correlation between the curves, and the arc-length ratio between the curves. We will consider any of these to be equally important, i.e., no additional weights. Each loss shall cover all intervals with weight \(=1\), except for the Long Stretch interval, where we will use a reduced weight.

There are \(4\) types of variables, \(7\) projects (two projects have consensus \(=0\), i.e., no weight) and \(2\times 3\) single losses, resulting in \(168\) losses to compute. The final weight for each loss is computed as: \(\omega_i=\omega^{(\text{project})}\times\omega^{(\text{vartype})}\times\omega^{(\text{interval})}\). For the phase Long Stretch, the weight for any loss will \(\frac{1}{2}\), and for the source code density we will chose \(\frac{1}{2}\), too. The weight of each project is based on the consensus of the ground truth. The ordinal scale for that is \([0,10]\), so that we will divide the score by \(10\) and use that as weight. Examples:

\begin{itemize}
\tightlist
\item
  \textbf{A} in Fire Drill in project \(p3\): \(\omega=0.6\times 1\times 1=0.6\) (consensus is \(6\) in project \(p3\))
\item
  \textbf{FREQ} in Long Stretch in project \(p7\): \(\omega=0.3\times 0.5\times 1=0.15\) and
\item
  \textbf{SCD} in Long Stretch in project \(p4\): \(\omega=0.8\times 0.5\times 0.5=0.2\).
\end{itemize}

In table \ref{tab:groundtruth-score} we show all projects with a consensus-score \(>0\), projects \(2\) and \(8\) are not included any longer.

\begin{Shaded}
\begin{Highlighting}[]
\NormalTok{ground\_truth}\SpecialCharTok{$}\NormalTok{consensus\_score }\OtherTok{\textless{}{-}}\NormalTok{ ground\_truth}\SpecialCharTok{$}\NormalTok{consensus}\SpecialCharTok{/}\DecValTok{10}
\NormalTok{weight\_vartype }\OtherTok{\textless{}{-}} \FunctionTok{c}\NormalTok{(}\AttributeTok{A =} \DecValTok{1}\NormalTok{, }\AttributeTok{CP =} \DecValTok{1}\NormalTok{, }\AttributeTok{FREQ =} \DecValTok{1}\NormalTok{, }\AttributeTok{SCD =} \FloatTok{0.5}\NormalTok{)}
\NormalTok{weight\_interval }\OtherTok{\textless{}{-}} \FunctionTok{c}\NormalTok{(}\AttributeTok{Begin =} \DecValTok{1}\NormalTok{, }\StringTok{\textasciigrave{}}\AttributeTok{Long Stretch}\StringTok{\textasciigrave{}} \OtherTok{=} \FloatTok{0.5}\NormalTok{, }\StringTok{\textasciigrave{}}\AttributeTok{Fire Drill}\StringTok{\textasciigrave{}} \OtherTok{=} \DecValTok{1}\NormalTok{, }\AttributeTok{Aftermath =} \DecValTok{1}\NormalTok{)}
\end{Highlighting}
\end{Shaded}

\begin{Shaded}
\begin{Highlighting}[]
\NormalTok{temp }\OtherTok{\textless{}{-}} \FunctionTok{expand.grid}\NormalTok{(weight\_interval, weight\_vartype, ground\_truth}\SpecialCharTok{$}\NormalTok{consensus\_score)}
\NormalTok{temp}\SpecialCharTok{$}\NormalTok{p }\OtherTok{\textless{}{-}}\NormalTok{ temp}\SpecialCharTok{$}\NormalTok{Var1 }\SpecialCharTok{*}\NormalTok{ temp}\SpecialCharTok{$}\NormalTok{Var2 }\SpecialCharTok{*}\NormalTok{ temp}\SpecialCharTok{$}\NormalTok{Var3}
\NormalTok{weight\_total }\OtherTok{\textless{}{-}} \FunctionTok{sum}\NormalTok{(temp}\SpecialCharTok{$}\NormalTok{p)}
\end{Highlighting}
\end{Shaded}

The sum of all weights combined is 31.85.

\begin{table}

\caption{\label{tab:groundtruth-score}Entire ground truth as of both raters}
\centering
\begin{tabular}[t]{llrr}
\toprule
  & project & consensus & consensus\_score\\
\midrule
1 & project\_1 & 1 & 0.1\\
3 & project\_3 & 6 & 0.6\\
4 & project\_4 & 8 & 0.8\\
5 & project\_5 & 1 & 0.1\\
6 & project\_6 & 2 & 0.2\\
\addlinespace
7 & project\_7 & 3 & 0.3\\
9 & project\_9 & 5 & 0.5\\
\bottomrule
\end{tabular}
\end{table}

\hypertarget{defining-the-losses}{%
\paragraph{Defining the losses}\label{defining-the-losses}}

For the optimization we will use mainly \textbf{\texttt{5}} classes:

\begin{itemize}
\tightlist
\item
  \texttt{srBTAW\_MultiVartype}: One instance globally, that manages all parameters across all instances of \texttt{srBTAW}.
\item
  \texttt{srBTAW}: One instance per variable-type, so here we'll end up with four instances.
\item
  \texttt{srBTAW\_LossLinearScalarizer}: A linear scalarizer that will take on all of the defined singular losses and compute and add them together according to their weight.
\item
  \texttt{srBTAW\_Loss2Curves}: Used for each of the \(168\) singular losses, and configured using a specific loss function, weight, and set of intervals where it ought to be used.
\item
  \texttt{TimeWarpRegularization}: One global instance for all \texttt{srBTAW} instances, to regularize extreme intervals. We chose a mild weight for this of just \(1\), which is small compared to the sum of all other weights (31.85).
\end{itemize}

\begin{Shaded}
\begin{Highlighting}[]
\NormalTok{p2\_smv }\OtherTok{\textless{}{-}}\NormalTok{ srBTAW\_MultiVartype}\SpecialCharTok{$}\FunctionTok{new}\NormalTok{()}

\NormalTok{p2\_vars }\OtherTok{\textless{}{-}} \FunctionTok{c}\NormalTok{(}\StringTok{"A"}\NormalTok{, }\StringTok{"CP"}\NormalTok{, }\StringTok{"FREQ"}\NormalTok{, }\StringTok{"SCD"}\NormalTok{)}
\NormalTok{p2\_inst }\OtherTok{\textless{}{-}} \FunctionTok{list}\NormalTok{()}
\ControlFlowTok{for}\NormalTok{ (name }\ControlFlowTok{in}\NormalTok{ p2\_vars) \{}
\NormalTok{  p2\_inst[[name]] }\OtherTok{\textless{}{-}}\NormalTok{ srBTAW}\SpecialCharTok{$}\FunctionTok{new}\NormalTok{(}
    \AttributeTok{theta\_b =} \FunctionTok{c}\NormalTok{(}\DecValTok{0}\NormalTok{, fd\_data\_boundaries, }\DecValTok{1}\NormalTok{),}
    \AttributeTok{gamma\_bed =} \FunctionTok{c}\NormalTok{(}\DecValTok{0}\NormalTok{, }\DecValTok{1}\NormalTok{, }\FunctionTok{sqrt}\NormalTok{(.Machine}\SpecialCharTok{$}\NormalTok{double.eps)),}
    \AttributeTok{lambda =} \FunctionTok{rep}\NormalTok{(}\FunctionTok{sqrt}\NormalTok{(.Machine}\SpecialCharTok{$}\NormalTok{double.eps), }\FunctionTok{length}\NormalTok{(p2\_vars)),}
    \AttributeTok{begin =} \DecValTok{0}\NormalTok{, }\AttributeTok{end =} \DecValTok{1}\NormalTok{, }\AttributeTok{openBegin =} \ConstantTok{FALSE}\NormalTok{, }\AttributeTok{openEnd =} \ConstantTok{FALSE}\NormalTok{,}
    \AttributeTok{useAmplitudeWarping =} \ConstantTok{TRUE}\NormalTok{,}
    \CommentTok{\# We allow these to be larger; however, the final result should be within [0,1]}
    \AttributeTok{lambda\_ymin =} \FunctionTok{rep}\NormalTok{(}\SpecialCharTok{{-}}\DecValTok{10}\NormalTok{, }\FunctionTok{length}\NormalTok{(p2\_vars)),}
    \AttributeTok{lambda\_ymax =} \FunctionTok{rep}\NormalTok{( }\DecValTok{10}\NormalTok{, }\FunctionTok{length}\NormalTok{(p2\_vars)),}
    \AttributeTok{isObjectiveLogarithmic =} \ConstantTok{TRUE}\NormalTok{,}
    \AttributeTok{paramNames =} \FunctionTok{c}\NormalTok{(}\StringTok{"v"}\NormalTok{,}
                   \FunctionTok{paste0}\NormalTok{(}\StringTok{"vtl\_"}\NormalTok{, }\FunctionTok{seq\_len}\NormalTok{(}\AttributeTok{length.out =} \FunctionTok{length}\NormalTok{(p2\_vars))),}
                   \FunctionTok{paste0}\NormalTok{(}\StringTok{"vty\_"}\NormalTok{, }\FunctionTok{seq\_len}\NormalTok{(}\AttributeTok{length.out =} \FunctionTok{length}\NormalTok{(p2\_vars)))))}
  
  \CommentTok{\# We can already add the WP:}
\NormalTok{  p2\_inst[[name]]}\SpecialCharTok{$}\FunctionTok{setSignal}\NormalTok{(}\AttributeTok{signal =}\NormalTok{ p1\_signals[[name]])}
\NormalTok{  p2\_smv}\SpecialCharTok{$}\FunctionTok{setSrbtaw}\NormalTok{(}\AttributeTok{varName =}\NormalTok{ name, }\AttributeTok{srbtaw =}\NormalTok{ p2\_inst[[name]])}
  
  \CommentTok{\# .. and also all the projects\textquotesingle{} signals:}
  \ControlFlowTok{for}\NormalTok{ (project }\ControlFlowTok{in}\NormalTok{ ground\_truth[ground\_truth}\SpecialCharTok{$}\NormalTok{consensus }\SpecialCharTok{\textgreater{}} \DecValTok{0}\NormalTok{, ]}\SpecialCharTok{$}\NormalTok{project) \{}
\NormalTok{    p2\_inst[[name]]}\SpecialCharTok{$}\FunctionTok{setSignal}\NormalTok{(}\AttributeTok{signal =}\NormalTok{ project\_signals[[project]][[name]])}
\NormalTok{  \}}
\NormalTok{\}}

\CommentTok{\# We call this there so there are parameters present.}
\FunctionTok{set.seed}\NormalTok{(}\DecValTok{1337}\NormalTok{)}
\NormalTok{p2\_smv}\SpecialCharTok{$}\FunctionTok{setParams}\NormalTok{(}\AttributeTok{params =}
  \StringTok{\textasciigrave{}}\AttributeTok{names\textless{}{-}}\StringTok{\textasciigrave{}}\NormalTok{(}\AttributeTok{x =} \FunctionTok{runif}\NormalTok{(}\AttributeTok{n =}\NormalTok{ p2\_smv}\SpecialCharTok{$}\FunctionTok{getNumParams}\NormalTok{()), }\AttributeTok{value =}\NormalTok{ p2\_smv}\SpecialCharTok{$}\FunctionTok{getParamNames}\NormalTok{()))}
\end{Highlighting}
\end{Shaded}

We can already initialize the linear scalarizer. This includes also to set up some progress-callback. Even with massive parallelization, this process will take its time so it will be good to know where we are approximately.

\begin{Shaded}
\begin{Highlighting}[]
\NormalTok{p2\_lls }\OtherTok{\textless{}{-}}\NormalTok{ srBTAW\_LossLinearScalarizer}\SpecialCharTok{$}\FunctionTok{new}\NormalTok{(}\AttributeTok{returnRaw =} \ConstantTok{FALSE}\NormalTok{, }\AttributeTok{computeParallel =} \ConstantTok{TRUE}\NormalTok{,}
  \AttributeTok{progressCallback =} \ControlFlowTok{function}\NormalTok{(what, step, total) \{}
    \CommentTok{\# if (step == total) \{ print(paste(what, step, total)) \}}
\NormalTok{  \})}

\ControlFlowTok{for}\NormalTok{ (name }\ControlFlowTok{in} \FunctionTok{names}\NormalTok{(p2\_inst)) \{}
\NormalTok{  p2\_inst[[name]]}\SpecialCharTok{$}\FunctionTok{setObjective}\NormalTok{(}\AttributeTok{obj =}\NormalTok{ p2\_lls)}
\NormalTok{\}}
\end{Highlighting}
\end{Shaded}

The basic infrastructure stands, so now it's time to instantiate all of the singular losses. First we define a helper-function to do the bulk-work, then we iterate all projects, variables and intervals.

\begin{Shaded}
\begin{Highlighting}[]
\CommentTok{\#\textquotesingle{} This function creates a singular loss that is a linear combination}
\CommentTok{\#\textquotesingle{} of an area{-}, correlation{-} and arclength{-}loss (all with same weight).}
\NormalTok{p2\_attach\_combined\_loss }\OtherTok{\textless{}{-}} \ControlFlowTok{function}\NormalTok{(project, vartype, intervals) \{}
\NormalTok{  weight\_p }\OtherTok{\textless{}{-}}\NormalTok{ ground\_truth[ground\_truth}\SpecialCharTok{$}\NormalTok{project }\SpecialCharTok{==}\NormalTok{ project, ]}\SpecialCharTok{$}\NormalTok{consensus\_score}
\NormalTok{  weight\_v }\OtherTok{\textless{}{-}}\NormalTok{ weight\_vartype[[vartype]]}
\NormalTok{  temp }\OtherTok{\textless{}{-}}\NormalTok{ weight\_interval[intervals]}
  \FunctionTok{stopifnot}\NormalTok{(}\FunctionTok{length}\NormalTok{(}\FunctionTok{unique}\NormalTok{(temp)) }\SpecialCharTok{==} \DecValTok{1}\NormalTok{)}
\NormalTok{  weight\_i }\OtherTok{\textless{}{-}} \FunctionTok{unique}\NormalTok{(temp)}
\NormalTok{  weight }\OtherTok{\textless{}{-}}\NormalTok{ weight\_p }\SpecialCharTok{*}\NormalTok{ weight\_v }\SpecialCharTok{*}\NormalTok{ weight\_i}

\NormalTok{  lossRss }\OtherTok{\textless{}{-}}\NormalTok{ srBTAW\_Loss\_Rss}\SpecialCharTok{$}\FunctionTok{new}\NormalTok{(}\AttributeTok{wpName =} \FunctionTok{paste0}\NormalTok{(}\StringTok{"p1\_"}\NormalTok{, vartype), }\AttributeTok{wcName =} \FunctionTok{paste}\NormalTok{(project,}
\NormalTok{    vartype, }\AttributeTok{sep =} \StringTok{"\_"}\NormalTok{), }\AttributeTok{weight =}\NormalTok{ weight, }\AttributeTok{intervals =}\NormalTok{ intervals, }\AttributeTok{continuous =} \ConstantTok{FALSE}\NormalTok{,}
    \AttributeTok{numSamples =} \FunctionTok{rep}\NormalTok{(}\DecValTok{500}\NormalTok{, }\FunctionTok{length}\NormalTok{(intervals)), }\AttributeTok{returnRaw =} \ConstantTok{TRUE}\NormalTok{)}

\NormalTok{  p2\_inst[[vartype]]}\SpecialCharTok{$}\FunctionTok{addLoss}\NormalTok{(}\AttributeTok{loss =}\NormalTok{ lossRss)}
\NormalTok{  p2\_lls}\SpecialCharTok{$}\FunctionTok{setObjective}\NormalTok{(}\AttributeTok{name =} \FunctionTok{paste}\NormalTok{(project, vartype, }\FunctionTok{paste}\NormalTok{(intervals, }\AttributeTok{collapse =} \StringTok{"\_"}\NormalTok{),}
    \StringTok{"rss"}\NormalTok{, }\AttributeTok{sep =} \StringTok{"\_"}\NormalTok{), }\AttributeTok{obj =}\NormalTok{ lossRss)}
\NormalTok{\}}
\end{Highlighting}
\end{Shaded}

Let's call our helper iteratively:

\begin{Shaded}
\begin{Highlighting}[]
\NormalTok{interval\_types }\OtherTok{\textless{}{-}} \FunctionTok{list}\NormalTok{(}\AttributeTok{A =} \FunctionTok{c}\NormalTok{(}\DecValTok{1}\NormalTok{, }\DecValTok{3}\NormalTok{, }\DecValTok{4}\NormalTok{), }\AttributeTok{B =} \DecValTok{2}\NormalTok{)}

\ControlFlowTok{for}\NormalTok{ (vartype }\ControlFlowTok{in}\NormalTok{ p2\_vars) \{}
  \ControlFlowTok{for}\NormalTok{ (project }\ControlFlowTok{in}\NormalTok{ ground\_truth[ground\_truth}\SpecialCharTok{$}\NormalTok{consensus }\SpecialCharTok{\textgreater{}} \DecValTok{0}\NormalTok{, ]}\SpecialCharTok{$}\NormalTok{project) \{}
    \ControlFlowTok{for}\NormalTok{ (intervals }\ControlFlowTok{in}\NormalTok{ interval\_types) \{}
      \FunctionTok{p2\_attach\_combined\_loss}\NormalTok{(}\AttributeTok{project =}\NormalTok{ project, }\AttributeTok{vartype =}\NormalTok{ vartype, }\AttributeTok{intervals =}\NormalTok{ intervals)}
\NormalTok{    \}}
\NormalTok{  \}}

  \CommentTok{\# Add one per variable{-}type:}
\NormalTok{  lossYtrans }\OtherTok{\textless{}{-}}\NormalTok{ YTransRegularization}\SpecialCharTok{$}\FunctionTok{new}\NormalTok{(}\AttributeTok{wpName =} \FunctionTok{paste0}\NormalTok{(}\StringTok{"p1\_"}\NormalTok{, vartype), }\AttributeTok{wcName =} \FunctionTok{paste}\NormalTok{(project,}
\NormalTok{    vartype, }\AttributeTok{sep =} \StringTok{"\_"}\NormalTok{), }\AttributeTok{intervals =} \FunctionTok{seq\_len}\NormalTok{(}\AttributeTok{length.out =} \DecValTok{4}\NormalTok{), }\AttributeTok{returnRaw =} \ConstantTok{TRUE}\NormalTok{,}
    \AttributeTok{weight =} \DecValTok{1}\NormalTok{, }\AttributeTok{use =} \StringTok{"tikhonov"}\NormalTok{)}

\NormalTok{  p2\_inst[[vartype]]}\SpecialCharTok{$}\FunctionTok{addLoss}\NormalTok{(}\AttributeTok{loss =}\NormalTok{ lossYtrans)}
\NormalTok{  p2\_lls}\SpecialCharTok{$}\FunctionTok{setObjective}\NormalTok{(}\AttributeTok{name =} \FunctionTok{paste}\NormalTok{(vartype, }\StringTok{"p2\_reg\_output"}\NormalTok{, }\AttributeTok{sep =} \StringTok{"\_"}\NormalTok{), }\AttributeTok{obj =}\NormalTok{ lossYtrans)}
\NormalTok{\}}
\end{Highlighting}
\end{Shaded}

Finally, we add the regularizer for extreme intervals:

\begin{Shaded}
\begin{Highlighting}[]
\NormalTok{p2\_lls}\SpecialCharTok{$}\FunctionTok{setObjective}\NormalTok{(}\AttributeTok{name =} \StringTok{"p2\_reg\_exint2"}\NormalTok{, }\AttributeTok{obj =}\NormalTok{ TimeWarpRegularization}\SpecialCharTok{$}\FunctionTok{new}\NormalTok{(}\AttributeTok{weight =} \FloatTok{0.25} \SpecialCharTok{*}
\NormalTok{  p2\_lls}\SpecialCharTok{$}\FunctionTok{getNumObjectives}\NormalTok{(), }\AttributeTok{use =} \StringTok{"exint2"}\NormalTok{, }\AttributeTok{returnRaw =} \ConstantTok{TRUE}\NormalTok{, }\AttributeTok{wpName =}\NormalTok{ p1\_signals}\SpecialCharTok{$}\NormalTok{A}\SpecialCharTok{$}\FunctionTok{getName}\NormalTok{(),}
  \AttributeTok{wcName =}\NormalTok{ project\_signals}\SpecialCharTok{$}\NormalTok{project\_1}\SpecialCharTok{$}\NormalTok{A}\SpecialCharTok{$}\FunctionTok{getName}\NormalTok{(), }\AttributeTok{intervals =} \FunctionTok{seq\_len}\NormalTok{(}\AttributeTok{length.out =} \FunctionTok{length}\NormalTok{(p2\_vars)))}\SpecialCharTok{$}\FunctionTok{setSrBtaw}\NormalTok{(}\AttributeTok{srbtaw =}\NormalTok{ p2\_inst}\SpecialCharTok{$}\NormalTok{A))}
\end{Highlighting}
\end{Shaded}

\hypertarget{fitting-the-pattern}{%
\paragraph{Fitting the pattern}\label{fitting-the-pattern}}

\begin{Shaded}
\begin{Highlighting}[]
\NormalTok{p2\_params }\OtherTok{\textless{}{-}} \FunctionTok{loadResultsOrCompute}\NormalTok{(}\AttributeTok{file =} \StringTok{"../results/p2\_params.rds"}\NormalTok{, }\AttributeTok{computeExpr =}\NormalTok{ \{}
\NormalTok{  cl }\OtherTok{\textless{}{-}}\NormalTok{ parallel}\SpecialCharTok{::}\FunctionTok{makePSOCKcluster}\NormalTok{(}\FunctionTok{min}\NormalTok{(}\DecValTok{64}\NormalTok{, parallel}\SpecialCharTok{::}\FunctionTok{detectCores}\NormalTok{()))}
\NormalTok{  tempf }\OtherTok{\textless{}{-}} \FunctionTok{tempfile}\NormalTok{()}
  \FunctionTok{saveRDS}\NormalTok{(}\AttributeTok{object =} \FunctionTok{list}\NormalTok{(}\AttributeTok{a =}\NormalTok{ p2\_smv, }\AttributeTok{b =}\NormalTok{ p2\_lls), }\AttributeTok{file =}\NormalTok{ tempf)}
\NormalTok{  parallel}\SpecialCharTok{::}\FunctionTok{clusterExport}\NormalTok{(cl, }\AttributeTok{varlist =} \FunctionTok{list}\NormalTok{(}\StringTok{"tempf"}\NormalTok{))}
  
\NormalTok{  res }\OtherTok{\textless{}{-}} \FunctionTok{doWithParallelClusterExplicit}\NormalTok{(}\AttributeTok{cl =}\NormalTok{ cl, }\AttributeTok{expr =}\NormalTok{ \{}
\NormalTok{    optimParallel}\SpecialCharTok{::}\FunctionTok{optimParallel}\NormalTok{(}
      \AttributeTok{par =}\NormalTok{ p2\_smv}\SpecialCharTok{$}\FunctionTok{getParams}\NormalTok{(),}
      \AttributeTok{method =} \StringTok{"L{-}BFGS{-}B"}\NormalTok{,}
      \AttributeTok{lower =} \FunctionTok{c}\NormalTok{(}
        \FunctionTok{rep}\NormalTok{(}\SpecialCharTok{{-}}\NormalTok{.Machine}\SpecialCharTok{$}\NormalTok{double.xmax, }\FunctionTok{length}\NormalTok{(p2\_vars)), }\CommentTok{\# v\_[vartype]}
        \FunctionTok{rep}\NormalTok{(}\FunctionTok{sqrt}\NormalTok{(.Machine}\SpecialCharTok{$}\NormalTok{double.eps), }\FunctionTok{length}\NormalTok{(p2\_vars)), }\CommentTok{\# vtl}
        \FunctionTok{rep}\NormalTok{(}\SpecialCharTok{{-}}\NormalTok{.Machine}\SpecialCharTok{$}\NormalTok{double.xmax, }\FunctionTok{length}\NormalTok{(p2\_vars) }\SpecialCharTok{*} \FunctionTok{length}\NormalTok{(weight\_vartype))), }\CommentTok{\# vty for each}
      \AttributeTok{upper =} \FunctionTok{c}\NormalTok{(}
        \FunctionTok{rep}\NormalTok{(.Machine}\SpecialCharTok{$}\NormalTok{double.xmax, }\FunctionTok{length}\NormalTok{(p2\_vars)),}
        \FunctionTok{rep}\NormalTok{(}\DecValTok{1}\NormalTok{, }\FunctionTok{length}\NormalTok{(p2\_vars)),}
        \FunctionTok{rep}\NormalTok{(.Machine}\SpecialCharTok{$}\NormalTok{double.xmax, }\FunctionTok{length}\NormalTok{(p2\_vars) }\SpecialCharTok{*} \FunctionTok{length}\NormalTok{(weight\_vartype))),}
      \AttributeTok{fn =} \ControlFlowTok{function}\NormalTok{(x) \{}
\NormalTok{        temp }\OtherTok{\textless{}{-}} \FunctionTok{readRDS}\NormalTok{(}\AttributeTok{file =}\NormalTok{ tempf)}
\NormalTok{        temp}\SpecialCharTok{$}\NormalTok{a}\SpecialCharTok{$}\FunctionTok{setParams}\NormalTok{(}\AttributeTok{params =}\NormalTok{ x)}
\NormalTok{        temp}\SpecialCharTok{$}\NormalTok{b}\SpecialCharTok{$}\FunctionTok{compute0}\NormalTok{()}
\NormalTok{      \},}
      \AttributeTok{parallel =} \FunctionTok{list}\NormalTok{(}\AttributeTok{cl =}\NormalTok{ cl, }\AttributeTok{forward =} \ConstantTok{FALSE}\NormalTok{, }\AttributeTok{loginfo =} \ConstantTok{TRUE}\NormalTok{)}
\NormalTok{    )}
\NormalTok{  \})}
\NormalTok{\})}

\NormalTok{p2\_fr }\OtherTok{\textless{}{-}}\NormalTok{ FitResult}\SpecialCharTok{$}\FunctionTok{new}\NormalTok{(}\StringTok{"a"}\NormalTok{)}
\NormalTok{p2\_fr}\SpecialCharTok{$}\FunctionTok{fromOptimParallel}\NormalTok{(p2\_params)}
\FunctionTok{format}\NormalTok{(p2\_fr}\SpecialCharTok{$}\FunctionTok{getBest}\NormalTok{(}\AttributeTok{paramName =} \StringTok{"loss"}\NormalTok{)[}
  \DecValTok{1}\NormalTok{, }\SpecialCharTok{!}\NormalTok{(p2\_fr}\SpecialCharTok{$}\FunctionTok{getParamNames}\NormalTok{() }\SpecialCharTok{\%in\%} \FunctionTok{c}\NormalTok{(}\StringTok{"duration"}\NormalTok{, }\StringTok{"begin"}\NormalTok{, }\StringTok{"end"}\NormalTok{))],}
  \AttributeTok{scientific =} \ConstantTok{FALSE}\NormalTok{, }\AttributeTok{digits =} \DecValTok{4}\NormalTok{, }\AttributeTok{nsmall =} \DecValTok{4}\NormalTok{)}
\end{Highlighting}
\end{Shaded}

\begin{verbatim}
##         v_A        v_CP      v_FREQ       v_SCD       vtl_1       vtl_2 
## "-0.014674" "-0.001527" "-0.008464" "-0.325218" " 0.486963" " 0.040402" 
##       vtl_3       vtl_4     vty_1_A    vty_1_CP  vty_1_FREQ   vty_1_SCD 
## " 0.493052" " 0.988596" " 0.055662" " 0.285855" " 0.413141" " 0.398693" 
##     vty_2_A    vty_2_CP  vty_2_FREQ   vty_2_SCD     vty_3_A    vty_3_CP 
## "-0.004618" " 0.023071" "-0.064090" "-0.149528" " 0.661491" "-0.333191" 
##  vty_3_FREQ   vty_3_SCD     vty_4_A    vty_4_CP  vty_4_FREQ   vty_4_SCD 
## " 0.462005" " 0.263016" "-1.048072" "-0.287820" "-1.437350" "-0.342797" 
##        loss 
## " 6.167237"
\end{verbatim}

\begin{figure}[ht!]
\includegraphics{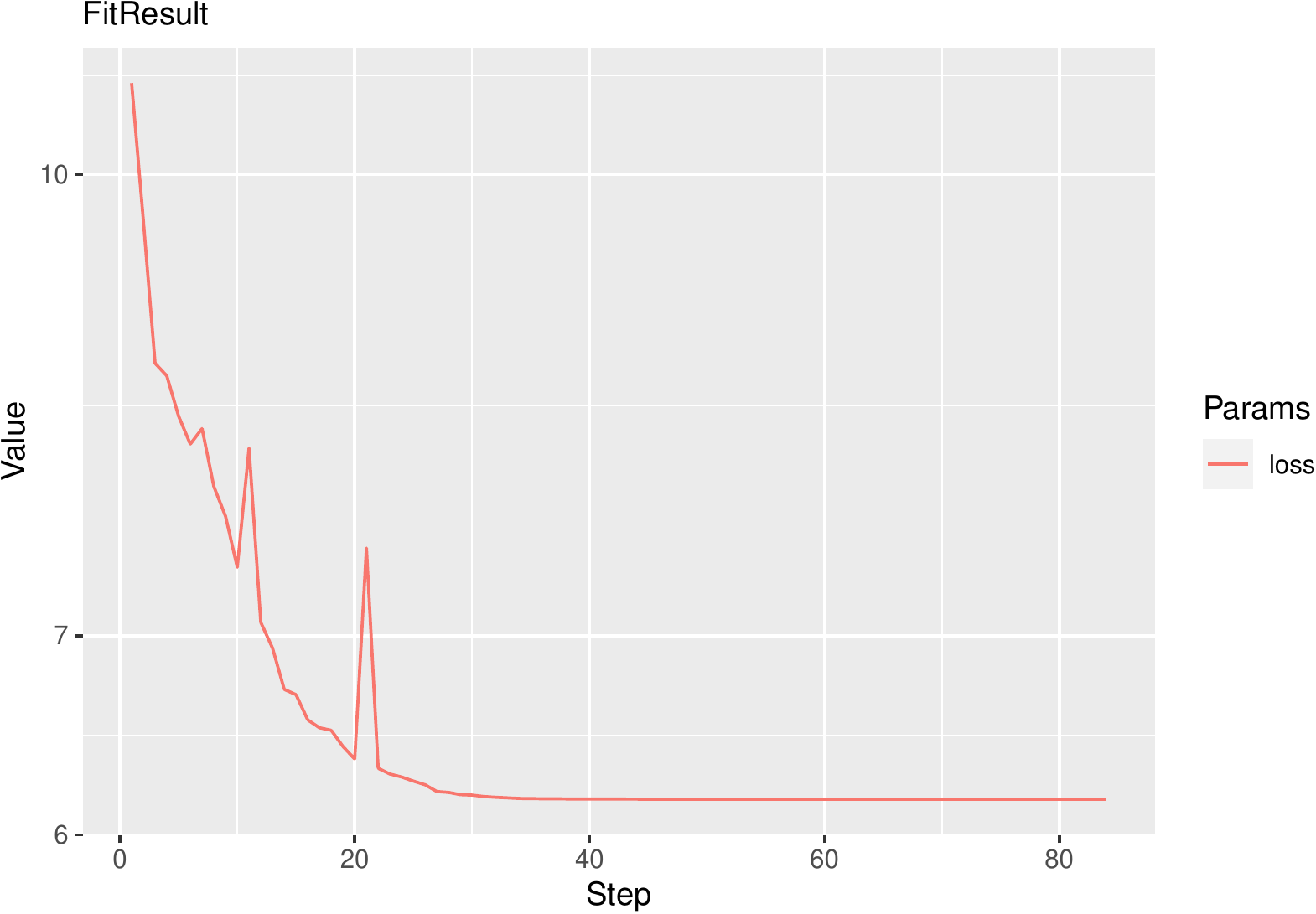} \caption{Neg. Log-loss of fitting pattern type II.}\label{fig:p2-params-fig}
\end{figure}

\hypertarget{inversing-the-parameters}{%
\paragraph{Inversing the parameters}\label{inversing-the-parameters}}

For this pattern, we have warped all the projects to the pattern, while the ultimate goal is to warp the pattern to all the projects (or, better, to warp each type of variable of the WP to the group of variables of the same type of all projects, according to their weight, which is determined by the consensus of the ground truth). So, if we know how to go from A to B, we can inverse the learned parameters and go from B to A, which means in our case that we have to apply the inverse parameters to the WP in order to obtain WP-prime.

As for y-translations (that is, \(v\), as well as all \(\bm{\vartheta}^{(y)}\)), the inversion is simple: we multiply these parameters with \(-1\). The explanation for that is straightforward: If, for example, we had to go down by \(-0.5\), to bring the data closer to the pattern, then that means that we have to lift the pattern by \(+0.5\) to achieve the inverse effect.

Inversing the the boundaries is simple, too, and is explained by how we take some portion of the WC (the source) and warp it to the corresponding interval of the WP (the target).

That's how we do it:

\begin{itemize}
\tightlist
\item
  Given are the WP's \textbf{original} boundaries, \(\bm{\theta}^{(b)}\), and the learned \(\bm{\vartheta}^{(l)}\). The goal is, for each \(q\)-th interval, to take what is in the WP's interval and warp it according to the learned length.
\item
  Given the boundaries-to-lengths operator, \(\mathsf{T}^{(l)}\), and the lengths-to-boundaries operator, \(\mathsf{T}^{(b)}\), we can convert between \(\bm{\theta}\) and \(\bm{\vartheta}\).
\item
  Start with a new instance of \texttt{SRBTW} (or \texttt{SRBTWBAW} for also warping y-translations) and set as \(\bm{\theta}^{(b)}=\mathsf{T}^{(b)}(\bm{\vartheta}^{(l)})\). The learned lengths will become the \textbf{target} intervals.
\item
  Add the variable that ought to be transformed as \textbf{WC}, and set \(\bm{\vartheta}^{(l)}=\mathsf{T}^{(l)}(\bm{\theta}^{(b)})\).
\item
  That will result in that we are taking what was \emph{originally} in each interval, and warp it to a new length.
\item
  The warped signal is then the \texttt{M}-function of the \texttt{SRBTW}/\texttt{SRBTWBAW}-instance.
\end{itemize}

Short example: Let's take the \texttt{SCD}-variable from the first pattern and warp it!

\begin{Shaded}
\begin{Highlighting}[]
\CommentTok{\# Transforming some learned lengths to new boundaries:}
\NormalTok{p2\_ex\_thetaB }\OtherTok{\textless{}{-}} \FunctionTok{c}\NormalTok{(}\DecValTok{0}\NormalTok{, }\FloatTok{0.3}\NormalTok{, }\FloatTok{0.5}\NormalTok{, }\FloatTok{0.7}\NormalTok{, }\DecValTok{1}\NormalTok{)}
\CommentTok{\# Transforming the original boundaries to lengths:}
\NormalTok{p2\_ex\_varthetaL }\OtherTok{\textless{}{-}} \FunctionTok{unname}\NormalTok{(}\FunctionTok{c}\NormalTok{(fd\_data\_boundaries[}\DecValTok{1}\NormalTok{], fd\_data\_boundaries[}\DecValTok{2}\NormalTok{] }\SpecialCharTok{{-}}\NormalTok{ fd\_data\_boundaries[}\DecValTok{1}\NormalTok{],}
\NormalTok{  fd\_data\_boundaries[}\DecValTok{3}\NormalTok{] }\SpecialCharTok{{-}}\NormalTok{ fd\_data\_boundaries[}\DecValTok{2}\NormalTok{], }\DecValTok{1} \SpecialCharTok{{-}}\NormalTok{ fd\_data\_boundaries[}\DecValTok{3}\NormalTok{]))}

\NormalTok{p2\_ex\_srbtw }\OtherTok{\textless{}{-}}\NormalTok{ SRBTW}\SpecialCharTok{$}\FunctionTok{new}\NormalTok{(}\AttributeTok{theta\_b =}\NormalTok{ p2\_ex\_thetaB, }\AttributeTok{gamma\_bed =} \FunctionTok{c}\NormalTok{(}\DecValTok{0}\NormalTok{, }\DecValTok{1}\NormalTok{, }\DecValTok{0}\NormalTok{), }\AttributeTok{wp =}\NormalTok{ p1\_signals}\SpecialCharTok{$}\NormalTok{SCD}\SpecialCharTok{$}\FunctionTok{get0Function}\NormalTok{(),}
  \AttributeTok{wc =}\NormalTok{ p1\_signals}\SpecialCharTok{$}\NormalTok{SCD}\SpecialCharTok{$}\FunctionTok{get0Function}\NormalTok{(), }\AttributeTok{lambda =} \FunctionTok{rep}\NormalTok{(}\DecValTok{0}\NormalTok{, }\DecValTok{4}\NormalTok{), }\AttributeTok{begin =} \DecValTok{0}\NormalTok{, }\AttributeTok{end =} \DecValTok{1}\NormalTok{)}

\NormalTok{p2\_ex\_srbtw}\SpecialCharTok{$}\FunctionTok{setParams}\NormalTok{(}\AttributeTok{vartheta\_l =}\NormalTok{ p2\_ex\_varthetaL)}
\end{Highlighting}
\end{Shaded}

\begin{figure}[ht!]
\includegraphics{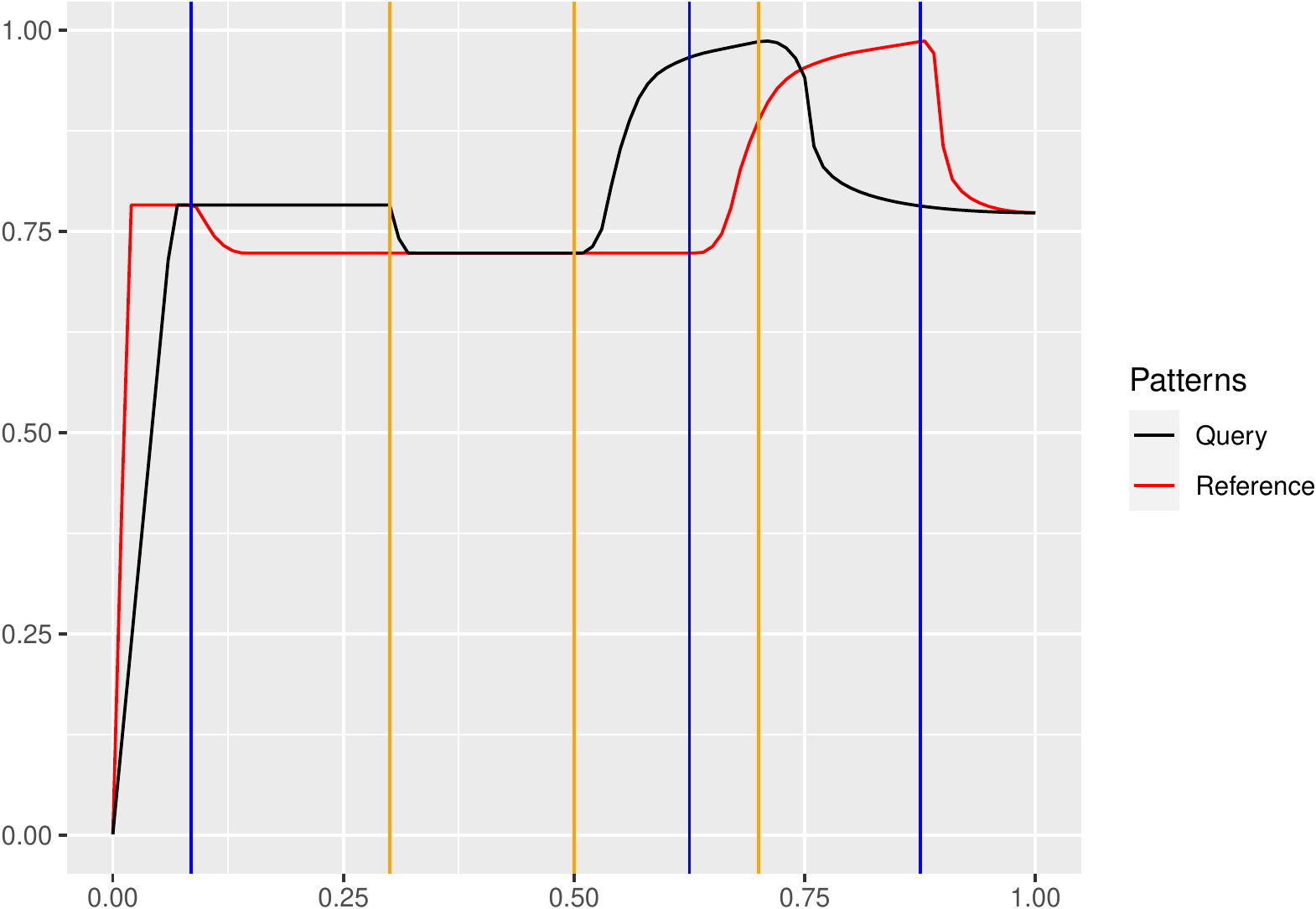} \caption{Warping the variable from within the blue to the orange intervals.}\label{fig:inverse-example-fig}
\end{figure}

In figure \ref{fig:inverse-example-fig} we can quite clearly see how the pattern warped from the blue intervals into the orange intervals

We have learned the following parameters from our optimization for pattern II:

\begin{Shaded}
\begin{Highlighting}[]
\NormalTok{p2\_best }\OtherTok{\textless{}{-}}\NormalTok{ p2\_fr}\SpecialCharTok{$}\FunctionTok{getBest}\NormalTok{(}\AttributeTok{paramName =} \StringTok{"loss"}\NormalTok{)[}\DecValTok{1}\NormalTok{, }\SpecialCharTok{!}\NormalTok{(p2\_fr}\SpecialCharTok{$}\FunctionTok{getParamNames}\NormalTok{() }\SpecialCharTok{\%in\%} \FunctionTok{c}\NormalTok{(}\StringTok{"begin"}\NormalTok{,}
  \StringTok{"end"}\NormalTok{, }\StringTok{"duration"}\NormalTok{))]}
\NormalTok{p2\_best}
\end{Highlighting}
\end{Shaded}

\begin{verbatim}
##          v_A         v_CP       v_FREQ        v_SCD        vtl_1        vtl_2 
## -0.014673620 -0.001527483 -0.008464488 -0.325217649  0.486962785  0.040401601 
##        vtl_3        vtl_4      vty_1_A     vty_1_CP   vty_1_FREQ    vty_1_SCD 
##  0.493052472  0.988595727  0.055661794  0.285855275  0.413141394  0.398693496 
##      vty_2_A     vty_2_CP   vty_2_FREQ    vty_2_SCD      vty_3_A     vty_3_CP 
## -0.004618406  0.023071288 -0.064090457 -0.149527947  0.661491448 -0.333190639 
##   vty_3_FREQ    vty_3_SCD      vty_4_A     vty_4_CP   vty_4_FREQ    vty_4_SCD 
##  0.462004516  0.263016381 -1.048071875 -0.287819945 -1.437350319 -0.342797057 
##         loss 
##  6.167237270
\end{verbatim}

All of the initial translations (\(v\)) are zero. The learned lengths converted to boundaries are:

\begin{Shaded}
\begin{Highlighting}[]
\CommentTok{\# Here, we transform the learned lengths to boundaries.}
\NormalTok{p2\_best\_varthetaL }\OtherTok{\textless{}{-}}\NormalTok{ p2\_best[}\FunctionTok{names}\NormalTok{(p2\_best) }\SpecialCharTok{\%in\%} \FunctionTok{paste0}\NormalTok{(}\StringTok{"vtl\_"}\NormalTok{, }\DecValTok{1}\SpecialCharTok{:}\DecValTok{4}\NormalTok{)]}\SpecialCharTok{/}\FunctionTok{sum}\NormalTok{(p2\_best[}\FunctionTok{names}\NormalTok{(p2\_best) }\SpecialCharTok{\%in\%}
  \FunctionTok{paste0}\NormalTok{(}\StringTok{"vtl\_"}\NormalTok{, }\DecValTok{1}\SpecialCharTok{:}\DecValTok{4}\NormalTok{)])}
\NormalTok{p2\_best\_varthetaL}
\end{Highlighting}
\end{Shaded}

\begin{verbatim}
##      vtl_1      vtl_2      vtl_3      vtl_4 
## 0.24238912 0.02011018 0.24542030 0.49208041
\end{verbatim}

\begin{Shaded}
\begin{Highlighting}[]
\NormalTok{p2\_best\_thetaB }\OtherTok{\textless{}{-}} \FunctionTok{unname}\NormalTok{(}\FunctionTok{c}\NormalTok{(}\DecValTok{0}\NormalTok{, p2\_best\_varthetaL[}\DecValTok{1}\NormalTok{], }\FunctionTok{sum}\NormalTok{(p2\_best\_varthetaL[}\DecValTok{1}\SpecialCharTok{:}\DecValTok{2}\NormalTok{]),}
  \FunctionTok{sum}\NormalTok{(p2\_best\_varthetaL[}\DecValTok{1}\SpecialCharTok{:}\DecValTok{3}\NormalTok{]), }\DecValTok{1}\NormalTok{))}
\NormalTok{p2\_best\_thetaB}
\end{Highlighting}
\end{Shaded}

\begin{verbatim}
## [1] 0.0000000 0.2423891 0.2624993 0.5079196 1.0000000
\end{verbatim}

The first two intervals are rather short, while the last two are comparatively long. Let's transform all of the pattern's variables according to the parameters:

\begin{Shaded}
\begin{Highlighting}[]
\NormalTok{p2\_signals }\OtherTok{\textless{}{-}} \FunctionTok{list}\NormalTok{()}

\ControlFlowTok{for}\NormalTok{ (vartype }\ControlFlowTok{in} \FunctionTok{names}\NormalTok{(weight\_vartype)) \{}
\NormalTok{  temp }\OtherTok{\textless{}{-}}\NormalTok{ SRBTWBAW}\SpecialCharTok{$}\FunctionTok{new}\NormalTok{(}\AttributeTok{theta\_b =} \FunctionTok{unname}\NormalTok{(p2\_best\_thetaB), }\AttributeTok{gamma\_bed =} \FunctionTok{c}\NormalTok{(}\DecValTok{0}\NormalTok{, }\DecValTok{1}\NormalTok{, }\DecValTok{0}\NormalTok{),}
    \AttributeTok{wp =}\NormalTok{ p1\_signals[[vartype]]}\SpecialCharTok{$}\FunctionTok{get0Function}\NormalTok{(), }\AttributeTok{wc =}\NormalTok{ p1\_signals[[vartype]]}\SpecialCharTok{$}\FunctionTok{get0Function}\NormalTok{(),}
    \AttributeTok{lambda =} \FunctionTok{rep}\NormalTok{(}\DecValTok{0}\NormalTok{, }\DecValTok{4}\NormalTok{), }\AttributeTok{begin =} \DecValTok{0}\NormalTok{, }\AttributeTok{end =} \DecValTok{1}\NormalTok{, }\AttributeTok{lambda\_ymin =} \FunctionTok{rep}\NormalTok{(}\DecValTok{0}\NormalTok{, }\DecValTok{4}\NormalTok{), }\AttributeTok{lambda\_ymax =} \FunctionTok{rep}\NormalTok{(}\DecValTok{1}\NormalTok{,}
      \DecValTok{4}\NormalTok{))  }\CommentTok{\# not important here}
  \CommentTok{\# That\textquotesingle{}s still the same (\textquotesingle{}p2\_ex\_varthetaL\textquotesingle{} is the original boundaries of}
  \CommentTok{\# Pattern I transformed to lengths):}
\NormalTok{  temp}\SpecialCharTok{$}\FunctionTok{setParams}\NormalTok{(}\AttributeTok{vartheta\_l =}\NormalTok{ p2\_ex\_varthetaL, }\AttributeTok{v =} \SpecialCharTok{{-}}\DecValTok{1} \SpecialCharTok{*}\NormalTok{ p2\_best[}\FunctionTok{paste0}\NormalTok{(}\StringTok{"v\_"}\NormalTok{, vartype)],}
    \AttributeTok{vartheta\_y =} \SpecialCharTok{{-}}\DecValTok{1} \SpecialCharTok{*}\NormalTok{ p2\_best[}\FunctionTok{paste0}\NormalTok{(}\StringTok{"vty\_"}\NormalTok{, }\DecValTok{1}\SpecialCharTok{:}\DecValTok{4}\NormalTok{, }\StringTok{"\_"}\NormalTok{, vartype)])}

\NormalTok{  p2\_signals[[vartype]] }\OtherTok{\textless{}{-}}\NormalTok{ Signal}\SpecialCharTok{$}\FunctionTok{new}\NormalTok{(}\AttributeTok{name =} \FunctionTok{paste0}\NormalTok{(}\StringTok{"p2\_"}\NormalTok{, vartype), }\AttributeTok{support =} \FunctionTok{c}\NormalTok{(}\DecValTok{0}\NormalTok{,}
    \DecValTok{1}\NormalTok{), }\AttributeTok{isWp =} \ConstantTok{TRUE}\NormalTok{, }\AttributeTok{func =} \FunctionTok{Vectorize}\NormalTok{(temp}\SpecialCharTok{$}\NormalTok{M))}
\NormalTok{\}}
\end{Highlighting}
\end{Shaded}

\begin{figure}[ht!]
\includegraphics{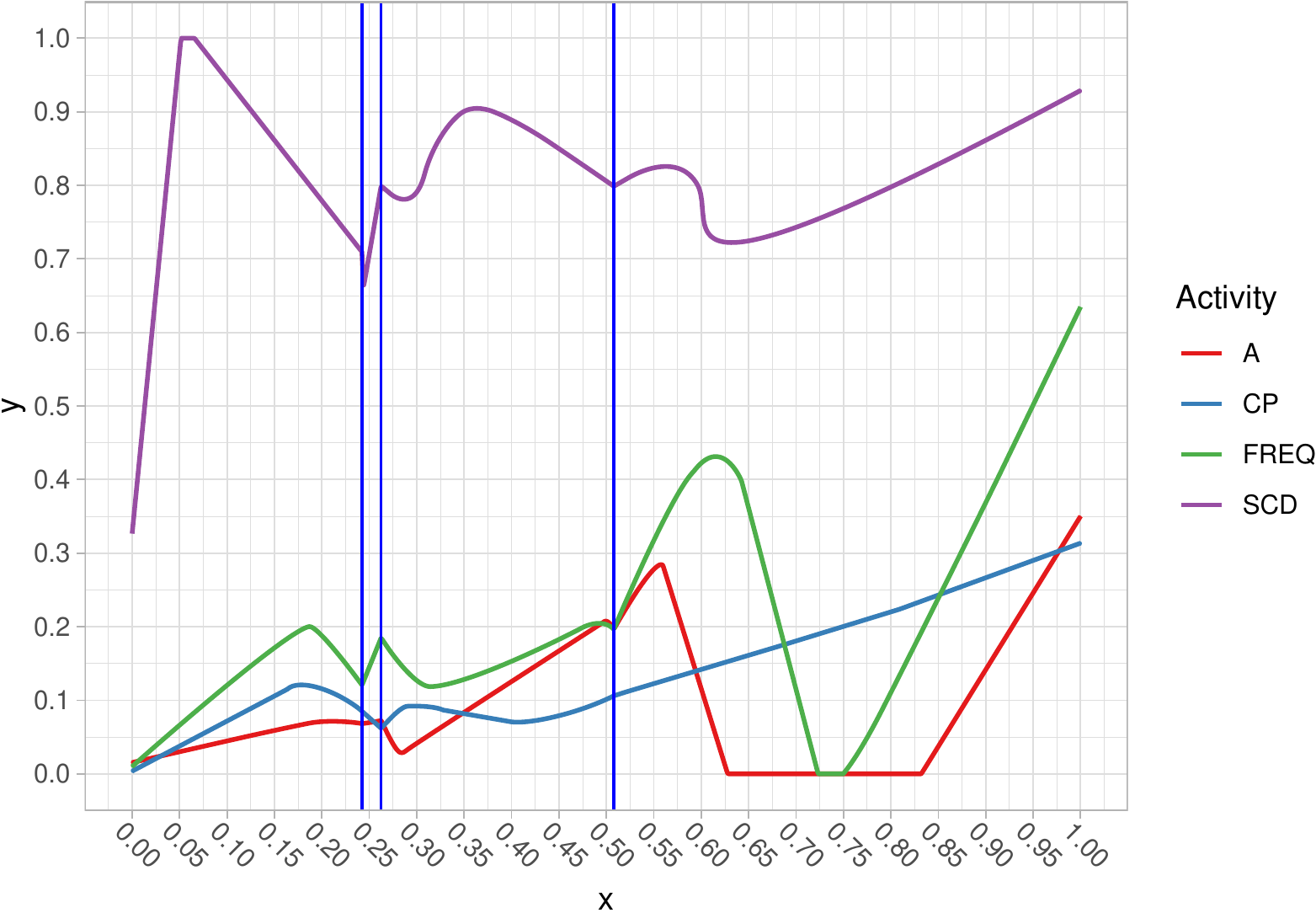} \caption{Second pattern as aligned by the ground truth.}\label{fig:p2-signals}
\end{figure}

The 2nd pattern, as derived from the ground truth, is shown in figure \ref{fig:p2-signals}.

While this worked I suppose it is fair to say that our initial pattern is hardly recognizable. Since we expected this, we planned for a third kind of pattern in section \ref{sec:pattern3}, that is purely evidence-based. It appears that, in order to match the ground truths we have at our disposal, projects register some kind of weak initial peak for the maintenance activities, that is followed by a somewhat uneventful second and third interval. Interestingly, the optimization seemed to have used to mostly straight lines in the Long Stretch phase to model linear declines and increases. The new Aftermath phase is the longest, so it is clear that the original pattern and its subdivision into phases is not a good mapping any longer. Instead of a sharp decline in the Aftermath, we now see an increase of all variables, without the chance of any decline before the last observed commit. We will check how this adapted pattern fares in section \ref{ssec:score-pattern2}.

\hypertarget{pattern-iii-averaging-the-ground-truth}{%
\subsubsection{Pattern III: Averaging the ground truth}\label{pattern-iii-averaging-the-ground-truth}}

We can produce a pattern by computing a weighted average over all available ground truth. As weight, we can use either rater's score, their mean or consensus (default).

\begin{Shaded}
\begin{Highlighting}[]
\NormalTok{gt\_weighted\_avg }\OtherTok{\textless{}{-}} \ControlFlowTok{function}\NormalTok{(vartype, }\AttributeTok{wtype =} \FunctionTok{c}\NormalTok{(}\StringTok{"consensus"}\NormalTok{, }\StringTok{"rater.a"}\NormalTok{, }\StringTok{"rater.b"}\NormalTok{,}
  \StringTok{"rater.mean"}\NormalTok{), }\AttributeTok{use\_signals =}\NormalTok{ project\_signals, }\AttributeTok{use\_ground\_truth =}\NormalTok{ ground\_truth) \{}
\NormalTok{  wtype }\OtherTok{\textless{}{-}} \FunctionTok{match.arg}\NormalTok{(wtype)}
\NormalTok{  gt }\OtherTok{\textless{}{-}}\NormalTok{ use\_ground\_truth[use\_ground\_truth[[wtype]] }\SpecialCharTok{\textgreater{}} \DecValTok{0}\NormalTok{, ]}
\NormalTok{  wTotal }\OtherTok{\textless{}{-}} \FunctionTok{sum}\NormalTok{(gt[[wtype]])}
\NormalTok{  proj }\OtherTok{\textless{}{-}}\NormalTok{ gt}\SpecialCharTok{$}\NormalTok{project}
\NormalTok{  weights }\OtherTok{\textless{}{-}} \StringTok{\textasciigrave{}}\AttributeTok{names\textless{}{-}}\StringTok{\textasciigrave{}}\NormalTok{(gt[[wtype]], gt}\SpecialCharTok{$}\NormalTok{project)}

\NormalTok{  funcs }\OtherTok{\textless{}{-}} \FunctionTok{lapply}\NormalTok{(use\_signals, }\ControlFlowTok{function}\NormalTok{(ps) ps[[vartype]]}\SpecialCharTok{$}\FunctionTok{get0Function}\NormalTok{())}

  \FunctionTok{Vectorize}\NormalTok{(}\ControlFlowTok{function}\NormalTok{(x) \{}
\NormalTok{    val }\OtherTok{\textless{}{-}} \DecValTok{0}
    \ControlFlowTok{for}\NormalTok{ (p }\ControlFlowTok{in}\NormalTok{ proj) \{}
\NormalTok{      val }\OtherTok{\textless{}{-}}\NormalTok{ val }\SpecialCharTok{+}\NormalTok{ weights[[p]] }\SpecialCharTok{*}\NormalTok{ funcs[[p]](x)}
\NormalTok{    \}}
\NormalTok{    val}\SpecialCharTok{/}\NormalTok{wTotal}
\NormalTok{  \})}
\NormalTok{\}}
\end{Highlighting}
\end{Shaded}

Now we can easily call above function to produce a weighted average of each signal:

\begin{Shaded}
\begin{Highlighting}[]
\NormalTok{p3\_avg\_signals }\OtherTok{\textless{}{-}} \FunctionTok{list}\NormalTok{()}

\ControlFlowTok{for}\NormalTok{ (vartype }\ControlFlowTok{in} \FunctionTok{names}\NormalTok{(weight\_vartype)) \{}
\NormalTok{  p3\_avg\_signals[[vartype]] }\OtherTok{\textless{}{-}}\NormalTok{ Signal}\SpecialCharTok{$}\FunctionTok{new}\NormalTok{(}\AttributeTok{name =} \FunctionTok{paste0}\NormalTok{(}\StringTok{"p3\_avg\_"}\NormalTok{, vartype), }\AttributeTok{support =} \FunctionTok{c}\NormalTok{(}\DecValTok{0}\NormalTok{,}
    \DecValTok{1}\NormalTok{), }\AttributeTok{isWp =} \ConstantTok{TRUE}\NormalTok{, }\AttributeTok{func =} \FunctionTok{gt\_weighted\_avg}\NormalTok{(}\AttributeTok{vartype =}\NormalTok{ vartype))}
\NormalTok{\}}
\end{Highlighting}
\end{Shaded}

\begin{figure}[ht!]
\includegraphics{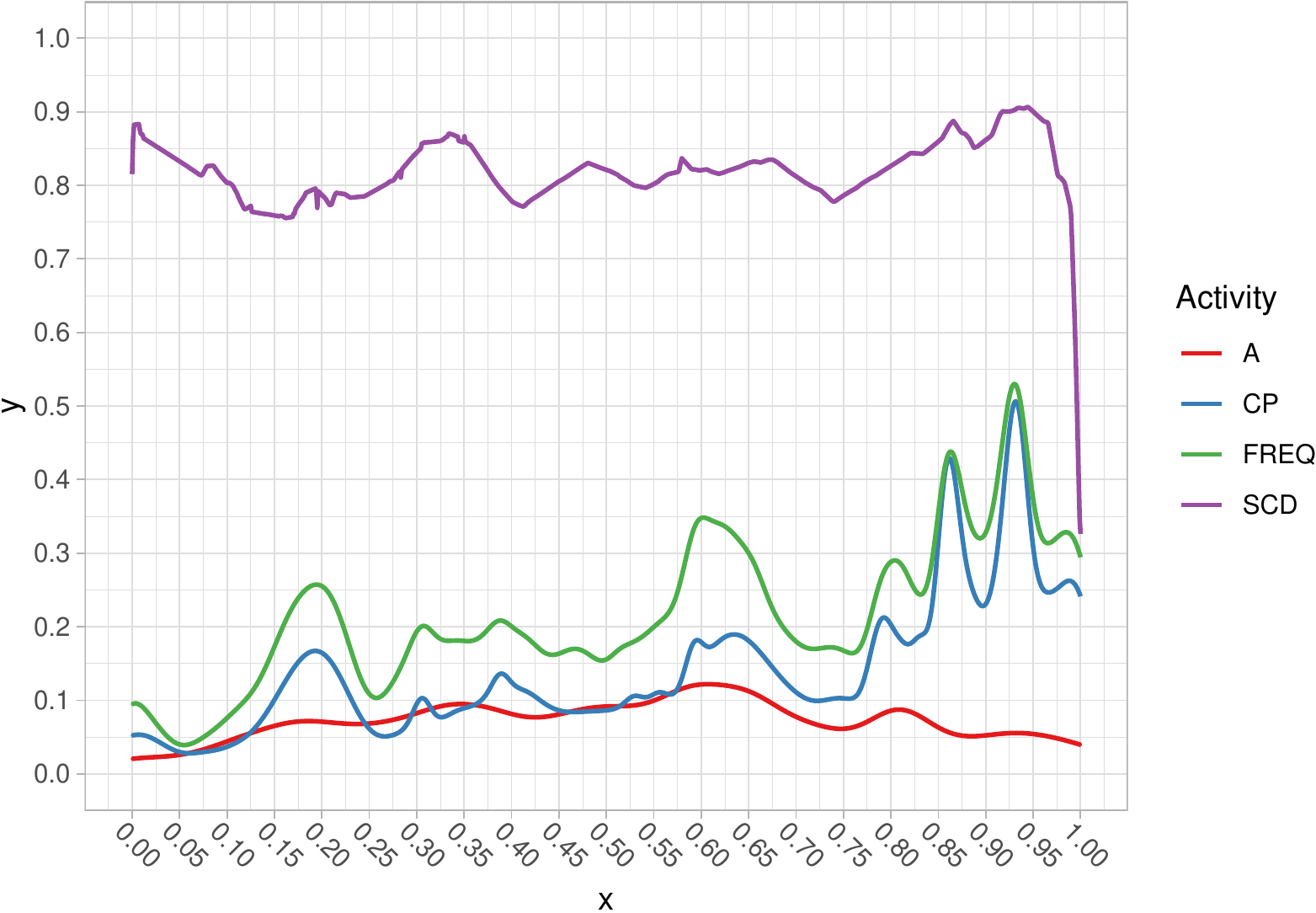} \caption{The third kind of pattern as weighted average over all ground truth.}\label{fig:p3-avg-signals}
\end{figure}

The 2nd pattern, as derived from the ground truth, is shown in figure \ref{fig:p3-avg-signals}.

\hypertarget{pattern-iii-b-evidence-based}{%
\subsubsection{\texorpdfstring{Pattern III (b): Evidence-based\label{sec:pattern3}}{Pattern III (b): Evidence-based}}\label{pattern-iii-b-evidence-based}}

A third kind of pattern is produced by starting with an empty warping pattern and having all available ground truth adapt to it. Empty means that we will start with a flat line located at \(0.5\) for each variable. Finally, the parameters are inversed. While we could do this the other way round, we have two reasons to do it this way, which is the same as we used for pattern II. First of all if the warping candidate was a perfectly flat line, it would be very difficult for the gradient to converge towards some alignment. Secondly, we want to use equidistantly-spaced boundaries (resulting in equal-length intervals) and using this approach, we can guarantee the interval lengths. To find the optimum amount of intervals, we try all values in a certain range and compute a fit, and then use an information criterion to decide which of the produced patterns provides the best trade-off between number of parameters and goodness-of-fit.

The process is the same as for pattern II: Using an instance of \texttt{srBTAW\_MultiVartype} that holds one instance of an \texttt{srBTAW} per variable-type. We will choose equidistantly-spaced boundaries over the WP, and start with just \(1\) interval, going up to some two-digit number. The best amount of parameters (intervals) is then determined using the Akaike Information Criterion (Akaike 1981), which is directly implemented in \texttt{srBTAW}. We either have to use continuous losses or make sure to \textbf{always} use the exact same amount of samples total. The amount per interval is determined by dividing by the number of intervals. This is important, as otherwise the information criterion will not work. We will do a single RSS-loss that covers all intervals. We will also use an instance of \texttt{TimeWarpRegularization} with the \texttt{exint2}-regularizer, as it scales with arbitrary many intervals (important!). I do not suppose that regularization for the y-values is needed, so we will not have this. This means that the resulting objective has just two losses.

For a set of equal-length number of intervals, we will fit such a multiple variable-type model. This also means we can do this in parallel. However, models with more intervals and hence more parameters will considerable take longer during gradient iterations. The more parameters, the fewer of these models should be fit simultaneously. We have access to 128-thread machine (of which about 125 thread can be used). Gradients are computed in parallel as well.

\hypertarget{preparation-1}{%
\paragraph{Preparation}\label{preparation-1}}

We define a single function that encapsulates the multiple variable-type model, losses and objectives and returns them, so that we can just fit them in a loop. The only configurable parameters is the amount of intervals.

\begin{Shaded}
\begin{Highlighting}[]
\NormalTok{p3\_prepare\_mvtypemodel }\OtherTok{\textless{}{-}} \ControlFlowTok{function}\NormalTok{(numIntervals) \{}
\NormalTok{  eps }\OtherTok{\textless{}{-}} \FunctionTok{sqrt}\NormalTok{(.Machine}\SpecialCharTok{$}\NormalTok{double.eps)}
\NormalTok{  p3\_smv }\OtherTok{\textless{}{-}}\NormalTok{ srBTAW\_MultiVartype}\SpecialCharTok{$}\FunctionTok{new}\NormalTok{()}
  
\NormalTok{  p3\_vars }\OtherTok{\textless{}{-}} \FunctionTok{c}\NormalTok{(}\StringTok{"A"}\NormalTok{, }\StringTok{"CP"}\NormalTok{, }\StringTok{"FREQ"}\NormalTok{, }\StringTok{"SCD"}\NormalTok{)}
\NormalTok{  p3\_inst }\OtherTok{\textless{}{-}} \FunctionTok{list}\NormalTok{()}
  
  \CommentTok{\# The objective:}
\NormalTok{  p3\_lls }\OtherTok{\textless{}{-}}\NormalTok{ srBTAW\_LossLinearScalarizer}\SpecialCharTok{$}\FunctionTok{new}\NormalTok{(}
    \AttributeTok{returnRaw =} \ConstantTok{FALSE}\NormalTok{, }\AttributeTok{computeParallel =} \ConstantTok{TRUE}\NormalTok{, }\AttributeTok{gradientParallel =} \ConstantTok{TRUE}\NormalTok{)}
  
  \ControlFlowTok{for}\NormalTok{ (name }\ControlFlowTok{in}\NormalTok{ p3\_vars) \{}
\NormalTok{    p3\_inst[[name]] }\OtherTok{\textless{}{-}}\NormalTok{ srBTAW}\SpecialCharTok{$}\FunctionTok{new}\NormalTok{(}
      \CommentTok{\# Always includes 0,1 {-} just as we need it! Works for values \textgreater{}= 1}
      \AttributeTok{theta\_b =} \FunctionTok{seq}\NormalTok{(}\AttributeTok{from =} \DecValTok{0}\NormalTok{, }\AttributeTok{to =} \DecValTok{1}\NormalTok{, }\AttributeTok{by =} \DecValTok{1} \SpecialCharTok{/}\NormalTok{ numIntervals),}
      \AttributeTok{gamma\_bed =} \FunctionTok{c}\NormalTok{(}\DecValTok{0}\NormalTok{, }\DecValTok{1}\NormalTok{, eps),}
      \AttributeTok{lambda =} \FunctionTok{rep}\NormalTok{(eps, numIntervals),}
      \AttributeTok{begin =} \DecValTok{0}\NormalTok{, }\AttributeTok{end =} \DecValTok{1}\NormalTok{, }\AttributeTok{openBegin =} \ConstantTok{FALSE}\NormalTok{, }\AttributeTok{openEnd =} \ConstantTok{FALSE}\NormalTok{,}
      \AttributeTok{useAmplitudeWarping =} \ConstantTok{TRUE}\NormalTok{,}
      \CommentTok{\# We allow these to be larger; however, the final result should be within [0,1]}
      \AttributeTok{lambda\_ymin =} \FunctionTok{rep}\NormalTok{(}\SpecialCharTok{{-}}\DecValTok{10}\NormalTok{, numIntervals),}
      \AttributeTok{lambda\_ymax =} \FunctionTok{rep}\NormalTok{( }\DecValTok{10}\NormalTok{, numIntervals),}
      \AttributeTok{isObjectiveLogarithmic =} \ConstantTok{TRUE}\NormalTok{,}
      \AttributeTok{paramNames =} \FunctionTok{c}\NormalTok{(}\StringTok{"v"}\NormalTok{,}
        \FunctionTok{paste0}\NormalTok{(}\StringTok{"vtl\_"}\NormalTok{, }\FunctionTok{seq\_len}\NormalTok{(}\AttributeTok{length.out =} \FunctionTok{length}\NormalTok{(p3\_vars))),}
        \FunctionTok{paste0}\NormalTok{(}\StringTok{"vty\_"}\NormalTok{, }\FunctionTok{seq\_len}\NormalTok{(}\AttributeTok{length.out =} \FunctionTok{length}\NormalTok{(p3\_vars)))))}
    
    \CommentTok{\# The WP is a flat line located at 0.5:}
\NormalTok{    p3\_inst[[name]]}\SpecialCharTok{$}\FunctionTok{setSignal}\NormalTok{(}\AttributeTok{signal =}\NormalTok{ Signal}\SpecialCharTok{$}\FunctionTok{new}\NormalTok{(}
      \AttributeTok{func =} \ControlFlowTok{function}\NormalTok{(x) .}\DecValTok{5}\NormalTok{, }\AttributeTok{isWp =} \ConstantTok{TRUE}\NormalTok{, }\AttributeTok{support =} \FunctionTok{c}\NormalTok{(}\DecValTok{0}\NormalTok{, }\DecValTok{1}\NormalTok{), }\AttributeTok{name =} \FunctionTok{paste0}\NormalTok{(}\StringTok{"p3\_"}\NormalTok{, name)))}
    
    \CommentTok{\# Set the common objective:}
\NormalTok{    p3\_inst[[name]]}\SpecialCharTok{$}\FunctionTok{setObjective}\NormalTok{(}\AttributeTok{obj =}\NormalTok{ p3\_lls)}
    
    \CommentTok{\# .. and also all the projects\textquotesingle{} signals:}
    \ControlFlowTok{for}\NormalTok{ (project }\ControlFlowTok{in}\NormalTok{ ground\_truth[ground\_truth}\SpecialCharTok{$}\NormalTok{consensus }\SpecialCharTok{\textgreater{}} \DecValTok{0}\NormalTok{, ]}\SpecialCharTok{$}\NormalTok{project) \{}
\NormalTok{      p3\_inst[[name]]}\SpecialCharTok{$}\FunctionTok{setSignal}\NormalTok{(}\AttributeTok{signal =}\NormalTok{ project\_signals[[project]][[name]])}
\NormalTok{    \}}
    
\NormalTok{    p3\_smv}\SpecialCharTok{$}\FunctionTok{setSrbtaw}\NormalTok{(}\AttributeTok{varName =}\NormalTok{ name, }\AttributeTok{srbtaw =}\NormalTok{ p3\_inst[[name]])}
\NormalTok{  \}}

  \CommentTok{\# We call this there so there are parameters present.}
  \FunctionTok{set.seed}\NormalTok{(}\DecValTok{1337} \SpecialCharTok{*}\NormalTok{ numIntervals)}
\NormalTok{  p3\_smv}\SpecialCharTok{$}\FunctionTok{setParams}\NormalTok{(}\AttributeTok{params =}
    \StringTok{\textasciigrave{}}\AttributeTok{names\textless{}{-}}\StringTok{\textasciigrave{}}\NormalTok{(}\AttributeTok{x =} \FunctionTok{runif}\NormalTok{(}\AttributeTok{n =}\NormalTok{ p3\_smv}\SpecialCharTok{$}\FunctionTok{getNumParams}\NormalTok{()), }\AttributeTok{value =}\NormalTok{ p3\_smv}\SpecialCharTok{$}\FunctionTok{getParamNames}\NormalTok{()))}
  
  \ControlFlowTok{for}\NormalTok{ (name }\ControlFlowTok{in}\NormalTok{ p3\_vars) \{}
    \CommentTok{\# Add RSS{-}loss per variable{-}pair:}
    \ControlFlowTok{for}\NormalTok{ (project }\ControlFlowTok{in}\NormalTok{ ground\_truth[ground\_truth}\SpecialCharTok{$}\NormalTok{consensus }\SpecialCharTok{\textgreater{}} \DecValTok{0}\NormalTok{, ]}\SpecialCharTok{$}\NormalTok{project) \{}
      \CommentTok{\# The RSS{-}loss:}
\NormalTok{      lossRss }\OtherTok{\textless{}{-}}\NormalTok{ srBTAW\_Loss\_Rss}\SpecialCharTok{$}\FunctionTok{new}\NormalTok{(}
        \AttributeTok{wpName =} \FunctionTok{paste0}\NormalTok{(}\StringTok{"p3\_"}\NormalTok{, name), }\AttributeTok{wcName =} \FunctionTok{paste}\NormalTok{(project, name, }\AttributeTok{sep =} \StringTok{"\_"}\NormalTok{),}
        \AttributeTok{weight =} \DecValTok{1}\NormalTok{, }\AttributeTok{intervals =} \FunctionTok{seq\_len}\NormalTok{(}\AttributeTok{length.out =}\NormalTok{ numIntervals), }\AttributeTok{continuous =} \ConstantTok{FALSE}\NormalTok{,}
        \AttributeTok{numSamples =} \FunctionTok{rep}\NormalTok{(}\FunctionTok{round}\NormalTok{(}\DecValTok{5000} \SpecialCharTok{/}\NormalTok{ numIntervals), numIntervals), }\AttributeTok{returnRaw =} \ConstantTok{TRUE}\NormalTok{)}
\NormalTok{      p3\_inst[[name]]}\SpecialCharTok{$}\FunctionTok{addLoss}\NormalTok{(}\AttributeTok{loss =}\NormalTok{ lossRss)}
\NormalTok{      p3\_lls}\SpecialCharTok{$}\FunctionTok{setObjective}\NormalTok{(}
        \AttributeTok{name =} \FunctionTok{paste}\NormalTok{(project, name, }\StringTok{"rss"}\NormalTok{, }\AttributeTok{sep =} \StringTok{"\_"}\NormalTok{), }\AttributeTok{obj =}\NormalTok{ lossRss)}
\NormalTok{    \}}
\NormalTok{  \}}
  
  \CommentTok{\# This has a much higher weight than we had for pattern II}
  \CommentTok{\# because we are using many more samples in the RSS{-}loss.}
\NormalTok{  p3\_lls}\SpecialCharTok{$}\FunctionTok{setObjective}\NormalTok{(}\AttributeTok{name =} \StringTok{"p3\_reg\_exint2"}\NormalTok{, }\AttributeTok{obj =}\NormalTok{ TimeWarpRegularization}\SpecialCharTok{$}\FunctionTok{new}\NormalTok{(}
    \AttributeTok{weight =}\NormalTok{ p3\_lls}\SpecialCharTok{$}\FunctionTok{getNumObjectives}\NormalTok{(), }\AttributeTok{use =} \StringTok{"exint2"}\NormalTok{, }\AttributeTok{returnRaw =} \ConstantTok{TRUE}\NormalTok{,}
    \AttributeTok{wpName =} \StringTok{"p3\_A"}\NormalTok{, }\AttributeTok{wcName =}\NormalTok{ project\_signals}\SpecialCharTok{$}\NormalTok{project\_1}\SpecialCharTok{$}\NormalTok{A}\SpecialCharTok{$}\FunctionTok{getName}\NormalTok{(),}
    \AttributeTok{intervals =} \FunctionTok{seq\_len}\NormalTok{(numIntervals)}
\NormalTok{  )}\SpecialCharTok{$}\FunctionTok{setSrBtaw}\NormalTok{(}\AttributeTok{srbtaw =}\NormalTok{ p3\_inst}\SpecialCharTok{$}\NormalTok{A))}
  
  \FunctionTok{list}\NormalTok{(}\AttributeTok{smv =}\NormalTok{ p3\_smv, }\AttributeTok{lls =}\NormalTok{ p3\_lls)}
\NormalTok{\}}
\end{Highlighting}
\end{Shaded}

Now we can compute these in parallel:

\begin{Shaded}
\begin{Highlighting}[]
\ControlFlowTok{for}\NormalTok{ (numIntervals }\ControlFlowTok{in} \FunctionTok{c}\NormalTok{(}\DecValTok{1}\SpecialCharTok{:}\DecValTok{16}\NormalTok{)) \{}
  \FunctionTok{loadResultsOrCompute}\NormalTok{(}
    \AttributeTok{file =} \FunctionTok{paste0}\NormalTok{(}\StringTok{"../results/p3{-}compute/i\_"}\NormalTok{, numIntervals, }\StringTok{".rds"}\NormalTok{),}
    \AttributeTok{computeExpr =}
\NormalTok{  \{}
\NormalTok{    p3\_vars }\OtherTok{\textless{}{-}} \FunctionTok{c}\NormalTok{(}\StringTok{"A"}\NormalTok{, }\StringTok{"CP"}\NormalTok{, }\StringTok{"FREQ"}\NormalTok{, }\StringTok{"SCD"}\NormalTok{)}
\NormalTok{    temp }\OtherTok{\textless{}{-}} \FunctionTok{p3\_prepare\_mvtypemodel}\NormalTok{(}\AttributeTok{numIntervals =}\NormalTok{ numIntervals)}
\NormalTok{    tempf }\OtherTok{\textless{}{-}} \FunctionTok{tempfile}\NormalTok{()}
    \FunctionTok{saveRDS}\NormalTok{(}\AttributeTok{object =}\NormalTok{ temp, }\AttributeTok{file =}\NormalTok{ tempf)}
    
    \CommentTok{\# It does not scale well beyond that.}
\NormalTok{    cl }\OtherTok{\textless{}{-}}\NormalTok{ parallel}\SpecialCharTok{::}\FunctionTok{makePSOCKcluster}\NormalTok{(}\FunctionTok{min}\NormalTok{(}\DecValTok{32}\NormalTok{, parallel}\SpecialCharTok{::}\FunctionTok{detectCores}\NormalTok{()))}
\NormalTok{    parallel}\SpecialCharTok{::}\FunctionTok{clusterExport}\NormalTok{(}\AttributeTok{cl =}\NormalTok{ cl, }\AttributeTok{varlist =} \FunctionTok{list}\NormalTok{(}\StringTok{"tempf"}\NormalTok{))}
    
\NormalTok{    optR }\OtherTok{\textless{}{-}} \FunctionTok{doWithParallelClusterExplicit}\NormalTok{(}\AttributeTok{cl =}\NormalTok{ cl, }\AttributeTok{expr =}\NormalTok{ \{}
\NormalTok{      optimParallel}\SpecialCharTok{::}\FunctionTok{optimParallel}\NormalTok{(}
        \AttributeTok{par =}\NormalTok{ temp}\SpecialCharTok{$}\NormalTok{smv}\SpecialCharTok{$}\FunctionTok{getParams}\NormalTok{(),}
        \AttributeTok{method =} \StringTok{"L{-}BFGS{-}B"}\NormalTok{,}
        \AttributeTok{lower =} \FunctionTok{c}\NormalTok{(}
          \FunctionTok{rep}\NormalTok{(}\SpecialCharTok{{-}}\NormalTok{.Machine}\SpecialCharTok{$}\NormalTok{double.xmax, }\FunctionTok{length}\NormalTok{(p3\_vars)), }\CommentTok{\# v\_[vartype]}
          \FunctionTok{rep}\NormalTok{(}\FunctionTok{sqrt}\NormalTok{(.Machine}\SpecialCharTok{$}\NormalTok{double.eps), }\FunctionTok{length}\NormalTok{(p3\_vars)), }\CommentTok{\# vtl}
          \FunctionTok{rep}\NormalTok{(}\SpecialCharTok{{-}}\NormalTok{.Machine}\SpecialCharTok{$}\NormalTok{double.xmax, }\FunctionTok{length}\NormalTok{(p3\_vars) }\SpecialCharTok{*} \FunctionTok{length}\NormalTok{(weight\_vartype))), }\CommentTok{\# vty for each}
        \AttributeTok{upper =} \FunctionTok{c}\NormalTok{(}
          \FunctionTok{rep}\NormalTok{(.Machine}\SpecialCharTok{$}\NormalTok{double.xmax, }\FunctionTok{length}\NormalTok{(p3\_vars)),}
          \FunctionTok{rep}\NormalTok{(}\DecValTok{1}\NormalTok{, }\FunctionTok{length}\NormalTok{(p3\_vars)),}
          \FunctionTok{rep}\NormalTok{(.Machine}\SpecialCharTok{$}\NormalTok{double.xmax, }\FunctionTok{length}\NormalTok{(p3\_vars) }\SpecialCharTok{*} \FunctionTok{length}\NormalTok{(weight\_vartype))),}
        \AttributeTok{fn =} \ControlFlowTok{function}\NormalTok{(x) \{}
\NormalTok{          temp }\OtherTok{\textless{}{-}} \FunctionTok{readRDS}\NormalTok{(}\AttributeTok{file =}\NormalTok{ tempf)}
\NormalTok{          temp}\SpecialCharTok{$}\NormalTok{smv}\SpecialCharTok{$}\FunctionTok{setParams}\NormalTok{(}\AttributeTok{params =}\NormalTok{ x)}
\NormalTok{          temp}\SpecialCharTok{$}\NormalTok{lls}\SpecialCharTok{$}\FunctionTok{compute0}\NormalTok{()}
\NormalTok{        \},}
        \AttributeTok{parallel =} \FunctionTok{list}\NormalTok{(}\AttributeTok{cl =}\NormalTok{ cl, }\AttributeTok{forward =} \ConstantTok{FALSE}\NormalTok{, }\AttributeTok{loginfo =} \ConstantTok{TRUE}\NormalTok{)}
\NormalTok{      )}
\NormalTok{    \})}
    \FunctionTok{list}\NormalTok{(}\AttributeTok{optR =}\NormalTok{ optR, }\AttributeTok{smv =}\NormalTok{ temp}\SpecialCharTok{$}\NormalTok{smv, }\AttributeTok{lls =}\NormalTok{ temp}\SpecialCharTok{$}\NormalTok{lls)}
\NormalTok{  \})}
\NormalTok{\}}
\end{Highlighting}
\end{Shaded}

\hypertarget{finding-the-best-fit}{%
\paragraph{Finding the best fit}\label{finding-the-best-fit}}

We will load all the results previously computed and compute an information criterion to compare fits, and then choose the best model.

\begin{Shaded}
\begin{Highlighting}[]
\NormalTok{p3\_params }\OtherTok{\textless{}{-}} \ConstantTok{NULL}
\ControlFlowTok{for}\NormalTok{ (tempPath }\ControlFlowTok{in}\NormalTok{ gtools}\SpecialCharTok{::}\FunctionTok{mixedsort}\NormalTok{(}
  \FunctionTok{Sys.glob}\NormalTok{(}\AttributeTok{paths =} \FunctionTok{paste0}\NormalTok{(}\FunctionTok{getwd}\NormalTok{(), }\StringTok{"/../results/p3{-}compute/i*.rds"}\NormalTok{)))}
\NormalTok{) \{}
\NormalTok{  temp }\OtherTok{\textless{}{-}} \FunctionTok{readRDS}\NormalTok{(}\AttributeTok{file =}\NormalTok{ tempPath)}
\NormalTok{  p3\_params }\OtherTok{\textless{}{-}} \FunctionTok{rbind}\NormalTok{(p3\_params, }\FunctionTok{data.frame}\NormalTok{(}
    \AttributeTok{numInt =}\NormalTok{ (temp}\SpecialCharTok{$}\NormalTok{lls}\SpecialCharTok{$}\FunctionTok{getNumParams}\NormalTok{() }\SpecialCharTok{{-}} \DecValTok{1}\NormalTok{) }\SpecialCharTok{/} \DecValTok{2}\NormalTok{,}
    \AttributeTok{numPar =}\NormalTok{ temp}\SpecialCharTok{$}\NormalTok{smv}\SpecialCharTok{$}\FunctionTok{getNumParams}\NormalTok{(),}
    \AttributeTok{numParSrBTAW =}\NormalTok{ temp}\SpecialCharTok{$}\NormalTok{lls}\SpecialCharTok{$}\FunctionTok{getNumParams}\NormalTok{(),}
    \CommentTok{\# This AIC would the original one!}
    \AttributeTok{AIC =} \DecValTok{2} \SpecialCharTok{*}\NormalTok{ temp}\SpecialCharTok{$}\NormalTok{smv}\SpecialCharTok{$}\FunctionTok{getNumParams}\NormalTok{() }\SpecialCharTok{{-}} \DecValTok{2} \SpecialCharTok{*} \FunctionTok{log}\NormalTok{(}\DecValTok{1} \SpecialCharTok{/} \FunctionTok{exp}\NormalTok{(temp}\SpecialCharTok{$}\NormalTok{optR}\SpecialCharTok{$}\NormalTok{value)),}
    \CommentTok{\# This AIC is based on the number of intervals, not parameters!}
    \AttributeTok{AIC1 =}\NormalTok{ (temp}\SpecialCharTok{$}\NormalTok{lls}\SpecialCharTok{$}\FunctionTok{getNumParams}\NormalTok{() }\SpecialCharTok{{-}} \DecValTok{1}\NormalTok{) }\SpecialCharTok{{-}} \DecValTok{2} \SpecialCharTok{*} \FunctionTok{log}\NormalTok{(}\DecValTok{1} \SpecialCharTok{/} \FunctionTok{exp}\NormalTok{(temp}\SpecialCharTok{$}\NormalTok{optR}\SpecialCharTok{$}\NormalTok{value)),}
    \CommentTok{\# This AIC is based on the amount of parameters per srBTAW instance:}
    \AttributeTok{AIC2 =} \DecValTok{2} \SpecialCharTok{*}\NormalTok{ temp}\SpecialCharTok{$}\NormalTok{lls}\SpecialCharTok{$}\FunctionTok{getNumParams}\NormalTok{() }\SpecialCharTok{{-}} \DecValTok{2} \SpecialCharTok{*} \FunctionTok{log}\NormalTok{(}\DecValTok{1} \SpecialCharTok{/} \FunctionTok{exp}\NormalTok{(temp}\SpecialCharTok{$}\NormalTok{optR}\SpecialCharTok{$}\NormalTok{value)),}
    \AttributeTok{logLoss =}\NormalTok{ temp}\SpecialCharTok{$}\NormalTok{optR}\SpecialCharTok{$}\NormalTok{value,}
    \AttributeTok{loss =} \FunctionTok{exp}\NormalTok{(temp}\SpecialCharTok{$}\NormalTok{optR}\SpecialCharTok{$}\NormalTok{value)}
\NormalTok{  ))}
\NormalTok{\}}
\end{Highlighting}
\end{Shaded}

In table \ref{tab:p3-params}, we show computed fits for various models, where the only difference is the number of intervals. Each interval comes with two degrees of freedom: its length and terminal y-translation. Recall that each computed fit concerns four variables. For example, the first model with just one interval per variable has nine parameters: All of the variables share the interval's length, the first parameter. Then, each variable has one \(v\)-parameter, the global y-translation. For each interval, we have one terminal y-translation. For example, the model with \(7\) intervals has \(7 + 4 + (4\times7)=39\) parameters.

We compute the AIC for each fit, which is formulated as in the following. The parameter \(k\) is the number of parameters in the model, i.e., as described, it refers to all the parameters in the \texttt{srBTAW\_MultiVartype}-model. The second AIC-alternative uses the parameter \(p\) instead, which refers to the number of variables per \texttt{srBTAW}-instance.

\[
\begin{aligned}
  \operatorname{AIC}=&\;2\times k - 2\times\log{(\mathcal{\hat{L}})}\;\text{, where}\;\mathcal{\hat{L}}\;\text{is the maximum log-likelihood of the model,}
  \\[1ex]
  \mathcal{\hat{L}}=&\;\frac{1}{\exp{\big(\;\text{lowest loss of the model}\;\big)}}\;\text{, since we use logarithmic losses.}
  \\[1em]
  \text{The alternatives}&\;\operatorname{AIC^1}\;\text{and}\;\operatorname{AIC^2}\;\text{ are defined as:}
  \\[1ex]
  \operatorname{AIC^1}=&\;k-2\times\log{(\mathcal{\hat{L}})}-1\;\text{, which is based on the number of intervals, and}
  \\[1ex]
  \operatorname{AIC^2}=&\;2\times p - 2\times\log{(\mathcal{\hat{L}})}\;\text{, where}\;p\;\text{is the amount of params per}\;\operatorname{srBTAW}\text{-instance.}
\end{aligned}
\]

\begin{table}

\caption{\label{tab:p3-params}Likelihood and Akaike information criteria (AIC) for computed models.}
\centering
\begin{tabular}[t]{rrrrrrrr}
\toprule
numInt & numPar & numParSrBTAW & AIC & AIC1 & AIC2 & logLoss & loss\\
\midrule
1 & 9 & 3 & 34.966 & 18.966 & 22.966 & 8.483 & 4831.189\\
2 & 14 & 5 & 44.052 & 20.052 & 26.052 & 8.026 & 3059.055\\
3 & 19 & 7 & 54.473 & 22.473 & 30.473 & 8.237 & 3776.854\\
4 & 24 & 9 & 63.694 & 23.694 & 33.694 & 7.847 & 2557.439\\
5 & 29 & 11 & 73.985 & 25.985 & 37.985 & 7.992 & 2958.219\\
\addlinespace
6 & 34 & 13 & 83.761 & 27.761 & 41.761 & 7.881 & 2645.198\\
7 & 39 & 15 & 94.783 & 30.783 & 46.783 & 8.392 & 4410.053\\
8 & 44 & 17 & 110.695 & 38.695 & 56.695 & 11.347 & 84745.003\\
9 & 49 & 19 & 113.807 & 33.807 & 53.807 & 7.904 & 2706.944\\
10 & 54 & 21 & 123.809 & 35.809 & 57.809 & 7.905 & 2709.809\\
\addlinespace
11 & 59 & 23 & 134.167 & 38.167 & 62.167 & 8.084 & 3240.653\\
12 & 64 & 25 & 144.350 & 40.350 & 66.350 & 8.175 & 3551.062\\
13 & 69 & 27 & 153.878 & 41.878 & 69.878 & 7.939 & 2804.353\\
14 & 74 & 29 & 164.122 & 44.122 & 74.122 & 8.061 & 3168.120\\
15 & 79 & 31 & 174.284 & 46.284 & 78.284 & 8.142 & 3435.957\\
\addlinespace
16 & 84 & 33 & 184.305 & 48.305 & 82.305 & 8.153 & 3472.327\\
\bottomrule
\end{tabular}
\end{table}

Comparing the results from table \ref{tab:p3-params}, it appears that no matter how we define the AIC, it is increasing with the number of parameters, and it does so faster than the loss reduces. So, picking a model by AIC is not terribly useful, as the results suggest we would to go with the \(1\)-interval model. The model with the lowest loss is the one with 4 intervals.

\hypertarget{create-pattern-from-best-fit}{%
\paragraph{Create pattern from best fit}\label{create-pattern-from-best-fit}}

This is the same process as for pattern II, as the parameters need inversion. We will reconstruct the warped signals according to the inversed parameters to produce the third pattern. According to the overview above, the best model (lowest loss, \textbf{not} AIC) is the one with \textbf{4} intervals. Its parameters are the following:

\begin{Shaded}
\begin{Highlighting}[]
\CommentTok{\# Let\textquotesingle{}s first define a function that inverses the params and reconstructs the pattern.}
\NormalTok{p3\_pattern\_from\_fit }\OtherTok{\textless{}{-}} \ControlFlowTok{function}\NormalTok{(whichNumIntervals) \{}
\NormalTok{  res }\OtherTok{\textless{}{-}} \FunctionTok{list}\NormalTok{()}
\NormalTok{  p3\_i }\OtherTok{\textless{}{-}} \FunctionTok{readRDS}\NormalTok{(}\AttributeTok{file =} \FunctionTok{paste0}\NormalTok{(}\FunctionTok{getwd}\NormalTok{(), }\StringTok{"/../results/p3{-}compute/i\_"}\NormalTok{, whichNumIntervals, }\StringTok{".rds"}\NormalTok{))}
  
  \CommentTok{\# FitResult:}
\NormalTok{  fr }\OtherTok{\textless{}{-}}\NormalTok{ FitResult}\SpecialCharTok{$}\FunctionTok{new}\NormalTok{(}\StringTok{"foo"}\NormalTok{)}
\NormalTok{  fr}\SpecialCharTok{$}\FunctionTok{fromOptimParallel}\NormalTok{(}\AttributeTok{optR =}\NormalTok{ p3\_i}\SpecialCharTok{$}\NormalTok{optR)}
\NormalTok{  res}\SpecialCharTok{$}\NormalTok{fr }\OtherTok{\textless{}{-}}\NormalTok{ fr}
  
  \CommentTok{\# Inversion:}
\NormalTok{  lambda }\OtherTok{\textless{}{-}}\NormalTok{ p3\_i}\SpecialCharTok{$}\NormalTok{smv}\SpecialCharTok{$}\NormalTok{.\_\_enclos\_env\_\_}\SpecialCharTok{$}\NormalTok{private}\SpecialCharTok{$}\NormalTok{instances}\SpecialCharTok{$}\NormalTok{A}\SpecialCharTok{$}\NormalTok{.\_\_enclos\_env\_\_}\SpecialCharTok{$}\NormalTok{private}\SpecialCharTok{$}\NormalTok{instances}\SpecialCharTok{$}\StringTok{\textasciigrave{}}\AttributeTok{p3\_A|project\_1\_A}\StringTok{\textasciigrave{}}\SpecialCharTok{$}\FunctionTok{getLambda}\NormalTok{()}
\NormalTok{  p3\_i\_varthetaL }\OtherTok{\textless{}{-}}\NormalTok{ p3\_i}\SpecialCharTok{$}\NormalTok{optR}\SpecialCharTok{$}\NormalTok{par[}\FunctionTok{grepl}\NormalTok{(}\AttributeTok{pattern =} \StringTok{"\^{}vtl\_"}\NormalTok{, }\AttributeTok{x =} \FunctionTok{names}\NormalTok{(p3\_i}\SpecialCharTok{$}\NormalTok{optR}\SpecialCharTok{$}\NormalTok{par))]}
  \ControlFlowTok{for}\NormalTok{ (q }\ControlFlowTok{in} \FunctionTok{seq\_len}\NormalTok{(}\AttributeTok{length.out =}\NormalTok{ whichNumIntervals)) \{}
    \ControlFlowTok{if}\NormalTok{ (p3\_i\_varthetaL[q] }\SpecialCharTok{\textless{}}\NormalTok{ lambda[q]) \{}
\NormalTok{      p3\_i\_varthetaL[q] }\OtherTok{\textless{}{-}}\NormalTok{ lambda[q]}
\NormalTok{    \}}
\NormalTok{  \}}
\NormalTok{  p3\_i\_varthetaL }\OtherTok{\textless{}{-}}\NormalTok{ p3\_i\_varthetaL }\SpecialCharTok{/} \FunctionTok{sum}\NormalTok{(p3\_i\_varthetaL)}
  
\NormalTok{  p3\_i\_thetaB }\OtherTok{\textless{}{-}} \FunctionTok{c}\NormalTok{(}\DecValTok{0}\NormalTok{)}
  \ControlFlowTok{for}\NormalTok{ (idx }\ControlFlowTok{in} \FunctionTok{seq\_len}\NormalTok{(}\AttributeTok{length.out =} \FunctionTok{length}\NormalTok{(p3\_i\_varthetaL))) \{}
\NormalTok{    p3\_i\_thetaB }\OtherTok{\textless{}{-}} \FunctionTok{c}\NormalTok{(p3\_i\_thetaB, }\FunctionTok{sum}\NormalTok{(p3\_i\_varthetaL[}\DecValTok{1}\SpecialCharTok{:}\NormalTok{idx]))}
\NormalTok{  \}}
\NormalTok{  p3\_i\_thetaB[}\FunctionTok{length}\NormalTok{(p3\_i\_thetaB)] }\OtherTok{\textless{}{-}} \DecValTok{1} \CommentTok{\# numeric stability}
  
\NormalTok{  p3\_i\_varthetaL}
\NormalTok{  p3\_i\_thetaB}
\NormalTok{  res}\SpecialCharTok{$}\NormalTok{varthetaL }\OtherTok{\textless{}{-}}\NormalTok{ p3\_i\_varthetaL}
\NormalTok{  res}\SpecialCharTok{$}\NormalTok{thetaB }\OtherTok{\textless{}{-}}\NormalTok{ p3\_i\_thetaB}
  
  \CommentTok{\# Signals:}
\NormalTok{  p3\_i\_numInt }\OtherTok{\textless{}{-}} \FunctionTok{length}\NormalTok{(p3\_i\_varthetaL)}
\NormalTok{  p3\_i\_signals }\OtherTok{\textless{}{-}} \FunctionTok{list}\NormalTok{()}
  
  \ControlFlowTok{for}\NormalTok{ (vartype }\ControlFlowTok{in} \FunctionTok{names}\NormalTok{(weight\_vartype)) \{}
\NormalTok{    emptySig }\OtherTok{\textless{}{-}}\NormalTok{ Signal}\SpecialCharTok{$}\FunctionTok{new}\NormalTok{(}
      \AttributeTok{isWp =} \ConstantTok{TRUE}\NormalTok{, }\CommentTok{\# does not matter here}
        \AttributeTok{func =} \ControlFlowTok{function}\NormalTok{(x) .}\DecValTok{5}\NormalTok{, }\AttributeTok{support =} \FunctionTok{c}\NormalTok{(}\DecValTok{0}\NormalTok{, }\DecValTok{1}\NormalTok{), }\AttributeTok{name =} \FunctionTok{paste0}\NormalTok{(}\StringTok{"p3\_"}\NormalTok{, vartype))}
    
\NormalTok{    temp }\OtherTok{\textless{}{-}}\NormalTok{ SRBTWBAW}\SpecialCharTok{$}\FunctionTok{new}\NormalTok{(}
      \AttributeTok{theta\_b =} \FunctionTok{unname}\NormalTok{(p3\_i\_thetaB), }\AttributeTok{gamma\_bed =} \FunctionTok{c}\NormalTok{(}\DecValTok{0}\NormalTok{, }\DecValTok{1}\NormalTok{, }\DecValTok{0}\NormalTok{),}
      \AttributeTok{wp =}\NormalTok{ emptySig}\SpecialCharTok{$}\FunctionTok{get0Function}\NormalTok{(), }\AttributeTok{wc =}\NormalTok{ emptySig}\SpecialCharTok{$}\FunctionTok{get0Function}\NormalTok{(),}
      \AttributeTok{lambda =} \FunctionTok{rep}\NormalTok{(}\DecValTok{0}\NormalTok{, p3\_i\_numInt), }\AttributeTok{begin =} \DecValTok{0}\NormalTok{, }\AttributeTok{end =} \DecValTok{1}\NormalTok{,}
      \AttributeTok{lambda\_ymin =} \FunctionTok{rep}\NormalTok{(}\DecValTok{0}\NormalTok{, p3\_i\_numInt), }\AttributeTok{lambda\_ymax =} \FunctionTok{rep}\NormalTok{(}\DecValTok{1}\NormalTok{, p3\_i\_numInt))}
    
    \CommentTok{\# Recall that originally we used equidistantly{-}spaced boundaries:}
\NormalTok{    temp}\SpecialCharTok{$}\FunctionTok{setParams}\NormalTok{(}\AttributeTok{vartheta\_l =} \FunctionTok{rep}\NormalTok{(}\DecValTok{1} \SpecialCharTok{/}\NormalTok{ p3\_i\_numInt, p3\_i\_numInt),}
                   \AttributeTok{v =} \SpecialCharTok{{-}}\DecValTok{1} \SpecialCharTok{*}\NormalTok{ p3\_i}\SpecialCharTok{$}\NormalTok{optR}\SpecialCharTok{$}\NormalTok{par[}\FunctionTok{paste0}\NormalTok{(}\StringTok{"v\_"}\NormalTok{, vartype)],}
                   \AttributeTok{vartheta\_y =} \SpecialCharTok{{-}}\DecValTok{1} \SpecialCharTok{*}\NormalTok{ p3\_i}\SpecialCharTok{$}\NormalTok{optR}\SpecialCharTok{$}\NormalTok{par[}\FunctionTok{paste0}\NormalTok{(}\StringTok{"vty\_"}\NormalTok{, }\DecValTok{1}\SpecialCharTok{:}\NormalTok{p3\_i\_numInt, }\StringTok{"\_"}\NormalTok{, vartype)])}
    
\NormalTok{    p3\_i\_signals[[vartype]] }\OtherTok{\textless{}{-}}\NormalTok{ Signal}\SpecialCharTok{$}\FunctionTok{new}\NormalTok{(}
      \AttributeTok{name =} \FunctionTok{paste0}\NormalTok{(}\StringTok{"p3\_"}\NormalTok{, vartype), }\AttributeTok{support =} \FunctionTok{c}\NormalTok{(}\DecValTok{0}\NormalTok{, }\DecValTok{1}\NormalTok{), }\AttributeTok{isWp =} \ConstantTok{TRUE}\NormalTok{, }\AttributeTok{func =} \FunctionTok{Vectorize}\NormalTok{(temp}\SpecialCharTok{$}\NormalTok{M))}
\NormalTok{  \}}
\NormalTok{  res}\SpecialCharTok{$}\NormalTok{signals }\OtherTok{\textless{}{-}}\NormalTok{ p3\_i\_signals}
  
  \CommentTok{\# Data:}
\NormalTok{  temp }\OtherTok{\textless{}{-}} \ConstantTok{NULL}
  \ControlFlowTok{for}\NormalTok{ (vartype }\ControlFlowTok{in} \FunctionTok{names}\NormalTok{(weight\_vartype)) \{}
\NormalTok{    f }\OtherTok{\textless{}{-}}\NormalTok{ p3\_i\_signals[[vartype]]}\SpecialCharTok{$}\FunctionTok{get0Function}\NormalTok{()}
\NormalTok{    x }\OtherTok{\textless{}{-}} \FunctionTok{seq}\NormalTok{(}\AttributeTok{from =} \DecValTok{0}\NormalTok{, }\AttributeTok{to =} \DecValTok{1}\NormalTok{, }\AttributeTok{length.out =} \FloatTok{1e3}\NormalTok{)}
\NormalTok{    y }\OtherTok{\textless{}{-}} \FunctionTok{f}\NormalTok{(x)}
    
\NormalTok{    temp }\OtherTok{\textless{}{-}} \FunctionTok{rbind}\NormalTok{(temp, }\FunctionTok{data.frame}\NormalTok{(}
      \AttributeTok{x =}\NormalTok{ x,}
      \AttributeTok{y =}\NormalTok{ y,}
      \AttributeTok{t =}\NormalTok{ vartype,}
      \AttributeTok{numInt =}\NormalTok{ whichNumIntervals}
\NormalTok{    ))}
\NormalTok{  \}}
\NormalTok{  res}\SpecialCharTok{$}\NormalTok{data }\OtherTok{\textless{}{-}}\NormalTok{ temp}
\NormalTok{  res}
\NormalTok{\}}
\end{Highlighting}
\end{Shaded}

\begin{Shaded}
\begin{Highlighting}[]
\NormalTok{p3\_best }\OtherTok{\textless{}{-}} \FunctionTok{readRDS}\NormalTok{(}\AttributeTok{file =} \FunctionTok{paste0}\NormalTok{(}\FunctionTok{getwd}\NormalTok{(), }\StringTok{"/../results/p3{-}compute/i\_"}\NormalTok{, p3\_params[}\FunctionTok{which.min}\NormalTok{(p3\_params}\SpecialCharTok{$}\NormalTok{loss),}
\NormalTok{  ]}\SpecialCharTok{$}\NormalTok{numInt, }\StringTok{".rds"}\NormalTok{))}
\NormalTok{p3\_best}\SpecialCharTok{$}\NormalTok{optR}\SpecialCharTok{$}\NormalTok{par}
\end{Highlighting}
\end{Shaded}

\begin{verbatim}
##           v_A          v_CP        v_FREQ         v_SCD         vtl_1 
##  0.4753518248  0.4449292077  0.3969243406 -0.3425136575  0.0486178486 
##         vtl_2         vtl_3         vtl_4       vty_1_A      vty_1_CP 
##  0.0160263762  0.3063930774  0.9919315307 -0.0085322067  0.0130265463 
##    vty_1_FREQ     vty_1_SCD       vty_2_A      vty_2_CP    vty_2_FREQ 
##  0.0423862173  0.0196908380 -0.0055657708 -0.0293088290 -0.0404751674 
##     vty_2_SCD       vty_3_A      vty_3_CP    vty_3_FREQ     vty_3_SCD 
##  0.0401942364 -0.0898785537 -0.0488678587 -0.1533263719  0.0002213948 
##       vty_4_A      vty_4_CP    vty_4_FREQ     vty_4_SCD 
##  0.0673046097 -0.1805227900 -0.1085880858 -0.0666110009
\end{verbatim}

First we have to inverse the parameters before we can reconstruct the signals:

\begin{Shaded}
\begin{Highlighting}[]
\NormalTok{p3\_best\_varthetaL }\OtherTok{\textless{}{-}}\NormalTok{ p3\_best}\SpecialCharTok{$}\NormalTok{optR}\SpecialCharTok{$}\NormalTok{par[}\FunctionTok{grepl}\NormalTok{(}\AttributeTok{pattern =} \StringTok{"\^{}vtl\_"}\NormalTok{, }\AttributeTok{x =} \FunctionTok{names}\NormalTok{(p3\_best}\SpecialCharTok{$}\NormalTok{optR}\SpecialCharTok{$}\NormalTok{par))]}
\NormalTok{p3\_best\_varthetaL }\OtherTok{\textless{}{-}}\NormalTok{ p3\_best\_varthetaL}\SpecialCharTok{/}\FunctionTok{sum}\NormalTok{(p3\_best\_varthetaL)}

\NormalTok{p3\_best\_thetaB }\OtherTok{\textless{}{-}} \FunctionTok{c}\NormalTok{(}\DecValTok{0}\NormalTok{)}
\ControlFlowTok{for}\NormalTok{ (idx }\ControlFlowTok{in} \FunctionTok{seq\_len}\NormalTok{(}\AttributeTok{length.out =} \FunctionTok{length}\NormalTok{(p3\_best\_varthetaL))) \{}
\NormalTok{  p3\_best\_thetaB }\OtherTok{\textless{}{-}} \FunctionTok{c}\NormalTok{(p3\_best\_thetaB, }\FunctionTok{sum}\NormalTok{(p3\_best\_varthetaL[}\DecValTok{1}\SpecialCharTok{:}\NormalTok{idx]))}
\NormalTok{\}}
\NormalTok{p3\_best\_thetaB[}\FunctionTok{length}\NormalTok{(p3\_best\_thetaB)] }\OtherTok{\textless{}{-}} \DecValTok{1}  \CommentTok{\# numeric stability}

\NormalTok{p3\_best\_varthetaL}
\end{Highlighting}
\end{Shaded}

\begin{verbatim}
##      vtl_1      vtl_2      vtl_3      vtl_4 
## 0.03567055 0.01175843 0.22479830 0.72777272
\end{verbatim}

\begin{Shaded}
\begin{Highlighting}[]
\NormalTok{p3\_best\_thetaB}
\end{Highlighting}
\end{Shaded}

\begin{verbatim}
## [1] 0.00000000 0.03567055 0.04742898 0.27222728 1.00000000
\end{verbatim}

\begin{Shaded}
\begin{Highlighting}[]
\NormalTok{p3\_best\_numInt }\OtherTok{\textless{}{-}} \FunctionTok{length}\NormalTok{(p3\_best\_varthetaL)}
\NormalTok{p3\_signals }\OtherTok{\textless{}{-}} \FunctionTok{list}\NormalTok{()}

\ControlFlowTok{for}\NormalTok{ (vartype }\ControlFlowTok{in} \FunctionTok{names}\NormalTok{(weight\_vartype)) \{}
\NormalTok{  emptySig }\OtherTok{\textless{}{-}}\NormalTok{ Signal}\SpecialCharTok{$}\FunctionTok{new}\NormalTok{(}
    \AttributeTok{isWp =} \ConstantTok{TRUE}\NormalTok{, }\CommentTok{\# does not matter here}
      \AttributeTok{func =} \ControlFlowTok{function}\NormalTok{(x) .}\DecValTok{5}\NormalTok{, }\AttributeTok{support =} \FunctionTok{c}\NormalTok{(}\DecValTok{0}\NormalTok{, }\DecValTok{1}\NormalTok{), }\AttributeTok{name =} \FunctionTok{paste0}\NormalTok{(}\StringTok{"p3\_"}\NormalTok{, vartype))}
  
\NormalTok{  temp }\OtherTok{\textless{}{-}}\NormalTok{ SRBTWBAW}\SpecialCharTok{$}\FunctionTok{new}\NormalTok{(}
    \AttributeTok{theta\_b =} \FunctionTok{unname}\NormalTok{(p3\_best\_thetaB), }\AttributeTok{gamma\_bed =} \FunctionTok{c}\NormalTok{(}\DecValTok{0}\NormalTok{, }\DecValTok{1}\NormalTok{, }\DecValTok{0}\NormalTok{),}
    \AttributeTok{wp =}\NormalTok{ emptySig}\SpecialCharTok{$}\FunctionTok{get0Function}\NormalTok{(), }\AttributeTok{wc =}\NormalTok{ emptySig}\SpecialCharTok{$}\FunctionTok{get0Function}\NormalTok{(),}
    \AttributeTok{lambda =} \FunctionTok{rep}\NormalTok{(}\DecValTok{0}\NormalTok{, p3\_best\_numInt), }\AttributeTok{begin =} \DecValTok{0}\NormalTok{, }\AttributeTok{end =} \DecValTok{1}\NormalTok{,}
    \AttributeTok{lambda\_ymin =} \FunctionTok{rep}\NormalTok{(}\DecValTok{0}\NormalTok{, p3\_best\_numInt), }\AttributeTok{lambda\_ymax =} \FunctionTok{rep}\NormalTok{(}\DecValTok{1}\NormalTok{, p3\_best\_numInt))}
  
  \CommentTok{\# Recall that originally we used equidistantly{-}spaced boundaries:}
\NormalTok{  temp}\SpecialCharTok{$}\FunctionTok{setParams}\NormalTok{(}\AttributeTok{vartheta\_l =} \FunctionTok{rep}\NormalTok{(}\DecValTok{1} \SpecialCharTok{/}\NormalTok{ p3\_best\_numInt, p3\_best\_numInt),}
                 \AttributeTok{v =} \SpecialCharTok{{-}}\DecValTok{1} \SpecialCharTok{*}\NormalTok{ p3\_best}\SpecialCharTok{$}\NormalTok{optR}\SpecialCharTok{$}\NormalTok{par[}\FunctionTok{paste0}\NormalTok{(}\StringTok{"v\_"}\NormalTok{, vartype)],}
                 \AttributeTok{vartheta\_y =} \SpecialCharTok{{-}}\DecValTok{1} \SpecialCharTok{*}\NormalTok{ p3\_best}\SpecialCharTok{$}\NormalTok{optR}\SpecialCharTok{$}\NormalTok{par[}\FunctionTok{paste0}\NormalTok{(}\StringTok{"vty\_"}\NormalTok{, }\DecValTok{1}\SpecialCharTok{:}\NormalTok{p3\_best\_numInt, }\StringTok{"\_"}\NormalTok{, vartype)])}
  
\NormalTok{  p3\_signals[[vartype]] }\OtherTok{\textless{}{-}}\NormalTok{ Signal}\SpecialCharTok{$}\FunctionTok{new}\NormalTok{(}
    \AttributeTok{name =} \FunctionTok{paste0}\NormalTok{(}\StringTok{"p3\_"}\NormalTok{, vartype), }\AttributeTok{support =} \FunctionTok{c}\NormalTok{(}\DecValTok{0}\NormalTok{, }\DecValTok{1}\NormalTok{), }\AttributeTok{isWp =} \ConstantTok{TRUE}\NormalTok{, }\AttributeTok{func =} \FunctionTok{Vectorize}\NormalTok{(temp}\SpecialCharTok{$}\NormalTok{M))}
\NormalTok{\}}
\end{Highlighting}
\end{Shaded}

\begin{figure}[ht!]
\includegraphics{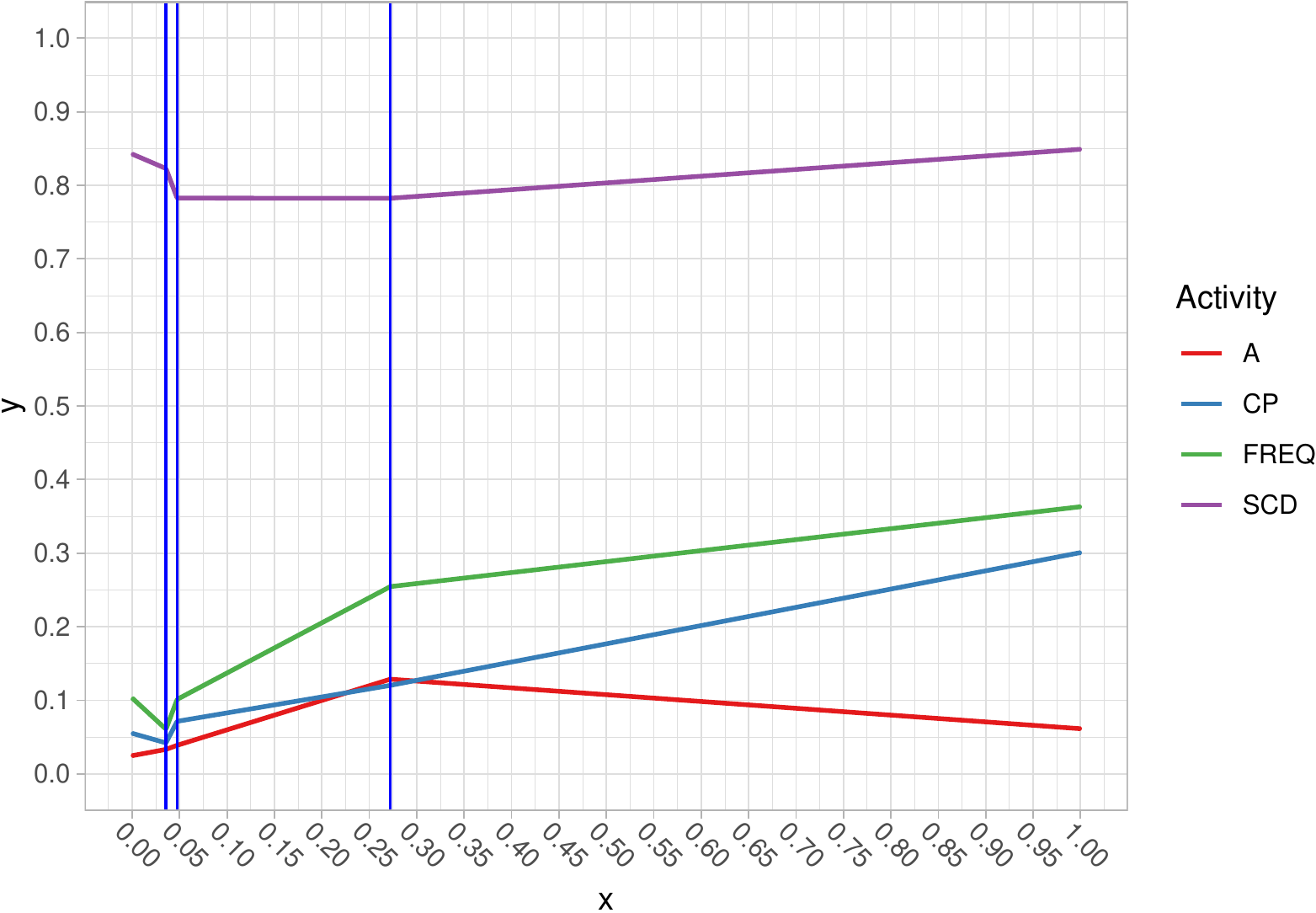} \caption{Pattern type III (b) pattern as aligned by the ground truth only.}\label{fig:p3-signals}
\end{figure}

The 2nd pattern, as derived from the ground truth, is shown in figure \ref{fig:p3-signals}.

\begin{figure}[ht!]
\includegraphics{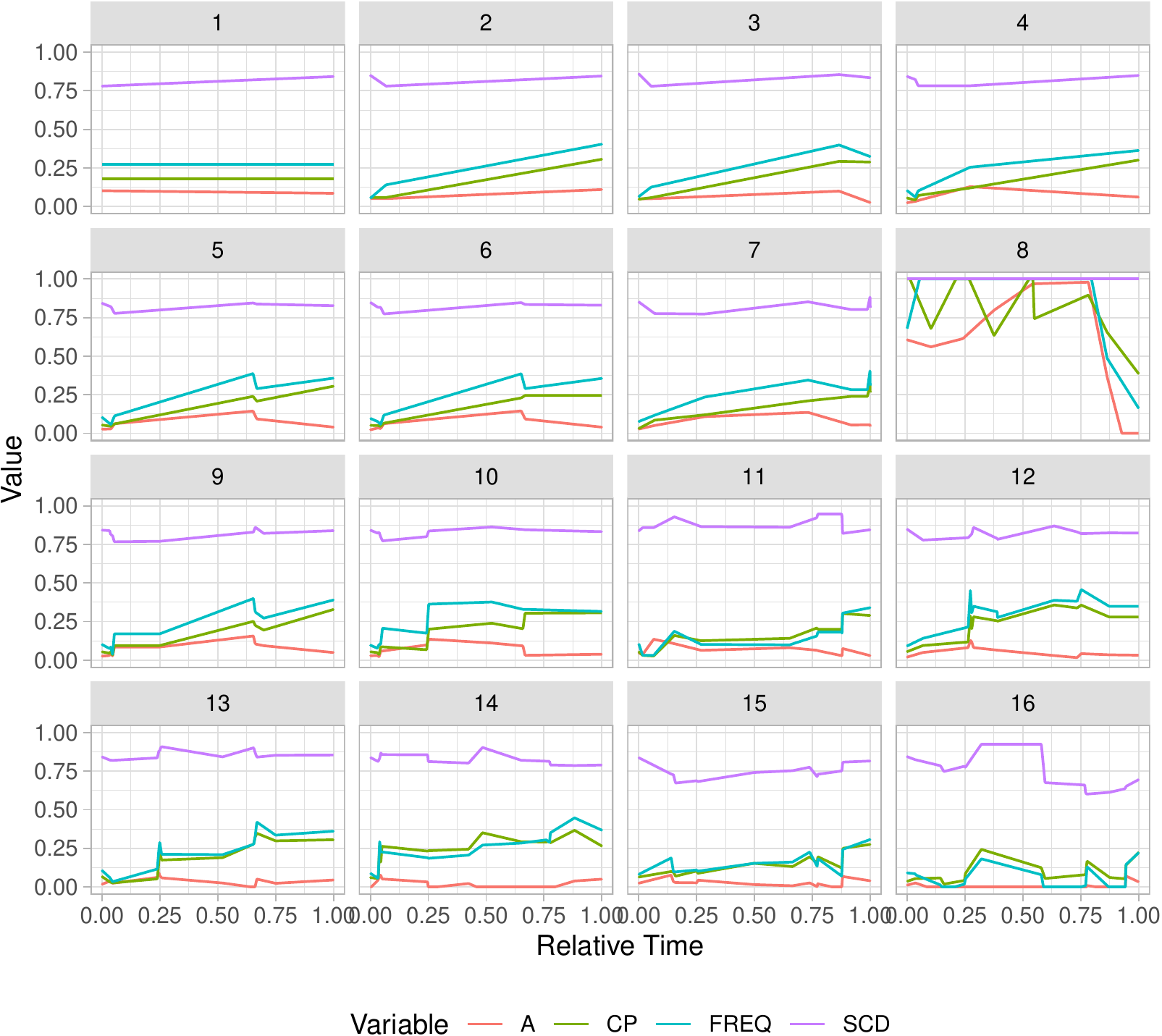} \caption{Computed pattern by number of intervals.}\label{fig:p3-all}
\end{figure}

Let's show all computed patterns in a grid:

In figure \ref{fig:p3-all} we can clearly observe how the pattern evolves with growing number or parameters. Almost all patterns with sufficiently many degrees of freedom have some crack at about one quarter of the projects' time, a second crack is observed at about three quarter's time. In all patterns, it appears that adaptive activities are the least common. All patterns started with randomized coefficients, and something must have gone wrong for pattern \(8\). From five and more intervals we can observe growing similarities with the weighted-average pattern, although it never comes really close. Even though we used a timewarp-regularizer with high weight, we frequently get extreme intervals.

\begin{figure}[ht!]
\includegraphics{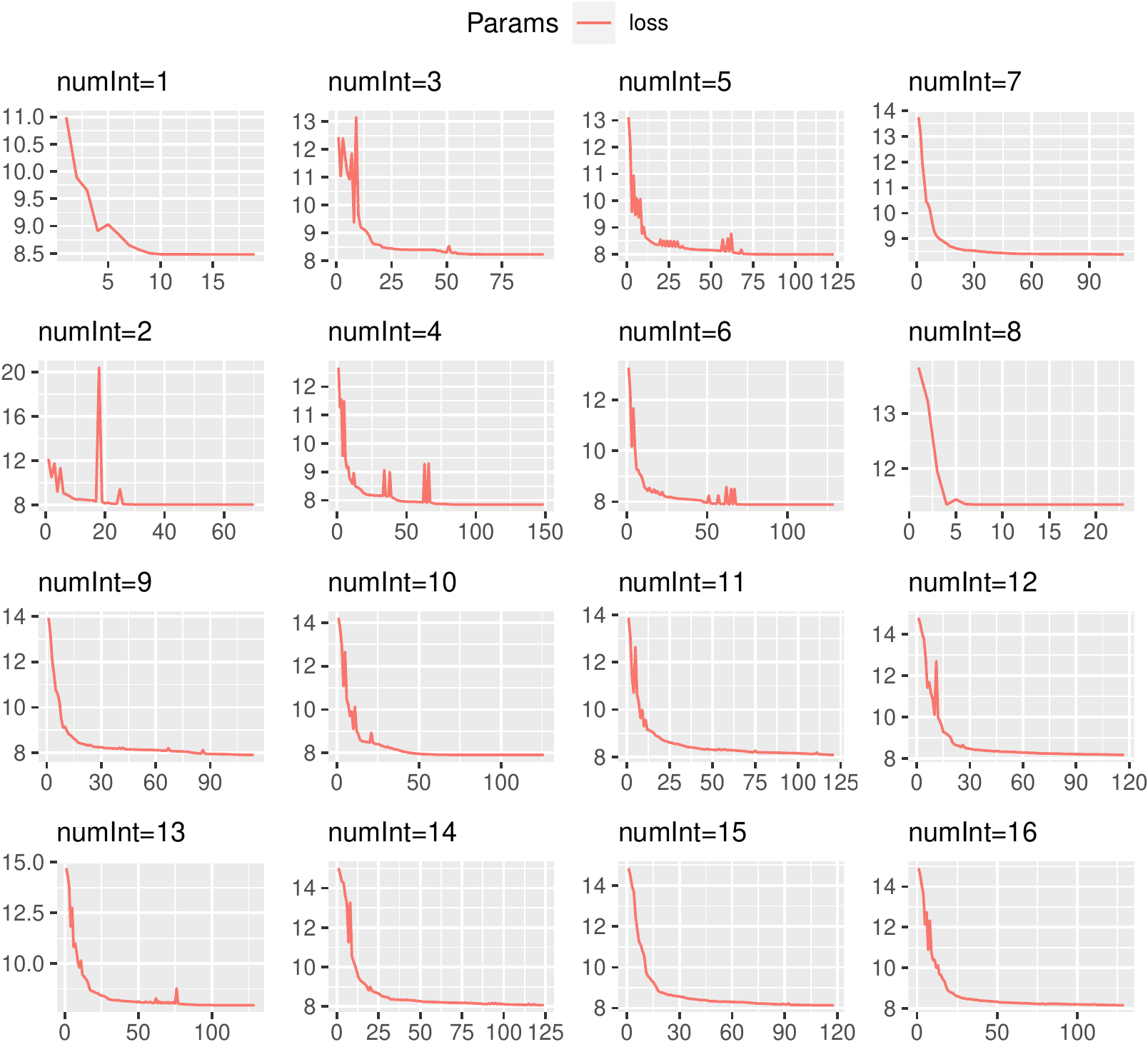} \caption{Losses for all computed pattern by number of intervals.}\label{fig:p3b-all-fr}
\end{figure}

In figure \ref{fig:p3b-all-fr} we can clearly see that all but the eighth pattern converged nicely (this was already visible in \ref{fig:p3-all}). The loss is logarithmic, so the progress is rather substantial. For example, going from \(\log{(14)}\approx1.2e6\) to \(\log{(8)}\approx3e3\) is a reduction by \(3\) (!) orders of magnitude.

\hypertarget{scoring-of-projects}{%
\subsection{Scoring of projects}\label{scoring-of-projects}}

The true main-purpose of our work is to take a pattern and check it against any project, with the goal of obtaining a score, or goodness-of-match so that we can determine if the AP in the pattern is present in the project. In the previous sections we have introduced a number of patterns that we are going to apply here.

How it works: Given some pattern that consists of one or arbitrary many signals, the pattern is added to a single instance of \texttt{srBTAW} as \textbf{Warping Pattern}. The project's signals are added as \textbf{Warping Candidates} to the same instance.

To compute a score, we need to define how to measure the distance between the WP and the WC (between each pair of signals and each interval). In the notebooks for sr-BTW we have previously defined some suitable losses with either \textbf{global} or \textbf{local} finite upper bounds. Currently, the Jensen--Shannon divergence (JSD), as well as the ratio-metrics (correlation, arc-lengths) have global upper bounds. For the JSD, it is \(\ln{(2)}\). Losses with local finite upper bound are, for example, the area between curves, the residual sum of squares, the Euclidean distance etc., basically any metric that has a limit within the rectangle demarcated by one or more intervals. For some of the patterns, we have used a combination of such losses with local bounds. In general, it is not necessary to fit a pattern with the same kinds of losses that are later on used for scoring, but it is recommended to avoid confusing may results.

\hypertarget{the-cost-of-alignment}{%
\subsubsection{The cost of alignment}\label{the-cost-of-alignment}}

When aligning a project to a pattern using boundary time warping, a deviation between the sections' lengths is introduced. Ideally, if the project would align with the pattern perfectly, there would be perfect agreement. The less good a project aligns with a pattern, the more time warping is required. However, the entire alignment needs to be assessed in conjunction with the scores -- the amount of required time warping alone is not sufficient to assess to overall goodness of fit.

During the optimization, we already used a regularizer for extreme intervals (\texttt{TimeWarpRegularization} with regularizer \texttt{exint2}).

\hypertarget{scoring-mechanisms}{%
\subsubsection{\texorpdfstring{Scoring mechanisms\label{sssec:score-mech}}{Scoring mechanisms}}\label{scoring-mechanisms}}

For scoring a single project, we first warp it to the pattern, then we measure the remaining distance. We only do time-warping of the projects to the pattern. We could compute a score for each interval. However, the ground truth does not yield this, so we will only compute a scores for entire signals, i.e., over all intervals. Once aligned, computing scores is cheap, so we will try a variety of scores and see what works best.

\begin{Shaded}
\begin{Highlighting}[]
\CommentTok{\# Function score\_variable\_alignment(..) has been moved to common{-}funcs.R!}
\end{Highlighting}
\end{Shaded}

We define a parallelized function to compute all scores of a project:

\begin{Shaded}
\begin{Highlighting}[]
\NormalTok{compute\_all\_scores }\OtherTok{\textless{}{-}} \ControlFlowTok{function}\NormalTok{(alignment, patternName) \{}
\NormalTok{  useScores }\OtherTok{\textless{}{-}} \FunctionTok{c}\NormalTok{(}\StringTok{"area"}\NormalTok{, }\StringTok{"corr"}\NormalTok{, }\StringTok{"jsd"}\NormalTok{, }\StringTok{"kl"}\NormalTok{, }\StringTok{"arclen"}\NormalTok{, }\StringTok{"sd"}\NormalTok{, }\StringTok{"var"}\NormalTok{, }\StringTok{"mae"}\NormalTok{, }\StringTok{"rmse"}\NormalTok{,}
    \StringTok{"RMS"}\NormalTok{, }\StringTok{"Kurtosis"}\NormalTok{, }\StringTok{"Peak"}\NormalTok{, }\StringTok{"ImpulseFactor"}\NormalTok{)}

  \StringTok{\textasciigrave{}}\AttributeTok{rownames\textless{}{-}}\StringTok{\textasciigrave{}}\NormalTok{(}\FunctionTok{doWithParallelCluster}\NormalTok{(}\AttributeTok{numCores =} \FunctionTok{length}\NormalTok{(alignment), }\AttributeTok{expr =}\NormalTok{ \{}
\NormalTok{    foreach}\SpecialCharTok{::}\FunctionTok{foreach}\NormalTok{(}\AttributeTok{projectName =} \FunctionTok{names}\NormalTok{(alignment), }\AttributeTok{.inorder =} \ConstantTok{TRUE}\NormalTok{, }\AttributeTok{.combine =}\NormalTok{ rbind,}
      \AttributeTok{.export =} \FunctionTok{c}\NormalTok{(}\StringTok{"score\_variable\_alignment"}\NormalTok{, }\StringTok{"weight\_vartype"}\NormalTok{)) }\SpecialCharTok{\%dopar\%}\NormalTok{ \{}
      \FunctionTok{source}\NormalTok{(}\StringTok{"./common{-}funcs.R"}\NormalTok{)}
      \FunctionTok{source}\NormalTok{(}\StringTok{"../models/modelsR6.R"}\NormalTok{)}
      \FunctionTok{source}\NormalTok{(}\StringTok{"../models/SRBTW{-}R6.R"}\NormalTok{)}

\NormalTok{      scores }\OtherTok{\textless{}{-}} \FunctionTok{c}\NormalTok{()}
      \ControlFlowTok{for}\NormalTok{ (score }\ControlFlowTok{in}\NormalTok{ useScores) \{}
\NormalTok{        temp }\OtherTok{\textless{}{-}} \FunctionTok{score\_variable\_alignment}\NormalTok{(}\AttributeTok{patternName =}\NormalTok{ patternName, }\AttributeTok{projectName =}\NormalTok{ projectName,}
          \AttributeTok{alignment =}\NormalTok{ alignment[[projectName]], }\AttributeTok{use =}\NormalTok{ score)}
\NormalTok{        scores }\OtherTok{\textless{}{-}} \FunctionTok{c}\NormalTok{(scores, }\StringTok{\textasciigrave{}}\AttributeTok{names\textless{}{-}}\StringTok{\textasciigrave{}}\NormalTok{(}\FunctionTok{c}\NormalTok{(}\FunctionTok{mean}\NormalTok{(temp), }\FunctionTok{prod}\NormalTok{(temp)), }\FunctionTok{c}\NormalTok{(}\FunctionTok{paste0}\NormalTok{(score,}
          \FunctionTok{c}\NormalTok{(}\StringTok{"\_m"}\NormalTok{, }\StringTok{"\_p"}\NormalTok{)))))}
\NormalTok{      \}}
      \StringTok{\textasciigrave{}}\AttributeTok{colnames\textless{}{-}}\StringTok{\textasciigrave{}}\NormalTok{(}\FunctionTok{matrix}\NormalTok{(}\AttributeTok{data =}\NormalTok{ scores, }\AttributeTok{nrow =} \DecValTok{1}\NormalTok{), }\FunctionTok{names}\NormalTok{(scores))}
\NormalTok{    \}}
\NormalTok{  \}), }\FunctionTok{sort}\NormalTok{(}\FunctionTok{names}\NormalTok{(alignment)))}
\NormalTok{\}}
\end{Highlighting}
\end{Shaded}

We also need to define a function for warping a project to the pattern:

\begin{Shaded}
\begin{Highlighting}[]
\CommentTok{\# Function time\_warp\_project(..) has been moved to common{-}funcs.R!}
\end{Highlighting}
\end{Shaded}

\hypertarget{pattern-i}{%
\subsubsection{Pattern I}\label{pattern-i}}

First we compute the alignment for all projects, then all scores.

\begin{Shaded}
\begin{Highlighting}[]
\FunctionTok{library}\NormalTok{(foreach)}

\NormalTok{p1\_align }\OtherTok{\textless{}{-}} \FunctionTok{loadResultsOrCompute}\NormalTok{(}\AttributeTok{file =} \StringTok{"../results/p1\_align.rds"}\NormalTok{, }\AttributeTok{computeExpr =}\NormalTok{ \{}
  \CommentTok{\# Let\textquotesingle{}s compute all projects in parallel!}
\NormalTok{  cl }\OtherTok{\textless{}{-}}\NormalTok{ parallel}\SpecialCharTok{::}\FunctionTok{makePSOCKcluster}\NormalTok{(}\FunctionTok{length}\NormalTok{(project\_signals))}
  \FunctionTok{unlist}\NormalTok{(}\FunctionTok{doWithParallelClusterExplicit}\NormalTok{(}\AttributeTok{cl =}\NormalTok{ cl, }\AttributeTok{expr =}\NormalTok{ \{}
\NormalTok{    foreach}\SpecialCharTok{::}\FunctionTok{foreach}\NormalTok{(}\AttributeTok{projectName =} \FunctionTok{names}\NormalTok{(project\_signals), }\AttributeTok{.inorder =} \ConstantTok{FALSE}\NormalTok{,}
      \AttributeTok{.packages =} \FunctionTok{c}\NormalTok{(}\StringTok{"parallel"}\NormalTok{)) }\SpecialCharTok{\%dopar\%}\NormalTok{ \{}
      \FunctionTok{source}\NormalTok{(}\StringTok{"./common{-}funcs.R"}\NormalTok{)}
      \FunctionTok{source}\NormalTok{(}\StringTok{"../models/modelsR6.R"}\NormalTok{)}
      \FunctionTok{source}\NormalTok{(}\StringTok{"../models/SRBTW{-}R6.R"}\NormalTok{)}

      \CommentTok{\# There are 5 objectives that can be computed in parallel!}
\NormalTok{      cl\_nested }\OtherTok{\textless{}{-}}\NormalTok{ parallel}\SpecialCharTok{::}\FunctionTok{makePSOCKcluster}\NormalTok{(}\DecValTok{5}\NormalTok{)}
      \StringTok{\textasciigrave{}}\AttributeTok{names\textless{}{-}}\StringTok{\textasciigrave{}}\NormalTok{(}\FunctionTok{list}\NormalTok{(}\FunctionTok{doWithParallelClusterExplicit}\NormalTok{(}\AttributeTok{cl =}\NormalTok{ cl\_nested, }\AttributeTok{expr =}\NormalTok{ \{}
\NormalTok{        temp }\OtherTok{\textless{}{-}} \FunctionTok{time\_warp\_project}\NormalTok{(}\AttributeTok{pattern =}\NormalTok{ p1\_signals, }\AttributeTok{project =}\NormalTok{ project\_signals[[projectName]])}
\NormalTok{        temp}\SpecialCharTok{$}\FunctionTok{fit}\NormalTok{(}\AttributeTok{verbose =} \ConstantTok{TRUE}\NormalTok{)}
\NormalTok{        temp  }\CommentTok{\# return the instance, it includes the FitResult}
\NormalTok{      \})), projectName)}
\NormalTok{    \}}
\NormalTok{  \}))}
\NormalTok{\})}
\end{Highlighting}
\end{Shaded}

\begin{Shaded}
\begin{Highlighting}[]
\NormalTok{p1\_scores }\OtherTok{\textless{}{-}} \FunctionTok{loadResultsOrCompute}\NormalTok{(}\AttributeTok{file =} \StringTok{"../results/p1\_scores.rds"}\NormalTok{, }\AttributeTok{computeExpr =}\NormalTok{ \{}
  \FunctionTok{as.data.frame}\NormalTok{(}\FunctionTok{compute\_all\_scores}\NormalTok{(}\AttributeTok{alignment =}\NormalTok{ p1\_align, }\AttributeTok{patternName =} \StringTok{"p1"}\NormalTok{))}
\NormalTok{\})}
\end{Highlighting}
\end{Shaded}

Recall that we are obtaining scores for each interval. To aggregate them we build the product and the mean in the following table, there is no weighing applied.

\begin{table}

\caption{\label{tab:p1-scores}Scores for the aligned projects with pattern I (p=product, m=mean).}
\centering
\begin{tabular}[t]{lrrrrrrrrr}
\toprule
  & pr\_1 & pr\_2 & pr\_3 & pr\_4 & pr\_5 & pr\_6 & pr\_7 & pr\_8 & pr\_9\\
\midrule
area\_m & 0.80 & 0.81 & 0.89 & 0.89 & 0.83 & 0.83 & 0.77 & 0.84 & 0.82\\
area\_p & 0.40 & 0.42 & 0.63 & 0.63 & 0.48 & 0.48 & 0.34 & 0.50 & 0.45\\
corr\_m & 0.50 & 0.64 & 0.67 & 0.58 & 0.59 & 0.52 & 0.55 & 0.54 & 0.54\\
corr\_p & 0.05 & 0.15 & 0.19 & 0.11 & 0.10 & 0.03 & 0.07 & 0.08 & 0.06\\
jsd\_m & 0.28 & 0.34 & 0.46 & 0.43 & 0.40 & 0.37 & 0.29 & 0.34 & 0.32\\
\addlinespace
jsd\_p & 0.00 & 0.01 & 0.03 & 0.02 & 0.02 & 0.01 & 0.00 & 0.01 & 0.01\\
kl\_m & 0.44 & 0.31 & 0.16 & 0.19 & 0.20 & 0.26 & 0.49 & 0.36 & 0.35\\
kl\_p & 0.01 & 0.00 & 0.00 & 0.00 & 0.00 & 0.00 & 0.01 & 0.00 & 0.00\\
arclen\_m & 0.47 & 0.52 & 0.43 & 0.51 & 0.57 & 0.85 & 0.54 & 0.53 & 0.50\\
arclen\_p & 0.04 & 0.07 & 0.03 & 0.06 & 0.10 & 0.50 & 0.06 & 0.08 & 0.06\\
\addlinespace
sd\_m & 0.65 & 0.70 & 0.77 & 0.74 & 0.76 & 0.75 & 0.69 & 0.70 & 0.68\\
sd\_p & 0.18 & 0.24 & 0.34 & 0.30 & 0.32 & 0.30 & 0.21 & 0.23 & 0.21\\
var\_m & 0.87 & 0.91 & 0.94 & 0.93 & 0.94 & 0.93 & 0.89 & 0.90 & 0.89\\
var\_p & 0.58 & 0.67 & 0.79 & 0.75 & 0.77 & 0.75 & 0.63 & 0.66 & 0.64\\
mae\_m & 0.80 & 0.81 & 0.89 & 0.89 & 0.83 & 0.83 & 0.77 & 0.84 & 0.82\\
\addlinespace
mae\_p & 0.40 & 0.42 & 0.62 & 0.63 & 0.48 & 0.48 & 0.34 & 0.50 & 0.45\\
rmse\_m & 0.74 & 0.76 & 0.83 & 0.81 & 0.78 & 0.80 & 0.71 & 0.76 & 0.76\\
rmse\_p & 0.30 & 0.33 & 0.48 & 0.43 & 0.37 & 0.40 & 0.24 & 0.33 & 0.32\\
RMS\_m & 0.66 & 0.73 & 0.50 & 0.60 & 0.52 & 0.76 & 0.58 & 0.69 & 0.50\\
RMS\_p & 0.14 & 0.24 & 0.03 & 0.06 & 0.04 & 0.28 & 0.06 & 0.18 & 0.04\\
\addlinespace
Kurtosis\_m & 0.40 & 0.49 & 0.39 & 0.31 & 0.24 & 0.39 & 0.35 & 0.35 & 0.13\\
Kurtosis\_p & 0.00 & 0.00 & 0.00 & 0.00 & 0.00 & 0.00 & 0.00 & 0.00 & 0.00\\
Peak\_m & 0.62 & 0.62 & 0.53 & 0.63 & 0.65 & 0.61 & 0.61 & 0.63 & 0.44\\
Peak\_p & 0.09 & 0.10 & 0.04 & 0.06 & 0.12 & 0.09 & 0.06 & 0.09 & 0.03\\
ImpulseFactor\_m & 0.60 & 0.59 & 0.54 & 0.66 & 0.48 & 0.61 & 0.57 & 0.54 & 0.64\\
\addlinespace
ImpulseFactor\_p & 0.11 & 0.11 & 0.05 & 0.11 & 0.05 & 0.08 & 0.06 & 0.06 & 0.16\\
\bottomrule
\end{tabular}
\end{table}

In table \ref{tab:p1-corr} the correlations of the scores with the ground truth as computed against the first pattern are shown.

\begin{table}

\caption{\label{tab:p1-corr}Correlation of the ground truth with all other scores for pattern I.}
\centering
\begin{tabular}[t]{lrlrlr}
\toprule
Score & Value & Score & Value & NA & NA\\
\midrule
area\_m & 0.6115889 & arclen\_p & -0.1684936 & RMS\_m & -0.5768954\\
area\_p & 0.6436132 & sd\_m & 0.3796681 & RMS\_p & -0.6222672\\
corr\_m & 0.2684177 & sd\_p & 0.3934965 & Kurtosis\_m & -0.3730592\\
corr\_p & 0.2605112 & var\_m & 0.3842355 & Kurtosis\_p & -0.6263601\\
jsd\_m & 0.5388574 & var\_p & 0.3900069 & Peak\_m & -0.4524193\\
\addlinespace
jsd\_p & 0.5938024 & mae\_m & 0.6101791 & Peak\_p & -0.7678158\\
kl\_m & -0.4227976 & mae\_p & 0.6423594 & ImpulseFactor\_m & 0.4717221\\
kl\_p & -0.2931612 & rmse\_m & 0.4837649 & ImpulseFactor\_p & 0.2300525\\
arclen\_m & -0.2546068 & rmse\_p & 0.5246423 & NA & NA\\
\bottomrule
\end{tabular}
\end{table}

Let's show a correlation matrix in figure \ref{fig:p1-corr-mat}:

\begin{figure}[ht!]
\includegraphics{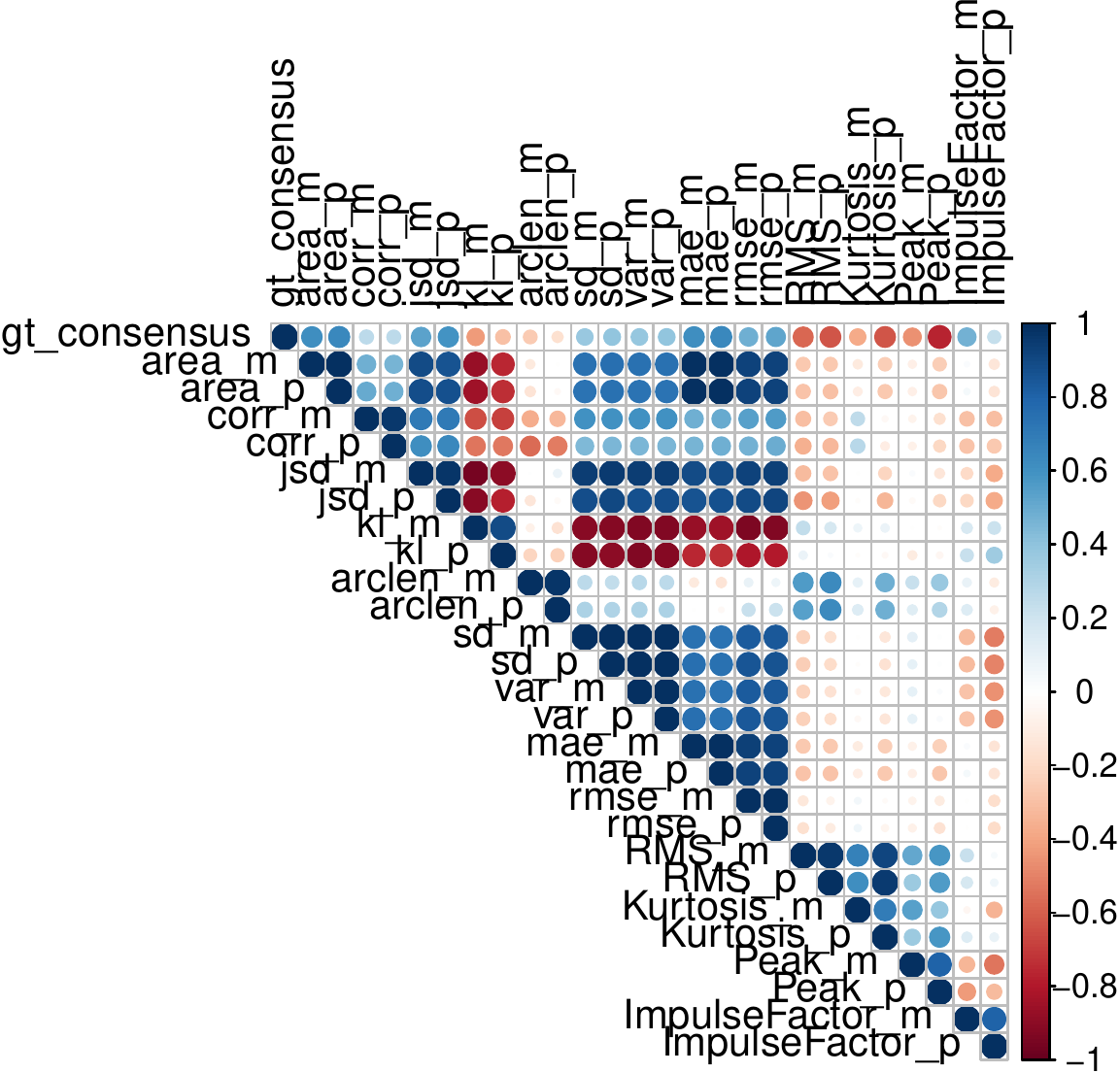} \caption{Correlation matrix for scores using pattern I.}\label{fig:p1-corr-mat}
\end{figure}

We appear to have mostly strongly negative correlations -- note that the measure using Kullback-Leibler is a divergence, not a similarity. Area and RMSE have a strong negative correlation, which suggests that whenever their score is high, the ground truth's consensus score is low. A high score for area or RMSE however means, that the distance between the signals is comparatively low, so we should have a good alignment, so how is this explained then?

Going back to table \ref{tab:groundtruth}, we will notice that projects 2, 3 and 5 have somewhat similar courses for their variables, yet their consensus scores are 0, 6 and 1, respectively. In other words, only project 3 has had a Fire Drill. We can hence conclude that the visual distance and the ground truth consensus are \textbf{not} proportional (at least not for the variables we chose to model). The visual similarity of a project to a pattern is just that; the score quantifies the deviation from the pattern, but it does not necessarily correlate with the ground truth. This however was our underlying assumption all along, hence the initial pattern. We deliberately chose to design patterns \emph{without} investigating any of the projects. Also, while we had access to the projects for some time now, the ground truth became available only very recently, after all modeling was done.

Nevertheless, it does not hurt to check out patterns II and III, as we would like to achieve better matches. Eventually, the goal with this approach is to improve correlations and to get more accurate scores. The final stage then could be to compute, for example, a weighted consensus, based on the projects that we have, or to create a linear model that can regress to a value close to the ground truth by considering all the different scores.

\hypertarget{pattern-ii}{%
\subsubsection{\texorpdfstring{Pattern II\label{ssec:score-pattern2}}{Pattern II}}\label{pattern-ii}}

The second pattern was produced by having it warp to all the ground truths simultaneously, using their weight.

\begin{Shaded}
\begin{Highlighting}[]
\FunctionTok{library}\NormalTok{(foreach)}

\NormalTok{p2\_align }\OtherTok{\textless{}{-}} \FunctionTok{loadResultsOrCompute}\NormalTok{(}\AttributeTok{file =} \StringTok{"../results/p2\_align.rds"}\NormalTok{, }\AttributeTok{computeExpr =}\NormalTok{ \{}
  \CommentTok{\# Let\textquotesingle{}s compute all projects in parallel!}
\NormalTok{  cl }\OtherTok{\textless{}{-}}\NormalTok{ parallel}\SpecialCharTok{::}\FunctionTok{makePSOCKcluster}\NormalTok{(}\FunctionTok{length}\NormalTok{(project\_signals))}
  \FunctionTok{unlist}\NormalTok{(}\FunctionTok{doWithParallelClusterExplicit}\NormalTok{(}\AttributeTok{cl =}\NormalTok{ cl, }\AttributeTok{expr =}\NormalTok{ \{}
\NormalTok{    foreach}\SpecialCharTok{::}\FunctionTok{foreach}\NormalTok{(}\AttributeTok{projectName =} \FunctionTok{names}\NormalTok{(project\_signals), }\AttributeTok{.inorder =} \ConstantTok{FALSE}\NormalTok{,}
      \AttributeTok{.packages =} \FunctionTok{c}\NormalTok{(}\StringTok{"parallel"}\NormalTok{)) }\SpecialCharTok{\%dopar\%}\NormalTok{ \{}
      \FunctionTok{source}\NormalTok{(}\StringTok{"./common{-}funcs.R"}\NormalTok{)}
      \FunctionTok{source}\NormalTok{(}\StringTok{"../models/modelsR6.R"}\NormalTok{)}
      \FunctionTok{source}\NormalTok{(}\StringTok{"../models/SRBTW{-}R6.R"}\NormalTok{)}

      \CommentTok{\# There are 5 objectives that can be computed in parallel!}
\NormalTok{      cl\_nested }\OtherTok{\textless{}{-}}\NormalTok{ parallel}\SpecialCharTok{::}\FunctionTok{makePSOCKcluster}\NormalTok{(}\DecValTok{5}\NormalTok{)}
      \StringTok{\textasciigrave{}}\AttributeTok{names\textless{}{-}}\StringTok{\textasciigrave{}}\NormalTok{(}\FunctionTok{list}\NormalTok{(}\FunctionTok{doWithParallelClusterExplicit}\NormalTok{(}\AttributeTok{cl =}\NormalTok{ cl\_nested, }\AttributeTok{expr =}\NormalTok{ \{}
\NormalTok{        temp }\OtherTok{\textless{}{-}} \FunctionTok{time\_warp\_project}\NormalTok{(}\AttributeTok{pattern =}\NormalTok{ p2\_signals, }\AttributeTok{project =}\NormalTok{ project\_signals[[projectName]])}
\NormalTok{        temp}\SpecialCharTok{$}\FunctionTok{fit}\NormalTok{(}\AttributeTok{verbose =} \ConstantTok{TRUE}\NormalTok{)}
\NormalTok{        temp  }\CommentTok{\# return the instance, it includes the FitResult}
\NormalTok{      \})), projectName)}
\NormalTok{    \}}
\NormalTok{  \}))}
\NormalTok{\})}
\end{Highlighting}
\end{Shaded}

\begin{Shaded}
\begin{Highlighting}[]
\NormalTok{p2\_scores }\OtherTok{\textless{}{-}} \FunctionTok{loadResultsOrCompute}\NormalTok{(}\AttributeTok{file =} \StringTok{"../results/p2\_scores.rds"}\NormalTok{, }\AttributeTok{computeExpr =}\NormalTok{ \{}
  \FunctionTok{as.data.frame}\NormalTok{(}\FunctionTok{compute\_all\_scores}\NormalTok{(}\AttributeTok{alignment =}\NormalTok{ p2\_align, }\AttributeTok{patternName =} \StringTok{"p2"}\NormalTok{))}
\NormalTok{\})}
\end{Highlighting}
\end{Shaded}

\begin{table}

\caption{\label{tab:p2-scores}Scores for the aligned projects with pattern II (p=product, m=mean).}
\centering
\begin{tabular}[t]{lrrrrrrrrr}
\toprule
  & pr\_1 & pr\_2 & pr\_3 & pr\_4 & pr\_5 & pr\_6 & pr\_7 & pr\_8 & pr\_9\\
\midrule
area\_m & 0.92 & 0.92 & 0.90 & 0.90 & 0.89 & 0.87 & 0.86 & 0.90 & 0.91\\
area\_p & 0.72 & 0.71 & 0.65 & 0.64 & 0.61 & 0.56 & 0.55 & 0.65 & 0.68\\
corr\_m & 0.69 & 0.61 & 0.55 & 0.68 & 0.58 & 0.52 & 0.59 & 0.69 & 0.58\\
corr\_p & 0.22 & 0.13 & 0.09 & 0.21 & 0.10 & 0.05 & 0.12 & 0.21 & 0.11\\
jsd\_m & 0.48 & 0.42 & 0.36 & 0.41 & 0.35 & 0.29 & 0.34 & 0.34 & 0.38\\
\addlinespace
jsd\_p & 0.05 & 0.03 & 0.02 & 0.02 & 0.01 & 0.00 & 0.01 & 0.01 & 0.02\\
kl\_m & 0.13 & 0.17 & 0.20 & 0.20 & 0.26 & 0.47 & 0.25 & 0.24 & 0.22\\
kl\_p & 0.00 & 0.00 & 0.00 & 0.00 & 0.00 & 0.01 & 0.00 & 0.00 & 0.00\\
arclen\_m & 0.53 & 0.54 & 0.48 & 0.57 & 0.59 & 0.73 & 0.51 & 0.58 & 0.57\\
arclen\_p & 0.05 & 0.07 & 0.04 & 0.06 & 0.11 & 0.24 & 0.04 & 0.09 & 0.08\\
\addlinespace
sd\_m & 0.86 & 0.83 & 0.81 & 0.84 & 0.81 & 0.76 & 0.76 & 0.80 & 0.82\\
sd\_p & 0.54 & 0.48 & 0.44 & 0.50 & 0.43 & 0.31 & 0.32 & 0.42 & 0.45\\
var\_m & 0.98 & 0.97 & 0.96 & 0.97 & 0.96 & 0.93 & 0.94 & 0.96 & 0.97\\
var\_p & 0.92 & 0.89 & 0.86 & 0.90 & 0.85 & 0.75 & 0.76 & 0.85 & 0.87\\
mae\_m & 0.92 & 0.92 & 0.90 & 0.90 & 0.89 & 0.87 & 0.86 & 0.90 & 0.91\\
\addlinespace
mae\_p & 0.72 & 0.71 & 0.64 & 0.64 & 0.61 & 0.56 & 0.55 & 0.65 & 0.68\\
rmse\_m & 0.89 & 0.88 & 0.86 & 0.86 & 0.85 & 0.81 & 0.81 & 0.85 & 0.87\\
rmse\_p & 0.62 & 0.59 & 0.55 & 0.56 & 0.53 & 0.42 & 0.42 & 0.53 & 0.57\\
RMS\_m & 0.62 & 0.66 & 0.71 & 0.65 & 0.83 & 0.62 & 0.58 & 0.64 & 0.83\\
RMS\_p & 0.11 & 0.14 & 0.21 & 0.15 & 0.45 & 0.14 & 0.11 & 0.13 & 0.46\\
\addlinespace
Kurtosis\_m & 0.42 & 0.29 & 0.22 & 0.30 & 0.33 & 0.11 & 0.30 & 0.39 & 0.37\\
Kurtosis\_p & 0.01 & 0.00 & 0.00 & 0.00 & 0.01 & 0.00 & 0.00 & 0.00 & 0.00\\
Peak\_m & 0.66 & 0.52 & 0.54 & 0.53 & 0.70 & 0.53 & 0.64 & 0.66 & 0.60\\
Peak\_p & 0.15 & 0.06 & 0.07 & 0.06 & 0.23 & 0.07 & 0.14 & 0.14 & 0.11\\
ImpulseFactor\_m & 0.66 & 0.63 & 0.57 & 0.51 & 0.62 & 0.61 & 0.77 & 0.55 & 0.55\\
\addlinespace
ImpulseFactor\_p & 0.17 & 0.13 & 0.07 & 0.02 & 0.09 & 0.08 & 0.34 & 0.06 & 0.06\\
\bottomrule
\end{tabular}
\end{table}

The correlation of just the ground truth with all scores is in table \ref{tab:p2-corr}.

\begin{table}

\caption{\label{tab:p2-corr}Correlation of the ground truth with all other scores for pattern II.}
\centering
\begin{tabular}[t]{lrlrlr}
\toprule
Score & Value & Score & Value & Score & Value\\
\midrule
area\_m & -0.0890189 & arclen\_p & -0.2557238 & RMS\_m & 0.1676109\\
area\_p & -0.1069064 & sd\_m & 0.1285950 & RMS\_p & 0.1289162\\
corr\_m & -0.0764780 & sd\_p & 0.1292517 & Kurtosis\_m & -0.2078491\\
corr\_p & -0.0362952 & var\_m & 0.1332358 & Kurtosis\_p & -0.3601837\\
jsd\_m & 0.0077401 & var\_p & 0.1339373 & Peak\_m & -0.4622136\\
\addlinespace
jsd\_p & -0.1128088 & mae\_m & -0.0908586 & Peak\_p & -0.4417711\\
kl\_m & -0.0882289 & mae\_p & -0.1087002 & ImpulseFactor\_m & -0.4271623\\
kl\_p & -0.1733968 & rmse\_m & 0.0336848 & ImpulseFactor\_p & -0.2815468\\
arclen\_m & -0.1780968 & rmse\_p & 0.0189570 & NA & NA\\
\bottomrule
\end{tabular}
\end{table}

\begin{figure}[ht!]
\includegraphics{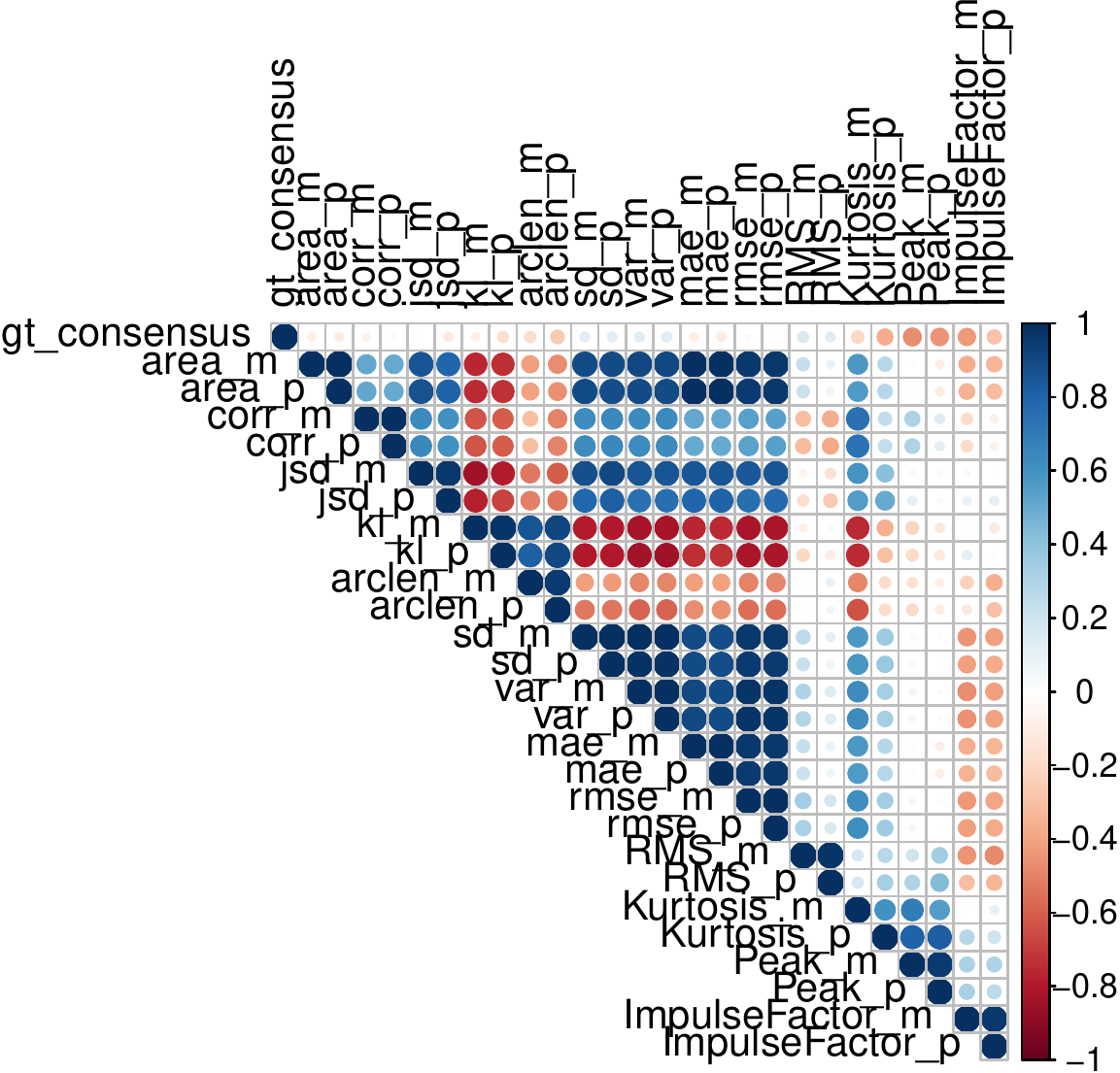} \caption{Correlation matrix for scores using pattern II.}\label{fig:p2-corr-mat}
\end{figure}

The correlation matrix looks as in figure \ref{fig:p2-corr-mat}.

With the second pattern we get much stronger correlations on average, meaning that the alignment of each project to the second pattern is better. While some correlations with the ground truth remain similar, we get strong positive correlations with the signal-measures \emph{Impulse-factor} (-0.427) and \emph{Peak} (-0.462) (Xi, Sun, and Krishnappa 2000). This however is explained by the double warping: First the pattern II was produced by time- and amplitude-warping it to the ground truth. Then, each project was time-warped to the that pattern. Simply put, this should result in some good alignment of all of the signals' peaks, explaining the high correlations.

\hypertarget{pattern-ii-without-alignment}{%
\subsubsection{Pattern II (without alignment)}\label{pattern-ii-without-alignment}}

Since pattern II was computed such that it warps to all projects, it already should correct for time- and amplitude warping. This gives us incentive to compute scores against the non-aligned projects:

\begin{Shaded}
\begin{Highlighting}[]
\NormalTok{p2\_no\_align }\OtherTok{\textless{}{-}} \FunctionTok{list}\NormalTok{()}

\ControlFlowTok{for}\NormalTok{ (project }\ControlFlowTok{in}\NormalTok{ ground\_truth}\SpecialCharTok{$}\NormalTok{project) \{}
\NormalTok{  temp }\OtherTok{\textless{}{-}}\NormalTok{ p2\_align[[project]]}\SpecialCharTok{$}\FunctionTok{clone}\NormalTok{()}
\NormalTok{  temp}\SpecialCharTok{$}\FunctionTok{setParams}\NormalTok{(}\AttributeTok{params =} \StringTok{\textasciigrave{}}\AttributeTok{names\textless{}{-}}\StringTok{\textasciigrave{}}\NormalTok{(p2\_ex\_varthetaL, temp}\SpecialCharTok{$}\FunctionTok{getParamNames}\NormalTok{()))}
\NormalTok{  p2\_no\_align[[project]] }\OtherTok{\textless{}{-}}\NormalTok{ temp}
\NormalTok{\}}
\end{Highlighting}
\end{Shaded}

\begin{Shaded}
\begin{Highlighting}[]
\NormalTok{p2\_no\_scores }\OtherTok{\textless{}{-}} \FunctionTok{loadResultsOrCompute}\NormalTok{(}\AttributeTok{file =} \StringTok{"../results/p2\_no\_scores.rds"}\NormalTok{, }\AttributeTok{computeExpr =}\NormalTok{ \{}
  \FunctionTok{as.data.frame}\NormalTok{(}\FunctionTok{compute\_all\_scores}\NormalTok{(}\AttributeTok{alignment =}\NormalTok{ p2\_no\_align, }\AttributeTok{patternName =} \StringTok{"p2"}\NormalTok{))}
\NormalTok{\})}
\end{Highlighting}
\end{Shaded}

\begin{table}

\caption{\label{tab:p2-no-scores}Scores for the non-aligned projects with pattern II (p=product, m=mean).}
\centering
\begin{tabular}[t]{lrrrrrrrrr}
\toprule
  & pr\_1 & pr\_2 & pr\_3 & pr\_4 & pr\_5 & pr\_6 & pr\_7 & pr\_8 & pr\_9\\
\midrule
area\_m & 0.78 & 0.88 & 0.89 & 0.88 & 0.85 & 0.86 & 0.89 & 0.83 & 0.90\\
area\_p & 0.36 & 0.61 & 0.63 & 0.59 & 0.52 & 0.53 & 0.62 & 0.48 & 0.64\\
corr\_m & 0.57 & 0.68 & 0.60 & 0.61 & 0.55 & 0.44 & 0.55 & 0.62 & 0.62\\
corr\_p & 0.10 & 0.21 & 0.10 & 0.13 & 0.09 & 0.03 & 0.08 & 0.13 & 0.14\\
jsd\_m & 0.32 & 0.42 & 0.36 & 0.33 & 0.31 & 0.26 & 0.31 & 0.35 & 0.37\\
\addlinespace
jsd\_p & 0.01 & 0.02 & 0.01 & 0.01 & 0.01 & 0.00 & 0.01 & 0.01 & 0.02\\
kl\_m & 0.29 & 0.19 & 0.25 & 0.31 & 0.29 & 0.47 & 0.27 & 0.22 & 0.20\\
kl\_p & 0.00 & 0.00 & 0.00 & 0.00 & 0.00 & 0.01 & 0.00 & 0.00 & 0.00\\
arclen\_m & 0.76 & 0.55 & 0.55 & 0.58 & 0.51 & 0.57 & 0.47 & 0.55 & 0.60\\
arclen\_p & 0.31 & 0.07 & 0.08 & 0.06 & 0.04 & 0.08 & 0.04 & 0.08 & 0.12\\
\addlinespace
sd\_m & 0.73 & 0.79 & 0.78 & 0.79 & 0.73 & 0.75 & 0.79 & 0.78 & 0.81\\
sd\_p & 0.27 & 0.37 & 0.37 & 0.39 & 0.28 & 0.30 & 0.38 & 0.35 & 0.43\\
var\_m & 0.92 & 0.95 & 0.95 & 0.95 & 0.92 & 0.93 & 0.95 & 0.94 & 0.96\\
var\_p & 0.71 & 0.80 & 0.81 & 0.82 & 0.71 & 0.74 & 0.81 & 0.79 & 0.86\\
mae\_m & 0.78 & 0.88 & 0.89 & 0.88 & 0.85 & 0.86 & 0.89 & 0.83 & 0.90\\
\addlinespace
mae\_p & 0.36 & 0.61 & 0.63 & 0.59 & 0.52 & 0.53 & 0.62 & 0.48 & 0.64\\
rmse\_m & 0.72 & 0.83 & 0.84 & 0.84 & 0.79 & 0.81 & 0.85 & 0.79 & 0.86\\
rmse\_p & 0.26 & 0.47 & 0.50 & 0.49 & 0.38 & 0.42 & 0.51 & 0.38 & 0.55\\
RMS\_m & 0.49 & 0.63 & 0.59 & 0.60 & 0.54 & 0.63 & 0.65 & 0.64 & 0.75\\
RMS\_p & 0.06 & 0.14 & 0.09 & 0.10 & 0.08 & 0.15 & 0.16 & 0.13 & 0.32\\
\addlinespace
Kurtosis\_m & 0.13 & 0.35 & 0.21 & 0.17 & 0.23 & 0.23 & 0.17 & 0.39 & 0.33\\
Kurtosis\_p & 0.00 & 0.00 & 0.00 & 0.00 & 0.00 & 0.00 & 0.00 & 0.00 & 0.00\\
Peak\_m & 0.53 & 0.66 & 0.52 & 0.53 & 0.64 & 0.60 & 0.54 & 0.66 & 0.70\\
Peak\_p & 0.07 & 0.15 & 0.06 & 0.06 & 0.14 & 0.11 & 0.07 & 0.14 & 0.23\\
ImpulseFactor\_m & 0.64 & 0.82 & 0.65 & 0.49 & 0.75 & 0.51 & 0.57 & 0.60 & 0.64\\
\addlinespace
ImpulseFactor\_p & 0.13 & 0.44 & 0.16 & 0.04 & 0.30 & 0.04 & 0.08 & 0.11 & 0.14\\
\bottomrule
\end{tabular}
\end{table}

The correlation of just the ground truth with all scores is in table \ref{tab:p2-no-corr}.

\begin{table}

\caption{\label{tab:p2-no-corr}Correlation of the ground truth with all other non-aligned scores for pattern II.}
\centering
\begin{tabular}[t]{lrlrlr}
\toprule
Score & Value & Score & Value & Score & Value\\
\midrule
area\_m & 0.5183212 & arclen\_p & -0.1900343 & RMS\_m & 0.2211204\\
area\_p & 0.5317444 & sd\_m & 0.5396011 & RMS\_p & 0.1957347\\
corr\_m & 0.0640167 & sd\_p & 0.5611052 & Kurtosis\_m & -0.4022877\\
corr\_p & -0.0668462 & var\_m & 0.5762125 & Kurtosis\_p & -0.1080604\\
jsd\_m & -0.0154419 & var\_p & 0.5825667 & Peak\_m & -0.4191859\\
\addlinespace
jsd\_p & -0.1826259 & mae\_m & 0.5186601 & Peak\_p & -0.2512678\\
kl\_m & 0.0640951 & mae\_p & 0.5321362 & ImpulseFactor\_m & -0.5075726\\
kl\_p & -0.1618328 & rmse\_m & 0.5857615 & ImpulseFactor\_p & -0.4730530\\
arclen\_m & -0.0765462 & rmse\_p & 0.6123484 & NA & NA\\
\bottomrule
\end{tabular}
\end{table}

The correlation matrix looks as in figure \ref{fig:p2-corr-mat}.

\begin{figure}[ht!]
\includegraphics{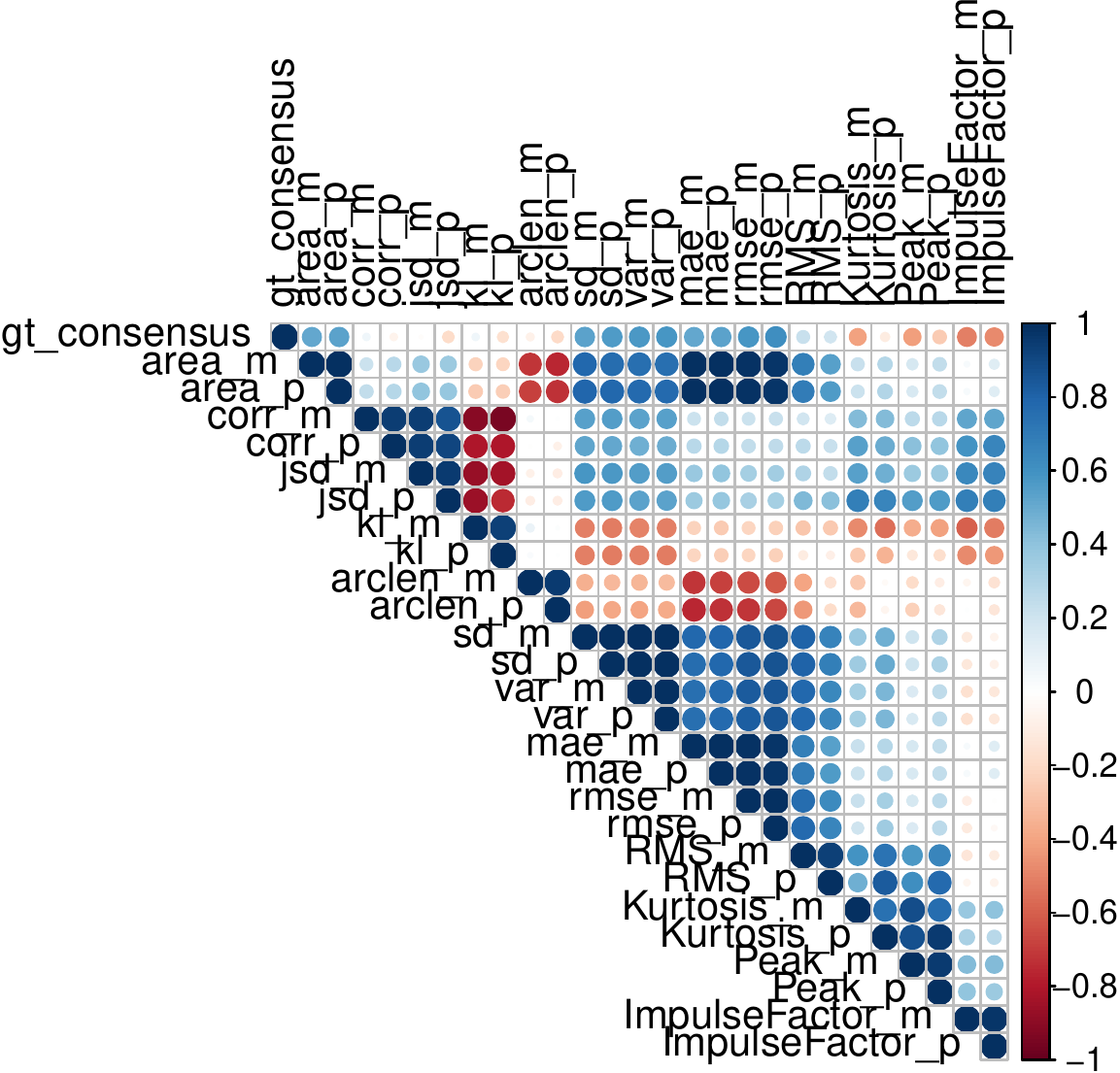} \caption{Correlation matrix for non-aligned scores using pattern II.}\label{fig:p2-no-corr-mat}
\end{figure}

While some correlations are weaker now, we observe more agreement between the scores, i.e., more correlations tend to be positive. \emph{Peak} and \emph{Impulse-factor} however have negative correlations now.

\hypertarget{pattern-iii-average}{%
\subsubsection{Pattern III (average)}\label{pattern-iii-average}}

The 3rd pattern that was produced as weighted average over all ground truth is scored in this section. \textbf{Note!} There is one important difference here: the weighted-average pattern does not have the same intervals as our initial pattern -- in fact, we cannot make any assumptions about any of the intervals. Therefore, we will compute this align with \textbf{ten equally-long} intervals. This amount was chosen arbitrarily as a trade-off between the time it takes to compute, and the resolution of the results. Adding more intervals increases both, computing time and resolution exponentially, however the latter much less.

\begin{Shaded}
\begin{Highlighting}[]
\FunctionTok{library}\NormalTok{(foreach)}

\NormalTok{p3\_avg\_align }\OtherTok{\textless{}{-}} \FunctionTok{loadResultsOrCompute}\NormalTok{(}\AttributeTok{file =} \StringTok{"../results/p3\_avg\_align.rds"}\NormalTok{, }\AttributeTok{computeExpr =}\NormalTok{ \{}
  \CommentTok{\# Let\textquotesingle{}s compute all projects in parallel!}
\NormalTok{  cl }\OtherTok{\textless{}{-}}\NormalTok{ parallel}\SpecialCharTok{::}\FunctionTok{makePSOCKcluster}\NormalTok{(}\FunctionTok{length}\NormalTok{(project\_signals))}
  \FunctionTok{unlist}\NormalTok{(}\FunctionTok{doWithParallelClusterExplicit}\NormalTok{(}\AttributeTok{cl =}\NormalTok{ cl, }\AttributeTok{expr =}\NormalTok{ \{}
\NormalTok{    foreach}\SpecialCharTok{::}\FunctionTok{foreach}\NormalTok{(}
      \AttributeTok{projectName =} \FunctionTok{names}\NormalTok{(project\_signals),}
      \AttributeTok{.inorder =} \ConstantTok{FALSE}\NormalTok{,}
      \AttributeTok{.packages =} \FunctionTok{c}\NormalTok{(}\StringTok{"parallel"}\NormalTok{)}
\NormalTok{    ) }\SpecialCharTok{\%dopar\%}\NormalTok{ \{}
      \FunctionTok{source}\NormalTok{(}\StringTok{"./common{-}funcs.R"}\NormalTok{)}
      \FunctionTok{source}\NormalTok{(}\StringTok{"../models/modelsR6.R"}\NormalTok{)}
      \FunctionTok{source}\NormalTok{(}\StringTok{"../models/SRBTW{-}R6.R"}\NormalTok{)}
      
\NormalTok{      cl\_nested }\OtherTok{\textless{}{-}}\NormalTok{ parallel}\SpecialCharTok{::}\FunctionTok{makePSOCKcluster}\NormalTok{(}\DecValTok{5}\NormalTok{)}
      \StringTok{\textasciigrave{}}\AttributeTok{names\textless{}{-}}\StringTok{\textasciigrave{}}\NormalTok{(}\FunctionTok{list}\NormalTok{(}\FunctionTok{doWithParallelClusterExplicit}\NormalTok{(}\AttributeTok{cl =}\NormalTok{ cl\_nested, }\AttributeTok{expr =}\NormalTok{ \{}
\NormalTok{        temp }\OtherTok{\textless{}{-}} \FunctionTok{time\_warp\_project}\NormalTok{(}
          \AttributeTok{thetaB =} \FunctionTok{seq}\NormalTok{(}\AttributeTok{from =} \DecValTok{0}\NormalTok{, }\AttributeTok{to =} \DecValTok{1}\NormalTok{, }\AttributeTok{by =} \FloatTok{0.1}\NormalTok{), }\CommentTok{\# important!}
          \AttributeTok{pattern =}\NormalTok{ p3\_avg\_signals, }\AttributeTok{project =}\NormalTok{ project\_signals[[projectName]])}
\NormalTok{        temp}\SpecialCharTok{$}\FunctionTok{fit}\NormalTok{(}\AttributeTok{verbose =} \ConstantTok{TRUE}\NormalTok{)}
\NormalTok{        temp }\CommentTok{\# return the instance, it includes the FitResult}
\NormalTok{      \})), projectName)}
\NormalTok{    \}}
\NormalTok{  \}))}
\NormalTok{\})}
\end{Highlighting}
\end{Shaded}

\begin{Shaded}
\begin{Highlighting}[]
\NormalTok{p3\_avg\_scores }\OtherTok{\textless{}{-}} \FunctionTok{loadResultsOrCompute}\NormalTok{(}\AttributeTok{file =} \StringTok{"../results/p3\_avg\_scores.rds"}\NormalTok{, }\AttributeTok{computeExpr =}\NormalTok{ \{}
  \FunctionTok{as.data.frame}\NormalTok{(}\FunctionTok{compute\_all\_scores}\NormalTok{(}\AttributeTok{alignment =}\NormalTok{ p3\_avg\_align, }\AttributeTok{patternName =} \StringTok{"p3\_avg"}\NormalTok{))}
\NormalTok{\})}
\end{Highlighting}
\end{Shaded}

\begin{table}

\caption{\label{tab:p3-avg-scores}Scores for the aligned projects with pattern III (average ground truth).}
\centering
\begin{tabular}[t]{lrrrrrrrrr}
\toprule
  & pr\_1 & pr\_2 & pr\_3 & pr\_4 & pr\_5 & pr\_6 & pr\_7 & pr\_8 & pr\_9\\
\midrule
area\_m & 0.93 & 0.90 & 0.88 & 0.94 & 0.94 & 0.93 & 0.89 & 0.93 & 0.92\\
area\_p & 0.73 & 0.66 & 0.60 & 0.77 & 0.79 & 0.74 & 0.62 & 0.75 & 0.71\\
corr\_m & 0.61 & 0.65 & 0.53 & 0.73 & 0.76 & 0.86 & 0.70 & 0.78 & 0.73\\
corr\_p & 0.13 & 0.14 & 0.07 & 0.28 & 0.32 & 0.55 & 0.22 & 0.37 & 0.27\\
jsd\_m & 0.45 & 0.43 & 0.38 & 0.56 & 0.52 & 0.54 & 0.39 & 0.53 & 0.48\\
\addlinespace
jsd\_p & 0.04 & 0.03 & 0.02 & 0.08 & 0.07 & 0.06 & 0.02 & 0.07 & 0.04\\
kl\_m & 0.14 & 0.16 & 0.19 & 0.10 & 0.09 & 0.12 & 0.18 & 0.09 & 0.16\\
kl\_p & 0.00 & 0.00 & 0.00 & 0.00 & 0.00 & 0.00 & 0.00 & 0.00 & 0.00\\
arclen\_m & 0.57 & 0.68 & 0.53 & 0.66 & 0.63 & 0.65 & 0.80 & 0.74 & 0.60\\
arclen\_p & 0.10 & 0.19 & 0.07 & 0.17 & 0.12 & 0.14 & 0.31 & 0.26 & 0.13\\
\addlinespace
sd\_m & 0.83 & 0.84 & 0.79 & 0.86 & 0.89 & 0.88 & 0.83 & 0.87 & 0.86\\
sd\_p & 0.48 & 0.50 & 0.38 & 0.54 & 0.61 & 0.60 & 0.47 & 0.58 & 0.55\\
var\_m & 0.97 & 0.97 & 0.95 & 0.98 & 0.98 & 0.98 & 0.97 & 0.98 & 0.98\\
var\_p & 0.88 & 0.90 & 0.81 & 0.90 & 0.94 & 0.93 & 0.88 & 0.93 & 0.91\\
mae\_m & 0.93 & 0.90 & 0.88 & 0.94 & 0.94 & 0.93 & 0.89 & 0.93 & 0.92\\
\addlinespace
mae\_p & 0.73 & 0.66 & 0.60 & 0.77 & 0.79 & 0.74 & 0.62 & 0.75 & 0.71\\
rmse\_m & 0.88 & 0.87 & 0.84 & 0.90 & 0.92 & 0.91 & 0.85 & 0.90 & 0.89\\
rmse\_p & 0.59 & 0.58 & 0.50 & 0.65 & 0.70 & 0.67 & 0.51 & 0.67 & 0.62\\
RMS\_m & 0.54 & 0.50 & 0.48 & 0.67 & 0.64 & 0.55 & 0.50 & 0.58 & 0.62\\
RMS\_p & 0.07 & 0.05 & 0.04 & 0.19 & 0.16 & 0.09 & 0.04 & 0.09 & 0.13\\
\addlinespace
Kurtosis\_m & 0.13 & 0.37 & 0.18 & 0.28 & 0.26 & 0.16 & 0.20 & 0.43 & 0.19\\
Kurtosis\_p & 0.00 & 0.00 & 0.00 & 0.00 & 0.00 & 0.00 & 0.00 & 0.01 & 0.00\\
Peak\_m & 0.53 & 0.71 & 0.57 & 0.60 & 0.68 & 0.58 & 0.47 & 0.72 & 0.54\\
Peak\_p & 0.07 & 0.20 & 0.10 & 0.13 & 0.20 & 0.11 & 0.03 & 0.21 & 0.08\\
ImpulseFactor\_m & 0.61 & 0.74 & 0.69 & 0.62 & 0.67 & 0.44 & 0.49 & 0.70 & 0.42\\
\addlinespace
ImpulseFactor\_p & 0.12 & 0.28 & 0.21 & 0.14 & 0.18 & 0.03 & 0.02 & 0.22 & 0.03\\
\bottomrule
\end{tabular}
\end{table}

The correlation of just the ground truth with all scores is in table \ref{tab:p3-avg-corr}.

\begin{table}

\caption{\label{tab:p3-avg-corr}Correlation of the ground truth with all other scores for pattern II.}
\centering
\begin{tabular}[t]{lrlrlr}
\toprule
Score & Value & Score & Value & Score & Value\\
\midrule
area\_m & -0.1308766 & arclen\_p & -0.2487774 & RMS\_m & 0.2993487\\
area\_p & -0.1294226 & sd\_m & -0.3115364 & RMS\_p & 0.3621570\\
corr\_m & -0.2236643 & sd\_p & -0.3275009 & Kurtosis\_m & -0.3521132\\
corr\_p & -0.1652319 & var\_m & -0.4254858 & Kurtosis\_p & -0.4821234\\
jsd\_m & 0.0300469 & var\_p & -0.4289845 & Peak\_m & -0.4520734\\
\addlinespace
jsd\_p & -0.0243270 & mae\_m & -0.1309105 & Peak\_p & -0.4502977\\
kl\_m & 0.2310304 & mae\_p & -0.1296077 & ImpulseFactor\_m & -0.2753840\\
kl\_p & 0.2108042 & rmse\_m & -0.2254399 & ImpulseFactor\_p & -0.2790409\\
arclen\_m & -0.2946963 & rmse\_p & -0.2372254 & NA & NA\\
\bottomrule
\end{tabular}
\end{table}

\begin{figure}[ht!]
\includegraphics{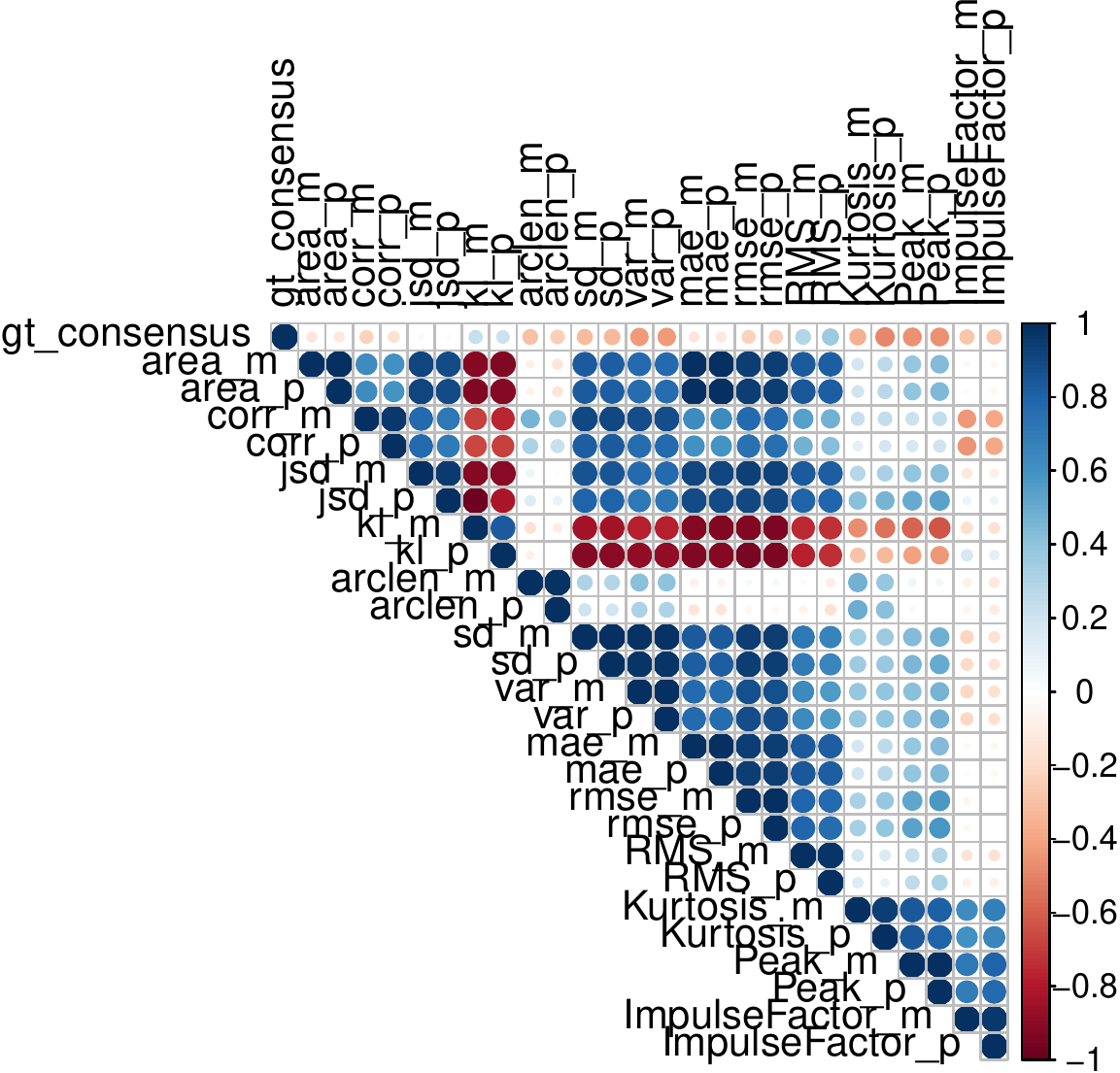} \caption{Correlation matrix for scores using pattern III (average).}\label{fig:p3-avg-corr-mat}
\end{figure}

The correlation matrix looks as in figure \ref{fig:p3-avg-corr-mat}.

I suppose that the most significant result here is the positive Jensen--Shannon divergence score correlation.

\hypertarget{pattern-iii-average-no-alignment}{%
\subsubsection{Pattern III (average, no alignment)}\label{pattern-iii-average-no-alignment}}

Before we go any further, I would also like to compute the scores for this data-driven pattern \textbf{without} having the projects aligned. After manually inspecting some of these alignments, it turns out that some are quite extreme. Since this pattern is a weighted average over all ground truth, not much alignment should be required.

\begin{Shaded}
\begin{Highlighting}[]
\CommentTok{\# We\textquotesingle{}ll have to mimic an \textquotesingle{}aligned\textquotesingle{} object, which is a list of}
\CommentTok{\# srBTAW\_MultiVartype instances. We can clone it and just undo the time}
\CommentTok{\# warping.}
\NormalTok{p3\_avg\_no\_align }\OtherTok{\textless{}{-}} \FunctionTok{list}\NormalTok{()}

\ControlFlowTok{for}\NormalTok{ (project }\ControlFlowTok{in}\NormalTok{ ground\_truth}\SpecialCharTok{$}\NormalTok{project) \{}
\NormalTok{  temp }\OtherTok{\textless{}{-}}\NormalTok{ p3\_avg\_align[[project]]}\SpecialCharTok{$}\FunctionTok{clone}\NormalTok{()}
\NormalTok{  temp}\SpecialCharTok{$}\FunctionTok{setParams}\NormalTok{(}\AttributeTok{params =} \StringTok{\textasciigrave{}}\AttributeTok{names\textless{}{-}}\StringTok{\textasciigrave{}}\NormalTok{(}\FunctionTok{rep}\NormalTok{(}\DecValTok{1}\SpecialCharTok{/}\NormalTok{temp}\SpecialCharTok{$}\FunctionTok{getNumParams}\NormalTok{(), temp}\SpecialCharTok{$}\FunctionTok{getNumParams}\NormalTok{()),}
\NormalTok{    temp}\SpecialCharTok{$}\FunctionTok{getParamNames}\NormalTok{()))}
\NormalTok{  p3\_avg\_no\_align[[project]] }\OtherTok{\textless{}{-}}\NormalTok{ temp}
\NormalTok{\}}
\end{Highlighting}
\end{Shaded}

\begin{Shaded}
\begin{Highlighting}[]
\NormalTok{p3\_avg\_no\_scores }\OtherTok{\textless{}{-}} \FunctionTok{loadResultsOrCompute}\NormalTok{(}\AttributeTok{file =} \StringTok{"../results/p3\_avg\_no\_scores.rds"}\NormalTok{,}
  \AttributeTok{computeExpr =}\NormalTok{ \{}
    \FunctionTok{as.data.frame}\NormalTok{(}\FunctionTok{compute\_all\_scores}\NormalTok{(}\AttributeTok{alignment =}\NormalTok{ p3\_avg\_no\_align, }\AttributeTok{patternName =} \StringTok{"p3\_avg"}\NormalTok{))}
\NormalTok{  \})}
\end{Highlighting}
\end{Shaded}

\begin{table}

\caption{\label{tab:p3-avg-no-scores}Scores for the non-aligned projects with pattern III (average ground truth).}
\centering
\begin{tabular}[t]{lrrrrrrrrr}
\toprule
  & pr\_1 & pr\_2 & pr\_3 & pr\_4 & pr\_5 & pr\_6 & pr\_7 & pr\_8 & pr\_9\\
\midrule
area\_m & 0.80 & 0.88 & 0.91 & 0.90 & 0.86 & 0.88 & 0.91 & 0.84 & 0.91\\
area\_p & 0.39 & 0.59 & 0.69 & 0.66 & 0.55 & 0.60 & 0.67 & 0.50 & 0.69\\
corr\_m & 0.66 & 0.50 & 0.71 & 0.82 & 0.49 & 0.58 & 0.73 & 0.59 & 0.77\\
corr\_p & 0.19 & 0.06 & 0.22 & 0.44 & 0.05 & 0.11 & 0.27 & 0.08 & 0.34\\
jsd\_m & 0.36 & 0.37 & 0.49 & 0.49 & 0.33 & 0.36 & 0.46 & 0.39 & 0.46\\
\addlinespace
jsd\_p & 0.01 & 0.01 & 0.04 & 0.04 & 0.01 & 0.01 & 0.03 & 0.02 & 0.03\\
kl\_m & 0.23 & 0.27 & 0.16 & 0.16 & 0.26 & 0.32 & 0.18 & 0.20 & 0.15\\
kl\_p & 0.00 & 0.00 & 0.00 & 0.00 & 0.00 & 0.00 & 0.00 & 0.00 & 0.00\\
arclen\_m & 0.75 & 0.61 & 0.71 & 0.67 & 0.58 & 0.63 & 0.65 & 0.71 & 0.76\\
arclen\_p & 0.26 & 0.13 & 0.23 & 0.16 & 0.10 & 0.15 & 0.15 & 0.21 & 0.29\\
\addlinespace
sd\_m & 0.75 & 0.77 & 0.83 & 0.87 & 0.75 & 0.80 & 0.84 & 0.80 & 0.84\\
sd\_p & 0.30 & 0.34 & 0.45 & 0.56 & 0.31 & 0.40 & 0.48 & 0.40 & 0.50\\
var\_m & 0.93 & 0.94 & 0.96 & 0.98 & 0.93 & 0.95 & 0.96 & 0.95 & 0.97\\
var\_p & 0.73 & 0.77 & 0.85 & 0.91 & 0.74 & 0.82 & 0.86 & 0.82 & 0.89\\
mae\_m & 0.80 & 0.88 & 0.91 & 0.90 & 0.86 & 0.88 & 0.91 & 0.84 & 0.91\\
\addlinespace
mae\_p & 0.39 & 0.59 & 0.69 & 0.66 & 0.55 & 0.60 & 0.67 & 0.50 & 0.69\\
rmse\_m & 0.73 & 0.82 & 0.87 & 0.89 & 0.80 & 0.84 & 0.88 & 0.80 & 0.88\\
rmse\_p & 0.27 & 0.45 & 0.57 & 0.61 & 0.41 & 0.50 & 0.58 & 0.40 & 0.60\\
RMS\_m & 0.46 & 0.43 & 0.50 & 0.55 & 0.39 & 0.48 & 0.56 & 0.47 & 0.54\\
RMS\_p & 0.02 & 0.03 & 0.06 & 0.09 & 0.02 & 0.05 & 0.09 & 0.03 & 0.07\\
\addlinespace
Kurtosis\_m & 0.14 & 0.11 & 0.21 & 0.12 & 0.12 & 0.10 & 0.32 & 0.37 & 0.37\\
Kurtosis\_p & 0.00 & 0.00 & 0.00 & 0.00 & 0.00 & 0.00 & 0.00 & 0.00 & 0.00\\
Peak\_m & 0.47 & 0.53 & 0.60 & 0.58 & 0.57 & 0.54 & 0.68 & 0.71 & 0.72\\
Peak\_p & 0.03 & 0.07 & 0.13 & 0.11 & 0.10 & 0.08 & 0.20 & 0.20 & 0.21\\
ImpulseFactor\_m & 0.49 & 0.71 & 0.69 & 0.37 & 0.78 & 0.37 & 0.74 & 0.76 & 0.69\\
\addlinespace
ImpulseFactor\_p & 0.05 & 0.17 & 0.22 & 0.01 & 0.34 & 0.01 & 0.29 & 0.30 & 0.20\\
\bottomrule
\end{tabular}
\end{table}

The correlation of just the ground truth with all scores is in table \ref{tab:p3-avg-corr}.

\begin{table}

\caption{\label{tab:p3-avg-no-corr}Correlation of the ground truth with all other scores for pattern II.}
\centering
\begin{tabular}[t]{lrlrlr}
\toprule
Score & Value & Score & Value & Score & Value\\
\midrule
area\_m & 0.6670842 & arclen\_p & 0.2514802 & RMS\_m & 0.7467092\\
area\_p & 0.6894208 & sd\_m & 0.8218241 & RMS\_p & 0.7740284\\
corr\_m & 0.8412487 & sd\_p & 0.8436768 & Kurtosis\_m & 0.0543283\\
corr\_p & 0.8835587 & var\_m & 0.7916103 & Kurtosis\_p & 0.0289034\\
jsd\_m & 0.8823326 & var\_p & 0.8025434 & Peak\_m & 0.1879883\\
\addlinespace
jsd\_p & 0.8724026 & mae\_m & 0.6670619 & Peak\_p & 0.1986320\\
kl\_m & -0.6623665 & mae\_p & 0.6893942 & ImpulseFactor\_m & -0.3763130\\
kl\_p & -0.7161612 & rmse\_m & 0.7436967 & ImpulseFactor\_p & -0.3217380\\
arclen\_m & 0.2914946 & rmse\_p & 0.7784040 & NA & NA\\
\bottomrule
\end{tabular}
\end{table}

\begin{figure}[ht!]
\includegraphics{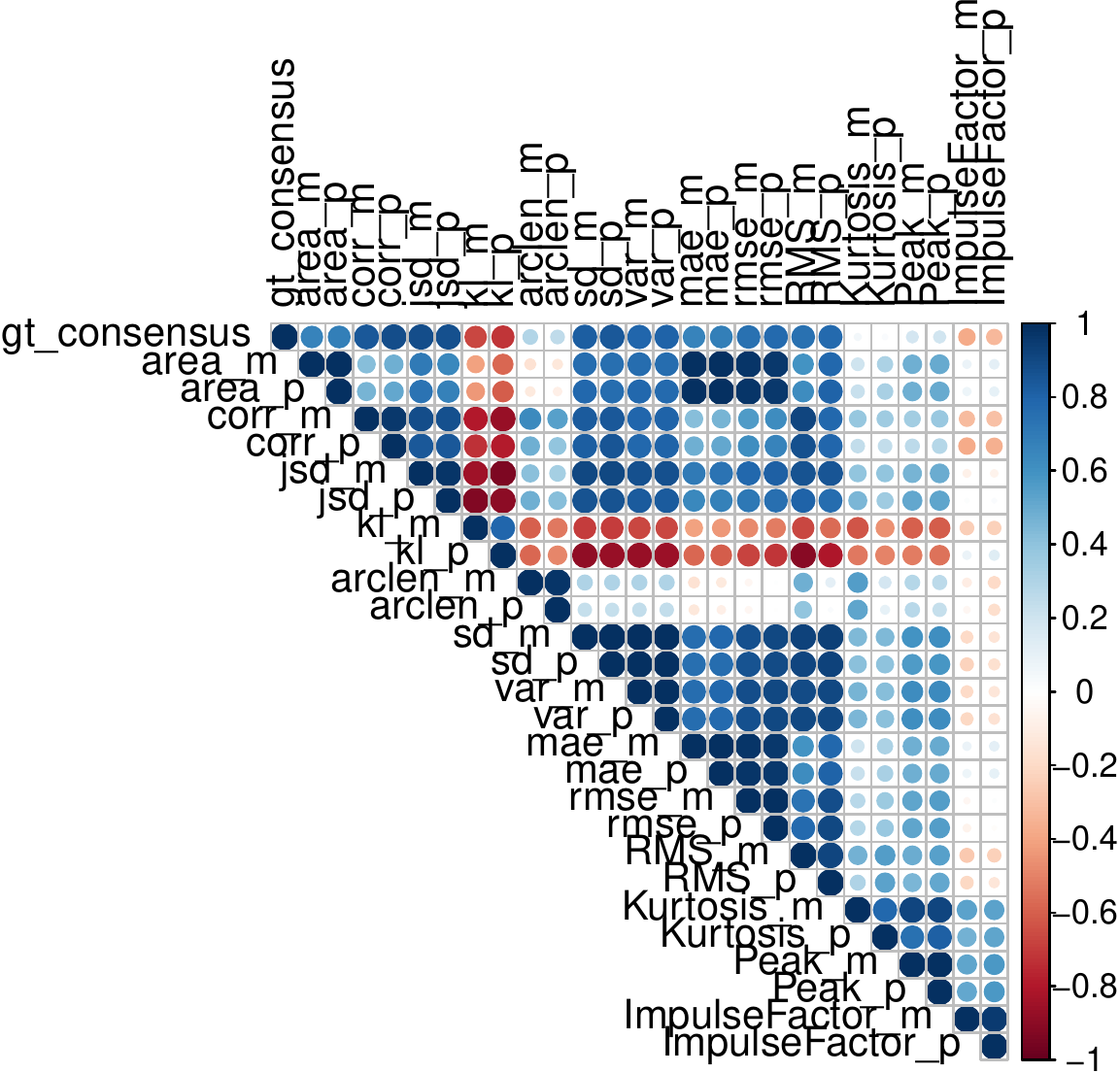} \caption{Correlation matrix for non-aligned scores using pattern III (average).}\label{fig:p3-avg-no-corr-mat}
\end{figure}

The correlation matrix looks as in figure \ref{fig:p3-avg-no-corr-mat}.

And here in figure \ref{fig:p3-avg-no-corr-mat} we got the result that was the most-expected. We get almost always positive correlations, and most of them range between medium and significant strength. If we look at the correlation for \texttt{sd\_p}, it is almost perfect with 0.844. The Jensen--Shannon divergence score (note: while called divergence, this is a score, the higher the better) of 0.882 is at a high level now. I mention this because this measure is less primitive than the others and tends to capture more properties of the signals. If this is high, it means we get a robust measure that ought to be usable stand-alone. The other low-level scores probably would need to be combined instead.

\hypertarget{linear-combination-of-scores}{%
\paragraph{Linear combination of scores}\label{linear-combination-of-scores}}

So far we have tested whether the calculated scores correlate with the scores of the ground truth, and we find some good examples. However, these scores are not scaled or translated in any way, so it is probably best to A) take multiple scores into account and B) create a regression model that makes these adjustments. We will test some linear combination of the scores \texttt{corr\_p}, \texttt{jsd\_m}, \texttt{RMS\_m} and \texttt{sd\_p}.

\begin{Shaded}
\begin{Highlighting}[]
\NormalTok{p3\_avg\_lm }\OtherTok{\textless{}{-}}\NormalTok{ stats}\SpecialCharTok{::}\FunctionTok{glm}\NormalTok{(}\AttributeTok{formula =}\NormalTok{ gt\_consensus }\SpecialCharTok{\textasciitilde{}}\NormalTok{ corr\_p }\SpecialCharTok{+}\NormalTok{ jsd\_m }\SpecialCharTok{+}\NormalTok{ RMS\_m }\SpecialCharTok{+}\NormalTok{ sd\_p, }\AttributeTok{data =}\NormalTok{ temp)}
\NormalTok{stats}\SpecialCharTok{::}\FunctionTok{coef}\NormalTok{(p3\_avg\_lm)}
\end{Highlighting}
\end{Shaded}

\begin{verbatim}
## (Intercept)      corr_p       jsd_m       RMS_m        sd_p 
##   0.1127856   1.4429241   2.3439285  -3.2697287   1.2561099
\end{verbatim}

\begin{Shaded}
\begin{Highlighting}[]
\FunctionTok{plot}\NormalTok{(p3\_avg\_lm, }\AttributeTok{ask =} \ConstantTok{FALSE}\NormalTok{, }\AttributeTok{which =} \DecValTok{1}\SpecialCharTok{:}\DecValTok{2}\NormalTok{)}
\end{Highlighting}
\end{Shaded}

\includegraphics{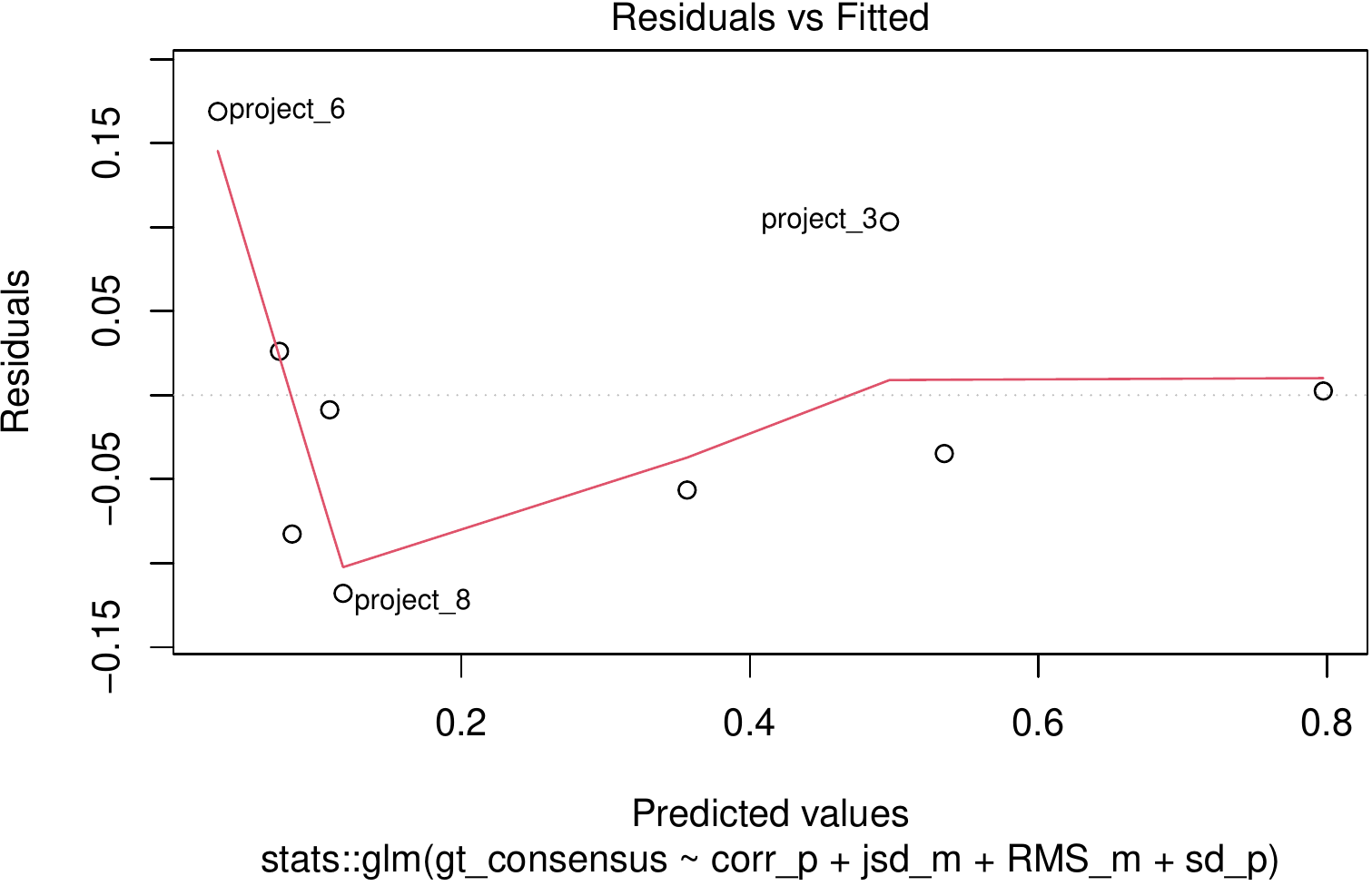} \includegraphics{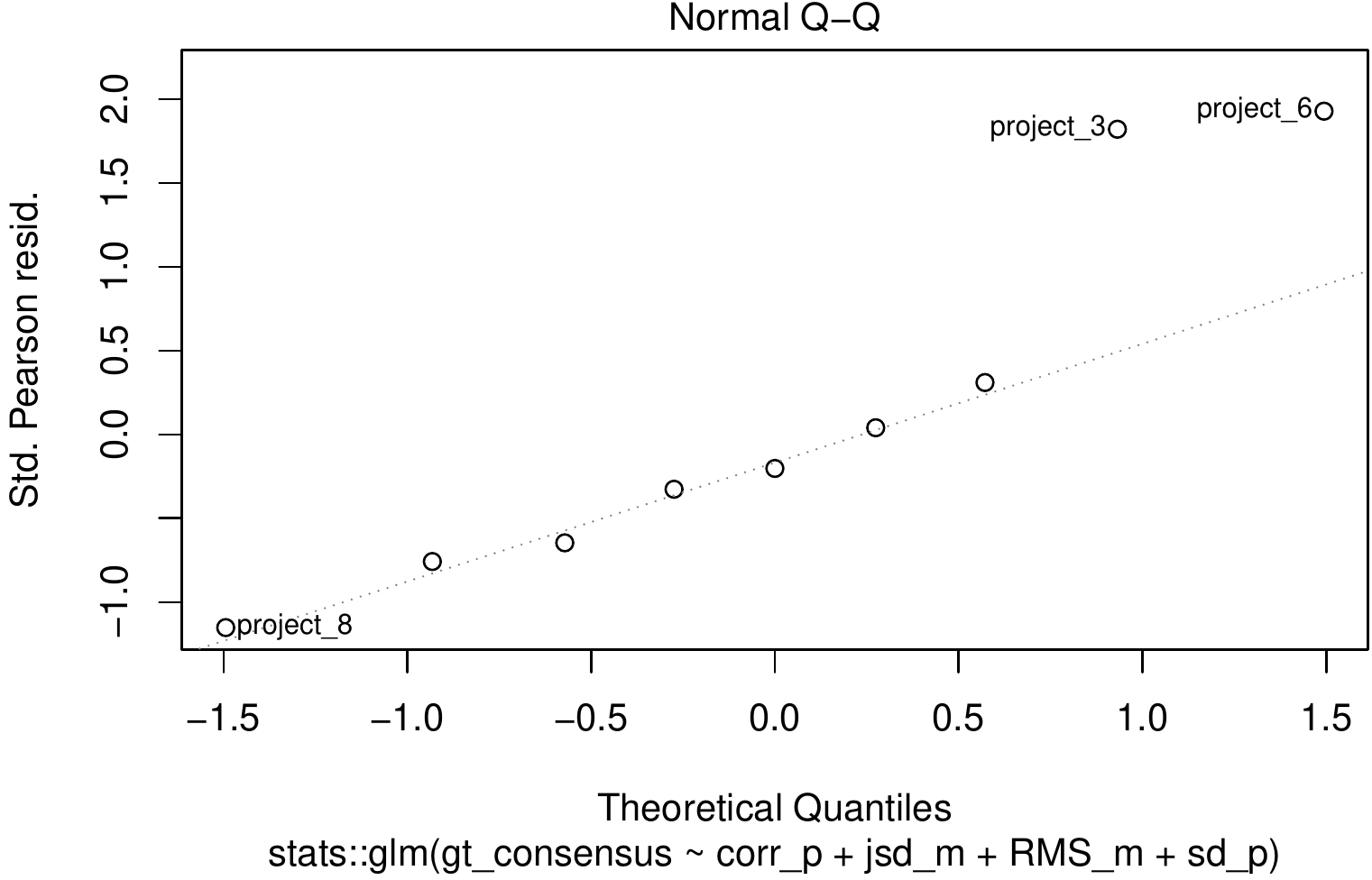}

Of course, this model should not be used to predict on the training data, but what we wanted to learn here is simply how to linearly combine the scores in order to get scores that are in the same range as the ground truth, we learn a re-scale so to say.

\begin{Shaded}
\begin{Highlighting}[]
\NormalTok{p3\_avg\_lm\_scores }\OtherTok{\textless{}{-}}\NormalTok{ stats}\SpecialCharTok{::}\FunctionTok{predict}\NormalTok{(p3\_avg\_lm, temp)}
\FunctionTok{round}\NormalTok{(p3\_avg\_lm\_scores }\SpecialCharTok{*} \DecValTok{10}\NormalTok{, }\DecValTok{3}\NormalTok{)}
\end{Highlighting}
\end{Shaded}

\begin{verbatim}
## project_1 project_2 project_3 project_4 project_5 project_6 project_7 project_8 
##     1.087     0.827     4.968     7.975     0.739     0.311     3.565     1.180 
## project_9 
##     5.348
\end{verbatim}

\begin{Shaded}
\begin{Highlighting}[]
\NormalTok{stats}\SpecialCharTok{::}\FunctionTok{cor}\NormalTok{(p3\_avg\_lm\_scores, ground\_truth}\SpecialCharTok{$}\NormalTok{consensus\_score)}
\end{Highlighting}
\end{Shaded}

\begin{verbatim}
## [1] 0.9485156
\end{verbatim}

This also increased the correlation to 0.949.

\hypertarget{finding-the-most-important-scores}{%
\paragraph{\texorpdfstring{Finding the most important scores\label{sssec:var-imp}}{Finding the most important scores}}\label{finding-the-most-important-scores}}

Previously, we have combined some hand-picked scores into a linear model, in order to mainly scale and translate them, in order to be able to obtain predictions in the range of the actual ground truth. However, this approach is not suitable for a model that shall generalize.

We have shown in figure \ref{fig:p3-avg-no-corr-mat} that most scores capture different properties of the alignment. It therefore makes sense to choose scores for a regression that, once combined, have the most predictive power in terms of accuracy (precision), and generalizability. There are quite many techniques and approaches to \emph{feature selection}. Here, we will attempt a \textbf{recursive feature elimination} (RFE), that uses partial least squares as estimator, a measure of \textbf{variable importance}, and using bootstrapped data (for example, Darst, Malecki, and Engelman (2018)).

Before we start, a few things are of importance. First, a feature selection should be done whenever a new model with prediction and generalization capabilities is to be trained, i.e., it is specific to the case (here process model and observed processes). Therefore, the features we select here are only a valid selection for a model that is to make predictions on scores as obtained from the pattern type III (average, no align). Secondly, the amount of data we have only suffices for running the suggested RFE, but the data is too scarce to obtain a robust model. Therefore, the results we will obtain here are practically only usable for giving an indication as to the relative importance of features (here: scores), but we must not deduce a genuine truth from them, nor can we conclude an absolute ordering that will hold for future runs of this approach, given different data. Thirdly, and that is specific to our case, we have separate scores for each interval (previously aggregated once using the mean, and once using the product), but the score is an aggregation across all variables (activities). Having one score represent the deviation over only one contiguous, aggregated interval, should be avoided in a real-world setting, and scores should be localized to each interval, to become their own feature. However, that requires even more data. Also, one would probably not aggregate scores across variables. We will examine this scenario more closely for issue-tracking data.

Let's start, we'll use the scores from \texttt{p3\_avg\_no\_scores}:

\begin{Shaded}
\begin{Highlighting}[]
\NormalTok{rfe\_data }\OtherTok{\textless{}{-}} \FunctionTok{cbind}\NormalTok{(p3\_avg\_no\_scores, }\FunctionTok{data.frame}\NormalTok{(}\AttributeTok{gt =}\NormalTok{ ground\_truth}\SpecialCharTok{$}\NormalTok{consensus))}
\end{Highlighting}
\end{Shaded}

The RFE is done via caret, with 3-times repeated, 10-fold cross validation as outer resampling method\footnote{\url{https://web.archive.org/web/20211120164401/https://topepo.github.io/caret/recursive-feature-elimination.html}}.

\begin{Shaded}
\begin{Highlighting}[]
\FunctionTok{library}\NormalTok{(caret, }\AttributeTok{quietly =} \ConstantTok{TRUE}\NormalTok{)}

\FunctionTok{set.seed}\NormalTok{(}\DecValTok{1337}\NormalTok{)}

\NormalTok{control }\OtherTok{\textless{}{-}}\NormalTok{ caret}\SpecialCharTok{::}\FunctionTok{trainControl}\NormalTok{(}\AttributeTok{method =} \StringTok{"repeatedcv"}\NormalTok{, }\AttributeTok{number =} \DecValTok{10}\NormalTok{, }\AttributeTok{repeats =} \DecValTok{3}\NormalTok{)}
\NormalTok{modelFit }\OtherTok{\textless{}{-}}\NormalTok{ caret}\SpecialCharTok{::}\FunctionTok{train}\NormalTok{(gt }\SpecialCharTok{\textasciitilde{}}\NormalTok{ ., }\AttributeTok{data =}\NormalTok{ rfe\_data, }\AttributeTok{method =} \StringTok{"pls"}\NormalTok{, }\AttributeTok{trControl =}\NormalTok{ control)}

\NormalTok{imp }\OtherTok{\textless{}{-}}\NormalTok{ caret}\SpecialCharTok{::}\FunctionTok{varImp}\NormalTok{(}\AttributeTok{object =}\NormalTok{ modelFit)}
\end{Highlighting}
\end{Shaded}

\begin{verbatim}
## 
## Attaching package: 'pls'
\end{verbatim}

\begin{verbatim}
## The following object is masked from 'package:caret':
## 
##     R2
\end{verbatim}

\begin{verbatim}
## The following object is masked from 'package:stats':
## 
##     loadings
\end{verbatim}

\begin{figure}[ht!]
\includegraphics{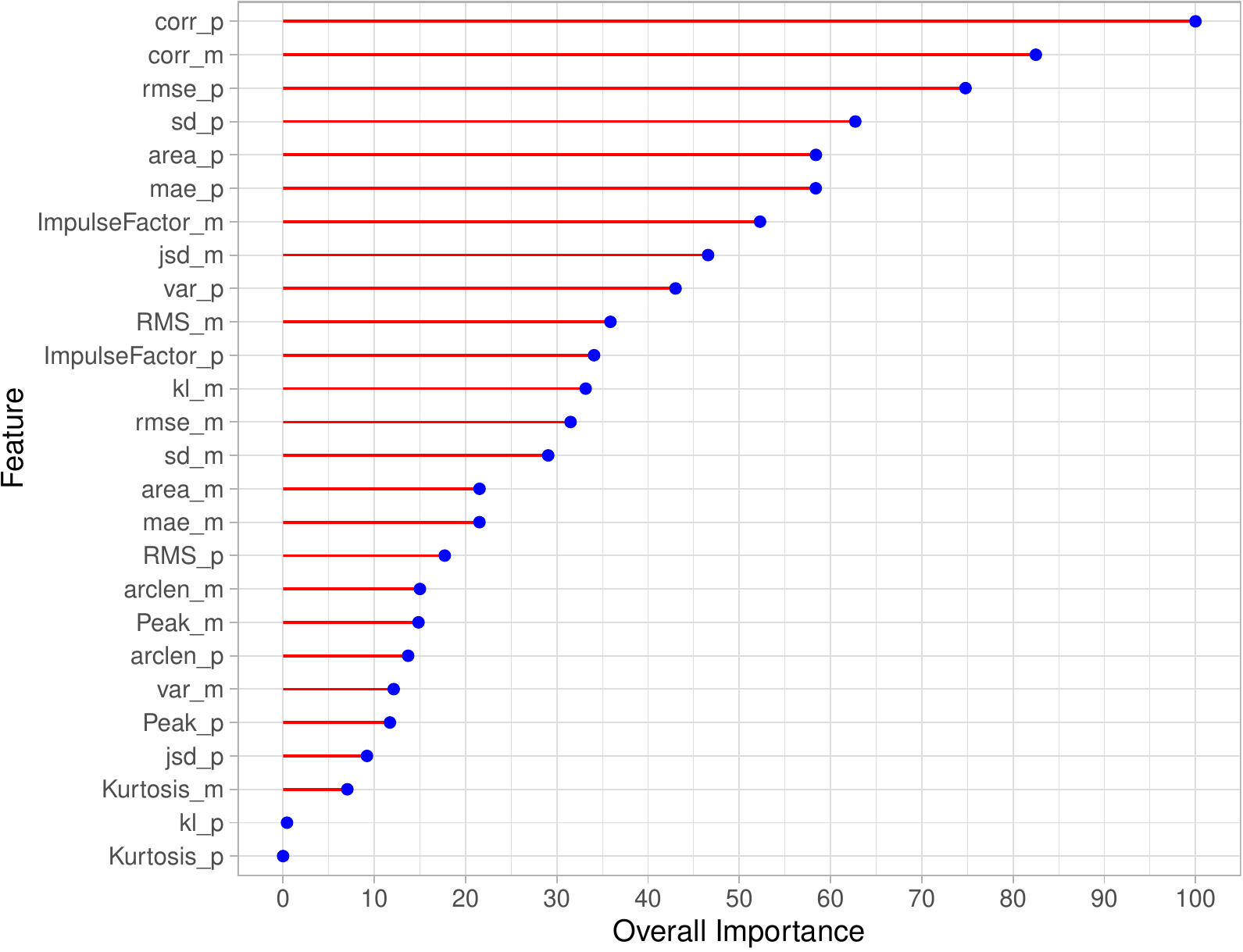} \caption{Plot of the relative variable importances of scores as computed against pattern III (average).}\label{fig:p3-avg-no-var-imp-plot}
\end{figure}

The relative variable importances are shown in table \ref{tab:p3-avg-no-varimp}. Unsurprisingly, correlation is the most important feature (score), as it expresses the degree to which two curves resemble each other (regardless of their absolute difference). This is then captured by the next most important score, the RMSE, closely followed by the standard deviation and area between curves. Note that area, RMSE, sd and variance are all highly correlated, so this does not come as a surprise. The next two places is the Impulsefactor and Jenson--Shannon divergence. We observe a somewhat healthy mix between mean- and product-measures. For values that tend to be comparatively tiny, such as the JSD or KL divergence, the product reduces variance too extreme (should have used log-sums), such that the score forfeits too much of its importance.

\begin{table}

\caption{\label{tab:p3-avg-no-varimp}Relative variable importances of scores as computed from all projects against pattern type III (average, no align).}
\centering
\begin{tabular}[t]{lr}
\toprule
  & Overall\\
\midrule
corr\_p & 100.0000000\\
corr\_m & 82.4997038\\
rmse\_p & 74.7951625\\
sd\_p & 62.7139275\\
area\_p & 58.3954015\\
\addlinespace
mae\_p & 58.3812807\\
ImpulseFactor\_m & 52.2749729\\
jsd\_m & 46.5783897\\
var\_p & 43.0166540\\
RMS\_m & 35.8758915\\
\addlinespace
ImpulseFactor\_p & 34.0850404\\
kl\_m & 33.1660758\\
rmse\_m & 31.5078958\\
sd\_m & 29.0562651\\
area\_m & 21.5288693\\
\addlinespace
mae\_m & 21.5241949\\
RMS\_p & 17.7170483\\
arclen\_m & 14.9978964\\
Peak\_m & 14.8411897\\
arclen\_p & 13.7074530\\
\addlinespace
var\_m & 12.1239044\\
Peak\_p & 11.7125090\\
jsd\_p & 9.1980319\\
Kurtosis\_m & 7.0437781\\
kl\_p & 0.4299751\\
\addlinespace
Kurtosis\_p & 0.0000000\\
\bottomrule
\end{tabular}
\end{table}

The final model as selected by the RFE approach has these properties:

\begin{Shaded}
\begin{Highlighting}[]
\NormalTok{modelFit}
\end{Highlighting}
\end{Shaded}

\begin{verbatim}
## Partial Least Squares 
## 
##  9 samples
## 26 predictors
## 
## No pre-processing
## Resampling: Cross-Validated (10 fold, repeated 3 times) 
## Summary of sample sizes: 8, 8, 8, 8, 8, 8, ... 
## Resampling results across tuning parameters:
## 
##   ncomp  RMSE      Rsquared  MAE     
##   1      1.515697  NaN       1.515697
##   2      1.317217  NaN       1.317217
##   3      1.497372  NaN       1.497372
## 
## RMSE was used to select the optimal model using the smallest value.
## The final value used for the model was ncomp = 2.
\end{verbatim}

Now if we were to make predictions with this model, the results would be these:

\begin{Shaded}
\begin{Highlighting}[]
\NormalTok{temp }\OtherTok{\textless{}{-}} \FunctionTok{predict}\NormalTok{(}\AttributeTok{object =}\NormalTok{ modelFit, }\AttributeTok{newdata =}\NormalTok{ p3\_avg\_no\_scores)}
\FunctionTok{round}\NormalTok{(}\AttributeTok{x =}\NormalTok{ temp, }\AttributeTok{digits =} \DecValTok{2}\NormalTok{)}
\end{Highlighting}
\end{Shaded}

\begin{verbatim}
## [1]  1.17  0.90  4.31  7.69 -0.47  3.14  3.96  0.20  5.11
\end{verbatim}

\begin{Shaded}
\begin{Highlighting}[]
\NormalTok{ground\_truth}\SpecialCharTok{$}\NormalTok{consensus}
\end{Highlighting}
\end{Shaded}

\begin{verbatim}
## [1] 1 0 6 8 1 2 3 0 5
\end{verbatim}

The correlation with the ground truth would then be:

\begin{Shaded}
\begin{Highlighting}[]
\FunctionTok{cor.test}\NormalTok{(ground\_truth}\SpecialCharTok{$}\NormalTok{consensus, temp)}
\end{Highlighting}
\end{Shaded}

\begin{verbatim}
## 
##  Pearson's product-moment correlation
## 
## data:  ground_truth$consensus and temp
## t = 6.9369, df = 7, p-value = 0.0002238
## alternative hypothesis: true correlation is not equal to 0
## 95 percent confidence interval:
##  0.7120974 0.9863926
## sample estimates:
##       cor 
## 0.9343479
\end{verbatim}

\hypertarget{pattern-as-confidence-surface}{%
\paragraph{Pattern as confidence surface}\label{pattern-as-confidence-surface}}

Exactly how it was done for issue-tracking data (refer for details to subsection \ref{sssec:inhomo-conf-interval}, as here we will only create and show the resulting pattern), we can produce a data-only pattern that features the weighted average for each variable, but also a confidence surface that takes the ground truth into account. Note that as the time of adding this section, we have access to a second batch of projects, including a ground truth. We will therefore include those here, as data-only patterns become better with each observation.

\begin{Shaded}
\begin{Highlighting}[]
\NormalTok{ground\_truth\_2nd\_batch }\OtherTok{\textless{}{-}} \FunctionTok{read.csv}\NormalTok{(}\AttributeTok{file =} \StringTok{"../data/ground{-}truth\_2nd{-}batch.csv"}\NormalTok{, }\AttributeTok{sep =} \StringTok{";"}\NormalTok{)}
\NormalTok{ground\_truth\_2nd\_batch}\SpecialCharTok{$}\NormalTok{consensus\_score }\OtherTok{\textless{}{-}}\NormalTok{ ground\_truth\_2nd\_batch}\SpecialCharTok{$}\NormalTok{consensus}\SpecialCharTok{/}\DecValTok{10}
\FunctionTok{rownames}\NormalTok{(ground\_truth\_2nd\_batch) }\OtherTok{\textless{}{-}} \FunctionTok{paste0}\NormalTok{((}\DecValTok{1} \SpecialCharTok{+} \FunctionTok{nrow}\NormalTok{(ground\_truth))}\SpecialCharTok{:}\NormalTok{(}\FunctionTok{nrow}\NormalTok{(ground\_truth) }\SpecialCharTok{+}
  \FunctionTok{nrow}\NormalTok{(ground\_truth\_2nd\_batch)))}

\NormalTok{temp }\OtherTok{\textless{}{-}} \FunctionTok{rbind}\NormalTok{(ground\_truth, ground\_truth\_2nd\_batch)}
\NormalTok{omega }\OtherTok{\textless{}{-}}\NormalTok{ temp}\SpecialCharTok{$}\NormalTok{consensus\_score}
\FunctionTok{names}\NormalTok{(omega) }\OtherTok{\textless{}{-}} \FunctionTok{paste0}\NormalTok{(}\StringTok{"Project"}\NormalTok{, }\DecValTok{1}\SpecialCharTok{:}\FunctionTok{length}\NormalTok{(omega))}
\end{Highlighting}
\end{Shaded}

The weighted averaged signals are stored in \texttt{p3\_avg\_signals}, so we need to compute two things: first, we need functions to delineate the confidence surface (\(\operatorname{CI}_{\text{upper}}(x)\) and \(\operatorname{CI}_{\text{upper}}(x)\)), and second, we need a function that produces the gradated surface within those bounds (\(\operatorname{CI}(x,y)\)). This is very similar to how it was done for issue-tracking, so most of the code is not shown here (check the source code for this notebook).

Let's load the new batch and do some required preprocessing:

\begin{Shaded}
\begin{Highlighting}[]
\NormalTok{student\_projects\_2nd\_batch }\OtherTok{\textless{}{-}} \FunctionTok{read.csv}\NormalTok{(}\AttributeTok{file =} \StringTok{"../data/student{-}projects\_2nd{-}batch.csv"}\NormalTok{,}
  \AttributeTok{sep =} \StringTok{";"}\NormalTok{)}
\NormalTok{student\_projects\_2nd\_batch}\SpecialCharTok{$}\NormalTok{label }\OtherTok{\textless{}{-}} \FunctionTok{as.factor}\NormalTok{(student\_projects\_2nd\_batch}\SpecialCharTok{$}\NormalTok{label)}
\NormalTok{student\_projects\_2nd\_batch}\SpecialCharTok{$}\NormalTok{project }\OtherTok{\textless{}{-}} \FunctionTok{as.factor}\NormalTok{(student\_projects\_2nd\_batch}\SpecialCharTok{$}\NormalTok{project)}
\NormalTok{student\_projects\_2nd\_batch}\SpecialCharTok{$}\NormalTok{AuthorTimeNormalized }\OtherTok{\textless{}{-}} \ConstantTok{NA\_real\_}

\ControlFlowTok{for}\NormalTok{ (pId }\ControlFlowTok{in} \FunctionTok{levels}\NormalTok{(student\_projects\_2nd\_batch}\SpecialCharTok{$}\NormalTok{project)) \{}
\NormalTok{  student\_projects\_2nd\_batch[student\_projects\_2nd\_batch}\SpecialCharTok{$}\NormalTok{project }\SpecialCharTok{==}\NormalTok{ pId, ]}\SpecialCharTok{$}\NormalTok{AuthorTimeNormalized }\OtherTok{\textless{}{-}}\NormalTok{ (student\_projects\_2nd\_batch[student\_projects\_2nd\_batch}\SpecialCharTok{$}\NormalTok{project }\SpecialCharTok{==}
\NormalTok{    pId, ]}\SpecialCharTok{$}\NormalTok{AuthorTimeUnixEpochMilliSecs }\SpecialCharTok{{-}} \FunctionTok{min}\NormalTok{(student\_projects\_2nd\_batch[student\_projects\_2nd\_batch}\SpecialCharTok{$}\NormalTok{project }\SpecialCharTok{==}
\NormalTok{    pId, ]}\SpecialCharTok{$}\NormalTok{AuthorTimeUnixEpochMilliSecs))}
\NormalTok{  student\_projects\_2nd\_batch[student\_projects\_2nd\_batch}\SpecialCharTok{$}\NormalTok{project }\SpecialCharTok{==}\NormalTok{ pId, ]}\SpecialCharTok{$}\NormalTok{AuthorTimeNormalized }\OtherTok{\textless{}{-}}\NormalTok{ (student\_projects\_2nd\_batch[student\_projects\_2nd\_batch}\SpecialCharTok{$}\NormalTok{project }\SpecialCharTok{==}
\NormalTok{    pId, ]}\SpecialCharTok{$}\NormalTok{AuthorTimeNormalized}\SpecialCharTok{/}\FunctionTok{max}\NormalTok{(student\_projects\_2nd\_batch[student\_projects\_2nd\_batch}\SpecialCharTok{$}\NormalTok{project }\SpecialCharTok{==}
\NormalTok{    pId, ]}\SpecialCharTok{$}\NormalTok{AuthorTimeNormalized))}
\NormalTok{\}}
\end{Highlighting}
\end{Shaded}

Below, we show the amount of commits in each project:

\begin{Shaded}
\begin{Highlighting}[]
\FunctionTok{table}\NormalTok{(student\_projects\_2nd\_batch}\SpecialCharTok{$}\NormalTok{project)}
\end{Highlighting}
\end{Shaded}

\begin{verbatim}
## 
## project_10 project_11 project_12 project_13 project_14 project_15 
##        158        542        136        192         62         66
\end{verbatim}

\begin{figure}[ht!]
\includegraphics{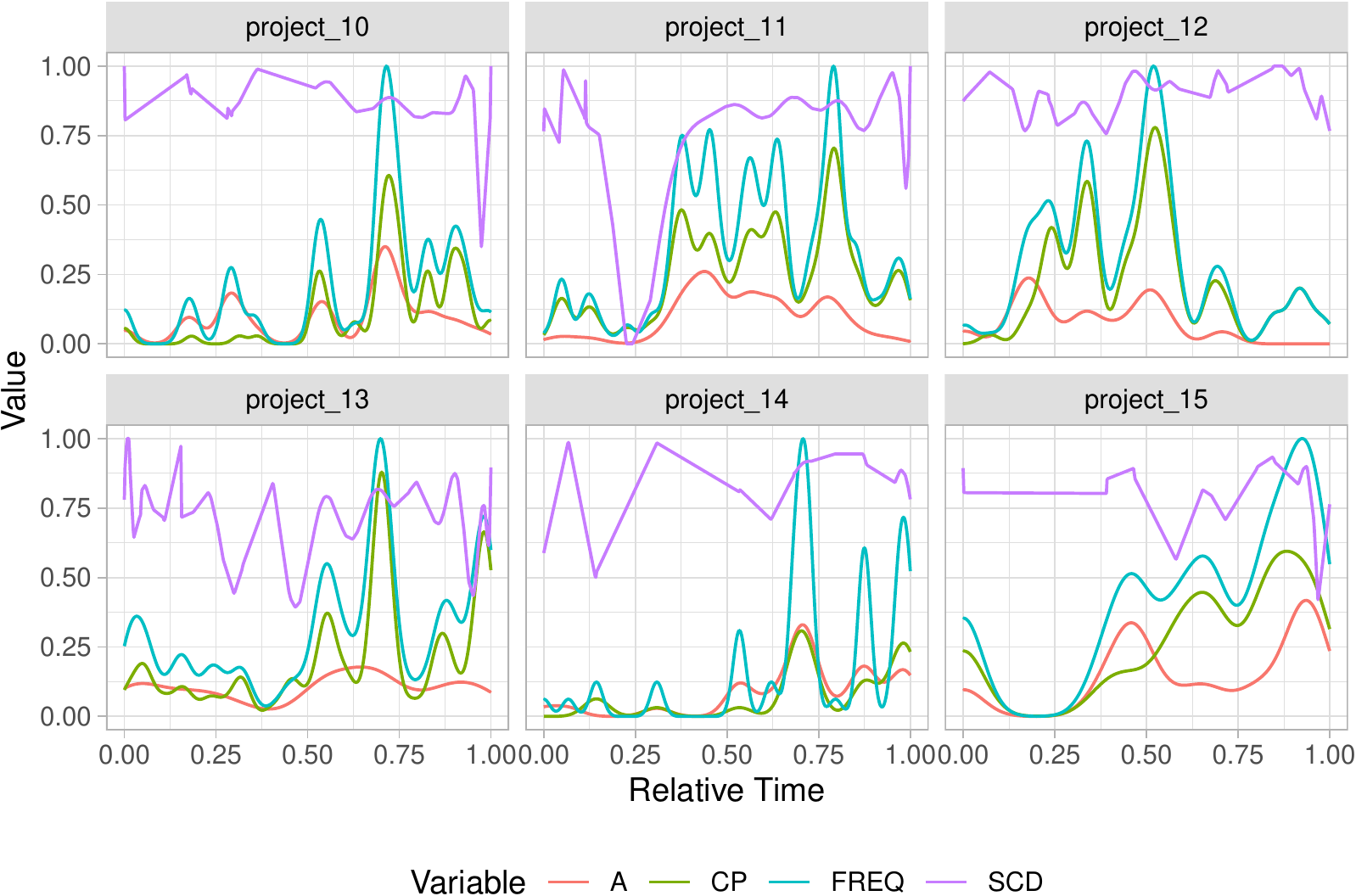} \caption{All variables over each project's time span (second batch of projects).}\label{fig:project-vars-2nd-batch}
\end{figure}

The second batch of projects and their variables are shown in figure \ref{fig:project-vars-2nd-batch}.

We will need to produce the weighted average for each variable, as we have done it previously for creating pattern type III. It needs to be redone, because we want to include the new projects.

\begin{Shaded}
\begin{Highlighting}[]
\NormalTok{p3\_avg\_signals\_all }\OtherTok{\textless{}{-}} \FunctionTok{list}\NormalTok{()}

\ControlFlowTok{for}\NormalTok{ (vartype }\ControlFlowTok{in} \FunctionTok{names}\NormalTok{(weight\_vartype)) \{}
\NormalTok{  p3\_avg\_signals\_all[[vartype]] }\OtherTok{\textless{}{-}}\NormalTok{ Signal}\SpecialCharTok{$}\FunctionTok{new}\NormalTok{(}\AttributeTok{name =} \FunctionTok{paste0}\NormalTok{(}\StringTok{"p3\_avg\_"}\NormalTok{, vartype),}
    \AttributeTok{support =} \FunctionTok{c}\NormalTok{(}\DecValTok{0}\NormalTok{, }\DecValTok{1}\NormalTok{), }\AttributeTok{isWp =} \ConstantTok{TRUE}\NormalTok{, }\AttributeTok{func =} \FunctionTok{gt\_weighted\_avg}\NormalTok{(}\AttributeTok{vartype =}\NormalTok{ vartype,}
      \AttributeTok{use\_signals =} \FunctionTok{append}\NormalTok{(project\_signals, project\_signals\_2nd\_batch), }\AttributeTok{use\_ground\_truth =} \FunctionTok{rbind}\NormalTok{(ground\_truth,}
\NormalTok{        ground\_truth\_2nd\_batch)))}
\NormalTok{\}}
\end{Highlighting}
\end{Shaded}

Next, we'll produce the lower and upper bounds vor each variable:

\begin{Shaded}
\begin{Highlighting}[]
\NormalTok{a\_ci\_upper\_p3avg }\OtherTok{\textless{}{-}} \ControlFlowTok{function}\NormalTok{(x) }\FunctionTok{CI\_bound\_p3avg}\NormalTok{(}\AttributeTok{x =}\NormalTok{ x, }\AttributeTok{funclist =}\NormalTok{ funclist\_A, }\AttributeTok{omega =}\NormalTok{ omega,}
  \AttributeTok{upper =} \ConstantTok{TRUE}\NormalTok{)}
\NormalTok{a\_ci\_lower\_p3avg }\OtherTok{\textless{}{-}} \ControlFlowTok{function}\NormalTok{(x) }\FunctionTok{CI\_bound\_p3avg}\NormalTok{(}\AttributeTok{x =}\NormalTok{ x, }\AttributeTok{funclist =}\NormalTok{ funclist\_A, }\AttributeTok{omega =}\NormalTok{ omega,}
  \AttributeTok{upper =} \ConstantTok{FALSE}\NormalTok{)}

\NormalTok{cp\_ci\_upper\_p3avg }\OtherTok{\textless{}{-}} \ControlFlowTok{function}\NormalTok{(x) }\FunctionTok{CI\_bound\_p3avg}\NormalTok{(}\AttributeTok{x =}\NormalTok{ x, }\AttributeTok{funclist =}\NormalTok{ funclist\_CP, }\AttributeTok{omega =}\NormalTok{ omega,}
  \AttributeTok{upper =} \ConstantTok{TRUE}\NormalTok{)}
\NormalTok{cp\_ci\_lower\_p3avg }\OtherTok{\textless{}{-}} \ControlFlowTok{function}\NormalTok{(x) }\FunctionTok{CI\_bound\_p3avg}\NormalTok{(}\AttributeTok{x =}\NormalTok{ x, }\AttributeTok{funclist =}\NormalTok{ funclist\_CP, }\AttributeTok{omega =}\NormalTok{ omega,}
  \AttributeTok{upper =} \ConstantTok{FALSE}\NormalTok{)}

\NormalTok{freq\_ci\_upper\_p3avg }\OtherTok{\textless{}{-}} \ControlFlowTok{function}\NormalTok{(x) }\FunctionTok{CI\_bound\_p3avg}\NormalTok{(}\AttributeTok{x =}\NormalTok{ x, }\AttributeTok{funclist =}\NormalTok{ funclist\_FREQ,}
  \AttributeTok{omega =}\NormalTok{ omega, }\AttributeTok{upper =} \ConstantTok{TRUE}\NormalTok{)}
\NormalTok{freq\_ci\_lower\_p3avg }\OtherTok{\textless{}{-}} \ControlFlowTok{function}\NormalTok{(x) }\FunctionTok{CI\_bound\_p3avg}\NormalTok{(}\AttributeTok{x =}\NormalTok{ x, }\AttributeTok{funclist =}\NormalTok{ funclist\_FREQ,}
  \AttributeTok{omega =}\NormalTok{ omega, }\AttributeTok{upper =} \ConstantTok{FALSE}\NormalTok{)}

\NormalTok{scd\_ci\_upper\_p3avg }\OtherTok{\textless{}{-}} \ControlFlowTok{function}\NormalTok{(x) }\FunctionTok{CI\_bound\_p3avg}\NormalTok{(}\AttributeTok{x =}\NormalTok{ x, }\AttributeTok{funclist =}\NormalTok{ funclist\_SCD,}
  \AttributeTok{omega =}\NormalTok{ omega, }\AttributeTok{upper =} \ConstantTok{TRUE}\NormalTok{)}
\NormalTok{scd\_ci\_lower\_p3avg }\OtherTok{\textless{}{-}} \ControlFlowTok{function}\NormalTok{(x) }\FunctionTok{CI\_bound\_p3avg}\NormalTok{(}\AttributeTok{x =}\NormalTok{ x, }\AttributeTok{funclist =}\NormalTok{ funclist\_SCD,}
  \AttributeTok{omega =}\NormalTok{ omega, }\AttributeTok{upper =} \ConstantTok{FALSE}\NormalTok{)}
\end{Highlighting}
\end{Shaded}

\begin{Shaded}
\begin{Highlighting}[]
\NormalTok{CI\_a\_p3avg }\OtherTok{\textless{}{-}} \ControlFlowTok{function}\NormalTok{(x, y) }\FunctionTok{CI\_p3avg}\NormalTok{(}\AttributeTok{x =}\NormalTok{ x, }\AttributeTok{y =}\NormalTok{ y, }\AttributeTok{funclist =}\NormalTok{ funclist\_A, }\AttributeTok{f\_ci\_upper =}\NormalTok{ a\_ci\_upper\_p3avg,}
  \AttributeTok{f\_ci\_lower =}\NormalTok{ a\_ci\_lower\_p3avg, }\AttributeTok{gbar =}\NormalTok{ p3\_avg\_signals\_all}\SpecialCharTok{$}\NormalTok{A}\SpecialCharTok{$}\FunctionTok{get0Function}\NormalTok{(), }\AttributeTok{omega =}\NormalTok{ omega)}
\NormalTok{CI\_cp\_p3avg }\OtherTok{\textless{}{-}} \ControlFlowTok{function}\NormalTok{(x, y) }\FunctionTok{CI\_p3avg}\NormalTok{(}\AttributeTok{x =}\NormalTok{ x, }\AttributeTok{y =}\NormalTok{ y, }\AttributeTok{funclist =}\NormalTok{ funclist\_CP, }\AttributeTok{f\_ci\_upper =}\NormalTok{ cp\_ci\_upper\_p3avg,}
  \AttributeTok{f\_ci\_lower =}\NormalTok{ cp\_ci\_lower\_p3avg, }\AttributeTok{gbar =}\NormalTok{ p3\_avg\_signals\_all}\SpecialCharTok{$}\NormalTok{CP}\SpecialCharTok{$}\FunctionTok{get0Function}\NormalTok{(),}
  \AttributeTok{omega =}\NormalTok{ omega)}
\NormalTok{CI\_freq\_p3avg }\OtherTok{\textless{}{-}} \ControlFlowTok{function}\NormalTok{(x, y) }\FunctionTok{CI\_p3avg}\NormalTok{(}\AttributeTok{x =}\NormalTok{ x, }\AttributeTok{y =}\NormalTok{ y, }\AttributeTok{funclist =}\NormalTok{ funclist\_FREQ,}
  \AttributeTok{f\_ci\_upper =}\NormalTok{ freq\_ci\_upper\_p3avg, }\AttributeTok{f\_ci\_lower =}\NormalTok{ freq\_ci\_lower\_p3avg, }\AttributeTok{gbar =}\NormalTok{ p3\_avg\_signals\_all}\SpecialCharTok{$}\NormalTok{FREQ}\SpecialCharTok{$}\FunctionTok{get0Function}\NormalTok{(),}
  \AttributeTok{omega =}\NormalTok{ omega)}
\NormalTok{CI\_scd\_p3avg }\OtherTok{\textless{}{-}} \ControlFlowTok{function}\NormalTok{(x, y) }\FunctionTok{CI\_p3avg}\NormalTok{(}\AttributeTok{x =}\NormalTok{ x, }\AttributeTok{y =}\NormalTok{ y, }\AttributeTok{funclist =}\NormalTok{ funclist\_SCD, }\AttributeTok{f\_ci\_upper =}\NormalTok{ scd\_ci\_upper\_p3avg,}
  \AttributeTok{f\_ci\_lower =}\NormalTok{ scd\_ci\_lower\_p3avg, }\AttributeTok{gbar =}\NormalTok{ p3\_avg\_signals\_all}\SpecialCharTok{$}\NormalTok{SCD}\SpecialCharTok{$}\FunctionTok{get0Function}\NormalTok{(),}
  \AttributeTok{omega =}\NormalTok{ omega)}

\FunctionTok{invisible}\NormalTok{(}\FunctionTok{loadResultsOrCompute}\NormalTok{(}\AttributeTok{file =} \StringTok{"../data/CI\_p3avg\_funcs\_SC.rds"}\NormalTok{, }\AttributeTok{computeExpr =}\NormalTok{ \{}
  \FunctionTok{list}\NormalTok{(}\AttributeTok{CI\_a\_p3avg =}\NormalTok{ CI\_a\_p3avg, }\AttributeTok{CI\_cp\_p3avg =}\NormalTok{ CI\_cp\_p3avg, }\AttributeTok{CI\_freq\_p3avg =}\NormalTok{ CI\_freq\_p3avg,}
    \AttributeTok{CI\_scd\_p3avg =}\NormalTok{ CI\_scd\_p3avg)}
\NormalTok{\}))}
\end{Highlighting}
\end{Shaded}

\begin{Shaded}
\begin{Highlighting}[]
\NormalTok{z\_a }\OtherTok{\textless{}{-}} \FunctionTok{loadResultsOrCompute}\NormalTok{(}\AttributeTok{file =} \StringTok{"../results/ci\_p3avg\_z\_a.rds"}\NormalTok{, }\AttributeTok{computeExpr =}\NormalTok{ \{}
  \FunctionTok{compute\_z\_p3avg}\NormalTok{(}\AttributeTok{varname =} \StringTok{"A"}\NormalTok{, }\AttributeTok{x =}\NormalTok{ x, }\AttributeTok{y =}\NormalTok{ y)}
\NormalTok{\})}
\NormalTok{z\_cp }\OtherTok{\textless{}{-}} \FunctionTok{loadResultsOrCompute}\NormalTok{(}\AttributeTok{file =} \StringTok{"../results/ci\_p3avg\_z\_cp.rds"}\NormalTok{, }\AttributeTok{computeExpr =}\NormalTok{ \{}
  \FunctionTok{compute\_z\_p3avg}\NormalTok{(}\AttributeTok{varname =} \StringTok{"CP"}\NormalTok{, }\AttributeTok{x =}\NormalTok{ x, }\AttributeTok{y =}\NormalTok{ y)}
\NormalTok{\})}
\NormalTok{z\_freq }\OtherTok{\textless{}{-}} \FunctionTok{loadResultsOrCompute}\NormalTok{(}\AttributeTok{file =} \StringTok{"../results/ci\_p3avg\_z\_freq.rds"}\NormalTok{, }\AttributeTok{computeExpr =}\NormalTok{ \{}
  \FunctionTok{compute\_z\_p3avg}\NormalTok{(}\AttributeTok{varname =} \StringTok{"FREQ"}\NormalTok{, }\AttributeTok{x =}\NormalTok{ x, }\AttributeTok{y =}\NormalTok{ y)}
\NormalTok{\})}
\NormalTok{z\_scd }\OtherTok{\textless{}{-}} \FunctionTok{loadResultsOrCompute}\NormalTok{(}\AttributeTok{file =} \StringTok{"../results/ci\_p3avg\_z\_scd.rds"}\NormalTok{, }\AttributeTok{computeExpr =}\NormalTok{ \{}
  \FunctionTok{compute\_z\_p3avg}\NormalTok{(}\AttributeTok{varname =} \StringTok{"SCD"}\NormalTok{, }\AttributeTok{x =}\NormalTok{ x, }\AttributeTok{y =}\NormalTok{ y)}
\NormalTok{\})}
\end{Highlighting}
\end{Shaded}

\begin{figure}[ht!]

{\centering \includegraphics{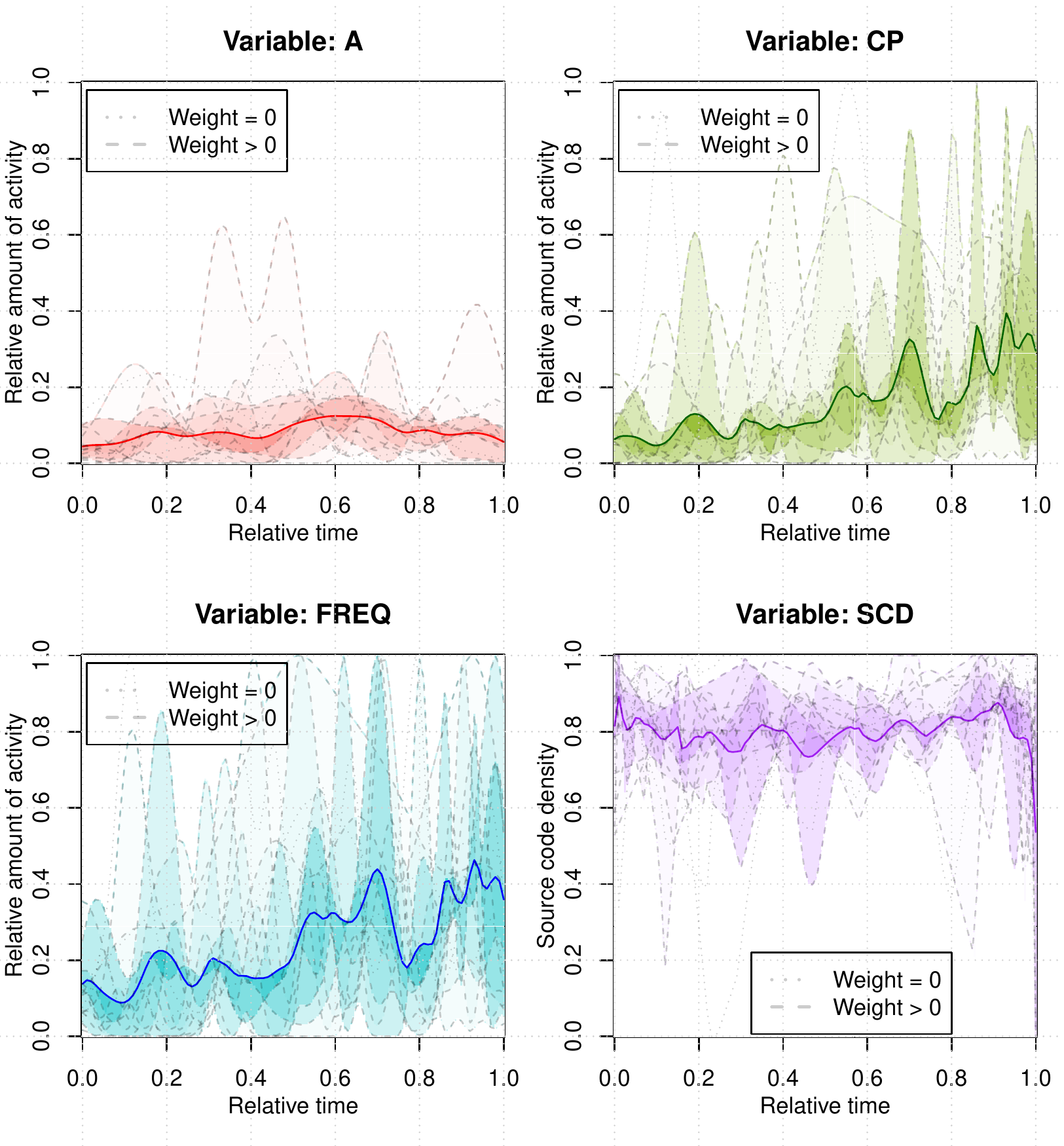} 

}

\caption{The empirical confidence intervals for the four variables as mined from the source code of all projects. Higher saturation of the color correlates with higher confidence. Projects with zero weight contribute to the CIs' boundaries, but not to the hyperplane.}\label{fig:p3-emp-cis-sc}
\end{figure}

In figure \ref{fig:p3-emp-cis-sc} we finally show the empirical confidence surfaces for all four variables, as computed over the first two batches of projects.

\hypertarget{pattern-iii-b}{%
\subsubsection{\texorpdfstring{Pattern III (b)\label{ssec:score-pattern3}}{Pattern III (b)}}\label{pattern-iii-b}}

The third and last pattern is based on the ground truth only. Starting with straight lines and equally long intervals, a pattern was generated and selected using a best trade-off between number of parameters and highest \emph{likelihood}, or by the lowest loss. Like pattern II, all of these patterns were produced with time warping applied, so that we do not need to align the projects to it. We have produced 16 patterns, so let's compute the scores for all of them.

\begin{Shaded}
\begin{Highlighting}[]
\NormalTok{p3b\_no\_scores }\OtherTok{\textless{}{-}} \FunctionTok{loadResultsOrCompute}\NormalTok{(}\AttributeTok{file =} \StringTok{"../results/p3b\_no\_scores.rds"}\NormalTok{, }\AttributeTok{computeExpr =}\NormalTok{ \{}
  \FunctionTok{unlist}\NormalTok{(}\FunctionTok{doWithParallelCluster}\NormalTok{(}\AttributeTok{numCores =} \FunctionTok{min}\NormalTok{(}\DecValTok{8}\NormalTok{, parallel}\SpecialCharTok{::}\FunctionTok{detectCores}\NormalTok{()), }\AttributeTok{expr =}\NormalTok{ \{}
\NormalTok{    foreach}\SpecialCharTok{::}\FunctionTok{foreach}\NormalTok{(}\AttributeTok{numIntervals =} \FunctionTok{seq\_len}\NormalTok{(}\AttributeTok{length.out =} \DecValTok{16}\NormalTok{), }\AttributeTok{.inorder =} \ConstantTok{FALSE}\NormalTok{) }\SpecialCharTok{\%dopar\%}
\NormalTok{      \{}
        \FunctionTok{source}\NormalTok{(}\StringTok{"./common{-}funcs.R"}\NormalTok{)}
        \FunctionTok{source}\NormalTok{(}\StringTok{"../models/modelsR6.R"}\NormalTok{)}
        \FunctionTok{source}\NormalTok{(}\StringTok{"../models/SRBTW{-}R6.R"}\NormalTok{)}

\NormalTok{        temp\_no\_align }\OtherTok{\textless{}{-}} \FunctionTok{list}\NormalTok{()}

        \ControlFlowTok{for}\NormalTok{ (project }\ControlFlowTok{in}\NormalTok{ ground\_truth}\SpecialCharTok{$}\NormalTok{project) \{}
\NormalTok{          inst }\OtherTok{\textless{}{-}}\NormalTok{ srBTAW}\SpecialCharTok{$}\FunctionTok{new}\NormalTok{(}\AttributeTok{theta\_b =} \FunctionTok{seq}\NormalTok{(}\DecValTok{0}\NormalTok{, }\DecValTok{1}\NormalTok{, }\AttributeTok{length.out =} \DecValTok{2}\NormalTok{), }\AttributeTok{gamma\_bed =} \FunctionTok{c}\NormalTok{(}\DecValTok{0}\NormalTok{,}
          \DecValTok{1}\NormalTok{, .Machine}\SpecialCharTok{$}\NormalTok{double.eps), }\AttributeTok{lambda =}\NormalTok{ .Machine}\SpecialCharTok{$}\NormalTok{double.eps, }\AttributeTok{begin =} \DecValTok{0}\NormalTok{,}
          \AttributeTok{end =} \DecValTok{1}\NormalTok{, }\AttributeTok{openBegin =} \ConstantTok{FALSE}\NormalTok{, }\AttributeTok{openEnd =} \ConstantTok{FALSE}\NormalTok{, }\AttributeTok{useAmplitudeWarping =} \ConstantTok{FALSE}\NormalTok{)}

          \ControlFlowTok{for}\NormalTok{ (vartype }\ControlFlowTok{in} \FunctionTok{names}\NormalTok{(weight\_vartype)) \{}
          \CommentTok{\# Set WP and WC:}
\NormalTok{          wp }\OtherTok{\textless{}{-}}\NormalTok{ p3b\_all[[}\FunctionTok{paste0}\NormalTok{(}\StringTok{"i\_"}\NormalTok{, numIntervals)]]}\SpecialCharTok{$}\NormalTok{signals[[vartype]]}
\NormalTok{          wc }\OtherTok{\textless{}{-}}\NormalTok{ project\_signals[[project]][[vartype]]}

\NormalTok{          inst}\SpecialCharTok{$}\FunctionTok{setSignal}\NormalTok{(}\AttributeTok{signal =}\NormalTok{ wp)}
\NormalTok{          inst}\SpecialCharTok{$}\FunctionTok{setSignal}\NormalTok{(}\AttributeTok{signal =}\NormalTok{ wc)}

\NormalTok{          srbtwbaw }\OtherTok{\textless{}{-}}\NormalTok{ inst}\SpecialCharTok{$}\NormalTok{.\_\_enclos\_env\_\_}\SpecialCharTok{$}\NormalTok{private}\SpecialCharTok{$}\FunctionTok{createInstance}\NormalTok{(}\AttributeTok{wp =}\NormalTok{ wp}\SpecialCharTok{$}\FunctionTok{get0Function}\NormalTok{(),}
            \AttributeTok{wc =}\NormalTok{ wc}\SpecialCharTok{$}\FunctionTok{get0Function}\NormalTok{())}
\NormalTok{          inst}\SpecialCharTok{$}\NormalTok{.\_\_enclos\_env\_\_}\SpecialCharTok{$}\NormalTok{private}\SpecialCharTok{$}\FunctionTok{addInstance}\NormalTok{(}\AttributeTok{instance =}\NormalTok{ srbtwbaw,}
            \AttributeTok{wpName =}\NormalTok{ wp}\SpecialCharTok{$}\FunctionTok{getName}\NormalTok{(), }\AttributeTok{wcName =}\NormalTok{ wc}\SpecialCharTok{$}\FunctionTok{getName}\NormalTok{())}
\NormalTok{          \}}

\NormalTok{          inst}\SpecialCharTok{$}\FunctionTok{setParams}\NormalTok{(}\AttributeTok{params =} \FunctionTok{c}\NormalTok{(}\AttributeTok{vtl\_1 =} \DecValTok{1}\NormalTok{))}
\NormalTok{          temp\_no\_align[[project]] }\OtherTok{\textless{}{-}}\NormalTok{ inst}
\NormalTok{        \}}

\NormalTok{        res }\OtherTok{\textless{}{-}} \StringTok{\textasciigrave{}}\AttributeTok{names\textless{}{-}}\StringTok{\textasciigrave{}}\NormalTok{(}\FunctionTok{list}\NormalTok{(}\FunctionTok{tryCatch}\NormalTok{(}\AttributeTok{expr =}\NormalTok{ \{}
          \CommentTok{\# compute\_all\_scores is already parallel!}
\NormalTok{          temp }\OtherTok{\textless{}{-}} \FunctionTok{as.data.frame}\NormalTok{(}\FunctionTok{compute\_all\_scores}\NormalTok{(}\AttributeTok{alignment =}\NormalTok{ temp\_no\_align,}
          \AttributeTok{patternName =} \StringTok{"p3"}\NormalTok{))}
          \FunctionTok{saveRDS}\NormalTok{(}\AttributeTok{object =}\NormalTok{ temp, }\AttributeTok{file =} \FunctionTok{paste0}\NormalTok{(}\StringTok{"../results/p3{-}compute/p3b\_no\_"}\NormalTok{,}
\NormalTok{          numIntervals, }\StringTok{".rds"}\NormalTok{))}
\NormalTok{          temp}
\NormalTok{        \}, }\AttributeTok{error =} \ControlFlowTok{function}\NormalTok{(cond) \{}
          \FunctionTok{list}\NormalTok{(}\AttributeTok{cond =}\NormalTok{ cond, }\AttributeTok{tb =} \FunctionTok{traceback}\NormalTok{())}
\NormalTok{        \})), }\FunctionTok{paste0}\NormalTok{(}\StringTok{"i\_"}\NormalTok{, numIntervals))}
\NormalTok{      \}}
\NormalTok{  \}), }\AttributeTok{recursive =} \ConstantTok{FALSE}\NormalTok{)}
\NormalTok{\})}
\end{Highlighting}
\end{Shaded}

\begin{figure}[ht!]
\includegraphics{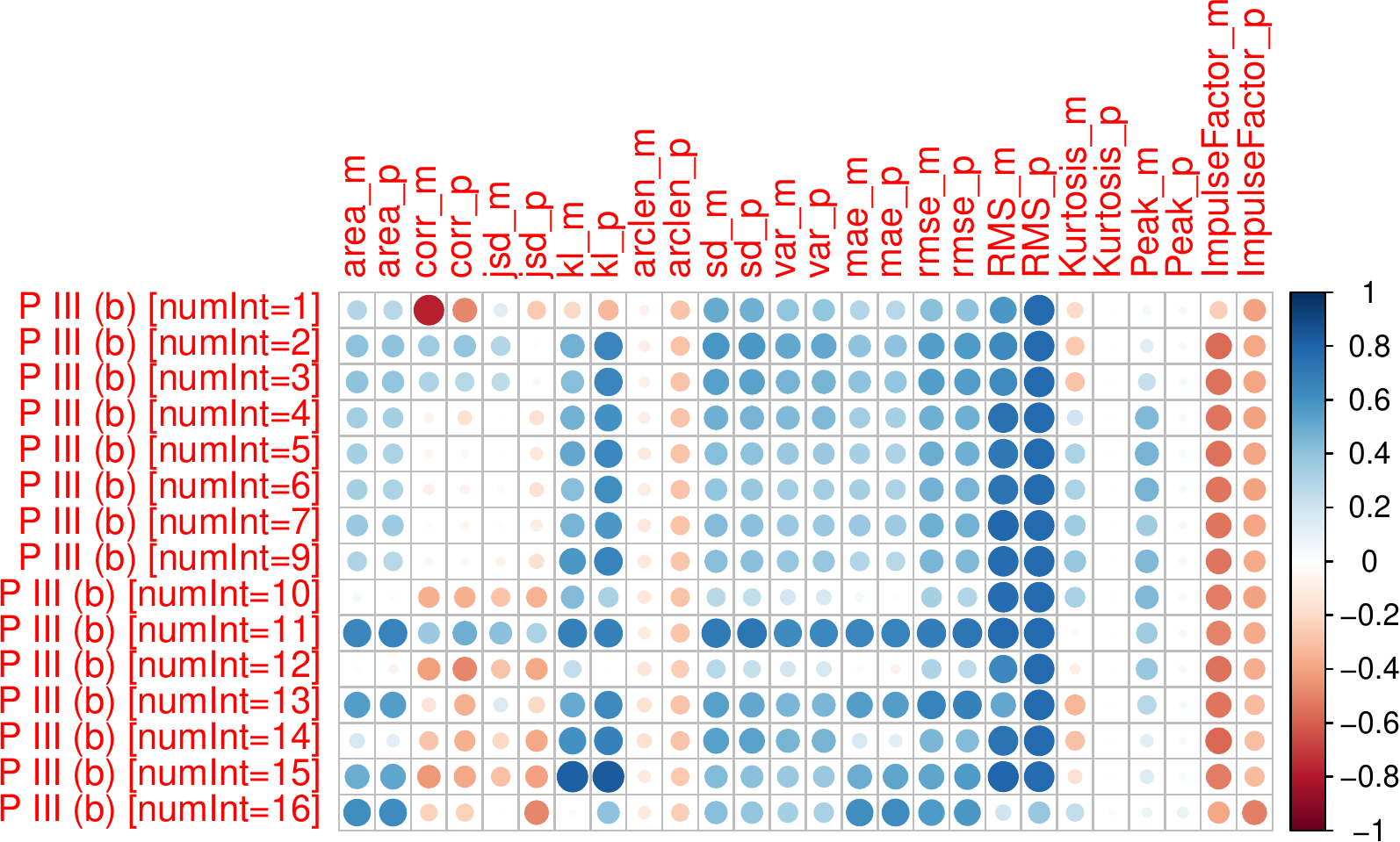} \caption{Correlation-scores for all 16 patterns (type III, b).}\label{fig:p3b-no-scores-corr}
\end{figure}

We will have a lot of results. In order to give an overview, we will show the correlation of the ground truth with the scores as obtained by scoring against each of the 16 patterns. The correlation plot in figure \ref{fig:p3b-no-scores-corr} should give us a good idea of how the scores changes with increasing number of parameters.

If \texttt{RMS} was to score to use, then the even the \(1\)-interval pattern will do, as the correlation for it is always strongly positive. In general, we can observe how some correlations get weaker, and some get stronger for patterns with higher number of parameters. There is no clear winner here, and the results are quite similar. What can say clearly is, that it is likely not worth to use highly parameterized models to detect the Fire Drill as it was present in our projects, as the manifestation is just not strong enough to warrant for patterns with high degrees of freedom. It is probably best, to use one of the other pattern types.

Let's also show an overview of the correlation with the ground truth for all of the other patterns:

\begin{figure}[ht!]
\includegraphics{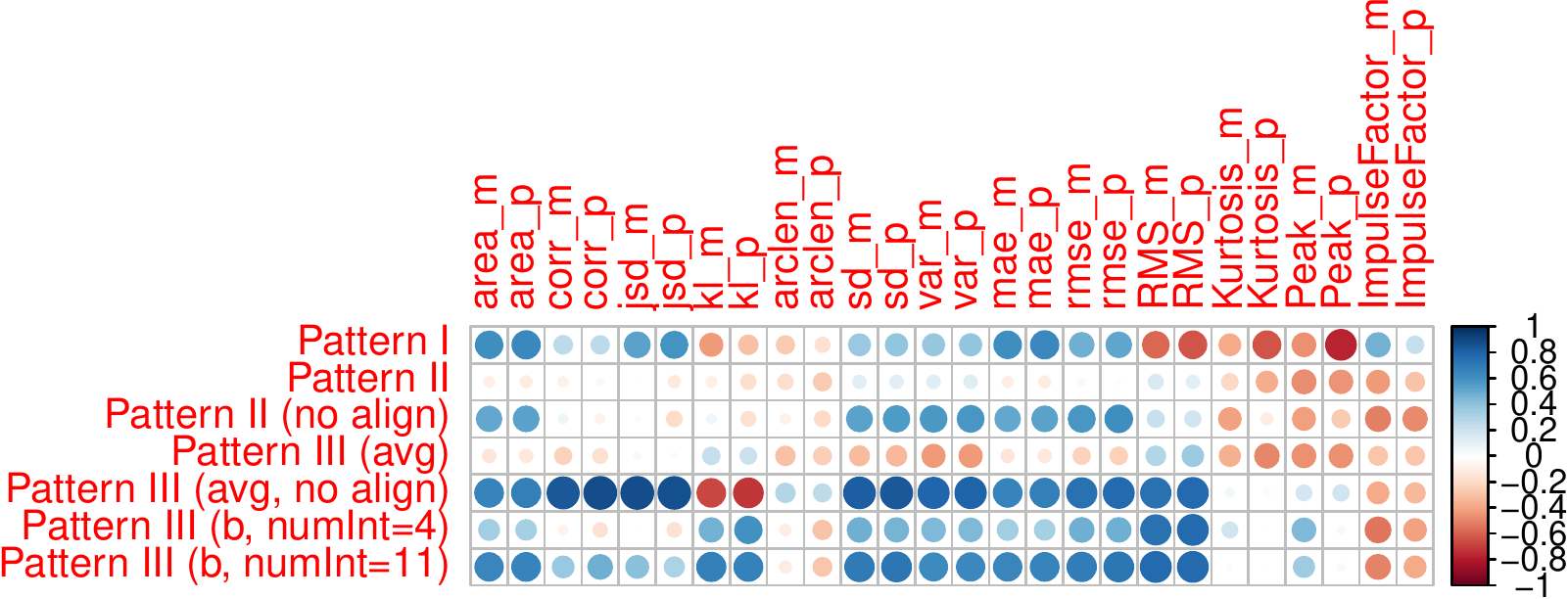} \caption{Overview of correlation-scores for all other types of patterns.}\label{fig:pall-corr}
\end{figure}

The correlation overview in figure \ref{fig:pall-corr} suggests that the no-alignment patterns have most of the strong positive-correlated scores. However, it appears that it was still worth adapting our initial pattern using the ground truth, as it is the only pattern with very high positive correlations for Peak and Impulse-factor. These however disappear, if we do not do the double warping.

\hypertarget{linear-combination-of-scores-1}{%
\paragraph{Linear combination of scores}\label{linear-combination-of-scores-1}}

The pattern that incurred the lowest loss was number \textbf{4}. We should however also test the pattern that had the highest correlations (positive \emph{or} negative) on average:

\begin{Shaded}
\begin{Highlighting}[]
\NormalTok{p3b\_corr\_all }\OtherTok{\textless{}{-}} \FunctionTok{apply}\NormalTok{(}\AttributeTok{X =}\NormalTok{ p3b\_corr, }\AttributeTok{MARGIN =} \DecValTok{1}\NormalTok{, }\AttributeTok{FUN =} \ControlFlowTok{function}\NormalTok{(row) }\FunctionTok{mean}\NormalTok{(}\FunctionTok{abs}\NormalTok{(row)))}
\end{Highlighting}
\end{Shaded}

\begin{table}

\caption{\label{tab:p3b-corr-all}Mean absolute correlation for all patterns of type III (b).}
\centering
\begin{tabular}[t]{lr}
\toprule
  & corr\\
\midrule
P III (b) [numInt=1] & 0.3303488\\
P III (b) [numInt=2] & 0.3960491\\
P III (b) [numInt=3] & 0.3836158\\
P III (b) [numInt=4] & 0.3516578\\
P III (b) [numInt=5] & 0.3421378\\
\addlinespace
P III (b) [numInt=6] & 0.3365299\\
P III (b) [numInt=7] & 0.3421056\\
P III (b) [numInt=9] & 0.3440612\\
P III (b) [numInt=10] & 0.2858276\\
P III (b) [numInt=11] & 0.4983041\\
\addlinespace
P III (b) [numInt=12] & 0.2555541\\
P III (b) [numInt=13] & 0.4151310\\
P III (b) [numInt=14] & 0.3560980\\
P III (b) [numInt=15] & 0.4242694\\
P III (b) [numInt=16] & 0.3373230\\
\bottomrule
\end{tabular}
\end{table}

In table \ref{tab:p3b-corr-all} we show the mean absolute correlation for the scores of all projects as computed against each pattern of type III (b).

\begin{Shaded}
\begin{Highlighting}[]
\NormalTok{p3b\_highest\_corr }\OtherTok{\textless{}{-}} \FunctionTok{names}\NormalTok{(}\FunctionTok{which.max}\NormalTok{(p3b\_corr\_all))}
\NormalTok{p3b\_highest\_corr}
\end{Highlighting}
\end{Shaded}

\begin{verbatim}
## [1] "P III (b) [numInt=11]"
\end{verbatim}

The pattern using \textbf{11} has the highest correlation.

Taking the best pattern of type III (b), we can also do a linear combination again. ``Best'' may refer to the pattern with the highest correlations over all scores or the one with the lowest loss when fitting to all projects.

\begin{Shaded}
\begin{Highlighting}[]
\CommentTok{\# Take the project with the highest mean absolute correlation.}
\NormalTok{temp }\OtherTok{\textless{}{-}} \FunctionTok{cbind}\NormalTok{(}\FunctionTok{data.frame}\NormalTok{(}\AttributeTok{gt\_consensus =}\NormalTok{ ground\_truth}\SpecialCharTok{$}\NormalTok{consensus\_score), p3b\_no\_scores[[}\FunctionTok{paste0}\NormalTok{(}\StringTok{"i\_"}\NormalTok{,}
  \FunctionTok{gsub}\NormalTok{(}\StringTok{"[\^{}0{-}9]"}\NormalTok{, }\StringTok{""}\NormalTok{, p3b\_highest\_corr))]])}
\NormalTok{p3b\_lm }\OtherTok{\textless{}{-}}\NormalTok{ stats}\SpecialCharTok{::}\FunctionTok{lm}\NormalTok{(}\AttributeTok{formula =}\NormalTok{ gt\_consensus }\SpecialCharTok{\textasciitilde{}}\NormalTok{ area\_p }\SpecialCharTok{+}\NormalTok{ corr\_p }\SpecialCharTok{+}\NormalTok{ jsd\_m }\SpecialCharTok{+}\NormalTok{ sd\_p }\SpecialCharTok{+}\NormalTok{ rmse\_p }\SpecialCharTok{+}
\NormalTok{  RMS\_p, }\AttributeTok{data =}\NormalTok{ temp)}
\NormalTok{stats}\SpecialCharTok{::}\FunctionTok{coef}\NormalTok{(p3b\_lm)}
\end{Highlighting}
\end{Shaded}

\begin{verbatim}
## (Intercept)      area_p      corr_p       jsd_m        sd_p      rmse_p 
##    2.616642  -38.644714    3.048282   12.414117  -18.992103   50.599212 
##       RMS_p 
##  -46.685270
\end{verbatim}

\begin{Shaded}
\begin{Highlighting}[]
\FunctionTok{plot}\NormalTok{(p3b\_lm, }\AttributeTok{ask =} \ConstantTok{FALSE}\NormalTok{, }\AttributeTok{which =} \DecValTok{1}\SpecialCharTok{:}\DecValTok{2}\NormalTok{)}
\end{Highlighting}
\end{Shaded}

\includegraphics{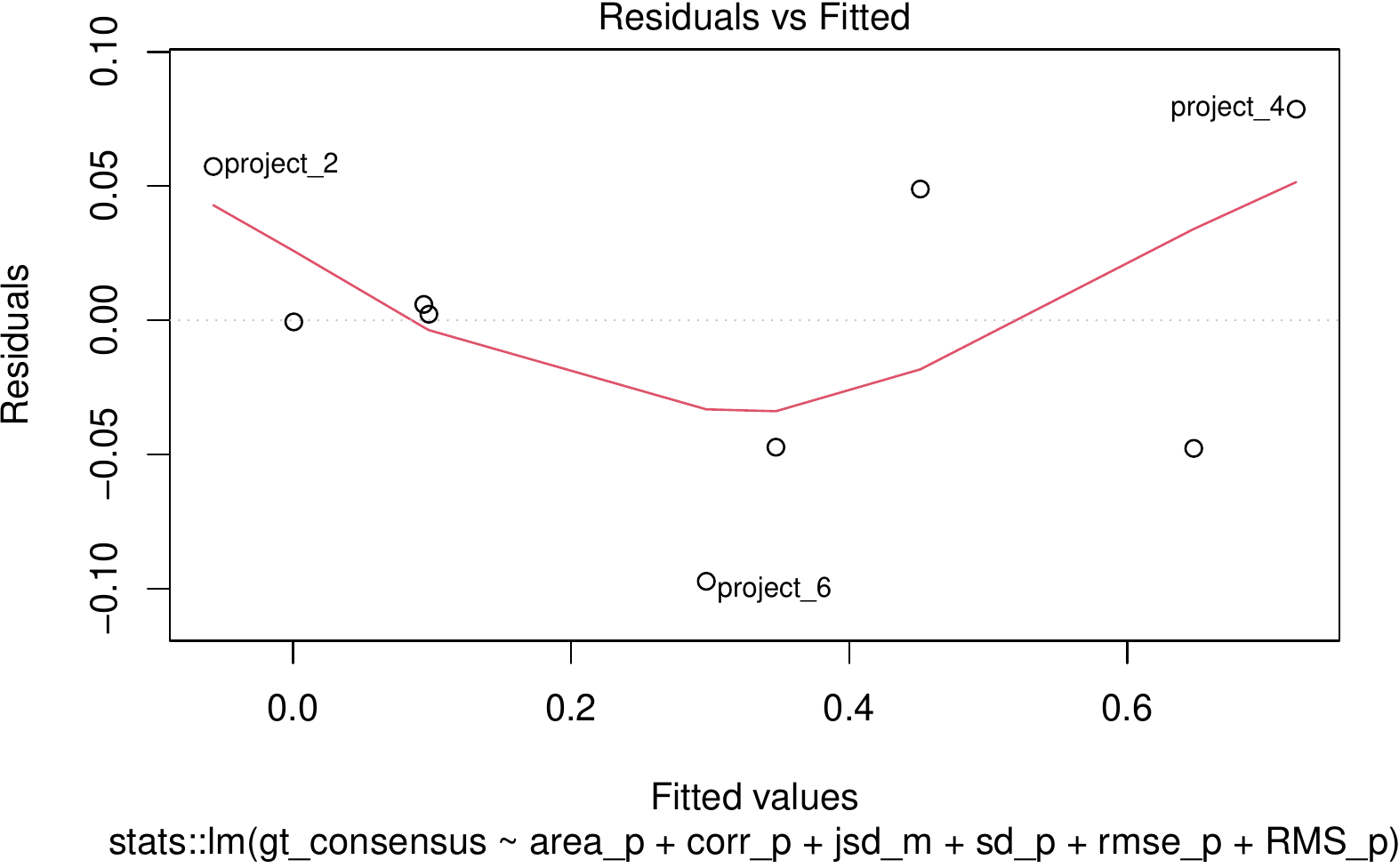} \includegraphics{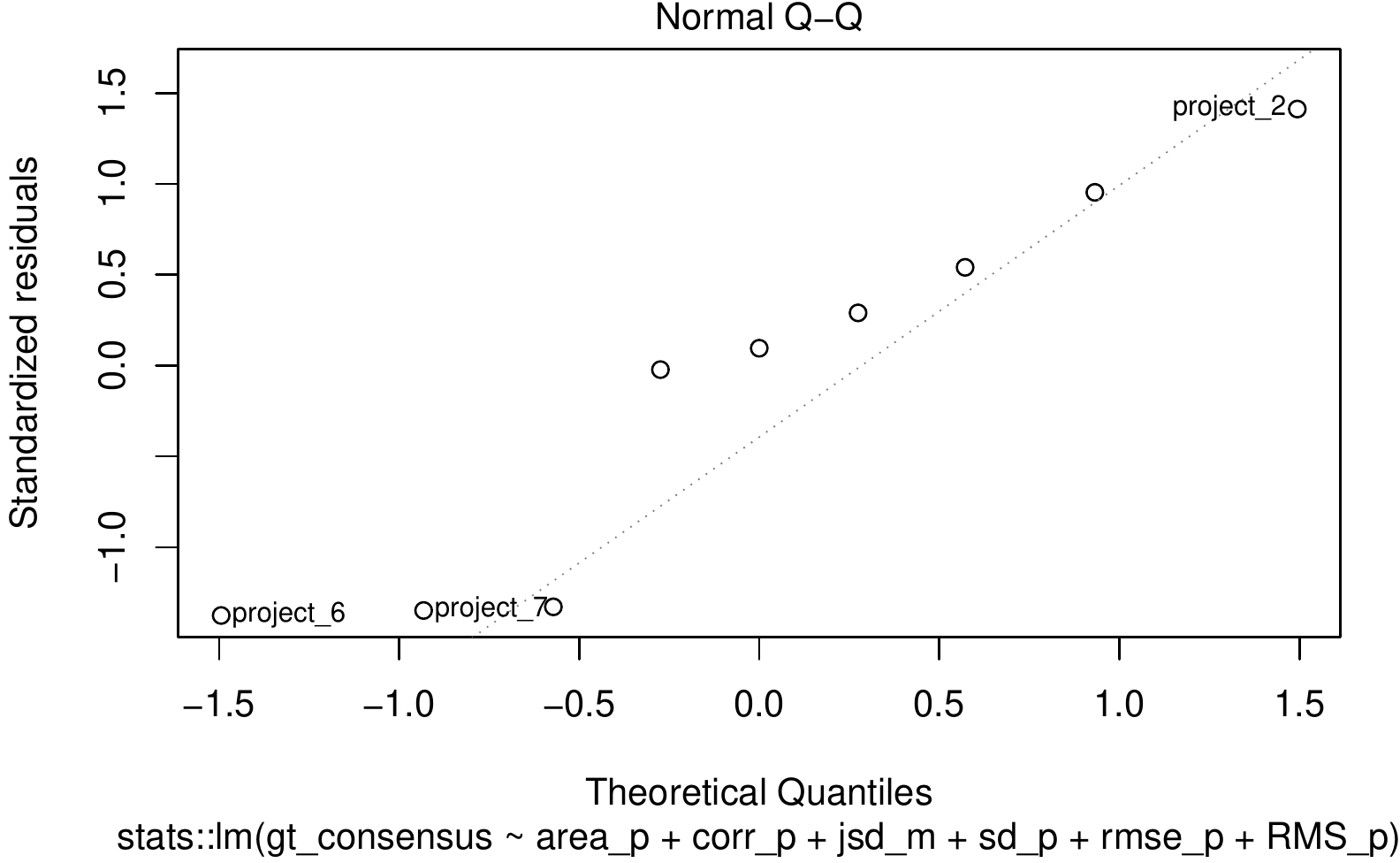}

\begin{Shaded}
\begin{Highlighting}[]
\NormalTok{p3b\_lm\_scores }\OtherTok{\textless{}{-}}\NormalTok{ stats}\SpecialCharTok{::}\FunctionTok{predict}\NormalTok{(p3b\_lm, temp)}
\FunctionTok{round}\NormalTok{(p3b\_lm\_scores }\SpecialCharTok{*} \DecValTok{10}\NormalTok{, }\DecValTok{3}\NormalTok{)}
\end{Highlighting}
\end{Shaded}

\begin{verbatim}
## project_1 project_2 project_3 project_4 project_5 project_6 project_7 project_8 
##     0.941    -0.573     6.477     7.214     0.978     2.972     3.473     0.006 
## project_9 
##     4.511
\end{verbatim}

\begin{Shaded}
\begin{Highlighting}[]
\NormalTok{stats}\SpecialCharTok{::}\FunctionTok{cor}\NormalTok{(p3b\_lm\_scores, ground\_truth}\SpecialCharTok{$}\NormalTok{consensus\_score)}
\end{Highlighting}
\end{Shaded}

\begin{verbatim}
## [1] 0.9798728
\end{verbatim}

With this linear model, we can report a high correlation with the consensus, too.

\hypertarget{automatic-calibration-of-the-continuous-process-models}{%
\subsection{\texorpdfstring{Automatic calibration of the continuous process models\label{sec:auto-calib}}{Automatic calibration of the continuous process models}}\label{automatic-calibration-of-the-continuous-process-models}}

This is a new section in the seventh iteration of this report.
``Automatic calibration'' refers to a new approach of gauging a continuous PM, so that it can compute true scores for any process, also scores that are \textbf{comparable} across PMs, given the same configuration (segments, objectives, regularizers) was used (we say that PMs using the same configuration are part of a \emph{family}).
This approach is based on the idea of \textbf{rectification of scores}\footnote{\url{https://github.com/MrShoenel/anti-pattern-models/blob/master/notebooks/rectify-score.md}}. The goal is to learn for every objective a \textbf{uniform} score, where uniform refers to both, a uniform distribution \emph{and} a codomain of \([0,1]\). With such scores, a normalizing linear scalarizer will produce a score with the \textbf{same} properties, and that means we will be able to compare the continuous process models we formulated so far!
This process has been more formalized in the meantime, since it has applications in multi-objective optimization (Hönel and Löwe 2022).
Even beyond that, we can use this score similar to a probabilistic measure of how well a process matches the PM, so in our case we can translate this information into the degree to which a Fire Drill is present in a project, according to the process model in question. We will also be able to compute the Brier- and Logarithmic scoring rules, and compare against the results that we got using the binary decision rule.

The plan for this new section is to demonstrate the following:

\begin{enumerate}
\def\labelenumi{\arabic{enumi}.}
\tightlist
\item
  Computation of available metrics/scores for all types of PMs (I, II, III (average), and III (b, numInt=11)).
\item
  Computation for each variable (e.g., FREQ), using a large number of random processes (\(\approx\) 1k--100k). Also, computing for each segment (we will be subdividing each PM into \(10\) equally long segments).
\item
  Scalarization of the scores, with and without weights (we will be able to choose a weight for each segment or score).
\item
  Perhaps make an attempt at using only the most important scores, or some automatic approach (e.g., RFE) to find the most important scores.
\item
  Computing the Brier- and Log-scoring rules so that we can compare to the binary decision rule's results.
\end{enumerate}

The reason for using segments is simple. First, the longer a segment is, the less explainability a score/metric has, as it captures too much of the characteristics. Second, we will be able to introduce weights for segments. Since it is difficult to choose weights a priori, we will only use two weighting schemes. In the first one, all weights are equal (calculate mean). In the second we will use a scheme to gradually (e.g., linearly or slightly exponentially) increase the weights towards the end of the project. This is simply based on the assumption that newer data is more important.

Example: In source code data, we have four variables. Using all \(13\) scores and ten segments requires to compute \(520\) single scores, \textbf{per} random process. Times four (for each kind of PM) equals \(2080\) computations. If we simulate \(10\)k random processes, we're looking at \(20.8\)m computations. This needs to be scaled reasonably so we get sufficiently well approximated results quickly.

\hypertarget{prerequisites-1}{%
\subsubsection{Prerequisites}\label{prerequisites-1}}

Before we can start, we should prepare this expensive task well. First is a function that takes a PM+P, a segment specification, and the name of the score to compute.

\begin{Shaded}
\begin{Highlighting}[]
\NormalTok{segment\_unit\_transform }\OtherTok{\textless{}{-}} \ControlFlowTok{function}\NormalTok{(func, from, to) \{}
\NormalTok{  ext }\OtherTok{\textless{}{-}}\NormalTok{ to }\SpecialCharTok{{-}}\NormalTok{ from}

  \ControlFlowTok{function}\NormalTok{(x) \{}
    \CommentTok{\# func has a support of [0, 1], and the returned function has a support of}
    \CommentTok{\# [from, to].}
    \FunctionTok{func}\NormalTok{(from }\SpecialCharTok{+}\NormalTok{ (x }\SpecialCharTok{*}\NormalTok{ ext))}
\NormalTok{  \}}
\NormalTok{\}}

\NormalTok{compute\_segment\_metric }\OtherTok{\textless{}{-}} \ControlFlowTok{function}\NormalTok{(PM, P, segment, }\AttributeTok{numSamples =} \DecValTok{1000}\NormalTok{, }\AttributeTok{use =} \FunctionTok{c}\NormalTok{(}\StringTok{"area"}\NormalTok{,}
  \StringTok{"corr"}\NormalTok{, }\StringTok{"jsd"}\NormalTok{, }\StringTok{"kl"}\NormalTok{, }\StringTok{"arclen"}\NormalTok{, }\StringTok{"sd"}\NormalTok{, }\StringTok{"var"}\NormalTok{, }\StringTok{"mae"}\NormalTok{, }\StringTok{"rmse"}\NormalTok{, }\StringTok{"RMS"}\NormalTok{, }\StringTok{"Kurtosis"}\NormalTok{,}
  \StringTok{"Peak"}\NormalTok{, }\StringTok{"ImpulseFactor"}\NormalTok{)) \{}

  \CommentTok{\# We\textquotesingle{}ll have to make this transform as all the scores we compute assume a}
  \CommentTok{\# support of [0,1] and compute over that.}
\NormalTok{  f1 }\OtherTok{\textless{}{-}} \FunctionTok{segment\_unit\_transform}\NormalTok{(PM, }\AttributeTok{from =}\NormalTok{ segment[}\DecValTok{1}\NormalTok{], }\AttributeTok{to =}\NormalTok{ segment[}\DecValTok{2}\NormalTok{])}
\NormalTok{  f2 }\OtherTok{\textless{}{-}} \FunctionTok{segment\_unit\_transform}\NormalTok{(P, }\AttributeTok{from =}\NormalTok{ segment[}\DecValTok{1}\NormalTok{], }\AttributeTok{to =}\NormalTok{ segment[}\DecValTok{2}\NormalTok{])}

  \CommentTok{\# Some metrics will require scaling after the segment{-}unit{-}transform}
\NormalTok{  ext }\OtherTok{\textless{}{-}}\NormalTok{ segment[}\DecValTok{2}\NormalTok{] }\SpecialCharTok{{-}}\NormalTok{ segment[}\DecValTok{1}\NormalTok{]}

\NormalTok{  diff }\OtherTok{\textless{}{-}} \ConstantTok{NA\_real\_}
  \ControlFlowTok{if}\NormalTok{ (use }\SpecialCharTok{==} \StringTok{"area"}\NormalTok{) \{}
    \CommentTok{\# This is integration and susceptible to scaling!}
\NormalTok{    diff }\OtherTok{\textless{}{-}}\NormalTok{ ext }\SpecialCharTok{*} \FunctionTok{area\_diff\_2\_functions}\NormalTok{(}\AttributeTok{f1 =}\NormalTok{ f1, }\AttributeTok{f2 =}\NormalTok{ f2)}\SpecialCharTok{$}\NormalTok{value}
\NormalTok{  \} }\ControlFlowTok{else} \ControlFlowTok{if}\NormalTok{ (use }\SpecialCharTok{==} \StringTok{"corr"}\NormalTok{) \{}
    \CommentTok{\# Returns a correlation in range [{-}1,1], but we want a difference,}
\NormalTok{    diff }\OtherTok{\textless{}{-}} \FunctionTok{stat\_diff\_2\_functions\_cor\_score}\NormalTok{(}\AttributeTok{allowReturnNA =} \ConstantTok{TRUE}\NormalTok{, }\AttributeTok{requiredSign =} \DecValTok{0}\NormalTok{,}
      \AttributeTok{numSamples =}\NormalTok{ numSamples)(}\AttributeTok{f1 =}\NormalTok{ f1, }\AttributeTok{f2 =}\NormalTok{ f2)}
    \CommentTok{\# so we\textquotesingle{}ll subtract it from 1 {-}\textgreater{} [0,2]}
\NormalTok{    diff }\OtherTok{\textless{}{-}} \DecValTok{1} \SpecialCharTok{{-}}\NormalTok{ (}\ControlFlowTok{if}\NormalTok{ (}\FunctionTok{is.na}\NormalTok{(diff))}
      \DecValTok{0} \ControlFlowTok{else}\NormalTok{ diff)  }\CommentTok{\# NA means no correlation +/{-}}
\NormalTok{  \} }\ControlFlowTok{else} \ControlFlowTok{if}\NormalTok{ (use }\SpecialCharTok{==} \StringTok{"jsd"}\NormalTok{) \{}
\NormalTok{    diff }\OtherTok{\textless{}{-}} \FunctionTok{stat\_diff\_2\_functions\_symmetric\_JSD\_sampled}\NormalTok{(}\AttributeTok{f1 =}\NormalTok{ f1, }\AttributeTok{f2 =}\NormalTok{ f2, }\AttributeTok{numSamples =}\NormalTok{ numSamples)}\SpecialCharTok{$}\NormalTok{value}
\NormalTok{  \} }\ControlFlowTok{else} \ControlFlowTok{if}\NormalTok{ (use }\SpecialCharTok{==} \StringTok{"kl"}\NormalTok{) \{}
\NormalTok{    diff }\OtherTok{\textless{}{-}} \FunctionTok{stat\_diff\_2\_functions\_symmetric\_KL\_sampled}\NormalTok{(}\AttributeTok{f1 =}\NormalTok{ f1, }\AttributeTok{f2 =}\NormalTok{ f2, }\AttributeTok{numSamples =}\NormalTok{ numSamples)}\SpecialCharTok{$}\NormalTok{value}
\NormalTok{  \} }\ControlFlowTok{else} \ControlFlowTok{if}\NormalTok{ (use }\SpecialCharTok{==} \StringTok{"arclen"}\NormalTok{) \{}
    \CommentTok{\# Returns a ration min/max {-}\textgreater{} (0,1], where 1 is ideal.  We should inverse}
    \CommentTok{\# that because we are returning a difference.}
\NormalTok{    diff }\OtherTok{\textless{}{-}} \DecValTok{1} \SpecialCharTok{{-}} \FunctionTok{stat\_diff\_2\_functions\_arclen\_score}\NormalTok{(}\AttributeTok{requiredSign =} \DecValTok{0}\NormalTok{, }\AttributeTok{numSamples =}\NormalTok{ numSamples)(}\AttributeTok{f1 =}\NormalTok{ f1,}
      \AttributeTok{f2 =}\NormalTok{ f2)}
\NormalTok{  \} }\ControlFlowTok{else} \ControlFlowTok{if}\NormalTok{ (use }\SpecialCharTok{\%in\%} \FunctionTok{c}\NormalTok{(}\StringTok{"sd"}\NormalTok{, }\StringTok{"var"}\NormalTok{, }\StringTok{"mae"}\NormalTok{, }\StringTok{"rmse"}\NormalTok{)) \{}
\NormalTok{    temp }\OtherTok{\textless{}{-}} \ControlFlowTok{switch}\NormalTok{(use, }\AttributeTok{sd =} \FunctionTok{stat\_diff\_2\_functions\_sd}\NormalTok{(}\AttributeTok{f1 =}\NormalTok{ f1, }\AttributeTok{f2 =}\NormalTok{ f2, }\AttributeTok{numSamples =}\NormalTok{ numSamples)}\SpecialCharTok{$}\NormalTok{value,}
      \AttributeTok{var =} \FunctionTok{stat\_diff\_2\_functions\_var}\NormalTok{(}\AttributeTok{f1 =}\NormalTok{ f1, }\AttributeTok{f2 =}\NormalTok{ f2, }\AttributeTok{numSamples =}\NormalTok{ numSamples)}\SpecialCharTok{$}\NormalTok{value,}
      \AttributeTok{mae =}\NormalTok{ ext }\SpecialCharTok{*} \FunctionTok{stat\_diff\_2\_functions\_mae}\NormalTok{(}\AttributeTok{f1 =}\NormalTok{ f1, }\AttributeTok{f2 =}\NormalTok{ f2, }\AttributeTok{numSamples =}\NormalTok{ numSamples)}\SpecialCharTok{$}\NormalTok{value,}
      \AttributeTok{rmse =}\NormalTok{ ext }\SpecialCharTok{*} \FunctionTok{stat\_diff\_2\_functions\_rmse}\NormalTok{(}\AttributeTok{f1 =}\NormalTok{ f1, }\AttributeTok{f2 =}\NormalTok{ f2, }\AttributeTok{numSamples =}\NormalTok{ numSamples)}\SpecialCharTok{$}\NormalTok{value,}
\NormalTok{      \{}
        \FunctionTok{stop}\NormalTok{(}\FunctionTok{paste0}\NormalTok{(}\StringTok{"Don\textquotesingle{}t know "}\NormalTok{, use, }\StringTok{"."}\NormalTok{))}
\NormalTok{      \})}
\NormalTok{    diff }\OtherTok{\textless{}{-}}\NormalTok{ temp}
\NormalTok{  \} }\ControlFlowTok{else} \ControlFlowTok{if}\NormalTok{ (use }\SpecialCharTok{\%in\%} \FunctionTok{c}\NormalTok{(}\StringTok{"RMS"}\NormalTok{, }\StringTok{"Kurtosis"}\NormalTok{, }\StringTok{"Peak"}\NormalTok{, }\StringTok{"ImpulseFactor"}\NormalTok{)) \{}
    \CommentTok{\# Returns a ratio min/max {-}\textgreater{} (0,1], where 1 is ideal.}
\NormalTok{    diff }\OtherTok{\textless{}{-}} \DecValTok{1} \SpecialCharTok{{-}} \FunctionTok{stat\_diff\_2\_functions\_signals\_score}\NormalTok{(}\AttributeTok{use =}\NormalTok{ use, }\AttributeTok{requiredSign =} \DecValTok{0}\NormalTok{,}
      \AttributeTok{numSamples =}\NormalTok{ numSamples)(}\AttributeTok{f1 =}\NormalTok{ f1, }\AttributeTok{f2 =}\NormalTok{ f2)}
\NormalTok{  \}}

\NormalTok{  diff}
\NormalTok{\}}
\end{Highlighting}
\end{Shaded}

We have previously used a function to get a smooth random process. We will be slightly adapting this function here to also include the process' derivative. We will not be working with derivative PMs in this notebook, but the very next when we use issue-tracking data.

It is important to note that we do not have any expectation as to how such a random process will look like, except for the fact that we expect it to run within the bounds of the process model, that is, we allow any random process to ``behave'' arbitrarily within the bounds of \([0,1]\).
In reality, however, we probably would have some expectation (or at least stricter limitations or some confidence intervals) as to how some process may look/behave.
In that case, it is perhaps better to construct a \emph{generative model} from these expectations, and then to uniformly sample from its quantile function (PPF) in order to generate random processes that match our expectations and are within some limitations.
However, that would require that we have such expectations. Those could come from another set of observations (not the observed projects), or we could perhaps craft some expert-designed expectations by reasoning about potential limits and confidence intervals, or by simply designing a set of plausible processes.
However here, we took the road where we do not have any of these, and thus allow any process to behave absolutely random within the bounds of our PMs. In other words, our expectation is that we do not have an expectation, and hence we randomly sample uniformly from all possibilities.

\begin{Shaded}
\begin{Highlighting}[]
\NormalTok{get\_smoothed\_curve }\OtherTok{\textless{}{-}} \ControlFlowTok{function}\NormalTok{(}\AttributeTok{seed =} \ConstantTok{NA}\NormalTok{, }\AttributeTok{npoints =} \DecValTok{15}\NormalTok{, }\AttributeTok{include\_deriv =} \ConstantTok{FALSE}\NormalTok{) \{}
  \ControlFlowTok{if}\NormalTok{ (}\SpecialCharTok{!}\FunctionTok{is.na}\NormalTok{(seed)) \{}
    \FunctionTok{set.seed}\NormalTok{(}\AttributeTok{seed =}\NormalTok{ seed)}
\NormalTok{  \}}

\NormalTok{  x }\OtherTok{\textless{}{-}} \FunctionTok{sort}\NormalTok{(}\FunctionTok{c}\NormalTok{(}\DecValTok{0}\NormalTok{, }\DecValTok{1}\NormalTok{, }\FunctionTok{runif}\NormalTok{(npoints }\SpecialCharTok{{-}} \DecValTok{2}\NormalTok{)))}
\NormalTok{  y }\OtherTok{\textless{}{-}} \FunctionTok{runif}\NormalTok{(}\FunctionTok{length}\NormalTok{(x))}
\NormalTok{  temp }\OtherTok{\textless{}{-}} \FunctionTok{loess.smooth}\NormalTok{(}\AttributeTok{x =}\NormalTok{ x, }\AttributeTok{y =}\NormalTok{ y, }\AttributeTok{span =} \FloatTok{0.35}\NormalTok{, }\AttributeTok{family =} \StringTok{"g"}\NormalTok{, }\AttributeTok{evaluation =} \DecValTok{1000}\NormalTok{)}
\NormalTok{  appr }\OtherTok{\textless{}{-}}\NormalTok{ stats}\SpecialCharTok{::}\FunctionTok{approxfun}\NormalTok{(}\AttributeTok{x =}\NormalTok{ ((temp}\SpecialCharTok{$}\NormalTok{x }\SpecialCharTok{{-}} \FunctionTok{min}\NormalTok{(temp}\SpecialCharTok{$}\NormalTok{x))}\SpecialCharTok{/}\NormalTok{(}\FunctionTok{max}\NormalTok{(temp}\SpecialCharTok{$}\NormalTok{x) }\SpecialCharTok{{-}} \FunctionTok{min}\NormalTok{(temp}\SpecialCharTok{$}\NormalTok{x))),}
    \AttributeTok{y =}\NormalTok{ temp}\SpecialCharTok{$}\NormalTok{y}\SpecialCharTok{/}\NormalTok{(}\FunctionTok{max}\NormalTok{(temp}\SpecialCharTok{$}\NormalTok{x) }\SpecialCharTok{{-}} \FunctionTok{min}\NormalTok{(temp}\SpecialCharTok{$}\NormalTok{x)), }\AttributeTok{yleft =}\NormalTok{ utils}\SpecialCharTok{::}\FunctionTok{head}\NormalTok{(temp}\SpecialCharTok{$}\NormalTok{y, }\DecValTok{1}\NormalTok{), }\AttributeTok{yright =}\NormalTok{ utils}\SpecialCharTok{::}\FunctionTok{tail}\NormalTok{(temp}\SpecialCharTok{$}\NormalTok{y,}
      \DecValTok{1}\NormalTok{))}

\NormalTok{  tempf }\OtherTok{\textless{}{-}} \FunctionTok{Vectorize}\NormalTok{(}\ControlFlowTok{function}\NormalTok{(x) \{}
    \CommentTok{\# Limit the resulting function to the bounding box of [0,1]}
    \FunctionTok{min}\NormalTok{(}\DecValTok{1}\NormalTok{, }\FunctionTok{max}\NormalTok{(}\DecValTok{0}\NormalTok{, }\FunctionTok{appr}\NormalTok{(x)))}
\NormalTok{  \})}

\NormalTok{  f1 }\OtherTok{\textless{}{-}} \ConstantTok{NULL}
  \ControlFlowTok{if}\NormalTok{ (include\_deriv) \{}
\NormalTok{    tempf1 }\OtherTok{\textless{}{-}} \FunctionTok{Vectorize}\NormalTok{(}\ControlFlowTok{function}\NormalTok{(x) \{}
\NormalTok{      ltol }\OtherTok{\textless{}{-}}\NormalTok{ x }\SpecialCharTok{\textless{}} \FunctionTok{sqrt}\NormalTok{(.Machine}\SpecialCharTok{$}\NormalTok{double.eps)}
\NormalTok{      rtol }\OtherTok{\textless{}{-}}\NormalTok{ x }\SpecialCharTok{\textgreater{}} \DecValTok{1} \SpecialCharTok{{-}} \FunctionTok{sqrt}\NormalTok{(.Machine}\SpecialCharTok{$}\NormalTok{double.eps)}
\NormalTok{      pracma}\SpecialCharTok{::}\FunctionTok{fderiv}\NormalTok{(}\AttributeTok{f =}\NormalTok{ tempf, }\AttributeTok{x =}\NormalTok{ x, }\AttributeTok{method =} \ControlFlowTok{if}\NormalTok{ (ltol)}
        \StringTok{"forward"} \ControlFlowTok{else} \ControlFlowTok{if}\NormalTok{ (rtol)}
        \StringTok{"backward"} \ControlFlowTok{else} \StringTok{"central"}\NormalTok{)}
\NormalTok{    \})}
\NormalTok{    x }\OtherTok{\textless{}{-}} \FunctionTok{seq}\NormalTok{(}\AttributeTok{from =} \DecValTok{0}\NormalTok{, }\AttributeTok{to =} \DecValTok{1}\NormalTok{, }\AttributeTok{length.out =} \DecValTok{2000}\NormalTok{)}
\NormalTok{    y }\OtherTok{\textless{}{-}} \FunctionTok{tempf1}\NormalTok{(x)}

\NormalTok{    f1 }\OtherTok{=}\NormalTok{ stats}\SpecialCharTok{::}\FunctionTok{approxfun}\NormalTok{(}\AttributeTok{x =}\NormalTok{ x, }\AttributeTok{y =}\NormalTok{ y, }\AttributeTok{yleft =}\NormalTok{ utils}\SpecialCharTok{::}\FunctionTok{head}\NormalTok{(y, }\DecValTok{1}\NormalTok{), }\AttributeTok{yright =}\NormalTok{ utils}\SpecialCharTok{::}\FunctionTok{tail}\NormalTok{(y,}
      \DecValTok{1}\NormalTok{))}
\NormalTok{  \}}

  \FunctionTok{list}\NormalTok{(}\AttributeTok{f0 =}\NormalTok{ tempf, }\AttributeTok{f1 =}\NormalTok{ f1)}
\NormalTok{\}}
\end{Highlighting}
\end{Shaded}

\hypertarget{computing-the-metrics}{%
\subsubsection{Computing the metrics}\label{computing-the-metrics}}

We have a handful of dimensions, and the goal is to obtain a data frame with the following columns:

\begin{itemize}
\tightlist
\item
  Process model (ID or name)
\item
  Variable name
\item
  Seed of random process
\item
  Segment index
\item
  Metric 1, \ldots, Metric 13 (one \emph{named} column per metric)
\end{itemize}

Each worker shall compute all metrics for one specific segment. It will therefore be given the PM's variable, the seed/random process, and the segment's index. This way, we can chunk each worker's workload into comfortably sized items.

\hypertarget{worker}{%
\paragraph{Worker}\label{worker}}

We will be making a grid of all combinations, and the worker will compute the results of that grid. Then finally, we will just be concatenating the grid with the results horizontally and store the result.

\begin{Shaded}
\begin{Highlighting}[]
\NormalTok{ac\_worker }\OtherTok{\textless{}{-}} \ControlFlowTok{function}\NormalTok{(PM, P, seg\_idx, }\AttributeTok{total\_segments =} \DecValTok{10}\NormalTok{) \{}
\NormalTok{  use }\OtherTok{\textless{}{-}} \FunctionTok{c}\NormalTok{(}\StringTok{"area"}\NormalTok{, }\StringTok{"corr"}\NormalTok{, }\StringTok{"jsd"}\NormalTok{, }\StringTok{"kl"}\NormalTok{, }\StringTok{"arclen"}\NormalTok{, }\StringTok{"sd"}\NormalTok{, }\StringTok{"var"}\NormalTok{, }\StringTok{"mae"}\NormalTok{, }\StringTok{"rmse"}\NormalTok{, }\StringTok{"RMS"}\NormalTok{,}
    \StringTok{"Kurtosis"}\NormalTok{, }\StringTok{"Peak"}\NormalTok{, }\StringTok{"ImpulseFactor"}\NormalTok{)}
\NormalTok{  df }\OtherTok{\textless{}{-}} \StringTok{\textasciigrave{}}\AttributeTok{colnames\textless{}{-}}\StringTok{\textasciigrave{}}\NormalTok{(}\FunctionTok{matrix}\NormalTok{(}\AttributeTok{nrow =} \DecValTok{1}\NormalTok{, }\AttributeTok{ncol =} \FunctionTok{length}\NormalTok{(use)), use)}

\NormalTok{  seg\_ext }\OtherTok{\textless{}{-}} \DecValTok{1}\SpecialCharTok{/}\NormalTok{total\_segments}

  \ControlFlowTok{for}\NormalTok{ (metric }\ControlFlowTok{in}\NormalTok{ use) \{}
\NormalTok{    df[}\DecValTok{1}\NormalTok{, metric] }\OtherTok{\textless{}{-}} \FunctionTok{compute\_segment\_metric}\NormalTok{(}\AttributeTok{PM =}\NormalTok{ PM, }\AttributeTok{P =}\NormalTok{ P, }\AttributeTok{use =}\NormalTok{ metric, }\AttributeTok{segment =} \FunctionTok{c}\NormalTok{((seg\_idx }\SpecialCharTok{{-}}
      \DecValTok{1}\NormalTok{) }\SpecialCharTok{*}\NormalTok{ seg\_ext, seg\_idx }\SpecialCharTok{*}\NormalTok{ seg\_ext))}
\NormalTok{  \}}
\NormalTok{  df}
\NormalTok{\}}
\end{Highlighting}
\end{Shaded}

\hypertarget{grid-for-the-random-processes}{%
\paragraph{Grid for the random processes}\label{grid-for-the-random-processes}}

Let's define the grid of parameters, which will eventually be concatenated with the results.

\begin{Shaded}
\begin{Highlighting}[]
\NormalTok{ac\_grid }\OtherTok{\textless{}{-}} \FunctionTok{loadResultsOrCompute}\NormalTok{(}\AttributeTok{file =} \StringTok{"../results/ac\_grid\_sc.rds"}\NormalTok{, }\AttributeTok{computeExpr =}\NormalTok{ \{}
  \FunctionTok{expand.grid}\NormalTok{(}\FunctionTok{list}\NormalTok{(}\AttributeTok{PM =} \FunctionTok{c}\NormalTok{(}\StringTok{"I"}\NormalTok{, }\StringTok{"II"}\NormalTok{, }\StringTok{"III(avg)"}\NormalTok{, }\StringTok{"III(b,11)"}\NormalTok{), }\AttributeTok{Var =} \FunctionTok{names}\NormalTok{(weight\_vartype),}
    \AttributeTok{Seed =} \FunctionTok{seq}\NormalTok{(}\AttributeTok{from =} \DecValTok{1}\NormalTok{, }\AttributeTok{length.out =} \DecValTok{10000}\NormalTok{)))}
\NormalTok{\})}

\FunctionTok{nrow}\NormalTok{(ac\_grid)}
\end{Highlighting}
\end{Shaded}

\begin{verbatim}
## [1] 160000
\end{verbatim}

Let's compute the grid (attention: this is expensive).

\begin{Shaded}
\begin{Highlighting}[]
\NormalTok{ac\_grid\_results }\OtherTok{\textless{}{-}} \FunctionTok{loadResultsOrCompute}\NormalTok{(}\AttributeTok{file =} \StringTok{"../results/ac\_grid\_results\_sc.rds"}\NormalTok{,}
  \AttributeTok{computeExpr =}\NormalTok{ \{}
    \FunctionTok{library}\NormalTok{(foreach)}

\NormalTok{    p3b11 }\OtherTok{\textless{}{-}}\NormalTok{ p3b\_all}\SpecialCharTok{$}\NormalTok{i\_11}

    \FunctionTok{doWithParallelCluster}\NormalTok{(}\AttributeTok{numCores =} \FunctionTok{min}\NormalTok{(}\DecValTok{123}\NormalTok{, parallel}\SpecialCharTok{::}\FunctionTok{detectCores}\NormalTok{()), }\AttributeTok{expr =}\NormalTok{ \{}
\NormalTok{      pb }\OtherTok{\textless{}{-}}\NormalTok{ utils}\SpecialCharTok{::}\FunctionTok{txtProgressBar}\NormalTok{(}\AttributeTok{min =} \DecValTok{1}\NormalTok{, }\AttributeTok{max =} \FunctionTok{nrow}\NormalTok{(ac\_grid), }\AttributeTok{style =} \DecValTok{3}\NormalTok{)}
\NormalTok{      progress }\OtherTok{\textless{}{-}} \ControlFlowTok{function}\NormalTok{(n) \{}
        \ControlFlowTok{if}\NormalTok{ (}\DecValTok{0} \SpecialCharTok{==}\NormalTok{ (n}\SpecialCharTok{\%\%}\DecValTok{25}\NormalTok{)) \{}
          \FunctionTok{print}\NormalTok{(n)}
\NormalTok{        \}}
\NormalTok{        utils}\SpecialCharTok{::}\FunctionTok{setTxtProgressBar}\NormalTok{(}\AttributeTok{pb =}\NormalTok{ pb, }\AttributeTok{value =}\NormalTok{ n)}
\NormalTok{      \}}

\NormalTok{      foreach}\SpecialCharTok{::}\FunctionTok{foreach}\NormalTok{(}\AttributeTok{grid\_idx =} \FunctionTok{rownames}\NormalTok{(ac\_grid), }\AttributeTok{.combine =}\NormalTok{ rbind, }\AttributeTok{.inorder =} \ConstantTok{FALSE}\NormalTok{,}
        \AttributeTok{.verbose =} \ConstantTok{TRUE}\NormalTok{, }\AttributeTok{.options.snow =} \FunctionTok{list}\NormalTok{(}\AttributeTok{progress =}\NormalTok{ progress)) }\SpecialCharTok{\%dopar\%}
\NormalTok{        \{}
          \FunctionTok{options}\NormalTok{(}\AttributeTok{warn =} \DecValTok{2}\NormalTok{)}
\NormalTok{          params }\OtherTok{\textless{}{-}}\NormalTok{ ac\_grid[grid\_idx, ]}
\NormalTok{          params}\SpecialCharTok{$}\NormalTok{PM }\OtherTok{\textless{}{-}} \FunctionTok{as.character}\NormalTok{(params}\SpecialCharTok{$}\NormalTok{PM)}
\NormalTok{          params}\SpecialCharTok{$}\NormalTok{Var }\OtherTok{\textless{}{-}} \FunctionTok{as.character}\NormalTok{(params}\SpecialCharTok{$}\NormalTok{Var)}

\NormalTok{          pm\_var }\OtherTok{\textless{}{-}} \ControlFlowTok{switch}\NormalTok{(params}\SpecialCharTok{$}\NormalTok{PM, }\AttributeTok{I =}\NormalTok{ p1\_signals, }\AttributeTok{II =}\NormalTok{ p2\_signals, }\StringTok{\textasciigrave{}}\AttributeTok{III(avg)}\StringTok{\textasciigrave{}} \OtherTok{=}\NormalTok{ p3\_avg\_signals,}
          \StringTok{\textasciigrave{}}\AttributeTok{III(b,11)}\StringTok{\textasciigrave{}} \OtherTok{=}\NormalTok{ p3b11}\SpecialCharTok{$}\NormalTok{signals, \{}
            \FunctionTok{stop}\NormalTok{(}\FunctionTok{paste0}\NormalTok{(}\StringTok{"Don\textquotesingle{}t know "}\NormalTok{, params}\SpecialCharTok{$}\NormalTok{PM, }\StringTok{"."}\NormalTok{))}
\NormalTok{          \})[[params}\SpecialCharTok{$}\NormalTok{Var]]}\SpecialCharTok{$}\FunctionTok{get0Function}\NormalTok{()}

\NormalTok{          p\_var }\OtherTok{\textless{}{-}} \FunctionTok{get\_smoothed\_curve}\NormalTok{(}\AttributeTok{seed =}\NormalTok{ params}\SpecialCharTok{$}\NormalTok{Seed, }\AttributeTok{include\_deriv =} \ConstantTok{FALSE}\NormalTok{)}\SpecialCharTok{$}\NormalTok{f0}

\NormalTok{          df }\OtherTok{\textless{}{-}} \ConstantTok{NULL}
          \ControlFlowTok{for}\NormalTok{ (seg\_idx }\ControlFlowTok{in} \DecValTok{1}\SpecialCharTok{:}\DecValTok{10}\NormalTok{) \{}
          \CommentTok{\# Important so we can correctly concatenate the results with the grid}
          \CommentTok{\# parameters!}
\NormalTok{          res }\OtherTok{\textless{}{-}} \FunctionTok{as.data.frame}\NormalTok{(}\FunctionTok{ac\_worker}\NormalTok{(}\AttributeTok{PM =}\NormalTok{ pm\_var, }\AttributeTok{P =}\NormalTok{ p\_var, }\AttributeTok{seg\_idx =}\NormalTok{ seg\_idx,}
            \AttributeTok{total\_segments =} \DecValTok{10}\NormalTok{))}
\NormalTok{          res}\SpecialCharTok{$}\NormalTok{GridIdx }\OtherTok{\textless{}{-}}\NormalTok{ grid\_idx}
\NormalTok{          res}\SpecialCharTok{$}\NormalTok{SegIdx }\OtherTok{\textless{}{-}}\NormalTok{ seg\_idx}
\NormalTok{          res }\OtherTok{\textless{}{-}} \FunctionTok{cbind}\NormalTok{(res, params)}
\NormalTok{          df }\OtherTok{\textless{}{-}} \ControlFlowTok{if}\NormalTok{ (}\FunctionTok{is.null}\NormalTok{(df))}
\NormalTok{            res }\ControlFlowTok{else} \FunctionTok{rbind}\NormalTok{(df, res)}
\NormalTok{          \}}
\NormalTok{          df}
\NormalTok{        \}}
\NormalTok{    \})}
\NormalTok{  \})}
\end{Highlighting}
\end{Shaded}

\hypertarget{approximating-marginal-cumulative-densities}{%
\subsubsection{\texorpdfstring{Approximating marginal cumulative densities\label{ssec:ac-approx-marginal-ecdfs}}{Approximating marginal cumulative densities}}\label{approximating-marginal-cumulative-densities}}

Now with the data at hand, the goal is to approximate the empirical cumulative distribution function (ECDF) for each objective, in each segment, for each variable, for each process model. For example, we will have one ECDF for the correlation of the A-variable in the first segment of the type-III (average) pattern.
All these ``address'' the specific ECDF. We will first write a helper function that can extract this.

\begin{Shaded}
\begin{Highlighting}[]
\NormalTok{ac\_extract\_data }\OtherTok{\textless{}{-}} \ControlFlowTok{function}\NormalTok{(pmName, varName, segIdx, metricName, ac\_grid, ac\_grid\_results) \{}
\NormalTok{  rows }\OtherTok{\textless{}{-}}\NormalTok{ ac\_grid[ac\_grid}\SpecialCharTok{$}\NormalTok{PM }\SpecialCharTok{==}\NormalTok{ pmName }\SpecialCharTok{\&}\NormalTok{ ac\_grid}\SpecialCharTok{$}\NormalTok{Var }\SpecialCharTok{==}\NormalTok{ varName, ]}
\NormalTok{  ac\_grid\_results[ac\_grid\_results}\SpecialCharTok{$}\NormalTok{SegIdx }\SpecialCharTok{==}\NormalTok{ segIdx }\SpecialCharTok{\&}\NormalTok{ ac\_grid\_results}\SpecialCharTok{$}\NormalTok{GridIdx }\SpecialCharTok{\%in\%}
    \FunctionTok{rownames}\NormalTok{(rows), metricName]}
\NormalTok{\}}
\end{Highlighting}
\end{Shaded}

So let's plot a few randomly picked examples:

\begin{figure}
\centering
\includegraphics{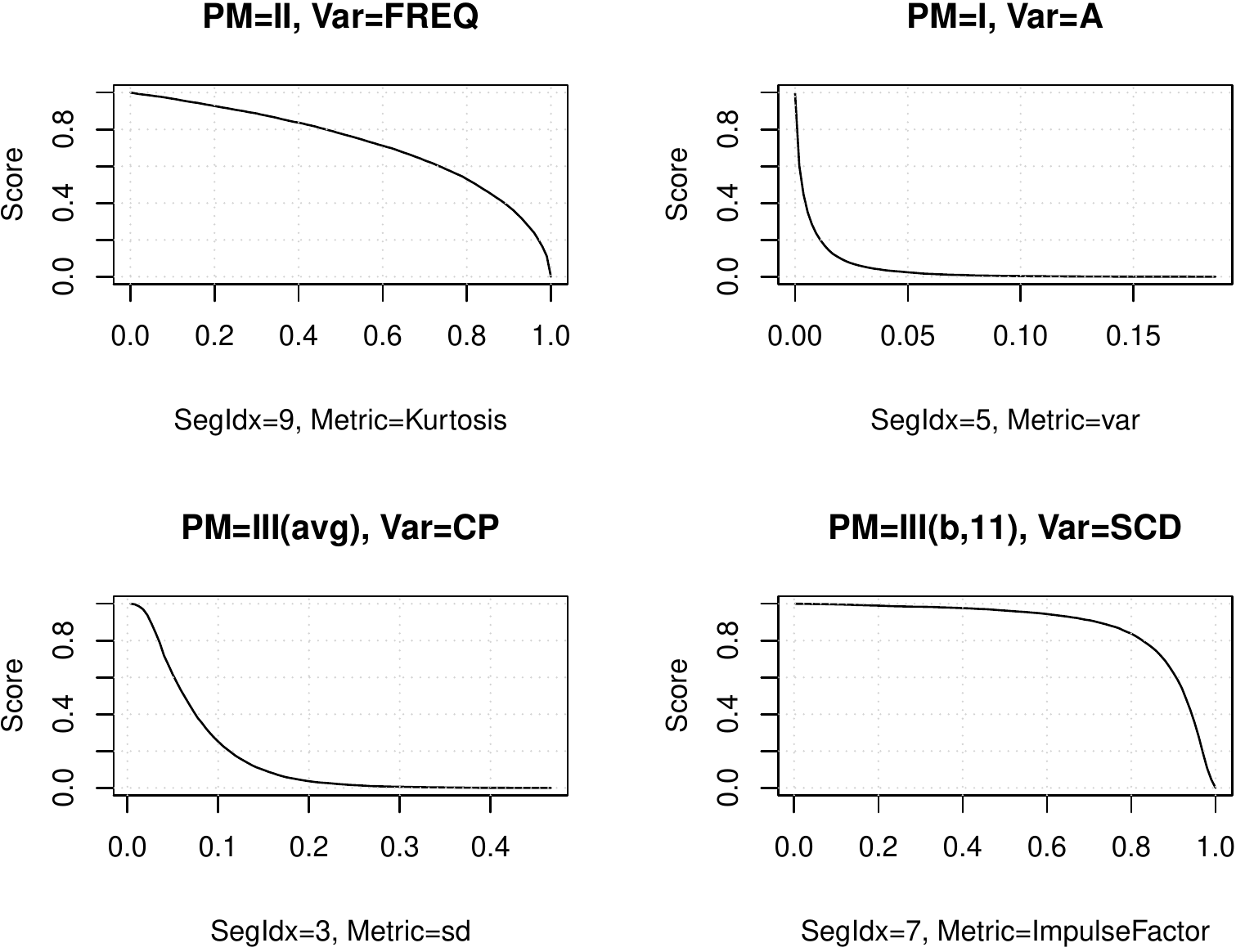}
\caption{\label{fig:ac-grid-example}Four randomly picked ECCDFs as simulated using the random processes of the automatic calibration.}
\end{figure}

So as we see in figure \ref{fig:ac-grid-example}, the ECCDFs can be quite diverse. Increasing distances are now non-linearly correlated with lower scores.
In probably all cases, the ECDF is \textbf{non-linear}, which is exactly what we are after, as it will then be used to rectify the corresponding objective later.
Below is a table of the last shown ECCDF (metric \texttt{ImpulseFactor} on variable \texttt{SCD} in segment \(7\) in process model type III(b, numInt=11)), where we linearly increase the distance by steps of \(0.05\), while the score clearly decreases non-linearly, especially towards the end.

\begin{Shaded}
\begin{Highlighting}[]
\NormalTok{tempf1 }\OtherTok{\textless{}{-}} \ControlFlowTok{function}\NormalTok{(x) }\DecValTok{1} \SpecialCharTok{{-}} \FunctionTok{tempf}\NormalTok{(x)}
\StringTok{\textasciigrave{}}\AttributeTok{names\textless{}{-}}\StringTok{\textasciigrave{}}\NormalTok{(}\FunctionTok{tempf1}\NormalTok{(}\FunctionTok{seq}\NormalTok{(}\DecValTok{0}\NormalTok{, }\FloatTok{0.5}\NormalTok{, }\AttributeTok{by =} \FloatTok{0.05}\NormalTok{)), }\FunctionTok{paste0}\NormalTok{(}\FunctionTok{seq}\NormalTok{(}\DecValTok{0}\NormalTok{, }\FloatTok{0.5}\NormalTok{, }\AttributeTok{by =} \FloatTok{0.05}\NormalTok{)))}
\end{Highlighting}
\end{Shaded}

\begin{verbatim}
##      0   0.05    0.1   0.15    0.2   0.25    0.3   0.35    0.4   0.45    0.5 
## 1.0000 0.9982 0.9962 0.9930 0.9898 0.9858 0.9836 0.9800 0.9762 0.9709 0.9632
\end{verbatim}

\begin{Shaded}
\begin{Highlighting}[]
\StringTok{\textasciigrave{}}\AttributeTok{names\textless{}{-}}\StringTok{\textasciigrave{}}\NormalTok{(}\FunctionTok{tempf1}\NormalTok{(}\FunctionTok{seq}\NormalTok{(}\FloatTok{0.55}\NormalTok{, }\DecValTok{1}\NormalTok{, }\AttributeTok{by =} \FloatTok{0.05}\NormalTok{)), }\FunctionTok{paste0}\NormalTok{(}\FunctionTok{seq}\NormalTok{(}\FloatTok{0.55}\NormalTok{, }\DecValTok{1}\NormalTok{, }\AttributeTok{by =} \FloatTok{0.05}\NormalTok{)))}
\end{Highlighting}
\end{Shaded}

\begin{verbatim}
##   0.55    0.6   0.65    0.7   0.75    0.8   0.85    0.9   0.95      1 
## 0.9549 0.9442 0.9298 0.9105 0.8814 0.8383 0.7622 0.6251 0.3557 0.0000
\end{verbatim}

\hypertarget{normality-and-uniformity-tests-of-the-objectives}{%
\paragraph{Normality and uniformity tests of the objectives}\label{normality-and-uniformity-tests-of-the-objectives}}

In fact, let's check if any of the scores is normally or uniformly distributed.
We will do three tests, the Shapiro--Wilk (Shapiro and Wilk 1965a) and the Kolmogorov--Smirnov (Stephens 1974) tests for normality, and the Kolmogorov--Smirnov test ofr uniformity.

\begin{Shaded}
\begin{Highlighting}[]
\NormalTok{ac\_distr\_test\_sc }\OtherTok{\textless{}{-}} \FunctionTok{loadResultsOrCompute}\NormalTok{(}\AttributeTok{file =} \StringTok{"../results/ac\_distr\_test\_sc.rds"}\NormalTok{,}
  \AttributeTok{computeExpr =}\NormalTok{ \{}
    \FunctionTok{library}\NormalTok{(foreach)}

\NormalTok{    temp.grid }\OtherTok{\textless{}{-}} \FunctionTok{expand.grid}\NormalTok{(}\FunctionTok{list}\NormalTok{(}\AttributeTok{PM =} \FunctionTok{levels}\NormalTok{(ac\_grid}\SpecialCharTok{$}\NormalTok{PM), }\AttributeTok{Var =} \FunctionTok{unique}\NormalTok{(ac\_grid}\SpecialCharTok{$}\NormalTok{Var),}
      \AttributeTok{Seg =} \DecValTok{1}\SpecialCharTok{:}\DecValTok{10}\NormalTok{, }\AttributeTok{Metric =} \FunctionTok{c}\NormalTok{(}\StringTok{"area"}\NormalTok{, }\StringTok{"corr"}\NormalTok{, }\StringTok{"jsd"}\NormalTok{, }\StringTok{"kl"}\NormalTok{, }\StringTok{"arclen"}\NormalTok{, }\StringTok{"sd"}\NormalTok{, }\StringTok{"var"}\NormalTok{,}
        \StringTok{"mae"}\NormalTok{, }\StringTok{"rmse"}\NormalTok{, }\StringTok{"RMS"}\NormalTok{, }\StringTok{"Kurtosis"}\NormalTok{, }\StringTok{"Peak"}\NormalTok{, }\StringTok{"ImpulseFactor"}\NormalTok{)))}

    \FunctionTok{as.data.frame}\NormalTok{(}\FunctionTok{doWithParallelCluster}\NormalTok{(}\AttributeTok{numCores =} \FunctionTok{min}\NormalTok{(parallel}\SpecialCharTok{::}\FunctionTok{detectCores}\NormalTok{(),}
      \DecValTok{32}\NormalTok{), }\AttributeTok{expr =}\NormalTok{ \{}
\NormalTok{      foreach}\SpecialCharTok{::}\FunctionTok{foreach}\NormalTok{(}\AttributeTok{rn =} \FunctionTok{rownames}\NormalTok{(temp.grid), }\AttributeTok{.combine =}\NormalTok{ rbind, }\AttributeTok{.inorder =} \ConstantTok{FALSE}\NormalTok{) }\SpecialCharTok{\%dopar\%}
\NormalTok{        \{}
          \FunctionTok{set.seed}\NormalTok{(}\DecValTok{1}\NormalTok{)}
\NormalTok{          row }\OtherTok{\textless{}{-}}\NormalTok{ temp.grid[rn, ]}
\NormalTok{          temp }\OtherTok{\textless{}{-}} \FunctionTok{ac\_extract\_data}\NormalTok{(}\AttributeTok{pmName =}\NormalTok{ row}\SpecialCharTok{$}\NormalTok{PM, }\AttributeTok{varName =}\NormalTok{ row}\SpecialCharTok{$}\NormalTok{Var, }\AttributeTok{segIdx =}\NormalTok{ row}\SpecialCharTok{$}\NormalTok{Seg,}
          \AttributeTok{metricName =}\NormalTok{ row}\SpecialCharTok{$}\NormalTok{Metric, }\AttributeTok{ac\_grid =}\NormalTok{ ac\_grid, }\AttributeTok{ac\_grid\_results =}\NormalTok{ ac\_grid\_results)}

\NormalTok{          samp }\OtherTok{\textless{}{-}} \FunctionTok{sample}\NormalTok{(}\AttributeTok{x =}\NormalTok{ temp, }\AttributeTok{size =} \DecValTok{5000}\NormalTok{)}
\NormalTok{          st }\OtherTok{\textless{}{-}} \ControlFlowTok{if}\NormalTok{ (}\FunctionTok{length}\NormalTok{(}\FunctionTok{unique}\NormalTok{(samp)) }\SpecialCharTok{==} \DecValTok{1}\NormalTok{)}
          \FunctionTok{list}\NormalTok{(}\AttributeTok{statistic =} \DecValTok{0}\NormalTok{, }\AttributeTok{p.value =} \DecValTok{0}\NormalTok{) }\ControlFlowTok{else} \FunctionTok{shapiro.test}\NormalTok{(samp)}

\NormalTok{          m }\OtherTok{\textless{}{-}} \FunctionTok{mean}\NormalTok{(temp)}
\NormalTok{          s }\OtherTok{\textless{}{-}} \FunctionTok{sd}\NormalTok{(temp)}
\NormalTok{          distFun\_norm }\OtherTok{\textless{}{-}} \ControlFlowTok{function}\NormalTok{(q) \{}
          \FunctionTok{pnorm}\NormalTok{(}\AttributeTok{q =}\NormalTok{ q, }\AttributeTok{mean =}\NormalTok{ m, }\AttributeTok{sd =}\NormalTok{ s)}
\NormalTok{          \}}
\NormalTok{          ks\_norm }\OtherTok{\textless{}{-}} \FunctionTok{ks.test}\NormalTok{(}\AttributeTok{x =}\NormalTok{ temp, }\AttributeTok{y =}\NormalTok{ distFun\_norm)}

\NormalTok{          ab }\OtherTok{\textless{}{-}} \FunctionTok{range}\NormalTok{(temp)}
\NormalTok{          distFun\_unif }\OtherTok{\textless{}{-}} \ControlFlowTok{function}\NormalTok{(q) \{}
          \FunctionTok{punif}\NormalTok{(}\AttributeTok{q =}\NormalTok{ q, }\AttributeTok{min =}\NormalTok{ ab[}\DecValTok{1}\NormalTok{], }\AttributeTok{max =}\NormalTok{ ab[}\DecValTok{2}\NormalTok{])}
\NormalTok{          \}}
\NormalTok{          ks\_unif }\OtherTok{\textless{}{-}} \FunctionTok{ks.test}\NormalTok{(}\AttributeTok{x =}\NormalTok{ temp, }\AttributeTok{y =}\NormalTok{ distFun\_unif)}

          \StringTok{\textasciigrave{}}\AttributeTok{colnames\textless{}{-}}\StringTok{\textasciigrave{}}\NormalTok{(}\AttributeTok{x =} \FunctionTok{matrix}\NormalTok{(}\AttributeTok{data =} \FunctionTok{c}\NormalTok{(st}\SpecialCharTok{$}\NormalTok{statistic, st}\SpecialCharTok{$}\NormalTok{p.value, ks\_norm}\SpecialCharTok{$}\NormalTok{statistic,}
\NormalTok{          ks\_norm}\SpecialCharTok{$}\NormalTok{p.value, ks\_unif}\SpecialCharTok{$}\NormalTok{statistic, ks\_unif}\SpecialCharTok{$}\NormalTok{p.value), }\AttributeTok{nrow =} \DecValTok{1}\NormalTok{),}
          \AttributeTok{value =} \FunctionTok{c}\NormalTok{(}\StringTok{"shap.W"}\NormalTok{, }\StringTok{"shap.pval"}\NormalTok{, }\StringTok{"ks\_norm.D"}\NormalTok{, }\StringTok{"ks\_norm.pval"}\NormalTok{,}
            \StringTok{"ks\_unif.D"}\NormalTok{, }\StringTok{"ks\_unif.pval"}\NormalTok{))}
\NormalTok{        \}}
\NormalTok{    \}))}
\NormalTok{  \})}
\end{Highlighting}
\end{Shaded}

Out of 2080 ECDFs, \textbf{3} are normally distributed, and \textbf{0} are uniformly distributed.
The highest p-values for either normally test are \ensuremath{3.293422\times 10^{-9}} and 0.1002497. The highest W-statistic was 0.9966326, and the lowest D-statistic was 0.0122334.
The highest p-value for the standard uniform test was 0.004937, and the lowest D-statistic was 0.0173265.
It is fair to say that nothing here is normally or uniformly distributed.
We now have reasonable evidence that justifies that conversion to linear behaving scores is required.

\hypertarget{calculating-scores}{%
\subsubsection{Calculating scores}\label{calculating-scores}}

In the previous sections we have simulated random continuous processes in order to approximate the marginal densities for each and every single objective.
Now we want to find out how each of our observed projects scores. So the first step is to compute these in the \emph{same} way as we did the random processes (we can reuse the function \texttt{ac\_worker()}), and then the next step will be to \emph{rectify} each score using the previously approximated densities. Only after that, we can continue to evaluate the PMs and see which one is best, and where the strong and weak parts are.

In order to calculate each project's scores, we need to calculate one objective per:

\begin{itemize}
\tightlist
\item
  Project (15),
\item
  Process model (4),
\item
  Variable (4),
\item
  Segment index (10), and
\item
  Metric (13) {[}this is not part of the grid, the \texttt{ac\_worker()} does this{]}.
\end{itemize}

That will leave us with \(15\times 4\times 4\times 10\times 13=31,200\) computations. It's probably best to generate a grid for that:

\begin{Shaded}
\begin{Highlighting}[]
\NormalTok{ac\_grid\_projects }\OtherTok{\textless{}{-}} \FunctionTok{expand.grid}\NormalTok{(}\FunctionTok{list}\NormalTok{(}\AttributeTok{Project =} \FunctionTok{c}\NormalTok{(}\FunctionTok{names}\NormalTok{(project\_signals), }\FunctionTok{names}\NormalTok{(project\_signals\_2nd\_batch)),}
  \AttributeTok{PM =} \FunctionTok{c}\NormalTok{(}\StringTok{"I"}\NormalTok{, }\StringTok{"II"}\NormalTok{, }\StringTok{"III(avg)"}\NormalTok{, }\StringTok{"III(b,11)"}\NormalTok{), }\AttributeTok{Var =} \FunctionTok{names}\NormalTok{(weight\_vartype), }\AttributeTok{SegIdx =} \DecValTok{1}\SpecialCharTok{:}\DecValTok{10}\NormalTok{))}

\FunctionTok{nrow}\NormalTok{(ac\_grid\_projects)}
\end{Highlighting}
\end{Shaded}

\begin{verbatim}
## [1] 2400
\end{verbatim}

Now we can compute the projects' scores:

\begin{Shaded}
\begin{Highlighting}[]
\NormalTok{ac\_grid\_projects\_results }\OtherTok{\textless{}{-}} \FunctionTok{loadResultsOrCompute}\NormalTok{(}\AttributeTok{file =} \StringTok{"../results/ac\_grid\_projects\_results\_sc.rds"}\NormalTok{,}
  \AttributeTok{computeExpr =}\NormalTok{ \{}
    \FunctionTok{library}\NormalTok{(foreach)}

\NormalTok{    p3b11 }\OtherTok{\textless{}{-}}\NormalTok{ p3b\_all}\SpecialCharTok{$}\NormalTok{i\_11}
\NormalTok{    project\_signals\_all }\OtherTok{\textless{}{-}} \FunctionTok{append}\NormalTok{(project\_signals, project\_signals\_2nd\_batch)}

    \FunctionTok{doWithParallelCluster}\NormalTok{(}\AttributeTok{numCores =} \FunctionTok{min}\NormalTok{(}\DecValTok{123}\NormalTok{, parallel}\SpecialCharTok{::}\FunctionTok{detectCores}\NormalTok{()), }\AttributeTok{expr =}\NormalTok{ \{}
\NormalTok{      pb }\OtherTok{\textless{}{-}}\NormalTok{ utils}\SpecialCharTok{::}\FunctionTok{txtProgressBar}\NormalTok{(}\AttributeTok{min =} \DecValTok{1}\NormalTok{, }\AttributeTok{max =} \FunctionTok{nrow}\NormalTok{(ac\_grid\_projects), }\AttributeTok{style =} \DecValTok{3}\NormalTok{)}
\NormalTok{      progress }\OtherTok{\textless{}{-}} \ControlFlowTok{function}\NormalTok{(n) \{}
        \ControlFlowTok{if}\NormalTok{ (}\DecValTok{0} \SpecialCharTok{==}\NormalTok{ (n}\SpecialCharTok{\%\%}\DecValTok{25}\NormalTok{)) \{}
          \FunctionTok{print}\NormalTok{(n)}
\NormalTok{        \}}
\NormalTok{        utils}\SpecialCharTok{::}\FunctionTok{setTxtProgressBar}\NormalTok{(}\AttributeTok{pb =}\NormalTok{ pb, }\AttributeTok{value =}\NormalTok{ n)}
\NormalTok{      \}}

\NormalTok{      foreach}\SpecialCharTok{::}\FunctionTok{foreach}\NormalTok{(}\AttributeTok{grid\_idx =} \FunctionTok{rownames}\NormalTok{(ac\_grid\_projects), }\AttributeTok{.combine =}\NormalTok{ rbind,}
        \AttributeTok{.inorder =} \ConstantTok{FALSE}\NormalTok{, }\AttributeTok{.verbose =} \ConstantTok{TRUE}\NormalTok{, }\AttributeTok{.options.snow =} \FunctionTok{list}\NormalTok{(}\AttributeTok{progress =}\NormalTok{ progress)) }\SpecialCharTok{\%dopar\%}
\NormalTok{        \{}
          \FunctionTok{options}\NormalTok{(}\AttributeTok{warn =} \DecValTok{2}\NormalTok{)}
\NormalTok{          params }\OtherTok{\textless{}{-}}\NormalTok{ ac\_grid\_projects[grid\_idx, ]}
\NormalTok{          params}\SpecialCharTok{$}\NormalTok{Project }\OtherTok{\textless{}{-}} \FunctionTok{as.character}\NormalTok{(params}\SpecialCharTok{$}\NormalTok{Project)}
\NormalTok{          params}\SpecialCharTok{$}\NormalTok{PM }\OtherTok{\textless{}{-}} \FunctionTok{as.character}\NormalTok{(params}\SpecialCharTok{$}\NormalTok{PM)}
\NormalTok{          params}\SpecialCharTok{$}\NormalTok{Var }\OtherTok{\textless{}{-}} \FunctionTok{as.character}\NormalTok{(params}\SpecialCharTok{$}\NormalTok{Var)}

\NormalTok{          pm\_var }\OtherTok{\textless{}{-}} \ControlFlowTok{switch}\NormalTok{(params}\SpecialCharTok{$}\NormalTok{PM, }\AttributeTok{I =}\NormalTok{ p1\_signals, }\AttributeTok{II =}\NormalTok{ p2\_signals, }\StringTok{\textasciigrave{}}\AttributeTok{III(avg)}\StringTok{\textasciigrave{}} \OtherTok{=}\NormalTok{ p3\_avg\_signals,}
          \StringTok{\textasciigrave{}}\AttributeTok{III(b,11)}\StringTok{\textasciigrave{}} \OtherTok{=}\NormalTok{ p3b11}\SpecialCharTok{$}\NormalTok{signals, \{}
            \FunctionTok{stop}\NormalTok{(}\FunctionTok{paste0}\NormalTok{(}\StringTok{"Don\textquotesingle{}t know "}\NormalTok{, params}\SpecialCharTok{$}\NormalTok{PM, }\StringTok{"."}\NormalTok{))}
\NormalTok{          \})[[params}\SpecialCharTok{$}\NormalTok{Var]]}\SpecialCharTok{$}\FunctionTok{get0Function}\NormalTok{()}

\NormalTok{          p\_var }\OtherTok{\textless{}{-}}\NormalTok{ project\_signals\_all[[params}\SpecialCharTok{$}\NormalTok{Project]][[params}\SpecialCharTok{$}\NormalTok{Var]]}\SpecialCharTok{$}\FunctionTok{get0Function}\NormalTok{()}

          \CommentTok{\# Important so we can correctly concatenate the results with the grid}
          \CommentTok{\# parameters!}
\NormalTok{          res }\OtherTok{\textless{}{-}} \FunctionTok{as.data.frame}\NormalTok{(}\FunctionTok{ac\_worker}\NormalTok{(}\AttributeTok{PM =}\NormalTok{ pm\_var, }\AttributeTok{P =}\NormalTok{ p\_var, }\AttributeTok{seg\_idx =}\NormalTok{ params}\SpecialCharTok{$}\NormalTok{SegIdx,}
          \AttributeTok{total\_segments =} \DecValTok{10}\NormalTok{))}

\NormalTok{          res}\SpecialCharTok{$}\NormalTok{grid\_idx }\OtherTok{\textless{}{-}}\NormalTok{ grid\_idx}
          \FunctionTok{cbind}\NormalTok{(res, params)}
\NormalTok{        \}}
\NormalTok{    \})}
\NormalTok{  \})}
\end{Highlighting}
\end{Shaded}

\hypertarget{rectification-of-raw-scores}{%
\paragraph{Rectification of raw scores}\label{rectification-of-raw-scores}}

Now that we have the raw scores, it is time to transform them using the marginal densities. Each single objective (ECDF) is reused 15 times (once for each project).
We will make a new temporary grid, where the number of rows in this grid corresponds to the number of ECDFs:

\begin{Shaded}
\begin{Highlighting}[]
\NormalTok{temp.Metric }\OtherTok{\textless{}{-}} \FunctionTok{c}\NormalTok{(}\StringTok{"area"}\NormalTok{, }\StringTok{"corr"}\NormalTok{, }\StringTok{"jsd"}\NormalTok{, }\StringTok{"kl"}\NormalTok{, }\StringTok{"arclen"}\NormalTok{, }\StringTok{"sd"}\NormalTok{, }\StringTok{"var"}\NormalTok{, }\StringTok{"mae"}\NormalTok{, }\StringTok{"rmse"}\NormalTok{,}
  \StringTok{"RMS"}\NormalTok{, }\StringTok{"Kurtosis"}\NormalTok{, }\StringTok{"Peak"}\NormalTok{, }\StringTok{"ImpulseFactor"}\NormalTok{)}
\NormalTok{temp.grid }\OtherTok{\textless{}{-}} \FunctionTok{expand.grid}\NormalTok{(}\FunctionTok{list}\NormalTok{(}\AttributeTok{PM =} \FunctionTok{c}\NormalTok{(}\StringTok{"I"}\NormalTok{, }\StringTok{"II"}\NormalTok{, }\StringTok{"III(avg)"}\NormalTok{, }\StringTok{"III(b,11)"}\NormalTok{), }\AttributeTok{Var =} \FunctionTok{names}\NormalTok{(weight\_vartype),}
  \AttributeTok{SegIdx =} \DecValTok{1}\SpecialCharTok{:}\DecValTok{10}\NormalTok{, }\AttributeTok{Metric =}\NormalTok{ temp.Metric))}

\FunctionTok{nrow}\NormalTok{(temp.grid)}
\end{Highlighting}
\end{Shaded}

\begin{verbatim}
## [1] 2080
\end{verbatim}

Now it's time to rectify the raw scores.
Also, and this \textbf{is important}, we are reversing the scores here into a notion where a score of \(0\) is the worst possible score, and \(1\) the best possible.
So far, we were using distances, and a lower distance was proportional with a low score, now we are inverting this.

\begin{Shaded}
\begin{Highlighting}[]
\NormalTok{ac\_grid\_projects\_results\_uniform }\OtherTok{\textless{}{-}} \FunctionTok{loadResultsOrCompute}\NormalTok{(}\AttributeTok{file =} \StringTok{"../results/ac\_grid\_projects\_results\_uniform\_sc.rds"}\NormalTok{,}
  \AttributeTok{computeExpr =}\NormalTok{ \{}
    \FunctionTok{doWithParallelCluster}\NormalTok{(}\AttributeTok{numCores =} \FunctionTok{min}\NormalTok{(}\DecValTok{32}\NormalTok{, parallel}\SpecialCharTok{::}\FunctionTok{detectCores}\NormalTok{()), }\AttributeTok{expr =}\NormalTok{ \{}
      \FunctionTok{library}\NormalTok{(foreach)}

\NormalTok{      project\_signals\_all }\OtherTok{\textless{}{-}} \FunctionTok{append}\NormalTok{(project\_signals, project\_signals\_2nd\_batch)}

\NormalTok{      foreach}\SpecialCharTok{::}\FunctionTok{foreach}\NormalTok{(}\AttributeTok{grid\_idx =} \FunctionTok{rownames}\NormalTok{(temp.grid), }\AttributeTok{.combine =}\NormalTok{ rbind, }\AttributeTok{.inorder =} \ConstantTok{FALSE}\NormalTok{,}
        \AttributeTok{.verbose =} \ConstantTok{TRUE}\NormalTok{) }\SpecialCharTok{\%dopar\%}\NormalTok{ \{}
        \FunctionTok{options}\NormalTok{(}\AttributeTok{warn =} \DecValTok{2}\NormalTok{)}
\NormalTok{        params }\OtherTok{\textless{}{-}}\NormalTok{ temp.grid[grid\_idx, ]}
\NormalTok{        params}\SpecialCharTok{$}\NormalTok{PM }\OtherTok{\textless{}{-}} \FunctionTok{as.character}\NormalTok{(params}\SpecialCharTok{$}\NormalTok{PM)}
\NormalTok{        params}\SpecialCharTok{$}\NormalTok{Var }\OtherTok{\textless{}{-}} \FunctionTok{as.character}\NormalTok{(params}\SpecialCharTok{$}\NormalTok{Var)}
\NormalTok{        params}\SpecialCharTok{$}\NormalTok{Metric }\OtherTok{\textless{}{-}} \FunctionTok{as.character}\NormalTok{(params}\SpecialCharTok{$}\NormalTok{Metric)}

\NormalTok{        temp.data }\OtherTok{\textless{}{-}} \FunctionTok{ac\_extract\_data}\NormalTok{(}\AttributeTok{pmName =}\NormalTok{ params}\SpecialCharTok{$}\NormalTok{PM, }\AttributeTok{varName =}\NormalTok{ params}\SpecialCharTok{$}\NormalTok{Var,}
          \AttributeTok{segIdx =}\NormalTok{ params}\SpecialCharTok{$}\NormalTok{SegIdx, }\AttributeTok{metricName =}\NormalTok{ params}\SpecialCharTok{$}\NormalTok{Metric, }\AttributeTok{ac\_grid =}\NormalTok{ ac\_grid,}
          \AttributeTok{ac\_grid\_results =}\NormalTok{ ac\_grid\_results)}
\NormalTok{        tempf }\OtherTok{\textless{}{-}}\NormalTok{ stats}\SpecialCharTok{::}\FunctionTok{ecdf}\NormalTok{(temp.data)}

\NormalTok{        res }\OtherTok{\textless{}{-}} \FunctionTok{matrix}\NormalTok{(}\AttributeTok{nrow =} \DecValTok{1}\NormalTok{, }\AttributeTok{ncol =} \FunctionTok{length}\NormalTok{(project\_signals\_all))}
        \ControlFlowTok{for}\NormalTok{ (i }\ControlFlowTok{in} \DecValTok{1}\SpecialCharTok{:}\FunctionTok{length}\NormalTok{(project\_signals\_all)) \{}
          \CommentTok{\# Attention! It is only here that we also transform a score into the}
          \CommentTok{\# notion of 1=best, 0=worst! Beware that the data for the ECDF is the}
          \CommentTok{\# empirical distribution of distances. So for any x where ECDF(x)=0}
          \CommentTok{\# is the best possible score, so we need to subtract that from 1 {-}{-}}
          \CommentTok{\# the more we subtract, the worse the score.}
\NormalTok{          res[}\DecValTok{1}\NormalTok{, i] }\OtherTok{\textless{}{-}} \DecValTok{1} \SpecialCharTok{{-}} \FunctionTok{tempf}\NormalTok{(ac\_grid\_projects\_results[ac\_grid\_projects\_results}\SpecialCharTok{$}\NormalTok{Project }\SpecialCharTok{==}
          \FunctionTok{names}\NormalTok{(project\_signals\_all)[i] }\SpecialCharTok{\&}\NormalTok{ ac\_grid\_projects\_results}\SpecialCharTok{$}\NormalTok{PM }\SpecialCharTok{==}
\NormalTok{          params}\SpecialCharTok{$}\NormalTok{PM }\SpecialCharTok{\&}\NormalTok{ ac\_grid\_projects\_results}\SpecialCharTok{$}\NormalTok{Var }\SpecialCharTok{==}\NormalTok{ params}\SpecialCharTok{$}\NormalTok{Var }\SpecialCharTok{\&}\NormalTok{ ac\_grid\_projects\_results}\SpecialCharTok{$}\NormalTok{SegIdx }\SpecialCharTok{==}
\NormalTok{          params}\SpecialCharTok{$}\NormalTok{SegIdx, params}\SpecialCharTok{$}\NormalTok{Metric])}
\NormalTok{        \}}
        \FunctionTok{cbind}\NormalTok{(}\StringTok{\textasciigrave{}}\AttributeTok{colnames\textless{}{-}}\StringTok{\textasciigrave{}}\NormalTok{(res, }\FunctionTok{names}\NormalTok{(project\_signals\_all)), params)}
\NormalTok{      \}}
\NormalTok{    \})}
\NormalTok{  \})}
\end{Highlighting}
\end{Shaded}

\hypertarget{non-weighted}{%
\paragraph{Non-weighted}\label{non-weighted}}

With the results from above, we can do many things, for example:

\begin{itemize}
\tightlist
\item
  Calculate a score between each pair of PM and P
\item
  Compare scores between PMs (e.g., which model has the highest average score and thus resembles the projects best?)

  \begin{itemize}
  \tightlist
  \item
    Compare Brier-score of PMs to the decision rule.
  \end{itemize}
\item
  Fitting the normalized linear scalarizer:

  \begin{itemize}
  \tightlist
  \item
    Compare scores across (groups of) metrics, segments, and variables to find out what is important and where for each PM.
  \item
    Prune overdetermined models (remove scores where the weight is (close to) zero in models that are specified using too many scores). This could perhaps also be understood as \emph{regularization} of an ill-posed problem, where \emph{ill-posed} refers to the fact of having too many coefficients in a regression model.
  \end{itemize}
\item
  Fit other kind of regression model to analyze what Brier-score or correlation is possible.
\item
  Fit other kind of model that allows examining the variable importance (e.g., a Random forest).
\end{itemize}

Here, we will take a first look at which PM appears to be the best. Since all PMs are in the same family, we can now perfectly compare them after the automatic calibration.
\emph{Same family} refers to having done the automatic calibration using the same segments, variables, and objectives. For example, the objective for process model type II, variable CP, segment 4, correlation, is the same across all PMs. The PM that returns the highest average score (across all projects) would then be the best for that single objective.
We have \(4*10*13=520\) objectives and the non-weighted approach is to simply add their scores together and divide by \(520\).

We will need a function that takes PM/P and returns that score:

\begin{Shaded}
\begin{Highlighting}[]
\NormalTok{ac\_pmp\_score }\OtherTok{\textless{}{-}} \ControlFlowTok{function}\NormalTok{(pmName, projName, results\_uniform, use\_metrics) \{}
\NormalTok{  temp }\OtherTok{\textless{}{-}}\NormalTok{ results\_uniform[results\_uniform}\SpecialCharTok{$}\NormalTok{PM }\SpecialCharTok{==}\NormalTok{ pmName }\SpecialCharTok{\&}\NormalTok{ results\_uniform}\SpecialCharTok{$}\NormalTok{Metric }\SpecialCharTok{\%in\%}
    \FunctionTok{as.character}\NormalTok{(use\_metrics), projName]}

  \FunctionTok{mean}\NormalTok{(temp)}
\NormalTok{\}}
\end{Highlighting}
\end{Shaded}

In order to compare PMs, we will find out the following: Which PM produces the highest/lowest score for the best/worst PM, and what is the Brier(MSE)-/Log-score and correlation for all projects.
Also, we will compute the Kullback--Leibler divergence between the distributions of the ground truth and the predictions.
The result is shown in table \ref{tab:ac-compare-pms-sc}.

\begin{Shaded}
\begin{Highlighting}[]
\NormalTok{as.density }\OtherTok{\textless{}{-}} \ControlFlowTok{function}\NormalTok{(x) \{}
\NormalTok{  temp }\OtherTok{\textless{}{-}}\NormalTok{ stats}\SpecialCharTok{::}\FunctionTok{density}\NormalTok{(x)}
  \StringTok{\textasciigrave{}}\AttributeTok{attributes\textless{}{-}}\StringTok{\textasciigrave{}}\NormalTok{(}\AttributeTok{x =}\NormalTok{ stats}\SpecialCharTok{::}\FunctionTok{approxfun}\NormalTok{(}\AttributeTok{x =}\NormalTok{ temp}\SpecialCharTok{$}\NormalTok{x, }\AttributeTok{y =}\NormalTok{ temp}\SpecialCharTok{$}\NormalTok{y, }\AttributeTok{yleft =} \DecValTok{0}\NormalTok{, }\AttributeTok{yright =} \DecValTok{0}\NormalTok{),}
    \AttributeTok{value =} \FunctionTok{list}\NormalTok{(}\AttributeTok{ab =} \FunctionTok{range}\NormalTok{(temp}\SpecialCharTok{$}\NormalTok{x)))}
\NormalTok{\}}

\NormalTok{kl\_div }\OtherTok{\textless{}{-}} \ControlFlowTok{function}\NormalTok{(vals1, vals2) \{}
\NormalTok{  f1 }\OtherTok{\textless{}{-}} \FunctionTok{as.density}\NormalTok{(vals1)}
\NormalTok{  f2 }\OtherTok{\textless{}{-}} \FunctionTok{as.density}\NormalTok{(vals2)}
\NormalTok{  r }\OtherTok{\textless{}{-}} \FunctionTok{range}\NormalTok{(}\FunctionTok{c}\NormalTok{(}\FunctionTok{attr}\NormalTok{(f1, }\StringTok{"ab"}\NormalTok{), }\FunctionTok{attr}\NormalTok{(f2, }\StringTok{"ab"}\NormalTok{)))}
\NormalTok{  x }\OtherTok{\textless{}{-}} \FunctionTok{seq}\NormalTok{(}\AttributeTok{from =}\NormalTok{ r[}\DecValTok{1}\NormalTok{], }\AttributeTok{to =}\NormalTok{ r[}\DecValTok{2}\NormalTok{], }\AttributeTok{length.out =} \DecValTok{1000}\NormalTok{)}
\NormalTok{  v1 }\OtherTok{\textless{}{-}} \FunctionTok{f1}\NormalTok{(x)}
\NormalTok{  v2 }\OtherTok{\textless{}{-}} \FunctionTok{f2}\NormalTok{(x)}
  \FunctionTok{unname}\NormalTok{(}\FunctionTok{suppressMessages}\NormalTok{(\{}
\NormalTok{    philentropy}\SpecialCharTok{::}\FunctionTok{distance}\NormalTok{(}\AttributeTok{method =} \StringTok{"kullback{-}leibler"}\NormalTok{, }\AttributeTok{unit =} \StringTok{"log"}\NormalTok{, }\AttributeTok{x =} \FunctionTok{rbind}\NormalTok{(v1}\SpecialCharTok{/}\FunctionTok{sum}\NormalTok{(v1),}
\NormalTok{      v2}\SpecialCharTok{/}\FunctionTok{sum}\NormalTok{(v2)))}
\NormalTok{  \}))}
\NormalTok{\}}
\end{Highlighting}
\end{Shaded}

\begin{Shaded}
\begin{Highlighting}[]
\NormalTok{temp }\OtherTok{\textless{}{-}} \FunctionTok{append}\NormalTok{(project\_signals, project\_signals\_2nd\_batch)}
\NormalTok{temp.gt }\OtherTok{\textless{}{-}} \FunctionTok{c}\NormalTok{(ground\_truth}\SpecialCharTok{$}\NormalTok{consensus\_score, ground\_truth\_2nd\_batch}\SpecialCharTok{$}\NormalTok{consensus\_score)}

\NormalTok{temp.df }\OtherTok{\textless{}{-}} \ConstantTok{NULL}

\ControlFlowTok{for}\NormalTok{ (pmName }\ControlFlowTok{in} \FunctionTok{c}\NormalTok{(}\StringTok{"I"}\NormalTok{, }\StringTok{"II"}\NormalTok{, }\StringTok{"III(avg)"}\NormalTok{, }\StringTok{"III(b,11)"}\NormalTok{)) \{}
\NormalTok{  temp.pred }\OtherTok{\textless{}{-}} \FunctionTok{data.frame}\NormalTok{(}\AttributeTok{pred =} \FunctionTok{sapply}\NormalTok{(}\AttributeTok{X =} \FunctionTok{names}\NormalTok{(temp), }\ControlFlowTok{function}\NormalTok{(pName) }\FunctionTok{ac\_pmp\_score}\NormalTok{(}\AttributeTok{use\_metrics =}\NormalTok{ temp.Metric,}
    \AttributeTok{pmName =}\NormalTok{ pmName, }\AttributeTok{projName =}\NormalTok{ pName, }\AttributeTok{results\_uniform =}\NormalTok{ ac\_grid\_projects\_results\_uniform)),}
    \AttributeTok{ground\_truth =}\NormalTok{ temp.gt)}

\NormalTok{  temp.df }\OtherTok{\textless{}{-}} \FunctionTok{rbind}\NormalTok{(temp.df, }\FunctionTok{data.frame}\NormalTok{(}\AttributeTok{PM =}\NormalTok{ pmName, }\AttributeTok{ScForBest =} \FunctionTok{ac\_pmp\_score}\NormalTok{(}\AttributeTok{use\_metrics =}\NormalTok{ temp.Metric,}
    \AttributeTok{pmName =}\NormalTok{ pmName, }\AttributeTok{projName =} \FunctionTok{paste0}\NormalTok{(}\StringTok{"project\_"}\NormalTok{, }\FunctionTok{which.max}\NormalTok{(temp.gt)), }\AttributeTok{results\_uniform =}\NormalTok{ ac\_grid\_projects\_results\_uniform),}
    \AttributeTok{ScForWorst =} \FunctionTok{ac\_pmp\_score}\NormalTok{(}\AttributeTok{use\_metrics =}\NormalTok{ temp.Metric, }\AttributeTok{pmName =}\NormalTok{ pmName, }\AttributeTok{projName =} \FunctionTok{paste0}\NormalTok{(}\StringTok{"project\_"}\NormalTok{,}
      \FunctionTok{which.min}\NormalTok{(temp.gt)), }\AttributeTok{results\_uniform =}\NormalTok{ ac\_grid\_projects\_results\_uniform),}
    \AttributeTok{ScMax =} \FunctionTok{max}\NormalTok{(temp.pred}\SpecialCharTok{$}\NormalTok{pred), }\AttributeTok{ScMin =} \FunctionTok{min}\NormalTok{(temp.pred}\SpecialCharTok{$}\NormalTok{pred), }\AttributeTok{ScAvg =} \FunctionTok{mean}\NormalTok{(temp.pred}\SpecialCharTok{$}\NormalTok{pred),}
    \AttributeTok{KLdiv =} \FunctionTok{kl\_div}\NormalTok{(temp.gt, temp.pred}\SpecialCharTok{$}\NormalTok{pred), }\AttributeTok{Corr =}\NormalTok{ stats}\SpecialCharTok{::}\FunctionTok{cor}\NormalTok{(temp.gt, temp.pred}\SpecialCharTok{$}\NormalTok{pred),}
    \AttributeTok{RMSE =} \FunctionTok{sqrt}\NormalTok{(Metrics}\SpecialCharTok{::}\FunctionTok{mse}\NormalTok{(}\AttributeTok{actual =}\NormalTok{ temp.gt, }\AttributeTok{predicted =}\NormalTok{ temp.pred}\SpecialCharTok{$}\NormalTok{pred)),}
    \AttributeTok{MSE =}\NormalTok{ Metrics}\SpecialCharTok{::}\FunctionTok{mse}\NormalTok{(}\AttributeTok{actual =}\NormalTok{ temp.gt, }\AttributeTok{predicted =}\NormalTok{ temp.pred}\SpecialCharTok{$}\NormalTok{pred), }\AttributeTok{Log =} \FunctionTok{mean}\NormalTok{(scoring}\SpecialCharTok{::}\FunctionTok{logscore}\NormalTok{(}\AttributeTok{object =}\NormalTok{ ground\_truth }\SpecialCharTok{\textasciitilde{}}
\NormalTok{      pred, }\AttributeTok{data =}\NormalTok{ temp.pred))))}
\NormalTok{\}}
\end{Highlighting}
\end{Shaded}

\begin{table}

\caption{\label{tab:ac-compare-pms-sc}Comparison of continuous PMs using source code for best/worst projects, as well as MSE- and Log-scores as average deviation from the ground truth consensus.}
\centering
\begin{tabular}[t]{lrrrrrrrrrr}
\toprule
PM & ScForBest & ScForWorst & ScMax & ScMin & ScAvg & KLdiv & Corr & RMSE & MSE & Log\\
\midrule
I & 0.4394 & 0.4951 & 0.5240 & 0.4227 & 0.4652 & 4.4682 & -0.369 & 0.3598 & 0.1294 & 0.6435\\
II & 0.5750 & 0.6383 & 0.6450 & 0.5160 & 0.5878 & 4.4264 & -0.169 & 0.4331 & 0.1876 & 0.8695\\
III(avg) & 0.5929 & 0.6370 & 0.7185 & 0.5745 & 0.6320 & 4.4184 & 0.392 & 0.4519 & 0.2042 & 0.9814\\
III(b,11) & 0.5772 & 0.6475 & 0.6475 & 0.5430 & 0.6069 & 4.4859 & -0.085 & 0.4447 & 0.1978 & 0.9162\\
\bottomrule
\end{tabular}
\end{table}

As we show in the technical report for issue-tracking, the lowest MSE for the binary decision rule is \(\approx0.14\) across all 15 projects.
The MSE of the ZeroR base model was \(\approx0.137\) across all projects.
While the first type of PM is undercutting this threshold, its results and those of the other PMs are likely unusable, as the score for the best project is consistently lower than the score for the worst project, where it should be the other way round (to become usable, we need to learn better weights).
This could indicate, for example, that the models' design of the expectation is off (i.e., the variables as designed do not reflect the manifestation of the phenomenon well), or that using equal weights for all objectives (or the choice of objectives, which in this case is all of them) is not good.
I suppose that more of the latter is true, since even the data-enhanced and especially the data-only models show this inconsistency, and those were designed using the data. So it is likely we are measuring the wrong things (perhaps in the wrong places), or we emphasizing not correctly.
Nonetheless, all these scores are close together, so picking a champion needs to be done with care. Also, the score from each PM is the unweighted aggregation across hundreds of objectives, and at this point we have yet to examine which objectives are the most important (next two sections).

In the non-weighted approach we assume to know nothing about the observations, implicitly assuming that a) all observations exhibit the sought-after phenomenon, and b) that they do so to a similar degree.
If this were true for our case, then the mean score of table \ref{tab:ac-compare-pms-sc} would give a perfectly usable score that we could pick a champion PM by (in any other case, the mean/best/worst scores are useless).
This is not quite true for our case, as our ground truth ranges from zero to one (full spectrum).
Therefore, a higher score is not just better, but rather a score that deviates less from the ground truth. So if we are fair and consider the ground truth, we need to measure the deviation from it using, e.g., the (R)MSE, correlation, or some divergence.
Going by these, the correlation is perhaps most important, followed by the (R)MSE to estimate each PM's bias.
The bias can be reduced when we additionally learning weights for all of the scores (see section \ref{ssec:ac-weights-optimized}).

\hypertarget{weighted-by-expert-decision-maker}{%
\paragraph{Weighted by expert decision maker}\label{weighted-by-expert-decision-maker}}

The scores as of table \ref{tab:ac-compare-pms-sc} for best/worst are close together, too close perhaps. This is due to the massive amount of objectives that were used. Also, all of the available objectives were used, across all segments and variables.
In reality however, an expert would almost certainly not configure a PM this way. Rather, they would choose specific objectives. We will simulate this scenario by choosing the metrics \texttt{area} and \texttt{corr}, as these are simple and comprehensible choices.
As for the weights, they will be the same for each metric, and only vary between segments. The weights will simply increase exponentially towards the end of the project.
The expression for the weights is \(\frac{1}{4}+\frac{3}{4}x^3\).

We have a large number of weights, so it is important to be able to ``address'' them correctly.
Therefore, we will use a grid as address register.
Let's define a weight grid:

\begin{Shaded}
\begin{Highlighting}[]
\NormalTok{ac\_weight\_grid\_expert }\OtherTok{\textless{}{-}} \FunctionTok{expand.grid}\NormalTok{(}\FunctionTok{list}\NormalTok{(}\AttributeTok{Var =} \FunctionTok{names}\NormalTok{(weight\_vartype), }\AttributeTok{Metric =} \FunctionTok{c}\NormalTok{(}\StringTok{"area"}\NormalTok{,}
  \StringTok{"corr"}\NormalTok{), }\AttributeTok{SegIdx =} \DecValTok{1}\SpecialCharTok{:}\DecValTok{10}\NormalTok{))}

\NormalTok{ac\_weight\_vector\_expert }\OtherTok{\textless{}{-}} \FunctionTok{sapply}\NormalTok{(}\AttributeTok{X =}\NormalTok{ ac\_weight\_grid\_expert}\SpecialCharTok{$}\NormalTok{SegIdx, }\AttributeTok{FUN =} \ControlFlowTok{function}\NormalTok{(si) \{}
  \FloatTok{0.25} \SpecialCharTok{+} \DecValTok{3} \SpecialCharTok{*}\NormalTok{ (si}\SpecialCharTok{/}\DecValTok{10}\NormalTok{)}\SpecialCharTok{\^{}}\DecValTok{3}\SpecialCharTok{/}\DecValTok{4}
\NormalTok{\})}
\end{Highlighting}
\end{Shaded}

Also, we will implement a variant of the function \texttt{ac\_pmp\_score()} that can handle weights:

\begin{Shaded}
\begin{Highlighting}[]
\NormalTok{ac\_pmp\_score\_weighted }\OtherTok{\textless{}{-}} \ControlFlowTok{function}\NormalTok{(pmName, projName, weightGrid, weightVector, results\_uniform) \{}
  \ControlFlowTok{if}\NormalTok{ (}\FunctionTok{length}\NormalTok{(weightVector) }\SpecialCharTok{!=} \FunctionTok{nrow}\NormalTok{(weightGrid)) \{}
    \FunctionTok{stop}\NormalTok{(}\StringTok{"Address register grid does not match weight vector."}\NormalTok{)}
\NormalTok{  \}}

\NormalTok{  res }\OtherTok{\textless{}{-}} \FunctionTok{c}\NormalTok{()}
  \ControlFlowTok{for}\NormalTok{ (i }\ControlFlowTok{in} \DecValTok{1}\SpecialCharTok{:}\FunctionTok{length}\NormalTok{(weightVector)) \{}
\NormalTok{    params }\OtherTok{\textless{}{-}}\NormalTok{ weightGrid[i, ]}
\NormalTok{    params}\SpecialCharTok{$}\NormalTok{Var }\OtherTok{\textless{}{-}} \FunctionTok{as.character}\NormalTok{(params}\SpecialCharTok{$}\NormalTok{Var)}
\NormalTok{    params}\SpecialCharTok{$}\NormalTok{Metric }\OtherTok{\textless{}{-}} \FunctionTok{as.character}\NormalTok{(params}\SpecialCharTok{$}\NormalTok{Metric)}

    \CommentTok{\# The following selection will result in a single value:}
\NormalTok{    score }\OtherTok{\textless{}{-}}\NormalTok{ results\_uniform[results\_uniform}\SpecialCharTok{$}\NormalTok{PM }\SpecialCharTok{==}\NormalTok{ pmName }\SpecialCharTok{\&}\NormalTok{ results\_uniform}\SpecialCharTok{$}\NormalTok{Var }\SpecialCharTok{==}
\NormalTok{      params}\SpecialCharTok{$}\NormalTok{Var }\SpecialCharTok{\&}\NormalTok{ results\_uniform}\SpecialCharTok{$}\NormalTok{Metric }\SpecialCharTok{==}\NormalTok{ params}\SpecialCharTok{$}\NormalTok{Metric }\SpecialCharTok{\&}\NormalTok{ results\_uniform}\SpecialCharTok{$}\NormalTok{SegIdx }\SpecialCharTok{==}
\NormalTok{      params}\SpecialCharTok{$}\NormalTok{SegIdx, projName]}

    \ControlFlowTok{if}\NormalTok{ (}\FunctionTok{length}\NormalTok{(score) }\SpecialCharTok{!=} \DecValTok{1}\NormalTok{) \{}
      \FunctionTok{stop}\NormalTok{()}
\NormalTok{    \}}

\NormalTok{    res }\OtherTok{\textless{}{-}} \FunctionTok{c}\NormalTok{(res, weightVector[i] }\SpecialCharTok{*}\NormalTok{ score)}
\NormalTok{  \}}

  \FunctionTok{sum}\NormalTok{(res)}\SpecialCharTok{/}\FunctionTok{sum}\NormalTok{(weightVector)}
\NormalTok{\}}
\end{Highlighting}
\end{Shaded}

The results are shown in table \ref{tab:ac-compare-pms-sc-expert}. Apparently, our expert managed to improve almost all process model types. All types except type III(b, numInt=11) do not exhibit the inconsistency for best/worst project any longer, and type I's MSE is now significantly lower than the ZeroR's of \(\approx0.137\) (still, it's quite bad).
PM type III(avg) has an acceptable spread for best/worst now, but its scores do not look good. However, its correlation improved significantly (actually, all correlations improved). If the expert-picked weights had resulted in a lower bias (i.e., lower (R)MSE), then we would have ended up with usable models.

Depending on the criteria to select a PM, going forward with type I (lowest MSE) or type III(avg) (highest correlation) would probably be the best choice, albeit with great care.
It appears our hypothetical expert made a choice that is slightly better than using all available objectives with same weight, as the MSE and all but the last two Log-scores for all PMs improved somewhat.

\begin{table}

\caption{\label{tab:ac-compare-pms-sc-expert}Comparison of continuous PMs using source code for best/worst projects, as well as MSE- and Log-scores as average deviation from the ground truth consensus, using expert-picked objectives and weights.}
\centering
\begin{tabular}[t]{lrrrrrrrrrr}
\toprule
PM & ScForBest & ScForWorst & ScMax & ScMin & ScAvg & KLdiv & Corr & RMSE & MSE & Log\\
\midrule
I & 0.4739 & 0.4220 & 0.5257 & 0.3765 & 0.4661 & 4.1636 & -0.1887 & 0.3598 & 0.1295 & 0.6376\\
II & 0.5840 & 0.5830 & 0.6265 & 0.4945 & 0.5673 & 4.4102 & -0.1080 & 0.4176 & 0.1744 & 0.8184\\
III(avg) & 0.6834 & 0.5799 & 0.7686 & 0.5799 & 0.6577 & 4.2681 & 0.6792 & 0.4598 & 0.2115 & 1.0356\\
III(b,11) & 0.6156 & 0.6424 & 0.6507 & 0.5341 & 0.6029 & 4.4345 & 0.0116 & 0.4400 & 0.1936 & 0.8963\\
\bottomrule
\end{tabular}
\end{table}

\hypertarget{weighted-by-optimization}{%
\paragraph{\texorpdfstring{Weighted by optimization\label{ssec:ac-weights-optimized}}{Weighted by optimization}}\label{weighted-by-optimization}}

Here we will attempt to find the best weights (variable importances, actually) by optimization.
The interesting thing is, that the weights have a \(1:1\) correlation to the principle of variable importance, since the objective is normalizing linear scalarizer, and and all objectives are scores with a linear co-domain of \([0,1]\).
A weight is specific to the variable (4), segment (10), and metric (13), so it is best to define an address grid again.

\begin{Shaded}
\begin{Highlighting}[]
\NormalTok{ac\_weight\_grid }\OtherTok{\textless{}{-}} \FunctionTok{expand.grid}\NormalTok{(}\FunctionTok{list}\NormalTok{(}\AttributeTok{Var =} \FunctionTok{names}\NormalTok{(weight\_vartype), }\AttributeTok{Metric =}\NormalTok{ temp.Metric,}
  \AttributeTok{SegIdx =} \DecValTok{1}\SpecialCharTok{:}\DecValTok{10}\NormalTok{))}

\FunctionTok{nrow}\NormalTok{(ac\_weight\_grid)}
\end{Highlighting}
\end{Shaded}

\begin{verbatim}
## [1] 520
\end{verbatim}

Now the objective for the optimization will search for weights that minimize the MSE between the weighted projects and the ground truth.
We will optimize the weights once for each process model, and then follow this up by a closer inspection of the champion model (the model with the lowest MSE).
We will run the optimization using a local search, facilitating a parallel version of the L-BFGS-B algorithm (Byrd et al. 1995) (Gerber and Furrer 2019), using a numerical gradient.

\begin{Shaded}
\begin{Highlighting}[]
\NormalTok{temp.names }\OtherTok{\textless{}{-}} \FunctionTok{names}\NormalTok{(}\FunctionTok{append}\NormalTok{(project\_signals, project\_signals\_2nd\_batch))}

\NormalTok{ac\_sc\_weights\_optim }\OtherTok{\textless{}{-}} \ControlFlowTok{function}\NormalTok{(pmName) \{}
  \FunctionTok{loadResultsOrCompute}\NormalTok{(}\AttributeTok{file =} \FunctionTok{paste0}\NormalTok{(}\StringTok{"../results/ac\_sc\_weights\_pm\_"}\NormalTok{, pmName, }\StringTok{".rds"}\NormalTok{),}
    \AttributeTok{computeExpr =}\NormalTok{ \{}
\NormalTok{      cl }\OtherTok{\textless{}{-}}\NormalTok{ parallel}\SpecialCharTok{::}\FunctionTok{makePSOCKcluster}\NormalTok{(}\FunctionTok{min}\NormalTok{(}\DecValTok{123}\NormalTok{, parallel}\SpecialCharTok{::}\FunctionTok{detectCores}\NormalTok{()))}
\NormalTok{      parallel}\SpecialCharTok{::}\FunctionTok{clusterExport}\NormalTok{(cl, }\AttributeTok{varlist =} \FunctionTok{list}\NormalTok{(}\StringTok{"temp.gt"}\NormalTok{, }\StringTok{"temp.names"}\NormalTok{, }\StringTok{"ac\_pmp\_score\_weighted"}\NormalTok{,}
        \StringTok{"ac\_weight\_grid"}\NormalTok{, }\StringTok{"ac\_grid\_projects\_results\_uniform"}\NormalTok{))}

      \FunctionTok{doWithParallelClusterExplicit}\NormalTok{(}\AttributeTok{cl =}\NormalTok{ cl, }\AttributeTok{expr =}\NormalTok{ \{}
\NormalTok{        optimParallel}\SpecialCharTok{::}\FunctionTok{optimParallel}\NormalTok{(}\AttributeTok{par =} \FunctionTok{rep}\NormalTok{(}\FloatTok{0.5}\NormalTok{, }\FunctionTok{nrow}\NormalTok{(ac\_weight\_grid)),}
          \AttributeTok{method =} \StringTok{"L{-}BFGS{-}B"}\NormalTok{, }\AttributeTok{lower =} \FunctionTok{rep}\NormalTok{(}\DecValTok{0}\NormalTok{, }\FunctionTok{nrow}\NormalTok{(ac\_weight\_grid)), }\AttributeTok{upper =} \FunctionTok{rep}\NormalTok{(}\DecValTok{1}\NormalTok{,}
          \FunctionTok{nrow}\NormalTok{(ac\_weight\_grid)), }\AttributeTok{fn =} \ControlFlowTok{function}\NormalTok{(x) \{}
\NormalTok{          Metrics}\SpecialCharTok{::}\FunctionTok{mse}\NormalTok{(}\AttributeTok{actual =}\NormalTok{ temp.gt, }\AttributeTok{predicted =} \FunctionTok{sapply}\NormalTok{(}\AttributeTok{X =}\NormalTok{ temp.names,}
            \AttributeTok{FUN =} \ControlFlowTok{function}\NormalTok{(projName) \{}
            \FunctionTok{ac\_pmp\_score\_weighted}\NormalTok{(}\AttributeTok{results\_uniform =}\NormalTok{ ac\_grid\_projects\_results\_uniform,}
              \AttributeTok{pmName =}\NormalTok{ pmName, }\AttributeTok{projName =}\NormalTok{ projName, }\AttributeTok{weightGrid =}\NormalTok{ ac\_weight\_grid,}
              \AttributeTok{weightVector =}\NormalTok{ x)}
\NormalTok{            \}))}
\NormalTok{          \}, }\AttributeTok{parallel =} \FunctionTok{list}\NormalTok{(}\AttributeTok{cl =}\NormalTok{ cl, }\AttributeTok{forward =} \ConstantTok{FALSE}\NormalTok{, }\AttributeTok{loginfo =} \ConstantTok{TRUE}\NormalTok{))}
\NormalTok{      \})}
\NormalTok{    \})}
\NormalTok{\}}
\end{Highlighting}
\end{Shaded}

\hypertarget{variable-importance-traditional-approach}{%
\subparagraph{Variable importance (traditional approach)}\label{variable-importance-traditional-approach}}

Assessing the variable importance can be done as we have done previously in section \ref{sssec:var-imp}.
So before we jump to the results of finding the weights (and thus the importance of each score), we attempt the traditional approach using a model like Random forest or partial least squares using process model type I:

\begin{Shaded}
\begin{Highlighting}[]
\NormalTok{ac\_sc\_weights\_varimp }\OtherTok{\textless{}{-}} \FunctionTok{loadResultsOrCompute}\NormalTok{(}\AttributeTok{file =} \StringTok{"../results/ac\_sc\_weights\_varimp.rds"}\NormalTok{,}
  \AttributeTok{computeExpr =}\NormalTok{ \{}
\NormalTok{    temp.gt }\OtherTok{\textless{}{-}} \FunctionTok{c}\NormalTok{(ground\_truth}\SpecialCharTok{$}\NormalTok{consensus\_score, ground\_truth\_2nd\_batch}\SpecialCharTok{$}\NormalTok{consensus\_score)}
\NormalTok{    projNames }\OtherTok{\textless{}{-}} \FunctionTok{names}\NormalTok{(}\FunctionTok{append}\NormalTok{(project\_signals, project\_signals\_2nd\_batch))}
\NormalTok{    templ }\OtherTok{\textless{}{-}} \FunctionTok{list}\NormalTok{()}

    \ControlFlowTok{for}\NormalTok{ (pmName }\ControlFlowTok{in} \FunctionTok{levels}\NormalTok{(ac\_grid}\SpecialCharTok{$}\NormalTok{PM)) \{}
\NormalTok{      temp }\OtherTok{\textless{}{-}} \FunctionTok{matrix}\NormalTok{(}\AttributeTok{nrow =} \FunctionTok{length}\NormalTok{(projNames), }\AttributeTok{ncol =} \DecValTok{1} \SpecialCharTok{+} \FunctionTok{nrow}\NormalTok{(ac\_weight\_grid))}
\NormalTok{      temp[, }\FunctionTok{ncol}\NormalTok{(temp)] }\OtherTok{\textless{}{-}}\NormalTok{ temp.gt}

      \ControlFlowTok{for}\NormalTok{ (i }\ControlFlowTok{in} \DecValTok{1}\SpecialCharTok{:}\FunctionTok{length}\NormalTok{(projNames)) \{}
        \ControlFlowTok{for}\NormalTok{ (j }\ControlFlowTok{in} \DecValTok{1}\SpecialCharTok{:}\FunctionTok{nrow}\NormalTok{(ac\_weight\_grid)) \{}
\NormalTok{          params }\OtherTok{\textless{}{-}}\NormalTok{ ac\_weight\_grid[j, ]}
\NormalTok{          params}\SpecialCharTok{$}\NormalTok{Var }\OtherTok{\textless{}{-}} \FunctionTok{as.character}\NormalTok{(params}\SpecialCharTok{$}\NormalTok{Var)}
\NormalTok{          params}\SpecialCharTok{$}\NormalTok{Metric }\OtherTok{\textless{}{-}} \FunctionTok{as.character}\NormalTok{(params}\SpecialCharTok{$}\NormalTok{Metric)}

\NormalTok{          temp[i, j] }\OtherTok{\textless{}{-}}\NormalTok{ ac\_grid\_projects\_results\_uniform[ac\_grid\_projects\_results\_uniform}\SpecialCharTok{$}\NormalTok{PM }\SpecialCharTok{==}
\NormalTok{          pmName }\SpecialCharTok{\&}\NormalTok{ ac\_grid\_projects\_results\_uniform}\SpecialCharTok{$}\NormalTok{Metric }\SpecialCharTok{==}\NormalTok{ params}\SpecialCharTok{$}\NormalTok{Metric }\SpecialCharTok{\&}
\NormalTok{          ac\_grid\_projects\_results\_uniform}\SpecialCharTok{$}\NormalTok{Var }\SpecialCharTok{==}\NormalTok{ params}\SpecialCharTok{$}\NormalTok{Var }\SpecialCharTok{\&}\NormalTok{ ac\_grid\_projects\_results\_uniform}\SpecialCharTok{$}\NormalTok{SegIdx }\SpecialCharTok{==}
\NormalTok{          params}\SpecialCharTok{$}\NormalTok{SegIdx, projNames[i]]}
\NormalTok{        \}}
\NormalTok{      \}}

\NormalTok{      temp }\OtherTok{\textless{}{-}} \FunctionTok{as.data.frame}\NormalTok{(temp)}
      \FunctionTok{colnames}\NormalTok{(temp) }\OtherTok{\textless{}{-}} \FunctionTok{c}\NormalTok{(}\FunctionTok{paste0}\NormalTok{(}\StringTok{"w"}\NormalTok{, }\DecValTok{1}\SpecialCharTok{:}\FunctionTok{nrow}\NormalTok{(ac\_weight\_grid)), }\StringTok{"gt"}\NormalTok{)}

\NormalTok{      templ[[pmName]] }\OtherTok{\textless{}{-}} \FunctionTok{doWithParallelCluster}\NormalTok{(}\AttributeTok{numCores =} \FunctionTok{min}\NormalTok{(}\DecValTok{8}\NormalTok{, parallel}\SpecialCharTok{::}\FunctionTok{detectCores}\NormalTok{()),}
        \AttributeTok{expr =}\NormalTok{ \{}
          \FunctionTok{library}\NormalTok{(caret, }\AttributeTok{quietly =} \ConstantTok{TRUE}\NormalTok{)}

          \FunctionTok{set.seed}\NormalTok{(}\DecValTok{1}\NormalTok{)}
\NormalTok{          control }\OtherTok{\textless{}{-}}\NormalTok{ caret}\SpecialCharTok{::}\FunctionTok{trainControl}\NormalTok{(}\AttributeTok{method =} \StringTok{"repeatedcv"}\NormalTok{, }\AttributeTok{number =} \DecValTok{10}\NormalTok{,}
          \AttributeTok{repeats =} \DecValTok{3}\NormalTok{)}
\NormalTok{          modelFit }\OtherTok{\textless{}{-}}\NormalTok{ caret}\SpecialCharTok{::}\FunctionTok{train}\NormalTok{(gt }\SpecialCharTok{\textasciitilde{}}\NormalTok{ ., }\AttributeTok{data =}\NormalTok{ temp, }\AttributeTok{method =} \StringTok{"rf"}\NormalTok{, }\AttributeTok{trControl =}\NormalTok{ control)}
\NormalTok{          imp }\OtherTok{\textless{}{-}}\NormalTok{ caret}\SpecialCharTok{::}\FunctionTok{varImp}\NormalTok{(}\AttributeTok{object =}\NormalTok{ modelFit)}

          \FunctionTok{list}\NormalTok{(}\AttributeTok{fit =}\NormalTok{ modelFit, }\AttributeTok{imp =}\NormalTok{ imp)}
\NormalTok{        \})}
\NormalTok{    \}}
\NormalTok{    templ}
\NormalTok{  \})}
\end{Highlighting}
\end{Shaded}

Let's take a look at the fitted model's performance (table \ref{tab:ac-sc-weights-varimp-mse})
The Random forest overfits to the data, for each expectation.
This is obvious as the explained variance is negative in all cases, i.e., it is worse than a random guess.
Therefore, it is hard to assess the validity of the obtained results for variable importance purposes.
It appears we are able to reach acceptable MSE's, albeit not \(\approx0\) for some reason. This model and the other fitted models cannot be used in prediction scenarios, as they do not generalize at all, the exercise was only to examine the variable importance.

\begin{table}

\caption{\label{tab:ac-sc-weights-varimp-mse}Best models of fitting of a Random forest to the calibrated data in order to assess the variable importance.}
\centering
\begin{tabular}[t]{lrrrrrrrrr}
\toprule
PM & mtry & RMSE & Rsq & MAE & RMSESD & RsqSD & MAESD & MSE & VarExpl\\
\midrule
I & 2 & 0.2754 & 1 & 0.2641 & 0.1690 & 0 & 0.1629 & 0.0759 & -12.383\\
II & 2 & 0.2811 & 1 & 0.2671 & 0.1740 & 0 & 0.1679 & 0.0790 & -13.020\\
III(avg) & 2 & 0.2664 & 1 & 0.2534 & 0.1797 & 0 & 0.1720 & 0.0710 & -8.831\\
III(b,11) & 2 & 0.2730 & 1 & 0.2593 & 0.1748 & 0 & 0.1692 & 0.0745 & -17.844\\
\bottomrule
\end{tabular}
\end{table}

\hypertarget{results-pm-vs.-pm}{%
\subparagraph{Results PM vs.~PM}\label{results-pm-vs.-pm}}

Now for the actual optimization of each type of process model.
An overview of the optimization process on a per-project basis is shown in table \ref{tab:ac-sc-weights-optim-overview}.

\begin{Shaded}
\begin{Highlighting}[]
\NormalTok{ac\_sc\_weights\_pm\_I }\OtherTok{\textless{}{-}} \FunctionTok{ac\_sc\_weights\_optim}\NormalTok{(}\AttributeTok{pmName =} \StringTok{"I"}\NormalTok{)}
\NormalTok{ac\_sc\_weights\_pm\_II }\OtherTok{\textless{}{-}} \FunctionTok{ac\_sc\_weights\_optim}\NormalTok{(}\AttributeTok{pmName =} \StringTok{"II"}\NormalTok{)}
\NormalTok{ac\_sc\_weights\_pm\_IIIavg }\OtherTok{\textless{}{-}} \FunctionTok{ac\_sc\_weights\_optim}\NormalTok{(}\AttributeTok{pmName =} \StringTok{"III(avg)"}\NormalTok{)}
\NormalTok{ac\_sc\_weights\_pm\_IIIb11 }\OtherTok{\textless{}{-}} \FunctionTok{ac\_sc\_weights\_optim}\NormalTok{(}\AttributeTok{pmName =} \StringTok{"III(b,11)"}\NormalTok{)}
\end{Highlighting}
\end{Shaded}

\begin{table}

\caption{\label{tab:ac-sc-weights-optim-overview}Overview of the results of optimizing the weights using the normalizing linear scalarizer objective, on a per-project basis.}
\centering
\begin{tabular}[t]{lrrrlrr}
\toprule
PM & RMSE & MSE & Num\_weights\_gt0 & Weights\_pruned & Num\_iter & Num\_grad\\
\midrule
I & 0.25558 & 0.06532 & 171 & 67.12\% & 120 & 120\\
II & 0.23726 & 0.05629 & 29 & 94.42\% & 68 & 68\\
III(avg) & 0.19310 & 0.03729 & 7 & 98.65\% & 105 & 105\\
III(b,11) & 0.19855 & 0.03942 & 8 & 98.46\% & 118 & 118\\
\bottomrule
\end{tabular}
\end{table}

The results from table \ref{tab:ac-sc-weights-optim-overview} show that in general, a large portion of the weights (between \(\approx67-98\)\%) was set to \(0\), and hence the corresponding objective is not used, or not important at all for assessing the goodness of fit, i.e., how well the pair PM/P matches (mutual resemblance).
This affords us to considerably \emph{prune} the given models.
With differing number of iterations of the optimizer, we can obtain optimized weights for all models.
We still observe significant variances among the models' MSEs, with type I having the largest and type III(avg) the lowest.
Also, all MSE are lower than the ones found using a Random forest. So, not only do we get pure variable importance information, we also get more precise models with this approach.

Albeit being the worst of the optimized models, type I performs surprisingly well, but requires comparatively many objectives. While it still uses more than \(171\) weights \(>0\), which is considerable more than any of the other models, and required the most iterations to find this optimum, the values obtained for MSE and RMSE appear to be already within a practically acceptable and usable range.

Model type II is a data-enhanced version of type I, and therefore already closer to the ground truth than type I.
This implies the expectation that its scores should be better (i.e., lower (R)MSE), while simultaneously requiring fewer weight(s)/adjustments, and fewer iterations to reach an optimum. All of these expectations were fulfilled, with the most significant being the considerably fewer objectives required (\(\approx17\)\%).

Now the real interesting models are the data-only models of type III, as they achieve the lowest (R)MSEs while simultaneously only requiring \(7\) resp. \(8\) weights (out of originally \(520\)) to do so.
Granted, both types of models were hard to optimize, given the amount of iterations required.
Recall that model type III(b, numInt=11) consists of piece-wise linear functions per variable, which means no perturbations. This could explain why it is among the best in terms of (R)MSE and number of weights as its nature means it is simple, yet effective.

In any case, the residual MSE also indicates that a PM cannot probably be optimized beyond what the optimization found. Thus, a residual MSE\(>0\) indicates an error in the model's formulation, and that even by using some best constellation of the weights, it is not possible to achieve perfect goodness of fit, which indicates a design flaw in either the model and/or the particular objectives used.
In other words, an MSE\(=0\) would mean that a given PM can perfectly assess the goodness of fit of the given processes, and that it was configured in a perfect way (i.e., using the right segments, variables, metrics, etc.) to do so.

\begin{table}

\caption{\label{tab:ac-compare-pms-sc-optim}Comparison of continuous PMs using source code for best/worst projects, as well as MSE- and Log-scores as average deviation from the ground truth consensus, using some best set of weights per PM as found by optimization.}
\centering
\begin{tabular}[t]{lrrrrrrrrrr}
\toprule
PM & ScForBest & ScForWorst & ScMax & ScMin & ScAvg & KLdiv & Corr & RMSE & MSE & Log\\
\midrule
I & 0.3875 & 0.2667 & 0.4052 & 0.2118 & 0.3039 & 2.7393 & 0.7205 & 0.2556 & 0.0653 & 0.3971\\
II & 0.5015 & 0.3052 & 0.5015 & 0.2064 & 0.3237 & 2.1529 & 0.7831 & 0.2373 & 0.0563 & 0.4002\\
III(avg) & 0.5828 & 0.1739 & 0.6926 & 0.1267 & 0.3330 & 0.6644 & 0.8206 & 0.1931 & 0.0373 & 0.4223\\
III(b,11) & 0.6917 & 0.2478 & 0.6917 & 0.2125 & 0.3364 & 1.9565 & 0.8616 & 0.1985 & 0.0394 & 0.3866\\
\bottomrule
\end{tabular}
\end{table}

Table \ref{tab:ac-compare-pms-sc-optim} shows the scores per optimized PM. We observe a significantly larger margin between the scores for the worst and best project, as well as significantly lower MSE.
Also, the score for the best is now consistently higher than the score for the worst project, so we were able to correct this phenomenon by finding more appropriate weights.
Going by only these values, the champion models is perhaps one of the type III models. While type III(avg) has a slightly lower MSE, type III(b, numInt=11) has a lower Log-score, which indicates that it produces less extreme outliers, as the Log-score is much more sensible to that than MSE.

Since the scores for the best/worst project are not \(1\) resp. \(0\), this indicates that either the models are flawed (that is, the design of the continuous expectation is off), the choice of objectives is inappropriate (to some degree), or that the ground truth is not accurate -- or any combination of these.
It could also be a hint at the fact that the observed projects do neither represent the worst nor best possible manifestation of the Fire Drill in source code which means that our ground truth is using too-extreme values (e.g., a \(10\) in the ground truth might in reality just be a \(6\), and likewise for a low value).
In summary, the conclusions that can be drawn from finding the optimal weights depend on the assumptions we are certain to make. So for example, if we are confident about our ground truth, it indicates a flaw in, e.g., the model (expectation) or the (choice of) objectives.
Or if we are confident in the expectation (that is, the design of the continuous model itself), then the flaw could be in the (choice of) objectives or the ground truth, and so on.
In any case, the fitted weights (actually, variable importances) allow us to reason about the shortcomings of the model (the expectation), its configuration (objectives), and the ground truth.

\hypertarget{results-in-detail}{%
\subparagraph{Results in detail}\label{results-in-detail}}

With each weight linked to a single objective (score), we have the ability to gain rich insights into which of these scores are most important.
Also, since each such score is tied to one variable, one segment, and one metric, we can find out more about any of these (for example, which is the most important variable in a PM, or which metrics are important and which may be discarded, and when/where).

The more scores that are used (i.e., weight \(>0\)), the more insights we can gain.
We will first look into the weights of model type III(b, numInt=11), as it is one of the champions.
Then we will inspect the detailed results of model type I, as it uses the most weights. It will allow us to get a better insight into the average importance of each segment, each variable, and each metric.
Of course, this could be further drilled down. However, this shall just be a demonstration of what is possible in terms of analyzing a fitted PM.
It is also worth noting that in this scenario we are using all available metrics, variables, and segments. In a more realistic scenario, an expert would probably only pick a few types of metrics. Also, as long as there are sufficiently many segments that are measured, it is of less importance what actually is measured, because with enough degrees of freedom, we can find optimal weights that will lead to an MSE that has enough utility.

\begin{table}

\caption{\label{tab:ac-sc-weights-detailed}All non-zero weights as optimized for PM type III(b, numInt=11), also showing which variable, metric, and segment they address.}
\centering
\begin{tabular}[t]{llrrl}
\toprule
Var & Metric & SegIdx & weight & weight\_percent\\
\midrule
SCD & corr & 4 & 1.0000000 & 25.81\%\\
SCD & Kurtosis & 9 & 0.9240681 & 23.85\%\\
SCD & arclen & 1 & 0.7747642 & 19.99\%\\
A & ImpulseFactor & 3 & 0.6624839 & 17.1\%\\
FREQ & arclen & 5 & 0.2450323 & 6.32\%\\
\addlinespace
CP & arclen & 5 & 0.1943053 & 5.01\%\\
A & ImpulseFactor & 2 & 0.0443235 & 1.14\%\\
SCD & Peak & 1 & 0.0298810 & 0.77\%\\
\bottomrule
\end{tabular}
\end{table}

Before the analysis, let's show all the weights that are \(>0\) for PM type III(b, numInt=11) in table \ref{tab:ac-sc-weights-detailed}.
The first obvious fact that we observe is that the not all segments are used (cannot be used). Segments \(\{6,7,8,10\}\) are not used (important) at all in this model, while segments \(\{1,5\}\) have been used twice to take measurements.
The first four most important weights account for \(\approx86.7\)\% of the total variable importance.
In terms of metrics, half of all measurements come from signal theory. The other half is one correlation and three arclen-ratios.
The least important variables in this type of model are \texttt{CP} and \texttt{FREQ} (one usage), followed by \texttt{A} (two usages), and trumped by \texttt{SCD} (four or half of all usages).
In conclusion, these results are a bit astounding, since they mean that with only \(8\) objectives, it is already quite possible to indicate the degree to which the Fire Drill is present in a project, without even inspecting \(40\)\% of it.

\begin{figure}
\centering
\includegraphics{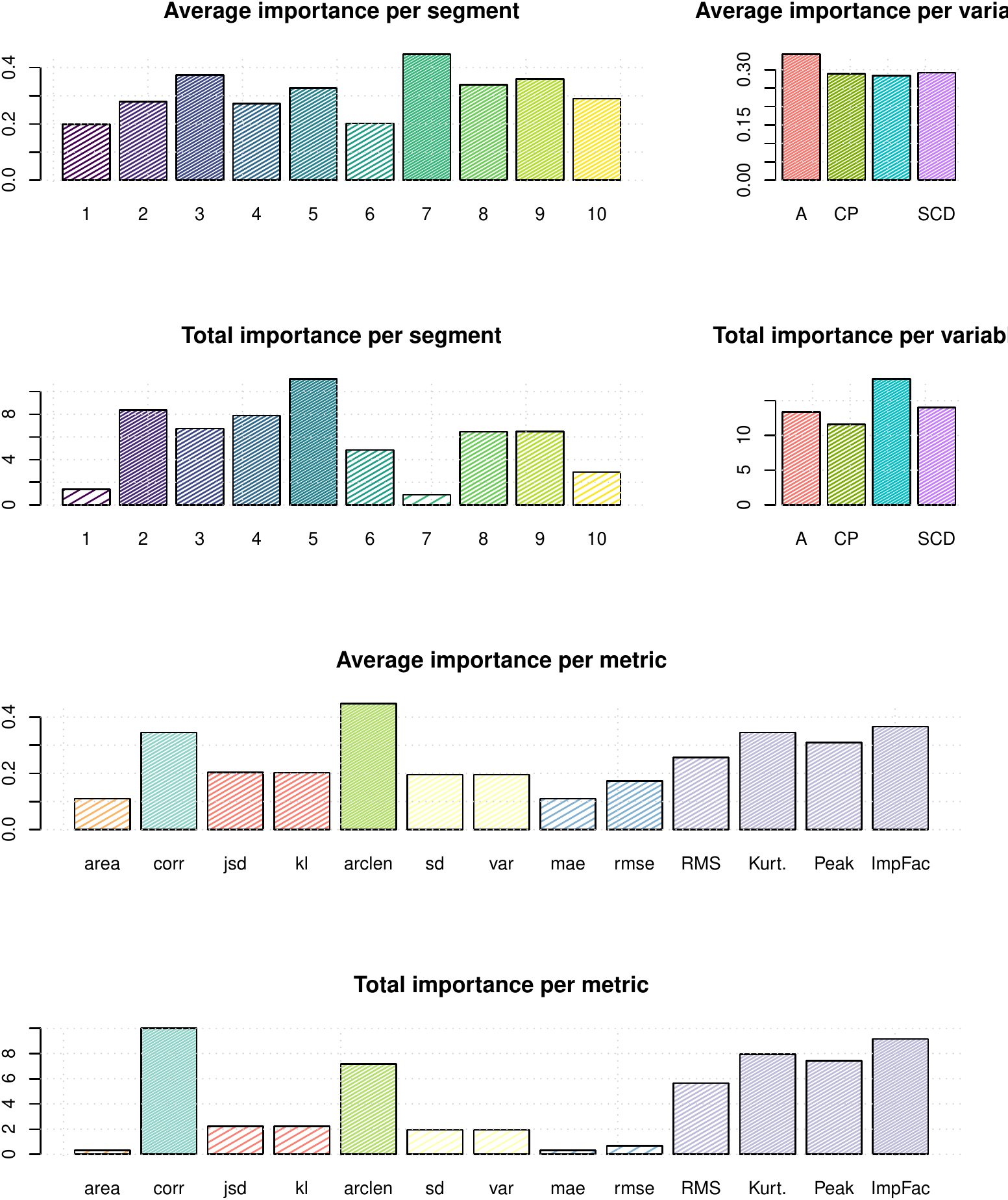}
\caption{\label{fig:ac-sc-weights-detailed-plot}Average and total importances of scores per segment, variable, and metric, for process model type I.}
\end{figure}

Let's look at the average importance per segment, variable, and metric averaged across \textbf{all} projects (figure \ref{fig:ac-sc-weights-detailed-plot}).

We may also show the most frequented segments, variables, and metrics:

\begin{figure}
\centering
\includegraphics{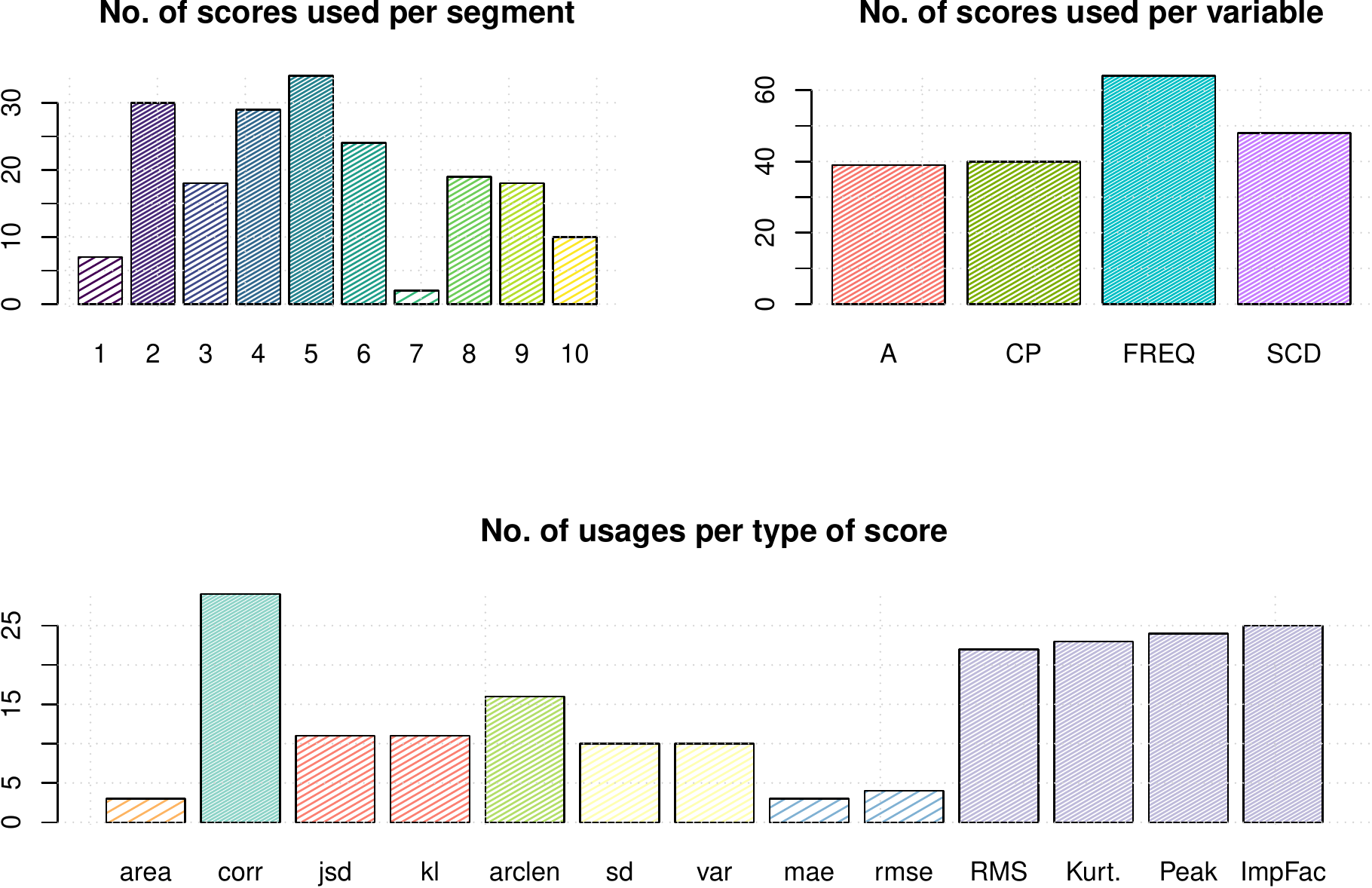}
\caption{\label{fig:ac-sc-weights-detailed-plot-counts}Number of score usages per segment, variable, and metric for process model type I.}
\end{figure}

The differences as of figures \ref{fig:ac-sc-weights-detailed-plot} and \ref{fig:ac-sc-weights-detailed-plot-counts} rely lie in the comparison of average importance and total importance/number of usages.
The average importance of each segments peaks at segment \(4\), is followed by the least important segment \(8\), only to be superseded by the absolute most important segment \(10\).
This means that it is more important to sample from a process in these segments, and that fluctuations in these segments are more sensible to the outcome (assessing the goodness of fit). This finding is also interesting w.r.t. the binary decision rule as used for issue-tracking data. It measures at three characteristic points in time of a project, namely at \(t_1=0.4\), \(t_2=0.85\), and \(t_{\mathrm{end}}=1\). If we look at the average importance of each segment, the choice for \(t_2\) is perhaps not ideal (under the assumptions that the source code variables quantify the same thing as the issue-tracking variables, which they do not).

The variables seem to be almost equally important on average. However, when we look at the total importance (that is, the sum of the weights per segment/metric/variable), this changes drastically, making the \texttt{FREQ} variable the most important.
Comparing average and total importance per type of metric is inconsistent, too, with the exception of the metrics \texttt{area}, \texttt{mae} (which is the same), and \texttt{rmse} being least important in either scenario.
Similarly, the signal theory metrics keep their importance between average and total.
A higher total than average importance could indicate that while a metric is suitable to quantify differences, comparatively many instance of it are required to accommodate the variance found in the observations.

\clearpage

\hypertarget{technical-report-detecting-the-fire-drill-anti-pattern-using-issue-tracking-data}{%
\section{\texorpdfstring{Technical Report: Detecting the Fire Drill anti-pattern using issue-tracking data\label{tr:fire-drill-issue-tracking-technical-report}}{Technical Report: Detecting the Fire Drill anti-pattern using issue-tracking data}}\label{technical-report-detecting-the-fire-drill-anti-pattern-using-issue-tracking-data}}

This is the self-contained technical report for detecting the Fire Drill anti-pattern using issue-tracking data.

\hypertarget{introduction-1}{%
\subsection{Introduction}\label{introduction-1}}

This is the complementary technical report for the paper/article tentatively entitled ``Multivariate Continuous Processes: Modeling, Instantiation, Goodness-of-fit, Forecasting''. Similar to the technical report for detecting the Fire Drill using source code, we import all projects' data and the ground truth. This notebook however is concerned with different and additional approaches, i.e., it is not just a repetition of the other technical report.

All complementary data and results can be found at Zenodo (Hönel, Pícha, et al. 2023). This notebook was written in a way that it can be run without any additional efforts to reproduce the outputs (using the pre-computed results). This notebook has a canonical URL\textsuperscript{\href{https://github.com/MrShoenel/anti-pattern-models/blob/master/notebooks/fire-drill-issue-tracking-technical-report.Rmd}{{[}Link{]}}} and can be read online as a rendered markdown\textsuperscript{\href{https://github.com/MrShoenel/anti-pattern-models/blob/master/notebooks/fire-drill-issue-tracking-technical-report.md}{{[}Link{]}}} version. All code can be found in this repository, too.

\hypertarget{importing-the-data}{%
\subsection{Importing the data}\label{importing-the-data}}

Here, we import the ground truth and the projects.

\hypertarget{ground-truth}{%
\subsubsection{Ground truth}\label{ground-truth}}

We have \(9\) projects conducted by students, and two raters have \textbf{independently}, i.e., without prior communication, assessed to what degree the AP is present in each project. This was done using a scale from zero to ten, where zero means that the AP was not present, and ten would indicate a strong manifestation The entire ground truth is shown in table \ref{tab:groundtruth}.

\begin{Shaded}
\begin{Highlighting}[]
\NormalTok{ground\_truth }\OtherTok{\textless{}{-}} \FunctionTok{read.csv}\NormalTok{(}\AttributeTok{file =} \StringTok{"../data/ground{-}truth.csv"}\NormalTok{, }\AttributeTok{sep =} \StringTok{";"}\NormalTok{)}
\NormalTok{ground\_truth}\SpecialCharTok{$}\NormalTok{consensus\_score }\OtherTok{\textless{}{-}}\NormalTok{ ground\_truth}\SpecialCharTok{$}\NormalTok{consensus}\SpecialCharTok{/}\DecValTok{10}
\end{Highlighting}
\end{Shaded}

\begin{table}

\caption{\label{tab:groundtruth}Entire ground truth as of both raters}
\centering
\begin{tabular}[t]{lrrrrr}
\toprule
project & rater.a & rater.b & consensus & rater.mean & consensus\_score\\
\midrule
project\_1 & 2 & 0 & 1 & 1.0 & 0.1\\
project\_2 & 0 & 0 & 0 & 0.0 & 0.0\\
project\_3 & 8 & 5 & 6 & 6.5 & 0.6\\
project\_4 & 8 & 6 & 8 & 7.0 & 0.8\\
project\_5 & 1 & 1 & 1 & 1.0 & 0.1\\
\addlinespace
project\_6 & 4 & 1 & 2 & 2.5 & 0.2\\
project\_7 & 2 & 3 & 3 & 2.5 & 0.3\\
project\_8 & 0 & 0 & 0 & 0.0 & 0.0\\
project\_9 & 1 & 4 & 5 & 2.5 & 0.5\\
\bottomrule
\end{tabular}
\end{table}

\hypertarget{project-data}{%
\subsubsection{Project data}\label{project-data}}

In this section we import the projects' \textbf{issue-tracking}-data. All projects' data will be normalized w.r.t. the time, i.e., each project will have a support of \([0,1]\). The variables are modeled as cumulative time spent on issues. Each variable in each project will be loaded into an instance of \texttt{Signal}.

\begin{Shaded}
\begin{Highlighting}[]
\FunctionTok{library}\NormalTok{(readxl)}

\NormalTok{load\_project\_issue\_data }\OtherTok{\textless{}{-}} \ControlFlowTok{function}\NormalTok{(pId) \{}
\NormalTok{  data }\OtherTok{\textless{}{-}} \FunctionTok{read\_excel}\NormalTok{(}\StringTok{"../data/FD\_issue{-}based\_detection.xlsx"}\NormalTok{, }\AttributeTok{sheet =}\NormalTok{ pId)}
\NormalTok{  data[}\FunctionTok{is.na}\NormalTok{(data)] }\OtherTok{\textless{}{-}} \DecValTok{0}

\NormalTok{  data}\SpecialCharTok{$}\NormalTok{req }\OtherTok{\textless{}{-}} \FunctionTok{as.numeric}\NormalTok{(data}\SpecialCharTok{$}\NormalTok{req)}
\NormalTok{  data}\SpecialCharTok{$}\NormalTok{dev }\OtherTok{\textless{}{-}} \FunctionTok{as.numeric}\NormalTok{(data}\SpecialCharTok{$}\NormalTok{dev)}
\NormalTok{  data}\SpecialCharTok{$}\NormalTok{desc }\OtherTok{\textless{}{-}} \FunctionTok{as.numeric}\NormalTok{(data}\SpecialCharTok{$}\NormalTok{desc)}

\NormalTok{  req\_cs }\OtherTok{\textless{}{-}} \FunctionTok{cumsum}\NormalTok{(data}\SpecialCharTok{$}\NormalTok{req)}\SpecialCharTok{/}\FunctionTok{sum}\NormalTok{(data}\SpecialCharTok{$}\NormalTok{req)}
\NormalTok{  dev\_cs }\OtherTok{\textless{}{-}} \FunctionTok{cumsum}\NormalTok{(data}\SpecialCharTok{$}\NormalTok{dev)}\SpecialCharTok{/}\FunctionTok{sum}\NormalTok{(data}\SpecialCharTok{$}\NormalTok{dev)}
\NormalTok{  desc\_cs }\OtherTok{\textless{}{-}} \FunctionTok{cumsum}\NormalTok{(data}\SpecialCharTok{$}\NormalTok{desc)}\SpecialCharTok{/}\FunctionTok{max}\NormalTok{(}\FunctionTok{cumsum}\NormalTok{(data}\SpecialCharTok{$}\NormalTok{dev))}
\NormalTok{  X }\OtherTok{\textless{}{-}} \FunctionTok{seq}\NormalTok{(}\AttributeTok{from =} \DecValTok{0}\NormalTok{, }\AttributeTok{to =} \DecValTok{1}\NormalTok{, }\AttributeTok{length.out =} \FunctionTok{length}\NormalTok{(req\_cs))}

\NormalTok{  signal\_req }\OtherTok{\textless{}{-}}\NormalTok{ Signal}\SpecialCharTok{$}\FunctionTok{new}\NormalTok{(}\AttributeTok{func =}\NormalTok{ stats}\SpecialCharTok{::}\FunctionTok{approxfun}\NormalTok{(}\AttributeTok{x =}\NormalTok{ X, }\AttributeTok{y =}\NormalTok{ req\_cs, }\AttributeTok{yleft =} \DecValTok{0}\NormalTok{,}
    \AttributeTok{yright =} \DecValTok{1}\NormalTok{), }\AttributeTok{name =} \StringTok{"REQ"}\NormalTok{, }\AttributeTok{support =} \FunctionTok{c}\NormalTok{(}\DecValTok{0}\NormalTok{, }\DecValTok{1}\NormalTok{), }\AttributeTok{isWp =} \ConstantTok{FALSE}\NormalTok{)}
\NormalTok{  signal\_dev }\OtherTok{\textless{}{-}}\NormalTok{ Signal}\SpecialCharTok{$}\FunctionTok{new}\NormalTok{(}\AttributeTok{func =}\NormalTok{ stats}\SpecialCharTok{::}\FunctionTok{approxfun}\NormalTok{(}\AttributeTok{x =}\NormalTok{ X, }\AttributeTok{y =}\NormalTok{ dev\_cs, }\AttributeTok{yleft =} \DecValTok{0}\NormalTok{,}
    \AttributeTok{yright =} \DecValTok{1}\NormalTok{), }\AttributeTok{name =} \StringTok{"DEV"}\NormalTok{, }\AttributeTok{support =} \FunctionTok{c}\NormalTok{(}\DecValTok{0}\NormalTok{, }\DecValTok{1}\NormalTok{), }\AttributeTok{isWp =} \ConstantTok{FALSE}\NormalTok{)}
\NormalTok{  signal\_desc }\OtherTok{\textless{}{-}}\NormalTok{ Signal}\SpecialCharTok{$}\FunctionTok{new}\NormalTok{(}\AttributeTok{func =}\NormalTok{ stats}\SpecialCharTok{::}\FunctionTok{approxfun}\NormalTok{(}\AttributeTok{x =}\NormalTok{ X, }\AttributeTok{y =}\NormalTok{ desc\_cs, }\AttributeTok{yleft =} \DecValTok{0}\NormalTok{,}
    \AttributeTok{yright =} \FunctionTok{max}\NormalTok{(desc\_cs)), }\AttributeTok{name =} \StringTok{"DESC"}\NormalTok{, }\AttributeTok{support =} \FunctionTok{c}\NormalTok{(}\DecValTok{0}\NormalTok{, }\DecValTok{1}\NormalTok{), }\AttributeTok{isWp =} \ConstantTok{FALSE}\NormalTok{)}

  \FunctionTok{list}\NormalTok{(}\AttributeTok{data =}\NormalTok{ data, }\AttributeTok{REQ =}\NormalTok{ signal\_req, }\AttributeTok{DEV =}\NormalTok{ signal\_dev, }\AttributeTok{DESC =}\NormalTok{ signal\_desc)}
\NormalTok{\}}
\end{Highlighting}
\end{Shaded}

\begin{figure}[ht!]

{\centering \includegraphics{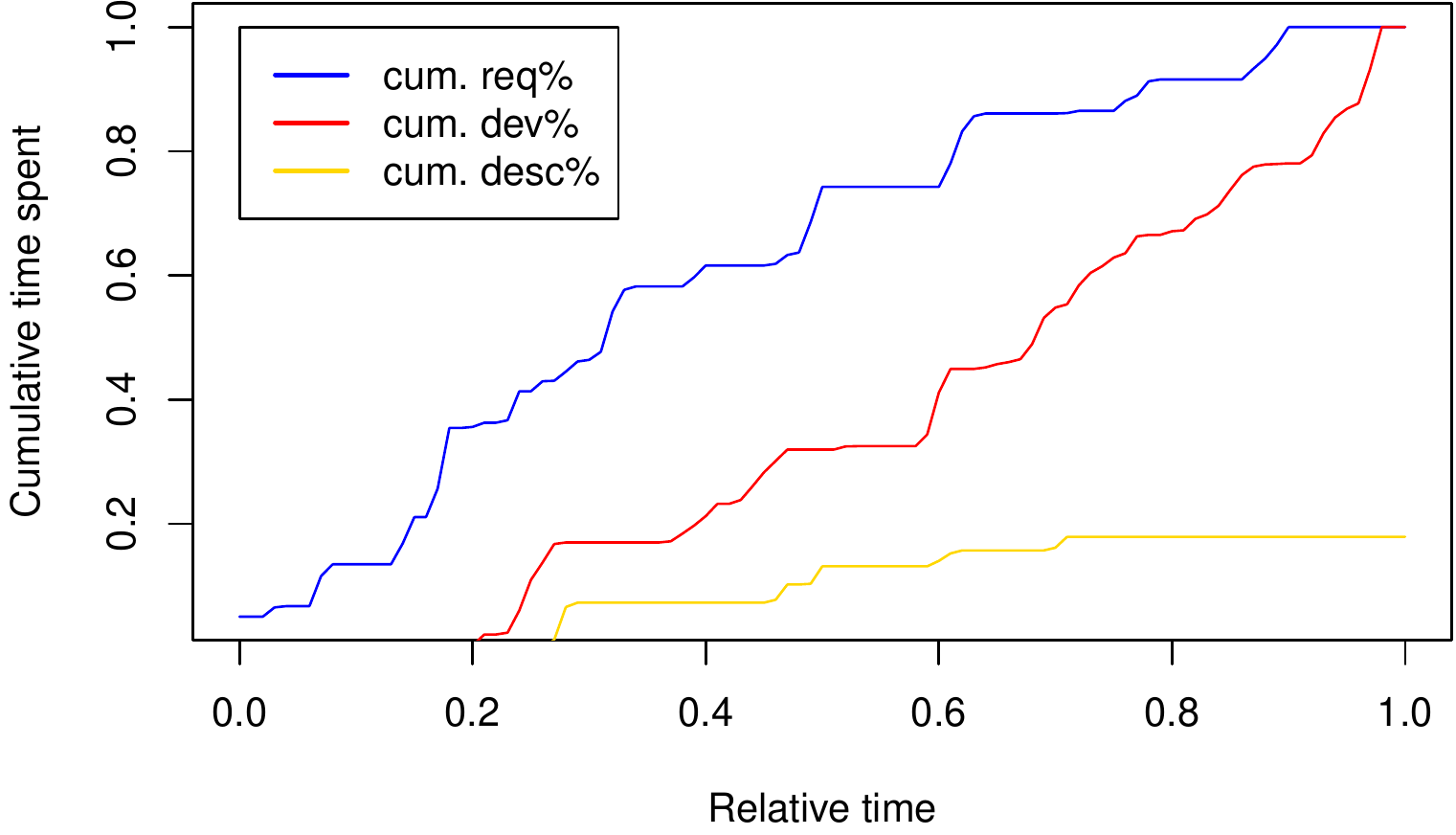} 

}

\caption{The three variables of the first project.}\label{fig:p1-example}
\end{figure}

Let's attempt to replicate the graphs of the first project (cf.~figure \ref{fig:p1-example}):

\begin{Shaded}
\begin{Highlighting}[]
\NormalTok{p3\_signals }\OtherTok{\textless{}{-}} \FunctionTok{load\_project\_issue\_data}\NormalTok{(}\AttributeTok{pId =} \StringTok{"Project3"}\NormalTok{)}
\NormalTok{req\_f }\OtherTok{\textless{}{-}}\NormalTok{ p3\_signals}\SpecialCharTok{$}\NormalTok{REQ}\SpecialCharTok{$}\FunctionTok{get0Function}\NormalTok{()}
\NormalTok{dev\_f }\OtherTok{\textless{}{-}}\NormalTok{ p3\_signals}\SpecialCharTok{$}\NormalTok{DEV}\SpecialCharTok{$}\FunctionTok{get0Function}\NormalTok{()}
\NormalTok{desc\_f }\OtherTok{\textless{}{-}}\NormalTok{ p3\_signals}\SpecialCharTok{$}\NormalTok{DESC}\SpecialCharTok{$}\FunctionTok{get0Function}\NormalTok{()}
\end{Highlighting}
\end{Shaded}

OK, that works well. It'll be the same for all projects, i.e., only two variables, time spent on requirements- and time spent on development-issues, is tracked. That means we will only be fitting two variables later.

\begin{figure}[ht!]
\includegraphics{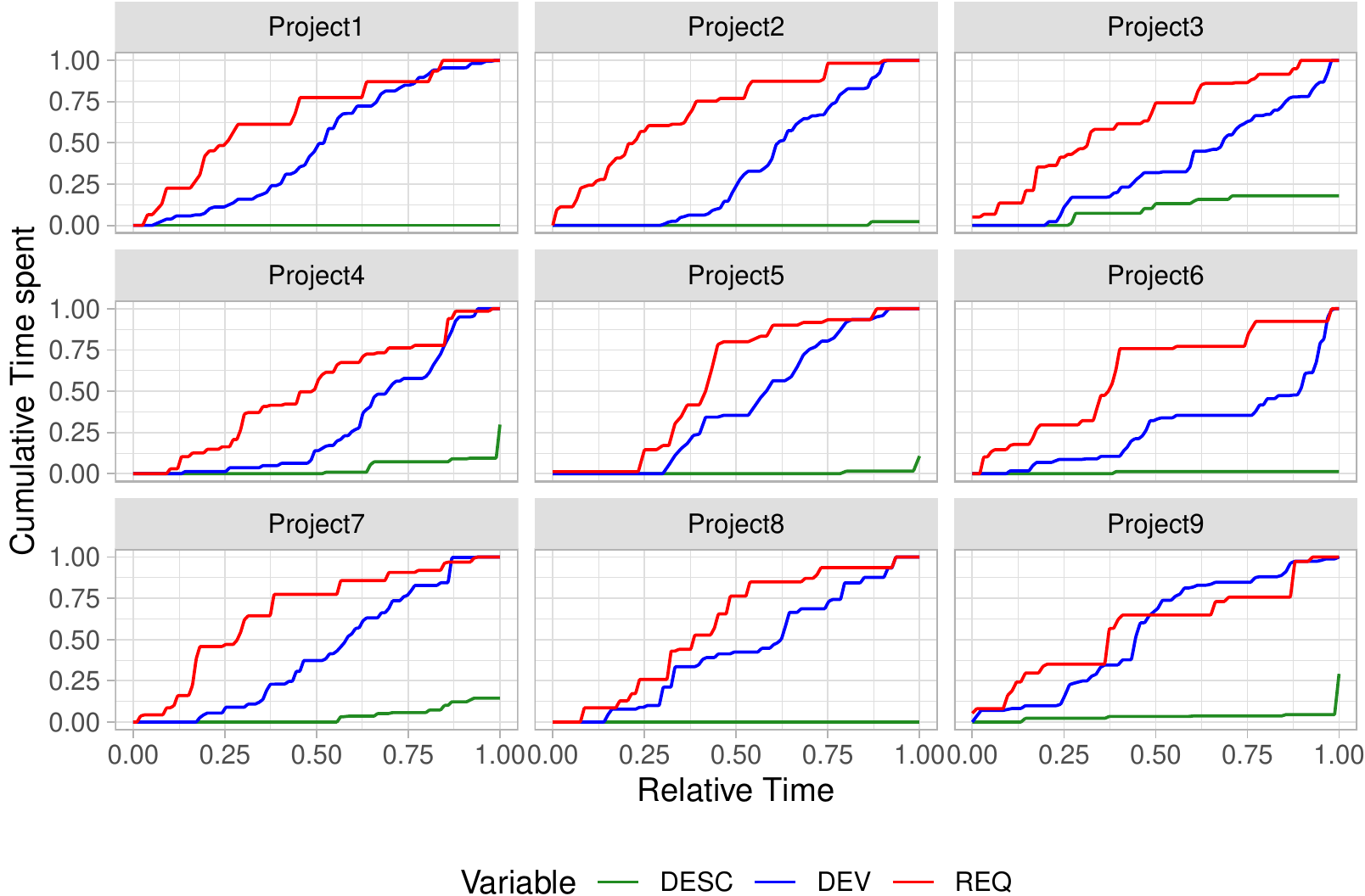} \caption{All variables over each project's time span.}\label{fig:project-it-vars}
\end{figure}

Let's load, store and visualize all projects (cf.~figure \ref{fig:project-it-vars}):

\begin{Shaded}
\begin{Highlighting}[]
\NormalTok{all\_signals }\OtherTok{\textless{}{-}} \FunctionTok{list}\NormalTok{()}
\ControlFlowTok{for}\NormalTok{ (pId }\ControlFlowTok{in} \FunctionTok{paste0}\NormalTok{(}\StringTok{"Project"}\NormalTok{, }\DecValTok{1}\SpecialCharTok{:}\DecValTok{9}\NormalTok{)) \{}
\NormalTok{  all\_signals[[pId]] }\OtherTok{\textless{}{-}} \FunctionTok{load\_project\_issue\_data}\NormalTok{(}\AttributeTok{pId =}\NormalTok{ pId)}
\NormalTok{\}}
\FunctionTok{invisible}\NormalTok{(}\FunctionTok{loadResultsOrCompute}\NormalTok{(}\AttributeTok{file =} \StringTok{"../results/project\_signals\_it.rds"}\NormalTok{, }\AttributeTok{computeExpr =}\NormalTok{ \{}
\NormalTok{  all\_signals}
\NormalTok{\}))}
\end{Highlighting}
\end{Shaded}

\hypertarget{patterns-for-scoring-the-projects-1}{%
\subsection{Patterns for scoring the projects}\label{patterns-for-scoring-the-projects-1}}

Here, we develop various patterns (process models) suitable for the detection of the Fire Drill anti-pattern using issue-tracking data.

\hypertarget{pattern-i-consensus-of-two-experts}{%
\subsubsection{Pattern I: Consensus of two experts}\label{pattern-i-consensus-of-two-experts}}

The initial pattern as defined for the detection of the Fire Drill AP is imported/created/defined here, and its variables and confidence intervals are modeled as continuous functions over time.

There are some values (x/y coordinates) for which we want to guarantee that the confidence intervals or the variables themselves pass through. Also, the two points in time \(t_1,t_2\) are defined to be at \(0.4\) and \(0.85\), respectively.

\begin{Shaded}
\begin{Highlighting}[]
\NormalTok{t\_1 }\OtherTok{\textless{}{-}} \FloatTok{0.4}
\NormalTok{t\_2 }\OtherTok{\textless{}{-}} \FloatTok{0.85}

\CommentTok{\# req(t\_1)}
\NormalTok{req\_t\_1 }\OtherTok{\textless{}{-}} \FloatTok{0.7}

\CommentTok{\# dev(t\_1), dev(t\_2)}
\NormalTok{dev\_t\_1 }\OtherTok{\textless{}{-}} \FloatTok{0.075}
\NormalTok{dev\_t\_2 }\OtherTok{\textless{}{-}} \FloatTok{0.4}
\end{Highlighting}
\end{Shaded}

This initial version of the pattern is not based on any data, observation or ground truth, but solely on two independent experts that reached a consensus for every value a priori any of the detection approaches.

\hypertarget{variable-requirements-analysis-planning}{%
\paragraph{Variable: Requirements, analysis, planning}\label{variable-requirements-analysis-planning}}

\begin{figure}[ht!]

{\centering \includegraphics{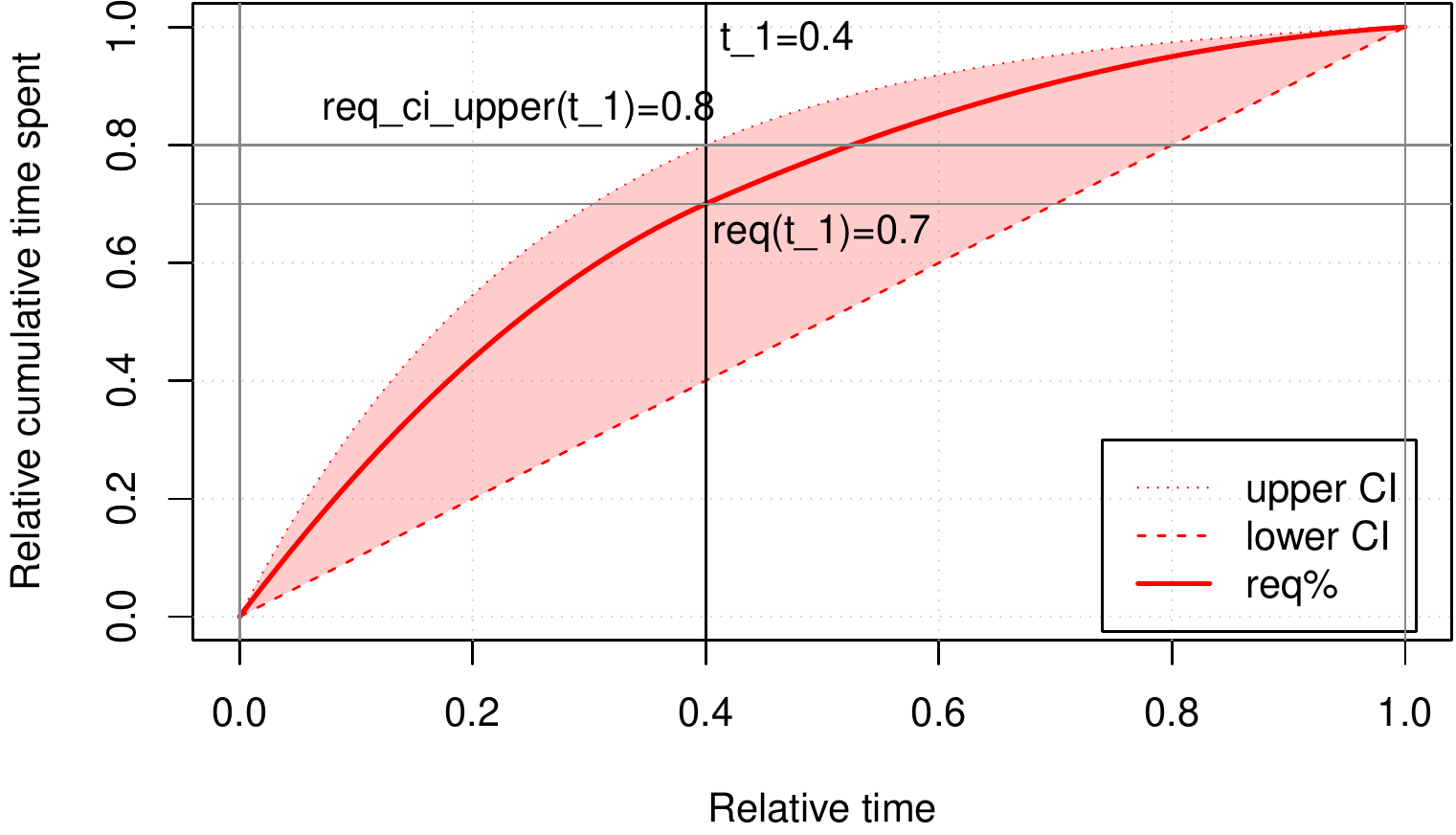} 

}

\caption{req\% and its lower- and upper confidence interval.}\label{fig:req-cis}
\end{figure}

The variable itself is not given, only its upper- and lower confidence-intervals (CI), where the latter simply is \(\operatorname{req}^{\text{CI}}_{\text{lower}}(x)=x\). The upper CI is given by the informal expression \(\operatorname{req}^{\text{CI}}_{\text{upper}}(x)=1.02261-1.02261\times\exp{(-3.811733\times x)}\). All together is shown in figure \ref{fig:req-cis}.

The variable itself is not given, as it was not important for the binary decision rule, whether or not a project's variable is within the confidence interval. It is still not important, what matters is that it runs through the confidence interval, and we will design it by fitting a polynomial through some inferred points from the plot. In some practical case however, the variable's course may be important, and while we will later use the variable to compute some kind of loss between it, the confidence interval and some project's variable, we only do this for demonstration purposes.

Let's first define the variable using some supporting x/y coordinates. It needs to be constrained such that it runs through 0,0 and 1,1:

\begin{Shaded}
\begin{Highlighting}[]
\NormalTok{req\_poly }\OtherTok{\textless{}{-}}\NormalTok{ cobs}\SpecialCharTok{::}\FunctionTok{cobs}\NormalTok{(}\AttributeTok{x =} \FunctionTok{seq}\NormalTok{(}\AttributeTok{from =} \DecValTok{0}\NormalTok{, }\AttributeTok{to =} \DecValTok{1}\NormalTok{, }\AttributeTok{by =} \FloatTok{0.1}\NormalTok{), }\AttributeTok{y =} \FunctionTok{c}\NormalTok{(}\DecValTok{0}\NormalTok{, }\FloatTok{0.25}\NormalTok{, }\FloatTok{0.425}\NormalTok{,}
  \FloatTok{0.475}\NormalTok{, }\FloatTok{0.7}\NormalTok{, }\FloatTok{0.8}\NormalTok{, }\FloatTok{0.85}\NormalTok{, }\FloatTok{0.9}\NormalTok{, }\FloatTok{0.95}\NormalTok{, }\FloatTok{0.975}\NormalTok{, }\DecValTok{1}\NormalTok{), }\AttributeTok{pointwise =} \FunctionTok{matrix}\NormalTok{(}\AttributeTok{data =} \FunctionTok{c}\NormalTok{(}\FunctionTok{c}\NormalTok{(}\DecValTok{0}\NormalTok{,}
  \DecValTok{0}\NormalTok{, }\DecValTok{0}\NormalTok{), }\FunctionTok{c}\NormalTok{(}\DecValTok{0}\NormalTok{, t\_1, req\_t\_1), }\FunctionTok{c}\NormalTok{(}\DecValTok{0}\NormalTok{, }\DecValTok{1}\NormalTok{, }\DecValTok{1}\NormalTok{)), }\AttributeTok{byrow =} \ConstantTok{TRUE}\NormalTok{, }\AttributeTok{ncol =} \DecValTok{3}\NormalTok{))}
\end{Highlighting}
\end{Shaded}

\begin{verbatim}
## Warning in .recacheSubclasses(def@className, def, env): undefined subclass
## "numericVector" of class "Mnumeric"; definition not updated
\end{verbatim}

\begin{verbatim}
## qbsks2():
##  Performing general knot selection ...
## 
##  Deleting unnecessary knots ...
\end{verbatim}

\begin{Shaded}
\begin{Highlighting}[]
\CommentTok{\# Now we can define the variable simply by predicting from the polynomial (btw.}
\CommentTok{\# this is vectorized automatically):}
\NormalTok{req }\OtherTok{\textless{}{-}} \ControlFlowTok{function}\NormalTok{(x) \{}
\NormalTok{  stats}\SpecialCharTok{::}\FunctionTok{predict}\NormalTok{(}\AttributeTok{object =}\NormalTok{ req\_poly, }\AttributeTok{z =}\NormalTok{ x)[, }\StringTok{"fit"}\NormalTok{]}
\NormalTok{\}}
\end{Highlighting}
\end{Shaded}

And now for the confidence intervals:

\begin{Shaded}
\begin{Highlighting}[]
\NormalTok{req\_ci\_lower }\OtherTok{\textless{}{-}} \ControlFlowTok{function}\NormalTok{(x) x}
\NormalTok{req\_ci\_upper }\OtherTok{\textless{}{-}} \ControlFlowTok{function}\NormalTok{(x) }\FloatTok{1.02261} \SpecialCharTok{{-}} \FloatTok{1.02261} \SpecialCharTok{*} \FunctionTok{exp}\NormalTok{(}\SpecialCharTok{{-}}\FloatTok{3.811733} \SpecialCharTok{*}\NormalTok{ x)}
\end{Highlighting}
\end{Shaded}

\hypertarget{variables-design-implementation-testing-bugfixing-and-descoping}{%
\paragraph{Variables: Design, implementation, testing, bugfixing and Descoping}\label{variables-design-implementation-testing-bugfixing-and-descoping}}

Again, the variable for design etc. is not given, but rather only its upper confidence interval. Its lower CI is simply always zero. The upper CI is given by the informal expression \(\operatorname{dev}^{\text{CI}}_{\text{upper}}(x)=0.07815904\times x+0.6222767\times x^2+0.2995643\times x^3\). The variable for de-scoping comes without confidence interval, and is defined by \(\operatorname{desc}(x)=0.01172386\times x + 0.0933415\times x^2 + 0.04493464\times x^3\).

First we will define/fit a polynomial that describes the variable for design etc., the same way we did for requirements etc. we do know that it should pass through the points \([t_1,\approx0.075]\), as well as \([t_2,\approx0.4]\).

\begin{Shaded}
\begin{Highlighting}[]
\NormalTok{dev\_poly }\OtherTok{\textless{}{-}}\NormalTok{ cobs}\SpecialCharTok{::}\FunctionTok{cobs}\NormalTok{(}\AttributeTok{x =} \FunctionTok{seq}\NormalTok{(}\AttributeTok{from =} \DecValTok{0}\NormalTok{, }\AttributeTok{to =} \DecValTok{1}\NormalTok{, }\AttributeTok{by =} \FloatTok{0.1}\NormalTok{), }\AttributeTok{y =} \FunctionTok{c}\NormalTok{(}\DecValTok{0}\NormalTok{, }\FloatTok{0.0175}\NormalTok{, }\FloatTok{0.035}\NormalTok{,}
  \FloatTok{0.055}\NormalTok{, dev\_t\_1, }\FloatTok{0.014}\NormalTok{, }\FloatTok{0.165}\NormalTok{, }\FloatTok{0.2}\NormalTok{, }\FloatTok{0.28}\NormalTok{, }\FloatTok{0.475}\NormalTok{, }\DecValTok{1}\NormalTok{), }\AttributeTok{print.warn =} \ConstantTok{FALSE}\NormalTok{, }\AttributeTok{print.mesg =} \ConstantTok{FALSE}\NormalTok{,}
  \AttributeTok{pointwise =} \FunctionTok{matrix}\NormalTok{(}\AttributeTok{data =} \FunctionTok{c}\NormalTok{(}\FunctionTok{c}\NormalTok{(}\DecValTok{0}\NormalTok{, t\_1, dev\_t\_1), }\FunctionTok{c}\NormalTok{(}\DecValTok{0}\NormalTok{, t\_2, dev\_t\_2), }\FunctionTok{c}\NormalTok{(}\DecValTok{0}\NormalTok{, }\DecValTok{1}\NormalTok{, }\DecValTok{1}\NormalTok{)),}
    \AttributeTok{byrow =} \ConstantTok{TRUE}\NormalTok{, }\AttributeTok{ncol =} \DecValTok{3}\NormalTok{))}

\CommentTok{\# Now we can define the variable simply by predicting from the polynomial (btw.}
\CommentTok{\# this is vectorized automatically):}
\NormalTok{dev }\OtherTok{\textless{}{-}} \ControlFlowTok{function}\NormalTok{(x) \{}
\NormalTok{  temp }\OtherTok{\textless{}{-}}\NormalTok{ stats}\SpecialCharTok{::}\FunctionTok{predict}\NormalTok{(}\AttributeTok{object =}\NormalTok{ dev\_poly, }\AttributeTok{z =}\NormalTok{ x)[, }\StringTok{"fit"}\NormalTok{]}
  \CommentTok{\# I cannot constrain the polynomial at 0,0 and it returns a very slight}
  \CommentTok{\# negative value there, so let\textquotesingle{}s do it this way:}
\NormalTok{  temp[temp }\SpecialCharTok{\textless{}} \DecValTok{0}\NormalTok{] }\OtherTok{\textless{}{-}} \DecValTok{0}
\NormalTok{  temp[temp }\SpecialCharTok{\textgreater{}} \DecValTok{1}\NormalTok{] }\OtherTok{\textless{}{-}} \DecValTok{1}
\NormalTok{  temp}
\NormalTok{\}}
\end{Highlighting}
\end{Shaded}

\begin{figure}[ht!]

{\centering \includegraphics{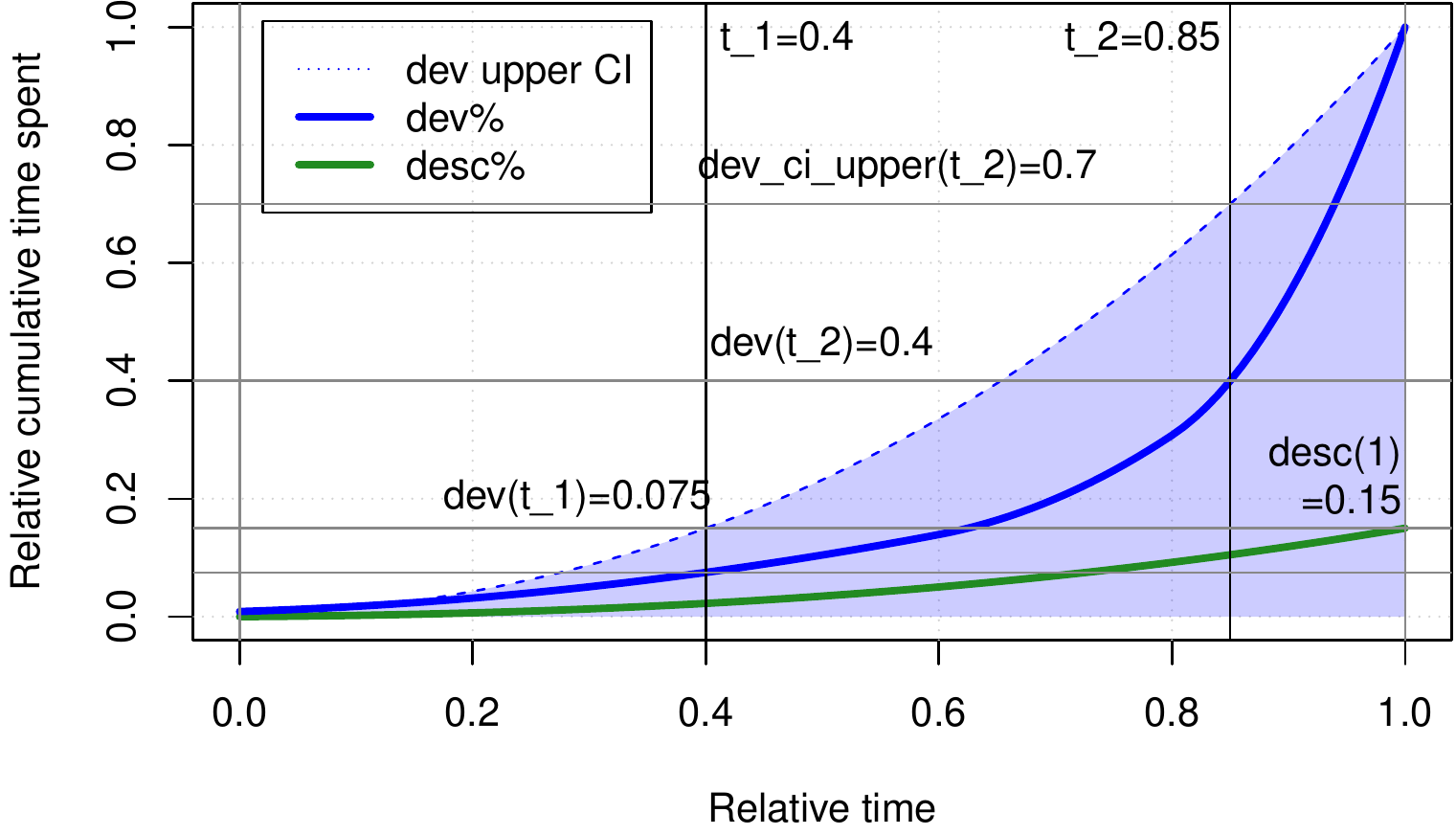} 

}

\caption{The variable dev\% and its upper confidence interval, as well as the variable desc\%.}\label{fig:dev-desc-cis}
\end{figure}

Next we define the upper confidence interval for the variable \texttt{DEV}, then the variable for de-scoping. All is shown in figure \ref{fig:dev-desc-cis}.

\begin{Shaded}
\begin{Highlighting}[]
\NormalTok{dev\_ci\_upper }\OtherTok{\textless{}{-}} \ControlFlowTok{function}\NormalTok{(x) }\FloatTok{0.07815904} \SpecialCharTok{*}\NormalTok{ x }\SpecialCharTok{+} \FloatTok{0.6222767} \SpecialCharTok{*}\NormalTok{ x}\SpecialCharTok{\^{}}\DecValTok{2} \SpecialCharTok{+} \FloatTok{0.2995643} \SpecialCharTok{*}\NormalTok{ x}\SpecialCharTok{\^{}}\DecValTok{3}
\NormalTok{desc }\OtherTok{\textless{}{-}} \ControlFlowTok{function}\NormalTok{(x) }\FloatTok{0.01172386} \SpecialCharTok{*}\NormalTok{ x }\SpecialCharTok{+} \FloatTok{0.0933415} \SpecialCharTok{*}\NormalTok{ x}\SpecialCharTok{\^{}}\DecValTok{2} \SpecialCharTok{+} \FloatTok{0.04493464} \SpecialCharTok{*}\NormalTok{ x}\SpecialCharTok{\^{}}\DecValTok{3}
\end{Highlighting}
\end{Shaded}

\hypertarget{pattern-ii-partial-adaptation-of-first-pattern}{%
\subsubsection{Pattern II: Partial adaptation of first pattern}\label{pattern-ii-partial-adaptation-of-first-pattern}}

We will be attempting three kinds of adaptations to the first pattern:

\begin{enumerate}
\def\labelenumi{\alph{enumi})}
\tightlist
\item
  Learn \(t_1,t_2\) from the data: There is not much to learn, but we could attempt to define these two thresholds as weighted average over the ground truth. Alternatively, we could formulate an optimization problem. We then use \emph{time warping} to alter the first pattern, \textbf{including} its confidence intervals.
\item
  Additionally to a (after learning \(t_1,t_2\)), we will apply \emph{amplitude warping} using \textbf{\texttt{srBTAW}}.
\item
  Take the first pattern and apply both, \emph{boundary time warping} \textbf{and} \emph{boundary amplitude warping}, to produce a pattern that is (hopefully) closest to all projects in the ground truth. This is the very same approach we attempted for adapting the pattern of type I that we defined for the Fire Drill in source code.
\end{enumerate}

\hypertarget{type-ii-a-adapt-type-i-using-thresholds-t_1t_2}{%
\paragraph{\texorpdfstring{Type II (a): Adapt type I using thresholds \(t_1,t_2\)\label{ssec:optim-t1t2}}{Type II (a): Adapt type I using thresholds t\_1,t\_2}}\label{type-ii-a-adapt-type-i-using-thresholds-t_1t_2}}

The two variables \texttt{REQ} and \texttt{DEV} in the first pattern describe the cumulative time spent on two distinct activities. It was designed with focus on the confidence intervals, and a binary decision rule, such that the actual variables' course was not of interest.

To find the optimal value for a threshold, we could look at when each project is closest to \(\operatorname{req}(t_1)\) and \(\operatorname{dev}(t_2)\) (in relative time), and then compute a weighted average over it. However, since we already modeled each project's variables as \emph{continuous-time stochastic process}, I suggest we use an optimization-based approach.

\begin{Shaded}
\begin{Highlighting}[]
\NormalTok{tempf }\OtherTok{\textless{}{-}}\NormalTok{ all\_signals}\SpecialCharTok{$}\NormalTok{Project5}\SpecialCharTok{$}\NormalTok{REQ}\SpecialCharTok{$}\FunctionTok{get0Function}\NormalTok{()}
\NormalTok{tempf1 }\OtherTok{\textless{}{-}} \ControlFlowTok{function}\NormalTok{(x) }\FunctionTok{abs}\NormalTok{(}\FunctionTok{tempf}\NormalTok{(x) }\SpecialCharTok{{-}}\NormalTok{ req\_t\_1)}
\NormalTok{optR }\OtherTok{\textless{}{-}} \FunctionTok{optimize}\NormalTok{(tempf1, }\AttributeTok{interval =} \FunctionTok{c}\NormalTok{(}\DecValTok{0}\NormalTok{, }\DecValTok{1}\NormalTok{))}
\NormalTok{optR}
\end{Highlighting}
\end{Shaded}

\begin{verbatim}
## $minimum
## [1] 0.6159966
## 
## $objective
## [1] 0.2
\end{verbatim}

\begin{figure}[ht!]

{\centering \includegraphics{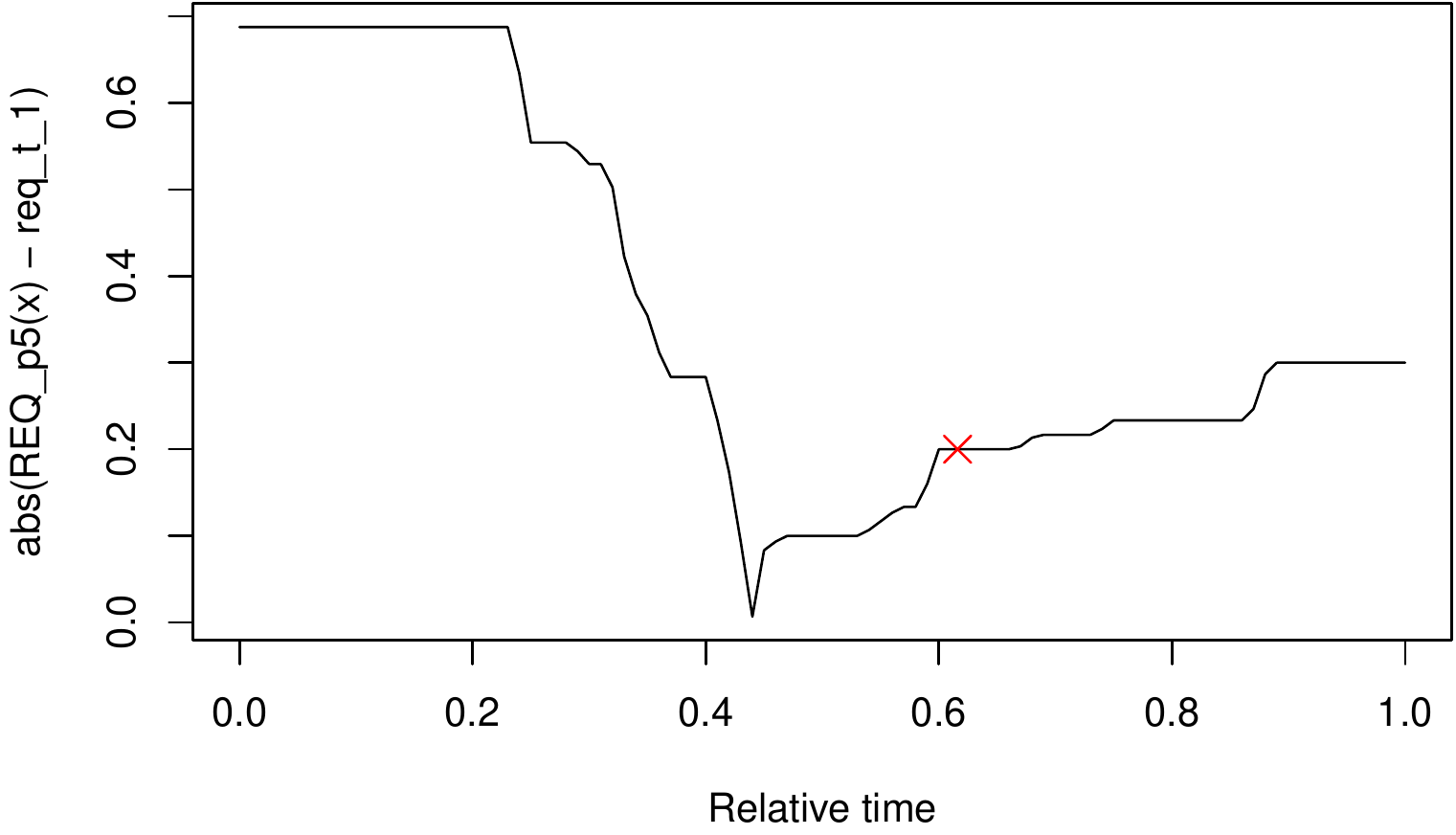} 

}

\caption{The non-optimal optimum found by gradient-based optimization in project 5.}\label{fig:t1t2-example-fig}
\end{figure}

We will find the optimum using \texttt{nlopt} and a global optimization, because we actually will have a global optimum by re-arranging each project's variables. Also, gradient-based methods do not work well because of the nature of the variables, having large horizontal plateaus. This can be seen in figure \ref{fig:t1t2-example-fig}. Approaches using the built-in \texttt{optim} do hence not work well, the problem is clearly demonstrated in the previous code chunk, resulting in an objective \(\gg0\) (which should ideally be \(0\)).

We want to find out when each project is closest to the previously defined thresholds. Each variable is a cumulative aggregation of the underlying values, which means that we have monotonic behavior.

\[
\begin{aligned}
  \min_{\hat{t}_1,\hat{t}_2\in R}&\;{\operatorname{req}(\hat{t}_1), \operatorname{dev}(\hat{t}_2)}\;\text{,}
  \\[1ex]
  \text{subject to}&\;0\leq\hat{t}_1,\hat{t}_2\leq1\;\text{, using}
  \\[1ex]
  \mathcal{L}_{\operatorname{req}}(x)=&\;\left\lVert\,\operatorname{req}(x)-\operatorname{req}(t_1)\,\right\rVert\;\text{, and}
  \\[1ex]
  \mathcal{L}_{\operatorname{dev}}(x)=&\;\left\lVert\,\operatorname{dev}(x)-\operatorname{dev}(t_2)\,\right\rVert\;\text{(quasi-convex loss functions).}
\end{aligned}
\]

The objective functions hence will be to find the global optimum (minimum), which occurs at \(y\approx0\). Since we have plateaus in our data, we will potentially have infinitely many global optima. However, we are satisfied with any that is \(\approx0\).

\[
\begin{aligned}
  \mathcal{O}_{\operatorname{dev}}(x)=&\;\underset{\hat{x}\in R}{arg\,min}\;{L_{\operatorname{dev}}(x)}\;\text{, and}
  \\[1ex]
  \mathcal{O}_{\operatorname{req}}(x)=&\;\underset{\hat{x}\in R}{arg\,min}\;{L_{\operatorname{req}}(x)}\text{.}
\end{aligned}
\]

\begin{Shaded}
\begin{Highlighting}[]
\FunctionTok{library}\NormalTok{(nloptr)}

\FunctionTok{set.seed}\NormalTok{(}\DecValTok{1}\NormalTok{)}

\NormalTok{t1t2\_opt }\OtherTok{\textless{}{-}} \FunctionTok{matrix}\NormalTok{(}\AttributeTok{ncol =} \DecValTok{4}\NormalTok{, }\AttributeTok{nrow =} \FunctionTok{length}\NormalTok{(all\_signals))}
\FunctionTok{rownames}\NormalTok{(t1t2\_opt) }\OtherTok{\textless{}{-}} \FunctionTok{names}\NormalTok{(all\_signals)}
\FunctionTok{colnames}\NormalTok{(t1t2\_opt) }\OtherTok{\textless{}{-}} \FunctionTok{c}\NormalTok{(}\StringTok{"req\_sol"}\NormalTok{, }\StringTok{"dev\_sol"}\NormalTok{, }\StringTok{"req\_obj"}\NormalTok{, }\StringTok{"dev\_obj"}\NormalTok{)}

\NormalTok{find\_global\_low }\OtherTok{\textless{}{-}} \ControlFlowTok{function}\NormalTok{(f) \{}
  \FunctionTok{nloptr}\NormalTok{(}\AttributeTok{x0 =} \FloatTok{0.5}\NormalTok{, }\AttributeTok{opts =} \FunctionTok{list}\NormalTok{(}\AttributeTok{maxeval =} \DecValTok{1000}\NormalTok{, }\AttributeTok{algorithm =} \StringTok{"NLOPT\_GN\_DIRECT\_L\_RAND"}\NormalTok{),}
    \AttributeTok{eval\_f =}\NormalTok{ f, }\AttributeTok{lb =} \DecValTok{0}\NormalTok{, }\AttributeTok{ub =} \DecValTok{1}\NormalTok{)}
\NormalTok{\}}

\ControlFlowTok{for}\NormalTok{ (pId }\ControlFlowTok{in} \FunctionTok{paste0}\NormalTok{(}\StringTok{"Project"}\NormalTok{, }\DecValTok{1}\SpecialCharTok{:}\DecValTok{9}\NormalTok{)) \{}
\NormalTok{  sig\_REQ }\OtherTok{\textless{}{-}}\NormalTok{ all\_signals[[pId]]}\SpecialCharTok{$}\NormalTok{REQ}\SpecialCharTok{$}\FunctionTok{get0Function}\NormalTok{()}
\NormalTok{  req\_abs }\OtherTok{\textless{}{-}} \ControlFlowTok{function}\NormalTok{(x) }\FunctionTok{abs}\NormalTok{(}\FunctionTok{sig\_REQ}\NormalTok{(x) }\SpecialCharTok{{-}}\NormalTok{ req\_t\_1)}

\NormalTok{  sig\_DEV }\OtherTok{\textless{}{-}}\NormalTok{ all\_signals[[pId]]}\SpecialCharTok{$}\NormalTok{DEV}\SpecialCharTok{$}\FunctionTok{get0Function}\NormalTok{()}
\NormalTok{  dev\_abs }\OtherTok{\textless{}{-}} \ControlFlowTok{function}\NormalTok{(x) }\FunctionTok{abs}\NormalTok{(}\FunctionTok{sig\_DEV}\NormalTok{(x) }\SpecialCharTok{{-}}\NormalTok{ dev\_t\_2)}

\NormalTok{  optRes\_REQ }\OtherTok{\textless{}{-}} \FunctionTok{find\_global\_low}\NormalTok{(}\AttributeTok{f =}\NormalTok{ req\_abs)}
\NormalTok{  optRes\_DEV }\OtherTok{\textless{}{-}} \FunctionTok{find\_global\_low}\NormalTok{(}\AttributeTok{f =}\NormalTok{ dev\_abs)}

\NormalTok{  t1t2\_opt[pId, ] }\OtherTok{\textless{}{-}} \FunctionTok{c}\NormalTok{(optRes\_REQ}\SpecialCharTok{$}\NormalTok{solution, optRes\_DEV}\SpecialCharTok{$}\NormalTok{solution, optRes\_REQ}\SpecialCharTok{$}\NormalTok{objective,}
\NormalTok{    optRes\_DEV}\SpecialCharTok{$}\NormalTok{objective)}
\NormalTok{\}}
\end{Highlighting}
\end{Shaded}

\begin{table}

\caption{\label{tab:t1t2-optvals}Optimum values for $t_1$ and $t_2$ for each project, together with the loss of each project's objective function at that offset (ideally 0).}
\centering
\begin{tabular}[t]{lrrrr}
\toprule
  & req\_sol & dev\_sol & req\_obj & dev\_obj\\
\midrule
Project1 & 0.4445887 & 0.4762108 & 0 & 0\\
Project2 & 0.3804348 & 0.5967236 & 0 & 0\\
Project3 & 0.4924851 & 0.5987420 & 0 & 0\\
Project4 & 0.6232323 & 0.6427128 & 0 & 0\\
Project5 & 0.4407407 & 0.5476190 & 0 & 0\\
\addlinespace
Project6 & 0.3973381 & 0.7788398 & 0 & 0\\
Project7 & 0.3780904 & 0.5306457 & 0 & 0\\
Project8 & 0.4775269 & 0.4451613 & 0 & 0\\
Project9 & 0.6582329 & 0.4357020 & 0 & 0\\
\bottomrule
\end{tabular}
\end{table}

From the table \ref{tab:t1t2-optvals}, we can take an example to demonstrate that the optimization indeed found the global optimum (or a value very close to it, usually with a deviation \(<1e^{-15}\)). For the picked example of project 5, we can clearly observe a value for \(x\) that results in a solution that is \(\approx0\).

\begin{figure}[ht!]

{\centering \includegraphics{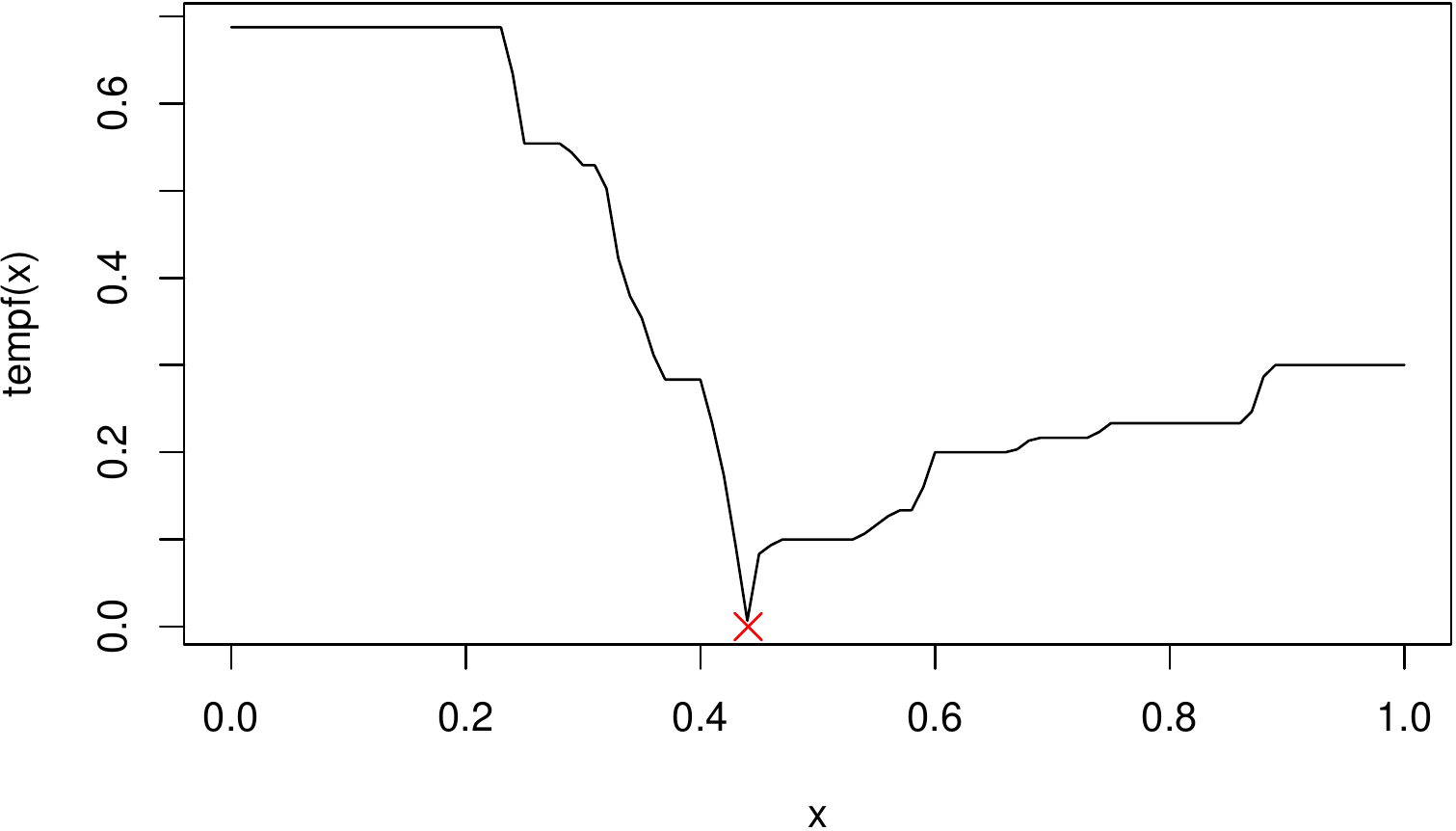} 

}

\caption{Example of finding the optimum for req($t_1$) in project 5.}\label{fig:test-test}
\end{figure}

Figure \ref{fig:test-test} depicts the optimal solution as found by using global optimization. Now what is left, is to calculate the weighted average for the optimized \(\bm{\hat{t}_1},\bm{\hat{t}_2}\). The weighted arithmetic mean is defined as:

\[
\begin{aligned}
  \text{weighted mean}=&\;\Big[\sum\bm{\omega}\Big]^{-1}\times\bm{\omega}^\top\cdot\bm{\hat{t}}\;\text{, where}
  \\[1ex]
  \bm{\omega}\dots&\;\text{weight vector that corresponds to the consensus-score, and}
  \\[1ex]
  \bm{\hat{t}}\dots&\;\text{vector with optimal values for either}\;t_1\;\text{or}\;t_2\;\text{(as learned earlier).}
\end{aligned}
\]

\begin{Shaded}
\begin{Highlighting}[]
\NormalTok{omega }\OtherTok{\textless{}{-}}\NormalTok{ ground\_truth}\SpecialCharTok{$}\NormalTok{consensus\_score}
\FunctionTok{names}\NormalTok{(omega) }\OtherTok{\textless{}{-}} \FunctionTok{paste0}\NormalTok{(}\StringTok{"Project"}\NormalTok{, }\DecValTok{1}\SpecialCharTok{:}\FunctionTok{length}\NormalTok{(omega))}

\NormalTok{t1\_wavg }\OtherTok{\textless{}{-}}\NormalTok{ t1t2\_opt[, }\DecValTok{1}\NormalTok{] }\SpecialCharTok{\%*\%}\NormalTok{ omega}\SpecialCharTok{/}\FunctionTok{sum}\NormalTok{(omega)}
\FunctionTok{print}\NormalTok{(}\FunctionTok{c}\NormalTok{(t\_1, t1\_wavg))}
\end{Highlighting}
\end{Shaded}

\begin{verbatim}
## [1] 0.4000000 0.5402389
\end{verbatim}

\begin{Shaded}
\begin{Highlighting}[]
\NormalTok{t2\_wavg }\OtherTok{\textless{}{-}}\NormalTok{ t1t2\_opt[, }\DecValTok{2}\NormalTok{] }\SpecialCharTok{\%*\%}\NormalTok{ omega}\SpecialCharTok{/}\FunctionTok{sum}\NormalTok{(omega)}
\FunctionTok{print}\NormalTok{(}\FunctionTok{c}\NormalTok{(t\_2, t2\_wavg))}
\end{Highlighting}
\end{Shaded}

\begin{verbatim}
## [1] 0.850000 0.580235
\end{verbatim}

Originally, \(t_1\) was guessed to be located at 0.4, with \(\operatorname{req}(t_1)=0.7\). The weighted average over the optimized values (where \(\operatorname{req}(\hat{t}_1)\approx0.7\)) suggests defining \(\hat{t}_1\)=0.54024.

\(t_2\) on the other hand was originally located at 0.85, with \(\operatorname{dev}(t_2)=0.4\). The weighted average over the optimized values (where \(\operatorname{dev}(\hat{t}_2)\approx0.4\)) suggests defining \(\hat{t}_2\)=0.58024.

Having learned these values, we can now adapt the pattern using time warping. For that, we have to instantiate the pattern using \texttt{srBTAW}, add the original boundaries and set them according to what we learned. Below, we define a function that can warp a single variable.

\begin{Shaded}
\begin{Highlighting}[]
\NormalTok{timewarp\_variable }\OtherTok{\textless{}{-}} \ControlFlowTok{function}\NormalTok{(f, t, t\_new) \{}
\NormalTok{  temp }\OtherTok{\textless{}{-}}\NormalTok{ SRBTW}\SpecialCharTok{$}\FunctionTok{new}\NormalTok{(}\AttributeTok{wp =}\NormalTok{ f, }\AttributeTok{wc =}\NormalTok{ f, }\AttributeTok{theta\_b =} \FunctionTok{c}\NormalTok{(}\DecValTok{0}\NormalTok{, t\_new, }\DecValTok{1}\NormalTok{), }\AttributeTok{gamma\_bed =} \FunctionTok{c}\NormalTok{(}\DecValTok{0}\NormalTok{,}
    \DecValTok{1}\NormalTok{, }\FunctionTok{sqrt}\NormalTok{(.Machine}\SpecialCharTok{$}\NormalTok{double.eps)), }\AttributeTok{lambda =} \FunctionTok{c}\NormalTok{(}\DecValTok{0}\NormalTok{, }\DecValTok{0}\NormalTok{), }\AttributeTok{begin =} \DecValTok{0}\NormalTok{, }\AttributeTok{end =} \DecValTok{1}\NormalTok{, }\AttributeTok{openBegin =} \ConstantTok{FALSE}\NormalTok{,}
    \AttributeTok{openEnd =} \ConstantTok{FALSE}\NormalTok{)}
\NormalTok{  temp}\SpecialCharTok{$}\FunctionTok{setParams}\NormalTok{(}\StringTok{\textasciigrave{}}\AttributeTok{names\textless{}{-}}\StringTok{\textasciigrave{}}\NormalTok{(}\FunctionTok{c}\NormalTok{(t, }\DecValTok{1} \SpecialCharTok{{-}}\NormalTok{ t), }\FunctionTok{c}\NormalTok{(}\StringTok{"vtl\_1"}\NormalTok{, }\StringTok{"vtl\_2"}\NormalTok{)))}
  \ControlFlowTok{function}\NormalTok{(x) }\FunctionTok{sapply}\NormalTok{(}\AttributeTok{X =}\NormalTok{ x, }\AttributeTok{FUN =}\NormalTok{ temp}\SpecialCharTok{$}\NormalTok{M)}
\NormalTok{\}}
\end{Highlighting}
\end{Shaded}

\begin{Shaded}
\begin{Highlighting}[]
\NormalTok{req\_p2a }\OtherTok{\textless{}{-}} \FunctionTok{timewarp\_variable}\NormalTok{(}\AttributeTok{f =}\NormalTok{ req, }\AttributeTok{t =}\NormalTok{ t\_1, }\AttributeTok{t\_new =}\NormalTok{ t1\_wavg)}
\NormalTok{req\_ci\_lower\_p2a }\OtherTok{\textless{}{-}} \FunctionTok{timewarp\_variable}\NormalTok{(}\AttributeTok{f =}\NormalTok{ req\_ci\_lower, }\AttributeTok{t =}\NormalTok{ t\_1, }\AttributeTok{t\_new =}\NormalTok{ t1\_wavg)}
\NormalTok{req\_ci\_upper\_p2a }\OtherTok{\textless{}{-}} \FunctionTok{timewarp\_variable}\NormalTok{(}\AttributeTok{f =}\NormalTok{ req\_ci\_upper, }\AttributeTok{t =}\NormalTok{ t\_1, }\AttributeTok{t\_new =}\NormalTok{ t1\_wavg)}
\end{Highlighting}
\end{Shaded}

\begin{figure}[ht!]

{\centering \includegraphics{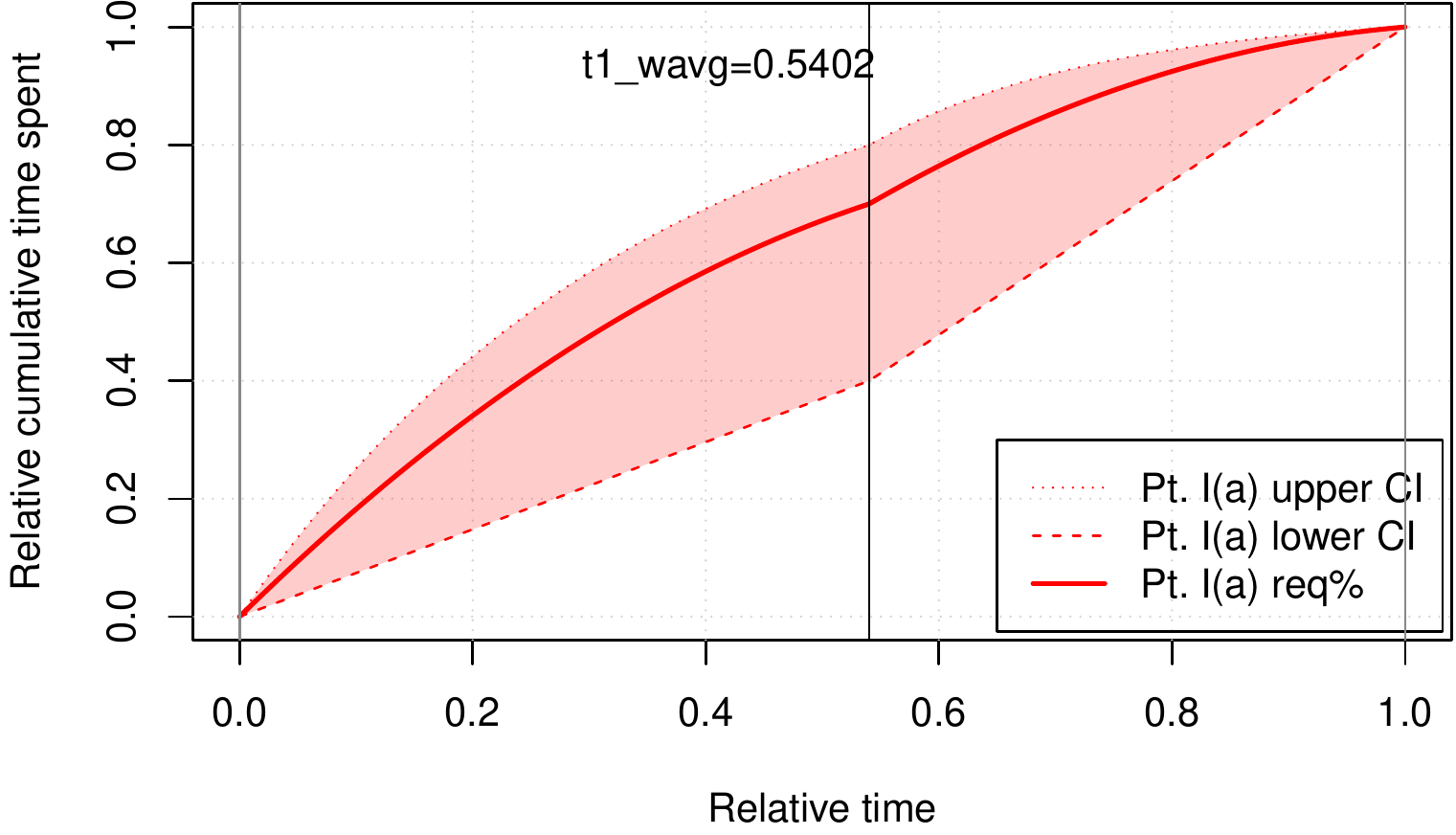} 

}

\caption{req\% and its lower- and upper confidence interval after time warping for pattern type I (a).}\label{fig:req-p1a-cis}
\end{figure}

Moving the boundary \(t_1\) farther behind, changes the variable \texttt{REQ} and its confidence interval slightly, as can be seen in figure \ref{fig:req-p1a-cis}. Next, we will adapt the remaining variables and their confidence intervals.

\begin{Shaded}
\begin{Highlighting}[]
\NormalTok{dev\_p2a }\OtherTok{\textless{}{-}} \FunctionTok{timewarp\_variable}\NormalTok{(}\AttributeTok{f =}\NormalTok{ dev, }\AttributeTok{t =}\NormalTok{ t\_2, }\AttributeTok{t\_new =}\NormalTok{ t2\_wavg)}
\NormalTok{dev\_ci\_upper\_p2a }\OtherTok{\textless{}{-}} \FunctionTok{timewarp\_variable}\NormalTok{(}\AttributeTok{f =}\NormalTok{ dev\_ci\_upper, }\AttributeTok{t =}\NormalTok{ t\_2, }\AttributeTok{t\_new =}\NormalTok{ t2\_wavg)}
\NormalTok{desc\_p2a }\OtherTok{\textless{}{-}} \FunctionTok{timewarp\_variable}\NormalTok{(}\AttributeTok{f =}\NormalTok{ desc, }\AttributeTok{t =}\NormalTok{ t\_2, }\AttributeTok{t\_new =}\NormalTok{ t2\_wavg)}
\end{Highlighting}
\end{Shaded}

\begin{figure}[ht!]

{\centering \includegraphics{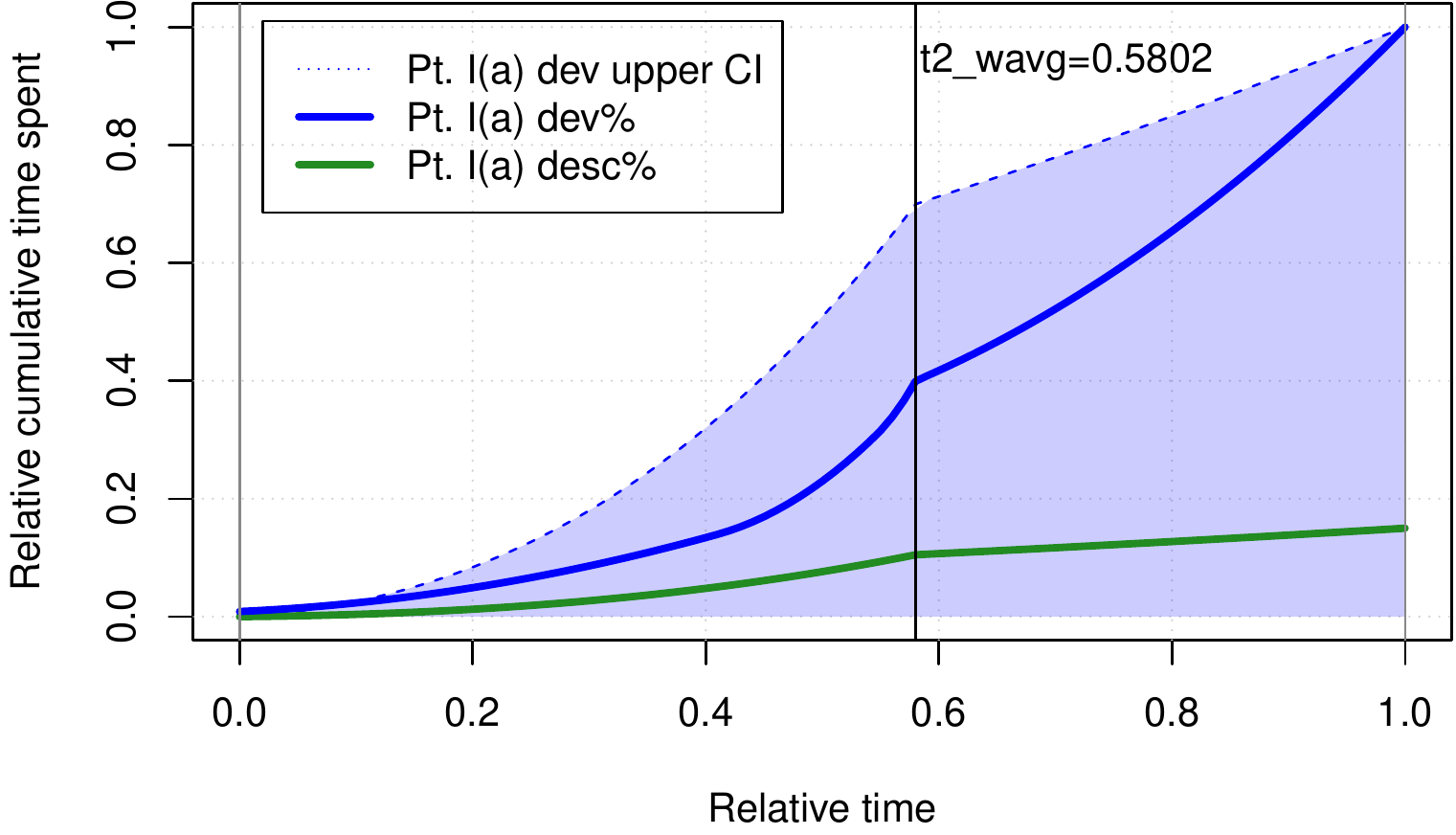} 

}

\caption{The variable dev\% and its upper confidence interval, as well as the variable desc\%, after time warping for pattern type I (a).}\label{fig:dev-desc-p1a-cis}
\end{figure}

Moving the boundary \(t_2\) was a more significant change for variables \texttt{DEV} and \texttt{DESC} than it was for \texttt{REQ}, as of figure \ref{fig:dev-desc-p1a-cis}.

\hypertarget{pattern-iii-averaging-the-ground-truth-1}{%
\subsubsection{Pattern III: Averaging the ground truth}\label{pattern-iii-averaging-the-ground-truth-1}}

This is the same approach we undertook for pattern type III (average) for the Fire Drill in source code. However, we will also learn an \textbf{empirical confidence interval}, which is later used for two additional detection methods. These methods have the advantage that they work over arbitrary (integration) intervals, making them also applicable for early detection of the process (i.e., not the entire process needs to be observed, and we can just attempt to detect what we have so far).

\begin{Shaded}
\begin{Highlighting}[]
\NormalTok{p3\_weighted\_var }\OtherTok{\textless{}{-}} \ControlFlowTok{function}\NormalTok{(name, omega) \{}
\NormalTok{  funcs }\OtherTok{\textless{}{-}} \FunctionTok{list}\NormalTok{()}
  \ControlFlowTok{for}\NormalTok{ (pId }\ControlFlowTok{in} \FunctionTok{names}\NormalTok{(all\_signals)) \{}
\NormalTok{    funcs[[pId]] }\OtherTok{\textless{}{-}}\NormalTok{ all\_signals[[pId]][[name]]}\SpecialCharTok{$}\FunctionTok{get0Function}\NormalTok{()}
\NormalTok{  \}}

  \ControlFlowTok{function}\NormalTok{(x) }\FunctionTok{sapply}\NormalTok{(}\AttributeTok{X =}\NormalTok{ x, }\AttributeTok{FUN =} \ControlFlowTok{function}\NormalTok{(x\_) \{}
\NormalTok{    omega }\SpecialCharTok{\%*\%} \FunctionTok{unlist}\NormalTok{(}\FunctionTok{lapply}\NormalTok{(funcs, }\ControlFlowTok{function}\NormalTok{(f) }\FunctionTok{f}\NormalTok{(x\_)))}\SpecialCharTok{/}\FunctionTok{sum}\NormalTok{(omega)}
\NormalTok{  \})}
\NormalTok{\}}
\end{Highlighting}
\end{Shaded}

\begin{Shaded}
\begin{Highlighting}[]
\NormalTok{req\_p3 }\OtherTok{\textless{}{-}} \FunctionTok{p3\_weighted\_var}\NormalTok{(}\AttributeTok{name =} \StringTok{"REQ"}\NormalTok{, }\AttributeTok{omega =}\NormalTok{ omega)}
\NormalTok{dev\_p3 }\OtherTok{\textless{}{-}} \FunctionTok{p3\_weighted\_var}\NormalTok{(}\AttributeTok{name =} \StringTok{"DEV"}\NormalTok{, }\AttributeTok{omega =}\NormalTok{ omega)}
\NormalTok{desc\_p3 }\OtherTok{\textless{}{-}} \FunctionTok{p3\_weighted\_var}\NormalTok{(}\AttributeTok{name =} \StringTok{"DESC"}\NormalTok{, }\AttributeTok{omega =}\NormalTok{ omega)}
\end{Highlighting}
\end{Shaded}

\begin{figure}[ht!]

{\centering \includegraphics{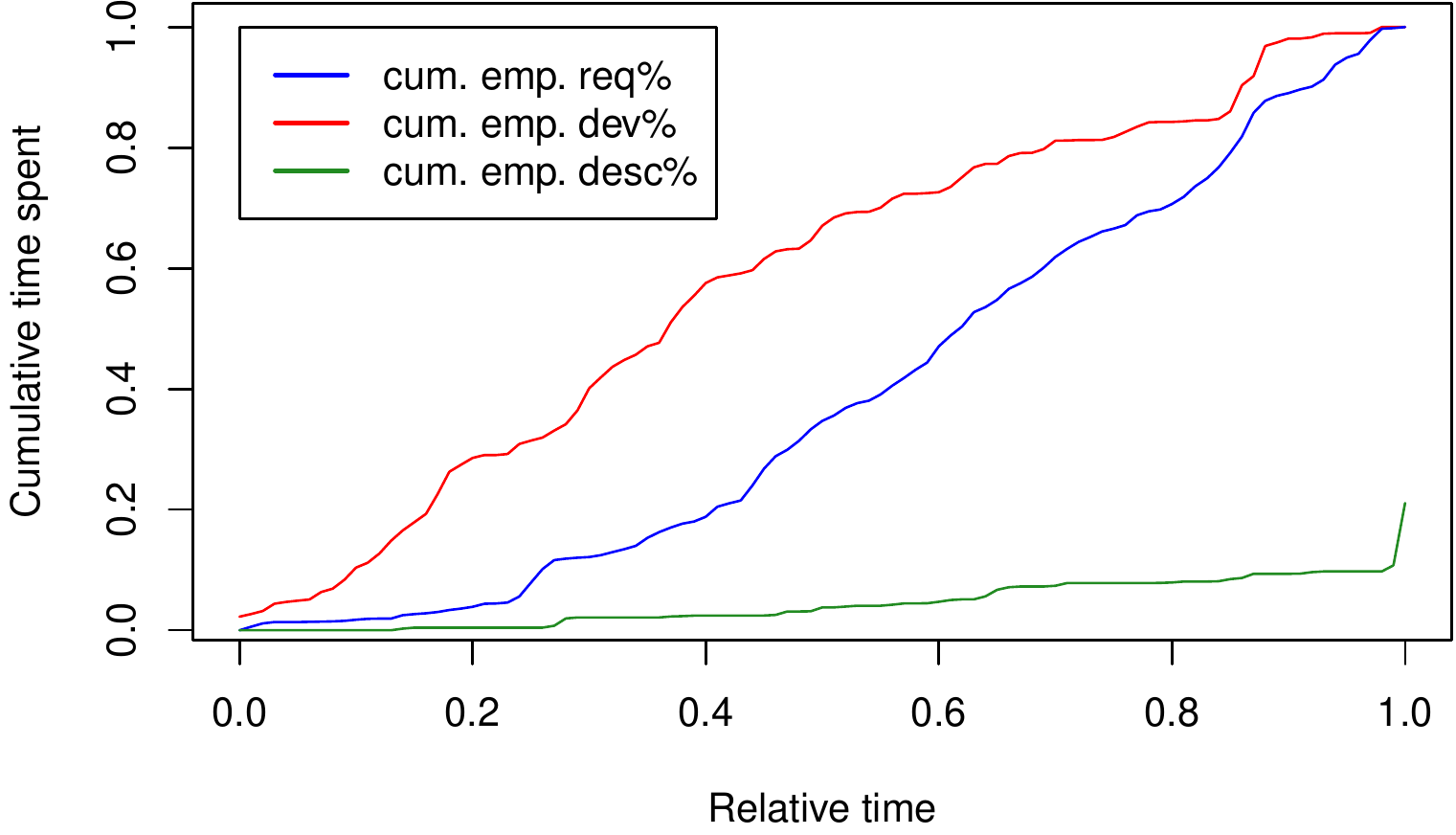} 

}

\caption{The weighted average for the three variables req\%, dev\% and desc\%.}\label{fig:avg-req-dev}
\end{figure}

The computed weighted average-variables for \texttt{REQ} and \texttt{DEV} are shown in figure \ref{fig:avg-req-dev}.

\hypertarget{determining-an-empirical-and-inhomogeneous-confidence-interval}{%
\paragraph{\texorpdfstring{Determining an empirical and inhomogeneous confidence interval\label{sssec:inhomo-conf-interval}}{Determining an empirical and inhomogeneous confidence interval}}\label{determining-an-empirical-and-inhomogeneous-confidence-interval}}

Also, we want to calculate an empirical confidence interval and -surface, based on the projects' data and the consensus of the ground truth. The boundary of the lower confidence interval is defined as the infimum of all signals (and the upper CI as the supremum of all signals):

\[
\begin{aligned}
  \bm{f}\dots&\;\text{vector of functions (here: project signals),}
  \\[1ex]
  \operatorname{CI}_{\text{upper}}(x)=&\;\sup{\Bigg(\forall\,f\in\bm{f}\;\bigg[\begin{cases}
    -\infty,&\text{if}\;\bm{\omega}_n=0,
    \\
    f_n(x),&\text{otherwise}
  \end{cases}\bigg]\;,\;\frown\;,\;\Big[\dots\Big]\Bigg)}\;\text{,}
  \\[1ex]
  \operatorname{CI}_{\text{upper}}(x)=&\;\inf{\Bigg(\forall\,f\in\bm{f}\;\bigg[\begin{cases}
    \infty,&\text{if}\;\bm{\omega}_n=0,
    \\
    f_n(x),&\text{otherwise}
  \end{cases}\bigg]\;,\;\frown\;,\;\Big[\dots\Big]\Bigg)}\;\text{.}
\end{aligned}
\]

\begin{Shaded}
\begin{Highlighting}[]
\NormalTok{funclist\_REQ }\OtherTok{\textless{}{-}} \FunctionTok{list}\NormalTok{()}
\NormalTok{funclist\_DEV }\OtherTok{\textless{}{-}} \FunctionTok{list}\NormalTok{()}
\NormalTok{funclist\_DESC }\OtherTok{\textless{}{-}} \FunctionTok{list}\NormalTok{()}
\ControlFlowTok{for}\NormalTok{ (pId }\ControlFlowTok{in} \FunctionTok{names}\NormalTok{(all\_signals)) \{}
\NormalTok{  funclist\_REQ[[pId]] }\OtherTok{\textless{}{-}}\NormalTok{ all\_signals[[pId]]}\SpecialCharTok{$}\NormalTok{REQ}\SpecialCharTok{$}\FunctionTok{get0Function}\NormalTok{()}
\NormalTok{  funclist\_DEV[[pId]] }\OtherTok{\textless{}{-}}\NormalTok{ all\_signals[[pId]]}\SpecialCharTok{$}\NormalTok{DEV}\SpecialCharTok{$}\FunctionTok{get0Function}\NormalTok{()}
\NormalTok{  funclist\_DESC[[pId]] }\OtherTok{\textless{}{-}}\NormalTok{ all\_signals[[pId]]}\SpecialCharTok{$}\NormalTok{DESC}\SpecialCharTok{$}\FunctionTok{get0Function}\NormalTok{()}
\NormalTok{\}}

\NormalTok{CI\_bound\_p3avg }\OtherTok{\textless{}{-}} \ControlFlowTok{function}\NormalTok{(x, funclist, omega, }\AttributeTok{upper =} \ConstantTok{TRUE}\NormalTok{) \{}
  \FunctionTok{sapply}\NormalTok{(}\AttributeTok{X =}\NormalTok{ x, }\AttributeTok{FUN =} \ControlFlowTok{function}\NormalTok{(x\_) \{}
\NormalTok{    val }\OtherTok{\textless{}{-}} \FunctionTok{unlist}\NormalTok{(}\FunctionTok{lapply}\NormalTok{(}\AttributeTok{X =} \FunctionTok{names}\NormalTok{(funclist), }\AttributeTok{FUN =} \ControlFlowTok{function}\NormalTok{(fname) \{}
      \ControlFlowTok{if}\NormalTok{ (omega[fname] }\SpecialCharTok{==} \DecValTok{0}\NormalTok{)}
\NormalTok{        (}\ControlFlowTok{if}\NormalTok{ (upper)}
          \SpecialCharTok{{-}}\ConstantTok{Inf} \ControlFlowTok{else} \ConstantTok{Inf}\NormalTok{) }\ControlFlowTok{else}\NormalTok{ funclist[[fname]](x\_)}
\NormalTok{    \}))}

    \ControlFlowTok{if}\NormalTok{ (upper)}
      \FunctionTok{max}\NormalTok{(val) }\ControlFlowTok{else} \FunctionTok{min}\NormalTok{(val)}
\NormalTok{  \})}
\NormalTok{\}}

\NormalTok{req\_ci\_upper\_p3avg }\OtherTok{\textless{}{-}} \ControlFlowTok{function}\NormalTok{(x) }\FunctionTok{CI\_bound\_p3avg}\NormalTok{(}\AttributeTok{x =}\NormalTok{ x, }\AttributeTok{funclist =}\NormalTok{ funclist\_REQ,}
  \AttributeTok{omega =}\NormalTok{ omega, }\AttributeTok{upper =} \ConstantTok{TRUE}\NormalTok{)}
\NormalTok{req\_ci\_lower\_p3avg }\OtherTok{\textless{}{-}} \ControlFlowTok{function}\NormalTok{(x) }\FunctionTok{CI\_bound\_p3avg}\NormalTok{(}\AttributeTok{x =}\NormalTok{ x, }\AttributeTok{funclist =}\NormalTok{ funclist\_REQ,}
  \AttributeTok{omega =}\NormalTok{ omega, }\AttributeTok{upper =} \ConstantTok{FALSE}\NormalTok{)}
\NormalTok{dev\_ci\_upper\_p3avg }\OtherTok{\textless{}{-}} \ControlFlowTok{function}\NormalTok{(x) }\FunctionTok{CI\_bound\_p3avg}\NormalTok{(}\AttributeTok{x =}\NormalTok{ x, }\AttributeTok{funclist =}\NormalTok{ funclist\_DEV,}
  \AttributeTok{omega =}\NormalTok{ omega, }\AttributeTok{upper =} \ConstantTok{TRUE}\NormalTok{)}
\NormalTok{dev\_ci\_lower\_p3avg }\OtherTok{\textless{}{-}} \ControlFlowTok{function}\NormalTok{(x) }\FunctionTok{CI\_bound\_p3avg}\NormalTok{(}\AttributeTok{x =}\NormalTok{ x, }\AttributeTok{funclist =}\NormalTok{ funclist\_DEV,}
  \AttributeTok{omega =}\NormalTok{ omega, }\AttributeTok{upper =} \ConstantTok{FALSE}\NormalTok{)}
\NormalTok{desc\_ci\_upper\_p3avg }\OtherTok{\textless{}{-}} \ControlFlowTok{function}\NormalTok{(x) }\FunctionTok{CI\_bound\_p3avg}\NormalTok{(}\AttributeTok{x =}\NormalTok{ x, }\AttributeTok{funclist =}\NormalTok{ funclist\_DESC,}
  \AttributeTok{omega =}\NormalTok{ omega, }\AttributeTok{upper =} \ConstantTok{TRUE}\NormalTok{)}
\NormalTok{desc\_ci\_lower\_p3avg }\OtherTok{\textless{}{-}} \ControlFlowTok{function}\NormalTok{(x) }\FunctionTok{CI\_bound\_p3avg}\NormalTok{(}\AttributeTok{x =}\NormalTok{ x, }\AttributeTok{funclist =}\NormalTok{ funclist\_DESC,}
  \AttributeTok{omega =}\NormalTok{ omega, }\AttributeTok{upper =} \ConstantTok{FALSE}\NormalTok{)}
\end{Highlighting}
\end{Shaded}

While the above expressions define the \emph{boundaries} of the lower and upper confidence intervals, we also need a function that interpolates in between. Recall that the CI of the first pattern was \textbf{homogeneous}, i.e., it provided no gradation, and was used for a binary decision rule. If we define a function that bases the strength of the confidence on the values of the ground truth of each project's variable, then it also means that with that pattern, all projects are included in the binary decision rule. Having gradation in the CI will allow us to make more probabilistic statements by computing some kind of score.

\begin{figure}[ht!]

{\centering \includegraphics{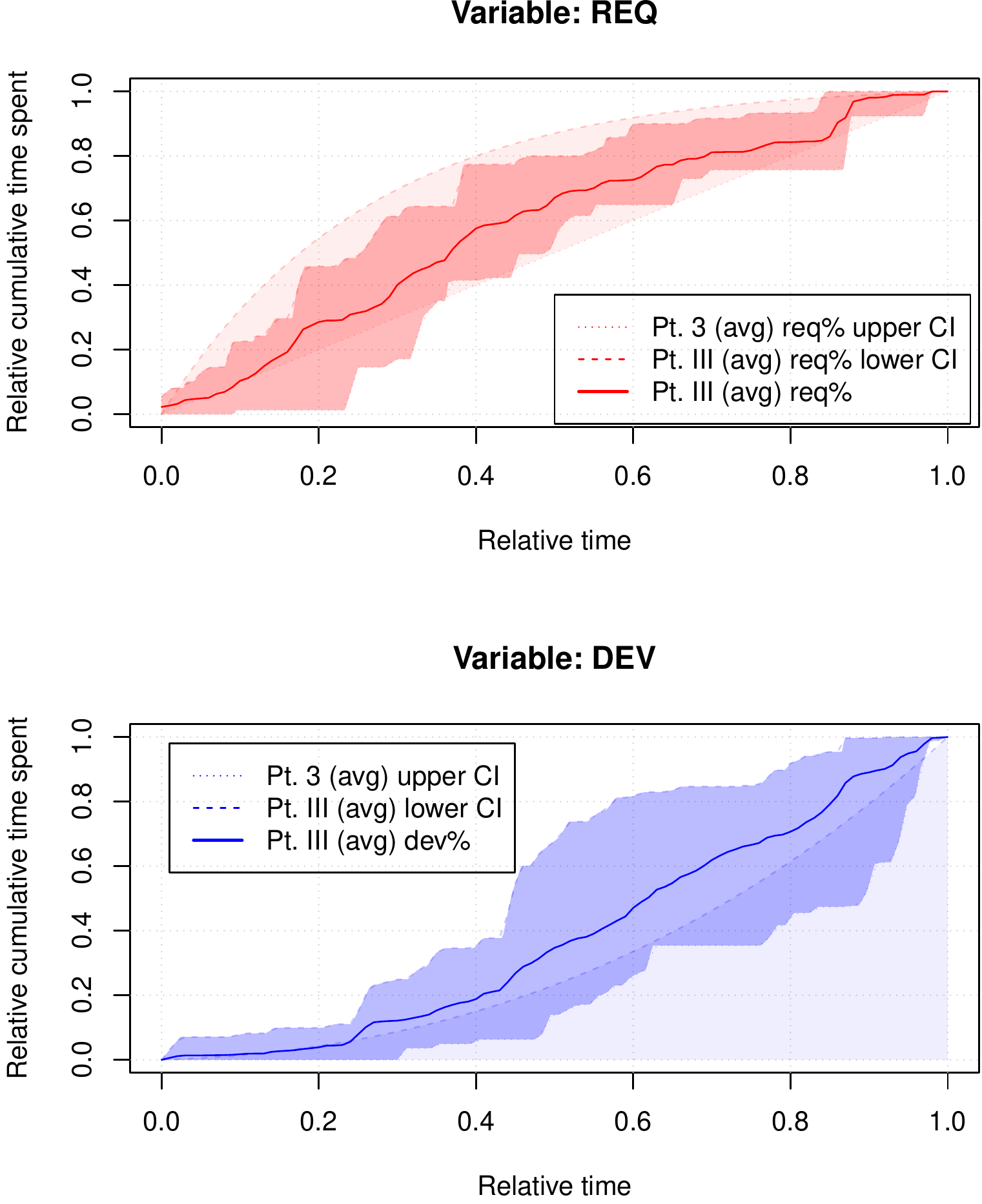} 

}

\caption{Empirical (average) req\% and dev\% and their lower- and upper empirical weighted confidence intervals (here without gradation).}\label{fig:req-dev-p3avg-cis}
\end{figure}

\begin{figure}[ht!]

{\centering \includegraphics{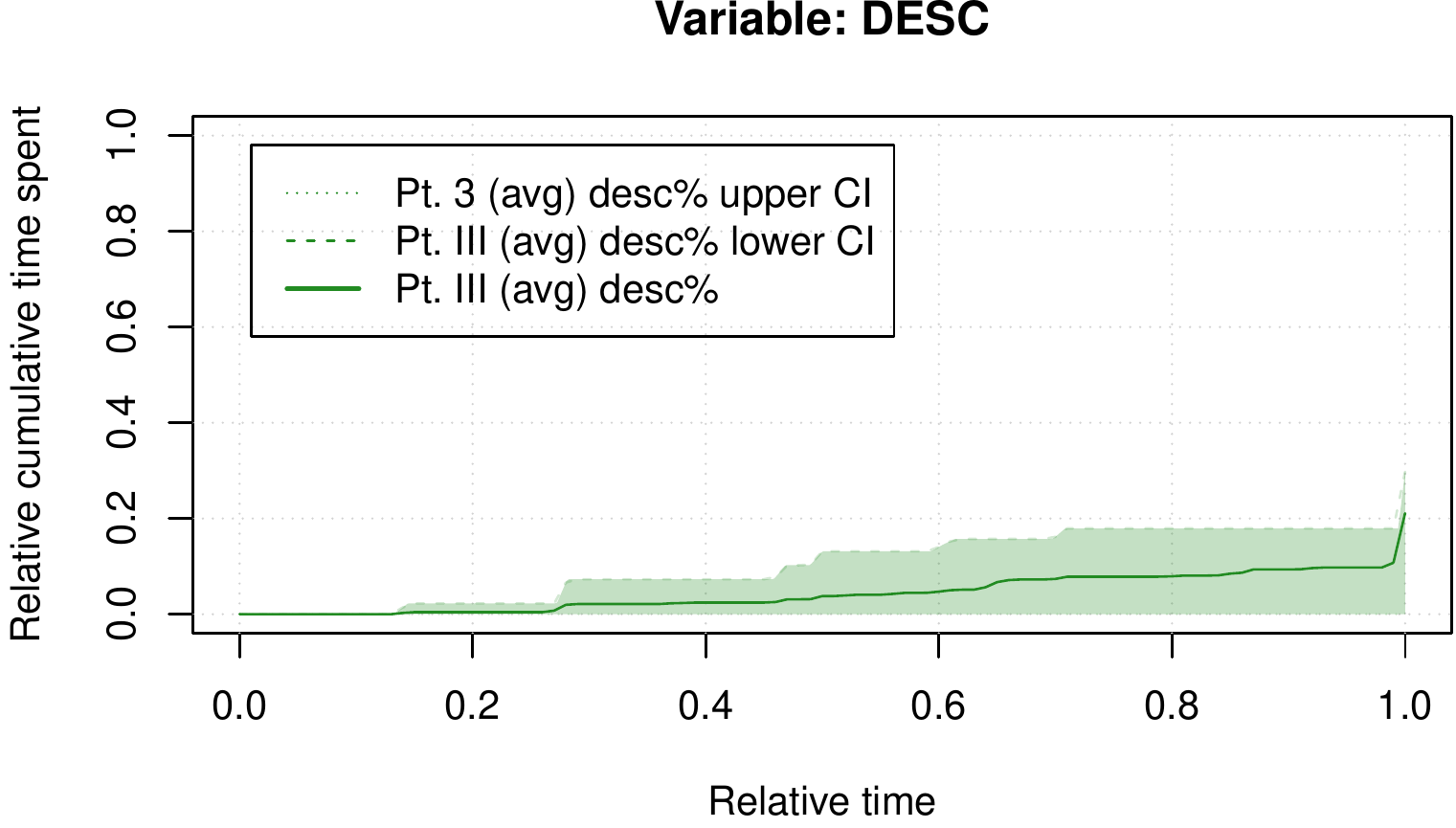} 

}

\caption{Empirical (average) desc\% and its lower- and upper empirical weighted confidence intervals (here without gradation).}\label{fig:desc-p3avg-cis}
\end{figure}

Figures \ref{fig:req-dev-p3avg-cis} and \ref{fig:desc-p3avg-cis} show the average variables and the empirical confidence intervals.

But first, we define a function \(f:R^2\mapsto R\) to compute a CI with gradation. For each x/y coordinate, it shall output a confidence based on the weights as of the ground truth, and all projects' variables that output a \(y\) that is smaller than (larger than) the given \(y\) shall be excluded.

\[
\begin{aligned}
  h_{\text{upper}}(f,x,y)=&\;\begin{cases}
        1,&\text{if}\;f(x)\geq y,
        \\
        0,&\text{otherwise,}
    \end{cases}
  \\
  \operatorname{CI}^+(x,y)=&\;\bm{\omega}^\top\cdot h_{\text{upper}}(\bm{f},x,y)\times\Big[\sum\bm{\omega}\Big]^{-1}\text{.}
\end{aligned}
\]

\(\operatorname{CI}^+\) is to be used for the upper confidence region, and likewise, we define \(\operatorname{CI}^-\) to be equivalent, but it uses \(h_{\text{lower}}\), that switches the condition to \(f(x)\leq y\). The decision on whether to use \(\operatorname{CI}^+\) or \(\operatorname{CI}^-\) depends on whether the given \(y\) is above or below the \emph{computed average variable}, i.e.,

\[
\begin{aligned}
  \bar{g}(x)\dots&\;\text{the computed average variable,}
  \\[1ex]
  \operatorname{CI}(x,y)=&\;\begin{cases}
    0,&\text{if}\;y>\operatorname{CI}_{\text{upper}}(x)\;\text{,}
    \\
    0,&\text{if}\;y<\operatorname{CI}_{\text{lower}}(x)\;\text{,}
    \\
    \bm{\omega}^\top\cdot\Bigg[\begin{cases}
      h_{\text{upper}}(\bm{f},x,y),&\text{if}\;\bar{g}(x)<y,
      \\
      h_{\text{lower}}(\bm{f},x,y),&\text{otherwise}
    \end{cases}\Bigg]\times\Big[\sum\bm{\omega}\Big]^{-1},&\text{otherwise}
  \end{cases}\;\text{.}
  \\
  =&\;\begin{cases}
    0,&\text{if}\;y>\operatorname{CI}_{\text{upper}}(x)\,\text{,}
    \\
    0,&\text{if}\;y<\operatorname{CI}_{\text{lower}}(x)\,\text{,}
    \\
    \bm{\omega}^\top\cdot\begin{cases}
      \operatorname{CI}^+(x,y),&\text{if}\;\bar{g}(x)<y,
      \\
      \operatorname{CI}^-(x,y),&\text{otherwise.}
    \end{cases}
  \end{cases}
\end{aligned}
\]

With this definition, we can compute a loss that is then based on a path that goes through this hyperplane. That path is a project's variable.

\begin{Shaded}
\begin{Highlighting}[]
\CommentTok{\# h\_p3avg \textless{}{-} function(funclist, x, y, upper = TRUE) \{ unlist(lapply(X =}
\CommentTok{\# funclist, FUN = function(f) \{ sapply(X = f(x), function(val) \{ if (val == 0)}
\CommentTok{\# val \textless{}{-} sqrt(.Machine$double.eps) if ((upper \&\& val \textgreater{}= y) || (!upper \&\& val \textless{}=}
\CommentTok{\# y)) val else 0 \}) \})) \}}

\CommentTok{\# We re{-}define this function to just indicate.}
\NormalTok{h\_p3avg }\OtherTok{\textless{}{-}} \ControlFlowTok{function}\NormalTok{(funclist, x, y, }\AttributeTok{upper =} \ConstantTok{TRUE}\NormalTok{, f\_ci) \{}
  \FunctionTok{unlist}\NormalTok{(}\FunctionTok{lapply}\NormalTok{(}\AttributeTok{X =}\NormalTok{ funclist, }\AttributeTok{FUN =} \ControlFlowTok{function}\NormalTok{(f) \{}
    \FunctionTok{sapply}\NormalTok{(}\AttributeTok{X =} \FunctionTok{f}\NormalTok{(x), }\ControlFlowTok{function}\NormalTok{(val) \{}
      \ControlFlowTok{if}\NormalTok{ (upper }\SpecialCharTok{\&\&}\NormalTok{ val }\SpecialCharTok{\textgreater{}=}\NormalTok{ y) \{}
        \DecValTok{1}
        \CommentTok{\# f\_ci(x) {-} val \textless{}{-}{-} this could be an alternative using the boundaries}
\NormalTok{      \} }\ControlFlowTok{else} \ControlFlowTok{if}\NormalTok{ (}\SpecialCharTok{!}\NormalTok{upper }\SpecialCharTok{\&\&}\NormalTok{ val }\SpecialCharTok{\textless{}=}\NormalTok{ y) \{}
        \DecValTok{1}
        \CommentTok{\# val {-} f\_ci(x)}
\NormalTok{      \} }\ControlFlowTok{else}\NormalTok{ \{}
        \DecValTok{0}
\NormalTok{      \}}
\NormalTok{    \})}
\NormalTok{  \}))}
\NormalTok{\}}

\NormalTok{h\_upper\_p3avg }\OtherTok{\textless{}{-}} \ControlFlowTok{function}\NormalTok{(funclist, x, y, f\_ci) }\FunctionTok{h\_p3avg}\NormalTok{(}\AttributeTok{funclist =}\NormalTok{ funclist, }\AttributeTok{x =}\NormalTok{ x,}
  \AttributeTok{y =}\NormalTok{ y, }\AttributeTok{upper =} \ConstantTok{TRUE}\NormalTok{, }\AttributeTok{f\_ci =}\NormalTok{ f\_ci)}
\NormalTok{h\_lower\_p3avg }\OtherTok{\textless{}{-}} \ControlFlowTok{function}\NormalTok{(funclist, x, y, f\_ci) }\FunctionTok{h\_p3avg}\NormalTok{(}\AttributeTok{funclist =}\NormalTok{ funclist, }\AttributeTok{x =}\NormalTok{ x,}
  \AttributeTok{y =}\NormalTok{ y, }\AttributeTok{upper =} \ConstantTok{FALSE}\NormalTok{, }\AttributeTok{f\_ci =}\NormalTok{ f\_ci)}

\NormalTok{CI\_p3avg }\OtherTok{\textless{}{-}} \ControlFlowTok{function}\NormalTok{(x, y, funclist, f\_ci\_upper, f\_ci\_lower, gbar, omega) \{}
  \FunctionTok{stopifnot}\NormalTok{(}\FunctionTok{length}\NormalTok{(x) }\SpecialCharTok{==} \FunctionTok{length}\NormalTok{(y))}

  \FunctionTok{sapply}\NormalTok{(}\AttributeTok{X =} \FunctionTok{seq\_len}\NormalTok{(}\AttributeTok{length.out =} \FunctionTok{length}\NormalTok{(x)), }\AttributeTok{FUN =} \ControlFlowTok{function}\NormalTok{(idx) \{}
\NormalTok{    xi }\OtherTok{\textless{}{-}}\NormalTok{ x[idx]}
\NormalTok{    yi }\OtherTok{\textless{}{-}}\NormalTok{ y[idx]}

    \ControlFlowTok{if}\NormalTok{ (yi }\SpecialCharTok{\textgreater{}} \FunctionTok{f\_ci\_upper}\NormalTok{(xi) }\SpecialCharTok{||}\NormalTok{ yi }\SpecialCharTok{\textless{}} \FunctionTok{f\_ci\_lower}\NormalTok{(xi)) \{}
      \FunctionTok{return}\NormalTok{(}\DecValTok{0}\NormalTok{)}
\NormalTok{    \}}

\NormalTok{    gbarval }\OtherTok{\textless{}{-}} \FunctionTok{gbar}\NormalTok{(xi)}
\NormalTok{    hval }\OtherTok{\textless{}{-}} \ControlFlowTok{if}\NormalTok{ (gbarval }\SpecialCharTok{\textless{}}\NormalTok{ yi) \{}
      \FunctionTok{h\_upper\_p3avg}\NormalTok{(}\AttributeTok{funclist =}\NormalTok{ funclist, }\AttributeTok{x =}\NormalTok{ xi, }\AttributeTok{y =}\NormalTok{ yi, }\AttributeTok{f\_ci =}\NormalTok{ f\_ci\_upper)}
\NormalTok{    \} }\ControlFlowTok{else}\NormalTok{ \{}
      \FunctionTok{h\_lower\_p3avg}\NormalTok{(}\AttributeTok{funclist =}\NormalTok{ funclist, }\AttributeTok{x =}\NormalTok{ xi, }\AttributeTok{y =}\NormalTok{ yi, }\AttributeTok{f\_ci =}\NormalTok{ f\_ci\_lower)}
\NormalTok{    \}}

\NormalTok{    omega }\SpecialCharTok{\%*\%}\NormalTok{ hval}\SpecialCharTok{/}\FunctionTok{sum}\NormalTok{(omega)}
\NormalTok{  \})}
\NormalTok{\}}

\NormalTok{CI\_req\_p3avg }\OtherTok{\textless{}{-}} \ControlFlowTok{function}\NormalTok{(x, y) }\FunctionTok{CI\_p3avg}\NormalTok{(}\AttributeTok{x =}\NormalTok{ x, }\AttributeTok{y =}\NormalTok{ y, }\AttributeTok{funclist =}\NormalTok{ funclist\_REQ, }\AttributeTok{f\_ci\_upper =}\NormalTok{ req\_ci\_upper\_p3avg,}
  \AttributeTok{f\_ci\_lower =}\NormalTok{ req\_ci\_lower\_p3avg, }\AttributeTok{gbar =}\NormalTok{ req\_p3, }\AttributeTok{omega =}\NormalTok{ omega)}
\NormalTok{CI\_dev\_p3avg }\OtherTok{\textless{}{-}} \ControlFlowTok{function}\NormalTok{(x, y) }\FunctionTok{CI\_p3avg}\NormalTok{(}\AttributeTok{x =}\NormalTok{ x, }\AttributeTok{y =}\NormalTok{ y, }\AttributeTok{funclist =}\NormalTok{ funclist\_DEV, }\AttributeTok{f\_ci\_upper =}\NormalTok{ dev\_ci\_upper\_p3avg,}
  \AttributeTok{f\_ci\_lower =}\NormalTok{ dev\_ci\_lower\_p3avg, }\AttributeTok{gbar =}\NormalTok{ dev\_p3, }\AttributeTok{omega =}\NormalTok{ omega)}
\NormalTok{CI\_desc\_p3avg }\OtherTok{\textless{}{-}} \ControlFlowTok{function}\NormalTok{(x, y) }\FunctionTok{CI\_p3avg}\NormalTok{(}\AttributeTok{x =}\NormalTok{ x, }\AttributeTok{y =}\NormalTok{ y, }\AttributeTok{funclist =}\NormalTok{ funclist\_DESC,}
  \AttributeTok{f\_ci\_upper =}\NormalTok{ desc\_ci\_upper\_p3avg, }\AttributeTok{f\_ci\_lower =}\NormalTok{ desc\_ci\_lower\_p3avg, }\AttributeTok{gbar =}\NormalTok{ desc\_p3,}
  \AttributeTok{omega =}\NormalTok{ omega)}

\FunctionTok{invisible}\NormalTok{(}\FunctionTok{loadResultsOrCompute}\NormalTok{(}\AttributeTok{file =} \StringTok{"../data/CI\_p3avg\_funcs.rds"}\NormalTok{, }\AttributeTok{computeExpr =}\NormalTok{ \{}
  \FunctionTok{list}\NormalTok{(}\AttributeTok{CI\_req\_p3avg =}\NormalTok{ CI\_req\_p3avg, }\AttributeTok{CI\_dev\_p3avg =}\NormalTok{ CI\_dev\_p3avg, }\AttributeTok{CI\_desc\_p3avg =}\NormalTok{ CI\_desc\_p3avg)}
\NormalTok{\}))}
\end{Highlighting}
\end{Shaded}

\begin{Shaded}
\begin{Highlighting}[]
\NormalTok{x }\OtherTok{\textless{}{-}} \FunctionTok{seq}\NormalTok{(}\DecValTok{0}\NormalTok{, }\DecValTok{1}\NormalTok{, }\AttributeTok{length.out =} \DecValTok{200}\NormalTok{)}
\NormalTok{y }\OtherTok{\textless{}{-}} \FunctionTok{seq}\NormalTok{(}\DecValTok{0}\NormalTok{, }\DecValTok{1}\NormalTok{, }\AttributeTok{length.out =} \DecValTok{200}\NormalTok{)}

\NormalTok{compute\_z\_p3avg }\OtherTok{\textless{}{-}} \ControlFlowTok{function}\NormalTok{(varname, x, y, }\AttributeTok{interp =} \ConstantTok{NA\_real\_}\NormalTok{) \{}
  \CommentTok{\# We cannot call outer because our functions are not properly vectorized. z}
  \CommentTok{\# \textless{}{-} outer(X = x, Y = y, FUN = CI\_req\_p3avg)}
\NormalTok{  f }\OtherTok{\textless{}{-}} \ControlFlowTok{if}\NormalTok{ (varname }\SpecialCharTok{==} \StringTok{"REQ"}\NormalTok{) \{}
\NormalTok{    CI\_req\_p3avg}
\NormalTok{  \} }\ControlFlowTok{else} \ControlFlowTok{if}\NormalTok{ (varname }\SpecialCharTok{==} \StringTok{"DEV"}\NormalTok{) \{}
\NormalTok{    CI\_dev\_p3avg}
\NormalTok{  \} }\ControlFlowTok{else}\NormalTok{ \{}
\NormalTok{    CI\_desc\_p3avg}
\NormalTok{  \}}

\NormalTok{  z }\OtherTok{\textless{}{-}} \FunctionTok{matrix}\NormalTok{(}\AttributeTok{nrow =} \FunctionTok{length}\NormalTok{(x), }\AttributeTok{ncol =} \FunctionTok{length}\NormalTok{(y))}
  \ControlFlowTok{for}\NormalTok{ (i }\ControlFlowTok{in} \DecValTok{1}\SpecialCharTok{:}\FunctionTok{length}\NormalTok{(x)) \{}
    \ControlFlowTok{for}\NormalTok{ (j }\ControlFlowTok{in} \DecValTok{1}\SpecialCharTok{:}\FunctionTok{length}\NormalTok{(y)) \{}
\NormalTok{      z[i, j] }\OtherTok{\textless{}{-}} \FunctionTok{f}\NormalTok{(}\AttributeTok{x =}\NormalTok{ x[i], }\AttributeTok{y =}\NormalTok{ y[j])}
\NormalTok{    \}}
\NormalTok{  \}}

\NormalTok{  res }\OtherTok{\textless{}{-}} \FunctionTok{list}\NormalTok{(}\AttributeTok{x =}\NormalTok{ x, }\AttributeTok{y =}\NormalTok{ y, }\AttributeTok{z =}\NormalTok{ z)}

  \ControlFlowTok{if}\NormalTok{ (}\SpecialCharTok{!}\FunctionTok{is.na}\NormalTok{(interp)) \{}
\NormalTok{    res }\OtherTok{\textless{}{-}}\NormalTok{ fields}\SpecialCharTok{::}\FunctionTok{interp.surface.grid}\NormalTok{(}\AttributeTok{obj =}\NormalTok{ res, }\AttributeTok{grid.list =} \FunctionTok{list}\NormalTok{(}\AttributeTok{x =} \FunctionTok{seq}\NormalTok{(}\AttributeTok{from =} \FunctionTok{min}\NormalTok{(x),}
      \AttributeTok{to =} \FunctionTok{max}\NormalTok{(x), }\AttributeTok{length.out =}\NormalTok{ interp), }\AttributeTok{y =} \FunctionTok{seq}\NormalTok{(}\AttributeTok{from =} \FunctionTok{min}\NormalTok{(y), }\AttributeTok{to =} \FunctionTok{max}\NormalTok{(y),}
      \AttributeTok{length.out =}\NormalTok{ interp)))}
\NormalTok{  \}}

\NormalTok{  res}
\NormalTok{\}}

\NormalTok{z\_req }\OtherTok{\textless{}{-}} \FunctionTok{loadResultsOrCompute}\NormalTok{(}\AttributeTok{file =} \StringTok{"../results/ci\_p3avg\_z\_req.rds"}\NormalTok{, }\AttributeTok{computeExpr =}\NormalTok{ \{}
  \FunctionTok{compute\_z\_p3avg}\NormalTok{(}\AttributeTok{varname =} \StringTok{"REQ"}\NormalTok{, }\AttributeTok{x =}\NormalTok{ x, }\AttributeTok{y =}\NormalTok{ y)}
\NormalTok{\})}
\NormalTok{z\_dev }\OtherTok{\textless{}{-}} \FunctionTok{loadResultsOrCompute}\NormalTok{(}\AttributeTok{file =} \StringTok{"../results/ci\_p3avg\_z\_dev.rds"}\NormalTok{, }\AttributeTok{computeExpr =}\NormalTok{ \{}
  \FunctionTok{compute\_z\_p3avg}\NormalTok{(}\AttributeTok{varname =} \StringTok{"DEV"}\NormalTok{, }\AttributeTok{x =}\NormalTok{ x, }\AttributeTok{y =}\NormalTok{ y)}
\NormalTok{\})}
\NormalTok{z\_desc }\OtherTok{\textless{}{-}} \FunctionTok{loadResultsOrCompute}\NormalTok{(}\AttributeTok{file =} \StringTok{"../results/ci\_p3avg\_z\_desc.rds"}\NormalTok{, }\AttributeTok{computeExpr =}\NormalTok{ \{}
  \FunctionTok{compute\_z\_p3avg}\NormalTok{(}\AttributeTok{varname =} \StringTok{"DESC"}\NormalTok{, }\AttributeTok{x =}\NormalTok{ x, }\AttributeTok{y =}\NormalTok{ y)}
\NormalTok{\})}
\end{Highlighting}
\end{Shaded}

\begin{figure}[ht!]

{\centering \includegraphics{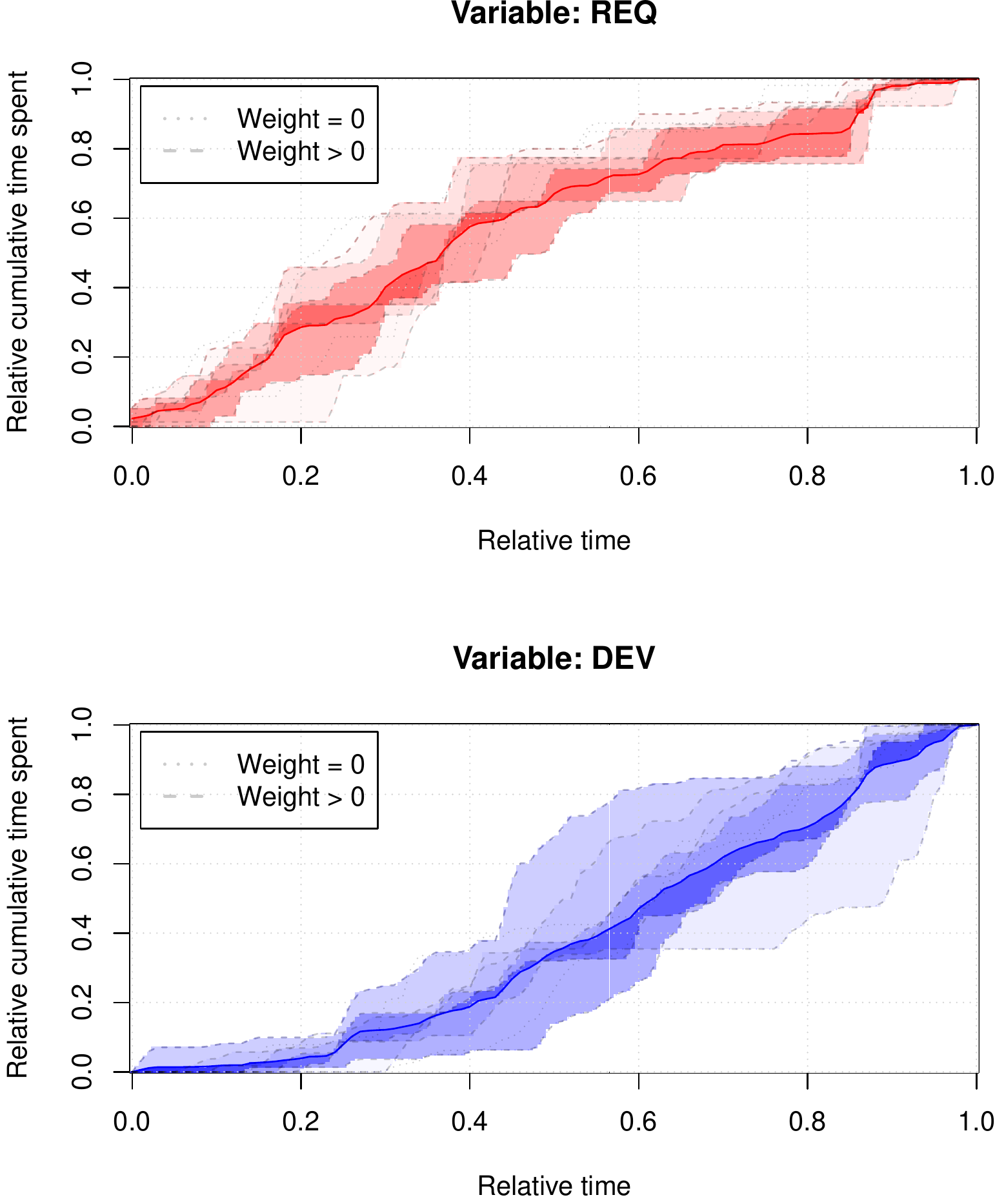} 

}

\caption{The empirical confidence intervals for the two variables req\% and dev\%. Higher saturation of the color correlates with higher confidence. Projects with zero weight contribute to the CIs' boundaries, but not to the hyperplane.}\label{fig:p3-emp-cis}
\end{figure}

\begin{figure}[ht!]

{\centering \includegraphics{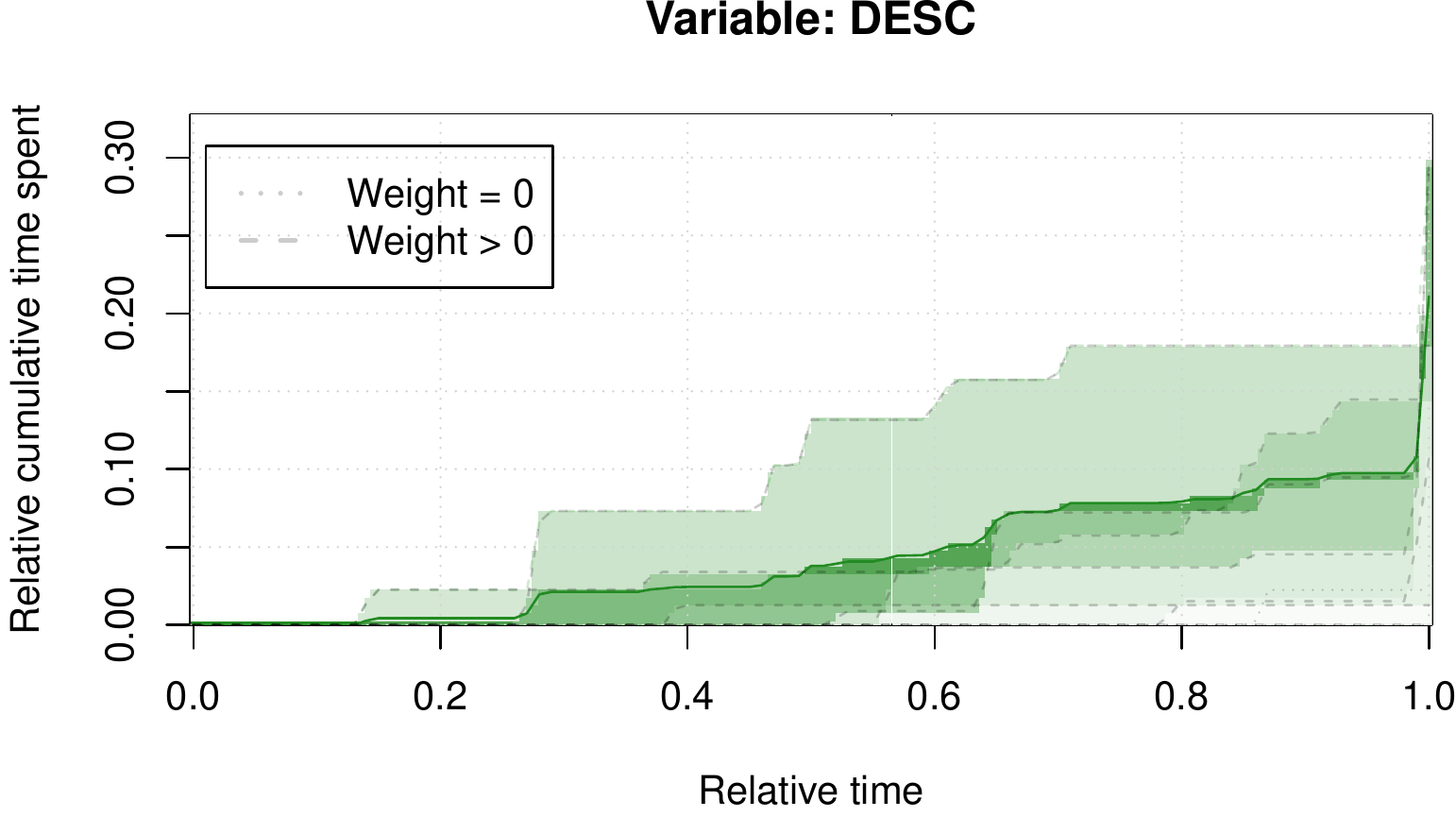} 

}

\caption{The empirical confidence intervals for the variable desc\%. Higher saturation of the color correlates with higher confidence.}\label{fig:p3-emp-desc-cis}
\end{figure}

Finally, we show the empirical confidence intervals in figures \ref{fig:p3-emp-cis} and \ref{fig:p3-emp-desc-cis}. Note that the minimum non-zero confidence is 0.038462, while the maximum is \(1\). Therefore, while we gradate the colors from \(0\) to \(1\), we slightly scale and transform the grid's non-zero values using the expression \(0.9\times z+0.1\), to improve the visibility.

\hypertarget{pattern-iv}{%
\subsubsection{\texorpdfstring{Pattern IV\label{ssec:pattern-iv}}{Pattern IV}}\label{pattern-iv}}

We had previously discussed the possibility of partial (early) detection. Some of the scores and losses use integration and custom intervals, and while these might be applicable in some cases without further changes, we need to be careful when dealing with variables that are normalized at project end. This is currently the case for the variables of pattern I.

Pattern IV explores the possibility of not using some pattern directly, but its \textbf{derivative} for all of the variables and the confidence intervals. Here we exploit the fact that a cumulative variable has monotonic behavior, depending on how it was designed even \emph{strictly monotonic} behavior. That means the slope at any point in time is \(>0\) (or at least \(\geq0\)) - resp. \(<0\) (or at least \(\leq0\)). The confidence intervals model the minimum and maximum extents we would expect, and their derivatives give us lower and upper bounds for the expected rate of change. Regardless of the actual value of the variable, we can thus transform our expectation of its value into an expectation of how it changes over time. Furthermore, the rate of change can be (numerically or analytically) computed for any other time-series variable, the procedure introduced here is not limited to cumulative and/or normalized variables.

Pattern type IV thus is a \textbf{Meta-Process model}, and can be applied to any of the previously introduced patterns (which are process models). Generically speaking, it supports the transformation of variables and confidence interval boundaries (but not inhomogeneous confidence interval surfaces). By using any other pattern (process model), pattern type IV gets \textbf{instantiated}. For the remainder of this notebook, we will be instantiating the pattern type IV using pattern I. While this helps our demonstration purposes, pattern I is partially far off the real-world data, which means that we must expect mediocre results. In practice, one should use this meta pattern only with well-designed and/or data-enhanced or data-only patterns.

\begin{figure}[ht!]

{\centering \includegraphics{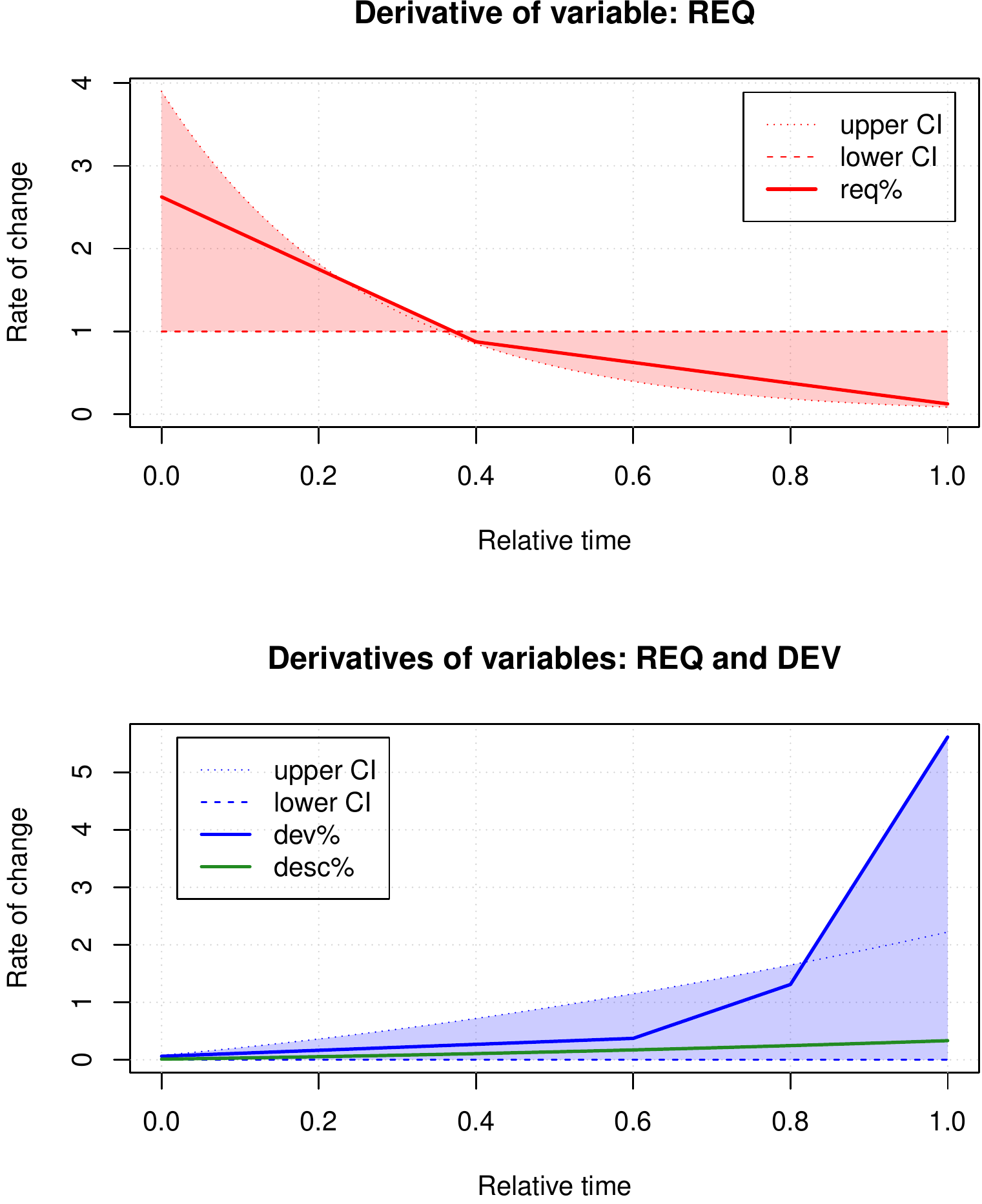} 

}

\caption{Derivatives of all variables and confidence intervals as of pattern I.}\label{fig:p4-req-dev}
\end{figure}

For creating the first derivatives, we can use either, analytical expressions (if available) or numeric methods, such finite difference approaches. In the following, we use both, as we also actually have analytical expressions for some of the curves. The first pattern, represented by its derivatives, is shown in figure \ref{fig:p4-req-dev}.

\begin{Shaded}
\begin{Highlighting}[]
\NormalTok{func\_d1 }\OtherTok{\textless{}{-}} \ControlFlowTok{function}\NormalTok{(f, x, }\AttributeTok{supp =} \FunctionTok{c}\NormalTok{(}\DecValTok{0}\NormalTok{, }\DecValTok{1}\NormalTok{)) \{}
  \FunctionTok{sapply}\NormalTok{(}\AttributeTok{X =}\NormalTok{ x, }\AttributeTok{FUN =} \ControlFlowTok{function}\NormalTok{(x\_) \{}
\NormalTok{    t }\OtherTok{\textless{}{-}} \FloatTok{1e{-}05}
\NormalTok{    m }\OtherTok{\textless{}{-}} \ControlFlowTok{if}\NormalTok{ (x\_ }\SpecialCharTok{\textless{}}\NormalTok{ (supp[}\DecValTok{1}\NormalTok{] }\SpecialCharTok{+}\NormalTok{ t))}
      \StringTok{"forward"} \ControlFlowTok{else} \ControlFlowTok{if}\NormalTok{ (x\_ }\SpecialCharTok{\textgreater{}}\NormalTok{ (supp[}\DecValTok{2}\NormalTok{] }\SpecialCharTok{{-}}\NormalTok{ t))}
      \StringTok{"backward"} \ControlFlowTok{else} \StringTok{"central"}
\NormalTok{    pracma}\SpecialCharTok{::}\FunctionTok{fderiv}\NormalTok{(}\AttributeTok{f =}\NormalTok{ f, }\AttributeTok{x =}\NormalTok{ x\_, }\AttributeTok{method =}\NormalTok{ m)}
\NormalTok{  \})}
\NormalTok{\}}

\NormalTok{req\_d1\_p4 }\OtherTok{\textless{}{-}} \ControlFlowTok{function}\NormalTok{(x) \{}
  \FunctionTok{func\_d1}\NormalTok{(}\AttributeTok{f =}\NormalTok{ req, }\AttributeTok{x =}\NormalTok{ x)}
\NormalTok{\}}
\NormalTok{req\_ci\_lower\_d1\_p4 }\OtherTok{\textless{}{-}} \ControlFlowTok{function}\NormalTok{(x) }\FunctionTok{rep}\NormalTok{(}\DecValTok{1}\NormalTok{, }\FunctionTok{length}\NormalTok{(x))}
\NormalTok{req\_ci\_upper\_d1\_p4 }\OtherTok{\textless{}{-}} \ControlFlowTok{function}\NormalTok{(x) }\FloatTok{3.89791628313} \SpecialCharTok{*} \FunctionTok{exp}\NormalTok{(}\SpecialCharTok{{-}}\NormalTok{(}\FloatTok{3.811733} \SpecialCharTok{*}\NormalTok{ x))}

\NormalTok{dev\_d1\_p4 }\OtherTok{\textless{}{-}} \ControlFlowTok{function}\NormalTok{(x) \{}
  \FunctionTok{func\_d1}\NormalTok{(}\AttributeTok{f =}\NormalTok{ dev, }\AttributeTok{x =}\NormalTok{ x)}
\NormalTok{\}}
\NormalTok{dev\_ci\_lower\_d1\_p4 }\OtherTok{\textless{}{-}} \ControlFlowTok{function}\NormalTok{(x) }\FunctionTok{rep}\NormalTok{(}\DecValTok{0}\NormalTok{, }\FunctionTok{length}\NormalTok{(x))}
\NormalTok{dev\_ci\_upper\_d1\_p4 }\OtherTok{\textless{}{-}} \ControlFlowTok{function}\NormalTok{(x) }\FloatTok{0.07815904} \SpecialCharTok{+}\NormalTok{ x }\SpecialCharTok{*}\NormalTok{ (}\FloatTok{0.8986929} \SpecialCharTok{*}\NormalTok{ x }\SpecialCharTok{+} \FloatTok{1.2445534}\NormalTok{)}

\CommentTok{\# For DESC, there are no confidence intervals}
\NormalTok{desc\_d1\_p4 }\OtherTok{\textless{}{-}} \ControlFlowTok{function}\NormalTok{(x) }\FloatTok{0.01172386} \SpecialCharTok{+}\NormalTok{ x }\SpecialCharTok{*}\NormalTok{ (}\FloatTok{0.13480392} \SpecialCharTok{*}\NormalTok{ x }\SpecialCharTok{+} \FloatTok{0.186683}\NormalTok{)}
\end{Highlighting}
\end{Shaded}

While we need the derivatives of any function that describes a variable or confidence interval over time, the derivatives of the confidence intervals now represent upper and lower bounds for the expected range of change. Furthermore, at any point the rate of change of the variable or either of the confidence intervals may exceed any of the other, and the curves can also cross. So, in order to define the upper and lower boundaries, we require the definition of a helper function that returns the minimum and/or maximum of either of these three functions for every \(x\):

\[
\begin{aligned}
  \operatorname{CI}_{\nabla \text{upper}}(\nabla f,\nabla f_{\text{lower}},\nabla f_{\text{upper}},x)=&\;\sup{\big(\nabla f(x),\nabla f_{\text{lower}}(x),\nabla f_{\text{upper}}(x)\big)}\;\text{.}
\end{aligned}
\]

Likewise, we define \(\operatorname{CI}_{\nabla\text{lower}}(\dots)\) using the infimum.

The next challenge lies in representing the data which, in this case, is cumulative time spent on issues. If we approximate functions in a \emph{zero-hold}-fashion as we did so far, then the gradients of these have extreme steps, and are likely unusable. In the following we attempt a few techniques to approximate a cumulative variable using LOESS-smoothing, constrained B-splines non-parametric regression quantiles, and fitting of orthogonal polynomials. The latter two result in smooth gradients if the number of knots or the degree of the polynomials are kept low. In the following subsections we use the signals of the 3rd project as an example.

\hypertarget{using-averaged-bins}{%
\paragraph{Using averaged bins}\label{using-averaged-bins}}

\begin{figure}[ht!]

{\centering \includegraphics{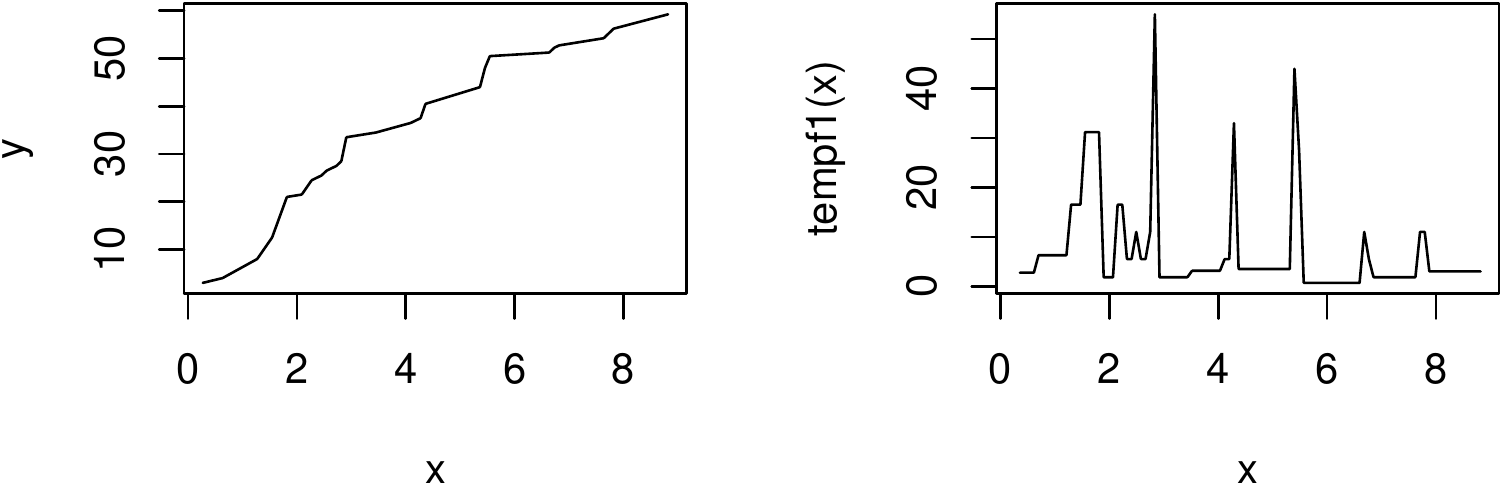} 

}

\caption{The average-bin signal and its gradient.}\label{fig:p4-averaged-bins-d1}
\end{figure}

In this approach we eliminate plateaus by aggregating all \(x\) that have the same \(y\), and moving \(x\) by the amount of values that have the same \(y\). Looking at the gradient in figure \ref{fig:p4-averaged-bins-d1}, it is still too extreme.

\begin{Shaded}
\begin{Highlighting}[]
\NormalTok{temp }\OtherTok{\textless{}{-}} \FunctionTok{table}\NormalTok{(p3\_signals}\SpecialCharTok{$}\NormalTok{data}\SpecialCharTok{$}\StringTok{\textasciigrave{}}\AttributeTok{cum req}\StringTok{\textasciigrave{}}\NormalTok{)}
\NormalTok{prev\_x }\OtherTok{\textless{}{-}} \DecValTok{0}
\NormalTok{x }\OtherTok{\textless{}{-}} \FunctionTok{c}\NormalTok{()}
\NormalTok{y }\OtherTok{\textless{}{-}} \FunctionTok{c}\NormalTok{()}
\ControlFlowTok{for}\NormalTok{ (i }\ControlFlowTok{in} \DecValTok{1}\SpecialCharTok{:}\FunctionTok{length}\NormalTok{(temp)) \{}
\NormalTok{  x }\OtherTok{\textless{}{-}} \FunctionTok{c}\NormalTok{(x, prev\_x }\SpecialCharTok{+}\NormalTok{ temp[[i]]}\SpecialCharTok{/}\FunctionTok{max}\NormalTok{(temp))}
\NormalTok{  prev\_x }\OtherTok{\textless{}{-}} \FunctionTok{tail}\NormalTok{(x, }\DecValTok{1}\NormalTok{)}
\NormalTok{  y }\OtherTok{\textless{}{-}} \FunctionTok{c}\NormalTok{(y, }\FunctionTok{as.numeric}\NormalTok{(}\FunctionTok{names}\NormalTok{(temp[i])))}
\NormalTok{\}}
\end{Highlighting}
\end{Shaded}

\hypertarget{eliminate-plateaus}{%
\paragraph{Eliminate plateaus}\label{eliminate-plateaus}}

The cumulative variables have large plateaus, and the following is a test to eliminate them. However, this additional manipulation step should be skipped, if possible.

\begin{figure}[ht!]

{\centering \includegraphics{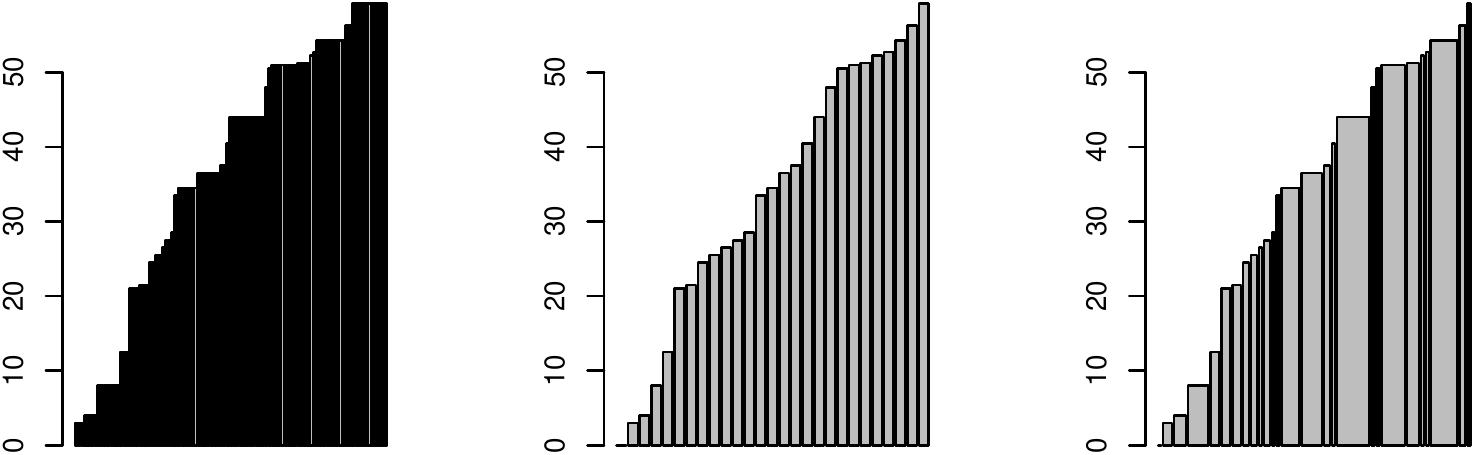} 

}

\caption{Transforming a signal into a non-plateau signal, using unequal widths.}\label{fig:p4-no-plateaus}
\end{figure}

In this test, we iterate over all values of the cumulative variable and only keep x/y pairs, when y increases. In figure \ref{fig:p4-no-plateaus} we compare the original cumulative variable against the one without plateaus, i.e., every \(x_t>x_{t-1}\) (note that the unequal spread of \(x\) in the first non-plateau plot is deliberately ignored in the figure).

\hypertarget{loess-smoothing}{%
\paragraph{LOESS smoothing}\label{loess-smoothing}}

First we define a helper function for smoothing signals using LOESS. It will also help us to scale and translate the resulting function into a specific support.

\begin{Shaded}
\begin{Highlighting}[]
\NormalTok{smooth\_signal\_loess }\OtherTok{\textless{}{-}} \ControlFlowTok{function}\NormalTok{(x, y, }\AttributeTok{support =} \FunctionTok{c}\NormalTok{(}\DecValTok{0}\NormalTok{, }\DecValTok{1}\NormalTok{), }\AttributeTok{span =}\NormalTok{ .}\DecValTok{35}\NormalTok{, }\AttributeTok{family =} \FunctionTok{c}\NormalTok{(}\StringTok{"s"}\NormalTok{, }\StringTok{"g"}\NormalTok{), }\AttributeTok{neval =} \FloatTok{1e4}\NormalTok{) \{}
\NormalTok{  temp }\OtherTok{\textless{}{-}} \ControlFlowTok{if}\NormalTok{ (span }\SpecialCharTok{\textless{}=} \DecValTok{0}\NormalTok{) \{}
    \FunctionTok{list}\NormalTok{(}\AttributeTok{x =}\NormalTok{ x, }\AttributeTok{y =}\NormalTok{ y)}
\NormalTok{  \} }\ControlFlowTok{else}\NormalTok{ \{}
    \FunctionTok{loess.smooth}\NormalTok{(}\AttributeTok{x =}\NormalTok{ x, }\AttributeTok{y =}\NormalTok{ y, }\AttributeTok{span =}\NormalTok{ span, }\AttributeTok{family =} \FunctionTok{match.arg}\NormalTok{(family), }\AttributeTok{evaluation =}\NormalTok{ neval)}
\NormalTok{  \}}
  
\NormalTok{  stats}\SpecialCharTok{::}\FunctionTok{approxfun}\NormalTok{(}
    \CommentTok{\# translate and scale to [0,1], then scale and translate to desired support.}
    \AttributeTok{x =}\NormalTok{ ((temp}\SpecialCharTok{$}\NormalTok{x }\SpecialCharTok{{-}} \FunctionTok{min}\NormalTok{(temp}\SpecialCharTok{$}\NormalTok{x)) }\SpecialCharTok{/}\NormalTok{ (}\FunctionTok{max}\NormalTok{(temp}\SpecialCharTok{$}\NormalTok{x) }\SpecialCharTok{{-}} \FunctionTok{min}\NormalTok{(temp}\SpecialCharTok{$}\NormalTok{x))) }\SpecialCharTok{*}\NormalTok{ (support[}\DecValTok{2}\NormalTok{] }\SpecialCharTok{{-}}\NormalTok{ support[}\DecValTok{1}\NormalTok{]) }\SpecialCharTok{+}\NormalTok{ support[}\DecValTok{1}\NormalTok{],}
    \CommentTok{\# scale y together with x, this is important}
    \AttributeTok{y =}\NormalTok{ temp}\SpecialCharTok{$}\NormalTok{y }\SpecialCharTok{/}\NormalTok{ (}\FunctionTok{max}\NormalTok{(temp}\SpecialCharTok{$}\NormalTok{x) }\SpecialCharTok{{-}} \FunctionTok{min}\NormalTok{(temp}\SpecialCharTok{$}\NormalTok{x)) }\SpecialCharTok{*}\NormalTok{ (support[}\DecValTok{2}\NormalTok{] }\SpecialCharTok{{-}}\NormalTok{ support[}\DecValTok{1}\NormalTok{]),}
    \AttributeTok{yleft =}\NormalTok{ utils}\SpecialCharTok{::}\FunctionTok{head}\NormalTok{(temp}\SpecialCharTok{$}\NormalTok{y, }\DecValTok{1}\NormalTok{),}
    \AttributeTok{yright =}\NormalTok{ utils}\SpecialCharTok{::}\FunctionTok{tail}\NormalTok{(temp}\SpecialCharTok{$}\NormalTok{y, }\DecValTok{1}\NormalTok{))}
\NormalTok{\}}
\end{Highlighting}
\end{Shaded}

\begin{verbatim}
## Warning: Removed 1 row(s) containing missing values (geom_path).
## Removed 1 row(s) containing missing values (geom_path).
## Removed 1 row(s) containing missing values (geom_path).
\end{verbatim}

\begin{figure}[ht!]

{\centering \includegraphics{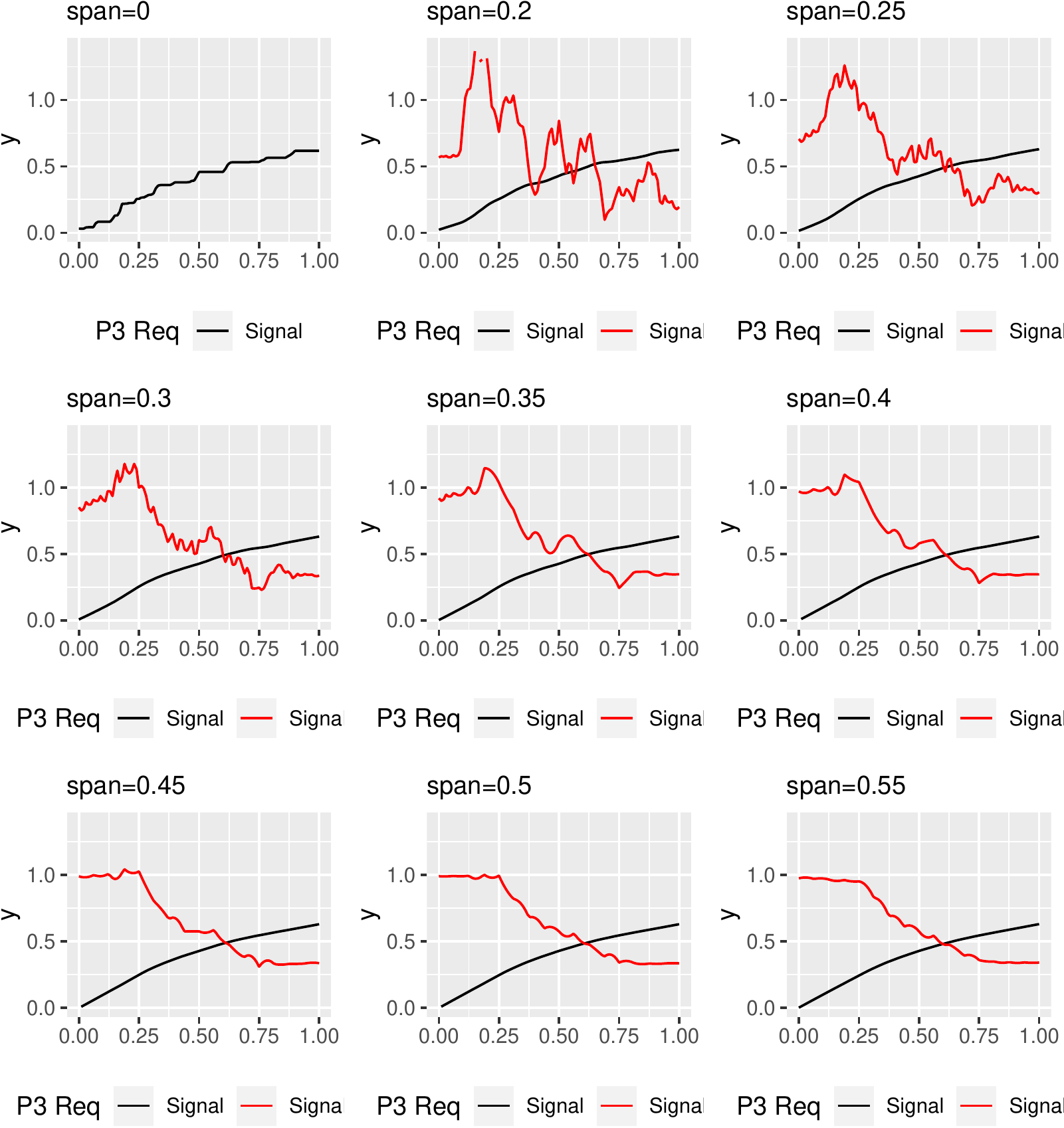} 

}

\caption{Increasing LOESS-smoothing of the non-plateau signal and its resulting gradient (span=0 means no smoothing).}\label{fig:p4-no-plateaus-smooth}
\end{figure}

In figure \ref{fig:p4-no-plateaus-smooth} we smooth raw project data, i.e., we do \textbf{not} use the non-plateau data or data from averaged bins, but rather the cumulative variable as-is, even \textbf{without normalization}, to demonstrate the advantages of this fourth pattern. In the top-left plot we focus on showing the raw signal. The extent of its gradient is well beyond the other (cf.~figure \ref{fig:p4-averaged-bins-d1}), smoothed versions. With increasing smoothing-span, both the signal and even more so its gradient become smooth. For our concern, which is the identification of a trend regarding the rate of change, I would say that a span in the range \([0.2,0.45]\) is probably most useful, so the default was selected to be \(0.35\), as it smoothes out many of the local peaks, while still preserving the important characteristics of the gradient (rate of change). However, everything \(\geq0.2\) appears to be applicable. Note that spans below that often result in errors (typically around \(0.15\) and below), so I do not recommend going below \(0.2\).

However, the smoothing-span should be chosen according to the intended application. For example, we can compute a score based on the area between the gradient and its expected value. Here, a less smooth gradient should be preferred as it conserves more details. If the application however were, e.g., to compute a correlation, and the curve used for comparison is smooth, then the span should be chosen accordingly higher (\(\geq0.4\)) to get usable results.

\begin{figure}[ht!]

{\centering \includegraphics{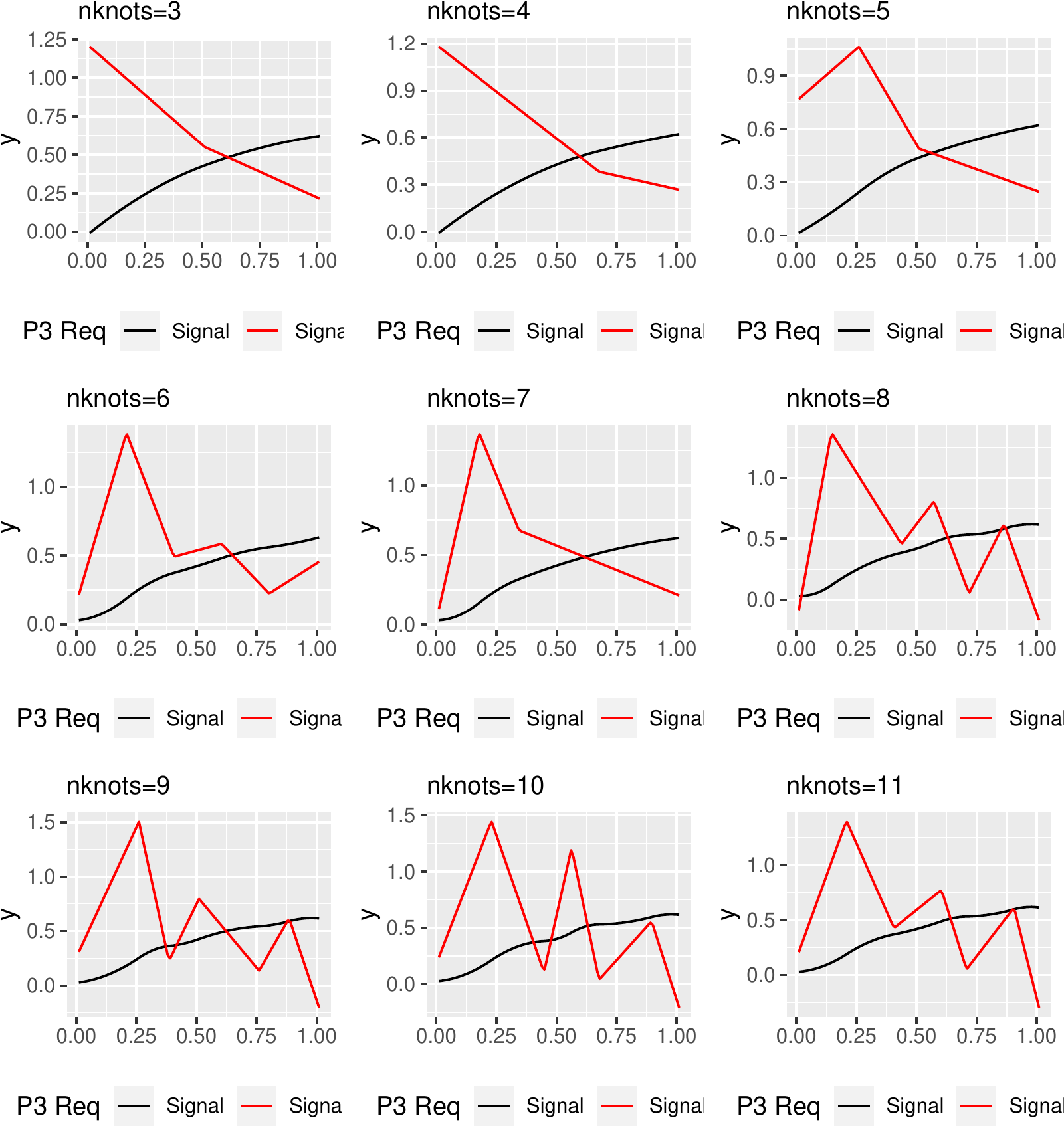} 

}

\caption{Fitted splines with different number of knots and their derivative.}\label{fig:p4-cobs-splines}
\end{figure}

\hypertarget{constrained-b-splines-non-parametric-regression-quantiles}{%
\paragraph{Constrained B-splines non-parametric regression quantiles}\label{constrained-b-splines-non-parametric-regression-quantiles}}

.. or COBS, estimates (fits) a smooth function using B-splines through some ``knots''. In figure \ref{fig:p4-cobs-splines} we are attempting fitting with \([3,11]\) knots. The number of knots should probably not exceed \(4\) or \(5\) for the gradient to be useful, as otherwise it gets too many modes. However, we can clearly see that there is no good compromise. Too few knots result in too smooth a signal and gradient, and everything with \(5\) knots or more is too rough an approximation, too.

\begin{Shaded}
\begin{Highlighting}[]
\FunctionTok{par}\NormalTok{(}\AttributeTok{mfrow =} \FunctionTok{c}\NormalTok{(}\DecValTok{4}\NormalTok{, }\DecValTok{4}\NormalTok{))}

\NormalTok{X }\OtherTok{\textless{}{-}} \DecValTok{1}\SpecialCharTok{:}\FunctionTok{nrow}\NormalTok{(p3\_signals}\SpecialCharTok{$}\NormalTok{data)}
\NormalTok{Y }\OtherTok{\textless{}{-}}\NormalTok{ p3\_signals}\SpecialCharTok{$}\NormalTok{data}\SpecialCharTok{$}\StringTok{\textasciigrave{}}\AttributeTok{cum req}\StringTok{\textasciigrave{}}\SpecialCharTok{/}\NormalTok{(}\FunctionTok{max}\NormalTok{(X) }\SpecialCharTok{{-}} \FunctionTok{min}\NormalTok{(X))}
\NormalTok{X }\OtherTok{\textless{}{-}}\NormalTok{ X}\SpecialCharTok{/}\NormalTok{(}\FunctionTok{max}\NormalTok{(X) }\SpecialCharTok{{-}} \FunctionTok{min}\NormalTok{(X))}
\NormalTok{templ }\OtherTok{\textless{}{-}} \FunctionTok{list}\NormalTok{()}

\ControlFlowTok{for}\NormalTok{ (i }\ControlFlowTok{in} \DecValTok{1}\SpecialCharTok{:}\DecValTok{9}\NormalTok{) \{}
\NormalTok{  templ[[i]] }\OtherTok{\textless{}{-}}\NormalTok{ (}\ControlFlowTok{function}\NormalTok{() \{}
\NormalTok{    temp }\OtherTok{\textless{}{-}}\NormalTok{ cobs}\SpecialCharTok{::}\FunctionTok{cobs}\NormalTok{(}\AttributeTok{x =}\NormalTok{ X, }\AttributeTok{y =}\NormalTok{ Y, }\AttributeTok{nknots =}\NormalTok{ i }\SpecialCharTok{+} \DecValTok{2}\NormalTok{, }\AttributeTok{print.mesg =} \ConstantTok{FALSE}\NormalTok{, }\AttributeTok{print.warn =} \ConstantTok{FALSE}\NormalTok{)}
\NormalTok{    Signal}\SpecialCharTok{$}\FunctionTok{new}\NormalTok{(}\AttributeTok{func =} \ControlFlowTok{function}\NormalTok{(x) stats}\SpecialCharTok{::}\FunctionTok{predict}\NormalTok{(}\AttributeTok{object =}\NormalTok{ temp, }\AttributeTok{z =}\NormalTok{ x)[, }\StringTok{"fit"}\NormalTok{],}
      \AttributeTok{name =} \StringTok{"P3 Req"}\NormalTok{, }\AttributeTok{support =} \FunctionTok{range}\NormalTok{(X), }\AttributeTok{isWp =} \ConstantTok{TRUE}\NormalTok{)}\SpecialCharTok{$}\FunctionTok{plot}\NormalTok{(}\AttributeTok{show1stDeriv =} \ConstantTok{TRUE}\NormalTok{) }\SpecialCharTok{+}
      \FunctionTok{labs}\NormalTok{(}\AttributeTok{subtitle =} \FunctionTok{paste0}\NormalTok{(}\StringTok{"nknots="}\NormalTok{, i }\SpecialCharTok{+} \DecValTok{2}\NormalTok{))}
\NormalTok{  \})()}
\NormalTok{\}}

\NormalTok{ggpubr}\SpecialCharTok{::}\FunctionTok{ggarrange}\NormalTok{(}\AttributeTok{plotlist =}\NormalTok{ templ, }\AttributeTok{ncol =} \DecValTok{3}\NormalTok{, }\AttributeTok{nrow =} \DecValTok{3}\NormalTok{)}
\end{Highlighting}
\end{Shaded}

\hypertarget{orthogonal-polynomials}{%
\paragraph{Orthogonal polynomials}\label{orthogonal-polynomials}}

\begin{figure}[ht!]

{\centering \includegraphics{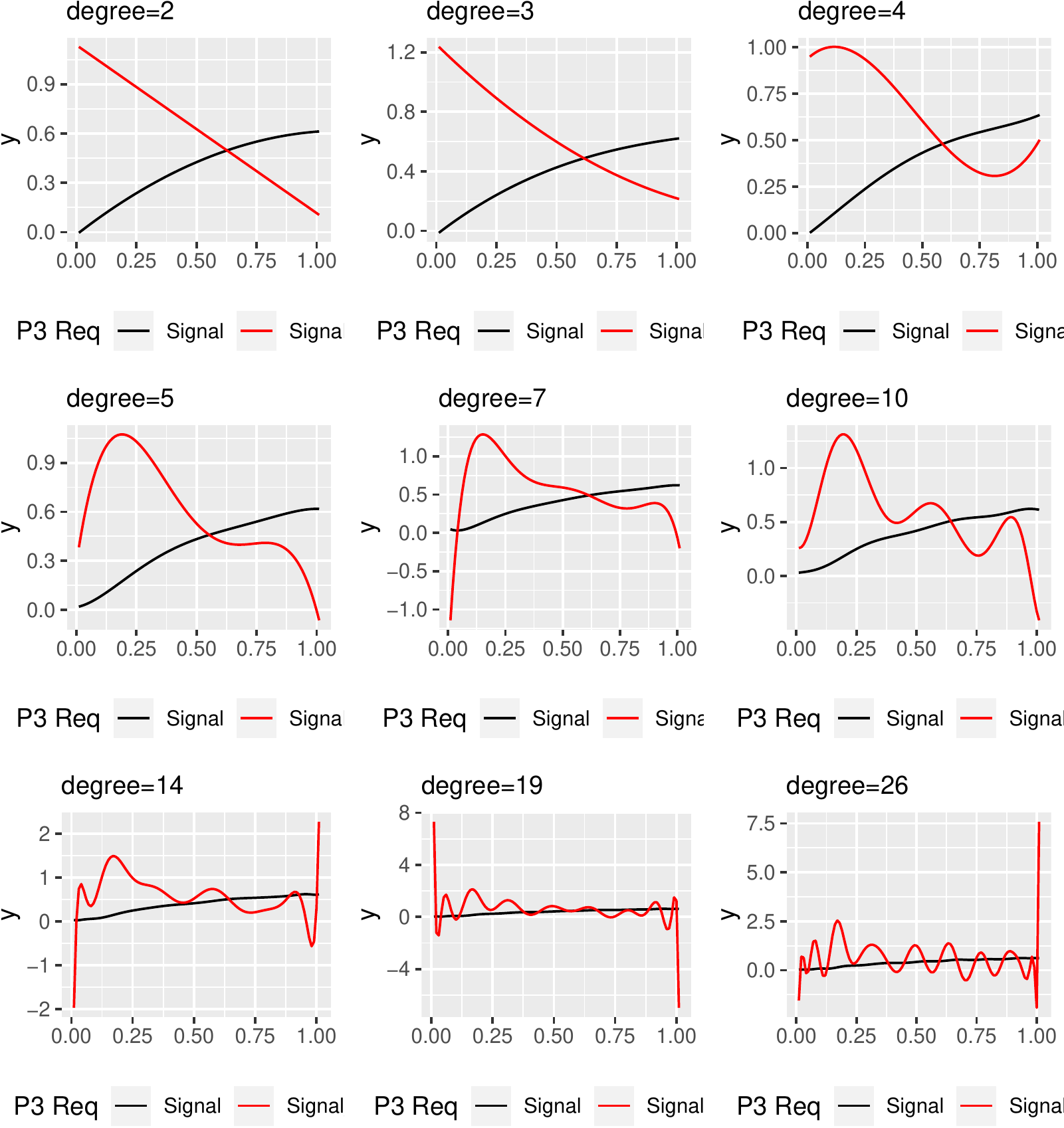} 

}

\caption{Fitted polynomials and their first derivative using varying degrees of freedom.}\label{fig:p4-polynomials-d1}
\end{figure}

Using a varying degree, we fit polynomials to the data in order to obtain a truly smooth curve. This technique always results in smooth approximations both for the signal and its gradient (unlike COBS). In figure \ref{fig:p4-polynomials-d1} we observe how the derivative captures more and more nuances of the signal, starting from a degree of approx. \(4\). For our tests with this pattern later, polynomials might be an alternative to LOESS-smoothed polynomials. I do recommend using these here polynomials of degree \(\geq3\), because the true purpose of all this is to estimate the non-linear trend for the variable and its rate of change.

\begin{Shaded}
\begin{Highlighting}[]
\FunctionTok{par}\NormalTok{(}\AttributeTok{mfrow =} \FunctionTok{c}\NormalTok{(}\DecValTok{4}\NormalTok{, }\DecValTok{4}\NormalTok{))}

\NormalTok{use\_degs }\OtherTok{\textless{}{-}} \FunctionTok{c}\NormalTok{(}\DecValTok{2}\NormalTok{, }\DecValTok{3}\NormalTok{, }\DecValTok{4}\NormalTok{, }\DecValTok{5}\NormalTok{, }\DecValTok{7}\NormalTok{, }\DecValTok{10}\NormalTok{, }\DecValTok{14}\NormalTok{, }\DecValTok{19}\NormalTok{, }\FunctionTok{length}\NormalTok{(stats}\SpecialCharTok{::}\FunctionTok{coef}\NormalTok{(}\FunctionTok{poly\_autofit\_max}\NormalTok{(}\AttributeTok{x =}\NormalTok{ X,}
  \AttributeTok{y =}\NormalTok{ Y))) }\SpecialCharTok{{-}} \DecValTok{1}\NormalTok{)}

\NormalTok{templ }\OtherTok{\textless{}{-}} \FunctionTok{list}\NormalTok{()}
\ControlFlowTok{for}\NormalTok{ (idx }\ControlFlowTok{in} \DecValTok{1}\SpecialCharTok{:}\FunctionTok{length}\NormalTok{(use\_degs)) \{}

\NormalTok{  templ[[idx]] }\OtherTok{\textless{}{-}}\NormalTok{ (}\ControlFlowTok{function}\NormalTok{() \{}
\NormalTok{    ud }\OtherTok{\textless{}{-}}\NormalTok{ use\_degs[idx]}
\NormalTok{    temp }\OtherTok{\textless{}{-}}\NormalTok{ stats}\SpecialCharTok{::}\FunctionTok{lm}\NormalTok{(Y }\SpecialCharTok{\textasciitilde{}} \FunctionTok{poly}\NormalTok{(}\AttributeTok{x =}\NormalTok{ X, }\AttributeTok{degree =}\NormalTok{ ud))}
\NormalTok{    Signal}\SpecialCharTok{$}\FunctionTok{new}\NormalTok{(}\AttributeTok{func =} \ControlFlowTok{function}\NormalTok{(x) stats}\SpecialCharTok{::}\FunctionTok{predict}\NormalTok{(temp, }\AttributeTok{newdata =} \FunctionTok{data.frame}\NormalTok{(}\AttributeTok{X =}\NormalTok{ x)),}
      \AttributeTok{name =} \StringTok{"P3 Req"}\NormalTok{, }\AttributeTok{support =} \FunctionTok{range}\NormalTok{(X), }\AttributeTok{isWp =} \ConstantTok{TRUE}\NormalTok{)}\SpecialCharTok{$}\FunctionTok{plot}\NormalTok{(}\AttributeTok{show1stDeriv =} \ConstantTok{TRUE}\NormalTok{) }\SpecialCharTok{+}
      \FunctionTok{labs}\NormalTok{(}\AttributeTok{subtitle =} \FunctionTok{paste0}\NormalTok{(}\StringTok{"degree="}\NormalTok{, ud))}
\NormalTok{  \})()}
\NormalTok{\}}

\NormalTok{ggpubr}\SpecialCharTok{::}\FunctionTok{ggarrange}\NormalTok{(}\AttributeTok{plotlist =}\NormalTok{ templ, }\AttributeTok{ncol =} \DecValTok{3}\NormalTok{, }\AttributeTok{nrow =} \DecValTok{3}\NormalTok{)}
\end{Highlighting}
\end{Shaded}

\clearpage

\hypertarget{assessing-the-goodness-of-fit}{%
\subsection{\texorpdfstring{Assessing the Goodness of Fit\label{sec:assess-gof}}{Assessing the Goodness of Fit}}\label{assessing-the-goodness-of-fit}}

In the technical report for detecting the Fire Drill in source code, we had previously introduced a plethora of methods to assess the goodness of fit. However, here we introduce additional methods that can exploit \textbf{confidence intervals}, both of homogeneous and inhomogeneous nature.

\hypertarget{score-based-on-ci-hyperplane}{%
\subsubsection{Score based on CI hyperplane}\label{score-based-on-ci-hyperplane}}

Simply put, this loss is based on the confidence interval's hyperplane, and calculates an absolute average confidence based on it. Each signal evaluated against the hyperplane is a slice of it. Since we are computing an average confidence, strictly speaking this is not a loss, but a score (higher is better).

\[
\begin{aligned}
  \mathit{L}^{\text{avgconf}}(x_1,x_2,f)=&\;\Bigg[\int_{x_1}^{x_2}\,\operatorname{CI}(x, f(x))\,dx\Bigg]\times(x_2-x_1)^{-1}\;\text{,}
  \\
  &\;\text{where}\;f\;\text{is the signal/variable and}\;x_2>x_1\text{.}
\end{aligned}
\]

We will do a full evaluation later, including creating a decision rule or learning how to scale the average weight to the consensus score, but let's take one project and test this.

\begin{Shaded}
\begin{Highlighting}[]
\NormalTok{L\_avgconf\_p3\_avg }\OtherTok{\textless{}{-}} \ControlFlowTok{function}\NormalTok{(x1, x2, f, CI) \{}
\NormalTok{  cubature}\SpecialCharTok{::}\FunctionTok{cubintegrate}\NormalTok{(}\AttributeTok{f =} \ControlFlowTok{function}\NormalTok{(x) \{}
    \FunctionTok{CI}\NormalTok{(}\AttributeTok{x =}\NormalTok{ x, }\AttributeTok{y =} \FunctionTok{f}\NormalTok{(x))}
\NormalTok{  \}, }\AttributeTok{lower =}\NormalTok{ x1, }\AttributeTok{upper =}\NormalTok{ x2)}\SpecialCharTok{$}\NormalTok{integral}\SpecialCharTok{/}\NormalTok{(x2 }\SpecialCharTok{{-}}\NormalTok{ x1)}
\NormalTok{\}}

\FunctionTok{invisible}\NormalTok{(}\FunctionTok{loadResultsOrCompute}\NormalTok{(}\AttributeTok{file =} \StringTok{"../results/ci\_p3avg\_Lavg{-}test.rds"}\NormalTok{, }\AttributeTok{computeExpr =}\NormalTok{ \{}
  \FunctionTok{c}\NormalTok{(}\AttributeTok{P2\_REQ =} \FunctionTok{L\_avgconf\_p3\_avg}\NormalTok{(}\AttributeTok{x1 =} \DecValTok{0}\NormalTok{, }\AttributeTok{x2 =} \DecValTok{1}\NormalTok{, }\AttributeTok{f =}\NormalTok{ all\_signals}\SpecialCharTok{$}\NormalTok{Project2}\SpecialCharTok{$}\NormalTok{REQ}\SpecialCharTok{$}\FunctionTok{get0Function}\NormalTok{(),}
    \AttributeTok{CI =}\NormalTok{ CI\_req\_p3avg), }\AttributeTok{P4\_REQ =} \FunctionTok{L\_avgconf\_p3\_avg}\NormalTok{(}\AttributeTok{x1 =} \DecValTok{0}\NormalTok{, }\AttributeTok{x2 =} \DecValTok{1}\NormalTok{, }\AttributeTok{f =}\NormalTok{ all\_signals}\SpecialCharTok{$}\NormalTok{Project4}\SpecialCharTok{$}\NormalTok{REQ}\SpecialCharTok{$}\FunctionTok{get0Function}\NormalTok{(),}
    \AttributeTok{CI =}\NormalTok{ CI\_req\_p3avg))}
\NormalTok{\}))}
\end{Highlighting}
\end{Shaded}

\begin{figure}[ht!]

{\centering \includegraphics{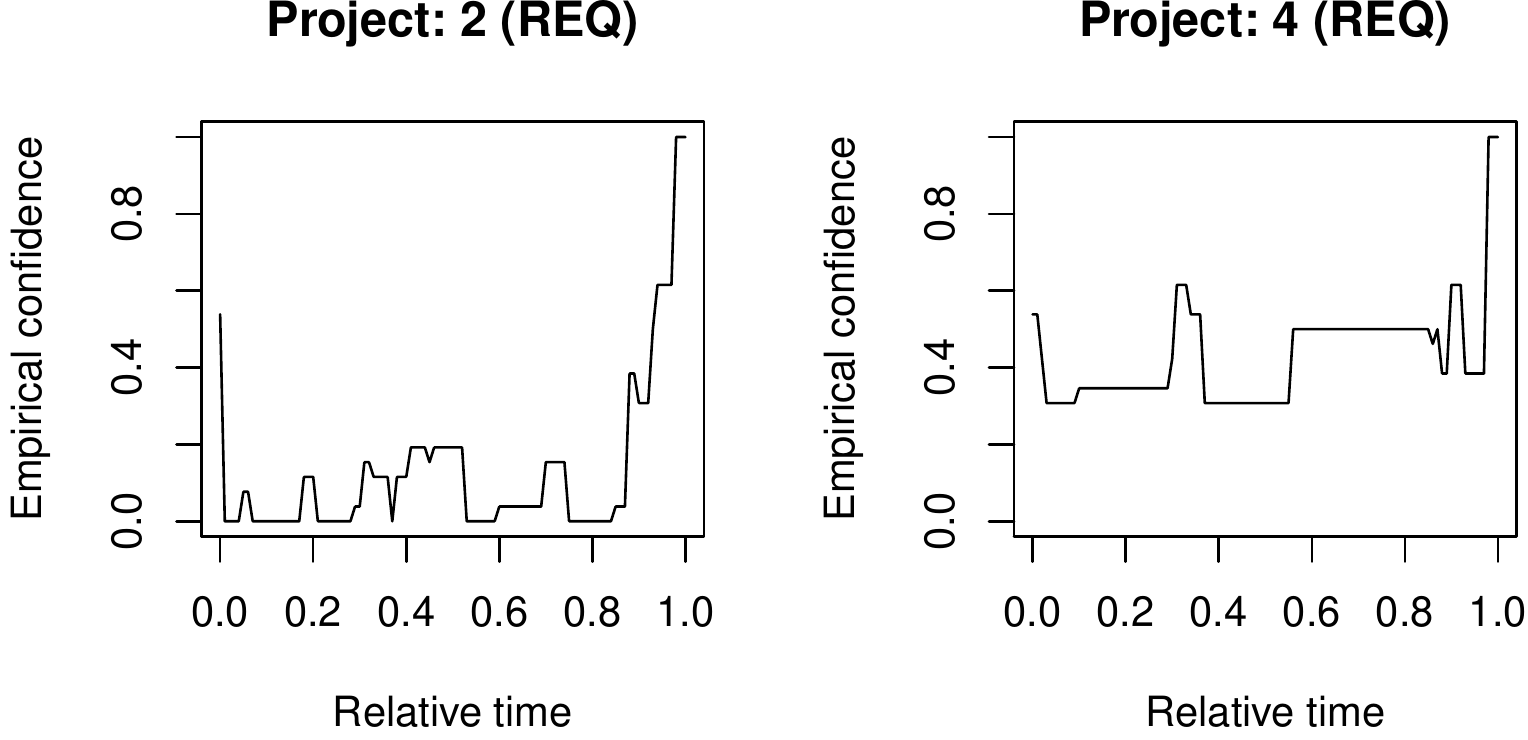} 

}

\caption{The confidence of req\% variable of projects 2 and 4 over relative project time.}\label{fig:p2p4-l3avgconf}
\end{figure}

Plotting \texttt{L\_avgconf\_p3\_avg} for the \texttt{REQ}-variable of both projects 2 and 4 gives us the function in figure \ref{fig:p2p4-l3avgconf}. We can clearly observe that the area under the latter is larger, and so will be the average.

From this example we can clearly see a difference in average confidence, which seems to be somewhat reconciling the projects' weights (consensus score). Let's try the next method, too.

\hypertarget{loss-based-on-distance-to-reference-variable}{%
\subsubsection{Loss based on distance to reference-variable}\label{loss-based-on-distance-to-reference-variable}}

We may choose to ignore the previously computed confidence hyperplane and compute a cost by, e.g., quantifying the distance between the previously computed average variable (or any other reference-variable) and another signal/variable. More precisely, we quantify the area between both variables, and compare it to the largest possible area, thereby obtaining an upper bound (with the lower bound being \(0\) obviously). Unlike the previous attempt, this function is a loss, that maps to the range \([0,1]\), where \(1\) means the largest possible distance (i.e., the variable compared is entirely outside (or collinear with) the confidence intervals). Hence this function is an attempt of measuring of dissimilarity.

This method requires three features: A reference-variable, and an upper- and a lower confidence interval. However, none of these are required to be of empirical nature. For example, we can even apply this method to the first of our patterns. A reference-variable may be generated as the average between the confidence intervals, etc. All this makes this method versatile. Its robustness comes from the fact any variable it scores, is confined to the boundaries of the confidence intervals.

\[
\begin{aligned}
  \mathit{L}^{\text{areadist}}(x_1,x_2,f)=&\;\int_{x_1}^{x_2}\,\left\lVert\,\bar{g}(x)-\overbrace{\min{\Big(\operatorname{CI}_{\text{upper}}(x),\;\max{\big(\operatorname{CI}_{\text{lower}}(x), f(x)\big)}\Big)}}^{\text{Confining of}\;f\;\text{to the confidence intervals.}}\,\right\rVert\,dx
  \\[1ex]
  &\;\times\Bigg[\;\overbrace{\int_{x_1}^{x_2}\,\sup{\Big(\bar{g}(x)-\operatorname{CI}_{\text{lower}}(x)\;,\;\operatorname{CI}_{\text{upper}}(x)-\bar{g}(x)\Big)\,dx}}^{\text{maximum possible area between}\;\bar{g}(x)\;\text{and either CI.}}\;\Bigg]^{-1}\;\text{.}
\end{aligned}
\]

This loss may also be alternatively defined using the following denominator:

\[
\begin{aligned}
  \mathit{L}^{\text{areadist2}}(x_1,x_2,f)=&\;\int_{x_1}^{x_2}\,\left\lVert\,\bar{g}(x)-\min{\Big(\operatorname{CI}_{\text{upper}}(x),\;\max{\big(\operatorname{CI}_{\text{lower}}(x), f(x)\big)}\Big)}\,\right\rVert\,dx
  \\[1ex]
  &\;\times\Bigg[\;\int_{x_1}^{x_2}\,\begin{cases}
    \operatorname{CI}_{\text{upper}}(x)-\bar{g}(x),&\text{if}\;f(x)\geq\bar{g}(x),
    \\
    \bar{g}(x)-\operatorname{CI}_{\text{lower}}(x),&\text{otherwise.}
  \end{cases}\,dx\;\Bigg]^{-1}\;\text{.}
\end{aligned}
\]

The difference is subtle but important, and corrects better for asymmetric confidence intervals. It now captures the maximum possible area, based on the currently valid confidence interval (at \(x\)). For the remainder, we will always be using the \textbf{second variant of this loss} because of this.

Again, we will do a full evaluation later, but let's just attempt computing this loss once.

\begin{Shaded}
\begin{Highlighting}[]
\NormalTok{L\_areadist\_p3\_avg }\OtherTok{\textless{}{-}} \ControlFlowTok{function}\NormalTok{(x1, x2, f, gbar, CI\_upper, CI\_lower, }\AttributeTok{use2ndVariant =} \ConstantTok{FALSE}\NormalTok{) \{}
\NormalTok{  int1 }\OtherTok{\textless{}{-}}\NormalTok{ cubature}\SpecialCharTok{::}\FunctionTok{cubintegrate}\NormalTok{(}\AttributeTok{f =} \ControlFlowTok{function}\NormalTok{(x) \{}
    \FunctionTok{abs}\NormalTok{(}\FunctionTok{gbar}\NormalTok{(x) }\SpecialCharTok{{-}} \FunctionTok{min}\NormalTok{(}\FunctionTok{CI\_upper}\NormalTok{(x), }\FunctionTok{max}\NormalTok{(}\FunctionTok{CI\_lower}\NormalTok{(x), }\FunctionTok{f}\NormalTok{(x))))}
\NormalTok{  \}, }\AttributeTok{lower =}\NormalTok{ x1, }\AttributeTok{upper =}\NormalTok{ x2)}\SpecialCharTok{$}\NormalTok{integral}

\NormalTok{  int2 }\OtherTok{\textless{}{-}}\NormalTok{ cubature}\SpecialCharTok{::}\FunctionTok{cubintegrate}\NormalTok{(}\AttributeTok{f =} \ControlFlowTok{function}\NormalTok{(x) \{}
\NormalTok{    gbarval }\OtherTok{\textless{}{-}} \FunctionTok{gbar}\NormalTok{(x)}
    \ControlFlowTok{if}\NormalTok{ (use2ndVariant) \{}
      \ControlFlowTok{if}\NormalTok{ (}\FunctionTok{f}\NormalTok{(x) }\SpecialCharTok{\textgreater{}=}\NormalTok{ gbarval) \{}
        \FunctionTok{CI\_upper}\NormalTok{(x) }\SpecialCharTok{{-}}\NormalTok{ gbarval}
\NormalTok{      \} }\ControlFlowTok{else}\NormalTok{ \{}
\NormalTok{        gbarval }\SpecialCharTok{{-}} \FunctionTok{CI\_lower}\NormalTok{(x)}
\NormalTok{      \}}
\NormalTok{    \} }\ControlFlowTok{else}\NormalTok{ \{}
      \FunctionTok{max}\NormalTok{(gbarval }\SpecialCharTok{{-}} \FunctionTok{CI\_lower}\NormalTok{(x), }\FunctionTok{CI\_upper}\NormalTok{(x) }\SpecialCharTok{{-}}\NormalTok{ gbarval)}
\NormalTok{    \}}
\NormalTok{  \}, }\AttributeTok{lower =}\NormalTok{ x1, }\AttributeTok{upper =}\NormalTok{ x2)}\SpecialCharTok{$}\NormalTok{integral}

  \FunctionTok{c}\NormalTok{(}\AttributeTok{area =}\NormalTok{ int1, }\AttributeTok{maxarea =}\NormalTok{ int2, }\AttributeTok{dist =}\NormalTok{ int1}\SpecialCharTok{/}\NormalTok{int2)}
\NormalTok{\}}

\FunctionTok{invisible}\NormalTok{(}\FunctionTok{loadResultsOrCompute}\NormalTok{(}\AttributeTok{file =} \StringTok{"../results/ci\_p3avg\_Larea{-}test.rds"}\NormalTok{, }\AttributeTok{computeExpr =}\NormalTok{ \{}
  \FunctionTok{list}\NormalTok{(}\AttributeTok{P2\_REQ =} \FunctionTok{L\_areadist\_p3\_avg}\NormalTok{(}\AttributeTok{x1 =} \DecValTok{0}\NormalTok{, }\AttributeTok{x2 =} \DecValTok{1}\NormalTok{, }\AttributeTok{f =}\NormalTok{ all\_signals}\SpecialCharTok{$}\NormalTok{Project2}\SpecialCharTok{$}\NormalTok{REQ}\SpecialCharTok{$}\FunctionTok{get0Function}\NormalTok{(),}
    \AttributeTok{use2ndVariant =} \ConstantTok{TRUE}\NormalTok{, }\AttributeTok{gbar =}\NormalTok{ req\_p3, }\AttributeTok{CI\_upper =}\NormalTok{ req\_ci\_upper\_p3avg, }\AttributeTok{CI\_lower =}\NormalTok{ req\_ci\_lower\_p3avg),}
    \AttributeTok{P4\_REQ =} \FunctionTok{L\_areadist\_p3\_avg}\NormalTok{(}\AttributeTok{x1 =} \DecValTok{0}\NormalTok{, }\AttributeTok{x2 =} \DecValTok{1}\NormalTok{, }\AttributeTok{f =}\NormalTok{ all\_signals}\SpecialCharTok{$}\NormalTok{Project4}\SpecialCharTok{$}\NormalTok{REQ}\SpecialCharTok{$}\FunctionTok{get0Function}\NormalTok{(),}
      \AttributeTok{use2ndVariant =} \ConstantTok{TRUE}\NormalTok{, }\AttributeTok{gbar =}\NormalTok{ req\_p3, }\AttributeTok{CI\_upper =}\NormalTok{ req\_ci\_upper\_p3avg, }\AttributeTok{CI\_lower =}\NormalTok{ req\_ci\_lower\_p3avg))}
\NormalTok{\}))}
\end{Highlighting}
\end{Shaded}

\begin{figure}[ht!]

{\centering \includegraphics{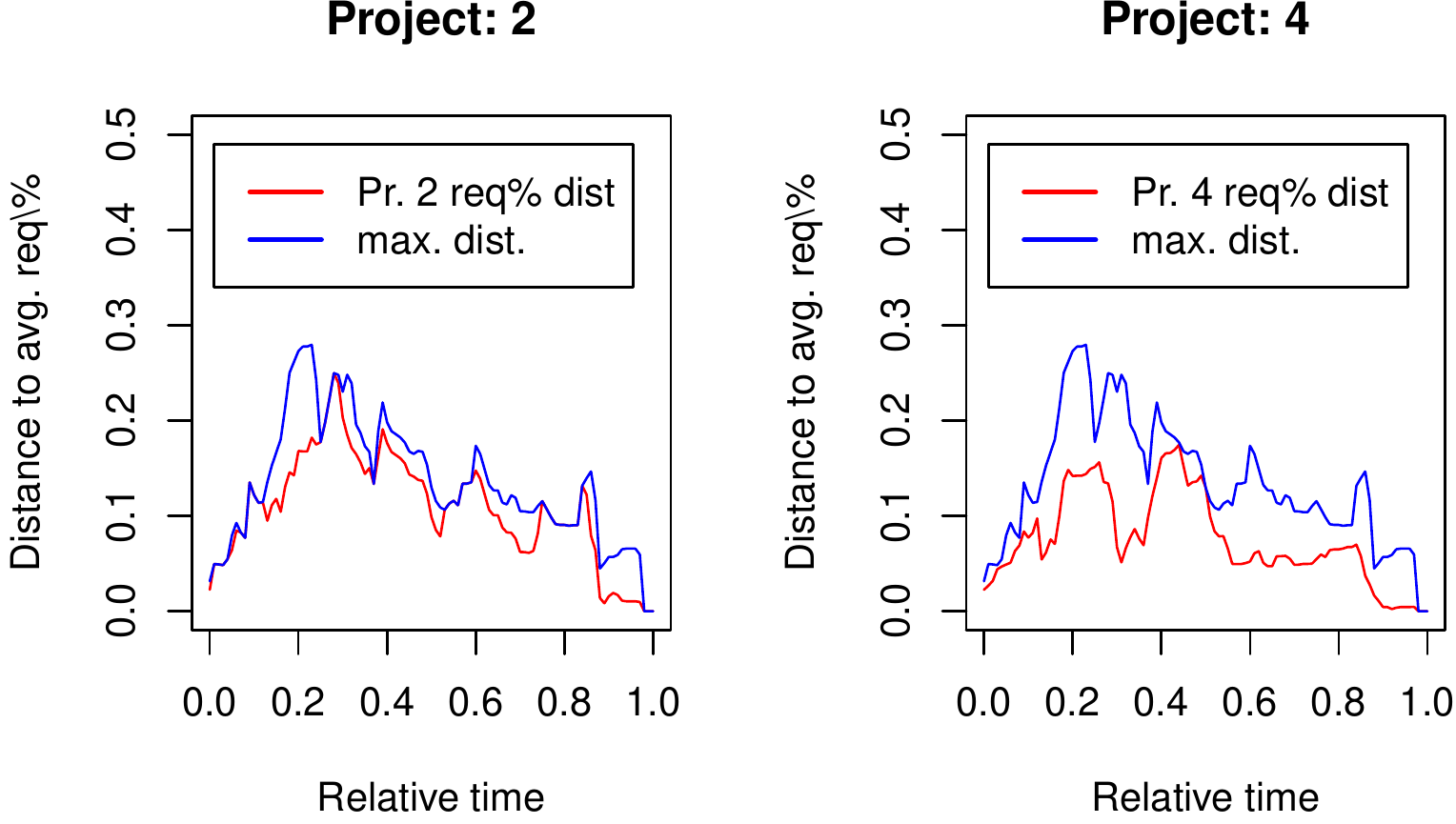} 

}

\caption{The req\% variable of projects 2 and 4, together with the maximum possible distance.}\label{fig:p2p4-lareadist}
\end{figure}

Let's show the maximum possible distance vs.~the distance of a project's variable in a plot (cf.~figure \ref{fig:p2p4-lareadist}). These figures clearly show the smaller distance of project 4 to the average. This is expected, as this project has the highest weight, so the average \texttt{REQ\%}-variable resembles this project most.

\hypertarget{loss-based-on-the-two-previous-approaches}{%
\subsubsection{Loss based on the two previous approaches}\label{loss-based-on-the-two-previous-approaches}}

This is an early-stadium idea. The essence is that for every \(x\), we have a vertical ``confidence-slice'' that we can integrate over and get an average confidence. Then, we obtain the confidence for the variable in question at the same \(x\). Both of these values can then be put into a relation. If we were to integrate this function, we would get the ratio between the variable's confidence and the average confidence, on average.

\begin{Shaded}
\begin{Highlighting}[]
\NormalTok{use\_x }\OtherTok{\textless{}{-}} \FloatTok{0.8}
\FunctionTok{dev\_p3}\NormalTok{(use\_x)}
\end{Highlighting}
\end{Shaded}

\begin{verbatim}
## [1] 0.706636
\end{verbatim}

\begin{Shaded}
\begin{Highlighting}[]
\NormalTok{cubature}\SpecialCharTok{::}\FunctionTok{cubintegrate}\NormalTok{(}\AttributeTok{f =} \ControlFlowTok{function}\NormalTok{(x) \{}
  \FunctionTok{CI\_dev\_p3avg}\NormalTok{(}\AttributeTok{x =}\NormalTok{ use\_x, }\AttributeTok{y =}\NormalTok{ x)}
\NormalTok{\}, }\AttributeTok{lower =} \FunctionTok{dev\_ci\_lower\_p3avg}\NormalTok{(use\_x), }\AttributeTok{upper =} \FunctionTok{dev\_ci\_upper\_p3avg}\NormalTok{(use\_x))}\SpecialCharTok{$}\NormalTok{integral}\SpecialCharTok{/}\NormalTok{(}\FunctionTok{dev\_ci\_upper\_p3avg}\NormalTok{(use\_x) }\SpecialCharTok{{-}}
  \FunctionTok{dev\_ci\_lower\_p3avg}\NormalTok{(use\_x))}
\end{Highlighting}
\end{Shaded}

\begin{verbatim}
## [1] 0.2650567
\end{verbatim}

\begin{Shaded}
\begin{Highlighting}[]
\FunctionTok{CI\_dev\_p3avg}\NormalTok{(}\AttributeTok{x =}\NormalTok{ use\_x, (all\_signals}\SpecialCharTok{$}\NormalTok{Project4}\SpecialCharTok{$}\NormalTok{DEV}\SpecialCharTok{$}\FunctionTok{get0Function}\NormalTok{())(use\_x))}
\end{Highlighting}
\end{Shaded}

\begin{verbatim}
## [1] 0.3846154
\end{verbatim}

At \texttt{x=0.8}, the average variable is at \(\approx0.71\), the average confidence for the slice is \(\approx0.19\), and the confidence of the evaluated variable (project 4, \texttt{DEV}) is at \(\approx0.22\). This means that the ratio is in the interval \((0,\infty)\). A ``perfect'' ratio of \(1.0\) would express that the tested variable is, on average, equal to the average confidence.

\hypertarget{f-divergences-as-objectives}{%
\subsubsection{\texorpdfstring{\emph{f}-divergences as objectives}{f-divergences as objectives}}\label{f-divergences-as-objectives}}

f-divergences describe the divergence between two continuous \emph{random} variables, not between two continuous non-random variables (processes).
As we will show, however, it is possible to mimic the most important properties of random variables, making f-divergences applicable to any continuous process.
The computation of any such divergence can be understood as an aggregation of all point-wise differences.
f-divergences are \emph{convex} functions that express by how much one quantity diverges from the other using a ratio. Therefore, a ratio of \(1\) results in a divergence of \(0\), and any other ratio in a divergence \(>0\).
Three properties are required to mimic random variables. First, for a continuous process \(f:\mathbb{R}\mapsto\mathbb{R}\), for all realizations \(x\in\mathcal{X}\), \(f(x)\geq0\). Second, the integral of \(f=1\), i.e., \(\int_{\mathcal{X}}\,f=1\). And third, \emph{Absolute continuity} between \(f\) and a second continuous process \(g\) is required, that is, the domains of \(f\) and \(g\) must coincide.
Absolute continuity, sharing the same support, is especially required in cases where the segments of the PM and the process are not the same, which might by a valid case for some objectives.

In many cases it might be feasible to fulfill all three properties.
The first one is perhaps just a translation of \(f\) by defining \(f'=f-\underset{f}{arg\,min}\). When designing a PM, this property can be conveniently considered.
The second property requires the first to hold. Then, \(f\) can perhaps be scaled by defining \(f''=f'\div\int\,f'\).
For fulfilling the third property, we suggest using a transform- and scale-operator, \(\mathsf{T}\) \eqref{eq:transf-op}.
Consider the example of the Kullback--Leibler divergence between distributions \(P,Q\), and their resp. densities \(p,q\) \eqref{eq:kl-div}.
Plugging in the transform operator \(\mathsf{T}\) allows us to define a ``relative'\,' variant of the divergence \eqref{eq:kl-div-rel}.
Another advantage is that integration intervals simplify to \([0,1]\) \eqref{eq:kl-div-rel-int}, which is beneficial to any measurement operator or objective.

\begin{align}
    \mathsf{T}(a,b,x)=&\;a+x(b-a)\label{eq:transf-op}\text{,}
    \\[1ex]
    D_{\text{KL}}\left(P\,\|\,Q\right)=&\;\int_{-\infty}^{\infty}\,p(x)\log{\left(\frac{p(x)}{q(x)}\right)}\,dx\label{eq:kl-div}\text{,}
    \\[1em]
    a_P,b_P,a_Q,b_Q\;\dots&\;\text{supports for}\;p,q\text{ (known a priori),}\nonumber
    \\[1ex]
    D_{\text{KL}}^{\text{rel}}\left(P\,\|\,Q\right)=&\;p\big(\mathsf{T}(a_P,b_P,x)\big)\log{\left(\frac{p\big(\mathsf{T}(a_P,b_P,x)\big)}{q\big(\mathsf{T}(a_Q,b_Q,x)\big)}\right)}\label{eq:kl-div-rel}\text{, where}
    \\[1ex]
    \operatorname{supp}\left(D_{\text{KL}}^{\text{rel}}\left(P\,\|\,Q\right)\right)=&\;[0,1]\nonumber\text{,}
    \\[1ex]
    %\operatorname{supp}\Big(D_{\text{KL}}^{\text{rel}}(P\,\|\,Q)\Big)=&\;[0,1]\;\text{, so the divergence}\;d_{PQ}\;\text{becomes}\nonumber
    %\\[1ex]
    d_{PQ}=&\;\int_0^1\,D_{\text{KL}}^{\text{rel}}\left(P\,\|\,Q\right)\,dx\label{eq:kl-div-rel-int}\text{.}
\end{align}

\hypertarget{m-dimensional-relative-continuous-pearson-sample-correlation-coefficient}{%
\subsubsection{\texorpdfstring{\(m\)-dimensional relative continuous Pearson sample correlation coefficient\label{ssec:m-dim-pearson}}{m-dimensional relative continuous Pearson sample correlation coefficient}}\label{m-dimensional-relative-continuous-pearson-sample-correlation-coefficient}}

First, here is the formula:

\[
\begin{aligned}
  f,g\;\dots&\,m\text{-dimensional continuous variables},
  \\[1ex]
  \bm{\mathit{S}}_f,\bm{\mathit{S}}_g=&\;a_{i,j},b_{i,j}\in \mathbb{R}^{m\times2}\,\text{, supports for}\;f,g\;\text{along each dimension,}
  \\[1ex]
  \overline{f}=\mathrm{E}[f]=&\;\left[\int_{a_{1,1}}^{a_{1,2}}\dots\int_{a_{m,1}}^{a_{m,2}}\,f(x_1,\dots,x_m)\,dx_1\dots dx_m\right]
  \\[0ex]
  &\;\;\times\prod_{i=1}^{m}\,(a_{i,1}-a_{i,2})^{-1}\,\text{, (analogously for}\;\overline{g}\;\text{using}\;\bm{\mathit{S}}_g\text{),}
  \\[1ex]
  =&\;\int_0^1\dots\int_0^1\,f(\mathsf{T}(a_{1,1},a_{1,2},x_1),\,\dots\,,\mathsf{T}(a_{m,1},a_{m,2},x_m))\,dx_1\,\dots\,dx_m\,\text{,}
  \\[1em]
  \operatorname{corr}(f,g)=&\frac{
    \splitfrac{
        \Big(f\big(\mathsf{T}(a_{1,1},a_{1,2},x_1),\,\dots,\,\mathsf{T}(a_{m,1},a_{m,2},x_m)\big)-\mathrm{E}[f]\Big)
    }{
        \times\Big(g\big(\mathsf{T}(b_{1,1},b_{1,2},x_1),\,\dots,\,\mathsf{T}(b_{m,1},b_{m,2},x_m)\big)-\mathrm{E}[g]\Big)
    }
  }{
    \splitfrac{
      \left[\int_0^1\dots\int_0^1\,\Big(f\big(\mathsf{T}(a_{1,1},a_{1,2},x_1),\,\dots\big)-\mathrm{E}[f]\Big)^2\,dx_1\,\dots\,dx_m\right]^{\frac{1}{2}}
  }{
      \times\left[\int_0^1\dots\int_0^1\,\Big(g\big(\mathsf{T}(b_{1,1},b_{1,2},x_1),\,\dots\big)-\mathrm{E}[g]\Big)^2\,dx_1\,\dots\,dx_m\right]^{\frac{1}{2}}
    }
  }\,\text{,}\label{eq:mdim-corr}
  \\[1em]
  c_{fg}=&\;\int_0^1\dots\int_0^1\,corr(f,g)\,dx_1\,\dots\,dx_m\,\text{.}
\end{aligned}
\]

\begin{figure}[ht!]

{\centering \includegraphics{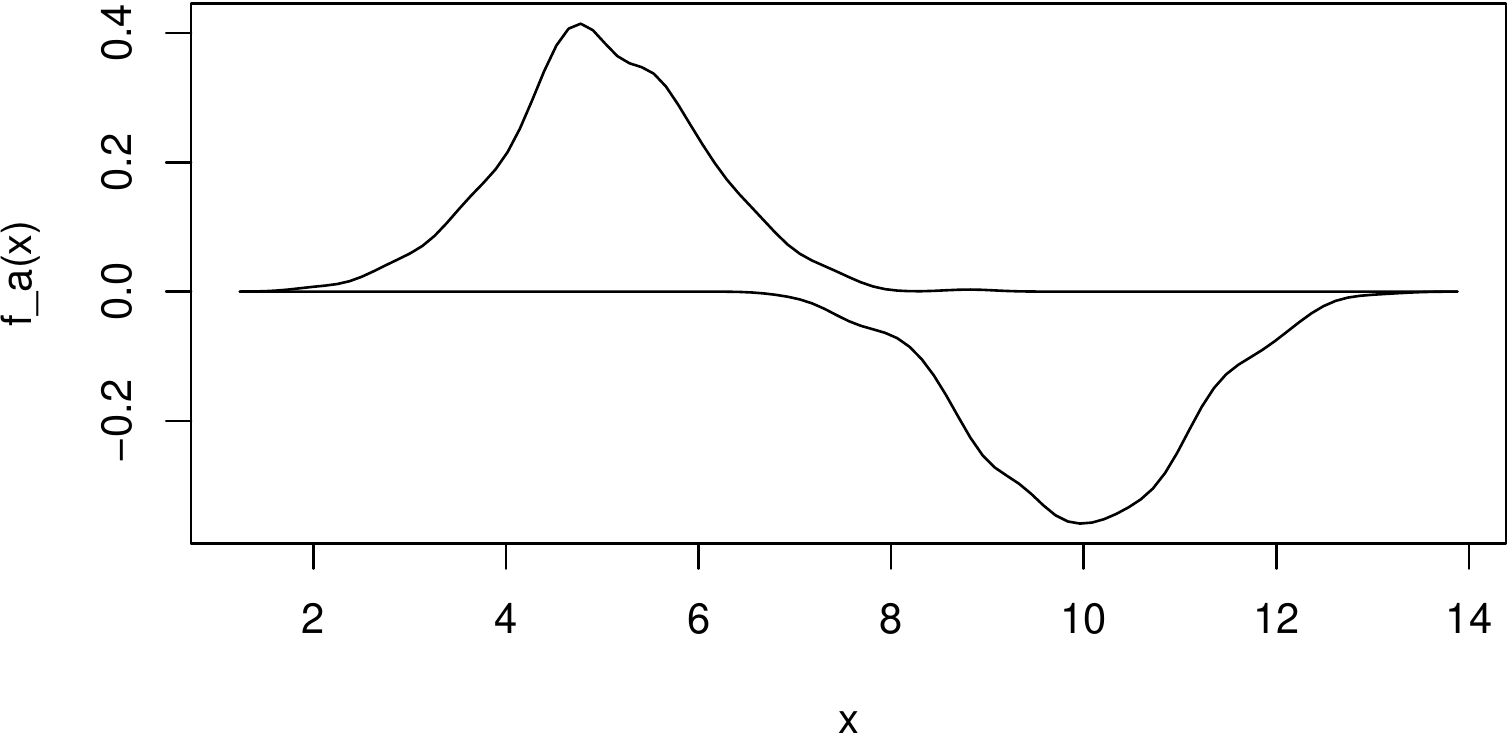} 

}

\caption{The densities of both normally-distributed variables.}\label{fig:dens-example-data}
\end{figure}

However first, we will implement a continuous relative 1D version to test our approach. Let's generate some sample data that should be highly correlated, shown in figure \ref{fig:dens-example-data}. See how we deliberately have different means in order to dislocate both variables.

\begin{Shaded}
\begin{Highlighting}[]
\FunctionTok{set.seed}\NormalTok{(}\DecValTok{1}\NormalTok{)}

\NormalTok{a }\OtherTok{\textless{}{-}} \FunctionTok{rnorm}\NormalTok{(}\DecValTok{500}\NormalTok{, }\AttributeTok{mean =} \DecValTok{5}\NormalTok{)}
\NormalTok{b }\OtherTok{\textless{}{-}} \FunctionTok{rnorm}\NormalTok{(}\DecValTok{500}\NormalTok{, }\AttributeTok{mean =} \DecValTok{10}\NormalTok{)}

\NormalTok{dens\_a }\OtherTok{\textless{}{-}} \FunctionTok{density}\NormalTok{(a)}
\NormalTok{dens\_b }\OtherTok{\textless{}{-}} \FunctionTok{density}\NormalTok{(b)}
\NormalTok{dens\_b}\SpecialCharTok{$}\NormalTok{y }\OtherTok{\textless{}{-}} \SpecialCharTok{{-}}\DecValTok{1} \SpecialCharTok{*}\NormalTok{ dens\_b}\SpecialCharTok{$}\NormalTok{y}

\NormalTok{f\_a }\OtherTok{\textless{}{-}} \FunctionTok{approxfun}\NormalTok{(}\AttributeTok{x =}\NormalTok{ dens\_a}\SpecialCharTok{$}\NormalTok{x, }\AttributeTok{y =}\NormalTok{ dens\_a}\SpecialCharTok{$}\NormalTok{y, }\AttributeTok{yleft =} \DecValTok{0}\NormalTok{, }\AttributeTok{yright =} \DecValTok{0}\NormalTok{)}
\NormalTok{f\_b }\OtherTok{\textless{}{-}} \FunctionTok{approxfun}\NormalTok{(}\AttributeTok{x =}\NormalTok{ dens\_b}\SpecialCharTok{$}\NormalTok{x, }\AttributeTok{y =}\NormalTok{ dens\_b}\SpecialCharTok{$}\NormalTok{y, }\AttributeTok{yleft =} \DecValTok{0}\NormalTok{, }\AttributeTok{yright =} \DecValTok{0}\NormalTok{)}
\end{Highlighting}
\end{Shaded}

As expected, the spatial difference in location does not influence the correlation:

\begin{Shaded}
\begin{Highlighting}[]
\FunctionTok{cor}\NormalTok{(dens\_a}\SpecialCharTok{$}\NormalTok{y, dens\_b}\SpecialCharTok{$}\NormalTok{y)}
\end{Highlighting}
\end{Shaded}

\begin{verbatim}
## [1] -0.9454304
\end{verbatim}

\begin{figure}[ht!]

{\centering \includegraphics{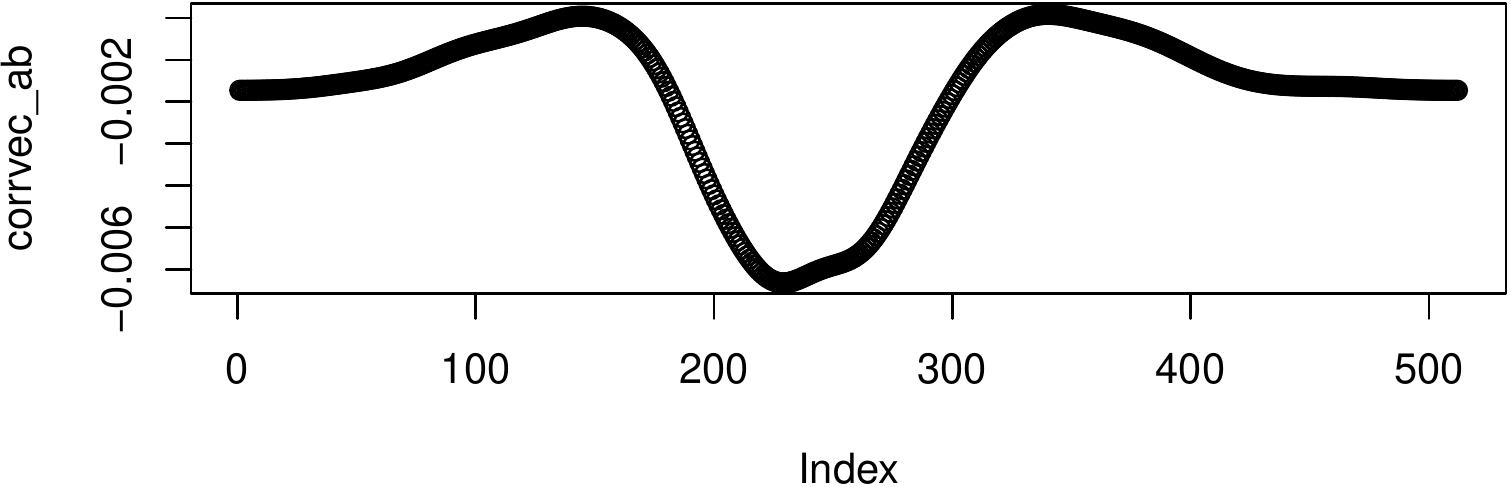} 

}

\caption{The one-dimensional correlation vector of the two discrete variables.}\label{fig:oned-corrvec}
\end{figure}

Before we go further, we want to manually implement the coefficient and visualize the correlation vector. It is shown in figure \ref{fig:oned-corrvec}.

\begin{Shaded}
\begin{Highlighting}[]
\NormalTok{corrvec\_ab }\OtherTok{\textless{}{-}}\NormalTok{ (dens\_a}\SpecialCharTok{$}\NormalTok{y }\SpecialCharTok{{-}} \FunctionTok{mean}\NormalTok{(dens\_a}\SpecialCharTok{$}\NormalTok{y)) }\SpecialCharTok{*}\NormalTok{ (dens\_b}\SpecialCharTok{$}\NormalTok{y }\SpecialCharTok{{-}} \FunctionTok{mean}\NormalTok{(dens\_b}\SpecialCharTok{$}\NormalTok{y))}\SpecialCharTok{/}\NormalTok{(}\FunctionTok{sqrt}\NormalTok{(}\FunctionTok{sum}\NormalTok{((dens\_a}\SpecialCharTok{$}\NormalTok{y }\SpecialCharTok{{-}}
  \FunctionTok{mean}\NormalTok{(dens\_a}\SpecialCharTok{$}\NormalTok{y))}\SpecialCharTok{\^{}}\DecValTok{2}\NormalTok{)) }\SpecialCharTok{*} \FunctionTok{sqrt}\NormalTok{(}\FunctionTok{sum}\NormalTok{((dens\_b}\SpecialCharTok{$}\NormalTok{y }\SpecialCharTok{{-}} \FunctionTok{mean}\NormalTok{(dens\_b}\SpecialCharTok{$}\NormalTok{y))}\SpecialCharTok{\^{}}\DecValTok{2}\NormalTok{)))}

\FunctionTok{sum}\NormalTok{(corrvec\_ab)}
\end{Highlighting}
\end{Shaded}

\begin{verbatim}
## [1] -0.9454304
\end{verbatim}

\hypertarget{d-continuous-relative-correlation}{%
\paragraph{1D continuous relative correlation}\label{d-continuous-relative-correlation}}

Now we define a continuous relative version of the Pearson sample correlation coefficient:

\begin{Shaded}
\begin{Highlighting}[]
\NormalTok{coef\_rel\_pearson\_1d }\OtherTok{\textless{}{-}} \ControlFlowTok{function}\NormalTok{(f, g, }\AttributeTok{supp\_f =} \FunctionTok{c}\NormalTok{(}\DecValTok{0}\NormalTok{, }\DecValTok{1}\NormalTok{), }\AttributeTok{supp\_g =} \FunctionTok{c}\NormalTok{(}\DecValTok{0}\NormalTok{, }\DecValTok{1}\NormalTok{)) \{}
  \CommentTok{\# sum[ (x\_i {-} bar\_x) x (y\_i {-} bar\_y) ] {-}{-}{-}{-}{-}{-}{-}{-}{-}{-}{-}{-}{-}{-}{-}{-}{-}{-}{-}{-}{-}{-}{-}{-}{-}{-}{-}{-}{-}{-}{-}{-}{-}{-}{-}{-}}
  \CommentTok{\# sqrt(sum[ (x\_i {-} bar\_x)\^{}2 ]) x sqrt(...)}

\NormalTok{  transform\_op }\OtherTok{\textless{}{-}} \ControlFlowTok{function}\NormalTok{(a, b, x) a }\SpecialCharTok{+}\NormalTok{ x }\SpecialCharTok{*}\NormalTok{ b }\SpecialCharTok{{-}}\NormalTok{ x }\SpecialCharTok{*}\NormalTok{ a}

  \CommentTok{\# Those work, too: bar\_f \textless{}{-} cubature::cubintegrate( f = f, lower = supp\_f[1],}
  \CommentTok{\# upper = supp\_f[2])$integral / (supp\_f[2] {-} supp\_f[1]) bar\_g \textless{}{-}}
  \CommentTok{\# cubature::cubintegrate( f = g, lower = supp\_g[1], upper =}
  \CommentTok{\# supp\_g[2])$integral / (supp\_g[2] {-} supp\_g[1])}

\NormalTok{  bar\_f }\OtherTok{\textless{}{-}}\NormalTok{ cubature}\SpecialCharTok{::}\FunctionTok{cubintegrate}\NormalTok{(}\AttributeTok{f =} \ControlFlowTok{function}\NormalTok{(x) }\FunctionTok{f}\NormalTok{(}\FunctionTok{transform\_op}\NormalTok{(supp\_f[}\DecValTok{1}\NormalTok{], supp\_f[}\DecValTok{2}\NormalTok{],}
\NormalTok{    x)), }\AttributeTok{lower =} \DecValTok{0}\NormalTok{, }\AttributeTok{upper =} \DecValTok{1}\NormalTok{)}\SpecialCharTok{$}\NormalTok{integral}
\NormalTok{  bar\_g }\OtherTok{\textless{}{-}}\NormalTok{ cubature}\SpecialCharTok{::}\FunctionTok{cubintegrate}\NormalTok{(}\AttributeTok{f =} \ControlFlowTok{function}\NormalTok{(x) }\FunctionTok{g}\NormalTok{(}\FunctionTok{transform\_op}\NormalTok{(supp\_g[}\DecValTok{1}\NormalTok{], supp\_g[}\DecValTok{2}\NormalTok{],}
\NormalTok{    x)), }\AttributeTok{lower =} \DecValTok{0}\NormalTok{, }\AttributeTok{upper =} \DecValTok{1}\NormalTok{)}\SpecialCharTok{$}\NormalTok{integral}

\NormalTok{  denom\_f }\OtherTok{\textless{}{-}} \FunctionTok{sqrt}\NormalTok{(cubature}\SpecialCharTok{::}\FunctionTok{cubintegrate}\NormalTok{(}\AttributeTok{f =} \ControlFlowTok{function}\NormalTok{(x) \{}
\NormalTok{    (}\FunctionTok{f}\NormalTok{(}\FunctionTok{transform\_op}\NormalTok{(supp\_f[}\DecValTok{1}\NormalTok{], supp\_f[}\DecValTok{2}\NormalTok{], x)) }\SpecialCharTok{{-}}\NormalTok{ bar\_f)}\SpecialCharTok{\^{}}\DecValTok{2}
\NormalTok{  \}, }\AttributeTok{lower =} \DecValTok{0}\NormalTok{, }\AttributeTok{upper =} \DecValTok{1}\NormalTok{)}\SpecialCharTok{$}\NormalTok{integral)}
\NormalTok{  denom\_g }\OtherTok{\textless{}{-}} \FunctionTok{sqrt}\NormalTok{(cubature}\SpecialCharTok{::}\FunctionTok{cubintegrate}\NormalTok{(}\AttributeTok{f =} \ControlFlowTok{function}\NormalTok{(x) \{}
\NormalTok{    (}\FunctionTok{g}\NormalTok{(}\FunctionTok{transform\_op}\NormalTok{(supp\_g[}\DecValTok{1}\NormalTok{], supp\_g[}\DecValTok{2}\NormalTok{], x)) }\SpecialCharTok{{-}}\NormalTok{ bar\_g)}\SpecialCharTok{\^{}}\DecValTok{2}
\NormalTok{  \}, }\AttributeTok{lower =} \DecValTok{0}\NormalTok{, }\AttributeTok{upper =} \DecValTok{1}\NormalTok{)}\SpecialCharTok{$}\NormalTok{integral)}

\NormalTok{  fnum }\OtherTok{\textless{}{-}} \ControlFlowTok{function}\NormalTok{(x) \{}
\NormalTok{    (}\FunctionTok{f}\NormalTok{(}\FunctionTok{transform\_op}\NormalTok{(}\AttributeTok{a =}\NormalTok{ supp\_f[}\DecValTok{1}\NormalTok{], }\AttributeTok{b =}\NormalTok{ supp\_f[}\DecValTok{2}\NormalTok{], }\AttributeTok{x =}\NormalTok{ x)) }\SpecialCharTok{{-}}\NormalTok{ bar\_f) }\SpecialCharTok{*}\NormalTok{ (}\FunctionTok{g}\NormalTok{(}\FunctionTok{transform\_op}\NormalTok{(}\AttributeTok{a =}\NormalTok{ supp\_g[}\DecValTok{1}\NormalTok{],}
      \AttributeTok{b =}\NormalTok{ supp\_g[}\DecValTok{2}\NormalTok{], }\AttributeTok{x =}\NormalTok{ x)) }\SpecialCharTok{{-}}\NormalTok{ bar\_g)}
\NormalTok{  \}}

\NormalTok{  numerator }\OtherTok{\textless{}{-}}\NormalTok{ cubature}\SpecialCharTok{::}\FunctionTok{cubintegrate}\NormalTok{(}\AttributeTok{f =}\NormalTok{ fnum, }\AttributeTok{lower =} \DecValTok{0}\NormalTok{, }\AttributeTok{upper =} \DecValTok{1}\NormalTok{)}\SpecialCharTok{$}\NormalTok{integral}

  \FunctionTok{list}\NormalTok{(}\AttributeTok{fnum =}\NormalTok{ fnum, }\AttributeTok{bar\_f =}\NormalTok{ bar\_f, }\AttributeTok{bar\_g =}\NormalTok{ bar\_g, }\AttributeTok{denom\_f =}\NormalTok{ denom\_f, }\AttributeTok{denom\_g =}\NormalTok{ denom\_g,}
    \AttributeTok{numerator =}\NormalTok{ numerator, }\AttributeTok{corr\_func =} \ControlFlowTok{function}\NormalTok{(x) \{}
      \FunctionTok{fnum}\NormalTok{(x)}\SpecialCharTok{/}\NormalTok{(denom\_f }\SpecialCharTok{*}\NormalTok{ denom\_g)}
\NormalTok{    \}, }\AttributeTok{corr\_fg =}\NormalTok{ numerator}\SpecialCharTok{/}\NormalTok{(denom\_f }\SpecialCharTok{*}\NormalTok{ denom\_g))}
\NormalTok{\}}
\end{Highlighting}
\end{Shaded}

\begin{figure}[ht!]

{\centering \includegraphics{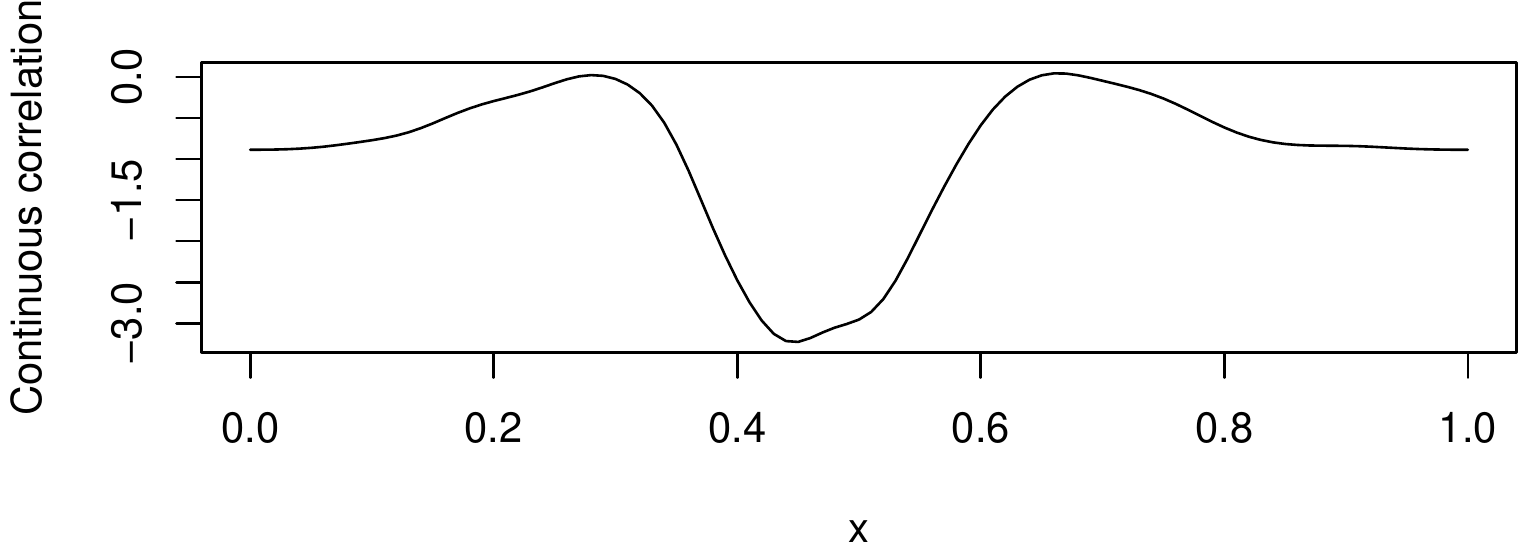} 

}

\caption{The one-dimensional correlation function of the two continuous variables.}\label{fig:oned-corrfunc}
\end{figure}

In figure \ref{fig:oned-corrfunc} we show the correlation of both continuous functions. Let's test using the previously generated data, density, and functions thereof:

\begin{verbatim}
## $bar_f
## [1] 0.120169
## 
## $bar_g
## [1] -0.1299947
## 
## $denom_f
## [1] 0.1366413
## 
## $denom_g
## [1] 0.1287124
## 
## $numerator
## [1] -0.01662643
## 
## $corr_fg
## [1] -0.9453587
\end{verbatim}

We can see that the results are very similar, except for some decimals. Next, we compare some individual intermittent results, that should be very similar (discrete vs.~continuous). Starting with the mean/expectation of either variable:

\begin{Shaded}
\begin{Highlighting}[]
\FunctionTok{c}\NormalTok{(}\FunctionTok{mean}\NormalTok{(dens\_a}\SpecialCharTok{$}\NormalTok{y), }\FunctionTok{mean}\NormalTok{(dens\_b}\SpecialCharTok{$}\NormalTok{y))}
\end{Highlighting}
\end{Shaded}

\begin{verbatim}
## [1]  0.1199343 -0.1297409
\end{verbatim}

\begin{Shaded}
\begin{Highlighting}[]
\FunctionTok{c}\NormalTok{(temp}\SpecialCharTok{$}\NormalTok{bar\_f, temp}\SpecialCharTok{$}\NormalTok{bar\_g)}
\end{Highlighting}
\end{Shaded}

\begin{verbatim}
## [1]  0.1201690 -0.1299947
\end{verbatim}

\begin{figure}[ht!]

{\centering \includegraphics{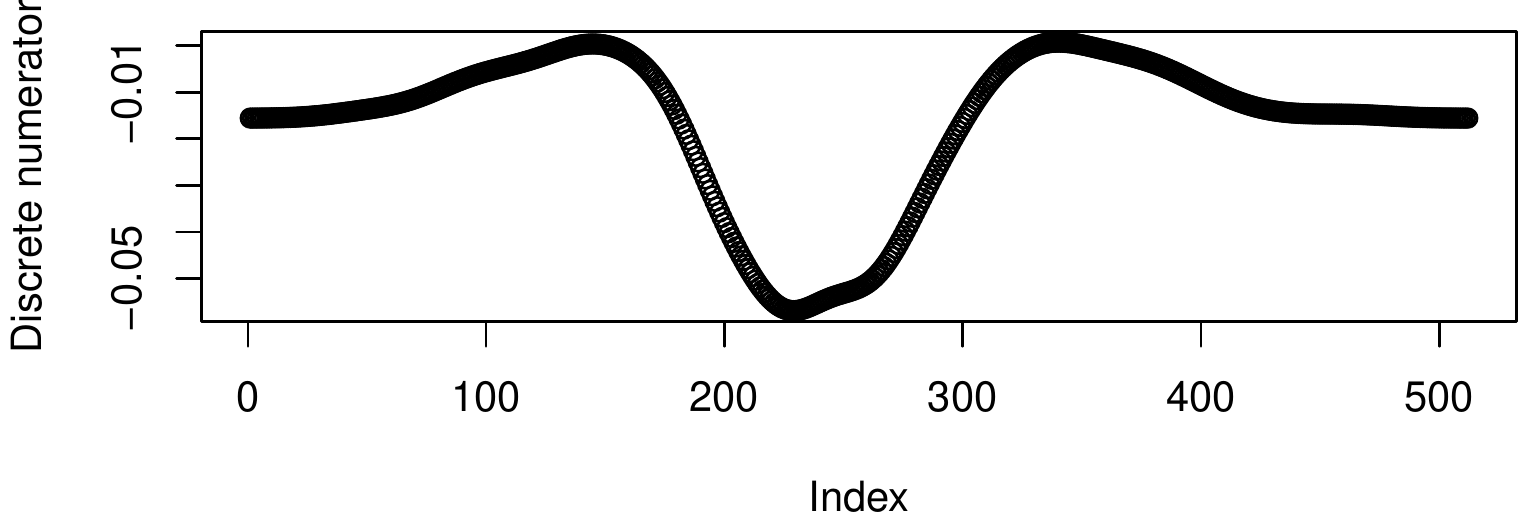} 

}

\caption{The discrete values that make up the numerator, plotted over all indexes.}\label{fig:disc-num}
\end{figure}

Check, very similar. Next, we compute and plot the result of the numerator (cf.~fig.~\ref{fig:disc-num}):

\begin{Shaded}
\begin{Highlighting}[]
\FunctionTok{sum}\NormalTok{((dens\_a}\SpecialCharTok{$}\NormalTok{y }\SpecialCharTok{{-}} \FunctionTok{mean}\NormalTok{(dens\_a}\SpecialCharTok{$}\NormalTok{y)) }\SpecialCharTok{*}\NormalTok{ (dens\_b}\SpecialCharTok{$}\NormalTok{y }\SpecialCharTok{{-}} \FunctionTok{mean}\NormalTok{(dens\_b}\SpecialCharTok{$}\NormalTok{y)))}
\end{Highlighting}
\end{Shaded}

\begin{verbatim}
## [1] -8.511955
\end{verbatim}

\begin{figure}[ht!]

{\centering \includegraphics{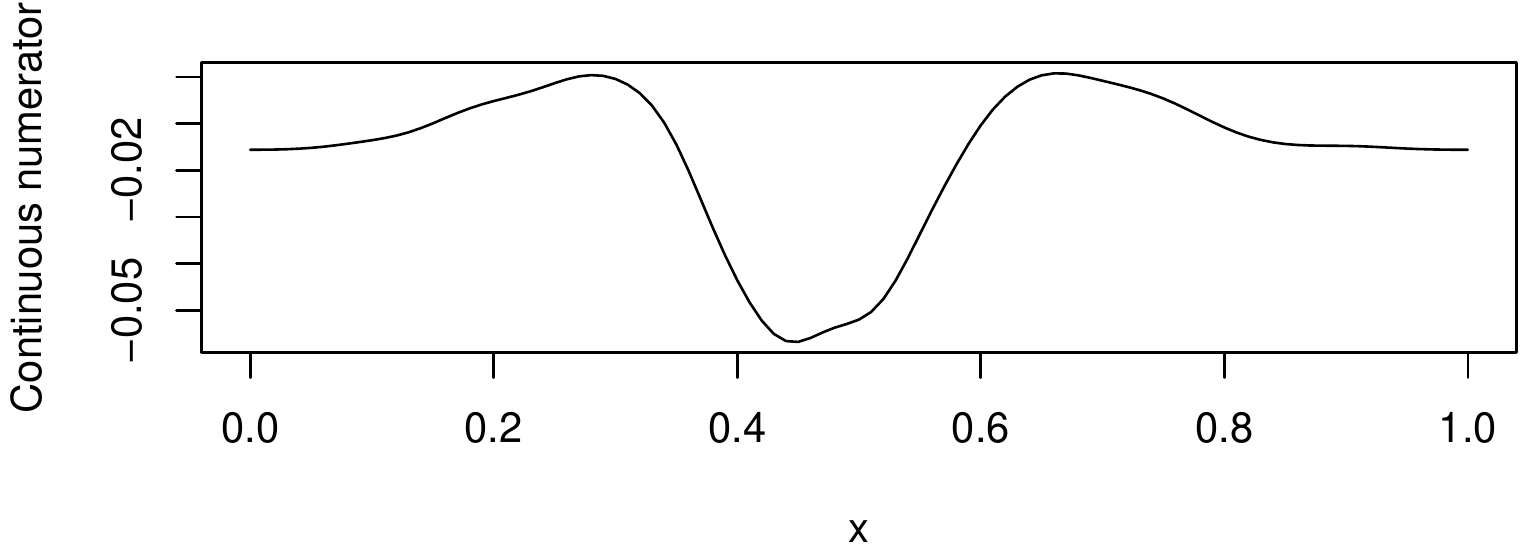} 

}

\caption{The relative function of the continuos numerator plotted over its support.}\label{fig:cont-num}
\end{figure}

The continuous version of the numerator is in the previously computed result. If we integrate it, we get the \textbf{mean} of the function (since the area under the curve over the interval \([0,1]\) is always a mean), not the sum. The function for the continuous numerator is shown in figure \ref{fig:cont-num}.

In order to get the same result as we got from the previous summation, we need to multiply by the number of elements in the discrete variable. The following result is very close to what we got from the summation:

\begin{Shaded}
\begin{Highlighting}[]
\NormalTok{cubature}\SpecialCharTok{::}\FunctionTok{cubintegrate}\NormalTok{(tempfff, }\DecValTok{0}\NormalTok{, }\DecValTok{1}\NormalTok{)}\SpecialCharTok{$}\NormalTok{integral }\SpecialCharTok{*} \FunctionTok{length}\NormalTok{(a)}
\end{Highlighting}
\end{Shaded}

\begin{verbatim}
## [1] -8.313215
\end{verbatim}

\begin{figure}[ht!]

{\centering \includegraphics{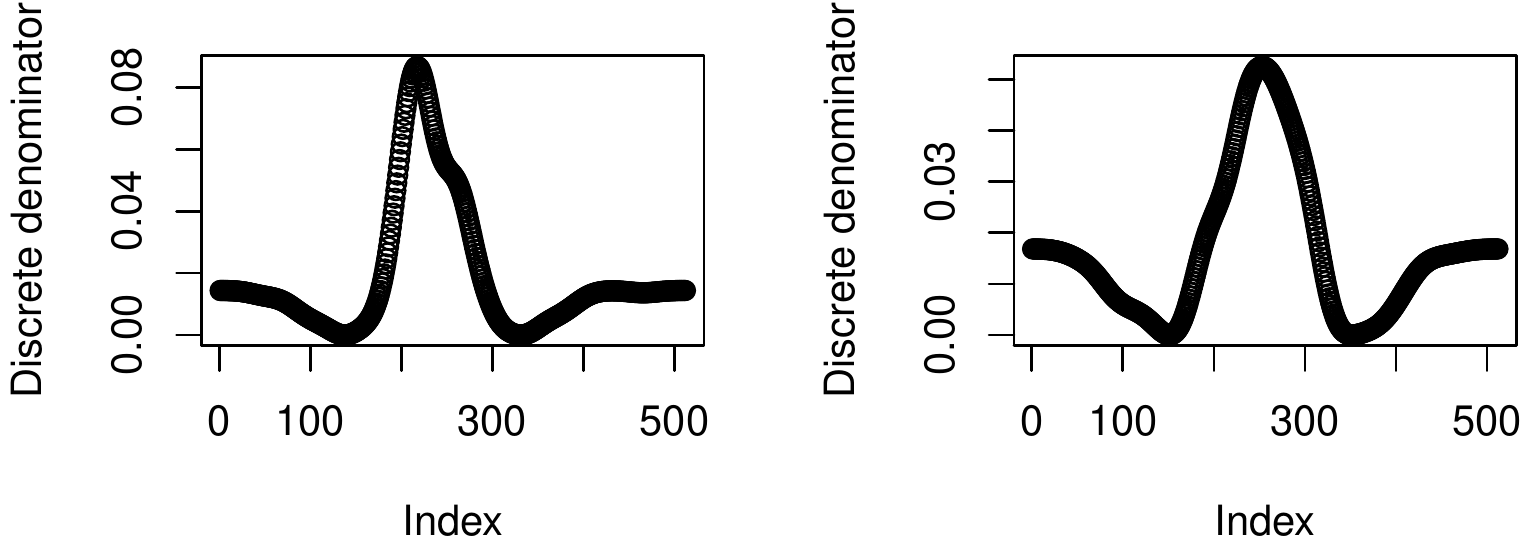} 

}

\caption{Plots of the discrete denominators for both data series.}\label{fig:disc-denoms}
\end{figure}

Let's check some values as computed in the denominator for the two variables (cf.~fig.~\ref{fig:disc-denoms}):

\begin{Shaded}
\begin{Highlighting}[]
\FunctionTok{c}\NormalTok{(}\FunctionTok{sqrt}\NormalTok{(}\FunctionTok{sum}\NormalTok{((dens\_a}\SpecialCharTok{$}\NormalTok{y }\SpecialCharTok{{-}} \FunctionTok{mean}\NormalTok{(dens\_a}\SpecialCharTok{$}\NormalTok{y))}\SpecialCharTok{\^{}}\DecValTok{2}\NormalTok{)), }\FunctionTok{sqrt}\NormalTok{(}\FunctionTok{sum}\NormalTok{((dens\_b}\SpecialCharTok{$}\NormalTok{y }\SpecialCharTok{{-}} \FunctionTok{mean}\NormalTok{(dens\_b}\SpecialCharTok{$}\NormalTok{y))}\SpecialCharTok{\^{}}\DecValTok{2}\NormalTok{)))}
\end{Highlighting}
\end{Shaded}

\begin{verbatim}
## [1] 3.091219 2.912528
\end{verbatim}

\begin{figure}[ht!]

{\centering \includegraphics{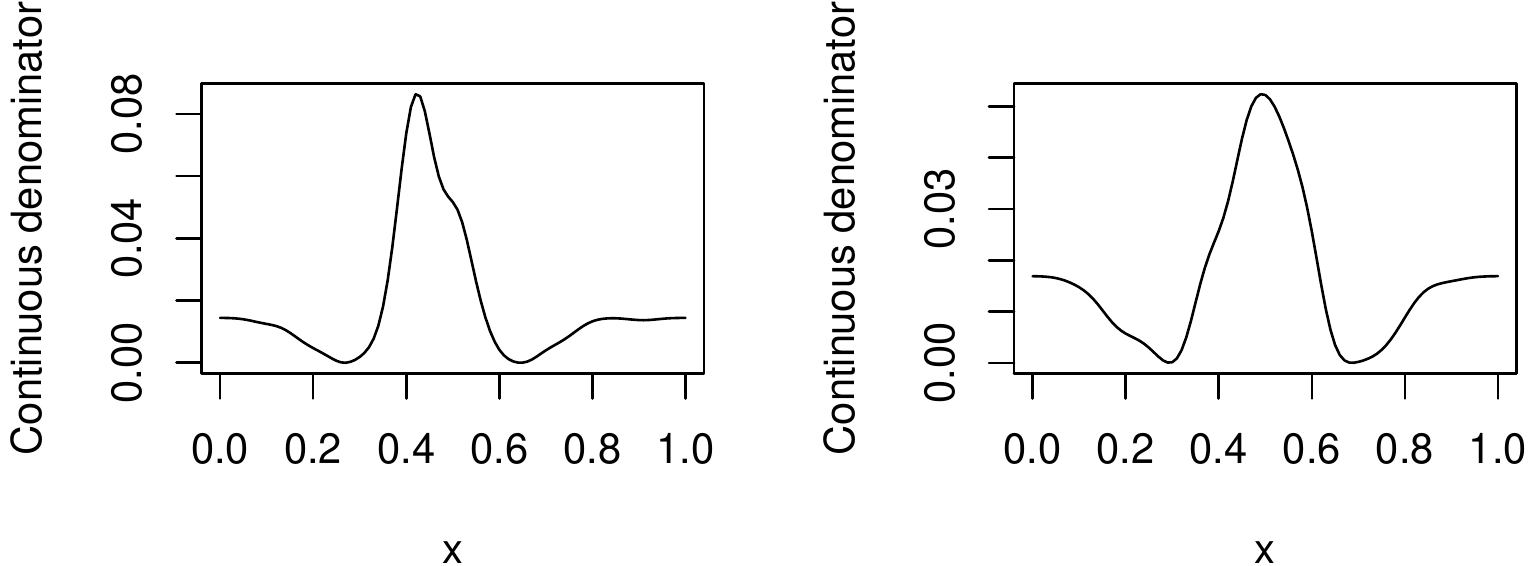} 

}

\caption{Plots of the continuous denominators for both data series.}\label{fig:cont-denoms}
\end{figure}

The continuous version of the denominators is the following (cf.~fig.~\ref{fig:cont-denoms}):

\begin{Shaded}
\begin{Highlighting}[]
\NormalTok{tempf\_a }\OtherTok{\textless{}{-}} \FunctionTok{Vectorize}\NormalTok{(}\ControlFlowTok{function}\NormalTok{(x) \{}
\NormalTok{  transform\_op }\OtherTok{\textless{}{-}} \ControlFlowTok{function}\NormalTok{(a, b, x) a }\SpecialCharTok{+}\NormalTok{ x }\SpecialCharTok{*}\NormalTok{ b }\SpecialCharTok{{-}}\NormalTok{ x }\SpecialCharTok{*}\NormalTok{ a}

\NormalTok{  (}\FunctionTok{f\_a}\NormalTok{(}\FunctionTok{transform\_op}\NormalTok{(}\FunctionTok{min}\NormalTok{(dens\_a}\SpecialCharTok{$}\NormalTok{x), }\FunctionTok{max}\NormalTok{(dens\_a}\SpecialCharTok{$}\NormalTok{x), }\AttributeTok{x =}\NormalTok{ x)) }\SpecialCharTok{{-}}\NormalTok{ temp}\SpecialCharTok{$}\NormalTok{bar\_f)}\SpecialCharTok{\^{}}\DecValTok{2}
\NormalTok{\})}
\NormalTok{tempf\_b }\OtherTok{\textless{}{-}} \FunctionTok{Vectorize}\NormalTok{(}\ControlFlowTok{function}\NormalTok{(x) \{}
\NormalTok{  transform\_op }\OtherTok{\textless{}{-}} \ControlFlowTok{function}\NormalTok{(a, b, x) a }\SpecialCharTok{+}\NormalTok{ x }\SpecialCharTok{*}\NormalTok{ b }\SpecialCharTok{{-}}\NormalTok{ x }\SpecialCharTok{*}\NormalTok{ a}
\NormalTok{  (}\FunctionTok{f\_b}\NormalTok{(}\FunctionTok{transform\_op}\NormalTok{(}\FunctionTok{min}\NormalTok{(dens\_b}\SpecialCharTok{$}\NormalTok{x), }\FunctionTok{max}\NormalTok{(dens\_b}\SpecialCharTok{$}\NormalTok{x), }\AttributeTok{x =}\NormalTok{ x)) }\SpecialCharTok{{-}}\NormalTok{ temp}\SpecialCharTok{$}\NormalTok{bar\_g)}\SpecialCharTok{\^{}}\DecValTok{2}
\NormalTok{\})}

\FunctionTok{c}\NormalTok{(}\FunctionTok{sqrt}\NormalTok{(cubature}\SpecialCharTok{::}\FunctionTok{cubintegrate}\NormalTok{(tempf\_a, }\DecValTok{0}\NormalTok{, }\DecValTok{1}\NormalTok{)}\SpecialCharTok{$}\NormalTok{integral }\SpecialCharTok{*} \FunctionTok{length}\NormalTok{(a)), }\FunctionTok{sqrt}\NormalTok{(cubature}\SpecialCharTok{::}\FunctionTok{cubintegrate}\NormalTok{(tempf\_b,}
  \DecValTok{0}\NormalTok{, }\DecValTok{1}\NormalTok{)}\SpecialCharTok{$}\NormalTok{integral }\SpecialCharTok{*} \FunctionTok{length}\NormalTok{(b)))}
\end{Highlighting}
\end{Shaded}

\begin{verbatim}
## [1] 3.055393 2.878096
\end{verbatim}

\hypertarget{d-continuous-relative-correlation-1}{%
\paragraph{2D continuous relative correlation}\label{d-continuous-relative-correlation-1}}

Finally, we'll implement the 2D version. Note that we even allow the support of \(y\) (the 2nd dimension) to depend on \(x\). Therefore, we pass in the support for the 2nd dimension as function. The following function works well, but the number of evaluations should be limited to get results in time (e.g., \(50\)). I tried \(1,000\) evaluations and got very precise results, but it ran for \(30\) minutes. With \(50\) evaluations, results are similarly close. Remember that we calculate correlations, and there it is often sufficient to have precision up to 2-3 decimals.

\begin{Shaded}
\begin{Highlighting}[]
\NormalTok{coef\_rel\_pearson\_2d }\OtherTok{\textless{}{-}} \ControlFlowTok{function}\NormalTok{(f, g, }\AttributeTok{supp\_f\_d1 =} \FunctionTok{c}\NormalTok{(}\DecValTok{0}\NormalTok{, }\DecValTok{1}\NormalTok{), }\AttributeTok{supp\_g\_d1 =} \FunctionTok{c}\NormalTok{(}\DecValTok{0}\NormalTok{, }\DecValTok{1}\NormalTok{), }\AttributeTok{supp\_f\_d2 =} \ControlFlowTok{function}\NormalTok{(x) }\FunctionTok{c}\NormalTok{(}\DecValTok{0}\NormalTok{,}
  \DecValTok{1}\NormalTok{), }\AttributeTok{supp\_g\_d2 =} \ControlFlowTok{function}\NormalTok{(x) }\FunctionTok{c}\NormalTok{(}\DecValTok{0}\NormalTok{, }\DecValTok{1}\NormalTok{), }\AttributeTok{maxEval =} \DecValTok{50}\NormalTok{) \{}
  \CommentTok{\# sum[ (x\_i {-} bar\_x) x (y\_i {-} bar\_y) ] {-}{-}{-}{-}{-}{-}{-}{-}{-}{-}{-}{-}{-}{-}{-}{-}{-}{-}{-}{-}{-}{-}{-}{-}{-}{-}{-}{-}{-}{-}{-}{-}{-}{-}{-}{-}}
  \CommentTok{\# sqrt(sum[ (x\_i {-} bar\_x)\^{}2 ]) x sqrt(...)}

\NormalTok{  transform\_op }\OtherTok{\textless{}{-}} \ControlFlowTok{function}\NormalTok{(a, b, x) a }\SpecialCharTok{+}\NormalTok{ x }\SpecialCharTok{*}\NormalTok{ b }\SpecialCharTok{{-}}\NormalTok{ x }\SpecialCharTok{*}\NormalTok{ a}

\NormalTok{  double\_int\_mean }\OtherTok{\textless{}{-}} \ControlFlowTok{function}\NormalTok{(func, supp\_d1, supp\_d2, }\AttributeTok{maxEval =} \DecValTok{50}\NormalTok{) \{}
\NormalTok{    cubature}\SpecialCharTok{::}\FunctionTok{cubintegrate}\NormalTok{(}\AttributeTok{f =} \ControlFlowTok{function}\NormalTok{(x) \{}
\NormalTok{      x1 }\OtherTok{\textless{}{-}} \FunctionTok{transform\_op}\NormalTok{(}\AttributeTok{a =}\NormalTok{ supp\_d1[}\DecValTok{1}\NormalTok{], }\AttributeTok{b =}\NormalTok{ supp\_d1[}\DecValTok{2}\NormalTok{], }\AttributeTok{x =}\NormalTok{ x)}
\NormalTok{      d2\_a }\OtherTok{\textless{}{-}} \FunctionTok{supp\_d2}\NormalTok{(x1)[}\DecValTok{1}\NormalTok{]}
\NormalTok{      d2\_b }\OtherTok{\textless{}{-}} \FunctionTok{supp\_d2}\NormalTok{(x1)[}\DecValTok{2}\NormalTok{]}

\NormalTok{      cubature}\SpecialCharTok{::}\FunctionTok{cubintegrate}\NormalTok{(}\AttributeTok{f =} \ControlFlowTok{function}\NormalTok{(y) \{}
\NormalTok{        y1 }\OtherTok{\textless{}{-}} \FunctionTok{transform\_op}\NormalTok{(}\AttributeTok{a =}\NormalTok{ d2\_a, }\AttributeTok{b =}\NormalTok{ d2\_b, }\AttributeTok{x =}\NormalTok{ y)}
        \FunctionTok{func}\NormalTok{(x1, y1)}
\NormalTok{      \}, }\AttributeTok{lower =} \DecValTok{0}\NormalTok{, }\AttributeTok{upper =} \DecValTok{1}\NormalTok{, }\AttributeTok{maxEval =}\NormalTok{ maxEval)}\SpecialCharTok{$}\NormalTok{integral}
\NormalTok{    \}, }\AttributeTok{lower =} \DecValTok{0}\NormalTok{, }\AttributeTok{upper =} \DecValTok{1}\NormalTok{, }\AttributeTok{maxEval =}\NormalTok{ maxEval)}\SpecialCharTok{$}\NormalTok{integral}
\NormalTok{  \}}

\NormalTok{  bar\_f }\OtherTok{\textless{}{-}} \FunctionTok{double\_int\_mean}\NormalTok{(}\AttributeTok{func =}\NormalTok{ f, }\AttributeTok{supp\_d1 =}\NormalTok{ supp\_f\_d1, }\AttributeTok{supp\_d2 =}\NormalTok{ supp\_f\_d2,}
    \AttributeTok{maxEval =}\NormalTok{ maxEval)}
\NormalTok{  bar\_g }\OtherTok{\textless{}{-}} \FunctionTok{double\_int\_mean}\NormalTok{(}\AttributeTok{func =}\NormalTok{ g, }\AttributeTok{supp\_d1 =}\NormalTok{ supp\_g\_d1, }\AttributeTok{supp\_d2 =}\NormalTok{ supp\_g\_d2,}
    \AttributeTok{maxEval =}\NormalTok{ maxEval)}


\NormalTok{  denom\_f }\OtherTok{\textless{}{-}} \FunctionTok{sqrt}\NormalTok{(}\FunctionTok{double\_int\_mean}\NormalTok{(}\AttributeTok{func =} \ControlFlowTok{function}\NormalTok{(x, y) \{}
\NormalTok{    (}\FunctionTok{f}\NormalTok{(x, y) }\SpecialCharTok{{-}}\NormalTok{ bar\_f)}\SpecialCharTok{\^{}}\DecValTok{2}
\NormalTok{  \}, }\AttributeTok{supp\_d1 =}\NormalTok{ supp\_f\_d1, }\AttributeTok{supp\_d2 =}\NormalTok{ supp\_f\_d2))}
\NormalTok{  denom\_g }\OtherTok{\textless{}{-}} \FunctionTok{sqrt}\NormalTok{(}\FunctionTok{double\_int\_mean}\NormalTok{(}\AttributeTok{func =} \ControlFlowTok{function}\NormalTok{(x, y) \{}
\NormalTok{    (}\FunctionTok{g}\NormalTok{(x, y) }\SpecialCharTok{{-}}\NormalTok{ bar\_g)}\SpecialCharTok{\^{}}\DecValTok{2}
\NormalTok{  \}, }\AttributeTok{supp\_d1 =}\NormalTok{ supp\_g\_d1, }\AttributeTok{supp\_d2 =}\NormalTok{ supp\_g\_d2))}

\NormalTok{  fnum }\OtherTok{\textless{}{-}} \ControlFlowTok{function}\NormalTok{(x, y) \{}
\NormalTok{    x1\_f }\OtherTok{\textless{}{-}} \FunctionTok{transform\_op}\NormalTok{(}\AttributeTok{a =}\NormalTok{ supp\_f\_d1[}\DecValTok{1}\NormalTok{], }\AttributeTok{b =}\NormalTok{ supp\_f\_d1[}\DecValTok{2}\NormalTok{], }\AttributeTok{x =}\NormalTok{ x)}
\NormalTok{    d2\_f\_a }\OtherTok{\textless{}{-}} \FunctionTok{supp\_f\_d2}\NormalTok{(x1\_f)[}\DecValTok{1}\NormalTok{]}
\NormalTok{    d2\_f\_b }\OtherTok{\textless{}{-}} \FunctionTok{supp\_f\_d2}\NormalTok{(x1\_f)[}\DecValTok{2}\NormalTok{]}
\NormalTok{    y1\_f }\OtherTok{\textless{}{-}} \FunctionTok{transform\_op}\NormalTok{(}\AttributeTok{a =}\NormalTok{ d2\_f\_a, }\AttributeTok{b =}\NormalTok{ d2\_f\_b, }\AttributeTok{x =}\NormalTok{ y)}

\NormalTok{    x1\_g }\OtherTok{\textless{}{-}} \FunctionTok{transform\_op}\NormalTok{(}\AttributeTok{a =}\NormalTok{ supp\_g\_d1[}\DecValTok{1}\NormalTok{], }\AttributeTok{b =}\NormalTok{ supp\_g\_d1[}\DecValTok{2}\NormalTok{], }\AttributeTok{x =}\NormalTok{ x)}
\NormalTok{    d2\_g\_a }\OtherTok{\textless{}{-}} \FunctionTok{supp\_g\_d2}\NormalTok{(x1\_g)[}\DecValTok{1}\NormalTok{]}
\NormalTok{    d2\_g\_b }\OtherTok{\textless{}{-}} \FunctionTok{supp\_g\_d2}\NormalTok{(x1\_g)[}\DecValTok{2}\NormalTok{]}
\NormalTok{    y1\_g }\OtherTok{\textless{}{-}} \FunctionTok{transform\_op}\NormalTok{(}\AttributeTok{a =}\NormalTok{ d2\_g\_a, }\AttributeTok{b =}\NormalTok{ d2\_g\_b, }\AttributeTok{x =}\NormalTok{ y)}

\NormalTok{    (}\FunctionTok{f}\NormalTok{(x1\_f, y1\_f) }\SpecialCharTok{{-}}\NormalTok{ bar\_f) }\SpecialCharTok{*}\NormalTok{ (}\FunctionTok{g}\NormalTok{(x1\_g, y1\_g) }\SpecialCharTok{{-}}\NormalTok{ bar\_g)}
\NormalTok{  \}}

\NormalTok{  numerator }\OtherTok{\textless{}{-}}\NormalTok{ cubature}\SpecialCharTok{::}\FunctionTok{cubintegrate}\NormalTok{(}\AttributeTok{f =} \ControlFlowTok{function}\NormalTok{(x) \{}
\NormalTok{    cubature}\SpecialCharTok{::}\FunctionTok{cubintegrate}\NormalTok{(}\AttributeTok{f =} \ControlFlowTok{function}\NormalTok{(y) \{}
      \FunctionTok{fnum}\NormalTok{(x, y)}
\NormalTok{    \}, }\AttributeTok{lower =} \DecValTok{0}\NormalTok{, }\AttributeTok{upper =} \DecValTok{1}\NormalTok{, }\AttributeTok{maxEval =}\NormalTok{ maxEval)}\SpecialCharTok{$}\NormalTok{integral}
\NormalTok{  \}, }\AttributeTok{lower =} \DecValTok{0}\NormalTok{, }\AttributeTok{upper =} \DecValTok{1}\NormalTok{, }\AttributeTok{maxEval =}\NormalTok{ maxEval)}\SpecialCharTok{$}\NormalTok{integral}

  \FunctionTok{list}\NormalTok{(}\AttributeTok{fnum =}\NormalTok{ fnum, }\AttributeTok{bar\_f =}\NormalTok{ bar\_f, }\AttributeTok{bar\_g =}\NormalTok{ bar\_g, }\AttributeTok{denom\_f =}\NormalTok{ denom\_f, }\AttributeTok{denom\_g =}\NormalTok{ denom\_g,}
    \AttributeTok{numerator =}\NormalTok{ numerator, }\AttributeTok{corr\_func =} \ControlFlowTok{function}\NormalTok{(x, y) \{}
      \FunctionTok{fnum}\NormalTok{(x, y)}\SpecialCharTok{/}\NormalTok{(denom\_f }\SpecialCharTok{*}\NormalTok{ denom\_g)}
\NormalTok{    \}, }\AttributeTok{corr\_fg =}\NormalTok{ numerator}\SpecialCharTok{/}\NormalTok{(denom\_f }\SpecialCharTok{*}\NormalTok{ denom\_g))}
\NormalTok{\}}
\end{Highlighting}
\end{Shaded}

The correlation of these two two-dimensional variables is \(\approx0.258\):

\begin{Shaded}
\begin{Highlighting}[]
\NormalTok{tempcorr }\OtherTok{\textless{}{-}} \FunctionTok{loadResultsOrCompute}\NormalTok{(}\AttributeTok{file =} \StringTok{"../results/2dcorr.rds"}\NormalTok{, }\AttributeTok{computeExpr =}\NormalTok{ \{}
  \FunctionTok{coef\_rel\_pearson\_2db}\NormalTok{(}\AttributeTok{f =}\NormalTok{ CI\_req\_p3avg, }\AttributeTok{g =}\NormalTok{ CI\_dev\_p3avg, }\AttributeTok{maxEval =} \DecValTok{250}\NormalTok{)}
\NormalTok{\})}
\NormalTok{tempcorr[}\SpecialCharTok{!}\NormalTok{(}\FunctionTok{names}\NormalTok{(tempcorr) }\SpecialCharTok{\%in\%} \FunctionTok{c}\NormalTok{(}\StringTok{"fnum"}\NormalTok{, }\StringTok{"corr\_func"}\NormalTok{))]}
\end{Highlighting}
\end{Shaded}

\begin{verbatim}
## $fbar
## [1] 0.06760212
## 
## $bar_g
## [1] 0.09045101
## 
## $denom_f
## [1] 0.1477128
## 
## $denom_g
## [1] 0.1574828
## 
## $numerator
## [1] 0.0008289947
## 
## $corr_fg
## [1] 0.03563694
\end{verbatim}

\begin{figure}[ht!]

{\centering \includegraphics{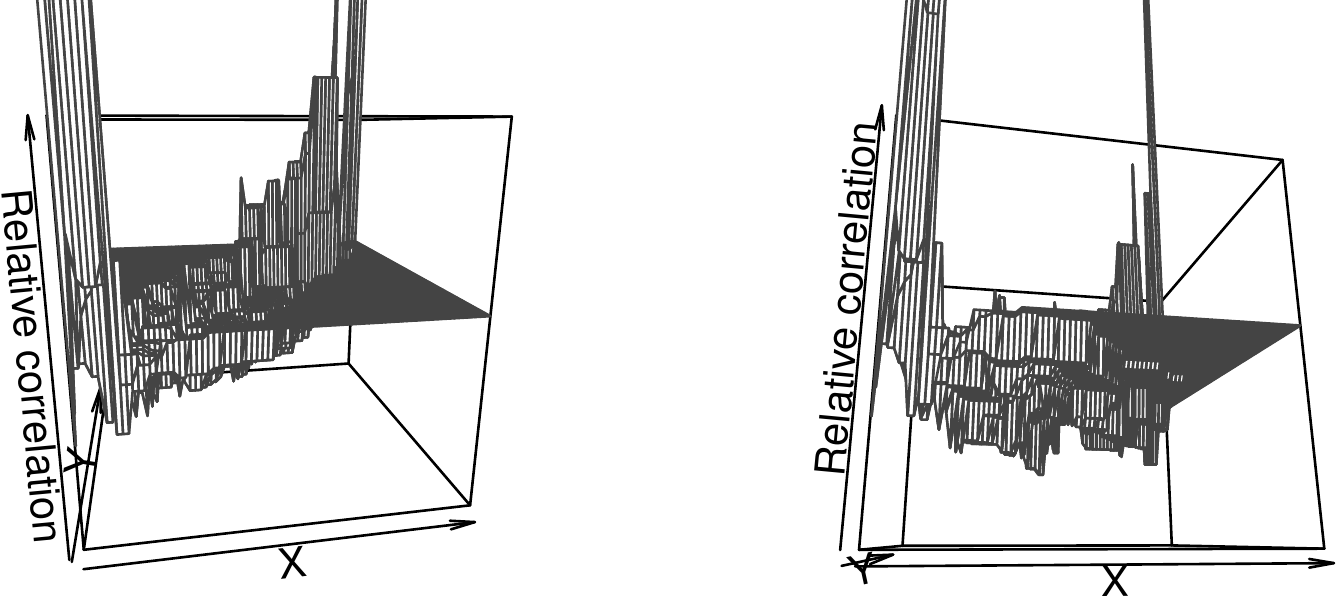} 

}

\caption{Correlation between the variables req\% and dev\%, plotted perspectively. On the left, we look at the correlation from above, while from the right, it is seen from underneath.}\label{fig:twod-correlation-req-dev-persp}
\end{figure}

In order to show the correlation in three dimensions, we'll compute a grid, cf.~fig.~\ref{fig:twod-correlation-req-dev-persp}:

\begin{Shaded}
\begin{Highlighting}[]
\NormalTok{tempgrid }\OtherTok{\textless{}{-}} \FunctionTok{loadResultsOrCompute}\NormalTok{(}\AttributeTok{file =} \StringTok{"../results/2dcorr\_grid.rds"}\NormalTok{, }\AttributeTok{computeExpr =}\NormalTok{ \{}
  \FunctionTok{outer}\NormalTok{(}\AttributeTok{X =} \FunctionTok{seq}\NormalTok{(}\DecValTok{0}\NormalTok{, }\DecValTok{1}\NormalTok{, }\AttributeTok{length.out =} \DecValTok{75}\NormalTok{), }\AttributeTok{Y =} \FunctionTok{seq}\NormalTok{(}\DecValTok{0}\NormalTok{, }\DecValTok{1}\NormalTok{, }\AttributeTok{length.out =} \DecValTok{75}\NormalTok{), }\AttributeTok{FUN =}\NormalTok{ tempcorr}\SpecialCharTok{$}\NormalTok{fnum)}
\NormalTok{\})}
\end{Highlighting}
\end{Shaded}

From this example we see there is clearly both, positive and negative correlation. Here is the same as a colorful contour plot (fig.~\ref{fig:twod-correlation-req-dev}) (correlation goes from blue/negative to red/positive, and no correlation is white):

\begin{figure}[ht!]

{\centering \includegraphics{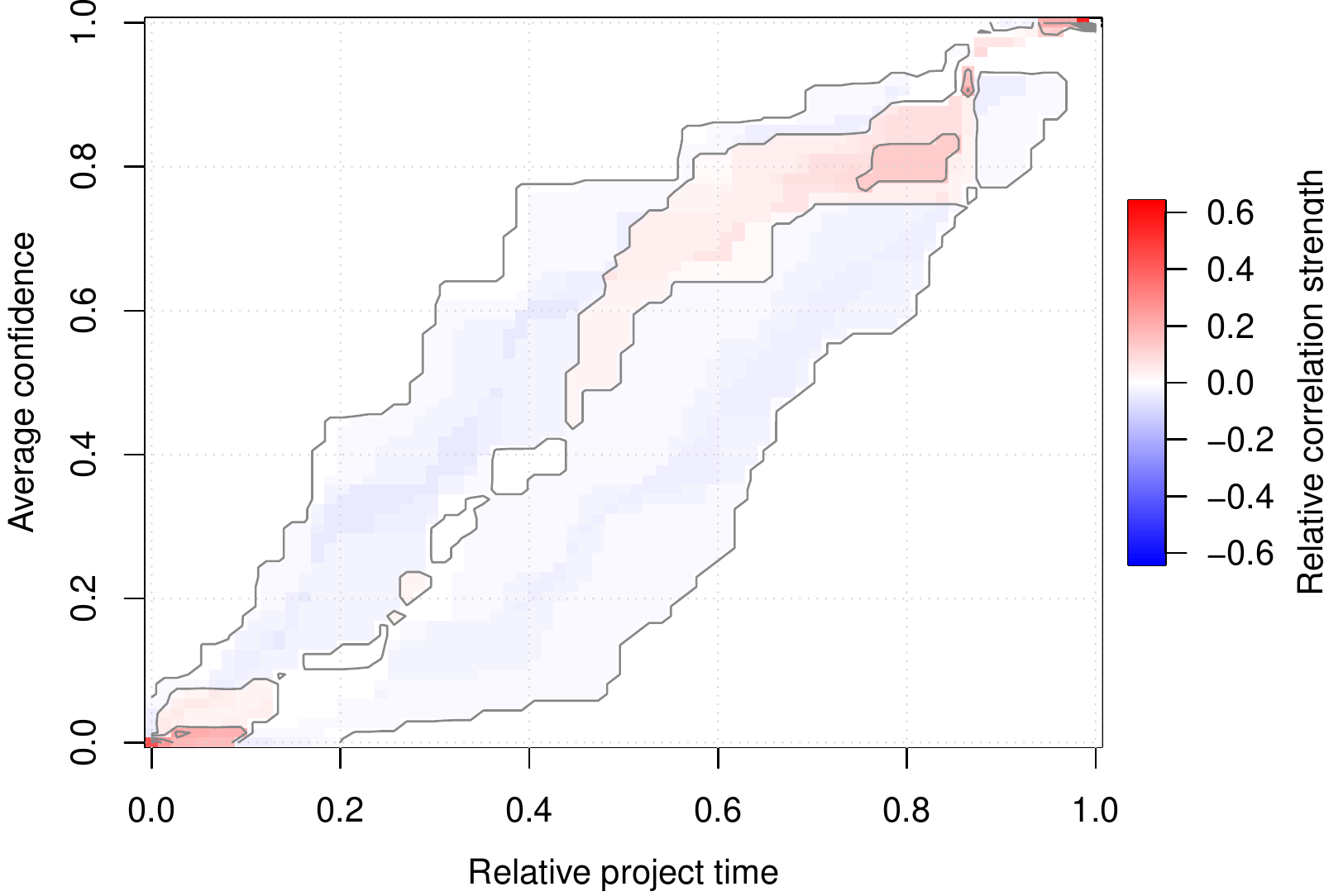} 

}

\caption{Correlation between the variables req\% and dev\%, plotted spatially.}\label{fig:twod-correlation-req-dev}
\end{figure}

\hypertarget{early-detection}{%
\subsection{Early detection}\label{early-detection}}

Many of the methods presented in the previous section \ref{sec:assess-gof} can also be used for early detection, by limiting (and re-scaling) the scores to the available interval. That means, that we would make a partial assessment of how well the process observed so far aligns with the process model in question. It only requires knowledge about at what point in time we are, i.e., \(t_{\text{now}}\). Scores would then be computed over the interval \([0,t_{\text{now}}]\).

The early detection can be split into two scenarios of interest:

\begin{enumerate}
\def\labelenumi{\arabic{enumi}.}
\tightlist
\item
  What happened since project begin until now? Scores may also be computed over an arbitrary interval \([t_{\text{begin}},t_{\text{end}}]\). More generally, the limits can be chosen arbitrarily for most scores, as long as \(t_{\text{begin}}<t_{\text{end}}\) and \(t_{\text{end}}\leq t_{\text{now}}\).
\item
  What will be the probable future, starting from now until a chosen point in the future, \(t_{\text{future}}\)? Here, \(t_{\text{end}}<t_{\text{future}}\) and the interval may be chosen arbitrarily. This scenario can be generically transformed into an instance of the first, by using any preferred prediction method to forecast the course of any variable, and then to compute the scores over, e.g., only the forecast interval, within \([t_0,t_{\text{forecast}}]\) etc.
\end{enumerate}

Early detection is widely applicable and can be, e.g., determined for each single variable separately. It may also be computed for derived variables, to make probabilistic statements about the expected \emph{rate of change}.

There is potentially a third scenario, where we would want to obtain point-in-time estimates, i.e., at any arbitrary point in time \(t\), we would like to obtain a total score. Furthermore, this would mean that we do \textbf{not} consider anything that happened before \(t\), nor after it. We may also not know the exact value for \(t\), i.e., how much of the total time has elapsed. In this case, we would probably require many more scores for trying to make up for the missing history. We will not consider this scenario further and only consider the first two scenarios that should cover most of the use cases.

\hypertarget{arbitrary-interval-scores}{%
\subsubsection{Arbitrary-interval scores}\label{arbitrary-interval-scores}}

Given some labeled training data (observations and assigned scores), we can attempt to learn the possibly non-linear relation for any interval in time, i.e., some \([t_{\text{begin}},t_{\text{end}}]\) (where \(t_{\text{end}}\leq t_{\text{now}}\)), the set of scores computed over that interval, \(\mathtt{S}_{t_{\text{begin}},t_{\text{end}}}\), and a ground truth, expressed as a function over continuous time, \(g(t_{\text{begin}},t_{\text{end}})\mapsto R\).

The ground truth may be a constant function, i.e., \(g(t_{\text{begin}},t_{\text{end}})\mapsto\mathrm{\text{constant}}\). In that case, the time span of the interval must be part of the input data, as we will then actually learn the relationship between the time span and the scores, i.e., some coefficients for the functions that take the time span delimiters, the scores over that interval, \(\mathtt{S}_{t_{\text{begin}},t_{\text{end}}}\), and output the constant total score (resp. a value close to it, as there will be residuals).

If the ground truth is not a constant function, but rather yields an exact or approximate total score for any arbitrarily chosen interval \([t_{\text{begin}},t_{\text{end}}]\), it is likely that we can skip using the time span as input variable. This kind of model is useful for when we do not know at what point in time we are at. If we had a good approximate idea, we could use a model that requires time as input and retrieve some average total score over an interval we are confident we are in. Otherwise, a model with non-constant ground truth is required.

The data we have available only provides us with a constant ground truth, i.e., for each project and at any point in time, it is constant. In section \ref{ssec:arbint-scores} we exemplary attempt to learn the non-linear relationship between arbitrarily chosen time spans and computed scores, and the constant ground truth.

\hypertarget{process-alignment}{%
\paragraph{Process alignment}\label{process-alignment}}

The procedure for computing scores over arbitrarily chosen intervals as just described requires approximate knowledge about when it was captured, since our ground truth is constant and the begin and end are input parameters to the trained model. The problem in a more generalized way is to align two (potentially multivariate) time series, where one of them is incomplete. Also, the alignment potentially has an open begin and/or end.

For these kind of tasks, Dynamic time warping (Giorgino 2009) works usually well. Depending on the nature of the process(es) and process model, we may even solve the problem using some optimization. For example, we modeled the issue-tracking data as cumulative processes. We could pose the essentially same optimization problem as earlier, where we attempted to find where the process model is \(0\) (see section \ref{ssec:optim-t1t2}), except that here would look for where the process model minus the observed process at \(t_{\text{begin}}\) equals zero (and the same for the end using \(t_{\text{end}}\)). However, this might not work for non-monotonic processes.

One could also attempt a more general way of mathematical optimization that computes an alignment between the process model and the partially observed process such that some loss is minimized the better start and end match. This approach is essentially a manual way of \textbf{one-interval} boundary time warping. For that, we have \textbf{\texttt{srBTAW}}, which allows us to use arbitrary intervals, losses, weight, etc. For example, an alignment may be computed using the correlation of the processes. DTW on the other hand uses the Euclidean distance as distance measure, and also requires discrete data, which greatly limits its flexibility.

In section \ref{ssec:proc-align-compute} we take an example and compute all of these.

\hypertarget{forecasting-within-vector-fields}{%
\subsubsection{Forecasting within Vector Fields}\label{forecasting-within-vector-fields}}

\begin{figure}[ht!]

{\centering \includegraphics{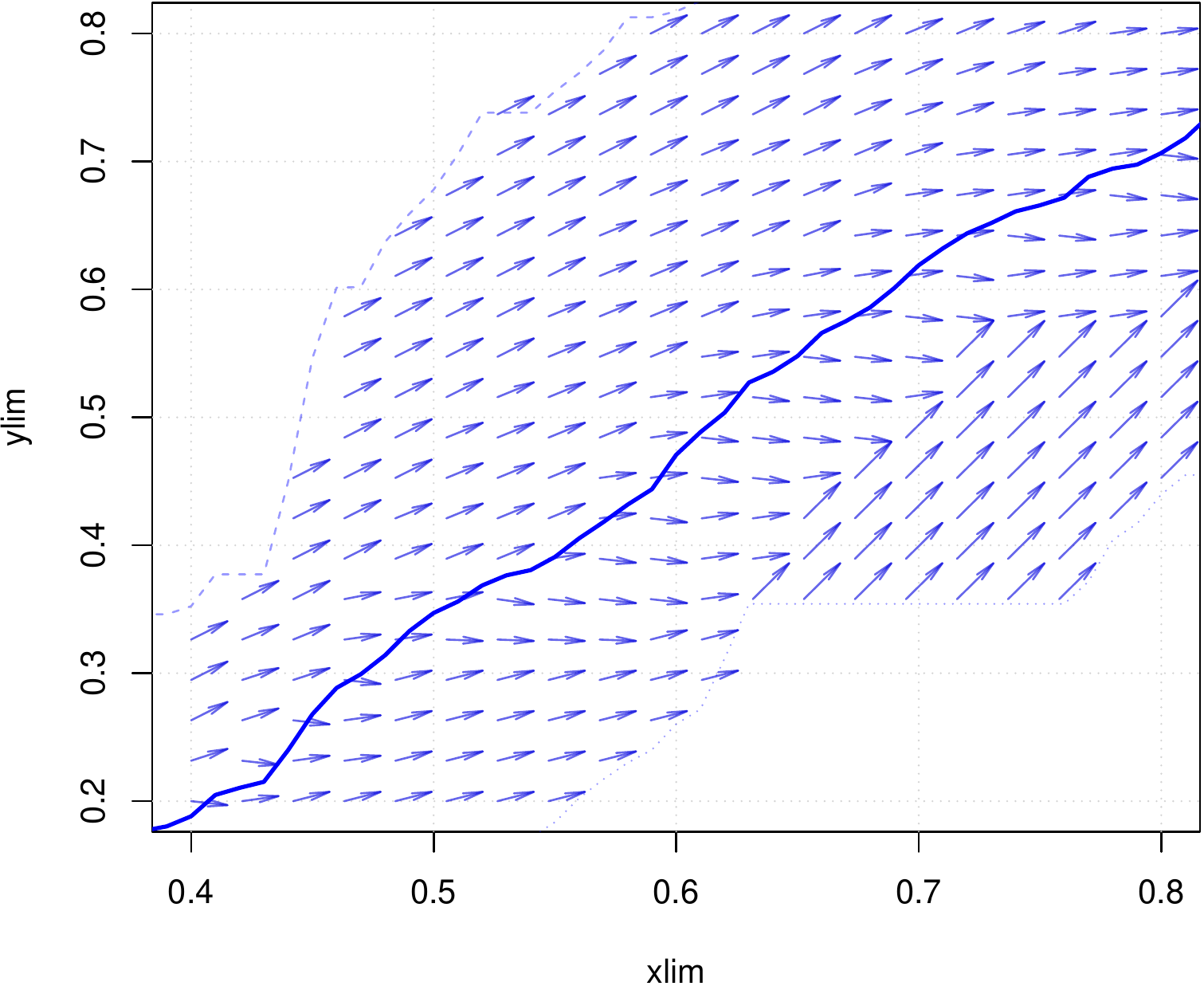} 

}

\caption{The inhomogeneous confidence interval of the dev\% variable and its vector field, pointing towards the largest increase in confidence for each pair of x/y coordinates. Here we use a non-smooth surface.}\label{fig:example-vectorfield}
\end{figure}

If we were to derive for an \emph{inhomogeneous} confidence interval, the result would be a vector field that, for each pair of coordinates \(x,y\), points into the direction of the largest change, which here means the direction into which the confidence would increase the most, on a cartesian coordinate system. This is exemplary shown in figure \ref{fig:example-vectorfield}. We suggest an operationalization in these ways:

\begin{itemize}
\tightlist
\item
  Using methods usually applied in time series forecasting, we have the means to determine the trends and probable path of a variable. Also, some of these methods will give as a confidence interval, which, over the vector field, is an area that may (partially) overlap.
\item
  We can determine the total and average confidence of the vector field that is affected by the overlap. This supports assumptions about whether we are currently in or headed into regions with lower or higher confidence (where the confidence represents the degree to which a process model is present).
\item
  We can determine the direction and strength of steepest increase of the confidence. The direction can be compared to the direction of the forecast. This may be exploited for making decisions that lead to heading away or into the confidence surface, depending on which choice is desirable. It also gives insights into the current projected trend.
\end{itemize}

In figure \ref{fig:example-vectorfield} we show an example of a vector field that is non-smooth. In the following examples however, we use smoothed version of x/y-slices (Green and Silverman 1993).

\begin{Shaded}
\begin{Highlighting}[]
\NormalTok{CI\_dev\_smooth\_slice\_x }\OtherTok{\textless{}{-}} \ControlFlowTok{function}\NormalTok{(y, }\AttributeTok{nsamp =} \DecValTok{100}\NormalTok{, }\AttributeTok{deriv =} \DecValTok{0}\NormalTok{) \{}
  \FunctionTok{stopifnot}\NormalTok{(}\FunctionTok{length}\NormalTok{(y) }\SpecialCharTok{==} \DecValTok{1}\NormalTok{)}

\NormalTok{  data\_x }\OtherTok{\textless{}{-}} \FunctionTok{seq}\NormalTok{(}\AttributeTok{from =} \DecValTok{0}\NormalTok{, }\AttributeTok{to =} \DecValTok{1}\NormalTok{, }\AttributeTok{length.out =}\NormalTok{ nsamp)}
\NormalTok{  data\_y }\OtherTok{\textless{}{-}} \FunctionTok{sapply}\NormalTok{(}\AttributeTok{X =}\NormalTok{ data\_x, }\AttributeTok{FUN =} \ControlFlowTok{function}\NormalTok{(xslice) \{}
    \FunctionTok{CI\_dev\_p3avg}\NormalTok{(}\AttributeTok{x =}\NormalTok{ xslice, }\AttributeTok{y =}\NormalTok{ y)}
\NormalTok{  \})}
\NormalTok{  pred\_smooth }\OtherTok{\textless{}{-}} \FunctionTok{suppressWarnings}\NormalTok{(}\AttributeTok{expr =}\NormalTok{ \{}
\NormalTok{    stats}\SpecialCharTok{::}\FunctionTok{smooth.spline}\NormalTok{(}\AttributeTok{x =}\NormalTok{ data\_x, }\AttributeTok{y =}\NormalTok{ data\_y)}
\NormalTok{  \})}

  \ControlFlowTok{function}\NormalTok{(x) \{}
\NormalTok{    vals }\OtherTok{\textless{}{-}}\NormalTok{ stats}\SpecialCharTok{::}\FunctionTok{predict}\NormalTok{(}\AttributeTok{object =}\NormalTok{ pred\_smooth, }\AttributeTok{deriv =}\NormalTok{ deriv, }\AttributeTok{x =}\NormalTok{ x)}\SpecialCharTok{$}\NormalTok{y}
    \ControlFlowTok{if}\NormalTok{ (deriv }\SpecialCharTok{==} \DecValTok{0}\NormalTok{) \{}
\NormalTok{      vals[vals }\SpecialCharTok{\textless{}} \DecValTok{0}\NormalTok{] }\OtherTok{\textless{}{-}} \DecValTok{0}
\NormalTok{      vals[vals }\SpecialCharTok{\textgreater{}} \DecValTok{1}\NormalTok{] }\OtherTok{\textless{}{-}} \DecValTok{1}
\NormalTok{    \}}
\NormalTok{    vals}
\NormalTok{  \}}
\NormalTok{\}}
\end{Highlighting}
\end{Shaded}

\begin{Shaded}
\begin{Highlighting}[]
\NormalTok{CI\_dev\_smooth\_slice\_y }\OtherTok{\textless{}{-}} \ControlFlowTok{function}\NormalTok{(x, }\AttributeTok{nsamp =} \DecValTok{100}\NormalTok{, }\AttributeTok{deriv =} \DecValTok{0}\NormalTok{) \{}
  \FunctionTok{stopifnot}\NormalTok{(}\FunctionTok{length}\NormalTok{(x) }\SpecialCharTok{==} \DecValTok{1}\NormalTok{)}

\NormalTok{  data\_x }\OtherTok{\textless{}{-}} \FunctionTok{seq}\NormalTok{(}\AttributeTok{from =} \DecValTok{0}\NormalTok{, }\AttributeTok{to =} \DecValTok{1}\NormalTok{, }\AttributeTok{length.out =}\NormalTok{ nsamp)}
\NormalTok{  data\_y }\OtherTok{\textless{}{-}} \FunctionTok{sapply}\NormalTok{(}\AttributeTok{X =}\NormalTok{ data\_x, }\AttributeTok{FUN =} \ControlFlowTok{function}\NormalTok{(yslice) \{}
    \FunctionTok{CI\_dev\_p3avg}\NormalTok{(}\AttributeTok{x =}\NormalTok{ x, }\AttributeTok{y =}\NormalTok{ yslice)}
\NormalTok{  \})}
\NormalTok{  pred\_smooth }\OtherTok{\textless{}{-}} \FunctionTok{suppressWarnings}\NormalTok{(}\AttributeTok{expr =}\NormalTok{ \{}
\NormalTok{    stats}\SpecialCharTok{::}\FunctionTok{smooth.spline}\NormalTok{(}\AttributeTok{x =}\NormalTok{ data\_x, }\AttributeTok{y =}\NormalTok{ data\_y)}
\NormalTok{  \})}

  \ControlFlowTok{function}\NormalTok{(y) \{}
\NormalTok{    vals }\OtherTok{\textless{}{-}}\NormalTok{ stats}\SpecialCharTok{::}\FunctionTok{predict}\NormalTok{(}\AttributeTok{object =}\NormalTok{ pred\_smooth, }\AttributeTok{deriv =}\NormalTok{ deriv, }\AttributeTok{x =}\NormalTok{ y)}\SpecialCharTok{$}\NormalTok{y}
    \ControlFlowTok{if}\NormalTok{ (deriv }\SpecialCharTok{==} \DecValTok{0}\NormalTok{) \{}
\NormalTok{      vals[vals }\SpecialCharTok{\textless{}} \DecValTok{0}\NormalTok{] }\OtherTok{\textless{}{-}} \DecValTok{0}
\NormalTok{      vals[vals }\SpecialCharTok{\textgreater{}} \DecValTok{1}\NormalTok{] }\OtherTok{\textless{}{-}} \DecValTok{1}
\NormalTok{    \}}
\NormalTok{    vals}
\NormalTok{  \}}
\NormalTok{\}}
\end{Highlighting}
\end{Shaded}

Ideally, the x-slice of \(y\) at \(x\) returns the same value as the y-slice of \(x\) at \(y\). This is however only approximately true for the smoothed slices, so we return the mean of these two values and the actual value, to obtain a final smoothed \(z\) value that is closest to the ground truth, i.e.,

\[
\begin{aligned}
  \operatorname{Slice}^X(y),\;\operatorname{Slice}^Y(x)\dots&\;\text{horizontal/vertical}\;x\text{/}y\text{-slice,}
  \\[1ex]
  \operatorname{Slice}_{\text{smooth}}^X(y),\;\operatorname{Slice}_{\text{smooth}}^Y(x)\dots&\;\text{smoothed versions, such that}
  \\[1ex]
  \operatorname{Slice}^X(y)\approx\operatorname{Slice}_{\text{smooth}}^X(y)\;\land&\;\operatorname{Slice}^Y(x)\approx\operatorname{Slice}_{\text{smooth}}^Y(x)\;\text{,}
  \\[1em]
  \operatorname{CI}_{\text{smooth}}(x,y)=&\;\Big(\operatorname{CI}(x,y)\;+\;\operatorname{Slice}_{\text{smooth}}^X(y)\;+\;\operatorname{Slice}_{\text{smooth}}^Y(x)\Big)\div3\;\text{.}
\end{aligned}
\]

\begin{Shaded}
\begin{Highlighting}[]
\NormalTok{CI\_dev\_smooth\_p3avg }\OtherTok{\textless{}{-}} \FunctionTok{Vectorize}\NormalTok{(}\ControlFlowTok{function}\NormalTok{(x, y, }\AttributeTok{nsamp =} \DecValTok{100}\NormalTok{) \{}
  \FunctionTok{stopifnot}\NormalTok{(}\FunctionTok{length}\NormalTok{(x) }\SpecialCharTok{==} \DecValTok{1} \SpecialCharTok{\&\&} \FunctionTok{length}\NormalTok{(y) }\SpecialCharTok{==} \DecValTok{1}\NormalTok{)}

\NormalTok{  xsl }\OtherTok{\textless{}{-}} \FunctionTok{CI\_dev\_smooth\_slice\_x}\NormalTok{(}\AttributeTok{y =}\NormalTok{ y, }\AttributeTok{nsamp =}\NormalTok{ nsamp)}
\NormalTok{  ysl }\OtherTok{\textless{}{-}} \FunctionTok{CI\_dev\_smooth\_slice\_y}\NormalTok{(}\AttributeTok{x =}\NormalTok{ x, }\AttributeTok{nsamp =}\NormalTok{ nsamp)}

  \FunctionTok{mean}\NormalTok{(}\FunctionTok{c}\NormalTok{(}\FunctionTok{xsl}\NormalTok{(}\AttributeTok{x =}\NormalTok{ x), }\FunctionTok{ysl}\NormalTok{(}\AttributeTok{y =}\NormalTok{ y), }\FunctionTok{CI\_dev\_p3avg}\NormalTok{(}\AttributeTok{x =}\NormalTok{ x, }\AttributeTok{y =}\NormalTok{ y)))}

\NormalTok{\}, }\AttributeTok{vectorize.args =} \FunctionTok{c}\NormalTok{(}\StringTok{"x"}\NormalTok{, }\StringTok{"y"}\NormalTok{))}
\end{Highlighting}
\end{Shaded}

The functions \texttt{CI\_dev\_smooth\_slice\_x()} and \texttt{CI\_dev\_smooth\_slice\_y()} can also return the derivative (gradient) of the slice, and we will use this when computing the direction and magnitude in each dimension. At every point \(x,y\), we can obtain now two vectors pointing into the steepest increase for either dimension, i.e., \(\overrightarrow{x},\overrightarrow{y}\).

\begin{Shaded}
\begin{Highlighting}[]
\NormalTok{arrow\_dir\_smooth }\OtherTok{\textless{}{-}} \ControlFlowTok{function}\NormalTok{(x, y) \{}
\NormalTok{  xsl }\OtherTok{\textless{}{-}} \FunctionTok{CI\_dev\_smooth\_slice\_x}\NormalTok{(}\AttributeTok{y =}\NormalTok{ y, }\AttributeTok{deriv =} \DecValTok{1}\NormalTok{)}
\NormalTok{  ysl }\OtherTok{\textless{}{-}} \FunctionTok{CI\_dev\_smooth\_slice\_y}\NormalTok{(}\AttributeTok{x =}\NormalTok{ x, }\AttributeTok{deriv =} \DecValTok{1}\NormalTok{)}

  \FunctionTok{c}\NormalTok{(}\AttributeTok{x =}\NormalTok{ x, }\AttributeTok{y =}\NormalTok{ y, }\AttributeTok{x1 =} \FunctionTok{xsl}\NormalTok{(}\AttributeTok{x =}\NormalTok{ x), }\AttributeTok{y1 =} \FunctionTok{ysl}\NormalTok{(}\AttributeTok{y =}\NormalTok{ y))}
\NormalTok{\}}
\end{Highlighting}
\end{Shaded}

\begin{figure}[ht!]

{\centering \includegraphics{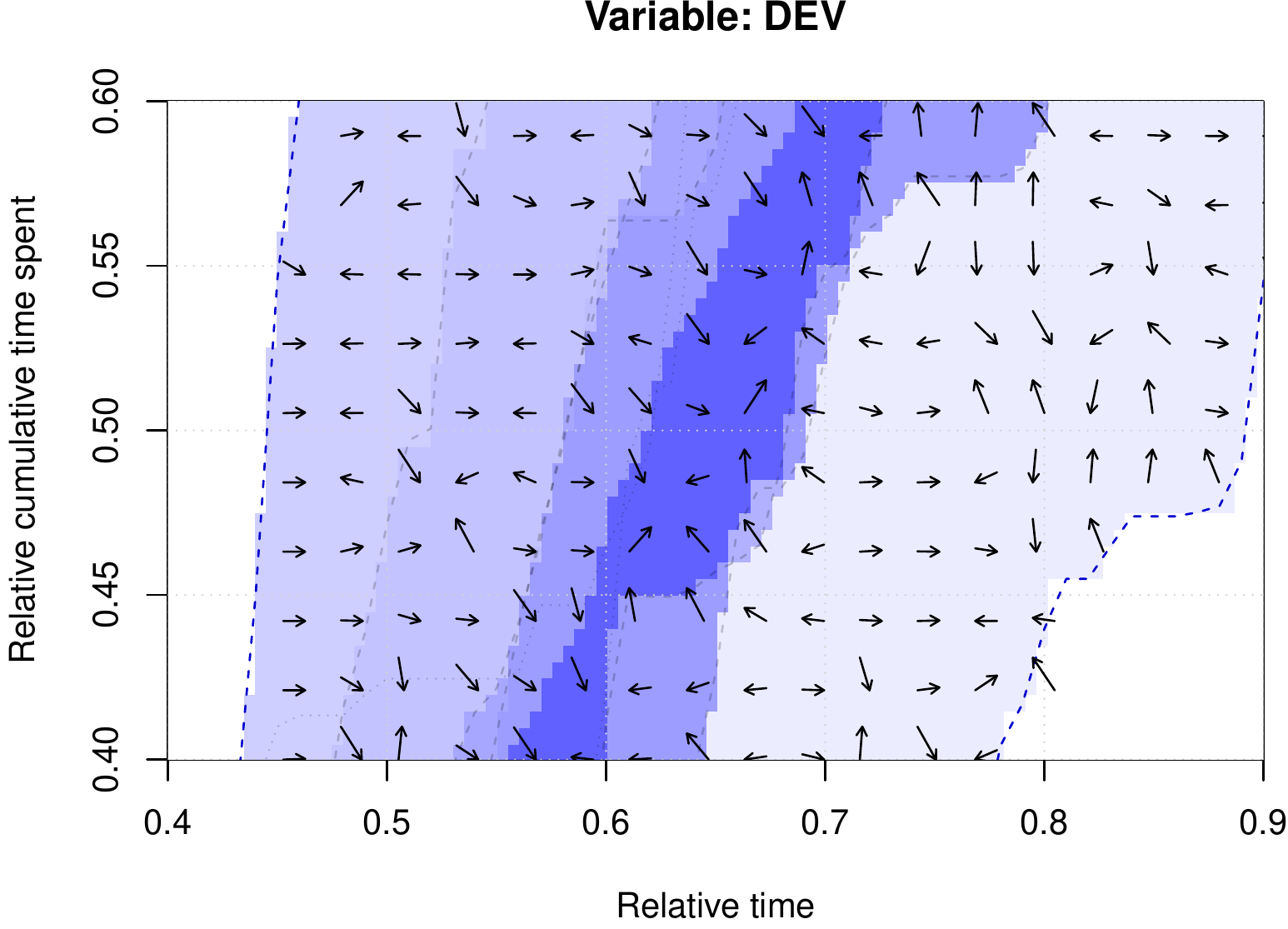} 

}

\caption{Example vector field computed using a smoothed surface. We can clearly observe how the arrows point now properly towards the direction with steepest increase.}\label{fig:example-vectorfield-smooth}
\end{figure}

Finally, we compute a vector field based on a smoothed confidence surface, shown in figure \ref{fig:example-vectorfield-smooth}. Note that all arrows have the same length in this figure, such that it does not depend on the magnitude of the steepest increase they are pointing towards.

We want to demonstrate the previously suggested methods using a section of the confidence surfaces in figures \ref{fig:example-vectorfield} and \ref{fig:example-vectorfield-smooth} and an example variable. We will be using the \texttt{DEV}-variable of project 5, and make a forecast from \(0.65\) to \(0.75\) in relative time. This also demonstrates another advantage of having modeled discrete variables in continuous time, as we are not required to make forecasts in discrete time either.

\hypertarget{average-confidence-in-overlapped-surface}{%
\paragraph{Average confidence in overlapped surface}\label{average-confidence-in-overlapped-surface}}

\begin{Shaded}
\begin{Highlighting}[]
\NormalTok{n\_periods }\OtherTok{\textless{}{-}} \DecValTok{10}
\NormalTok{p5\_signals }\OtherTok{\textless{}{-}}\NormalTok{ all\_signals}\SpecialCharTok{$}\NormalTok{Project5}
\NormalTok{pr5\_dev }\OtherTok{\textless{}{-}}\NormalTok{ p5\_signals}\SpecialCharTok{$}\NormalTok{DEV}\SpecialCharTok{$}\FunctionTok{get0Function}\NormalTok{()}
\NormalTok{forecast\_y }\OtherTok{\textless{}{-}} \FunctionTok{sapply}\NormalTok{(}\AttributeTok{X =} \FunctionTok{seq}\NormalTok{(}\AttributeTok{from =} \DecValTok{0}\NormalTok{, }\AttributeTok{to =} \FloatTok{0.65}\NormalTok{, }\AttributeTok{length.out =} \DecValTok{65}\NormalTok{), }\AttributeTok{FUN =}\NormalTok{ pr5\_dev)}

\CommentTok{\# Fit an EMA to the data and produce 80\% and 95\% confidence intervals:}
\NormalTok{fch }\OtherTok{\textless{}{-}}\NormalTok{ forecast}\SpecialCharTok{::}\FunctionTok{holt}\NormalTok{(}\AttributeTok{y =}\NormalTok{ forecast\_y, }\AttributeTok{h =} \DecValTok{10}\NormalTok{)}
\end{Highlighting}
\end{Shaded}

\begin{verbatim}
## Registered S3 method overwritten by 'quantmod':
##   method            from
##   as.zoo.data.frame zoo
\end{verbatim}

\begin{Shaded}
\begin{Highlighting}[]
\NormalTok{fch\_x }\OtherTok{\textless{}{-}} \FunctionTok{c}\NormalTok{(}\DecValTok{65}\NormalTok{, }\FunctionTok{seq}\NormalTok{(}\AttributeTok{from =} \DecValTok{66}\NormalTok{, }\AttributeTok{to =} \DecValTok{65} \SpecialCharTok{+}\NormalTok{ n\_periods, }\AttributeTok{by =} \DecValTok{1}\NormalTok{))}\SpecialCharTok{/}\DecValTok{100}

\NormalTok{fch\_mean }\OtherTok{\textless{}{-}}\NormalTok{ stats}\SpecialCharTok{::}\FunctionTok{approxfun}\NormalTok{(}\AttributeTok{x =}\NormalTok{ fch\_x, }\AttributeTok{y =} \FunctionTok{c}\NormalTok{(}\FunctionTok{pr5\_dev}\NormalTok{(}\FloatTok{0.65}\NormalTok{), }\FunctionTok{as.vector}\NormalTok{(fch}\SpecialCharTok{$}\NormalTok{mean)))}
\NormalTok{fch\_80\_upper }\OtherTok{\textless{}{-}}\NormalTok{ stats}\SpecialCharTok{::}\FunctionTok{approxfun}\NormalTok{(}\AttributeTok{x =}\NormalTok{ fch\_x, }\AttributeTok{y =} \FunctionTok{c}\NormalTok{(}\FunctionTok{pr5\_dev}\NormalTok{(}\FloatTok{0.65}\NormalTok{), }\FunctionTok{as.vector}\NormalTok{(fch}\SpecialCharTok{$}\NormalTok{upper[,}
  \DecValTok{1}\NormalTok{])), }\AttributeTok{yleft =} \DecValTok{0}\NormalTok{, }\AttributeTok{yright =} \DecValTok{0}\NormalTok{)}
\NormalTok{fch\_80\_lower }\OtherTok{\textless{}{-}}\NormalTok{ stats}\SpecialCharTok{::}\FunctionTok{approxfun}\NormalTok{(}\AttributeTok{x =}\NormalTok{ fch\_x, }\AttributeTok{y =} \FunctionTok{c}\NormalTok{(}\FunctionTok{pr5\_dev}\NormalTok{(}\FloatTok{0.65}\NormalTok{), }\FunctionTok{as.vector}\NormalTok{(fch}\SpecialCharTok{$}\NormalTok{lower[,}
  \DecValTok{1}\NormalTok{])), }\AttributeTok{yleft =} \DecValTok{0}\NormalTok{, }\AttributeTok{yright =} \DecValTok{0}\NormalTok{)}
\NormalTok{fch\_95\_upper }\OtherTok{\textless{}{-}}\NormalTok{ stats}\SpecialCharTok{::}\FunctionTok{approxfun}\NormalTok{(}\AttributeTok{x =}\NormalTok{ fch\_x, }\AttributeTok{y =} \FunctionTok{c}\NormalTok{(}\FunctionTok{pr5\_dev}\NormalTok{(}\FloatTok{0.65}\NormalTok{), }\FunctionTok{as.vector}\NormalTok{(fch}\SpecialCharTok{$}\NormalTok{upper[,}
  \DecValTok{2}\NormalTok{])), }\AttributeTok{yleft =} \DecValTok{0}\NormalTok{, }\AttributeTok{yright =} \DecValTok{0}\NormalTok{)}
\NormalTok{fch\_95\_lower }\OtherTok{\textless{}{-}}\NormalTok{ stats}\SpecialCharTok{::}\FunctionTok{approxfun}\NormalTok{(}\AttributeTok{x =}\NormalTok{ fch\_x, }\AttributeTok{y =} \FunctionTok{c}\NormalTok{(}\FunctionTok{pr5\_dev}\NormalTok{(}\FloatTok{0.65}\NormalTok{), }\FunctionTok{as.vector}\NormalTok{(fch}\SpecialCharTok{$}\NormalTok{lower[,}
  \DecValTok{2}\NormalTok{])), }\AttributeTok{yleft =} \DecValTok{0}\NormalTok{, }\AttributeTok{yright =} \DecValTok{0}\NormalTok{)}
\end{Highlighting}
\end{Shaded}

\begin{figure}[ht!]

{\centering \includegraphics{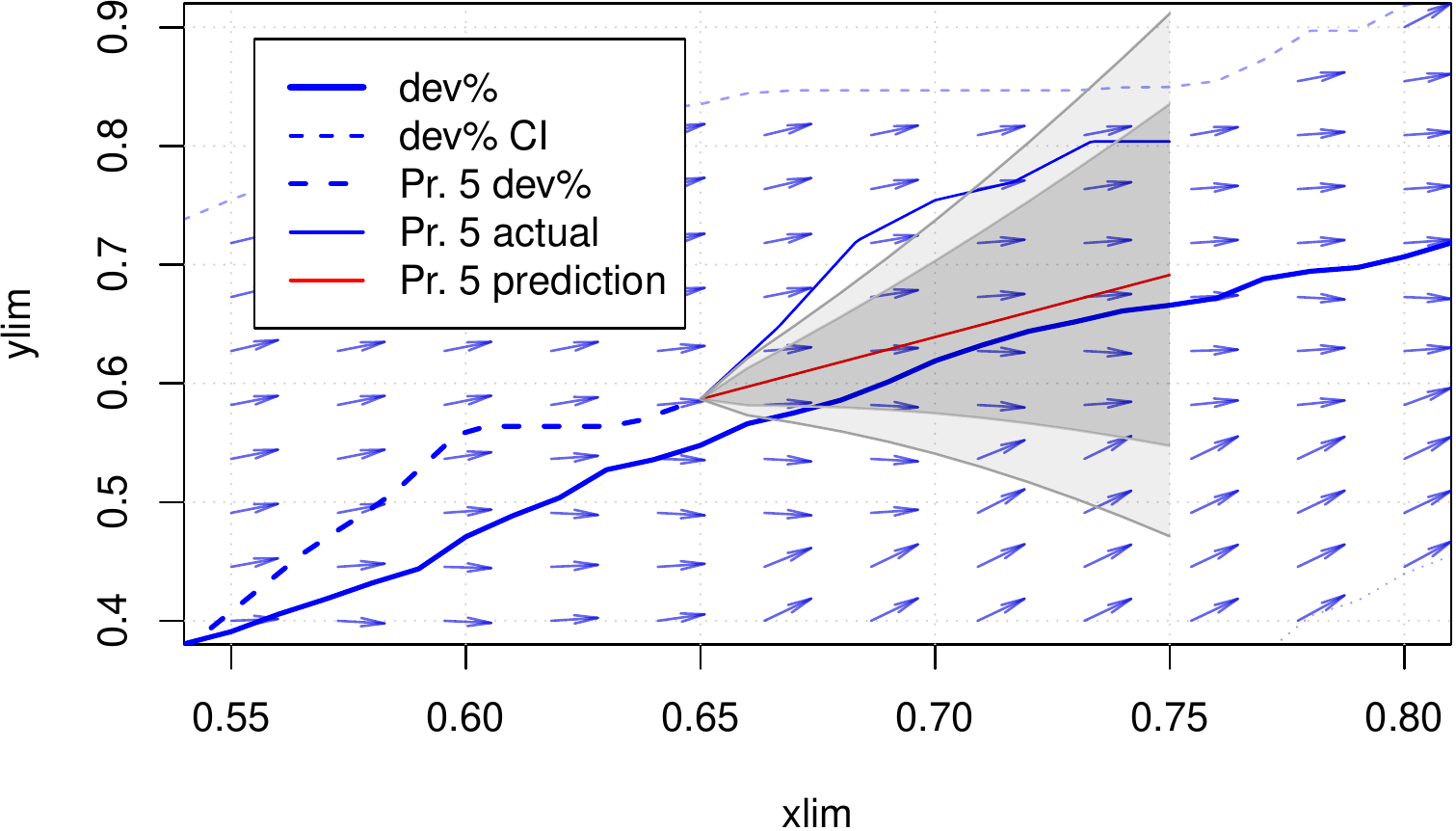} 

}

\caption{Overlap of empirical confidence surface and project 5 (variable: dev\%). The 80\% and 95\% prediction confidence intervals are shown in light- and darkgray.}\label{fig:ed-avg-conf}
\end{figure}

In figure \ref{fig:ed-avg-conf} we use the smoothed confidence surface again. We show a section of the variable \texttt{DEV} and its vector field, pointing towards the steepest increase of confidence at every \(x,y\). Then, the course of project 5's \texttt{DEV} variable is shown until \(0.65\), and forecast until \(0.75\) (red line). The prediction confidence intervals are shown in lightgray (80\%) and darkgray (95\%).

In order to calculate the average confidence in the area that overlaps with the vector field, we need a double integral. The upper 95\% confidence interval is partially outside the vector field. Generally, we must not integrate areas that do not overlap. Therefore, we define a function \(\operatorname{CI}_{\text{overlap}}(x,y)\) to help with that:

\[
\begin{aligned}
  f_{\text{lower}}(x),f_{\text{upper}}(x)\dots&\;\text{functions for the lower/upper prediction confidence intervals,}
  \\[1ex]
  \operatorname{CI}_{\text{lower}}(x),\operatorname{CI}_{\text{upper}}(x)\dots&\;\text{lower/upper confidence boundaries for confidence surface,}
  \\[1ex]
  \operatorname{CI}_{\text{overlap}}(x,y)=&\;\begin{cases}
    0,&\text{if}\;f_{\text{lower}}(x)<\operatorname{CI}_{\text{lower}}(x),
    \\
    0,&\text{if}\;f_{\text{upper}}(x)>\operatorname{CI}_{\text{upper}}(x),
    \\
    \operatorname{CI}(x,y),&\text{otherwise,}
  \end{cases}
  \\[1em]
  \text{average confidence}\;=&\;\int_a^b\bigg[\int_{f_{\text{lower}}(x)}^{f_{\text{upper}}(x)}\,\operatorname{CI}_{\text{overlap}}(x,y)\times\big(f_{\text{upper}}(x)-f_{\text{lower}}(x)\big)^{-1}\,dy\bigg]\times(b-a)^{-1}\,dx\;\text{.}
\end{aligned}
\]

The approximate value for 80\% / 95\% prediction confidence intervals for the concrete example is calculated as:

\begin{Shaded}
\begin{Highlighting}[]
\NormalTok{average\_confidence }\OtherTok{\textless{}{-}} \ControlFlowTok{function}\NormalTok{(f\_low, f\_upp, CI\_low, CI\_upp, CI, lower, upper, }\AttributeTok{maxEval =} \DecValTok{15}\NormalTok{) \{}
\NormalTok{  cubature}\SpecialCharTok{::}\FunctionTok{cubintegrate}\NormalTok{(}\AttributeTok{f =} \FunctionTok{Vectorize}\NormalTok{(}\ControlFlowTok{function}\NormalTok{(x) \{}
\NormalTok{    l }\OtherTok{\textless{}{-}} \FunctionTok{f\_low}\NormalTok{(x)}
\NormalTok{    u }\OtherTok{\textless{}{-}} \FunctionTok{f\_upp}\NormalTok{(x)}
\NormalTok{    cubature}\SpecialCharTok{::}\FunctionTok{cubintegrate}\NormalTok{(}\AttributeTok{f =} \FunctionTok{Vectorize}\NormalTok{(}\ControlFlowTok{function}\NormalTok{(y) \{}
      \ControlFlowTok{if}\NormalTok{ (l }\SpecialCharTok{\textless{}} \FunctionTok{CI\_low}\NormalTok{(x) }\SpecialCharTok{||}\NormalTok{ u }\SpecialCharTok{\textgreater{}} \FunctionTok{CI\_upp}\NormalTok{(x))}
        \DecValTok{0} \ControlFlowTok{else} \FunctionTok{CI}\NormalTok{(}\AttributeTok{x =}\NormalTok{ x, }\AttributeTok{y =}\NormalTok{ y)}
\NormalTok{    \}), }\AttributeTok{lower =}\NormalTok{ l, }\AttributeTok{upper =}\NormalTok{ u, }\AttributeTok{maxEval =}\NormalTok{ maxEval)}\SpecialCharTok{$}\NormalTok{integral}\SpecialCharTok{/}\NormalTok{(u }\SpecialCharTok{{-}}\NormalTok{ l)}
\NormalTok{  \}), }\AttributeTok{lower =}\NormalTok{ lower, }\AttributeTok{upper =}\NormalTok{ upper, }\AttributeTok{maxEval =}\NormalTok{ maxEval)}\SpecialCharTok{$}\NormalTok{integral}\SpecialCharTok{/}\NormalTok{(upper }\SpecialCharTok{{-}}\NormalTok{ lower)}
\NormalTok{\}}
\end{Highlighting}
\end{Shaded}

\begin{Shaded}
\begin{Highlighting}[]
\FunctionTok{invisible}\NormalTok{(}\FunctionTok{loadResultsOrCompute}\NormalTok{(}\AttributeTok{file =} \StringTok{"../results/ed\_avgconf\_80.rds"}\NormalTok{, }\AttributeTok{computeExpr =}\NormalTok{ \{}
  \FunctionTok{average\_confidence}\NormalTok{(}
    \AttributeTok{f\_low =}\NormalTok{ fch\_80\_lower, }\AttributeTok{f\_upp =}\NormalTok{ fch\_80\_upper,}
    \AttributeTok{CI\_low =}\NormalTok{ dev\_ci\_lower\_p3avg, }\AttributeTok{CI\_upp =}\NormalTok{ dev\_ci\_upper\_p3avg,}
    \AttributeTok{CI =}\NormalTok{ CI\_dev\_smooth\_p3avg, }\CommentTok{\# CI\_dev\_p3avg,}
    \AttributeTok{lower =} \FloatTok{0.65}\NormalTok{, }\AttributeTok{upper =} \FloatTok{0.75}\NormalTok{)}
\NormalTok{\}))}
\end{Highlighting}
\end{Shaded}

\begin{Shaded}
\begin{Highlighting}[]
\FunctionTok{invisible}\NormalTok{(}\FunctionTok{loadResultsOrCompute}\NormalTok{(}\AttributeTok{file =} \StringTok{"../results/ed\_avgconf\_95.rds"}\NormalTok{, }\AttributeTok{computeExpr =}\NormalTok{ \{}
  \FunctionTok{average\_confidence}\NormalTok{(}
    \AttributeTok{f\_low =}\NormalTok{ fch\_95\_lower, }\AttributeTok{f\_upp =}\NormalTok{ fch\_95\_upper,}
    \AttributeTok{CI\_low =}\NormalTok{ dev\_ci\_lower\_p3avg, }\AttributeTok{CI\_upp =}\NormalTok{ dev\_ci\_upper\_p3avg,}
    \AttributeTok{CI =}\NormalTok{ CI\_dev\_smooth\_p3avg, }\CommentTok{\# CI\_dev\_p3avg,}
    \AttributeTok{lower =} \FloatTok{0.65}\NormalTok{, }\AttributeTok{upper =} \FloatTok{0.75}\NormalTok{)}
\NormalTok{\}))}
\end{Highlighting}
\end{Shaded}

As expected, the average confidence in the overlap of the 95\% prediction interval is a little lower.

\hypertarget{average-direction-of-steepest-confidence-increase}{%
\paragraph{Average direction of steepest confidence increase}\label{average-direction-of-steepest-confidence-increase}}

Similar to finding the average confidence in the overlapped surface, we may also determine the direction and strength of the steepest increase in confidence. For every point that is part of the overlapped area, we need to sum up the gradient in x- and y-direction. The resulting combined vector then gives the direction of the steepest increase.

This is quite similar to how we computed the average confidence, and requires the following double integral:

\[
\begin{aligned}
  x\text{/}y\;\text{total change}\;=&\;\int_a^b\bigg[\int_{f_{\text{lower}}(x)}^{f_{\text{upper}}(x)}\,\bigg\{\frac{\partial}{\partial\,x}\,\operatorname{CI}_{\text{overlap}}(x,y)\;,\;\frac{\partial}{\partial\,y}\,\operatorname{CI}_{\text{overlap}}(x,y)\bigg\}^\top\,dy\bigg]\,dx\;\text{,}
  \\[1ex]
  &\;\text{or using individual slices and an overlap-helper:}
  \\[1em]
  \operatorname{overlap}(x,y)=&\;\begin{cases}
    0,&\text{if}\;f_{\text{lower}}(x)<\operatorname{CI}_{\text{lower}}(x),
    \\
    0,&\text{if}\;f_{\text{upper}}(x)>\operatorname{CI}_{\text{upper}}(x),
    \\
    1,&\text{otherwise,}
  \end{cases}
  \\[1ex]
  x\;\text{total change}\;=&\;\int_a^b\bigg[\int_{f_{\text{lower}}(x)}^{f_{\text{upper}}(x)}\,\operatorname{overlap}(x,y)\times\frac{\partial}{\partial\,x}\,\operatorname{Slice}_{\text{smooth}}^X(y)\,dy\bigg]\,dx\;\text{, also similarly for}\;y\text{.}
\end{aligned}
\]

\begin{Shaded}
\begin{Highlighting}[]
\NormalTok{total\_change }\OtherTok{\textless{}{-}} \ControlFlowTok{function}\NormalTok{(}\AttributeTok{axis =} \FunctionTok{c}\NormalTok{(}\StringTok{"x"}\NormalTok{, }\StringTok{"y"}\NormalTok{), f\_low, f\_upp, CI\_low, CI\_upp, lower,}
\NormalTok{  upper, }\AttributeTok{maxEval =} \DecValTok{15}\NormalTok{) \{}
\NormalTok{  use\_x }\OtherTok{\textless{}{-}} \FunctionTok{match.arg}\NormalTok{(axis) }\SpecialCharTok{==} \StringTok{"x"}

\NormalTok{  cubature}\SpecialCharTok{::}\FunctionTok{cubintegrate}\NormalTok{(}\AttributeTok{f =} \FunctionTok{Vectorize}\NormalTok{(}\ControlFlowTok{function}\NormalTok{(x) \{}
\NormalTok{    l }\OtherTok{\textless{}{-}} \FunctionTok{f\_low}\NormalTok{(x)}
\NormalTok{    u }\OtherTok{\textless{}{-}} \FunctionTok{f\_upp}\NormalTok{(x)}

\NormalTok{    cubature}\SpecialCharTok{::}\FunctionTok{cubintegrate}\NormalTok{(}\AttributeTok{f =} \ControlFlowTok{function}\NormalTok{(y) \{}
      \FunctionTok{sapply}\NormalTok{(}\AttributeTok{X =}\NormalTok{ y, }\AttributeTok{FUN =} \ControlFlowTok{function}\NormalTok{(y\_) \{}
        \ControlFlowTok{if}\NormalTok{ (l }\SpecialCharTok{\textless{}} \FunctionTok{CI\_low}\NormalTok{(y\_) }\SpecialCharTok{||}\NormalTok{ u }\SpecialCharTok{\textgreater{}} \FunctionTok{CI\_upp}\NormalTok{(y\_)) \{}
          \FunctionTok{return}\NormalTok{(}\DecValTok{0}\NormalTok{)}
\NormalTok{        \}}
\NormalTok{        val }\OtherTok{\textless{}{-}} \FunctionTok{arrow\_dir\_smooth}\NormalTok{(}\AttributeTok{x =}\NormalTok{ x, }\AttributeTok{y =}\NormalTok{ y\_)}
        \ControlFlowTok{if}\NormalTok{ (use\_x)}
\NormalTok{          val[}\StringTok{"x1"}\NormalTok{] }\ControlFlowTok{else}\NormalTok{ val[}\StringTok{"y1"}\NormalTok{]}
\NormalTok{      \})}
\NormalTok{    \}, }\AttributeTok{lower =}\NormalTok{ l, }\AttributeTok{upper =}\NormalTok{ u, }\AttributeTok{maxEval =}\NormalTok{ maxEval)}\SpecialCharTok{$}\NormalTok{integral}\SpecialCharTok{/}\NormalTok{(u }\SpecialCharTok{{-}}\NormalTok{ l)}
\NormalTok{  \}), }\AttributeTok{lower =}\NormalTok{ lower, }\AttributeTok{upper =}\NormalTok{ upper, }\AttributeTok{maxEval =}\NormalTok{ maxEval)}\SpecialCharTok{$}\NormalTok{integral}\SpecialCharTok{/}\NormalTok{(upper }\SpecialCharTok{{-}}\NormalTok{ lower)}
\NormalTok{\}}
\end{Highlighting}
\end{Shaded}

\begin{Shaded}
\begin{Highlighting}[]
\NormalTok{ed\_tc\_x }\OtherTok{\textless{}{-}} \FunctionTok{loadResultsOrCompute}\NormalTok{(}\AttributeTok{file =} \StringTok{"../results/ed\_total\_change\_x.rds"}\NormalTok{, }\AttributeTok{computeExpr =}\NormalTok{ \{}
  \FunctionTok{total\_change}\NormalTok{(}\AttributeTok{f\_low =}\NormalTok{ fch\_95\_lower, }\AttributeTok{f\_upp =}\NormalTok{ fch\_95\_upper, }\AttributeTok{CI\_low =}\NormalTok{ dev\_ci\_lower\_p3avg,}
    \AttributeTok{CI\_upp =}\NormalTok{ dev\_ci\_upper\_p3avg, }\AttributeTok{lower =} \FloatTok{0.65}\NormalTok{, }\AttributeTok{upper =} \FloatTok{0.75}\NormalTok{, }\AttributeTok{axis =} \StringTok{"x"}\NormalTok{)}
\NormalTok{\})}
\NormalTok{ed\_tc\_x}
\end{Highlighting}
\end{Shaded}

\begin{verbatim}
## [1] 2.015401
\end{verbatim}

\begin{Shaded}
\begin{Highlighting}[]
\NormalTok{ed\_tc\_y }\OtherTok{\textless{}{-}} \FunctionTok{loadResultsOrCompute}\NormalTok{(}\AttributeTok{file =} \StringTok{"../results/ed\_total\_change\_y.rds"}\NormalTok{, }\AttributeTok{computeExpr =}\NormalTok{ \{}
  \FunctionTok{total\_change}\NormalTok{(}\AttributeTok{f\_low =}\NormalTok{ fch\_95\_lower, }\AttributeTok{f\_upp =}\NormalTok{ fch\_95\_upper, }\AttributeTok{CI\_low =}\NormalTok{ dev\_ci\_lower\_p3avg,}
    \AttributeTok{CI\_upp =}\NormalTok{ dev\_ci\_upper\_p3avg, }\AttributeTok{lower =} \FloatTok{0.65}\NormalTok{, }\AttributeTok{upper =} \FloatTok{0.75}\NormalTok{, }\AttributeTok{axis =} \StringTok{"y"}\NormalTok{)}
\NormalTok{\})}
\NormalTok{ed\_tc\_y}
\end{Highlighting}
\end{Shaded}

\begin{verbatim}
## [1] -1.5785
\end{verbatim}

\begin{figure}[ht!]

{\centering \includegraphics{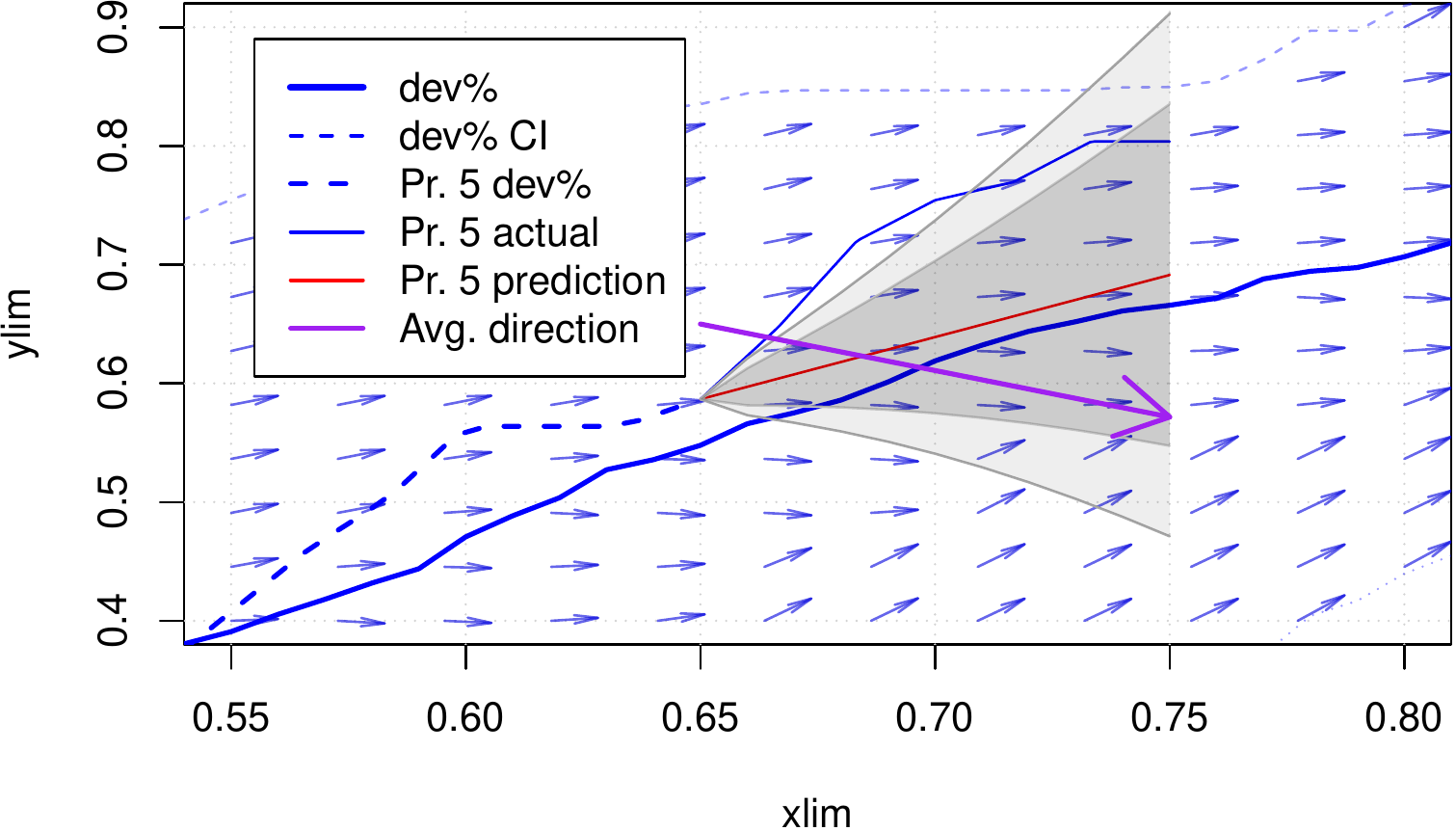} 

}

\caption{The average direction towards the steepest increase in confidence for the dev\% variable in its confidence surface that is overlapped by the 95\% confidence interval of the prediction. Note that the direction may change if computed over the 80\% confidence interval.}\label{fig:ed-avg-dir}
\end{figure}

These results mean that the slope of the average change is \(\approx\)-0.78 in the 95\% confidence interval of the prediction that overlaps with the vector field of the \texttt{DEV}-variable's confidence surface. This is shown in figure \ref{fig:ed-avg-dir}.

Now we can use this information to calculate, e.g., an angle between the predicted variable and the average direction in the field. Depending on the case, one usually wants to either follow into the average direction of steepest increase or diverge from it. In our case of predicting the presence of the Fire Drill anti-pattern, we would want to move (stay) away from it.

\begin{Shaded}
\begin{Highlighting}[]
\NormalTok{matlib}\SpecialCharTok{::}\FunctionTok{angle}\NormalTok{(}\FunctionTok{c}\NormalTok{(}\FloatTok{0.1}\NormalTok{, }\FunctionTok{fch\_mean}\NormalTok{(}\FloatTok{0.75}\NormalTok{) }\SpecialCharTok{{-}} \FunctionTok{fch\_mean}\NormalTok{(}\FloatTok{0.65}\NormalTok{)), }\FunctionTok{c}\NormalTok{(ed\_tc\_x, ed\_tc\_y))}
\end{Highlighting}
\end{Shaded}

\begin{verbatim}
##         [,1]
## [1,] 84.3715
\end{verbatim}

The angle between the projected (predicted) \texttt{DEV}-variable of project 5 and the average direction of steepest increase in confidence is \(\approx84.4^{\circ}\).

\hypertarget{correlation-of-scores}{%
\subsection{Correlation of scores}\label{correlation-of-scores}}

This is a new section in the fifth iteration of this report. Similar to computing the correlation of scores against all models using source code data, we will do this here for each and every pattern based on issue-tracking data. We will be using the function \texttt{score\_variable\_alignment} (cf.~section \ref{sssec:score-mech}) from the technical report using source code data. We do not have any phases in the process models for issue-tracking data. Therefore, no alignment is required. However, we still have to wrap each PM in an alignment, and will do so by defining one closed interval, without warping or amplitude correction.

As a preparation for the pseudo-alignment, we need to present each project's signals as instances of \texttt{Signal}, as well as each pattern's signals then later. This has been done at the beginning of the report, and is stored in \texttt{all\_signals}. It only requires some slight modifications before we can continue.

Also, for the running the projects against type-IV patterns, we'll need to compute their gradients. Recall that the projects' signals were created using a zero-order hold, creating a cumulative quantity that increases step-wise, which results in extreme gradients. In section \ref{ssec:pattern-iv} we have previously examined which kind of smoothing is perhaps most applicable in order to obtain relatively smooth gradients. The recommended solution was to use LOESS smoothing with a span of \texttt{0.35}, and not to go lower than \texttt{0.2}. A lower span preserves more details, and the results below were computed with a value of \texttt{0.35}. We have previously run these computations with the value \texttt{0.2}, with no significant different results.

\begin{Shaded}
\begin{Highlighting}[]
\NormalTok{use\_span }\OtherTok{\textless{}{-}} \FloatTok{0.35}

\NormalTok{time\_warp\_wrapper }\OtherTok{\textless{}{-}} \ControlFlowTok{function}\NormalTok{(pattern, }\AttributeTok{derive =} \ConstantTok{FALSE}\NormalTok{, }\AttributeTok{use\_signals =}\NormalTok{ all\_signals,}
  \AttributeTok{use\_ground\_truth =}\NormalTok{ ground\_truth) \{}
\NormalTok{  signals }\OtherTok{\textless{}{-}} \FunctionTok{list}\NormalTok{()}

  \ControlFlowTok{for}\NormalTok{ (project }\ControlFlowTok{in} \FunctionTok{paste0}\NormalTok{(}\StringTok{"Project"}\NormalTok{, }\FunctionTok{rownames}\NormalTok{(use\_ground\_truth))) \{}
\NormalTok{    temp }\OtherTok{\textless{}{-}} \FunctionTok{list}\NormalTok{()}
    \ControlFlowTok{if}\NormalTok{ (derive) \{}
\NormalTok{      temp[[}\StringTok{"REQ"}\NormalTok{]] }\OtherTok{\textless{}{-}}\NormalTok{ Signal}\SpecialCharTok{$}\FunctionTok{new}\NormalTok{(}\AttributeTok{name =} \FunctionTok{paste0}\NormalTok{(project, }\StringTok{"\_REQ"}\NormalTok{), }\AttributeTok{support =} \FunctionTok{c}\NormalTok{(}\DecValTok{0}\NormalTok{,}
        \DecValTok{1}\NormalTok{), }\AttributeTok{isWp =} \ConstantTok{FALSE}\NormalTok{, }\AttributeTok{func =} \FunctionTok{smooth\_signal\_loess}\NormalTok{(}\AttributeTok{x =} \DecValTok{1}\SpecialCharTok{:}\FunctionTok{nrow}\NormalTok{(use\_signals[[project]]}\SpecialCharTok{$}\NormalTok{data),}
        \AttributeTok{y =} \FunctionTok{cumsum}\NormalTok{(use\_signals[[project]]}\SpecialCharTok{$}\NormalTok{data}\SpecialCharTok{$}\NormalTok{req), }\AttributeTok{span =}\NormalTok{ use\_span, }\AttributeTok{family =} \StringTok{"g"}\NormalTok{))}
\NormalTok{      temp[[}\StringTok{"DEV"}\NormalTok{]] }\OtherTok{\textless{}{-}}\NormalTok{ Signal}\SpecialCharTok{$}\FunctionTok{new}\NormalTok{(}\AttributeTok{name =} \FunctionTok{paste0}\NormalTok{(project, }\StringTok{"\_DEV"}\NormalTok{), }\AttributeTok{support =} \FunctionTok{c}\NormalTok{(}\DecValTok{0}\NormalTok{,}
        \DecValTok{1}\NormalTok{), }\AttributeTok{isWp =} \ConstantTok{FALSE}\NormalTok{, }\AttributeTok{func =} \FunctionTok{smooth\_signal\_loess}\NormalTok{(}\AttributeTok{x =} \DecValTok{1}\SpecialCharTok{:}\FunctionTok{nrow}\NormalTok{(use\_signals[[project]]}\SpecialCharTok{$}\NormalTok{data),}
        \AttributeTok{y =} \FunctionTok{cumsum}\NormalTok{(use\_signals[[project]]}\SpecialCharTok{$}\NormalTok{data}\SpecialCharTok{$}\NormalTok{dev), }\AttributeTok{span =}\NormalTok{ use\_span, }\AttributeTok{family =} \StringTok{"g"}\NormalTok{))}
\NormalTok{      temp[[}\StringTok{"DESC"}\NormalTok{]] }\OtherTok{\textless{}{-}}\NormalTok{ Signal}\SpecialCharTok{$}\FunctionTok{new}\NormalTok{(}\AttributeTok{name =} \FunctionTok{paste0}\NormalTok{(project, }\StringTok{"\_DESC"}\NormalTok{), }\AttributeTok{support =} \FunctionTok{c}\NormalTok{(}\DecValTok{0}\NormalTok{,}
        \DecValTok{1}\NormalTok{), }\AttributeTok{isWp =} \ConstantTok{FALSE}\NormalTok{, }\AttributeTok{func =} \FunctionTok{smooth\_signal\_loess}\NormalTok{(}\AttributeTok{x =} \DecValTok{1}\SpecialCharTok{:}\FunctionTok{nrow}\NormalTok{(use\_signals[[project]]}\SpecialCharTok{$}\NormalTok{data),}
        \AttributeTok{y =} \FunctionTok{cumsum}\NormalTok{(use\_signals[[project]]}\SpecialCharTok{$}\NormalTok{data}\SpecialCharTok{$}\NormalTok{desc), }\AttributeTok{span =}\NormalTok{ use\_span, }\AttributeTok{family =} \StringTok{"g"}\NormalTok{))}
\NormalTok{    \} }\ControlFlowTok{else}\NormalTok{ \{}
\NormalTok{      temp[[}\StringTok{"REQ"}\NormalTok{]] }\OtherTok{\textless{}{-}}\NormalTok{ use\_signals[[project]]}\SpecialCharTok{$}\NormalTok{REQ}\SpecialCharTok{$}\FunctionTok{clone}\NormalTok{()}
\NormalTok{      temp}\SpecialCharTok{$}\NormalTok{REQ}\SpecialCharTok{$}\NormalTok{.\_\_enclos\_env\_\_}\SpecialCharTok{$}\NormalTok{private}\SpecialCharTok{$}\NormalTok{name }\OtherTok{\textless{}{-}} \FunctionTok{paste0}\NormalTok{(project, }\StringTok{"\_REQ"}\NormalTok{)}
\NormalTok{      temp[[}\StringTok{"DEV"}\NormalTok{]] }\OtherTok{\textless{}{-}}\NormalTok{ use\_signals[[project]]}\SpecialCharTok{$}\NormalTok{DEV}\SpecialCharTok{$}\FunctionTok{clone}\NormalTok{()}
\NormalTok{      temp}\SpecialCharTok{$}\NormalTok{DEV}\SpecialCharTok{$}\NormalTok{.\_\_enclos\_env\_\_}\SpecialCharTok{$}\NormalTok{private}\SpecialCharTok{$}\NormalTok{name }\OtherTok{\textless{}{-}} \FunctionTok{paste0}\NormalTok{(project, }\StringTok{"\_DEV"}\NormalTok{)}
\NormalTok{      temp[[}\StringTok{"DESC"}\NormalTok{]] }\OtherTok{\textless{}{-}}\NormalTok{ use\_signals[[project]]}\SpecialCharTok{$}\NormalTok{DESC}\SpecialCharTok{$}\FunctionTok{clone}\NormalTok{()}
\NormalTok{      temp}\SpecialCharTok{$}\NormalTok{DESC}\SpecialCharTok{$}\NormalTok{.\_\_enclos\_env\_\_}\SpecialCharTok{$}\NormalTok{private}\SpecialCharTok{$}\NormalTok{name }\OtherTok{\textless{}{-}} \FunctionTok{paste0}\NormalTok{(project, }\StringTok{"\_DESC"}\NormalTok{)}
\NormalTok{    \}}
\NormalTok{    signals[[project]] }\OtherTok{\textless{}{-}} \FunctionTok{time\_warp\_project}\NormalTok{(}\AttributeTok{pattern =}\NormalTok{ pattern, }\AttributeTok{project =}\NormalTok{ temp,}
      \AttributeTok{thetaB =} \FunctionTok{c}\NormalTok{(}\DecValTok{0}\NormalTok{, }\DecValTok{1}\NormalTok{))}
\NormalTok{  \}}

\NormalTok{  signals}
\NormalTok{\}}
\end{Highlighting}
\end{Shaded}

\hypertarget{pattern-i-iia-and-iii}{%
\subsubsection{Pattern I, II(a), and III}\label{pattern-i-iia-and-iii}}

We'll first have to assemble a list of signals that define each pattern, before we can wrap it in an empty alignment and calculate the various scores.

\begin{Shaded}
\begin{Highlighting}[]
\NormalTok{p1\_it\_signals }\OtherTok{\textless{}{-}} \FunctionTok{loadResultsOrCompute}\NormalTok{(}\AttributeTok{file =} \StringTok{"../data/p1\_it\_signals.rds"}\NormalTok{, }\AttributeTok{computeExpr =}\NormalTok{ \{}
  \FunctionTok{list}\NormalTok{(}\AttributeTok{REQ =}\NormalTok{ Signal}\SpecialCharTok{$}\FunctionTok{new}\NormalTok{(}\AttributeTok{name =} \StringTok{"p1\_it\_REQ"}\NormalTok{, }\AttributeTok{func =}\NormalTok{ req, }\AttributeTok{support =} \FunctionTok{c}\NormalTok{(}\DecValTok{0}\NormalTok{, }\DecValTok{1}\NormalTok{), }\AttributeTok{isWp =} \ConstantTok{TRUE}\NormalTok{),}
    \AttributeTok{DEV =}\NormalTok{ Signal}\SpecialCharTok{$}\FunctionTok{new}\NormalTok{(}\AttributeTok{name =} \StringTok{"p1\_it\_DEV"}\NormalTok{, }\AttributeTok{func =}\NormalTok{ dev, }\AttributeTok{support =} \FunctionTok{c}\NormalTok{(}\DecValTok{0}\NormalTok{, }\DecValTok{1}\NormalTok{), }\AttributeTok{isWp =} \ConstantTok{TRUE}\NormalTok{),}
    \AttributeTok{DESC =}\NormalTok{ Signal}\SpecialCharTok{$}\FunctionTok{new}\NormalTok{(}\AttributeTok{name =} \StringTok{"p1\_it\_DESC"}\NormalTok{, }\AttributeTok{func =}\NormalTok{ desc, }\AttributeTok{support =} \FunctionTok{c}\NormalTok{(}\DecValTok{0}\NormalTok{, }\DecValTok{1}\NormalTok{), }\AttributeTok{isWp =} \ConstantTok{TRUE}\NormalTok{))}
\NormalTok{\})}

\NormalTok{p1\_it\_projects }\OtherTok{\textless{}{-}} \FunctionTok{time\_warp\_wrapper}\NormalTok{(}\AttributeTok{pattern =}\NormalTok{ p1\_it\_signals, }\AttributeTok{derive =} \ConstantTok{FALSE}\NormalTok{)}
\end{Highlighting}
\end{Shaded}

\begin{Shaded}
\begin{Highlighting}[]
\NormalTok{p1\_it\_scores }\OtherTok{\textless{}{-}} \FunctionTok{loadResultsOrCompute}\NormalTok{(}\AttributeTok{file =} \StringTok{"../results/p1\_it\_scores.rds"}\NormalTok{, }\AttributeTok{computeExpr =}\NormalTok{ \{}
  \FunctionTok{as.data.frame}\NormalTok{(}\FunctionTok{compute\_all\_scores\_it}\NormalTok{(}\AttributeTok{alignment =}\NormalTok{ p1\_it\_projects, }\AttributeTok{patternName =} \StringTok{"p1\_it"}\NormalTok{,}
    \AttributeTok{vartypes =} \FunctionTok{names}\NormalTok{(p1\_it\_signals)))}
\NormalTok{\})}
\end{Highlighting}
\end{Shaded}

Table \ref{tab:p1-it-scores} shows the correlations for each variable. This is different to how we did it for source code data. Each variable is shown separately in order not to conceal more extreme correlations that would otherwise be inaccessible due to aggregation.

\begin{table}

\caption{\label{tab:p1-it-scores}Scores for the aligned projects with pattern I (issue-tracking; p=product, m=mean).}
\centering
\begin{tabular}[t]{lrrrrrrrrr}
\toprule
  & pr\_1 & pr\_2 & pr\_3 & pr\_4 & pr\_5 & pr\_6 & pr\_7 & pr\_8 & pr\_9\\
\midrule
REQ\_area & 0.96 & 0.98 & 0.93 & 0.81 & 0.86 & 0.91 & 0.96 & 0.89 & 0.87\\
DEV\_area & 0.70 & 0.81 & 0.83 & 0.87 & 0.75 & 0.91 & 0.76 & 0.74 & 0.64\\
DESC\_area & 0.95 & 0.95 & 0.95 & 0.98 & 0.96 & 0.96 & 0.98 & 0.95 & 0.97\\
REQ\_corr & 1.00 & 1.00 & 0.99 & 0.98 & 0.97 & 0.98 & 0.99 & 0.98 & 0.98\\
DEV\_corr & 0.90 & 0.94 & 0.95 & 0.97 & 0.92 & 0.97 & 0.93 & 0.93 & 0.87\\
\addlinespace
DESC\_corr & 0.50 & 0.87 & 0.94 & 0.96 & 0.84 & 0.87 & 0.98 & 0.50 & 0.83\\
REQ\_jsd & 0.86 & 0.92 & 0.75 & 0.57 & 0.40 & 0.73 & 0.78 & 0.58 & 0.72\\
DEV\_jsd & 0.42 & 0.43 & 0.52 & 0.53 & 0.40 & 0.57 & 0.45 & 0.46 & 0.38\\
DESC\_jsd & 0.52 & 0.58 & 0.67 & 0.78 & 0.59 & 0.58 & 0.80 & 0.52 & 0.63\\
REQ\_kl & 0.00 & 0.00 & 0.02 & 0.06 & 0.16 & 0.02 & 0.01 & 0.06 & 0.02\\
\addlinespace
DEV\_kl & 0.15 & 0.14 & 0.09 & 0.08 & 0.15 & 0.06 & 0.12 & 0.12 & 0.18\\
DESC\_kl & 0.08 & 0.06 & 0.03 & 0.01 & 0.06 & 0.06 & 0.01 & 0.08 & 0.04\\
REQ\_arclen & 0.83 & 0.84 & 0.85 & 0.84 & 0.83 & 0.82 & 0.82 & 0.80 & 0.83\\
DEV\_arclen & 0.98 & 0.95 & 0.94 & 0.93 & 0.95 & 0.91 & 0.92 & 0.89 & 0.94\\
DESC\_arclen & 0.98 & 1.00 & 0.90 & 0.81 & 0.94 & 0.99 & 0.93 & 0.98 & 0.81\\
\addlinespace
REQ\_sd & 0.94 & 0.96 & 0.93 & 0.86 & 0.76 & 0.88 & 0.92 & 0.84 & 0.88\\
DEV\_sd & 0.67 & 0.71 & 0.81 & 0.78 & 0.68 & 0.89 & 0.71 & 0.75 & 0.66\\
DESC\_sd & 0.94 & 0.95 & 0.95 & 0.98 & 0.95 & 0.94 & 0.98 & 0.94 & 0.95\\
REQ\_var & 1.00 & 1.00 & 1.00 & 0.98 & 0.94 & 0.99 & 0.99 & 0.97 & 0.99\\
DEV\_var & 0.89 & 0.92 & 0.96 & 0.95 & 0.90 & 0.99 & 0.92 & 0.94 & 0.89\\
\addlinespace
DESC\_var & 1.00 & 1.00 & 1.00 & 1.00 & 1.00 & 1.00 & 1.00 & 1.00 & 1.00\\
REQ\_mae & 0.96 & 0.98 & 0.93 & 0.81 & 0.86 & 0.91 & 0.96 & 0.89 & 0.87\\
DEV\_mae & 0.70 & 0.81 & 0.83 & 0.87 & 0.75 & 0.91 & 0.76 & 0.74 & 0.64\\
DESC\_mae & 0.95 & 0.95 & 0.95 & 0.98 & 0.96 & 0.96 & 0.98 & 0.95 & 0.97\\
REQ\_rmse & 0.95 & 0.97 & 0.92 & 0.79 & 0.79 & 0.88 & 0.94 & 0.85 & 0.85\\
\addlinespace
DEV\_rmse & 0.63 & 0.73 & 0.79 & 0.80 & 0.67 & 0.88 & 0.69 & 0.69 & 0.57\\
DESC\_rmse & 0.93 & 0.94 & 0.94 & 0.98 & 0.94 & 0.94 & 0.98 & 0.93 & 0.96\\
REQ\_RMS & 0.95 & 1.00 & 0.93 & 0.85 & 0.69 & 0.92 & 0.90 & 0.78 & 0.97\\
DEV\_RMS & 0.62 & 0.60 & 0.76 & 0.65 & 0.60 & 0.95 & 0.63 & 0.68 & 0.64\\
DESC\_RMS & 0.00 & 0.18 & 0.61 & 0.95 & 0.23 & 0.14 & 0.91 & 0.00 & 0.50\\
\addlinespace
REQ\_Kurtosis & 0.89 & 0.98 & 0.99 & 0.93 & 0.63 & 0.84 & 0.73 & 0.79 & 0.73\\
DEV\_Kurtosis & 0.94 & 0.74 & 0.83 & 0.68 & 0.76 & 0.85 & 0.84 & 1.00 & 0.93\\
DESC\_Kurtosis & 0.00 & 0.01 & 0.34 & 0.47 & 0.22 & 0.00 & 0.56 & 0.00 & 0.34\\
REQ\_Peak & 1.00 & 1.00 & 0.95 & 1.00 & 0.99 & 1.00 & 1.00 & 1.00 & 0.95\\
DEV\_Peak & 0.99 & 0.99 & 0.99 & 0.99 & 0.99 & 0.99 & 0.99 & 0.99 & 0.99\\
\addlinespace
DESC\_Peak & 0.00 & 0.15 & 0.84 & 0.50 & 0.71 & 0.08 & 0.96 & 0.00 & 0.51\\
REQ\_ImpulseFactor & 0.95 & 0.98 & 0.96 & 0.73 & 0.83 & 0.87 & 0.97 & 0.85 & 0.86\\
DEV\_ImpulseFactor & 0.40 & 0.54 & 0.54 & 0.63 & 0.46 & 0.70 & 0.45 & 0.44 & 0.35\\
DESC\_ImpulseFactor & 0.00 & 0.42 & 0.57 & 0.32 & 0.11 & 0.52 & 0.75 & 0.00 & 0.33\\
\bottomrule
\end{tabular}
\end{table}

The correlation of just the ground truth with all scores is in table \ref{tab:p1-it-corr}.

\begin{verbatim}
## Warning in stats::cor(ground_truth$consensus, p1_it_scores): the standard
## deviation is zero
\end{verbatim}

\begin{table}

\caption{\label{tab:p1-it-corr}Correlation of the ground truth with all other scores for pattern I (issue-tracking).}
\centering
\begin{tabular}[t]{lrlrlr}
\toprule
Score & Value & Score & Value & Score & Value\\
\midrule
REQ\_area & -0.5187100 & DEV\_arclen & 0.0259164 & DESC\_rmse & 0.5599840\\
DEV\_area & 0.2119023 & DESC\_arclen & -0.8739954 & REQ\_RMS & 0.1460863\\
DESC\_area & 0.4882235 & REQ\_sd & 0.0289615 & DEV\_RMS & 0.1058164\\
REQ\_corr & -0.1641480 & DEV\_sd & 0.2512036 & DESC\_RMS & 0.8152119\\
DEV\_corr & 0.2042772 & DESC\_sd & 0.5686608 & REQ\_Kurtosis & 0.2631592\\
\addlinespace
DESC\_corr & 0.5812074 & REQ\_var & 0.1419239 & DEV\_Kurtosis & -0.3622383\\
REQ\_jsd & -0.1277921 & DEV\_var & 0.2840502 & DESC\_Kurtosis & 0.7484784\\
DEV\_jsd & 0.3730029 & DESC\_var & 0.6121042 & REQ\_Peak & -0.4799340\\
DESC\_jsd & 0.7200972 & REQ\_mae & -0.5187037 & DEV\_Peak & NA\\
REQ\_kl & -0.0827761 & DEV\_mae & 0.2119031 & DESC\_Peak & 0.5326076\\
\addlinespace
DEV\_kl & -0.3500757 & DESC\_mae & 0.4882361 & REQ\_ImpulseFactor & -0.4222044\\
DESC\_kl & -0.7618455 & REQ\_rmse & -0.3374198 & DEV\_ImpulseFactor & 0.2564932\\
REQ\_arclen & 0.4237075 & DEV\_rmse & 0.2219994 & DESC\_ImpulseFactor & 0.3848855\\
\bottomrule
\end{tabular}
\end{table}

\begin{figure}[ht!]
\includegraphics{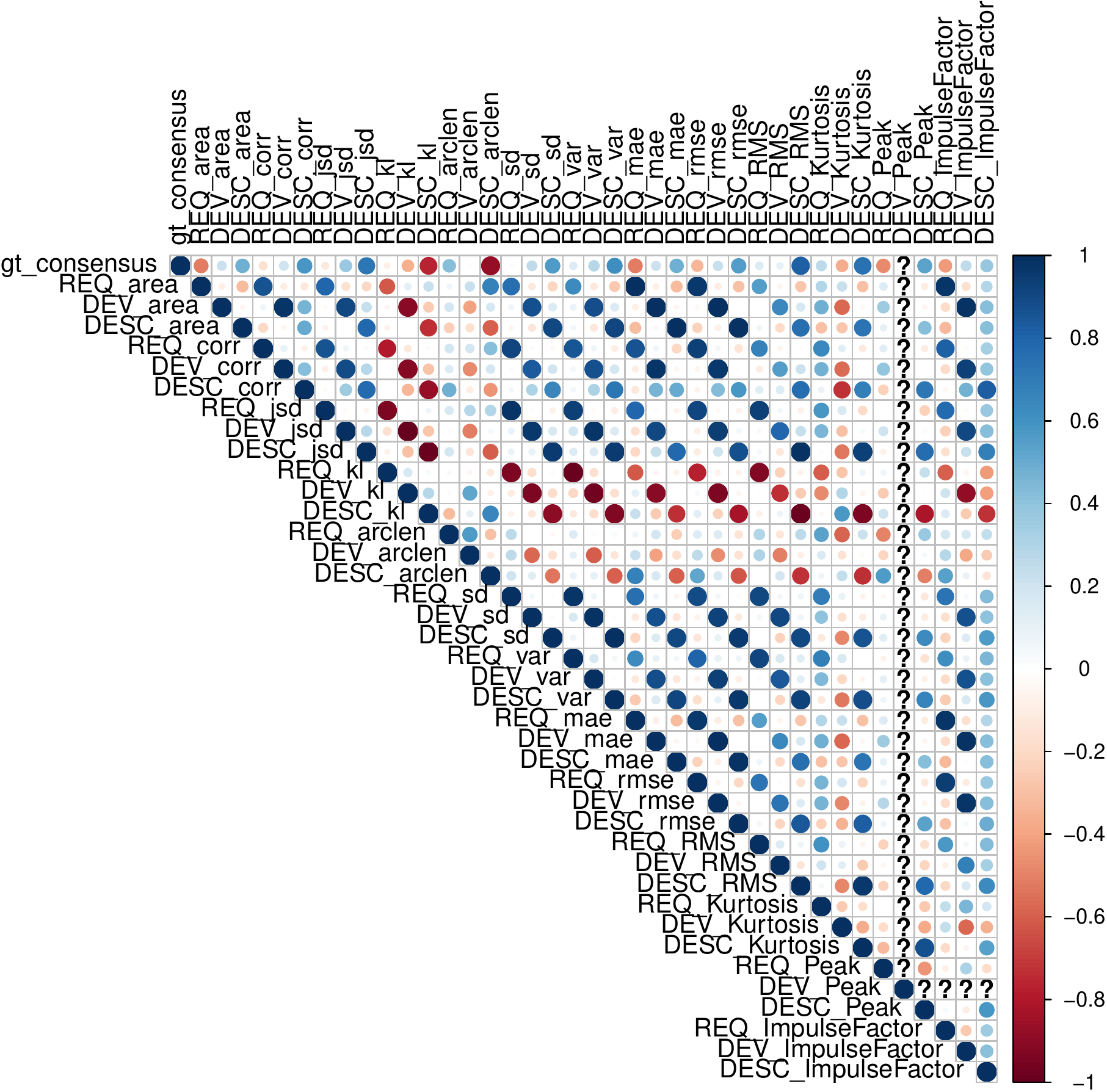} \caption{Correlation matrix for scores using pattern I (issue-tracking).}\label{fig:p1-it-corr-mat}
\end{figure}

The correlation matrix looks as in figure \ref{fig:p1-it-corr-mat}.

\begin{verbatim}
## Warning in stats::cor(temp): the standard deviation is zero
\end{verbatim}

From now on, we'll only compute the correlation scores for each variable without showing intermediate results. We'll then later build a larger matrix with results from all patterns.

\begin{Shaded}
\begin{Highlighting}[]
\NormalTok{p2a\_it\_signals }\OtherTok{\textless{}{-}} \FunctionTok{list}\NormalTok{(}\AttributeTok{REQ =}\NormalTok{ Signal}\SpecialCharTok{$}\FunctionTok{new}\NormalTok{(}\AttributeTok{name =} \StringTok{"p2a\_it\_REQ"}\NormalTok{, }\AttributeTok{func =}\NormalTok{ req\_p2a, }\AttributeTok{support =} \FunctionTok{c}\NormalTok{(}\DecValTok{0}\NormalTok{,}
  \DecValTok{1}\NormalTok{), }\AttributeTok{isWp =} \ConstantTok{TRUE}\NormalTok{), }\AttributeTok{DEV =}\NormalTok{ Signal}\SpecialCharTok{$}\FunctionTok{new}\NormalTok{(}\AttributeTok{name =} \StringTok{"p2a\_it\_DEV"}\NormalTok{, }\AttributeTok{func =}\NormalTok{ dev\_p2a, }\AttributeTok{support =} \FunctionTok{c}\NormalTok{(}\DecValTok{0}\NormalTok{,}
  \DecValTok{1}\NormalTok{), }\AttributeTok{isWp =} \ConstantTok{TRUE}\NormalTok{), }\AttributeTok{DESC =}\NormalTok{ Signal}\SpecialCharTok{$}\FunctionTok{new}\NormalTok{(}\AttributeTok{name =} \StringTok{"p2a\_it\_DESC"}\NormalTok{, }\AttributeTok{func =}\NormalTok{ desc\_p2a, }\AttributeTok{support =} \FunctionTok{c}\NormalTok{(}\DecValTok{0}\NormalTok{,}
  \DecValTok{1}\NormalTok{), }\AttributeTok{isWp =} \ConstantTok{TRUE}\NormalTok{))}

\NormalTok{p2a\_it\_projects }\OtherTok{\textless{}{-}} \FunctionTok{time\_warp\_wrapper}\NormalTok{(}\AttributeTok{pattern =}\NormalTok{ p2a\_it\_signals, }\AttributeTok{derive =} \ConstantTok{FALSE}\NormalTok{)}
\end{Highlighting}
\end{Shaded}

\begin{Shaded}
\begin{Highlighting}[]
\NormalTok{p2a\_it\_scores }\OtherTok{\textless{}{-}} \FunctionTok{loadResultsOrCompute}\NormalTok{(}\AttributeTok{file =} \StringTok{"../results/p2a\_it\_scores.rds"}\NormalTok{, }\AttributeTok{computeExpr =}\NormalTok{ \{}
  \FunctionTok{as.data.frame}\NormalTok{(}\FunctionTok{compute\_all\_scores\_it}\NormalTok{(}\AttributeTok{alignment =}\NormalTok{ p2a\_it\_projects, }\AttributeTok{patternName =} \StringTok{"p2a\_it"}\NormalTok{,}
    \AttributeTok{vartypes =} \FunctionTok{names}\NormalTok{(p2a\_it\_signals)))}
\NormalTok{\})}
\end{Highlighting}
\end{Shaded}

\begin{Shaded}
\begin{Highlighting}[]
\NormalTok{p3\_it\_signals }\OtherTok{\textless{}{-}} \FunctionTok{list}\NormalTok{(}\AttributeTok{REQ =}\NormalTok{ Signal}\SpecialCharTok{$}\FunctionTok{new}\NormalTok{(}\AttributeTok{name =} \StringTok{"p3\_it\_REQ"}\NormalTok{, }\AttributeTok{func =}\NormalTok{ req\_p3, }\AttributeTok{support =} \FunctionTok{c}\NormalTok{(}\DecValTok{0}\NormalTok{,}
  \DecValTok{1}\NormalTok{), }\AttributeTok{isWp =} \ConstantTok{TRUE}\NormalTok{), }\AttributeTok{DEV =}\NormalTok{ Signal}\SpecialCharTok{$}\FunctionTok{new}\NormalTok{(}\AttributeTok{name =} \StringTok{"p3\_it\_DEV"}\NormalTok{, }\AttributeTok{func =}\NormalTok{ dev\_p3, }\AttributeTok{support =} \FunctionTok{c}\NormalTok{(}\DecValTok{0}\NormalTok{,}
  \DecValTok{1}\NormalTok{), }\AttributeTok{isWp =} \ConstantTok{TRUE}\NormalTok{), }\AttributeTok{DESC =}\NormalTok{ Signal}\SpecialCharTok{$}\FunctionTok{new}\NormalTok{(}\AttributeTok{name =} \StringTok{"p3\_it\_DESC"}\NormalTok{, }\AttributeTok{func =}\NormalTok{ desc\_p3, }\AttributeTok{support =} \FunctionTok{c}\NormalTok{(}\DecValTok{0}\NormalTok{,}
  \DecValTok{1}\NormalTok{), }\AttributeTok{isWp =} \ConstantTok{TRUE}\NormalTok{))}

\NormalTok{p3\_it\_projects }\OtherTok{\textless{}{-}} \FunctionTok{time\_warp\_wrapper}\NormalTok{(}\AttributeTok{pattern =}\NormalTok{ p3\_it\_signals, }\AttributeTok{derive =} \ConstantTok{FALSE}\NormalTok{)}
\end{Highlighting}
\end{Shaded}

\begin{Shaded}
\begin{Highlighting}[]
\NormalTok{p3\_it\_scores }\OtherTok{\textless{}{-}} \FunctionTok{loadResultsOrCompute}\NormalTok{(}\AttributeTok{file =} \StringTok{"../results/p3\_it\_scores.rds"}\NormalTok{, }\AttributeTok{computeExpr =}\NormalTok{ \{}
  \FunctionTok{as.data.frame}\NormalTok{(}\FunctionTok{compute\_all\_scores\_it}\NormalTok{(}\AttributeTok{alignment =}\NormalTok{ p3\_it\_projects, }\AttributeTok{patternName =} \StringTok{"p3\_it"}\NormalTok{,}
    \AttributeTok{vartypes =} \FunctionTok{names}\NormalTok{(p3\_it\_signals)))}
\NormalTok{\})}
\end{Highlighting}
\end{Shaded}

Let's attempt to aggregate the correlation scores from the first three patterns, and show a plot.

\begin{figure}[ht!]

{\centering \includegraphics{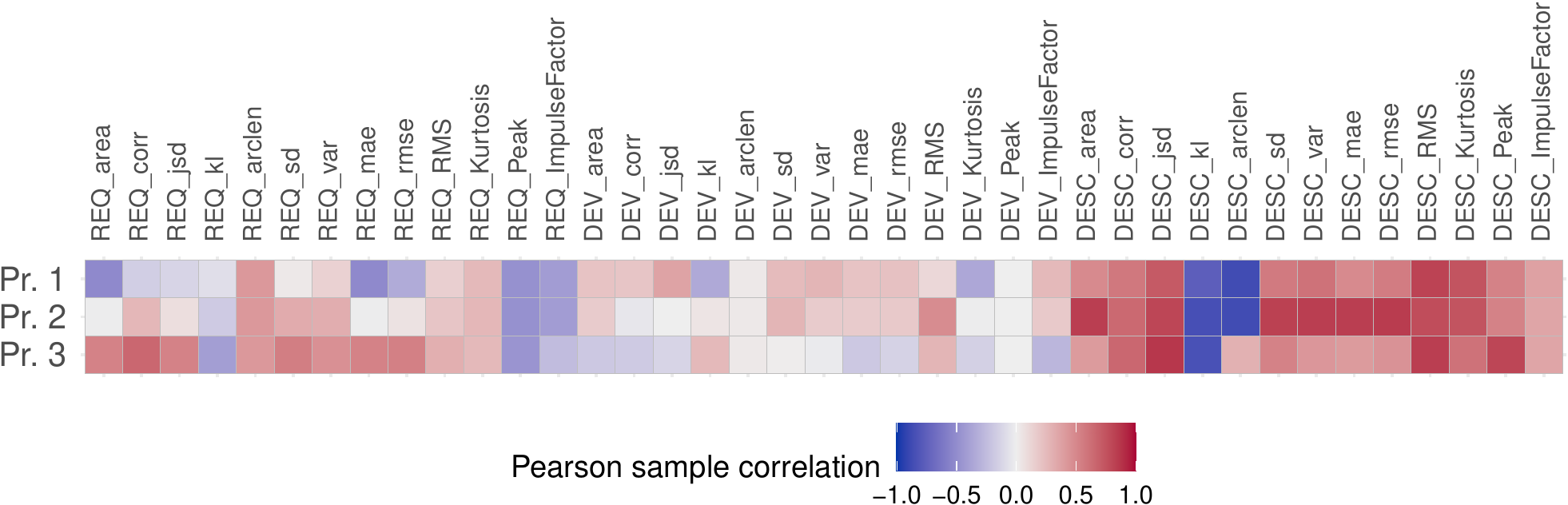} 

}

\caption{Table with correlation scores for the first three types of patterns, calculated against each variable separately.}\label{fig:p1-p2a-p3-corr-scores}
\end{figure}

In \ref{fig:p1-p2a-p3-corr-scores} we show now the correlation scores per variable and score against each of the three patterns I, II(a), and III.

\hypertarget{derivative-of-pattern-i-iia-and-iii}{%
\subsubsection{Derivative of Pattern I, II(a), and III}\label{derivative-of-pattern-i-iia-and-iii}}

In the following, we'll basically repeat calculating the correlation scores of against each process model, but this time we will use \textbf{derivative} models and processes. That is, we will compute the scores based on the \emph{rate of change}.

For every cumulative quantity we should apply some smoothing in order to get usable gradients. This applies to all projects in all comparisons, as well as to process model III, since its variables are weighted averages of such quantities.

\begin{Shaded}
\begin{Highlighting}[]
\NormalTok{get\_smoothed\_signal\_d1 }\OtherTok{\textless{}{-}} \ControlFlowTok{function}\NormalTok{(name, func, use\_span) \{}
\NormalTok{  xvals }\OtherTok{\textless{}{-}} \FunctionTok{seq}\NormalTok{(}\AttributeTok{from =} \DecValTok{0}\NormalTok{, }\AttributeTok{to =} \DecValTok{1}\NormalTok{, }\AttributeTok{length.out =} \DecValTok{10000}\NormalTok{)}
\NormalTok{  func }\OtherTok{\textless{}{-}} \FunctionTok{smooth\_signal\_loess}\NormalTok{(}\AttributeTok{x =}\NormalTok{ xvals, }\AttributeTok{y =} \FunctionTok{func}\NormalTok{(xvals), }\AttributeTok{support =} \FunctionTok{c}\NormalTok{(}\DecValTok{0}\NormalTok{, }\DecValTok{1}\NormalTok{), }\AttributeTok{family =} \StringTok{"g"}\NormalTok{,}
    \AttributeTok{span =}\NormalTok{ use\_span)}

\NormalTok{  temp }\OtherTok{\textless{}{-}}\NormalTok{ Signal}\SpecialCharTok{$}\FunctionTok{new}\NormalTok{(}\AttributeTok{name =}\NormalTok{ name, }\AttributeTok{isWp =} \ConstantTok{TRUE}\NormalTok{, }\AttributeTok{support =} \FunctionTok{c}\NormalTok{(}\DecValTok{0}\NormalTok{, }\DecValTok{1}\NormalTok{), }\AttributeTok{func =}\NormalTok{ func)}
\NormalTok{  Signal}\SpecialCharTok{$}\FunctionTok{new}\NormalTok{(}\AttributeTok{name =}\NormalTok{ name, }\AttributeTok{isWp =} \ConstantTok{TRUE}\NormalTok{, }\AttributeTok{support =} \FunctionTok{c}\NormalTok{(}\DecValTok{0}\NormalTok{, }\DecValTok{1}\NormalTok{), }\AttributeTok{func =}\NormalTok{ temp}\SpecialCharTok{$}\FunctionTok{get1stOrderPd}\NormalTok{())}
\NormalTok{\}}
\end{Highlighting}
\end{Shaded}

The derivatives of the projects I, II(a) and III look like in the following figure \ref{fig:p4of-p1-p2a-p3}:

\begin{figure}[ht!]

{\centering \includegraphics{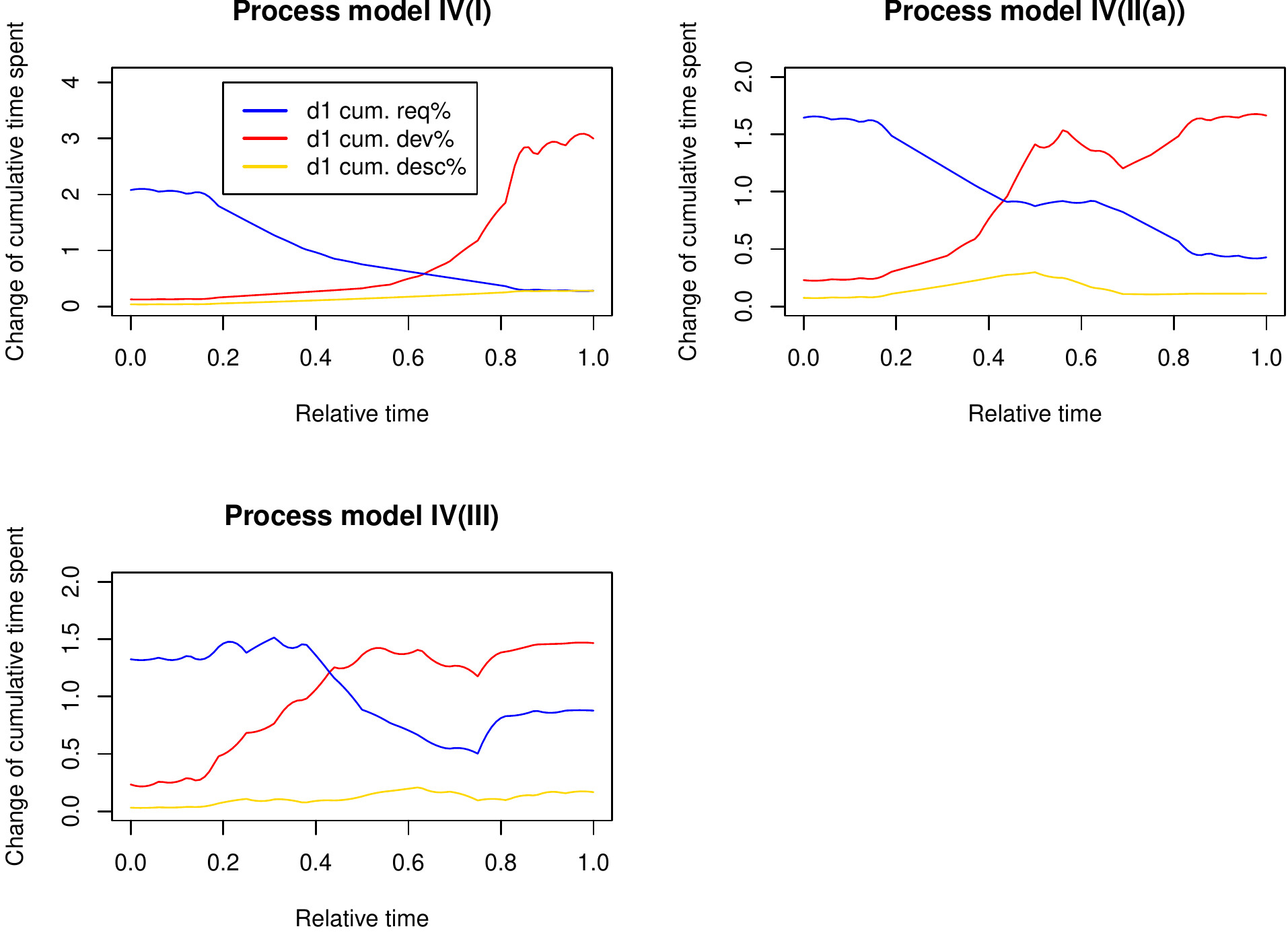} 

}

\caption{Derivative process models type I, II(a) and III, using a smoothing-span of 0.35.}\label{fig:p4of-p1-p2a-p3}
\end{figure}

Let's compute the scores next.

\begin{Shaded}
\begin{Highlighting}[]
\CommentTok{\# We\textquotesingle{}ll use this custom range to compute scores for area and sd/var/mae/rmse}
\NormalTok{use\_yrange }\OtherTok{\textless{}{-}} \FunctionTok{c}\NormalTok{(}\SpecialCharTok{{-}}\DecValTok{1}\NormalTok{, }\DecValTok{10}\NormalTok{)}

\NormalTok{p4p1\_it\_scores }\OtherTok{\textless{}{-}} \FunctionTok{loadResultsOrCompute}\NormalTok{(}\AttributeTok{file =} \StringTok{"../results/p4p1\_it\_scores.rds"}\NormalTok{, }\AttributeTok{computeExpr =}\NormalTok{ \{}
  \FunctionTok{as.data.frame}\NormalTok{(}\FunctionTok{compute\_all\_scores\_it}\NormalTok{(}\AttributeTok{useYRange =}\NormalTok{ use\_yrange, }\AttributeTok{alignment =}\NormalTok{ p4p1\_it\_projects,}
    \AttributeTok{patternName =} \StringTok{"p4p1\_it"}\NormalTok{, }\AttributeTok{vartypes =} \FunctionTok{names}\NormalTok{(p4p1\_it\_signals)))}
\NormalTok{\})}

\NormalTok{p4p2a\_it\_scores }\OtherTok{\textless{}{-}} \FunctionTok{loadResultsOrCompute}\NormalTok{(}\AttributeTok{file =} \StringTok{"../results/p4p2a\_it\_scores.rds"}\NormalTok{,}
  \AttributeTok{computeExpr =}\NormalTok{ \{}
    \FunctionTok{as.data.frame}\NormalTok{(}\FunctionTok{compute\_all\_scores\_it}\NormalTok{(}\AttributeTok{useYRange =}\NormalTok{ use\_yrange, }\AttributeTok{alignment =}\NormalTok{ p4p2a\_it\_projects,}
      \AttributeTok{patternName =} \StringTok{"p4p2a\_it"}\NormalTok{, }\AttributeTok{vartypes =} \FunctionTok{names}\NormalTok{(p4p2a\_it\_signals)))}
\NormalTok{  \})}

\NormalTok{p4p3\_it\_scores }\OtherTok{\textless{}{-}} \FunctionTok{loadResultsOrCompute}\NormalTok{(}\AttributeTok{file =} \StringTok{"../results/p4p3\_it\_scores.rds"}\NormalTok{, }\AttributeTok{computeExpr =}\NormalTok{ \{}
  \FunctionTok{as.data.frame}\NormalTok{(}\FunctionTok{compute\_all\_scores\_it}\NormalTok{(}\AttributeTok{useYRange =}\NormalTok{ use\_yrange, }\AttributeTok{alignment =}\NormalTok{ p4p3\_it\_projects,}
    \AttributeTok{patternName =} \StringTok{"p4p3\_it"}\NormalTok{, }\AttributeTok{vartypes =} \FunctionTok{names}\NormalTok{(p4p3\_it\_signals)))}
\NormalTok{\})}
\end{Highlighting}
\end{Shaded}

\begin{Shaded}
\begin{Highlighting}[]
\NormalTok{p4All\_it\_corr }\OtherTok{\textless{}{-}} \FunctionTok{matrix}\NormalTok{(}\AttributeTok{nrow =} \DecValTok{3}\NormalTok{, }\AttributeTok{ncol =} \FunctionTok{length}\NormalTok{(var\_types) }\SpecialCharTok{*} \FunctionTok{length}\NormalTok{(score\_types))}

\NormalTok{i }\OtherTok{\textless{}{-}} \DecValTok{0}
\NormalTok{c\_names }\OtherTok{\textless{}{-}} \ConstantTok{NULL}
\ControlFlowTok{for}\NormalTok{ (vt }\ControlFlowTok{in}\NormalTok{ var\_types) \{}
  \ControlFlowTok{for}\NormalTok{ (pIdx }\ControlFlowTok{in} \DecValTok{1}\SpecialCharTok{:}\DecValTok{3}\NormalTok{) \{}
\NormalTok{    p\_data }\OtherTok{\textless{}{-}} \ControlFlowTok{if}\NormalTok{ (pIdx }\SpecialCharTok{==} \DecValTok{1}\NormalTok{) \{}
\NormalTok{      p4p1\_it\_scores}
\NormalTok{    \} }\ControlFlowTok{else} \ControlFlowTok{if}\NormalTok{ (pIdx }\SpecialCharTok{==} \DecValTok{2}\NormalTok{) \{}
\NormalTok{      p4p2a\_it\_scores}
\NormalTok{    \} }\ControlFlowTok{else}\NormalTok{ \{}
\NormalTok{      p4p3\_it\_scores}
\NormalTok{    \}}

\NormalTok{    p4All\_it\_corr[pIdx, (i }\SpecialCharTok{*} \FunctionTok{length}\NormalTok{(score\_types) }\SpecialCharTok{+} \DecValTok{1}\NormalTok{)}\SpecialCharTok{:}\NormalTok{(i }\SpecialCharTok{*} \FunctionTok{length}\NormalTok{(score\_types) }\SpecialCharTok{+}
      \FunctionTok{length}\NormalTok{(score\_types))] }\OtherTok{\textless{}{-}}\NormalTok{ stats}\SpecialCharTok{::}\FunctionTok{cor}\NormalTok{(}\AttributeTok{x =}\NormalTok{ ground\_truth}\SpecialCharTok{$}\NormalTok{consensus\_score,}
      \AttributeTok{y =}\NormalTok{ p\_data[, }\FunctionTok{grepl}\NormalTok{(}\AttributeTok{pattern =} \FunctionTok{paste0}\NormalTok{(}\StringTok{"\^{}"}\NormalTok{, vt), }\AttributeTok{x =} \FunctionTok{colnames}\NormalTok{(p\_data))])}
\NormalTok{  \}}

\NormalTok{  i }\OtherTok{\textless{}{-}}\NormalTok{ i }\SpecialCharTok{+} \DecValTok{1}

\NormalTok{  c\_names }\OtherTok{\textless{}{-}} \FunctionTok{c}\NormalTok{(c\_names, }\FunctionTok{paste0}\NormalTok{(vt, }\StringTok{"\_"}\NormalTok{, score\_types))}
\NormalTok{\}}

\FunctionTok{colnames}\NormalTok{(p4All\_it\_corr) }\OtherTok{\textless{}{-}}\NormalTok{ c\_names}
\FunctionTok{rownames}\NormalTok{(p4All\_it\_corr) }\OtherTok{\textless{}{-}} \FunctionTok{paste0}\NormalTok{(}\StringTok{"Pr. "}\NormalTok{, }\DecValTok{1}\SpecialCharTok{:}\DecValTok{3}\NormalTok{)}
\NormalTok{p4All\_it\_corr[}\FunctionTok{is.na}\NormalTok{(p4All\_it\_corr)] }\OtherTok{\textless{}{-}} \FunctionTok{sqrt}\NormalTok{(.Machine}\SpecialCharTok{$}\NormalTok{double.eps)}
\end{Highlighting}
\end{Shaded}

\begin{figure}[ht!]

{\centering \includegraphics{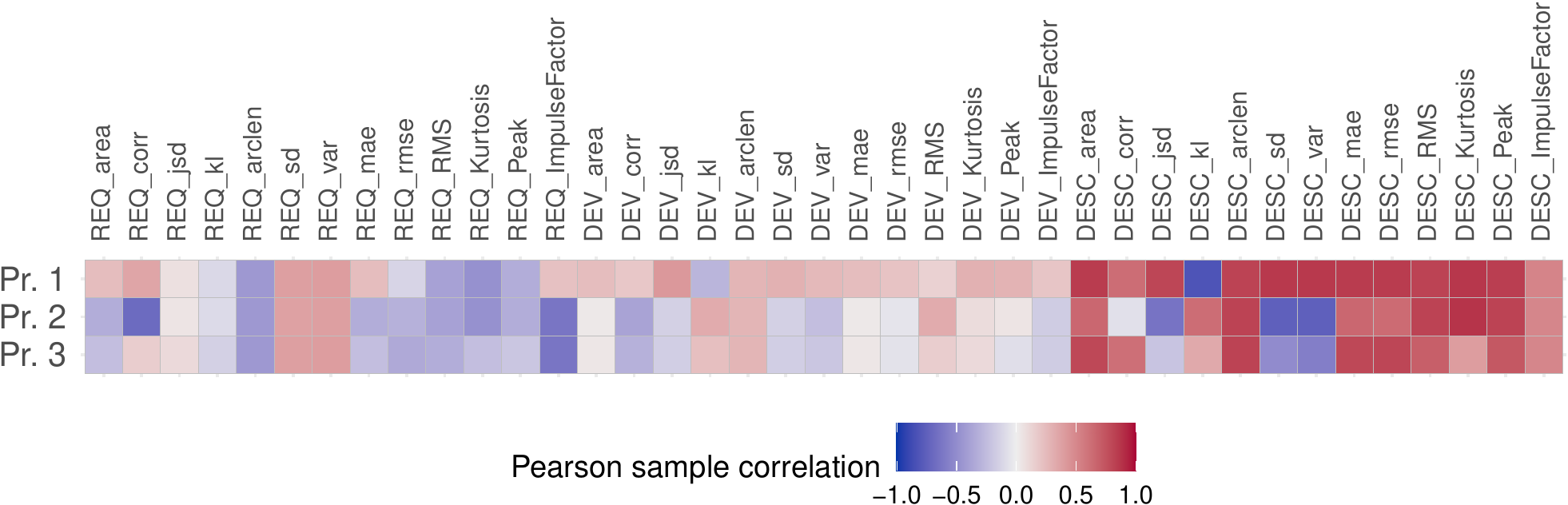} 

}

\caption{Table with correlation scores for the first three types of patterns, calculated against each variable separately.}\label{fig:p4of-p1-p2a-p3-corr-scores}
\end{figure}

In figure \ref{fig:p4of-p1-p2a-p3-corr-scores} we show now the correlation scores per variable and score against each of the three patterns I, II(a), and III.

\hypertarget{scoring-of-projects-first-batch}{%
\subsection{Scoring of projects (first batch)}\label{scoring-of-projects-first-batch}}

In the technical report for detecting the Fire Drill using source code data, we already explored a wide range of possible patterns and scoring mechanisms. All of them are based on comparing/scoring the variables (process model vs.~process). Some of these we will apply here, too, but our focus is first on detection mechanisms that can facilitate the \textbf{confidence intervals}.

\hypertarget{pattern-i-1}{%
\subsubsection{Pattern I}\label{pattern-i-1}}

This is the expert-guess of how the Fire Drill would manifest in issue-tracking data. The pattern was conceived with precise values for the confidence intervals at certain points (\(t_1,t_2\)), and the variables were not of importance. It was only used thus far using a binary decision rule.

\hypertarget{binary-detection-decision-rule}{%
\paragraph{Binary detection decision rule}\label{binary-detection-decision-rule}}

In this section, we will only replicate earlier results by applying the existing rule. It is further formulated using indicators and thresholds as:

\[
\begin{aligned}
  I_1 =&\;\operatorname{req}(t_1) < y_1 \land \operatorname{req}(t_1) > y_2,
  \\[1ex]
  I_2 =&\;\operatorname{dev}(t_1) < y_3 \land \operatorname{dev}(t_2) < y_4,
  \\[1ex]
  I_3 =&\;\operatorname{desc}(1) > y_5,
  \\[1em]
  \operatorname{detect}^{\text{binary}}(I_1,I_2,I_3) =&\;\begin{cases}
    1,&\text{if}\;I_1 \land (I_2 \lor I_3),
    \\
    0,&\text{otherwise}
  \end{cases}\;\text{, using the threshold values}
  \\[1ex]
  \bm{y}=&\;\{0.8,0.4,0.15,0.7,0.15\}^\top\;\text{.}
\end{aligned}
\]

This can be encapsulated in a single function:

\begin{Shaded}
\begin{Highlighting}[]
\NormalTok{p1\_dr }\OtherTok{\textless{}{-}} \ControlFlowTok{function}\NormalTok{(projName, }\AttributeTok{y =} \FunctionTok{c}\NormalTok{(}\FloatTok{0.8}\NormalTok{, }\FloatTok{0.4}\NormalTok{, }\FloatTok{0.15}\NormalTok{, }\FloatTok{0.7}\NormalTok{, }\FloatTok{0.15}\NormalTok{), }\AttributeTok{signals =}\NormalTok{ all\_signals) \{}
\NormalTok{  req }\OtherTok{\textless{}{-}}\NormalTok{ signals[[projName]]}\SpecialCharTok{$}\NormalTok{REQ}\SpecialCharTok{$}\FunctionTok{get0Function}\NormalTok{()}
\NormalTok{  dev }\OtherTok{\textless{}{-}}\NormalTok{ signals[[projName]]}\SpecialCharTok{$}\NormalTok{DEV}\SpecialCharTok{$}\FunctionTok{get0Function}\NormalTok{()}
\NormalTok{  desc }\OtherTok{\textless{}{-}}\NormalTok{ signals[[projName]]}\SpecialCharTok{$}\NormalTok{DESC}\SpecialCharTok{$}\FunctionTok{get0Function}\NormalTok{()}

\NormalTok{  I1 }\OtherTok{\textless{}{-}} \FunctionTok{req}\NormalTok{(t\_1) }\SpecialCharTok{\textless{}}\NormalTok{ y[}\DecValTok{1}\NormalTok{] }\SpecialCharTok{\&\&} \FunctionTok{req}\NormalTok{(t\_1) }\SpecialCharTok{\textgreater{}}\NormalTok{ y[}\DecValTok{2}\NormalTok{]}
\NormalTok{  I2 }\OtherTok{\textless{}{-}} \FunctionTok{dev}\NormalTok{(t\_1) }\SpecialCharTok{\textless{}}\NormalTok{ y[}\DecValTok{3}\NormalTok{] }\SpecialCharTok{\&\&} \FunctionTok{dev}\NormalTok{(t\_2) }\SpecialCharTok{\textless{}}\NormalTok{ y[}\DecValTok{4}\NormalTok{]}
\NormalTok{  I3 }\OtherTok{\textless{}{-}} \FunctionTok{desc}\NormalTok{(}\DecValTok{1}\NormalTok{) }\SpecialCharTok{\textgreater{}}\NormalTok{ y[}\DecValTok{5}\NormalTok{]}

\NormalTok{  I1 }\SpecialCharTok{\&\&}\NormalTok{ (I2 }\SpecialCharTok{||}\NormalTok{ I3)}
\NormalTok{\}}
\end{Highlighting}
\end{Shaded}

\begin{Shaded}
\begin{Highlighting}[]
\NormalTok{temp }\OtherTok{\textless{}{-}} \FunctionTok{sapply}\NormalTok{(}\AttributeTok{X =} \FunctionTok{names}\NormalTok{(all\_signals), }\AttributeTok{FUN =}\NormalTok{ p1\_dr)}
\NormalTok{p1\_detect }\OtherTok{\textless{}{-}} \FunctionTok{data.frame}\NormalTok{(}\AttributeTok{detect =}\NormalTok{ temp, }\AttributeTok{ground\_truth =}\NormalTok{ ground\_truth}\SpecialCharTok{$}\NormalTok{consensus, }\AttributeTok{correct =}\NormalTok{ (temp }\SpecialCharTok{\&}
\NormalTok{  ground\_truth}\SpecialCharTok{$}\NormalTok{consensus }\SpecialCharTok{\textgreater{}=} \DecValTok{5}\NormalTok{) }\SpecialCharTok{|}\NormalTok{ (}\SpecialCharTok{!}\NormalTok{temp }\SpecialCharTok{\&}\NormalTok{ ground\_truth}\SpecialCharTok{$}\NormalTok{consensus }\SpecialCharTok{\textless{}} \DecValTok{5}\NormalTok{))}
\end{Highlighting}
\end{Shaded}

\begin{table}

\caption{\label{tab:p1-bin-detect}Binary detection using a decision rule based on homogeneous confidence intervals of pattern I.}
\centering
\begin{tabular}[t]{llrl}
\toprule
  & detect & ground\_truth & correct\\
\midrule
Project1 & FALSE & 1 & TRUE\\
Project2 & FALSE & 0 & TRUE\\
Project3 & TRUE & 6 & TRUE\\
Project4 & TRUE & 8 & TRUE\\
Project5 & FALSE & 1 & TRUE\\
\addlinespace
Project6 & TRUE & 2 & FALSE\\
Project7 & FALSE & 3 & TRUE\\
Project8 & FALSE & 0 & TRUE\\
Project9 & TRUE & 5 & TRUE\\
\bottomrule
\end{tabular}
\end{table}

In table \ref{tab:p1-bin-detect} we show the results of the binary detection, which is based on the manually defined homogeneous confidence intervals.

\hypertarget{manually-adjusting-the-rule}{%
\subparagraph{Manually adjusting the rule}\label{manually-adjusting-the-rule}}

Through manual inspection, we found out that the decision rule can be re-calibrated in order to achieve 100\% accuracy on the projects. If the thresholds were \(y_1=0.73\) (instead of \(0.8\)) and \(y_4=0.69\) (instead of \$0.7), not only would the detection results align perfectly with the ground truth assessment (eliminating both false positives), but also \(I_1\) no longer be detected in all projects, failing in projects 2, 6 and 7.

\hypertarget{automatically-adjusting-the-rule}{%
\subparagraph{Automatically adjusting the rule}\label{automatically-adjusting-the-rule}}

In order to automatically adjust the thresholds, we could use a global search. This should work well because the problem is tiny and can be quickly evaluated. First, we define a function to return the accuracy given some thresholds. This function can then be used in the optimizer:

\begin{Shaded}
\begin{Highlighting}[]
\NormalTok{eval\_thresholds }\OtherTok{\textless{}{-}}\NormalTok{ (}\ControlFlowTok{function}\NormalTok{() \{}
\NormalTok{  temp }\OtherTok{\textless{}{-}} \FunctionTok{append}\NormalTok{(all\_signals, all\_signals\_2nd\_batch)}
\NormalTok{  gt }\OtherTok{\textless{}{-}} \FunctionTok{c}\NormalTok{(ground\_truth}\SpecialCharTok{$}\NormalTok{consensus }\SpecialCharTok{\textgreater{}=} \DecValTok{5}\NormalTok{, ground\_truth\_2nd\_batch}\SpecialCharTok{$}\NormalTok{consensus }\SpecialCharTok{\textgreater{}=} \DecValTok{5}\NormalTok{)}
  \ControlFlowTok{function}\NormalTok{(x) \{}
\NormalTok{    det }\OtherTok{\textless{}{-}} \FunctionTok{sapply}\NormalTok{(}\AttributeTok{X =} \FunctionTok{names}\NormalTok{(temp), }\AttributeTok{FUN =} \ControlFlowTok{function}\NormalTok{(pName) \{}
      \FunctionTok{p1\_dr}\NormalTok{(}\AttributeTok{projName =}\NormalTok{ pName, }\AttributeTok{signals =}\NormalTok{ temp, }\AttributeTok{y =}\NormalTok{ x)}
\NormalTok{    \})}
    \FunctionTok{sum}\NormalTok{(det }\SpecialCharTok{==}\NormalTok{ gt)}\SpecialCharTok{/}\FunctionTok{length}\NormalTok{(gt)}
\NormalTok{  \}}
\NormalTok{\})()}
\end{Highlighting}
\end{Shaded}

Using the above function and the manually re-calibrated thresholds, we get an accuracy of \texttt{1}:

\begin{Shaded}
\begin{Highlighting}[]
\FunctionTok{eval\_thresholds}\NormalTok{(}\FunctionTok{c}\NormalTok{(}\FloatTok{0.73}\NormalTok{, }\FloatTok{0.4}\NormalTok{, }\FloatTok{0.15}\NormalTok{, }\FloatTok{0.69}\NormalTok{, }\FloatTok{0.15}\NormalTok{))}
\end{Highlighting}
\end{Shaded}

\begin{verbatim}
## [1] 1
\end{verbatim}

Now let's try to find thresholds using a global search:

\begin{Shaded}
\begin{Highlighting}[]
\FunctionTok{set.seed}\NormalTok{(}\DecValTok{1337}\NormalTok{)}

\NormalTok{x0 }\OtherTok{\textless{}{-}} \FunctionTok{c}\NormalTok{(}\FloatTok{0.8}\NormalTok{, }\FloatTok{0.4}\NormalTok{, }\FloatTok{0.15}\NormalTok{, }\FloatTok{0.7}\NormalTok{, }\FloatTok{0.15}\NormalTok{)}

\FunctionTok{nloptr}\NormalTok{(}
  \AttributeTok{x0 =}\NormalTok{ x0,}
  \AttributeTok{eval\_f =} \ControlFlowTok{function}\NormalTok{(x) \{}
    \DecValTok{1} \SpecialCharTok{{-}} \FunctionTok{eval\_thresholds}\NormalTok{(}\AttributeTok{x=}\NormalTok{x)}
\NormalTok{  \},}
  \AttributeTok{lb =} \FunctionTok{c}\NormalTok{(}\FloatTok{0.7}\NormalTok{, }\FloatTok{0.3}\NormalTok{, }\FloatTok{0.05}\NormalTok{, }\FloatTok{0.6}\NormalTok{, }\FloatTok{0.05}\NormalTok{),}
  \AttributeTok{ub =} \FunctionTok{c}\NormalTok{(}\FloatTok{0.9}\NormalTok{, }\FloatTok{0.5}\NormalTok{, }\FloatTok{0.25}\NormalTok{, }\FloatTok{0.8}\NormalTok{, }\FloatTok{0.25}\NormalTok{),}
  \AttributeTok{opts =} \FunctionTok{list}\NormalTok{(}
    \AttributeTok{stopval =} \DecValTok{0}\NormalTok{, }\CommentTok{\# any solution with accuracy=1 is optimal!}
    \AttributeTok{algorithm =} \StringTok{"NLOPT\_GN\_DIRECT\_L\_RAND"}\NormalTok{),}
\NormalTok{)}
\end{Highlighting}
\end{Shaded}

\begin{verbatim}
## 
## Call:
## nloptr(x0 = x0, eval_f = function(x) {    1 - eval_thresholds(x = x)
## }, lb = c(0.7, 0.3, 0.05, 0.6, 0.05), ub = c(0.9, 0.5, 0.25, 
##     0.8, 0.25), opts = list(stopval = 0, algorithm = "NLOPT_GN_DIRECT_L_RAND"))
## 
## 
## 
## Minimization using NLopt version 2.7.1 
## 
## NLopt solver status: 3 ( NLOPT_FTOL_REACHED: Optimization stopped because 
## ftol_rel or ftol_abs (above) was reached. )
## 
## Number of Iterations....: 5 
## Termination conditions:  stopval: 0 
## Number of inequality constraints:  0 
## Number of equality constraints:    0 
## Optimal value of objective function:  0 
## Optimal value of controls: 0.8 0.4 0.08333333 0.6333333 0.15
\end{verbatim}

Re-running this global search produces different optimal solutions, such as \(\bm{y}=\{0.8,0.4,0.15,0.666,0.15\}^\top\), \(\bm{y}=\{0.8,0.4,0.0833,0.633,0.15\}^\top\), \(\bm{y}=\{0.8,0.333,0.083,0.633,0.15\}^\top\), or \(\bm{y}=\{0.722,0.1,0.5,0.1,0.177\}^\top\).

As an experiment, we could also force to find an optimal solution that is maximally far away from our manually found solution (\(\bm{x}_0\)), by adding some regularization:

\begin{Shaded}
\begin{Highlighting}[]
\NormalTok{res }\OtherTok{\textless{}{-}} \FunctionTok{loadResultsOrCompute}\NormalTok{(}\AttributeTok{file =} \StringTok{"../results/p1\_optim\_y.rds"}\NormalTok{, }\AttributeTok{computeExpr =}\NormalTok{ \{}
  \FunctionTok{set.seed}\NormalTok{(}\DecValTok{1339}\NormalTok{)}

\NormalTok{  req\_max }\OtherTok{\textless{}{-}} \FunctionTok{sqrt}\NormalTok{(}\FunctionTok{sum}\NormalTok{((}\FunctionTok{c}\NormalTok{(}\DecValTok{0}\NormalTok{, }\DecValTok{1}\NormalTok{, }\DecValTok{1}\NormalTok{, }\DecValTok{0}\NormalTok{, }\DecValTok{1}\NormalTok{) }\SpecialCharTok{{-}}\NormalTok{ x0)}\SpecialCharTok{\^{}}\DecValTok{2}\NormalTok{)) }\CommentTok{\# \textasciitilde{}1.7141}
  
\NormalTok{  res }\OtherTok{\textless{}{-}}\NormalTok{ nloptr}\SpecialCharTok{::}\FunctionTok{nloptr}\NormalTok{(}
    \AttributeTok{x0 =}\NormalTok{ x0,}
    \AttributeTok{eval\_f =} \ControlFlowTok{function}\NormalTok{(x) \{}
\NormalTok{      acc }\OtherTok{\textless{}{-}} \FunctionTok{eval\_thresholds}\NormalTok{(}\AttributeTok{x=}\NormalTok{x)}
\NormalTok{      reg }\OtherTok{\textless{}{-}} \FunctionTok{sqrt}\NormalTok{(}\FunctionTok{sum}\NormalTok{((x }\SpecialCharTok{{-}}\NormalTok{ x0)}\SpecialCharTok{\^{}}\DecValTok{2}\NormalTok{)) }\SpecialCharTok{/}\NormalTok{ req\_max }\CommentTok{\# the max is \textasciitilde{}1.7131 using the bounds 0,1}
      \CommentTok{\# We want to maximize accuracy + reg (increase distance)!}
      \SpecialCharTok{{-}}\FunctionTok{log}\NormalTok{(reg }\SpecialCharTok{+} \DecValTok{2}\SpecialCharTok{*}\NormalTok{acc) }\CommentTok{\# accuracy is twice as important!}
\NormalTok{    \},}
    \AttributeTok{lb =} \FunctionTok{rep}\NormalTok{(}\DecValTok{0}\NormalTok{, }\DecValTok{5}\NormalTok{),}
    \AttributeTok{ub =} \FunctionTok{rep}\NormalTok{(}\DecValTok{1}\NormalTok{, }\DecValTok{5}\NormalTok{),}
    \AttributeTok{opts =} \FunctionTok{list}\NormalTok{(}
      \AttributeTok{maxeval =} \FloatTok{1e4}\NormalTok{,}
      \AttributeTok{stopval =} \SpecialCharTok{{-}}\FunctionTok{log}\NormalTok{(}\DecValTok{3} \SpecialCharTok{{-}} \FunctionTok{sqrt}\NormalTok{(.Machine}\SpecialCharTok{$}\NormalTok{double.eps)), }\CommentTok{\# \textasciitilde{}{-}1.0986 but we\textquotesingle{}re not gonna get there}
      \AttributeTok{algorithm =} \StringTok{"NLOPT\_GN\_DIRECT\_L\_RAND"}\NormalTok{),}
\NormalTok{  )}
\NormalTok{\})}

\NormalTok{res}
\end{Highlighting}
\end{Shaded}

\begin{verbatim}
## 
## Call:
## nloptr::nloptr(x0 = x0, eval_f = function(x) {
##     acc <- eval_thresholds(x = x)    reg <- sqrt(sum((x - x0)^2))/req_max
##     -log(reg + 2 * acc)
## }, lb = rep(0, 5), ub = rep(1, 5), opts = list(maxeval = 10000, 
##     stopval = -log(3 - sqrt(.Machine$double.eps)), algorithm = "NLOPT_GN_DIRECT_L_RAND"))
## 
## 
## 
## Minimization using NLopt version 2.4.2 
## 
## NLopt solver status: 5 ( NLOPT_MAXEVAL_REACHED: Optimization stopped because 
## maxeval (above) was reached. )
## 
## Number of Iterations....: 10000 
## Termination conditions:  maxeval: 10000  stopval: -1.09861228370106 
## Number of inequality constraints:  0 
## Number of equality constraints:    0 
## Current value of objective function:  -0.991004514452949 
## Current value of controls: 1 8.379398e-16 1 2.982802e-16 0.179224
\end{verbatim}

So I have re-run this a couple of times, and it appears that there is a small solution space that is maximally far away, where the first four values for \(\bm{y}\) are \(\approx\{1,0,1,0\}\), and \(y_5\) gets pushed towards \(0.18\), but ultimately needs to be lower than that (\(\approx 0.179\)).

\hypertarget{calculating-scoring-rules}{%
\subparagraph{Calculating scoring rules}\label{calculating-scoring-rules}}

The decision rule is not a probabilistic classifier, and hence does not output class probabilities. By definition, indicators are of binary nature.
Scoring rules are applicable to tasks in which predictions must assign probabilities to a set of mutually exclusive discrete outcomes or classes.
The set of possible outcomes can be either binary or categorical in nature, and the probabilities assigned to this set of outcomes must sum to one (where each individual probability is in the range of 0 to 1).
We can apply scoring rules if we inverse the relation classifier and ground truth. In our case, the classifier outputs binary predictions (probabilities \(0,1\)), and the ground truth (the consensus score) is of probabilistic nature.

In the following, we will compute two scoring rules:

\begin{itemize}
\tightlist
\item
  Brier score (Brier et al. 1950): This is equivalent to the MSE: \(\frac{1}{n}\sum_{i=1}^{N}\,\left(f_i-o_i\right)^2\), where \(f_i\) is the probability of the ground truth, and \(o_i\) is the classification result (either \(0\) or \(1\)). A Brier score of \(0\) would hence be a perfect result, and \(1\) the worst possible.
\item
  Logarithmic score: \(\frac{1}{N}\sum_{i=1}^{N}\,-\log{\left(1-\left|f_i-o_i\right|\right)}\). Like the Brier score, \(0\) would be perfect, but there is no limit, so \(+\infty\) is possible.
\end{itemize}

\begin{Shaded}
\begin{Highlighting}[]
\NormalTok{temp }\OtherTok{\textless{}{-}}\NormalTok{ p1\_detect[, ]}
\NormalTok{temp}\SpecialCharTok{$}\NormalTok{ground\_truth }\OtherTok{\textless{}{-}}\NormalTok{ temp}\SpecialCharTok{$}\NormalTok{ground\_truth}\SpecialCharTok{/}\DecValTok{10}
\NormalTok{temp}\SpecialCharTok{$}\NormalTok{zero\_r }\OtherTok{\textless{}{-}} \FloatTok{0.5}
\NormalTok{temp}\SpecialCharTok{$}\NormalTok{detect\_prob }\OtherTok{\textless{}{-}} \FunctionTok{sapply}\NormalTok{(temp}\SpecialCharTok{$}\NormalTok{detect, }\ControlFlowTok{function}\NormalTok{(d) }\ControlFlowTok{if}\NormalTok{ (d) }\DecValTok{1} \ControlFlowTok{else} \DecValTok{0}\NormalTok{)}

\StringTok{\textasciigrave{}}\AttributeTok{names\textless{}{-}}\StringTok{\textasciigrave{}}\NormalTok{(}\FunctionTok{c}\NormalTok{(}\FunctionTok{mean}\NormalTok{(scoring}\SpecialCharTok{::}\FunctionTok{brierscore}\NormalTok{(}\AttributeTok{object =}\NormalTok{ detect\_prob }\SpecialCharTok{\textasciitilde{}}\NormalTok{ ground\_truth, }\AttributeTok{data =}\NormalTok{ temp)),}
  \FunctionTok{mean}\NormalTok{(scoring}\SpecialCharTok{::}\FunctionTok{logscore}\NormalTok{(}\AttributeTok{object =}\NormalTok{ detect\_prob }\SpecialCharTok{\textasciitilde{}}\NormalTok{ ground\_truth, }\AttributeTok{data =}\NormalTok{ temp)), stats}\SpecialCharTok{::}\FunctionTok{cor}\NormalTok{(temp}\SpecialCharTok{$}\NormalTok{ground\_truth,}
\NormalTok{    temp}\SpecialCharTok{$}\NormalTok{detect\_prob)), }\FunctionTok{c}\NormalTok{(}\StringTok{"Brier score"}\NormalTok{, }\StringTok{"Log score"}\NormalTok{, }\StringTok{"Correlation"}\NormalTok{))}
\end{Highlighting}
\end{Shaded}

\begin{verbatim}
## Brier score   Log score Correlation 
##   0.1333333   0.4004389   0.7864978
\end{verbatim}

These results indicate that the binary decision rule does not have perfect scores.
If we relax the probabilities it outputs to \(0.75\) resp. \(0.25\), the Brier- and Log-score deteriorate even more (\(\approx0.156\) and \(\approx0.445\)), so doing that is not in favor of the decision rule.

Also, we need to compute a baseline, using a so-called Zero-Rule, which represents the simplest possible regression model we can suggest \emph{without} any prior knowledge. Since we try to predict probabilities, the ZeroR regression is just to predict a probability of \(0.5\).

\begin{Shaded}
\begin{Highlighting}[]
\NormalTok{Metrics}\SpecialCharTok{::}\FunctionTok{mse}\NormalTok{(temp}\SpecialCharTok{$}\NormalTok{ground\_truth, temp}\SpecialCharTok{$}\NormalTok{zero\_r)}
\end{Highlighting}
\end{Shaded}

\begin{verbatim}
## [1] 0.1166667
\end{verbatim}

For the first batch of projects, ZeroR outperforms our decision rule in terms of MSE.

\hypertarget{automatic-calibration-of-the-decision-rule}{%
\subparagraph{Automatic calibration of the decision rule}\label{automatic-calibration-of-the-decision-rule}}

We have previously done the automatic calibration of the continuous PMs based on source code, see section \ref{sec:auto-calib}. Note that this section (``calibration'') is conceptually different from the previous sections, where we \emph{adjusted} the decision rule.
The decision rule can be calibrated as well, but since it is discrete, things will work a bit differently:

\begin{itemize}
\tightlist
\item
  \texttt{REQ} is expected to be within a threshold at \(t_1\). However, the threshold is only a logical delimiter for the decision rule. For the automatic calibration, only the \emph{expectation} of \texttt{REQ(t\_1)} is important (which is \(0.7\)). Since the threshold is supposed to be a confidence interval, it would have an effect on the calibration. However, the threshold is homogeneous, so there is no such effect.
\item
  \texttt{DEV} is to be measured twice, once at \(t_1\) and once at \(t_2\). Again, we have a threshold/confidence interval, but only the expectation at both points (\(0.075,0.4\)) is of importance.
\item
  \texttt{DESC} is measured at \(t=1\). The variables as it was drawn in this report is slightly misleading, as we are not drawing the variable, but its upper confidence boundary. De-scoping is \emph{always} negative (bad), so that our expectation for this activity is \textbf{\(0\)}. In other words, any de-scoping is bad, and \(I_3\) would be positive \emph{above} a value of \(0.15\).
\end{itemize}

\begin{figure}
\centering
\includegraphics{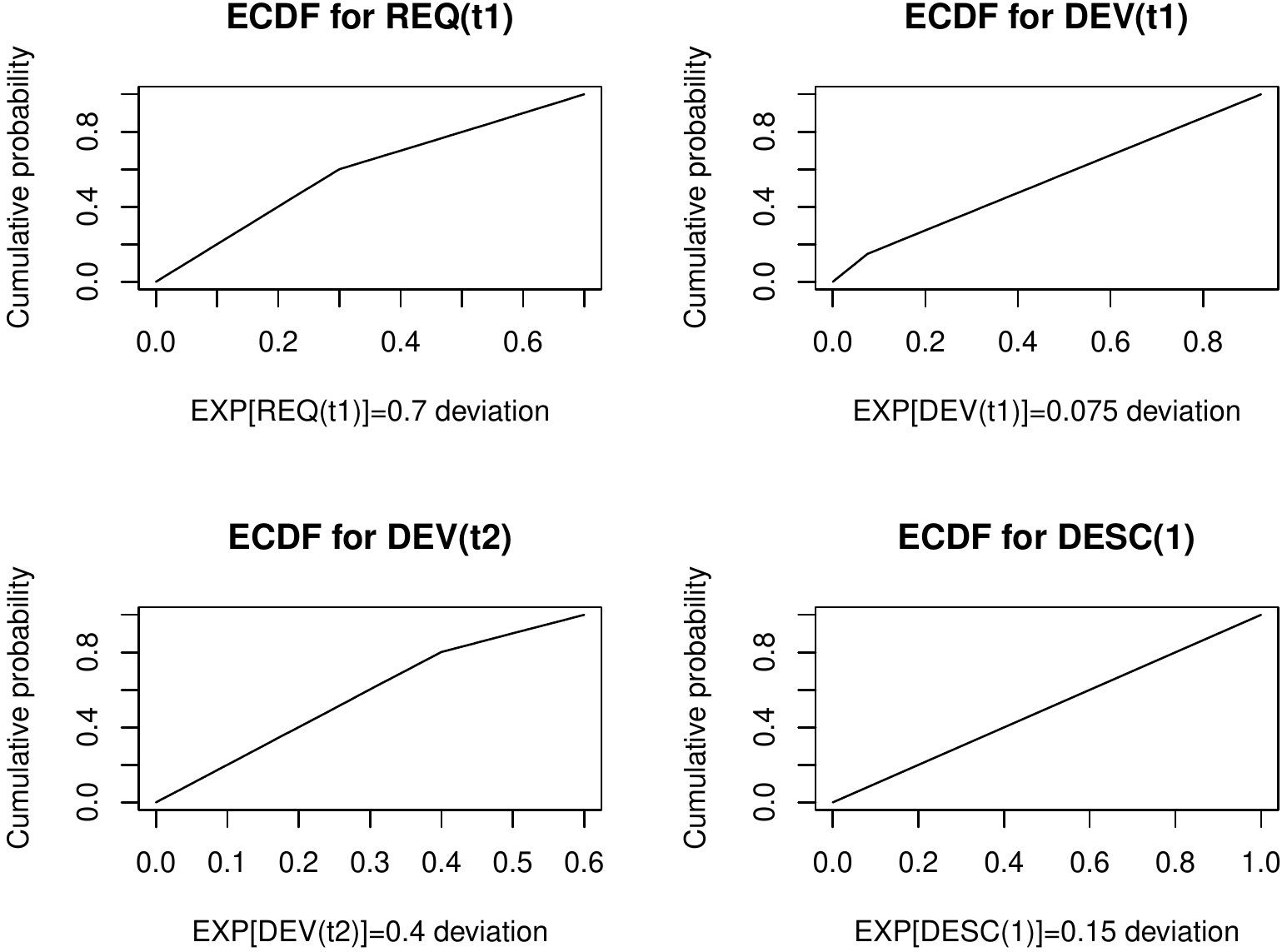}
\caption{\label{fig:ac-calib-dec-rule}The automatically calibrated scores (losses) of the decision rule.}
\end{figure}

The objective for the calibration is always the same, it is the absolute difference of any measurement from the expectation.
The issue-tracking models are normalized. Therefore any measurement must be in the range \texttt{{[}0,1{]}}. We will simulate uniformly distributed points in that range.
Let's do the calibration for the four points \texttt{REQ(t\_1)}, \texttt{DEV(t\_1)}, \texttt{DEV(t\_2)}, and \texttt{DESC(1)} (the empirical distributions are shown in figure \ref{fig:ac-calib-dec-rule}):

\begin{Shaded}
\begin{Highlighting}[]
\FunctionTok{set.seed}\NormalTok{(}\DecValTok{1}\NormalTok{)}

\NormalTok{temp }\OtherTok{\textless{}{-}} \FunctionTok{req}\NormalTok{(t\_1)}
\NormalTok{ac\_itp1\_reqt1\_ecdf }\OtherTok{\textless{}{-}}\NormalTok{ stats}\SpecialCharTok{::}\FunctionTok{ecdf}\NormalTok{(}\FunctionTok{abs}\NormalTok{(temp }\SpecialCharTok{{-}}\NormalTok{ stats}\SpecialCharTok{::}\FunctionTok{runif}\NormalTok{(}\AttributeTok{n =} \FloatTok{2e+05}\NormalTok{)))}
\NormalTok{temp }\OtherTok{\textless{}{-}} \FunctionTok{dev}\NormalTok{(t\_1)}
\NormalTok{ac\_itp1\_devt1\_ecdf }\OtherTok{\textless{}{-}}\NormalTok{ stats}\SpecialCharTok{::}\FunctionTok{ecdf}\NormalTok{(}\FunctionTok{abs}\NormalTok{(temp }\SpecialCharTok{{-}}\NormalTok{ stats}\SpecialCharTok{::}\FunctionTok{runif}\NormalTok{(}\AttributeTok{n =} \FloatTok{2e+05}\NormalTok{)))}
\NormalTok{temp }\OtherTok{\textless{}{-}} \FunctionTok{dev}\NormalTok{(t\_2)}
\NormalTok{ac\_itp1\_devt2\_ecdf }\OtherTok{\textless{}{-}}\NormalTok{ stats}\SpecialCharTok{::}\FunctionTok{ecdf}\NormalTok{(}\FunctionTok{abs}\NormalTok{(temp }\SpecialCharTok{{-}}\NormalTok{ stats}\SpecialCharTok{::}\FunctionTok{runif}\NormalTok{(}\AttributeTok{n =} \FloatTok{2e+05}\NormalTok{)))}
\CommentTok{\# All have same likelihood since the expectation for DESC(1) = 0 and the}
\CommentTok{\# distance is the L1{-}norm:}
\NormalTok{ac\_itp1\_desc1\_ecdf }\OtherTok{\textless{}{-}} \ControlFlowTok{function}\NormalTok{(x) x}
\end{Highlighting}
\end{Shaded}

Since we are calibrating around single points, the ECDFs in figure \ref{fig:ac-calib-dec-rule} are all piece-wise linear, with a single break at the expectation. If the expectation were \(0.5\), there would be no break. An expectation \(\neq 0.5\) however indicates a different uniform probability before and after it.
In this simple case we could have constructed these ECDFs manually instead of random sampling around each expectation.

Next we will define a function that uses the estimated ECDFs and gives us a probabilistic score from the automatically calibrated decision rule.
The binary decision rule is a logical concatenation of the three indicators: \(I_1\land(I_2\lor I_3)\). We have now transformed the rule into four independent objectives, and each objective describes the deviation from an expected value using a uniformly scaled and uniformly distributed score in the range \([0,1]\).
As I see it, we have the justification for two distinct setups for the weights. Either, all weights are the same, with each deviation from the corresponding deviation equally important.
In the other case, we see that the first indicator must be true for the binary decision rule to indicate detection (one half of the entire condition). We could therefore reason to give twice the weight to the objective for \texttt{REQ(t\_1)}. In the following, we will try both.

\begin{Shaded}
\begin{Highlighting}[]
\NormalTok{expectations }\OtherTok{\textless{}{-}} \FunctionTok{c}\NormalTok{(}
  \StringTok{"reqt1"} \OtherTok{=} \FunctionTok{unname}\NormalTok{(}\FunctionTok{req}\NormalTok{(t\_1)), }\StringTok{"devt1"} \OtherTok{=} \FunctionTok{unname}\NormalTok{(}\FunctionTok{dev}\NormalTok{(t\_1)),}
  \CommentTok{\# Note the expectation for DESC(1) = 0!}
  \StringTok{"devt2"} \OtherTok{=} \FunctionTok{unname}\NormalTok{(}\FunctionTok{dev}\NormalTok{(t\_2)), }\StringTok{"desc1"} \OtherTok{=} \DecValTok{0}\NormalTok{)}

\NormalTok{ac\_p1\_dr }\OtherTok{\textless{}{-}} \ControlFlowTok{function}\NormalTok{(}
\NormalTok{  projName, }\AttributeTok{signals =}\NormalTok{ all\_signals,}
  \AttributeTok{w\_reqt1 =} \DecValTok{1}\NormalTok{, }\AttributeTok{w\_devt1 =} \DecValTok{1}\NormalTok{, }\AttributeTok{w\_devt2 =} \DecValTok{1}\NormalTok{, }\AttributeTok{w\_desc1 =} \DecValTok{1}\NormalTok{)}
\NormalTok{\{}
\NormalTok{  p\_req }\OtherTok{\textless{}{-}}\NormalTok{ signals[[projName]]}\SpecialCharTok{$}\NormalTok{REQ}\SpecialCharTok{$}\FunctionTok{get0Function}\NormalTok{()}
\NormalTok{  p\_dev }\OtherTok{\textless{}{-}}\NormalTok{ signals[[projName]]}\SpecialCharTok{$}\NormalTok{DEV}\SpecialCharTok{$}\FunctionTok{get0Function}\NormalTok{()}
\NormalTok{  p\_desc }\OtherTok{\textless{}{-}}\NormalTok{ signals[[projName]]}\SpecialCharTok{$}\NormalTok{DESC}\SpecialCharTok{$}\FunctionTok{get0Function}\NormalTok{()}
  
\NormalTok{  w\_l1 }\OtherTok{\textless{}{-}}\NormalTok{ w\_reqt1 }\SpecialCharTok{+}\NormalTok{ w\_devt1 }\SpecialCharTok{+}\NormalTok{ w\_devt2 }\SpecialCharTok{+}\NormalTok{ w\_desc1}
  
\NormalTok{  (}
    \CommentTok{\# Note: The objective must be given the absolute difference between}
    \CommentTok{\# the expectation and the project\textquotesingle{}s value!}
    \CommentTok{\# Note the objective return the distance, not the similarity.}
    \CommentTok{\# Therefore, we have to subtract the value from 1!}
\NormalTok{    w\_reqt1 }\SpecialCharTok{*}\NormalTok{ (}\DecValTok{1} \SpecialCharTok{{-}} \FunctionTok{ac\_itp1\_reqt1\_ecdf}\NormalTok{(}\FunctionTok{abs}\NormalTok{(expectations[}\StringTok{"reqt1"}\NormalTok{] }\SpecialCharTok{{-}} \FunctionTok{p\_req}\NormalTok{(t\_1)))) }\SpecialCharTok{+}
\NormalTok{    w\_devt1 }\SpecialCharTok{*}\NormalTok{ (}\DecValTok{1} \SpecialCharTok{{-}} \FunctionTok{ac\_itp1\_devt1\_ecdf}\NormalTok{(}\FunctionTok{abs}\NormalTok{(expectations[}\StringTok{"devt1"}\NormalTok{] }\SpecialCharTok{{-}} \FunctionTok{p\_dev}\NormalTok{(t\_1)))) }\SpecialCharTok{+}
\NormalTok{    w\_devt2 }\SpecialCharTok{*}\NormalTok{ (}\DecValTok{1} \SpecialCharTok{{-}} \FunctionTok{ac\_itp1\_devt2\_ecdf}\NormalTok{(}\FunctionTok{abs}\NormalTok{(expectations[}\StringTok{"devt2"}\NormalTok{] }\SpecialCharTok{{-}} \FunctionTok{p\_dev}\NormalTok{(t\_2)))) }\SpecialCharTok{+}
\NormalTok{    w\_desc1 }\SpecialCharTok{*}\NormalTok{ (}\DecValTok{1} \SpecialCharTok{{-}} \FunctionTok{ac\_itp1\_desc1\_ecdf}\NormalTok{(}\FunctionTok{abs}\NormalTok{(expectations[}\StringTok{"desc1"}\NormalTok{] }\SpecialCharTok{{-}} \FunctionTok{p\_desc}\NormalTok{(}\DecValTok{1}\NormalTok{))))}
\NormalTok{  ) }\SpecialCharTok{/}\NormalTok{ w\_l1}
\NormalTok{\}}
\end{Highlighting}
\end{Shaded}

Now we can compute the scores from the automatically calibrated decision rule, using either set of weights.
Also, we shore the calibrated score for each of the objectives in table \ref{tab:ac-it-dr-scores}.

\begin{Shaded}
\begin{Highlighting}[]
\NormalTok{temp }\OtherTok{\textless{}{-}} \FunctionTok{append}\NormalTok{(all\_signals, all\_signals\_2nd\_batch)}

\NormalTok{ac\_it\_dr\_scores }\OtherTok{\textless{}{-}} \FunctionTok{as.data.frame}\NormalTok{(}\StringTok{\textasciigrave{}}\AttributeTok{colnames\textless{}{-}}\StringTok{\textasciigrave{}}\NormalTok{(}\StringTok{\textasciigrave{}}\AttributeTok{rownames\textless{}{-}}\StringTok{\textasciigrave{}}\NormalTok{(}\FunctionTok{matrix}\NormalTok{(}\AttributeTok{ncol =} \DecValTok{7}\NormalTok{, }\AttributeTok{nrow =} \FunctionTok{length}\NormalTok{(temp),}
  \AttributeTok{byrow =} \ConstantTok{FALSE}\NormalTok{, }\AttributeTok{data =} \FunctionTok{c}\NormalTok{(}\FunctionTok{sapply}\NormalTok{(}\AttributeTok{X =} \FunctionTok{names}\NormalTok{(temp), }\ControlFlowTok{function}\NormalTok{(pName) }\FunctionTok{ac\_p1\_dr}\NormalTok{(}\AttributeTok{projName =}\NormalTok{ pName,}
    \AttributeTok{signals =}\NormalTok{ temp)), }\FunctionTok{sapply}\NormalTok{(}\AttributeTok{X =} \FunctionTok{names}\NormalTok{(temp), }\ControlFlowTok{function}\NormalTok{(pName) }\FunctionTok{ac\_p1\_dr}\NormalTok{(}\AttributeTok{projName =}\NormalTok{ pName,}
    \AttributeTok{signals =}\NormalTok{ temp, }\AttributeTok{w\_reqt1 =} \DecValTok{2}\NormalTok{)), }\FunctionTok{sapply}\NormalTok{(}\AttributeTok{X =} \FunctionTok{names}\NormalTok{(temp), }\ControlFlowTok{function}\NormalTok{(pName) \{}
\NormalTok{    p\_req }\OtherTok{\textless{}{-}}\NormalTok{ temp[[pName]]}\SpecialCharTok{$}\NormalTok{REQ}\SpecialCharTok{$}\FunctionTok{get0Function}\NormalTok{()}
    \DecValTok{1} \SpecialCharTok{{-}} \FunctionTok{ac\_itp1\_reqt1\_ecdf}\NormalTok{(}\FunctionTok{abs}\NormalTok{(expectations[}\StringTok{"reqt1"}\NormalTok{] }\SpecialCharTok{{-}} \FunctionTok{p\_req}\NormalTok{(t\_1)))}
\NormalTok{  \}), }\FunctionTok{sapply}\NormalTok{(}\AttributeTok{X =} \FunctionTok{names}\NormalTok{(temp), }\ControlFlowTok{function}\NormalTok{(pName) \{}
\NormalTok{    p\_dev }\OtherTok{\textless{}{-}}\NormalTok{ temp[[pName]]}\SpecialCharTok{$}\NormalTok{DEV}\SpecialCharTok{$}\FunctionTok{get0Function}\NormalTok{()}
    \DecValTok{1} \SpecialCharTok{{-}} \FunctionTok{ac\_itp1\_devt1\_ecdf}\NormalTok{(}\FunctionTok{abs}\NormalTok{(expectations[}\StringTok{"devt1"}\NormalTok{] }\SpecialCharTok{{-}} \FunctionTok{p\_dev}\NormalTok{(t\_1)))}
\NormalTok{  \}), }\FunctionTok{sapply}\NormalTok{(}\AttributeTok{X =} \FunctionTok{names}\NormalTok{(temp), }\ControlFlowTok{function}\NormalTok{(pName) \{}
\NormalTok{    p\_dev }\OtherTok{\textless{}{-}}\NormalTok{ temp[[pName]]}\SpecialCharTok{$}\NormalTok{DEV}\SpecialCharTok{$}\FunctionTok{get0Function}\NormalTok{()}
    \DecValTok{1} \SpecialCharTok{{-}} \FunctionTok{ac\_itp1\_devt2\_ecdf}\NormalTok{(}\FunctionTok{abs}\NormalTok{(expectations[}\StringTok{"devt2"}\NormalTok{] }\SpecialCharTok{{-}} \FunctionTok{p\_dev}\NormalTok{(t\_2)))}
\NormalTok{  \}), }\FunctionTok{sapply}\NormalTok{(}\AttributeTok{X =} \FunctionTok{names}\NormalTok{(temp), }\ControlFlowTok{function}\NormalTok{(pName) \{}
\NormalTok{    p\_desc }\OtherTok{\textless{}{-}}\NormalTok{ temp[[pName]]}\SpecialCharTok{$}\NormalTok{DESC}\SpecialCharTok{$}\FunctionTok{get0Function}\NormalTok{()}
    \DecValTok{1} \SpecialCharTok{{-}} \FunctionTok{ac\_itp1\_desc1\_ecdf}\NormalTok{(}\FunctionTok{abs}\NormalTok{(expectations[}\StringTok{"desc1"}\NormalTok{] }\SpecialCharTok{{-}} \FunctionTok{p\_desc}\NormalTok{(}\DecValTok{1}\NormalTok{)))}
\NormalTok{  \}), }\FunctionTok{c}\NormalTok{(ground\_truth}\SpecialCharTok{$}\NormalTok{consensus\_score, ground\_truth\_2nd\_batch}\SpecialCharTok{$}\NormalTok{consensus\_score))),}
  \FunctionTok{names}\NormalTok{(temp)), }\FunctionTok{c}\NormalTok{(}\StringTok{"all\_equal"}\NormalTok{, }\StringTok{"pref\_reqt1"}\NormalTok{, }\StringTok{"reqt1\_ecdf"}\NormalTok{, }\StringTok{"devt1\_ecdf"}\NormalTok{, }\StringTok{"devt2\_ecdf"}\NormalTok{,}
  \StringTok{"desc1\_ecdf"}\NormalTok{, }\StringTok{"ground\_truth"}\NormalTok{)))}
\end{Highlighting}
\end{Shaded}

\begin{table}

\caption{\label{tab:ac-it-dr-scores}Scores as calculated by the automatically calibrated decision rule, once using equal weights, and once giving preference to the objective for REQ(t1).}
\centering
\begin{tabular}[t]{lrrrrrrr}
\toprule
  & all\_equal & pref\_reqt1 & reqt1\_ecdf & devt1\_ecdf & devt2\_ecdf & desc1\_ecdf & ground\_truth\\
\midrule
Project1 & 0.6550 & 0.6890 & 0.8249 & 0.7500 & 0.0452 & 1.0000 & 0.1\\
Project2 & 0.7538 & 0.7820 & 0.8952 & 0.9737 & 0.1686 & 0.9775 & 0.0\\
Project3 & 0.6906 & 0.7188 & 0.8314 & 0.7880 & 0.3222 & 0.8208 & 0.6\\
Project4 & 0.5816 & 0.5511 & 0.4292 & 0.9655 & 0.2293 & 0.7024 & 0.8\\
Project5 & 0.5376 & 0.5166 & 0.4326 & 0.7606 & 0.0638 & 0.8933 & 0.1\\
\addlinespace
Project6 & 0.9283 & 0.9294 & 0.9339 & 0.9401 & 0.8520 & 0.9873 & 0.2\\
Project7 & 0.6578 & 0.6966 & 0.8516 & 0.7700 & 0.1543 & 0.8553 & 0.3\\
Project8 & 0.6055 & 0.6149 & 0.6526 & 0.6479 & 0.1214 & 1.0000 & 0.0\\
Project9 & 0.5766 & 0.6320 & 0.8536 & 0.6486 & 0.0964 & 0.7078 & 0.5\\
Project10 & 0.6300 & 0.6512 & 0.7357 & 0.7861 & 0.0605 & 0.9378 & 0.2\\
\addlinespace
Project11 & 0.6409 & 0.7051 & 0.9617 & 0.6365 & 0.0672 & 0.8984 & 0.0\\
Project12 & 0.6252 & 0.6830 & 0.9140 & 0.5221 & 0.0648 & 1.0000 & 0.2\\
Project13 & 0.5955 & 0.6741 & 0.9886 & 0.8311 & 0.1432 & 0.4190 & 1.0\\
Project14 & 0.7619 & 0.7539 & 0.7220 & 0.9953 & 0.4009 & 0.9294 & 0.1\\
Project15 & 0.6344 & 0.6401 & 0.6629 & 0.7578 & 0.1257 & 0.9913 & 0.1\\
\bottomrule
\end{tabular}
\end{table}

The scores from the automatically calibrated decision rule are shown in table \ref{tab:ac-it-dr-scores}.

We should also examine the scores more in detail for both configurations of weights, and find out what were the scores for the best/worst projects (table \ref{tab:ac-it-dr-scores-2-confs}).

\begin{Shaded}
\begin{Highlighting}[]
\NormalTok{temp.gt }\OtherTok{\textless{}{-}} \FunctionTok{c}\NormalTok{(ground\_truth}\SpecialCharTok{$}\NormalTok{consensus\_score, ground\_truth\_2nd\_batch}\SpecialCharTok{$}\NormalTok{consensus\_score)}

\NormalTok{temp.df }\OtherTok{\textless{}{-}} \ConstantTok{NULL}

\ControlFlowTok{for}\NormalTok{ (confName }\ControlFlowTok{in} \FunctionTok{c}\NormalTok{(}\StringTok{"all\_equal"}\NormalTok{, }\StringTok{"pref\_reqt1"}\NormalTok{)) \{}
\NormalTok{  temp.weight }\OtherTok{\textless{}{-}} \ControlFlowTok{if}\NormalTok{ (confName }\SpecialCharTok{==} \StringTok{"all\_equal"}\NormalTok{)}
    \DecValTok{1} \ControlFlowTok{else} \DecValTok{2}
\NormalTok{  temp.pred }\OtherTok{\textless{}{-}} \FunctionTok{data.frame}\NormalTok{(}\AttributeTok{pred =} \FunctionTok{sapply}\NormalTok{(}\AttributeTok{X =} \FunctionTok{names}\NormalTok{(temp), }\ControlFlowTok{function}\NormalTok{(pName) }\FunctionTok{ac\_p1\_dr}\NormalTok{(}\AttributeTok{projName =}\NormalTok{ pName,}
    \AttributeTok{signals =}\NormalTok{ temp, }\AttributeTok{w\_reqt1 =}\NormalTok{ temp.weight)), }\AttributeTok{ground\_truth =}\NormalTok{ temp.gt)}

\NormalTok{  temp.df }\OtherTok{\textless{}{-}} \FunctionTok{rbind}\NormalTok{(temp.df, }\FunctionTok{data.frame}\NormalTok{(}\AttributeTok{Config =}\NormalTok{ confName, }\AttributeTok{ScForBest =} \FunctionTok{ac\_p1\_dr}\NormalTok{(}\AttributeTok{signals =}\NormalTok{ temp,}
    \AttributeTok{projName =} \FunctionTok{paste0}\NormalTok{(}\StringTok{"Project"}\NormalTok{, }\FunctionTok{which.max}\NormalTok{(temp.gt))), }\AttributeTok{ScForWorst =} \FunctionTok{ac\_p1\_dr}\NormalTok{(}\AttributeTok{signals =}\NormalTok{ temp,}
    \AttributeTok{projName =} \FunctionTok{paste0}\NormalTok{(}\StringTok{"Project"}\NormalTok{, }\FunctionTok{which.min}\NormalTok{(temp.gt))), }\AttributeTok{ScMax =} \FunctionTok{max}\NormalTok{(temp.pred}\SpecialCharTok{$}\NormalTok{pred),}
    \AttributeTok{ScMin =} \FunctionTok{min}\NormalTok{(temp.pred}\SpecialCharTok{$}\NormalTok{pred), }\AttributeTok{ScAvg =} \FunctionTok{mean}\NormalTok{(temp.pred}\SpecialCharTok{$}\NormalTok{pred), }\AttributeTok{KLdiv =} \FunctionTok{kl\_div}\NormalTok{(temp.gt,}
\NormalTok{      temp.pred}\SpecialCharTok{$}\NormalTok{pred), }\AttributeTok{RMSE =} \FunctionTok{sqrt}\NormalTok{(Metrics}\SpecialCharTok{::}\FunctionTok{mse}\NormalTok{(}\AttributeTok{actual =}\NormalTok{ temp.gt, }\AttributeTok{predicted =}\NormalTok{ temp.pred}\SpecialCharTok{$}\NormalTok{pred)),}
    \AttributeTok{MSE =}\NormalTok{ Metrics}\SpecialCharTok{::}\FunctionTok{mse}\NormalTok{(}\AttributeTok{actual =}\NormalTok{ temp.gt, }\AttributeTok{predicted =}\NormalTok{ temp.pred}\SpecialCharTok{$}\NormalTok{pred), }\AttributeTok{Log =} \FunctionTok{mean}\NormalTok{(scoring}\SpecialCharTok{::}\FunctionTok{logscore}\NormalTok{(}\AttributeTok{object =}\NormalTok{ ground\_truth }\SpecialCharTok{\textasciitilde{}}
\NormalTok{      pred, }\AttributeTok{data =}\NormalTok{ temp.pred))))}
\NormalTok{\}}

\FunctionTok{rownames}\NormalTok{(temp.df) }\OtherTok{\textless{}{-}} \ConstantTok{NULL}
\end{Highlighting}
\end{Shaded}

\begin{table}

\caption{\label{tab:ac-it-dr-scores-2-confs}Scores and scoring rules for two configurations as calculated by the decision rule for two different configurations.}
\centering
\begin{tabular}[t]{lrrrrrrrrr}
\toprule
temp.df[, 1] & ScForBest & ScForWorst & ScMax & ScMin & ScAvg & KLdiv & RMSE & MSE & Log\\
\midrule
all\_equal & 0.5955 & 0.7538 & 0.9283 & 0.5376 & 0.6583 & 3.9791 & 0.5045 & 0.2546 & 1.1159\\
pref\_reqt1 & 0.5955 & 0.7538 & 0.9294 & 0.5166 & 0.6825 & 3.9068 & 0.5198 & 0.2702 & 1.1685\\
\bottomrule
\end{tabular}
\end{table}

Next, we will check the correlation with the ground truth, first for all-equal weights, then using the preference for \texttt{REQ}.

\begin{Shaded}
\begin{Highlighting}[]
\NormalTok{stats}\SpecialCharTok{::}\FunctionTok{cor.test}\NormalTok{(}\AttributeTok{x =}\NormalTok{ ac\_it\_dr\_scores[, }\DecValTok{1}\NormalTok{], }\FunctionTok{c}\NormalTok{(ground\_truth}\SpecialCharTok{$}\NormalTok{consensus\_score, ground\_truth\_2nd\_batch}\SpecialCharTok{$}\NormalTok{consensus\_score))}
\end{Highlighting}
\end{Shaded}

\begin{verbatim}
## 
##  Pearson's product-moment correlation
## 
## data:  ac_it_dr_scores[, 1] and c(ground_truth$consensus_score, ground_truth_2nd_batch$consensus_score)
## t = -0.97559, df = 13, p-value = 0.3471
## alternative hypothesis: true correlation is not equal to 0
## 95 percent confidence interval:
##  -0.6821772  0.2898574
## sample estimates:
##        cor 
## -0.2611874
\end{verbatim}

\begin{Shaded}
\begin{Highlighting}[]
\NormalTok{stats}\SpecialCharTok{::}\FunctionTok{cor.test}\NormalTok{(}\AttributeTok{x =}\NormalTok{ ac\_it\_dr\_scores[, }\DecValTok{2}\NormalTok{], }\FunctionTok{c}\NormalTok{(ground\_truth}\SpecialCharTok{$}\NormalTok{consensus\_score, ground\_truth\_2nd\_batch}\SpecialCharTok{$}\NormalTok{consensus\_score))}
\end{Highlighting}
\end{Shaded}

\begin{verbatim}
## 
##  Pearson's product-moment correlation
## 
## data:  ac_it_dr_scores[, 2] and c(ground_truth$consensus_score, ground_truth_2nd_batch$consensus_score)
## t = -0.72494, df = 13, p-value = 0.4813
## alternative hypothesis: true correlation is not equal to 0
## 95 percent confidence interval:
##  -0.6443192  0.3505422
## sample estimates:
##        cor 
## -0.1971168
\end{verbatim}

There is almost no correlation between these scores and the ground truth, the large p-values support the null hypothesis (no correlation). Let's check the Brier-scoring rule (we will check the MSE as we do not have discrete outcomes).

\begin{Shaded}
\begin{Highlighting}[]
\StringTok{\textasciigrave{}}\AttributeTok{names\textless{}{-}}\StringTok{\textasciigrave{}}\NormalTok{(}\FunctionTok{c}\NormalTok{(Metrics}\SpecialCharTok{::}\FunctionTok{mse}\NormalTok{(ac\_it\_dr\_scores}\SpecialCharTok{$}\NormalTok{all\_equal, ac\_it\_dr\_scores}\SpecialCharTok{$}\NormalTok{ground\_truth),}
\NormalTok{  Metrics}\SpecialCharTok{::}\FunctionTok{mse}\NormalTok{(ac\_it\_dr\_scores}\SpecialCharTok{$}\NormalTok{pref\_reqt1, ac\_it\_dr\_scores}\SpecialCharTok{$}\NormalTok{ground\_truth)), }\FunctionTok{c}\NormalTok{(}\StringTok{"MSE (all\_equal)"}\NormalTok{,}
  \StringTok{"MSE (pref\_reqt1)"}\NormalTok{))}
\end{Highlighting}
\end{Shaded}

\begin{verbatim}
##  MSE (all_equal) MSE (pref_reqt1) 
##        0.2545587        0.2701644
\end{verbatim}

These results are much worse than those from the scoring rule of the original decision rule. However, some manual tests indicate that the weights we picked might not have been optimal. We will therefore attempt to optimize them. This is now a simple task since we can define the loss to be the Brier score.

\begin{Shaded}
\begin{Highlighting}[]
\NormalTok{eval\_f }\OtherTok{\textless{}{-}} \ControlFlowTok{function}\NormalTok{(x) \{}
\NormalTok{  actual }\OtherTok{\textless{}{-}} \DecValTok{1}\SpecialCharTok{/}\FunctionTok{sum}\NormalTok{(x) }\SpecialCharTok{*}\NormalTok{ (x[}\DecValTok{1}\NormalTok{] }\SpecialCharTok{*}\NormalTok{ ac\_it\_dr\_scores}\SpecialCharTok{$}\NormalTok{reqt1\_ecdf }\SpecialCharTok{+}\NormalTok{ x[}\DecValTok{2}\NormalTok{] }\SpecialCharTok{*}\NormalTok{ ac\_it\_dr\_scores}\SpecialCharTok{$}\NormalTok{devt1\_ecdf }\SpecialCharTok{+}
\NormalTok{    x[}\DecValTok{3}\NormalTok{] }\SpecialCharTok{*}\NormalTok{ ac\_it\_dr\_scores}\SpecialCharTok{$}\NormalTok{devt2\_ecdf }\SpecialCharTok{+}\NormalTok{ x[}\DecValTok{4}\NormalTok{] }\SpecialCharTok{*}\NormalTok{ ac\_it\_dr\_scores}\SpecialCharTok{$}\NormalTok{desc1\_ecdf)}

  \CommentTok{\# Use this instead to maximize the correlation: 1 {-} stats::cor(x = actual, y}
  \CommentTok{\# = ac\_it\_dr\_scores$ground\_truth)}
\NormalTok{  Metrics}\SpecialCharTok{::}\FunctionTok{mse}\NormalTok{(}\AttributeTok{actual =}\NormalTok{ actual, }\AttributeTok{predicted =}\NormalTok{ ac\_it\_dr\_scores}\SpecialCharTok{$}\NormalTok{ground\_truth)}
\NormalTok{\}}

\NormalTok{res }\OtherTok{\textless{}{-}}\NormalTok{ nloptr}\SpecialCharTok{::}\FunctionTok{nloptr}\NormalTok{(}\AttributeTok{x0 =} \FunctionTok{rep}\NormalTok{(}\FloatTok{0.5}\NormalTok{, }\DecValTok{4}\NormalTok{), }\AttributeTok{lb =} \FunctionTok{rep}\NormalTok{(}\DecValTok{0}\NormalTok{, }\DecValTok{4}\NormalTok{), }\AttributeTok{ub =} \FunctionTok{rep}\NormalTok{(}\DecValTok{1}\NormalTok{, }\DecValTok{4}\NormalTok{), }\AttributeTok{eval\_f =}\NormalTok{ eval\_f,}
  \AttributeTok{eval\_grad\_f =} \ControlFlowTok{function}\NormalTok{(x) pracma}\SpecialCharTok{::}\FunctionTok{grad}\NormalTok{(}\AttributeTok{f =}\NormalTok{ eval\_f, }\AttributeTok{x0 =}\NormalTok{ x), }\AttributeTok{opts =} \FunctionTok{list}\NormalTok{(}\AttributeTok{maxeval =} \DecValTok{100}\NormalTok{,}
    \AttributeTok{algorithm =} \StringTok{"NLOPT\_LD\_TNEWTON"}\NormalTok{))}
\NormalTok{res}
\end{Highlighting}
\end{Shaded}

\begin{verbatim}
## 
## Call:
## 
## nloptr::nloptr(x0 = rep(0.5, 4), eval_f = eval_f, eval_grad_f = function(x) pracma::grad(f = eval_f, 
##     x0 = x), lb = rep(0, 4), ub = rep(1, 4), opts = list(maxeval = 100, 
##     algorithm = "NLOPT_LD_TNEWTON"))
## 
## 
## Minimization using NLopt version 2.7.1 
## 
## NLopt solver status: 1 ( NLOPT_SUCCESS: Generic success return value. )
## 
## Number of Iterations....: 20 
## Termination conditions:  maxeval: 100 
## Number of inequality constraints:  0 
## Number of equality constraints:    0 
## Optimal value of objective function:  0.110940310213263 
## Optimal value of controls: 0.1599596 0.1150134 1 0
\end{verbatim}

This Brier/MSE score is the best so far, the value achieved by the binary decision rule across all projects previously was \(0.14\) (with a Log-score of \(\approx0.431\) and a correlation of \(\approx0.696\)).
The weights found by the optimization are shown in table \ref{tab:ac-it-dr-opt-scores}.

\begin{table}

\caption{\label{tab:ac-it-dr-opt-scores}Relative importance of the scores as found by optimization and used in the automatically calibrated decision rule.}
\centering
\begin{tabular}[t]{rrrr}
\toprule
reqt1 & devt1 & devt2 & desc1\\
\midrule
0.1599596 & 0.1150134 & 1 & 0\\
\bottomrule
\end{tabular}
\end{table}

It appears, for example, that we should not put any weight on the indicator for the \texttt{DESC}-variable at \(t=1\).
Let's use the weights (see table \ref{tab:ac-it-dr-opt-scores-with-weights}):

\begin{Shaded}
\begin{Highlighting}[]
\NormalTok{temp }\OtherTok{\textless{}{-}} \FunctionTok{append}\NormalTok{(all\_signals, all\_signals\_2nd\_batch)}

\NormalTok{temp }\OtherTok{\textless{}{-}} \StringTok{\textasciigrave{}}\AttributeTok{colnames\textless{}{-}}\StringTok{\textasciigrave{}}\NormalTok{(}\StringTok{\textasciigrave{}}\AttributeTok{rownames\textless{}{-}}\StringTok{\textasciigrave{}}\NormalTok{(}\FunctionTok{matrix}\NormalTok{(}\AttributeTok{ncol =} \DecValTok{2}\NormalTok{, }\AttributeTok{nrow =} \FunctionTok{length}\NormalTok{(temp), }\AttributeTok{byrow =} \ConstantTok{FALSE}\NormalTok{,}
  \AttributeTok{data =} \FunctionTok{c}\NormalTok{(}\FunctionTok{sapply}\NormalTok{(}\AttributeTok{X =} \FunctionTok{names}\NormalTok{(temp), }\ControlFlowTok{function}\NormalTok{(pName) }\FunctionTok{ac\_p1\_dr}\NormalTok{(}\AttributeTok{projName =}\NormalTok{ pName, }\AttributeTok{signals =}\NormalTok{ temp,}
    \AttributeTok{w\_reqt1 =}\NormalTok{ res}\SpecialCharTok{$}\NormalTok{solution[}\DecValTok{1}\NormalTok{], }\AttributeTok{w\_devt1 =}\NormalTok{ res}\SpecialCharTok{$}\NormalTok{solution[}\DecValTok{2}\NormalTok{], }\AttributeTok{w\_devt2 =}\NormalTok{ res}\SpecialCharTok{$}\NormalTok{solution[}\DecValTok{3}\NormalTok{],}
    \AttributeTok{w\_desc1 =}\NormalTok{ res}\SpecialCharTok{$}\NormalTok{solution[}\DecValTok{4}\NormalTok{])), }\FunctionTok{c}\NormalTok{(ground\_truth}\SpecialCharTok{$}\NormalTok{consensus\_score, ground\_truth\_2nd\_batch}\SpecialCharTok{$}\NormalTok{consensus\_score))),}
  \FunctionTok{names}\NormalTok{(temp)), }\FunctionTok{c}\NormalTok{(}\StringTok{"optimized"}\NormalTok{, }\StringTok{"ground\_truth"}\NormalTok{))}
\end{Highlighting}
\end{Shaded}

\begin{table}

\caption{\label{tab:ac-it-dr-opt-scores-with-weights}Using the optimized weights for the decision rule to demonstrate the deviation from the ground truth.}
\centering
\begin{tabular}[t]{lrr}
\toprule
  & optimized & ground\_truth\\
\midrule
Project1 & 0.2065690 & 0.1\\
Project2 & 0.3324114 & 0.0\\
Project3 & 0.4281273 & 0.6\\
Project4 & 0.3208072 & 0.8\\
Project5 & 0.1729124 & 0.1\\
\addlinespace
Project6 & 0.8702079 & 0.2\\
Project7 & 0.2973469 & 0.3\\
Project8 & 0.2355321 & 0.0\\
Project9 & 0.2412459 & 0.5\\
Project10 & 0.2106481 & 0.2\\
\addlinespace
Project11 & 0.2307607 & 0.0\\
Project12 & 0.2125678 & 0.2\\
Project13 & 0.3113479 & 1.0\\
Project14 & 0.4947850 & 0.1\\
Project15 & 0.2501160 & 0.1\\
\bottomrule
\end{tabular}
\end{table}

Brier-score (MSE) and correlation are:

\begin{Shaded}
\begin{Highlighting}[]
\StringTok{\textasciigrave{}}\AttributeTok{names\textless{}{-}}\StringTok{\textasciigrave{}}\NormalTok{(}\FunctionTok{c}\NormalTok{(Metrics}\SpecialCharTok{::}\FunctionTok{mse}\NormalTok{(}\FunctionTok{c}\NormalTok{(ground\_truth}\SpecialCharTok{$}\NormalTok{consensus\_score, ground\_truth\_2nd\_batch}\SpecialCharTok{$}\NormalTok{consensus\_score),}
\NormalTok{  temp[, }\DecValTok{1}\NormalTok{]), }\FunctionTok{cor}\NormalTok{(}\FunctionTok{c}\NormalTok{(ground\_truth}\SpecialCharTok{$}\NormalTok{consensus\_score, ground\_truth\_2nd\_batch}\SpecialCharTok{$}\NormalTok{consensus\_score),}
\NormalTok{  temp[, }\DecValTok{1}\NormalTok{])), }\FunctionTok{c}\NormalTok{(}\StringTok{"Brier{-}score (MSE)"}\NormalTok{, }\StringTok{"Correlation"}\NormalTok{))}
\end{Highlighting}
\end{Shaded}

\begin{verbatim}
## Brier-score (MSE)       Correlation 
##        0.11094031        0.07623258
\end{verbatim}

Note that we can optimize using other objectives, too. For example, if we were to optimize for the correlation between the scores and the ground truth, we can bump the correlation significantly to \(\approx0.197\), while the Brier-score would plummet to \(\approx0.34\) (however, still far off from the correlation of the original decision rule).
So while we can optimize the Brier-score, it is ultimately up to the decision maker to pick an objective that will maximize the utility of the model.
This short demonstration showed us that we can completely ignore the \texttt{desc1} score, and still achieve a lower Brier score than the original decision rule. Given the assumption that the discrete binary decision rule and its continuous counter part measure the same thing, then this would mean that the indicator for \texttt{DESC} is redundant.

So far, we have attempted to learn a weight vector where each weight \(\geq0\) (however the sum of all weights must be \(>0\) for the normalizing linear weighted scalarizer to work).
The reason behind this was that all objectives from the automatic calibration have the properties of being uniformly distributed in the range \([0,1]\). A normalizing linear combination \(l(x)\) of such scores has the exact same properties as well, i.e.,

\[
\begin{aligned}
  \bm{s}\dots\;&\text{vector of score functions, where}\;\forall\,s\in\bm{s},\,s\mapsto\mathcal{U}_{[0,1]}\text{,}
  \\[1ex]
  \bm{\omega}\dots\;&\text{weight vector, where}\;\forall\,\omega\in\bm{\omega},\,\omega>0\text{,}
  \\[1ex]
  l(x)=&\;\left\lVert\,\bm{\omega}\,\right\rVert_1^{-1}\times\left(\sum_{i=1}^{N}\,\omega_i\times s_i(x)\right)\text{, scalarizer with same properties as}\;\forall\,s\text{.}
\end{aligned}
\]

This means, that even after adapting the weights, the scalarizer \(l(x)\) still \(\mapsto\mathcal{U}_{[0,1]}\), which still allows to compare calibrated models (of the same family) against each other.
Furthermore, if the weights were adapted this way, we gain insights into the \emph{relative importance} of each objective (here: score). In our case, we learned for example that the objective \texttt{devt2} is the most important, while we should completely ignore objective \texttt{desc1} and put only marginal importance into the objectives \texttt{reqt1} and \texttt{devt1} (see table \ref{tab:ac-it-dr-opt-scores}).

In other words, the normalizing linear scalarizer is its own objective, but we may use any other objective to build a regression model that learns coefficients for our features (here: the scores).
For example, in linear least squares (Bates and Maechler 2019), we learn an intercept and a coefficient (multiplier) for each score. However, the resulting fitted model will not that clearly allow us to reason about the relative importance of each score.

Linear least squares, that is

\[
\begin{aligned}
  \hat{\bm{\beta}}=&\;\underset{\bm{\beta}}{arg\,min}\,\left\lVert\,\bm{y}-\bm{X}\,\right\rVert^2
  \\[1ex]
  =&\;\left(\bm{X}^\top\bm{X}\right)^{-1}\,\bm{X}^\top\bm{y}\text{,}
\end{aligned}
\]

is another kind of objective that will lead to a model of which the predictions will ``better'' resemble the ground truth. ``Better'' because the choice of objective is subjective. For example, previously we chose to optimize the Brier-score or correlation.
Here is the solution of linear least squares, including an intercept term:

\begin{Shaded}
\begin{Highlighting}[]
\NormalTok{m }\OtherTok{\textless{}{-}} \FunctionTok{cbind}\NormalTok{(}\DecValTok{1}\NormalTok{, }\FunctionTok{as.matrix}\NormalTok{(ac\_it\_dr\_scores[, }\FunctionTok{c}\NormalTok{(}\StringTok{"reqt1\_ecdf"}\NormalTok{, }\StringTok{"devt1\_ecdf"}\NormalTok{, }\StringTok{"devt2\_ecdf"}\NormalTok{,}
  \StringTok{"desc1\_ecdf"}\NormalTok{)]))}
\FunctionTok{solve}\NormalTok{(}\FunctionTok{t}\NormalTok{(m) }\SpecialCharTok{\%*\%}\NormalTok{ m) }\SpecialCharTok{\%*\%} \FunctionTok{t}\NormalTok{(m) }\SpecialCharTok{\%*\%}\NormalTok{ ac\_it\_dr\_scores}\SpecialCharTok{$}\NormalTok{ground\_truth}
\end{Highlighting}
\end{Shaded}

\begin{verbatim}
##                  [,1]
##             2.1333307
## reqt1_ecdf -0.2069818
## devt1_ecdf -0.2130925
## devt2_ecdf  0.3356454
## desc1_ecdf -1.8178072
\end{verbatim}

This should be equivalent to a linear model:

\begin{Shaded}
\begin{Highlighting}[]
\NormalTok{temp.lm }\OtherTok{\textless{}{-}}\NormalTok{ stats}\SpecialCharTok{::}\FunctionTok{lm}\NormalTok{(}\AttributeTok{data =}\NormalTok{ ac\_it\_dr\_scores, }\AttributeTok{formula =}\NormalTok{ ground\_truth }\SpecialCharTok{\textasciitilde{}}\NormalTok{ reqt1\_ecdf }\SpecialCharTok{+}
\NormalTok{  devt1\_ecdf }\SpecialCharTok{+}\NormalTok{ devt2\_ecdf }\SpecialCharTok{+}\NormalTok{ desc1\_ecdf)}

\NormalTok{stats}\SpecialCharTok{::}\FunctionTok{coef}\NormalTok{(temp.lm)}
\end{Highlighting}
\end{Shaded}

\begin{verbatim}
## (Intercept)  reqt1_ecdf  devt1_ecdf  devt2_ecdf  desc1_ecdf 
##   2.1333307  -0.2069818  -0.2130925   0.3356454  -1.8178072
\end{verbatim}

\begin{figure}
\centering
\includegraphics{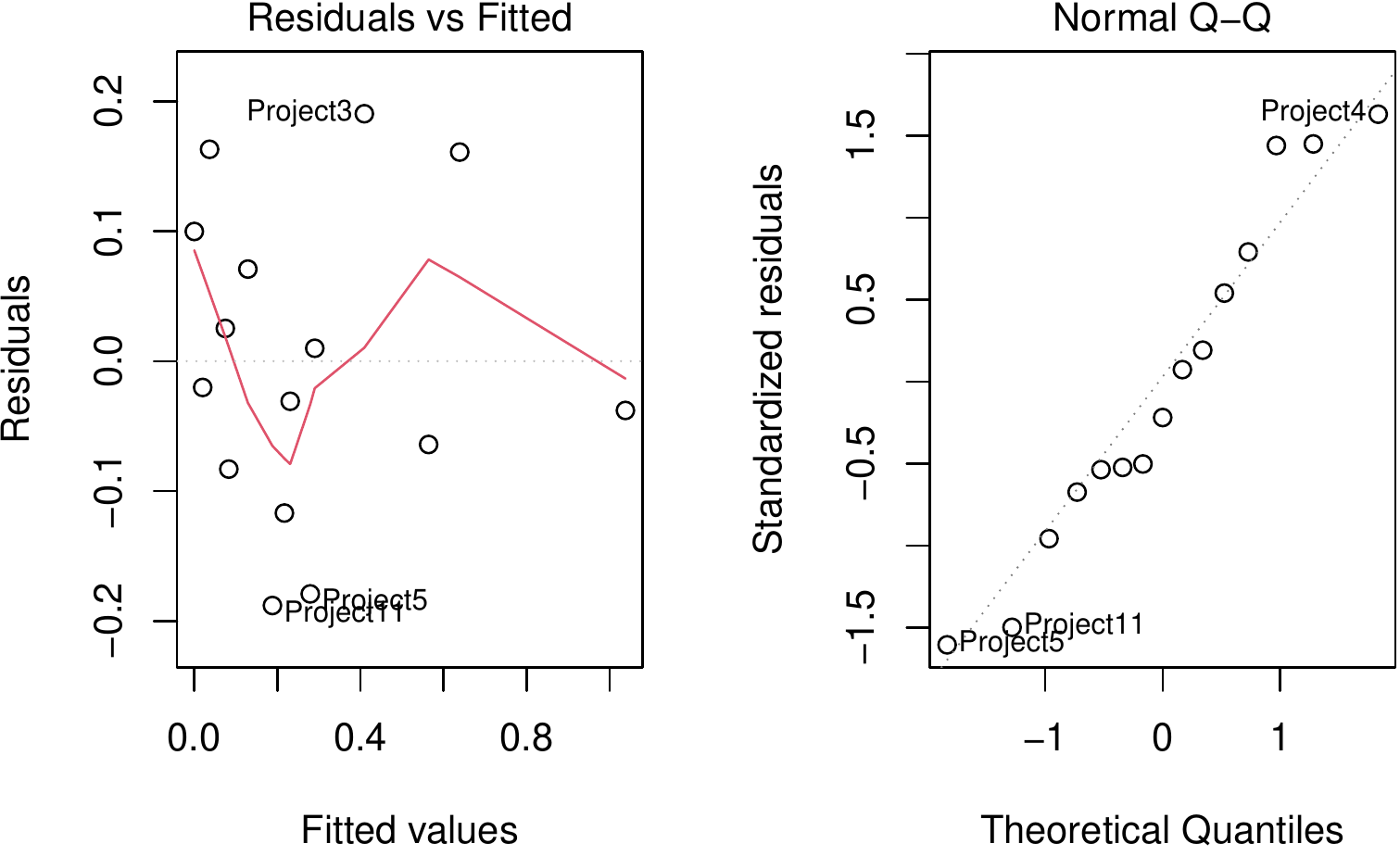}
\caption{\label{fig:ac-it-lm-fit}Residuals and QQ-plot of the automatically calibrated decision rule that was fitted using a linear model.}
\end{figure}

\begin{Shaded}
\begin{Highlighting}[]
\StringTok{\textasciigrave{}}\AttributeTok{names\textless{}{-}}\StringTok{\textasciigrave{}}\NormalTok{(}\FunctionTok{c}\NormalTok{(Metrics}\SpecialCharTok{::}\FunctionTok{mse}\NormalTok{(ac\_it\_dr\_scores}\SpecialCharTok{$}\NormalTok{ground\_truth, stats}\SpecialCharTok{::}\FunctionTok{predict}\NormalTok{(temp.lm, ac\_it\_dr\_scores)),}
  \FunctionTok{cor}\NormalTok{(ac\_it\_dr\_scores}\SpecialCharTok{$}\NormalTok{ground\_truth, stats}\SpecialCharTok{::}\FunctionTok{predict}\NormalTok{(temp.lm, ac\_it\_dr\_scores))),}
  \FunctionTok{c}\NormalTok{(}\StringTok{"Brier{-}score (MSE)"}\NormalTok{, }\StringTok{"Correlation"}\NormalTok{))}
\end{Highlighting}
\end{Shaded}

\begin{verbatim}
## Brier-score (MSE)       Correlation 
##        0.01330897        0.92153060
\end{verbatim}

This fitted model has exceptional good values for the Brier-score and correlation (see fig.~\ref{fig:ac-it-lm-fit}), but we cannot use it in reality as it is almost certainly overfit due to the scarcity of the data.
Also, because of the intercept, it is much harder to make any valid statements about the variable importance of our scores.

\hypertarget{average-distance-to-reference-pattern-type-i}{%
\paragraph{Average distance to reference (pattern type I)}\label{average-distance-to-reference-pattern-type-i}}

As a bonus, to demonstrate the versatility and robustness of this method, we will score the projects against the first pattern, and the second pattern type II (a), which is supposed to be a slight improvement over type I. We will use the variables and confidence intervals as they were defined there, i.e., those are neither data-based nor data-enhanced. For the following tests, the variables \texttt{REQ} and \texttt{DEV} will be considered, as \texttt{DESC} does not have non-empirical confidence intervals.

Pattern I's CIs for the \texttt{REQ} variable align quite well with the boundaries of the empirical CIs, which means that most projects lie, by a large degree, within the expert-guessed CIs, so this method should work well. For the \texttt{DEV} variable's CIs in pattern I however, only a fraction of the projects are within the guessed CI, the averaged-variable appears even to be completely outside of the empirical CI. Therefore, we cannot expect the method to work well in any way in this scenario (cf.~figure \ref{fig:req-dev-p3avg-cis}).

\begin{figure}[ht!]

{\centering \includegraphics{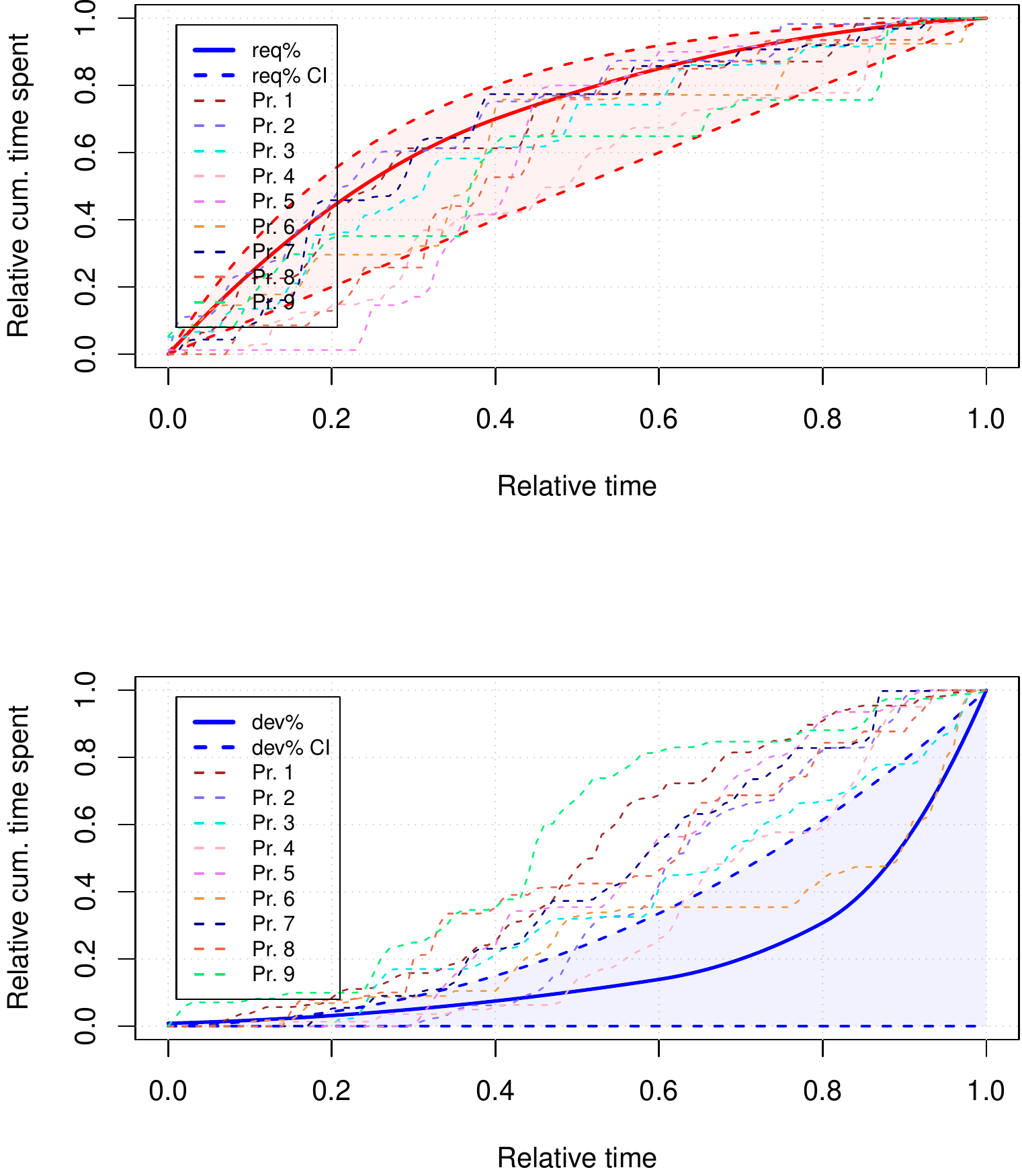} 

}

\caption{All projects plotted against the two variables req\% and dev\% of the first pattern.}\label{fig:p1-avg-area-scores}
\end{figure}

In figure \ref{fig:p1-avg-area-scores} we show the first pattern and its confidence intervals, plotted against all projects' variables. The method used here to compute a score calculates the area between the pattern's variable (here: \texttt{REQ} or \texttt{DEV}) and a project's variable, where the upper bound would be the confidence intervals. While the first variant of this loss uses as max distance the CI that is farther away, the 2nd variant uses the CI based on whether the signal is above or below the compared-to variable. If we look at figure \ref{fig:p1-avg-area-scores}, it becomes apparent why that is better: Consider, e.g., project 9 and the \texttt{DEV} variable. It is completely above the upper CI. The are between the upper CI and \texttt{DEV} and the lower CI and \texttt{DEV} are differently large, with the latter being definitely larger. In variant 1, due to the confinement, the area-distance, the are between the variable and the signal would be approximately equivalent to the area between \texttt{DEV} and the upper CI. That is then put into relation to the maximum area, which is the one between \texttt{DEV} and the lower CI. This results in a distance \(\ll1\), even though it could not be worse. Variant 2 however considers, for all realizations of \(X\), which CI should be used, and correctly determines the distance as \(\approx1\) (note: it might be slightly less or more, due to numerical error).

\begin{table}

\caption{\label{tab:p1-avg-area-scores}The average distance of the variables REQ and DEV of each project to the reference-variables REQ/DEV as defined by pattern I.}
\centering
\begin{tabular}[t]{lrr}
\toprule
  & REQ & DEV\\
\midrule
Project1 & 0.2282387 & 1.0025387\\
Project2 & 0.2055580 & 0.9217657\\
Project3 & 0.3492357 & 0.9801568\\
Project4 & 0.8581838 & 0.8096948\\
Project5 & 0.6717589 & 0.9942192\\
\addlinespace
Project6 & 0.5131105 & 0.5068238\\
Project7 & 0.2535267 & 0.9983062\\
Project8 & 0.5540218 & 0.9995019\\
Project9 & 0.6359345 & 1.0106332\\
\bottomrule
\end{tabular}
\end{table}

\begin{Shaded}
\begin{Highlighting}[]
\FunctionTok{cor}\NormalTok{(}\AttributeTok{x =}\NormalTok{ ground\_truth}\SpecialCharTok{$}\NormalTok{consensus\_score, }\AttributeTok{y =}\NormalTok{ p1\_avg\_area\_scores[, }\StringTok{"REQ"}\NormalTok{])}
\end{Highlighting}
\end{Shaded}

\begin{verbatim}
## [1] 0.4789826
\end{verbatim}

\begin{Shaded}
\begin{Highlighting}[]
\FunctionTok{cor}\NormalTok{(}\AttributeTok{x =}\NormalTok{ ground\_truth}\SpecialCharTok{$}\NormalTok{consensus\_score, }\AttributeTok{y =}\NormalTok{ p1\_avg\_area\_scores[, }\StringTok{"DEV"}\NormalTok{])}
\end{Highlighting}
\end{Shaded}

\begin{verbatim}
## [1] -0.09021552
\end{verbatim}

I think it is fair to say that the first pattern and its confidence intervals do not work well with this detection approach. While we do get a moderate correlation for \texttt{REQ}, it is positive, when it should have been negative.

As expected, \texttt{DEV} is quite unusable, as the correlations are low, albeit negative.

\hypertarget{average-distance-to-reference-pattern-type-ii-a}{%
\paragraph{Average distance to reference (pattern type II (a))}\label{average-distance-to-reference-pattern-type-ii-a}}

Let's check the next pattern: In type II (a), we have adapted the thresholds \(t_1,t_2\) according to the ground truth. We can see, that the CIs of this align already much better with the projects, esp.~for the \texttt{DEV} variable.

\begin{figure}[ht!]

{\centering \includegraphics{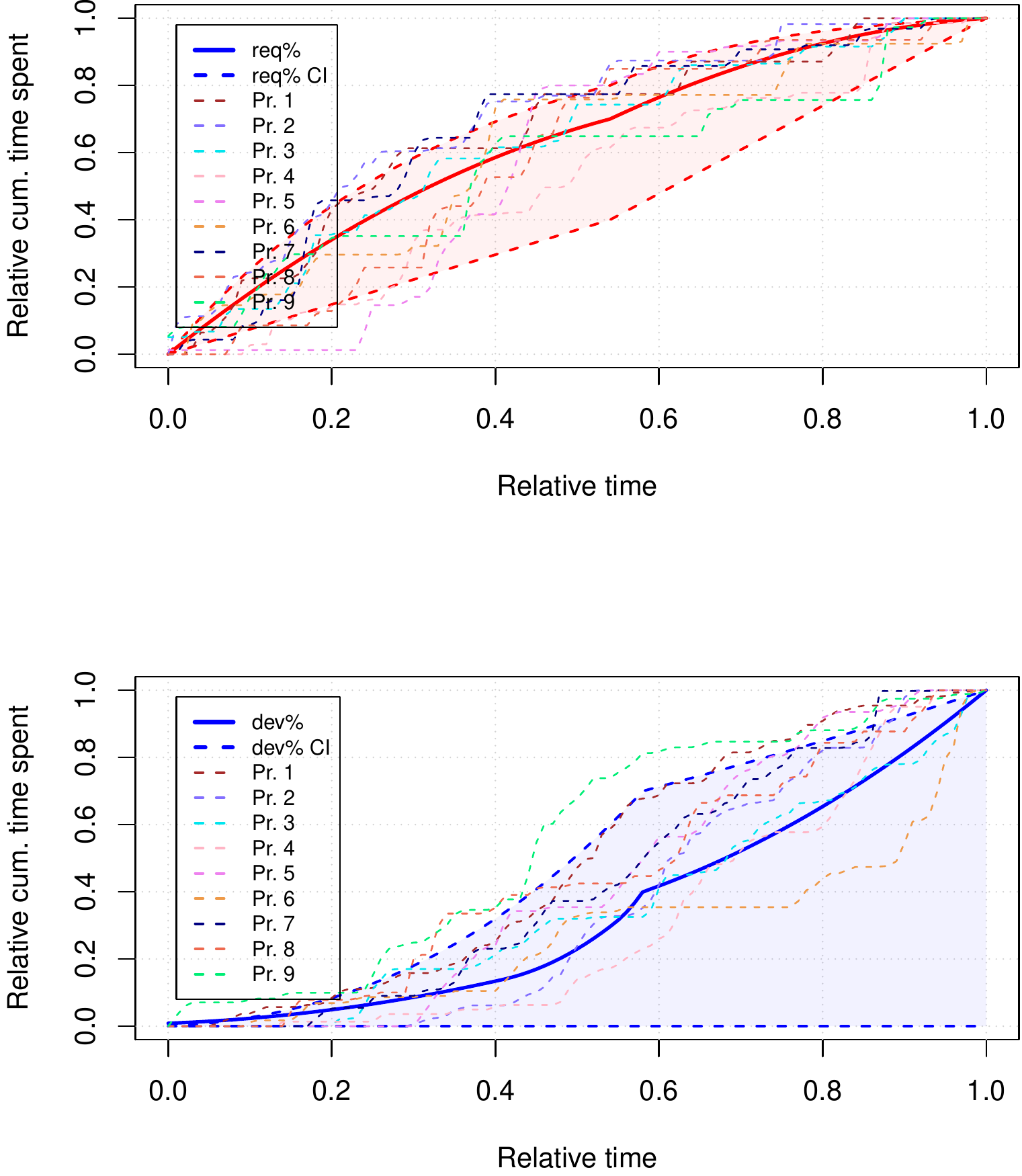} 

}

\caption{All projects plotted against the two variables req\% and dev\% of pattern type II (a).}\label{fig:p2a-avg-area-scores}
\end{figure}

\begin{table}

\caption{\label{tab:p2a-avg-area-scores}The average distance of the variables REQ and DEV of each project to the reference-variables $\operatorname{REQ},\operatorname{DEV}$ as taken from pattern I and adjusted by the optimized $t_1,t_2$ thresholds and timewarping.}
\centering
\begin{tabular}[t]{lrr}
\toprule
  & REQ & DEV\\
\midrule
Project1 & 0.4523361 & 0.9013424\\
Project2 & 0.8687620 & 0.4740080\\
Project3 & 0.1920389 & 0.1698280\\
Project4 & 0.5830514 & 0.2945775\\
Project5 & 0.7912241 & 0.6853936\\
\addlinespace
Project6 & 0.4024728 & 0.2748405\\
Project7 & 0.6703912 & 0.5923427\\
Project8 & 0.5822121 & 0.6677470\\
Project9 & 0.4021386 & 1.0050354\\
\bottomrule
\end{tabular}
\end{table}

\begin{Shaded}
\begin{Highlighting}[]
\FunctionTok{cor}\NormalTok{(}\AttributeTok{x =}\NormalTok{ ground\_truth}\SpecialCharTok{$}\NormalTok{consensus\_score, }\AttributeTok{y =}\NormalTok{ p2a\_avg\_area\_scores[, }\StringTok{"REQ"}\NormalTok{])}
\end{Highlighting}
\end{Shaded}

\begin{verbatim}
## [1] -0.4976161
\end{verbatim}

\begin{Shaded}
\begin{Highlighting}[]
\FunctionTok{cor}\NormalTok{(}\AttributeTok{x =}\NormalTok{ ground\_truth}\SpecialCharTok{$}\NormalTok{consensus\_score, }\AttributeTok{y =}\NormalTok{ p2a\_avg\_area\_scores[, }\StringTok{"DEV"}\NormalTok{])}
\end{Highlighting}
\end{Shaded}

\begin{verbatim}
## [1] -0.3558058
\end{verbatim}

Now we can observe quite an improvement for both variables. The correlation for \texttt{REQ} has increased by more than \(1\), so it is moderate now and has the right sign. As for \texttt{DEV}, the correlation is almost four times as strong.

\hypertarget{pattern-iii-average-1}{%
\subsubsection{\texorpdfstring{Pattern III (average)\label{ssec:score-p3avg}}{Pattern III (average)}}\label{pattern-iii-average-1}}

The third kind of pattern is based on data only, all the variables, confidence intervals and the strength thereof are based on the nine projects and the weight, which is the same as their consensus score.

\hypertarget{scoring-based-on-the-confidence-intervals}{%
\paragraph{Scoring based on the confidence intervals}\label{scoring-based-on-the-confidence-intervals}}

We have calculated gradated confidence intervals, which means two things. First, we cannot apply a binary detection rule any longer, as the boundaries of the intervals include each project, only the weight is different. Second, when calculating a score, we will obtain a continuous measure, of which we can calculate a correlation to the consensus score of the ground truth, or, e.g., fit a linear model for scaling these scores.

\begin{Shaded}
\begin{Highlighting}[]
\NormalTok{p3\_avg\_ci\_scores\_compute }\OtherTok{\textless{}{-}} \ControlFlowTok{function}\NormalTok{(pId, }\AttributeTok{x1 =} \DecValTok{0}\NormalTok{, }\AttributeTok{x2 =} \DecValTok{1}\NormalTok{, }\AttributeTok{signals =}\NormalTok{ all\_signals) \{}
\NormalTok{  req }\OtherTok{\textless{}{-}}\NormalTok{ signals[[pId]]}\SpecialCharTok{$}\NormalTok{REQ}\SpecialCharTok{$}\FunctionTok{get0Function}\NormalTok{()}
\NormalTok{  dev }\OtherTok{\textless{}{-}}\NormalTok{ signals[[pId]]}\SpecialCharTok{$}\NormalTok{DEV}\SpecialCharTok{$}\FunctionTok{get0Function}\NormalTok{()}
\NormalTok{  desc }\OtherTok{\textless{}{-}}\NormalTok{ signals[[pId]]}\SpecialCharTok{$}\NormalTok{DESC}\SpecialCharTok{$}\FunctionTok{get0Function}\NormalTok{()}

  \StringTok{\textasciigrave{}}\AttributeTok{rownames\textless{}{-}}\StringTok{\textasciigrave{}}\NormalTok{(}\FunctionTok{data.frame}\NormalTok{(}\AttributeTok{REQ =} \FunctionTok{L\_avgconf\_p3\_avg}\NormalTok{(}\AttributeTok{x1 =}\NormalTok{ x1, }\AttributeTok{x2 =}\NormalTok{ x2, }\AttributeTok{f =}\NormalTok{ req, }\AttributeTok{CI =}\NormalTok{ CI\_req\_p3avg),}
    \AttributeTok{DEV =} \FunctionTok{L\_avgconf\_p3\_avg}\NormalTok{(}\AttributeTok{x1 =}\NormalTok{ x1, }\AttributeTok{x2 =}\NormalTok{ x2, }\AttributeTok{f =}\NormalTok{ dev, }\AttributeTok{CI =}\NormalTok{ CI\_dev\_p3avg), }\AttributeTok{DESC =} \FunctionTok{L\_avgconf\_p3\_avg}\NormalTok{(}\AttributeTok{x1 =}\NormalTok{ x1,}
      \AttributeTok{x2 =}\NormalTok{ x2, }\AttributeTok{f =}\NormalTok{ desc, }\AttributeTok{CI =}\NormalTok{ CI\_desc\_p3avg), }\AttributeTok{Project =}\NormalTok{ pId), pId)}
\NormalTok{\}}
\end{Highlighting}
\end{Shaded}

\begin{Shaded}
\begin{Highlighting}[]
\NormalTok{p3\_avg\_ci\_scores }\OtherTok{\textless{}{-}} \FunctionTok{loadResultsOrCompute}\NormalTok{(}\AttributeTok{file =} \StringTok{"../results/p3\_avg\_ci\_scores.rds"}\NormalTok{,}
  \AttributeTok{computeExpr =}\NormalTok{ \{}
    \FunctionTok{doWithParallelCluster}\NormalTok{(}\AttributeTok{numCores =} \FunctionTok{length}\NormalTok{(all\_signals), }\AttributeTok{expr =}\NormalTok{ \{}
      \FunctionTok{library}\NormalTok{(foreach)}

\NormalTok{      foreach}\SpecialCharTok{::}\FunctionTok{foreach}\NormalTok{(}\AttributeTok{pId =} \FunctionTok{names}\NormalTok{(all\_signals), }\AttributeTok{.inorder =} \ConstantTok{TRUE}\NormalTok{, }\AttributeTok{.combine =}\NormalTok{ rbind) }\SpecialCharTok{\%dopar\%}
\NormalTok{        \{}
          \FunctionTok{p3\_avg\_ci\_scores\_compute}\NormalTok{(}\AttributeTok{pId =}\NormalTok{ pId, }\AttributeTok{x1 =} \DecValTok{0}\NormalTok{, }\AttributeTok{x2 =} \DecValTok{1}\NormalTok{)}
\NormalTok{        \}}
\NormalTok{    \})}
\NormalTok{  \})}
\end{Highlighting}
\end{Shaded}

Table \ref{tab:p3-avg-ci-scores} shows the computed scores.

\begin{table}

\caption{\label{tab:p3-avg-ci-scores}The average confidence of the variables REQ, DEV and DESC of each project as integrated over the confidence intervals' hyperplane.}
\centering
\begin{tabular}[t]{lrrr}
\toprule
  & REQ & DEV & DESC\\
\midrule
Project1 & 0.1729721 & 0.1364909 & 0.0000000\\
Project2 & 0.1069433 & 0.1491474 & 0.0002014\\
Project3 & 0.3122518 & 0.1695559 & 0.0235139\\
Project4 & 0.2379591 & 0.1543914 & 0.0140107\\
Project5 & 0.1074743 & 0.1425569 & 0.0003490\\
\addlinespace
Project6 & 0.2019605 & 0.0579568 & 0.0009256\\
Project7 & 0.1715675 & 0.1737240 & 0.0131727\\
Project8 & 0.1415105 & 0.1708401 & 0.0000000\\
Project9 & 0.2201456 & 0.1436310 & 0.0117499\\
\bottomrule
\end{tabular}
\end{table}

Let's test the correlation between either kind of score and the ground truth consensus score. The null hypothesis of this test states that both samples have no correlation.

\begin{Shaded}
\begin{Highlighting}[]
\FunctionTok{cor.test}\NormalTok{(}\AttributeTok{x =}\NormalTok{ ground\_truth}\SpecialCharTok{$}\NormalTok{consensus\_score, }\AttributeTok{y =}\NormalTok{ p3\_avg\_ci\_scores[, }\StringTok{"REQ"}\NormalTok{])}
\end{Highlighting}
\end{Shaded}

\begin{verbatim}
## 
##  Pearson's product-moment correlation
## 
## data:  ground_truth$consensus_score and p3_avg_ci_scores[, "REQ"]
## t = 3.9034, df = 7, p-value = 0.005873
## alternative hypothesis: true correlation is not equal to 0
## 95 percent confidence interval:
##  0.3634519 0.9626722
## sample estimates:
##       cor 
## 0.8277696
\end{verbatim}

For the variable \texttt{REQ} we get a significant correlation of \(\approx0.83\), and there is no significant evidence for the null hypothesis, so must reject it.

\begin{Shaded}
\begin{Highlighting}[]
\FunctionTok{cor.test}\NormalTok{(}\AttributeTok{x =}\NormalTok{ ground\_truth}\SpecialCharTok{$}\NormalTok{consensus\_score, }\AttributeTok{y =}\NormalTok{ p3\_avg\_ci\_scores[, }\StringTok{"DEV"}\NormalTok{])}
\end{Highlighting}
\end{Shaded}

\begin{verbatim}
## 
##  Pearson's product-moment correlation
## 
## data:  ground_truth$consensus_score and p3_avg_ci_scores[, "DEV"]
## t = 0.4571, df = 7, p-value = 0.6614
## alternative hypothesis: true correlation is not equal to 0
## 95 percent confidence interval:
##  -0.5568336  0.7496134
## sample estimates:
##      cor 
## 0.170246
\end{verbatim}

For the variable \texttt{DEV} however, the correlation is quite low, \(\approx0.17\). Also, we have significant evidence for accepting the null hypothesis (no correlation).

\begin{Shaded}
\begin{Highlighting}[]
\FunctionTok{cor.test}\NormalTok{(}\AttributeTok{x =}\NormalTok{ ground\_truth}\SpecialCharTok{$}\NormalTok{consensus\_score, }\AttributeTok{y =}\NormalTok{ p3\_avg\_ci\_scores[, }\StringTok{"DESC"}\NormalTok{])}
\end{Highlighting}
\end{Shaded}

\begin{verbatim}
## 
##  Pearson's product-moment correlation
## 
## data:  ground_truth$consensus_score and p3_avg_ci_scores[, "DESC"]
## t = 4.2844, df = 7, p-value = 0.003636
## alternative hypothesis: true correlation is not equal to 0
## 95 percent confidence interval:
##  0.4293062 0.9679894
## sample estimates:
##       cor 
## 0.8508428
\end{verbatim}

Looks like we are getting some strong positive correlation for the variable \texttt{DESC} of \(\approx0.85\). There is almost no evidence at all for accepting the null hypothesis.

\hypertarget{scoring-based-on-the-distance-to-average}{%
\paragraph{Scoring based on the distance to average}\label{scoring-based-on-the-distance-to-average}}

Here we compute the distance of each project's variables to the previously averaged variables. This approach does not rely on inhomogeneous confidence intervals, and only considers the intervals' boundaries to integrate some distance. Ideally, the distance is \(0\), and in the worst case it is \(1\). If this ought to be used as score, we probably would want to compute it as \(\mathit{S}^{\text{areadist}}=1-\mathit{L}^{\text{areadist}}\).

\begin{Shaded}
\begin{Highlighting}[]
\NormalTok{p3\_avg\_area\_scores\_compute }\OtherTok{\textless{}{-}} \ControlFlowTok{function}\NormalTok{(pId, }\AttributeTok{x1 =} \DecValTok{0}\NormalTok{, }\AttributeTok{x2 =} \DecValTok{1}\NormalTok{, }\AttributeTok{signals =}\NormalTok{ all\_signals) \{}
\NormalTok{  req }\OtherTok{\textless{}{-}}\NormalTok{ signals[[pId]]}\SpecialCharTok{$}\NormalTok{REQ}\SpecialCharTok{$}\FunctionTok{get0Function}\NormalTok{()}
\NormalTok{  dev }\OtherTok{\textless{}{-}}\NormalTok{ signals[[pId]]}\SpecialCharTok{$}\NormalTok{DEV}\SpecialCharTok{$}\FunctionTok{get0Function}\NormalTok{()}
\NormalTok{  desc }\OtherTok{\textless{}{-}}\NormalTok{ signals[[pId]]}\SpecialCharTok{$}\NormalTok{DESC}\SpecialCharTok{$}\FunctionTok{get0Function}\NormalTok{()}
  
  \StringTok{\textasciigrave{}}\AttributeTok{rownames\textless{}{-}}\StringTok{\textasciigrave{}}\NormalTok{(}\FunctionTok{data.frame}\NormalTok{(}
    \AttributeTok{REQ =} \FunctionTok{L\_areadist\_p3\_avg}\NormalTok{(}
      \AttributeTok{x1 =}\NormalTok{ x1, }\AttributeTok{x2 =}\NormalTok{ x2, }\AttributeTok{f =}\NormalTok{ req, }\AttributeTok{gbar =}\NormalTok{ req\_p3, }\AttributeTok{use2ndVariant =} \ConstantTok{TRUE}\NormalTok{,}
      \AttributeTok{CI\_upper =}\NormalTok{ req\_ci\_upper\_p3avg, }\AttributeTok{CI\_lower =}\NormalTok{ req\_ci\_lower\_p3avg)[}\StringTok{"dist"}\NormalTok{],}
    
    \AttributeTok{DEV =} \FunctionTok{L\_areadist\_p3\_avg}\NormalTok{(}
      \AttributeTok{x1 =}\NormalTok{ x1, }\AttributeTok{x2 =}\NormalTok{ x2, }\AttributeTok{f =}\NormalTok{ dev, }\AttributeTok{gbar =}\NormalTok{ dev\_p3, }\AttributeTok{use2ndVariant =} \ConstantTok{TRUE}\NormalTok{,}
      \AttributeTok{CI\_upper =}\NormalTok{ dev\_ci\_upper\_p3avg, }\AttributeTok{CI\_lower =}\NormalTok{ dev\_ci\_lower\_p3avg)[}\StringTok{"dist"}\NormalTok{],}
    
    \AttributeTok{DESC =} \FunctionTok{L\_areadist\_p3\_avg}\NormalTok{(}
      \AttributeTok{x1 =}\NormalTok{ x1, }\AttributeTok{x2 =}\NormalTok{ x2, }\AttributeTok{f =}\NormalTok{ desc, }\AttributeTok{gbar =}\NormalTok{ desc\_p3, }\AttributeTok{use2ndVariant =} \ConstantTok{TRUE}\NormalTok{,}
      \AttributeTok{CI\_upper =}\NormalTok{ desc\_ci\_upper\_p3avg, }\AttributeTok{CI\_lower =}\NormalTok{ desc\_ci\_lower\_p3avg)[}\StringTok{"dist"}\NormalTok{],}
    
    \AttributeTok{Project =}\NormalTok{ pId}
\NormalTok{  ), pId)}
\NormalTok{\}}
\end{Highlighting}
\end{Shaded}

\begin{Shaded}
\begin{Highlighting}[]
\NormalTok{p3\_avg\_area\_scores }\OtherTok{\textless{}{-}} \FunctionTok{loadResultsOrCompute}\NormalTok{(}\AttributeTok{file =} \StringTok{"../results/p3\_avg\_area\_scores.rds"}\NormalTok{,}
  \AttributeTok{computeExpr =}\NormalTok{ \{}
    \FunctionTok{doWithParallelCluster}\NormalTok{(}\AttributeTok{numCores =} \FunctionTok{length}\NormalTok{(all\_signals), }\AttributeTok{expr =}\NormalTok{ \{}
      \FunctionTok{library}\NormalTok{(foreach)}

\NormalTok{      foreach}\SpecialCharTok{::}\FunctionTok{foreach}\NormalTok{(}\AttributeTok{pId =} \FunctionTok{names}\NormalTok{(all\_signals), }\AttributeTok{.inorder =} \ConstantTok{TRUE}\NormalTok{, }\AttributeTok{.combine =}\NormalTok{ rbind) }\SpecialCharTok{\%dopar\%}
\NormalTok{        \{}
          \FunctionTok{p3\_avg\_area\_scores\_compute}\NormalTok{(}\AttributeTok{pId =}\NormalTok{ pId)}
\NormalTok{        \}}
\NormalTok{    \})}
\NormalTok{  \})}
\end{Highlighting}
\end{Shaded}

Table \ref{tab:p3-avg-area-scores} shows the computed scores.

\begin{table}

\caption{\label{tab:p3-avg-area-scores}The average distance of the variables REQ, DEV and DESC of each project to the previously averaged reference-variables $\overline{\operatorname{REQ}},\overline{\operatorname{DEV}},\overline{\operatorname{DESC}}$.}
\centering
\begin{tabular}[t]{lrrr}
\toprule
  & REQ & DEV & DESC\\
\midrule
Project1 & 0.6745385 & 0.6031728 & 1.0000000\\
Project2 & 0.8944326 & 0.4870020 & 0.9293380\\
Project3 & 0.3962041 & 0.2562217 & 0.9876595\\
Project4 & 0.6750274 & 0.6539538 & 0.3154989\\
Project5 & 0.9296596 & 0.4517074 & 0.9089733\\
\addlinespace
Project6 & 0.4607223 & 0.7177923 & 0.8203402\\
Project7 & 0.7656557 & 0.3180007 & 0.4546803\\
Project8 & 0.5561467 & 0.3651845 & 1.0000000\\
Project9 & 0.5192891 & 0.9612175 & 0.4064163\\
\bottomrule
\end{tabular}
\end{table}

As for the correlation tests, ideally, we get negative correlations, as the computed score is \textbf{lower} the less distance we find between the area of the average variable and a project's variable, hence the relation must be antiproportional.

\begin{Shaded}
\begin{Highlighting}[]
\FunctionTok{cor.test}\NormalTok{(}\AttributeTok{x =}\NormalTok{ ground\_truth}\SpecialCharTok{$}\NormalTok{consensus\_score, }\AttributeTok{y =}\NormalTok{ p3\_avg\_area\_scores[, }\StringTok{"REQ"}\NormalTok{])}
\end{Highlighting}
\end{Shaded}

\begin{verbatim}
## 
##  Pearson's product-moment correlation
## 
## data:  ground_truth$consensus_score and p3_avg_area_scores[, "REQ"]
## t = -1.208, df = 7, p-value = 0.2663
## alternative hypothesis: true correlation is not equal to 0
## 95 percent confidence interval:
##  -0.8460845  0.3435348
## sample estimates:
##        cor 
## -0.4153482
\end{verbatim}

For the variable \texttt{REQ} we get a weaker, yet moderate correlation of \(\approx-0.42\). However, there is evidence for the null hypothesis, which suggests that there is no statistical significant correlation. This means, we will have to use this with care, if at all.

\begin{Shaded}
\begin{Highlighting}[]
\FunctionTok{cor.test}\NormalTok{(}\AttributeTok{x =}\NormalTok{ ground\_truth}\SpecialCharTok{$}\NormalTok{consensus\_score, }\AttributeTok{y =}\NormalTok{ p3\_avg\_area\_scores[, }\StringTok{"DEV"}\NormalTok{])}
\end{Highlighting}
\end{Shaded}

\begin{verbatim}
## 
##  Pearson's product-moment correlation
## 
## data:  ground_truth$consensus_score and p3_avg_area_scores[, "DEV"]
## t = 0.5941, df = 7, p-value = 0.5711
## alternative hypothesis: true correlation is not equal to 0
## 95 percent confidence interval:
##  -0.5208081  0.7710272
## sample estimates:
##       cor 
## 0.2190938
\end{verbatim}

The correlation for the variable \texttt{DEV} is poor again. Also, it is positive, which means that the scores calculated using this method are proportional, when they should not be. The p-value is significant, so there is most likely no significant correlation for this variable and the ground truth.

\begin{Shaded}
\begin{Highlighting}[]
\FunctionTok{cor.test}\NormalTok{(}\AttributeTok{x =}\NormalTok{ ground\_truth}\SpecialCharTok{$}\NormalTok{consensus\_score, }\AttributeTok{y =}\NormalTok{ p3\_avg\_area\_scores[, }\StringTok{"DESC"}\NormalTok{])}
\end{Highlighting}
\end{Shaded}

\begin{verbatim}
## 
##  Pearson's product-moment correlation
## 
## data:  ground_truth$consensus_score and p3_avg_area_scores[, "DESC"]
## t = -2.3941, df = 7, p-value = 0.04788
## alternative hypothesis: true correlation is not equal to 0
## 95 percent confidence interval:
##  -0.92355184 -0.01235339
## sample estimates:
##        cor 
## -0.6709704
\end{verbatim}

The correlation for \texttt{DESC} is substantial, and also it is negative like it should be. The p-value is below the significance level \(0.05\), suggesting correlation. We might be able to use this score.

\hypertarget{linear-combination-of-the-two-methods}{%
\paragraph{Linear combination of the two methods}\label{linear-combination-of-the-two-methods}}

\begin{Shaded}
\begin{Highlighting}[]
\NormalTok{temp }\OtherTok{\textless{}{-}} \FunctionTok{data.frame}\NormalTok{(}\AttributeTok{gt\_consensus =}\NormalTok{ ground\_truth}\SpecialCharTok{$}\NormalTok{consensus\_score, }\AttributeTok{ci\_req =}\NormalTok{ p3\_avg\_ci\_scores}\SpecialCharTok{$}\NormalTok{REQ,}
  \AttributeTok{area\_req =}\NormalTok{ p3\_avg\_area\_scores}\SpecialCharTok{$}\NormalTok{REQ, }\AttributeTok{ci\_dev =}\NormalTok{ p3\_avg\_ci\_scores}\SpecialCharTok{$}\NormalTok{DEV, }\AttributeTok{area\_dev =}\NormalTok{ p3\_avg\_area\_scores}\SpecialCharTok{$}\NormalTok{DEV,}
  \AttributeTok{ci\_desc =}\NormalTok{ p3\_avg\_ci\_scores}\SpecialCharTok{$}\NormalTok{DESC, }\AttributeTok{area\_desc =}\NormalTok{ p3\_avg\_area\_scores}\SpecialCharTok{$}\NormalTok{DESC)}

\CommentTok{\# ci\_req + area\_desc gives us \textasciitilde{}0.951 already!  ci\_req + ci\_dev + area\_desc}
\CommentTok{\# gives \textasciitilde{}0.962}
\NormalTok{p3\_avg\_lm }\OtherTok{\textless{}{-}}\NormalTok{ stats}\SpecialCharTok{::}\FunctionTok{lm}\NormalTok{(}\AttributeTok{formula =}\NormalTok{ gt\_consensus }\SpecialCharTok{\textasciitilde{}}\NormalTok{ ci\_req }\SpecialCharTok{+}\NormalTok{ ci\_dev }\SpecialCharTok{+}\NormalTok{ area\_desc, }\AttributeTok{data =}\NormalTok{ temp)}
\NormalTok{stats}\SpecialCharTok{::}\FunctionTok{coef}\NormalTok{(p3\_avg\_lm)}
\end{Highlighting}
\end{Shaded}

\begin{verbatim}
## (Intercept)      ci_req      ci_dev   area_desc 
## -0.02917711  3.01066887  0.83917207 -0.47825651
\end{verbatim}

\begin{figure}
\centering
\includegraphics{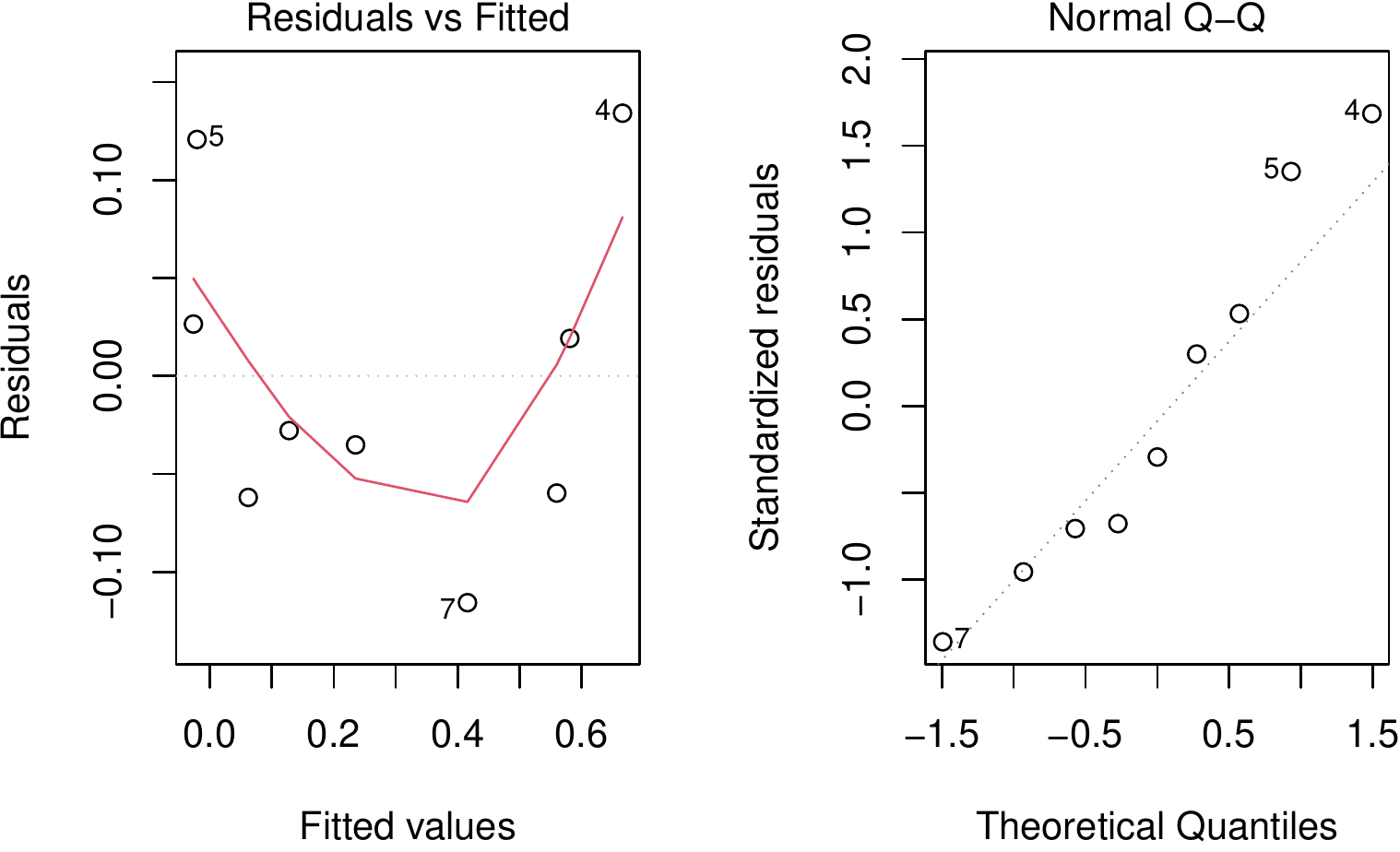}
\caption{\label{fig:ac-it-lm-fit2}Residuals and QQ-plot of the automatically calibrated decision rule that was fitted using a linear model.}
\end{figure}

Using the approximate coefficients of the linear model (figure \ref{fig:ac-it-lm-fit2}), we can define the detector as follows:

\[
\begin{aligned}
  x_1,x_2,\operatorname{req},\operatorname{dev},\operatorname{desc}\dots&\;\text{lower/upper integration interval and project signals,}
  \\[1ex]
  \bm{\tau}=&\;\Big\{\mathit{L}^{\text{avgconf}}(x_1,x_2,\operatorname{req}),\;\mathit{L}^{\text{avgconf}}(x_1,x_2,\operatorname{dev}),\;\mathit{L}^{\text{areadist2}}(x_1,x_2,\operatorname{desc})\Big\}\;\text{,}
  \\[1ex]
  \operatorname{detect}^{\text{ci+area}}(\bm{\tau})=&\;-0.029 + 3.011\times\bm{\tau}_1 + 0.839\times\bm{\tau}_2 - 0.478\times\bm{\tau}_3\;\text{.}
\end{aligned}
\]

\begin{Shaded}
\begin{Highlighting}[]
\NormalTok{p3\_avg\_lm\_scores }\OtherTok{\textless{}{-}}\NormalTok{ stats}\SpecialCharTok{::}\FunctionTok{predict}\NormalTok{(p3\_avg\_lm, temp)}
\CommentTok{\# Since we are attempting a regression to positive scores, we set any negative}
\CommentTok{\# predictions to 0. Same goes for \textgreater{}1.}
\NormalTok{p3\_avg\_lm\_scores[p3\_avg\_lm\_scores }\SpecialCharTok{\textless{}} \DecValTok{0}\NormalTok{] }\OtherTok{\textless{}{-}} \DecValTok{0}
\NormalTok{p3\_avg\_lm\_scores[p3\_avg\_lm\_scores }\SpecialCharTok{\textgreater{}} \DecValTok{1}\NormalTok{] }\OtherTok{\textless{}{-}} \DecValTok{1}

\FunctionTok{round}\NormalTok{(p3\_avg\_lm\_scores }\SpecialCharTok{*} \DecValTok{10}\NormalTok{, }\DecValTok{3}\NormalTok{)}
\end{Highlighting}
\end{Shaded}

\begin{verbatim}
##     1     2     3     4     5     6     7     8     9 
## 1.279 0.000 5.808 6.659 0.000 2.352 4.157 0.620 5.598
\end{verbatim}

\begin{Shaded}
\begin{Highlighting}[]
\NormalTok{stats}\SpecialCharTok{::}\FunctionTok{cor}\NormalTok{(p3\_avg\_lm\_scores, ground\_truth}\SpecialCharTok{$}\NormalTok{consensus\_score)}
\end{Highlighting}
\end{Shaded}

\begin{verbatim}
## [1] 0.960387
\end{verbatim}

With this linear combination of only three scores (out of six), we were able to significantly boost the detection to \(\approx0.96\), which implies that combining both methods is of worth for a detection using (inhomogeneous) confidence intervals only. If we only combine the scores of the variable \texttt{REQ} into a model, we still achieve a correlation of \(\approx0.88\). This should probably be preferred to keep the degrees of freedom low, countering overfitting. Using four or more scores goes beyond \(0.97\).

\hypertarget{variable-importance-most-important-scores}{%
\paragraph{\texorpdfstring{Variable importance: most important scores\label{sssec:var-imp-it}}{Variable importance: most important scores}}\label{variable-importance-most-important-scores}}

In the report for source code data (section \ref{sssec:var-imp}), we have previously determined the most important features. We'll do the same computation here. The results will then allow us to compare against the relative importances as determined by source code data.

\begin{Shaded}
\begin{Highlighting}[]
\NormalTok{rfe\_data\_it }\OtherTok{\textless{}{-}} \FunctionTok{cbind}\NormalTok{(p3\_it\_scores, }\FunctionTok{data.frame}\NormalTok{(}\AttributeTok{gt =}\NormalTok{ ground\_truth}\SpecialCharTok{$}\NormalTok{consensus))}
\end{Highlighting}
\end{Shaded}

\begin{Shaded}
\begin{Highlighting}[]
\FunctionTok{library}\NormalTok{(caret, }\AttributeTok{quietly =} \ConstantTok{TRUE}\NormalTok{)}
\FunctionTok{library}\NormalTok{(pls, }\AttributeTok{quietly =} \ConstantTok{TRUE}\NormalTok{)}
\FunctionTok{library}\NormalTok{(randomForest, }\AttributeTok{quietly =} \ConstantTok{TRUE}\NormalTok{)}

\FunctionTok{set.seed}\NormalTok{(}\DecValTok{1337}\NormalTok{)}

\NormalTok{control }\OtherTok{\textless{}{-}}\NormalTok{ caret}\SpecialCharTok{::}\FunctionTok{trainControl}\NormalTok{(}\AttributeTok{method =} \StringTok{"repeatedcv"}\NormalTok{, }\AttributeTok{number =} \DecValTok{10}\NormalTok{, }\AttributeTok{repeats =} \DecValTok{3}\NormalTok{)}
\NormalTok{modelFit\_it\_all }\OtherTok{\textless{}{-}}\NormalTok{ caret}\SpecialCharTok{::}\FunctionTok{train}\NormalTok{(gt }\SpecialCharTok{\textasciitilde{}}\NormalTok{ ., }\AttributeTok{data =}\NormalTok{ rfe\_data\_it, }\AttributeTok{method =} \StringTok{"pls"}\NormalTok{, }\AttributeTok{trControl =}\NormalTok{ control)}

\NormalTok{imp\_it\_all }\OtherTok{\textless{}{-}}\NormalTok{ caret}\SpecialCharTok{::}\FunctionTok{varImp}\NormalTok{(}\AttributeTok{object =}\NormalTok{ modelFit\_it\_all)}
\end{Highlighting}
\end{Shaded}

\begin{figure}[ht!]

{\centering \includegraphics{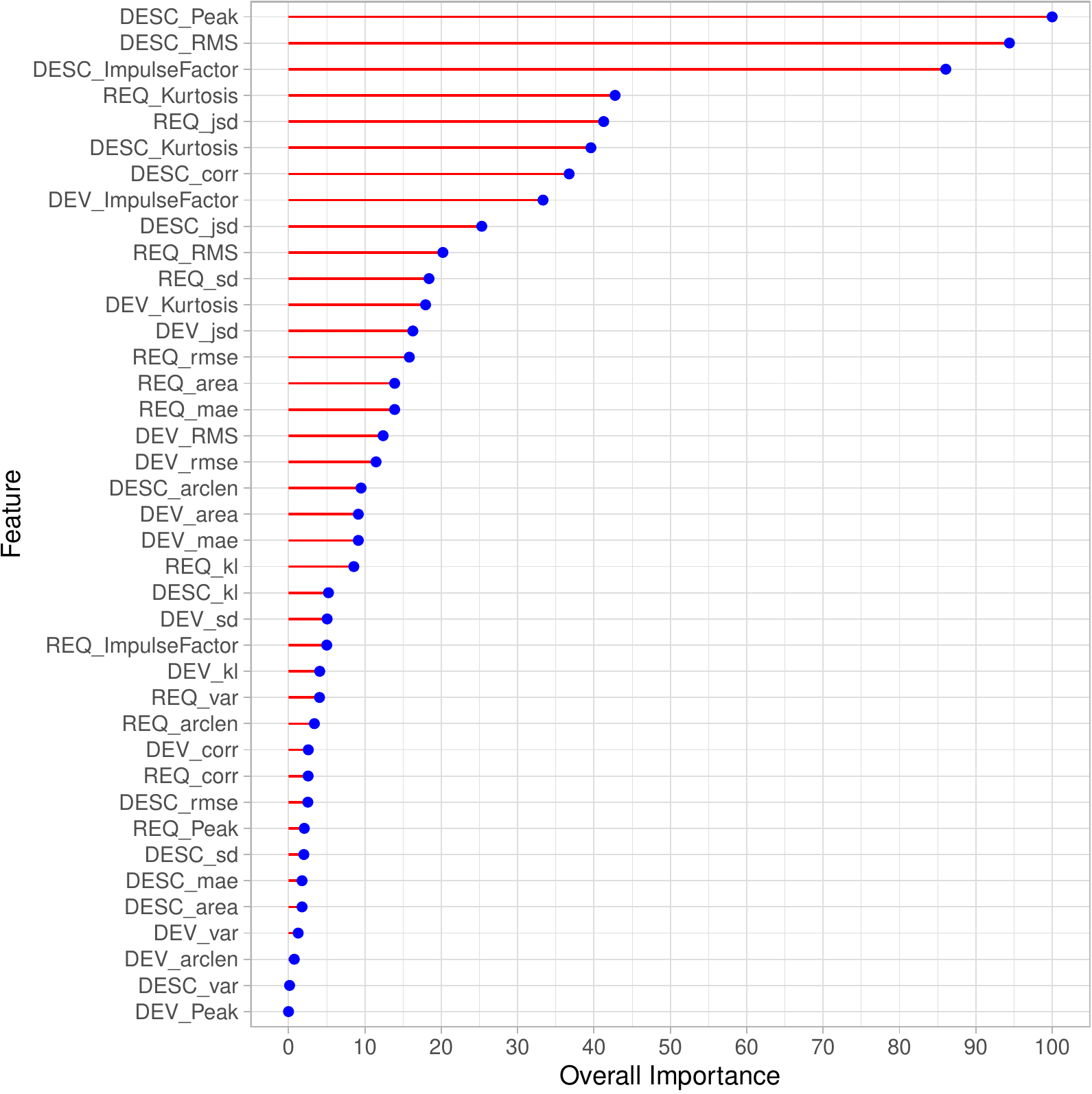} 

}

\caption{Normalized variable-importance for all features (activities) for predicting the ground truth, relative to each other and expressed in percent.}\label{fig:var-imp-all-features}
\end{figure}

The \texttt{DESC}-activity with its three most important and dominating features Peak, RMS, and ImpulseFactor dominates the field, which could be due to, e.g., the \texttt{DESC}-activity being less characteristic than the others, ergo we get a more coherent predictor. Also, the first three features are quite similar to each other. The resulting model uses \textbf{3} components.

For issue-tracking data we do have the scores for each variable (activity) separately, so we will attempt to also compute importances on a per-activity basis. This will also allow us to compare importances across activities. The expectation is that different activities will have a different order and magnitude of importances, since each activity has its own characteristics.

\begin{Shaded}
\begin{Highlighting}[]
\NormalTok{temp }\OtherTok{\textless{}{-}} \FunctionTok{loadResultsOrCompute}\NormalTok{(}\AttributeTok{file =} \StringTok{"../results/modelFit\_it\_3.rds"}\NormalTok{, }\AttributeTok{computeExpr =}\NormalTok{ \{}
  \FunctionTok{set.seed}\NormalTok{(}\DecValTok{48879}\NormalTok{)}

\NormalTok{  temp }\OtherTok{\textless{}{-}} \FunctionTok{colnames}\NormalTok{(rfe\_data\_it)}
\NormalTok{  cols\_req }\OtherTok{\textless{}{-}}\NormalTok{ temp[}\FunctionTok{grep}\NormalTok{(}\AttributeTok{pattern =} \StringTok{"\^{}(req\_)|(gt)"}\NormalTok{, }\AttributeTok{ignore.case =} \ConstantTok{TRUE}\NormalTok{, }\AttributeTok{x =}\NormalTok{ temp)]}
\NormalTok{  cols\_dev }\OtherTok{\textless{}{-}}\NormalTok{ temp[}\FunctionTok{grep}\NormalTok{(}\AttributeTok{pattern =} \StringTok{"\^{}(dev\_)|(gt)"}\NormalTok{, }\AttributeTok{ignore.case =} \ConstantTok{TRUE}\NormalTok{, }\AttributeTok{x =}\NormalTok{ temp)]}
\NormalTok{  cols\_desc }\OtherTok{\textless{}{-}}\NormalTok{ temp[}\FunctionTok{grep}\NormalTok{(}\AttributeTok{pattern =} \StringTok{"\^{}(desc\_)|(gt)"}\NormalTok{, }\AttributeTok{ignore.case =} \ConstantTok{TRUE}\NormalTok{, }\AttributeTok{x =}\NormalTok{ temp)]}

  \FunctionTok{list}\NormalTok{(}\AttributeTok{req =}\NormalTok{ caret}\SpecialCharTok{::}\FunctionTok{train}\NormalTok{(gt }\SpecialCharTok{\textasciitilde{}}\NormalTok{ ., }\AttributeTok{data =}\NormalTok{ rfe\_data\_it[cols\_req], }\AttributeTok{method =} \StringTok{"pls"}\NormalTok{,}
    \AttributeTok{trControl =}\NormalTok{ control), }\AttributeTok{dev =}\NormalTok{ caret}\SpecialCharTok{::}\FunctionTok{train}\NormalTok{(gt }\SpecialCharTok{\textasciitilde{}}\NormalTok{ ., }\AttributeTok{data =}\NormalTok{ rfe\_data\_it[cols\_dev],}
    \AttributeTok{method =} \StringTok{"pls"}\NormalTok{, }\AttributeTok{trControl =}\NormalTok{ control), }\AttributeTok{desc =}\NormalTok{ caret}\SpecialCharTok{::}\FunctionTok{train}\NormalTok{(gt }\SpecialCharTok{\textasciitilde{}}\NormalTok{ ., }\AttributeTok{data =}\NormalTok{ rfe\_data\_it[cols\_desc],}
    \AttributeTok{method =} \StringTok{"pls"}\NormalTok{, }\AttributeTok{trControl =}\NormalTok{ control))}
\NormalTok{\})}

\NormalTok{imp\_it\_req }\OtherTok{\textless{}{-}}\NormalTok{ caret}\SpecialCharTok{::}\FunctionTok{varImp}\NormalTok{(}\AttributeTok{object =}\NormalTok{ temp}\SpecialCharTok{$}\NormalTok{req)}
\NormalTok{imp\_it\_dev }\OtherTok{\textless{}{-}}\NormalTok{ caret}\SpecialCharTok{::}\FunctionTok{varImp}\NormalTok{(}\AttributeTok{object =}\NormalTok{ temp}\SpecialCharTok{$}\NormalTok{dev)}
\NormalTok{imp\_it\_desc }\OtherTok{\textless{}{-}}\NormalTok{ caret}\SpecialCharTok{::}\FunctionTok{varImp}\NormalTok{(}\AttributeTok{object =}\NormalTok{ temp}\SpecialCharTok{$}\NormalTok{desc)}
\end{Highlighting}
\end{Shaded}

\begin{figure}[ht!]

{\centering \includegraphics{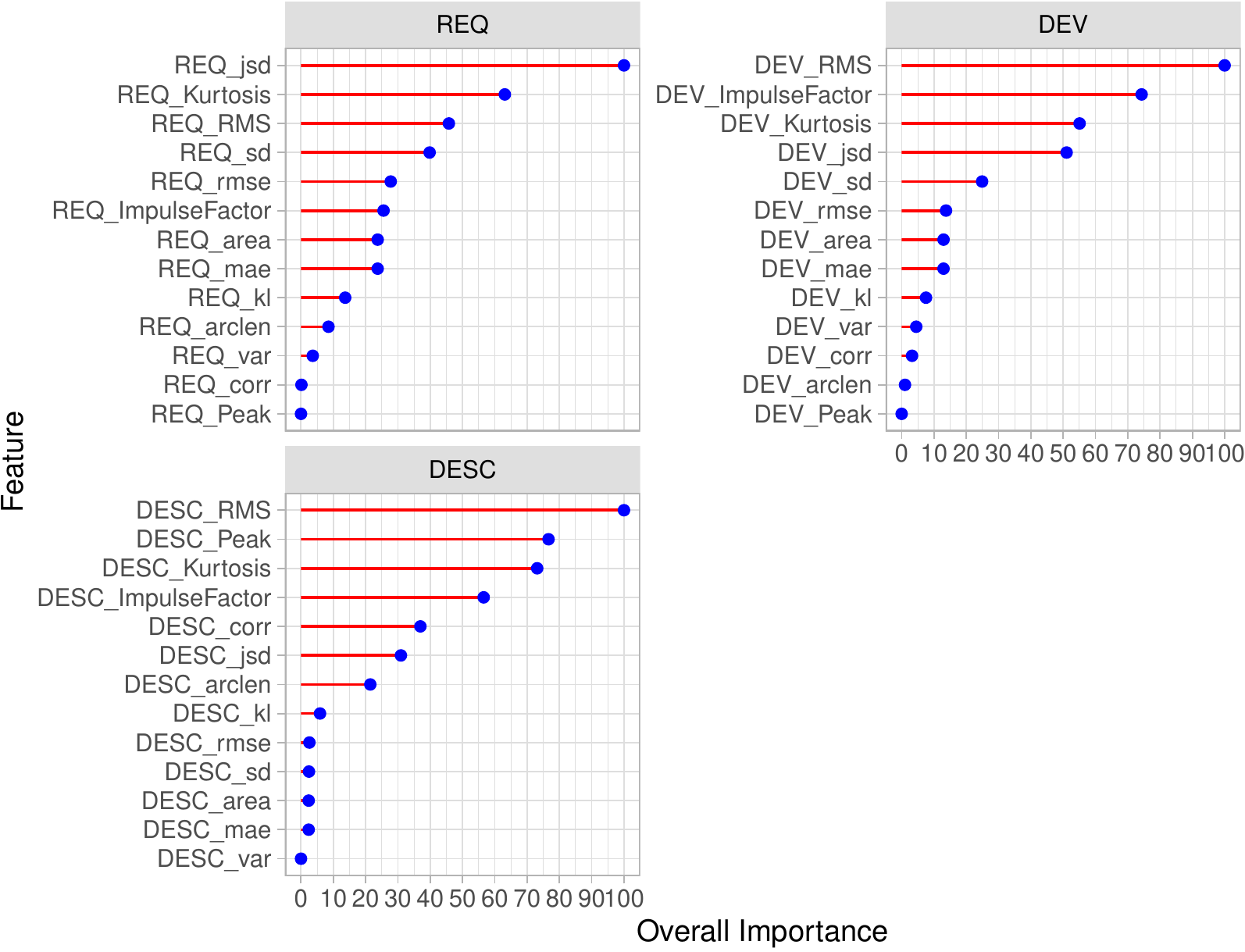} 

}

\caption{Variable-importance for all features (activities) separately for predicting the ground truth on a per-activity basis (normalized and relative).}\label{fig:var-imp-all-features-sep}
\end{figure}

When comparing figures \ref{fig:var-imp-all-features} and \ref{fig:var-imp-all-features-sep}, we observe that the importance of features varies if we were to predict a ground truth on a per-activity basis. We also observe some features being important in all cases, such as RMS, or the Jensen--Shannon divergence.

\hypertarget{arbitrary-interval-scores-computing}{%
\paragraph{\texorpdfstring{Arbitrary-interval scores computing\label{ssec:arbint-scores}}{Arbitrary-interval scores computing}}\label{arbitrary-interval-scores-computing}}

Up to now, we presented two methods to compute losses/scores over the closed interval \([0,1]\). We then have fitted a linear model to weigh, scale and translate the scores. We could evaluate how well this linear model performs if we select arbitrary intervals and compute the scores, but the expected is that smaller and earlier intervals would perform poorly. We will compute some the scores for some arbitrary intervals and then evaluate this. Then, we will attempt to learn a non-linear mapping that should correct for these residuals.

\begin{Shaded}
\begin{Highlighting}[]
\NormalTok{p3\_avg\_area\_scores\_arb\_int }\OtherTok{\textless{}{-}} \FunctionTok{loadResultsOrCompute}\NormalTok{(}\AttributeTok{file =} \StringTok{"../results/p3\_avg\_area\_scores\_arb\_int.rds"}\NormalTok{,}
  \AttributeTok{computeExpr =}\NormalTok{ \{}
    \FunctionTok{doWithParallelCluster}\NormalTok{(}\AttributeTok{numCores =} \FunctionTok{min}\NormalTok{(}\DecValTok{123}\NormalTok{, parallel}\SpecialCharTok{::}\FunctionTok{detectCores}\NormalTok{()), }\AttributeTok{expr =}\NormalTok{ \{}
      \FunctionTok{library}\NormalTok{(foreach)}

\NormalTok{      tempm }\OtherTok{\textless{}{-}} \FunctionTok{matrix}\NormalTok{(}\AttributeTok{ncol =} \DecValTok{3}\NormalTok{, }\AttributeTok{byrow =} \ConstantTok{TRUE}\NormalTok{, }\AttributeTok{data =} \FunctionTok{sapply}\NormalTok{(}\AttributeTok{X =} \FunctionTok{seq\_len}\NormalTok{(}\DecValTok{400} \SpecialCharTok{*}
        \FunctionTok{length}\NormalTok{(all\_signals)), }\AttributeTok{FUN =} \ControlFlowTok{function}\NormalTok{(x) \{}
        \FunctionTok{set.seed}\NormalTok{(}\FunctionTok{bitwXor}\NormalTok{(}\DecValTok{1337}\NormalTok{, x))}

\NormalTok{        t }\OtherTok{\textless{}{-}} \FunctionTok{rep}\NormalTok{(}\ConstantTok{NA\_real\_}\NormalTok{, }\DecValTok{2}\NormalTok{)}
        \ControlFlowTok{if}\NormalTok{ (x }\SpecialCharTok{\textless{}=} \DecValTok{400}\NormalTok{) \{}
\NormalTok{          t }\OtherTok{\textless{}{-}} \FunctionTok{c}\NormalTok{(}\DecValTok{0}\NormalTok{, }\FunctionTok{runif}\NormalTok{(}\AttributeTok{n =} \DecValTok{1}\NormalTok{, }\AttributeTok{min =} \FloatTok{0.01}\NormalTok{))}
\NormalTok{        \} }\ControlFlowTok{else} \ControlFlowTok{if}\NormalTok{ (x }\SpecialCharTok{\textless{}=} \DecValTok{3200}\NormalTok{) \{}
          \ControlFlowTok{while}\NormalTok{ (}\ConstantTok{TRUE}\NormalTok{) \{}
\NormalTok{          t }\OtherTok{\textless{}{-}} \FunctionTok{sort}\NormalTok{(}\FunctionTok{runif}\NormalTok{(}\AttributeTok{n =} \DecValTok{2}\NormalTok{))}
          \ControlFlowTok{if}\NormalTok{ ((t[}\DecValTok{2}\NormalTok{] }\SpecialCharTok{{-}}\NormalTok{ t[}\DecValTok{1}\NormalTok{]) }\SpecialCharTok{\textgreater{}} \FloatTok{0.01}\NormalTok{) \{}
            \ControlFlowTok{break}
\NormalTok{          \}}
\NormalTok{          \}}
\NormalTok{        \} }\ControlFlowTok{else}\NormalTok{ \{}
\NormalTok{          t }\OtherTok{\textless{}{-}} \FunctionTok{c}\NormalTok{(}\FunctionTok{runif}\NormalTok{(}\AttributeTok{n =} \DecValTok{1}\NormalTok{, }\AttributeTok{max =} \FloatTok{0.99}\NormalTok{), }\DecValTok{1}\NormalTok{)}
\NormalTok{        \}}

\NormalTok{        pId }\OtherTok{\textless{}{-}}\NormalTok{ x}\SpecialCharTok{\%\%}\FunctionTok{length}\NormalTok{(all\_signals)}
        \ControlFlowTok{if}\NormalTok{ (pId }\SpecialCharTok{==} \DecValTok{0}\NormalTok{) \{}
\NormalTok{          pId }\OtherTok{\textless{}{-}} \FunctionTok{length}\NormalTok{(all\_signals)}
\NormalTok{        \}}

        \FunctionTok{c}\NormalTok{(t, pId)}
\NormalTok{      \}))}
\NormalTok{      tempm[, }\DecValTok{3}\NormalTok{] }\OtherTok{\textless{}{-}} \FunctionTok{sample}\NormalTok{(tempm[, }\DecValTok{3}\NormalTok{])  }\CommentTok{\# randomize projects}

\NormalTok{      temp }\OtherTok{\textless{}{-}}\NormalTok{ foreach}\SpecialCharTok{::}\FunctionTok{foreach}\NormalTok{(}\AttributeTok{permIdx =} \DecValTok{1}\SpecialCharTok{:}\FunctionTok{nrow}\NormalTok{(tempm), }\AttributeTok{.inorder =} \ConstantTok{TRUE}\NormalTok{, }\AttributeTok{.combine =}\NormalTok{ rbind) }\SpecialCharTok{\%dopar\%}
\NormalTok{        \{}
\NormalTok{          p }\OtherTok{\textless{}{-}}\NormalTok{ tempm[permIdx, ]}
\NormalTok{          pId }\OtherTok{\textless{}{-}} \FunctionTok{paste0}\NormalTok{(}\StringTok{"Project"}\NormalTok{, p[}\DecValTok{3}\NormalTok{])}

\NormalTok{          temp\_ci }\OtherTok{\textless{}{-}} \FunctionTok{p3\_avg\_ci\_scores\_compute}\NormalTok{(}\AttributeTok{pId =}\NormalTok{ pId, }\AttributeTok{x1 =}\NormalTok{ p[}\DecValTok{1}\NormalTok{], }\AttributeTok{x2 =}\NormalTok{ p[}\DecValTok{2}\NormalTok{])}
          \FunctionTok{colnames}\NormalTok{(temp\_ci) }\OtherTok{\textless{}{-}} \FunctionTok{paste0}\NormalTok{(}\FunctionTok{colnames}\NormalTok{(temp\_ci), }\StringTok{"\_ci"}\NormalTok{)}
\NormalTok{          temp\_ci}\SpecialCharTok{$}\NormalTok{begin }\OtherTok{\textless{}{-}}\NormalTok{ p[}\DecValTok{1}\NormalTok{]}
\NormalTok{          temp\_ci}\SpecialCharTok{$}\NormalTok{end }\OtherTok{\textless{}{-}}\NormalTok{ p[}\DecValTok{2}\NormalTok{]}

\NormalTok{          temp\_area }\OtherTok{\textless{}{-}} \FunctionTok{p3\_avg\_area\_scores\_compute}\NormalTok{(}\AttributeTok{pId =}\NormalTok{ pId, }\AttributeTok{x1 =}\NormalTok{ p[}\DecValTok{1}\NormalTok{], }\AttributeTok{x2 =}\NormalTok{ p[}\DecValTok{2}\NormalTok{])}
          \FunctionTok{colnames}\NormalTok{(temp\_area) }\OtherTok{\textless{}{-}} \FunctionTok{paste0}\NormalTok{(}\FunctionTok{colnames}\NormalTok{(temp\_area), }\StringTok{"\_area"}\NormalTok{)}
          \StringTok{\textasciigrave{}}\AttributeTok{rownames\textless{}{-}}\StringTok{\textasciigrave{}}\NormalTok{(}\FunctionTok{cbind}\NormalTok{(temp\_ci, temp\_area), }\ConstantTok{NULL}\NormalTok{)}
\NormalTok{        \}}

\NormalTok{      temp}\SpecialCharTok{$}\NormalTok{gt }\OtherTok{\textless{}{-}} \FunctionTok{sapply}\NormalTok{(}\AttributeTok{X =}\NormalTok{ temp}\SpecialCharTok{$}\NormalTok{Project\_ci, }\AttributeTok{FUN =} \ControlFlowTok{function}\NormalTok{(p) \{}
\NormalTok{        ground\_truth[ground\_truth}\SpecialCharTok{$}\NormalTok{project }\SpecialCharTok{==} \FunctionTok{paste0}\NormalTok{(}\StringTok{"project\_"}\NormalTok{, }\FunctionTok{substr}\NormalTok{(p,}
          \FunctionTok{nchar}\NormalTok{(p), }\FunctionTok{nchar}\NormalTok{(p))), ]}\SpecialCharTok{$}\NormalTok{consensus}
\NormalTok{      \})}
\NormalTok{      temp}
\NormalTok{    \})}
\NormalTok{  \})}
\end{Highlighting}
\end{Shaded}

Now that we have produced some labeled training data, we can attempt to learn a mapping between the intervals and scores, and the ground truth. We use a Random forest, and as outer resampling method a ten times repeated ten-fold cross validation. We apply z-standardization to the data.

\begin{Shaded}
\begin{Highlighting}[]
\NormalTok{p3\_avg\_arb\_int\_pred }\OtherTok{\textless{}{-}} \FunctionTok{loadResultsOrCompute}\NormalTok{(}\AttributeTok{file =} \StringTok{"../results/p3\_avg\_arb\_int\_pred.rds"}\NormalTok{,}
  \AttributeTok{computeExpr =}\NormalTok{ \{}
    \FunctionTok{library}\NormalTok{(caret)}
    \FunctionTok{set.seed}\NormalTok{(}\DecValTok{1337}\NormalTok{)}

\NormalTok{    temp }\OtherTok{\textless{}{-}}\NormalTok{ p3\_avg\_area\_scores\_arb\_int[}\FunctionTok{complete.cases}\NormalTok{(p3\_avg\_area\_scores\_arb\_int),}
\NormalTok{      ]}
\NormalTok{    temp }\OtherTok{\textless{}{-}}\NormalTok{ temp[, }\FunctionTok{c}\NormalTok{(}\StringTok{"gt"}\NormalTok{, }\StringTok{"begin"}\NormalTok{, }\StringTok{"end"}\NormalTok{, }\StringTok{"REQ\_ci"}\NormalTok{, }\StringTok{"DEV\_ci"}\NormalTok{, }\StringTok{"DESC\_ci"}\NormalTok{, }\StringTok{"REQ\_area"}\NormalTok{,}
      \StringTok{"DEV\_area"}\NormalTok{, }\StringTok{"DESC\_area"}\NormalTok{)]}

\NormalTok{    inTraining }\OtherTok{\textless{}{-}}\NormalTok{ caret}\SpecialCharTok{::}\FunctionTok{createDataPartition}\NormalTok{(temp}\SpecialCharTok{$}\NormalTok{gt, }\AttributeTok{p =} \FloatTok{0.8}\NormalTok{, }\AttributeTok{list =} \ConstantTok{FALSE}\NormalTok{)}
\NormalTok{    training }\OtherTok{\textless{}{-}}\NormalTok{ temp[inTraining, ]}
\NormalTok{    testing }\OtherTok{\textless{}{-}}\NormalTok{ temp[}\SpecialCharTok{{-}}\NormalTok{inTraining, ]}

\NormalTok{    fitControl }\OtherTok{\textless{}{-}}\NormalTok{ caret}\SpecialCharTok{::}\FunctionTok{trainControl}\NormalTok{(}\AttributeTok{method =} \StringTok{"repeatedcv"}\NormalTok{, }\AttributeTok{number =} \DecValTok{10}\NormalTok{, }\AttributeTok{repeats =} \DecValTok{10}\NormalTok{)}

    \FunctionTok{doWithParallelCluster}\NormalTok{(}\AttributeTok{expr =}\NormalTok{ \{}
      \FunctionTok{list}\NormalTok{(}\AttributeTok{fit =}\NormalTok{ caret}\SpecialCharTok{::}\FunctionTok{train}\NormalTok{(gt }\SpecialCharTok{\textasciitilde{}}\NormalTok{ ., }\AttributeTok{data =}\NormalTok{ training, }\AttributeTok{method =} \StringTok{"ranger"}\NormalTok{, }\AttributeTok{trControl =}\NormalTok{ fitControl,}
        \AttributeTok{preProcess =} \FunctionTok{c}\NormalTok{(}\StringTok{"center"}\NormalTok{, }\StringTok{"scale"}\NormalTok{)), }\AttributeTok{train =}\NormalTok{ training, }\AttributeTok{test =}\NormalTok{ testing)}
\NormalTok{    \})}
\NormalTok{  \})}
\end{Highlighting}
\end{Shaded}

\begin{verbatim}
## Random Forest 
## 
## 2801 samples
##    8 predictor
## 
## Pre-processing: centered (8), scaled (8) 
## Resampling: Cross-Validated (10 fold, repeated 10 times) 
## Summary of sample sizes: 2519, 2521, 2522, 2520, 2521, 2522, ... 
## Resampling results across tuning parameters:
## 
##   mtry  splitrule   RMSE       Rsquared   MAE      
##   2     variance    0.4824646  0.9695206  0.2245037
##   2     extratrees  0.4951226  0.9695519  0.2548989
##   5     variance    0.4400933  0.9728952  0.1645185
##   5     extratrees  0.3902164  0.9793062  0.1637342
##   8     variance    0.4715037  0.9684441  0.1617345
##   8     extratrees  0.3681580  0.9810770  0.1433879
## 
## Tuning parameter 'min.node.size' was held constant at a value of 5
## RMSE was used to select the optimal model using the smallest value.
## The final values used for the model were mtry = 8, splitrule = extratrees
##  and min.node.size = 5.
\end{verbatim}

\begin{verbatim}
##      corr       MAE      RMSE 
## 0.9920896 0.1222479 0.3472935
\end{verbatim}

With a correlation of \(\approx0.99\), an MAE of \(\approx0.12\) and RMSE of \(\approx0.35\) I suppose we already have a predictor that is quite usable. Remember that we trained on the ground truth in the range \([0,10]\), so these mean deviations are probably already acceptable for some use cases. Also, I only produced 400 random ranges/scores per project, so in total we have 3600 records in the data, and the split was made at \(0.8\) (during each fold of the cross-validation, the default split of \(0.75\) was used). With about one quarter of the amount of that data, we get similarly good results. Also, this model is likely overfitting, because when I leave out \texttt{begin} and \texttt{end} of each interval, the results are similarly good. However, the whole purpose of this model fitting here is to merely demonstrate that one could fit this kind of model and then make accurate predictions over arbitrary time intervals.

So, if we take the MAE, then these results mean that for any arbitrarily chosen interval, the deviation from the computed scores \(\mathit{L}^{\text{avgconf}}\) and \(\mathit{L}^{\text{areadist2}}\) is less than \(\pm\approx0.29\), and that on a scale of \([0,10]\). Since this is \(<0.5\), it means that we should get an even better result when rounding (at least for the MAE), and indeed:

\begin{Shaded}
\begin{Highlighting}[]
\NormalTok{Metrics}\SpecialCharTok{::}\FunctionTok{mae}\NormalTok{(p3\_avg\_arb\_int\_pred}\SpecialCharTok{$}\NormalTok{test}\SpecialCharTok{$}\NormalTok{gt, }\FunctionTok{round}\NormalTok{(stats}\SpecialCharTok{::}\FunctionTok{predict}\NormalTok{(}\AttributeTok{object =}\NormalTok{ p3\_avg\_arb\_int\_pred}\SpecialCharTok{$}\NormalTok{fit,}
  \AttributeTok{newdata =}\NormalTok{ p3\_avg\_arb\_int\_pred}\SpecialCharTok{$}\NormalTok{test)))}
\end{Highlighting}
\end{Shaded}

\begin{verbatim}
## [1] 0.06294707
\end{verbatim}

While it is not shown here, we can convert the rounded prediction to a factor and then show a confusion matrix. The diagonal in this case is well filled, i.e., only few mismatches exist. All of the mismatches are also in the next neighboring cell, which means that not a single prediction is more off than a single level.

If we run the same routine as a classification task, then we typically achieve \(\approx99\)\% accuracy, and very high Kappa of \(>0.98\) (\(1\) is perfect). However, we cannot train for levels of ground truth currently not present in our data, which means that this kind of model will not generalize well to new data. However, all this was just a demonstration for the case of arbitrary-interval scores computing.

\begin{Shaded}
\begin{Highlighting}[]
\NormalTok{p3\_avg\_arb\_int\_pred\_cls }\OtherTok{\textless{}{-}} \FunctionTok{loadResultsOrCompute}\NormalTok{(}\AttributeTok{file =} \StringTok{"../results/p3\_avg\_arb\_int\_pred\_cls.rds"}\NormalTok{,}
  \AttributeTok{computeExpr =}\NormalTok{ \{}
    \FunctionTok{library}\NormalTok{(caret)}
    \FunctionTok{set.seed}\NormalTok{(}\DecValTok{1337}\NormalTok{)}

\NormalTok{    temp }\OtherTok{\textless{}{-}}\NormalTok{ p3\_avg\_area\_scores\_arb\_int[}\FunctionTok{complete.cases}\NormalTok{(p3\_avg\_area\_scores\_arb\_int),}
\NormalTok{      ]}
\NormalTok{    temp }\OtherTok{\textless{}{-}}\NormalTok{ temp[, }\FunctionTok{c}\NormalTok{(}\StringTok{"gt"}\NormalTok{, }\StringTok{"begin"}\NormalTok{, }\StringTok{"end"}\NormalTok{, }\StringTok{"REQ\_ci"}\NormalTok{, }\StringTok{"DEV\_ci"}\NormalTok{, }\StringTok{"DESC\_ci"}\NormalTok{, }\StringTok{"REQ\_area"}\NormalTok{,}
      \StringTok{"DEV\_area"}\NormalTok{, }\StringTok{"DESC\_area"}\NormalTok{)]}
\NormalTok{    temp}\SpecialCharTok{$}\NormalTok{gt }\OtherTok{\textless{}{-}} \FunctionTok{factor}\NormalTok{(}\AttributeTok{ordered =} \ConstantTok{TRUE}\NormalTok{, }\AttributeTok{x =} \FunctionTok{as.character}\NormalTok{(temp}\SpecialCharTok{$}\NormalTok{gt), }\AttributeTok{levels =} \FunctionTok{sort}\NormalTok{(}\FunctionTok{unique}\NormalTok{(}\FunctionTok{as.character}\NormalTok{(temp}\SpecialCharTok{$}\NormalTok{gt))))}

\NormalTok{    inTraining }\OtherTok{\textless{}{-}}\NormalTok{ caret}\SpecialCharTok{::}\FunctionTok{createDataPartition}\NormalTok{(temp}\SpecialCharTok{$}\NormalTok{gt, }\AttributeTok{p =} \FloatTok{0.8}\NormalTok{, }\AttributeTok{list =} \ConstantTok{FALSE}\NormalTok{)}
\NormalTok{    training }\OtherTok{\textless{}{-}}\NormalTok{ temp[inTraining, ]}
\NormalTok{    testing }\OtherTok{\textless{}{-}}\NormalTok{ temp[}\SpecialCharTok{{-}}\NormalTok{inTraining, ]}

\NormalTok{    fitControl }\OtherTok{\textless{}{-}}\NormalTok{ caret}\SpecialCharTok{::}\FunctionTok{trainControl}\NormalTok{(}\AttributeTok{method =} \StringTok{"repeatedcv"}\NormalTok{, }\AttributeTok{number =} \DecValTok{10}\NormalTok{, }\AttributeTok{repeats =} \DecValTok{10}\NormalTok{)}

    \FunctionTok{doWithParallelCluster}\NormalTok{(}\AttributeTok{numCores =} \DecValTok{10}\NormalTok{, }\AttributeTok{expr =}\NormalTok{ \{}
      \FunctionTok{list}\NormalTok{(}\AttributeTok{fit =}\NormalTok{ caret}\SpecialCharTok{::}\FunctionTok{train}\NormalTok{(gt }\SpecialCharTok{\textasciitilde{}}\NormalTok{ ., }\AttributeTok{data =}\NormalTok{ training, }\AttributeTok{method =} \StringTok{"ranger"}\NormalTok{, }\AttributeTok{trControl =}\NormalTok{ fitControl,}
        \AttributeTok{preProcess =} \FunctionTok{c}\NormalTok{(}\StringTok{"center"}\NormalTok{, }\StringTok{"scale"}\NormalTok{)), }\AttributeTok{train =}\NormalTok{ training, }\AttributeTok{test =}\NormalTok{ testing)}
\NormalTok{    \})}
\NormalTok{  \})}
\end{Highlighting}
\end{Shaded}

\begin{verbatim}
## Confusion Matrix and Statistics
## 
##           Reference
## Prediction   0   1   2   3   5   6   8
##          0 153   2   0   0   0   0   0
##          1   0 153   1   1   0   0   0
##          2   1   0  77   0   0   0   0
##          3   0   0   0  75   0   2   0
##          5   0   0   0   0  78   0   0
##          6   0   0   0   0   0  77   0
##          8   2   0   0   0   0   0  76
## 
## Overall Statistics
##                                           
##                Accuracy : 0.9871          
##                  95% CI : (0.9757, 0.9941)
##     No Information Rate : 0.2235          
##     P-Value [Acc > NIR] : < 2.2e-16       
##                                           
##                   Kappa : 0.9846          
##                                           
##  Mcnemar's Test P-Value : NA              
## 
## Statistics by Class:
## 
##                      Class: 0 Class: 1 Class: 2 Class: 3 Class: 5 Class: 6
## Sensitivity            0.9808   0.9871   0.9872   0.9868   1.0000   0.9747
## Specificity            0.9963   0.9963   0.9984   0.9968   1.0000   1.0000
## Pos Pred Value         0.9871   0.9871   0.9872   0.9740   1.0000   1.0000
## Neg Pred Value         0.9945   0.9963   0.9984   0.9984   1.0000   0.9968
## Prevalence             0.2235   0.2221   0.1117   0.1089   0.1117   0.1132
## Detection Rate         0.2192   0.2192   0.1103   0.1074   0.1117   0.1103
## Detection Prevalence   0.2221   0.2221   0.1117   0.1103   0.1117   0.1103
## Balanced Accuracy      0.9885   0.9917   0.9928   0.9918   1.0000   0.9873
##                      Class: 8
## Sensitivity            1.0000
## Specificity            0.9968
## Pos Pred Value         0.9744
## Neg Pred Value         1.0000
## Prevalence             0.1089
## Detection Rate         0.1089
## Detection Prevalence   0.1117
## Balanced Accuracy      0.9984
\end{verbatim}

\hypertarget{process-alignment-dtw-optimization-srbtaw}{%
\paragraph{\texorpdfstring{Process alignment: DTW, Optimization, srBTAW\label{ssec:proc-align-compute}}{Process alignment: DTW, Optimization, srBTAW}}\label{process-alignment-dtw-optimization-srbtaw}}

In the previous subsection we learned a predictive model that requires as input the start and end over which some scores were obtained (it is required since the available ground truth is constant). In a real-world scenario we might not know where exactly in the process we are. In such cases, it might be helpful to find the best alignment of some observed process with the process model. As an example, we will be using the processes from project three, and slice out the interval \([0.25,0.45]\).

\begin{Shaded}
\begin{Highlighting}[]
\NormalTok{slice\_supp }\OtherTok{\textless{}{-}} \FunctionTok{c}\NormalTok{(}\FloatTok{0.25}\NormalTok{, }\FloatTok{0.45}\NormalTok{)}
\NormalTok{use\_p }\OtherTok{\textless{}{-}}\NormalTok{ all\_signals}\SpecialCharTok{$}\NormalTok{Project3}

\NormalTok{get\_slice\_func }\OtherTok{\textless{}{-}} \ControlFlowTok{function}\NormalTok{(varname) \{}
\NormalTok{  use\_func }\OtherTok{\textless{}{-}}\NormalTok{ use\_p[[varname]]}\SpecialCharTok{$}\FunctionTok{get0Function}\NormalTok{()}
  \ControlFlowTok{function}\NormalTok{(x) }\FunctionTok{sapply}\NormalTok{(}\AttributeTok{X =}\NormalTok{ x, }\AttributeTok{FUN =} \ControlFlowTok{function}\NormalTok{(x\_) \{}
    \FunctionTok{use\_func}\NormalTok{((slice\_supp[}\DecValTok{2}\NormalTok{] }\SpecialCharTok{{-}}\NormalTok{ slice\_supp[}\DecValTok{1}\NormalTok{]) }\SpecialCharTok{*}\NormalTok{ x\_ }\SpecialCharTok{+}\NormalTok{ slice\_supp[}\DecValTok{1}\NormalTok{])}
\NormalTok{  \})}
\NormalTok{\}}

\CommentTok{\# Re{-}define the slices to support [0,1], to make the problem harder and more}
\CommentTok{\# realistic.}
\NormalTok{slice\_req }\OtherTok{\textless{}{-}}\NormalTok{ Signal}\SpecialCharTok{$}\FunctionTok{new}\NormalTok{(}\AttributeTok{name =} \StringTok{"REQ\_WP"}\NormalTok{, }\AttributeTok{func =} \FunctionTok{get\_slice\_func}\NormalTok{(}\StringTok{"REQ"}\NormalTok{), }\AttributeTok{support =} \FunctionTok{c}\NormalTok{(}\DecValTok{0}\NormalTok{,}
  \DecValTok{1}\NormalTok{), }\AttributeTok{isWp =} \ConstantTok{TRUE}\NormalTok{)}
\NormalTok{slice\_dev }\OtherTok{\textless{}{-}}\NormalTok{ Signal}\SpecialCharTok{$}\FunctionTok{new}\NormalTok{(}\AttributeTok{name =} \StringTok{"DEV\_WP"}\NormalTok{, }\AttributeTok{func =} \FunctionTok{get\_slice\_func}\NormalTok{(}\StringTok{"DEV"}\NormalTok{), }\AttributeTok{support =} \FunctionTok{c}\NormalTok{(}\DecValTok{0}\NormalTok{,}
  \DecValTok{1}\NormalTok{), }\AttributeTok{isWp =} \ConstantTok{TRUE}\NormalTok{)}
\NormalTok{slice\_desc }\OtherTok{\textless{}{-}}\NormalTok{ Signal}\SpecialCharTok{$}\FunctionTok{new}\NormalTok{(}\AttributeTok{name =} \StringTok{"DESC\_WP"}\NormalTok{, }\AttributeTok{func =} \FunctionTok{get\_slice\_func}\NormalTok{(}\StringTok{"DESC"}\NormalTok{), }\AttributeTok{support =} \FunctionTok{c}\NormalTok{(}\DecValTok{0}\NormalTok{,}
  \DecValTok{1}\NormalTok{), }\AttributeTok{isWp =} \ConstantTok{TRUE}\NormalTok{)}
\end{Highlighting}
\end{Shaded}

\hypertarget{optimization-based-approach}{%
\subparagraph{Optimization-based approach}\label{optimization-based-approach}}

Similar to how we found the optimum in section \ref{ssec:optim-t1t2}, we can pose an optimization problem:

\[
\begin{aligned}
  {[}a,b{]}\dots&\;\text{support of the observed process (here:}\,[0.25,0.45]\text{),}
  \\[1ex]
  \mathcal{L}_{\text{begin}}(t_{\text{begin}})=&\;\left\lVert\,\operatorname{\overline{req}}(t_{\text{begin}})-\operatorname{req}(a)\,\right\rVert+\left\lVert\,\operatorname{\overline{dev}}(t_{\text{begin}})-\operatorname{dev}(a)\,\right\rVert+\left\lVert\,\operatorname{\overline{desc}}(t_{\text{begin}})-\operatorname{desc}(a)\,\right\rVert\,\text{,}
  \\[1ex]
  \mathcal{L}_{\text{end}}(t_{\text{end}})=&\;\left\lVert\,\operatorname{\overline{req}}(t_{\text{end}})-\operatorname{req}(b)\,\right\rVert+\left\lVert\,\operatorname{\overline{dev}}(t_{\text{end}})-\operatorname{dev}(b)\,\right\rVert+\left\lVert\,\operatorname{\overline{desc}}(t_{\text{end}})-\operatorname{desc}(b)\,\right\rVert\,\text{,}
  \\[1ex]
  \mathcal{L}(t_{\text{begin}},t_{\text{end}})=&\;\mathcal{L}_{\text{begin}}(t_{\text{begin}})+\mathcal{L}_{\text{end}}(t_{\text{end}})\,\text{,}
  \\[1ex]
  \min_{t_{\text{begin}},t_{\text{end}}\in R}&\;\mathcal{L}(t_{\text{begin}},t_{\text{end}})\,\text{,}
  \\[1ex]
  \text{subject to}&\;0\leq t_{\text{begin}}<t_{\text{end}}\leq1\;\text{.}
\end{aligned}
\]

\begin{Shaded}
\begin{Highlighting}[]
\NormalTok{req\_ab }\OtherTok{\textless{}{-}}\NormalTok{ (slice\_req}\SpecialCharTok{$}\FunctionTok{get0Function}\NormalTok{())(}\FunctionTok{c}\NormalTok{(}\DecValTok{0}\NormalTok{, }\DecValTok{1}\NormalTok{))}
\NormalTok{dev\_ab }\OtherTok{\textless{}{-}}\NormalTok{ (slice\_dev}\SpecialCharTok{$}\FunctionTok{get0Function}\NormalTok{())(}\FunctionTok{c}\NormalTok{(}\DecValTok{0}\NormalTok{, }\DecValTok{1}\NormalTok{))}
\NormalTok{desc\_ab }\OtherTok{\textless{}{-}}\NormalTok{ (slice\_desc}\SpecialCharTok{$}\FunctionTok{get0Function}\NormalTok{())(}\FunctionTok{c}\NormalTok{(}\DecValTok{0}\NormalTok{, }\DecValTok{1}\NormalTok{))}

\FunctionTok{set.seed}\NormalTok{(}\DecValTok{1}\NormalTok{)}

\NormalTok{res }\OtherTok{\textless{}{-}} \FunctionTok{nloptr}\NormalTok{(}\AttributeTok{x0 =} \FunctionTok{c}\NormalTok{(}\FloatTok{0.5}\NormalTok{, }\FloatTok{0.5}\NormalTok{), }\AttributeTok{eval\_f =} \ControlFlowTok{function}\NormalTok{(x) \{}
\NormalTok{  a }\OtherTok{\textless{}{-}}\NormalTok{ x[}\DecValTok{1}\NormalTok{]}
\NormalTok{  b }\OtherTok{\textless{}{-}}\NormalTok{ x[}\DecValTok{2}\NormalTok{]}
  \FunctionTok{abs}\NormalTok{(}\FunctionTok{req\_p3}\NormalTok{(a) }\SpecialCharTok{{-}}\NormalTok{ req\_ab[}\DecValTok{1}\NormalTok{]) }\SpecialCharTok{+} \FunctionTok{abs}\NormalTok{(}\FunctionTok{dev\_p3}\NormalTok{(a) }\SpecialCharTok{{-}}\NormalTok{ dev\_ab[}\DecValTok{1}\NormalTok{]) }\SpecialCharTok{+} \FunctionTok{abs}\NormalTok{(}\FunctionTok{desc\_p3}\NormalTok{(a) }\SpecialCharTok{{-}}\NormalTok{ desc\_ab[}\DecValTok{1}\NormalTok{]) }\SpecialCharTok{+}
    \FunctionTok{abs}\NormalTok{(}\FunctionTok{req\_p3}\NormalTok{(b) }\SpecialCharTok{{-}}\NormalTok{ req\_ab[}\DecValTok{2}\NormalTok{]) }\SpecialCharTok{+} \FunctionTok{abs}\NormalTok{(}\FunctionTok{dev\_p3}\NormalTok{(b) }\SpecialCharTok{{-}}\NormalTok{ dev\_ab[}\DecValTok{2}\NormalTok{]) }\SpecialCharTok{+} \FunctionTok{abs}\NormalTok{(}\FunctionTok{desc\_p3}\NormalTok{(b) }\SpecialCharTok{{-}}
\NormalTok{    desc\_ab[}\DecValTok{2}\NormalTok{])}
\NormalTok{\}, }\AttributeTok{lb =} \FunctionTok{c}\NormalTok{(}\DecValTok{0}\NormalTok{, }\DecValTok{0}\NormalTok{), }\AttributeTok{ub =} \FunctionTok{c}\NormalTok{(}\DecValTok{1}\NormalTok{, }\DecValTok{1}\NormalTok{), }\AttributeTok{opts =} \FunctionTok{list}\NormalTok{(}\AttributeTok{maxeval =} \DecValTok{1000}\NormalTok{, }\AttributeTok{algorithm =} \StringTok{"NLOPT\_LN\_BOBYQA"}\NormalTok{))}

\FunctionTok{c}\NormalTok{(res}\SpecialCharTok{$}\NormalTok{solution, res}\SpecialCharTok{$}\NormalTok{objective)}
\end{Highlighting}
\end{Shaded}

\begin{verbatim}
## [1] 0.30356210 0.45695785 0.09384887
\end{verbatim}

We found the parameters 0.3035621 and 0.4569579 for the overall begin an end, which is quite close.

\hypertarget{dtw-based-approach}{%
\subparagraph{DTW-based approach}\label{dtw-based-approach}}

Here, we use open-end/-begin DTW to find the offset per variable, then for all variables combined.

\begin{Shaded}
\begin{Highlighting}[]
\NormalTok{dtwRes }\OtherTok{\textless{}{-}} \FunctionTok{loadResultsOrCompute}\NormalTok{(}\AttributeTok{file =} \StringTok{"../results/proc\_align\_dtw.rds"}\NormalTok{, }\AttributeTok{computeExpr =}\NormalTok{ \{}
  \FunctionTok{library}\NormalTok{(dtw)}
\NormalTok{  X }\OtherTok{\textless{}{-}} \FunctionTok{seq}\NormalTok{(}\DecValTok{0}\NormalTok{, }\DecValTok{1}\NormalTok{, }\AttributeTok{length.out =} \DecValTok{1000}\NormalTok{)}

\NormalTok{  dtwRes\_req }\OtherTok{\textless{}{-}}\NormalTok{ dtw}\SpecialCharTok{::}\FunctionTok{dtw}\NormalTok{(}\AttributeTok{x =} \FunctionTok{req\_p3}\NormalTok{(X), }\AttributeTok{y =}\NormalTok{ (slice\_req}\SpecialCharTok{$}\FunctionTok{get0Function}\NormalTok{())(X), }\AttributeTok{keep.internals =} \ConstantTok{TRUE}\NormalTok{)}
\NormalTok{  dtwEx\_req }\OtherTok{\textless{}{-}} \FunctionTok{extract\_signal\_from\_window}\NormalTok{(dtwRes\_req, }\AttributeTok{window =}\NormalTok{ X)}

\NormalTok{  dtwRes\_dev }\OtherTok{\textless{}{-}}\NormalTok{ dtw}\SpecialCharTok{::}\FunctionTok{dtw}\NormalTok{(}\AttributeTok{x =} \FunctionTok{dev\_p3}\NormalTok{(X), }\AttributeTok{y =}\NormalTok{ (slice\_dev}\SpecialCharTok{$}\FunctionTok{get0Function}\NormalTok{())(X), }\AttributeTok{keep.internals =} \ConstantTok{TRUE}\NormalTok{)}
\NormalTok{  dtwEx\_dev }\OtherTok{\textless{}{-}} \FunctionTok{extract\_signal\_from\_window}\NormalTok{(dtwRes\_dev, }\AttributeTok{window =}\NormalTok{ X)}

\NormalTok{  dtwRes\_desc }\OtherTok{\textless{}{-}}\NormalTok{ dtw}\SpecialCharTok{::}\FunctionTok{dtw}\NormalTok{(}\AttributeTok{x =} \FunctionTok{desc\_p3}\NormalTok{(X), }\AttributeTok{y =}\NormalTok{ (slice\_desc}\SpecialCharTok{$}\FunctionTok{get0Function}\NormalTok{())(X), }\AttributeTok{keep.internals =} \ConstantTok{TRUE}\NormalTok{)}
\NormalTok{  dtwEx\_desc }\OtherTok{\textless{}{-}} \FunctionTok{extract\_signal\_from\_window}\NormalTok{(dtwRes\_desc, }\AttributeTok{window =}\NormalTok{ X)}

\NormalTok{  dtwRes\_ALL }\OtherTok{\textless{}{-}}\NormalTok{ dtw}\SpecialCharTok{::}\FunctionTok{dtw}\NormalTok{(}\AttributeTok{x =} \FunctionTok{matrix}\NormalTok{(}\AttributeTok{ncol =} \DecValTok{3}\NormalTok{, }\AttributeTok{data =} \FunctionTok{c}\NormalTok{(}\FunctionTok{req\_p3}\NormalTok{(X), }\FunctionTok{dev\_p3}\NormalTok{(X), }\FunctionTok{desc\_p3}\NormalTok{(X))),}
    \AttributeTok{y =} \FunctionTok{matrix}\NormalTok{(}\AttributeTok{ncol =} \DecValTok{3}\NormalTok{, }\AttributeTok{data =} \FunctionTok{c}\NormalTok{((slice\_req}\SpecialCharTok{$}\FunctionTok{get0Function}\NormalTok{())(X), (slice\_dev}\SpecialCharTok{$}\FunctionTok{get0Function}\NormalTok{())(X),}
\NormalTok{      (slice\_desc}\SpecialCharTok{$}\FunctionTok{get0Function}\NormalTok{())(X))), }\AttributeTok{keep.internals =} \ConstantTok{TRUE}\NormalTok{)}
\NormalTok{  dtwEx\_ALL }\OtherTok{\textless{}{-}} \FunctionTok{extract\_signal\_from\_window}\NormalTok{(dtwRes\_ALL, }\AttributeTok{window =}\NormalTok{ X)}

  \FunctionTok{c}\NormalTok{(}\AttributeTok{begin\_req =}\NormalTok{ dtwEx\_req}\SpecialCharTok{$}\NormalTok{start\_rel, }\AttributeTok{end\_req =}\NormalTok{ dtwEx\_req}\SpecialCharTok{$}\NormalTok{end\_rel, }\AttributeTok{begin\_dev =}\NormalTok{ dtwEx\_dev}\SpecialCharTok{$}\NormalTok{start\_rel,}
    \AttributeTok{end\_dev =}\NormalTok{ dtwEx\_dev}\SpecialCharTok{$}\NormalTok{end\_rel, }\AttributeTok{begin\_desc =}\NormalTok{ dtwEx\_desc}\SpecialCharTok{$}\NormalTok{start\_rel, }\AttributeTok{end\_desc =}\NormalTok{ dtwEx\_desc}\SpecialCharTok{$}\NormalTok{end\_rel,}
    \AttributeTok{begin\_ALL =}\NormalTok{ dtwEx\_ALL}\SpecialCharTok{$}\NormalTok{start\_rel, }\AttributeTok{end\_ALL =}\NormalTok{ dtwEx\_ALL}\SpecialCharTok{$}\NormalTok{end\_rel)}
\NormalTok{\})}

\NormalTok{dtwRes}
\end{Highlighting}
\end{Shaded}

\begin{verbatim}
##  begin_req    end_req  begin_dev    end_dev begin_desc   end_desc  begin_ALL 
## 0.30330330 0.45045045 0.26326326 0.45645646 0.08008008 0.69869870 0.30130130 
##    end_ALL 
## 0.46846847
\end{verbatim}

Here we get a couple results, depending on the variable. The overall result is marginally worse compared to the one we got from the optimization. Here it becomes apparent that a custom objective would be needed to detect the true begin and end, which is not possible with DTW.

\hypertarget{srbtaw-based-approach}{%
\subparagraph{srBTAW-based approach}\label{srbtaw-based-approach}}

Here we attempt to align the partially observed process with the process model using two different approaches. First, we attempt to use the RSS loss, which should give us similar results to what DTW achieved for the overall begin and end. Then, we use an objective to compute the correlation between processes.

\begin{Shaded}
\begin{Highlighting}[]
\NormalTok{proc\_align\_srbtaw }\OtherTok{\textless{}{-}} \ControlFlowTok{function}\NormalTok{(}\AttributeTok{use =} \FunctionTok{c}\NormalTok{(}\StringTok{"rss"}\NormalTok{, }\StringTok{"logratio"}\NormalTok{)) \{}
\NormalTok{  use }\OtherTok{\textless{}{-}} \FunctionTok{match.arg}\NormalTok{(use)}

\NormalTok{  srbtaw }\OtherTok{\textless{}{-}}\NormalTok{ srBTAW}\SpecialCharTok{$}\FunctionTok{new}\NormalTok{(}\AttributeTok{theta\_b =} \FunctionTok{seq}\NormalTok{(}\DecValTok{0}\NormalTok{, }\DecValTok{1}\NormalTok{, }\AttributeTok{by =} \FloatTok{0.1}\NormalTok{), }\AttributeTok{gamma\_bed =} \FunctionTok{c}\NormalTok{(}\DecValTok{0}\NormalTok{, }\DecValTok{1}\NormalTok{, .Machine}\SpecialCharTok{$}\NormalTok{double.eps),}
    \AttributeTok{lambda =} \FunctionTok{rep}\NormalTok{(.Machine}\SpecialCharTok{$}\NormalTok{double.eps, }\DecValTok{10}\NormalTok{), }\AttributeTok{begin =} \DecValTok{0}\NormalTok{, }\AttributeTok{end =} \DecValTok{1}\NormalTok{, }\AttributeTok{openBegin =} \ConstantTok{TRUE}\NormalTok{,}
    \AttributeTok{openEnd =} \ConstantTok{TRUE}\NormalTok{)}
\NormalTok{  srbtaw}\SpecialCharTok{$}\FunctionTok{setParams}\NormalTok{(}\StringTok{\textasciigrave{}}\AttributeTok{names\textless{}{-}}\StringTok{\textasciigrave{}}\NormalTok{(}\FunctionTok{c}\NormalTok{(}\DecValTok{0}\NormalTok{, }\DecValTok{1}\NormalTok{, }\FunctionTok{rep}\NormalTok{(}\DecValTok{1}\SpecialCharTok{/}\DecValTok{10}\NormalTok{, }\DecValTok{10}\NormalTok{)), srbtaw}\SpecialCharTok{$}\FunctionTok{getParamNames}\NormalTok{()))}

\NormalTok{  srbtaw}\SpecialCharTok{$}\FunctionTok{setSignal}\NormalTok{(}\AttributeTok{signal =}\NormalTok{ Signal}\SpecialCharTok{$}\FunctionTok{new}\NormalTok{(}\AttributeTok{name =} \StringTok{"REQ"}\NormalTok{, }\AttributeTok{func =}\NormalTok{ req\_p3, }\AttributeTok{support =} \FunctionTok{c}\NormalTok{(}\DecValTok{0}\NormalTok{,}
    \DecValTok{1}\NormalTok{), }\AttributeTok{isWp =} \ConstantTok{FALSE}\NormalTok{))}
\NormalTok{  srbtaw}\SpecialCharTok{$}\FunctionTok{setSignal}\NormalTok{(slice\_req)}
\NormalTok{  srbtaw}\SpecialCharTok{$}\FunctionTok{setSignal}\NormalTok{(}\AttributeTok{signal =}\NormalTok{ Signal}\SpecialCharTok{$}\FunctionTok{new}\NormalTok{(}\AttributeTok{name =} \StringTok{"DEV"}\NormalTok{, }\AttributeTok{func =}\NormalTok{ dev\_p3, }\AttributeTok{support =} \FunctionTok{c}\NormalTok{(}\DecValTok{0}\NormalTok{,}
    \DecValTok{1}\NormalTok{), }\AttributeTok{isWp =} \ConstantTok{FALSE}\NormalTok{))}
\NormalTok{  srbtaw}\SpecialCharTok{$}\FunctionTok{setSignal}\NormalTok{(slice\_dev)}
\NormalTok{  srbtaw}\SpecialCharTok{$}\FunctionTok{setSignal}\NormalTok{(}\AttributeTok{signal =}\NormalTok{ Signal}\SpecialCharTok{$}\FunctionTok{new}\NormalTok{(}\AttributeTok{name =} \StringTok{"DESC"}\NormalTok{, }\AttributeTok{func =}\NormalTok{ desc\_p3, }\AttributeTok{support =} \FunctionTok{c}\NormalTok{(}\DecValTok{0}\NormalTok{,}
    \DecValTok{1}\NormalTok{), }\AttributeTok{isWp =} \ConstantTok{FALSE}\NormalTok{))}
\NormalTok{  srbtaw}\SpecialCharTok{$}\FunctionTok{setSignal}\NormalTok{(slice\_desc)}

\NormalTok{  obj }\OtherTok{\textless{}{-}}\NormalTok{ srBTAW\_LossLinearScalarizer}\SpecialCharTok{$}\FunctionTok{new}\NormalTok{(}\AttributeTok{computeParallel =} \ConstantTok{TRUE}\NormalTok{, }\AttributeTok{gradientParallel =} \ConstantTok{TRUE}\NormalTok{,}
    \AttributeTok{returnRaw =} \ConstantTok{TRUE}\NormalTok{)}

  \ControlFlowTok{if}\NormalTok{ (use }\SpecialCharTok{==} \StringTok{"rss"}\NormalTok{) \{}
\NormalTok{    loss\_req }\OtherTok{\textless{}{-}}\NormalTok{ srBTAW\_Loss\_Rss}\SpecialCharTok{$}\FunctionTok{new}\NormalTok{(}\AttributeTok{wpName =} \StringTok{"REQ\_WP"}\NormalTok{, }\AttributeTok{wcName =} \StringTok{"REQ"}\NormalTok{, }\AttributeTok{intervals =} \DecValTok{1}\SpecialCharTok{:}\DecValTok{10}\NormalTok{,}
      \AttributeTok{returnRaw =} \ConstantTok{FALSE}\NormalTok{)}
\NormalTok{    srbtaw}\SpecialCharTok{$}\FunctionTok{addLoss}\NormalTok{(}\AttributeTok{loss =}\NormalTok{ loss\_req)}
\NormalTok{    loss\_dev }\OtherTok{\textless{}{-}}\NormalTok{ srBTAW\_Loss\_Rss}\SpecialCharTok{$}\FunctionTok{new}\NormalTok{(}\AttributeTok{wpName =} \StringTok{"DEV\_WP"}\NormalTok{, }\AttributeTok{wcName =} \StringTok{"DEV"}\NormalTok{, }\AttributeTok{intervals =} \DecValTok{1}\SpecialCharTok{:}\DecValTok{10}\NormalTok{,}
      \AttributeTok{returnRaw =} \ConstantTok{FALSE}\NormalTok{)}
\NormalTok{    srbtaw}\SpecialCharTok{$}\FunctionTok{addLoss}\NormalTok{(}\AttributeTok{loss =}\NormalTok{ loss\_dev)}
\NormalTok{    loss\_desc }\OtherTok{\textless{}{-}}\NormalTok{ srBTAW\_Loss\_Rss}\SpecialCharTok{$}\FunctionTok{new}\NormalTok{(}\AttributeTok{wpName =} \StringTok{"DESC\_WP"}\NormalTok{, }\AttributeTok{wcName =} \StringTok{"DESC"}\NormalTok{, }\AttributeTok{intervals =} \DecValTok{1}\SpecialCharTok{:}\DecValTok{10}\NormalTok{,}
      \AttributeTok{returnRaw =} \ConstantTok{FALSE}\NormalTok{)}
\NormalTok{    srbtaw}\SpecialCharTok{$}\FunctionTok{addLoss}\NormalTok{(}\AttributeTok{loss =}\NormalTok{ loss\_desc)}

\NormalTok{    obj}\SpecialCharTok{$}\FunctionTok{setObjective}\NormalTok{(}\AttributeTok{name =} \StringTok{"Loss\_REQ"}\NormalTok{, }\AttributeTok{obj =}\NormalTok{ loss\_req)}
\NormalTok{    obj}\SpecialCharTok{$}\FunctionTok{setObjective}\NormalTok{(}\AttributeTok{name =} \StringTok{"Loss\_DEV"}\NormalTok{, }\AttributeTok{obj =}\NormalTok{ loss\_dev)}
\NormalTok{    obj}\SpecialCharTok{$}\FunctionTok{setObjective}\NormalTok{(}\AttributeTok{name =} \StringTok{"Loss\_DESC"}\NormalTok{, }\AttributeTok{obj =}\NormalTok{ loss\_desc)}
\NormalTok{  \} }\ControlFlowTok{else}\NormalTok{ \{}
\NormalTok{    loss\_req2 }\OtherTok{\textless{}{-}}\NormalTok{ srBTAW\_Loss2Curves}\SpecialCharTok{$}\FunctionTok{new}\NormalTok{(}\AttributeTok{use =} \StringTok{"logratio"}\NormalTok{, }\AttributeTok{srbtaw =}\NormalTok{ srbtaw, }\AttributeTok{wpName =} \StringTok{"REQ\_WP"}\NormalTok{,}
      \AttributeTok{wcName =} \StringTok{"REQ"}\NormalTok{, }\AttributeTok{intervals =} \DecValTok{1}\SpecialCharTok{:}\DecValTok{10}\NormalTok{)}
\NormalTok{    srbtaw}\SpecialCharTok{$}\FunctionTok{addLoss}\NormalTok{(}\AttributeTok{loss =}\NormalTok{ loss\_req2)}

\NormalTok{    loss\_dev2 }\OtherTok{\textless{}{-}}\NormalTok{ srBTAW\_Loss2Curves}\SpecialCharTok{$}\FunctionTok{new}\NormalTok{(}\AttributeTok{use =} \StringTok{"logratio"}\NormalTok{, }\AttributeTok{srbtaw =}\NormalTok{ srbtaw, }\AttributeTok{wpName =} \StringTok{"DEV\_WP"}\NormalTok{,}
      \AttributeTok{wcName =} \StringTok{"DEV"}\NormalTok{, }\AttributeTok{intervals =} \DecValTok{1}\SpecialCharTok{:}\DecValTok{10}\NormalTok{)}
\NormalTok{    srbtaw}\SpecialCharTok{$}\FunctionTok{addLoss}\NormalTok{(}\AttributeTok{loss =}\NormalTok{ loss\_dev2)}

\NormalTok{    loss\_desc2 }\OtherTok{\textless{}{-}}\NormalTok{ srBTAW\_Loss2Curves}\SpecialCharTok{$}\FunctionTok{new}\NormalTok{(}\AttributeTok{use =} \StringTok{"logratio"}\NormalTok{, }\AttributeTok{srbtaw =}\NormalTok{ srbtaw, }\AttributeTok{wpName =} \StringTok{"DESC\_WP"}\NormalTok{,}
      \AttributeTok{wcName =} \StringTok{"DESC"}\NormalTok{, }\AttributeTok{intervals =} \DecValTok{1}\SpecialCharTok{:}\DecValTok{10}\NormalTok{)}
\NormalTok{    srbtaw}\SpecialCharTok{$}\FunctionTok{addLoss}\NormalTok{(}\AttributeTok{loss =}\NormalTok{ loss\_desc2)}

\NormalTok{    obj}\SpecialCharTok{$}\FunctionTok{setObjective}\NormalTok{(}\AttributeTok{name =} \StringTok{"Loss\_REQ2"}\NormalTok{, }\AttributeTok{obj =}\NormalTok{ loss\_req2)}
\NormalTok{    obj}\SpecialCharTok{$}\FunctionTok{setObjective}\NormalTok{(}\AttributeTok{name =} \StringTok{"Loss\_DEV2"}\NormalTok{, }\AttributeTok{obj =}\NormalTok{ loss\_dev2)}
\NormalTok{    obj}\SpecialCharTok{$}\FunctionTok{setObjective}\NormalTok{(}\AttributeTok{name =} \StringTok{"Loss\_DESC2"}\NormalTok{, }\AttributeTok{obj =}\NormalTok{ loss\_desc2)}
\NormalTok{  \}}

\NormalTok{  reg }\OtherTok{\textless{}{-}}\NormalTok{ TimeWarpRegularization}\SpecialCharTok{$}\FunctionTok{new}\NormalTok{(}\AttributeTok{weight =} \DecValTok{1}\SpecialCharTok{/}\DecValTok{2}\NormalTok{, }\AttributeTok{use =} \StringTok{"exint2"}\NormalTok{, }\AttributeTok{wpName =} \StringTok{"REQ\_WP"}\NormalTok{,}
    \AttributeTok{wcName =} \StringTok{"REQ"}\NormalTok{, }\AttributeTok{returnRaw =}\NormalTok{ use }\SpecialCharTok{==} \StringTok{"logratio"}\NormalTok{, }\AttributeTok{intervals =} \DecValTok{1}\SpecialCharTok{:}\DecValTok{10}\NormalTok{)}

\NormalTok{  srbtaw}\SpecialCharTok{$}\FunctionTok{addLoss}\NormalTok{(}\AttributeTok{loss =}\NormalTok{ reg)}
\NormalTok{  obj}\SpecialCharTok{$}\FunctionTok{setObjective}\NormalTok{(}\AttributeTok{name =} \StringTok{"reg\_exint2"}\NormalTok{, }\AttributeTok{obj =}\NormalTok{ reg)}
\NormalTok{  srbtaw}\SpecialCharTok{$}\FunctionTok{setObjective}\NormalTok{(}\AttributeTok{obj =}\NormalTok{ obj)}
\NormalTok{  srbtaw}\SpecialCharTok{$}\FunctionTok{setIsObjectiveLogarithmic}\NormalTok{(}\AttributeTok{val =} \ConstantTok{TRUE}\NormalTok{)}
\NormalTok{\}}
\end{Highlighting}
\end{Shaded}

\begin{Shaded}
\begin{Highlighting}[]
\NormalTok{align\_srbtaw\_rss }\OtherTok{\textless{}{-}} \FunctionTok{loadResultsOrCompute}\NormalTok{(}\AttributeTok{file =} \StringTok{"../results/proc\_align\_srbtaw\_rss.rds"}\NormalTok{,}
  \AttributeTok{computeExpr =}\NormalTok{ \{}
\NormalTok{    srbtaw }\OtherTok{\textless{}{-}} \FunctionTok{proc\_align\_srbtaw}\NormalTok{(}\AttributeTok{use =} \StringTok{"rss"}\NormalTok{)}
    \FunctionTok{compute\_proc\_align}\NormalTok{(}\AttributeTok{srbtaw =}\NormalTok{ srbtaw)}
\NormalTok{  \})}
\NormalTok{align\_srbtaw\_rss}\SpecialCharTok{$}\NormalTok{res}\SpecialCharTok{$}\NormalTok{par[}\FunctionTok{c}\NormalTok{(}\StringTok{"b"}\NormalTok{, }\StringTok{"e"}\NormalTok{)]}
\end{Highlighting}
\end{Shaded}

\begin{verbatim}
##         b         e 
## 0.4555679 0.3067260
\end{verbatim}

\begin{Shaded}
\begin{Highlighting}[]
\NormalTok{align\_srbtaw\_logratio }\OtherTok{\textless{}{-}} \FunctionTok{loadResultsOrCompute}\NormalTok{(}\AttributeTok{file =} \StringTok{"../results/proc\_align\_srbtaw\_logratio.rds"}\NormalTok{,}
  \AttributeTok{computeExpr =}\NormalTok{ \{}
\NormalTok{    srbtaw }\OtherTok{\textless{}{-}} \FunctionTok{proc\_align\_srbtaw}\NormalTok{(}\AttributeTok{use =} \StringTok{"logratio"}\NormalTok{)}
    \FunctionTok{compute\_proc\_align}\NormalTok{(}\AttributeTok{srbtaw =}\NormalTok{ srbtaw)}
\NormalTok{  \})}
\NormalTok{align\_srbtaw\_logratio}\SpecialCharTok{$}\NormalTok{res}\SpecialCharTok{$}\NormalTok{par[}\FunctionTok{c}\NormalTok{(}\StringTok{"b"}\NormalTok{, }\StringTok{"e"}\NormalTok{)]}
\end{Highlighting}
\end{Shaded}

\begin{verbatim}
##         b         e 
## 0.2492560 0.5264296
\end{verbatim}

\begin{figure}[ht!]

{\centering \includegraphics{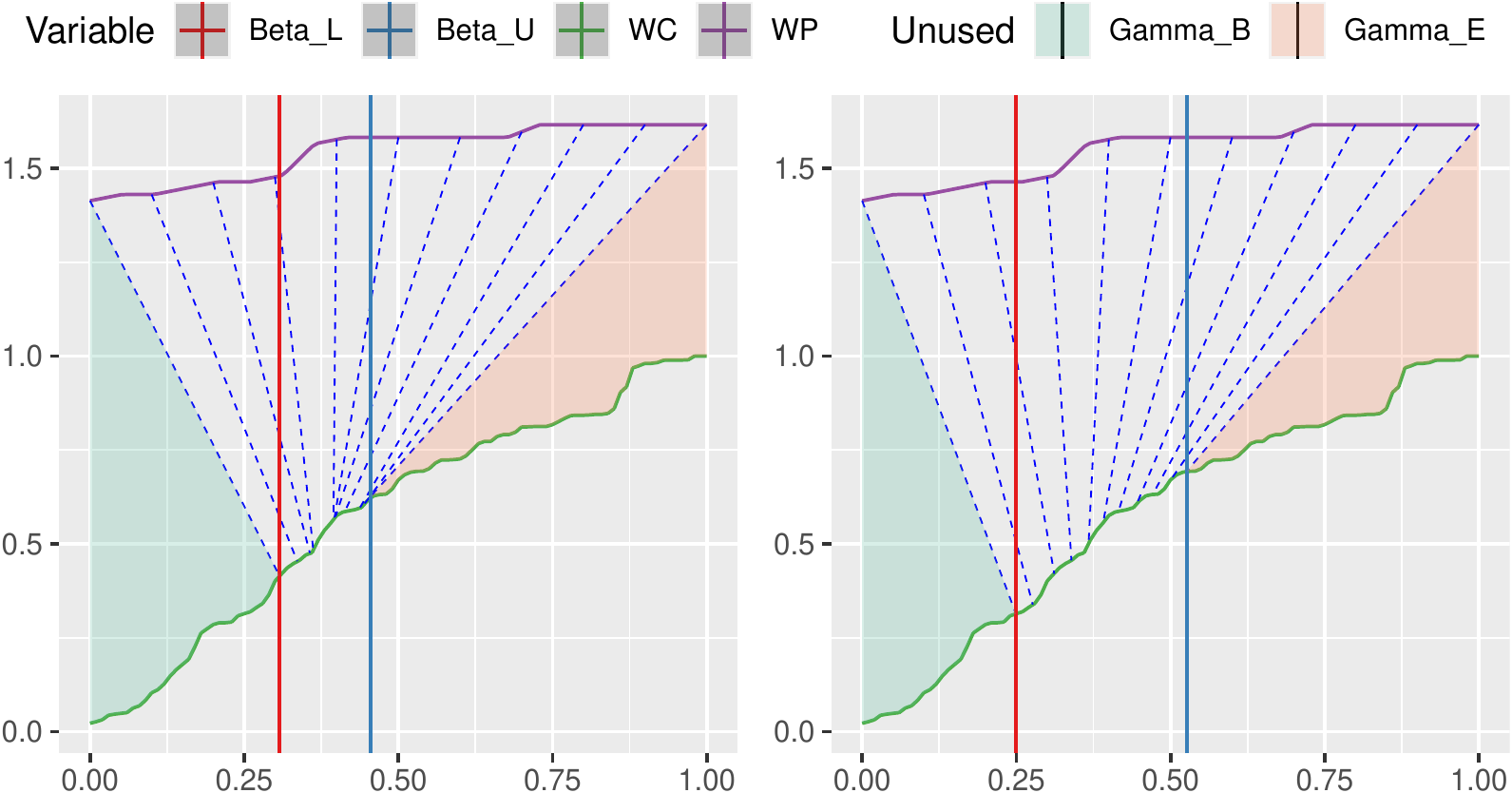} 

}

\caption{Open alignment of the observed process using srBTAW and 10 intervals. On the left using RSS as loss, and on the right using correlation.}\label{fig:proc-align-srbtaw}
\end{figure}

As expected, the RSS-based result is very similar to the overall result as obtained by DTW (note that the mix-up of begin and end does not matter to srBTAW, is it regularizes this using its internal representation \(\beta_l,\beta_u\)). The correlation-based loss, surprisingly, manages to discover the true begin of \(\approx0.25\). It is likely possible to find a combination of losses, weights, etc., that can find the offsets of some process better, but that is a bit out of scope here. In fig.~\ref{fig:proc-align-srbtaw} we show the time warping as computed by the two objectives (RSS and correlation).

For demonstration purposes, we can compute the loss surface of a the model that was previously fit using RSS. Here, we keep the vector \(\bm{\vartheta}^{(l)}\) constant, and only update \(b,e\).

\begin{figure}[ht!]

{\centering \includegraphics{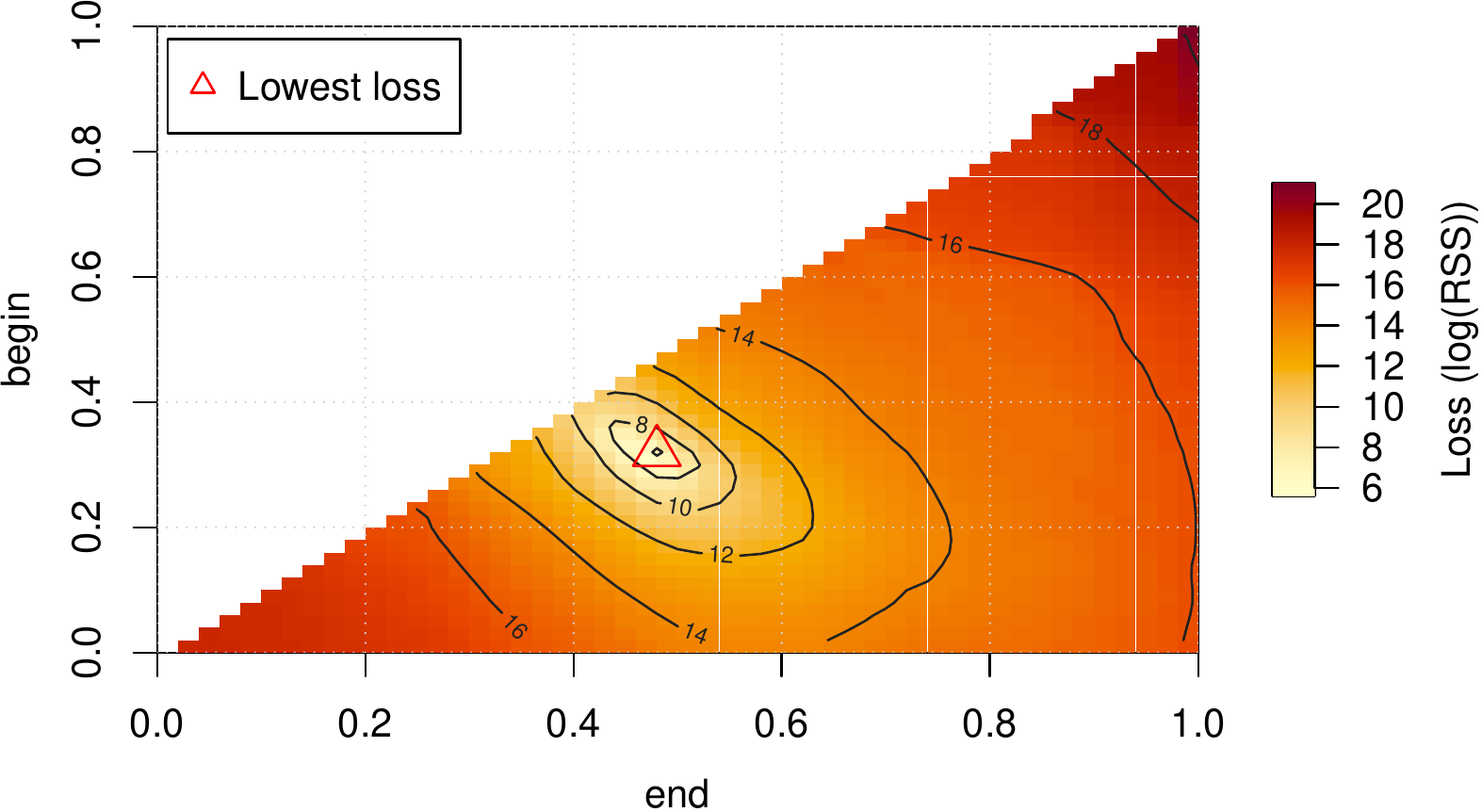} 

}

\caption{The loss surface for ten-interval time warping using srBTAW of the parameters begin and end. The lowest loss is marked with a triangle}\label{fig:proc-align-srbtaw-loss-surface}
\end{figure}

In figure \ref{fig:proc-align-srbtaw-loss-surface} we show the loss surface of the parameters begin and end of a one-interval alignment using srBTAW and RSS-losses.

\begin{Shaded}
\begin{Highlighting}[]
\NormalTok{use\_resol }\OtherTok{\textless{}{-}} \DecValTok{50}
\NormalTok{minStep }\OtherTok{\textless{}{-}} \DecValTok{1}\SpecialCharTok{/}\NormalTok{use\_resol}
\NormalTok{use\_begin }\OtherTok{\textless{}{-}} \FunctionTok{seq}\NormalTok{(}\DecValTok{0}\NormalTok{, }\DecValTok{1} \SpecialCharTok{{-}}\NormalTok{ minStep, }\AttributeTok{by =}\NormalTok{ minStep)}
\NormalTok{use\_par }\OtherTok{\textless{}{-}}\NormalTok{ align\_srbtaw\_rss}\SpecialCharTok{$}\NormalTok{res}\SpecialCharTok{$}\NormalTok{par}

\NormalTok{align\_srbtaw\_rss\_surface }\OtherTok{\textless{}{-}} \FunctionTok{loadResultsOrCompute}\NormalTok{(}\AttributeTok{file =} \StringTok{"../results/align\_srbtaw\_rss\_surface.rds"}\NormalTok{,}
  \AttributeTok{computeExpr =}\NormalTok{ \{}
    \FunctionTok{library}\NormalTok{(foreach)}

\NormalTok{    cl }\OtherTok{\textless{}{-}}\NormalTok{ parallel}\SpecialCharTok{::}\FunctionTok{makePSOCKcluster}\NormalTok{(}\FunctionTok{min}\NormalTok{(}\DecValTok{123}\NormalTok{, parallel}\SpecialCharTok{::}\FunctionTok{detectCores}\NormalTok{()))}
\NormalTok{    parallel}\SpecialCharTok{::}\FunctionTok{clusterExport}\NormalTok{(cl, }\AttributeTok{varlist =} \FunctionTok{list}\NormalTok{(}\StringTok{"slice\_supp"}\NormalTok{))}

    \FunctionTok{doWithParallelClusterExplicit}\NormalTok{(}\AttributeTok{cl =}\NormalTok{ cl, }\AttributeTok{expr =}\NormalTok{ \{}
\NormalTok{      foreach}\SpecialCharTok{::}\FunctionTok{foreach}\NormalTok{(}\AttributeTok{begin =}\NormalTok{ use\_begin, }\AttributeTok{.inorder =} \ConstantTok{TRUE}\NormalTok{, }\AttributeTok{.combine =}\NormalTok{ cbind) }\SpecialCharTok{\%dopar\%}
\NormalTok{        \{}
          \FunctionTok{source}\NormalTok{(}\AttributeTok{file =} \StringTok{"../helpers.R"}\NormalTok{)}
          \FunctionTok{source}\NormalTok{(}\AttributeTok{file =} \StringTok{"./common{-}funcs.R"}\NormalTok{)}
          \FunctionTok{source}\NormalTok{(}\AttributeTok{file =} \StringTok{"../models/modelsR6.R"}\NormalTok{)}
          \FunctionTok{source}\NormalTok{(}\AttributeTok{file =} \StringTok{"../models/SRBTW{-}R6.R"}\NormalTok{)}

\NormalTok{          end }\OtherTok{\textless{}{-}}\NormalTok{ begin }\SpecialCharTok{+}\NormalTok{ minStep}
\NormalTok{          res }\OtherTok{\textless{}{-}} \FunctionTok{c}\NormalTok{()}
          \ControlFlowTok{while}\NormalTok{ (}\ConstantTok{TRUE}\NormalTok{) \{}
          \ControlFlowTok{if}\NormalTok{ (end }\SpecialCharTok{\textgreater{}} \DecValTok{1}\NormalTok{)}
            \ControlFlowTok{break}

\NormalTok{          srbtaw }\OtherTok{\textless{}{-}} \FunctionTok{temp\_get\_srbtaw}\NormalTok{()}
\NormalTok{          params }\OtherTok{\textless{}{-}} \FunctionTok{c}\NormalTok{(use\_par)}
\NormalTok{          params[}\StringTok{"b"}\NormalTok{] }\OtherTok{\textless{}{-}}\NormalTok{ begin}
\NormalTok{          params[}\StringTok{"e"}\NormalTok{] }\OtherTok{\textless{}{-}}\NormalTok{ end}
          \CommentTok{\# srbtaw$setParams(\textasciigrave{}names\textless{}{-}\textasciigrave{}(c(begin, end, 1),}
          \CommentTok{\# srbtaw$getParamNames()))}
\NormalTok{          srbtaw}\SpecialCharTok{$}\FunctionTok{setParams}\NormalTok{(}\AttributeTok{params =}\NormalTok{ params)}
\NormalTok{          res }\OtherTok{\textless{}{-}} \FunctionTok{c}\NormalTok{(res, srbtaw}\SpecialCharTok{$}\FunctionTok{getObjective}\NormalTok{()}\SpecialCharTok{$}\FunctionTok{compute0}\NormalTok{())}

\NormalTok{          end }\OtherTok{\textless{}{-}}\NormalTok{ end }\SpecialCharTok{+}\NormalTok{ minStep}
\NormalTok{          \}}

          \CommentTok{\# pad left:}
          \FunctionTok{c}\NormalTok{(}\FunctionTok{rep}\NormalTok{(}\ConstantTok{NA\_real\_}\NormalTok{, use\_resol }\SpecialCharTok{{-}} \FunctionTok{length}\NormalTok{(res)), res)}
\NormalTok{        \}}
\NormalTok{    \})}
\NormalTok{  \})}
\end{Highlighting}
\end{Shaded}

\hypertarget{pattern-iv-1}{%
\subsubsection{Pattern IV}\label{pattern-iv-1}}

This pattern emerged only recently, but we have some options of scoring the projects against it. Let me start by emphasizing again how the concrete tests here are performed using an instantiation of type IV using pattern I, and how improper type I actually fits the data. The tests here, in the best case, should be run for all of the previously introduced patterns, and our expectation is that the data-only pattern would perform best, with the data-enhanced one coming in at second place. We mostly chose type I as we have analytical expressions for some of the variables.

Again, we have (derived) confidence intervals for the variables \texttt{REQ} and \texttt{DEV}. Then we have derivatives for all variables. Scoring based on CIs are hence applicable, but only for the first two variables. Other than that, any method based on comparing two variables is applicable. For our evaluations, we will use LOESS-smoothing with different smoothing-spans for either score, that are applicable.

\begin{figure}[ht!]

{\centering \includegraphics{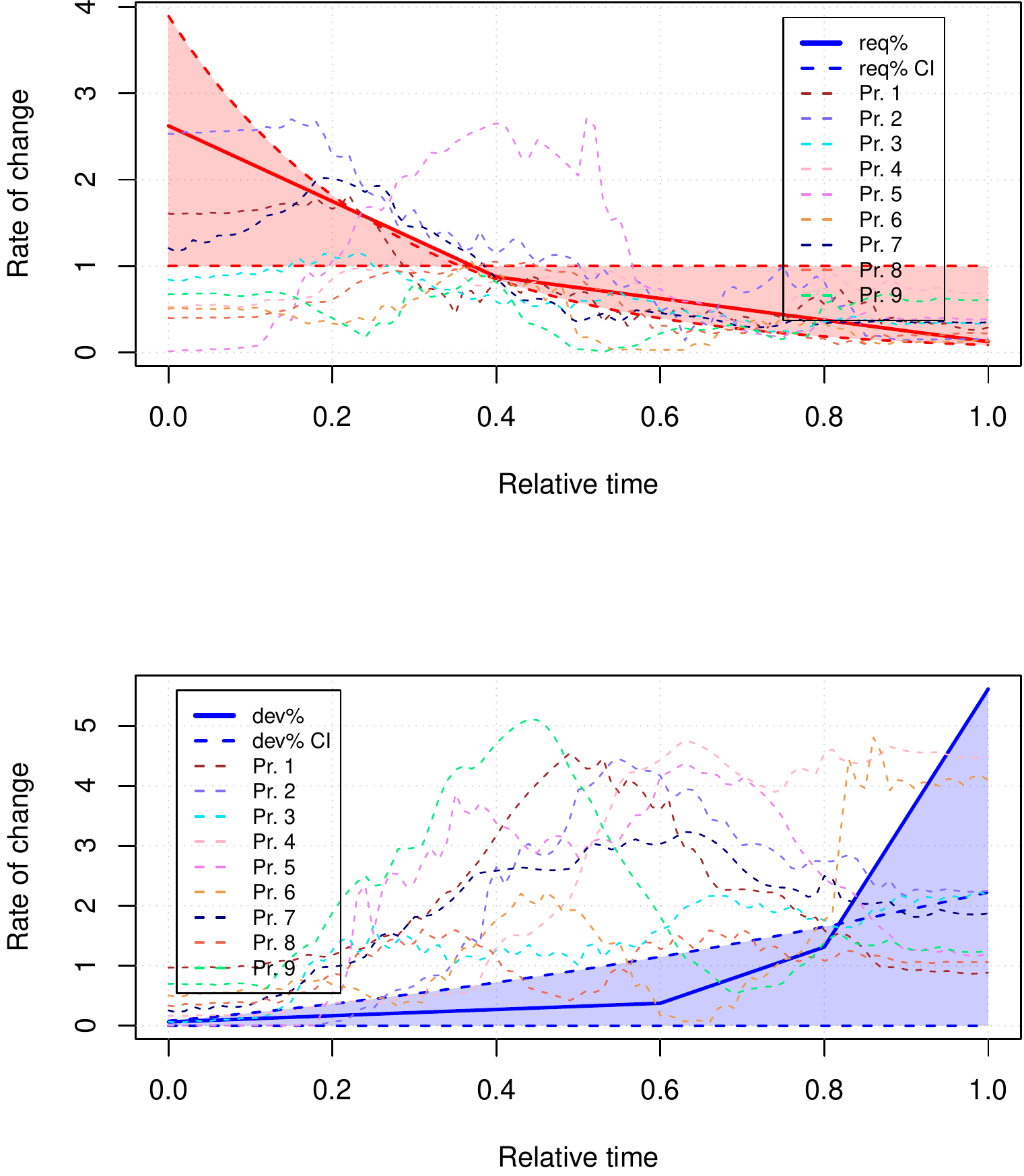} 

}

\caption{All projects' derivative variables plotted against the derivatives of the two variables req\% and dev\% of pattern type I. The selected smoothing-span was 0.3.}\label{fig:p4-deriv-signals}
\end{figure}

In figure \ref{fig:p4-deriv-signals} we show all projects against the fourth pattern, as smoothed using LOESS and derived. It is interesting to see that at \(\approx0.37\), where the confidence surface is the smallest for \texttt{REQ} and the upper confidence interval's rate of change becomes \(0\), most projects also have a turning point.

\hypertarget{scoring-based-on-the-distance-to-reference}{%
\paragraph{Scoring based on the distance to reference}\label{scoring-based-on-the-distance-to-reference}}

This method is applicable because we only do have the confidence intervals' boundaries, and a homogeneous surface. The derivative of the respective variable is the expectation for the rate of change, and the confidence intervals demarcate a minimum and maximum expectation. This applies only to the variables \texttt{REQ} and \texttt{DEV}. Since we compute the distance in terms of the area between curves, the gradients should use a lower smoothing-span to be detail-preserving. However, we will try a few to see what works best.

\begin{Shaded}
\begin{Highlighting}[]
\NormalTok{p4\_compute\_areadist }\OtherTok{\textless{}{-}} \ControlFlowTok{function}\NormalTok{(span, }\AttributeTok{intFrom =} \DecValTok{0}\NormalTok{, }\AttributeTok{intTo =} \DecValTok{1}\NormalTok{) \{}
\NormalTok{  d1\_vals }\OtherTok{\textless{}{-}} \ControlFlowTok{function}\NormalTok{(f, f\_low, f\_upp, x, }\AttributeTok{useMax =} \ConstantTok{TRUE}\NormalTok{) \{}
    \FunctionTok{sapply}\NormalTok{(}\AttributeTok{X =}\NormalTok{ x, }\AttributeTok{FUN =} \ControlFlowTok{function}\NormalTok{(x\_) \{}
\NormalTok{      temp }\OtherTok{\textless{}{-}} \FunctionTok{c}\NormalTok{(}\FunctionTok{f}\NormalTok{(x\_), }\FunctionTok{f\_low}\NormalTok{(x\_), }\FunctionTok{f\_upp}\NormalTok{(x\_))}
      \ControlFlowTok{if}\NormalTok{ (useMax) }\FunctionTok{max}\NormalTok{(temp) }\ControlFlowTok{else} \FunctionTok{min}\NormalTok{(temp)}
\NormalTok{    \})}
\NormalTok{  \}}
  
  \FunctionTok{doWithParallelCluster}\NormalTok{(}\AttributeTok{numCores =} \FunctionTok{length}\NormalTok{(all\_signals), }\AttributeTok{expr =}\NormalTok{ \{}
    \FunctionTok{library}\NormalTok{(foreach)}
    
\NormalTok{    foreach}\SpecialCharTok{::}\FunctionTok{foreach}\NormalTok{(}
      \AttributeTok{pId =} \FunctionTok{names}\NormalTok{(all\_signals),}
      \AttributeTok{.inorder =} \ConstantTok{TRUE}\NormalTok{,}
      \AttributeTok{.combine =}\NormalTok{ rbind,}
      \AttributeTok{.packages =} \FunctionTok{c}\NormalTok{(}\StringTok{"cobs"}\NormalTok{),}
      \AttributeTok{.export =} \FunctionTok{c}\NormalTok{(}\StringTok{"all\_signals"}\NormalTok{, }\StringTok{"L\_areadist\_p3\_avg"}\NormalTok{,}
                  \StringTok{"func\_d1"}\NormalTok{, }\StringTok{"req"}\NormalTok{, }\StringTok{"dev"}\NormalTok{, }\StringTok{"req\_d1\_p4"}\NormalTok{, }\StringTok{"dev\_d1\_p4"}\NormalTok{,}
                  \StringTok{"req\_ci\_upper\_d1\_p4"}\NormalTok{, }\StringTok{"dev\_ci\_upper\_d1\_p4"}\NormalTok{,}
                  \StringTok{"req\_ci\_lower\_d1\_p4"}\NormalTok{, }\StringTok{"dev\_ci\_lower\_d1\_p4"}\NormalTok{,}
                  \StringTok{"smooth\_signal\_loess"}\NormalTok{, }\StringTok{"req\_poly"}\NormalTok{, }\StringTok{"dev\_poly"}\NormalTok{)}
\NormalTok{    ) }\SpecialCharTok{\%dopar\%}\NormalTok{ \{}
\NormalTok{      X }\OtherTok{\textless{}{-}} \DecValTok{1}\SpecialCharTok{:}\FunctionTok{nrow}\NormalTok{(all\_signals[[pId]]}\SpecialCharTok{$}\NormalTok{data)}
\NormalTok{      Y\_req }\OtherTok{\textless{}{-}} \FunctionTok{cumsum}\NormalTok{(all\_signals[[pId]]}\SpecialCharTok{$}\NormalTok{data}\SpecialCharTok{$}\NormalTok{req)}
\NormalTok{      Y\_dev }\OtherTok{\textless{}{-}} \FunctionTok{cumsum}\NormalTok{(all\_signals[[pId]]}\SpecialCharTok{$}\NormalTok{data}\SpecialCharTok{$}\NormalTok{dev)}
      
\NormalTok{      req\_tempf }\OtherTok{\textless{}{-}} \FunctionTok{smooth\_signal\_loess}\NormalTok{(}\AttributeTok{x =}\NormalTok{ X, }\AttributeTok{y =}\NormalTok{ Y\_req, }\AttributeTok{span =}\NormalTok{ span, }\AttributeTok{neval =} \FloatTok{1e3}\NormalTok{)}
\NormalTok{      dev\_tempf }\OtherTok{\textless{}{-}} \FunctionTok{smooth\_signal\_loess}\NormalTok{(}\AttributeTok{x =}\NormalTok{ X, }\AttributeTok{y =}\NormalTok{ Y\_dev, }\AttributeTok{span =}\NormalTok{ span, }\AttributeTok{neval =} \FloatTok{1e3}\NormalTok{)}
      
\NormalTok{      req\_d1 }\OtherTok{\textless{}{-}} \ControlFlowTok{function}\NormalTok{(x) }\FunctionTok{func\_d1}\NormalTok{(}\AttributeTok{f =}\NormalTok{ req\_tempf, }\AttributeTok{x =}\NormalTok{ x)}
\NormalTok{      dev\_d1 }\OtherTok{\textless{}{-}} \ControlFlowTok{function}\NormalTok{(x) }\FunctionTok{func\_d1}\NormalTok{(}\AttributeTok{f =}\NormalTok{ dev\_tempf, }\AttributeTok{x =}\NormalTok{ x)}
      
\NormalTok{      req\_upper\_tempf }\OtherTok{\textless{}{-}} \ControlFlowTok{function}\NormalTok{(x) }\FunctionTok{d1\_vals}\NormalTok{(}
        \AttributeTok{f =}\NormalTok{ req\_d1\_p4, }\AttributeTok{f\_low =}\NormalTok{ req\_ci\_lower\_d1\_p4, }\AttributeTok{f\_upp =}\NormalTok{ req\_ci\_upper\_d1\_p4, }\AttributeTok{x =}\NormalTok{ x, }\AttributeTok{useMax =} \ConstantTok{TRUE}\NormalTok{)}
\NormalTok{      req\_lower\_tempf }\OtherTok{\textless{}{-}} \ControlFlowTok{function}\NormalTok{(x) }\FunctionTok{d1\_vals}\NormalTok{(}
        \AttributeTok{f =}\NormalTok{ req\_d1\_p4, }\AttributeTok{f\_low =}\NormalTok{ req\_ci\_lower\_d1\_p4, }\AttributeTok{f\_upp =}\NormalTok{ req\_ci\_upper\_d1\_p4, }\AttributeTok{x =}\NormalTok{ x, }\AttributeTok{useMax =} \ConstantTok{FALSE}\NormalTok{)}
      
\NormalTok{      dev\_upper\_tempf }\OtherTok{\textless{}{-}} \ControlFlowTok{function}\NormalTok{(x) }\FunctionTok{d1\_vals}\NormalTok{(}
        \AttributeTok{f =}\NormalTok{ dev\_d1\_p4, }\AttributeTok{f\_low =}\NormalTok{ dev\_ci\_lower\_d1\_p4, }\AttributeTok{f\_upp =}\NormalTok{ dev\_ci\_upper\_d1\_p4, }\AttributeTok{x =}\NormalTok{ x, }\AttributeTok{useMax =} \ConstantTok{TRUE}\NormalTok{)}
\NormalTok{      dev\_lower\_tempf }\OtherTok{\textless{}{-}} \ControlFlowTok{function}\NormalTok{(x) }\FunctionTok{d1\_vals}\NormalTok{(}
        \AttributeTok{f =}\NormalTok{ dev\_d1\_p4, }\AttributeTok{f\_low =}\NormalTok{ dev\_ci\_lower\_d1\_p4, }\AttributeTok{f\_upp =}\NormalTok{ dev\_ci\_upper\_d1\_p4, }\AttributeTok{x =}\NormalTok{ x, }\AttributeTok{useMax =} \ConstantTok{FALSE}\NormalTok{)}
      
      \StringTok{\textasciigrave{}}\AttributeTok{rownames\textless{}{-}}\StringTok{\textasciigrave{}}\NormalTok{(}\FunctionTok{data.frame}\NormalTok{(}
        \AttributeTok{REQ =} \FunctionTok{L\_areadist\_p3\_avg}\NormalTok{(}
          \AttributeTok{x1 =}\NormalTok{ intFrom, }\AttributeTok{x2 =}\NormalTok{ intTo, }\AttributeTok{f =}\NormalTok{ req\_d1, }\AttributeTok{gbar =}\NormalTok{ req\_d1\_p4, }\AttributeTok{use2ndVariant =} \ConstantTok{TRUE}\NormalTok{,}
          \AttributeTok{CI\_upper =}\NormalTok{ req\_upper\_tempf, }\AttributeTok{CI\_lower =}\NormalTok{ req\_lower\_tempf)[}\StringTok{"dist"}\NormalTok{],}
        
        \AttributeTok{DEV =} \FunctionTok{L\_areadist\_p3\_avg}\NormalTok{(}
          \AttributeTok{x1 =}\NormalTok{ intFrom, }\AttributeTok{x2 =}\NormalTok{ intTo, }\AttributeTok{f =}\NormalTok{ dev\_d1, }\AttributeTok{gbar =}\NormalTok{ dev\_d1\_p4, }\AttributeTok{use2ndVariant =} \ConstantTok{TRUE}\NormalTok{,}
          \AttributeTok{CI\_upper =}\NormalTok{ dev\_upper\_tempf, }\AttributeTok{CI\_lower =}\NormalTok{ dev\_lower\_tempf)[}\StringTok{"dist"}\NormalTok{],}
        
        \AttributeTok{span =}\NormalTok{ span,}
        \AttributeTok{begin =}\NormalTok{ intFrom,}
        \AttributeTok{end =}\NormalTok{ intTo}
\NormalTok{      ), pId)}
\NormalTok{    \}}
\NormalTok{  \})}
\NormalTok{\}}
\end{Highlighting}
\end{Shaded}

\begin{table}

\caption{\label{tab:p4-area-scores-corr}Impact of the smoothing-span on the correlation of the computed distance vs. the ground truth.}
\centering
\begin{tabular}[t]{rrrrr}
\toprule
span & corr\_REQ & corr\_DEV & corr\_REQ\_pval & corr\_DEV\_pval\\
\midrule
0.2 & 0.3466692 & -0.6169631 & 0.3607302 & 0.0767463\\
0.3 & 0.3319356 & -0.3295731 & 0.3828371 & 0.3864396\\
0.4 & 0.3309750 & -0.3387297 & 0.3843000 & 0.3725654\\
0.5 & 0.2902033 & -0.2217038 & 0.4487374 & 0.5664418\\
0.6 & 0.3033020 & -0.1205962 & 0.4275460 & 0.7572814\\
\addlinespace
0.7 & 0.3719289 & -0.0141294 & 0.3243096 & 0.9712207\\
0.8 & 0.4423170 & -0.0130786 & 0.2332020 & 0.9733603\\
0.9 & 0.4903329 & -0.0500595 & 0.1802188 & 0.8982322\\
1.0 & 0.4405066 & -0.0870196 & 0.2353474 & 0.8238400\\
\bottomrule
\end{tabular}
\end{table}

In table \ref{tab:p4-area-scores-corr} we can observe a clear impact of the smoothing-span on the correlation of the computed distance vs.~the ground truth. Note that ideally, the correlation is negative, as a higher distance corresponds to a lower score.

The correlations for the \texttt{REQ} variable get slightly stronger with increasing smoothing-spans. However, the correlations are positive when they should not be. This increase is perhaps explained by the gradually increasing area overlaps. However, since it affects all projects and all of them have a similar overlap with the confidence interval, the distance gets lower, hence resulting in higher correlations.

As for the \texttt{DEV} variable, with increasing span the correlations get lower. That is because most of the projects run outside the confidence intervals of the pattern, and with increasing span, those parts that were inside, are getting more and more outside, as the peaks are smoothed out, hence resulting in a lower correlation. The correlation for the smallest span is close to being acceptable, if we consider the p-value. Also, all the correlations are negative, like they should be.

\begin{table}

\caption{\label{tab:p4-area-scores-partial-corr}Correlation of computed scores decline with more time considered.}
\centering
\begin{tabular}[t]{rrrrrr}
\toprule
span & end & corr\_REQ & corr\_DEV & corr\_REQ\_pval & corr\_DEV\_pval\\
\midrule
0.2 & 0.33 & 0.4458399 & -0.1813007 & 0.2290578 & 0.6406250\\
0.2 & 0.50 & 0.4378943 & -0.1222024 & 0.2384618 & 0.7541285\\
0.2 & 0.67 & 0.3789016 & -0.0569737 & 0.3145935 & 0.8842477\\
0.3 & 0.33 & 0.4603568 & -0.4740913 & 0.2124080 & 0.1972918\\
0.3 & 0.50 & 0.4473922 & -0.0069892 & 0.2272445 & 0.9857623\\
\addlinespace
0.3 & 0.67 & 0.4197792 & 0.1514301 & 0.2606651 & 0.6973434\\
0.4 & 0.33 & 0.4555741 & 0.1585221 & 0.2178173 & 0.6837478\\
0.4 & 0.50 & 0.4323705 & 0.3837292 & 0.2451201 & 0.3079534\\
0.4 & 0.67 & 0.3995959 & 0.3455309 & 0.2866377 & 0.3624157\\
0.6 & 0.33 & 0.4902852 & -0.6729866 & 0.1802676 & 0.0469613\\
\addlinespace
0.6 & 0.50 & 0.4654046 & -0.6729866 & 0.2067803 & 0.0469613\\
0.6 & 0.67 & 0.3972392 & 0.2682183 & 0.2897542 & 0.4852965\\
0.8 & 0.33 & 0.5502693 & 0.0000000 & 0.1247540 & 0.0000000\\
0.8 & 0.50 & 0.5372084 & 0.0000000 & 0.1358324 & 0.0000000\\
0.8 & 0.67 & 0.4508288 & 0.2782687 & 0.2232583 & 0.4684327\\
\addlinespace
1.0 & 0.33 & 0.4772620 & 0.0000000 & 0.1938905 & 0.0000000\\
1.0 & 0.50 & 0.4521476 & 0.0000000 & 0.2217388 & 0.0000000\\
1.0 & 0.67 & 0.3827532 & 0.3352792 & 0.3092902 & 0.3777657\\
\bottomrule
\end{tabular}
\end{table}

In table \ref{tab:p4-area-scores-partial-corr} we can clearly observe how the correlation of the computed score declines in almost every group of spans (esp.~for the \texttt{REQ} variable), the more time we consider, i.e., in a consistent manner the scores decline, and it is the same phenomenon for every smoothing-span. This is expected since pattern IV is the derivative of pattern I, which itself is only a poor reconciliation of how the Fire Drill apparently manifests in the data, which means that the more discrepancy we consider, the less good the results get. Again, we should consider computing this correlation-based score when using a data-enhanced or data-only pattern as actually, a typical moderate correlation with the scores of \(\approx-0.4\) (or better) as in table \ref{tab:p4-area-scores-partial-corr} for the \texttt{DEV}-variable can be expected to substantially increase with a more suitable pattern.

\hypertarget{correlation-between-curves}{%
\paragraph{Correlation between curves}\label{correlation-between-curves}}

Here we will compare each project's variable with the corresponding variable from the fourth pattern. We will take samples from either, pattern and variable, at the same \(x\), and then compute the sample correlation. This is a very simple but efficient test. Also, it is subject to user-defined intervals, i.e., we do not have to sample from project to project end. Note that this would probably not work for the first pattern, as it normalizes the variables at project end. Pattern IV however uses the rate of change, which is not affected by that.

We will be using LOESS-smoothed curved with a somewhat higher smoothing-span of \(0.6\). Also, since this test does not consider confidence intervals, we can also compare the \texttt{DESC} variable.

\begin{Shaded}
\begin{Highlighting}[]
\NormalTok{N }\OtherTok{\textless{}{-}} \DecValTok{200}
\NormalTok{use\_span }\OtherTok{\textless{}{-}} \FloatTok{0.6}
\NormalTok{X\_samp }\OtherTok{\textless{}{-}} \FunctionTok{seq}\NormalTok{(}\AttributeTok{from =} \DecValTok{0}\NormalTok{, }\AttributeTok{to =} \DecValTok{1}\NormalTok{, }\AttributeTok{length.out =}\NormalTok{ N)}

\NormalTok{p4\_samples }\OtherTok{\textless{}{-}} \FunctionTok{list}\NormalTok{(}\AttributeTok{REQ =} \FunctionTok{req\_d1\_p4}\NormalTok{(X\_samp), }\AttributeTok{DEV =} \FunctionTok{dev\_d1\_p4}\NormalTok{(X\_samp), }\AttributeTok{DESC =} \FunctionTok{desc\_d1\_p4}\NormalTok{(X\_samp))}

\NormalTok{p4\_corr }\OtherTok{\textless{}{-}} \ConstantTok{NULL}
\ControlFlowTok{for}\NormalTok{ (pId }\ControlFlowTok{in} \FunctionTok{names}\NormalTok{(all\_signals)) \{}

\NormalTok{  X }\OtherTok{\textless{}{-}} \DecValTok{1}\SpecialCharTok{:}\FunctionTok{nrow}\NormalTok{(all\_signals[[pId]]}\SpecialCharTok{$}\NormalTok{data)}
\NormalTok{  Y\_req }\OtherTok{\textless{}{-}} \FunctionTok{cumsum}\NormalTok{(all\_signals[[pId]]}\SpecialCharTok{$}\NormalTok{data}\SpecialCharTok{$}\NormalTok{req)}
\NormalTok{  Y\_dev }\OtherTok{\textless{}{-}} \FunctionTok{cumsum}\NormalTok{(all\_signals[[pId]]}\SpecialCharTok{$}\NormalTok{data}\SpecialCharTok{$}\NormalTok{dev)}
\NormalTok{  Y\_desc }\OtherTok{\textless{}{-}} \FunctionTok{cumsum}\NormalTok{(all\_signals[[pId]]}\SpecialCharTok{$}\NormalTok{data}\SpecialCharTok{$}\NormalTok{desc)}

\NormalTok{  loess\_req }\OtherTok{\textless{}{-}} \FunctionTok{smooth\_signal\_loess}\NormalTok{(}\AttributeTok{x =}\NormalTok{ X, }\AttributeTok{y =}\NormalTok{ Y\_req, }\AttributeTok{span =}\NormalTok{ use\_span)}
\NormalTok{  req\_tempf\_d1 }\OtherTok{\textless{}{-}} \ControlFlowTok{function}\NormalTok{(x) }\FunctionTok{func\_d1}\NormalTok{(}\AttributeTok{f =}\NormalTok{ loess\_req, }\AttributeTok{x =}\NormalTok{ x)}
\NormalTok{  loess\_dev }\OtherTok{\textless{}{-}} \FunctionTok{smooth\_signal\_loess}\NormalTok{(}\AttributeTok{x =}\NormalTok{ X, }\AttributeTok{y =}\NormalTok{ Y\_dev, }\AttributeTok{span =}\NormalTok{ use\_span)}
\NormalTok{  dev\_tempf\_d1 }\OtherTok{\textless{}{-}} \ControlFlowTok{function}\NormalTok{(x) }\FunctionTok{func\_d1}\NormalTok{(}\AttributeTok{f =}\NormalTok{ loess\_dev, }\AttributeTok{x =}\NormalTok{ x)}
\NormalTok{  loess\_desc }\OtherTok{\textless{}{-}} \FunctionTok{tryCatch}\NormalTok{(\{}
    \FunctionTok{smooth\_signal\_loess}\NormalTok{(}\AttributeTok{x =}\NormalTok{ X, }\AttributeTok{y =}\NormalTok{ Y\_desc, }\AttributeTok{span =}\NormalTok{ use\_span)}
\NormalTok{  \}, }\AttributeTok{error =} \ControlFlowTok{function}\NormalTok{(cond) }\ControlFlowTok{function}\NormalTok{(x) }\FunctionTok{rep}\NormalTok{(}\DecValTok{0}\NormalTok{, }\FunctionTok{length}\NormalTok{(x)))}
\NormalTok{  desc\_tempf\_d1 }\OtherTok{\textless{}{-}} \ControlFlowTok{function}\NormalTok{(x) }\FunctionTok{func\_d1}\NormalTok{(}\AttributeTok{f =}\NormalTok{ loess\_desc, }\AttributeTok{x =}\NormalTok{ x)}

\NormalTok{  p4\_corr }\OtherTok{\textless{}{-}} \FunctionTok{suppressWarnings}\NormalTok{(}\AttributeTok{expr =}\NormalTok{ \{}
    \FunctionTok{rbind}\NormalTok{(p4\_corr, }\StringTok{\textasciigrave{}}\AttributeTok{rownames\textless{}{-}}\StringTok{\textasciigrave{}}\NormalTok{(}\FunctionTok{data.frame}\NormalTok{(}\AttributeTok{REQ =}\NormalTok{ stats}\SpecialCharTok{::}\FunctionTok{cor}\NormalTok{(p4\_samples}\SpecialCharTok{$}\NormalTok{REQ, }\FunctionTok{req\_tempf\_d1}\NormalTok{(X\_samp)),}
      \AttributeTok{DEV =}\NormalTok{ stats}\SpecialCharTok{::}\FunctionTok{cor}\NormalTok{(p4\_samples}\SpecialCharTok{$}\NormalTok{DEV, }\FunctionTok{dev\_tempf\_d1}\NormalTok{(X\_samp)), }\AttributeTok{DESC =}\NormalTok{ stats}\SpecialCharTok{::}\FunctionTok{cor}\NormalTok{(p4\_samples}\SpecialCharTok{$}\NormalTok{DESC,}
        \FunctionTok{desc\_tempf\_d1}\NormalTok{(X\_samp))), pId))}
\NormalTok{  \})}
\NormalTok{\}}
\end{Highlighting}
\end{Shaded}

\begin{table}

\caption{\label{tab:p4-corr}Correlation scores of the derivatives with the derived project signals.}
\centering
\begin{tabular}[t]{lrrr}
\toprule
  & REQ & DEV & DESC\\
\midrule
Project1 & 0.9586322 & -0.2892431 & 0.0000000\\
Project2 & 0.9522143 & 0.4056002 & 0.0000000\\
Project3 & 0.9378174 & 0.7004125 & -0.7225524\\
Project4 & 0.5991196 & 0.6427211 & 0.7798444\\
Project5 & 0.3398163 & 0.2029798 & 0.0000000\\
\addlinespace
Project6 & 0.7018211 & 0.3541736 & 0.0679762\\
Project7 & 0.9340890 & 0.2999956 & 0.9620032\\
Project8 & 0.6612469 & 0.5963677 & 0.0000000\\
Project9 & 0.6951769 & -0.5883952 & -0.8020752\\
\bottomrule
\end{tabular}
\end{table}

\begin{Shaded}
\begin{Highlighting}[]
\FunctionTok{c}\NormalTok{(}
  \AttributeTok{REQ =} \FunctionTok{cor}\NormalTok{(ground\_truth}\SpecialCharTok{$}\NormalTok{consensus\_score, p4\_corr}\SpecialCharTok{$}\NormalTok{REQ),}
  \AttributeTok{DEV =} \FunctionTok{cor}\NormalTok{(ground\_truth}\SpecialCharTok{$}\NormalTok{consensus\_score, p4\_corr}\SpecialCharTok{$}\NormalTok{DEV),}
  \AttributeTok{DESC =} \FunctionTok{cor}\NormalTok{(ground\_truth}\SpecialCharTok{$}\NormalTok{consensus\_score, p4\_corr}\SpecialCharTok{$}\NormalTok{DESC, }\AttributeTok{use =} \StringTok{"pa"}\NormalTok{),}
  
  \AttributeTok{REQ\_pval =} \FunctionTok{cor.test}\NormalTok{(ground\_truth}\SpecialCharTok{$}\NormalTok{consensus\_score, p4\_corr}\SpecialCharTok{$}\NormalTok{REQ)}\SpecialCharTok{$}\NormalTok{p.value,}
  \AttributeTok{DEV\_pval =} \FunctionTok{cor.test}\NormalTok{(ground\_truth}\SpecialCharTok{$}\NormalTok{consensus\_score, p4\_corr}\SpecialCharTok{$}\NormalTok{DEV)}\SpecialCharTok{$}\NormalTok{p.value,}
  \AttributeTok{DESC\_pval =} \FunctionTok{cor.test}\NormalTok{(ground\_truth}\SpecialCharTok{$}\NormalTok{consensus\_score, p4\_corr}\SpecialCharTok{$}\NormalTok{DESC, }\AttributeTok{use =} \StringTok{"pa"}\NormalTok{)}\SpecialCharTok{$}\NormalTok{p.value}
\NormalTok{)}
\end{Highlighting}
\end{Shaded}

\begin{verbatim}
##          REQ          DEV         DESC     REQ_pval     DEV_pval    DESC_pval 
## -0.038684284  0.122120875  0.006891493  0.921291235  0.754288535  0.985961321
\end{verbatim}

The correlations in table \ref{tab:p4-corr} are varying, but their correlation with the ground truth is poor, and in the case of \texttt{REQ} and \texttt{DESC} even negative. That means that computing the correlation of the derived data with the derived project signals is not a suitable detector using pattern IV, which is the derivative of the expert-designed pattern I. However, this might work with a data-enhanced or data-only pattern. Our previous attempt using the distance in areas was much better.

\hypertarget{scoring-of-projects-2nd-batch}{%
\subsection{Scoring of projects (2nd batch)}\label{scoring-of-projects-2nd-batch}}

In the meantime, we got a new batch of projects (IDs \texttt{{[}10,15{]}}). We want to use the regression model based on the previous batch of projects to calculate a degree to which the Fire Drill is present. To goal is to find out whether the model built on previous observations can generalize to new, previously unseen data. The model was fit on scores as computed against pattern III (average), and it used two kinds of scores: a) the average confidence as computed by the path of each variable it takes through the confidence surface, and b) the average distance to the reference variable (the weighted average of previous projects).

So, in order to obtain predictions, we need to do n things:

\begin{enumerate}
\def\labelenumi{\arabic{enumi}.}
\tightlist
\item
  Load the new projects -- the file \texttt{FD\_issue-based\_detection.xlsx} was updated to include the new projects. We'll use the function \texttt{load\_project\_issue\_data()} to instantiate the signals.
\item
  Compute the two kinds of scores; we'll run the \textbf{new} projects against the pattern that is based on the \textbf{old} projects.
\item
  Use the scores in the previously fit regression model and predict a ground truth.
\end{enumerate}

\hypertarget{ground-truth-1}{%
\subsubsection{Ground truth}\label{ground-truth-1}}

\begin{Shaded}
\begin{Highlighting}[]
\NormalTok{ground\_truth\_2nd\_batch }\OtherTok{\textless{}{-}} \FunctionTok{read.csv}\NormalTok{(}\AttributeTok{file =} \StringTok{"../data/ground{-}truth\_2nd{-}batch.csv"}\NormalTok{, }\AttributeTok{sep =} \StringTok{";"}\NormalTok{)}
\NormalTok{ground\_truth\_2nd\_batch}\SpecialCharTok{$}\NormalTok{consensus\_score }\OtherTok{\textless{}{-}}\NormalTok{ ground\_truth\_2nd\_batch}\SpecialCharTok{$}\NormalTok{consensus}\SpecialCharTok{/}\DecValTok{10}
\FunctionTok{rownames}\NormalTok{(ground\_truth\_2nd\_batch) }\OtherTok{\textless{}{-}} \FunctionTok{paste0}\NormalTok{((}\DecValTok{1} \SpecialCharTok{+} \FunctionTok{nrow}\NormalTok{(ground\_truth))}\SpecialCharTok{:}\NormalTok{(}\FunctionTok{nrow}\NormalTok{(ground\_truth) }\SpecialCharTok{+}
  \FunctionTok{nrow}\NormalTok{(ground\_truth\_2nd\_batch)))}
\end{Highlighting}
\end{Shaded}

We have ground truth available for the second batch. Again, the two raters assessed it independently first, and only later found a consensus. The a priori agreement between both raters was better this time, as the quadratic weighted Kappa increased substantially to a value of \textbf{0.929} (correlation is \textbf{0.975}). Table \ref{tab:groundtruth-2nd-batch} shows the ground truth for the new projects \texttt{{[}10-15{]}}.

For the entirety of \textbf{both batches}, the quadratic weighted Kappa is \textbf{0.813}, and the Pearson correlation of both raters' assessments is \textbf{0.815}.

\begin{table}

\caption{\label{tab:groundtruth-2nd-batch}Entire ground truth as of both raters}
\centering
\begin{tabular}[t]{llrrrrr}
\toprule
  & project & rater.a & rater.b & consensus & rater.mean & consensus\_score\\
\midrule
10 & project\_10 & 2 & 3 & 2 & 2.5 & 0.2\\
11 & project\_11 & 0 & 1 & 0 & 0.5 & 0.0\\
12 & project\_12 & 1 & 2 & 2 & 1.5 & 0.2\\
13 & project\_13 & 8 & 10 & 10 & 9.0 & 1.0\\
14 & project\_14 & 1 & 0 & 1 & 0.5 & 0.1\\
\addlinespace
15 & project\_15 & 1 & 1 & 1 & 1.0 & 0.1\\
\bottomrule
\end{tabular}
\end{table}

\hypertarget{ground-truth-inter-rater-agreement}{%
\subsubsection{Ground truth inter-rater agreement}\label{ground-truth-inter-rater-agreement}}

We have previously reported the quadratic weighted Kappa, as well as the Pearson correlation.
We should note, however, that the proposed scale by Landis and Koch (1977) was chosen rather arbitrarily, and that each Kappa \emph{agreement coefficient} is a point estimate associated with a probability distribution and a margin of error (Klein 2018).
It is therefore recommended to properly \emph{benchmark} raw Kappa coefficients.
Gwet suggests to compute the probability of a Kappa to fall into a certain range by integrating a standard normal distribution (where the Kappa coefficient is the mean and its associated error is the standard deviation) (Gwet 2014).
We should probably also report Gwet's coefficient called ``AC\textsubscript{$1$}'', which outperforms other coefficients in terms of having reasonably small biases for estimating the true inter-rater reliability (Gwet 2008).
It is especially applicable in the presence of high agreement because of its comparatively low bias.

Also, so far, we had two raters. In the meantime, however, a third rater gave their assessment for all \(15\) projects.
In order to assess inter-rater agreement of three or more raters, we will have to use a Kappa or coefficient that can handle at least three raters (Gwet's can, for example).

\begin{Shaded}
\begin{Highlighting}[]
\CommentTok{\# Let\textquotesingle{}s put them together.}
\NormalTok{ground\_truth\_all }\OtherTok{\textless{}{-}} \FunctionTok{rbind}\NormalTok{(ground\_truth, ground\_truth\_2nd\_batch)}
\CommentTok{\# Add the 3rd rater to it:}
\NormalTok{ground\_truth\_all}\SpecialCharTok{$}\NormalTok{rater.c }\OtherTok{\textless{}{-}} \FunctionTok{c}\NormalTok{(}\DecValTok{0}\NormalTok{, }\DecValTok{0}\NormalTok{, }\DecValTok{10}\NormalTok{, }\DecValTok{8}\NormalTok{, }\DecValTok{0}\NormalTok{, }\DecValTok{2}\NormalTok{, }\DecValTok{4}\NormalTok{, }\DecValTok{2}\NormalTok{, }\DecValTok{2}\NormalTok{, }\DecValTok{1}\NormalTok{, }\DecValTok{0}\NormalTok{, }\DecValTok{0}\NormalTok{, }\DecValTok{6}\NormalTok{, }\DecValTok{0}\NormalTok{, }\DecValTok{0}\NormalTok{)}
\end{Highlighting}
\end{Shaded}

In the following, we will report three things:

\begin{itemize}
\tightlist
\item
  Pair-wise Pearson correlation (rater A vs.~B, A vs.~C, B vs.~C)
\item
  Percentage agreement between all three raters
\item
  Various other Kappas: Gwet's AC\textsubscript{$1$}, Fleiss' Kappa, Krippendorff's Alpha, and Conger's Kappa.
\end{itemize}

For the last point, we will also benchmark these Kappas according to Landis and Koch's scale, as well as Fleiss' scale (Altman's scale is almost identical with Landis + Koch).
Note that the various Kappas and coefficients (except for Krippendorff's) are designed to work with nominal data only.
All of the ones we are going to test support, however, weight matrices. That means, we can make them work with ordinal data (i.e., ordered nominal data).
We will therefore first convert our ground truth, which is in the range \([0,10]\) to a factor/string, and then use linear weights (our ground truth uses a linear Likert scale).
We can therefore treat our Likert-scaled data as ordinal. Since the items are spaced linearly, so will be the weights.
We should note, however, that there are are various other methods for assessing inter-rater agreement on Likert scales (James, Demaree, and Wolf 1984; ONeill 2017) but here, however, we will treat it as an equally-spaced ordinal scale, which is equivalent but more straightforward.
The weights then assign an order and absolute distance between items to it. The bigger the differences between two raters' assessments, the bigger the difference in weight will be.

\hypertarget{pair-wise-pearson-correlation}{%
\paragraph{Pair-wise Pearson correlation}\label{pair-wise-pearson-correlation}}

Let's see how well the individual raters' assessments correlate with each other.

\begin{Shaded}
\begin{Highlighting}[]
\FunctionTok{c}\NormalTok{(stats}\SpecialCharTok{::}\FunctionTok{cor}\NormalTok{(ground\_truth\_all}\SpecialCharTok{$}\NormalTok{rater.a, ground\_truth\_all}\SpecialCharTok{$}\NormalTok{rater.b), stats}\SpecialCharTok{::}\FunctionTok{cor}\NormalTok{(ground\_truth\_all}\SpecialCharTok{$}\NormalTok{rater.a,}
\NormalTok{  ground\_truth\_all}\SpecialCharTok{$}\NormalTok{rater.c), stats}\SpecialCharTok{::}\FunctionTok{cor}\NormalTok{(ground\_truth\_all}\SpecialCharTok{$}\NormalTok{rater.b, ground\_truth\_all}\SpecialCharTok{$}\NormalTok{rater.c))}
\end{Highlighting}
\end{Shaded}

\begin{verbatim}
## [1] 0.8152014 0.8966555 0.7534160
\end{verbatim}

These are quite significant correlations. The p-values for these combinations are all almost zero:

\begin{Shaded}
\begin{Highlighting}[]
\FunctionTok{c}\NormalTok{(stats}\SpecialCharTok{::}\FunctionTok{cor.test}\NormalTok{(ground\_truth\_all}\SpecialCharTok{$}\NormalTok{rater.a, ground\_truth\_all}\SpecialCharTok{$}\NormalTok{rater.b)}\SpecialCharTok{$}\NormalTok{p.value, stats}\SpecialCharTok{::}\FunctionTok{cor.test}\NormalTok{(ground\_truth\_all}\SpecialCharTok{$}\NormalTok{rater.a,}
\NormalTok{  ground\_truth\_all}\SpecialCharTok{$}\NormalTok{rater.c)}\SpecialCharTok{$}\NormalTok{p.value, stats}\SpecialCharTok{::}\FunctionTok{cor.test}\NormalTok{(ground\_truth\_all}\SpecialCharTok{$}\NormalTok{rater.b,}
\NormalTok{  ground\_truth\_all}\SpecialCharTok{$}\NormalTok{rater.c)}\SpecialCharTok{$}\NormalTok{p.value)}
\end{Highlighting}
\end{Shaded}

\begin{verbatim}
## [1] 2.129309e-04 5.983635e-06 1.181933e-03
\end{verbatim}

The null hypothesis is that there is no correlation, but we clearly have to reject it in all cases, as there is no statistical significant evidence for it.

\hypertarget{percentage-agreement}{%
\paragraph{Percentage agreement}\label{percentage-agreement}}

This is roughly the same as one minus the mean of the mean absolute error among the pairs of raters divided by the range of the ground truth:

\begin{Shaded}
\begin{Highlighting}[]
\NormalTok{(temp }\OtherTok{\textless{}{-}} \DecValTok{1} \SpecialCharTok{{-}} \FunctionTok{mean}\NormalTok{(}\FunctionTok{c}\NormalTok{(Metrics}\SpecialCharTok{::}\FunctionTok{mae}\NormalTok{(ground\_truth\_all}\SpecialCharTok{$}\NormalTok{rater.a, ground\_truth\_all}\SpecialCharTok{$}\NormalTok{rater.b),}
\NormalTok{  Metrics}\SpecialCharTok{::}\FunctionTok{mae}\NormalTok{(ground\_truth\_all}\SpecialCharTok{$}\NormalTok{rater.a, ground\_truth\_all}\SpecialCharTok{$}\NormalTok{rater.c), Metrics}\SpecialCharTok{::}\FunctionTok{mae}\NormalTok{(ground\_truth\_all}\SpecialCharTok{$}\NormalTok{rater.b,}
\NormalTok{    ground\_truth\_all}\SpecialCharTok{$}\NormalTok{rater.c)))}\SpecialCharTok{/}\DecValTok{10}\NormalTok{)}
\end{Highlighting}
\end{Shaded}

\begin{verbatim}
## [1] 0.8622222
\end{verbatim}

This means that we have a percentual agreement between all three raters of about 86.22\%.
This quite good, but we have to proceed to the next method of benchmarking to get recognizable results.

\hypertarget{benchmarking-of-fleiss-kappa-gwets-ac1ac2-krippendorffs-alpha-and-congers-kappa}{%
\paragraph{Benchmarking of Fleiss' Kappa, Gwet's AC1/AC2, Krippendorff's Alpha, and Conger's Kappa}\label{benchmarking-of-fleiss-kappa-gwets-ac1ac2-krippendorffs-alpha-and-congers-kappa}}

Let's compute a few Kappas and agreement coefficients.
Table \ref{tab:gt-kappas} shows an overview. The percentage agreement is almost always the same and equal to what we computed previously manually.
The standard error seems to be quite similar, too.
But we see differences for the coefficient itself, with the largest being Gwet's AC\textsubscript{$1$}/AC\textsubscript{$2$} (the remaining others being pretty similar).
This is the one we should report, as Gwet has shown that it outperforms other Kappas/coefficients in the presence of high agreement, which is the case here (Gwet 2008).

\begin{Shaded}
\begin{Highlighting}[]
\NormalTok{ratings }\OtherTok{\textless{}{-}} \FunctionTok{cbind}\NormalTok{(ground\_truth\_all[, }\FunctionTok{c}\NormalTok{(}\StringTok{"rater.a"}\NormalTok{, }\StringTok{"rater.b"}\NormalTok{)], }\FunctionTok{data.frame}\NormalTok{(}\AttributeTok{rater.c =}\NormalTok{ ground\_truth\_all}\SpecialCharTok{$}\NormalTok{rater.c))}

\NormalTok{temp }\OtherTok{\textless{}{-}} \FunctionTok{rbind}\NormalTok{(irrCAC}\SpecialCharTok{::}\FunctionTok{fleiss.kappa.raw}\NormalTok{(}\AttributeTok{ratings =}\NormalTok{ ratings, }\AttributeTok{weights =} \StringTok{"linear"}\NormalTok{, }\AttributeTok{categ.labels =} \FunctionTok{paste0}\NormalTok{(}\DecValTok{0}\SpecialCharTok{:}\DecValTok{10}\NormalTok{))}\SpecialCharTok{$}\NormalTok{est,}
\NormalTok{  irrCAC}\SpecialCharTok{::}\FunctionTok{gwet.ac1.raw}\NormalTok{(}\AttributeTok{ratings =}\NormalTok{ ratings, }\AttributeTok{weights =} \StringTok{"linear"}\NormalTok{, }\AttributeTok{categ.labels =} \FunctionTok{paste0}\NormalTok{(}\DecValTok{0}\SpecialCharTok{:}\DecValTok{10}\NormalTok{))}\SpecialCharTok{$}\NormalTok{est,}
\NormalTok{  irrCAC}\SpecialCharTok{::}\FunctionTok{krippen.alpha.raw}\NormalTok{(}\AttributeTok{ratings =}\NormalTok{ ratings, }\AttributeTok{weights =} \StringTok{"linear"}\NormalTok{, }\AttributeTok{categ.labels =} \FunctionTok{paste0}\NormalTok{(}\DecValTok{0}\SpecialCharTok{:}\DecValTok{10}\NormalTok{))}\SpecialCharTok{$}\NormalTok{est,}
\NormalTok{  irrCAC}\SpecialCharTok{::}\FunctionTok{conger.kappa.raw}\NormalTok{(}\AttributeTok{ratings =}\NormalTok{ ratings, }\AttributeTok{weights =} \StringTok{"linear"}\NormalTok{, }\AttributeTok{categ.labels =} \FunctionTok{paste0}\NormalTok{(}\DecValTok{0}\SpecialCharTok{:}\DecValTok{10}\NormalTok{))}\SpecialCharTok{$}\NormalTok{est)}
\end{Highlighting}
\end{Shaded}

\begin{table}

\caption{\label{tab:gt-kappas}Fleiss' Kappa, Gwet's AC1/AC2, Krippendorff's Alpha, and Conger's Kappa. The table shows the percent agreement (pa), percent chance agreement (pe), the coefficient's/Kappa's value, its standard error, the confidence interval, and p-value.}
\centering
\begin{tabular}[t]{lrrrrlrl}
\toprule
coeff.name & pa & pe & coeff.val & coeff.se & conf.int & p.value & w.name\\
\midrule
Fleiss' Kappa & 0.8622222 & 0.7009383 & 0.53930 & 0.07785 & (0.372,0.706) & 7.0e-06 & linear\\
AC2 & 0.8622222 & 0.5676049 & 0.68136 & 0.07187 & (0.527,0.835) & 2.0e-07 & linear\\
Krippendorff's Alpha & 0.8652840 & 0.7009383 & 0.54954 & 0.07785 & (0.383,0.717) & 5.7e-06 & linear\\
Conger's Kappa & 0.8622222 & 0.6983704 & 0.54322 & 0.07711 & (0.378,0.709) & 5.8e-06 & linear\\
\bottomrule
\end{tabular}
\end{table}

Let's compute some benchmarks.
They are shown in tables \ref{tab:bench-landis-koch} and \ref{tab:bench-fleiss}.
What these tables show is the amount of probability with with the computed agreement coefficient falls into the defined range.
For example, about \(82.18\)\% of the computed coefficient falls into the \emph{Substantial}-category.
Also, the Kappa is \(>99.99\)\% associated with the category \emph{Moderate} (\(\geq0.4\)) or better.
These results mean that the majority of the Kappa is substantial, and we have a small (but not insignificant) portion being in the almost perfect category (\(\approx5\)\%). Another \(\approx12.9\)\% are in the moderate category.
In table \ref{tab:bench-fleiss}, about \(1/6\) falls into the category \emph{Excellent}, and the remaining \(5/6\) into the next best category \emph{Intermediate to Good}.

\begin{Shaded}
\begin{Highlighting}[]
\NormalTok{ac1 }\OtherTok{\textless{}{-}}\NormalTok{ irrCAC}\SpecialCharTok{::}\FunctionTok{gwet.ac1.raw}\NormalTok{(}\AttributeTok{ratings =}\NormalTok{ ratings, }\AttributeTok{weights =} \StringTok{"linear"}\NormalTok{, }\AttributeTok{categ.labels =} \FunctionTok{paste0}\NormalTok{(}\DecValTok{0}\SpecialCharTok{:}\DecValTok{10}\NormalTok{))}\SpecialCharTok{$}\NormalTok{est}

\NormalTok{temp1 }\OtherTok{\textless{}{-}}\NormalTok{ irrCAC}\SpecialCharTok{::}\FunctionTok{landis.koch.bf}\NormalTok{(}\AttributeTok{coeff =}\NormalTok{ ac1}\SpecialCharTok{$}\NormalTok{coeff.val, }\AttributeTok{se =}\NormalTok{ ac1}\SpecialCharTok{$}\NormalTok{coeff.se)}
\NormalTok{temp2 }\OtherTok{\textless{}{-}}\NormalTok{ irrCAC}\SpecialCharTok{::}\FunctionTok{fleiss.bf}\NormalTok{(}\AttributeTok{coeff =}\NormalTok{ ac1}\SpecialCharTok{$}\NormalTok{coeff.val, }\AttributeTok{se =}\NormalTok{ ac1}\SpecialCharTok{$}\NormalTok{coeff.se)}
\end{Highlighting}
\end{Shaded}

\begin{table}

\caption{\label{tab:bench-landis-koch}Landis-Koch benchmark (cumulative probabilities) for Gwet's AC1/AC2 agreement coefficient for all three raters.}
\centering
\begin{tabular}[t]{lllr}
\toprule
  & Landis-Koch & CumProb & Prob\\
\midrule
(0.8 to 1) & Almost Perfect & 0.04939 & 0.04939\\
(0.6 to 0.8) & Substantial & 0.87119 & 0.82180\\
(0.4 to 0.6) & Moderate & 0.99995 & 0.12876\\
(0.2 to 0.4) & Fair & 1 & 0.00005\\
(0 to 0.2) & Slight & 1 & 0.00000\\
\addlinespace
(-1 to 0) & Poor & 1 & 0.00000\\
\bottomrule
\end{tabular}
\end{table}

\begin{table}

\caption{\label{tab:bench-fleiss}Fleiss's benchmark (cumulative probabilities) for Gwet's AC1/AC2 agreement coefficient for all three raters.}
\centering
\begin{tabular}[t]{lllr}
\toprule
  & Fleiss & CumProb & Prob\\
\midrule
(0.75 to 1) & Excellent & 0.16977 & 0.16977\\
(0.4 to 0.75) & Intermediate to Good & 0.99995 & 0.83018\\
(-1 to 0.4) & Poor & 1 & 0.00005\\
\bottomrule
\end{tabular}
\end{table}

Tables \ref{tab:bench-landis-koch} and \ref{tab:bench-fleiss} show the benchmarks for Landis and Koch (1977) and Fleiss (1981), respectively.
There are two more common benchmarks, the one by Cicchetti and Sparrow (1981) and one by Regier et al. (2013).
These are shown in tables \ref{tab:bench-cicc} and \ref{tab:bench-regier}, respectively.
Note that the benchmark by Regier et al. (2013) uses the same levels as the one by Landis and Koch (1977), but only uses different labels.
We will compute these, too, to get a better idea of the goodness of the measured agreement.

\begin{Shaded}
\begin{Highlighting}[]
\NormalTok{bench\_cicc }\OtherTok{\textless{}{-}} \FunctionTok{c}\NormalTok{(}\AttributeTok{start =} \DecValTok{0}\NormalTok{, }\AttributeTok{Poor =} \FloatTok{0.4}\NormalTok{, }\AttributeTok{Fair =} \FloatTok{0.6}\NormalTok{, }\AttributeTok{Good =} \FloatTok{0.75}\NormalTok{, }\AttributeTok{Excellent =} \DecValTok{1}\NormalTok{)}
\NormalTok{bench\_regier }\OtherTok{\textless{}{-}} \FunctionTok{c}\NormalTok{(}\AttributeTok{start =} \DecValTok{0}\NormalTok{, }\AttributeTok{Unacceptable =} \FloatTok{0.2}\NormalTok{, }\AttributeTok{Questionable =} \FloatTok{0.4}\NormalTok{, }\AttributeTok{Good =} \FloatTok{0.6}\NormalTok{,}
  \StringTok{\textasciigrave{}}\AttributeTok{Very Good}\StringTok{\textasciigrave{}} \OtherTok{=} \FloatTok{0.8}\NormalTok{, }\AttributeTok{Excellent =} \DecValTok{1}\NormalTok{)}

\NormalTok{compute\_benchmark }\OtherTok{\textless{}{-}} \ControlFlowTok{function}\NormalTok{(benchvec, name, cdf) \{}
\NormalTok{  df }\OtherTok{\textless{}{-}} \ConstantTok{NULL}
  \ControlFlowTok{for}\NormalTok{ (idx }\ControlFlowTok{in} \DecValTok{2}\SpecialCharTok{:}\FunctionTok{length}\NormalTok{(benchvec)) \{}
\NormalTok{    a }\OtherTok{\textless{}{-}}\NormalTok{ benchvec[idx }\SpecialCharTok{{-}} \DecValTok{1}\NormalTok{]}
\NormalTok{    b }\OtherTok{\textless{}{-}}\NormalTok{ benchvec[idx]}
\NormalTok{    df }\OtherTok{\textless{}{-}} \FunctionTok{rbind}\NormalTok{(df, }\FunctionTok{data.frame}\NormalTok{(}\AttributeTok{X1 =} \FunctionTok{names}\NormalTok{(benchvec)[idx], }\AttributeTok{X2 =} \FunctionTok{round}\NormalTok{(}\FunctionTok{cdf}\NormalTok{(b) }\SpecialCharTok{{-}}
      \FunctionTok{cdf}\NormalTok{(a), }\DecValTok{5}\NormalTok{), }\AttributeTok{stringsAsFactors =} \ConstantTok{FALSE}\NormalTok{))}
\NormalTok{  \}}
  \FunctionTok{rownames}\NormalTok{(}\AttributeTok{x =}\NormalTok{ df) }\OtherTok{\textless{}{-}} \FunctionTok{sapply}\NormalTok{(}\AttributeTok{X =} \DecValTok{2}\SpecialCharTok{:}\FunctionTok{length}\NormalTok{(benchvec), }\AttributeTok{FUN =} \ControlFlowTok{function}\NormalTok{(idx) \{}
    \FunctionTok{paste0}\NormalTok{(}\StringTok{"("}\NormalTok{, benchvec[idx }\SpecialCharTok{{-}} \DecValTok{1}\NormalTok{], }\StringTok{" to "}\NormalTok{, benchvec[idx], }\StringTok{")"}\NormalTok{)}
\NormalTok{  \})}
  \FunctionTok{colnames}\NormalTok{(}\AttributeTok{x =}\NormalTok{ df) }\OtherTok{\textless{}{-}} \FunctionTok{c}\NormalTok{(name, }\StringTok{"Prob"}\NormalTok{)}
\NormalTok{  df}\SpecialCharTok{$}\NormalTok{CumProb }\OtherTok{\textless{}{-}} \FunctionTok{rev}\NormalTok{(}\FunctionTok{cumsum}\NormalTok{(}\FunctionTok{rev}\NormalTok{(df}\SpecialCharTok{$}\NormalTok{Prob)))}
\NormalTok{  df[}\FunctionTok{order}\NormalTok{(}\FunctionTok{nrow}\NormalTok{(df)}\SpecialCharTok{:}\DecValTok{1}\NormalTok{), ]}
\NormalTok{\}}
\end{Highlighting}
\end{Shaded}

\begin{table}

\caption{\label{tab:bench-cicc}Cicchetti et al.'s benchmark (cumulative probabilities) for Gwet's AC1/AC2 agreement coefficient for all three raters.}
\centering
\begin{tabular}[t]{llrr}
\toprule
  & Cicchetti & Prob & CumProb\\
\midrule
(0.75 to 1) & Excellent & 0.16977 & 0.16977\\
(0.6 to 0.75) & Good & 0.70142 & 0.87119\\
(0.4 to 0.6) & Fair & 0.12876 & 0.99995\\
(0 to 0.4) & Poor & 0.00005 & 1.00000\\
\bottomrule
\end{tabular}
\end{table}

\begin{table}

\caption{\label{tab:bench-regier}Regier et al.'s benchmark (cumulative probabilities) for Gwet's AC1/AC2 agreement coefficient for all three raters.}
\centering
\begin{tabular}[t]{llrr}
\toprule
  & Regier & Prob & CumProb\\
\midrule
(0.8 to 1) & Excellent & 0.04939 & 0.04939\\
(0.6 to 0.8) & Very Good & 0.82180 & 0.87119\\
(0.4 to 0.6) & Good & 0.12876 & 0.99995\\
(0.2 to 0.4) & Questionable & 0.00005 & 1.00000\\
(0 to 0.2) & Unacceptable & 0.00000 & 1.00000\\
\bottomrule
\end{tabular}
\end{table}

We can also visualize the probability. It is computed using a standard normal distribution (Klein 2018).
Equation \ref{eq:kappa-bench} shows how to compute the probability for an agreement coefficient \(\kappa\) to be in the range \([a,b]\).
This is also called \emph{interval membership probability} (Gwet 2014).
Note that the square root of the variance (\(\sqrt{\operatorname{Var}}\)) is the same as the standard deviation.
Instead of this formula, we can also use a parameterized PDF (or CDF) directly (see equation \ref{eq:kappa-bench-param}).
For example:

\begin{Shaded}
\begin{Highlighting}[]
\NormalTok{tempf }\OtherTok{\textless{}{-}} \ControlFlowTok{function}\NormalTok{(q) }\FunctionTok{pnorm}\NormalTok{(}\AttributeTok{q =}\NormalTok{ q, }\AttributeTok{mean =}\NormalTok{ ac1}\SpecialCharTok{$}\NormalTok{coeff.val, }\AttributeTok{sd =}\NormalTok{ ac1}\SpecialCharTok{$}\NormalTok{coeff.se)}

\CommentTok{\# This is equivalent to the Landis{-}Koch benchmark of \textquotesingle{}Almost Perfect\textquotesingle{} (see}
\CommentTok{\# previous table)}
\FunctionTok{c}\NormalTok{(}\FunctionTok{tempf}\NormalTok{(}\DecValTok{1}\NormalTok{) }\SpecialCharTok{{-}} \FunctionTok{tempf}\NormalTok{(}\FloatTok{0.8}\NormalTok{))}
\end{Highlighting}
\end{Shaded}

\begin{verbatim}
## [1] 0.04938933
\end{verbatim}

\begin{align}
    \Phi\dots&\;\text{CDF of the standard normal distribution}\nonumber,
    \\[1ex]
    Pr(a\leq\kappa\leq b)=&\;\Phi\left(\frac{\kappa-a}{\sqrt{\operatorname{Var}(\kappa)}}\right)-\Phi\left(\frac{\kappa-b}{\sqrt{\operatorname{Var}(\kappa)}}\right)\label{eq:kappa-bench},
    \\[1ex]
    =&\;\int_{\begin{cases}
    a&\text{, if}\;a>0,
    \\
    -\infty&\text{, otherwise}
    \end{cases}}^{\begin{cases}
    b&\text{, if}\;b<1,
    \\
    \infty&\text{, otherwise}
    \end{cases}}\,\left[\frac{1}{\sqrt{\operatorname{Var}(\kappa)}\sqrt{2\pi}}e^{-\frac{1}{2}\left(\frac{x-\kappa}{\sqrt{\operatorname{Var}(\kappa)}}\right)^2}\right]\,dx\label{eq:kappa-bench-param}.
\end{align}

\begin{figure}[ht!]
\includegraphics{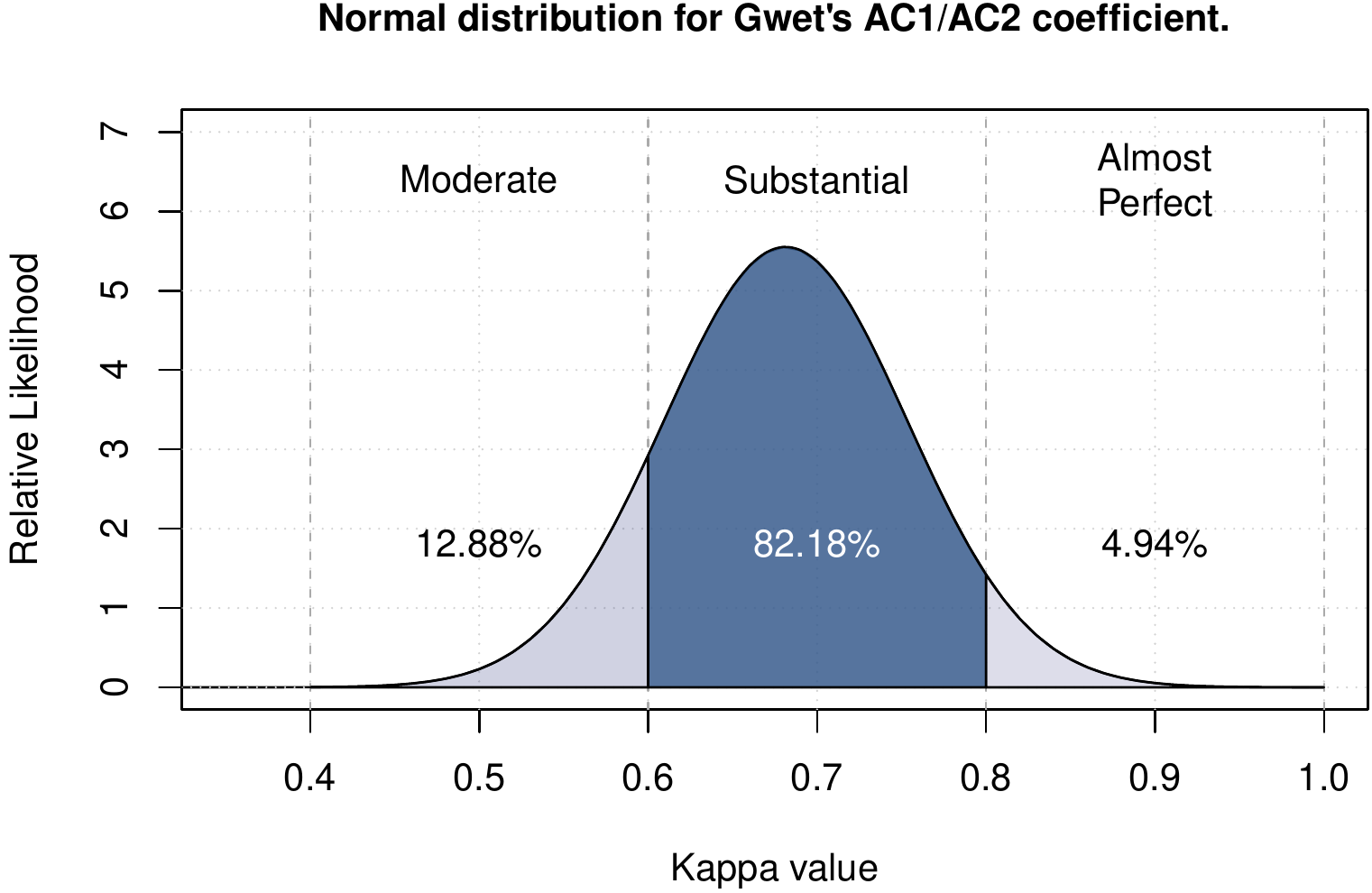} \caption{Normal distribution for Gwet's AC1/AC2 coefficient. Shown are the integrated areas and the opacity indicates the probability for the coefficient to be in the given range. The mean is at approx. 0.681 and the standard deviation is about 0.072.}\label{fig:gwet-ac1ac2-pdf}
\end{figure}

Figure \ref{fig:gwet-ac1ac2-pdf} shows the three top-most categories according to Landis and Koch (1977) and their integral.

\hypertarget{loading-the-new-projects}{%
\subsubsection{Loading the new projects}\label{loading-the-new-projects}}

\begin{Shaded}
\begin{Highlighting}[]
\FunctionTok{library}\NormalTok{(readxl)}

\NormalTok{all\_signals\_2nd\_batch }\OtherTok{\textless{}{-}} \FunctionTok{list}\NormalTok{()}

\ControlFlowTok{for}\NormalTok{ (pId }\ControlFlowTok{in} \FunctionTok{paste0}\NormalTok{(}\StringTok{"Project"}\NormalTok{, }\DecValTok{10}\SpecialCharTok{:}\DecValTok{15}\NormalTok{)) \{}
\NormalTok{  all\_signals\_2nd\_batch[[pId]] }\OtherTok{\textless{}{-}} \FunctionTok{load\_project\_issue\_data}\NormalTok{(}\AttributeTok{pId =}\NormalTok{ pId)}
\NormalTok{\}}
\end{Highlighting}
\end{Shaded}

Let's see how they look (sanity check):

\begin{figure}[ht!]
\includegraphics{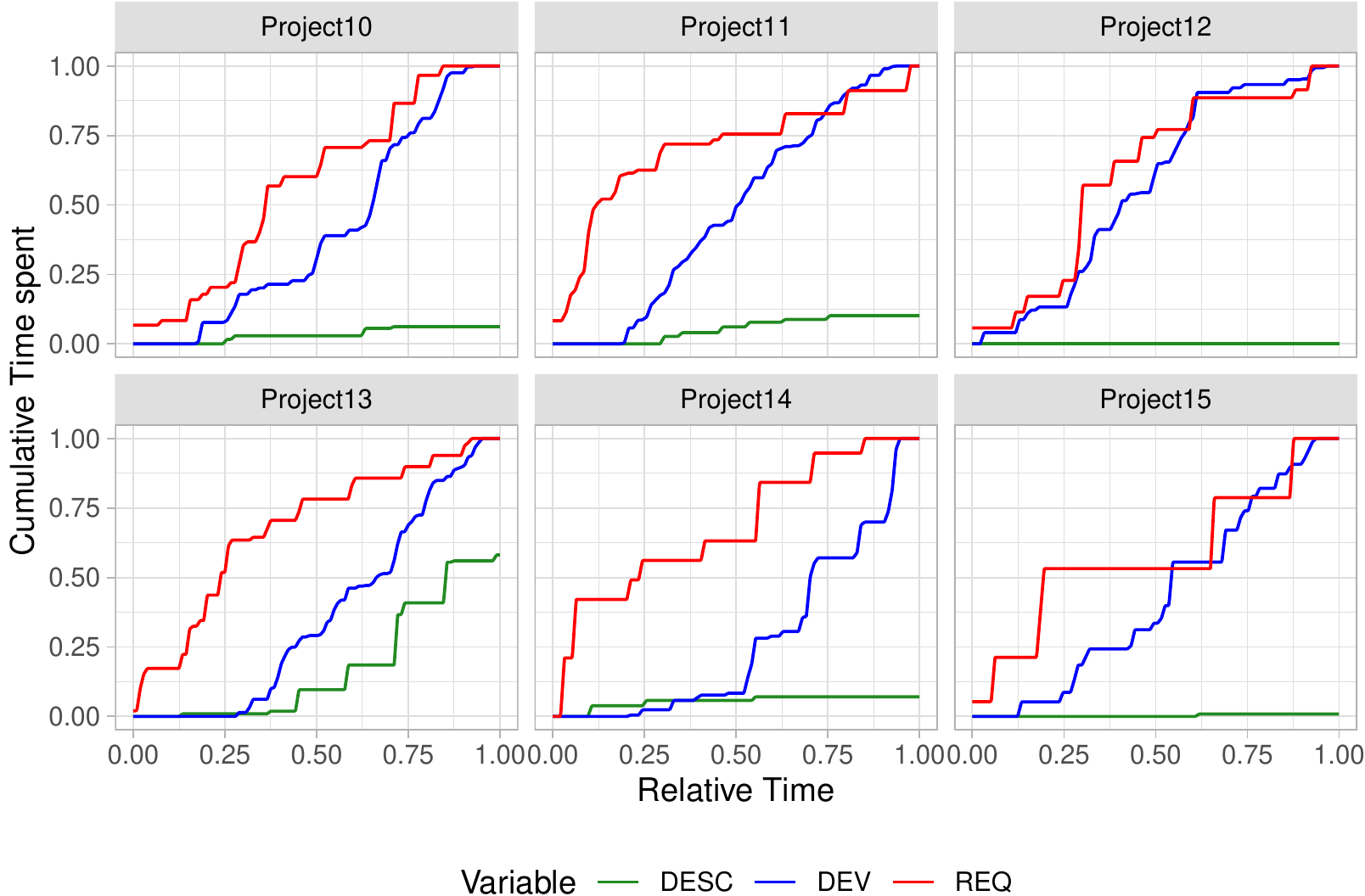} \caption{The second batch of projects [10-15]. All variables over each project's time span.}\label{fig:project-it-vars-2nd-batch}
\end{figure}

The six new projects in figure \ref{fig:project-it-vars-2nd-batch} look like they do in the excel, the import was successful.

\hypertarget{evaluation-of-the-binary-decision-rule}{%
\subsubsection{\texorpdfstring{Evaluation of the binary decision rule\label{ssec:eval-dr-2nd-batch}}{Evaluation of the binary decision rule}}\label{evaluation-of-the-binary-decision-rule}}

Here we will evaluate the existing binary decision rule to find out whether it can classify the second batch of projects correctly.

\begin{Shaded}
\begin{Highlighting}[]
\NormalTok{temp }\OtherTok{\textless{}{-}} \FunctionTok{sapply}\NormalTok{(}\AttributeTok{X =} \FunctionTok{names}\NormalTok{(all\_signals\_2nd\_batch), }\AttributeTok{FUN =} \ControlFlowTok{function}\NormalTok{(pName) \{}
  \FunctionTok{p1\_dr}\NormalTok{(}\AttributeTok{projName =}\NormalTok{ pName, }\AttributeTok{signals =}\NormalTok{ all\_signals\_2nd\_batch)}
\NormalTok{\})}
\NormalTok{p1\_detect\_2nd\_batch }\OtherTok{\textless{}{-}} \FunctionTok{data.frame}\NormalTok{(}\AttributeTok{detect =}\NormalTok{ temp, }\AttributeTok{ground\_truth =}\NormalTok{ ground\_truth\_2nd\_batch}\SpecialCharTok{$}\NormalTok{consensus,}
  \AttributeTok{correct =}\NormalTok{ (temp }\SpecialCharTok{\&}\NormalTok{ ground\_truth\_2nd\_batch}\SpecialCharTok{$}\NormalTok{consensus }\SpecialCharTok{\textgreater{}=} \DecValTok{5}\NormalTok{) }\SpecialCharTok{|}\NormalTok{ (}\SpecialCharTok{!}\NormalTok{temp }\SpecialCharTok{\&}\NormalTok{ ground\_truth\_2nd\_batch}\SpecialCharTok{$}\NormalTok{consensus }\SpecialCharTok{\textless{}}
    \DecValTok{5}\NormalTok{))}
\end{Highlighting}
\end{Shaded}

The results are shown in table \ref{tab:p1-bin-detect-2nd-batch}. The rule classifies \textbf{5} / \textbf{6} as correct. The precision is \textbf{0.75}, and the recall is \textbf{1}.

\begin{table}

\caption{\label{tab:p1-bin-detect-2nd-batch}Binary detection using the previously defined decision rule based on homogeneous confidence intervals of pattern I, for the second batch of projects.}
\centering
\begin{tabular}[t]{llrl}
\toprule
  & detect & ground\_truth & correct\\
\midrule
Project10 & FALSE & 2 & TRUE\\
Project11 & FALSE & 0 & TRUE\\
Project12 & FALSE & 2 & TRUE\\
Project13 & TRUE & 10 & TRUE\\
Project14 & TRUE & 1 & FALSE\\
\addlinespace
Project15 & FALSE & 1 & TRUE\\
\bottomrule
\end{tabular}
\end{table}

\hypertarget{calculating-scoring-rules-1}{%
\paragraph{Calculating scoring rules}\label{calculating-scoring-rules-1}}

Also, let's calculate the Brier- and Log-scoring rules for the second batch, and all projects together.

\begin{Shaded}
\begin{Highlighting}[]
\NormalTok{temp }\OtherTok{\textless{}{-}}\NormalTok{ p1\_detect\_2nd\_batch[, ]}
\NormalTok{temp}\SpecialCharTok{$}\NormalTok{ground\_truth }\OtherTok{\textless{}{-}}\NormalTok{ temp}\SpecialCharTok{$}\NormalTok{ground\_truth}\SpecialCharTok{/}\DecValTok{10}
\NormalTok{temp}\SpecialCharTok{$}\NormalTok{zero\_r }\OtherTok{\textless{}{-}} \FloatTok{0.5}
\NormalTok{temp}\SpecialCharTok{$}\NormalTok{detect\_prob }\OtherTok{\textless{}{-}} \FunctionTok{sapply}\NormalTok{(temp}\SpecialCharTok{$}\NormalTok{detect, }\ControlFlowTok{function}\NormalTok{(d) }\ControlFlowTok{if}\NormalTok{ (d) }\DecValTok{1} \ControlFlowTok{else} \DecValTok{0}\NormalTok{)}

\StringTok{\textasciigrave{}}\AttributeTok{names\textless{}{-}}\StringTok{\textasciigrave{}}\NormalTok{(}\FunctionTok{c}\NormalTok{(}\FunctionTok{mean}\NormalTok{(scoring}\SpecialCharTok{::}\FunctionTok{brierscore}\NormalTok{(}\AttributeTok{object =}\NormalTok{ detect\_prob }\SpecialCharTok{\textasciitilde{}}\NormalTok{ ground\_truth, }\AttributeTok{data =}\NormalTok{ temp)),}
  \FunctionTok{mean}\NormalTok{(scoring}\SpecialCharTok{::}\FunctionTok{logscore}\NormalTok{(}\AttributeTok{object =}\NormalTok{ detect\_prob }\SpecialCharTok{\textasciitilde{}}\NormalTok{ ground\_truth, }\AttributeTok{data =}\NormalTok{ temp)), stats}\SpecialCharTok{::}\FunctionTok{cor}\NormalTok{(temp}\SpecialCharTok{$}\NormalTok{ground\_truth,}
\NormalTok{    temp}\SpecialCharTok{$}\NormalTok{detect\_prob)), }\FunctionTok{c}\NormalTok{(}\StringTok{"Brier score"}\NormalTok{, }\StringTok{"Log score"}\NormalTok{, }\StringTok{"Correlation"}\NormalTok{))}
\end{Highlighting}
\end{Shaded}

\begin{verbatim}
## Brier score   Log score Correlation 
##   0.1500000   0.4757055   0.5980579
\end{verbatim}

ZeroR:

\begin{Shaded}
\begin{Highlighting}[]
\NormalTok{Metrics}\SpecialCharTok{::}\FunctionTok{mse}\NormalTok{(temp}\SpecialCharTok{$}\NormalTok{ground\_truth, temp}\SpecialCharTok{$}\NormalTok{zero\_r)}
\end{Highlighting}
\end{Shaded}

\begin{verbatim}
## [1] 0.1666667
\end{verbatim}

These results are a little worse than for the first batch (\(\approx0.133\) and \(\approx0.4\)).
Also for the second batch, the binary decision rule for the first time is slightly better than ZeroR.
Let's compute the scoring rules for all projects together, the results should be somewhere in between:

\begin{Shaded}
\begin{Highlighting}[]
\NormalTok{temp }\OtherTok{\textless{}{-}} \FunctionTok{rbind}\NormalTok{(p1\_detect, p1\_detect\_2nd\_batch)}
\NormalTok{temp}\SpecialCharTok{$}\NormalTok{ground\_truth }\OtherTok{\textless{}{-}}\NormalTok{ temp}\SpecialCharTok{$}\NormalTok{ground\_truth}\SpecialCharTok{/}\DecValTok{10}
\NormalTok{temp}\SpecialCharTok{$}\NormalTok{zero\_r }\OtherTok{\textless{}{-}} \FloatTok{0.5}
\NormalTok{temp}\SpecialCharTok{$}\NormalTok{detect\_prob }\OtherTok{\textless{}{-}} \FunctionTok{sapply}\NormalTok{(temp}\SpecialCharTok{$}\NormalTok{detect, }\ControlFlowTok{function}\NormalTok{(d) }\ControlFlowTok{if}\NormalTok{ (d) }\DecValTok{1} \ControlFlowTok{else} \DecValTok{0}\NormalTok{)}

\StringTok{\textasciigrave{}}\AttributeTok{names\textless{}{-}}\StringTok{\textasciigrave{}}\NormalTok{(}\FunctionTok{c}\NormalTok{(}\FunctionTok{mean}\NormalTok{(scoring}\SpecialCharTok{::}\FunctionTok{brierscore}\NormalTok{(}\AttributeTok{object =}\NormalTok{ detect\_prob }\SpecialCharTok{\textasciitilde{}}\NormalTok{ ground\_truth, }\AttributeTok{data =}\NormalTok{ temp)),}
  \FunctionTok{mean}\NormalTok{(scoring}\SpecialCharTok{::}\FunctionTok{logscore}\NormalTok{(}\AttributeTok{object =}\NormalTok{ detect\_prob }\SpecialCharTok{\textasciitilde{}}\NormalTok{ ground\_truth, }\AttributeTok{data =}\NormalTok{ temp)), stats}\SpecialCharTok{::}\FunctionTok{cor}\NormalTok{(temp}\SpecialCharTok{$}\NormalTok{ground\_truth,}
\NormalTok{    temp}\SpecialCharTok{$}\NormalTok{detect\_prob)), }\FunctionTok{c}\NormalTok{(}\StringTok{"Brier score"}\NormalTok{, }\StringTok{"Log score"}\NormalTok{, }\StringTok{"Correlation"}\NormalTok{))}
\end{Highlighting}
\end{Shaded}

\begin{verbatim}
## Brier score   Log score Correlation 
##   0.1400000   0.4305455   0.6962230
\end{verbatim}

We can use these results later for comparing against scores as computed by the continuous process models.

Also, computing ZeroR for all projects:

\begin{Shaded}
\begin{Highlighting}[]
\NormalTok{Metrics}\SpecialCharTok{::}\FunctionTok{mse}\NormalTok{(temp}\SpecialCharTok{$}\NormalTok{ground\_truth, temp}\SpecialCharTok{$}\NormalTok{zero\_r)}
\end{Highlighting}
\end{Shaded}

\begin{verbatim}
## [1] 0.1366667
\end{verbatim}

Again, ZeroR beats the decision rule.

\hypertarget{computing-the-scores}{%
\subsubsection{Computing the scores}\label{computing-the-scores}}

We will start by computing the scores that are based on the gradated confidence intervals. We can reuse the function \texttt{p3\_avg\_ci\_scores\_compute()}.

\begin{Shaded}
\begin{Highlighting}[]
\NormalTok{p3\_avg\_ci\_2nd\_batch\_scores }\OtherTok{\textless{}{-}} \FunctionTok{loadResultsOrCompute}\NormalTok{(}\AttributeTok{file =} \StringTok{"../results/p3\_avg\_ci\_2nd\_batch\_scores.rds"}\NormalTok{,}
  \AttributeTok{computeExpr =}\NormalTok{ \{}
    \FunctionTok{doWithParallelCluster}\NormalTok{(}\AttributeTok{numCores =} \FunctionTok{length}\NormalTok{(all\_signals\_2nd\_batch), }\AttributeTok{expr =}\NormalTok{ \{}
      \FunctionTok{library}\NormalTok{(foreach)}

\NormalTok{      foreach}\SpecialCharTok{::}\FunctionTok{foreach}\NormalTok{(}\AttributeTok{pId =} \FunctionTok{names}\NormalTok{(all\_signals\_2nd\_batch), }\AttributeTok{.inorder =} \ConstantTok{TRUE}\NormalTok{,}
        \AttributeTok{.combine =}\NormalTok{ rbind) }\SpecialCharTok{\%dopar\%}\NormalTok{ \{}
        \FunctionTok{p3\_avg\_ci\_scores\_compute}\NormalTok{(}\AttributeTok{pId =}\NormalTok{ pId, }\AttributeTok{x1 =} \DecValTok{0}\NormalTok{, }\AttributeTok{x2 =} \DecValTok{1}\NormalTok{, }\AttributeTok{signals =}\NormalTok{ all\_signals\_2nd\_batch)}
\NormalTok{      \}}
\NormalTok{    \})}
\NormalTok{  \})}
\end{Highlighting}
\end{Shaded}

Table \ref{tab:p3-avg-ci-2nd-batch-scores} shows the computed scores.

\begin{table}

\caption{\label{tab:p3-avg-ci-2nd-batch-scores}The average confidence of the variables REQ, DEV and DESC of each of the 2nd batch's projects as integrated over the confidence intervals' hyperplane (as computed from the first batch of projects).}
\centering
\begin{tabular}[t]{lrrrl}
\toprule
  & REQ & DEV & DESC & Project\\
\midrule
Project10 & 0.3559307 & 0.4116880 & 0.5283078 & Project10\\
Project11 & 0.2046314 & 0.3029024 & 0.4360296 & Project11\\
Project12 & 0.2950464 & 0.1413461 & 0.4061248 & Project12\\
Project13 & 0.1992798 & 0.4388567 & 0.2785768 & Project13\\
Project14 & 0.2197666 & 0.2707810 & 0.3170778 & Project14\\
\addlinespace
Project15 & 0.2804880 & 0.4226075 & 0.4064899 & Project15\\
\bottomrule
\end{tabular}
\end{table}

The correlation with the ground truth of these scores is:

\begin{Shaded}
\begin{Highlighting}[]
\FunctionTok{cor.test}\NormalTok{(}\AttributeTok{x =}\NormalTok{ ground\_truth\_2nd\_batch}\SpecialCharTok{$}\NormalTok{consensus\_score, }\AttributeTok{y =}\NormalTok{ p3\_avg\_ci\_2nd\_batch\_scores[,}
  \StringTok{"REQ"}\NormalTok{])}
\end{Highlighting}
\end{Shaded}

\begin{verbatim}
## 
##  Pearson's product-moment correlation
## 
## data:  ground_truth_2nd_batch$consensus_score and p3_avg_ci_2nd_batch_scores[, "REQ"]
## t = -0.6509, df = 4, p-value = 0.5506
## alternative hypothesis: true correlation is not equal to 0
## 95 percent confidence interval:
##  -0.8959986  0.6704848
## sample estimates:
##        cor 
## -0.3094729
\end{verbatim}

\begin{Shaded}
\begin{Highlighting}[]
\FunctionTok{cor.test}\NormalTok{(}\AttributeTok{x =}\NormalTok{ ground\_truth\_2nd\_batch}\SpecialCharTok{$}\NormalTok{consensus\_score, }\AttributeTok{y =}\NormalTok{ p3\_avg\_ci\_2nd\_batch\_scores[,}
  \StringTok{"DEV"}\NormalTok{])}
\end{Highlighting}
\end{Shaded}

\begin{verbatim}
## 
##  Pearson's product-moment correlation
## 
## data:  ground_truth_2nd_batch$consensus_score and p3_avg_ci_2nd_batch_scores[, "DEV"]
## t = 0.91889, df = 4, p-value = 0.4102
## alternative hypothesis: true correlation is not equal to 0
## 95 percent confidence interval:
##  -0.5960122  0.9180114
## sample estimates:
##       cor 
## 0.4174883
\end{verbatim}

\begin{Shaded}
\begin{Highlighting}[]
\FunctionTok{cor.test}\NormalTok{(}\AttributeTok{x =}\NormalTok{ ground\_truth\_2nd\_batch}\SpecialCharTok{$}\NormalTok{consensus\_score, }\AttributeTok{y =}\NormalTok{ p3\_avg\_ci\_2nd\_batch\_scores[,}
  \StringTok{"DESC"}\NormalTok{])}
\end{Highlighting}
\end{Shaded}

\begin{verbatim}
## 
##  Pearson's product-moment correlation
## 
## data:  ground_truth_2nd_batch$consensus_score and p3_avg_ci_2nd_batch_scores[, "DESC"]
## t = -1.435, df = 4, p-value = 0.2246
## alternative hypothesis: true correlation is not equal to 0
## 95 percent confidence interval:
##  -0.9466541  0.4338556
## sample estimates:
##        cor 
## -0.5829694
\end{verbatim}

While these correlations seem significant, the p-values indicate evidence for the null hypothesis, which means no correlation. Now that we have these scores, we'll compute scores for the average distance to the respective reference variable.

\begin{Shaded}
\begin{Highlighting}[]
\NormalTok{p3\_avg\_area\_2nd\_batch\_scores }\OtherTok{\textless{}{-}} \FunctionTok{loadResultsOrCompute}\NormalTok{(}\AttributeTok{file =} \StringTok{"../results/p3\_avg\_area\_2nd\_batch\_scores.rds"}\NormalTok{,}
  \AttributeTok{computeExpr =}\NormalTok{ \{}
    \FunctionTok{doWithParallelCluster}\NormalTok{(}\AttributeTok{numCores =} \FunctionTok{length}\NormalTok{(all\_signals\_2nd\_batch), }\AttributeTok{expr =}\NormalTok{ \{}
      \FunctionTok{library}\NormalTok{(foreach)}

\NormalTok{      foreach}\SpecialCharTok{::}\FunctionTok{foreach}\NormalTok{(}\AttributeTok{pId =} \FunctionTok{names}\NormalTok{(all\_signals\_2nd\_batch), }\AttributeTok{.inorder =} \ConstantTok{TRUE}\NormalTok{,}
        \AttributeTok{.combine =}\NormalTok{ rbind) }\SpecialCharTok{\%dopar\%}\NormalTok{ \{}
        \FunctionTok{p3\_avg\_area\_scores\_compute}\NormalTok{(}\AttributeTok{pId =}\NormalTok{ pId, }\AttributeTok{signals =}\NormalTok{ all\_signals\_2nd\_batch)}
\NormalTok{      \}}
\NormalTok{    \})}
\NormalTok{  \})}
\end{Highlighting}
\end{Shaded}

Table \ref{tab:p3-avg-area-2nd-batch-scores} shows the computed scores.

\begin{table}

\caption{\label{tab:p3-avg-area-2nd-batch-scores}The average distance of the variables REQ, DEV and DESC of each of the second batch's projects to the previously averaged reference-variables $\overline{\operatorname{REQ}},\overline{\operatorname{DEV}},\overline{\operatorname{DESC}}$ as computed over tethe first batch.}
\centering
\begin{tabular}[t]{lrrrl}
\toprule
  & REQ & DEV & DESC & Project\\
\midrule
Project10 & 0.3960715 & 0.3542538 & 0.2625533 & Project10\\
Project11 & 0.7154275 & 0.6566564 & 0.2399364 & Project11\\
Project12 & 0.5015237 & 0.8966831 & 1.0000000 & Project12\\
Project13 & 0.7542608 & 0.3181776 & 0.8678727 & Project13\\
Project14 & 0.6904980 & 0.6757826 & 0.3152234 & Project14\\
\addlinespace
Project15 & 0.6500128 & 0.3035216 & 0.9224892 & Project15\\
\bottomrule
\end{tabular}
\end{table}

\hypertarget{predicting-the-ground-truth}{%
\subsubsection{Predicting the ground truth}\label{predicting-the-ground-truth}}

Now we have all the scores of the new projects computed against the pattern as generated from the previous projects.

\begin{Shaded}
\begin{Highlighting}[]
\NormalTok{temp }\OtherTok{\textless{}{-}} \FunctionTok{data.frame}\NormalTok{(}\AttributeTok{gt\_consensus =}\NormalTok{ ground\_truth\_2nd\_batch}\SpecialCharTok{$}\NormalTok{consensus\_score, }\AttributeTok{ci\_req =}\NormalTok{ p3\_avg\_ci\_2nd\_batch\_scores}\SpecialCharTok{$}\NormalTok{REQ,}
  \AttributeTok{area\_req =}\NormalTok{ p3\_avg\_area\_2nd\_batch\_scores}\SpecialCharTok{$}\NormalTok{REQ, }\AttributeTok{ci\_dev =}\NormalTok{ p3\_avg\_ci\_2nd\_batch\_scores}\SpecialCharTok{$}\NormalTok{DEV,}
  \AttributeTok{area\_dev =}\NormalTok{ p3\_avg\_area\_2nd\_batch\_scores}\SpecialCharTok{$}\NormalTok{DEV, }\AttributeTok{ci\_desc =}\NormalTok{ p3\_avg\_ci\_2nd\_batch\_scores}\SpecialCharTok{$}\NormalTok{DESC,}
  \AttributeTok{area\_desc =}\NormalTok{ p3\_avg\_area\_2nd\_batch\_scores}\SpecialCharTok{$}\NormalTok{DESC)}

\NormalTok{p3\_avg\_lm\_2nd\_batch\_scores }\OtherTok{\textless{}{-}}\NormalTok{ stats}\SpecialCharTok{::}\FunctionTok{predict}\NormalTok{(}\AttributeTok{object =}\NormalTok{ p3\_avg\_lm, temp)}
\CommentTok{\# Since we are attempting a regression to positive scores, we set any negative}
\CommentTok{\# predictions to 0. Same goes for \textgreater{}1.}
\NormalTok{p3\_avg\_lm\_2nd\_batch\_scores[p3\_avg\_lm\_2nd\_batch\_scores }\SpecialCharTok{\textless{}} \DecValTok{0}\NormalTok{] }\OtherTok{\textless{}{-}} \DecValTok{0}
\NormalTok{p3\_avg\_lm\_2nd\_batch\_scores[p3\_avg\_lm\_2nd\_batch\_scores }\SpecialCharTok{\textgreater{}} \DecValTok{1}\NormalTok{] }\OtherTok{\textless{}{-}} \DecValTok{1}

\StringTok{\textasciigrave{}}\AttributeTok{names\textless{}{-}}\StringTok{\textasciigrave{}}\NormalTok{(}\FunctionTok{round}\NormalTok{(p3\_avg\_lm\_2nd\_batch\_scores }\SpecialCharTok{*} \DecValTok{10}\NormalTok{, }\DecValTok{3}\NormalTok{), }\DecValTok{10}\SpecialCharTok{:}\DecValTok{15}\NormalTok{)}
\end{Highlighting}
\end{Shaded}

\begin{verbatim}
##     10     11     12     13     14     15 
## 10.000  7.263  4.995  5.240  7.089  7.287
\end{verbatim}

\begin{Shaded}
\begin{Highlighting}[]
\NormalTok{stats}\SpecialCharTok{::}\FunctionTok{cor}\NormalTok{(p3\_avg\_lm\_2nd\_batch\_scores, ground\_truth\_2nd\_batch}\SpecialCharTok{$}\NormalTok{consensus\_score)}
\end{Highlighting}
\end{Shaded}

\begin{verbatim}
## [1] -0.4500864
\end{verbatim}

These results mean that the prediction is anti-proportional to the ground truth, and weak at the same time. It is actually worse to have a negative correlation than having no correlation. This can only be interpreted as our previously trained regression model not being able to generalize to new, unseen data. This is expected, since a) we did not have sufficient training data, and b) the previous regression model being overfit on the previous data. However, and I wrote that before, the main point of these regression models was translation and scaling of the scores, not prediction. For that, much more data is required (both, instances and features, as the selected features cannot capture all of the important aspects of the deviation process and process model). Furthermore, these regression models (including the one we built for the source code data) were built using hand-picked features. Ideally, we would have a great number of different features, paired with something like a recursive feature elimination, in order to come up with a regression model that has reliable generalizability. Therefore, I am inclined not to repeat this exercise with the source code data, as we will have the same problem there.

\hypertarget{predicting-using-the-best-rfe-model}{%
\subsubsection{Predicting using the best RFE model}\label{predicting-using-the-best-rfe-model}}

We have previously computed the variable importances in section \ref{sssec:var-imp-it}, using the first batch of projects. We will not attempt this for the second batch. Rather, we will use the best model as found by the recursive feature elimination and use it to make predictions on the second batch. Previously, we evaluated the binary decision rule on the new projects (cf.~section \ref{ssec:eval-dr-2nd-batch}).

\begin{Shaded}
\begin{Highlighting}[]
\NormalTok{p3\_it\_2nd\_batch\_scores }\OtherTok{\textless{}{-}} \FunctionTok{loadResultsOrCompute}\NormalTok{(}\AttributeTok{file =} \StringTok{"../results/p3\_it\_2nd\_batch\_scores.rds"}\NormalTok{,}
  \AttributeTok{computeExpr =}\NormalTok{ \{}
\NormalTok{    p3\_it\_projects\_2nd\_batch }\OtherTok{\textless{}{-}} \FunctionTok{time\_warp\_wrapper}\NormalTok{(}\AttributeTok{pattern =}\NormalTok{ p3\_it\_signals, }\AttributeTok{derive =} \ConstantTok{FALSE}\NormalTok{,}
      \AttributeTok{use\_signals =}\NormalTok{ all\_signals\_2nd\_batch, }\AttributeTok{use\_ground\_truth =}\NormalTok{ ground\_truth\_2nd\_batch)}

    \FunctionTok{as.data.frame}\NormalTok{(}\FunctionTok{compute\_all\_scores\_it}\NormalTok{(}\AttributeTok{alignment =}\NormalTok{ p3\_it\_projects\_2nd\_batch,}
      \AttributeTok{patternName =} \StringTok{"p3\_it"}\NormalTok{, }\AttributeTok{vartypes =} \FunctionTok{names}\NormalTok{(p3\_it\_signals)))}
\NormalTok{  \})}
\end{Highlighting}
\end{Shaded}

\begin{Shaded}
\begin{Highlighting}[]
\CommentTok{\# Note that these predictions are already scaled!}
\NormalTok{p3\_avg\_rfe\_2nd\_batch\_scores }\OtherTok{\textless{}{-}}\NormalTok{ stats}\SpecialCharTok{::}\FunctionTok{predict}\NormalTok{(}\AttributeTok{object =}\NormalTok{ modelFit\_it\_all, p3\_it\_2nd\_batch\_scores)}
\CommentTok{\# Since we are attempting a regression to positive scores, we set any negative}
\CommentTok{\# predictions to 0. Same goes for \textgreater{}1.}
\NormalTok{p3\_avg\_rfe\_2nd\_batch\_scores[p3\_avg\_rfe\_2nd\_batch\_scores }\SpecialCharTok{\textless{}} \DecValTok{0}\NormalTok{] }\OtherTok{\textless{}{-}} \DecValTok{0}
\NormalTok{p3\_avg\_rfe\_2nd\_batch\_scores[p3\_avg\_rfe\_2nd\_batch\_scores }\SpecialCharTok{\textgreater{}} \DecValTok{10}\NormalTok{] }\OtherTok{\textless{}{-}} \DecValTok{10}

\StringTok{\textasciigrave{}}\AttributeTok{names\textless{}{-}}\StringTok{\textasciigrave{}}\NormalTok{(}\FunctionTok{round}\NormalTok{(p3\_avg\_rfe\_2nd\_batch\_scores, }\DecValTok{4}\NormalTok{), }\FunctionTok{rownames}\NormalTok{(ground\_truth\_2nd\_batch))}
\end{Highlighting}
\end{Shaded}

\begin{verbatim}
##     10     11     12     13     14     15 
## 4.1674 4.9388 1.5291 0.4735 4.0357 0.0000
\end{verbatim}

\begin{Shaded}
\begin{Highlighting}[]
\NormalTok{stats}\SpecialCharTok{::}\FunctionTok{cor}\NormalTok{(p3\_avg\_rfe\_2nd\_batch\_scores, ground\_truth\_2nd\_batch}\SpecialCharTok{$}\NormalTok{consensus)}
\end{Highlighting}
\end{Shaded}

\begin{verbatim}
## [1] -0.5208951
\end{verbatim}

These results are similarly worse to those obtained from the previous section, where we predicted using the linear model. The only conclusion we can now draw confidently is that we have insufficient training data in order to train a model that can generalize to new data (likely both, instances and feature granularity). However, the features we found to be most important have a good chance to be actually the most important ones. If we were to repeat the variable-importance experiment with a lot more data (projects), we might get a similar ranking and relative importance. It is only that currently, the amount of data does not suffice to generalize from the observations, even if we have a good selection and weighting of features.

\hypertarget{automatic-calibration-of-the-continuous-process-models-1}{%
\subsection{Automatic calibration of the continuous process models}\label{automatic-calibration-of-the-continuous-process-models-1}}

We have previously attempted the automatic calibration of the continuous process models that use source code data (section \ref{sec:auto-calib}), and now we will do the same here for the models using issue-tracking data.
We will reuse the functions defined for source code data as much as possible here.
A notable difference is of course the process models. Here, we have one less variable, but six instead of four models (I, II(a), III, as well as their resp. derivative models).

Another notable difference is how we simulate the random processes. While for source code, we could not make any assumptions how a random process may look, for issue tracking we can make and guarantee these:

\begin{itemize}
\tightlist
\item
  Processes for issue-tracking data are monotonically increasing. Therefore, when we fit a curve through some random points, we will order them smallest to largest.
\item
  Furthermore, all processes start at \(0,0\) (x/y). The variables \texttt{REQ} and \texttt{DEV} also always end at \(1,1\).
\item
  While \(DESC\) does never end at \(y=1\) (it cannot as of the normalization through whatever amount \texttt{DEV} accumulated), we can conservably limit it to \(y=\frac{2}{3}\).
\end{itemize}

These assumptions, which will hold, should allow us to simulate random processes that much closer resemble the actual processes as we model them for issue tracking data. Therefore, we expect better outcomes than we did for source code data.

\hypertarget{computing-the-metrics-1}{%
\subsubsection{Computing the metrics}\label{computing-the-metrics-1}}

We will need an extra grid for issue-tracking data, before we can proceed.

\begin{Shaded}
\begin{Highlighting}[]
\NormalTok{ac\_grid\_it }\OtherTok{\textless{}{-}} \FunctionTok{loadResultsOrCompute}\NormalTok{(}\AttributeTok{file =} \StringTok{"../results/ac\_grid\_it.rds"}\NormalTok{, }\AttributeTok{computeExpr =}\NormalTok{ \{}
  \FunctionTok{expand.grid}\NormalTok{(}\FunctionTok{list}\NormalTok{(}\AttributeTok{PM =} \FunctionTok{c}\NormalTok{(}\StringTok{"I"}\NormalTok{, }\StringTok{"II(a)"}\NormalTok{, }\StringTok{"III"}\NormalTok{, }\StringTok{"IV(I)"}\NormalTok{, }\StringTok{"IV(II(a))"}\NormalTok{, }\StringTok{"IV(III)"}\NormalTok{),}
    \AttributeTok{Var =} \FunctionTok{c}\NormalTok{(}\StringTok{"REQ"}\NormalTok{, }\StringTok{"DEV"}\NormalTok{, }\StringTok{"DESC"}\NormalTok{), }\AttributeTok{Seed =} \FunctionTok{seq}\NormalTok{(}\AttributeTok{from =} \DecValTok{1}\NormalTok{, }\AttributeTok{length.out =} \DecValTok{10000}\NormalTok{)))}
\NormalTok{\})}

\FunctionTok{nrow}\NormalTok{(ac\_grid\_it)}
\end{Highlighting}
\end{Shaded}

\begin{verbatim}
## [1] 180000
\end{verbatim}

Let's compute the full grid (warning -- expensive):

\begin{Shaded}
\begin{Highlighting}[]
\NormalTok{ac\_grid\_results\_it }\OtherTok{\textless{}{-}} \FunctionTok{loadResultsOrCompute}\NormalTok{(}\AttributeTok{file =} \StringTok{"../results/ac\_grid\_results\_it.rds"}\NormalTok{,}
  \AttributeTok{computeExpr =}\NormalTok{ \{}
    \FunctionTok{library}\NormalTok{(foreach)}

\NormalTok{    cl }\OtherTok{\textless{}{-}}\NormalTok{ parallel}\SpecialCharTok{::}\FunctionTok{makePSOCKcluster}\NormalTok{(}\FunctionTok{min}\NormalTok{(}\DecValTok{123}\NormalTok{, parallel}\SpecialCharTok{::}\FunctionTok{detectCores}\NormalTok{()))}
\NormalTok{    parallel}\SpecialCharTok{::}\FunctionTok{clusterExport}\NormalTok{(cl, }\AttributeTok{varlist =} \FunctionTok{list}\NormalTok{(}\StringTok{"req\_poly"}\NormalTok{, }\StringTok{"dev\_poly"}\NormalTok{))}

    \FunctionTok{doWithParallelClusterExplicit}\NormalTok{(}\AttributeTok{cl =}\NormalTok{ cl, }\AttributeTok{stopCl =} \ConstantTok{TRUE}\NormalTok{, }\AttributeTok{expr =}\NormalTok{ \{}
\NormalTok{      pb }\OtherTok{\textless{}{-}}\NormalTok{ utils}\SpecialCharTok{::}\FunctionTok{txtProgressBar}\NormalTok{(}\AttributeTok{min =} \DecValTok{1}\NormalTok{, }\AttributeTok{max =} \FunctionTok{nrow}\NormalTok{(ac\_grid\_it), }\AttributeTok{style =} \DecValTok{3}\NormalTok{)}
\NormalTok{      progress }\OtherTok{\textless{}{-}} \ControlFlowTok{function}\NormalTok{(n) \{}
        \ControlFlowTok{if}\NormalTok{ (}\DecValTok{0} \SpecialCharTok{==}\NormalTok{ (n}\SpecialCharTok{\%\%}\DecValTok{25}\NormalTok{)) \{}
          \FunctionTok{print}\NormalTok{(n)}
\NormalTok{        \}}
\NormalTok{        utils}\SpecialCharTok{::}\FunctionTok{setTxtProgressBar}\NormalTok{(}\AttributeTok{pb =}\NormalTok{ pb, }\AttributeTok{value =}\NormalTok{ n)}
\NormalTok{      \}}

\NormalTok{      foreach}\SpecialCharTok{::}\FunctionTok{foreach}\NormalTok{(}\AttributeTok{grid\_idx =} \FunctionTok{rownames}\NormalTok{(ac\_grid\_it), }\AttributeTok{.combine =}\NormalTok{ rbind, }\AttributeTok{.inorder =} \ConstantTok{FALSE}\NormalTok{,}
        \AttributeTok{.verbose =} \ConstantTok{TRUE}\NormalTok{, }\AttributeTok{.packages =} \FunctionTok{c}\NormalTok{(}\StringTok{"cobs"}\NormalTok{), }\AttributeTok{.options.snow =} \FunctionTok{list}\NormalTok{(}\AttributeTok{progress =}\NormalTok{ progress)) }\SpecialCharTok{\%dopar\%}
\NormalTok{        \{}
          \FunctionTok{options}\NormalTok{(}\AttributeTok{warn =} \DecValTok{2}\NormalTok{)}
\NormalTok{          params }\OtherTok{\textless{}{-}}\NormalTok{ ac\_grid\_it[grid\_idx, ]}
\NormalTok{          params}\SpecialCharTok{$}\NormalTok{PM }\OtherTok{\textless{}{-}} \FunctionTok{as.character}\NormalTok{(params}\SpecialCharTok{$}\NormalTok{PM)}
\NormalTok{          params}\SpecialCharTok{$}\NormalTok{Var }\OtherTok{\textless{}{-}} \FunctionTok{as.character}\NormalTok{(params}\SpecialCharTok{$}\NormalTok{Var)}

\NormalTok{          pm\_var }\OtherTok{\textless{}{-}} \ControlFlowTok{switch}\NormalTok{(params}\SpecialCharTok{$}\NormalTok{PM, }\AttributeTok{I =}\NormalTok{ p1\_it\_signals, }\StringTok{\textasciigrave{}}\AttributeTok{II(a)}\StringTok{\textasciigrave{}} \OtherTok{=}\NormalTok{ p2a\_it\_signals,}
          \AttributeTok{III =}\NormalTok{ p3\_it\_signals, }\StringTok{\textasciigrave{}}\AttributeTok{IV(I)}\StringTok{\textasciigrave{}} \OtherTok{=}\NormalTok{ p4p1\_it\_signals, }\StringTok{\textasciigrave{}}\AttributeTok{IV(II(a))}\StringTok{\textasciigrave{}} \OtherTok{=}\NormalTok{ p4p2a\_it\_signals,}
          \StringTok{\textasciigrave{}}\AttributeTok{IV(III)}\StringTok{\textasciigrave{}} \OtherTok{=}\NormalTok{ p4p3\_it\_signals, \{}
            \FunctionTok{stop}\NormalTok{(}\FunctionTok{paste0}\NormalTok{(}\StringTok{"Don\textquotesingle{}t know "}\NormalTok{, params}\SpecialCharTok{$}\NormalTok{PM, }\StringTok{"."}\NormalTok{))}
\NormalTok{          \})[[params}\SpecialCharTok{$}\NormalTok{Var]]}\SpecialCharTok{$}\FunctionTok{get0Function}\NormalTok{()}

\NormalTok{          include\_deriv }\OtherTok{\textless{}{-}}\NormalTok{ params}\SpecialCharTok{$}\NormalTok{PM }\SpecialCharTok{\%in\%} \FunctionTok{c}\NormalTok{(}\StringTok{"IV(I)"}\NormalTok{, }\StringTok{"IV(II(a))"}\NormalTok{, }\StringTok{"IV(III)"}\NormalTok{)}
\NormalTok{          smooth\_curve }\OtherTok{\textless{}{-}} \FunctionTok{get\_smoothed\_curve}\NormalTok{(}\AttributeTok{seed =}\NormalTok{ params}\SpecialCharTok{$}\NormalTok{Seed, }\AttributeTok{include\_deriv =}\NormalTok{ include\_deriv,}
          \AttributeTok{issue\_tracking =} \ConstantTok{TRUE}\NormalTok{, }\AttributeTok{desc\_var =}\NormalTok{ params}\SpecialCharTok{$}\NormalTok{Var }\SpecialCharTok{==} \StringTok{"DESC"}\NormalTok{)}
\NormalTok{          p\_var }\OtherTok{\textless{}{-}} \ControlFlowTok{if}\NormalTok{ (include\_deriv)}
\NormalTok{          smooth\_curve}\SpecialCharTok{$}\NormalTok{f1 }\ControlFlowTok{else}\NormalTok{ smooth\_curve}\SpecialCharTok{$}\NormalTok{f0}

\NormalTok{          df }\OtherTok{\textless{}{-}} \ConstantTok{NULL}
          \ControlFlowTok{for}\NormalTok{ (seg\_idx }\ControlFlowTok{in} \DecValTok{1}\SpecialCharTok{:}\DecValTok{10}\NormalTok{) \{}
          \CommentTok{\# Important so we can correctly concatenate the results with the grid}
          \CommentTok{\# parameters!}
\NormalTok{          res }\OtherTok{\textless{}{-}} \FunctionTok{as.data.frame}\NormalTok{(}\FunctionTok{ac\_worker}\NormalTok{(}\AttributeTok{PM =}\NormalTok{ pm\_var, }\AttributeTok{P =}\NormalTok{ p\_var, }\AttributeTok{seg\_idx =}\NormalTok{ seg\_idx,}
            \AttributeTok{total\_segments =} \DecValTok{10}\NormalTok{))}
\NormalTok{          res}\SpecialCharTok{$}\NormalTok{GridIdx }\OtherTok{\textless{}{-}}\NormalTok{ grid\_idx}
\NormalTok{          res}\SpecialCharTok{$}\NormalTok{SegIdx }\OtherTok{\textless{}{-}}\NormalTok{ seg\_idx}
\NormalTok{          res }\OtherTok{\textless{}{-}} \FunctionTok{cbind}\NormalTok{(res, params)}
\NormalTok{          df }\OtherTok{\textless{}{-}} \ControlFlowTok{if}\NormalTok{ (}\FunctionTok{is.null}\NormalTok{(df))}
\NormalTok{            res }\ControlFlowTok{else} \FunctionTok{rbind}\NormalTok{(df, res)}
\NormalTok{          \}}
\NormalTok{          df}
\NormalTok{        \}}
\NormalTok{    \})}
\NormalTok{  \})}
\end{Highlighting}
\end{Shaded}

\hypertarget{approximating-marginal-cumulative-densities-1}{%
\subsubsection{Approximating marginal cumulative densities}\label{approximating-marginal-cumulative-densities-1}}

\begin{figure}
\centering
\includegraphics{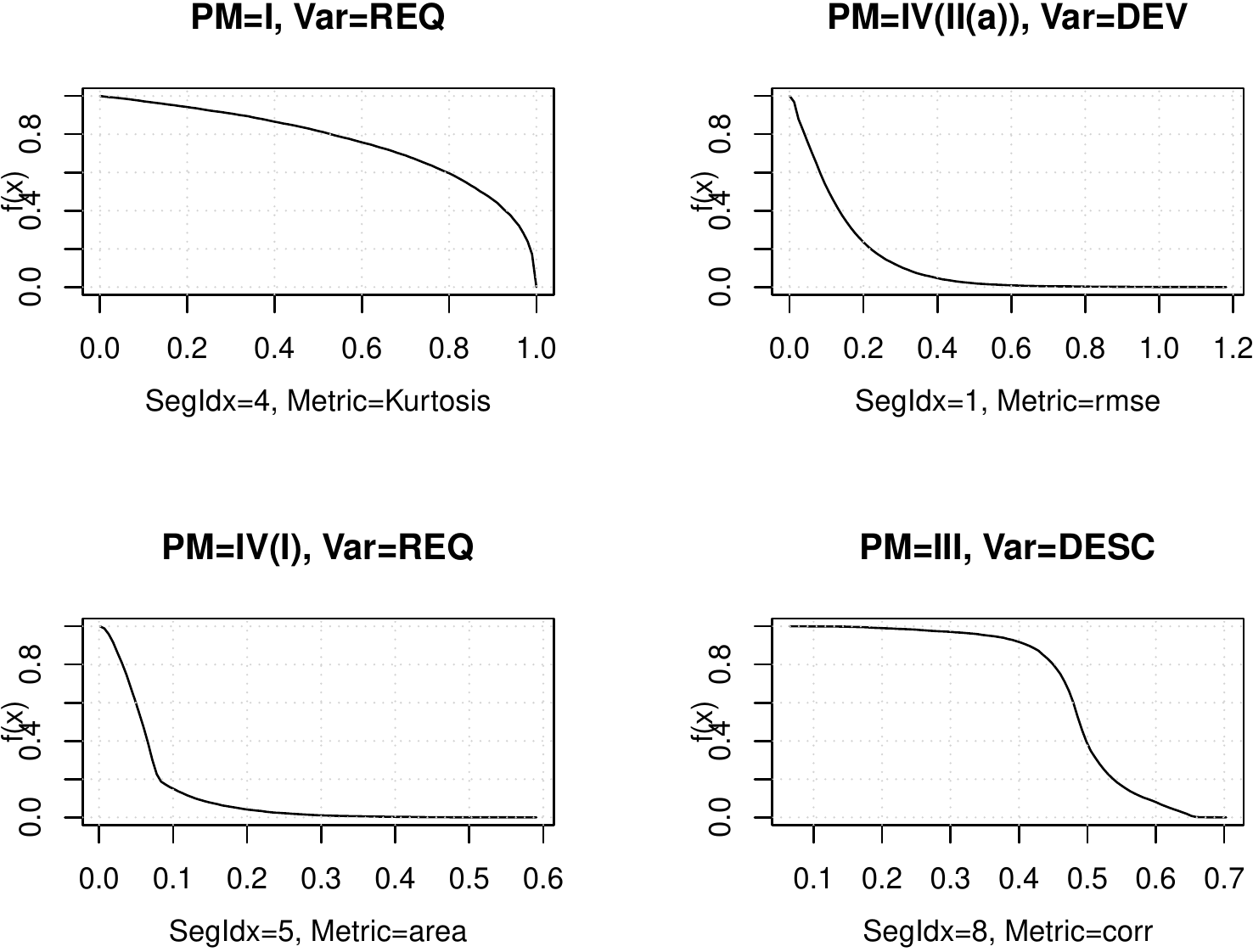}
\caption{\label{fig:ac-grid-example-it}Four randomly picked ECCDFs as simulated using the random processes of the automatic calibration for issue-tracking data.}
\end{figure}

Now with the simulated random processes, we can approximate the marginal (cumulative) densities for each objective. This is the same as we did previously using source code data (cf.~section \ref{ssec:ac-approx-marginal-ecdfs}), and we will reuse some of the previously defined functions there.
Let's show some randomly picked examples in figure \ref{fig:ac-grid-example-it}. Similarly to source code data, the ECCDFs are all clearly non-linear.

\hypertarget{normality-and-uniformity-tests-of-the-objectives-1}{%
\paragraph{Normality and uniformity tests of the objectives}\label{normality-and-uniformity-tests-of-the-objectives-1}}

In fact, let's check if any of the scores is normally or uniformly distributed.
We will do two tests, the Shapiro--Wilk and the Kolmogorov--Smirnov tests for normality.

\begin{Shaded}
\begin{Highlighting}[]
\NormalTok{ac\_distr\_test\_it }\OtherTok{\textless{}{-}} \FunctionTok{loadResultsOrCompute}\NormalTok{(}\AttributeTok{file =} \StringTok{"../results/ac\_distr\_test\_it.rds"}\NormalTok{,}
  \AttributeTok{computeExpr =}\NormalTok{ \{}
\NormalTok{    temp.grid }\OtherTok{\textless{}{-}} \FunctionTok{expand.grid}\NormalTok{(}\FunctionTok{list}\NormalTok{(}\AttributeTok{PM =} \FunctionTok{levels}\NormalTok{(ac\_grid\_it}\SpecialCharTok{$}\NormalTok{PM), }\AttributeTok{Var =} \FunctionTok{unique}\NormalTok{(ac\_grid\_it}\SpecialCharTok{$}\NormalTok{Var),}
      \AttributeTok{Seg =} \DecValTok{1}\SpecialCharTok{:}\DecValTok{10}\NormalTok{, }\AttributeTok{Metric =} \FunctionTok{c}\NormalTok{(}\StringTok{"area"}\NormalTok{, }\StringTok{"corr"}\NormalTok{, }\StringTok{"jsd"}\NormalTok{, }\StringTok{"kl"}\NormalTok{, }\StringTok{"arclen"}\NormalTok{, }\StringTok{"sd"}\NormalTok{, }\StringTok{"var"}\NormalTok{,}
        \StringTok{"mae"}\NormalTok{, }\StringTok{"rmse"}\NormalTok{, }\StringTok{"RMS"}\NormalTok{, }\StringTok{"Kurtosis"}\NormalTok{, }\StringTok{"Peak"}\NormalTok{, }\StringTok{"ImpulseFactor"}\NormalTok{)))}

    \FunctionTok{as.data.frame}\NormalTok{(}\FunctionTok{doWithParallelCluster}\NormalTok{(}\AttributeTok{numCores =} \FunctionTok{min}\NormalTok{(parallel}\SpecialCharTok{::}\FunctionTok{detectCores}\NormalTok{(),}
      \DecValTok{32}\NormalTok{), }\AttributeTok{expr =}\NormalTok{ \{}
\NormalTok{      foreach}\SpecialCharTok{::}\FunctionTok{foreach}\NormalTok{(}\AttributeTok{rn =} \FunctionTok{rownames}\NormalTok{(temp.grid), }\AttributeTok{.combine =}\NormalTok{ rbind, }\AttributeTok{.inorder =} \ConstantTok{FALSE}\NormalTok{) }\SpecialCharTok{\%dopar\%}
\NormalTok{        \{}
          \FunctionTok{set.seed}\NormalTok{(}\DecValTok{1}\NormalTok{)}
\NormalTok{          row }\OtherTok{\textless{}{-}}\NormalTok{ temp.grid[rn, ]}
\NormalTok{          temp }\OtherTok{\textless{}{-}} \FunctionTok{ac\_extract\_data}\NormalTok{(}\AttributeTok{pmName =}\NormalTok{ row}\SpecialCharTok{$}\NormalTok{PM, }\AttributeTok{varName =}\NormalTok{ row}\SpecialCharTok{$}\NormalTok{Var, }\AttributeTok{segIdx =}\NormalTok{ row}\SpecialCharTok{$}\NormalTok{Seg,}
          \AttributeTok{metricName =}\NormalTok{ row}\SpecialCharTok{$}\NormalTok{Metric, }\AttributeTok{ac\_grid =}\NormalTok{ ac\_grid\_it, }\AttributeTok{ac\_grid\_results =}\NormalTok{ ac\_grid\_results\_it)}

\NormalTok{          samp }\OtherTok{\textless{}{-}} \FunctionTok{sample}\NormalTok{(}\AttributeTok{x =}\NormalTok{ temp, }\AttributeTok{size =} \DecValTok{5000}\NormalTok{)}
\NormalTok{          st }\OtherTok{\textless{}{-}} \ControlFlowTok{if}\NormalTok{ (}\FunctionTok{length}\NormalTok{(}\FunctionTok{unique}\NormalTok{(samp)) }\SpecialCharTok{==} \DecValTok{1}\NormalTok{)}
          \FunctionTok{list}\NormalTok{(}\AttributeTok{statistic =} \DecValTok{0}\NormalTok{, }\AttributeTok{p.value =} \DecValTok{0}\NormalTok{) }\ControlFlowTok{else} \FunctionTok{shapiro.test}\NormalTok{(samp)}

\NormalTok{          m }\OtherTok{\textless{}{-}} \FunctionTok{mean}\NormalTok{(temp)}
\NormalTok{          s }\OtherTok{\textless{}{-}} \FunctionTok{sd}\NormalTok{(temp)}
\NormalTok{          distFun\_norm }\OtherTok{\textless{}{-}} \ControlFlowTok{function}\NormalTok{(q) \{}
          \FunctionTok{pnorm}\NormalTok{(}\AttributeTok{q =}\NormalTok{ q, }\AttributeTok{mean =}\NormalTok{ m, }\AttributeTok{sd =}\NormalTok{ s)}
\NormalTok{          \}}
\NormalTok{          ks\_norm }\OtherTok{\textless{}{-}} \FunctionTok{ks.test}\NormalTok{(}\AttributeTok{x =}\NormalTok{ temp, }\AttributeTok{y =}\NormalTok{ distFun\_norm)}

\NormalTok{          ab }\OtherTok{\textless{}{-}} \FunctionTok{range}\NormalTok{(temp)}
\NormalTok{          distFun\_unif }\OtherTok{\textless{}{-}} \ControlFlowTok{function}\NormalTok{(q) \{}
          \FunctionTok{punif}\NormalTok{(}\AttributeTok{q =}\NormalTok{ q, }\AttributeTok{min =}\NormalTok{ ab[}\DecValTok{1}\NormalTok{], }\AttributeTok{max =}\NormalTok{ ab[}\DecValTok{2}\NormalTok{])}
\NormalTok{          \}}
\NormalTok{          ks\_unif }\OtherTok{\textless{}{-}} \FunctionTok{ks.test}\NormalTok{(}\AttributeTok{x =}\NormalTok{ temp, }\AttributeTok{y =}\NormalTok{ distFun\_unif)}

          \StringTok{\textasciigrave{}}\AttributeTok{colnames\textless{}{-}}\StringTok{\textasciigrave{}}\NormalTok{(}\AttributeTok{x =} \FunctionTok{matrix}\NormalTok{(}\AttributeTok{data =} \FunctionTok{c}\NormalTok{(st}\SpecialCharTok{$}\NormalTok{statistic, st}\SpecialCharTok{$}\NormalTok{p.value, ks\_norm}\SpecialCharTok{$}\NormalTok{statistic,}
\NormalTok{          ks\_norm}\SpecialCharTok{$}\NormalTok{p.value, ks\_unif}\SpecialCharTok{$}\NormalTok{statistic, ks\_unif}\SpecialCharTok{$}\NormalTok{p.value), }\AttributeTok{nrow =} \DecValTok{1}\NormalTok{),}
          \AttributeTok{value =} \FunctionTok{c}\NormalTok{(}\StringTok{"shap.W"}\NormalTok{, }\StringTok{"shap.pval"}\NormalTok{, }\StringTok{"ks\_norm.D"}\NormalTok{, }\StringTok{"ks\_norm.pval"}\NormalTok{,}
            \StringTok{"ks\_unif.D"}\NormalTok{, }\StringTok{"ks\_unif.pval"}\NormalTok{))}
\NormalTok{        \}}
\NormalTok{    \}))}
\NormalTok{  \})}
\end{Highlighting}
\end{Shaded}

Out of 2340 ECDFs, \textbf{0} are normally distributed, and \textbf{0} are uniformly distributed.
The highest p-values for either normally test are \ensuremath{2.6938311\times 10^{-13}} and 0.0022923. The highest W-statistic was 0.9942279, and the lowest D-statistic was 0.0184002.
The highest p-value for the standard uniform test was \ensuremath{3.6761199\times 10^{-9}}, and the lowest D-statistic was 0.0317132.
It is fair to say that nothing here is normally or uniformly distributed.
We now have reasonable evidence that justifies that conversion to linear behaving scores is required.

\hypertarget{calculating-scores-1}{%
\subsubsection{Calculating scores}\label{calculating-scores-1}}

This is the same approach as for source code data, with only minor differences.
In order to calculate each project's scores, we need to calculate one objective per:

\begin{itemize}
\tightlist
\item
  Project (15),
\item
  Process model (6),
\item
  Variable (3),
\item
  Segment index (10), and
\item
  Metric (13) {[}this is not part of the grid, the \texttt{ac\_worker()} does this{]}.
\end{itemize}

That will leave us with \(15\times 6\times 3\times 10\times 13=35,100\) computations. It's probably best to generate a grid for that:

\begin{Shaded}
\begin{Highlighting}[]
\NormalTok{ac\_grid\_projects\_it }\OtherTok{\textless{}{-}} \FunctionTok{expand.grid}\NormalTok{(}\FunctionTok{list}\NormalTok{(}\AttributeTok{Project =} \FunctionTok{c}\NormalTok{(}\FunctionTok{names}\NormalTok{(all\_signals), }\FunctionTok{names}\NormalTok{(all\_signals\_2nd\_batch)),}
  \AttributeTok{PM =} \FunctionTok{c}\NormalTok{(}\StringTok{"I"}\NormalTok{, }\StringTok{"II(a)"}\NormalTok{, }\StringTok{"III"}\NormalTok{, }\StringTok{"IV(I)"}\NormalTok{, }\StringTok{"IV(II(a))"}\NormalTok{, }\StringTok{"IV(III)"}\NormalTok{), }\AttributeTok{Var =} \FunctionTok{c}\NormalTok{(}\StringTok{"REQ"}\NormalTok{,}
    \StringTok{"DEV"}\NormalTok{, }\StringTok{"DESC"}\NormalTok{), }\AttributeTok{SegIdx =} \DecValTok{1}\SpecialCharTok{:}\DecValTok{10}\NormalTok{))}

\FunctionTok{nrow}\NormalTok{(ac\_grid\_projects\_it)}
\end{Highlighting}
\end{Shaded}

\begin{verbatim}
## [1] 2700
\end{verbatim}

Now we can compute the projects' scores:

\begin{Shaded}
\begin{Highlighting}[]
\NormalTok{ac\_grid\_projects\_results\_it }\OtherTok{\textless{}{-}} \FunctionTok{loadResultsOrCompute}\NormalTok{(}\AttributeTok{file =} \StringTok{"../results/ac\_grid\_projects\_results\_it.rds"}\NormalTok{,}
  \AttributeTok{computeExpr =}\NormalTok{ \{}
    \FunctionTok{library}\NormalTok{(foreach)}

\NormalTok{    cl }\OtherTok{\textless{}{-}}\NormalTok{ parallel}\SpecialCharTok{::}\FunctionTok{makePSOCKcluster}\NormalTok{(}\FunctionTok{min}\NormalTok{(}\DecValTok{123}\NormalTok{, parallel}\SpecialCharTok{::}\FunctionTok{detectCores}\NormalTok{()))}
\NormalTok{    parallel}\SpecialCharTok{::}\FunctionTok{clusterExport}\NormalTok{(cl, }\AttributeTok{varlist =} \FunctionTok{list}\NormalTok{(}\StringTok{"req\_poly"}\NormalTok{, }\StringTok{"dev\_poly"}\NormalTok{))}
\NormalTok{    project\_signals\_all }\OtherTok{\textless{}{-}} \FunctionTok{append}\NormalTok{(all\_signals, all\_signals\_2nd\_batch)}

    \FunctionTok{doWithParallelClusterExplicit}\NormalTok{(}\AttributeTok{cl =}\NormalTok{ cl, }\AttributeTok{expr =}\NormalTok{ \{}
\NormalTok{      pb }\OtherTok{\textless{}{-}}\NormalTok{ utils}\SpecialCharTok{::}\FunctionTok{txtProgressBar}\NormalTok{(}\AttributeTok{min =} \DecValTok{1}\NormalTok{, }\AttributeTok{max =} \FunctionTok{nrow}\NormalTok{(ac\_grid\_projects\_it),}
        \AttributeTok{style =} \DecValTok{3}\NormalTok{)}
\NormalTok{      progress }\OtherTok{\textless{}{-}} \ControlFlowTok{function}\NormalTok{(n) \{}
        \ControlFlowTok{if}\NormalTok{ (}\DecValTok{0} \SpecialCharTok{==}\NormalTok{ (n}\SpecialCharTok{\%\%}\DecValTok{25}\NormalTok{)) \{}
          \FunctionTok{print}\NormalTok{(n)}
\NormalTok{        \}}
\NormalTok{        utils}\SpecialCharTok{::}\FunctionTok{setTxtProgressBar}\NormalTok{(}\AttributeTok{pb =}\NormalTok{ pb, }\AttributeTok{value =}\NormalTok{ n)}
\NormalTok{      \}}

\NormalTok{      foreach}\SpecialCharTok{::}\FunctionTok{foreach}\NormalTok{(}\AttributeTok{grid\_idx =} \FunctionTok{rownames}\NormalTok{(ac\_grid\_projects\_it), }\AttributeTok{.combine =}\NormalTok{ rbind,}
        \AttributeTok{.inorder =} \ConstantTok{FALSE}\NormalTok{, }\AttributeTok{.verbose =} \ConstantTok{TRUE}\NormalTok{, }\AttributeTok{.packages =} \FunctionTok{c}\NormalTok{(}\StringTok{"cobs"}\NormalTok{, }\StringTok{"R6"}\NormalTok{), }\AttributeTok{.options.snow =} \FunctionTok{list}\NormalTok{(}\AttributeTok{progress =}\NormalTok{ progress)) }\SpecialCharTok{\%dopar\%}
\NormalTok{        \{}
          \FunctionTok{options}\NormalTok{(}\AttributeTok{warn =} \DecValTok{2}\NormalTok{)}
          \FunctionTok{source}\NormalTok{(}\StringTok{"../models/SRBTW{-}R6.R"}\NormalTok{)}
\NormalTok{          params }\OtherTok{\textless{}{-}}\NormalTok{ ac\_grid\_projects\_it[grid\_idx, ]}
\NormalTok{          params}\SpecialCharTok{$}\NormalTok{Project }\OtherTok{\textless{}{-}} \FunctionTok{as.character}\NormalTok{(params}\SpecialCharTok{$}\NormalTok{Project)}
\NormalTok{          params}\SpecialCharTok{$}\NormalTok{PM }\OtherTok{\textless{}{-}} \FunctionTok{as.character}\NormalTok{(params}\SpecialCharTok{$}\NormalTok{PM)}
\NormalTok{          params}\SpecialCharTok{$}\NormalTok{Var }\OtherTok{\textless{}{-}} \FunctionTok{as.character}\NormalTok{(params}\SpecialCharTok{$}\NormalTok{Var)}

\NormalTok{          pm\_var }\OtherTok{\textless{}{-}} \ControlFlowTok{switch}\NormalTok{(params}\SpecialCharTok{$}\NormalTok{PM, }\AttributeTok{I =}\NormalTok{ p1\_it\_signals, }\StringTok{\textasciigrave{}}\AttributeTok{II(a)}\StringTok{\textasciigrave{}} \OtherTok{=}\NormalTok{ p2a\_it\_signals,}
          \AttributeTok{III =}\NormalTok{ p3\_it\_signals, }\StringTok{\textasciigrave{}}\AttributeTok{IV(I)}\StringTok{\textasciigrave{}} \OtherTok{=}\NormalTok{ p4p1\_it\_signals, }\StringTok{\textasciigrave{}}\AttributeTok{IV(II(a))}\StringTok{\textasciigrave{}} \OtherTok{=}\NormalTok{ p4p2a\_it\_signals,}
          \StringTok{\textasciigrave{}}\AttributeTok{IV(III)}\StringTok{\textasciigrave{}} \OtherTok{=}\NormalTok{ p4p3\_it\_signals, \{}
            \FunctionTok{stop}\NormalTok{(}\FunctionTok{paste0}\NormalTok{(}\StringTok{"Don\textquotesingle{}t know "}\NormalTok{, params}\SpecialCharTok{$}\NormalTok{PM, }\StringTok{"."}\NormalTok{))}
\NormalTok{          \})[[params}\SpecialCharTok{$}\NormalTok{Var]]}\SpecialCharTok{$}\FunctionTok{get0Function}\NormalTok{()}

\NormalTok{          tempf }\OtherTok{\textless{}{-}}\NormalTok{ project\_signals\_all[[params}\SpecialCharTok{$}\NormalTok{Project]][[params}\SpecialCharTok{$}\NormalTok{Var]]}\SpecialCharTok{$}\FunctionTok{get0Function}\NormalTok{()}
\NormalTok{          use\_deriv }\OtherTok{\textless{}{-}}\NormalTok{ params}\SpecialCharTok{$}\NormalTok{PM }\SpecialCharTok{\%in\%} \FunctionTok{c}\NormalTok{(}\StringTok{"IV(I)"}\NormalTok{, }\StringTok{"IV(II(a))"}\NormalTok{, }\StringTok{"IV(III)"}\NormalTok{)}
\NormalTok{          p\_var }\OtherTok{\textless{}{-}} \ControlFlowTok{if}\NormalTok{ (use\_deriv) \{}
          \CommentTok{\# If using the derivative, we approximate a somewhat smooth function}
          \CommentTok{\# of the variable.  The recommended span is \textgreater{}= 0.2, and we\textquotesingle{}ll use the}
          \CommentTok{\# value 0.35 as this was also the span used for smoothing the}
          \CommentTok{\# projects\textquotesingle{} variables.}
          \FunctionTok{get\_smoothed\_signal\_d1}\NormalTok{(}\AttributeTok{name =} \StringTok{"foo"}\NormalTok{, }\AttributeTok{func =}\NormalTok{ tempf, }\AttributeTok{use\_span =} \FloatTok{0.35}\NormalTok{)}\SpecialCharTok{$}\FunctionTok{get0Function}\NormalTok{()}
\NormalTok{          \} }\ControlFlowTok{else}\NormalTok{ \{}
\NormalTok{          tempf}
\NormalTok{          \}}

          \CommentTok{\# Important so we can correctly concatenate the results with the grid}
          \CommentTok{\# parameters!}
\NormalTok{          res }\OtherTok{\textless{}{-}} \FunctionTok{as.data.frame}\NormalTok{(}\FunctionTok{ac\_worker}\NormalTok{(}\AttributeTok{PM =}\NormalTok{ pm\_var, }\AttributeTok{P =}\NormalTok{ p\_var, }\AttributeTok{seg\_idx =}\NormalTok{ params}\SpecialCharTok{$}\NormalTok{SegIdx,}
          \AttributeTok{total\_segments =} \DecValTok{10}\NormalTok{))}
\NormalTok{          res}\SpecialCharTok{$}\NormalTok{grid\_idx }\OtherTok{\textless{}{-}}\NormalTok{ grid\_idx}
          \FunctionTok{cbind}\NormalTok{(res, params)}
\NormalTok{        \}}
\NormalTok{    \})}
\NormalTok{  \})}
\end{Highlighting}
\end{Shaded}

\hypertarget{rectification-of-raw-scores-1}{%
\paragraph{Rectification of raw scores}\label{rectification-of-raw-scores-1}}

Now that we have the raw scores, it is time to transform them using the marginal densities. Each single objective (ECDF) is reused 15 times (once for each project).
We will make a new temporary grid, where the number of rows in this grid corresponds to the number of ECDFs:

\begin{Shaded}
\begin{Highlighting}[]
\NormalTok{temp.Metric }\OtherTok{\textless{}{-}} \FunctionTok{c}\NormalTok{(}\StringTok{"area"}\NormalTok{, }\StringTok{"corr"}\NormalTok{, }\StringTok{"jsd"}\NormalTok{, }\StringTok{"kl"}\NormalTok{, }\StringTok{"arclen"}\NormalTok{, }\StringTok{"sd"}\NormalTok{, }\StringTok{"var"}\NormalTok{, }\StringTok{"mae"}\NormalTok{, }\StringTok{"rmse"}\NormalTok{,}
  \StringTok{"RMS"}\NormalTok{, }\StringTok{"Kurtosis"}\NormalTok{, }\StringTok{"Peak"}\NormalTok{, }\StringTok{"ImpulseFactor"}\NormalTok{)}
\NormalTok{temp.grid }\OtherTok{\textless{}{-}} \FunctionTok{expand.grid}\NormalTok{(}\FunctionTok{list}\NormalTok{(}\AttributeTok{PM =} \FunctionTok{c}\NormalTok{(}\StringTok{"I"}\NormalTok{, }\StringTok{"II(a)"}\NormalTok{, }\StringTok{"III"}\NormalTok{, }\StringTok{"IV(I)"}\NormalTok{, }\StringTok{"IV(II(a))"}\NormalTok{, }\StringTok{"IV(III)"}\NormalTok{),}
  \AttributeTok{Var =} \FunctionTok{c}\NormalTok{(}\StringTok{"REQ"}\NormalTok{, }\StringTok{"DEV"}\NormalTok{, }\StringTok{"DESC"}\NormalTok{), }\AttributeTok{SegIdx =} \DecValTok{1}\SpecialCharTok{:}\DecValTok{10}\NormalTok{, }\AttributeTok{Metric =}\NormalTok{ temp.Metric))}

\FunctionTok{nrow}\NormalTok{(temp.grid)}
\end{Highlighting}
\end{Shaded}

\begin{verbatim}
## [1] 2340
\end{verbatim}

Now it's time to rectify the raw scores:

\begin{Shaded}
\begin{Highlighting}[]
\NormalTok{ac\_grid\_projects\_results\_uniform\_it }\OtherTok{\textless{}{-}} \FunctionTok{loadResultsOrCompute}\NormalTok{(}\AttributeTok{file =} \StringTok{"../results/ac\_grid\_projects\_results\_uniform\_it.rds"}\NormalTok{,}
  \AttributeTok{computeExpr =}\NormalTok{ \{}
    \FunctionTok{doWithParallelCluster}\NormalTok{(}\AttributeTok{numCores =} \FunctionTok{min}\NormalTok{(}\DecValTok{32}\NormalTok{, parallel}\SpecialCharTok{::}\FunctionTok{detectCores}\NormalTok{()), }\AttributeTok{expr =}\NormalTok{ \{}
      \FunctionTok{library}\NormalTok{(foreach)}

\NormalTok{      project\_signals\_all }\OtherTok{\textless{}{-}} \FunctionTok{append}\NormalTok{(all\_signals, all\_signals\_2nd\_batch)}

\NormalTok{      foreach}\SpecialCharTok{::}\FunctionTok{foreach}\NormalTok{(}\AttributeTok{grid\_idx =} \FunctionTok{rownames}\NormalTok{(temp.grid), }\AttributeTok{.combine =}\NormalTok{ rbind, }\AttributeTok{.inorder =} \ConstantTok{FALSE}\NormalTok{,}
        \AttributeTok{.verbose =} \ConstantTok{TRUE}\NormalTok{) }\SpecialCharTok{\%dopar\%}\NormalTok{ \{}
        \FunctionTok{options}\NormalTok{(}\AttributeTok{warn =} \DecValTok{2}\NormalTok{)}
\NormalTok{        params }\OtherTok{\textless{}{-}}\NormalTok{ temp.grid[grid\_idx, ]}
\NormalTok{        params}\SpecialCharTok{$}\NormalTok{PM }\OtherTok{\textless{}{-}} \FunctionTok{as.character}\NormalTok{(params}\SpecialCharTok{$}\NormalTok{PM)}
\NormalTok{        params}\SpecialCharTok{$}\NormalTok{Var }\OtherTok{\textless{}{-}} \FunctionTok{as.character}\NormalTok{(params}\SpecialCharTok{$}\NormalTok{Var)}
\NormalTok{        params}\SpecialCharTok{$}\NormalTok{Metric }\OtherTok{\textless{}{-}} \FunctionTok{as.character}\NormalTok{(params}\SpecialCharTok{$}\NormalTok{Metric)}

\NormalTok{        temp.data }\OtherTok{\textless{}{-}} \FunctionTok{ac\_extract\_data}\NormalTok{(}\AttributeTok{pmName =}\NormalTok{ params}\SpecialCharTok{$}\NormalTok{PM, }\AttributeTok{varName =}\NormalTok{ params}\SpecialCharTok{$}\NormalTok{Var,}
          \AttributeTok{segIdx =}\NormalTok{ params}\SpecialCharTok{$}\NormalTok{SegIdx, }\AttributeTok{metricName =}\NormalTok{ params}\SpecialCharTok{$}\NormalTok{Metric, }\AttributeTok{ac\_grid =}\NormalTok{ ac\_grid\_it,}
          \AttributeTok{ac\_grid\_results =}\NormalTok{ ac\_grid\_results\_it)}
\NormalTok{        tempf }\OtherTok{\textless{}{-}}\NormalTok{ stats}\SpecialCharTok{::}\FunctionTok{ecdf}\NormalTok{(temp.data)}

\NormalTok{        res }\OtherTok{\textless{}{-}} \FunctionTok{matrix}\NormalTok{(}\AttributeTok{nrow =} \DecValTok{1}\NormalTok{, }\AttributeTok{ncol =} \FunctionTok{length}\NormalTok{(project\_signals\_all))}
        \ControlFlowTok{for}\NormalTok{ (i }\ControlFlowTok{in} \DecValTok{1}\SpecialCharTok{:}\FunctionTok{length}\NormalTok{(project\_signals\_all)) \{}
\NormalTok{          res[}\DecValTok{1}\NormalTok{, i] }\OtherTok{\textless{}{-}} \DecValTok{1} \SpecialCharTok{{-}} \FunctionTok{tempf}\NormalTok{(ac\_grid\_projects\_results\_it[ac\_grid\_projects\_results\_it}\SpecialCharTok{$}\NormalTok{Project }\SpecialCharTok{==}
          \FunctionTok{names}\NormalTok{(project\_signals\_all)[i] }\SpecialCharTok{\&}\NormalTok{ ac\_grid\_projects\_results\_it}\SpecialCharTok{$}\NormalTok{PM }\SpecialCharTok{==}
\NormalTok{          params}\SpecialCharTok{$}\NormalTok{PM }\SpecialCharTok{\&}\NormalTok{ ac\_grid\_projects\_results\_it}\SpecialCharTok{$}\NormalTok{Var }\SpecialCharTok{==}\NormalTok{ params}\SpecialCharTok{$}\NormalTok{Var }\SpecialCharTok{\&}\NormalTok{ ac\_grid\_projects\_results\_it}\SpecialCharTok{$}\NormalTok{SegIdx }\SpecialCharTok{==}
\NormalTok{          params}\SpecialCharTok{$}\NormalTok{SegIdx, params}\SpecialCharTok{$}\NormalTok{Metric])}
\NormalTok{        \}}
        \FunctionTok{cbind}\NormalTok{(}\StringTok{\textasciigrave{}}\AttributeTok{colnames\textless{}{-}}\StringTok{\textasciigrave{}}\NormalTok{(res, }\FunctionTok{names}\NormalTok{(project\_signals\_all)), params)}
\NormalTok{      \}}
\NormalTok{    \})}
\NormalTok{  \})}
\end{Highlighting}
\end{Shaded}

\hypertarget{non-weighted-1}{%
\paragraph{Non-weighted}\label{non-weighted-1}}

Let's have a look at how the calibrated models behave, without further fitting of weights.
This will give us insights into which (family of) models is best suitable for predicting the ground truth (given the configuration).

\begin{Shaded}
\begin{Highlighting}[]
\NormalTok{temp }\OtherTok{\textless{}{-}} \FunctionTok{append}\NormalTok{(all\_signals, all\_signals\_2nd\_batch)}
\NormalTok{temp.gt }\OtherTok{\textless{}{-}} \FunctionTok{c}\NormalTok{(ground\_truth}\SpecialCharTok{$}\NormalTok{consensus\_score, ground\_truth\_2nd\_batch}\SpecialCharTok{$}\NormalTok{consensus\_score)}

\NormalTok{temp.df }\OtherTok{\textless{}{-}} \ConstantTok{NULL}

\ControlFlowTok{for}\NormalTok{ (pmName }\ControlFlowTok{in} \FunctionTok{c}\NormalTok{(}\StringTok{"I"}\NormalTok{, }\StringTok{"II(a)"}\NormalTok{, }\StringTok{"III"}\NormalTok{, }\StringTok{"IV(I)"}\NormalTok{, }\StringTok{"IV(II(a))"}\NormalTok{, }\StringTok{"IV(III)"}\NormalTok{)) \{}
\NormalTok{  temp.pred }\OtherTok{\textless{}{-}} \FunctionTok{data.frame}\NormalTok{(}\AttributeTok{pred =} \FunctionTok{sapply}\NormalTok{(}\AttributeTok{X =} \FunctionTok{names}\NormalTok{(temp), }\ControlFlowTok{function}\NormalTok{(pName) }\FunctionTok{ac\_pmp\_score}\NormalTok{(}\AttributeTok{use\_metrics =}\NormalTok{ temp.Metric,}
    \AttributeTok{pmName =}\NormalTok{ pmName, }\AttributeTok{projName =}\NormalTok{ pName, }\AttributeTok{results\_uniform =}\NormalTok{ ac\_grid\_projects\_results\_uniform\_it)),}
    \AttributeTok{ground\_truth =}\NormalTok{ temp.gt)}

\NormalTok{  temp.df }\OtherTok{\textless{}{-}} \FunctionTok{rbind}\NormalTok{(temp.df, }\FunctionTok{data.frame}\NormalTok{(}\AttributeTok{PM =}\NormalTok{ pmName, }\AttributeTok{ScForWorst =} \FunctionTok{ac\_pmp\_score}\NormalTok{(}\AttributeTok{use\_metrics =}\NormalTok{ temp.Metric,}
    \AttributeTok{pmName =}\NormalTok{ pmName, }\AttributeTok{projName =} \FunctionTok{paste0}\NormalTok{(}\StringTok{"Project"}\NormalTok{, }\FunctionTok{which.min}\NormalTok{(temp.gt)), }\AttributeTok{results\_uniform =}\NormalTok{ ac\_grid\_projects\_results\_uniform\_it),}
    \AttributeTok{ScForBest =} \FunctionTok{ac\_pmp\_score}\NormalTok{(}\AttributeTok{use\_metrics =}\NormalTok{ temp.Metric, }\AttributeTok{pmName =}\NormalTok{ pmName, }\AttributeTok{projName =} \FunctionTok{paste0}\NormalTok{(}\StringTok{"Project"}\NormalTok{,}
      \FunctionTok{which.max}\NormalTok{(temp.gt)), }\AttributeTok{results\_uniform =}\NormalTok{ ac\_grid\_projects\_results\_uniform\_it),}
    \AttributeTok{ScMax =} \FunctionTok{max}\NormalTok{(temp.pred}\SpecialCharTok{$}\NormalTok{pred), }\AttributeTok{ScMin =} \FunctionTok{min}\NormalTok{(temp.pred}\SpecialCharTok{$}\NormalTok{pred), }\AttributeTok{ScAvg =} \FunctionTok{mean}\NormalTok{(temp.pred}\SpecialCharTok{$}\NormalTok{pred),}
    \AttributeTok{KLdiv =} \FunctionTok{kl\_div}\NormalTok{(temp.gt, temp.pred}\SpecialCharTok{$}\NormalTok{pred), }\AttributeTok{Corr =}\NormalTok{ stats}\SpecialCharTok{::}\FunctionTok{cor}\NormalTok{(temp.gt, temp.pred}\SpecialCharTok{$}\NormalTok{pred),}
    \AttributeTok{RMSE =} \FunctionTok{sqrt}\NormalTok{(Metrics}\SpecialCharTok{::}\FunctionTok{mse}\NormalTok{(}\AttributeTok{actual =}\NormalTok{ temp.gt, }\AttributeTok{predicted =}\NormalTok{ temp.pred}\SpecialCharTok{$}\NormalTok{pred)),}
    \AttributeTok{MSE =}\NormalTok{ Metrics}\SpecialCharTok{::}\FunctionTok{mse}\NormalTok{(}\AttributeTok{actual =}\NormalTok{ temp.gt, }\AttributeTok{predicted =}\NormalTok{ temp.pred}\SpecialCharTok{$}\NormalTok{pred), }\AttributeTok{Log =} \FunctionTok{mean}\NormalTok{(scoring}\SpecialCharTok{::}\FunctionTok{logscore}\NormalTok{(}\AttributeTok{object =}\NormalTok{ ground\_truth }\SpecialCharTok{\textasciitilde{}}
\NormalTok{      pred, }\AttributeTok{data =}\NormalTok{ temp.pred))))}
\NormalTok{\}}
\end{Highlighting}
\end{Shaded}

\begin{table}

\caption{\label{tab:ac-compare-pms-it}Comparison of continuous PMs using issue-tracking data for best/worst projects, as well as MSE- and Log-scores as average deviation from the ground truth consensus.}
\centering
\begin{tabular}[t]{lrrrrrrrrrr}
\toprule
PM & ScForWorst & ScForBest & ScMax & ScMin & ScAvg & KLdiv & Corr & RMSE & MSE & Log\\
\midrule
I & 0.4587 & 0.4928 & 0.4928 & 0.3578 & 0.4340 & 4.0769 & 0.4718 & 0.3181 & 0.1012 & 0.5744\\
II(a) & 0.4189 & 0.4880 & 0.5122 & 0.3535 & 0.4218 & 4.0012 & 0.5326 & 0.3092 & 0.0956 & 0.5544\\
III & 0.4311 & 0.4579 & 0.5792 & 0.3763 & 0.4503 & 3.8621 & 0.6080 & 0.3148 & 0.0991 & 0.6160\\
IV(I) & 0.7080 & 0.6848 & 0.7393 & 0.5867 & 0.6764 & 4.3515 & 0.2838 & 0.4896 & 0.2397 & 1.0867\\
IV(II(a)) & 0.7461 & 0.6890 & 0.7864 & 0.6050 & 0.6993 & 4.3042 & 0.1454 & 0.5122 & 0.2624 & 1.1648\\
\addlinespace
IV(III) & 0.7009 & 0.6829 & 0.7908 & 0.6373 & 0.7059 & 4.3786 & 0.3862 & 0.5108 & 0.2610 & 1.1869\\
\bottomrule
\end{tabular}
\end{table}

Table \ref{tab:ac-compare-pms-it} shows the results of the calibrated models. It is worth noting that we are looking at two families of models: the ordinary PMs, and the derivative PMs are in on family each.
What we can see directly is that the derivative models all suffer from the inconsistency that the score for the best project is lower than the score for the worst project. The ordinary models do not have that problem. Also the MSE is much larger for derivative models.
Another observation is that the score for best/worst is not necessarily the lowest/highest score predicted.
The mean score in each family tells us which the best model is. Here, it is type III, and that is no big surprise, as it was derived from data, so it should resemble the observed projects best.
However, looking at the spread of best/worst scores, it seems that there is room for improvement in terms of selecting the right objectives (weights).

\hypertarget{weighted-by-optimization-1}{%
\paragraph{\texorpdfstring{Weighted by optimization\label{sssec:weighted-by-opt}}{Weighted by optimization}}\label{weighted-by-optimization-1}}

\begin{Shaded}
\begin{Highlighting}[]
\NormalTok{ac\_weight\_grid\_it }\OtherTok{\textless{}{-}} \FunctionTok{expand.grid}\NormalTok{(}\FunctionTok{list}\NormalTok{(}\AttributeTok{Var =} \FunctionTok{c}\NormalTok{(}\StringTok{"REQ"}\NormalTok{, }\StringTok{"DEV"}\NormalTok{, }\StringTok{"DESC"}\NormalTok{), }\AttributeTok{Metric =}\NormalTok{ temp.Metric,}
  \AttributeTok{SegIdx =} \DecValTok{1}\SpecialCharTok{:}\DecValTok{10}\NormalTok{))}

\FunctionTok{nrow}\NormalTok{(ac\_weight\_grid\_it)}
\end{Highlighting}
\end{Shaded}

\begin{verbatim}
## [1] 390
\end{verbatim}

Now the objective for the optimization will search for weights that minimize the MSE between the weighted projects and the ground truth.
We will optimize the weights once for each process model, and then follow this up by a closer inspection of the champion model (the model with the lowest MSE).

\begin{Shaded}
\begin{Highlighting}[]
\NormalTok{temp.names }\OtherTok{\textless{}{-}} \FunctionTok{names}\NormalTok{(}\FunctionTok{append}\NormalTok{(all\_signals, all\_signals\_2nd\_batch))}

\NormalTok{ac\_it\_weights\_optim }\OtherTok{\textless{}{-}} \ControlFlowTok{function}\NormalTok{(pmName, }\AttributeTok{useLog =} \ConstantTok{FALSE}\NormalTok{) \{}
  \FunctionTok{loadResultsOrCompute}\NormalTok{(}\AttributeTok{file =} \FunctionTok{paste0}\NormalTok{(}\StringTok{"../results/ac\_it\_weights\_pm\_"}\NormalTok{, pmName, (}\ControlFlowTok{if}\NormalTok{ (useLog)}
    \StringTok{"\_Log"}\NormalTok{), }\StringTok{".rds"}\NormalTok{), }\AttributeTok{computeExpr =}\NormalTok{ \{}
\NormalTok{    cl }\OtherTok{\textless{}{-}}\NormalTok{ parallel}\SpecialCharTok{::}\FunctionTok{makePSOCKcluster}\NormalTok{(}\FunctionTok{min}\NormalTok{(}\DecValTok{123}\NormalTok{, parallel}\SpecialCharTok{::}\FunctionTok{detectCores}\NormalTok{()))}
\NormalTok{    parallel}\SpecialCharTok{::}\FunctionTok{clusterExport}\NormalTok{(cl, }\AttributeTok{varlist =} \FunctionTok{list}\NormalTok{(}\StringTok{"temp.gt"}\NormalTok{, }\StringTok{"temp.names"}\NormalTok{, }\StringTok{"ac\_pmp\_score\_weighted"}\NormalTok{,}
      \StringTok{"ac\_weight\_grid\_it"}\NormalTok{, }\StringTok{"ac\_grid\_projects\_results\_uniform\_it"}\NormalTok{))}

    \FunctionTok{doWithParallelClusterExplicit}\NormalTok{(}\AttributeTok{cl =}\NormalTok{ cl, }\AttributeTok{expr =}\NormalTok{ \{}
\NormalTok{      optimParallel}\SpecialCharTok{::}\FunctionTok{optimParallel}\NormalTok{(}\AttributeTok{par =} \FunctionTok{rep}\NormalTok{(}\FloatTok{0.5}\NormalTok{, }\FunctionTok{nrow}\NormalTok{(ac\_weight\_grid\_it)),}
        \AttributeTok{method =} \StringTok{"L{-}BFGS{-}B"}\NormalTok{, }\AttributeTok{lower =} \FunctionTok{rep}\NormalTok{(}\DecValTok{0}\NormalTok{, }\FunctionTok{nrow}\NormalTok{(ac\_weight\_grid\_it)), }\AttributeTok{upper =} \FunctionTok{rep}\NormalTok{(}\DecValTok{1}\NormalTok{,}
          \FunctionTok{nrow}\NormalTok{(ac\_weight\_grid\_it)), }\AttributeTok{fn =} \ControlFlowTok{function}\NormalTok{(x) \{}
          \ControlFlowTok{if}\NormalTok{ (useLog) \{}
\NormalTok{          temp.pred }\OtherTok{\textless{}{-}} \FunctionTok{data.frame}\NormalTok{(}\AttributeTok{ground\_truth =}\NormalTok{ temp.gt, }\AttributeTok{pred =} \FunctionTok{sapply}\NormalTok{(}\AttributeTok{X =}\NormalTok{ temp.names,}
            \AttributeTok{FUN =} \ControlFlowTok{function}\NormalTok{(projName) \{}
            \FunctionTok{ac\_pmp\_score\_weighted}\NormalTok{(}\AttributeTok{pmName =}\NormalTok{ pmName, }\AttributeTok{projName =}\NormalTok{ projName,}
              \AttributeTok{weightGrid =}\NormalTok{ ac\_weight\_grid\_it, }\AttributeTok{results\_uniform =}\NormalTok{ ac\_grid\_projects\_results\_uniform\_it,}
              \AttributeTok{weightVector =}\NormalTok{ x)}
\NormalTok{            \}))}

          \FunctionTok{return}\NormalTok{(}\FunctionTok{mean}\NormalTok{(scoring}\SpecialCharTok{::}\FunctionTok{logscore}\NormalTok{(}\AttributeTok{object =}\NormalTok{ ground\_truth }\SpecialCharTok{\textasciitilde{}}\NormalTok{ pred, }\AttributeTok{data =}\NormalTok{ temp.pred)))}
\NormalTok{          \} }\ControlFlowTok{else}\NormalTok{ \{}
          \FunctionTok{return}\NormalTok{(Metrics}\SpecialCharTok{::}\FunctionTok{mse}\NormalTok{(}\AttributeTok{actual =}\NormalTok{ temp.gt, }\AttributeTok{predicted =} \FunctionTok{sapply}\NormalTok{(}\AttributeTok{X =}\NormalTok{ temp.names,}
            \AttributeTok{FUN =} \ControlFlowTok{function}\NormalTok{(projName) \{}
            \FunctionTok{ac\_pmp\_score\_weighted}\NormalTok{(}\AttributeTok{pmName =}\NormalTok{ pmName, }\AttributeTok{projName =}\NormalTok{ projName,}
              \AttributeTok{weightGrid =}\NormalTok{ ac\_weight\_grid\_it, }\AttributeTok{results\_uniform =}\NormalTok{ ac\_grid\_projects\_results\_uniform\_it,}
              \AttributeTok{weightVector =}\NormalTok{ x)}
\NormalTok{            \})))}
\NormalTok{          \}}
\NormalTok{        \}, }\AttributeTok{parallel =} \FunctionTok{list}\NormalTok{(}\AttributeTok{cl =}\NormalTok{ cl, }\AttributeTok{forward =} \ConstantTok{FALSE}\NormalTok{, }\AttributeTok{loginfo =} \ConstantTok{TRUE}\NormalTok{))}
\NormalTok{    \})}
\NormalTok{  \})}
\NormalTok{\}}
\end{Highlighting}
\end{Shaded}

\hypertarget{results-pm-vs.-pm-1}{%
\subparagraph{Results PM vs.~PM}\label{results-pm-vs.-pm-1}}

Now for the actual optimization of each type of process model.
An overview of the optimization process on a per-project basis is shown in table \ref{tab:ac-it-weights-optim-overview}.

\begin{Shaded}
\begin{Highlighting}[]
\NormalTok{ac\_it\_weights\_pm\_I }\OtherTok{\textless{}{-}} \FunctionTok{ac\_it\_weights\_optim}\NormalTok{(}\AttributeTok{pmName =} \StringTok{"I"}\NormalTok{)}
\NormalTok{ac\_it\_weights\_pm\_IIa }\OtherTok{\textless{}{-}} \FunctionTok{ac\_it\_weights\_optim}\NormalTok{(}\AttributeTok{pmName =} \StringTok{"II(a)"}\NormalTok{)}
\NormalTok{ac\_it\_weights\_pm\_III }\OtherTok{\textless{}{-}} \FunctionTok{ac\_it\_weights\_optim}\NormalTok{(}\AttributeTok{pmName =} \StringTok{"III"}\NormalTok{)}
\NormalTok{ac\_it\_weights\_pm\_IV\_I }\OtherTok{\textless{}{-}} \FunctionTok{ac\_it\_weights\_optim}\NormalTok{(}\AttributeTok{pmName =} \StringTok{"IV(I)"}\NormalTok{)}
\NormalTok{ac\_it\_weights\_pm\_IV\_IIa }\OtherTok{\textless{}{-}} \FunctionTok{ac\_it\_weights\_optim}\NormalTok{(}\AttributeTok{pmName =} \StringTok{"IV(II(a))"}\NormalTok{)}
\NormalTok{ac\_it\_weights\_pm\_IV\_III }\OtherTok{\textless{}{-}} \FunctionTok{ac\_it\_weights\_optim}\NormalTok{(}\AttributeTok{pmName =} \StringTok{"IV(III)"}\NormalTok{)}

\CommentTok{\# Also, load the optimization results using the Log{-}score as objective:}
\NormalTok{ac\_it\_weights\_pm\_I\_Log }\OtherTok{\textless{}{-}} \FunctionTok{ac\_it\_weights\_optim}\NormalTok{(}\AttributeTok{pmName =} \StringTok{"I"}\NormalTok{, }\AttributeTok{useLog =} \ConstantTok{TRUE}\NormalTok{)}
\NormalTok{ac\_it\_weights\_pm\_IIa\_Log }\OtherTok{\textless{}{-}} \FunctionTok{ac\_it\_weights\_optim}\NormalTok{(}\AttributeTok{pmName =} \StringTok{"II(a)"}\NormalTok{, }\AttributeTok{useLog =} \ConstantTok{TRUE}\NormalTok{)}
\NormalTok{ac\_it\_weights\_pm\_III\_Log }\OtherTok{\textless{}{-}} \FunctionTok{ac\_it\_weights\_optim}\NormalTok{(}\AttributeTok{pmName =} \StringTok{"III"}\NormalTok{, }\AttributeTok{useLog =} \ConstantTok{TRUE}\NormalTok{)}
\NormalTok{ac\_it\_weights\_pm\_IV\_I\_Log }\OtherTok{\textless{}{-}} \FunctionTok{ac\_it\_weights\_optim}\NormalTok{(}\AttributeTok{pmName =} \StringTok{"IV(I)"}\NormalTok{, }\AttributeTok{useLog =} \ConstantTok{TRUE}\NormalTok{)}
\NormalTok{ac\_it\_weights\_pm\_IV\_IIa\_Log }\OtherTok{\textless{}{-}} \FunctionTok{ac\_it\_weights\_optim}\NormalTok{(}\AttributeTok{pmName =} \StringTok{"IV(II(a))"}\NormalTok{, }\AttributeTok{useLog =} \ConstantTok{TRUE}\NormalTok{)}
\NormalTok{ac\_it\_weights\_pm\_IV\_III\_Log }\OtherTok{\textless{}{-}} \FunctionTok{ac\_it\_weights\_optim}\NormalTok{(}\AttributeTok{pmName =} \StringTok{"IV(III)"}\NormalTok{, }\AttributeTok{useLog =} \ConstantTok{TRUE}\NormalTok{)}
\end{Highlighting}
\end{Shaded}

\begin{table}

\caption{\label{tab:ac-it-weights-optim-overview}Overview of the results of optimizing the weights using the normalizing linear scalarizer objective, on a per-project basis, for issue-tracking process models. Shown are results once minimized using the MSE, and once using the Log-score.}
\centering
\begin{tabular}[t]{llrrrlrr}
\toprule
PM & Objective & Solution & RMSE & Num\_w\_gt0 & Perc\_w\_pruned & Iter & Iter\_grad\\
\midrule
I & MSE & 0.01197 & 0.10938 & 12 & 96.92\% & 120 & 120\\
II(a) & MSE & 0.01794 & 0.13394 & 13 & 96.67\% & 110 & 110\\
III & MSE & 0.00957 & 0.09783 & 10 & 97.44\% & 93 & 93\\
IV(I) & MSE & 0.04139 & 0.20344 & 10 & 97.44\% & 127 & 127\\
IV(II(a)) & MSE & 0.10014 & 0.31645 & 120 & 69.23\% & 30 & 30\\
\addlinespace
IV(III) & MSE & 0.03862 & 0.19651 & 6 & 98.46\% & 115 & 115\\
I & Log-Score & 0.09588 & NA & 6 & 98.46\% & 139 & 139\\
II(a) & Log-Score & 0.08874 & NA & 7 & 98.21\% & 144 & 144\\
III & Log-Score & 0.09507 & NA & 6 & 98.46\% & 104 & 104\\
IV(I) & Log-Score & 0.27298 & NA & 4 & 98.97\% & 102 & 102\\
\addlinespace
IV(II(a)) & Log-Score & 0.27320 & NA & 2 & 99.49\% & 44 & 44\\
IV(III) & Log-Score & 0.25981 & NA & 3 & 99.23\% & 42 & 42\\
\bottomrule
\end{tabular}
\end{table}

Table \ref{tab:ac-it-weights-optim-overview} gives on overview of the optimization results, which are quite diverse.
All but model type IV(II(a)) could be pruned considerably (by \(\approx97\)\% or more), which means that about \(6-12\) objectives is enough to predict with a small deviation.
Again, we observe a substantial difference between the families of ordinary and derived models, with the former performing considerably better.
Running the minimization with the Log-score results in significantly smaller models. However, this was just a test, and we should not derive any conclusions from the Log-score results. This is because technically, the Log-score assumes mutually-exclusive, discrete classes (which we do not have).
So the test was whether or not some optimization can be reached using another objective.

\begin{table}

\caption{\label{tab:ac-compare-pms-it-optim}Comparison of continuous PMs using issue-tracking data for best/worst projects, as well as MSE- and Log-scores as average deviation from the ground truth consensus, using some best set of weights per PM as found by optimization.}
\centering
\begin{tabular}[t]{llrrrrrrrrrr}
\toprule
PM & Obj & ScForBest & ScForWorst & ScMax & ScMin & ScAvg & KLdiv & Corr & RMSE & MSE & Log\\
\midrule
I & MSE & 0.7841 & 0.1542 & 0.7841 & 0.0687 & 0.2752 & 0.1569 & 0.9500 & 0.1094 & 0.0120 & 0.3123\\
II(a) & MSE & 0.7500 & 0.1822 & 0.7500 & 0.0926 & 0.2798 & 0.6052 & 0.9410 & 0.1339 & 0.0179 & 0.3084\\
III & MSE & 0.7882 & 0.1326 & 0.7882 & 0.0551 & 0.2883 & 0.1457 & 0.9756 & 0.0978 & 0.0096 & 0.3224\\
IV(I) & MSE & 0.7114 & 0.1675 & 0.7114 & 0.1264 & 0.3409 & 0.5680 & 0.7800 & 0.2034 & 0.0414 & 0.3992\\
IV(II(a)) & MSE & 0.6294 & 0.3878 & 0.7217 & 0.1790 & 0.4614 & 0.9389 & 0.5017 & 0.3164 & 0.1001 & 0.6435\\
\addlinespace
IV(III) & MSE & 0.7089 & 0.2951 & 0.7089 & 0.0002 & 0.3450 & 0.5149 & 0.8062 & 0.1965 & 0.0386 & 0.4072\\
I & Log & 0.5616 & 0.0000 & 0.5616 & 0.0000 & 0.0875 & 1.3685 & 0.5898 & 0.3080 & 0.0949 & 0.0959\\
II(a) & Log & 0.6327 & 0.0000 & 0.6327 & 0.0000 & 0.0931 & 0.9971 & 0.6035 & 0.3018 & 0.0911 & 0.0887\\
III & Log & 0.6006 & 0.0000 & 0.6006 & 0.0000 & 0.0890 & 0.4912 & 0.5294 & 0.3175 & 0.1008 & 0.0951\\
IV(I) & Log & 0.5710 & 0.0532 & 0.6605 & 0.0385 & 0.2213 & 0.2230 & 0.4119 & 0.2869 & 0.0823 & 0.2730\\
\addlinespace
IV(II(a)) & Log & 0.6930 & 0.1874 & 0.7138 & 0.0000 & 0.2292 & 0.1079 & 0.4590 & 0.2873 & 0.0825 & 0.2732\\
IV(III) & Log & 0.4960 & 0.0935 & 0.5282 & 0.0000 & 0.2020 & 0.3785 & 0.4246 & 0.2882 & 0.0831 & 0.2598\\
\bottomrule
\end{tabular}
\end{table}

The results of table \ref{tab:ac-compare-pms-it-optim} demonstrate that we can drastically improve the models by optimizing the variable importance.
Going by lowest MSE and widest spread, the clear winner is perhaps again type III.
Again, the family of derivative models has much higher MSEs, but we were able to correct the inconsistency for best/worst projects.

In general we can say that the family of derivative models performs worse than their ordinary counter part. However, the performance of the derivative models is perhaps still acceptable in some scenarios, and from these results we can say that they work and are operationalizable, even if not the best.
An MSE \(<\approx0.1\) means that the models is perhaps already usable in practice.

We also want to check what results are achievable using an ordinary Random forest.
For that, we are going to use the champion model (type III).

\begin{Shaded}
\begin{Highlighting}[]
\NormalTok{ac\_it\_weights\_varimp }\OtherTok{\textless{}{-}} \FunctionTok{loadResultsOrCompute}\NormalTok{(}\AttributeTok{file =} \StringTok{"../results/ac\_it\_weights\_varimp.rds"}\NormalTok{,}
  \AttributeTok{computeExpr =}\NormalTok{ \{}
\NormalTok{    temp.gt }\OtherTok{\textless{}{-}} \FunctionTok{c}\NormalTok{(ground\_truth}\SpecialCharTok{$}\NormalTok{consensus\_score, ground\_truth\_2nd\_batch}\SpecialCharTok{$}\NormalTok{consensus\_score)}
\NormalTok{    projNames }\OtherTok{\textless{}{-}} \FunctionTok{names}\NormalTok{(}\FunctionTok{append}\NormalTok{(all\_signals, all\_signals\_2nd\_batch))}
\NormalTok{    templ }\OtherTok{\textless{}{-}} \FunctionTok{list}\NormalTok{()}

    \ControlFlowTok{for}\NormalTok{ (pmName }\ControlFlowTok{in} \FunctionTok{levels}\NormalTok{(ac\_grid\_it}\SpecialCharTok{$}\NormalTok{PM)) \{}
\NormalTok{      temp }\OtherTok{\textless{}{-}} \FunctionTok{matrix}\NormalTok{(}\AttributeTok{nrow =} \FunctionTok{length}\NormalTok{(projNames), }\AttributeTok{ncol =} \DecValTok{1} \SpecialCharTok{+} \FunctionTok{nrow}\NormalTok{(ac\_weight\_grid\_it))}
\NormalTok{      temp[, }\FunctionTok{ncol}\NormalTok{(temp)] }\OtherTok{\textless{}{-}}\NormalTok{ temp.gt}

      \ControlFlowTok{for}\NormalTok{ (i }\ControlFlowTok{in} \DecValTok{1}\SpecialCharTok{:}\FunctionTok{length}\NormalTok{(projNames)) \{}
        \ControlFlowTok{for}\NormalTok{ (j }\ControlFlowTok{in} \DecValTok{1}\SpecialCharTok{:}\FunctionTok{nrow}\NormalTok{(ac\_weight\_grid\_it)) \{}
\NormalTok{          params }\OtherTok{\textless{}{-}}\NormalTok{ ac\_weight\_grid\_it[j, ]}
\NormalTok{          params}\SpecialCharTok{$}\NormalTok{Var }\OtherTok{\textless{}{-}} \FunctionTok{as.character}\NormalTok{(params}\SpecialCharTok{$}\NormalTok{Var)}
\NormalTok{          params}\SpecialCharTok{$}\NormalTok{Metric }\OtherTok{\textless{}{-}} \FunctionTok{as.character}\NormalTok{(params}\SpecialCharTok{$}\NormalTok{Metric)}

\NormalTok{          temp[i, j] }\OtherTok{\textless{}{-}}\NormalTok{ ac\_grid\_projects\_results\_uniform\_it[ac\_grid\_projects\_results\_uniform\_it}\SpecialCharTok{$}\NormalTok{PM }\SpecialCharTok{==}
\NormalTok{          pmName }\SpecialCharTok{\&}\NormalTok{ ac\_grid\_projects\_results\_uniform\_it}\SpecialCharTok{$}\NormalTok{Metric }\SpecialCharTok{==}\NormalTok{ params}\SpecialCharTok{$}\NormalTok{Metric }\SpecialCharTok{\&}
\NormalTok{          ac\_grid\_projects\_results\_uniform\_it}\SpecialCharTok{$}\NormalTok{Var }\SpecialCharTok{==}\NormalTok{ params}\SpecialCharTok{$}\NormalTok{Var }\SpecialCharTok{\&}\NormalTok{ ac\_grid\_projects\_results\_uniform\_it}\SpecialCharTok{$}\NormalTok{SegIdx }\SpecialCharTok{==}
\NormalTok{          params}\SpecialCharTok{$}\NormalTok{SegIdx, projNames[i]]}
\NormalTok{        \}}
\NormalTok{      \}}

\NormalTok{      temp }\OtherTok{\textless{}{-}} \FunctionTok{as.data.frame}\NormalTok{(temp)}
      \FunctionTok{colnames}\NormalTok{(temp) }\OtherTok{\textless{}{-}} \FunctionTok{c}\NormalTok{(}\FunctionTok{paste0}\NormalTok{(}\StringTok{"w"}\NormalTok{, }\DecValTok{1}\SpecialCharTok{:}\FunctionTok{nrow}\NormalTok{(ac\_weight\_grid\_it)), }\StringTok{"gt"}\NormalTok{)}

\NormalTok{      templ[[pmName]] }\OtherTok{\textless{}{-}} \FunctionTok{doWithParallelCluster}\NormalTok{(}\AttributeTok{numCores =} \FunctionTok{min}\NormalTok{(}\DecValTok{8}\NormalTok{, parallel}\SpecialCharTok{::}\FunctionTok{detectCores}\NormalTok{()),}
        \AttributeTok{expr =}\NormalTok{ \{}
          \FunctionTok{library}\NormalTok{(caret, }\AttributeTok{quietly =} \ConstantTok{TRUE}\NormalTok{)}

          \FunctionTok{set.seed}\NormalTok{(}\DecValTok{1}\NormalTok{)}
\NormalTok{          control }\OtherTok{\textless{}{-}}\NormalTok{ caret}\SpecialCharTok{::}\FunctionTok{trainControl}\NormalTok{(}\AttributeTok{method =} \StringTok{"repeatedcv"}\NormalTok{, }\AttributeTok{number =} \DecValTok{10}\NormalTok{,}
          \AttributeTok{repeats =} \DecValTok{3}\NormalTok{)}
\NormalTok{          modelFit }\OtherTok{\textless{}{-}}\NormalTok{ caret}\SpecialCharTok{::}\FunctionTok{train}\NormalTok{(gt }\SpecialCharTok{\textasciitilde{}}\NormalTok{ ., }\AttributeTok{data =}\NormalTok{ temp, }\AttributeTok{method =} \StringTok{"rf"}\NormalTok{, }\AttributeTok{trControl =}\NormalTok{ control)}
\NormalTok{          imp }\OtherTok{\textless{}{-}}\NormalTok{ caret}\SpecialCharTok{::}\FunctionTok{varImp}\NormalTok{(}\AttributeTok{object =}\NormalTok{ modelFit)}

          \FunctionTok{list}\NormalTok{(}\AttributeTok{fit =}\NormalTok{ modelFit, }\AttributeTok{imp =}\NormalTok{ imp)}
\NormalTok{        \})}
\NormalTok{    \}}
\NormalTok{    templ}
\NormalTok{  \})}
\end{Highlighting}
\end{Shaded}

Let's take a look at the fitted model's performance (table \ref{tab:ac-it-weights-varimp-mse}).
The results are only ever so slightly better than those obtained using the source code PMs (table \ref{tab:ac-sc-weights-varimp-mse}).
Four out of six explained variances are still negative, with the other only being barely positive, meaning that the Random forest overfits every time.
Therefore, it is hard to assess the validity of the obtained results for the variable importance.
The RMSEs are slightly better, but similar.
The MSE to beat was \(\approx0.0096\), and the lowest MSE achieved by a Random forest was \(\approx0.058\), which is an order of magnitude larger. So, clearly, our variable importance approach is better.

\begin{table}

\caption{\label{tab:ac-it-weights-varimp-mse}Best models of fitting of a Random forest to the calibrated data in order to assess the variable importance.}
\centering
\begin{tabular}[t]{llrrrrrrrrr}
\toprule
  & PM & mtry & RMSE & Rsq & MAE & RMSESD & RsqSD & MAESD & MSE & VarExpl\\
\midrule
1 & I & 2 & 0.2542 & 1 & 0.2429 & 0.1880 & 0 & 0.1835 & 0.0646 & -10.795\\
2 & II(a) & 2 & 0.2512 & 1 & 0.2387 & 0.1748 & 0 & 0.1703 & 0.0631 & -8.531\\
3 & III & 2 & 0.2528 & 1 & 0.2405 & 0.1853 & 0 & 0.1803 & 0.0639 & -2.752\\
31 & IV(I) & 390 & 0.2420 & 1 & 0.2299 & 0.1764 & 0 & 0.1730 & 0.0586 & 2.692\\
21 & IV(II(a)) & 196 & 0.2520 & 1 & 0.2397 & 0.1764 & 0 & 0.1731 & 0.0635 & 6.450\\
\addlinespace
32 & IV(III) & 390 & 0.2522 & 1 & 0.2410 & 0.1838 & 0 & 0.1778 & 0.0636 & -3.738\\
\bottomrule
\end{tabular}
\end{table}

\hypertarget{detailed-results-for-model-type-i}{%
\subparagraph{Detailed results for model type I}\label{detailed-results-for-model-type-i}}

If we inspect the metrics used in model (I/MSE), we can make some very interesting observations (table \ref{tab:ac-it-details-pm-i}).
Using only twelve objectives in total (5/3/4 usages for \texttt{REQ}/\texttt{DEV}/\texttt{DESC}), we see that the by far most important objective is the \texttt{ImpulseFactor} on the \texttt{DESC} variable in segment \(9\). This seems plausible, as the de-scoping is an important indicator for a project going awry, but only towards project end.
The impulse factor divides the process' peak by its average and is therefore sensitive to perturbations of the entire signal and not just its extremes. A difference between \texttt{Peak} and \texttt{ImpulseFactor} could indicate that in fact differences between process and model exist.
The next two most (and equally) important objectives are the standard deviation and the variance in segment \(4\).
Apart from the \texttt{arclen} in segment one and the \texttt{jsd} and \texttt{kl} in segment \(10\), the rest of the objectives are signal metrics.

Also, segments \(\{2,3,6\}\) were not used and deemed unimportant. In other words, using only \(7\)/\(10\) segments, the model is still able to detect the presence (or absence) of the Fire Drill with great confidence.

\begin{table}

\caption{\label{tab:ac-it-details-pm-i}The metrics used in PM I. All but one metric are metrics from signal theory.}
\centering
\begin{tabular}[t]{llrrrr}
\toprule
Var & Metric & SegIdx & weight & weight\_rel & weight\_cum\\
\midrule
REQ & arclen & 1 & 0.0832654 & 0.0348751 & 0.0348751\\
REQ & sd & 4 & 0.2616520 & 0.1095911 & 0.1444662\\
REQ & var & 4 & 0.2616520 & 0.1095911 & 0.2540573\\
REQ & RMS & 4 & 0.1874265 & 0.0785023 & 0.3325596\\
DEV & RMS & 4 & 0.0299558 & 0.0125468 & 0.3451064\\
\addlinespace
REQ & Peak & 5 & 0.1926650 & 0.0806964 & 0.4258028\\
DESC & Peak & 7 & 0.0187210 & 0.0078412 & 0.4336439\\
DESC & Kurtosis & 8 & 0.0323242 & 0.0135388 & 0.4471827\\
DESC & ImpulseFactor & 8 & 0.0139573 & 0.0058459 & 0.4530286\\
DESC & ImpulseFactor & 9 & 1.0000000 & 0.4188430 & 0.8718716\\
\addlinespace
DEV & jsd & 10 & 0.1561476 & 0.0654013 & 0.9372729\\
DEV & kl & 10 & 0.1497628 & 0.0627271 & 1.0000000\\
\bottomrule
\end{tabular}
\end{table}

\hypertarget{detailed-results-ordinary-vs.-derivative-models}{%
\subparagraph{Detailed results ordinary vs.~derivative models}\label{detailed-results-ordinary-vs.-derivative-models}}

Let's compare the performance of all ordinary PMs vs their derivative counterparts.

\begin{figure}
\centering
\includegraphics{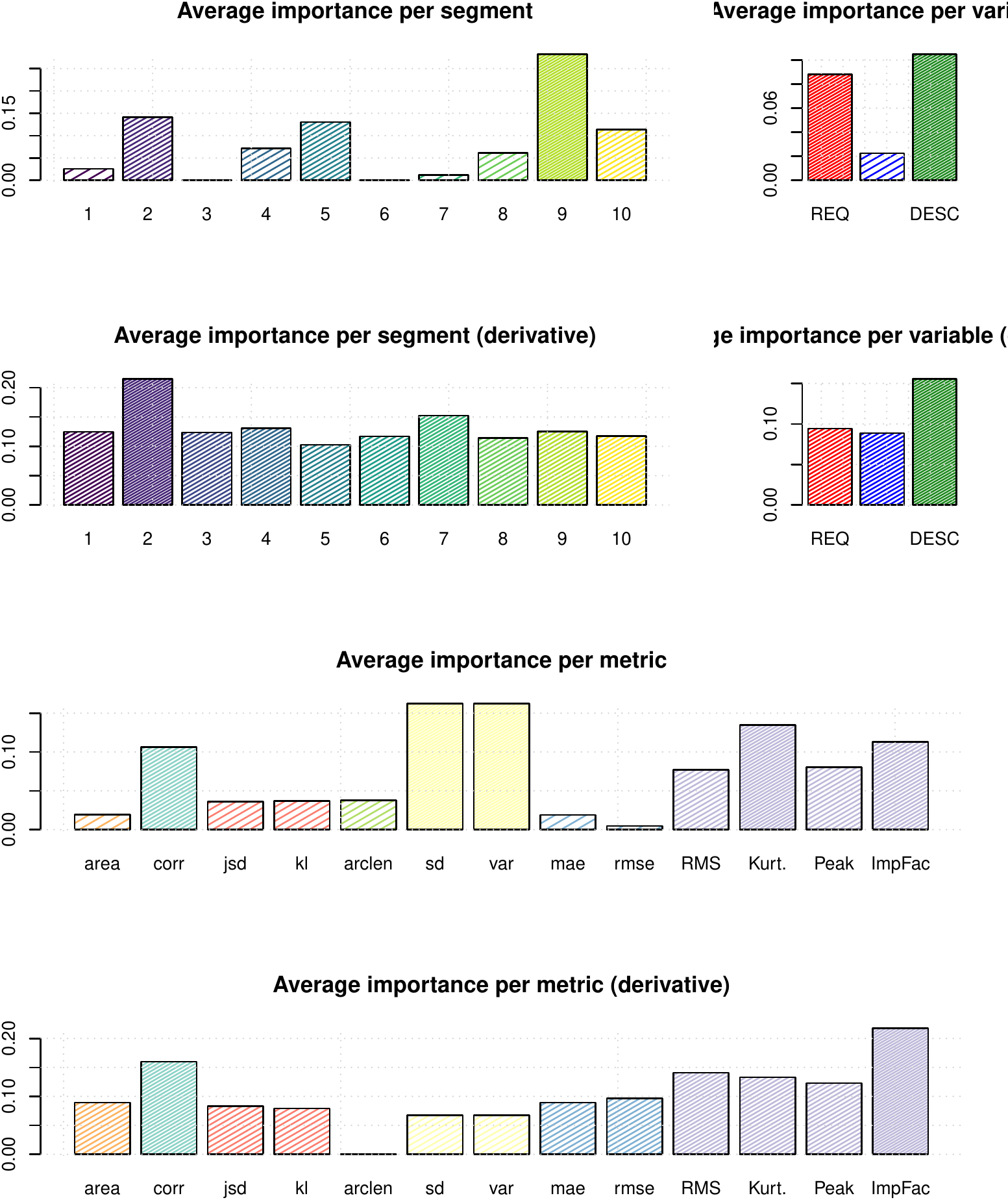}
\caption{\label{fig:ac-it-weights-detailed-plot}Average and total importances of scores per segment, variable, and metric, for the families of ordinary and derivative process models (optimized using MSE).}
\end{figure}

In figure \ref{fig:ac-it-weights-detailed-plot} we compare importances between the two families of ordinary and derivative process models.
It appears that the latter type of models make a more even use across all segments, while ordinary models focus their attention more on some segments than on others.
In both cases, the \texttt{DESC} variable is the most important, followed by \texttt{REQ}.
Correlation and signal metrics are important for both families, with varying importance for other kinds of metrics.
This is interesting, as these metrics are comparatively cheap to compute.

For the ordinary models it does not seem important by how much PM/P actually differ in absolute terms. In other words, it appears that it is more important to quantify deviation and resemblance by metrics other than \texttt{area}/\texttt{mae} or \texttt{rmse}. Instead, it makes apparently more sense to compare PM/P as signals (ratios), as well as by how their \emph{shape} diverges (\texttt{corr}, \texttt{sd}/\texttt{var} are sensitive to these).

\hypertarget{comparing-process-models}{%
\subsubsection{Comparing process models}\label{comparing-process-models}}

\begin{figure}[ht!]

{\centering \includegraphics[width=0.95\linewidth]{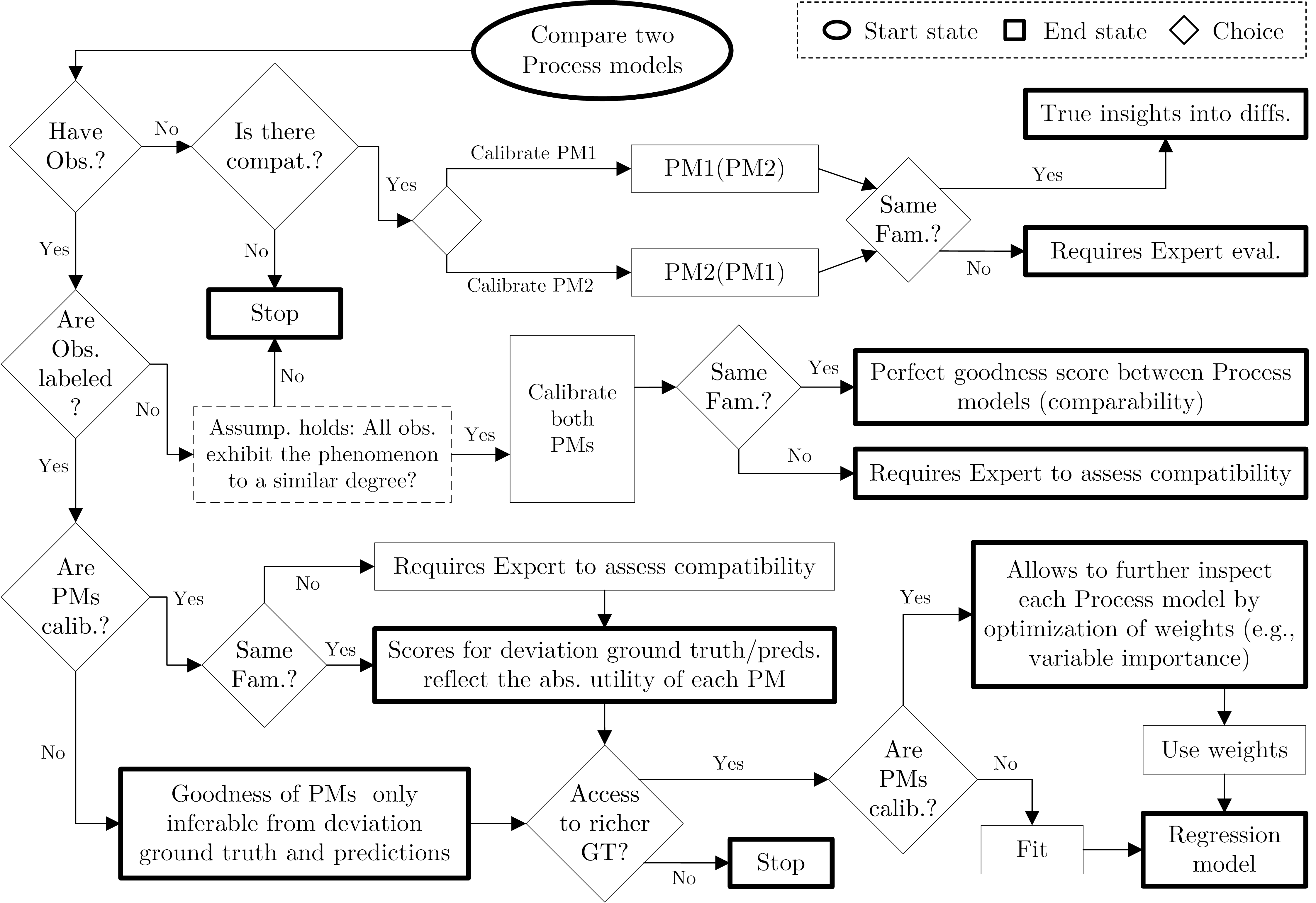} 

}

\caption{Workflow for examining differences between two PMs. Automatic calibration allows to gain detailed insights into each PM and the used configuration.}\label{fig:compare-pms}
\end{figure}

Figure \ref{fig:compare-pms} summarizes the workflow for comparing two or more process models.
This may be done with or with (labeled) observations.
Without observations but some compatibility, one of the PMs' configurations can be calibrated, in order to quantify how and by how much the other PM deviates from it.
Even without a ground truth, we can learn something about the PMs if we were to calibrate them, and that is which of the two better resembles the observations, given the chosen configuration.
This however is only a valid procedure under the assumptions that a) all observations exhibit the sought-after phenomenon, and b) they all do so to a similar degree.
If both PMs are in the same family, then we will obtain a perfect goodness score that allows to choose a champion PM within the family.
The champion PM is the one that incurs the lowest loss (or highest scores) across all unlabeled observations, which means that the champion resembles the observations best, given the same configuration as used within the family.

Another interesting case arises from having access to a somewhat richer corpus of ground truth, as we can then learn the variable importances (here: how important each calibrated score is) of a PM (see section \ref{sssec:weighted-by-opt}).
These allow us to inspect in detail, which the scores of importance are. For example, one might learn that a certain score or segment is irrelevant for capturing the phenomenon the PM describes.
Variable importances can also be further facilitated in a regression model for future observations.

\hypertarget{qualitative-evaluation-of-the-raters-notes}{%
\subsection{Qualitative Evaluation of the Rater's Notes}\label{qualitative-evaluation-of-the-raters-notes}}

Although this is not specific to issue tracking (or source code for that matter), we add a section about a qualitative evaluation of the three raters' findings for each project.
The analysis is already done and is based on commit \texttt{7a4d8}\footnote{\url{https://github.com/MrShoenel/anti-pattern-models/blob/master/notebooks/comm-class-models.Rmd}}.
What we will do here, is to load the data into a table so that we can produce nice figures for the upcoming paper (Hönel, Picha, et al. 2023).
It is the proper successor to our previous pilot study (Picha et al. 2022).

\hypertarget{list-of-symptoms-and-consequences}{%
\subsubsection{List of Symptoms and Consequences}\label{list-of-symptoms-and-consequences}}

The first data structure we'll employ is a list with associations Symptoms \& Consequence =\textgreater{} Empirical Instance(s).
The empirical instances will be put into a dataframe next.

\begin{Shaded}
\begin{Highlighting}[]
\NormalTok{qe\_sc }\OtherTok{\textless{}{-}} \FunctionTok{list}\NormalTok{(}
  \AttributeTok{SC1 =} \FunctionTok{paste0}\NormalTok{(}\StringTok{"E"}\NormalTok{, }\DecValTok{1}\SpecialCharTok{:}\DecValTok{3}\NormalTok{),}
  \AttributeTok{SC2 =} \FunctionTok{paste0}\NormalTok{(}\StringTok{"E"}\NormalTok{, }\DecValTok{4}\NormalTok{),}
  \AttributeTok{SC3 =} \FunctionTok{paste0}\NormalTok{(}\StringTok{"E"}\NormalTok{, }\DecValTok{5}\NormalTok{),}
  \AttributeTok{SC4 =} \FunctionTok{paste0}\NormalTok{(}\StringTok{"E"}\NormalTok{, }\DecValTok{6}\SpecialCharTok{:}\DecValTok{7}\NormalTok{),}
  \AttributeTok{SC5 =} \FunctionTok{paste0}\NormalTok{(}\StringTok{"E"}\NormalTok{, }\DecValTok{8}\NormalTok{),}
  \AttributeTok{SC6 =} \FunctionTok{paste0}\NormalTok{(}\StringTok{"E"}\NormalTok{, }\DecValTok{9}\SpecialCharTok{:}\DecValTok{11}\NormalTok{),}
  \AttributeTok{SC7 =} \FunctionTok{c}\NormalTok{(), }\CommentTok{\# No concrete instances for this one!}
  
  \CommentTok{\# ESC1, ESC2, and ESC3 are new, empirical symptoms and consequences!}
  \AttributeTok{ESC1 =} \FunctionTok{paste0}\NormalTok{(}\StringTok{"E"}\NormalTok{, }\DecValTok{12}\SpecialCharTok{:}\DecValTok{16}\NormalTok{),}
  \AttributeTok{ESC2 =} \FunctionTok{paste0}\NormalTok{(}\StringTok{"E"}\NormalTok{, }\DecValTok{17}\SpecialCharTok{:}\DecValTok{27}\NormalTok{),}
  \AttributeTok{ESC3 =} \FunctionTok{paste0}\NormalTok{(}\StringTok{"E"}\NormalTok{, }\DecValTok{28}\SpecialCharTok{:}\DecValTok{31}\NormalTok{))}
\end{Highlighting}
\end{Shaded}

\hypertarget{aggregations-for-empirical-instances}{%
\subsubsection{Aggregations for Empirical Instances}\label{aggregations-for-empirical-instances}}

\begin{Shaded}
\begin{Highlighting}[]
\NormalTok{make\_row }\OtherTok{\textless{}{-}} \ControlFlowTok{function}\NormalTok{(idx, obs, obs\_agg) \{}
  \StringTok{\textasciigrave{}}\AttributeTok{rownames\textless{}{-}}\StringTok{\textasciigrave{}}\NormalTok{(}\FunctionTok{data.frame}\NormalTok{(}
    \AttributeTok{E =}\NormalTok{ idx,}
    
    \AttributeTok{obs =} \FunctionTok{paste0}\NormalTok{(obs, }\AttributeTok{collapse =} \StringTok{","}\NormalTok{),}
    \AttributeTok{num =} \FunctionTok{length}\NormalTok{(obs),}
    \AttributeTok{avg =} \FunctionTok{round}\NormalTok{(}\FunctionTok{mean}\NormalTok{(obs), }\DecValTok{2}\NormalTok{),}
    \AttributeTok{sum =} \FunctionTok{sum}\NormalTok{(obs),}
    
    \AttributeTok{agg =} \FunctionTok{paste0}\NormalTok{(obs\_agg, }\AttributeTok{collapse =} \StringTok{","}\NormalTok{),}
    \AttributeTok{numAgg =} \FunctionTok{length}\NormalTok{(obs\_agg),}
    \AttributeTok{meanAgg =} \FunctionTok{round}\NormalTok{(}\FunctionTok{mean}\NormalTok{(obs\_agg), }\DecValTok{2}\NormalTok{),}
    \AttributeTok{sumAgg =} \FunctionTok{sum}\NormalTok{(obs\_agg),}
    
    \AttributeTok{stringsAsFactors =} \ConstantTok{FALSE}
\NormalTok{  ), }\FunctionTok{paste0}\NormalTok{(}\StringTok{"E"}\NormalTok{, idx))}
\NormalTok{\}}

\NormalTok{split\_obs }\OtherTok{\textless{}{-}} \ControlFlowTok{function}\NormalTok{(obs) }\FunctionTok{as.integer}\NormalTok{(}\FunctionTok{strsplit}\NormalTok{(}\AttributeTok{x =}\NormalTok{ obs, }\AttributeTok{split =} \StringTok{","}\NormalTok{)[[}\DecValTok{1}\NormalTok{]])}
\end{Highlighting}
\end{Shaded}

\begin{Shaded}
\begin{Highlighting}[]
\NormalTok{qe\_ei }\OtherTok{\textless{}{-}}\NormalTok{ dplyr}\SpecialCharTok{::}\FunctionTok{bind\_rows}\NormalTok{(}\FunctionTok{list}\NormalTok{(}
  \CommentTok{\# [1,A,1], [3,B,2], [5,A,1], [7,B,1], [9,B,2], [13,A,1], [14,A,1], [15,A,2]}
  \FunctionTok{make\_row}\NormalTok{( }\DecValTok{1}\NormalTok{, }\FunctionTok{c}\NormalTok{(}\DecValTok{1}\NormalTok{,}\DecValTok{2}\NormalTok{,}\DecValTok{1}\NormalTok{,}\DecValTok{1}\NormalTok{,}\DecValTok{2}\NormalTok{,}\DecValTok{1}\NormalTok{,}\DecValTok{1}\NormalTok{,}\DecValTok{2}\NormalTok{), }\FunctionTok{c}\NormalTok{(}\DecValTok{1}\NormalTok{,}\DecValTok{2}\NormalTok{,}\DecValTok{1}\NormalTok{,}\DecValTok{1}\NormalTok{,}\DecValTok{2}\NormalTok{,}\DecValTok{1}\NormalTok{,}\DecValTok{1}\NormalTok{,}\DecValTok{2}\NormalTok{)),}
  \CommentTok{\# [1,A,0], [15,A,0]}
  \FunctionTok{make\_row}\NormalTok{( }\DecValTok{2}\NormalTok{, }\FunctionTok{c}\NormalTok{(}\DecValTok{0}\NormalTok{,}\DecValTok{0}\NormalTok{), }\FunctionTok{c}\NormalTok{(}\DecValTok{0}\NormalTok{,}\DecValTok{0}\NormalTok{)),}
  \CommentTok{\# [3,C,1], ([13,A,3], [13,B,2], [13,C,4]) {-}\textgreater{} [13,ABC,5], [15,C,1]}
  \FunctionTok{make\_row}\NormalTok{( }\DecValTok{3}\NormalTok{, }\FunctionTok{c}\NormalTok{(}\DecValTok{1}\NormalTok{,}\DecValTok{3}\NormalTok{,}\DecValTok{2}\NormalTok{,}\DecValTok{4}\NormalTok{,}\DecValTok{1}\NormalTok{), }\FunctionTok{c}\NormalTok{(}\DecValTok{1}\NormalTok{,}\DecValTok{5}\NormalTok{,}\DecValTok{1}\NormalTok{)),}
  \CommentTok{\# [3,C,1], [6,A,1], [14,A,3]}
  \FunctionTok{make\_row}\NormalTok{( }\DecValTok{4}\NormalTok{, }\FunctionTok{c}\NormalTok{(}\DecValTok{1}\NormalTok{,}\DecValTok{1}\NormalTok{,}\DecValTok{3}\NormalTok{), }\FunctionTok{c}\NormalTok{(}\DecValTok{1}\NormalTok{,}\DecValTok{1}\NormalTok{,}\DecValTok{3}\NormalTok{)),}
  \CommentTok{\# [13,A,3]}
  \FunctionTok{make\_row}\NormalTok{( }\DecValTok{5}\NormalTok{, }\FunctionTok{c}\NormalTok{(}\DecValTok{3}\NormalTok{), }\FunctionTok{c}\NormalTok{(}\DecValTok{3}\NormalTok{)),}
  \CommentTok{\# [3,C,2]}
  \FunctionTok{make\_row}\NormalTok{( }\DecValTok{6}\NormalTok{, }\FunctionTok{c}\NormalTok{(}\DecValTok{2}\NormalTok{), }\FunctionTok{c}\NormalTok{(}\DecValTok{2}\NormalTok{)),}
  \CommentTok{\# [12,A,1]}
  \FunctionTok{make\_row}\NormalTok{( }\DecValTok{7}\NormalTok{, }\FunctionTok{c}\NormalTok{(}\DecValTok{1}\NormalTok{), }\FunctionTok{c}\NormalTok{(}\DecValTok{1}\NormalTok{)),}
  \CommentTok{\# ([4,B,1], [4,C,2]) {-}\textgreater{} [4,BC,3], [6,C,2], [7,B,2], ([13,A,4], [13,C,2]) {-}\textgreater{} [13,AC,5]}
  \FunctionTok{make\_row}\NormalTok{( }\DecValTok{8}\NormalTok{, }\FunctionTok{c}\NormalTok{(}\DecValTok{1}\NormalTok{,}\DecValTok{2}\NormalTok{,}\DecValTok{2}\NormalTok{,}\DecValTok{2}\NormalTok{,}\DecValTok{4}\NormalTok{,}\DecValTok{2}\NormalTok{), }\FunctionTok{c}\NormalTok{(}\DecValTok{3}\NormalTok{,}\DecValTok{2}\NormalTok{,}\DecValTok{2}\NormalTok{,}\DecValTok{5}\NormalTok{)),}
  \CommentTok{\# [1,A,1], [3,C,4], ([5,A,1], [5,B,1]) {-}\textgreater{} [5,AB,2], [9,A,1], [10,B,2], [12,A,3],}
  \CommentTok{\# ([13,A,4], [13,B,4]) {-}\textgreater{} [13,AB,5]}
  \FunctionTok{make\_row}\NormalTok{( }\DecValTok{9}\NormalTok{, }\FunctionTok{c}\NormalTok{(}\DecValTok{1}\NormalTok{,}\DecValTok{4}\NormalTok{,}\DecValTok{1}\NormalTok{,}\DecValTok{1}\NormalTok{,}\DecValTok{1}\NormalTok{,}\DecValTok{2}\NormalTok{,}\DecValTok{3}\NormalTok{,}\DecValTok{4}\NormalTok{,}\DecValTok{4}\NormalTok{), }\FunctionTok{c}\NormalTok{(}\DecValTok{1}\NormalTok{,}\DecValTok{4}\NormalTok{,}\DecValTok{2}\NormalTok{,}\DecValTok{1}\NormalTok{,}\DecValTok{2}\NormalTok{,}\DecValTok{3}\NormalTok{,}\DecValTok{5}\NormalTok{)),}
  \CommentTok{\# [3,C,4], [7,C,3], [9,C,3], [10,C,1], [14,A,1], [15,B,1]}
  \FunctionTok{make\_row}\NormalTok{(}\DecValTok{10}\NormalTok{, }\FunctionTok{c}\NormalTok{(}\DecValTok{4}\NormalTok{,}\DecValTok{3}\NormalTok{,}\DecValTok{3}\NormalTok{,}\DecValTok{1}\NormalTok{,}\DecValTok{1}\NormalTok{,}\DecValTok{1}\NormalTok{), }\FunctionTok{c}\NormalTok{(}\DecValTok{4}\NormalTok{,}\DecValTok{3}\NormalTok{,}\DecValTok{3}\NormalTok{,}\DecValTok{1}\NormalTok{,}\DecValTok{1}\NormalTok{,}\DecValTok{1}\NormalTok{)),}
  \CommentTok{\# [3,C,4], [4,C,2], [7,C,1]}
  \FunctionTok{make\_row}\NormalTok{(}\DecValTok{11}\NormalTok{, }\FunctionTok{c}\NormalTok{(}\DecValTok{4}\NormalTok{,}\DecValTok{2}\NormalTok{,}\DecValTok{1}\NormalTok{), }\FunctionTok{c}\NormalTok{(}\DecValTok{4}\NormalTok{,}\DecValTok{2}\NormalTok{,}\DecValTok{1}\NormalTok{)),}
  \CommentTok{\# ([3,A,4], [3,B,4]) {-}\textgreater{} [3,AB,5], ([4,A,3], [4,B,4], [4,C,3]) {-}\textgreater{} [4,ABC,5], [5,C,1],}
  \CommentTok{\# ([6,A,3], [6,B,3]) {-}\textgreater{} [6,AB,4], ([7,A,3], [7,C,3]) {-}\textgreater{} [7,AC,4],}
  \CommentTok{\# ([10,A,3], [10,B,3], [10,C,2]) {-}\textgreater{} [10,ABC,5], [11,A,1], [12,A,1], [13,C,1], [14,A,1],}
  \CommentTok{\# ([15,A,1], [15,C,1]) {-}\textgreater{} [15,AC,2]}
  \FunctionTok{make\_row}\NormalTok{(}\DecValTok{12}\NormalTok{, }\FunctionTok{c}\NormalTok{(}\DecValTok{4}\NormalTok{,}\DecValTok{4}\NormalTok{,}\DecValTok{3}\NormalTok{,}\DecValTok{4}\NormalTok{,}\DecValTok{3}\NormalTok{,}\DecValTok{1}\NormalTok{,}\DecValTok{3}\NormalTok{,}\DecValTok{3}\NormalTok{,}\DecValTok{3}\NormalTok{,}\DecValTok{3}\NormalTok{,}\DecValTok{3}\NormalTok{,}\DecValTok{3}\NormalTok{,}\DecValTok{2}\NormalTok{,}\DecValTok{1}\NormalTok{,}\DecValTok{1}\NormalTok{,}\DecValTok{1}\NormalTok{,}\DecValTok{1}\NormalTok{,}\DecValTok{1}\NormalTok{,}\DecValTok{1}\NormalTok{), }\FunctionTok{c}\NormalTok{(}\DecValTok{5}\NormalTok{,}\DecValTok{5}\NormalTok{,}\DecValTok{1}\NormalTok{,}\DecValTok{4}\NormalTok{,}\DecValTok{4}\NormalTok{,}\DecValTok{5}\NormalTok{,}\DecValTok{1}\NormalTok{,}\DecValTok{1}\NormalTok{,}\DecValTok{1}\NormalTok{,}\DecValTok{1}\NormalTok{,}\DecValTok{2}\NormalTok{)),}
  \CommentTok{\# [3,C,3], ([10,A,3], [10,B,3]) {-}\textgreater{} [10,AB,4], ([13,A,3], [13,C,4]) {-}\textgreater{} [13,AC,5]}
  \FunctionTok{make\_row}\NormalTok{(}\DecValTok{13}\NormalTok{, }\FunctionTok{c}\NormalTok{(}\DecValTok{3}\NormalTok{,}\DecValTok{3}\NormalTok{,}\DecValTok{3}\NormalTok{,}\DecValTok{3}\NormalTok{,}\DecValTok{4}\NormalTok{), }\FunctionTok{c}\NormalTok{(}\DecValTok{3}\NormalTok{,}\DecValTok{4}\NormalTok{,}\DecValTok{5}\NormalTok{)),}
  \CommentTok{\# [3,A,2]}
  \FunctionTok{make\_row}\NormalTok{(}\DecValTok{14}\NormalTok{, }\FunctionTok{c}\NormalTok{(}\DecValTok{2}\NormalTok{), }\FunctionTok{c}\NormalTok{(}\DecValTok{2}\NormalTok{)),}
  \CommentTok{\# [3,A,3], [4,C,1]}
  \FunctionTok{make\_row}\NormalTok{(}\DecValTok{15}\NormalTok{, }\FunctionTok{c}\NormalTok{(}\DecValTok{3}\NormalTok{,}\DecValTok{1}\NormalTok{), }\FunctionTok{c}\NormalTok{(}\DecValTok{3}\NormalTok{,}\DecValTok{1}\NormalTok{)),}
  \CommentTok{\# [9,B,2]}
  \FunctionTok{make\_row}\NormalTok{(}\DecValTok{16}\NormalTok{, }\FunctionTok{c}\NormalTok{(}\DecValTok{2}\NormalTok{), }\FunctionTok{c}\NormalTok{(}\DecValTok{2}\NormalTok{)),}
  \CommentTok{\# ([3,C,3], [3,A,3]) {-}\textgreater{} [3,AC,4], [4,C,3], [6,A,2], [12,A,1]}
  \FunctionTok{make\_row}\NormalTok{(}\DecValTok{17}\NormalTok{, }\FunctionTok{c}\NormalTok{(}\DecValTok{3}\NormalTok{,}\DecValTok{3}\NormalTok{,}\DecValTok{3}\NormalTok{,}\DecValTok{2}\NormalTok{,}\DecValTok{1}\NormalTok{), }\FunctionTok{c}\NormalTok{(}\DecValTok{4}\NormalTok{,}\DecValTok{3}\NormalTok{,}\DecValTok{2}\NormalTok{,}\DecValTok{1}\NormalTok{)),}
  \CommentTok{\# [3,C,2]}
  \FunctionTok{make\_row}\NormalTok{(}\DecValTok{18}\NormalTok{, }\FunctionTok{c}\NormalTok{(}\DecValTok{2}\NormalTok{), }\FunctionTok{c}\NormalTok{(}\DecValTok{2}\NormalTok{)),}
  \CommentTok{\# [3,C,1], [4,A,2]}
  \FunctionTok{make\_row}\NormalTok{(}\DecValTok{19}\NormalTok{, }\FunctionTok{c}\NormalTok{(}\DecValTok{1}\NormalTok{,}\DecValTok{2}\NormalTok{), }\FunctionTok{c}\NormalTok{(}\DecValTok{1}\NormalTok{,}\DecValTok{2}\NormalTok{)),}
  \CommentTok{\# ([6,A,2], [6,B,2]) {-}\textgreater{} [6,AB,3], [12,A,1], [15,A,1]}
  \FunctionTok{make\_row}\NormalTok{(}\DecValTok{20}\NormalTok{, }\FunctionTok{c}\NormalTok{(}\DecValTok{2}\NormalTok{,}\DecValTok{2}\NormalTok{,}\DecValTok{1}\NormalTok{,}\DecValTok{1}\NormalTok{), }\FunctionTok{c}\NormalTok{(}\DecValTok{3}\NormalTok{,}\DecValTok{1}\NormalTok{,}\DecValTok{1}\NormalTok{)),}
  \CommentTok{\# ([6,B,2], [6,C,2]) {-}\textgreater{} [6,BC,3], ([7,A,1], [7,B,1]) {-}\textgreater{} [7,AB,2], [9,A,1], [12,C,1],}
  \CommentTok{\# ([13,A,3], [13,B,2]) {-}\textgreater{} [13,AB,4]}
  \FunctionTok{make\_row}\NormalTok{(}\DecValTok{21}\NormalTok{, }\FunctionTok{c}\NormalTok{(}\DecValTok{2}\NormalTok{,}\DecValTok{2}\NormalTok{,}\DecValTok{1}\NormalTok{,}\DecValTok{1}\NormalTok{,}\DecValTok{1}\NormalTok{,}\DecValTok{1}\NormalTok{,}\DecValTok{3}\NormalTok{,}\DecValTok{2}\NormalTok{), }\FunctionTok{c}\NormalTok{(}\DecValTok{3}\NormalTok{,}\DecValTok{2}\NormalTok{,}\DecValTok{1}\NormalTok{,}\DecValTok{1}\NormalTok{,}\DecValTok{4}\NormalTok{)),}
  \CommentTok{\# [6,C,1], [9,A,1], [13,C,2]}
  \FunctionTok{make\_row}\NormalTok{(}\DecValTok{22}\NormalTok{, }\FunctionTok{c}\NormalTok{(}\DecValTok{1}\NormalTok{,}\DecValTok{1}\NormalTok{,}\DecValTok{2}\NormalTok{), }\FunctionTok{c}\NormalTok{(}\DecValTok{1}\NormalTok{,}\DecValTok{1}\NormalTok{,}\DecValTok{2}\NormalTok{)),}
  \CommentTok{\# [7,C,4], [13,A,4]}
  \FunctionTok{make\_row}\NormalTok{(}\DecValTok{23}\NormalTok{, }\FunctionTok{c}\NormalTok{(}\DecValTok{4}\NormalTok{,}\DecValTok{4}\NormalTok{), }\FunctionTok{c}\NormalTok{(}\DecValTok{4}\NormalTok{,}\DecValTok{4}\NormalTok{)),}
  \CommentTok{\# [9,C,1], [11,A,1], [13,A,1]}
  \FunctionTok{make\_row}\NormalTok{(}\DecValTok{24}\NormalTok{, }\FunctionTok{c}\NormalTok{(}\DecValTok{1}\NormalTok{,}\DecValTok{1}\NormalTok{,}\DecValTok{1}\NormalTok{), }\FunctionTok{c}\NormalTok{(}\DecValTok{1}\NormalTok{,}\DecValTok{1}\NormalTok{,}\DecValTok{1}\NormalTok{)),}
  \CommentTok{\# [10,C,1]}
  \FunctionTok{make\_row}\NormalTok{(}\DecValTok{25}\NormalTok{, }\FunctionTok{c}\NormalTok{(}\DecValTok{1}\NormalTok{), }\FunctionTok{c}\NormalTok{(}\DecValTok{1}\NormalTok{)),}
  \CommentTok{\# ([12,A,3], [12,B,1]) {-}\textgreater{} [12,AB,3], ([15,A,1], [15,C,1]) {-}\textgreater{} [15,AC,2]}
  \FunctionTok{make\_row}\NormalTok{(}\DecValTok{26}\NormalTok{, }\FunctionTok{c}\NormalTok{(}\DecValTok{3}\NormalTok{,}\DecValTok{1}\NormalTok{,}\DecValTok{1}\NormalTok{,}\DecValTok{1}\NormalTok{), }\FunctionTok{c}\NormalTok{(}\DecValTok{3}\NormalTok{,}\DecValTok{2}\NormalTok{)),}
  \CommentTok{\# [13,B,4], [15,A,2]}
  \FunctionTok{make\_row}\NormalTok{(}\DecValTok{27}\NormalTok{, }\FunctionTok{c}\NormalTok{(}\DecValTok{4}\NormalTok{,}\DecValTok{2}\NormalTok{), }\FunctionTok{c}\NormalTok{(}\DecValTok{4}\NormalTok{,}\DecValTok{2}\NormalTok{)),}
  \CommentTok{\# [3,C,1]}
  \FunctionTok{make\_row}\NormalTok{(}\DecValTok{28}\NormalTok{, }\FunctionTok{c}\NormalTok{(}\DecValTok{1}\NormalTok{), }\FunctionTok{c}\NormalTok{(}\DecValTok{1}\NormalTok{)),}
  \CommentTok{\# [4,C,1]}
  \FunctionTok{make\_row}\NormalTok{(}\DecValTok{29}\NormalTok{, }\FunctionTok{c}\NormalTok{(}\DecValTok{1}\NormalTok{), }\FunctionTok{c}\NormalTok{(}\DecValTok{1}\NormalTok{)),}
  \CommentTok{\# [3,C,1], [7,C,4], [10,A,1], [15,A,1]}
  \FunctionTok{make\_row}\NormalTok{(}\DecValTok{30}\NormalTok{, }\FunctionTok{c}\NormalTok{(}\DecValTok{1}\NormalTok{,}\DecValTok{4}\NormalTok{,}\DecValTok{1}\NormalTok{,}\DecValTok{1}\NormalTok{), }\FunctionTok{c}\NormalTok{(}\DecValTok{1}\NormalTok{,}\DecValTok{4}\NormalTok{,}\DecValTok{1}\NormalTok{,}\DecValTok{1}\NormalTok{)),}
  \CommentTok{\# [4,B,1]}
  \FunctionTok{make\_row}\NormalTok{(}\DecValTok{31}\NormalTok{, }\FunctionTok{c}\NormalTok{(}\DecValTok{1}\NormalTok{), }\FunctionTok{c}\NormalTok{(}\DecValTok{1}\NormalTok{))}
\NormalTok{))}
\end{Highlighting}
\end{Shaded}

Table \ref{tab:empirical-fd-instances} shows, for each empirical instance of a symptom/consequence, how often it has been observed and how severe it was.
The columns ending in \texttt{\_agg} represent the count, sum (total severity), and average severity of the aggregated instances.
Aggregation happens if two or more raters observed the same instance.

\begin{table}

\caption{\label{tab:empirical-fd-instances}Empirical instances of Fire Drill symptoms and consequences as found in the 15 projects.}
\centering
\begin{tabular}[t]{lrlrrrlrrr}
\toprule
  & E & obs & num & avg & sum & agg & numAgg & meanAgg & sumAgg\\
\midrule
E1 & 1 & 1,2,1,1,2,1,1, ... & 8 & 1.38 & 11 & 1,2,1,1,2,1,1, ... & 8 & 1.38 & 11\\
E2 & 2 & 0,0 & 2 & 0.00 & 0 & 0,0 & 2 & 0.00 & 0\\
E3 & 3 & 1,3,2,4,1 & 5 & 2.20 & 11 & 1,5,1 & 3 & 2.33 & 7\\
E4 & 4 & 1,1,3 & 3 & 1.67 & 5 & 1,1,3 & 3 & 1.67 & 5\\
E5 & 5 & 3 & 1 & 3.00 & 3 & 3 & 1 & 3.00 & 3\\
\addlinespace
E6 & 6 & 2 & 1 & 2.00 & 2 & 2 & 1 & 2.00 & 2\\
E7 & 7 & 1 & 1 & 1.00 & 1 & 1 & 1 & 1.00 & 1\\
E8 & 8 & 1,2,2,2,4,2 & 6 & 2.17 & 13 & 3,2,2,5 & 4 & 3.00 & 12\\
E9 & 9 & 1,4,1,1,1,2,3, ... & 9 & 2.33 & 21 & 1,4,2,1,2,3,5 & 7 & 2.57 & 18\\
E10 & 10 & 4,3,3,1,1,1 & 6 & 2.17 & 13 & 4,3,3,1,1,1 & 6 & 2.17 & 13\\
\addlinespace
E11 & 11 & 4,2,1 & 3 & 2.33 & 7 & 4,2,1 & 3 & 2.33 & 7\\
E12 & 12 & 4,4,3,4,3,1,3, ... & 19 & 2.37 & 45 & 5,5,1,4,4,5,1, ... & 11 & 2.73 & 30\\
E13 & 13 & 3,3,3,3,4 & 5 & 3.20 & 16 & 3,4,5 & 3 & 4.00 & 12\\
E14 & 14 & 2 & 1 & 2.00 & 2 & 2 & 1 & 2.00 & 2\\
E15 & 15 & 3,1 & 2 & 2.00 & 4 & 3,1 & 2 & 2.00 & 4\\
\addlinespace
E16 & 16 & 2 & 1 & 2.00 & 2 & 2 & 1 & 2.00 & 2\\
E17 & 17 & 3,3,3,2,1 & 5 & 2.40 & 12 & 4,3,2,1 & 4 & 2.50 & 10\\
E18 & 18 & 2 & 1 & 2.00 & 2 & 2 & 1 & 2.00 & 2\\
E19 & 19 & 1,2 & 2 & 1.50 & 3 & 1,2 & 2 & 1.50 & 3\\
E20 & 20 & 2,2,1,1 & 4 & 1.50 & 6 & 3,1,1 & 3 & 1.67 & 5\\
\addlinespace
E21 & 21 & 2,2,1,1,1,1,3, ... & 8 & 1.62 & 13 & 3,2,1,1,4 & 5 & 2.20 & 11\\
E22 & 22 & 1,1,2 & 3 & 1.33 & 4 & 1,1,2 & 3 & 1.33 & 4\\
E23 & 23 & 4,4 & 2 & 4.00 & 8 & 4,4 & 2 & 4.00 & 8\\
E24 & 24 & 1,1,1 & 3 & 1.00 & 3 & 1,1,1 & 3 & 1.00 & 3\\
E25 & 25 & 1 & 1 & 1.00 & 1 & 1 & 1 & 1.00 & 1\\
\addlinespace
E26 & 26 & 3,1,1,1 & 4 & 1.50 & 6 & 3,2 & 2 & 2.50 & 5\\
E27 & 27 & 4,2 & 2 & 3.00 & 6 & 4,2 & 2 & 3.00 & 6\\
E28 & 28 & 1 & 1 & 1.00 & 1 & 1 & 1 & 1.00 & 1\\
E29 & 29 & 1 & 1 & 1.00 & 1 & 1 & 1 & 1.00 & 1\\
E30 & 30 & 1,4,1,1 & 4 & 1.75 & 7 & 1,4,1,1 & 4 & 1.75 & 7\\
\addlinespace
E31 & 31 & 1 & 1 & 1.00 & 1 & 1 & 1 & 1.00 & 1\\
\bottomrule
\end{tabular}
\end{table}

\hypertarget{connect-symptoms-consequences-and-empirical-instances}{%
\subsubsection{Connect Symptoms, Consequences, and Empirical Instances}\label{connect-symptoms-consequences-and-empirical-instances}}

We are now prepared to produce an aggregated table.

\begin{Shaded}
\begin{Highlighting}[]
\NormalTok{emp\_obs\_df }\OtherTok{\textless{}{-}} \ControlFlowTok{function}\NormalTok{(rns) \{}
\NormalTok{  df }\OtherTok{\textless{}{-}} \ConstantTok{NULL}
  \ControlFlowTok{for}\NormalTok{ (rn }\ControlFlowTok{in}\NormalTok{ rns) \{}
\NormalTok{    all\_obs\_agg }\OtherTok{\textless{}{-}} \FunctionTok{split\_obs}\NormalTok{(qe\_ei[rn,]}\SpecialCharTok{$}\NormalTok{agg)}
\NormalTok{    df }\OtherTok{\textless{}{-}} \FunctionTok{rbind}\NormalTok{(df, }\FunctionTok{data.frame}\NormalTok{(}
      \AttributeTok{E =}\NormalTok{ rn,}
      \AttributeTok{num =} \FunctionTok{length}\NormalTok{(all\_obs\_agg),}
      \AttributeTok{total =} \FunctionTok{sum}\NormalTok{(all\_obs\_agg),}
      \AttributeTok{mean =} \ControlFlowTok{if}\NormalTok{ (}\FunctionTok{length}\NormalTok{(all\_obs\_agg) }\SpecialCharTok{==} \DecValTok{0}\NormalTok{) NA\_real }\ControlFlowTok{else} \FunctionTok{mean}\NormalTok{(all\_obs\_agg)}
\NormalTok{    ))}
\NormalTok{  \}}
\NormalTok{  df}
\NormalTok{\}}

\NormalTok{aggregate\_e }\OtherTok{\textless{}{-}} \ControlFlowTok{function}\NormalTok{(rns) \{}
\NormalTok{  all\_obs }\OtherTok{\textless{}{-}} \FunctionTok{do.call}\NormalTok{(}\StringTok{"c"}\NormalTok{, }\FunctionTok{lapply}\NormalTok{(}\AttributeTok{X =}\NormalTok{ rns, }\ControlFlowTok{function}\NormalTok{(rn) \{}
    \FunctionTok{split\_obs}\NormalTok{(qe\_ei[rn,]}\SpecialCharTok{$}\NormalTok{obs)}
\NormalTok{  \}))}
\NormalTok{  all\_obs\_agg }\OtherTok{\textless{}{-}} \FunctionTok{do.call}\NormalTok{(}\StringTok{"c"}\NormalTok{, }\FunctionTok{lapply}\NormalTok{(}\AttributeTok{X =}\NormalTok{ rns, }\ControlFlowTok{function}\NormalTok{(rn) \{}
    \FunctionTok{split\_obs}\NormalTok{(qe\_ei[rn,]}\SpecialCharTok{$}\NormalTok{agg)}
\NormalTok{  \}))}
  
\NormalTok{  temp }\OtherTok{\textless{}{-}} \ControlFlowTok{if}\NormalTok{ (}\FunctionTok{length}\NormalTok{(rns) }\SpecialCharTok{==} \DecValTok{0}\NormalTok{) }\ConstantTok{NULL} \ControlFlowTok{else} \FunctionTok{emp\_obs\_df}\NormalTok{(}\AttributeTok{rns =}\NormalTok{ rns) }\SpecialCharTok{\%\textgreater{}\%} \FunctionTok{arrange}\NormalTok{(}\SpecialCharTok{{-}}\NormalTok{mean)}
  
  \FunctionTok{data.frame}\NormalTok{(}
    \AttributeTok{numObs =} \FunctionTok{length}\NormalTok{(all\_obs),}
    \AttributeTok{numAgg =} \FunctionTok{length}\NormalTok{(all\_obs\_agg),}
    \AttributeTok{min =} \ControlFlowTok{if}\NormalTok{ (}\FunctionTok{length}\NormalTok{(all\_obs\_agg) }\SpecialCharTok{==} \DecValTok{0}\NormalTok{) }\ConstantTok{NA\_real\_} \ControlFlowTok{else} \FunctionTok{min}\NormalTok{(all\_obs\_agg),}
    \AttributeTok{mean =} \ControlFlowTok{if}\NormalTok{ (}\FunctionTok{length}\NormalTok{(all\_obs\_agg) }\SpecialCharTok{==} \DecValTok{0}\NormalTok{) }\ConstantTok{NA\_real\_} \ControlFlowTok{else} \FunctionTok{round}\NormalTok{(}\FunctionTok{mean}\NormalTok{(all\_obs\_agg), }\DecValTok{2}\NormalTok{),}
    \CommentTok{\#median = if (length(all\_obs\_agg) == 0) NA\_real\_ else round(median(all\_obs\_agg), 2),}
    \AttributeTok{max =} \ControlFlowTok{if}\NormalTok{ (}\FunctionTok{length}\NormalTok{(all\_obs\_agg) }\SpecialCharTok{==} \DecValTok{0}\NormalTok{) }\ConstantTok{NA\_real\_} \ControlFlowTok{else} \FunctionTok{max}\NormalTok{(all\_obs\_agg),}
    \AttributeTok{total =} \FunctionTok{sum}\NormalTok{(all\_obs\_agg),}
    \AttributeTok{severity =} \ControlFlowTok{if}\NormalTok{ (}\FunctionTok{is.null}\NormalTok{(temp)) }\StringTok{""} \ControlFlowTok{else} \FunctionTok{paste0}\NormalTok{(}\FunctionTok{sapply}\NormalTok{(}\AttributeTok{X =} \FunctionTok{rownames}\NormalTok{(temp), }\AttributeTok{FUN =} \ControlFlowTok{function}\NormalTok{(rn) \{}
\NormalTok{      row }\OtherTok{\textless{}{-}}\NormalTok{ temp[rn,]}
      \FunctionTok{paste0}\NormalTok{(row}\SpecialCharTok{$}\NormalTok{E, }\StringTok{"("}\NormalTok{, row}\SpecialCharTok{$}\NormalTok{num, }\StringTok{"; "}\NormalTok{, }\FunctionTok{round}\NormalTok{(row}\SpecialCharTok{$}\NormalTok{mean, }\DecValTok{2}\NormalTok{), }\StringTok{")"}\NormalTok{)}
\NormalTok{    \}), }\AttributeTok{collapse =} \StringTok{" \textgreater{}= "}\NormalTok{)}
\NormalTok{  )}
\NormalTok{\}}

\NormalTok{sc\_agg }\OtherTok{\textless{}{-}}\NormalTok{ dplyr}\SpecialCharTok{::}\FunctionTok{bind\_rows}\NormalTok{(}\FunctionTok{lapply}\NormalTok{(}\AttributeTok{X =} \FunctionTok{names}\NormalTok{(qe\_sc), }\AttributeTok{FUN =} \ControlFlowTok{function}\NormalTok{(sc) \{}
  \FunctionTok{cbind}\NormalTok{(}\FunctionTok{data.frame}\NormalTok{(}
    \AttributeTok{SC =}\NormalTok{ sc,}
    \AttributeTok{numE =} \FunctionTok{length}\NormalTok{(qe\_sc[[sc]])}
\NormalTok{  ), }\FunctionTok{aggregate\_e}\NormalTok{(qe\_sc[[sc]]))}
\NormalTok{\}))}
\NormalTok{sc\_agg}\SpecialCharTok{$}\NormalTok{SC }\OtherTok{\textless{}{-}} \FunctionTok{factor}\NormalTok{(}\AttributeTok{x =}\NormalTok{ sc\_agg}\SpecialCharTok{$}\NormalTok{SC, }\AttributeTok{levels =}\NormalTok{ sc\_agg}\SpecialCharTok{$}\NormalTok{SC, }\AttributeTok{ordered =} \ConstantTok{TRUE}\NormalTok{)}
\end{Highlighting}
\end{Shaded}

Table \ref{tab:sc-ei-overview} shows, for each of the Fire Drill's symptoms/consequences, how many types of empirical instances were observed, how often they occurred, as well as the average severity.
The last column orders the associated empirical instances from highest to lowest total severity (only shown interactively).

\begin{table}

\caption{\label{tab:sc-ei-overview}Association between (empirical) symptoms and consequences, and empirical instances of concrete observations.}
\centering
\begin{tabular}[t]{lrrrrrrr}
\toprule
SC & numE & numObs & numAgg & min & mean & max & total\\
\midrule
SC1 & 3 & 15 & 13 & 0 & 1.38 & 5 & 18\\
SC2 & 1 & 3 & 3 & 1 & 1.67 & 3 & 5\\
SC3 & 1 & 1 & 1 & 3 & 3.00 & 3 & 3\\
SC4 & 2 & 2 & 2 & 1 & 1.50 & 2 & 3\\
SC5 & 1 & 6 & 4 & 2 & 3.00 & 5 & 12\\
\addlinespace
SC6 & 3 & 18 & 16 & 1 & 2.38 & 5 & 38\\
SC7 & 0 & 0 & 0 & NA & NA & NA & 0\\
ESC1 & 5 & 28 & 18 & 1 & 2.78 & 5 & 50\\
ESC2 & 11 & 35 & 28 & 1 & 2.07 & 4 & 58\\
ESC3 & 4 & 7 & 7 & 1 & 1.43 & 4 & 10\\
\bottomrule
\end{tabular}
\end{table}

\begin{figure}
\centering
\includegraphics{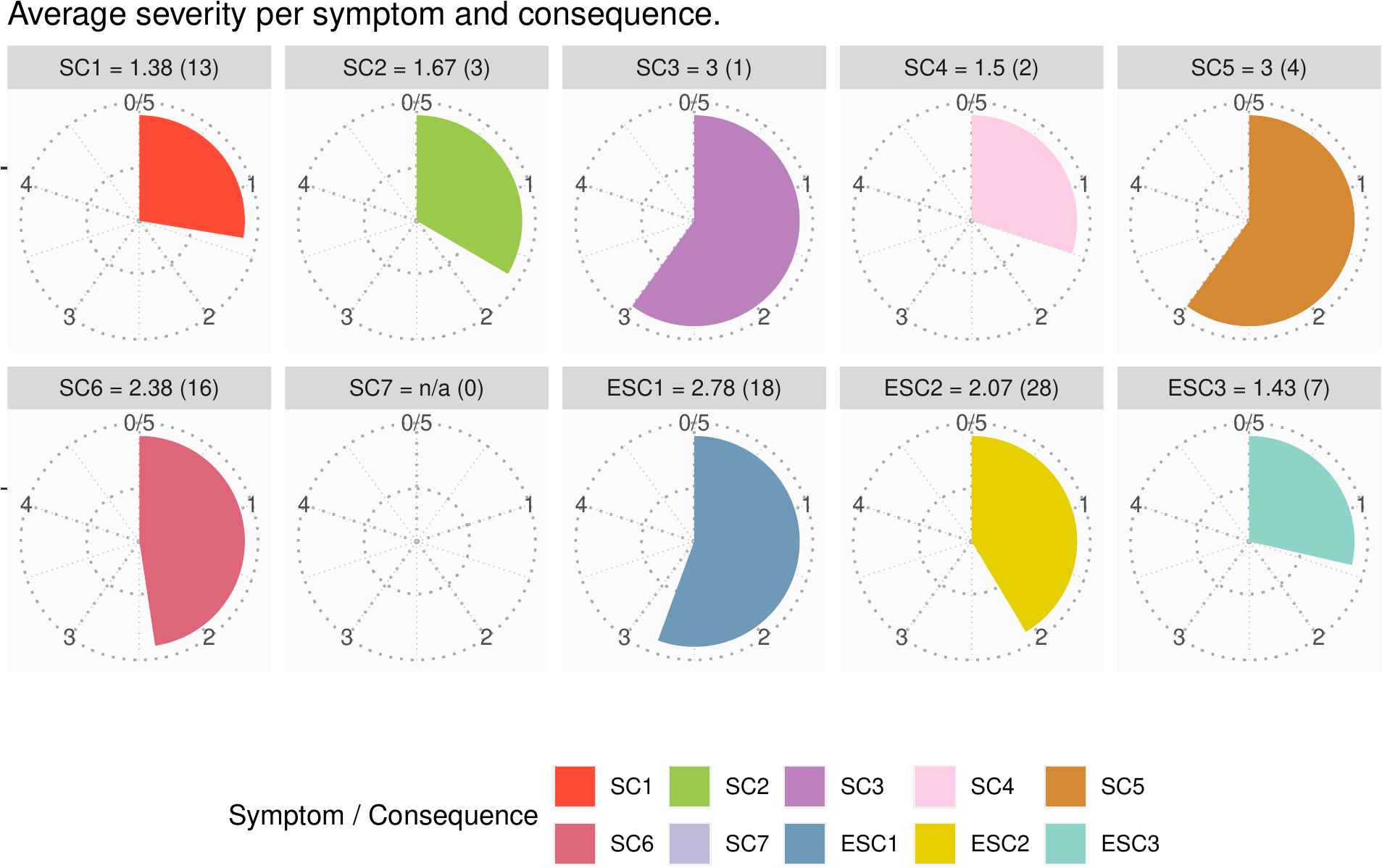}
\caption{\label{fig:fd-sc-mean}Average severity for each (empirical) Symptom / Consequence.}
\end{figure}

\begin{figure}
\centering
\includegraphics{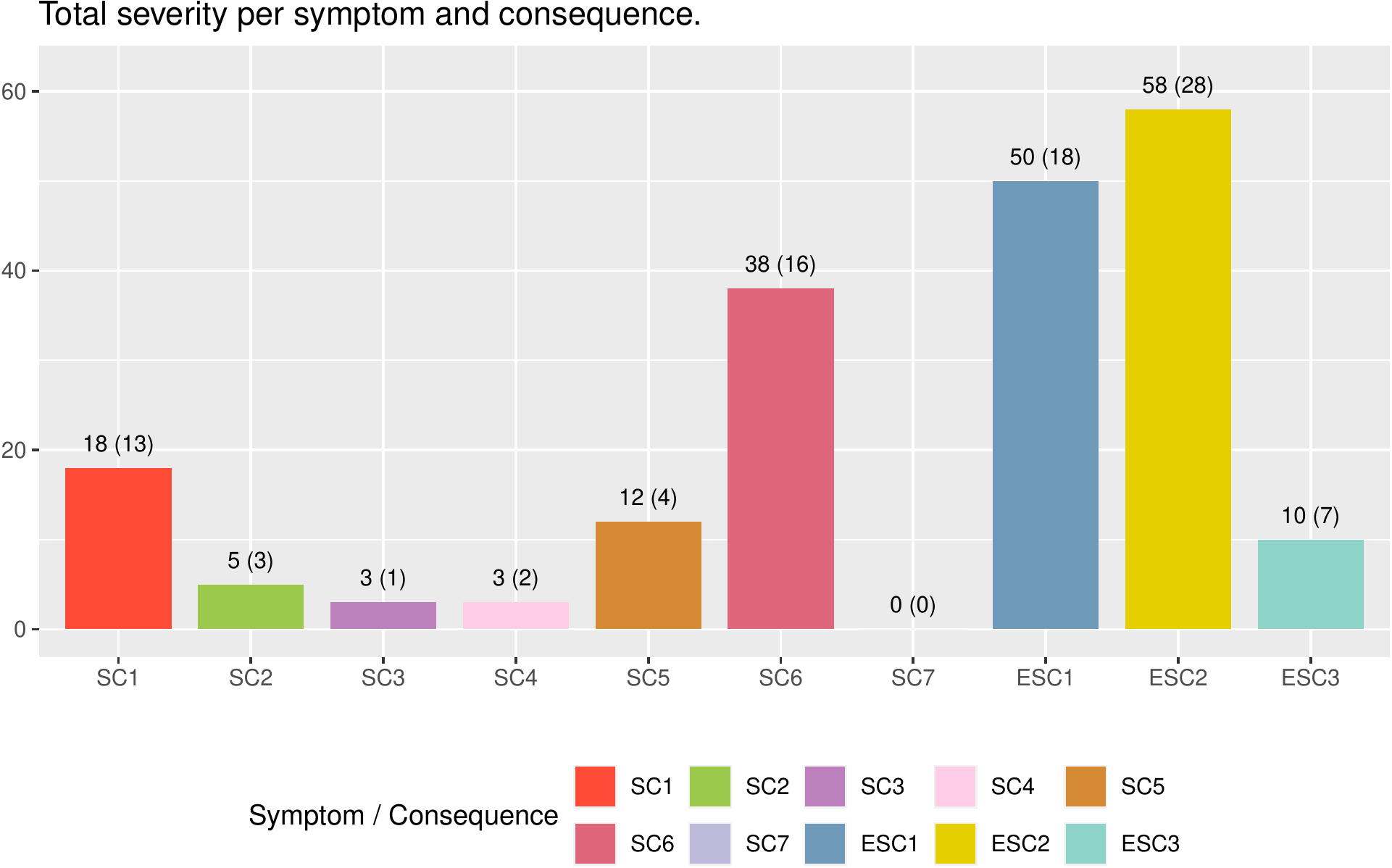}
\caption{\label{fig:fd-sc-total}Total severity and number of (aggregated) observations for each (empirical) Symptom / Consequence.}
\end{figure}

Figures \ref{fig:fd-sc-mean} and \ref{fig:fd-sc-total} show the average (out of maximal \(5\)) and total severity per symptom / consequence.

\clearpage

\hypertarget{technical-report-building-robust-regression-models-for-scoring-the-presence-of-the-fire-drill-anti-pattern}{%
\section{\texorpdfstring{Technical Report: Building Robust Regression Models for Scoring the Presence of the Fire Drill Anti-pattern\label{tr:fire-drill-rob-reg-technical-report}}{Technical Report: Building Robust Regression Models for Scoring the Presence of the Fire Drill Anti-pattern}}\label{technical-report-building-robust-regression-models-for-scoring-the-presence-of-the-fire-drill-anti-pattern}}

This is the self-contained technical report for the no-pattern detection of Fire Drill anti-pattern.

\hypertarget{introduction-2}{%
\subsection{\texorpdfstring{Introduction\label{tr:fire-drill-rob-reg-technical-report}}{Introduction}}\label{introduction-2}}

This technical report was added in the eighth iteration of the compilation.
So far, we have developed an unsupervised approach for properly scoring the presence of the anti-pattern.
However, here we will attempt to find a robust regression model in a supervised way. In other words, the regression model to be found will be the surrogate scoring mechanism.
The challenge lies in the scarcity of our data, as it has many more features than data points. It is therefore essential to obtain a regression model that, even when trained on only few instances, delivers a low generalization error.

All complementary data and results can be found at Zenodo (Hönel, Pícha, et al. 2023). This notebook was written in a way that it can be run without any additional efforts to reproduce the outputs (using the pre-computed results). This notebook has a canonical URL\textsuperscript{\href{https://github.com/MrShoenel/anti-pattern-models/blob/master/notebooks/fire-drill-issue-tracking-technical-report.Rmd}{{[}Link{]}}} and can be read online as a rendered markdown\textsuperscript{\href{https://github.com/MrShoenel/anti-pattern-models/blob/master/notebooks/fire-drill-issue-tracking-technical-report.md}{{[}Link{]}}} version. All code can be found in this repository, too.

\hypertarget{overview-of-the-approach}{%
\subsection{Overview of the Approach}\label{overview-of-the-approach}}

We will be importing the two initial process models/patterns for the Fire Drill anti-pattern in source code and issue-tracking data.
These patterns have four and three activities/metrics, respectively.
Both, patterns and projects, are formulated \emph{continuously}, that is, as curves over time. Also, both have been normalized along both, \(x\)- and \(y\)-axes (having a domain/range of \([0,1]\)).

As for the features, we will try two approaches.
First we will sample ten equidistantly-spaced points and compute the distance between these points of the process model and a project.
The second approach will subdivide the time axis into ten equally long segments, of which we will calculate the area between curves and the correlation.

Then, we will use a standard machine learning approach with repeated cross-validation in order to find out whether the generalization error on some holdout data will converge with increasing size of the training set.
We will attempt standard pre-processing, such as removing (near-)zero variance features, highly correlated features, or dimensionality reduction of the feature space.

\hypertarget{slight-adaption-of-the-source-code-pattern}{%
\subsection{Slight Adaption of the Source Code Pattern}\label{slight-adaption-of-the-source-code-pattern}}

The pattern used for source code is slightly erroneous with regard to how the source code density is modeled. It does not currently properly account for how the Fire Drill is described.
Also, we want to change the pattern to use smooth CDFs for the maintenance activities, as the plan is to only measure correlation and area between curves. Using CDFs has a few advantages, such as no normalization and better suitability for the chosen metrics.
We will also change the \texttt{FREQ} activity and represent it as a mixture of \texttt{A} and \texttt{CP}, such that all three densities become true densities (integrating to \(\approx1\)).

\hypertarget{modeling-a-cp-and-freq}{%
\subsubsection{\texorpdfstring{Modeling \texttt{A}, \texttt{CP}, and \texttt{FREQ}}{Modeling A, CP, and FREQ}}\label{modeling-a-cp-and-freq}}

The process for \texttt{A}, \texttt{CP}, and \texttt{FREQ} is the following (both, for the pattern and projects):

\begin{itemize}
\tightlist
\item
  Rejection Sampling:

  \begin{itemize}
  \tightlist
  \item
    If pattern, sample from under the modeled curve.
  \item
    If project, sample from the KDE of the activity's curve.
  \end{itemize}
\item
  Estimate new KDE from these samples for \texttt{A} and \texttt{CP}.
\item
  Compose a mixture density for \texttt{FREQ} as \(f(x)=r_A f_A(x) + r_{CP} f_{CP}(x)\), where \(r_a,r_{CP}\) correspond to the ratios of these activities.

  \begin{itemize}
  \tightlist
  \item
    We will also compose an ordinary density called \texttt{FREQ\_nm} (non-mixture) that just captures all commit when they happened. This is going to be used later in the no-pattern approach.
  \end{itemize}
\item
  Perform rejection sampling from the three densities, using large sample sizes, so that we can obtain a smooth ECDF.
\end{itemize}

This process will leave us with three separate smooth ECDFs.

\hypertarget{modeling-scd}{%
\subsubsection{\texorpdfstring{Modeling \texttt{SCD}}{Modeling SCD}}\label{modeling-scd}}

The source code density can move quite unpredictably, such that it does not make a lot of sense obtaining curves and measuring their correlation.
Instead, we will only look at the difference between the pattern's modeled density and that of each project.
Therefore, we will simply interpolate linearly between the projects' commits.

As for the pattern, the curve will be smooth and needs to be redesigned.
Note that the Fire Drill description says the following: ``only analytical or documentational artefacts for a long time''. The current pattern does not account for that.
Instead, the current pattern pretty much reduces all activities and the \texttt{SCD} during what we called the ``Long Stretch''.
While we mostly stick with how the activities were modeled, the \texttt{SCD}, however, requires a more accommodating redesign.
It should pick up pace almost immediately, followed by a slight increase until the Fire Drill starts.
Then, we should see a rapid increase that peaks at \(1\) and a normalization after the Fire Drill ends (go back to previous level).
We also modify the other activities and add a slight slope to them, but these are rather cosmetic changes.

\hypertarget{importing-the-raw-svg-points}{%
\subsubsection{Importing the Raw SVG Points}\label{importing-the-raw-svg-points}}

Please note that after sketching the SVG, we rasterize it to points using a tool called ``PathToPoints''\footnote{\url{https://github.com/MrShoenel/PathToPoints}}.
In the CSV, we will find four different data series (attention: of different length), one corresponding to each activity/metric.
Since \texttt{FREQ} will be a mixture, we are not modelling it explicitly.

Paths are associated with colors: Coral is \texttt{SCD}, Gold is \texttt{A}, and Green is \texttt{CP}.

\begin{Shaded}
\begin{Highlighting}[]
\NormalTok{temp }\OtherTok{\textless{}{-}} \FunctionTok{read.csv}\NormalTok{(}\AttributeTok{file =} \StringTok{"../data/Fire{-}Drill\_second{-}guess.csv"}\NormalTok{, }\AttributeTok{header =} \ConstantTok{TRUE}\NormalTok{, }\AttributeTok{sep =} \StringTok{";"}\NormalTok{)}
\NormalTok{assocs }\OtherTok{\textless{}{-}} \FunctionTok{c}\NormalTok{(}\AttributeTok{coral =} \StringTok{"SCD"}\NormalTok{, }\AttributeTok{gold =} \StringTok{"A"}\NormalTok{, }\AttributeTok{green =} \StringTok{"CP"}\NormalTok{)}
\NormalTok{assoc\_inv }\OtherTok{\textless{}{-}} \FunctionTok{names}\NormalTok{(assocs)}
\FunctionTok{names}\NormalTok{(assoc\_inv) }\OtherTok{\textless{}{-}} \FunctionTok{unname}\NormalTok{(assocs)}

\CommentTok{\# Y\textquotesingle{}s are negative.}
\ControlFlowTok{for}\NormalTok{ (assoc }\ControlFlowTok{in} \FunctionTok{names}\NormalTok{(assocs)) \{}
\NormalTok{  temp[, }\FunctionTok{paste0}\NormalTok{(assoc, }\StringTok{"\_y"}\NormalTok{)] }\OtherTok{\textless{}{-}} \SpecialCharTok{{-}}\DecValTok{1} \SpecialCharTok{*}\NormalTok{ temp[, }\FunctionTok{paste0}\NormalTok{(assoc, }\StringTok{"\_y"}\NormalTok{)]}
\NormalTok{\}}

\CommentTok{\# Next, we gotta normalize them all together, so we have to determine the}
\CommentTok{\# y{-}extent.}
\NormalTok{y\_ext }\OtherTok{\textless{}{-}} \FunctionTok{range}\NormalTok{(}\FunctionTok{c}\NormalTok{(temp}\SpecialCharTok{$}\NormalTok{coral\_y, temp}\SpecialCharTok{$}\NormalTok{gold\_y, temp}\SpecialCharTok{$}\NormalTok{green\_y), }\AttributeTok{na.rm =} \ConstantTok{TRUE}\NormalTok{)}
\ControlFlowTok{for}\NormalTok{ (assoc }\ControlFlowTok{in} \FunctionTok{names}\NormalTok{(assocs)) \{}
\NormalTok{  temp[, }\FunctionTok{paste0}\NormalTok{(assoc, }\StringTok{"\_y"}\NormalTok{)] }\OtherTok{\textless{}{-}}\NormalTok{ (temp[, }\FunctionTok{paste0}\NormalTok{(assoc, }\StringTok{"\_y"}\NormalTok{)] }\SpecialCharTok{{-}}\NormalTok{ y\_ext[}\DecValTok{1}\NormalTok{])}\SpecialCharTok{/}\NormalTok{(y\_ext[}\DecValTok{2}\NormalTok{] }\SpecialCharTok{{-}}
\NormalTok{    y\_ext[}\DecValTok{1}\NormalTok{])}
\NormalTok{\}}
\end{Highlighting}
\end{Shaded}

\begin{Shaded}
\begin{Highlighting}[]
\NormalTok{templ }\OtherTok{\textless{}{-}} \FunctionTok{list}\NormalTok{()}

\ControlFlowTok{for}\NormalTok{ (assoc }\ControlFlowTok{in} \FunctionTok{names}\NormalTok{(assocs)) \{}
\NormalTok{  data }\OtherTok{\textless{}{-}}\NormalTok{ temp[, }\FunctionTok{paste0}\NormalTok{(assoc, }\FunctionTok{c}\NormalTok{(}\StringTok{"\_x"}\NormalTok{, }\StringTok{"\_y"}\NormalTok{))]}
\NormalTok{  data }\OtherTok{\textless{}{-}}\NormalTok{ data[stats}\SpecialCharTok{::}\FunctionTok{complete.cases}\NormalTok{(data), ]}
\NormalTok{  tempf }\OtherTok{\textless{}{-}}\NormalTok{ stats}\SpecialCharTok{::}\FunctionTok{approxfun}\NormalTok{(}\AttributeTok{x =}\NormalTok{ data[, }\FunctionTok{paste0}\NormalTok{(assoc, }\StringTok{"\_x"}\NormalTok{)], }\AttributeTok{y =}\NormalTok{ data[, }\FunctionTok{paste0}\NormalTok{(assoc,}
    \StringTok{"\_y"}\NormalTok{)], }\AttributeTok{yleft =} \DecValTok{0}\NormalTok{, }\AttributeTok{yright =} \DecValTok{0}\NormalTok{)}
  \CommentTok{\# Let\textquotesingle{}s sample from that function linearly}
\NormalTok{  use\_y }\OtherTok{\textless{}{-}} \FunctionTok{tempf}\NormalTok{(}\FunctionTok{seq}\NormalTok{(}\AttributeTok{from =} \FunctionTok{min}\NormalTok{(data[, }\FunctionTok{paste0}\NormalTok{(assoc, }\StringTok{"\_x"}\NormalTok{)]), }\AttributeTok{to =} \FunctionTok{max}\NormalTok{(data[, }\FunctionTok{paste0}\NormalTok{(assoc,}
    \StringTok{"\_x"}\NormalTok{)]), }\AttributeTok{length.out =} \DecValTok{1000}\NormalTok{))}
\NormalTok{  templ[[assocs[assoc]]] }\OtherTok{\textless{}{-}}\NormalTok{ stats}\SpecialCharTok{::}\FunctionTok{approxfun}\NormalTok{(}\AttributeTok{x =} \FunctionTok{seq}\NormalTok{(}\AttributeTok{from =} \DecValTok{0}\NormalTok{, }\AttributeTok{to =} \DecValTok{1}\NormalTok{, }\AttributeTok{length.out =} \DecValTok{1000}\NormalTok{),}
    \AttributeTok{y =}\NormalTok{ use\_y, }\AttributeTok{yleft =}\NormalTok{ use\_y[}\DecValTok{1}\NormalTok{], }\AttributeTok{yright =}\NormalTok{ use\_y[}\DecValTok{2}\NormalTok{])}
\NormalTok{\}}
\end{Highlighting}
\end{Shaded}

Now we have a list with with \texttt{A}, \texttt{CP}, and \texttt{SCD}.
Next, we will do the rejection sampling from the former two in order to create \texttt{FREQ}.
As for the mixture ratio, we will assume that the \textbf{adaptive activities make up for \(40\)\%} of all activities.
So the mixture will be \(0.4/0.6\).

\begin{Shaded}
\begin{Highlighting}[]
\NormalTok{use\_x }\OtherTok{\textless{}{-}} \FunctionTok{seq}\NormalTok{(}\AttributeTok{from =} \DecValTok{0}\NormalTok{, }\AttributeTok{to =} \DecValTok{1}\NormalTok{, }\AttributeTok{length.out =} \FloatTok{1e+07}\NormalTok{)}


\NormalTok{use\_y }\OtherTok{\textless{}{-}}\NormalTok{ stats}\SpecialCharTok{::}\FunctionTok{runif}\NormalTok{(}\AttributeTok{n =} \FunctionTok{length}\NormalTok{(use\_x), }\AttributeTok{min =} \DecValTok{0}\NormalTok{, }\AttributeTok{max =} \FunctionTok{max}\NormalTok{(temp[[}\FunctionTok{paste0}\NormalTok{(assoc\_inv[}\StringTok{"A"}\NormalTok{],}
  \StringTok{"\_y"}\NormalTok{)]], }\AttributeTok{na.rm =} \ConstantTok{TRUE}\NormalTok{))}
\NormalTok{tempdens\_A }\OtherTok{\textless{}{-}}\NormalTok{ stats}\SpecialCharTok{::}\FunctionTok{density}\NormalTok{(}\AttributeTok{x =}\NormalTok{ use\_x[use\_y }\SpecialCharTok{\textless{}=}\NormalTok{ templ}\SpecialCharTok{$}\FunctionTok{A}\NormalTok{(use\_x)], }\AttributeTok{bw =} \StringTok{"SJ"}\NormalTok{, }\AttributeTok{cut =} \ConstantTok{TRUE}\NormalTok{)}
\NormalTok{tempf\_A }\OtherTok{\textless{}{-}}\NormalTok{ stats}\SpecialCharTok{::}\FunctionTok{approxfun}\NormalTok{(}\AttributeTok{x =}\NormalTok{ tempdens\_A}\SpecialCharTok{$}\NormalTok{x, }\AttributeTok{y =}\NormalTok{ tempdens\_A}\SpecialCharTok{$}\NormalTok{y, }\AttributeTok{yleft =} \DecValTok{0}\NormalTok{, }\AttributeTok{yright =} \DecValTok{0}\NormalTok{)}
\NormalTok{tempecdf\_A }\OtherTok{\textless{}{-}} \FunctionTok{make\_smooth\_ecdf}\NormalTok{(}\AttributeTok{values =}\NormalTok{ use\_x[use\_y }\SpecialCharTok{\textless{}=}\NormalTok{ templ}\SpecialCharTok{$}\FunctionTok{A}\NormalTok{(use\_x)], }\AttributeTok{slope =} \DecValTok{0}\NormalTok{,}
  \AttributeTok{inverse =} \ConstantTok{FALSE}\NormalTok{)}

\NormalTok{use\_y }\OtherTok{\textless{}{-}}\NormalTok{ stats}\SpecialCharTok{::}\FunctionTok{runif}\NormalTok{(}\AttributeTok{n =} \FunctionTok{length}\NormalTok{(use\_x), }\AttributeTok{min =} \DecValTok{0}\NormalTok{, }\AttributeTok{max =} \FunctionTok{max}\NormalTok{(temp[[}\FunctionTok{paste0}\NormalTok{(assoc\_inv[}\StringTok{"CP"}\NormalTok{],}
  \StringTok{"\_y"}\NormalTok{)]], }\AttributeTok{na.rm =} \ConstantTok{TRUE}\NormalTok{))}
\NormalTok{tempdens\_CP }\OtherTok{\textless{}{-}}\NormalTok{ stats}\SpecialCharTok{::}\FunctionTok{density}\NormalTok{(}\AttributeTok{x =}\NormalTok{ use\_x[use\_y }\SpecialCharTok{\textless{}=}\NormalTok{ templ}\SpecialCharTok{$}\FunctionTok{CP}\NormalTok{(use\_x)], }\AttributeTok{bw =} \StringTok{"SJ"}\NormalTok{, }\AttributeTok{cut =} \ConstantTok{TRUE}\NormalTok{)}
\NormalTok{tempf\_CP }\OtherTok{\textless{}{-}}\NormalTok{ stats}\SpecialCharTok{::}\FunctionTok{approxfun}\NormalTok{(}\AttributeTok{x =}\NormalTok{ tempdens\_CP}\SpecialCharTok{$}\NormalTok{x, }\AttributeTok{y =}\NormalTok{ tempdens\_CP}\SpecialCharTok{$}\NormalTok{y, }\AttributeTok{yleft =} \DecValTok{0}\NormalTok{, }\AttributeTok{yright =} \DecValTok{0}\NormalTok{)}
\NormalTok{tempecdf\_CP }\OtherTok{\textless{}{-}} \FunctionTok{make\_smooth\_ecdf}\NormalTok{(}\AttributeTok{values =}\NormalTok{ use\_x[use\_y }\SpecialCharTok{\textless{}=}\NormalTok{ templ}\SpecialCharTok{$}\FunctionTok{CP}\NormalTok{(use\_x)], }\AttributeTok{slope =} \DecValTok{0}\NormalTok{,}
  \AttributeTok{inverse =} \ConstantTok{FALSE}\NormalTok{)}

\FunctionTok{c}\NormalTok{(cubature}\SpecialCharTok{::}\FunctionTok{cubintegrate}\NormalTok{(}\AttributeTok{f =}\NormalTok{ tempf\_A, }\DecValTok{0}\NormalTok{, }\DecValTok{1}\NormalTok{)}\SpecialCharTok{$}\NormalTok{integral, cubature}\SpecialCharTok{::}\FunctionTok{cubintegrate}\NormalTok{(}\AttributeTok{f =}\NormalTok{ tempf\_CP,}
  \DecValTok{0}\NormalTok{, }\DecValTok{1}\NormalTok{)}\SpecialCharTok{$}\NormalTok{integral)  }\CommentTok{\# Those should be very close to 1}
\end{Highlighting}
\end{Shaded}

\begin{verbatim}
## [1] 1.000987 1.000835
\end{verbatim}

\begin{Shaded}
\begin{Highlighting}[]
\NormalTok{pattern\_sc }\OtherTok{\textless{}{-}} \FunctionTok{list}\NormalTok{(}\AttributeTok{SCD =}\NormalTok{ templ}\SpecialCharTok{$}\NormalTok{SCD, }\AttributeTok{A =}\NormalTok{ tempf\_A, }\AttributeTok{CP =}\NormalTok{ tempf\_CP, }\AttributeTok{FREQ =} \ControlFlowTok{function}\NormalTok{(x) }\FloatTok{0.4} \SpecialCharTok{*}
  \FunctionTok{tempf\_A}\NormalTok{(x) }\SpecialCharTok{+} \FloatTok{0.6} \SpecialCharTok{*} \FunctionTok{tempf\_CP}\NormalTok{(x))}

\NormalTok{cubature}\SpecialCharTok{::}\FunctionTok{cubintegrate}\NormalTok{(}\AttributeTok{f =}\NormalTok{ pattern\_sc}\SpecialCharTok{$}\NormalTok{FREQ, }\DecValTok{0}\NormalTok{, }\DecValTok{1}\NormalTok{)}\SpecialCharTok{$}\NormalTok{integral}
\end{Highlighting}
\end{Shaded}

\begin{verbatim}
## [1] 1.000896
\end{verbatim}

\begin{Shaded}
\begin{Highlighting}[]
\CommentTok{\# We\textquotesingle{}ll need it in the next step.}
\NormalTok{use\_y }\OtherTok{\textless{}{-}}\NormalTok{ stats}\SpecialCharTok{::}\FunctionTok{runif}\NormalTok{(}\AttributeTok{n =} \FunctionTok{length}\NormalTok{(use\_x), }\AttributeTok{min =} \DecValTok{0}\NormalTok{, }\AttributeTok{max =} \SpecialCharTok{{-}}\FloatTok{1.01} \SpecialCharTok{*}\NormalTok{ nloptr}\SpecialCharTok{::}\FunctionTok{nloptr}\NormalTok{(}\AttributeTok{x0 =} \FloatTok{0.5}\NormalTok{,}
  \AttributeTok{lb =} \DecValTok{0}\NormalTok{, }\AttributeTok{ub =} \DecValTok{1}\NormalTok{, }\AttributeTok{eval\_f =} \ControlFlowTok{function}\NormalTok{(x) }\SpecialCharTok{{-}}\NormalTok{pattern\_sc}\SpecialCharTok{$}\FunctionTok{FREQ}\NormalTok{(x), }\AttributeTok{opts =} \FunctionTok{list}\NormalTok{(}\AttributeTok{algorithm =} \StringTok{"NLOPT\_GN\_DIRECT\_L\_RAND"}\NormalTok{,}
    \AttributeTok{xtol\_rel =} \FloatTok{0.001}\NormalTok{, }\AttributeTok{maxeval =} \FloatTok{1e+05}\NormalTok{))}\SpecialCharTok{$}\NormalTok{objective)}
\NormalTok{tempdens\_FREQ }\OtherTok{\textless{}{-}}\NormalTok{ stats}\SpecialCharTok{::}\FunctionTok{density}\NormalTok{(}\AttributeTok{x =}\NormalTok{ use\_x[use\_y }\SpecialCharTok{\textless{}=}\NormalTok{ pattern\_sc}\SpecialCharTok{$}\FunctionTok{FREQ}\NormalTok{(use\_x)], }\AttributeTok{bw =} \StringTok{"SJ"}\NormalTok{,}
  \AttributeTok{cut =} \ConstantTok{TRUE}\NormalTok{)}
\NormalTok{tempf\_FREQ }\OtherTok{\textless{}{-}}\NormalTok{ stats}\SpecialCharTok{::}\FunctionTok{approxfun}\NormalTok{(}\AttributeTok{x =}\NormalTok{ tempdens\_FREQ}\SpecialCharTok{$}\NormalTok{x, }\AttributeTok{y =}\NormalTok{ tempdens\_FREQ}\SpecialCharTok{$}\NormalTok{y, }\AttributeTok{yleft =} \DecValTok{0}\NormalTok{,}
  \AttributeTok{yright =} \DecValTok{0}\NormalTok{)}
\NormalTok{tempecdf\_FREQ }\OtherTok{\textless{}{-}} \FunctionTok{make\_smooth\_ecdf}\NormalTok{(}\AttributeTok{values =}\NormalTok{ use\_x[use\_y }\SpecialCharTok{\textless{}=}\NormalTok{ pattern\_sc}\SpecialCharTok{$}\FunctionTok{FREQ}\NormalTok{(use\_x)],}
  \AttributeTok{slope =} \DecValTok{0}\NormalTok{, }\AttributeTok{inverse =} \ConstantTok{FALSE}\NormalTok{)}

\NormalTok{cubature}\SpecialCharTok{::}\FunctionTok{cubintegrate}\NormalTok{(}\AttributeTok{f =}\NormalTok{ tempf\_FREQ, }\DecValTok{0}\NormalTok{, }\DecValTok{1}\NormalTok{)}\SpecialCharTok{$}\NormalTok{integral}
\end{Highlighting}
\end{Shaded}

\begin{verbatim}
## [1] 1.000908
\end{verbatim}

\begin{figure}[ht!]
\includegraphics{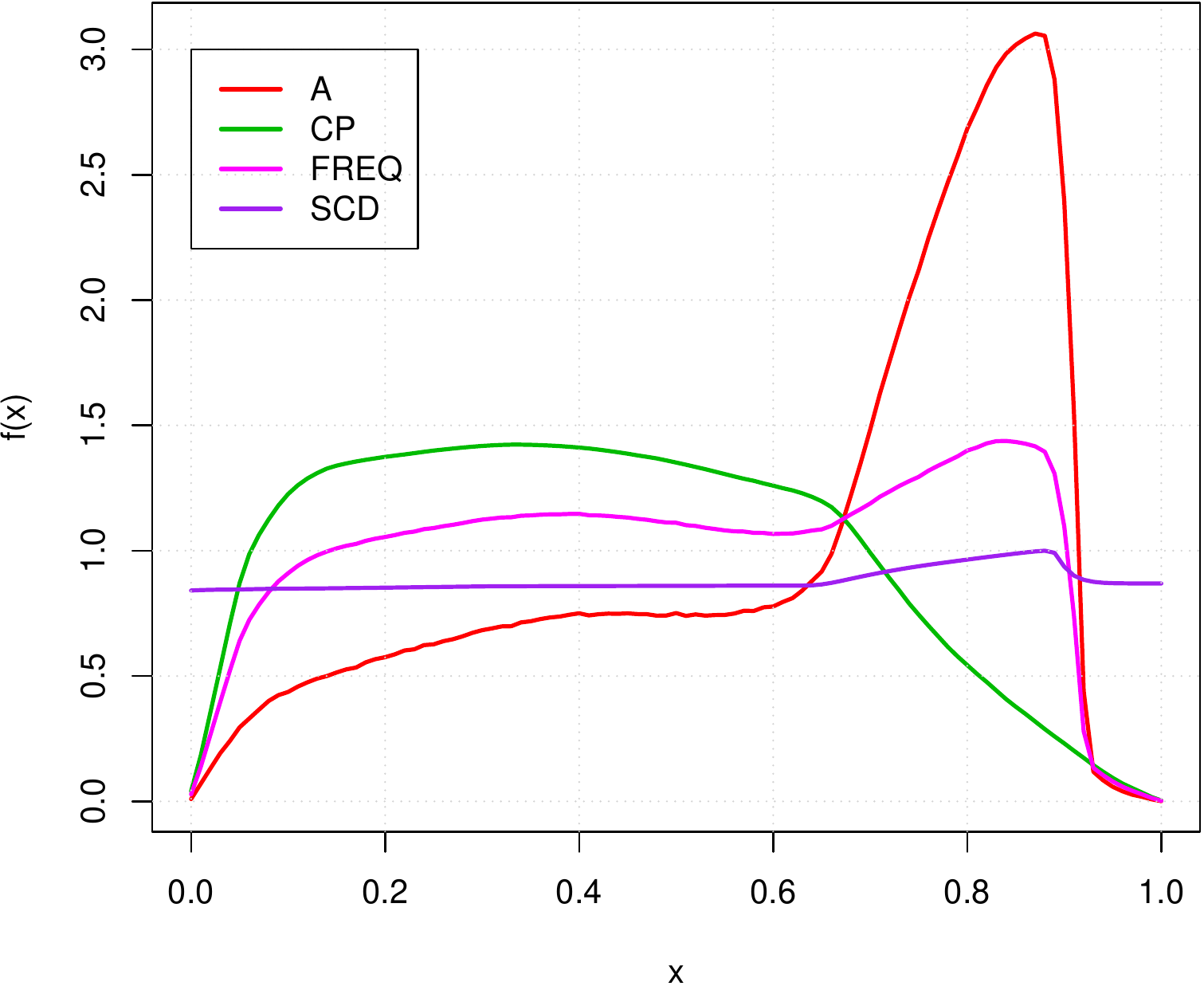} \caption{The second guess of the Fire Drill anti-pattern in source code.}\label{fig:fd-sc-2nd-guess}
\end{figure}

Now it's time to print the new pattern for Source Code!
Note that in figure \ref{fig:fd-sc-2nd-guess} the curves for the variables \texttt{A}, \texttt{CP}, and \texttt{FREQ} are now all proper probability densities (i.e., each of them integrates to \(1\)).
The variable \texttt{SCD} is not a density and peaks at \(1\), because that is the maximum value for the source code density (it is ratio, actually).

I have to admit this pattern looks quite nice. Let's keep it and produce an \emph{additional} version using CDFs (we will keep and test both).

Now, the last step is to transform the three maintenance activities into \textbf{cumulative} densities.
We will perform rejection sampling from these KDE-PDFs and estimate the ECDFs.

\begin{Shaded}
\begin{Highlighting}[]
\NormalTok{pattern\_sc\_cdf }\OtherTok{\textless{}{-}} \FunctionTok{append}\NormalTok{(pattern\_sc, }\FunctionTok{list}\NormalTok{())}

\NormalTok{temp\_x }\OtherTok{\textless{}{-}} \FunctionTok{seq}\NormalTok{(}\AttributeTok{from =} \DecValTok{0}\NormalTok{, }\AttributeTok{to =} \DecValTok{1}\NormalTok{, }\AttributeTok{length.out =} \DecValTok{5000}\NormalTok{)}

\ControlFlowTok{for}\NormalTok{ (vname }\ControlFlowTok{in} \FunctionTok{names}\NormalTok{(pattern\_sc\_cdf)) \{}
  \ControlFlowTok{if}\NormalTok{ (vname }\SpecialCharTok{==} \StringTok{"SCD"}\NormalTok{) \{}
    \ControlFlowTok{next}  \CommentTok{\# Not this one..}
\NormalTok{  \}}

  \CommentTok{\# Let\textquotesingle{}s replace the pattern\textquotesingle{}s activities with approximate ECDFs:}
\NormalTok{  pattern\_sc\_cdf[[vname]] }\OtherTok{\textless{}{-}}\NormalTok{ stats}\SpecialCharTok{::}\FunctionTok{approxfun}\NormalTok{(}\AttributeTok{x =}\NormalTok{ temp\_x, }\AttributeTok{y =} \FunctionTok{get}\NormalTok{(}\FunctionTok{paste0}\NormalTok{(}\StringTok{"tempecdf\_"}\NormalTok{,}
\NormalTok{    vname))(temp\_x), }\AttributeTok{yleft =} \DecValTok{0}\NormalTok{, }\AttributeTok{yright =} \DecValTok{1}\NormalTok{)}
\NormalTok{\}}
\end{Highlighting}
\end{Shaded}

\begin{figure}[ht!]
\includegraphics{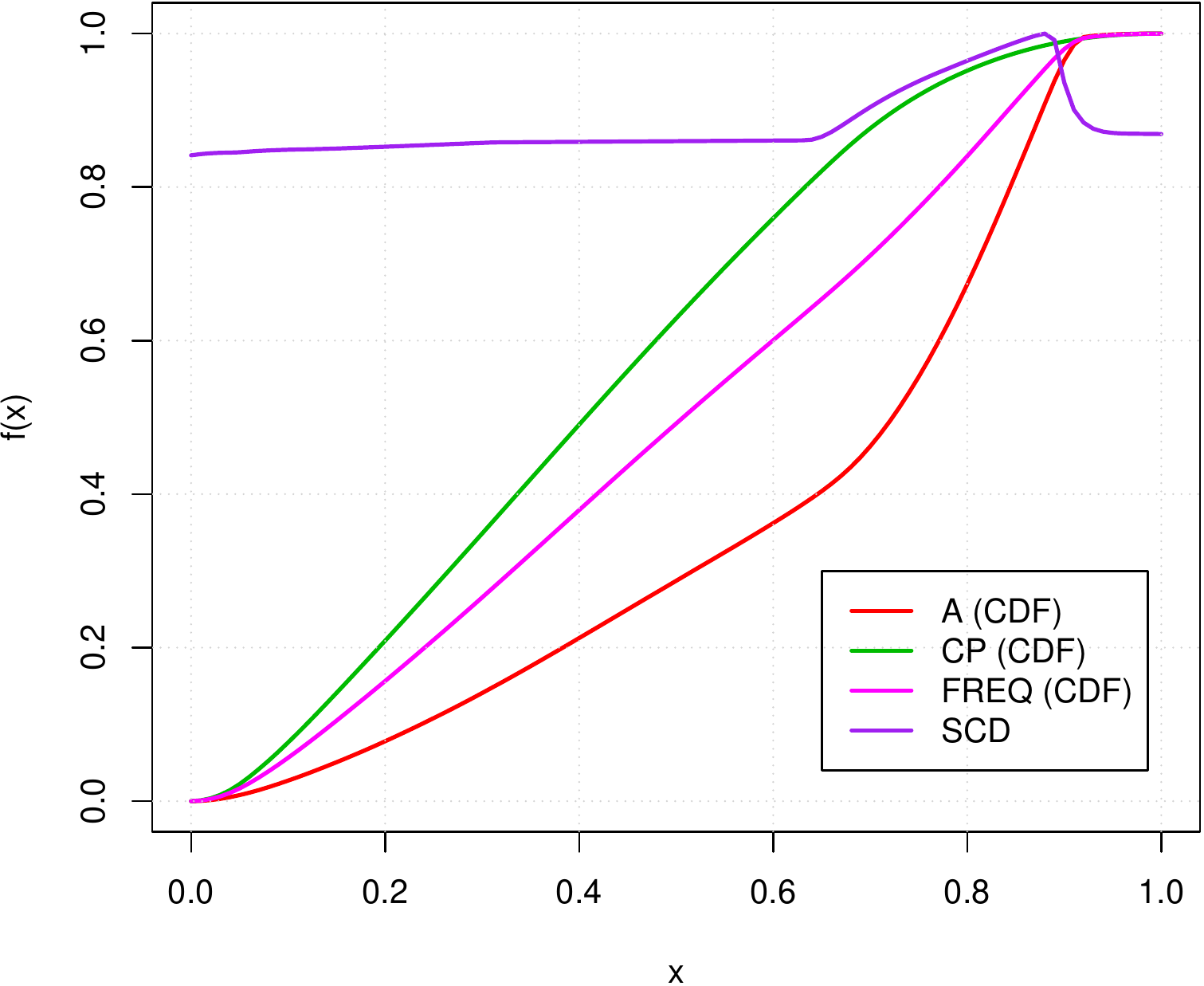} \caption{The second guess of the Fire Drill anti-pattern in source code, where the maintenance activities have been converted to cumulative densities.}\label{fig:fd-sc-2nd-guess-ecdf}
\end{figure}

In figure \ref{fig:fd-sc-2nd-guess-ecdf} we show the final CDF-version of the improved pattern for source code.
The three activities \texttt{A}, \texttt{CP}, and \texttt{FREQ} have now been converted to cumulative densities (no change to \texttt{SCD}).

\hypertarget{importing-and-preparing-the-data}{%
\subsection{Importing and preparing the Data}\label{importing-and-preparing-the-data}}

We have a total of \(15\) projects and a ground truth for each.

\hypertarget{load-the-ground-truth}{%
\subsubsection{Load the Ground Truth}\label{load-the-ground-truth}}

Here, we will simply join the two batches.

\begin{Shaded}
\begin{Highlighting}[]
\NormalTok{ground\_truth\_all }\OtherTok{\textless{}{-}} \FunctionTok{rbind}\NormalTok{(}\FunctionTok{read.csv}\NormalTok{(}\AttributeTok{file =} \StringTok{"../data/ground{-}truth.csv"}\NormalTok{, }\AttributeTok{sep =} \StringTok{";"}\NormalTok{),}
  \FunctionTok{read.csv}\NormalTok{(}\AttributeTok{file =} \StringTok{"../data/ground{-}truth\_2nd{-}batch.csv"}\NormalTok{, }\AttributeTok{sep =} \StringTok{";"}\NormalTok{))}
\end{Highlighting}
\end{Shaded}

\hypertarget{importing-the-issue-tracking-pattern}{%
\subsubsection{Importing the Issue-Tracking Pattern}\label{importing-the-issue-tracking-pattern}}

Here we load the issue-tracking pattern. Note that there is only one and we will not be using any variations.

\begin{Shaded}
\begin{Highlighting}[]
\NormalTok{pattern\_it }\OtherTok{\textless{}{-}} \FunctionTok{readRDS}\NormalTok{(}\AttributeTok{file =} \StringTok{"../results/pattern\_I\_it.rds"}\NormalTok{)}
\end{Highlighting}
\end{Shaded}

\hypertarget{importing-the-project-data-for-source-code}{%
\subsubsection{Importing the Project Data for Source Code}\label{importing-the-project-data-for-source-code}}

We cannot just import the projects as they were pre-processed previously for source code data.
Instead, we will load the raw data and transform the projects similar to how we produced the patterns.
Also, for source code, there will be two sets of projects: an ordinary one and a CDF-transformed one.

Let's first import and prepare the source code data:

\begin{Shaded}
\begin{Highlighting}[]
\NormalTok{temp }\OtherTok{\textless{}{-}} \FunctionTok{rbind}\NormalTok{(}\FunctionTok{read.csv}\NormalTok{(}\AttributeTok{file =} \StringTok{"../data/student{-}projects.csv"}\NormalTok{, }\AttributeTok{sep =} \StringTok{";"}\NormalTok{), }\FunctionTok{read.csv}\NormalTok{(}\AttributeTok{file =} \StringTok{"../data/student{-}projects\_2nd{-}batch.csv"}\NormalTok{,}
  \AttributeTok{sep =} \StringTok{";"}\NormalTok{))}
\NormalTok{temp}\SpecialCharTok{$}\NormalTok{label }\OtherTok{\textless{}{-}} \FunctionTok{factor}\NormalTok{(}\AttributeTok{x =} \FunctionTok{toupper}\NormalTok{(temp}\SpecialCharTok{$}\NormalTok{label), }\AttributeTok{levels =} \FunctionTok{sort}\NormalTok{(}\FunctionTok{toupper}\NormalTok{(}\FunctionTok{unique}\NormalTok{(temp}\SpecialCharTok{$}\NormalTok{label))))}
\NormalTok{temp}\SpecialCharTok{$}\NormalTok{project }\OtherTok{\textless{}{-}} \FunctionTok{factor}\NormalTok{(}\AttributeTok{x =}\NormalTok{ temp}\SpecialCharTok{$}\NormalTok{project, }\AttributeTok{levels =} \FunctionTok{sort}\NormalTok{(}\FunctionTok{unique}\NormalTok{(temp}\SpecialCharTok{$}\NormalTok{project)))}
\NormalTok{temp}\SpecialCharTok{$}\NormalTok{AuthorTimeNormalized }\OtherTok{\textless{}{-}} \ConstantTok{NA\_real\_}

\ControlFlowTok{for}\NormalTok{ (pId }\ControlFlowTok{in} \FunctionTok{levels}\NormalTok{(temp}\SpecialCharTok{$}\NormalTok{project)) \{}
\NormalTok{  temp[temp}\SpecialCharTok{$}\NormalTok{project }\SpecialCharTok{==}\NormalTok{ pId, ]}\SpecialCharTok{$}\NormalTok{AuthorTimeNormalized }\OtherTok{\textless{}{-}}\NormalTok{ (temp[temp}\SpecialCharTok{$}\NormalTok{project }\SpecialCharTok{==}\NormalTok{ pId,}
\NormalTok{    ]}\SpecialCharTok{$}\NormalTok{AuthorTimeUnixEpochMilliSecs }\SpecialCharTok{{-}} \FunctionTok{min}\NormalTok{(temp[temp}\SpecialCharTok{$}\NormalTok{project }\SpecialCharTok{==}\NormalTok{ pId, ]}\SpecialCharTok{$}\NormalTok{AuthorTimeUnixEpochMilliSecs))}
\NormalTok{  temp[temp}\SpecialCharTok{$}\NormalTok{project }\SpecialCharTok{==}\NormalTok{ pId, ]}\SpecialCharTok{$}\NormalTok{AuthorTimeNormalized }\OtherTok{\textless{}{-}}\NormalTok{ (temp[temp}\SpecialCharTok{$}\NormalTok{project }\SpecialCharTok{==}\NormalTok{ pId,}
\NormalTok{    ]}\SpecialCharTok{$}\NormalTok{AuthorTimeNormalized}\SpecialCharTok{/}\FunctionTok{max}\NormalTok{(temp[temp}\SpecialCharTok{$}\NormalTok{project }\SpecialCharTok{==}\NormalTok{ pId, ]}\SpecialCharTok{$}\NormalTok{AuthorTimeNormalized))}
\NormalTok{\}}
\end{Highlighting}
\end{Shaded}

\begin{Shaded}
\begin{Highlighting}[]
\NormalTok{projects\_sc }\OtherTok{\textless{}{-}} \FunctionTok{list}\NormalTok{()}
\NormalTok{projects\_sc\_cdf }\OtherTok{\textless{}{-}} \FunctionTok{list}\NormalTok{()}


\NormalTok{get\_densities }\OtherTok{\textless{}{-}} \ControlFlowTok{function}\NormalTok{(data, }\AttributeTok{x\_samples =} \DecValTok{10000}\NormalTok{) \{}
  \FunctionTok{suppressWarnings}\NormalTok{(\{}
\NormalTok{    use\_x }\OtherTok{\textless{}{-}} \FunctionTok{seq}\NormalTok{(}\AttributeTok{from =} \DecValTok{0}\NormalTok{, }\AttributeTok{to =} \DecValTok{1}\NormalTok{, }\AttributeTok{length.out =}\NormalTok{ x\_samples)}
\NormalTok{    tempdens }\OtherTok{\textless{}{-}}\NormalTok{ stats}\SpecialCharTok{::}\FunctionTok{density}\NormalTok{(}\AttributeTok{x =}\NormalTok{ data, }\AttributeTok{bw =} \StringTok{"SJ"}\NormalTok{, }\AttributeTok{cut =} \ConstantTok{TRUE}\NormalTok{)}
\NormalTok{    tempdens\_f }\OtherTok{\textless{}{-}}\NormalTok{ stats}\SpecialCharTok{::}\FunctionTok{approxfun}\NormalTok{(}\AttributeTok{x =}\NormalTok{ tempdens}\SpecialCharTok{$}\NormalTok{x, }\AttributeTok{y =}\NormalTok{ tempdens}\SpecialCharTok{$}\NormalTok{y, }\AttributeTok{yleft =} \DecValTok{0}\NormalTok{,}
      \AttributeTok{yright =} \DecValTok{0}\NormalTok{)}
\NormalTok{    use\_y }\OtherTok{\textless{}{-}}\NormalTok{ stats}\SpecialCharTok{::}\FunctionTok{runif}\NormalTok{(}\AttributeTok{n =}\NormalTok{ x\_samples, }\AttributeTok{min =} \DecValTok{0}\NormalTok{, }\AttributeTok{max =} \FunctionTok{max}\NormalTok{(tempdens}\SpecialCharTok{$}\NormalTok{y))}

\NormalTok{    temp\_samples }\OtherTok{\textless{}{-}}\NormalTok{ use\_x[use\_y }\SpecialCharTok{\textless{}=} \FunctionTok{tempdens\_f}\NormalTok{(use\_x)]}
\NormalTok{    tempdens }\OtherTok{\textless{}{-}}\NormalTok{ stats}\SpecialCharTok{::}\FunctionTok{density}\NormalTok{(}\AttributeTok{x =}\NormalTok{ temp\_samples, }\AttributeTok{bw =} \StringTok{"SJ"}\NormalTok{, }\AttributeTok{cut =} \ConstantTok{TRUE}\NormalTok{)}

    \FunctionTok{list}\NormalTok{(}\AttributeTok{PDF =}\NormalTok{ stats}\SpecialCharTok{::}\FunctionTok{approxfun}\NormalTok{(}\AttributeTok{x =}\NormalTok{ tempdens}\SpecialCharTok{$}\NormalTok{x, }\AttributeTok{y =}\NormalTok{ tempdens}\SpecialCharTok{$}\NormalTok{y, }\AttributeTok{yleft =} \DecValTok{0}\NormalTok{, }\AttributeTok{yright =} \DecValTok{0}\NormalTok{),}
      \AttributeTok{CDF =} \FunctionTok{make\_smooth\_ecdf}\NormalTok{(}\AttributeTok{values =}\NormalTok{ temp\_samples, }\AttributeTok{slope =} \DecValTok{0}\NormalTok{, }\AttributeTok{inverse =} \ConstantTok{FALSE}\NormalTok{))}
\NormalTok{  \})}
\NormalTok{\}}

\NormalTok{make\_mixture }\OtherTok{\textless{}{-}} \ControlFlowTok{function}\NormalTok{(pdf1, pdf2, pdf1\_ratio, }\AttributeTok{x\_samples =} \DecValTok{10000}\NormalTok{) \{}
\NormalTok{  use\_x }\OtherTok{\textless{}{-}} \FunctionTok{seq}\NormalTok{(}\AttributeTok{from =} \DecValTok{0}\NormalTok{, }\AttributeTok{to =} \DecValTok{1}\NormalTok{, }\AttributeTok{length.out =}\NormalTok{ x\_samples)}
\NormalTok{  tempf }\OtherTok{\textless{}{-}} \ControlFlowTok{function}\NormalTok{(x) pdf1\_ratio }\SpecialCharTok{*} \FunctionTok{pdf1}\NormalTok{(x) }\SpecialCharTok{+}\NormalTok{ (}\DecValTok{1} \SpecialCharTok{{-}}\NormalTok{ pdf1\_ratio) }\SpecialCharTok{*} \FunctionTok{pdf2}\NormalTok{(x)}

\NormalTok{  use\_y }\OtherTok{\textless{}{-}}\NormalTok{ stats}\SpecialCharTok{::}\FunctionTok{runif}\NormalTok{(}\AttributeTok{n =} \FunctionTok{length}\NormalTok{(use\_x), }\AttributeTok{min =} \DecValTok{0}\NormalTok{, }\AttributeTok{max =} \SpecialCharTok{{-}}\FloatTok{1.01} \SpecialCharTok{*}\NormalTok{ nloptr}\SpecialCharTok{::}\FunctionTok{nloptr}\NormalTok{(}\AttributeTok{x0 =} \FloatTok{0.5}\NormalTok{,}
    \AttributeTok{lb =} \DecValTok{0}\NormalTok{, }\AttributeTok{ub =} \DecValTok{1}\NormalTok{, }\AttributeTok{eval\_f =} \ControlFlowTok{function}\NormalTok{(x) }\SpecialCharTok{{-}}\FunctionTok{tempf}\NormalTok{(x), }\AttributeTok{opts =} \FunctionTok{list}\NormalTok{(}\AttributeTok{algorithm =} \StringTok{"NLOPT\_GN\_DIRECT\_L\_RAND"}\NormalTok{,}
      \AttributeTok{xtol\_rel =} \FloatTok{0.001}\NormalTok{, }\AttributeTok{maxeval =} \FloatTok{1e+05}\NormalTok{))}\SpecialCharTok{$}\NormalTok{objective)}

  \FunctionTok{list}\NormalTok{(}\AttributeTok{PDF =}\NormalTok{ tempf, }\AttributeTok{CDF =} \FunctionTok{make\_smooth\_ecdf}\NormalTok{(}\AttributeTok{values =}\NormalTok{ use\_x[use\_y }\SpecialCharTok{\textless{}=} \FunctionTok{tempf}\NormalTok{(use\_x)],}
    \AttributeTok{slope =} \DecValTok{0}\NormalTok{, }\AttributeTok{inverse =} \ConstantTok{FALSE}\NormalTok{))}
\NormalTok{\}}


\ControlFlowTok{for}\NormalTok{ (lvl }\ControlFlowTok{in} \FunctionTok{levels}\NormalTok{(temp}\SpecialCharTok{$}\NormalTok{project)) \{}
\NormalTok{  df }\OtherTok{\textless{}{-}}\NormalTok{ temp[temp}\SpecialCharTok{$}\NormalTok{project }\SpecialCharTok{==}\NormalTok{ lvl, ]}
\NormalTok{  templ }\OtherTok{\textless{}{-}} \FunctionTok{list}\NormalTok{()}
\NormalTok{  templ\_cdf }\OtherTok{\textless{}{-}} \FunctionTok{list}\NormalTok{()}
\NormalTok{  templ[[}\StringTok{"df"}\NormalTok{]] }\OtherTok{\textless{}{-}}\NormalTok{ df[, ]}

\NormalTok{  temp\_A }\OtherTok{\textless{}{-}} \FunctionTok{get\_densities}\NormalTok{(}\AttributeTok{data =}\NormalTok{ df[df}\SpecialCharTok{$}\NormalTok{label }\SpecialCharTok{==} \StringTok{"A"}\NormalTok{, ]}\SpecialCharTok{$}\NormalTok{AuthorTimeNormalized)}
\NormalTok{  temp\_C }\OtherTok{\textless{}{-}} \FunctionTok{get\_densities}\NormalTok{(}\AttributeTok{data =}\NormalTok{ df[df}\SpecialCharTok{$}\NormalTok{label }\SpecialCharTok{==} \StringTok{"C"}\NormalTok{, ]}\SpecialCharTok{$}\NormalTok{AuthorTimeNormalized)}
\NormalTok{  temp\_P }\OtherTok{\textless{}{-}} \FunctionTok{get\_densities}\NormalTok{(}\AttributeTok{data =}\NormalTok{ df[df}\SpecialCharTok{$}\NormalTok{label }\SpecialCharTok{==} \StringTok{"P"}\NormalTok{, ]}\SpecialCharTok{$}\NormalTok{AuthorTimeNormalized)}
\NormalTok{  temp\_CP }\OtherTok{\textless{}{-}} \FunctionTok{get\_densities}\NormalTok{(}\AttributeTok{data =}\NormalTok{ df[df}\SpecialCharTok{$}\NormalTok{label }\SpecialCharTok{\%in\%} \FunctionTok{c}\NormalTok{(}\StringTok{"C"}\NormalTok{, }\StringTok{"P"}\NormalTok{), ]}\SpecialCharTok{$}\NormalTok{AuthorTimeNormalized)}
\NormalTok{  acp\_ratio }\OtherTok{\textless{}{-}} \FunctionTok{nrow}\NormalTok{(df[df}\SpecialCharTok{$}\NormalTok{label }\SpecialCharTok{==} \StringTok{"A"}\NormalTok{, ])}\SpecialCharTok{/}\FunctionTok{nrow}\NormalTok{(df)}
\NormalTok{  temp\_FREQ }\OtherTok{\textless{}{-}} \FunctionTok{make\_mixture}\NormalTok{(}\AttributeTok{pdf1 =}\NormalTok{ temp\_A}\SpecialCharTok{$}\NormalTok{PDF, }\AttributeTok{pdf2 =}\NormalTok{ temp\_CP}\SpecialCharTok{$}\NormalTok{PDF, }\AttributeTok{pdf1\_ratio =}\NormalTok{ acp\_ratio)}
  \CommentTok{\# We will use this later on}
\NormalTok{  temp\_FREQ\_no\_mixture }\OtherTok{\textless{}{-}} \FunctionTok{get\_densities}\NormalTok{(}\AttributeTok{data =}\NormalTok{ df}\SpecialCharTok{$}\NormalTok{AuthorTimeNormalized)}

\NormalTok{  templ[[}\StringTok{"A"}\NormalTok{]] }\OtherTok{\textless{}{-}}\NormalTok{ temp\_A}\SpecialCharTok{$}\NormalTok{PDF}
\NormalTok{  templ[[}\StringTok{"C"}\NormalTok{]] }\OtherTok{\textless{}{-}}\NormalTok{ temp\_C}\SpecialCharTok{$}\NormalTok{PDF}
\NormalTok{  templ[[}\StringTok{"P"}\NormalTok{]] }\OtherTok{\textless{}{-}}\NormalTok{ temp\_P}\SpecialCharTok{$}\NormalTok{PDF}
\NormalTok{  templ[[}\StringTok{"CP"}\NormalTok{]] }\OtherTok{\textless{}{-}}\NormalTok{ temp\_CP}\SpecialCharTok{$}\NormalTok{PDF}
\NormalTok{  templ[[}\StringTok{"FREQ"}\NormalTok{]] }\OtherTok{\textless{}{-}}\NormalTok{ temp\_FREQ}\SpecialCharTok{$}\NormalTok{PDF}
\NormalTok{  templ[[}\StringTok{"FREQ\_nm"}\NormalTok{]] }\OtherTok{\textless{}{-}}\NormalTok{ temp\_FREQ\_no\_mixture}\SpecialCharTok{$}\NormalTok{PDF}

\NormalTok{  templ\_cdf[[}\StringTok{"A"}\NormalTok{]] }\OtherTok{\textless{}{-}}\NormalTok{ temp\_A}\SpecialCharTok{$}\NormalTok{CDF}
\NormalTok{  templ\_cdf[[}\StringTok{"C"}\NormalTok{]] }\OtherTok{\textless{}{-}}\NormalTok{ temp\_C}\SpecialCharTok{$}\NormalTok{CDF}
\NormalTok{  templ\_cdf[[}\StringTok{"P"}\NormalTok{]] }\OtherTok{\textless{}{-}}\NormalTok{ temp\_P}\SpecialCharTok{$}\NormalTok{CDF}
\NormalTok{  templ\_cdf[[}\StringTok{"CP"}\NormalTok{]] }\OtherTok{\textless{}{-}}\NormalTok{ temp\_CP}\SpecialCharTok{$}\NormalTok{CDF}
\NormalTok{  templ\_cdf[[}\StringTok{"FREQ"}\NormalTok{]] }\OtherTok{\textless{}{-}}\NormalTok{ temp\_FREQ}\SpecialCharTok{$}\NormalTok{CDF}
\NormalTok{  templ\_cdf[[}\StringTok{"FREQ\_nm"}\NormalTok{]] }\OtherTok{\textless{}{-}}\NormalTok{ temp\_FREQ\_no\_mixture}\SpecialCharTok{$}\NormalTok{CDF}

\NormalTok{  templ[[}\StringTok{"SCD"}\NormalTok{]] }\OtherTok{\textless{}{-}} \FunctionTok{suppressWarnings}\NormalTok{(\{}
\NormalTok{    stats}\SpecialCharTok{::}\FunctionTok{approxfun}\NormalTok{(}\AttributeTok{x =}\NormalTok{ df}\SpecialCharTok{$}\NormalTok{AuthorTimeNormalized, }\AttributeTok{y =}\NormalTok{ df}\SpecialCharTok{$}\NormalTok{Density, }\AttributeTok{rule =} \DecValTok{2}\NormalTok{)}
\NormalTok{  \})}
\NormalTok{  templ\_cdf[[}\StringTok{"SCD"}\NormalTok{]] }\OtherTok{\textless{}{-}}\NormalTok{ templ[[}\StringTok{"SCD"}\NormalTok{]]  }\CommentTok{\# it\textquotesingle{}s the same because there is no CDF of it}

\NormalTok{  projects\_sc[[lvl]] }\OtherTok{\textless{}{-}}\NormalTok{ templ}
\NormalTok{  projects\_sc\_cdf[[lvl]] }\OtherTok{\textless{}{-}}\NormalTok{ templ\_cdf}
\NormalTok{\}}
\end{Highlighting}
\end{Shaded}

\begin{figure}[ht!]
\includegraphics{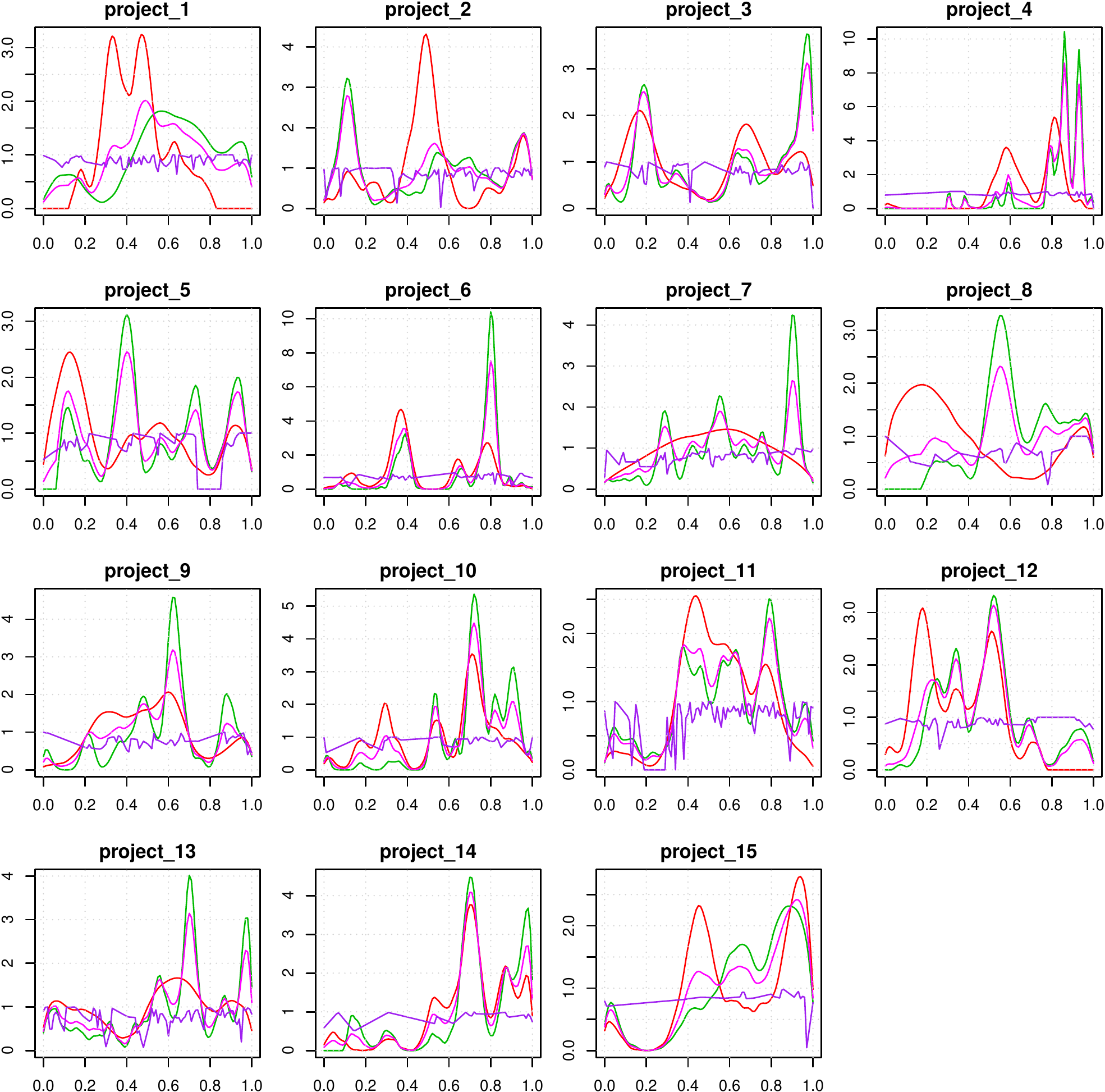} \caption{The projects in source code with the new transform applied. It is obvious that the densities are now exceeding $1.0$ since they are no longer normalized.}\label{fig:project-vars-new}
\end{figure}

In figure \ref{fig:project-vars-new} we show the \(15\) projects with the new transform applied.

\hypertarget{importing-the-project-data-for-issue-tracking}{%
\subsubsection{Importing the Project Data for Issue-tracking}\label{importing-the-project-data-for-issue-tracking}}

This case is more straightforward, as we do not change the way this was done before.
Also, there will only be one version, not two like we have for source code.

\begin{Shaded}
\begin{Highlighting}[]
\NormalTok{projects\_it }\OtherTok{\textless{}{-}} \FunctionTok{list}\NormalTok{()}
\NormalTok{projects\_it }\OtherTok{\textless{}{-}} \FunctionTok{append}\NormalTok{(projects\_it, }\FunctionTok{readRDS}\NormalTok{(}\AttributeTok{file =} \StringTok{"../results/project\_signals\_it.rds"}\NormalTok{))}
\NormalTok{projects\_it }\OtherTok{\textless{}{-}} \FunctionTok{append}\NormalTok{(projects\_it, }\FunctionTok{readRDS}\NormalTok{(}\AttributeTok{file =} \StringTok{"../results/project\_signals\_2nd\_batch\_it.rds"}\NormalTok{))}
\end{Highlighting}
\end{Shaded}

\hypertarget{creating-the-datasets}{%
\subsection{Creating the Datasets}\label{creating-the-datasets}}

For each pattern, we will create a dataset that contains the deviation for each project against each pattern.
Since we have three patterns, we will have three datasets.

Since the patterns and projects are modeled as curves, we will compute segment-wise features. For that, each project/pattern is subdivided into ten equally long intervals. Then, a distance metric is compute for each interval and activity.
We will compute \textbf{two metrics}: The area between curves and the \(2\)-dimensional relative continuous Pearson sample correlation coefficient (see section \ref{ssec:m-dim-pearson}).
This means that we will get a total of \(2\times 10\times 4=80\) features for source code patterns and \(60\) for issue-tracking patterns (because IT has only three activities/variables).
This means that we will get a lot more features than data points.
Also, our data is not balanced. For these reasons, we will have to use oversampling and synthetically inflate our dataset.
Note that this is OK for the intended purpose of finding an upper bound of required training instances for obtaining robust regression models.
In practice, it would of course be better just to label additional instances.

\hypertarget{function-for-computing-a-distance}{%
\subsubsection{Function for computing a Distance}\label{function-for-computing-a-distance}}

The following function will be used to compute the distance metrics for area between curves and correlation.

\begin{Shaded}
\begin{Highlighting}[]
\NormalTok{compute\_distance }\OtherTok{\textless{}{-}} \ControlFlowTok{function}\NormalTok{(f\_pattern, f\_project, }\AttributeTok{metric =} \FunctionTok{c}\NormalTok{(}\StringTok{"area"}\NormalTok{, }\StringTok{"corr"}\NormalTok{), }\AttributeTok{interval =} \DecValTok{1}\SpecialCharTok{:}\DecValTok{10}\NormalTok{,}
  \AttributeTok{num\_samples =} \DecValTok{1000}\NormalTok{) \{}
\NormalTok{  metric }\OtherTok{\textless{}{-}} \FunctionTok{match.arg}\NormalTok{(metric)}

\NormalTok{  interval\_ext }\OtherTok{\textless{}{-}} \FunctionTok{c}\NormalTok{(interval}\SpecialCharTok{/}\DecValTok{10} \SpecialCharTok{{-}} \FloatTok{0.1}\NormalTok{, interval}\SpecialCharTok{/}\DecValTok{10}\NormalTok{)}
\NormalTok{  use\_x }\OtherTok{\textless{}{-}} \FunctionTok{seq}\NormalTok{(}\AttributeTok{from =}\NormalTok{ interval\_ext[}\DecValTok{1}\NormalTok{], }\AttributeTok{to =}\NormalTok{ interval\_ext[}\DecValTok{2}\NormalTok{], }\AttributeTok{length.out =}\NormalTok{ num\_samples)}

\NormalTok{  v1 }\OtherTok{\textless{}{-}} \FunctionTok{f\_pattern}\NormalTok{(use\_x)}
\NormalTok{  v2 }\OtherTok{\textless{}{-}} \FunctionTok{f\_project}\NormalTok{(use\_x)}

  \ControlFlowTok{if}\NormalTok{ (metric }\SpecialCharTok{==} \StringTok{"area"}\NormalTok{) \{}
    \CommentTok{\# It\textquotesingle{}s the same as MAE for large samples:}
    \FunctionTok{return}\NormalTok{(Metrics}\SpecialCharTok{::}\FunctionTok{mae}\NormalTok{(}\AttributeTok{actual =}\NormalTok{ v1, }\AttributeTok{predicted =}\NormalTok{ v2))}
\NormalTok{  \}}
\NormalTok{  temp }\OtherTok{\textless{}{-}} \FunctionTok{suppressWarnings}\NormalTok{(\{}
\NormalTok{    stats}\SpecialCharTok{::}\FunctionTok{cor}\NormalTok{(}\AttributeTok{x =}\NormalTok{ v1, }\AttributeTok{y =}\NormalTok{ v2, }\AttributeTok{method =} \StringTok{"pearson"}\NormalTok{)}
\NormalTok{  \})}
  \ControlFlowTok{if}\NormalTok{ (}\FunctionTok{is.na}\NormalTok{(temp))}
    \DecValTok{0} \ControlFlowTok{else}\NormalTok{ temp}
\NormalTok{\}}
\end{Highlighting}
\end{Shaded}

\hypertarget{create-datasets-for-source-code}{%
\subsubsection{Create Datasets for Source Code}\label{create-datasets-for-source-code}}

Here we will create both datasets (ordinary and CDF) in the same loop.

\begin{Shaded}
\begin{Highlighting}[]
\NormalTok{grid }\OtherTok{\textless{}{-}} \FunctionTok{expand.grid}\NormalTok{(}\FunctionTok{list}\NormalTok{(}\AttributeTok{distance =} \FunctionTok{c}\NormalTok{(}\StringTok{"area"}\NormalTok{, }\StringTok{"corr"}\NormalTok{), }\AttributeTok{interval =} \DecValTok{1}\SpecialCharTok{:}\DecValTok{10}\NormalTok{, }\AttributeTok{activity =} \FunctionTok{c}\NormalTok{(}\StringTok{"A"}\NormalTok{,}
  \StringTok{"CP"}\NormalTok{, }\StringTok{"FREQ"}\NormalTok{, }\StringTok{"SCD"}\NormalTok{)))}
\NormalTok{grid}\SpecialCharTok{$}\NormalTok{distance }\OtherTok{\textless{}{-}} \FunctionTok{as.character}\NormalTok{(grid}\SpecialCharTok{$}\NormalTok{distance)}
\NormalTok{grid}\SpecialCharTok{$}\NormalTok{activity }\OtherTok{\textless{}{-}} \FunctionTok{as.character}\NormalTok{(grid}\SpecialCharTok{$}\NormalTok{activity)}

\NormalTok{dataset\_sc }\OtherTok{\textless{}{-}} \StringTok{\textasciigrave{}}\AttributeTok{colnames\textless{}{-}}\StringTok{\textasciigrave{}}\NormalTok{(}\AttributeTok{x =} \FunctionTok{matrix}\NormalTok{(}\AttributeTok{nrow =} \DecValTok{0}\NormalTok{, }\AttributeTok{ncol =} \FunctionTok{nrow}\NormalTok{(grid)), }\AttributeTok{value =} \FunctionTok{sapply}\NormalTok{(}\AttributeTok{X =} \FunctionTok{rownames}\NormalTok{(grid),}
  \AttributeTok{FUN =} \ControlFlowTok{function}\NormalTok{(rn) \{}
\NormalTok{    r }\OtherTok{\textless{}{-}}\NormalTok{ grid[rn, ]}
    \FunctionTok{paste}\NormalTok{(r}\SpecialCharTok{$}\NormalTok{activity, r}\SpecialCharTok{$}\NormalTok{interval, r}\SpecialCharTok{$}\NormalTok{distance, }\AttributeTok{sep =} \StringTok{"\_"}\NormalTok{)}
\NormalTok{  \}))}
\NormalTok{dataset\_sc\_cdf }\OtherTok{\textless{}{-}}\NormalTok{ dataset\_sc[, ]}


\ControlFlowTok{for}\NormalTok{ (pname }\ControlFlowTok{in} \FunctionTok{names}\NormalTok{(projects\_sc)) \{}
\NormalTok{  newrow\_sc }\OtherTok{\textless{}{-}} \StringTok{\textasciigrave{}}\AttributeTok{colnames\textless{}{-}}\StringTok{\textasciigrave{}}\NormalTok{(}\AttributeTok{x =} \FunctionTok{matrix}\NormalTok{(}\AttributeTok{ncol =} \FunctionTok{ncol}\NormalTok{(dataset\_sc)), }\AttributeTok{value =} \FunctionTok{colnames}\NormalTok{(dataset\_sc))}
\NormalTok{  newrow\_sc\_cdf }\OtherTok{\textless{}{-}} \StringTok{\textasciigrave{}}\AttributeTok{colnames\textless{}{-}}\StringTok{\textasciigrave{}}\NormalTok{(}\AttributeTok{x =} \FunctionTok{matrix}\NormalTok{(}\AttributeTok{ncol =} \FunctionTok{ncol}\NormalTok{(dataset\_sc)), }\AttributeTok{value =} \FunctionTok{colnames}\NormalTok{(dataset\_sc))}

  \ControlFlowTok{for}\NormalTok{ (rn }\ControlFlowTok{in} \FunctionTok{rownames}\NormalTok{(grid)) \{}
\NormalTok{    row }\OtherTok{\textless{}{-}}\NormalTok{ grid[rn, ]}
\NormalTok{    tempf\_pattern }\OtherTok{\textless{}{-}}\NormalTok{ pattern\_sc[[row}\SpecialCharTok{$}\NormalTok{activity]]}
\NormalTok{    tempf\_pattern\_cdf }\OtherTok{\textless{}{-}}\NormalTok{ pattern\_sc\_cdf[[row}\SpecialCharTok{$}\NormalTok{activity]]}
\NormalTok{    tempf\_project }\OtherTok{\textless{}{-}}\NormalTok{ projects\_sc[[pname]][[row}\SpecialCharTok{$}\NormalTok{activity]]}
\NormalTok{    tempf\_project\_cdf }\OtherTok{\textless{}{-}}\NormalTok{ projects\_sc\_cdf[[pname]][[row}\SpecialCharTok{$}\NormalTok{activity]]}

\NormalTok{    feat\_name }\OtherTok{\textless{}{-}} \FunctionTok{paste}\NormalTok{(row}\SpecialCharTok{$}\NormalTok{activity, row}\SpecialCharTok{$}\NormalTok{interval, row}\SpecialCharTok{$}\NormalTok{distance, }\AttributeTok{sep =} \StringTok{"\_"}\NormalTok{)}
\NormalTok{    newrow\_sc[}\DecValTok{1}\NormalTok{, feat\_name] }\OtherTok{\textless{}{-}} \FunctionTok{compute\_distance}\NormalTok{(}\AttributeTok{f\_pattern =}\NormalTok{ tempf\_pattern, }\AttributeTok{f\_project =}\NormalTok{ tempf\_project,}
      \AttributeTok{metric =}\NormalTok{ row}\SpecialCharTok{$}\NormalTok{distance, }\AttributeTok{interval =}\NormalTok{ row}\SpecialCharTok{$}\NormalTok{interval)}
\NormalTok{    newrow\_sc\_cdf[}\DecValTok{1}\NormalTok{, feat\_name] }\OtherTok{\textless{}{-}} \FunctionTok{compute\_distance}\NormalTok{(}\AttributeTok{f\_pattern =}\NormalTok{ tempf\_pattern\_cdf,}
      \AttributeTok{f\_project =}\NormalTok{ tempf\_project\_cdf, }\AttributeTok{metric =}\NormalTok{ row}\SpecialCharTok{$}\NormalTok{distance, }\AttributeTok{interval =}\NormalTok{ row}\SpecialCharTok{$}\NormalTok{interval)}
\NormalTok{  \}}

\NormalTok{  dataset\_sc }\OtherTok{\textless{}{-}} \FunctionTok{rbind}\NormalTok{(dataset\_sc, newrow\_sc)}
\NormalTok{  dataset\_sc\_cdf }\OtherTok{\textless{}{-}} \FunctionTok{rbind}\NormalTok{(dataset\_sc\_cdf, newrow\_sc\_cdf)}
\NormalTok{\}}

\NormalTok{dataset\_sc }\OtherTok{\textless{}{-}} \FunctionTok{as.data.frame}\NormalTok{(dataset\_sc)}
\NormalTok{dataset\_sc\_cdf }\OtherTok{\textless{}{-}} \FunctionTok{as.data.frame}\NormalTok{(dataset\_sc\_cdf)}
\end{Highlighting}
\end{Shaded}

\hypertarget{create-dataset-for-issue-tracking}{%
\subsubsection{Create Dataset for Issue-tracking}\label{create-dataset-for-issue-tracking}}

This is very similar to how we created the datasets for source code, mainly the activities will differ and there is only one version.

\begin{Shaded}
\begin{Highlighting}[]
\NormalTok{grid }\OtherTok{\textless{}{-}} \FunctionTok{expand.grid}\NormalTok{(}\FunctionTok{list}\NormalTok{(}\AttributeTok{distance =} \FunctionTok{c}\NormalTok{(}\StringTok{"area"}\NormalTok{, }\StringTok{"corr"}\NormalTok{), }\AttributeTok{interval =} \DecValTok{1}\SpecialCharTok{:}\DecValTok{10}\NormalTok{, }\AttributeTok{activity =} \FunctionTok{c}\NormalTok{(}\StringTok{"REQ"}\NormalTok{,}
  \StringTok{"DEV"}\NormalTok{, }\StringTok{"DESC"}\NormalTok{)))}
\NormalTok{grid}\SpecialCharTok{$}\NormalTok{distance }\OtherTok{\textless{}{-}} \FunctionTok{as.character}\NormalTok{(grid}\SpecialCharTok{$}\NormalTok{distance)}
\NormalTok{grid}\SpecialCharTok{$}\NormalTok{activity }\OtherTok{\textless{}{-}} \FunctionTok{as.character}\NormalTok{(grid}\SpecialCharTok{$}\NormalTok{activity)}

\NormalTok{dataset\_it }\OtherTok{\textless{}{-}} \StringTok{\textasciigrave{}}\AttributeTok{colnames\textless{}{-}}\StringTok{\textasciigrave{}}\NormalTok{(}\AttributeTok{x =} \FunctionTok{matrix}\NormalTok{(}\AttributeTok{nrow =} \DecValTok{0}\NormalTok{, }\AttributeTok{ncol =} \FunctionTok{nrow}\NormalTok{(grid)), }\AttributeTok{value =} \FunctionTok{sapply}\NormalTok{(}\AttributeTok{X =} \FunctionTok{rownames}\NormalTok{(grid),}
  \AttributeTok{FUN =} \ControlFlowTok{function}\NormalTok{(rn) \{}
\NormalTok{    r }\OtherTok{\textless{}{-}}\NormalTok{ grid[rn, ]}
    \FunctionTok{paste}\NormalTok{(r}\SpecialCharTok{$}\NormalTok{activity, r}\SpecialCharTok{$}\NormalTok{interval, r}\SpecialCharTok{$}\NormalTok{distance, }\AttributeTok{sep =} \StringTok{"\_"}\NormalTok{)}
\NormalTok{  \}))}

\ControlFlowTok{for}\NormalTok{ (pname }\ControlFlowTok{in} \FunctionTok{names}\NormalTok{(projects\_it)) \{}
\NormalTok{  newrow\_it }\OtherTok{\textless{}{-}} \StringTok{\textasciigrave{}}\AttributeTok{colnames\textless{}{-}}\StringTok{\textasciigrave{}}\NormalTok{(}\AttributeTok{x =} \FunctionTok{matrix}\NormalTok{(}\AttributeTok{ncol =} \FunctionTok{ncol}\NormalTok{(dataset\_it)), }\AttributeTok{value =} \FunctionTok{colnames}\NormalTok{(dataset\_it))}

  \ControlFlowTok{for}\NormalTok{ (rn }\ControlFlowTok{in} \FunctionTok{rownames}\NormalTok{(grid)) \{}
\NormalTok{    row }\OtherTok{\textless{}{-}}\NormalTok{ grid[rn, ]}
\NormalTok{    tempf\_pattern }\OtherTok{\textless{}{-}}\NormalTok{ pattern\_it[[row}\SpecialCharTok{$}\NormalTok{activity]]}
\NormalTok{    tempf\_project }\OtherTok{\textless{}{-}}\NormalTok{ projects\_it[[pname]][[row}\SpecialCharTok{$}\NormalTok{activity]]}\SpecialCharTok{$}\FunctionTok{get0Function}\NormalTok{()}

\NormalTok{    feat\_name }\OtherTok{\textless{}{-}} \FunctionTok{paste}\NormalTok{(row}\SpecialCharTok{$}\NormalTok{activity, row}\SpecialCharTok{$}\NormalTok{interval, row}\SpecialCharTok{$}\NormalTok{distance, }\AttributeTok{sep =} \StringTok{"\_"}\NormalTok{)}
\NormalTok{    newrow\_it[}\DecValTok{1}\NormalTok{, feat\_name] }\OtherTok{\textless{}{-}} \FunctionTok{compute\_distance}\NormalTok{(}\AttributeTok{f\_pattern =}\NormalTok{ tempf\_pattern, }\AttributeTok{f\_project =}\NormalTok{ tempf\_project,}
      \AttributeTok{metric =}\NormalTok{ row}\SpecialCharTok{$}\NormalTok{distance, }\AttributeTok{interval =}\NormalTok{ row}\SpecialCharTok{$}\NormalTok{interval)}
\NormalTok{  \}}

\NormalTok{  dataset\_it }\OtherTok{\textless{}{-}} \FunctionTok{rbind}\NormalTok{(dataset\_it, newrow\_it)}
\NormalTok{\}}

\NormalTok{dataset\_it }\OtherTok{\textless{}{-}} \FunctionTok{as.data.frame}\NormalTok{(dataset\_it)}
\end{Highlighting}
\end{Shaded}

\hypertarget{adaptive-training-for-robust-regression-models}{%
\subsection{Adaptive Training For Robust Regression Models}\label{adaptive-training-for-robust-regression-models}}

In this section, we will train a few state-of-the-art models \emph{adaptively}, which means that we will keep adding labeled training data in order to hopefully observe some convergence towards a generalization error on validation data that will allow us to estimate the minimum amount of training data required to achieve a certain error.
We will repeat fitting the same model many times, using repeated \(k\)-fold cross validation and bootstrapping, in order to obtain reliable results.
As for the models, it has been shown that the Random forest is the average top performer (Delgado et al. 2014), so we will pick it.
Furthermore, we will attempt to fit these models: GLM (generalized linear model), GBM (gradient boosting machine), XGB (extreme gradient boosting), SVM-Poly (SVM with polynomial kernel), avNNET (model averaged neural network), rPART2 (recursive partitioning of rules).

\hypertarget{preparation-of-the-datasets}{%
\subsubsection{Preparation of the Datasets}\label{preparation-of-the-datasets}}

\begin{figure}[ht!]
\includegraphics{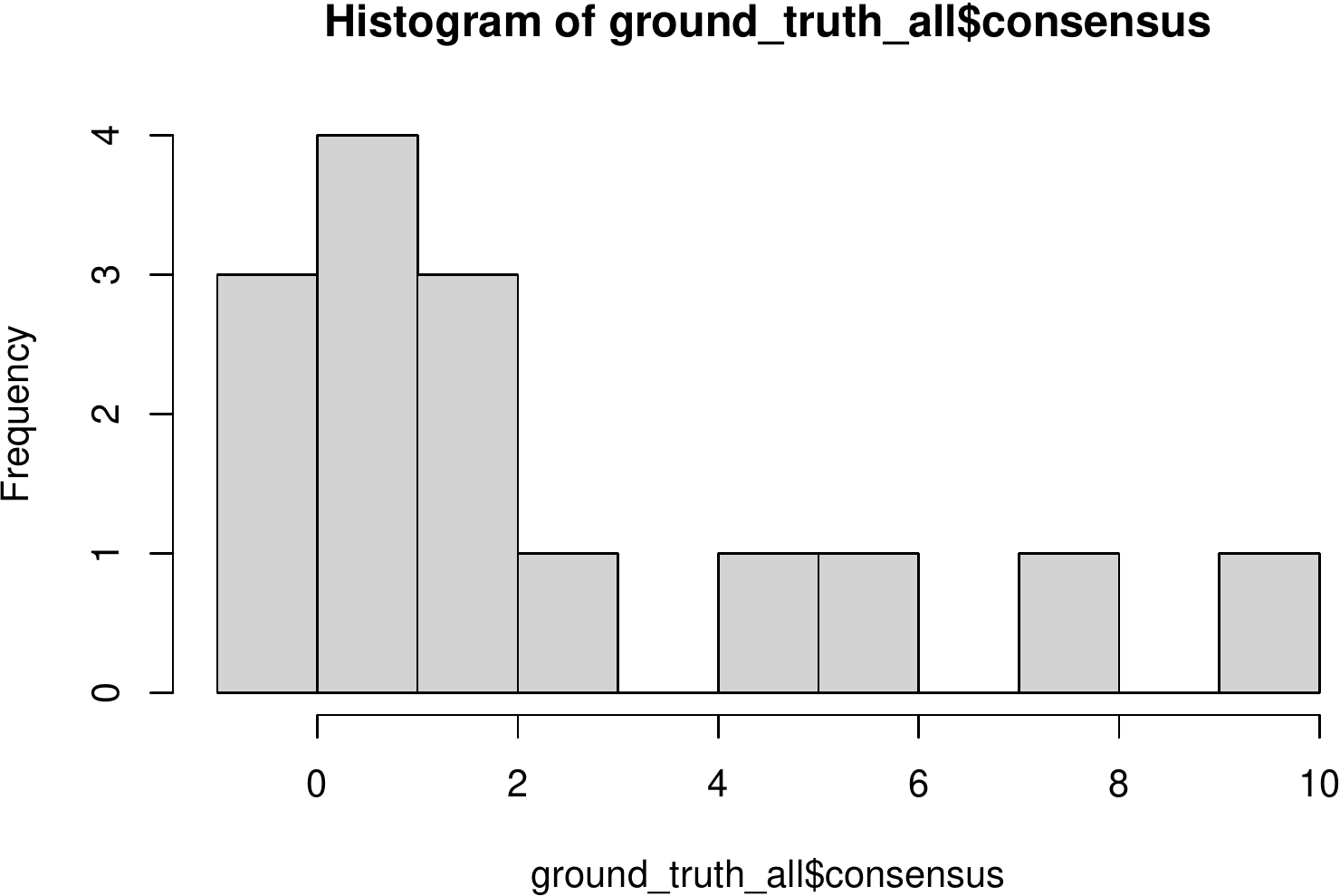} \caption{Histogram of the ground truth consensus for all 15 projects.}\label{fig:gt-label-hist}
\end{figure}

Our datasets are problematic in that we have only few instances but many features. Also, we do not have an example for every possible ground truth.
For example, we have no project with a score of \texttt{4} or \texttt{7} (see figure \ref{fig:gt-label-hist}).
Through a number of measures, however, we can alleviate all these problems to a satisfactory degree.

The first we will do is to synthetically oversample our dataset using SMOTE for regression (Torgo et al. 2013; Branco, Ribeiro, and Torgo 2016).
SMOTE is a well-established technique that has been proven to significantly increase model performance.
Since our label is numeric (we do regression), we will have to oversample instances such that new data points are generated that can be imputed and fill up the numeric labels.
While ordinary SMOTE cannot oversample unseen classes, the technique works differently for regression and allows us to generate synthetic samples with a previously unseen numeric label.
To reduce the amount of features, we will eliminate those with zero or near-zero variance.
Then, optionally, we further reduce the feature space's dimensionality by applying a principal component analysis (PCA).

\begin{Shaded}
\begin{Highlighting}[]
\CommentTok{\#\textquotesingle{} This function creates new instances for a given numeric label.}
\NormalTok{oversample\_y }\OtherTok{\textless{}{-}} \ControlFlowTok{function}\NormalTok{(dataset, num, lab) \{}
\NormalTok{  new\_ds }\OtherTok{\textless{}{-}} \ConstantTok{NULL}
\NormalTok{  i }\OtherTok{\textless{}{-}} \DecValTok{0}
  \ControlFlowTok{while}\NormalTok{ (}\ConstantTok{TRUE}\NormalTok{) \{}
    \FunctionTok{set.seed}\NormalTok{(i)}
\NormalTok{    i }\OtherTok{\textless{}{-}}\NormalTok{ i }\SpecialCharTok{+} \DecValTok{1}
\NormalTok{    temp }\OtherTok{\textless{}{-}}\NormalTok{ UBL}\SpecialCharTok{::}\FunctionTok{SmoteRegress}\NormalTok{(}\AttributeTok{form =}\NormalTok{ gt }\SpecialCharTok{\textasciitilde{}}\NormalTok{ ., }\AttributeTok{dat =}\NormalTok{ dataset, }\AttributeTok{thr.rel =} \DecValTok{1}\SpecialCharTok{/}\DecValTok{15}\NormalTok{)}
\NormalTok{    temp}\SpecialCharTok{$}\NormalTok{gt }\OtherTok{\textless{}{-}} \FunctionTok{round}\NormalTok{(temp}\SpecialCharTok{$}\NormalTok{gt)}
\NormalTok{    temp }\OtherTok{\textless{}{-}}\NormalTok{ temp[temp}\SpecialCharTok{$}\NormalTok{gt }\SpecialCharTok{==}\NormalTok{ lab, ]}
    \ControlFlowTok{if}\NormalTok{ (}\FunctionTok{nrow}\NormalTok{(temp) }\SpecialCharTok{\textgreater{}} \DecValTok{0}\NormalTok{) \{}
\NormalTok{      new\_ds }\OtherTok{\textless{}{-}} \FunctionTok{rbind}\NormalTok{(new\_ds, }\FunctionTok{head}\NormalTok{(temp, }\DecValTok{1}\NormalTok{))}
\NormalTok{    \}}
    \ControlFlowTok{if}\NormalTok{ (}\FunctionTok{is.data.frame}\NormalTok{(new\_ds) }\SpecialCharTok{\&\&} \FunctionTok{nrow}\NormalTok{(new\_ds) }\SpecialCharTok{==}\NormalTok{ num) \{}
      \ControlFlowTok{break}
\NormalTok{    \}}
\NormalTok{  \}}
\NormalTok{  new\_ds}
\NormalTok{\}}
\end{Highlighting}
\end{Shaded}

We will also create another short function that will balance all numeric labels in a given dataset.

\begin{Shaded}
\begin{Highlighting}[]
\NormalTok{balance\_num\_labels }\OtherTok{\textless{}{-}} \ControlFlowTok{function}\NormalTok{(dataset, }\AttributeTok{num =} \DecValTok{10}\NormalTok{) \{}
\NormalTok{  new\_ds }\OtherTok{\textless{}{-}} \ConstantTok{NULL}

  \ControlFlowTok{for}\NormalTok{ (i }\ControlFlowTok{in} \DecValTok{0}\SpecialCharTok{:}\DecValTok{10}\NormalTok{) \{}
    \CommentTok{\# All possible numeric labels}
\NormalTok{    has\_num\_rows }\OtherTok{\textless{}{-}} \FunctionTok{nrow}\NormalTok{(dataset[dataset}\SpecialCharTok{$}\NormalTok{gt }\SpecialCharTok{==}\NormalTok{ i, ])}
\NormalTok{    req\_rows }\OtherTok{\textless{}{-}}\NormalTok{ num }\SpecialCharTok{{-}}\NormalTok{ has\_num\_rows}
    \ControlFlowTok{if}\NormalTok{ (req\_rows }\SpecialCharTok{\textgreater{}} \DecValTok{0}\NormalTok{) \{}
\NormalTok{      new\_ds }\OtherTok{\textless{}{-}} \FunctionTok{suppressWarnings}\NormalTok{(\{}
        \FunctionTok{rbind}\NormalTok{(new\_ds, }\FunctionTok{oversample\_y}\NormalTok{(}\AttributeTok{dataset =}\NormalTok{ dataset, }\AttributeTok{num =}\NormalTok{ req\_rows, }\AttributeTok{lab =}\NormalTok{ i))}
\NormalTok{      \})}
\NormalTok{    \}}
\NormalTok{  \}}

  \ControlFlowTok{if}\NormalTok{ (}\SpecialCharTok{!}\FunctionTok{is.null}\NormalTok{(new\_ds))}
    \FunctionTok{rbind}\NormalTok{(dataset, new\_ds) }\ControlFlowTok{else}\NormalTok{ dataset}
\NormalTok{\}}
\end{Highlighting}
\end{Shaded}

And here we create the oversampled datasets.

\begin{Shaded}
\begin{Highlighting}[]
\NormalTok{min\_rows }\OtherTok{\textless{}{-}} \DecValTok{15}  \CommentTok{\# Number of rows we want to end up with per numeric label}

\NormalTok{dataset\_sc\_oversampled }\OtherTok{\textless{}{-}} \FunctionTok{cbind}\NormalTok{(dataset\_sc, }\FunctionTok{data.frame}\NormalTok{(}\AttributeTok{gt =}\NormalTok{ ground\_truth\_all}\SpecialCharTok{$}\NormalTok{consensus))}
\NormalTok{dataset\_sc\_oversampled }\OtherTok{\textless{}{-}} \FunctionTok{balance\_num\_labels}\NormalTok{(}\AttributeTok{dataset =}\NormalTok{ dataset\_sc\_oversampled, }\AttributeTok{num =}\NormalTok{ min\_rows)}

\NormalTok{dataset\_sc\_cdf\_oversampled }\OtherTok{\textless{}{-}} \FunctionTok{cbind}\NormalTok{(dataset\_sc\_cdf, }\FunctionTok{data.frame}\NormalTok{(}\AttributeTok{gt =}\NormalTok{ ground\_truth\_all}\SpecialCharTok{$}\NormalTok{consensus))}
\NormalTok{dataset\_sc\_cdf\_oversampled }\OtherTok{\textless{}{-}} \FunctionTok{balance\_num\_labels}\NormalTok{(}\AttributeTok{dataset =}\NormalTok{ dataset\_sc\_cdf\_oversampled,}
  \AttributeTok{num =}\NormalTok{ min\_rows)}

\NormalTok{dataset\_it\_oversampled }\OtherTok{\textless{}{-}} \FunctionTok{cbind}\NormalTok{(dataset\_it, }\FunctionTok{data.frame}\NormalTok{(}\AttributeTok{gt =}\NormalTok{ ground\_truth\_all}\SpecialCharTok{$}\NormalTok{consensus))}
\NormalTok{dataset\_it\_oversampled }\OtherTok{\textless{}{-}} \FunctionTok{balance\_num\_labels}\NormalTok{(}\AttributeTok{dataset =}\NormalTok{ dataset\_it\_oversampled, }\AttributeTok{num =}\NormalTok{ min\_rows)}
\end{Highlighting}
\end{Shaded}

So, as we see in the following overview for all three datasets, we have the same amount of instances for each numeric label (\(0\) through \(10\)):

\begin{Shaded}
\begin{Highlighting}[]
\FunctionTok{rbind}\NormalTok{(}\FunctionTok{table}\NormalTok{(dataset\_sc\_oversampled}\SpecialCharTok{$}\NormalTok{gt), }\FunctionTok{table}\NormalTok{(dataset\_sc\_cdf\_oversampled}\SpecialCharTok{$}\NormalTok{gt), }\FunctionTok{table}\NormalTok{(dataset\_it\_oversampled}\SpecialCharTok{$}\NormalTok{gt))}
\end{Highlighting}
\end{Shaded}

\begin{verbatim}
##       0  1  2  3  4  5  6  7  8  9 10
## [1,] 15 15 15 15 15 15 15 15 15 15 15
## [2,] 15 15 15 15 15 15 15 15 15 15 15
## [3,] 15 15 15 15 15 15 15 15 15 15 15
\end{verbatim}

\hypertarget{adaptive-training}{%
\subsubsection{Adaptive Training}\label{adaptive-training}}

We define a function that can train one of the chosen models adaptively (using an outer grid search).
This function mainly handles models that are trainable with caret, but there are two exceptions:
When the selected model is \texttt{gbm}, we use a custom grid to accommodate small and very small sample sizes, as the default grid warrants for much larger training sets.
When the selected model is \texttt{rankModel}, we will actually test a new model of our own that uses quantile normalization of the inputs and an inverse rank transform of the outputs, with some non-linearity in between\footnote{See the Rank Model repository at \url{https://github.com/MrShoenel/R-rank-model}.}.
This model has been proven efficient with very small sample sizes.

\begin{Shaded}
\begin{Highlighting}[]
\NormalTok{adaptive\_training\_caret }\OtherTok{\textless{}{-}} \ControlFlowTok{function}\NormalTok{(}
\NormalTok{    org\_dataset,}
\NormalTok{    seeds, }\CommentTok{\# number of seeds equals number of repeats}
    \AttributeTok{model\_type =} \FunctionTok{c}\NormalTok{(}\StringTok{"avNNet"}\NormalTok{, }\StringTok{"gbm"}\NormalTok{, }\StringTok{"glm"}\NormalTok{, }\StringTok{"M5"}\NormalTok{, }\StringTok{"nnet"}\NormalTok{, }\StringTok{"rankModel"}\NormalTok{, }\StringTok{"rf"}\NormalTok{, }\StringTok{"svmPoly"}\NormalTok{, }\StringTok{"treebag"}\NormalTok{, }\StringTok{"xgbTree"}\NormalTok{),}
    \AttributeTok{num\_train =} \DecValTok{5}\NormalTok{,}
    \AttributeTok{num\_valid =} \DecValTok{50}\NormalTok{,}
    \AttributeTok{do\_pca =} \ConstantTok{TRUE}\NormalTok{,}
    \AttributeTok{return\_num\_comps =} \ConstantTok{NULL}\NormalTok{,}
    \AttributeTok{num\_caret\_repeats =} \DecValTok{10}\NormalTok{,}
    \AttributeTok{return\_instance =} \ConstantTok{FALSE}
\NormalTok{) \{}
\NormalTok{  model\_type }\OtherTok{\textless{}{-}} \FunctionTok{match.arg}\NormalTok{(model\_type)}
  
\NormalTok{  cn }\OtherTok{\textless{}{-}} \FunctionTok{colnames}\NormalTok{(org\_dataset)}
\NormalTok{  cn\_x }\OtherTok{\textless{}{-}}\NormalTok{ cn[cn }\SpecialCharTok{!=} \StringTok{"gt"}\NormalTok{]}
\NormalTok{  pre\_proc\_method }\OtherTok{\textless{}{-}} \FunctionTok{c}\NormalTok{(}\StringTok{"nzv"}\NormalTok{)}
  \ControlFlowTok{if}\NormalTok{ (do\_pca) \{}
\NormalTok{    pre\_proc\_method }\OtherTok{\textless{}{-}} \FunctionTok{c}\NormalTok{(}\StringTok{"pca"}\NormalTok{, pre\_proc\_method)}
\NormalTok{  \}}
\NormalTok{  pre\_proc\_method }\OtherTok{\textless{}{-}} \FunctionTok{c}\NormalTok{(pre\_proc\_method, }\StringTok{"center"}\NormalTok{, }\StringTok{"scale"}\NormalTok{)}
  
\NormalTok{  res }\OtherTok{\textless{}{-}} \ConstantTok{NULL}

  
  \ControlFlowTok{for}\NormalTok{ (seed }\ControlFlowTok{in}\NormalTok{ seeds) \{}
    \FunctionTok{set.seed}\NormalTok{(seed)}
\NormalTok{    idx }\OtherTok{\textless{}{-}} \FunctionTok{sample}\NormalTok{(}\AttributeTok{x =} \FunctionTok{rownames}\NormalTok{(org\_dataset), }\AttributeTok{size =} \FunctionTok{nrow}\NormalTok{(org\_dataset))}
    
\NormalTok{    df\_train }\OtherTok{\textless{}{-}}\NormalTok{ org\_dataset[idx[}\DecValTok{1}\SpecialCharTok{:}\NormalTok{num\_train],]}
    \CommentTok{\# Fit the pre{-}processor on the training data, then...}
\NormalTok{    pre\_proc }\OtherTok{\textless{}{-}}\NormalTok{ caret}\SpecialCharTok{::}\FunctionTok{preProcess}\NormalTok{(}\AttributeTok{x =}\NormalTok{ df\_train[, cn\_x], }\AttributeTok{method =}\NormalTok{ pre\_proc\_method, }\AttributeTok{thresh =} \FloatTok{0.975}\NormalTok{)}
    
    \CommentTok{\# Transform training AND validation data with it:}
\NormalTok{    df\_train }\OtherTok{\textless{}{-}}\NormalTok{ stats}\SpecialCharTok{::}\FunctionTok{predict}\NormalTok{(pre\_proc, }\AttributeTok{newdata =}\NormalTok{ df\_train)}
\NormalTok{    df\_valid }\OtherTok{\textless{}{-}}\NormalTok{ org\_dataset[idx[(num\_train }\SpecialCharTok{+} \DecValTok{1}\NormalTok{)}\SpecialCharTok{:}\NormalTok{(num\_train }\SpecialCharTok{+}\NormalTok{ num\_valid)],]}
\NormalTok{    df\_valid }\OtherTok{\textless{}{-}}\NormalTok{ stats}\SpecialCharTok{::}\FunctionTok{predict}\NormalTok{(pre\_proc, }\AttributeTok{newdata =}\NormalTok{ df\_valid)}
    
    \CommentTok{\# Colnames change to PC1, PC2, ..., if PCA, otherwise they will be a subset}
    \CommentTok{\# of the original column names (because of (N)ZV):}
\NormalTok{    use\_cn\_x }\OtherTok{\textless{}{-}} \ControlFlowTok{if}\NormalTok{ (do\_pca) }\FunctionTok{paste0}\NormalTok{(}\StringTok{"PC"}\NormalTok{, }\DecValTok{1}\SpecialCharTok{:}\NormalTok{(}\FunctionTok{ncol}\NormalTok{(df\_train) }\SpecialCharTok{{-}} \DecValTok{1}\NormalTok{)) }\ControlFlowTok{else} \FunctionTok{colnames}\NormalTok{(df\_train)}
\NormalTok{    use\_cn\_x }\OtherTok{\textless{}{-}}\NormalTok{ use\_cn\_x[use\_cn\_x }\SpecialCharTok{!=} \StringTok{"gt"}\NormalTok{]}
    
    \ControlFlowTok{if}\NormalTok{ (}\SpecialCharTok{!}\FunctionTok{is.null}\NormalTok{(return\_num\_comps) }\SpecialCharTok{\&\&}\NormalTok{ return\_num\_comps) \{}
      \ControlFlowTok{if}\NormalTok{ (}\SpecialCharTok{!}\NormalTok{do\_pca) \{}
        \FunctionTok{stop}\NormalTok{(}\StringTok{"Can only return number of components if doing PCA!"}\NormalTok{)}
\NormalTok{      \}}
      \FunctionTok{return}\NormalTok{(}\FunctionTok{length}\NormalTok{(use\_cn\_x))}
\NormalTok{    \}}
    
\NormalTok{    pred\_train }\OtherTok{\textless{}{-}} \FunctionTok{c}\NormalTok{()}
\NormalTok{    pred\_valid }\OtherTok{\textless{}{-}} \FunctionTok{c}\NormalTok{()}
    \FunctionTok{tryCatch}\NormalTok{(\{}
      \ControlFlowTok{if}\NormalTok{ (model\_type }\SpecialCharTok{==} \StringTok{"rankModel"}\NormalTok{) \{}
        \CommentTok{\# Those need sourcing again.}
        \FunctionTok{source}\NormalTok{(}\AttributeTok{file =} \StringTok{"../helpers\_rank{-}model.R"}\NormalTok{)}
        \FunctionTok{source}\NormalTok{(}\AttributeTok{file =} \StringTok{"../models/rank{-}model.R"}\NormalTok{)}
        
        \CommentTok{\# We\textquotesingle{}re gonna have a small grid, so that each model is fit a few times.}
\NormalTok{        use\_grid }\OtherTok{\textless{}{-}} \FunctionTok{expand.grid}\NormalTok{(}\FunctionTok{list}\NormalTok{(}
          \AttributeTok{cdf\_type =} \FunctionTok{c}\NormalTok{(}\StringTok{"auto"}\NormalTok{, }\StringTok{"gauss"}\NormalTok{, }\StringTok{"ecdf"}\NormalTok{),}
          \AttributeTok{attempt =} \DecValTok{1}\SpecialCharTok{:}\DecValTok{5} \CommentTok{\# This is basically a dummy (well, it\textquotesingle{}s the number of repeats).}
\NormalTok{        ))}
\NormalTok{        use\_grid}\SpecialCharTok{$}\NormalTok{cdf\_type }\OtherTok{\textless{}{-}} \FunctionTok{as.character}\NormalTok{(use\_grid}\SpecialCharTok{$}\NormalTok{cdf\_type)}
        
\NormalTok{        best\_rm }\OtherTok{\textless{}{-}} \FloatTok{1e30}
        \ControlFlowTok{for}\NormalTok{ (rn }\ControlFlowTok{in} \FunctionTok{rownames}\NormalTok{(use\_grid)) \{}
\NormalTok{          row }\OtherTok{\textless{}{-}}\NormalTok{ use\_grid[rn,]}
\NormalTok{          temp1 }\OtherTok{\textless{}{-}} \FunctionTok{create\_model}\NormalTok{(}\AttributeTok{df\_train =}\NormalTok{ df\_train, }\AttributeTok{x\_cols =}\NormalTok{ use\_cn\_x, }\AttributeTok{y\_col =} \StringTok{"gt"}\NormalTok{, }\AttributeTok{cdf\_type =}\NormalTok{ row}\SpecialCharTok{$}\NormalTok{cdf\_type)}
\NormalTok{          tempf }\OtherTok{\textless{}{-}} \ControlFlowTok{function}\NormalTok{(x) }\FunctionTok{model\_loss}\NormalTok{(}\AttributeTok{model =}\NormalTok{ temp1, }\AttributeTok{x =}\NormalTok{ x, }\AttributeTok{df =}\NormalTok{ df\_train, }\AttributeTok{y\_col =} \StringTok{"gt"}\NormalTok{)}
          
          \CommentTok{\# Let\textquotesingle{}s fit the model!}
\NormalTok{          num\_params }\OtherTok{\textless{}{-}} \DecValTok{2} \SpecialCharTok{+} \DecValTok{3} \SpecialCharTok{*} \FunctionTok{length}\NormalTok{(use\_cn\_x)}
\NormalTok{          opt\_res }\OtherTok{\textless{}{-}}\NormalTok{ nloptr}\SpecialCharTok{::}\FunctionTok{nloptr}\NormalTok{(}
            \AttributeTok{x0 =} \FunctionTok{runif}\NormalTok{(}\AttributeTok{n =}\NormalTok{ num\_params),}
            \AttributeTok{eval\_f =}\NormalTok{ tempf,}
            \AttributeTok{eval\_grad\_f =} \ControlFlowTok{function}\NormalTok{(x) pracma}\SpecialCharTok{::}\FunctionTok{grad}\NormalTok{(}\AttributeTok{f =}\NormalTok{ tempf, }\AttributeTok{x0 =}\NormalTok{ x),}
            \AttributeTok{lb =} \FunctionTok{rep}\NormalTok{(}\SpecialCharTok{{-}}\FloatTok{1e3}\NormalTok{, num\_params),}
            \AttributeTok{ub =} \FunctionTok{rep}\NormalTok{( }\FloatTok{1e3}\NormalTok{, num\_params),}
            \AttributeTok{opts =} \FunctionTok{list}\NormalTok{(}\AttributeTok{algorithm =} \StringTok{"NLOPT\_LD\_TNEWTON"}\NormalTok{, }\AttributeTok{xtol\_rel=}\FloatTok{5e{-}4}\NormalTok{, }\AttributeTok{maxeval=}\DecValTok{250}\NormalTok{))}
          
\NormalTok{          temp\_pred\_valid }\OtherTok{\textless{}{-}} \FunctionTok{temp1}\NormalTok{(}\AttributeTok{x =}\NormalTok{ opt\_res}\SpecialCharTok{$}\NormalTok{solution, }\AttributeTok{df =}\NormalTok{ df\_valid)}
\NormalTok{          temp\_best\_rm }\OtherTok{\textless{}{-}}\NormalTok{ Metrics}\SpecialCharTok{::}\FunctionTok{rmse}\NormalTok{(}\AttributeTok{actual =}\NormalTok{ df\_valid}\SpecialCharTok{$}\NormalTok{gt, }\AttributeTok{predicted =}\NormalTok{ temp\_pred\_valid)}
          \ControlFlowTok{if}\NormalTok{ (temp\_best\_rm }\SpecialCharTok{\textless{}}\NormalTok{ best\_rm) \{}
\NormalTok{            best\_rm }\OtherTok{\textless{}{-}}\NormalTok{ temp\_best\_rm}
\NormalTok{            pred\_valid }\OtherTok{\textless{}{-}}\NormalTok{ temp\_pred\_valid}
\NormalTok{            pred\_train }\OtherTok{\textless{}{-}} \FunctionTok{temp1}\NormalTok{(}\AttributeTok{x =}\NormalTok{ opt\_res}\SpecialCharTok{$}\NormalTok{solution, }\AttributeTok{df =}\NormalTok{ df\_train)}
\NormalTok{          \}}
\NormalTok{        \}}
\NormalTok{      \} }\ControlFlowTok{else} \ControlFlowTok{if}\NormalTok{ (model\_type }\SpecialCharTok{==} \StringTok{"nnet"}\NormalTok{) \{}
\NormalTok{        use\_grid }\OtherTok{\textless{}{-}} \FunctionTok{expand.grid}\NormalTok{(}\FunctionTok{list}\NormalTok{(}
          \AttributeTok{hidden1 =} \FunctionTok{c}\NormalTok{(}\FunctionTok{seq}\NormalTok{(}\AttributeTok{from =} \DecValTok{1}\NormalTok{, }\AttributeTok{to =} \DecValTok{25}\NormalTok{, }\AttributeTok{by =} \DecValTok{3}\NormalTok{), }\DecValTok{35}\NormalTok{, }\DecValTok{50}\NormalTok{),}
          \AttributeTok{hidden2 =} \FunctionTok{c}\NormalTok{(}\ConstantTok{NA\_real\_}\NormalTok{, }\FunctionTok{seq}\NormalTok{(}\AttributeTok{from =} \DecValTok{1}\NormalTok{, }\AttributeTok{to =} \DecValTok{16}\NormalTok{, }\AttributeTok{by =} \DecValTok{3}\NormalTok{)),}
          \AttributeTok{act.fct =} \FunctionTok{c}\NormalTok{(}\StringTok{"sigmoid"}\NormalTok{, }\StringTok{"tanh"}\NormalTok{)}
\NormalTok{        ))}
        
\NormalTok{        best\_instance }\OtherTok{\textless{}{-}} \ConstantTok{NULL}
\NormalTok{        best\_nnet }\OtherTok{\textless{}{-}} \FloatTok{1e30}
        \ControlFlowTok{for}\NormalTok{ (rn }\ControlFlowTok{in} \FunctionTok{rownames}\NormalTok{(use\_grid)) \{}
          \FunctionTok{tryCatch}\NormalTok{(\{}
\NormalTok{            row }\OtherTok{\textless{}{-}}\NormalTok{ use\_grid[rn,]}
\NormalTok{            act.fct }\OtherTok{\textless{}{-}} \ControlFlowTok{if}\NormalTok{ (row}\SpecialCharTok{$}\NormalTok{act.fct }\SpecialCharTok{==} \StringTok{"tanh"}\NormalTok{) }\StringTok{"tanh"} \ControlFlowTok{else} \ControlFlowTok{function}\NormalTok{(x) }\DecValTok{1} \SpecialCharTok{/}\NormalTok{ (}\DecValTok{1} \SpecialCharTok{+} \FunctionTok{exp}\NormalTok{(}\SpecialCharTok{{-}}\NormalTok{x)) }\CommentTok{\# sigmoid}
\NormalTok{            hidden }\OtherTok{\textless{}{-}}\NormalTok{ row}\SpecialCharTok{$}\NormalTok{hidden1}
            \ControlFlowTok{if}\NormalTok{ (}\SpecialCharTok{!}\FunctionTok{is.na}\NormalTok{(row}\SpecialCharTok{$}\NormalTok{hidden2)) \{}
\NormalTok{              hidden }\OtherTok{\textless{}{-}} \FunctionTok{c}\NormalTok{(hidden, row}\SpecialCharTok{$}\NormalTok{hidden2)}
\NormalTok{            \}}
            
\NormalTok{            temp1 }\OtherTok{\textless{}{-}}\NormalTok{ neuralnet}\SpecialCharTok{::}\FunctionTok{neuralnet}\NormalTok{(}
              \AttributeTok{formula =}\NormalTok{ gt}\SpecialCharTok{\textasciitilde{}}\NormalTok{., }\AttributeTok{data =}\NormalTok{ df\_train, }\AttributeTok{act.fct =}\NormalTok{ act.fct,}
              \AttributeTok{hidden =}\NormalTok{ hidden, }\AttributeTok{threshold =} \FloatTok{5e{-}3}\NormalTok{, }\AttributeTok{linear.output =} \ConstantTok{TRUE}\NormalTok{, }\AttributeTok{rep =} \DecValTok{3}\NormalTok{, }\AttributeTok{lifesign =} \StringTok{"full"}\NormalTok{, }\AttributeTok{lifesign.step =} \DecValTok{1000}\NormalTok{)}
            
\NormalTok{            temp\_pred\_valid }\OtherTok{\textless{}{-}}\NormalTok{ stats}\SpecialCharTok{::}\FunctionTok{predict}\NormalTok{(temp1, df\_valid)}
\NormalTok{            temp\_best\_nnet }\OtherTok{\textless{}{-}}\NormalTok{ Metrics}\SpecialCharTok{::}\FunctionTok{rmse}\NormalTok{(}\AttributeTok{actual =}\NormalTok{ df\_valid}\SpecialCharTok{$}\NormalTok{gt, }\AttributeTok{predicted =}\NormalTok{ temp\_pred\_valid)}
            \ControlFlowTok{if}\NormalTok{ (temp\_best\_nnet }\SpecialCharTok{\textless{}}\NormalTok{ best\_nnet) \{}
\NormalTok{              best\_nnet }\OtherTok{\textless{}{-}}\NormalTok{ temp\_best\_nnet}
\NormalTok{              pred\_valid }\OtherTok{\textless{}{-}}\NormalTok{ temp\_pred\_valid}
\NormalTok{              pred\_train }\OtherTok{\textless{}{-}}\NormalTok{ stats}\SpecialCharTok{::}\FunctionTok{predict}\NormalTok{(temp1, df\_train)}
\NormalTok{              best\_instance }\OtherTok{\textless{}{-}}\NormalTok{ temp1}
\NormalTok{            \}}
\NormalTok{          \}, }\AttributeTok{error =} \ControlFlowTok{function}\NormalTok{(cond) \{}
            \FunctionTok{print}\NormalTok{(}\FunctionTok{paste}\NormalTok{(}\StringTok{"The model did not converge:"}\NormalTok{, cond))}
            \CommentTok{\# Do nothing}
\NormalTok{          \})}
\NormalTok{        \}}
        
        \ControlFlowTok{if}\NormalTok{ (best\_nnet }\SpecialCharTok{\textgreater{}=} \FloatTok{1e30}\NormalTok{) \{}
          \FunctionTok{stop}\NormalTok{(}\StringTok{"NNET fit not possible."}\NormalTok{)}
\NormalTok{        \}}
        \ControlFlowTok{if}\NormalTok{ (return\_instance) \{}
          \FunctionTok{return}\NormalTok{(}\FunctionTok{list}\NormalTok{(}
            \StringTok{"best\_instance"} \OtherTok{=}\NormalTok{ best\_instance,}
            \StringTok{"df\_train"} \OtherTok{=}\NormalTok{ df\_train,}
            \StringTok{"df\_valid"} \OtherTok{=}\NormalTok{ df\_valid}
\NormalTok{          ))}
\NormalTok{        \}}
\NormalTok{      \} }\ControlFlowTok{else} \ControlFlowTok{if}\NormalTok{ (model\_type }\SpecialCharTok{==} \StringTok{"gbm"}\NormalTok{) \{}
\NormalTok{        use\_grid }\OtherTok{\textless{}{-}} \FunctionTok{expand.grid}\NormalTok{(}\FunctionTok{list}\NormalTok{(}
          \AttributeTok{n.trees =} \FunctionTok{c}\NormalTok{(}\DecValTok{2}\SpecialCharTok{:}\DecValTok{10}\NormalTok{, }\FunctionTok{seq}\NormalTok{(}\AttributeTok{from =} \DecValTok{12}\NormalTok{, }\AttributeTok{to =} \DecValTok{30}\NormalTok{, }\AttributeTok{by =} \DecValTok{2}\NormalTok{), }\FunctionTok{seq}\NormalTok{(}\AttributeTok{from =} \DecValTok{35}\NormalTok{, }\AttributeTok{to =} \DecValTok{80}\NormalTok{, }\AttributeTok{by =} \DecValTok{5}\NormalTok{), }\DecValTok{90}\NormalTok{, }\DecValTok{100}\NormalTok{),}
          \AttributeTok{n.minobsinnode =} \DecValTok{1}\SpecialCharTok{:}\DecValTok{10}\NormalTok{,}
          \AttributeTok{bag.fraction =} \FunctionTok{seq}\NormalTok{(}\AttributeTok{from =} \FloatTok{0.1}\NormalTok{, }\AttributeTok{to =} \FloatTok{0.7}\NormalTok{, }\AttributeTok{by =} \FloatTok{0.2}\NormalTok{)}
\NormalTok{        ))}
        
\NormalTok{        best\_gbm }\OtherTok{\textless{}{-}} \FloatTok{1e30}
        \ControlFlowTok{for}\NormalTok{ (rn }\ControlFlowTok{in} \FunctionTok{rownames}\NormalTok{(use\_grid)) \{}
\NormalTok{          row }\OtherTok{\textless{}{-}}\NormalTok{ use\_grid[rn,]}
          \FunctionTok{tryCatch}\NormalTok{(\{}
\NormalTok{            temp1 }\OtherTok{\textless{}{-}}\NormalTok{ gbm}\SpecialCharTok{::}\FunctionTok{gbm}\NormalTok{(}
              \AttributeTok{formula =}\NormalTok{ gt}\SpecialCharTok{\textasciitilde{}}\NormalTok{., }\AttributeTok{distribution =} \StringTok{"gaussian"}\NormalTok{, }\AttributeTok{data =}\NormalTok{ df\_train, }\AttributeTok{n.cores =} \DecValTok{1}\NormalTok{,}
              \AttributeTok{n.trees =}\NormalTok{ row}\SpecialCharTok{$}\NormalTok{n.trees, }\AttributeTok{n.minobsinnode =}\NormalTok{ row}\SpecialCharTok{$}\NormalTok{n.minobsinnode,}
              \AttributeTok{cv.folds =}\NormalTok{ num\_caret\_repeats, }\AttributeTok{train.fraction =} \DecValTok{1} \SpecialCharTok{{-}}\NormalTok{ (}\DecValTok{1} \SpecialCharTok{/} \FunctionTok{nrow}\NormalTok{(df\_train))) }\CommentTok{\# Let\textquotesingle{}s do repeated, nested LOOCV}
            
\NormalTok{            temp\_pred\_valid }\OtherTok{\textless{}{-}}\NormalTok{ stats}\SpecialCharTok{::}\FunctionTok{predict}\NormalTok{(temp1, df\_valid)}
\NormalTok{            temp\_best\_gbm }\OtherTok{\textless{}{-}}\NormalTok{ Metrics}\SpecialCharTok{::}\FunctionTok{rmse}\NormalTok{(}\AttributeTok{actual =}\NormalTok{ df\_valid}\SpecialCharTok{$}\NormalTok{gt, }\AttributeTok{predicted =}\NormalTok{ temp\_pred\_valid)}
            \ControlFlowTok{if}\NormalTok{ (temp\_best\_gbm }\SpecialCharTok{\textless{}}\NormalTok{ best\_gbm) \{}
\NormalTok{              best\_gbm }\OtherTok{\textless{}{-}}\NormalTok{ temp\_best\_gbm}
\NormalTok{              pred\_valid }\OtherTok{\textless{}{-}}\NormalTok{ temp\_pred\_valid}
\NormalTok{              pred\_train }\OtherTok{\textless{}{-}}\NormalTok{ stats}\SpecialCharTok{::}\FunctionTok{predict}\NormalTok{(temp1, df\_train)}
\NormalTok{            \}}
\NormalTok{          \}, }\AttributeTok{error =} \ControlFlowTok{function}\NormalTok{(cond) \{}
            \CommentTok{\# Do nothing, silently swallow error (it\textquotesingle{}s handled by having NA{-}predictions)}
\NormalTok{          \})}
\NormalTok{        \}}
        
        \ControlFlowTok{if}\NormalTok{ (best\_gbm }\SpecialCharTok{\textgreater{}=} \FloatTok{1e30}\NormalTok{) \{}
          \FunctionTok{stop}\NormalTok{(}\StringTok{"GBM fit not possible."}\NormalTok{)}
\NormalTok{        \}}
\NormalTok{      \} }\ControlFlowTok{else}\NormalTok{ \{}
\NormalTok{        temp1 }\OtherTok{\textless{}{-}}\NormalTok{ caret}\SpecialCharTok{::}\FunctionTok{train}\NormalTok{(}
          \AttributeTok{x =}\NormalTok{ df\_train[, use\_cn\_x], }\AttributeTok{y =}\NormalTok{ df\_train[, }\StringTok{"gt"}\NormalTok{], }\AttributeTok{method =}\NormalTok{ model\_type,}
          \AttributeTok{trControl =}\NormalTok{ caret}\SpecialCharTok{::}\FunctionTok{trainControl}\NormalTok{(}\AttributeTok{method =} \StringTok{"LOOCV"}\NormalTok{, }\AttributeTok{number =}\NormalTok{ num\_caret\_repeats))}
\NormalTok{        pred\_train }\OtherTok{\textless{}{-}}\NormalTok{ kernlab}\SpecialCharTok{::}\FunctionTok{predict}\NormalTok{(temp1, df\_train[, use\_cn\_x])}
\NormalTok{        pred\_valid }\OtherTok{\textless{}{-}}\NormalTok{ kernlab}\SpecialCharTok{::}\FunctionTok{predict}\NormalTok{(temp1, df\_valid[, use\_cn\_x])}
\NormalTok{      \}}
\NormalTok{    \}, }\AttributeTok{error =} \ControlFlowTok{function}\NormalTok{(cond) \{}
      \FunctionTok{print}\NormalTok{(}\FunctionTok{paste}\NormalTok{(}\StringTok{"Training failed:"}\NormalTok{, cond))}
\NormalTok{      pred\_train }\OtherTok{\textless{}{-}} \FunctionTok{rep}\NormalTok{(}\ConstantTok{NA\_real\_}\NormalTok{, }\FunctionTok{nrow}\NormalTok{(df\_train))}
\NormalTok{      pred\_valid }\OtherTok{\textless{}{-}} \FunctionTok{rep}\NormalTok{(}\ConstantTok{NA\_real\_}\NormalTok{, }\FunctionTok{nrow}\NormalTok{(df\_valid))}
\NormalTok{    \})}
    
    
\NormalTok{    res }\OtherTok{\textless{}{-}} \FunctionTok{rbind}\NormalTok{(res, }\FunctionTok{data.frame}\NormalTok{(}
      \AttributeTok{seed =}\NormalTok{ seed,}
      \AttributeTok{do\_pca =}\NormalTok{ do\_pca,}
      \AttributeTok{num\_train =}\NormalTok{ num\_train,}
      \AttributeTok{num\_valid =}\NormalTok{ num\_valid,}
      \AttributeTok{model\_type =}\NormalTok{ model\_type,}
      \AttributeTok{num\_comps =} \ControlFlowTok{if}\NormalTok{ (do\_pca) }\FunctionTok{length}\NormalTok{(use\_cn\_x) }\ControlFlowTok{else} \ConstantTok{NA\_real\_}\NormalTok{,}
      
      \AttributeTok{mae\_train =}\NormalTok{ Metrics}\SpecialCharTok{::}\FunctionTok{mae}\NormalTok{(}\AttributeTok{actual =}\NormalTok{ df\_train}\SpecialCharTok{$}\NormalTok{gt, }\AttributeTok{predicted =}\NormalTok{ pred\_train),}
      \AttributeTok{mae\_valid =}\NormalTok{ Metrics}\SpecialCharTok{::}\FunctionTok{mae}\NormalTok{(}\AttributeTok{actual =}\NormalTok{ df\_valid}\SpecialCharTok{$}\NormalTok{gt, }\AttributeTok{predicted =}\NormalTok{ pred\_valid),}
      \AttributeTok{rmse\_train =}\NormalTok{ Metrics}\SpecialCharTok{::}\FunctionTok{rmse}\NormalTok{(}\AttributeTok{actual =}\NormalTok{ df\_train}\SpecialCharTok{$}\NormalTok{gt, }\AttributeTok{predicted =}\NormalTok{ pred\_train),}
      \AttributeTok{rmse\_valid =}\NormalTok{ Metrics}\SpecialCharTok{::}\FunctionTok{rmse}\NormalTok{(}\AttributeTok{actual =}\NormalTok{ df\_valid}\SpecialCharTok{$}\NormalTok{gt, }\AttributeTok{predicted =}\NormalTok{ pred\_valid),}
      \AttributeTok{sd\_train =}\NormalTok{ stats}\SpecialCharTok{::}\FunctionTok{sd}\NormalTok{(}\AttributeTok{x =}\NormalTok{ df\_train}\SpecialCharTok{$}\NormalTok{gt }\SpecialCharTok{{-}}\NormalTok{ pred\_train),}
      \AttributeTok{sd\_valid =}\NormalTok{ stats}\SpecialCharTok{::}\FunctionTok{sd}\NormalTok{(}\AttributeTok{x =}\NormalTok{ df\_valid}\SpecialCharTok{$}\NormalTok{gt }\SpecialCharTok{{-}}\NormalTok{ pred\_valid)}
\NormalTok{    ))}
\NormalTok{  \}}
  
\NormalTok{  res}
\NormalTok{\}}
\end{Highlighting}
\end{Shaded}

And here we compute the grid (test each model on each dataset).
Attention: The following is expensive.

\begin{Shaded}
\begin{Highlighting}[]
\NormalTok{grid }\OtherTok{\textless{}{-}} \FunctionTok{expand.grid}\NormalTok{(}\FunctionTok{list}\NormalTok{(}\AttributeTok{seed =} \DecValTok{1}\SpecialCharTok{:}\DecValTok{25}\NormalTok{, }\AttributeTok{num\_train =} \DecValTok{2}\SpecialCharTok{:}\DecValTok{50}\NormalTok{, }\AttributeTok{do\_pca =} \FunctionTok{c}\NormalTok{(}\ConstantTok{TRUE}\NormalTok{, }\ConstantTok{FALSE}\NormalTok{)))}
\NormalTok{model\_types }\OtherTok{\textless{}{-}} \FunctionTok{c}\NormalTok{(}\StringTok{"avNNet"}\NormalTok{, }\StringTok{"gbm"}\NormalTok{, }\StringTok{"glm"}\NormalTok{, }\StringTok{"M5"}\NormalTok{, }\StringTok{"rf"}\NormalTok{, }\StringTok{"svmPoly"}\NormalTok{, }\StringTok{"treebag"}\NormalTok{, }\StringTok{"rankModel"}\NormalTok{)}
\NormalTok{use\_datasets }\OtherTok{\textless{}{-}} \FunctionTok{c}\NormalTok{(}\StringTok{"dataset\_it\_oversampled"}\NormalTok{, }\StringTok{"dataset\_sc\_oversampled"}\NormalTok{, }\StringTok{"dataset\_sc\_cdf\_oversampled"}\NormalTok{)}


\FunctionTok{library}\NormalTok{(foreach)}
\NormalTok{res\_grid }\OtherTok{\textless{}{-}} \FunctionTok{loadResultsOrCompute}\NormalTok{(}\AttributeTok{file =} \StringTok{"../results/rob{-}reg\_res{-}grid.rds"}\NormalTok{, }\AttributeTok{computeExpr =}\NormalTok{ \{}
\NormalTok{  res }\OtherTok{\textless{}{-}} \ConstantTok{NULL}

  \ControlFlowTok{for}\NormalTok{ (use\_dataset }\ControlFlowTok{in}\NormalTok{ use\_datasets) \{}
\NormalTok{    dataset }\OtherTok{\textless{}{-}} \FunctionTok{get}\NormalTok{(use\_dataset)}

    \ControlFlowTok{for}\NormalTok{ (model\_type }\ControlFlowTok{in}\NormalTok{ model\_types) \{}

\NormalTok{      the\_file }\OtherTok{\textless{}{-}} \FunctionTok{paste0}\NormalTok{(}\StringTok{"../results/rob{-}reg\_res{-}grid\_"}\NormalTok{, model\_type, }\StringTok{"\_"}\NormalTok{, use\_dataset,}
        \StringTok{".rds"}\NormalTok{)}
\NormalTok{      temp }\OtherTok{\textless{}{-}} \FunctionTok{as.data.frame}\NormalTok{(}\FunctionTok{loadResultsOrCompute}\NormalTok{(}\AttributeTok{file =}\NormalTok{ the\_file, }\AttributeTok{computeExpr =}\NormalTok{ \{}
\NormalTok{        temp1 }\OtherTok{\textless{}{-}} \FunctionTok{doWithParallelCluster}\NormalTok{(}\AttributeTok{numCores =} \FunctionTok{min}\NormalTok{(parallel}\SpecialCharTok{::}\FunctionTok{detectCores}\NormalTok{(),}
          \DecValTok{123}\NormalTok{), }\AttributeTok{expr =}\NormalTok{ \{}
\NormalTok{          foreach}\SpecialCharTok{::}\FunctionTok{foreach}\NormalTok{(}\AttributeTok{rn =} \FunctionTok{rownames}\NormalTok{(grid), }\AttributeTok{.combine =}\NormalTok{ rbind, }\AttributeTok{.inorder =} \ConstantTok{FALSE}\NormalTok{,}
          \AttributeTok{.verbose =} \ConstantTok{TRUE}\NormalTok{, }\AttributeTok{.export =} \FunctionTok{c}\NormalTok{(}\StringTok{"adaptive\_training\_caret"}\NormalTok{)) }\SpecialCharTok{\%dopar\%}
\NormalTok{          \{}
            \FunctionTok{tryCatch}\NormalTok{(\{}
\NormalTok{            row }\OtherTok{\textless{}{-}}\NormalTok{ grid[rn, ]}
            \FunctionTok{adaptive\_training\_caret}\NormalTok{(}\AttributeTok{org\_dataset =}\NormalTok{ dataset, }\AttributeTok{seeds =}\NormalTok{ row}\SpecialCharTok{$}\NormalTok{seed,}
              \AttributeTok{model\_type =}\NormalTok{ model\_type, }\AttributeTok{num\_train =}\NormalTok{ row}\SpecialCharTok{$}\NormalTok{num\_train, }\AttributeTok{do\_pca =}\NormalTok{ row}\SpecialCharTok{$}\NormalTok{do\_pca)}
\NormalTok{            \}, }\AttributeTok{error =} \ControlFlowTok{function}\NormalTok{(cond) \{}
            \ConstantTok{NULL}  \CommentTok{\# Return an empty result that will not disturb .combine}
\NormalTok{            \})}
\NormalTok{          \}}
\NormalTok{        \})}
        \FunctionTok{saveRDS}\NormalTok{(}\AttributeTok{object =}\NormalTok{ temp1, }\AttributeTok{file =}\NormalTok{ the\_file)}
\NormalTok{        temp1}
\NormalTok{      \}))}
\NormalTok{      temp}\SpecialCharTok{$}\NormalTok{dataset }\OtherTok{\textless{}{-}} \FunctionTok{rep}\NormalTok{(use\_dataset, }\FunctionTok{nrow}\NormalTok{(temp))}

\NormalTok{      res }\OtherTok{\textless{}{-}} \FunctionTok{rbind}\NormalTok{(res, temp)}
\NormalTok{    \}}
\NormalTok{  \}}

\NormalTok{  res}\SpecialCharTok{$}\NormalTok{dataset }\OtherTok{\textless{}{-}} \FunctionTok{factor}\NormalTok{(res}\SpecialCharTok{$}\NormalTok{dataset)}
\NormalTok{  res}
\NormalTok{\})}
\end{Highlighting}
\end{Shaded}

\begin{figure}[ht!]
\includegraphics{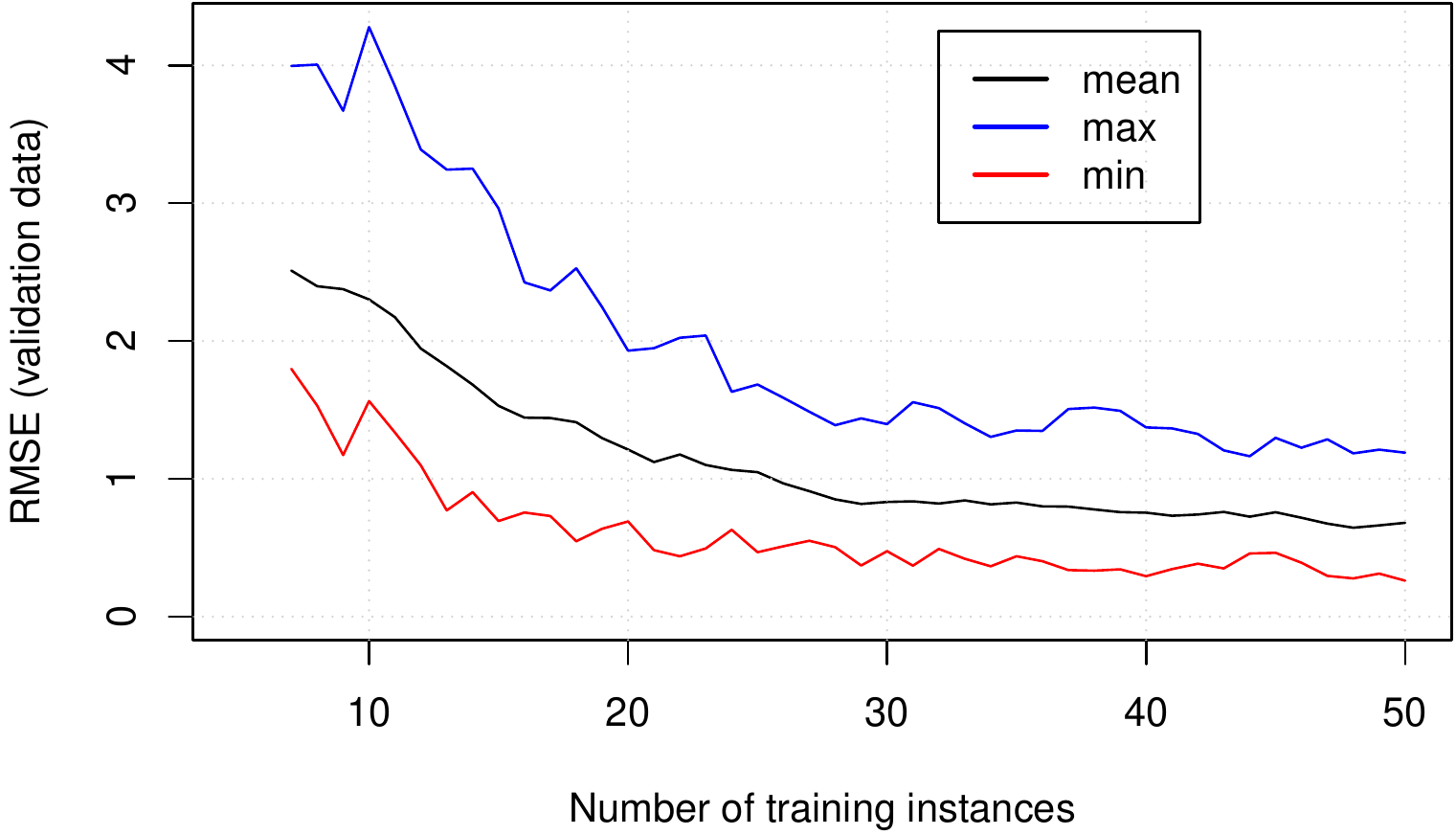} \caption{Robust regression using a Gradient Boosting Machine, PCA, and the oversampled issue-tracking dataset.}\label{fig:rob-reg-gbm1}
\end{figure}

Let's show some results for the Gradient Boosting Machine (figure \ref{fig:rob-reg-gbm1}).
It appears that the GBM needs at least \(7\) training examples to work at all.
With \(\approx25\) training examples, the expected generalization error (using the RMSE) is \(\approx1\), and with about \(15\) examples it is \(\approx1.5\), both of which might be acceptable in a real-world scenario.

\begin{figure}[ht!]
\includegraphics{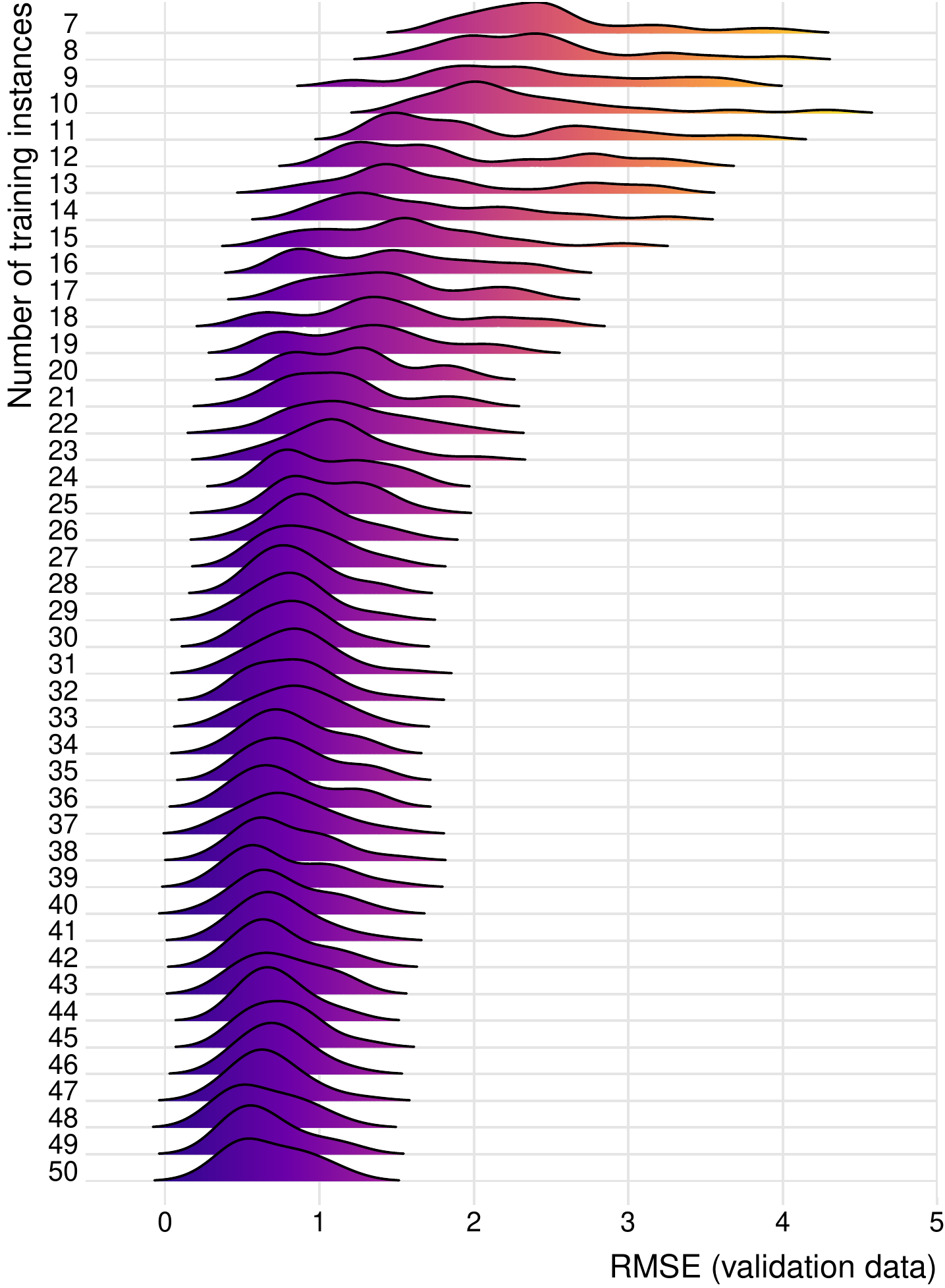} \caption{Ridge plot of distributions of RMSEs on validation data, dependent on the number of training samples.}\label{fig:rob-reg-gbm2}
\end{figure}

In figure \ref{fig:rob-reg-gbm2} we show the distribution of validation errors depending on the number of training instances (same setup as in figure \ref{fig:rob-reg-gbm1}).

We will also extract the number of components in case PCA was applied.

\begin{Shaded}
\begin{Highlighting}[]
\CommentTok{\# Do a PCA{-}only grid:}
\NormalTok{grid }\OtherTok{\textless{}{-}} \FunctionTok{expand.grid}\NormalTok{(}\FunctionTok{list}\NormalTok{(}
  \AttributeTok{seed =} \DecValTok{1}\SpecialCharTok{:}\DecValTok{25}\NormalTok{,}
  \AttributeTok{num\_train =} \DecValTok{2}\SpecialCharTok{:}\DecValTok{50}\NormalTok{,}
  \AttributeTok{use\_dataset =}\NormalTok{ use\_datasets}
\NormalTok{))}
\NormalTok{grid}\SpecialCharTok{$}\NormalTok{use\_dataset }\OtherTok{\textless{}{-}} \FunctionTok{as.character}\NormalTok{(grid}\SpecialCharTok{$}\NormalTok{use\_dataset)}

\NormalTok{res\_grid\_comps }\OtherTok{\textless{}{-}} \FunctionTok{loadResultsOrCompute}\NormalTok{(}\AttributeTok{file =} \StringTok{"../results/rob{-}reg\_res{-}grid\_comps.rds"}\NormalTok{, }\AttributeTok{computeExpr =}\NormalTok{ \{}
\NormalTok{  cl }\OtherTok{\textless{}{-}}\NormalTok{ parallel}\SpecialCharTok{::}\FunctionTok{makePSOCKcluster}\NormalTok{(}\FunctionTok{min}\NormalTok{(}\DecValTok{32}\NormalTok{, parallel}\SpecialCharTok{::}\FunctionTok{detectCores}\NormalTok{()))}
\NormalTok{  parallel}\SpecialCharTok{::}\FunctionTok{clusterExport}\NormalTok{(}\AttributeTok{cl =}\NormalTok{ cl, }\AttributeTok{varlist =}\NormalTok{ use\_datasets)}
  
  \FunctionTok{doWithParallelClusterExplicit}\NormalTok{(}\AttributeTok{cl =}\NormalTok{ cl, }\AttributeTok{stopCl =} \ConstantTok{TRUE}\NormalTok{, }\AttributeTok{expr =}\NormalTok{ \{}
\NormalTok{    foreach}\SpecialCharTok{::}\FunctionTok{foreach}\NormalTok{(}
      \AttributeTok{rn =} \FunctionTok{rownames}\NormalTok{(grid),}
      \AttributeTok{.combine =}\NormalTok{ rbind,}
      \AttributeTok{.inorder =} \ConstantTok{FALSE}\NormalTok{,}
      \AttributeTok{.export =} \FunctionTok{c}\NormalTok{(}\StringTok{"adaptive\_training\_caret"}\NormalTok{)}
\NormalTok{    ) }\SpecialCharTok{\%dopar\%}\NormalTok{ \{}
\NormalTok{      row }\OtherTok{\textless{}{-}}\NormalTok{ grid[rn,]}
\NormalTok{      row}\SpecialCharTok{$}\NormalTok{num\_comps }\OtherTok{\textless{}{-}} \FunctionTok{tryCatch}\NormalTok{(\{}
        \FunctionTok{adaptive\_training\_caret}\NormalTok{(}
          \AttributeTok{org\_dataset =} \FunctionTok{get}\NormalTok{(row}\SpecialCharTok{$}\NormalTok{use\_dataset), }\AttributeTok{seeds =}\NormalTok{ row}\SpecialCharTok{$}\NormalTok{seed, }\AttributeTok{model\_type =}\NormalTok{ model\_types[}\DecValTok{1}\NormalTok{], }\CommentTok{\# irrelevant}
          \AttributeTok{num\_train =}\NormalTok{ row}\SpecialCharTok{$}\NormalTok{num\_train, }\AttributeTok{do\_pca =} \ConstantTok{TRUE}\NormalTok{, }\AttributeTok{return\_num\_comps =} \ConstantTok{TRUE}\NormalTok{)}
\NormalTok{      \}, }\AttributeTok{error =} \ControlFlowTok{function}\NormalTok{(cond) }\ConstantTok{NA\_real\_}\NormalTok{)}
\NormalTok{      row}
\NormalTok{    \}}
\NormalTok{  \})}
\NormalTok{\})}
\FunctionTok{table}\NormalTok{(res\_grid\_comps}\SpecialCharTok{$}\NormalTok{num\_comps)}
\end{Highlighting}
\end{Shaded}

\begin{verbatim}
## 
##   2   3   4   5   6   7   8   9  10  11  12 
## 182 121 182 249 271 368 628 536 609 432  92
\end{verbatim}

\begin{figure}
\centering
\includegraphics{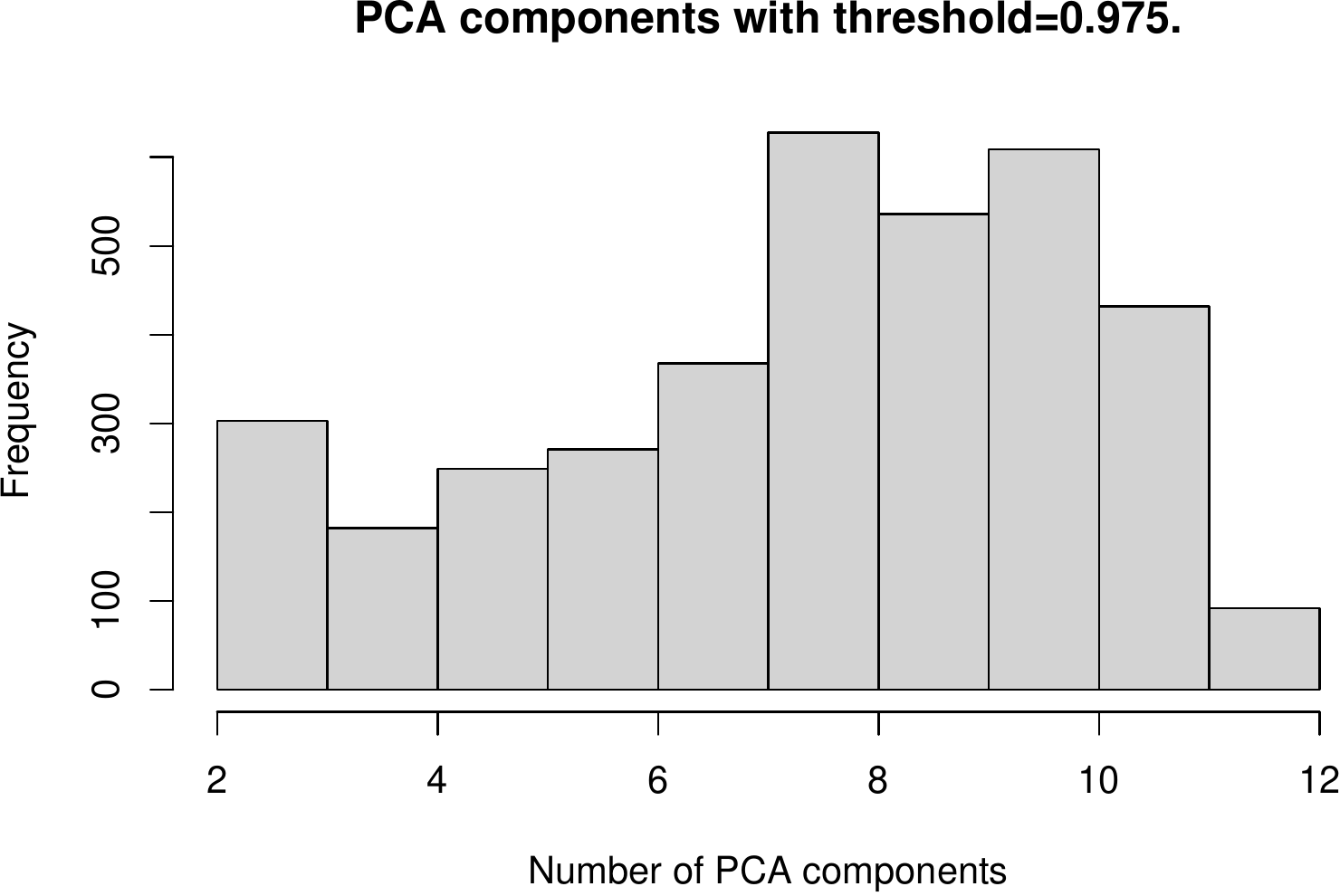}
\caption{\label{fig:num-comps-hist}Histogram over the number of components used in the grid search (threshold=0.975).}
\end{figure}

Above we see a table with how often a certain number of components was used, and in figure \ref{fig:num-comps-hist} it is illustrated in a histogram.

\hypertarget{analysis}{%
\subsubsection{Analysis}\label{analysis}}

Here we are going to analyze some of the results that we got (e.g., best pattern or model, expectations, lower/upper bounds, etc.).

\hypertarget{pca-vs.-non-pca}{%
\paragraph{PCA vs.~non-PCA}\label{pca-vs.-non-pca}}

Here we want to find out whether applying principal components analysis is useful or not.

\begin{table}

\caption{\label{tab:pca-comps-table}Mean, median, standard deviation, min, and max of the validation error, grouped by conditionally applying PCA.}
\centering
\begin{tabular}[t]{lrrrrr}
\toprule
do\_pca & RMSE\_mean & RMSE\_median & RMSE\_sd & RMSE\_min & RMSE\_max\\
\midrule
FALSE & 4.124120e+00 & 2.339129 & 3.823950e+01 & 0.0000000 & 2.993918e+03\\
TRUE & 3.147114e+39 & 1.829596 & 5.349446e+41 & 0.1541162 & 9.092957e+43\\
\bottomrule
\end{tabular}
\end{table}

\begin{figure}
\centering
\includegraphics{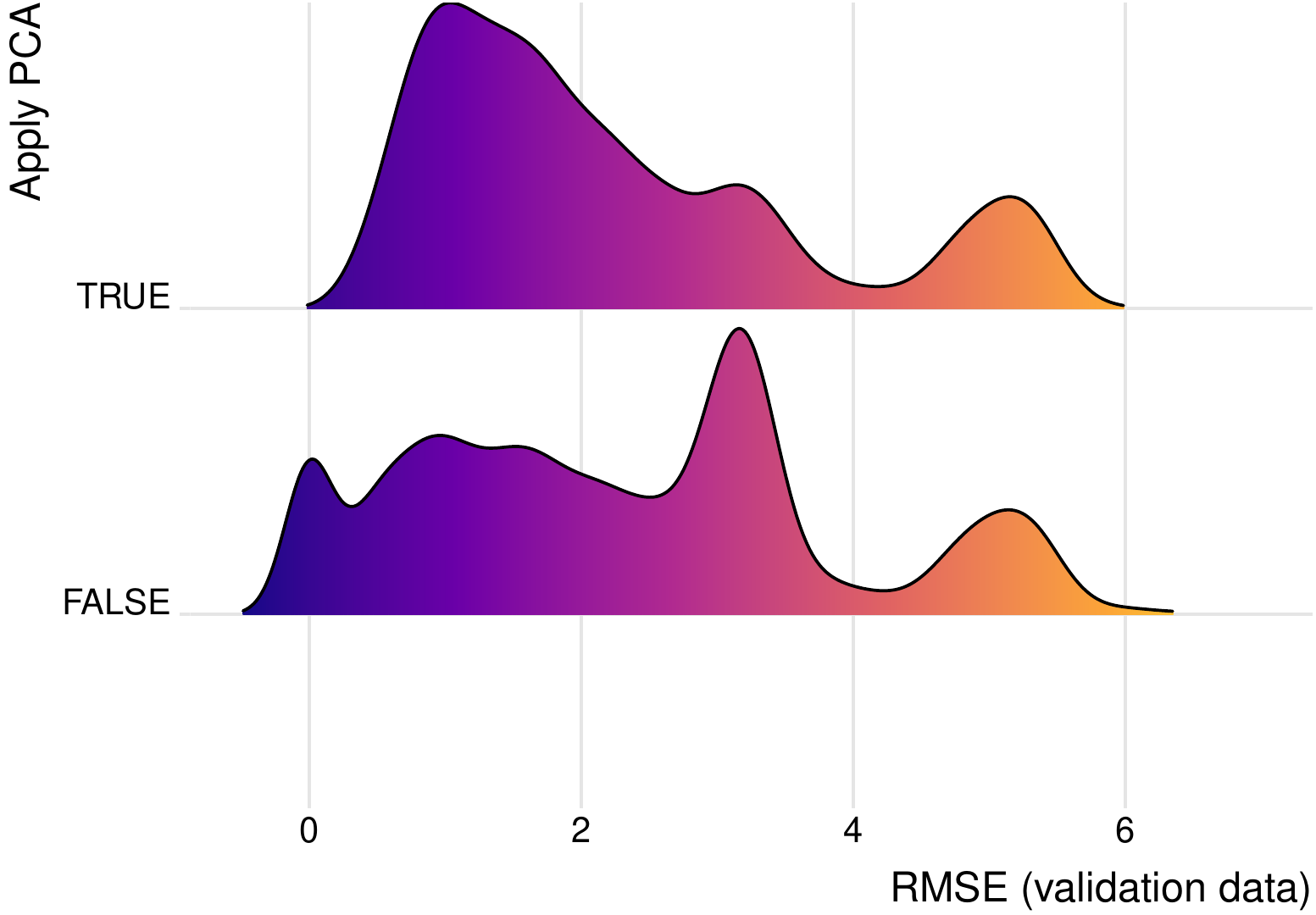}
\caption{\label{fig:pca-comps-ridge}Distributions of the validation error conditionally applying PCA.}
\end{figure}

Let's also show the distributions. In figure \ref{fig:pca-comps-ridge} we observe that applying PCA results in more smaller errors which are also more steadily distributed.
While the lowest possible validation error as of above table \ref{tab:pca-comps-table} is \(0\) for when not using PCA, this is likely due to the models overfitting, since without PCA we end up with many features and some models introduce extra degrees of freedom for each feature.
We shall therefore conclude that applying PCA is favorable.

\hypertarget{variable-importance-for-assessing-the-pattern}{%
\subsection{Variable Importance for Assessing the Pattern}\label{variable-importance-for-assessing-the-pattern}}

Obtaining variable importance allows us to understand which features are more or less important. Since in our case features are tied to a metric (area, correlation), a data kind (source code, issue tracking), an activity/metric (e.g., \texttt{REQ}, \texttt{SCD}), and a segment (one of ten equally long consecutive project phases), we can facilitate variable importance to learn about any of these (how important they are to predict a precise severity of the Fire Drill's presence).
We could potentially also (alternatively) learn about the quality of the ground truth. However, at this point we must assume that it is perfect as we have no evidence indicating otherwise.

Let's create a function that can give us a proper dataframe from the variable importance.

\begin{Shaded}
\begin{Highlighting}[]
\NormalTok{create\_variable\_importance }\OtherTok{\textless{}{-}} \ControlFlowTok{function}\NormalTok{(dataset) \{}
  \CommentTok{\# Note that PLS is very sensitive to (un{-})scaled data, so we need to make}
  \CommentTok{\# sure the data is properly pre{-}processed.}
\NormalTok{  cn }\OtherTok{\textless{}{-}} \FunctionTok{colnames}\NormalTok{(dataset)}
\NormalTok{  cn\_x }\OtherTok{\textless{}{-}}\NormalTok{ cn[cn }\SpecialCharTok{!=} \StringTok{"gt"}\NormalTok{]}
\NormalTok{  pre\_proc\_method }\OtherTok{\textless{}{-}} \FunctionTok{c}\NormalTok{(}\StringTok{"nzv"}\NormalTok{, }\StringTok{"center"}\NormalTok{, }\StringTok{"scale"}\NormalTok{)}
\NormalTok{  pre\_proc }\OtherTok{\textless{}{-}}\NormalTok{ caret}\SpecialCharTok{::}\FunctionTok{preProcess}\NormalTok{(}\AttributeTok{x =}\NormalTok{ dataset[, cn\_x], }\AttributeTok{method =}\NormalTok{ pre\_proc\_method)}
\NormalTok{  dataset }\OtherTok{\textless{}{-}}\NormalTok{ stats}\SpecialCharTok{::}\FunctionTok{predict}\NormalTok{(pre\_proc, }\AttributeTok{newdata =}\NormalTok{ dataset)}

\NormalTok{  model }\OtherTok{\textless{}{-}}\NormalTok{ caret}\SpecialCharTok{::}\FunctionTok{train}\NormalTok{(}\AttributeTok{x =}\NormalTok{ dataset[, cn\_x], }\AttributeTok{y =}\NormalTok{ dataset[, ]}\SpecialCharTok{$}\NormalTok{gt, }\AttributeTok{method =} \StringTok{"pls"}\NormalTok{,}
    \AttributeTok{trControl =}\NormalTok{ caret}\SpecialCharTok{::}\FunctionTok{trainControl}\NormalTok{(}\AttributeTok{method =} \StringTok{"LOOCV"}\NormalTok{, }\AttributeTok{number =} \DecValTok{100}\NormalTok{))}

\NormalTok{  vi }\OtherTok{\textless{}{-}}\NormalTok{ caret}\SpecialCharTok{::}\FunctionTok{varImp}\NormalTok{(model)}
\NormalTok{  df }\OtherTok{\textless{}{-}}\NormalTok{ vi}\SpecialCharTok{$}\NormalTok{importance}
\NormalTok{  res }\OtherTok{\textless{}{-}} \ConstantTok{NULL}

  \ControlFlowTok{for}\NormalTok{ (rn }\ControlFlowTok{in} \FunctionTok{rownames}\NormalTok{(df)) \{}
\NormalTok{    sp }\OtherTok{\textless{}{-}} \FunctionTok{strsplit}\NormalTok{(}\AttributeTok{x =}\NormalTok{ rn, }\AttributeTok{split =} \StringTok{"\_"}\NormalTok{)[[}\DecValTok{1}\NormalTok{]]}
\NormalTok{    res }\OtherTok{\textless{}{-}} \FunctionTok{rbind}\NormalTok{(res, }\FunctionTok{data.frame}\NormalTok{(}\AttributeTok{variable =}\NormalTok{ sp[}\DecValTok{1}\NormalTok{], }\AttributeTok{segment =} \FunctionTok{as.numeric}\NormalTok{(sp[}\DecValTok{2}\NormalTok{]),}
      \AttributeTok{metric =}\NormalTok{ sp[}\DecValTok{3}\NormalTok{], }\AttributeTok{vi =}\NormalTok{ df[rn, ]))}
\NormalTok{  \}}
\NormalTok{  res}\SpecialCharTok{$}\NormalTok{vi\_rel }\OtherTok{\textless{}{-}}\NormalTok{ res}\SpecialCharTok{$}\NormalTok{vi}\SpecialCharTok{/}\FunctionTok{sum}\NormalTok{(res}\SpecialCharTok{$}\NormalTok{vi)}
  \FunctionTok{list}\NormalTok{(}\AttributeTok{res =}\NormalTok{ res[}\FunctionTok{order}\NormalTok{(}\SpecialCharTok{{-}}\NormalTok{res}\SpecialCharTok{$}\NormalTok{vi), ], }\AttributeTok{vi =}\NormalTok{ vi)}
\NormalTok{\}}
\end{Highlighting}
\end{Shaded}

Let's compute the variable importance for each of our datasets:

\begin{Shaded}
\begin{Highlighting}[]
\NormalTok{vi\_sc }\OtherTok{\textless{}{-}} \FunctionTok{create\_variable\_importance}\NormalTok{(}\AttributeTok{dataset =} \FunctionTok{cbind}\NormalTok{(dataset\_sc, }\FunctionTok{data.frame}\NormalTok{(}\AttributeTok{gt =}\NormalTok{ ground\_truth\_all}\SpecialCharTok{$}\NormalTok{consensus)))}
\NormalTok{vi\_sc\_cdf }\OtherTok{\textless{}{-}} \FunctionTok{create\_variable\_importance}\NormalTok{(}\AttributeTok{dataset =} \FunctionTok{cbind}\NormalTok{(dataset\_sc\_cdf, }\FunctionTok{data.frame}\NormalTok{(}\AttributeTok{gt =}\NormalTok{ ground\_truth\_all}\SpecialCharTok{$}\NormalTok{consensus)))}
\NormalTok{vi\_it }\OtherTok{\textless{}{-}} \FunctionTok{create\_variable\_importance}\NormalTok{(}\AttributeTok{dataset =} \FunctionTok{cbind}\NormalTok{(dataset\_it, }\FunctionTok{data.frame}\NormalTok{(}\AttributeTok{gt =}\NormalTok{ ground\_truth\_all}\SpecialCharTok{$}\NormalTok{consensus)))}
\end{Highlighting}
\end{Shaded}

\begin{figure}[ht!]
\includegraphics{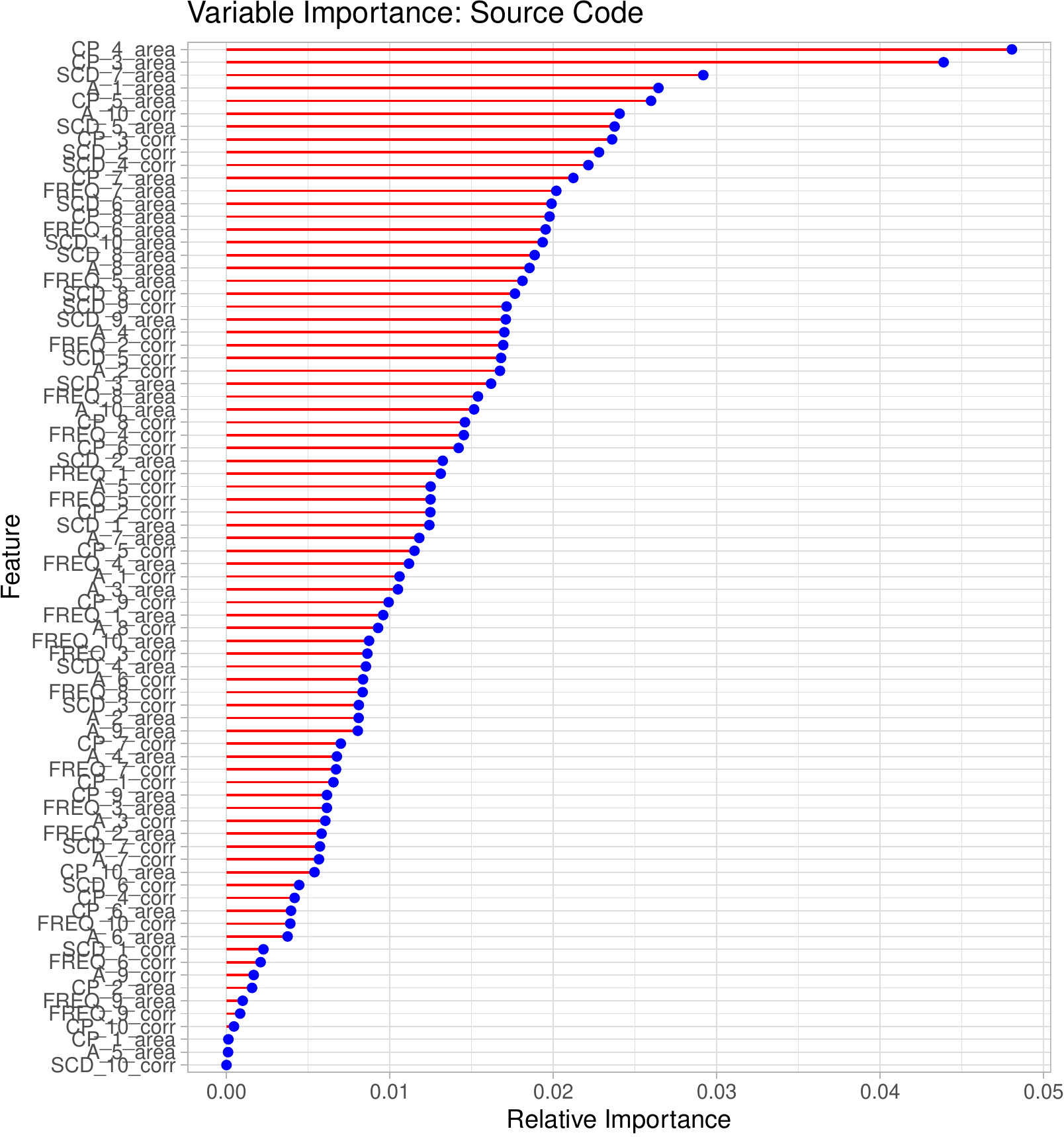} \caption{Variable importance for the source code dataset determined by partial least squares and cross validation.}\label{fig:varimp-sc}
\end{figure}

\begin{figure}[ht!]
\includegraphics{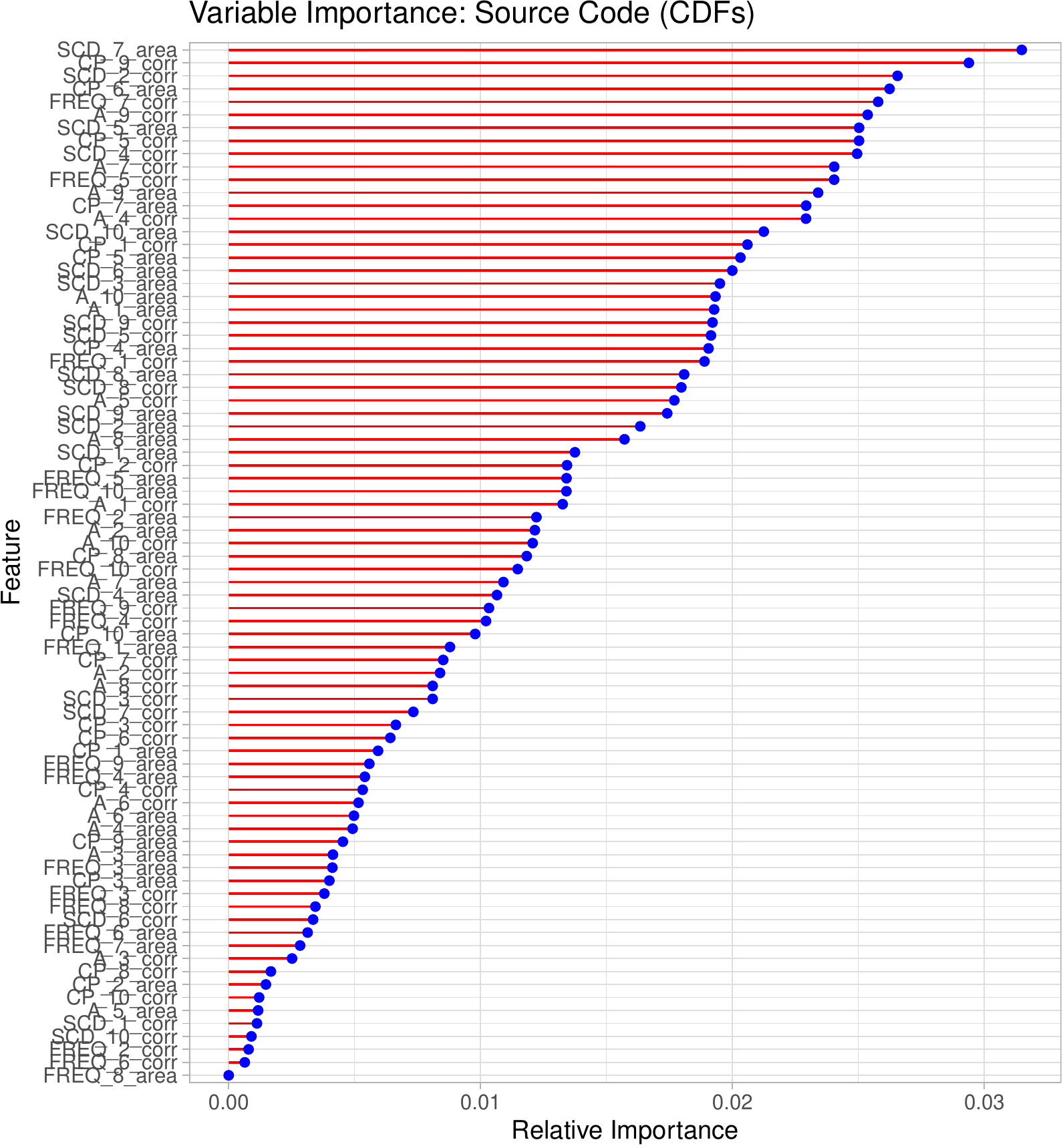} \caption{Variable importance for the source code (CDF) dataset determined by partial least squares and cross validation.}\label{fig:varimp-sc-cdf}
\end{figure}

\begin{figure}[ht!]
\includegraphics{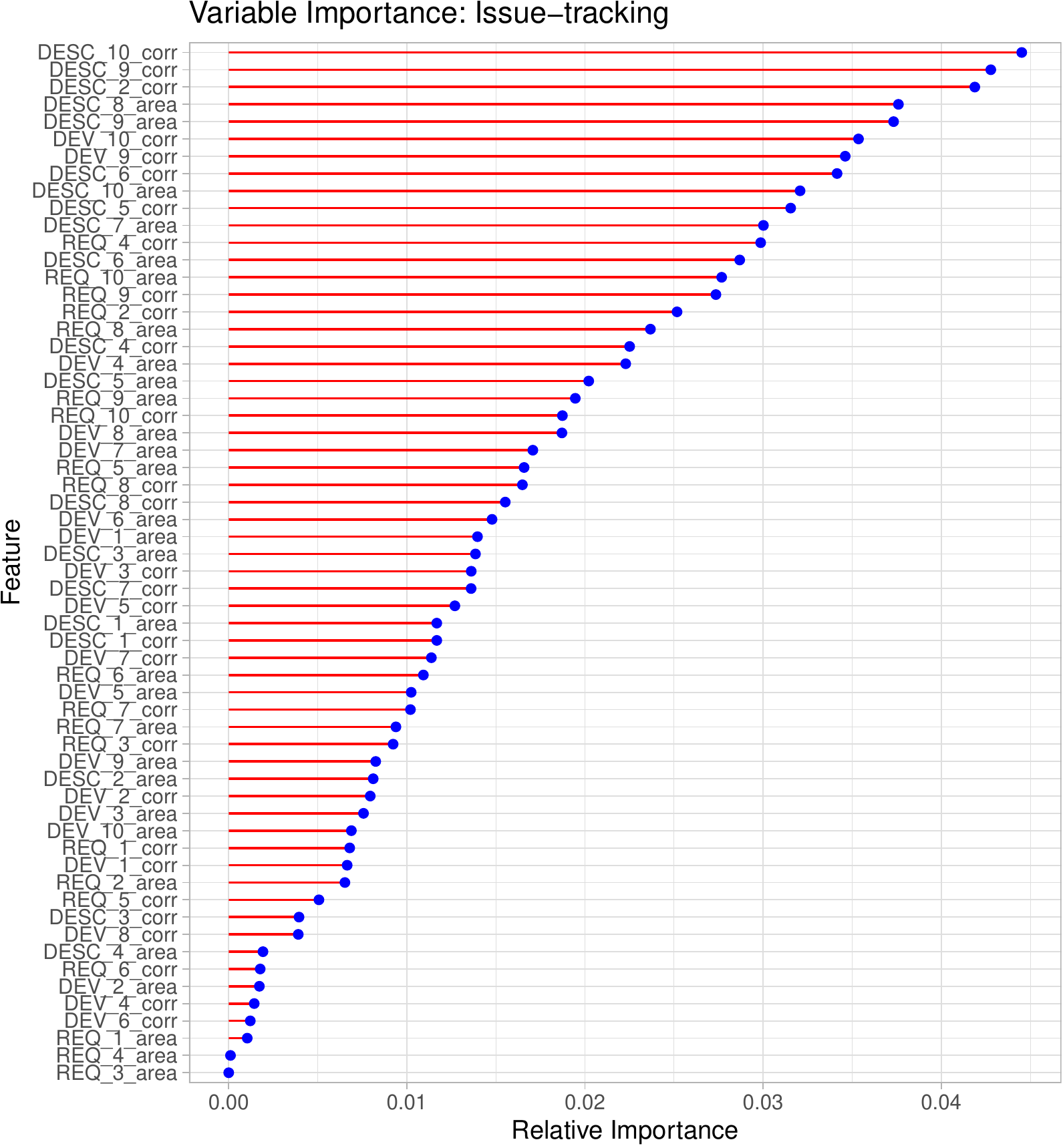} \caption{Variable importance for the issue-tracking dataset determined by partial least squares and cross validation.}\label{fig:varimp-it}
\end{figure}

\begin{figure}[ht!]
\includegraphics{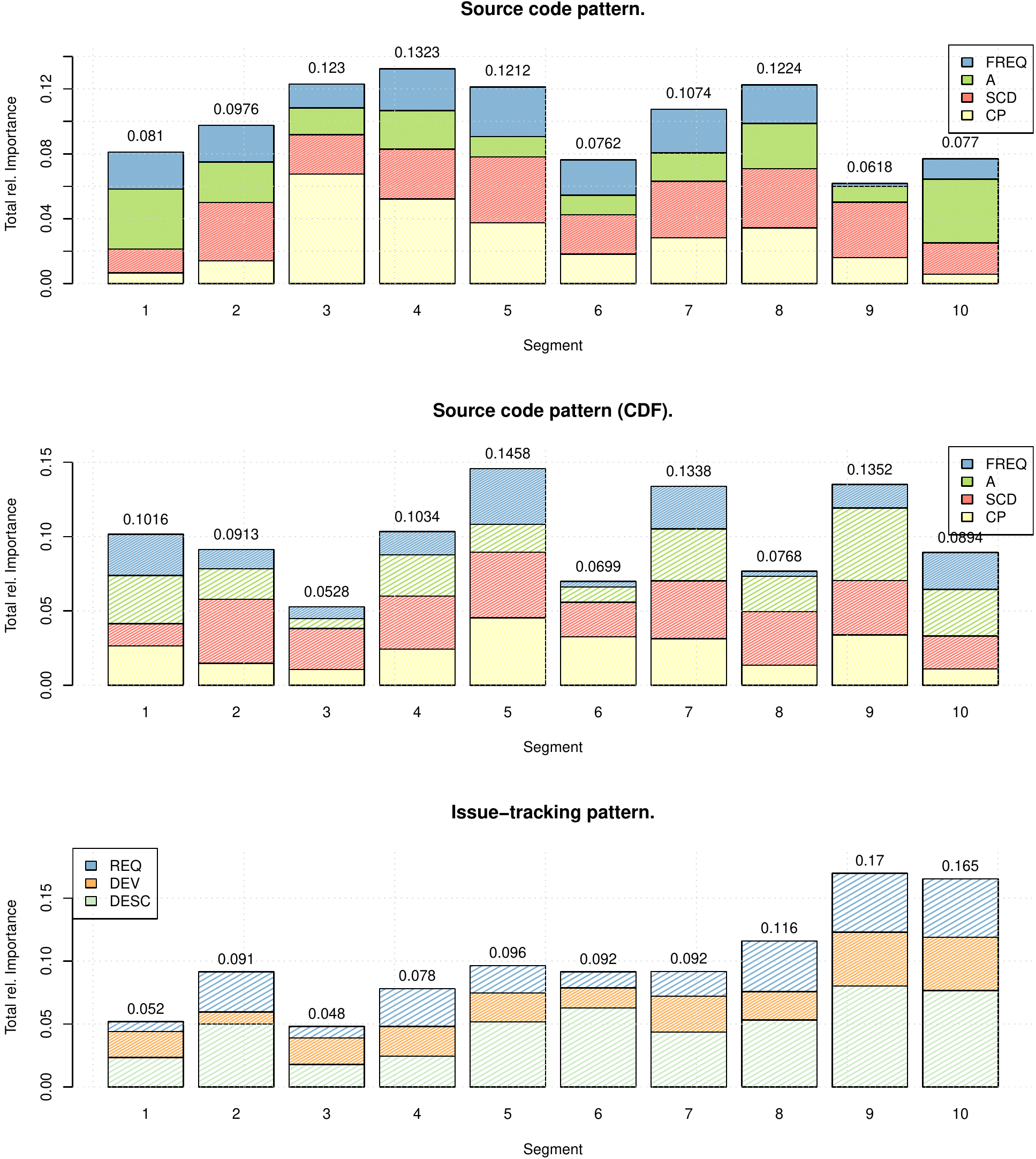} \caption{Variable importance per pattern, activity, and segment.}\label{fig:varimp-all}
\end{figure}

The variable importance for each dataset is shown in figures \ref{fig:varimp-sc}, \ref{fig:varimp-sc-cdf}, and \ref{fig:varimp-it}.
Figure \ref{fig:varimp-all} shows a more detailed drill-down.

\hypertarget{analysis-of-most-important-segments-metrics-activities-etc.}{%
\subsubsection{Analysis of most important Segments, Metrics, Activities, etc.}\label{analysis-of-most-important-segments-metrics-activities-etc.}}

\clearpage

\hypertarget{pattern-less-detection}{%
\subsection{Pattern-less detection}\label{pattern-less-detection}}

Let's try something radically different and simplified. From the beginning, we had used patterns so that we could measure the differences between a project and each pattern.
That was useful for the research case of developing a scoring mechanism.
If we do simple regression, this is not needed. Using a (constant) pattern can be thought of applying a (constant) non-linear transform to each activity.
For example, consider we have a pattern activity \(f(t)\) (where \(t\) is time) and a project activity \(g(t)\). What we have done so far was to produce \(g'(t)\) as, for example, \(g'=g-f\). When we used automatic calibration, the goal was to uniformly sample (create random instances) of project activities in order to estimate the probability density of \(g'(t)\).
Since \(f(t)\) was a constant function (i.e., the same \(f\) was used for each project), this transform is redundant if everything we want is simply a regression model.

We will therefore attempt the following:
We will model each activity, both for source code and issue-tracking, as probability distributions (PDFs).
Since there is nothing to compare these distributions to, we will simply create features by integrating the PDF on equally-long segments (e.g., \([0.0,0.1]; \dots; [0.9,1.0]\)), which is equivalent to \(\operatorname{CDF}(b)-\operatorname{CDF}(a)\) for some segment \([a,b]\).
Each feature will therefore express the amount of activity happening in the corresponding segment.
As for the source code density, we will simply take the average of it in each segment.
We will have to re-design the issue-tracking activities. The plan is to do this very similar to how we re-designed the source code activities in this report, using rejection sampling to obtain somewhat smooth densities.
We can reuse our method for adaptively computing grid results, we'll only have to prepare new datasets.

Let's define a shortcut function for rejection sampling (using \(x/y\) data to approximate a function from first):

\begin{Shaded}
\begin{Highlighting}[]
\NormalTok{rejection\_sampling }\OtherTok{\textless{}{-}} \ControlFlowTok{function}\NormalTok{(x\_data, y\_data, }\AttributeTok{xlim =} \FunctionTok{c}\NormalTok{(}\DecValTok{0}\NormalTok{, }\DecValTok{1}\NormalTok{), }\AttributeTok{num\_x =} \FloatTok{1e+05}\NormalTok{) \{}
\NormalTok{  tempf }\OtherTok{\textless{}{-}}\NormalTok{ stats}\SpecialCharTok{::}\FunctionTok{approxfun}\NormalTok{(}\AttributeTok{x =}\NormalTok{ x\_data, }\AttributeTok{y =}\NormalTok{ y\_data, }\AttributeTok{rule =} \DecValTok{2}\NormalTok{)}
\NormalTok{  use\_x }\OtherTok{\textless{}{-}} \FunctionTok{seq}\NormalTok{(}\AttributeTok{from =}\NormalTok{ xlim[}\DecValTok{1}\NormalTok{], }\AttributeTok{to =}\NormalTok{ xlim[}\DecValTok{2}\NormalTok{], }\AttributeTok{length.out =}\NormalTok{ num\_x)}
\NormalTok{  use\_y }\OtherTok{\textless{}{-}}\NormalTok{ stats}\SpecialCharTok{::}\FunctionTok{runif}\NormalTok{(}\AttributeTok{n =} \FunctionTok{length}\NormalTok{(use\_x), }\AttributeTok{min =} \DecValTok{0}\NormalTok{, }\AttributeTok{max =} \FunctionTok{max}\NormalTok{(y\_data))}
\NormalTok{  tempdens }\OtherTok{\textless{}{-}}\NormalTok{ stats}\SpecialCharTok{::}\FunctionTok{density}\NormalTok{(}\AttributeTok{x =}\NormalTok{ use\_x[use\_y }\SpecialCharTok{\textless{}=} \FunctionTok{tempf}\NormalTok{(use\_x)], }\AttributeTok{bw =} \StringTok{"SJ"}\NormalTok{, }\AttributeTok{cut =} \ConstantTok{TRUE}\NormalTok{)}

  \FunctionTok{list}\NormalTok{(}\AttributeTok{func =}\NormalTok{ tempf, }\AttributeTok{PDF =}\NormalTok{ stats}\SpecialCharTok{::}\FunctionTok{approxfun}\NormalTok{(}\AttributeTok{x =}\NormalTok{ tempdens}\SpecialCharTok{$}\NormalTok{x, }\AttributeTok{y =}\NormalTok{ tempdens}\SpecialCharTok{$}\NormalTok{y, }\AttributeTok{yleft =} \DecValTok{0}\NormalTok{,}
    \AttributeTok{yright =} \DecValTok{0}\NormalTok{), }\AttributeTok{CDF =} \FunctionTok{make\_smooth\_ecdf}\NormalTok{(}\AttributeTok{values =}\NormalTok{ use\_x[use\_y }\SpecialCharTok{\textless{}=} \FunctionTok{tempf}\NormalTok{(use\_x)],}
    \AttributeTok{slope =} \DecValTok{0}\NormalTok{, }\AttributeTok{inverse =} \ConstantTok{FALSE}\NormalTok{))}
\NormalTok{\}}
\end{Highlighting}
\end{Shaded}

Let's create a function to compute the Jensen--Shannon divergence between two PDFs on a segment:

\begin{Shaded}
\begin{Highlighting}[]
\NormalTok{kl\_div\_segment }\OtherTok{\textless{}{-}} \ControlFlowTok{function}\NormalTok{(p, q, }\AttributeTok{ext =} \FunctionTok{c}\NormalTok{(}\DecValTok{0}\NormalTok{, }\FloatTok{0.1}\NormalTok{), }\AttributeTok{xtol =} \FloatTok{1e{-}20}\NormalTok{) \{}
\NormalTok{  cubature}\SpecialCharTok{::}\FunctionTok{cubintegrate}\NormalTok{(}\AttributeTok{f =} \ControlFlowTok{function}\NormalTok{(x) \{}
\NormalTok{    p\_ }\OtherTok{\textless{}{-}} \FunctionTok{p}\NormalTok{(x)}
\NormalTok{    q\_ }\OtherTok{\textless{}{-}} \FunctionTok{q}\NormalTok{(x)}
    \ControlFlowTok{if}\NormalTok{ (}\FunctionTok{abs}\NormalTok{(p\_) }\SpecialCharTok{\textless{}}\NormalTok{ xtol }\SpecialCharTok{||} \FunctionTok{abs}\NormalTok{(q\_) }\SpecialCharTok{\textless{}}\NormalTok{ xtol)}
      \DecValTok{0} \ControlFlowTok{else}\NormalTok{ p\_ }\SpecialCharTok{*} \FunctionTok{log}\NormalTok{(p\_}\SpecialCharTok{/}\NormalTok{q\_)}
\NormalTok{  \}, }\AttributeTok{lower =}\NormalTok{ ext[}\DecValTok{1}\NormalTok{], }\AttributeTok{upper =}\NormalTok{ ext[}\DecValTok{2}\NormalTok{])}\SpecialCharTok{$}\NormalTok{integral}
\NormalTok{\}}

\NormalTok{jsd\_segment }\OtherTok{\textless{}{-}} \ControlFlowTok{function}\NormalTok{(p, q, }\AttributeTok{ext =} \FunctionTok{c}\NormalTok{(}\DecValTok{0}\NormalTok{, }\FloatTok{0.1}\NormalTok{), }\AttributeTok{xtol =} \FloatTok{1e{-}20}\NormalTok{) \{}
\NormalTok{  cubature}\SpecialCharTok{::}\FunctionTok{cubintegrate}\NormalTok{(}\AttributeTok{f =} \ControlFlowTok{function}\NormalTok{(x) \{}
\NormalTok{    p\_ }\OtherTok{\textless{}{-}} \FunctionTok{p}\NormalTok{(x)}
\NormalTok{    q\_ }\OtherTok{\textless{}{-}} \FunctionTok{q}\NormalTok{(x)}
\NormalTok{    m\_ }\OtherTok{\textless{}{-}} \FloatTok{0.5} \SpecialCharTok{*}\NormalTok{ (p\_ }\SpecialCharTok{+}\NormalTok{ q\_)}
    \ControlFlowTok{if}\NormalTok{ (}\FunctionTok{abs}\NormalTok{(p\_) }\SpecialCharTok{\textless{}}\NormalTok{ xtol }\SpecialCharTok{||} \FunctionTok{abs}\NormalTok{(q\_) }\SpecialCharTok{\textless{}}\NormalTok{ xtol }\SpecialCharTok{||} \FunctionTok{abs}\NormalTok{(m\_) }\SpecialCharTok{\textless{}}\NormalTok{ xtol)}
      \DecValTok{0} \ControlFlowTok{else} \FloatTok{0.5} \SpecialCharTok{*}\NormalTok{ p\_ }\SpecialCharTok{*} \FunctionTok{log}\NormalTok{(p\_}\SpecialCharTok{/}\NormalTok{m\_) }\SpecialCharTok{+} \FloatTok{0.5} \SpecialCharTok{*}\NormalTok{ q\_ }\SpecialCharTok{*} \FunctionTok{log}\NormalTok{(q\_}\SpecialCharTok{/}\NormalTok{m\_)}
\NormalTok{  \}, }\AttributeTok{lower =}\NormalTok{ ext[}\DecValTok{1}\NormalTok{], }\AttributeTok{upper =}\NormalTok{ ext[}\DecValTok{2}\NormalTok{])}\SpecialCharTok{$}\NormalTok{integral}
\NormalTok{\}}
\end{Highlighting}
\end{Shaded}

\hypertarget{create-dataset-for-source-code}{%
\subsubsection{Create Dataset for Source Code}\label{create-dataset-for-source-code}}

Source code projects are already present as PDFs, so we can just integrate for the three maintenance activities (actually, we'll use the CDFs directly).
For the source code density, we'll draw some samples per segment and average the result, that's it.

Also, here we will be using the \textbf{non-mixture} version of the \texttt{FREQ} activity. The mixture as used previously, simply is a linearly weighted convex combination of the \texttt{A}- and \texttt{CP}-densities.
In the following, when attempting to adaptively train a regression model and learning from the features, we expect that using a non-mixture version will increase the entropy.

\begin{Shaded}
\begin{Highlighting}[]
\NormalTok{dataset\_np\_sc }\OtherTok{\textless{}{-}} \FunctionTok{loadResultsOrCompute}\NormalTok{(}\AttributeTok{file =} \StringTok{"../data/rob{-}reg\_dataset\_np\_sc.csv"}\NormalTok{,}
  \AttributeTok{computeExpr =}\NormalTok{ \{}
\NormalTok{    grid }\OtherTok{\textless{}{-}} \FunctionTok{expand.grid}\NormalTok{(}\FunctionTok{list}\NormalTok{(}\AttributeTok{interval =} \DecValTok{1}\SpecialCharTok{:}\DecValTok{10}\NormalTok{, }\AttributeTok{activity =} \FunctionTok{c}\NormalTok{(}\StringTok{"A"}\NormalTok{, }\StringTok{"CP"}\NormalTok{, }\StringTok{"FREQ\_nm"}\NormalTok{,}
      \StringTok{"A\_vs\_CP"}\NormalTok{, }\StringTok{"A\_vs\_FREQ\_nm"}\NormalTok{, }\StringTok{"CP\_vs\_FREQ\_nm"}\NormalTok{)  }\CommentTok{\# Let\textquotesingle{}s skip SCD and only use the activities for now}
\NormalTok{))}
\NormalTok{    grid}\SpecialCharTok{$}\NormalTok{activity }\OtherTok{\textless{}{-}} \FunctionTok{as.character}\NormalTok{(grid}\SpecialCharTok{$}\NormalTok{activity)}

\NormalTok{    dataset\_np\_sc }\OtherTok{\textless{}{-}} \StringTok{\textasciigrave{}}\AttributeTok{colnames\textless{}{-}}\StringTok{\textasciigrave{}}\NormalTok{(}\AttributeTok{x =} \FunctionTok{matrix}\NormalTok{(}\AttributeTok{nrow =} \DecValTok{0}\NormalTok{, }\AttributeTok{ncol =} \FunctionTok{nrow}\NormalTok{(grid)), }\AttributeTok{value =} \FunctionTok{sapply}\NormalTok{(}\AttributeTok{X =} \FunctionTok{rownames}\NormalTok{(grid),}
      \AttributeTok{FUN =} \ControlFlowTok{function}\NormalTok{(rn) \{}
\NormalTok{        r }\OtherTok{\textless{}{-}}\NormalTok{ grid[rn, ]}
        \FunctionTok{paste}\NormalTok{(r}\SpecialCharTok{$}\NormalTok{activity, r}\SpecialCharTok{$}\NormalTok{interval, }\AttributeTok{sep =} \StringTok{"\_"}\NormalTok{)}
\NormalTok{      \}))}

    \ControlFlowTok{for}\NormalTok{ (pname }\ControlFlowTok{in} \FunctionTok{names}\NormalTok{(projects\_sc)) \{}
\NormalTok{      newrow\_sc }\OtherTok{\textless{}{-}} \StringTok{\textasciigrave{}}\AttributeTok{colnames\textless{}{-}}\StringTok{\textasciigrave{}}\NormalTok{(}\AttributeTok{x =} \FunctionTok{matrix}\NormalTok{(}\AttributeTok{ncol =} \FunctionTok{ncol}\NormalTok{(dataset\_np\_sc)), }\AttributeTok{value =} \FunctionTok{colnames}\NormalTok{(dataset\_np\_sc))}

      \ControlFlowTok{for}\NormalTok{ (rn }\ControlFlowTok{in} \FunctionTok{rownames}\NormalTok{(grid)) \{}
\NormalTok{        row }\OtherTok{\textless{}{-}}\NormalTok{ grid[rn, ]}
\NormalTok{        interval\_ext }\OtherTok{\textless{}{-}} \FunctionTok{c}\NormalTok{(row}\SpecialCharTok{$}\NormalTok{interval}\SpecialCharTok{/}\DecValTok{10} \SpecialCharTok{{-}} \FloatTok{0.1}\NormalTok{, row}\SpecialCharTok{$}\NormalTok{interval}\SpecialCharTok{/}\DecValTok{10}\NormalTok{)}
\NormalTok{        feat\_name }\OtherTok{\textless{}{-}} \FunctionTok{paste}\NormalTok{(row}\SpecialCharTok{$}\NormalTok{activity, row}\SpecialCharTok{$}\NormalTok{interval, }\AttributeTok{sep =} \StringTok{"\_"}\NormalTok{)}
\NormalTok{        is\_versus }\OtherTok{\textless{}{-}} \FunctionTok{grepl}\NormalTok{(}\AttributeTok{pattern =} \StringTok{"\_vs\_"}\NormalTok{, }\AttributeTok{x =}\NormalTok{ row}\SpecialCharTok{$}\NormalTok{activity)}

        \ControlFlowTok{if}\NormalTok{ (is\_versus) \{}
          \CommentTok{\# Calculate symmetric divergence of two activities on the segment}
\NormalTok{          sp }\OtherTok{\textless{}{-}} \FunctionTok{strsplit}\NormalTok{(}\AttributeTok{x =}\NormalTok{ row}\SpecialCharTok{$}\NormalTok{activity, }\AttributeTok{split =} \StringTok{"\_vs\_"}\NormalTok{)[[}\DecValTok{1}\NormalTok{]]}
\NormalTok{          pdf\_p }\OtherTok{\textless{}{-}}\NormalTok{ projects\_sc[[pname]][[sp[}\DecValTok{1}\NormalTok{]]]}
\NormalTok{          pdf\_q }\OtherTok{\textless{}{-}}\NormalTok{ projects\_sc[[pname]][[sp[}\DecValTok{2}\NormalTok{]]]}
\NormalTok{          newrow\_sc[}\DecValTok{1}\NormalTok{, feat\_name] }\OtherTok{\textless{}{-}} \FunctionTok{jsd\_segment}\NormalTok{(}\AttributeTok{p =}\NormalTok{ pdf\_p, }\AttributeTok{q =}\NormalTok{ pdf\_q, }\AttributeTok{ext =}\NormalTok{ interval\_ext)}
\NormalTok{        \} }\ControlFlowTok{else}\NormalTok{ \{}
          \CommentTok{\# Integrate the PDF of the activity: newrow\_sc[1, feat\_name] \textless{}{-}}
          \CommentTok{\# cubature::cubintegrate( f = projects\_sc[[pname]][[row$activity]],}
          \CommentTok{\# lower = interval\_ext[1], upper = interval\_ext[2])$integral}
          \CommentTok{\# Actually, let\textquotesingle{}s use the CDF directly:}
\NormalTok{          cdf }\OtherTok{\textless{}{-}}\NormalTok{ projects\_sc\_cdf[[pname]][[row}\SpecialCharTok{$}\NormalTok{activity]]}
\NormalTok{          newrow\_sc[}\DecValTok{1}\NormalTok{, feat\_name] }\OtherTok{\textless{}{-}} \FunctionTok{cdf}\NormalTok{(interval\_ext[}\DecValTok{2}\NormalTok{]) }\SpecialCharTok{{-}} \FunctionTok{cdf}\NormalTok{(interval\_ext[}\DecValTok{1}\NormalTok{])}
\NormalTok{        \}}
\NormalTok{      \}}

\NormalTok{      dataset\_np\_sc }\OtherTok{\textless{}{-}} \FunctionTok{rbind}\NormalTok{(dataset\_np\_sc, newrow\_sc)}
\NormalTok{    \}}

    \FunctionTok{as.data.frame}\NormalTok{(dataset\_np\_sc)}
\NormalTok{  \})}
\end{Highlighting}
\end{Shaded}

\hypertarget{create-dataset-for-issue-tracking-1}{%
\subsubsection{Create Dataset for Issue-Tracking}\label{create-dataset-for-issue-tracking-1}}

Here we load the original data and transform it into a density. Issue-tracking data has one major difference compared to source code data.
In the latter, we know the timely accumulation of certain activities, but we do not know how much time was spent on an activity.
In issue-tracking data, we do know that. Therefore, we will use the following procedure for creating the issue-tracking data:

\begin{enumerate}
\def\labelenumi{\arabic{enumi}.}
\tightlist
\item
  Use normalized time as \(x\) and the activity's spent time as \(y\).
\item
  Filter \(x\) by using the condition \(y>0\).
\item
  Estimate a density over \textbf{\(x\)}, using the duration of each activity (\(y\)) as weights. We will have to normalize the non-zero activities to sum up to \(1\) first.
\item
  Perform rejection sampling on that density to obtain a smooth CDF which will be used to estimate the amount of each activity in each segment.
\end{enumerate}

\begin{Shaded}
\begin{Highlighting}[]
\NormalTok{rejection\_sampling\_issue\_tracking }\OtherTok{\textless{}{-}} \ControlFlowTok{function}\NormalTok{(use\_x, use\_y) \{}
  \CommentTok{\# Apply filter:}
\NormalTok{  temp\_x }\OtherTok{\textless{}{-}}\NormalTok{ use\_x[use\_y }\SpecialCharTok{\textgreater{}} \DecValTok{0}\NormalTok{]}
\NormalTok{  temp\_y }\OtherTok{\textless{}{-}}\NormalTok{ use\_y[use\_y }\SpecialCharTok{\textgreater{}} \DecValTok{0}\NormalTok{]}

  \FunctionTok{tryCatch}\NormalTok{(\{}
\NormalTok{    temp\_dens }\OtherTok{\textless{}{-}} \FunctionTok{suppressWarnings}\NormalTok{(\{}
\NormalTok{      stats}\SpecialCharTok{::}\FunctionTok{density}\NormalTok{(}\AttributeTok{x =}\NormalTok{ temp\_x, }\AttributeTok{weights =}\NormalTok{ temp\_y}\SpecialCharTok{/}\FunctionTok{sum}\NormalTok{(temp\_y), }\AttributeTok{bw =} \StringTok{"SJ"}\NormalTok{, }\AttributeTok{cut =} \ConstantTok{TRUE}\NormalTok{)}
\NormalTok{    \})}
\NormalTok{    use\_x }\OtherTok{\textless{}{-}}\NormalTok{ temp\_dens}\SpecialCharTok{$}\NormalTok{x}
\NormalTok{    use\_y }\OtherTok{\textless{}{-}}\NormalTok{ temp\_dens}\SpecialCharTok{$}\NormalTok{y}
\NormalTok{  \}, }\AttributeTok{error =} \ControlFlowTok{function}\NormalTok{(cond) \{}
    \ControlFlowTok{if}\NormalTok{ (}\ConstantTok{FALSE}\NormalTok{) \{}
      \FunctionTok{print}\NormalTok{(}\FunctionTok{paste0}\NormalTok{(}\StringTok{"Cannot estimate density for project "}\NormalTok{, pname, }\StringTok{" and activity "}\NormalTok{,}
\NormalTok{        activity))}
\NormalTok{    \}}
\NormalTok{  \})}

  \FunctionTok{rejection\_sampling}\NormalTok{(}\AttributeTok{x\_data =}\NormalTok{ use\_x, }\AttributeTok{y\_data =}\NormalTok{ use\_y, }\AttributeTok{xlim =} \FunctionTok{range}\NormalTok{(use\_x))}
\NormalTok{\}}
\end{Highlighting}
\end{Shaded}

\begin{Shaded}
\begin{Highlighting}[]
\NormalTok{dataset\_np\_it }\OtherTok{\textless{}{-}} \FunctionTok{loadResultsOrCompute}\NormalTok{(}\AttributeTok{file =} \StringTok{"../data/rob{-}reg\_dataset\_np\_it.csv"}\NormalTok{,}
  \AttributeTok{computeExpr =}\NormalTok{ \{}
    \FunctionTok{library}\NormalTok{(readxl)}

\NormalTok{    grid }\OtherTok{\textless{}{-}} \FunctionTok{expand.grid}\NormalTok{(}\FunctionTok{list}\NormalTok{(}\AttributeTok{interval =} \DecValTok{1}\SpecialCharTok{:}\DecValTok{10}\NormalTok{, }\AttributeTok{activity =} \FunctionTok{c}\NormalTok{(}\StringTok{"REQ"}\NormalTok{, }\StringTok{"DEV"}\NormalTok{, }\StringTok{"DESC"}\NormalTok{,}
      \StringTok{"REQ\_vs\_DEV"}\NormalTok{, }\StringTok{"REQ\_vs\_DESC"}\NormalTok{, }\StringTok{"DEV\_vs\_DESC"}\NormalTok{)))}
\NormalTok{    grid}\SpecialCharTok{$}\NormalTok{activity }\OtherTok{\textless{}{-}} \FunctionTok{as.character}\NormalTok{(grid}\SpecialCharTok{$}\NormalTok{activity)}

\NormalTok{    dataset\_np\_it }\OtherTok{\textless{}{-}} \StringTok{\textasciigrave{}}\AttributeTok{colnames\textless{}{-}}\StringTok{\textasciigrave{}}\NormalTok{(}\AttributeTok{x =} \FunctionTok{matrix}\NormalTok{(}\AttributeTok{nrow =} \DecValTok{0}\NormalTok{, }\AttributeTok{ncol =} \FunctionTok{nrow}\NormalTok{(grid)), }\AttributeTok{value =} \FunctionTok{sapply}\NormalTok{(}\AttributeTok{X =} \FunctionTok{rownames}\NormalTok{(grid),}
      \AttributeTok{FUN =} \ControlFlowTok{function}\NormalTok{(rn) \{}
\NormalTok{        r }\OtherTok{\textless{}{-}}\NormalTok{ grid[rn, ]}
        \FunctionTok{paste}\NormalTok{(r}\SpecialCharTok{$}\NormalTok{activity, r}\SpecialCharTok{$}\NormalTok{interval, }\AttributeTok{sep =} \StringTok{"\_"}\NormalTok{)}
\NormalTok{      \}))}

    \ControlFlowTok{for}\NormalTok{ (pname }\ControlFlowTok{in} \FunctionTok{names}\NormalTok{(projects\_it)) \{}
\NormalTok{      temp }\OtherTok{\textless{}{-}} \FunctionTok{read\_excel}\NormalTok{(}\StringTok{"../data/FD\_issue{-}based\_detection.xlsx"}\NormalTok{, }\AttributeTok{sheet =}\NormalTok{ pname)}
\NormalTok{      newrow\_it }\OtherTok{\textless{}{-}} \StringTok{\textasciigrave{}}\AttributeTok{colnames\textless{}{-}}\StringTok{\textasciigrave{}}\NormalTok{(}\AttributeTok{x =} \FunctionTok{matrix}\NormalTok{(}\AttributeTok{ncol =} \FunctionTok{ncol}\NormalTok{(dataset\_np\_it)), }\AttributeTok{value =} \FunctionTok{colnames}\NormalTok{(dataset\_np\_it))}

      \ControlFlowTok{for}\NormalTok{ (activity }\ControlFlowTok{in} \FunctionTok{c}\NormalTok{(}\StringTok{"REQ"}\NormalTok{, }\StringTok{"DEV"}\NormalTok{, }\StringTok{"DESC"}\NormalTok{)) \{}
\NormalTok{        use\_y }\OtherTok{\textless{}{-}} \FunctionTok{as.numeric}\NormalTok{(temp[[}\FunctionTok{tolower}\NormalTok{(activity)]])}
\NormalTok{        use\_y[}\FunctionTok{is.na}\NormalTok{(use\_y)] }\OtherTok{\textless{}{-}} \DecValTok{0}
\NormalTok{        cdf }\OtherTok{\textless{}{-}} \FunctionTok{rejection\_sampling\_issue\_tracking}\NormalTok{(}\AttributeTok{use\_x =}\NormalTok{ temp}\SpecialCharTok{$}\StringTok{\textasciigrave{}}\AttributeTok{time\%}\StringTok{\textasciigrave{}}\NormalTok{, }\AttributeTok{use\_y =}\NormalTok{ use\_y)}\SpecialCharTok{$}\NormalTok{CDF}

        \ControlFlowTok{for}\NormalTok{ (idx }\ControlFlowTok{in} \DecValTok{1}\SpecialCharTok{:}\DecValTok{10}\NormalTok{) \{}
\NormalTok{          interval\_ext }\OtherTok{\textless{}{-}} \FunctionTok{c}\NormalTok{(idx}\SpecialCharTok{/}\DecValTok{10} \SpecialCharTok{{-}} \FloatTok{0.1}\NormalTok{, idx}\SpecialCharTok{/}\DecValTok{10}\NormalTok{)}
\NormalTok{          feat\_name }\OtherTok{\textless{}{-}} \FunctionTok{paste}\NormalTok{(activity, idx, }\AttributeTok{sep =} \StringTok{"\_"}\NormalTok{)}
\NormalTok{          newrow\_it[}\DecValTok{1}\NormalTok{, feat\_name] }\OtherTok{\textless{}{-}} \FunctionTok{cdf}\NormalTok{(interval\_ext[}\DecValTok{2}\NormalTok{]) }\SpecialCharTok{{-}} \FunctionTok{cdf}\NormalTok{(interval\_ext[}\DecValTok{1}\NormalTok{])}
\NormalTok{        \}}
\NormalTok{      \}}

      \ControlFlowTok{for}\NormalTok{ (divergence }\ControlFlowTok{in} \FunctionTok{c}\NormalTok{(}\StringTok{"REQ\_vs\_DEV"}\NormalTok{, }\StringTok{"REQ\_vs\_DESC"}\NormalTok{, }\StringTok{"DEV\_vs\_DESC"}\NormalTok{)) \{}
\NormalTok{        sp }\OtherTok{\textless{}{-}} \FunctionTok{strsplit}\NormalTok{(}\AttributeTok{x =}\NormalTok{ divergence, }\AttributeTok{split =} \StringTok{"\_vs\_"}\NormalTok{)[[}\DecValTok{1}\NormalTok{]]}

\NormalTok{        use\_y\_p }\OtherTok{\textless{}{-}} \FunctionTok{as.numeric}\NormalTok{(temp[[}\FunctionTok{tolower}\NormalTok{(sp[}\DecValTok{1}\NormalTok{])]])}
\NormalTok{        use\_y\_p[}\FunctionTok{is.na}\NormalTok{(use\_y\_p)] }\OtherTok{\textless{}{-}} \DecValTok{0}
\NormalTok{        pdf\_p }\OtherTok{\textless{}{-}} \FunctionTok{rejection\_sampling\_issue\_tracking}\NormalTok{(}\AttributeTok{use\_x =}\NormalTok{ temp}\SpecialCharTok{$}\StringTok{\textasciigrave{}}\AttributeTok{time\%}\StringTok{\textasciigrave{}}\NormalTok{,}
          \AttributeTok{use\_y =}\NormalTok{ use\_y\_p)}\SpecialCharTok{$}\NormalTok{PDF}

\NormalTok{        use\_y\_q }\OtherTok{\textless{}{-}} \FunctionTok{as.numeric}\NormalTok{(temp[[}\FunctionTok{tolower}\NormalTok{(sp[}\DecValTok{2}\NormalTok{])]])}
\NormalTok{        use\_y\_q[}\FunctionTok{is.na}\NormalTok{(use\_y\_q)] }\OtherTok{\textless{}{-}} \DecValTok{0}
\NormalTok{        pdf\_q }\OtherTok{\textless{}{-}} \FunctionTok{rejection\_sampling\_issue\_tracking}\NormalTok{(}\AttributeTok{use\_x =}\NormalTok{ temp}\SpecialCharTok{$}\StringTok{\textasciigrave{}}\AttributeTok{time\%}\StringTok{\textasciigrave{}}\NormalTok{,}
          \AttributeTok{use\_y =}\NormalTok{ use\_y\_q)}\SpecialCharTok{$}\NormalTok{PDF}

        \ControlFlowTok{for}\NormalTok{ (idx }\ControlFlowTok{in} \DecValTok{1}\SpecialCharTok{:}\DecValTok{10}\NormalTok{) \{}
\NormalTok{          interval\_ext }\OtherTok{\textless{}{-}} \FunctionTok{c}\NormalTok{(idx}\SpecialCharTok{/}\DecValTok{10} \SpecialCharTok{{-}} \FloatTok{0.1}\NormalTok{, idx}\SpecialCharTok{/}\DecValTok{10}\NormalTok{)}
\NormalTok{          feat\_name }\OtherTok{\textless{}{-}} \FunctionTok{paste}\NormalTok{(divergence, idx, }\AttributeTok{sep =} \StringTok{"\_"}\NormalTok{)}
\NormalTok{          newrow\_it[}\DecValTok{1}\NormalTok{, feat\_name] }\OtherTok{\textless{}{-}} \FunctionTok{jsd\_segment}\NormalTok{(}\AttributeTok{p =}\NormalTok{ pdf\_p, }\AttributeTok{q =}\NormalTok{ pdf\_q, }\AttributeTok{ext =}\NormalTok{ interval\_ext)}
\NormalTok{        \}}
\NormalTok{      \}}

\NormalTok{      dataset\_np\_it }\OtherTok{\textless{}{-}} \FunctionTok{rbind}\NormalTok{(dataset\_np\_it, newrow\_it)}
\NormalTok{    \}}

    \FunctionTok{as.data.frame}\NormalTok{(dataset\_np\_it)}
\NormalTok{  \})}
\end{Highlighting}
\end{Shaded}

\hypertarget{oversampling}{%
\subsubsection{Oversampling}\label{oversampling}}

The non-pattern datasets have the same problem as the ones using a pattern: the data is scarce and we have many features.
Therefore, we will go through the same procedure and oversample them.

\begin{Shaded}
\begin{Highlighting}[]
\NormalTok{dataset\_np\_sc\_oversampled }\OtherTok{\textless{}{-}} \FunctionTok{loadResultsOrCompute}\NormalTok{(}\AttributeTok{file =} \StringTok{"../data/rob{-}reg\_dataset\_np\_sc\_oversampled.csv"}\NormalTok{,}
  \AttributeTok{computeExpr =}\NormalTok{ \{}
\NormalTok{    temp }\OtherTok{\textless{}{-}} \FunctionTok{cbind}\NormalTok{(dataset\_np\_sc, }\FunctionTok{data.frame}\NormalTok{(}\AttributeTok{gt =}\NormalTok{ ground\_truth\_all}\SpecialCharTok{$}\NormalTok{consensus))}
    \FunctionTok{balance\_num\_labels}\NormalTok{(}\AttributeTok{dataset =}\NormalTok{ temp, }\AttributeTok{num =}\NormalTok{ min\_rows)}
\NormalTok{  \})}

\NormalTok{dataset\_np\_it\_oversampled }\OtherTok{\textless{}{-}} \FunctionTok{loadResultsOrCompute}\NormalTok{(}\AttributeTok{file =} \StringTok{"../data/rob{-}reg\_dataset\_np\_it\_oversampled.csv"}\NormalTok{,}
  \AttributeTok{computeExpr =}\NormalTok{ \{}
\NormalTok{    temp }\OtherTok{\textless{}{-}} \FunctionTok{cbind}\NormalTok{(dataset\_np\_it, }\FunctionTok{data.frame}\NormalTok{(}\AttributeTok{gt =}\NormalTok{ ground\_truth\_all}\SpecialCharTok{$}\NormalTok{consensus))}
    \FunctionTok{balance\_num\_labels}\NormalTok{(}\AttributeTok{dataset =}\NormalTok{ temp, }\AttributeTok{num =}\NormalTok{ min\_rows)}
\NormalTok{  \})}
\end{Highlighting}
\end{Shaded}

\hypertarget{adaptive-training-1}{%
\subsubsection{Adaptive Training}\label{adaptive-training-1}}

And here we compute the grid (test each model on each dataset).

\textbf{Note}: We have computed the following actually three times. The first time using both type of features (amount of activity per segment, pair-wise symmetric divergence between activities), and then once using only one type of feature.
There are three folders in the results-folder: \texttt{np\_both}, \texttt{np\_amount}, and \texttt{np\_divergence} holding the results for each complete run.
We tested all three combinations to find out whether fewer features are sufficient for a low generalization error.

Attention: The following is expensive.

\begin{Shaded}
\begin{Highlighting}[]
\NormalTok{compute\_np\_grid\_adaptive }\OtherTok{\textless{}{-}} \ControlFlowTok{function}\NormalTok{(}\AttributeTok{type =} \FunctionTok{c}\NormalTok{(}\StringTok{"both"}\NormalTok{, }\StringTok{"amount"}\NormalTok{, }\StringTok{"divergence"}\NormalTok{)) \{}
\NormalTok{  type }\OtherTok{\textless{}{-}} \FunctionTok{match.arg}\NormalTok{(type)}

\NormalTok{  grid }\OtherTok{\textless{}{-}} \FunctionTok{expand.grid}\NormalTok{(}\FunctionTok{list}\NormalTok{(}\AttributeTok{seed =} \DecValTok{1}\SpecialCharTok{:}\DecValTok{50}\NormalTok{, }\AttributeTok{num\_train =} \DecValTok{2}\SpecialCharTok{:}\DecValTok{50}\NormalTok{, }\AttributeTok{do\_pca =} \FunctionTok{c}\NormalTok{(}\ConstantTok{TRUE}\NormalTok{, }\ConstantTok{FALSE}\NormalTok{)))}
\NormalTok{  model\_types }\OtherTok{\textless{}{-}} \FunctionTok{c}\NormalTok{(}\StringTok{"gbm"}\NormalTok{, }\StringTok{"glm"}\NormalTok{, }\StringTok{"nnet"}\NormalTok{, }\StringTok{"rf"}\NormalTok{, }\StringTok{"svmPoly"}\NormalTok{, }\StringTok{"treebag"}\NormalTok{, }\StringTok{"rankModel"}\NormalTok{)  }\CommentTok{\#, \textquotesingle{}avNNet\textquotesingle{}, \textquotesingle{}M5\textquotesingle{})}
\NormalTok{  use\_datasets }\OtherTok{\textless{}{-}} \FunctionTok{c}\NormalTok{(}\StringTok{"dataset\_np\_it\_oversampled"}\NormalTok{, }\StringTok{"dataset\_np\_sc\_oversampled"}\NormalTok{)}


  \FunctionTok{library}\NormalTok{(foreach)}
  \FunctionTok{loadResultsOrCompute}\NormalTok{(}\AttributeTok{file =} \FunctionTok{paste0}\NormalTok{(}\StringTok{"../results/np\_"}\NormalTok{, type, }\StringTok{"/rob{-}reg{-}np\_res{-}grid.rds"}\NormalTok{),}
    \AttributeTok{computeExpr =}\NormalTok{ \{}
\NormalTok{      res }\OtherTok{\textless{}{-}} \ConstantTok{NULL}

      \ControlFlowTok{for}\NormalTok{ (use\_dataset }\ControlFlowTok{in}\NormalTok{ use\_datasets) \{}
\NormalTok{        dataset }\OtherTok{\textless{}{-}} \FunctionTok{get}\NormalTok{(use\_dataset)}
        \CommentTok{\# Let\textquotesingle{}s select the right columns. If type = \textquotesingle{}both\textquotesingle{}, then use all.}
        \ControlFlowTok{if}\NormalTok{ (type }\SpecialCharTok{==} \StringTok{"amount"}\NormalTok{) \{}
          \CommentTok{\# Take out the \textquotesingle{}\_vs\_\textquotesingle{}{-}columns:}
\NormalTok{          cn }\OtherTok{\textless{}{-}} \FunctionTok{colnames}\NormalTok{(dataset)}
          \CommentTok{\# This also keeps the \textquotesingle{}gt\textquotesingle{}{-}column}
\NormalTok{          dataset }\OtherTok{\textless{}{-}}\NormalTok{ dataset[, cn[}\SpecialCharTok{!}\FunctionTok{grepl}\NormalTok{(}\AttributeTok{pattern =} \StringTok{"\_vs\_"}\NormalTok{, }\AttributeTok{x =}\NormalTok{ cn)]]}
\NormalTok{        \} }\ControlFlowTok{else} \ControlFlowTok{if}\NormalTok{ (type }\SpecialCharTok{==} \StringTok{"divergence"}\NormalTok{) \{}
\NormalTok{          cn }\OtherTok{\textless{}{-}} \FunctionTok{colnames}\NormalTok{(dataset)}
          \CommentTok{\# Only take \textquotesingle{}\_vs\_\textquotesingle{}{-}columns and re{-}append the \textquotesingle{}gt\textquotesingle{}{-}column}
\NormalTok{          dataset }\OtherTok{\textless{}{-}}\NormalTok{ dataset[, }\FunctionTok{c}\NormalTok{(cn[}\FunctionTok{grepl}\NormalTok{(}\AttributeTok{pattern =} \StringTok{"\_vs\_"}\NormalTok{, }\AttributeTok{x =}\NormalTok{ cn)], }\StringTok{"gt"}\NormalTok{)]}
\NormalTok{        \}}

        \ControlFlowTok{for}\NormalTok{ (model\_type }\ControlFlowTok{in}\NormalTok{ model\_types) \{}

\NormalTok{          the\_file }\OtherTok{\textless{}{-}} \FunctionTok{paste0}\NormalTok{(}\StringTok{"../results/np\_"}\NormalTok{, type, }\StringTok{"/rob{-}reg{-}np\_res{-}grid\_"}\NormalTok{,}
\NormalTok{          model\_type, }\StringTok{"\_"}\NormalTok{, use\_dataset, }\StringTok{".rds"}\NormalTok{)}
\NormalTok{          temp }\OtherTok{\textless{}{-}} \FunctionTok{as.data.frame}\NormalTok{(}\FunctionTok{loadResultsOrCompute}\NormalTok{(}\AttributeTok{file =}\NormalTok{ the\_file, }\AttributeTok{computeExpr =}\NormalTok{ \{}
\NormalTok{          temp1 }\OtherTok{\textless{}{-}} \FunctionTok{doWithParallelCluster}\NormalTok{(}\AttributeTok{numCores =} \FunctionTok{min}\NormalTok{(parallel}\SpecialCharTok{::}\FunctionTok{detectCores}\NormalTok{(),}
            \DecValTok{123}\NormalTok{), }\AttributeTok{expr =}\NormalTok{ \{}
\NormalTok{            foreach}\SpecialCharTok{::}\FunctionTok{foreach}\NormalTok{(}\AttributeTok{rn =} \FunctionTok{rownames}\NormalTok{(grid), }\AttributeTok{.combine =}\NormalTok{ rbind, }\AttributeTok{.inorder =} \ConstantTok{FALSE}\NormalTok{,}
            \AttributeTok{.verbose =} \ConstantTok{TRUE}\NormalTok{, }\AttributeTok{.export =} \FunctionTok{c}\NormalTok{(}\StringTok{"adaptive\_training\_caret"}\NormalTok{)) }\SpecialCharTok{\%dopar\%}
\NormalTok{            \{}
              \FunctionTok{tryCatch}\NormalTok{(\{}
\NormalTok{              row }\OtherTok{\textless{}{-}}\NormalTok{ grid[rn, ]}
              \FunctionTok{adaptive\_training\_caret}\NormalTok{(}\AttributeTok{org\_dataset =}\NormalTok{ dataset, }\AttributeTok{seeds =}\NormalTok{ row}\SpecialCharTok{$}\NormalTok{seed,}
                \AttributeTok{model\_type =}\NormalTok{ model\_type, }\AttributeTok{num\_train =}\NormalTok{ row}\SpecialCharTok{$}\NormalTok{num\_train,}
                \AttributeTok{do\_pca =}\NormalTok{ row}\SpecialCharTok{$}\NormalTok{do\_pca, }\AttributeTok{num\_caret\_repeats =} \DecValTok{25}\NormalTok{)}
\NormalTok{              \}, }\AttributeTok{error =} \ControlFlowTok{function}\NormalTok{(cond) \{}
              \ConstantTok{NULL}  \CommentTok{\# Return an empty result that will not disturb .combine}
\NormalTok{              \})}
\NormalTok{            \}}
\NormalTok{          \})}
          \FunctionTok{saveRDS}\NormalTok{(}\AttributeTok{object =}\NormalTok{ temp1, }\AttributeTok{file =}\NormalTok{ the\_file)}
\NormalTok{          temp1}
\NormalTok{          \}))}
\NormalTok{          temp}\SpecialCharTok{$}\NormalTok{dataset }\OtherTok{\textless{}{-}} \FunctionTok{rep}\NormalTok{(use\_dataset, }\FunctionTok{nrow}\NormalTok{(temp))}

\NormalTok{          res }\OtherTok{\textless{}{-}} \FunctionTok{rbind}\NormalTok{(res, temp)}
\NormalTok{        \}}
\NormalTok{      \}}

\NormalTok{      res}\SpecialCharTok{$}\NormalTok{dataset }\OtherTok{\textless{}{-}} \FunctionTok{factor}\NormalTok{(res}\SpecialCharTok{$}\NormalTok{dataset)}
\NormalTok{      res}
\NormalTok{    \})}
\NormalTok{\}}
\end{Highlighting}
\end{Shaded}

\begin{Shaded}
\begin{Highlighting}[]
\NormalTok{res\_grid\_np\_both }\OtherTok{\textless{}{-}} \FunctionTok{compute\_np\_grid\_adaptive}\NormalTok{(}\AttributeTok{type =} \StringTok{"b"}\NormalTok{)}
\NormalTok{res\_grid\_np\_amount }\OtherTok{\textless{}{-}} \FunctionTok{compute\_np\_grid\_adaptive}\NormalTok{(}\AttributeTok{type =} \StringTok{"a"}\NormalTok{)}
\NormalTok{res\_grid\_np\_divergence }\OtherTok{\textless{}{-}} \FunctionTok{compute\_np\_grid\_adaptive}\NormalTok{(}\AttributeTok{type =} \StringTok{"d"}\NormalTok{)}
\end{Highlighting}
\end{Shaded}

\begin{figure}
\centering
\includegraphics{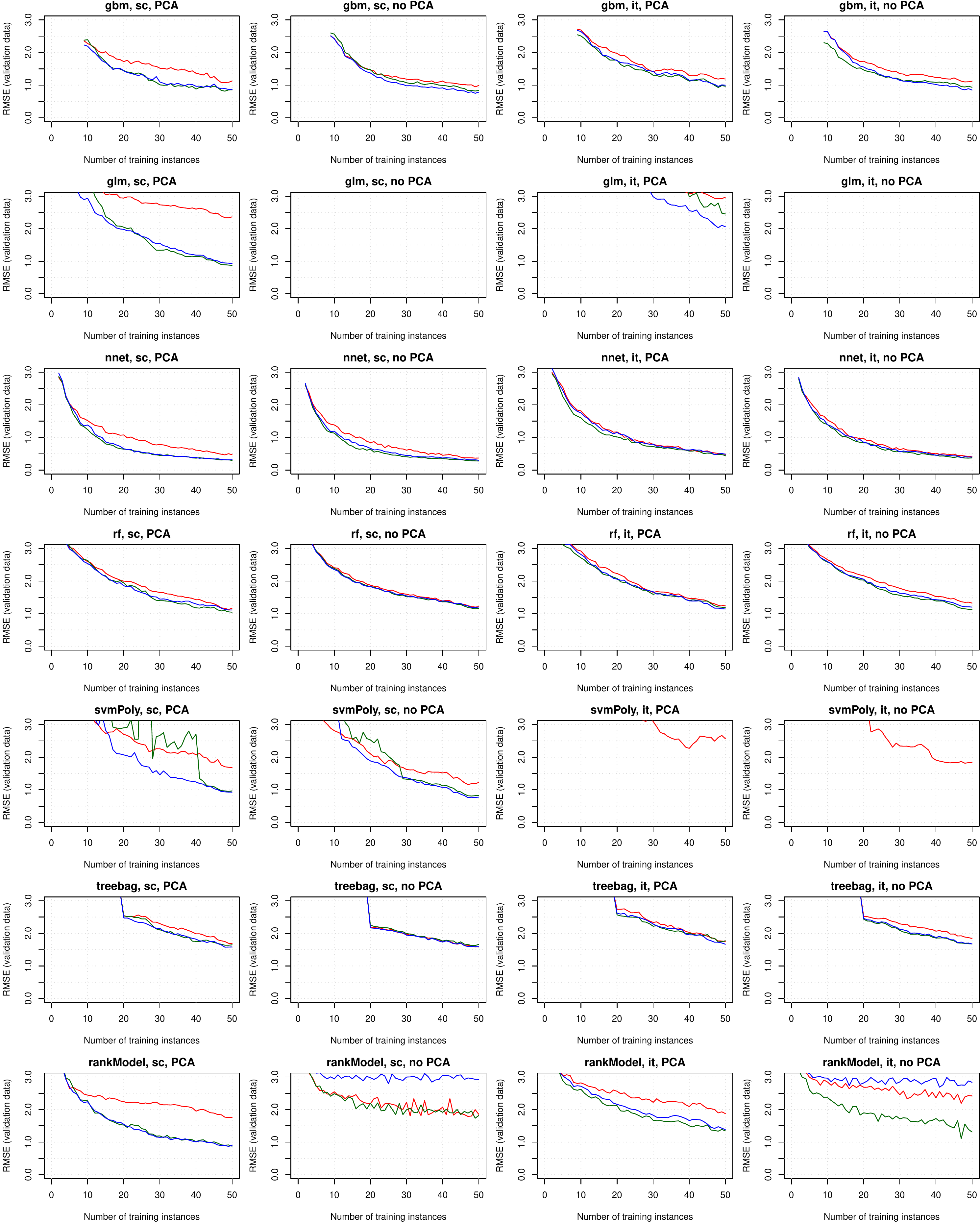}
\caption{\label{fig:np-overview}Comparing the expected validation error of the trained models of the No-pattern approach. Amount-only features are red, divergence-only are green, and both are blue.}
\end{figure}

\clearpage

Figure \ref{fig:np-overview} compares all of the seven trained models across both data types, source code and issue-tracking, with and without applied PCA, and using different feature sets (amount only, divergence only, both).
The visual inspection tells us that the best model is the neural network for source code, and the one not using PCA being a little better (``nnet, sc, no PCA'').
That model has, for the number of training instances of \(15\), an expectation \(<1\), regardless of the feature set used.
The best feature set by a small margin is ``divergence'', followed by ``both''.

\hypertarget{pca-vs.-non-pca-1}{%
\subsubsection{PCA vs.~non-PCA}\label{pca-vs.-non-pca-1}}

We will inspect the best model (neural network) and check whether it makes a difference to apply PCA or not when using both types of features (amount and divergence).
Table \ref{tab:nnet-pca-comps-table} shows the effects of conditionally applying PCA.
It appears that the expected average validation error is lower and has also lower variance when \textbf{\emph{not}} using PCA.
Simplified, more features is better, but the difference is not large (but still significant).

\begin{table}

\caption{\label{tab:nnet-pca-comps-table}Mean, median, standard deviation, min, and max of the RMSE validation error, grouped by the type of data (it=issue-tracking, sc=source code) conditionally applying PCA for feature set "both".}
\centering
\begin{tabular}[t]{llrrrrr}
\toprule
dataset & do\_pca & RMSE\_mean & RMSE\_median & RMSE\_sd & RMSE\_min & RMSE\_max\\
\midrule
np\_it\_oversampled & FALSE & 0.9177586 & 0.7117307 & 0.6464355 & 0.1575782 & 4.361813\\
np\_it\_oversampled & TRUE & 1.1258226 & 0.9113579 & 0.7432705 & 0.1732961 & 4.653067\\
np\_sc\_oversampled & FALSE & 0.7500831 & 0.5211888 & 0.5834025 & 0.1519206 & 4.361813\\
np\_sc\_oversampled & TRUE & 0.8095111 & 0.5071958 & 0.6947143 & 0.1643221 & 4.576710\\
\bottomrule
\end{tabular}
\end{table}

\hypertarget{confidence-intervals}{%
\subsubsection{Confidence Intervals}\label{confidence-intervals}}

We show some results for this model in table \ref{tab:nnet-best-model}.
Using all features, only \(14\) instances are required to achieve a generalization error (RMSE) of \(\leq1\) (with a standard deviation of \(\approx0.31\)).
The error falls to \(\approx0.9\) for \(15\) instances, and for \(17\) instances, the error plus one standard deviation is even below \(1\).

\begin{figure}
\centering
\includegraphics{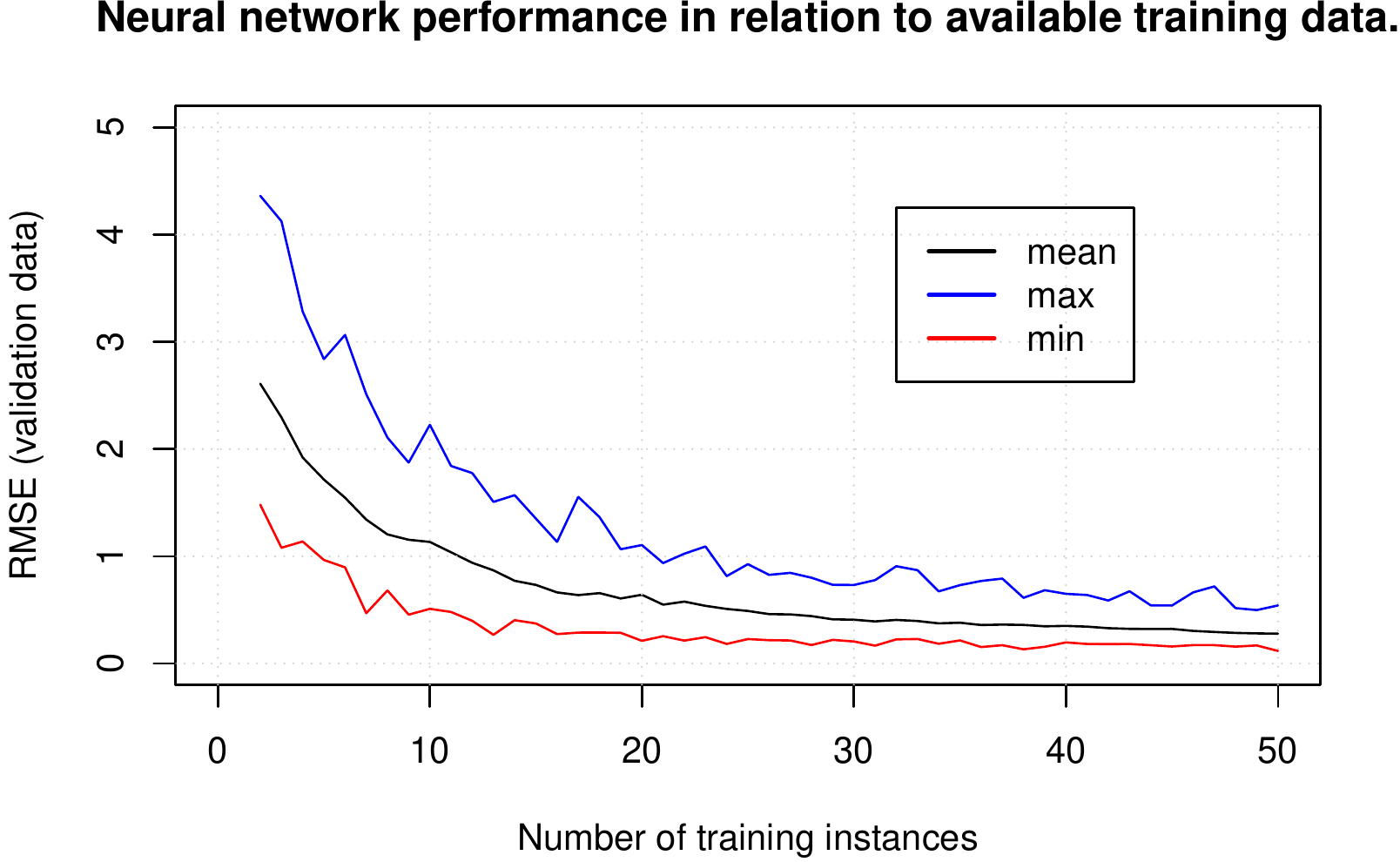}
\caption{\label{fig:nnet-expected-gen-err}The expected validation error for the best-performing model and dataset (source code, no PCA, divergence-features).}
\end{figure}

Before we evaluate the confidence intervals, let's take a look at the expected validation error, and the observed maximum and minimum (figure \ref{fig:nnet-expected-gen-err}).
We observe that the error steadily decreases with more and more training data being added.
The best results were obtained when \textbf{\emph{no}} PCA was applied and the dataset was using source code data (without pattern) and both type of features (amount and divergence).
The expected error on the withheld validation data falls below \(1\) for about \(15\) or more training instances.
Examining the confidence intervals will allow us to understand the likelihood for the RMSE to deviate from the expectation.

\begin{table}

\caption{\label{tab:nnet-best-model}Normality test and p-value, mean, median, standard deviation, min, and max of the RMSE validation error, for the neural network model, using the source code data, amount- and divergence-features combined, and no PCA. The error reduces steadily with the number of training instances.}
\centering
\begin{tabular}[t]{rllrrrrrr}
\toprule
num\_train & Unimodal? & Normal? & p.value & Mean & Median & Sd & Min & Max\\
\midrule
2 & TRUE & FALSE & 0.0002933 & 2.6560 & 2.6108 & 0.4871 & 1.7428 & 4.3618\\
3 & TRUE & FALSE & 0.0012234 & 2.2683 & 2.2408 & 0.4260 & 1.5189 & 3.8886\\
4 & TRUE & TRUE & 0.9306687 & 2.0016 & 2.0126 & 0.4518 & 1.0477 & 3.2353\\
5 & TRUE & FALSE & 0.0068398 & 1.7502 & 1.6614 & 0.4007 & 1.1662 & 3.0969\\
6 & TRUE & FALSE & 0.0016281 & 1.6044 & 1.5644 & 0.3794 & 1.0437 & 2.7770\\
\addlinespace
7 & FALSE & FALSE & 0.0013964 & 1.4480 & 1.4060 & 0.4012 & 0.8714 & 2.8000\\
8 & TRUE & FALSE & 0.0122055 & 1.2948 & 1.2541 & 0.2881 & 0.6515 & 2.1895\\
9 & TRUE & TRUE & 0.3837350 & 1.1780 & 1.1753 & 0.2670 & 0.5358 & 1.6948\\
10 & TRUE & TRUE & 0.3680971 & 1.1947 & 1.1936 & 0.3093 & 0.6545 & 1.9583\\
11 & TRUE & TRUE & 0.4238983 & 1.0958 & 1.0944 & 0.3191 & 0.4599 & 1.9639\\
\addlinespace
12 & TRUE & TRUE & 0.1660170 & 1.0188 & 1.0387 & 0.2722 & 0.3069 & 1.7309\\
13 & TRUE & TRUE & 0.6085910 & 0.9233 & 0.9128 & 0.2800 & 0.2821 & 1.4495\\
14 & TRUE & FALSE & 0.0467554 & 0.9408 & 0.8998 & 0.2832 & 0.4935 & 1.5325\\
15 & TRUE & TRUE & 0.1011428 & 0.8406 & 0.8464 & 0.2792 & 0.3806 & 1.5476\\
16 & TRUE & TRUE & 0.4112078 & 0.8324 & 0.7839 & 0.2437 & 0.3146 & 1.4020\\
\addlinespace
17 & TRUE & TRUE & 0.1865716 & 0.7487 & 0.7479 & 0.2225 & 0.3617 & 1.1923\\
18 & TRUE & TRUE & 0.4609954 & 0.7508 & 0.7364 & 0.1856 & 0.4262 & 1.1680\\
19 & TRUE & TRUE & 0.0938250 & 0.7126 & 0.6776 & 0.2349 & 0.3424 & 1.2776\\
20 & FALSE & FALSE & 0.0231780 & 0.6655 & 0.6561 & 0.1955 & 0.2776 & 1.0030\\
21 & TRUE & TRUE & 0.1231393 & 0.6167 & 0.5710 & 0.2056 & 0.2484 & 1.0391\\
\addlinespace
22 & FALSE & FALSE & 0.0038020 & 0.6263 & 0.5528 & 0.2191 & 0.2562 & 1.1417\\
23 & TRUE & TRUE & 0.0515424 & 0.6096 & 0.5671 & 0.2082 & 0.2487 & 1.0090\\
24 & FALSE & FALSE & 0.0022813 & 0.5617 & 0.4978 & 0.1853 & 0.2449 & 1.0093\\
25 & TRUE & FALSE & 0.0008305 & 0.5522 & 0.5034 & 0.2220 & 0.2266 & 1.2933\\
26 & TRUE & TRUE & 0.2186786 & 0.5231 & 0.4967 & 0.1488 & 0.2299 & 0.9234\\
\addlinespace
27 & TRUE & FALSE & 0.0151417 & 0.5254 & 0.5092 & 0.1742 & 0.1802 & 1.0202\\
28 & TRUE & TRUE & 0.0511006 & 0.4811 & 0.4603 & 0.1506 & 0.2499 & 0.8991\\
29 & TRUE & TRUE & 0.1096857 & 0.4641 & 0.4531 & 0.1502 & 0.2331 & 0.8483\\
30 & TRUE & FALSE & 0.0220127 & 0.4485 & 0.4285 & 0.1614 & 0.1887 & 0.9925\\
31 & TRUE & FALSE & 0.0000899 & 0.4475 & 0.4152 & 0.1733 & 0.2314 & 1.0600\\
\addlinespace
32 & TRUE & FALSE & 0.0027620 & 0.4059 & 0.3674 & 0.1452 & 0.1573 & 0.8154\\
33 & TRUE & TRUE & 0.0901130 & 0.4125 & 0.3808 & 0.1351 & 0.1585 & 0.7335\\
34 & TRUE & FALSE & 0.0122786 & 0.4067 & 0.3659 & 0.1352 & 0.1605 & 0.7748\\
35 & TRUE & FALSE & 0.0011656 & 0.4039 & 0.3752 & 0.1486 & 0.1929 & 0.8576\\
36 & FALSE & FALSE & 0.0002820 & 0.3985 & 0.3687 & 0.1444 & 0.2258 & 0.9291\\
\addlinespace
37 & TRUE & FALSE & 0.0005344 & 0.4104 & 0.4018 & 0.1563 & 0.1598 & 1.0016\\
38 & TRUE & TRUE & 0.3144747 & 0.3944 & 0.3862 & 0.1296 & 0.1599 & 0.7915\\
39 & TRUE & FALSE & 0.0002961 & 0.3953 & 0.3445 & 0.1657 & 0.1717 & 0.8912\\
40 & TRUE & TRUE & 0.4392768 & 0.3684 & 0.3566 & 0.0993 & 0.1874 & 0.6489\\
41 & TRUE & FALSE & 0.0024674 & 0.3817 & 0.3564 & 0.1308 & 0.2018 & 0.7052\\
\bottomrule
\end{tabular}
\end{table}

Table \ref{tab:nnet-best-model} also shows whether the generalization error is normally distributed (Shapiro and Wilk 1965b; Royston 1982), which it is in many cases (and in some almost).
The advantage of a normally distributed generalization error allows us to make statements with regard to the standard deviation, additionally to the expected error.
Taking the case of \(17\) training instances again, the expected generalization error is \(\approx0.75\) and the standard deviation is \(\approx0.25\).
Since the generalization error is normally distributed, we expect the error to deviate less than or equal to \(0.25\) in about \textbf{\(\approx68\)\%} of all cases, since for a normal distribution, cases that are up to one standard deviation away from the mean account for \(\approx68\)\% of all cases (three sigma rule; Pukelsheim (1994)).
In other words, a neural network trained on \(17\) instances will deliver predictions with a probability of \(\approx68\)\% that are off by \(\leq1\).

Another interesting example from table \ref{tab:nnet-best-model} is the one using \(26\) instances.
The validation error there is unimodal, normally distributed, and has an expectation of \(0.497\) and standard deviation of \(0.153\).
This means, according to the Three Sigma rule which we can apply in this case, that we can expect \(99.7\)\% of all predicted values to be below \(\approx0.956\) (mean plus three standard deviations).
In other words, we can practically guarantee to predict an error that is always less than \(1\).
With about \(68\)\% certainty, the error will be less than \(\approx0.65\). With about \(95\)\% certainty, it will be below \(\approx0.803\).
I'd say these are pretty good expectations for a model trained on only \(26\) instances.

Table \ref{tab:nnet-best-model-confidence-intervals} shows some common confidence intervals (where ``Low'' and ``High'' demarcate the lowest and highest expected deviation, according to the mean and standard deviation of the validation error and the fact that it is normally distributed).
For example, with a confidence of \(99\)\%, the lowest expected error is just \(0.1115\), and the highest is only \(1.3960\).
Note that the error cannot be negative and that this happens because it is not a perfect normal distribution.
So, for example, the confidence is the probability that \(Pr(\mu-a\leq X\leq\mu+a)\). If we use two times the standard deviation, then this will cover \(95.45\)\% of all values.

The table also shows lows and high according to Chebyshev's inequality (Tchébychef 1867).
Chebyshev's inequality is useful in cases when the actual distribution is not known. While the lows are lower and the highs are higher, we still get upper and lower bounds, even for cases where the generalization error is \emph{not} normally distributed.

\begin{table}

\caption{\label{tab:nnet-best-model-confidence-intervals}Confidence intervals of the best neural network model trained on $17$ instances. The table includes values according to the 68-95-99.7\%-rule and Chebyshev's inequality (cut off at 0).}
\centering
\begin{tabular}[t]{rrrrrrr}
\toprule
Confidence & Low & Mean1 & High & Chebyshev\_Low & Mean2 & Chebyshev\_High\\
\midrule
50.00 & 0.5986 & 0.75 & 0.8987 & 0.4340 & 0.75 & 1.0633\\
55.00 & 0.5806 & 0.75 & 0.9167 & 0.4170 & 0.75 & 1.0803\\
60.00 & 0.5614 & 0.75 & 0.9359 & 0.3969 & 0.75 & 1.1005\\
65.00 & 0.5407 & 0.75 & 0.9566 & 0.3726 & 0.75 & 1.1248\\
68.27 & 0.5262 & 0.75 & 0.9712 & 0.3537 & 0.75 & 1.1437\\
\addlinespace
70.00 & 0.5181 & 0.75 & 0.9793 & 0.3424 & 0.75 & 1.1549\\
75.00 & 0.4927 & 0.75 & 1.0046 & 0.3037 & 0.75 & 1.1937\\
80.00 & 0.4635 & 0.75 & 1.0338 & 0.2512 & 0.75 & 1.2462\\
85.00 & 0.4284 & 0.75 & 1.0690 & 0.1742 & 0.75 & 1.3231\\
90.00 & 0.3827 & 0.75 & 1.1146 & 0.0451 & 0.75 & 1.4523\\
\addlinespace
95.00 & 0.3126 & 0.75 & 1.1847 & 0.0000 & 0.75 & 1.7437\\
95.45 & 0.3037 & 0.75 & 1.1937 & 0.0000 & 0.75 & 1.7917\\
97.50 & 0.2500 & 0.75 & 1.2474 & 0.0000 & 0.75 & 2.1558\\
98.00 & 0.2311 & 0.75 & 1.2663 & 0.0000 & 0.75 & 2.3219\\
99.00 & 0.1756 & 0.75 & 1.3218 & 0.0000 & 0.75 & 2.9736\\
\addlinespace
99.50 & 0.1241 & 0.75 & 1.3732 & 0.0000 & 0.75 & 3.8952\\
99.73 & 0.0812 & 0.75 & 1.4161 & 0.0000 & 0.75 & 5.0306\\
\bottomrule
\end{tabular}
\end{table}

Table \ref{tab:confidence-intervals-inequalities} demonstrates four different inequalities.
For \(17\) training instances and a hypothetical RMSE (the \texttt{diff} column), it shows the confidence for the three sigma rule, the Vysochanskij--Petunin inequality, Gauss' inequality, and Chebyshev's inequality.
The order of the inequalities reflects their tightness. The more assumptions we can make about the distribution of the sample, the tighter we can select the bounds.
There are various other \textbf{\emph{concentration inequalities}}, such as Markov's, Paley--Zygmund's, Cantelli's, or Chernoff's bounds.
A disadvantage with Chebyshev's inequality is that it only works for more than one standard deviation (\(\approx1.63\) standard deviations for Vysochanskij--Petunin's inequality).
While Gauss' inequality works in that case, it requires to determine the mode of the sample, and that the sample is unimodal.

Therefore, in practice, I suggest to try one inequality after the other and take the one that has its requirements satisfied first (testing the tight ones first).

\begin{table}

\caption{\label{tab:confidence-intervals-inequalities}Four different inequalities giving and their confidence for a given deviation (for the best neural network model trained on $17$ instances). The Three Sigma rule has the tightest bounds, while Chebyshev's inequality has the loosest. The mean of the data is $0.749$ and the standard deviation is $0.222$.}
\centering
\begin{tabular}[t]{rrrrr}
\toprule
diff & ThreeSigmaR & VysoPetunin & GaussIneq & Chebyshev\\
\midrule
0.0 & 0.00000 & 0.00000 & 0.00000 & 0.00000\\
0.1 & 34.68925 & 0.00000 & 22.65391 & 0.00000\\
0.2 & 63.12935 & 0.00000 & 45.30782 & 0.00000\\
0.3 & 82.24536 & 0.00000 & 67.92495 & 44.99580\\
0.4 & 92.77909 & 86.24895 & 81.95778 & 69.06014\\
\addlinespace
0.5 & 97.53758 & 91.19933 & 88.45298 & 80.19849\\
0.6 & 99.29969 & 93.88842 & 91.98124 & 86.24895\\
0.7 & 99.83456 & 95.50986 & 94.10866 & 89.89719\\
0.8 & 99.96764 & 96.56224 & 95.48945 & 92.26503\\
0.9 & 99.99477 & 97.28374 & 96.43610 & 93.88842\\
\addlinespace
1.0 & 99.99930 & 97.79983 & 97.11325 & 95.04962\\
1.1 & 99.99992 & 98.18168 & 97.61425 & 95.90878\\
1.2 & 99.99999 & 98.47211 & 97.99531 & 96.56224\\
1.3 & 100.00000 & 98.69813 & 98.29186 & 97.07078\\
1.4 & 100.00000 & 98.87747 & 98.52717 & 97.47430\\
\addlinespace
1.5 & 100.00000 & 99.02215 & 98.71700 & 97.79983\\
1.6 & 100.00000 & 99.14056 & 98.87236 & 98.06626\\
1.7 & 100.00000 & 99.23870 & 99.00112 & 98.28707\\
1.8 & 100.00000 & 99.32094 & 99.10903 & 98.47211\\
1.9 & 100.00000 & 99.39054 & 99.20034 & 98.62870\\
\addlinespace
2.0 & 100.00000 & 99.44996 & 99.27831 & 98.76241\\
\bottomrule
\end{tabular}
\end{table}

\begin{figure}
\centering
\includegraphics{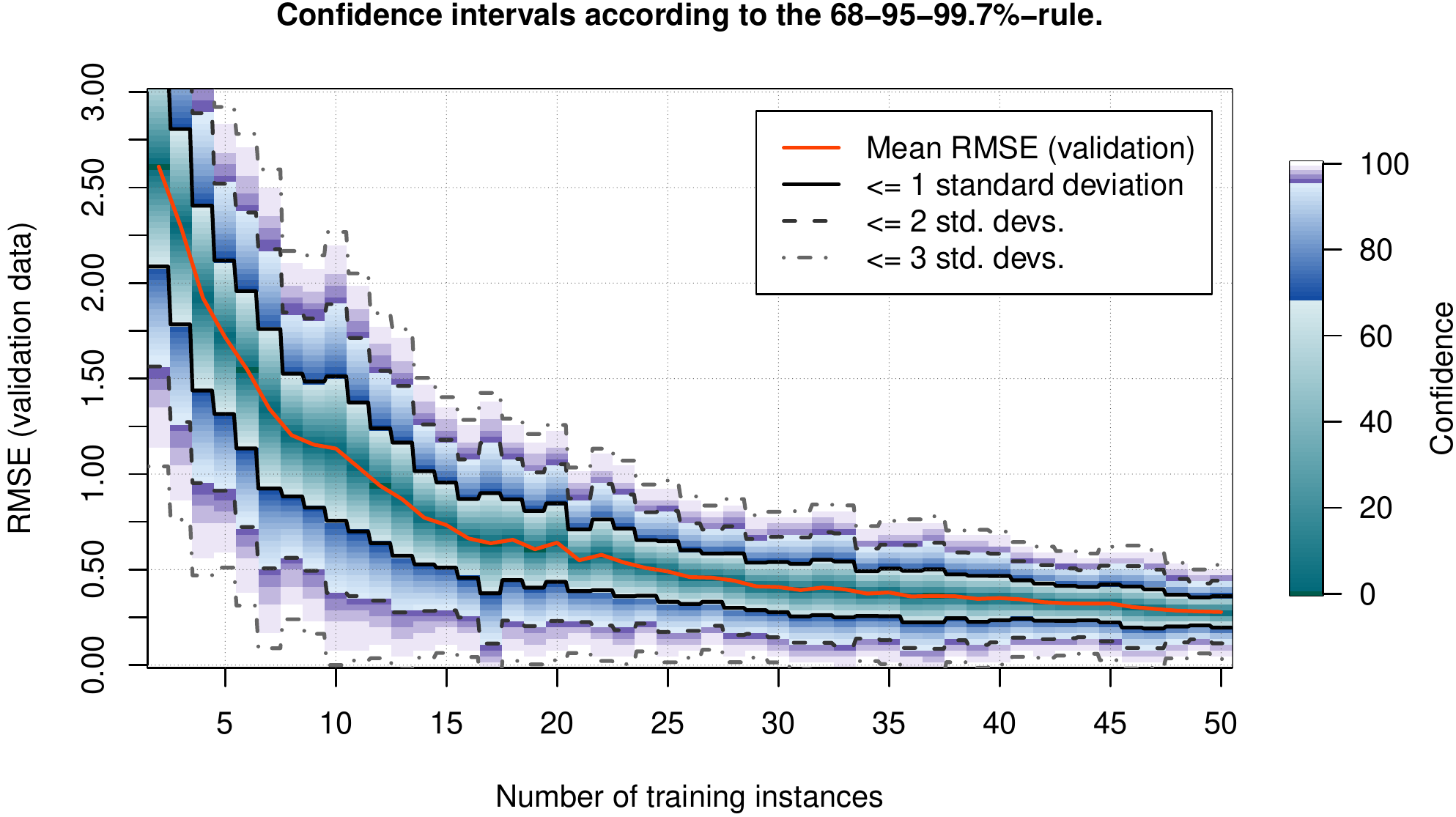}
\caption{\label{fig:nnet-rmse-cis}Continuous confidence of the neural network predictor, with regard to number of training instances. Shown are the values according to the 68-95-99.7\%-rule (assuming a normal distribution for every generalization error). The mean RMSE was determined using 50 models' predictions on validation data. The three color gradients correspond to the three sigmas.}
\end{figure}

Figure \ref{fig:nnet-rmse-cis} shows the confidence interval of the generalization error per number of training instances.
It is obvious that with a lower generalization error, the size of the confidence interval reduces as well (i.e., less variance and more stable predictions).

\begin{figure}
\centering
\includegraphics{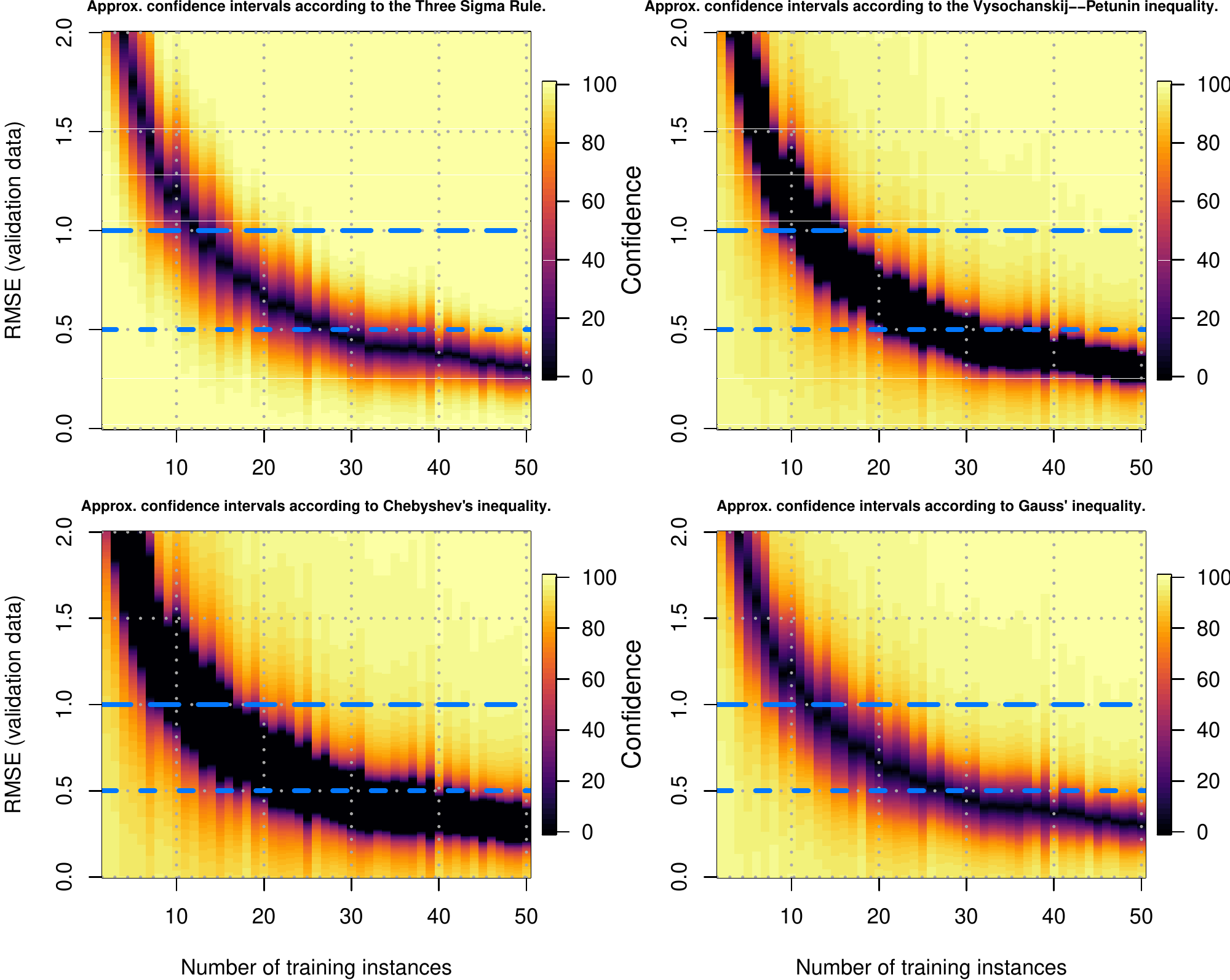}
\caption{\label{fig:nnet-rmse-cis-chebyshev}Continuous confidence of the neural network predictor, with regard to number of training instances. Shown are the values according to four different inequalities.}
\end{figure}

Figure \ref{fig:nnet-rmse-cis-chebyshev} shows the same confidence intervals for the generalization error, according to Chebyshev's inequality (ignoring that some errors are, in fact, normally distributed).
Note that this figure contains a wide area where the confidence is pictured as unknown.
This is because Chebyshev's inequality is only useful for \(k>1\), that is, more than one standard deviations. For \(k\leq1\), we cannot guarantee that any of the values will be within the mean plus/minus up to one standard deviation using Chebyshev's rule, so its utility lies more in obtaining upper bounds for when the distribution is not known.
For the sake of the figure, I have implemented it as \(\frac{1}{\max{(1,k^2)}}\). You see that we have a wide dark area in which we cannot guarantee anything (that it will contain values).
This would then be essentially the same as Selberg's generalization (Selberg 1940).

There is also the Vysochanskij--Petunin inequality, but it only gives bounds for more than \(\sqrt{(\frac{8}{3})}\) standard deviations (Vysochanskij and Petunin 1980).
It is defined as \(\frac{4}{9\lambda^2}\), so for the minimum lambda it includes \(\approx\frac{5}{6}\) of all values (and is not applicable for probabilities below that).

\begin{Shaded}
\begin{Highlighting}[]
\NormalTok{compute\_inequalities }\OtherTok{\textless{}{-}} \ControlFlowTok{function}\NormalTok{(num\_train, rmse, }\AttributeTok{use =} \FunctionTok{c}\NormalTok{(}\StringTok{"threesigma"}\NormalTok{, }\StringTok{"vyso"}\NormalTok{, }\StringTok{"chebyshev"}\NormalTok{,}
  \StringTok{"gauss"}\NormalTok{)) \{}
\NormalTok{  vals }\OtherTok{\textless{}{-}}\NormalTok{ temp[temp}\SpecialCharTok{$}\NormalTok{num\_train }\SpecialCharTok{==}\NormalTok{ num\_train, ]}\SpecialCharTok{$}\NormalTok{rmse\_valid}
\NormalTok{  mean\_ }\OtherTok{\textless{}{-}} \FunctionTok{mean}\NormalTok{(vals)}
\NormalTok{  sd\_ }\OtherTok{\textless{}{-}} \FunctionTok{sd}\NormalTok{(vals)}
\NormalTok{  use }\OtherTok{\textless{}{-}} \FunctionTok{match.arg}\NormalTok{(use)}

  \ControlFlowTok{if}\NormalTok{ (use }\SpecialCharTok{==} \StringTok{"threesigma"}\NormalTok{) \{}
    \CommentTok{\# Three Sigma rule}
\NormalTok{    temp }\OtherTok{\textless{}{-}} \DecValTok{200} \SpecialCharTok{*} \FunctionTok{abs}\NormalTok{(}\FloatTok{0.5} \SpecialCharTok{{-}} \FunctionTok{pnorm}\NormalTok{(}\AttributeTok{q =}\NormalTok{ rmse, }\AttributeTok{mean =}\NormalTok{ mean\_, }\AttributeTok{sd =}\NormalTok{ sd\_))}
\NormalTok{    temp}
    \CommentTok{\# if (temp \textless{} 99.9) temp else NA\_real\_}
\NormalTok{  \} }\ControlFlowTok{else} \ControlFlowTok{if}\NormalTok{ (use }\SpecialCharTok{==} \StringTok{"vyso"}\NormalTok{) \{}
    \CommentTok{\# Vysochanskij–Petunin inequality}
\NormalTok{    min\_ }\OtherTok{\textless{}{-}} \FunctionTok{sqrt}\NormalTok{(}\DecValTok{8}\SpecialCharTok{/}\DecValTok{3}\NormalTok{)}
\NormalTok{    diff\_ }\OtherTok{\textless{}{-}} \FunctionTok{abs}\NormalTok{(mean\_ }\SpecialCharTok{{-}}\NormalTok{ rmse)}
\NormalTok{    lambda }\OtherTok{\textless{}{-}}\NormalTok{ sd\_}\SpecialCharTok{/}\NormalTok{diff\_}
    \CommentTok{\# if (diff\_ \textgreater{}= min\_ * sd\_) 100 * (1 {-} 4 / 9 * lambda\^{}2) else NA\_real\_}
    \FunctionTok{max}\NormalTok{(}\DecValTok{0}\NormalTok{, }\DecValTok{100} \SpecialCharTok{*}\NormalTok{ (}\DecValTok{1} \SpecialCharTok{{-}} \DecValTok{4}\SpecialCharTok{/}\DecValTok{9} \SpecialCharTok{*}\NormalTok{ lambda}\SpecialCharTok{\^{}}\DecValTok{2}\NormalTok{))}
\NormalTok{  \} }\ControlFlowTok{else} \ControlFlowTok{if}\NormalTok{ (use }\SpecialCharTok{==} \StringTok{"chebyshev"}\NormalTok{) \{}
    \CommentTok{\# Chebyshev:}
\NormalTok{    diff }\OtherTok{\textless{}{-}} \FunctionTok{abs}\NormalTok{(mean\_ }\SpecialCharTok{{-}}\NormalTok{ rmse)}
\NormalTok{    k2 }\OtherTok{\textless{}{-}}\NormalTok{ (diff}\SpecialCharTok{/}\NormalTok{sd\_)}\SpecialCharTok{\^{}}\DecValTok{2}
    \DecValTok{100} \SpecialCharTok{*}\NormalTok{ (}\DecValTok{1} \SpecialCharTok{{-}} \DecValTok{1}\SpecialCharTok{/}\FunctionTok{max}\NormalTok{(}\DecValTok{1}\NormalTok{, k2))}
    \CommentTok{\# if (k2 \textless{} 1) NA\_real\_ else 100 * (1 {-} 1 / k2)}
\NormalTok{  \} }\ControlFlowTok{else} \ControlFlowTok{if}\NormalTok{ (use }\SpecialCharTok{==} \StringTok{"gauss"}\NormalTok{) \{}
    \CommentTok{\# Gauss\textquotesingle{}s inequality:}
\NormalTok{    k }\OtherTok{\textless{}{-}} \FunctionTok{abs}\NormalTok{(mean\_ }\SpecialCharTok{{-}}\NormalTok{ rmse)}
\NormalTok{    modes }\OtherTok{\textless{}{-}}\NormalTok{ LaplacesDemon}\SpecialCharTok{::}\FunctionTok{Modes}\NormalTok{(}\AttributeTok{x =}\NormalTok{ vals)}\SpecialCharTok{$}\NormalTok{modes}
    \ControlFlowTok{if}\NormalTok{ (}\FunctionTok{length}\NormalTok{(modes) }\SpecialCharTok{!=} \DecValTok{1}\NormalTok{) \{}
      \CommentTok{\# return(NA\_real\_) \# Only works for unimodal data; fall back to}
      \CommentTok{\# Chebyshev:}
\NormalTok{      k2 }\OtherTok{\textless{}{-}}\NormalTok{ (k}\SpecialCharTok{/}\NormalTok{sd\_)}\SpecialCharTok{\^{}}\DecValTok{2}
      \DecValTok{100} \SpecialCharTok{*}\NormalTok{ (}\DecValTok{1} \SpecialCharTok{{-}} \DecValTok{1}\SpecialCharTok{/}\FunctionTok{max}\NormalTok{(}\DecValTok{1}\NormalTok{, k2))}
\NormalTok{    \}}
\NormalTok{    mode\_ }\OtherTok{\textless{}{-}}\NormalTok{ modes[}\DecValTok{1}\NormalTok{]}
\NormalTok{    r }\OtherTok{\textless{}{-}} \FunctionTok{sqrt}\NormalTok{((mean\_ }\SpecialCharTok{{-}}\NormalTok{ mode\_)}\SpecialCharTok{\^{}}\DecValTok{2} \SpecialCharTok{+}\NormalTok{ sd\_}\SpecialCharTok{\^{}}\DecValTok{2}\NormalTok{)}
    \DecValTok{100} \SpecialCharTok{*}\NormalTok{ (}\DecValTok{1} \SpecialCharTok{{-}}\NormalTok{ (}\ControlFlowTok{if}\NormalTok{ (k }\SpecialCharTok{\textgreater{}=} \DecValTok{2} \SpecialCharTok{*}\NormalTok{ r}\SpecialCharTok{/}\FunctionTok{sqrt}\NormalTok{(}\DecValTok{3}\NormalTok{))}
\NormalTok{      ((}\DecValTok{2} \SpecialCharTok{*}\NormalTok{ r)}\SpecialCharTok{/}\NormalTok{(}\DecValTok{3} \SpecialCharTok{*}\NormalTok{ k))}\SpecialCharTok{\^{}}\DecValTok{2} \ControlFlowTok{else}\NormalTok{ (}\DecValTok{1} \SpecialCharTok{{-}}\NormalTok{ k}\SpecialCharTok{/}\NormalTok{(r }\SpecialCharTok{*} \FunctionTok{sqrt}\NormalTok{(}\DecValTok{3}\NormalTok{)))))}
\NormalTok{  \}}
\NormalTok{\}}
\end{Highlighting}
\end{Shaded}

\hypertarget{variable-importance}{%
\subsubsection{Variable Importance}\label{variable-importance}}

Here we will estimate the variable importance, similar to how it was done previously.
The main difference is that instead of only relying on partial least squares, we will rely on the averaged variable importance of a handful of models, in order to get more robust results.

\begin{Shaded}
\begin{Highlighting}[]
\NormalTok{vi\_np\_sc }\OtherTok{\textless{}{-}} \FunctionTok{loadResultsOrCompute}\NormalTok{(}\AttributeTok{file =} \StringTok{"../results/rob{-}reg\_vi\_np\_sc.rds"}\NormalTok{, }\AttributeTok{computeExpr =}\NormalTok{ \{}
  \FunctionTok{doWithParallelCluster}\NormalTok{(}\AttributeTok{expr =}\NormalTok{ \{}
\NormalTok{    templ }\OtherTok{\textless{}{-}} \FunctionTok{list}\NormalTok{()}
    \ControlFlowTok{for}\NormalTok{ (method }\ControlFlowTok{in} \FunctionTok{c}\NormalTok{(}\StringTok{"gamboost"}\NormalTok{, }\StringTok{"nnet"}\NormalTok{, }\StringTok{"pls"}\NormalTok{, }\StringTok{"plsRglm"}\NormalTok{, }\StringTok{"treebag"}\NormalTok{)) \{}
\NormalTok{      templ[[method]] }\OtherTok{\textless{}{-}} \FunctionTok{create\_variable\_importance}\NormalTok{(}\AttributeTok{dataset =}\NormalTok{ dataset\_np\_sc\_oversampled,}
        \AttributeTok{method =}\NormalTok{ method)}
\NormalTok{    \}}
\NormalTok{    templ}
\NormalTok{  \})}
\NormalTok{\})}

\NormalTok{vi\_np\_it }\OtherTok{\textless{}{-}} \FunctionTok{loadResultsOrCompute}\NormalTok{(}\AttributeTok{file =} \StringTok{"../results/rob{-}reg\_vi\_np\_it.rds"}\NormalTok{, }\AttributeTok{computeExpr =}\NormalTok{ \{}
  \FunctionTok{doWithParallelCluster}\NormalTok{(}\AttributeTok{expr =}\NormalTok{ \{}
\NormalTok{    templ }\OtherTok{\textless{}{-}} \FunctionTok{list}\NormalTok{()}
    \ControlFlowTok{for}\NormalTok{ (method }\ControlFlowTok{in} \FunctionTok{c}\NormalTok{(}\StringTok{"gamboost"}\NormalTok{, }\StringTok{"nnet"}\NormalTok{, }\StringTok{"pls"}\NormalTok{, }\StringTok{"plsRglm"}\NormalTok{, }\StringTok{"treebag"}\NormalTok{)) \{}
\NormalTok{      templ[[method]] }\OtherTok{\textless{}{-}} \FunctionTok{create\_variable\_importance}\NormalTok{(}\AttributeTok{dataset =}\NormalTok{ dataset\_np\_it\_oversampled,}
        \AttributeTok{method =}\NormalTok{ method)}
\NormalTok{    \}}
\NormalTok{    templ}
\NormalTok{  \})}
\NormalTok{\})}
\end{Highlighting}
\end{Shaded}

\begin{figure}
\centering
\includegraphics{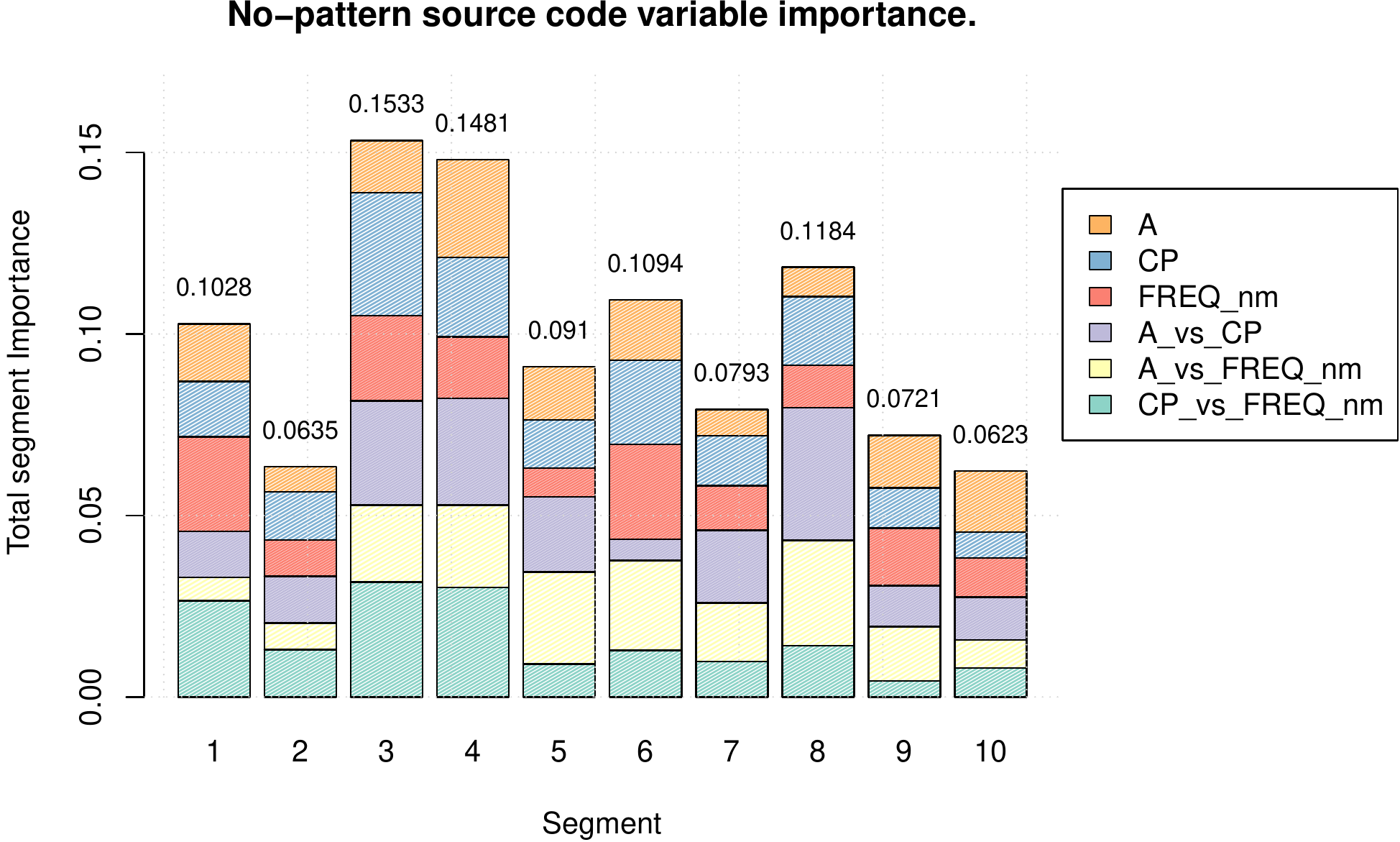}
\caption{\label{fig:varimp-np-sc}Variable importance for the no-pattern source code dataset estimated using 100 repeats and averaged over five models (Boosted Generalized Additive Model (Hofner, Boccuto, and Göker 2015), Neural network (Venables and Ripley 2002), (Generalized Linear) Partial Least Squares (Bertrand and Maumy-Bertrand 2014), and Bagged CART (Peters and Hothorn 2019)).}
\end{figure}

\begin{table}

\caption{\label{tab:varimp-np-sc-detailed}Detailed variable importance (scaled to percent) for the no-pattern source code dataset, including means and sums across segments and features.}
\centering
\begin{tabular}[t]{lrrrrrrrrrrr}
\toprule
  & Seg1 & Seg2 & Seg3 & Seg4 & Seg5 & Seg6 & Seg7 & Seg8 & Seg9 & Seg10 & sum\\
\midrule
A & 1.58 & 0.69 & 1.44 & 2.70 & 1.47 & 1.67 & 0.73 & 0.81 & 1.44 & 1.68 & 14.20\\
CP & 1.53 & 1.32 & 3.39 & 2.19 & 1.33 & 2.31 & 1.38 & 1.89 & 1.10 & 0.72 & 17.16\\
FREQ\_nm & 2.60 & 1.00 & 2.35 & 1.69 & 0.79 & 2.62 & 1.23 & 1.17 & 1.59 & 1.08 & 16.12\\
A\_vs\_CP & 1.27 & 1.28 & 2.87 & 2.95 & 2.08 & 0.59 & 2.01 & 3.65 & 1.13 & 1.18 & 19.00\\
A\_vs\_FREQ\_nm & 0.64 & 0.74 & 2.12 & 2.27 & 2.53 & 2.47 & 1.61 & 2.91 & 1.49 & 0.77 & 17.55\\
\addlinespace
CP\_vs\_FREQ\_nm & 2.65 & 1.31 & 3.16 & 3.02 & 0.91 & 1.28 & 0.98 & 1.41 & 0.45 & 0.80 & 15.97\\
mean & 1.71 & 1.06 & 2.55 & 2.47 & 1.52 & 1.82 & 1.32 & 1.97 & 1.20 & 1.04 & NA\\
sum & 10.28 & 6.35 & 15.33 & 14.81 & 9.10 & 10.94 & 7.93 & 11.84 & 7.21 & 6.23 & 200.00\\
\bottomrule
\end{tabular}
\end{table}

Table \ref{tab:varimp-np-sc-detailed} shows the variable importance from figure \ref{fig:varimp-np-sc}, including summaries for average and total importance across segments and features.

\begin{figure}
\centering
\includegraphics{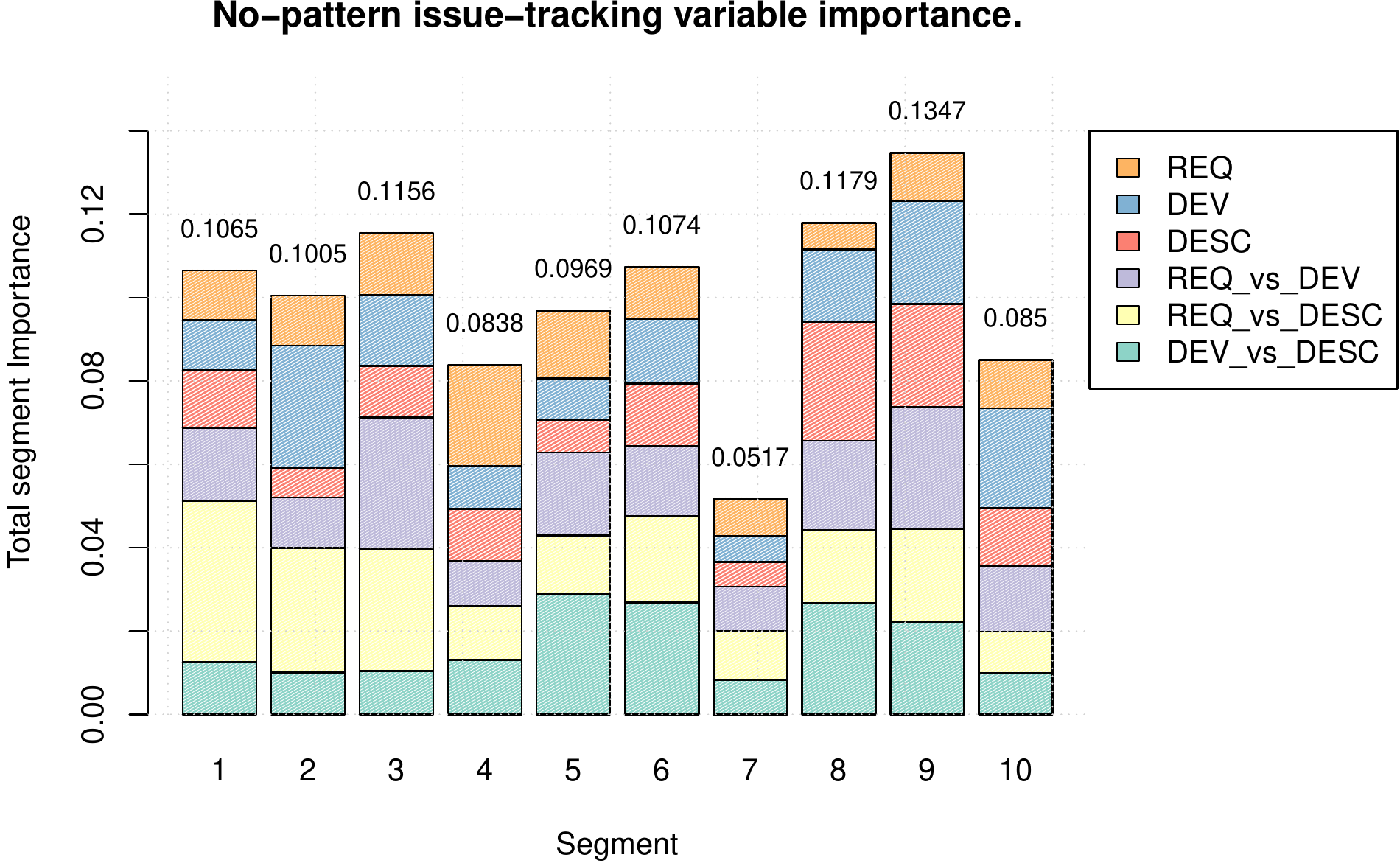}
\caption{\label{fig:varimp-np-it}Variable importance for the no-pattern issue-tracking dataset estimated using 100 repeats and averaged over the same five models as used for source code.}
\end{figure}

\begin{table}

\caption{\label{tab:varimp-np-it-detailed}Detailed variable importance (scaled to percent) for the no-pattern issue-tracking dataset, including means and sums across segments and features.}
\centering
\begin{tabular}[t]{lrrrrrrrrrrr}
\toprule
  & Seg1 & Seg2 & Seg3 & Seg4 & Seg5 & Seg6 & Seg7 & Seg8 & Seg9 & Seg10 & sum\\
\midrule
REQ & 1.19 & 1.20 & 1.50 & 2.42 & 1.63 & 1.25 & 0.89 & 0.64 & 1.15 & 1.16 & 13.02\\
DEV & 1.20 & 2.93 & 1.70 & 1.02 & 1.00 & 1.55 & 0.62 & 1.74 & 2.48 & 2.39 & 16.63\\
DESC & 1.39 & 0.72 & 1.24 & 1.26 & 0.78 & 1.50 & 0.60 & 2.84 & 2.47 & 1.39 & 14.19\\
REQ\_vs\_DEV & 1.76 & 1.21 & 3.14 & 1.07 & 1.99 & 1.68 & 1.06 & 2.15 & 2.91 & 1.58 & 18.55\\
REQ\_vs\_DESC & 3.86 & 2.99 & 2.94 & 1.31 & 1.42 & 2.07 & 1.17 & 1.75 & 2.23 & 0.99 & 20.72\\
\addlinespace
DEV\_vs\_DESC & 1.25 & 1.00 & 1.04 & 1.30 & 2.88 & 2.68 & 0.83 & 2.67 & 2.23 & 1.00 & 16.89\\
mean & 1.78 & 1.68 & 1.93 & 1.40 & 1.62 & 1.79 & 0.86 & 1.97 & 2.24 & 1.42 & NA\\
sum & 10.65 & 10.05 & 11.56 & 8.38 & 9.69 & 10.74 & 5.17 & 11.79 & 13.47 & 8.50 & 200.00\\
\bottomrule
\end{tabular}
\end{table}

Figure \ref{fig:varimp-np-it} and table \ref{tab:varimp-np-it-detailed} show the same thing, only for issue-tracking data.

In figure \ref{fig:varimp-np-sc} it becomes evident that the divergence features are slightly more important than the amount features (with \(\approx\) 52.52\%), when \textbf{both} types of features are used in a regression model.
The Fire Drill is a phenomenon that relies on a sensible balance between activities and disturbances to this balance can indicate it.
Also, the source code activities may not capture the true activities with high precision, so reliance on the divergence between activities is a feature with stronger predictive power on average.
Still, it is important \emph{how much} of each activity happened in each segment.

Table \ref{tab:varimp-np-sc-detailed} shows the total importance per feature.
The table shows that the divergence features are the most important, esp.~\texttt{A\_vs\_CP}, which, in theory, should be the strongest indicator of a Fire Drill, as it can give indication to a disturbed balance between two main development activities (that should exhibit a certain balance otherwise).
However, the relative importance between features is perhaps not significantly different (i.e., there is no feature with extreme low or high relative importance).

In figure \ref{fig:varimp-np-sc} we can already see the total importance per segment.
We can see that the two most important segments are \(\{3,4\}\).
The segments before and after are less important and there is almost a somewhat linearly declining trend, at least towards project end.
That could indicate that a Fire Drill may very well be detectable (predictable with high confidence) in earlier stages of a project.

\hypertarget{decision-tree}{%
\subsubsection{Decision Tree}\label{decision-tree}}

Another kind of regression model is a decision tree. The goal here is to construct an approximate, yet useful tree.
Today, we have \emph{no quantitative} understanding of the Fire Drill and a decision tree can provide us with one such perspective.

In the following, we build a tree based on the ``amount''-features only. The reason behind this is simple: these features are plainly understandable and straightforward to comprehend.
Also, we do not need to z-standardize them, which is almost required if we want to interpret the resulting tree.

\begin{Shaded}
\begin{Highlighting}[]
\FunctionTok{set.seed}\NormalTok{(}\DecValTok{1}\NormalTok{)}
\NormalTok{temp }\OtherTok{\textless{}{-}}\NormalTok{ dataset\_np\_sc\_oversampled[, }\SpecialCharTok{!}\FunctionTok{grepl}\NormalTok{(}\AttributeTok{pattern =} \StringTok{"\_vs\_"}\NormalTok{, }\AttributeTok{x =} \FunctionTok{colnames}\NormalTok{(dataset\_np\_sc\_oversampled))]}
\NormalTok{temp }\OtherTok{\textless{}{-}} \FunctionTok{cbind}\NormalTok{(}\DecValTok{100} \SpecialCharTok{*}\NormalTok{ temp[, }\FunctionTok{colnames}\NormalTok{(temp) }\SpecialCharTok{!=} \StringTok{"gt"}\NormalTok{], }\FunctionTok{data.frame}\NormalTok{(}\AttributeTok{gt =}\NormalTok{ temp}\SpecialCharTok{$}\NormalTok{gt))}

\NormalTok{cn }\OtherTok{\textless{}{-}} \FunctionTok{colnames}\NormalTok{(temp)}
\NormalTok{cn\_x }\OtherTok{\textless{}{-}}\NormalTok{ cn[cn }\SpecialCharTok{!=} \StringTok{"gt"}\NormalTok{]}
\NormalTok{cn\_x }\OtherTok{\textless{}{-}} \FunctionTok{gsub}\NormalTok{(}\AttributeTok{pattern =} \StringTok{"\_nm"}\NormalTok{, }\AttributeTok{replacement =} \StringTok{"nm"}\NormalTok{, }\AttributeTok{x =}\NormalTok{ cn\_x, }\AttributeTok{ignore.case =} \ConstantTok{TRUE}\NormalTok{)}
\FunctionTok{colnames}\NormalTok{(temp) }\OtherTok{\textless{}{-}} \FunctionTok{c}\NormalTok{(cn\_x, }\StringTok{"gt"}\NormalTok{)}

\CommentTok{\# cn \textless{}{-} colnames(temp) cn\_x \textless{}{-} cn[cn != \textquotesingle{}gt\textquotesingle{}] pre\_proc\_method \textless{}{-} c(\textquotesingle{}nzv\textquotesingle{},}
\CommentTok{\# \textquotesingle{}center\textquotesingle{}, \textquotesingle{}scale\textquotesingle{}) pre\_proc \textless{}{-} caret::preProcess(x = temp[, cn\_x], method =}
\CommentTok{\# pre\_proc\_method) temp \textless{}{-} stats::predict(pre\_proc, newdata = temp)}

\CommentTok{\# build the initial decision tree}
\NormalTok{tree }\OtherTok{\textless{}{-}} \FunctionTok{rpart}\NormalTok{(gt }\SpecialCharTok{\textasciitilde{}}\NormalTok{ ., }\AttributeTok{data =}\NormalTok{ temp, }\AttributeTok{method =} \StringTok{"anova"}\NormalTok{, }\AttributeTok{control =} \FunctionTok{rpart.control}\NormalTok{(}\AttributeTok{minsplit =} \DecValTok{3}\NormalTok{,}
  \AttributeTok{cp =} \FloatTok{1e{-}06}\NormalTok{))}

\CommentTok{\# identify best cp value to use}
\NormalTok{best }\OtherTok{\textless{}{-}}\NormalTok{ tree}\SpecialCharTok{$}\NormalTok{cptable[}\FunctionTok{which.min}\NormalTok{(tree}\SpecialCharTok{$}\NormalTok{cptable[, }\StringTok{"xerror"}\NormalTok{]), }\StringTok{"CP"}\NormalTok{]}

\CommentTok{\# produce a pruned tree based on the best cp value}
\NormalTok{pruned\_tree }\OtherTok{\textless{}{-}} \FunctionTok{prune}\NormalTok{(tree, }\AttributeTok{cp =}\NormalTok{ best)}
\end{Highlighting}
\end{Shaded}

\begin{figure}[ht!]
\includegraphics{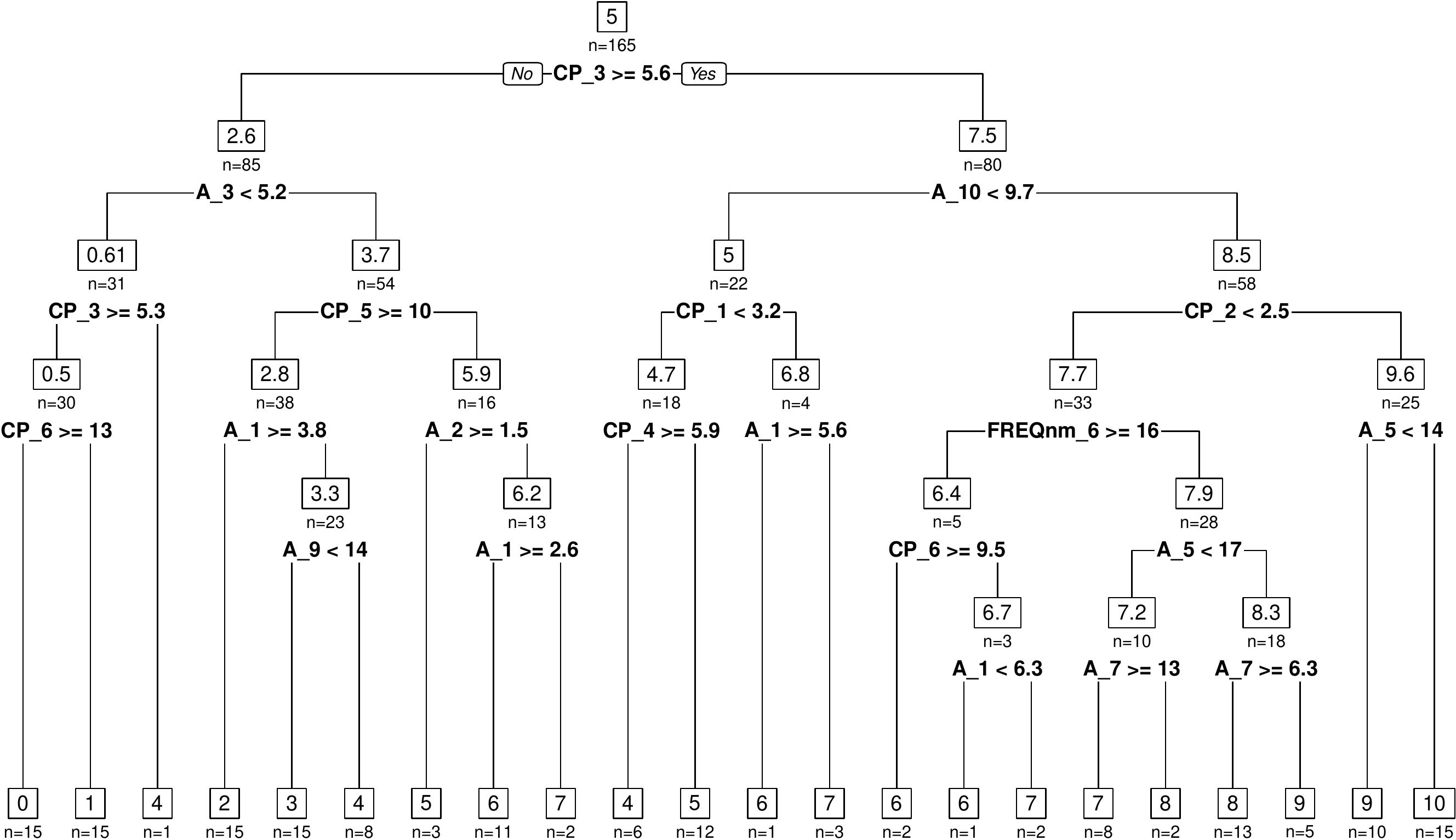} \caption{A decision tree using source code data and amount-type features only.}\label{fig:dec-tree-amount-feats}
\end{figure}

The tree is shown in figure \ref{fig:dec-tree-amount-feats}. The values have been scaled to reflect percent.
The leaf nodes are the predicted Fire Drill severity and the tree can be traversed comfortably manually.
The root node has a severity of \(5\), which is similar to the \emph{ZeroR} regression rule, in which we predict the average without further information.
If, for example, we observe that in the 3rd segment, more than \(5.8\)\% of the total \texttt{CP}-activity have happened, then we proceed to the next node, which corresponds to a severity of \(7.5\).
We can proceed with the drill down to obtain more and more accurate estimates.

\hypertarget{weighted-mixtures-from-data}{%
\subsubsection{Weighted Mixtures from Data}\label{weighted-mixtures-from-data}}

We have previously estimated densities for each project and activity.
Using the ground truth, we can produce a mixture using the projects' weights.

The main purpose of this exercise is to obtain a density, per activity, that is indicative of the Fire Drill behavior for this activity.
It lets us better understand the Fire Drill in terms of these activities, since it is a weighted mixture of the observed instances.
It could also be then used as a pattern.
Like the decision tree, this is another kind of quantitative understanding of the Fire Drill.

\begin{figure}
\centering
\includegraphics{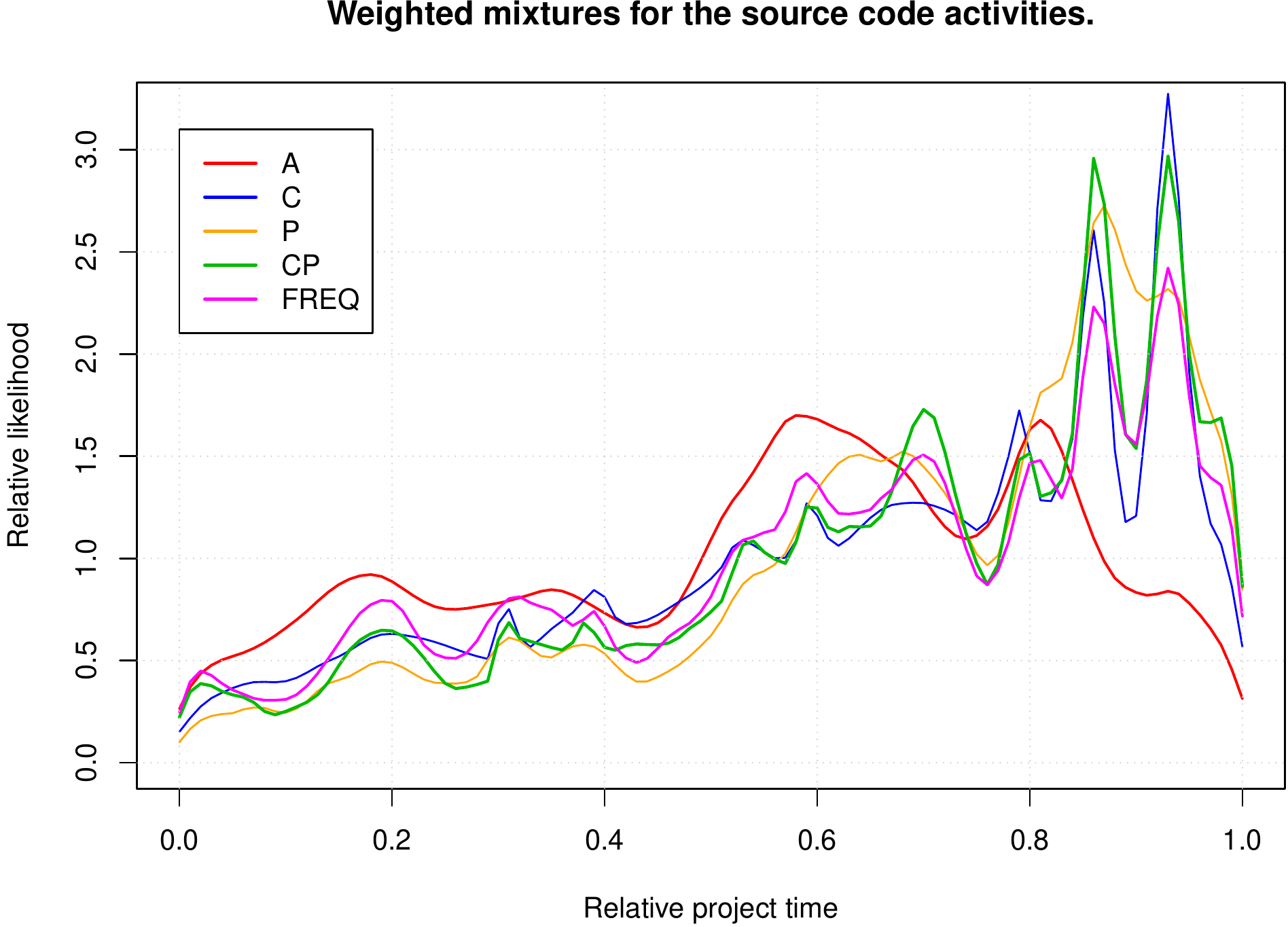}
\caption{\label{fig:weighted-mixt-sc}The weighted mixtures as of the source code activities. For FREQ, the non-mixture version is shown.}
\end{figure}

Figure \ref{fig:weighted-mixt-sc} shows the mixtures for source code.
It appears that we have most of the adaptive activities in the 2nd half of the project, with a decline then towards the end.
Corrective and perfective activities (Swanson 1976; Mockus and Votta 2000) steadily increase and peak towards the end.
My interpretation of this plot, after having seen the raters' notes, would be a late-stage rush with adding new features, followed by switch of dedication of time towards making them work as the project approaches the deadline.

In figure \ref{fig:weighted-mixt-sc} I had split \texttt{CP} into corrective and perfective again.
While the Fire Drill does not make a distinction between these two activities, we can perfectly do it again.
It appears that both activities exhibit very similar behavior: Up until \(\approx75\) \% of the time, they show a somewhat linear increase.
After that, they spike, and the decline only happens close to project end in both cases.
Now the spikes do not look exactly the same.
For corrective, it appears to have been carried out in three consecutive phases, each of which contained more corrective activity as the predecessing phase.
It could hint at some sort of ``panic'' that grew towards project end.
Perfective activities have basically only have a single bulk in that time. However, we can also discern another concentration right before it. Interestingly, that concentration ends (normalizes), and right after that, we get into highly active phases of both, corrective and perfective.
While the accumulation of perfective activities looks more homogeneous, we could perhaps still see this as what is perhaps three spikes.
The first one lasts a little longer than its corrective counter part, and the second and third seem to overlap with (perhaps accommodate) corrective activities.

\begin{Shaded}
\begin{Highlighting}[]
\NormalTok{weighted\_mixture\_sc }\OtherTok{\textless{}{-}} \ControlFlowTok{function}\NormalTok{(x, }\AttributeTok{activity =} \FunctionTok{c}\NormalTok{(}\StringTok{"A"}\NormalTok{, }\StringTok{"C"}\NormalTok{, }\StringTok{"P"}\NormalTok{, }\StringTok{"CP"}\NormalTok{, }\StringTok{"FREQ\_nm"}\NormalTok{)) \{}
\NormalTok{  activity }\OtherTok{\textless{}{-}} \FunctionTok{match.arg}\NormalTok{(activity)}
\NormalTok{  s }\OtherTok{\textless{}{-}} \FunctionTok{sum}\NormalTok{(ground\_truth\_all}\SpecialCharTok{$}\NormalTok{consensus)}
\NormalTok{  y }\OtherTok{\textless{}{-}} \DecValTok{0}

  \ControlFlowTok{for}\NormalTok{ (pidx }\ControlFlowTok{in} \DecValTok{1}\SpecialCharTok{:}\DecValTok{15}\NormalTok{) \{}
\NormalTok{    p }\OtherTok{\textless{}{-}} \FunctionTok{paste0}\NormalTok{(}\StringTok{"project\_"}\NormalTok{, pidx)}
\NormalTok{    w }\OtherTok{\textless{}{-}}\NormalTok{ ground\_truth\_all[ground\_truth\_all}\SpecialCharTok{$}\NormalTok{project }\SpecialCharTok{==}\NormalTok{ p, ]}\SpecialCharTok{$}\NormalTok{consensus}\SpecialCharTok{/}\NormalTok{s}
\NormalTok{    y }\OtherTok{\textless{}{-}}\NormalTok{ y }\SpecialCharTok{+}\NormalTok{ w }\SpecialCharTok{*}\NormalTok{ projects\_sc[[p]][[activity]](x)}
\NormalTok{  \}}
\NormalTok{  y}
\NormalTok{\}}
\end{Highlighting}
\end{Shaded}

\begin{figure}
\centering
\includegraphics{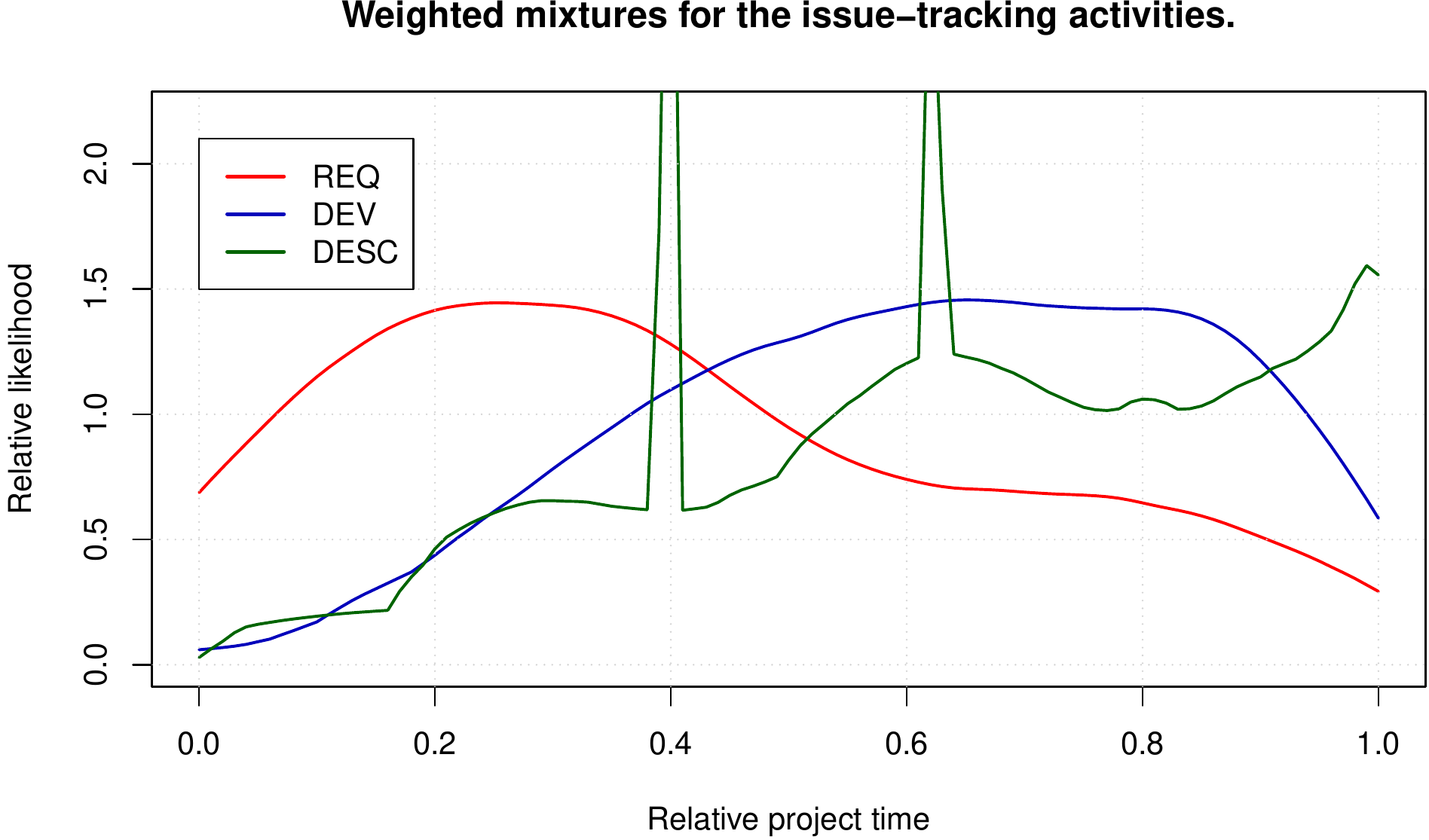}
\caption{\label{fig:weighted-mixt-it}The weighted mixtures as of the issue-tracking activities.}
\end{figure}

For issue-tracking, we previously only used the CDF from oversampling directly.
Therefore, we have to redo this task and take the PDF then.
The result is shown in figure \ref{fig:weighted-mixt-it}.
Time spent on \texttt{REQ} is much more likely in the first half of a project and declines to almost zero towards project end.
\texttt{DEV} begins at almost zero, shows a slow start, and reaches a somewhat wide peak in the second half of a project. It meets its decline only close to the end of a project.
Descoping, \texttt{DESC}, almost linearly increases from project begin towards project end.
Therefore, the more time has passed, the more likely descoping is.

\begin{Shaded}
\begin{Highlighting}[]
\NormalTok{weighted\_mixture\_it }\OtherTok{\textless{}{-}}\NormalTok{ (}\ControlFlowTok{function}\NormalTok{() \{}
  \FunctionTok{library}\NormalTok{(readxl)}

\NormalTok{  plist }\OtherTok{\textless{}{-}} \FunctionTok{list}\NormalTok{()}

  \ControlFlowTok{for}\NormalTok{ (pname }\ControlFlowTok{in} \FunctionTok{names}\NormalTok{(projects\_it)) \{}
\NormalTok{    temp }\OtherTok{\textless{}{-}} \FunctionTok{read\_excel}\NormalTok{(}\StringTok{"../data/FD\_issue{-}based\_detection.xlsx"}\NormalTok{, }\AttributeTok{sheet =}\NormalTok{ pname)}
\NormalTok{    templ }\OtherTok{\textless{}{-}} \FunctionTok{list}\NormalTok{()}

    \ControlFlowTok{for}\NormalTok{ (activity }\ControlFlowTok{in} \FunctionTok{c}\NormalTok{(}\StringTok{"REQ"}\NormalTok{, }\StringTok{"DEV"}\NormalTok{, }\StringTok{"DESC"}\NormalTok{)) \{}
\NormalTok{      use\_y }\OtherTok{\textless{}{-}} \FunctionTok{as.numeric}\NormalTok{(temp[[}\FunctionTok{tolower}\NormalTok{(activity)]])}
\NormalTok{      use\_y[}\FunctionTok{is.na}\NormalTok{(use\_y)] }\OtherTok{\textless{}{-}} \DecValTok{0}
\NormalTok{      templ[[activity]] }\OtherTok{\textless{}{-}} \FunctionTok{rejection\_sampling\_issue\_tracking}\NormalTok{(}\AttributeTok{use\_x =}\NormalTok{ temp}\SpecialCharTok{$}\StringTok{\textasciigrave{}}\AttributeTok{time\%}\StringTok{\textasciigrave{}}\NormalTok{,}
        \AttributeTok{use\_y =}\NormalTok{ use\_y)}
\NormalTok{    \}}

\NormalTok{    plist[[pname]] }\OtherTok{\textless{}{-}}\NormalTok{ templ}
\NormalTok{  \}}


  \ControlFlowTok{function}\NormalTok{(x, }\AttributeTok{activity =} \FunctionTok{c}\NormalTok{(}\StringTok{"REQ"}\NormalTok{, }\StringTok{"DEV"}\NormalTok{, }\StringTok{"DESC"}\NormalTok{)) \{}
\NormalTok{    activity }\OtherTok{\textless{}{-}} \FunctionTok{match.arg}\NormalTok{(activity)}
\NormalTok{    s }\OtherTok{\textless{}{-}} \FunctionTok{sum}\NormalTok{(ground\_truth\_all}\SpecialCharTok{$}\NormalTok{consensus)}
\NormalTok{    y }\OtherTok{\textless{}{-}} \DecValTok{0}

    \ControlFlowTok{for}\NormalTok{ (pidx }\ControlFlowTok{in} \DecValTok{1}\SpecialCharTok{:}\DecValTok{15}\NormalTok{) \{}
\NormalTok{      w }\OtherTok{\textless{}{-}}\NormalTok{ ground\_truth\_all[ground\_truth\_all}\SpecialCharTok{$}\NormalTok{project }\SpecialCharTok{==} \FunctionTok{paste0}\NormalTok{(}\StringTok{"project\_"}\NormalTok{,}
\NormalTok{        pidx), ]}\SpecialCharTok{$}\NormalTok{consensus}\SpecialCharTok{/}\NormalTok{s}
\NormalTok{      y }\OtherTok{\textless{}{-}}\NormalTok{ y }\SpecialCharTok{+}\NormalTok{ w }\SpecialCharTok{*}\NormalTok{ plist[[}\FunctionTok{paste0}\NormalTok{(}\StringTok{"Project"}\NormalTok{, pidx)]][[activity]]}\SpecialCharTok{$}\FunctionTok{PDF}\NormalTok{(x)}
\NormalTok{    \}}
\NormalTok{    y}
\NormalTok{  \}}
\NormalTok{\})()}
\end{Highlighting}
\end{Shaded}

\hypertarget{unsupervised-learning}{%
\subsection{Unsupervised Learning}\label{unsupervised-learning}}

Let us also try unsupervised learning: Clustering of similar projects.
The goal here is to see, whether similar projects (that is, with a similar ground truth) end up in the same clusters or are at least close to each other in a lower-dimensional space.

\hypertarget{d-and-3d-embeddings-using-t-distributed-stochastic-neighbor-embeddings}{%
\subsubsection{2D and 3D embeddings using t-distributed Stochastic Neighbor Embeddings}\label{d-and-3d-embeddings-using-t-distributed-stochastic-neighbor-embeddings}}

When we cluster, we must specify the number of clusters ahead of time.
t-SNE on the other hand, reduces the dimensionality of our observations (Maaten and Hinton 2008).
If we choose to produce low-dimensional observations (i.e., 2D or 3D), then we can comfortably plot the result.
The expectation is that we can observe clusters of similar observations (here: projects with a similar ground truth).

\begin{figure}[ht!]
\includegraphics{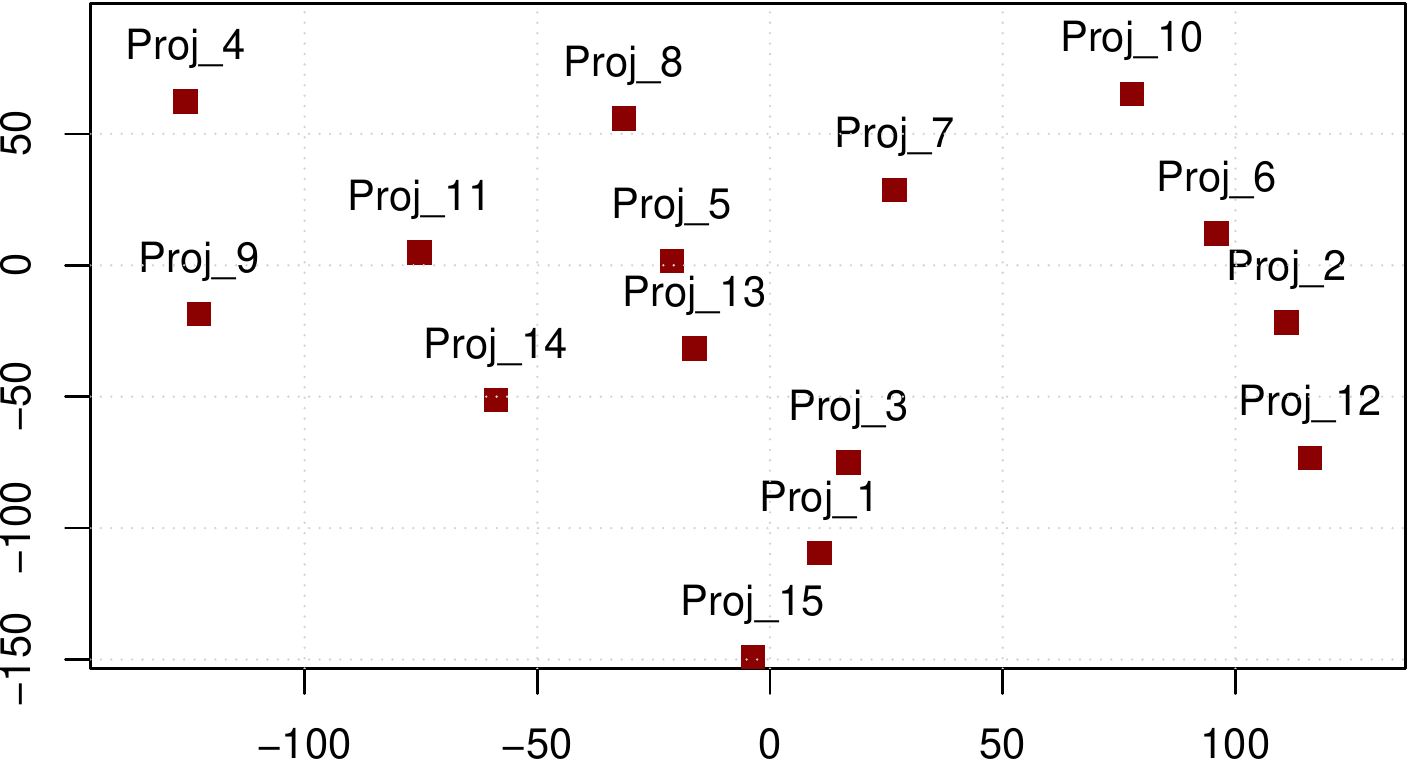} \caption{2-dimensional t-SNE embedding of all projects.}\label{fig:tsne-first-test}
\end{figure}

Figure \ref{fig:tsne-first-test} shows a 2D-embedding of our \(15\) projects.
It is problematic to have so few projects and so many expected clusters (ideally, we expect each severity to be a cluster, which means \(11\) clusters).
Therefore, we could try to either use the oversampled data, or to group projects into severity ranges in order to reduce the amount of expected clusters.

\begin{Shaded}
\begin{Highlighting}[]
\FunctionTok{set.seed}\NormalTok{(}\DecValTok{1}\NormalTok{)}
\NormalTok{temp }\OtherTok{\textless{}{-}}\NormalTok{ Rtsne}\SpecialCharTok{::}\FunctionTok{Rtsne}\NormalTok{(}\AttributeTok{X =}\NormalTok{ dataset\_np\_sc[, ], }\AttributeTok{dims =} \DecValTok{2}\NormalTok{, }\AttributeTok{perplexity =} \DecValTok{3}\NormalTok{, }\AttributeTok{pca =} \ConstantTok{TRUE}\NormalTok{,}
  \AttributeTok{check\_duplicates =} \ConstantTok{FALSE}\NormalTok{, }\AttributeTok{theta =} \DecValTok{0}\NormalTok{)}

\FunctionTok{plot}\NormalTok{(}\AttributeTok{x =}\NormalTok{ temp}\SpecialCharTok{$}\NormalTok{Y[, }\DecValTok{1}\NormalTok{], }\AttributeTok{y =}\NormalTok{ temp}\SpecialCharTok{$}\NormalTok{Y[, }\DecValTok{2}\NormalTok{], }\AttributeTok{col =} \FunctionTok{rgb}\NormalTok{(}\AttributeTok{red =} \DecValTok{0}\NormalTok{, }\AttributeTok{green =} \DecValTok{0}\NormalTok{, }\AttributeTok{blue =} \DecValTok{0}\NormalTok{, }\AttributeTok{alpha =} \DecValTok{0}\NormalTok{),}
  \AttributeTok{xlab =} \StringTok{""}\NormalTok{, }\AttributeTok{ylab =} \StringTok{""}\NormalTok{, }\AttributeTok{xlim =} \FunctionTok{c}\NormalTok{(}\FunctionTok{min}\NormalTok{(temp}\SpecialCharTok{$}\NormalTok{Y[, }\DecValTok{1}\NormalTok{]) }\SpecialCharTok{{-}} \DecValTok{10}\NormalTok{, }\FunctionTok{max}\NormalTok{(temp}\SpecialCharTok{$}\NormalTok{Y[, }\DecValTok{1}\NormalTok{]) }\SpecialCharTok{+} \DecValTok{10}\NormalTok{),}
  \AttributeTok{ylim =} \FunctionTok{c}\NormalTok{(}\FunctionTok{min}\NormalTok{(temp}\SpecialCharTok{$}\NormalTok{Y[, }\DecValTok{2}\NormalTok{]) }\SpecialCharTok{{-}} \DecValTok{15}\NormalTok{, }\FunctionTok{max}\NormalTok{(temp}\SpecialCharTok{$}\NormalTok{Y[, }\DecValTok{2}\NormalTok{]) }\SpecialCharTok{+} \DecValTok{5}\NormalTok{))}
\FunctionTok{text}\NormalTok{(}\AttributeTok{x =}\NormalTok{ temp}\SpecialCharTok{$}\NormalTok{Y[, }\DecValTok{1}\NormalTok{], }\AttributeTok{y =}\NormalTok{ temp}\SpecialCharTok{$}\NormalTok{Y[, }\DecValTok{2}\NormalTok{], }\AttributeTok{labels =} \FunctionTok{paste0}\NormalTok{(}\StringTok{"Proj\_"}\NormalTok{, }\DecValTok{1}\SpecialCharTok{:}\DecValTok{15}\NormalTok{))}
\FunctionTok{points}\NormalTok{(}\AttributeTok{x =}\NormalTok{ temp}\SpecialCharTok{$}\NormalTok{Y[, }\DecValTok{1}\NormalTok{], }\AttributeTok{y =}\NormalTok{ temp}\SpecialCharTok{$}\NormalTok{Y[, }\DecValTok{2}\NormalTok{] }\SpecialCharTok{{-}} \DecValTok{20}\NormalTok{, }\AttributeTok{pch =} \DecValTok{15}\NormalTok{, }\AttributeTok{cex =} \FloatTok{1.3}\NormalTok{, }\AttributeTok{col =} \StringTok{"darkred"}\NormalTok{)}
\FunctionTok{grid}\NormalTok{()}
\end{Highlighting}
\end{Shaded}

\begin{Shaded}
\begin{Highlighting}[]
\FunctionTok{set.seed}\NormalTok{(}\DecValTok{1}\NormalTok{)}

\NormalTok{dataset }\OtherTok{\textless{}{-}}\NormalTok{ dataset\_np\_sc\_oversampled[, ]}
\NormalTok{dataset}\SpecialCharTok{$}\NormalTok{gt }\OtherTok{\textless{}{-}} \FunctionTok{round}\NormalTok{(dataset}\SpecialCharTok{$}\NormalTok{gt)}\SpecialCharTok{/}\DecValTok{10} \SpecialCharTok{*} \DecValTok{11}  \CommentTok{\# inflate by 1 for visual reasons only}

\NormalTok{cn }\OtherTok{\textless{}{-}} \FunctionTok{colnames}\NormalTok{(dataset)}
\NormalTok{cn\_x }\OtherTok{\textless{}{-}}\NormalTok{ cn[cn }\SpecialCharTok{!=} \StringTok{"gt"}\NormalTok{]}
\NormalTok{pre\_proc\_method }\OtherTok{\textless{}{-}} \FunctionTok{c}\NormalTok{(}\StringTok{"nzv"}\NormalTok{, }\StringTok{"center"}\NormalTok{, }\StringTok{"scale"}\NormalTok{)}
\NormalTok{pre\_proc }\OtherTok{\textless{}{-}}\NormalTok{ caret}\SpecialCharTok{::}\FunctionTok{preProcess}\NormalTok{(}\AttributeTok{x =}\NormalTok{ dataset[, cn\_x], }\AttributeTok{method =}\NormalTok{ pre\_proc\_method)}
\NormalTok{temp }\OtherTok{\textless{}{-}}\NormalTok{ stats}\SpecialCharTok{::}\FunctionTok{predict}\NormalTok{(pre\_proc, }\AttributeTok{newdata =}\NormalTok{ dataset)}

\CommentTok{\# pal \textless{}{-} viridis::viridis(11) pal \textless{}{-} RColorBrewer::brewer.pal(n = 11, name =}
\CommentTok{\# \textquotesingle{}Set3\textquotesingle{}) pal \textless{}{-} viridis::inferno(n = 11, begin = 0, end = .95) pal \textless{}{-}}
\CommentTok{\# ggsci::pal\_d3(palette = \textquotesingle{}category20\textquotesingle{})(n=11) pal \textless{}{-} ggsci::pal\_igv()(n = 11)}
\CommentTok{\# pal \textless{}{-} ggsci::pal\_rickandmorty()(n = 11) pal \textless{}{-} ggsci::pal\_simpsons()(n = 11)}
\CommentTok{\# pal \textless{}{-} ggsci::pal\_ucscgb()(n = 11)}
\NormalTok{pal }\OtherTok{\textless{}{-}} \FunctionTok{c}\NormalTok{((ggsci}\SpecialCharTok{::}\FunctionTok{pal\_material}\NormalTok{(}\AttributeTok{palette =} \StringTok{"green"}\NormalTok{, }\AttributeTok{n =} \DecValTok{3}\NormalTok{))(}\AttributeTok{n =} \DecValTok{3}\NormalTok{)[}\DecValTok{2}\NormalTok{], (ggsci}\SpecialCharTok{::}\FunctionTok{pal\_material}\NormalTok{(}\AttributeTok{palette =} \StringTok{"blue"}\NormalTok{,}
  \AttributeTok{n =} \DecValTok{7}\NormalTok{))(}\AttributeTok{n =} \DecValTok{7}\NormalTok{)[}\FunctionTok{c}\NormalTok{(}\DecValTok{2}\NormalTok{, }\DecValTok{4}\NormalTok{, }\DecValTok{6}\NormalTok{)], (ggsci}\SpecialCharTok{::}\FunctionTok{pal\_material}\NormalTok{(}\AttributeTok{palette =} \StringTok{"purple"}\NormalTok{, }\AttributeTok{n =} \DecValTok{9}\NormalTok{))(}\AttributeTok{n =} \DecValTok{9}\NormalTok{)[}\FunctionTok{c}\NormalTok{(}\DecValTok{2}\NormalTok{,}
  \DecValTok{4}\NormalTok{, }\DecValTok{6}\NormalTok{, }\DecValTok{8}\NormalTok{)], (ggsci}\SpecialCharTok{::}\FunctionTok{pal\_material}\NormalTok{(}\AttributeTok{palette =} \StringTok{"orange"}\NormalTok{, }\AttributeTok{n =} \DecValTok{6}\NormalTok{))(}\AttributeTok{n =} \DecValTok{6}\NormalTok{)[}\FunctionTok{c}\NormalTok{(}\DecValTok{2}\NormalTok{, }\DecValTok{4}\NormalTok{, }\DecValTok{6}\NormalTok{)])}

\NormalTok{tsne\_2d }\OtherTok{\textless{}{-}}\NormalTok{ Rtsne}\SpecialCharTok{::}\FunctionTok{Rtsne}\NormalTok{(}\AttributeTok{X =}\NormalTok{ temp[, }\StringTok{"gt"} \SpecialCharTok{!=} \FunctionTok{colnames}\NormalTok{(temp)], }\AttributeTok{dims =} \DecValTok{2}\NormalTok{, }\AttributeTok{perplexity =} \DecValTok{30}\NormalTok{,}
  \AttributeTok{pca =} \ConstantTok{TRUE}\NormalTok{, }\AttributeTok{check\_duplicates =} \ConstantTok{FALSE}\NormalTok{, }\AttributeTok{theta =} \FloatTok{0.05}\NormalTok{, }\AttributeTok{max\_iter =} \DecValTok{2000}\NormalTok{)}
\NormalTok{tsne\_3d }\OtherTok{\textless{}{-}}\NormalTok{ Rtsne}\SpecialCharTok{::}\FunctionTok{Rtsne}\NormalTok{(}\AttributeTok{X =}\NormalTok{ temp[, }\StringTok{"gt"} \SpecialCharTok{!=} \FunctionTok{colnames}\NormalTok{(temp)], }\AttributeTok{dims =} \DecValTok{3}\NormalTok{, }\AttributeTok{perplexity =} \DecValTok{30}\NormalTok{,}
  \AttributeTok{pca =} \ConstantTok{TRUE}\NormalTok{, }\AttributeTok{check\_duplicates =} \ConstantTok{FALSE}\NormalTok{, }\AttributeTok{theta =} \FloatTok{0.05}\NormalTok{, }\AttributeTok{max\_iter =} \DecValTok{2000}\NormalTok{)}
\end{Highlighting}
\end{Shaded}

\begin{figure}[ht!]
\includegraphics{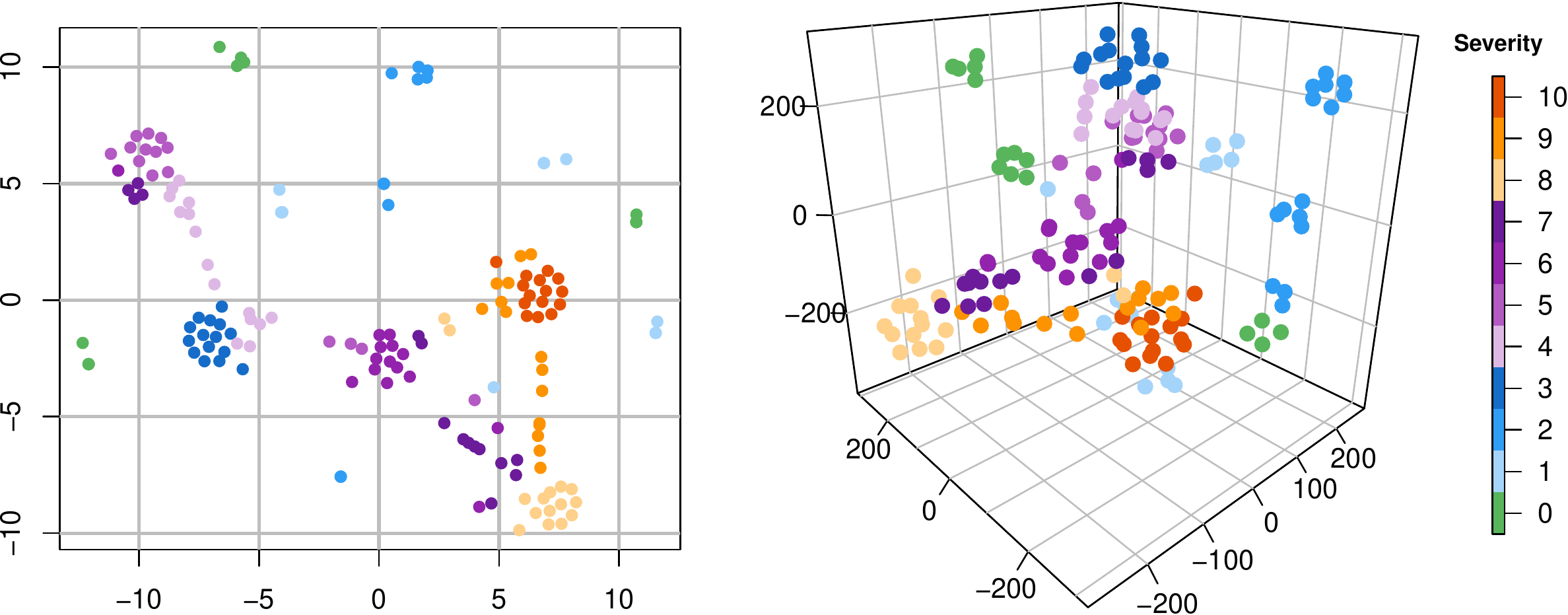} \caption{A 2- and 3-dimensional embedding of the oversampled projects (using z-standardized source code features).}\label{fig:tsne-dual-embedding}
\end{figure}

Figure \ref{fig:tsne-dual-embedding} shows two embeddings. I have asked my colleague Angelos to interpret these, without telling him any details about the underlying dataset.

He kindly responded:

\emph{Starting with the 2D projection, Observation 1 (O1) is about the distortion in the vertical space. I believe if there had been more space to project your points, I would have used a square (or as square as possible) grid. Following that, many points are overlapping in the 2D embedding and require interactivity in order to see them in comparison to 3D, which was clear in that case.}

\emph{O2 is related to groups of points forming a kind of circular motive. I think this happens when the data instances/records are quite similar or even identical.}

\emph{O3 is the most important observation because it seems that as you reduce the severity, there is a linear (or time-dependent) connection between your points. For example, the orange points are between dark red and light orange. The same is true for the purple points, but this is more clearly visible from the 3D projection. I am not sure why though; it could be that the 3D is ``cheating here'' due to the perspective from which you have taken the screenshot. I would definitely need to interact with both charts to understand more about what is going on.}

\textbf{Comment Sebastian}: That linear trend appears to correspond to the linear Likert-scale used to assign the severity.

\emph{The same goes for O4, where 2D is split into three areas while 3D shows two. Also, the overplotting issue is still evident here, especially for the 2D embedding. If I had no colors, I would argue that almost nothing is easy to observe from the plots except for the very distant and tight clusters. The remaining points in the middle would have been considered one cluster, in my opinion.}

\emph{Something deceiving in the 3D plot is marked as O2, with the green points appearing closer to the dark red compared to the 2D embedding. I guess it is because of the position of the ``camera,'' so interaction is key here.}

The second observation follows from the synthetic oversampling of the data, which will result in very similar data points.
I actually had to disable checking for duplicates when computing the t-SNE on the oversampled datasets for it to work.

\begin{figure}[ht!]
\includegraphics{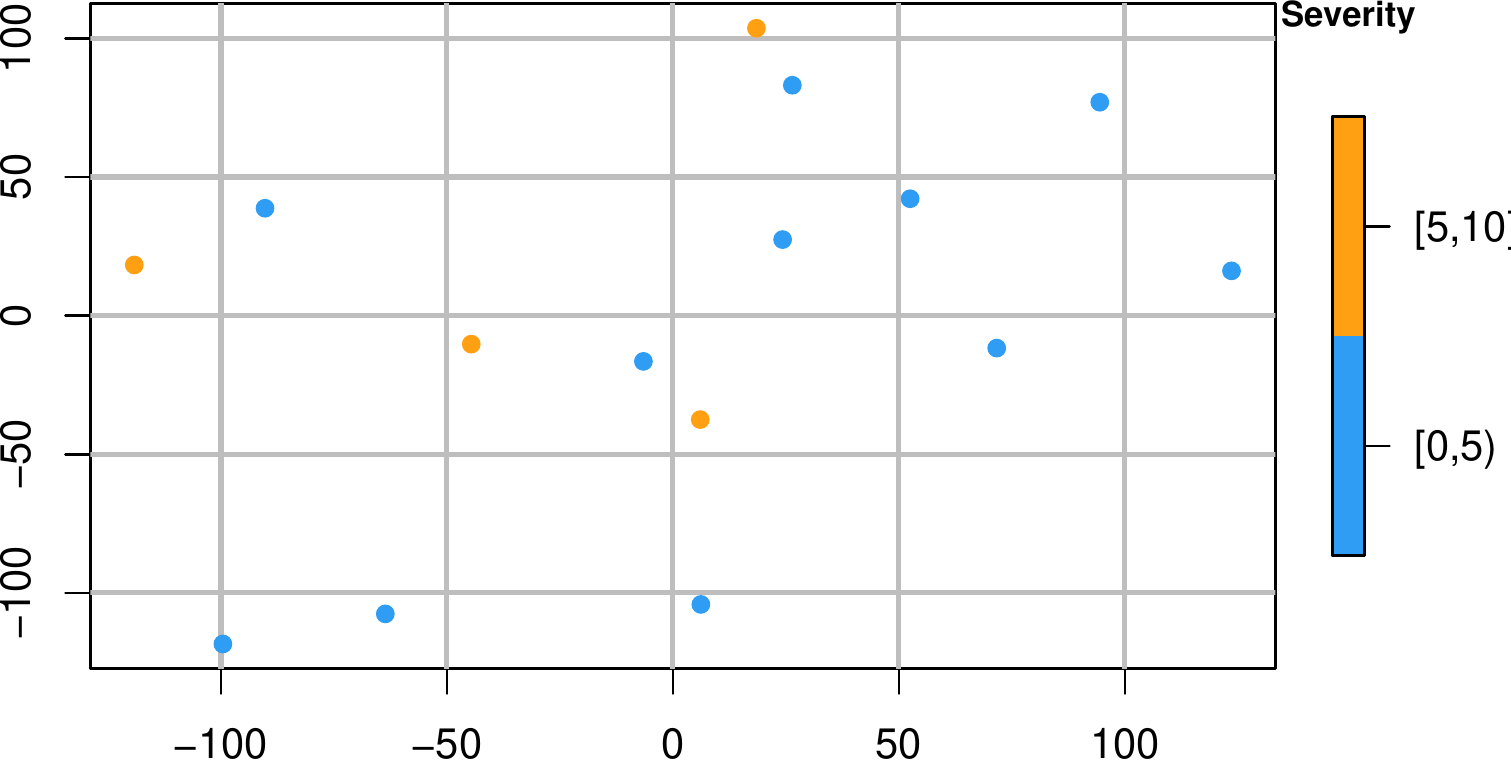} \caption{2D t-SNE embedding of the projects using source code data. The expected clusters have been reduced to two.}\label{fig:tsne-severity-groups}
\end{figure}

So, what we could try, is to reduce the amount of expected clusters and use the original, non-oversampled data.
Figure \ref{fig:tsne-severity-groups} shows the result of this. There I have reduced the projects' cluster to two, similar to how the decision rule does it (ground truth less than \(5\) means ``no Fire Drill'').
Still, this plot is not meaningful and we cannot observe any useful clusters.
The reason could quite well be that we do not have sufficient data. It could also very well be that the projects' underlying data does not reflect the ground truth all too well.
Further investigation and/or more data is needed.

\hypertarget{k-means-clustering}{%
\subsubsection{k-Means Clustering}\label{k-means-clustering}}

\begin{figure}[ht!]
\includegraphics{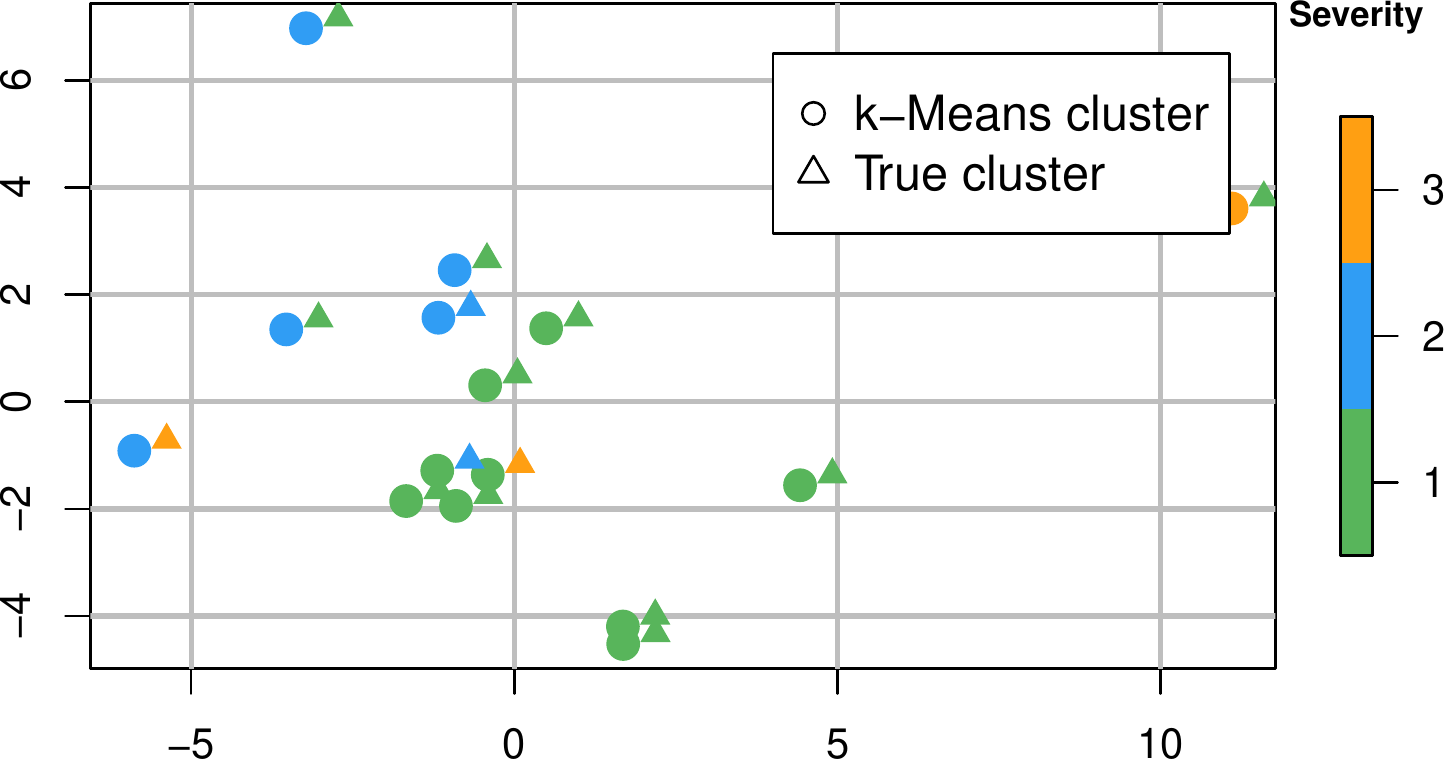} \caption{k-Means clustering of projects using 3 clusters and the first two principal components.}\label{fig:kmeans-embedding}
\end{figure}

Here we use k-Means clustering of the projects using three clusters (severity \(<4\), severity \(<7\), and severity \(>7\)),
In figure \ref{fig:kmeans-embedding}, the \(x/y\)-coordinates are obtained from the first two principal components to produce an embedding.
With a circle, the k-Means-assigned cluster for each project is shown. Next to each circle, the true cluster (using a triangle) is plotted.
Only four out of \(15\) projects are assigned a correct cluster, three are in the wrong, directly neighboring cluster (e.g., assigned to severity cluster \(2\) instead of \(1\), or \(1\) instead of \(0\)), and eight are in the most distant cluster (i.e., \(0\) instead of \(2\) or vice versa).

\clearpage

\hypertarget{technical-report-self-regularizing-boundary-time-warping-and-boundary-amplitude-warping}{%
\section{\texorpdfstring{Technical Report: Self-Regularizing Boundary Time Warping and Boundary Amplitude Warping\label{tr:srBTAW-technical-report}}{Technical Report: Self-Regularizing Boundary Time Warping and Boundary Amplitude Warping}}\label{technical-report-self-regularizing-boundary-time-warping-and-boundary-amplitude-warping}}

This is the self-contained technical report for self-regularizing Boundary Time Warping and Boundary Amplitude Warping.

\hypertarget{introduction-3}{%
\subsection{Introduction}\label{introduction-3}}

In this notebook, we will go back to our initial Fire Drill (FD) problem, and design models to accommodate it. The goal is to bring together everything we learned from all the notebooks in between, and to come up with solutions specifically designed to address the matter.

Self-regularizing boundary time and amplitude warping, \textbf{\texttt{srBTAW}}, is a model that allows for rectifying signals. Similar to Dynamic time warping (DTW), one signal is aligned to another. However, srBTAW is not limited to the time-domain only, and can optionally correct the amplitude of a signal, too. The suggested models have built-in self-regularization, meaning that they will somewhat automatically correct extreme parameters without requiring additional regularization.

Among others, srBTAW has the following features:

\begin{itemize}
\tightlist
\item
  srBTAW is an optimization algorithm, and not based on dynamic programming. While DTW finds some best alignment for each point of discrete signals, srBTAW finds some best alignment for whole intervals. Warping for srBTAW means to linearly affine transform intervals.
\item
  It is designed to foremost work with continuous signals, although it can be used for discrete signals just as well, too.
\item
  Closed-, half- or fully-open matching: It is up to the user whether a signal must be matched from the start, to the end, both, or neither of them.
\item
  Custom objectives: Unlike DTW, srBTAW allows for arbitrary user-defined objectives to compute, e.g., distances or similarities between (intervals of) signals. DTW almost always makes the stipulation of computing distances using the Euclidean distance. In this notebook, we demonstrate using losses from various domains, such as information- or signal-theory.
\item
  srBTAW is multivariate: both the pattern (reference) and candidate (query) may consist of one or arbitrary many separate signals. Each of them may have their own support.
\item
  For DTW the underlying assumption is that the query-signal is a time-distorted version of the reference signal, and that there are no portions of the query that have a higher amplitude. In other words, DTW assumes only degrees of freedom on the time axis and can actually not support alignment of signals with different amplitudes well. Furthermore, DTW performs poor for very dissimilar signals. This is not a problem for srBTAW because of its interval-alignments.
\end{itemize}

srBTAW was developed to solve some concrete problems in the project management domain, and this notebook uses some of the use cases to evolve the models. However, srBTAW can work with many various types of time series-related problems, such as classification, motif discovery etc.

All complementary data and results can be found at Zenodo (Hönel, Pícha, et al. 2023). This notebook was written in a way that it can be run without any additional efforts to reproduce the outputs (using the pre-computed results). This notebook has a canonical URL\textsuperscript{\href{https://github.com/MrShoenel/anti-pattern-models/blob/master/notebooks/srBTAW-technical-report.Rmd}{{[}Link{]}}} and can be read online as a rendered markdown\textsuperscript{\href{https://github.com/MrShoenel/anti-pattern-models/blob/master/notebooks/srBTAW-technical-report.md}{{[}Link{]}}} version (attention: the math is probably not rendered correctly there!). All code can be found in this repository, too.

\hypertarget{problem-description}{%
\subsection{Problem description}\label{problem-description}}

Initially I thought the answer would be using dynamic time Warping, but it is not suitable for an analytical solution (actually, closed-form expression). What makes our problem a bit special are the following:

\begin{itemize}
\tightlist
\item
  \textbf{Reference} and \textbf{Query} signal have the same support: The FD is an AP that spans the entire project, start to finish. Matching must be attempted at the very beginning of the data, and is concluded at the very end. The query signal is scaled such that the extent of its support is the same as the reference's.

  \begin{itemize}
  \tightlist
  \item
    As we will see later, this constraint is optionally relaxed by allowing to the query-signal to cover a proper sub-support of the reference using open begin- and/or -end time warping.
  \end{itemize}
\item
  All data has to be used: We can apply local time warping to the query signal, but nothing can be left out. The warping may even result in a length of zero of some intervals, but they must not be negative.
\end{itemize}

\hypertarget{optimization-goals}{%
\subsubsection{Optimization goals}\label{optimization-goals}}

We have \textbf{3} optimization goals for the joint paper/article (each of them has a dedicated section in this notebook):

\begin{itemize}
\tightlist
\item
  Fitting of project data to a defined AP (here: the Fire Drill)
\item
  Validate selection of sub-models and scores; also: calibration of scores
\item
  Averaging the reference pattern over all known ground truths
\end{itemize}

\hypertarget{optimization-goal-i-fitting-of-project-data-to-a-defined-ap}{%
\subsection{Optimization goal I: Fitting of project data to a defined AP}\label{optimization-goal-i-fitting-of-project-data-to-a-defined-ap}}

This is the primary problem, and we should start with a visualization:

\includegraphics{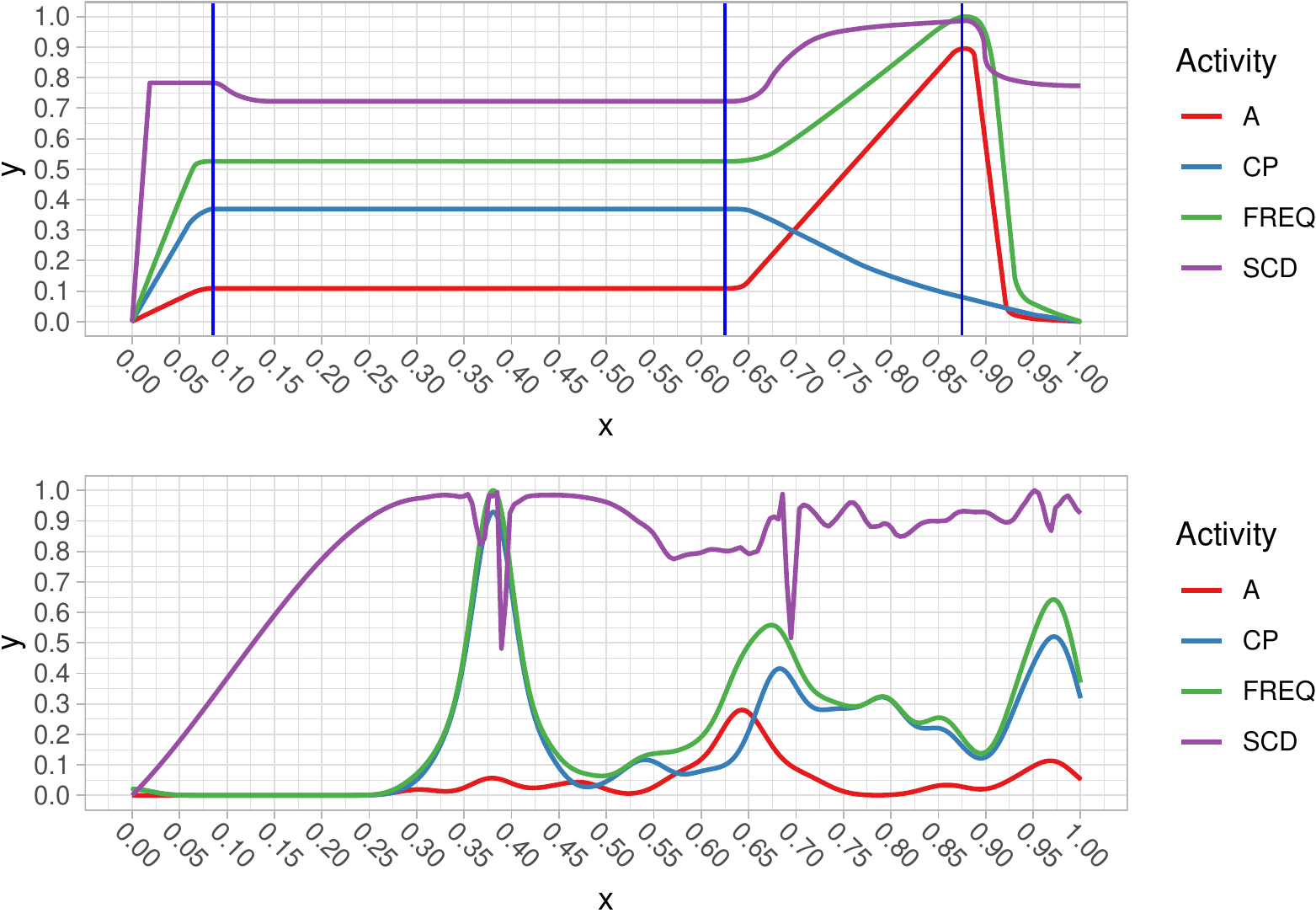}

The first plot is the reference pattern, the second is a random project. All data has already been scaled and normalized, for example, both supports are the same. We are going to use an entirely new approach to warp the project data from the 2nd plot the reference data from the first plot.

\hypertarget{approach}{%
\subsubsection{Approach}\label{approach}}

In order to understand this problem, we need to make some statements:

\begin{itemize}
\tightlist
\item
  The reference pattern is subdivided into intervals by some boundaries (blue in the above plot). The data in each such interval is considered \textbf{constant}. That means, if we were to sample from that interval, these samples would always be identical. If we were to look at each interval as a function, then it will always show the same behavior. Each interval has its own \textbf{support}.
\item
  The first interval starts at \(0\) and the last interval ends at \(1\) (we always expect that all data has been scaled and translated to \([0,1]\) prior to this, to keep things simple).
\item
  Any and all variables are strictly \textbf{positive}, with a co-domain of \([0,1]\) as well. We require this for reasons of simplification, although negative values would be no problem (but require extra effort at this point).
\item
  An interval's \textbf{length} is never negative. In practice, one may even want to define a minimum length for all or for each individual interval.
\item
  An interval's \textbf{offset} is the sum of the lengths of its preceding intervals. All intervals' lengths sum up to \(1\).
\end{itemize}

To find the sub-support of the query that best matches a variable (or many) in an interval, the query signal is translated and scaled. Another way to think about this is that the query signal is translated and only valid in the support that starts with the translate, and ends there plus the length of the current interval. I mention this because this is how we will begin.

In the above plot, there are \textbf{\(4\)} intervals, and let's say we index them with the parameter \(q\). If we were to describe the reference- and query-signals through the two functions \(r,f\), and we were given some boundary \(b_q\) that delimits the reference-interval, for \(q=1\) we would get:

\[
\begin{aligned}
  r=&\;\mathbb{R}\mapsto\mathbb{R}^m\;\land\;m>0\text{,}
  \\[1ex]
  r_q=&\;r(\cdot)\;\text{, reference-signal for interval}\;q,
  \\[1ex]
  \text{supp}(r_q)=&\;[0,b_q)\;\text{, the support of}\;r_q.
\end{aligned}
\]

During optimization, the goal is to find the best length for each interval. Assuming a given vector that contains these four lengths, \(\bm{\vartheta}^{(l)}\), we get:

\[
\begin{aligned}
  f=&\;\mathbb{R}\mapsto\mathbb{R}^n\;\land\;n>0\text{,}
  \\[1ex]
  f_q=&\;f(\cdot)\;\text{, query-signal for interval}\;q,
  \\[1ex]
  \text{supp}(f_q)=&\;\big\{\;x\in\mathbb{R}\;\rvert\;x\geq0\;\land\;x<\bm{\vartheta}^{(l)}_q\;\big\}\;\text{, the support of}\;f_q.
\end{aligned}
\]

The supports will almost always differ (especially for \(q>1\)). For the discrete case this may not be important, because the model may choose to always take the same amount of equidistantly-spaced samples from the reference- and query-intervals, regardless of the support. This is essentially the same as scaling and translating the query-support to be the same as the reference support. This is what would be required for the continuous case.

The result of the previous definition is an \textbf{unchanged} query-signal; however, we specify its support that \emph{corresponds} to the support of the current reference interval. In general, to translate (and scale) an interval from one offset and length to another offset and length, the expression is:

\[
\begin{aligned}
  f(x)\dots&\;\text{function to be translated,}
  \\[1ex]
  f'(x)=&\;f\Bigg(\frac{(x - t_b) \times (s_e - s_b)}{t_e - t_b} + s_b\Bigg)\;\text{, where}
  \\[1ex]
  s_b,s_e,t_b,t_e\dots&\;\text{begin (b) and end (e) of the source- (s) and target-intervals/supports (t).}
\end{aligned}
\]

We will do some testing with the pattern and data we have, focusing on the \textbf{adaptive} (\textbf{A}) variable.

However, notice how we have defined \(r,f\) to map to dimensionalities of \(1\) or greater, effectively making them vector-valued:

\begin{itemize}
\tightlist
\item
  A loss between both functions is concerning a specific \(q\)-th interval, and also usually concerning a specific index in each function, e.g., \(\mathcal{L}_q=(r_q^{(13)}(x)-f_q^{(4)}(x))^2\). Each such loss may have its \textbf{individual weight} assigned.
\item
  This concept of being multivariate can be extended to having \textbf{multiple signals} per \emph{Warping Pattern/Candidate}, too, by simple concatenation of the output of two or more such vector-valued functions, e.g., \(f(x)=f_a^{(1,2,3)}\frown f_b^{(7,11,13)}\frown\dots\), where \(f_a,f_b\) are individual vector-valued/multivariate Warping Candidates.
\item
  The dimensionalities of \(r,f\) do not necessarily have to be the same: Imagine the case where we compute some loss of one variable of the Warping Pattern over the same variable of one or more Warping Candidates. Also, as in the previous example, not all of a signal's variables have to be used.
\end{itemize}

\begin{verbatim}
## Scale for 'y' is already present. Adding another scale for 'y', which will
## replace the existing scale.
\end{verbatim}

\includegraphics{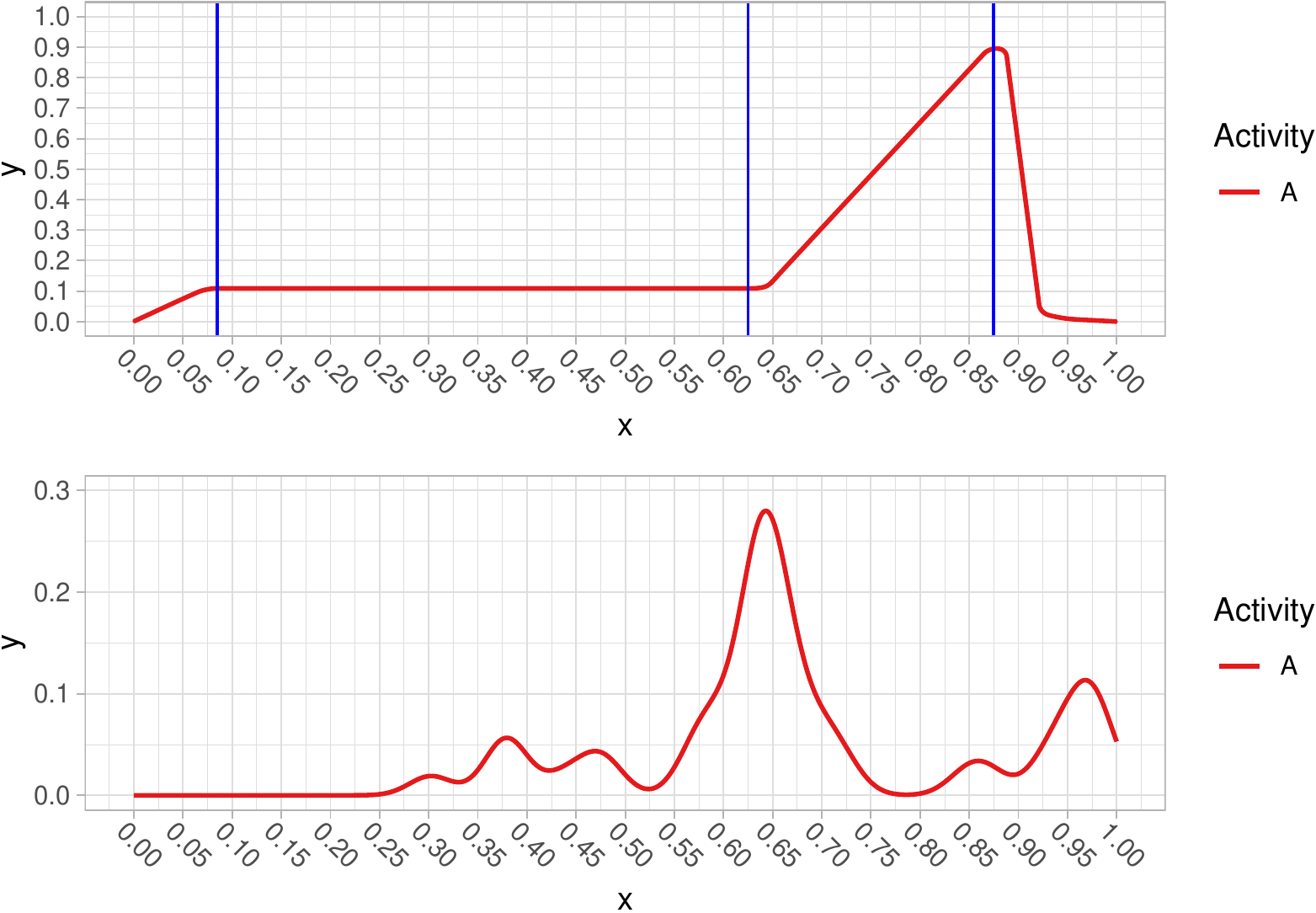}

We will do a test with \(f_q\) and \(f'_q\), and set \(q\) to \(1\) and then to \(3\). In the first case, no re-definition of \(f\) is required. We only need to define where to query-intervals should be (thus their length).

\begin{Shaded}
\begin{Highlighting}[]
\CommentTok{\# Use these lengths for q=3:}
\NormalTok{vt\_q3 }\OtherTok{\textless{}{-}} \FunctionTok{c}\NormalTok{(}\FloatTok{0.15}\NormalTok{, }\FloatTok{0.45}\NormalTok{, }\FloatTok{0.1}\NormalTok{, }\FloatTok{0.3}\NormalTok{)}

\NormalTok{a\_ref\_q3 }\OtherTok{\textless{}{-}}\NormalTok{ a\_ref[a\_ref }\SpecialCharTok{\textgreater{}=}\NormalTok{ bounds[}\DecValTok{2}\NormalTok{] }\SpecialCharTok{\&}\NormalTok{ a\_ref }\SpecialCharTok{\textless{}}\NormalTok{ bounds[}\DecValTok{3}\NormalTok{], ]}
\NormalTok{a\_query\_q3 }\OtherTok{\textless{}{-}}\NormalTok{ a\_query[a\_query }\SpecialCharTok{\textgreater{}=}\NormalTok{ (vt\_q3[}\DecValTok{1}\NormalTok{] }\SpecialCharTok{+}\NormalTok{ vt\_q3[}\DecValTok{2}\NormalTok{]) }\SpecialCharTok{\&}\NormalTok{ a\_query }\SpecialCharTok{\textless{}}\NormalTok{ (}\DecValTok{1} \SpecialCharTok{{-}}\NormalTok{ vt\_q3[}\DecValTok{4}\NormalTok{]),}
\NormalTok{  ]}

\NormalTok{r\_q3 }\OtherTok{\textless{}{-}}\NormalTok{ stats}\SpecialCharTok{::}\FunctionTok{approxfun}\NormalTok{(}\AttributeTok{x =}\NormalTok{ a\_ref\_q3}\SpecialCharTok{$}\NormalTok{x, }\AttributeTok{y =}\NormalTok{ a\_ref\_q3}\SpecialCharTok{$}\NormalTok{y)}
\NormalTok{f\_q3 }\OtherTok{\textless{}{-}}\NormalTok{ stats}\SpecialCharTok{::}\FunctionTok{approxfun}\NormalTok{(}\AttributeTok{x =}\NormalTok{ a\_query\_q3}\SpecialCharTok{$}\NormalTok{x, }\AttributeTok{y =}\NormalTok{ a\_query\_q3}\SpecialCharTok{$}\NormalTok{y)}

\FunctionTok{ggplot}\NormalTok{() }\SpecialCharTok{+} \FunctionTok{stat\_function}\NormalTok{(}\AttributeTok{fun =}\NormalTok{ r\_q3, }\FunctionTok{aes}\NormalTok{(}\AttributeTok{color =} \StringTok{"r\_q3"}\NormalTok{)) }\SpecialCharTok{+} \FunctionTok{stat\_function}\NormalTok{(}\AttributeTok{fun =}\NormalTok{ f\_q3,}
  \FunctionTok{aes}\NormalTok{(}\AttributeTok{color =} \StringTok{"f\_q3"}\NormalTok{)) }\SpecialCharTok{+} \FunctionTok{xlim}\NormalTok{(}\FloatTok{0.6}\NormalTok{, }\FloatTok{0.9}\NormalTok{)}
\end{Highlighting}
\end{Shaded}

\includegraphics{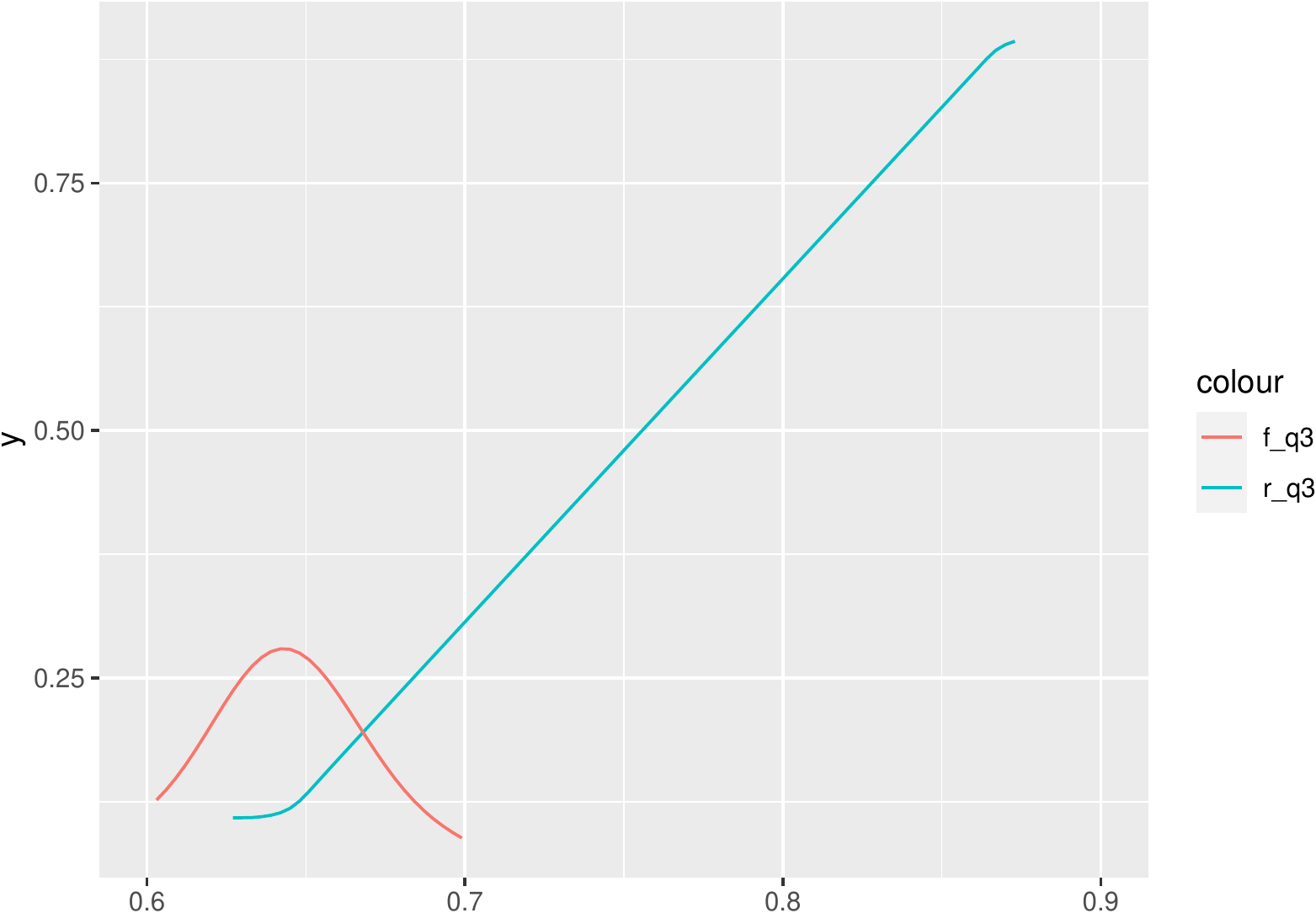}

And now we re-define \(f\) such that it has the same support as \texttt{r\_q3}:

\begin{Shaded}
\begin{Highlighting}[]
\NormalTok{f\_prime\_proto }\OtherTok{\textless{}{-}} \ControlFlowTok{function}\NormalTok{(x, source\_b, source\_e, target\_b, target\_e) \{}
  \FunctionTok{f}\NormalTok{((((x }\SpecialCharTok{{-}}\NormalTok{ target\_b) }\SpecialCharTok{*}\NormalTok{ (source\_e }\SpecialCharTok{{-}}\NormalTok{ source\_b))}\SpecialCharTok{/}\NormalTok{(target\_e }\SpecialCharTok{{-}}\NormalTok{ target\_b)) }\SpecialCharTok{+}\NormalTok{ source\_b)}
\NormalTok{\}}

\NormalTok{f\_q3\_prime }\OtherTok{\textless{}{-}} \ControlFlowTok{function}\NormalTok{(x) \{}
  \CommentTok{\# These two are where the signal currently is (translation source range)}
\NormalTok{  phi\_q }\OtherTok{\textless{}{-}}\NormalTok{ vt\_q3[}\DecValTok{1}\NormalTok{] }\SpecialCharTok{+}\NormalTok{ vt\_q3[}\DecValTok{2}\NormalTok{]}
\NormalTok{  vartheta\_q }\OtherTok{\textless{}{-}}\NormalTok{ vt\_q3[}\DecValTok{3}\NormalTok{]}
  \CommentTok{\# These two is where we want the signal to be (translation target range)}
  \CommentTok{\# theta\_q \textless{}{-} bounds[3] {-} bounds[2]}

\NormalTok{  source\_b }\OtherTok{\textless{}{-}}\NormalTok{ phi\_q}
\NormalTok{  source\_e }\OtherTok{\textless{}{-}}\NormalTok{ phi\_q }\SpecialCharTok{+}\NormalTok{ vartheta\_q}
\NormalTok{  target\_b }\OtherTok{\textless{}{-}}\NormalTok{ bounds[}\DecValTok{2}\NormalTok{]}
\NormalTok{  target\_e }\OtherTok{\textless{}{-}}\NormalTok{ bounds[}\DecValTok{3}\NormalTok{]}

  \FunctionTok{f\_prime\_proto}\NormalTok{(x, }\AttributeTok{target\_b =}\NormalTok{ target\_b, }\AttributeTok{target\_e =}\NormalTok{ target\_e, }\AttributeTok{source\_b =}\NormalTok{ source\_b,}
    \AttributeTok{source\_e =}\NormalTok{ source\_e)}
\NormalTok{\}}

\FunctionTok{curve}\NormalTok{(f\_q3\_prime, }\DecValTok{0}\NormalTok{, }\DecValTok{1}\NormalTok{)}
\end{Highlighting}
\end{Shaded}

\includegraphics{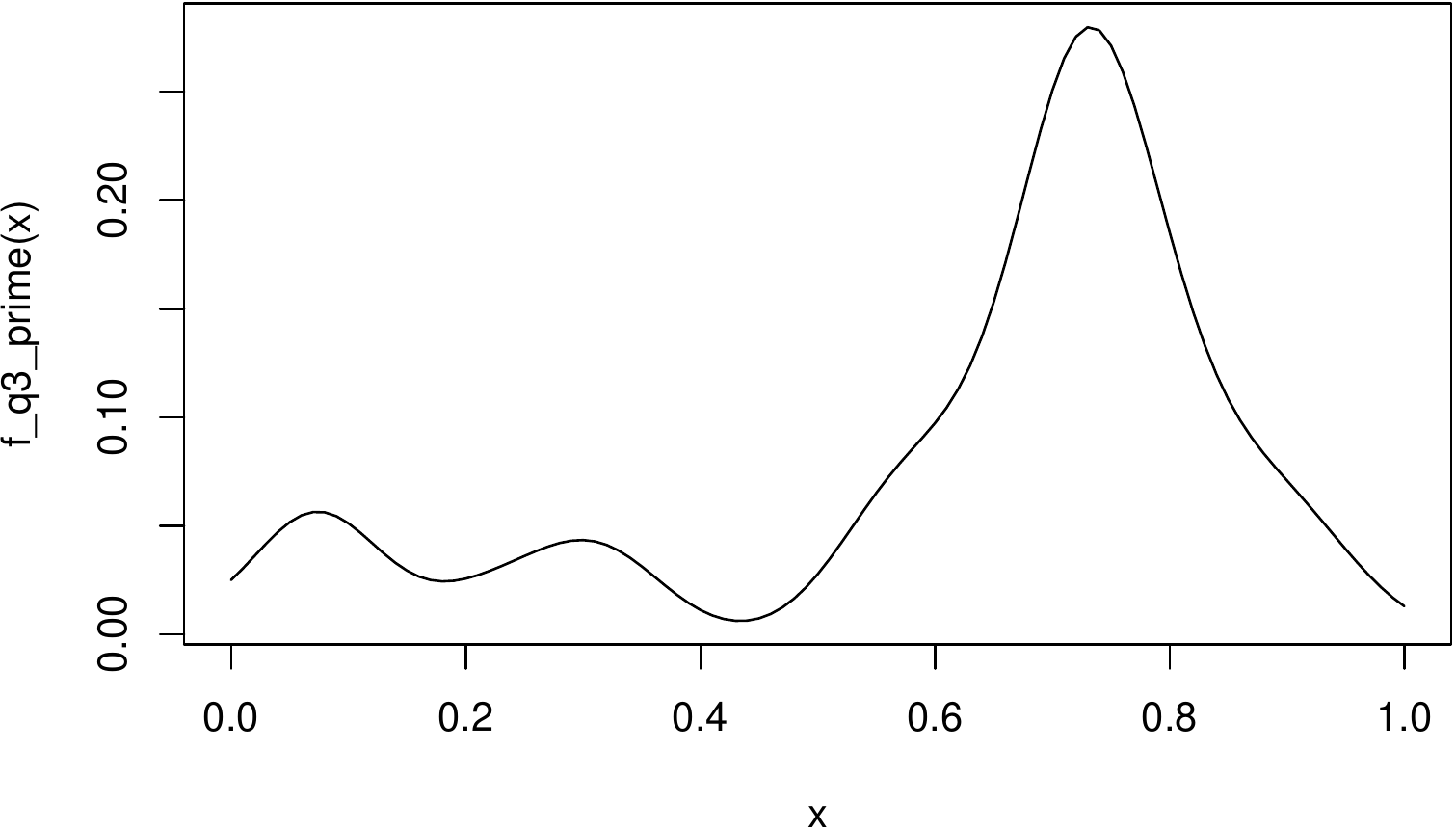}

\begin{Shaded}
\begin{Highlighting}[]
\FunctionTok{ggplot}\NormalTok{() }\SpecialCharTok{+} \FunctionTok{stat\_function}\NormalTok{(}\AttributeTok{fun =}\NormalTok{ r\_q3, }\FunctionTok{aes}\NormalTok{(}\AttributeTok{color =} \StringTok{"r\_q3"}\NormalTok{)) }\SpecialCharTok{+} \FunctionTok{stat\_function}\NormalTok{(}\AttributeTok{fun =}\NormalTok{ f\_q3\_prime,}
  \FunctionTok{aes}\NormalTok{(}\AttributeTok{color =} \StringTok{"f\_q3\_prime"}\NormalTok{)) }\SpecialCharTok{+} \FunctionTok{geom\_vline}\NormalTok{(}\AttributeTok{xintercept =}\NormalTok{ bounds[}\DecValTok{2}\NormalTok{]) }\SpecialCharTok{+} \FunctionTok{geom\_vline}\NormalTok{(}\AttributeTok{xintercept =}\NormalTok{ bounds[}\DecValTok{3}\NormalTok{]) }\SpecialCharTok{+}
  \FunctionTok{xlim}\NormalTok{(bounds[}\DecValTok{2}\SpecialCharTok{:}\DecValTok{3}\NormalTok{])}
\end{Highlighting}
\end{Shaded}

\includegraphics{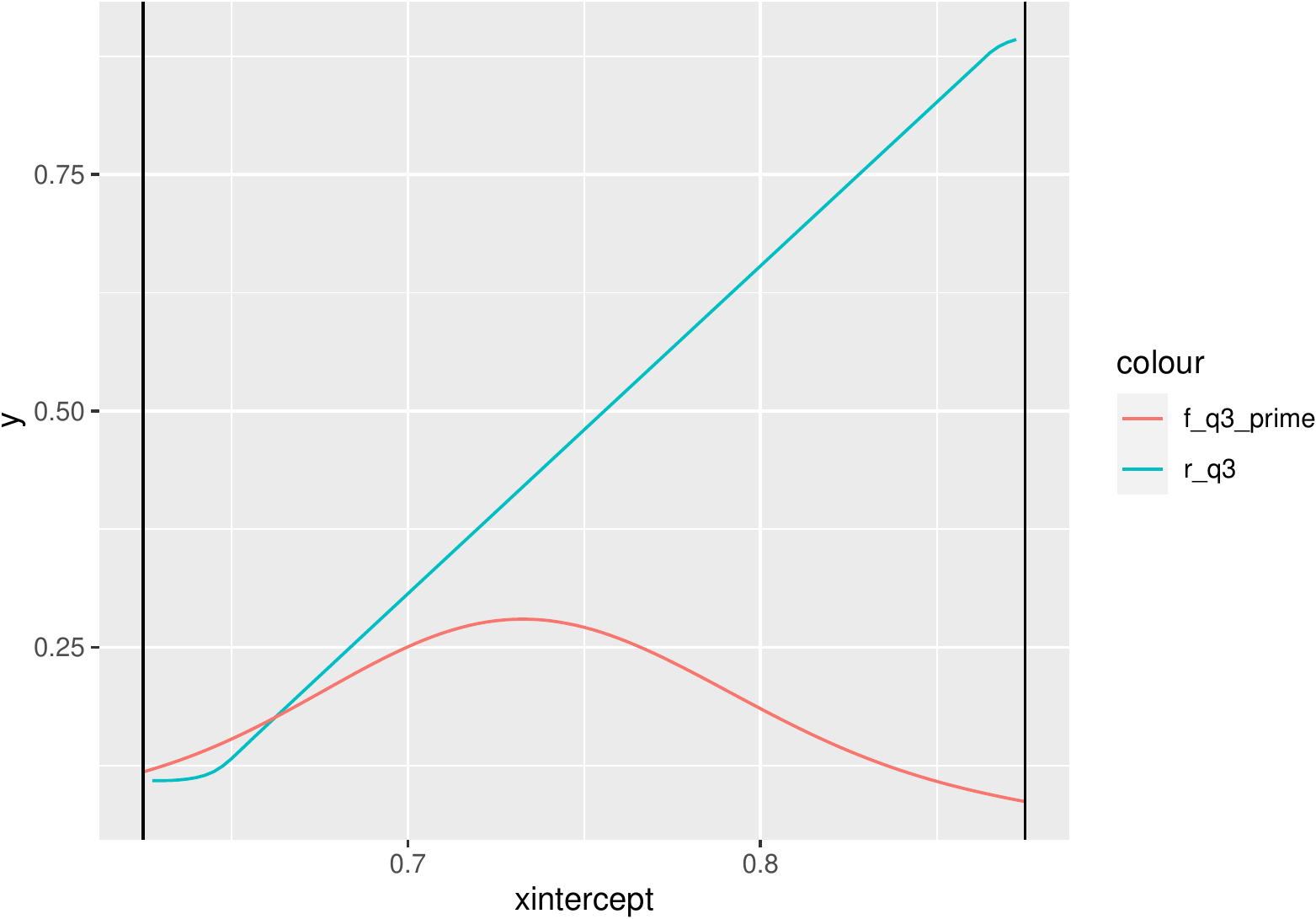}

Another test, squeezing \(f\) into the interval \([0.5,1]\):

\begin{Shaded}
\begin{Highlighting}[]
\NormalTok{f\_q3\_test }\OtherTok{\textless{}{-}} \ControlFlowTok{function}\NormalTok{(x) \{}
  \FunctionTok{f\_prime\_proto}\NormalTok{(x, }\AttributeTok{source\_b =} \DecValTok{0}\NormalTok{, }\AttributeTok{source\_e =} \DecValTok{1}\NormalTok{, }\AttributeTok{target\_b =} \FloatTok{0.5}\NormalTok{, }\AttributeTok{target\_e =} \DecValTok{1}\NormalTok{)}
\NormalTok{\}}

\NormalTok{f\_q3\_test2 }\OtherTok{\textless{}{-}} \ControlFlowTok{function}\NormalTok{(x) \{}
  \FunctionTok{f\_prime\_proto}\NormalTok{(x, }\AttributeTok{source\_b =} \FloatTok{0.55}\NormalTok{, }\AttributeTok{source\_e =} \FloatTok{0.7}\NormalTok{, }\AttributeTok{target\_b =} \DecValTok{0}\NormalTok{, }\AttributeTok{target\_e =} \DecValTok{1}\NormalTok{)}
\NormalTok{\}}

\FunctionTok{curve}\NormalTok{(f, }\DecValTok{0}\NormalTok{, }\DecValTok{1}\NormalTok{)}
\end{Highlighting}
\end{Shaded}

\includegraphics{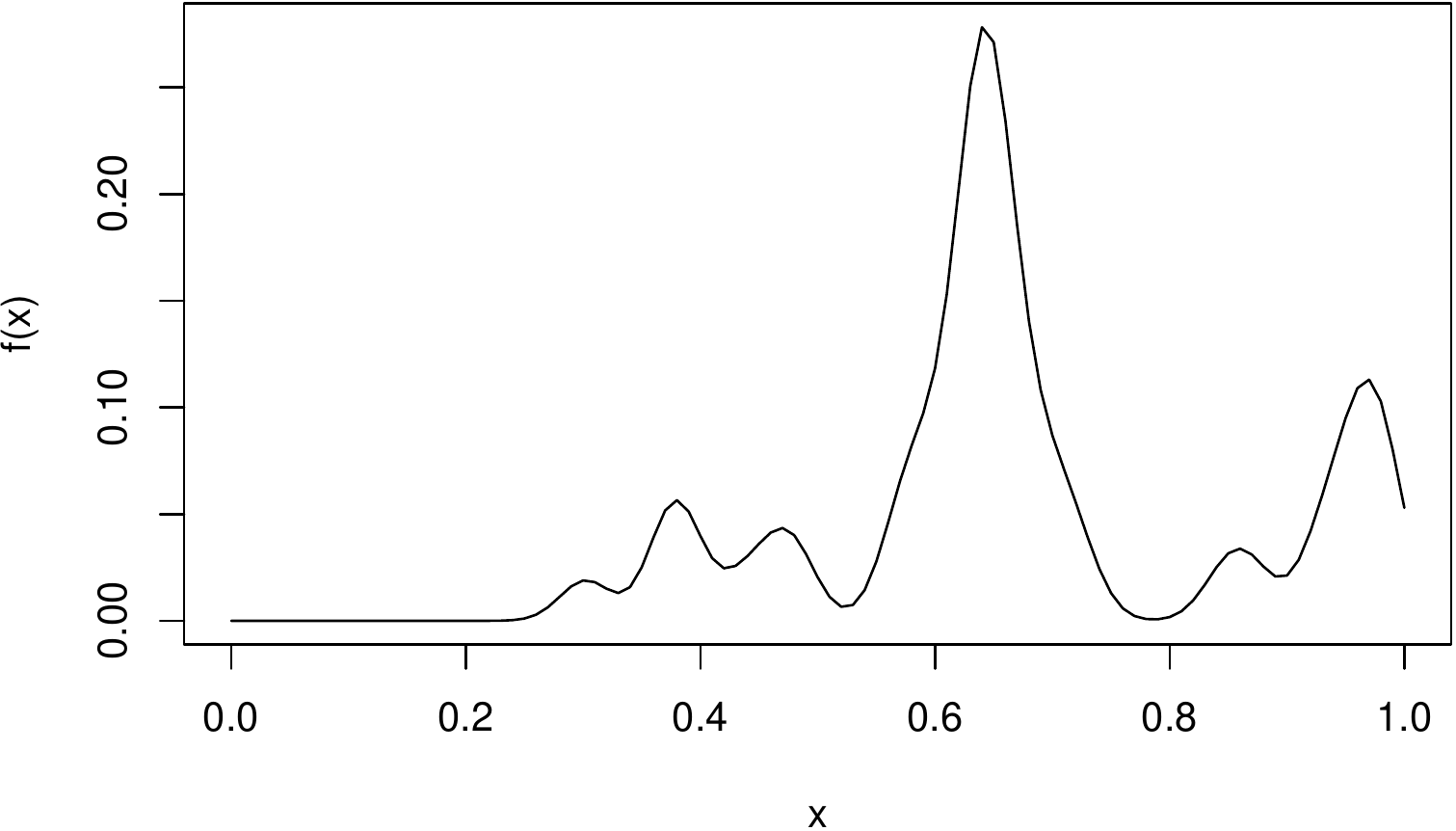}

\begin{Shaded}
\begin{Highlighting}[]
\FunctionTok{curve}\NormalTok{(f\_q3\_test, }\DecValTok{0}\NormalTok{, }\DecValTok{1}\NormalTok{)}
\end{Highlighting}
\end{Shaded}

\includegraphics{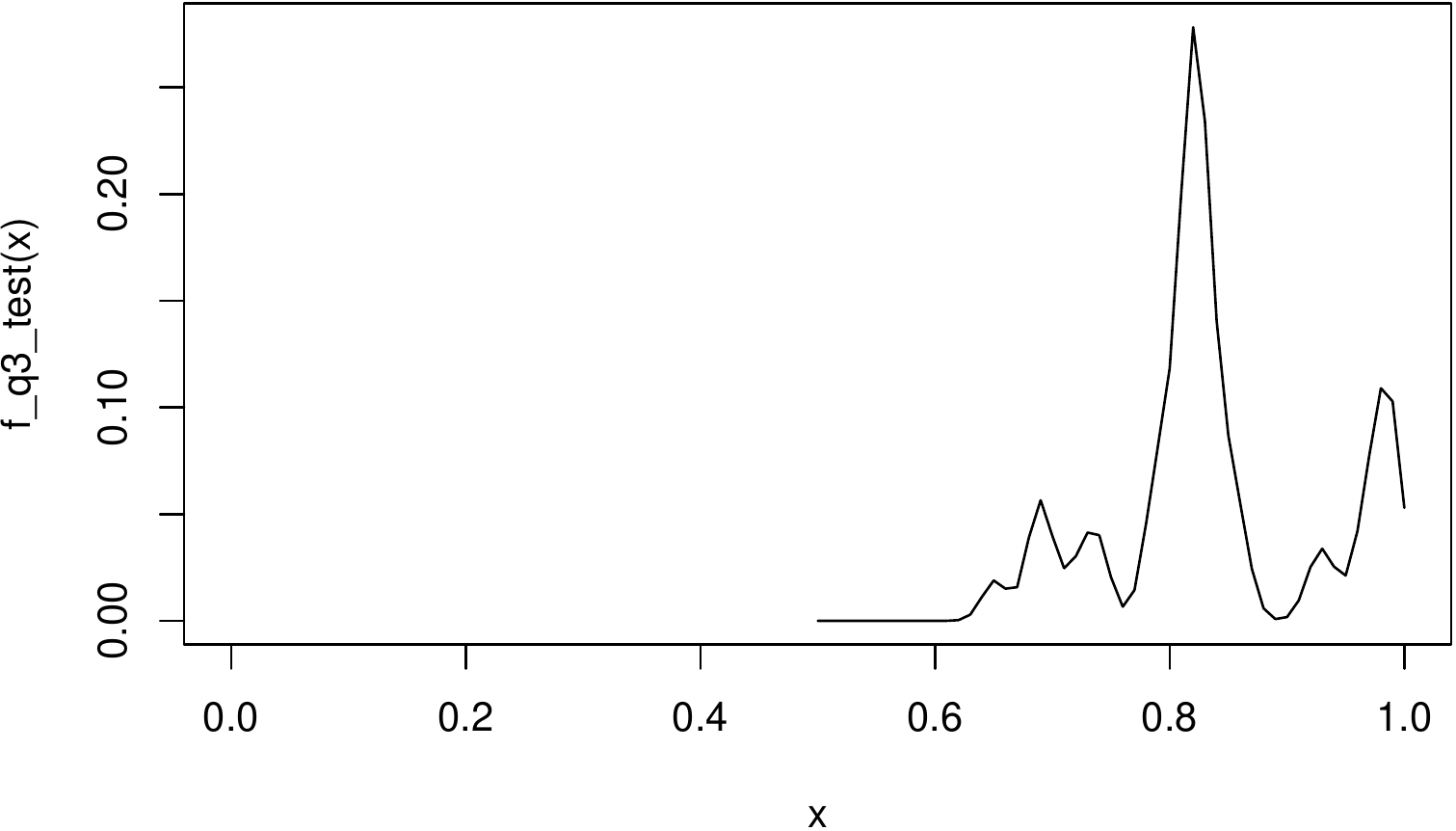}

\begin{Shaded}
\begin{Highlighting}[]
\FunctionTok{curve}\NormalTok{(f\_q3\_test2, }\DecValTok{0}\NormalTok{, }\DecValTok{1}\NormalTok{)}
\end{Highlighting}
\end{Shaded}

\includegraphics{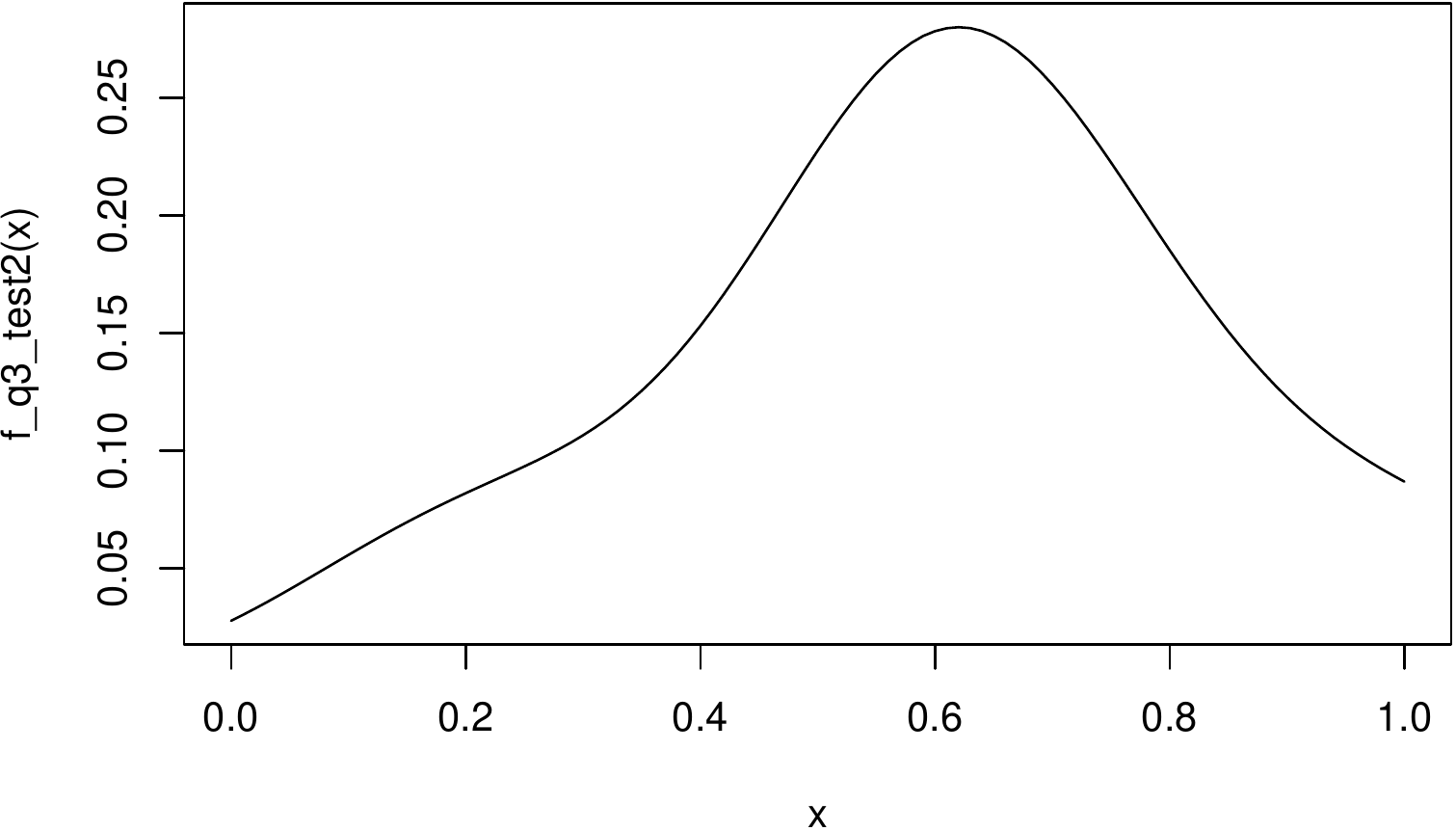}

\hypertarget{sub-model-formulation}{%
\subsubsection{Sub-Model formulation}\label{sub-model-formulation}}

In practice, we want to re-define \(f\), such that we scale and translate it in a way that its support and the reference support perfectly overlap. This will be done using above expression to derive \(f\) into \(f'\). Also, we want our model to satisfy additional constraints, without having to use regularization or (in-)equality constraints. These constraints are:

\begin{itemize}
\tightlist
\item
  All intervals must have a length greater than \(0\) (or any arbitrary positive number).
\item
  The lengths of all intervals must sum up to \(1\).
\end{itemize}

During unconstrained optimization, there is no way to enforce these constraints, but we can build these into our model! To satisfy the first constraint, we will replace using any \(l_q\) directly by \(\max{(\bm{\lambda}_q,l_q)}\) (where \(\bm{\lambda}_q\) is the lower bound for interval \(q\) and \(>0\)). This could however lead to the sum of all such lengths being greater than the extent of the support. The solution to the second constraint will satisfy this, by normalizing all lengths using the sum of all lengths. This way, lengths are converted to ratios. Later, using the \emph{target-extent}, these ratios are scaled back to actual lengths.

We define the new model with the built-in constraints (\(m_q^c\)):

\[
\begin{aligned}
  \bm{\theta}^{(b)}\;\dots&\;\text{ordered vector of boundaries, where the first and last boundary together}
  \\[0ex]
  &\;\text{represent the support of the signal they subdivide,}
  \\[1ex]
  b\dots&\;\text{the begin (absolute offset) of the first source-interval (usually}\;0\text{),}
  \\[1ex]
  e\dots&\;\text{the end (absolute offset) of the last source-interval (usually}\;1\text{), also}\;e>b\text{,}
  \\[1ex]
  \gamma_b,\gamma_e,\gamma_d\dots&\;\text{global min/max for}\;b,e\;\text{and min-distance between them,}
  \\[1ex]
  &\;\text{where}\;\gamma_d\geq 0\;\land\;\gamma_b+\gamma_d\leq\gamma_e\;\text{,}
  \\[1ex]
  \beta_l=&\;\min{\Big(\gamma_e-\gamma_d,\max{\big(\gamma_b, \min{(b,e)}\big)}\Big)}\;\text{, lower boundary as of}\;b,e\text{,}
  \\[1ex]
  \beta_u=&\;\max{\Big(\gamma_b+\gamma_d,\min{\big(\gamma_e, \max{(b,e)}\big)}\Big)}\;\text{, upper boundary as of}\;b,e\text{,}
  \\[1ex]
  \bm{\lambda}\dots&\;\text{vector with minimum lengths for each interval,}\;\forall\,\bm{\lambda}_q\geq 0\text{,}
  \\[1em]
  l_q\dots&\;\text{the length of the }q\text{-th source-interval,}
  \\[1ex]
  l'_q=&\;\max{\big(\bm{\lambda}_q,\left\lVert\,l_q\,\right\rVert\big)}\;\text{, where}\;\left\lVert\,l_q\,\right\rVert\equiv\max{\big(-l_q,l_q\big)}\equiv\mathcal{R}(2\times l_q)-l_q\text{,}
  \\[0ex]
  &\;\text{(the corrected (positive) interval-length that is strictly}\;\geq\bm{\lambda}_q\text{),}
  \\[1ex]
  \psi=&\;\sum_{i=1}^{\max{(Q)}}\,l'_i\equiv \bm{l'}^\top\hat{\bm{u}}\;\text{, the sum used to normalize each and every}\;q\text{-th}\;l\text{,}
  \\[0ex]
  &\;\text{(}\hat{\bm{u}}\;\text{is the unit-vector),}
  \\[1ex]
  l_q^{(c)}=&\;\frac{l'_q}{\psi}\times(\beta_u-\beta_l)\;\text{, corrected, normalized and re-scaled version of}\;l_q\text{,}
  \\[1ex]
  \phi_q\equiv&\;\begin{cases}
    0,&\text{if}\;q=1,
    \\
    \sum_{i=1}^{q-1}\,l_i^{(c)},&\text{otherwise}
  \end{cases}\;,
  \\[1ex]
  &\;\text{(the sum of the lengths of all corrected preceding intervals).}
\end{aligned}
\]

With all these, we can now formulate the sub-model using the previously used parameters \(s_b,s_e,t_b,t_e\) for translation and scaling of one interval (the source) to another interval (the target):

\[
\begin{aligned}
  s_b^{(q)}=&\;\beta_l+\phi_q\;\text{, and}
  \\[1ex]
  s_e^{(q)}=&\;s_b^{(q)}+l_q^{(c)}\;\text{, begin- and end-offset of the }q\text{-th source-interval,}
  \\[1ex]
  t_b^{(q)}=&\;\bm{\theta}^{(b)}_q\;\text{, and}
  \\[1ex]
  t_e^{(q)}=&\;\bm{\theta}^{(b)}_{q+1}\;\text{, begin- and end-offset of the }q\text{-th target-interval,}
  \\[1ex]
  \delta_q^{(t)}=&\;t_e^{(q)}-t_b^{(q)}\;\text{, the extent of the }q\text{-th target-interval,}
  \\[1ex]
  m_q^c\Big(f,x,t_b^{q},l_q^{(c)},\delta_q^{(t)},s_b^{(q)}\Big)=&\;f\Bigg(\frac{\Big(x-t_b^{(q)}\Big)\times l_q^{(c)}}{\delta_q^{(t)}}+s_b^{(q)}\Bigg)\;\text{, with all non-constants expanded:}
  \\[1ex]
  =&\;f\Bigg(\overbrace{\Bigg[\overbrace{\min{\Big(\gamma_e-\gamma_d,\max{\big(\gamma_b, \min{(b,e)}\big)}\Big)}}^{\beta_l}+\overbrace{\begin{cases}
    0,&\text{if}\;q=1,
    \\
    \sum_{i=1}^{q-1}\,l_i^{(c)},&\text{otherwise}
  \end{cases}}^{\phi_q}\Bigg]}^{s_b^{(q)}}
  \\[0ex]
  &\;\;\;\;\;\;+\Big(x-t_b^{(q)}\Big)\times{\delta_q^{(t)}}^{-1}\times \overbrace{\max{\Big(\bm{\lambda}_q,\mathcal{R}(2\times l_q)-l_q\Big)}\times\Bigg[\sum_{i=1}^{\max{(Q)}}\,l'_i\Bigg]^{-1}}^{\text{recall}\;l_q^{(c)}=\frac{l'_q}{\psi}\times(\beta_u-\beta_l)\text{, here we got}\;l'_q\times\psi^{-1}}
  \\[0ex]
  &\;\;\;\;\;\;\times\bigg(\overbrace{\max{\Big(\gamma_b+\gamma_d,\min{\big(\gamma_e, \max{(b,e)}\big)}\Big)}}^{\beta_u}-\overbrace{\min{\Big(\gamma_e-\gamma_d,\max{\big(\gamma_b, \min{(b,e)}\big)}\Big)}}^{\beta_l}\bigg)\Bigg)\;\text{.}
\end{aligned}
\]

The model \(m_q^c\) now satisfies these properties:

\begin{itemize}
\tightlist
\item
  Each interval has a length greater than or equal to \(\bm{\lambda}_q\) (which must be strictly positive; a length of zero however is allowed).
\item
  The first interval begins at \(\beta_l\), the last interval ends at \(\beta_u\) (these parameters are the absolute begin/end of the source-intervals, and hence apply to where onto the query will be mapped to). These parameters can either be constant or learned during optimization, effectively allowing open/closed begin- and/or -end time warping.
\item
  Each interval begins exactly after its predecessor, such that there are no overlaps. Intervals are seamlessly strung together.
\item
  The sum of the lengths of all intervals is normalized and then re-scaled using \(\phi\), considering the constraints of \(\gamma_d\) and the minimum length of each \(q\)-th interval (using its corresponding \(\bm{\lambda}_q\)).
\end{itemize}

\hypertarget{reference-vs.-query}{%
\paragraph{Reference vs.~Query}\label{reference-vs.-query}}

Thus far, we have made the difference between these two notions. The implicit assumption thus far was, that the reference is constant, i.e., once given as a pattern, it is never altered, and the query is warped to it. I came across the important difference while thinking about closed- vs.~open -begin and/or -end time warping: In the closed case, the \textbf{whole} query is warped onto the \textbf{whole} reference, i.e., both signals have to be used to their full extent. If, however, the begin or end (or both) are open, only a portion of the query is used to match still the entire reference. So, not only is the reference constant, it also has to be matched wholly, i.e., all of its reference-intervals need to be used. This raises two important points.

First, regarding the differences between both signals during warping, we should propose more appropriate names. I suggest \textbf{``Warping Pattern''} for the constant signal that has to be used in its entirety, and \textbf{``Warping Candidate''} for the signal that is translated and scaled to be closest to the Warping Pattern.

Second, in some cases it may be useful \textbf{not} to use the entire Warping Pattern. In this case, both signals can simply be swapped with each other, and a non-closed model is fit. After fitting, the warping path is inverted. Also, this is how DTW works, one signal has to be mapped in full. This flipping case is useful to find a sub-support of the Warping Pattern within the Warping Candidate.

\hypertarget{minmax-as-ramp-function}{%
\paragraph{Min/Max as ramp function}\label{minmax-as-ramp-function}}

Note that we define \(\min,\max\) in terms of the ramp-function and its derivative, the Heaviside step function.

\[
\begin{aligned}
  \max{(x,y)}\equiv&\;\mathcal{R}(y-x)+x\;\text{, with gradient}
  \\[1ex]
  \nabla\,\mathcal{R}(y-x)+x=\Bigg[\frac{\partial\,\mathcal{R}(\cdot)}{\partial\,x},\frac{\partial\,\mathcal{R}(\cdot)}{\partial\,y}\Bigg]=&\;\Bigg[\mathcal{H}(x-y)\;,\;\mathcal{H}(y-x)\Bigg]\;\text{, also, we define}\;\min\;\text{as:}
  \\[1ex]
  \min{(x,y)}=&\;\,y-\mathcal{R}(y-x)\;\text{, with gradient}
  \\[1ex]
  \nabla\,y-\mathcal{R}(y-x)=\Bigg[\frac{\partial\,\mathcal{R}(\cdot)}{\partial\,x},\frac{\partial\,\mathcal{R}(\cdot)}{\partial\,y}\Bigg]=&\;\Bigg[\mathcal{H}(y-x)\;,\;\mathcal{H}(x-y)\Bigg]\;\text{.}
\end{aligned}
\]

\hypertarget{overall-model-formulation}{%
\subsubsection{Overall model formulation}\label{overall-model-formulation}}

Now that we have sub-models that are able to transform a section of the query-signal for a given a section of the reference-signal, the logical next step is to define the entire (or overall) model in terms of all intervals. The following is in large parts identical to how we defined it in the notebook \emph{``Boundary Time Warping (final, update)''}.

It is important to once more recall the mechanics of this suggested model:

\begin{itemize}
\tightlist
\item
  The reference signal is segmented into one or more intervals, each with a length \(>0\). This segmentation is done using a given vector of \emph{reference-boundaries}.

  \begin{itemize}
  \tightlist
  \item
    These intervals are constant, and never changed afterwards. For each variable in each reference-interval, we have a constant segment of the corresponding reference-signal (continuous case), or a constant vector of data.
  \end{itemize}
\item
  Each \(q\)-th reference-interval is associated with a \(q\)-th sub-model. Each such sub-model has access to the whole query-signal and its gradient, and the goal is to translate and scale the query-signal to \textbf{best-fit} the current reference interval (this requires a loss).

  \begin{itemize}
  \tightlist
  \item
    The sub-models are designed in a way that the intervals they cover cannot overlap and connect seamlessly, among other criteria (see above).
  \end{itemize}
\end{itemize}

The input to the overall model is, among the signals, a set of reference- and query-boundaries, which are transformed into vectors of interval lengths. Albeit defined above with the sub-models, the parameters \(b,e,\gamma_b,\gamma_e,\gamma_d\) are parameters of the overall model. In practice, it may be more or less convenient to work with boundaries, and internally the overall model uses lengths/ratios, and an absolute offset for the first interval. Both of these concepts are convertible to each other. In the following, we start with a definition that uses boundaries, and then transform them to lengths and an offset.

\[
\begin{aligned}
  \mathsf{T}^{(l)}\Big(\bm{\tau}^{(b)}\Big)=&\;\Big\{\;l_1=\bm{\tau}^{(b)}_2-\bm{\tau}^{(b)}_1,\;\dots,\;l_{\left\lVert\,\bm{\tau}\,\right\rVert-1}=\bm{\tau}^{(b)}_{\left\lVert\,\bm{\tau}\,\right\rVert}-\bm{\tau}^{(b)}_{\left\lVert\,\bm{\tau}\,\right\rVert-1}\;\Big\}\;\text{,}
  \\[0ex]
  &\;\text{(boundaries-to-lengths transform operator),}
  \\[1ex]
  \mathcal{X}^{(\text{WP})}=&\;\Big[\min{\bm{\theta}^{(b)}}\,,\,\max{\bm{\theta}^{(b)}}\Big]\;\text{, the support of the Warping Pattern's signal,}
  \\[1ex]
  \bm{\vartheta}^{(l)}=\mathsf{T}^{(l)}\Big(\bm{\vartheta}^{(b)}\Big)\;\dots&\;\text{the query-intervals as lengths and boundaries,}
  \\[1ex]
  \mathcal{X}^{(\text{WC})}=&\;\Big[\beta_l\,,\,\beta_u\Big]\;\text{, the support of the Warping Candidate's signal,}
  \\[1ex]
  q\in Q;\;Q=&\;\Big\{1,\;\dots\;,\left\lVert\,\bm{\theta}^{(b)}\,\right\rVert-1\Big\}\;\text{, where}\;Q\;\text{is ordered low to high, i.e.,}\;q_i\prec q_{i+1}\text{,}
  \\[0ex]
  &\;\text{(boundaries converted to interval-indices),}
  \\[1ex]
  \mathcal{I}^{(\text{WP})},\mathcal{I}^{(\text{WC})}\;\dots&\;\text{the set of Warping Pattern- and Candidate-intervals with length}\;[\max{Q}]\text{,}
  \\[0ex]
  &\;\text{with each Warping Candidate-interval delimited by}
  \\[0ex]
  \mathbf{x}_q^{(\text{WC})}\subset\mathcal{X}^{(\text{WC})}=&\;\Big[\bm{\vartheta}^{(b)}_q\,,\,\bm{\vartheta}^{(b)}_{q+1}\Big)\;\text{, for the first}\;Q-1\;\text{intervals,}
  \\[0ex]
  =&\;\Big[\bm{\vartheta}^{(b)}_q\,,\,\bm{\vartheta}^{(b)}_{q+1}\Big]\;\text{, for the last interval; }
  \\[0ex]
  &\;\text{proper sub-supports for model}\;m_q\;\text{and its interval}\;\mathbf{x}_q^{(\text{WC})}\text{, such that}
  \\[0ex]
  &\;\mathbf{x}_q^{(\text{WC})}\prec\mathbf{x}_{q+1}^{(\text{WC})}\;\text{, i.e., sub-supports are ordered,}
  \\[1ex]
  \mathbf{y}_q^{(\text{WP})}=&\;r\Big(I_q^{(\text{WP})}\Big)\;\text{, the reference data for the}\;q\text{-th interval,}
  \\[1ex]
  \mathbf{y}=&\;\Big\{\;\mathbf{y}_1^{(\text{WP})}\,\frown\,\dots\,\frown\,\mathbf{y}_q^{(\text{WP})}\;\Big\},\;\forall\,q\in Q\;\text{.}
\end{aligned}
\]

Likewise, we can obtain \(\hat{\mathbf{y}}_q^{(\text{WC})}\) and concatenate those by calling each sub-model. During optimization, The vector of query-interval-lengths, \(\bm{\vartheta}^{(l)}\), is altered. If the model allows either or both, open begin or end, then these parameters are altered, too.

\[
\begin{aligned}
  \mathbf{\hat{y}}=\mathsf{M}\Big(r,f,\bm{\theta}^{(b)},\bm{\vartheta}^{(l)},&b,e,\gamma_b,\gamma_e,\gamma_d,\bm{\lambda}\Big)=
  \\[1ex]
  \Big\{\;&m^c_1(\cdot)\,\frown\,\dots\,\frown\,m^c_q\Big(f,\mathbf{x}^{(\text{WP})}_q,t_b^{q},l_q^{(c)},\delta_q^{(t)},s_b^{(q)}\Big)\;\Big\},\;\forall\,q\in Q
\end{aligned}
\]

Instead of returning a discrete vector for each sub-model, \(\mathbf{\hat{y}}_q\), in practice we probably want to return a tuple with the sub-model function, the support of it, as well as the relative length of its interval, i.e., \(\langle m_q^c,\text{supp}(q),l_q^{\text{rel}} \rangle\). The advantage is, that any loss function can decide how to sample from this function, or even do numerical integration.

\hypertarget{gradient-of-the-model}{%
\subsubsection{Gradient of the model}\label{gradient-of-the-model}}

\emph{Attention}: I had previously fixed an error in the model formulation, so that some parts of this subsection are unfortunately outdated. Still, we keep them for historical reasons.

We have previously shown the sub-model with built-in constraints, \(m_q^c\). We have quite many \(\min,\max\)-expressions and dynamically-long sums in our model, and for a symbolic solution, we suggest to expand all terms to create a sum, where each term can be derived for separately. Before we get there, however, it makes sense to derive some of the recurring expressions, so that we can later just plug them in.

It is important to create a fully-expanded version of the model \(m_q^c\), to better understand its gradient. Fully expanded, the model \(m_q^c\) is given as:

\[
\begin{aligned}
  m_q^c(\dots)=f\Bigg(s_b^{(q)}\;+\;&\bigg(\delta_q
  \\[1ex]
  &\times\bigg(x-\overbrace{\min{\Big(\gamma_e-\gamma_d,\max{\big(\gamma_b, \min{(b,e)}\big)}\Big)}}^{\beta_l}-\overbrace{\Big(\max{(\bm{\lambda}_1,l_1)}+\;\dots\;+\max{(\bm{\lambda}_{q-1},l_{q-1})}\Big)}^{\phi_q\;\text{(sum of all preceding intervals' lengths;}\;0\;\text{if}\;q=1\text{)}}\bigg)
  \\[1ex]
  &\times\overbrace{\Bigg(\overbrace{\max{\Big(\gamma_b+\gamma_d,\min{\big(\gamma_e, \max{(b,e)}\big)}\Big)}}^{\beta_u}-\overbrace{\min{\Big(\gamma_e-\gamma_d,\max{\big(\gamma_b, \min{(b,e)}\big)}\Big)}}^{\beta_l}\Bigg)}^{\phi^{(e)}\;\text{(extent of}\;\phi\text{)}}
  \\[1ex]
  &\times\overbrace{\max{\Big(\gamma_d,\;\max{(\bm{\lambda}_1,l_1)}+\;\dots\;+\max{(\bm{\lambda}_{\max{(Q)}},l_{\max{(Q)}})}\Big)}}^{\phi^{(s)}\;\text{(scale of}\;\phi\text{; sum of all intervals' minimum lengths)}}
  \\[1ex]
  &\times\max{\big(\bm{\lambda}_q,l_q\big)}^{-1}\bigg)\Bigg)
\end{aligned}
\]

In the remainder of this section, we will step-wise produce a gradient of the open begin- and -end model \(m_{q=4}^c\) (meaning that there is a gradient for \(b,e\), too). The reason to use \(q=4\) is simple: This guarantees that \(\phi_q\) is not zero (and that we can effortlessly see what would happen if \(q=1\)), and the \(\phi\) is not zero, either. Recall the overall model, and how it is composed of these sub-models. If we have four intervals, we get for sub-models. The difference in these sub-models lies in \(\phi_q\), which is different for every \(q\)-th sub-model (\(\phi\) is the same).

The gradient of the overall model has ``categories'', meaning that partial derivation for some of the parameters is very similar:

\begin{itemize}
\tightlist
\item
  Derivation for \(b\) or \(e\): This is category 1, and affects all terms that use \(\beta_l,\beta_u\).
\item
  Derivation for any \(l_r,r<q\): Category 2, this mainly affects \(\phi\) and \(\phi_q\).
\item
  Derivation for the current \(l_q\): Category 3, affects the denominator and \(\phi\).
\end{itemize}

\hypertarget{expanding-the-denominator}{%
\paragraph{Expanding the denominator}\label{expanding-the-denominator}}

The denominator for the product/fraction given to \(f\) is \(\max{(\bm{\lambda_q},l_q)^{-1}}\) (the last factor in the above expansion). It can only be differentiated for \(l_q\), as all \(\bm{\lambda}\) are constant:

\[
\begin{aligned}
  \max{(\bm{\lambda}_q,l_q)}^{-1}=&\;\Big(\mathcal{R}(l_q-\bm{\lambda}_q)+\bm{\lambda}_q\Big)^{-1}\;\text{, with derivative for}\;l_q
  \\[1ex]
  \frac{\partial\,\max{(\bm{\lambda}_q,l_q)}^{-1}}{\partial\,l_q}=&\;\frac{\mathcal{H}(l_q-\bm{\lambda}_q)}{\Big(\mathcal{R}(l_q-\bm{\lambda}_q)+\bm{\lambda}_q\Big)^2}
\end{aligned}
\]

\hypertarget{expanding-beta_lbeta_u}{%
\paragraph{\texorpdfstring{Expanding \(\beta_l,\beta_u\)}{Expanding \textbackslash beta\_l,\textbackslash beta\_u}}\label{expanding-beta_lbeta_u}}

We also should expand \(\beta_l,\beta_u\) and do the substitutions for our \(\min,\max\) definitions, so that we can derive our model without having to check conditions later:

\[
\begin{aligned}
  \beta_l=&\;\overbrace{\min{\Big(\gamma_e-\gamma_d,\overbrace{\max{\big(\gamma_b, \overbrace{\min{(b,e)}}^{s_2=e-\mathcal{R}(e-b)}\big)}\Big)}^{s_1=\mathcal{R}(s_2-\gamma_b)+\gamma_b}}}^{\beta_l=s_1-\mathcal{R}(s_1-(\gamma_e-\gamma_d))}\;\text{, rearranged as}
  \\[1ex]
  =&\;\overbrace{\mathcal{R}(\overbrace{e-\mathcal{R}(e-b)}^{s_2}-\gamma_b)+\gamma_b}^{s_1}-\mathcal{R}(\overbrace{\mathcal{R}(\overbrace{e-\mathcal{R}(e-b)}^{s_2}-\gamma_b)+\gamma_b}^{s_1}-(\gamma_e-\gamma_d))\;\text{, with gradient}
  \\[1ex]
  \bigg[\frac{\partial\,\beta_l}{\partial\,b},\frac{\partial\,\beta_l}{\partial\,e}\bigg]=&\bigg[\;-\mathcal{H}(e-b)\times\mathcal{H}\big(-\mathcal{R}(e-b)-\gamma_b+e\big)\times\Big(\mathcal{H}\big(\mathcal{R}(-\mathcal{R}(e-b)-\gamma_b+e)+\gamma_b-\gamma_e-\gamma_d\big)-1\Big),
  \\[0ex]
  &\;\;\;\;\Big(\mathcal{H}\big(-\mathcal{R}(e-b)-\gamma_b+e\big)-1\Big)\times\Big(e\times\mathcal{H}\Big(\mathcal{R}\big(-\mathcal{R}(e-b)-\gamma_b+e\big)+\gamma_b-\gamma_e+\gamma_d\Big)-1\Big)\;\bigg]\;\text{.}
\end{aligned}
\]

Since we get a lot of Heaviside step functions in these products, evaluation can be stopped early if the result of any is \(0\). Here we do the same for \(\beta_u\):

\[
\begin{aligned}
  \beta_u=&\;\overbrace{\max{\Big(\gamma_b+\gamma_d,\overbrace{\min{\big(\gamma_e,\overbrace{\max{(b,e)}\big)}^{s_2=\mathcal{R}(e-b)+b}}\Big)}^{s_1=s_2-\mathcal{R}(s_2-\gamma_e)}}}^{\beta_u=\mathcal{R}(s_1-(\gamma_b+\gamma_d))+(\gamma_b+\gamma_d)}\;\text{, rearranged as}
  \\[1ex]
  =&\;\mathcal{R}(\overbrace{\overbrace{\mathcal{R}(e-b)+b}^{s_2}-\mathcal{R}(\overbrace{\mathcal{R}(e-b)+b}^{s_2}-\gamma_e)}^{s_1}-(\gamma_b+\gamma_d))+(\gamma_b+\gamma_d)\;\text{, with gradient}
  \\[1ex]
  \bigg[\frac{\partial\,\beta_u}{\partial\,b},\frac{\partial\,\beta_u}{\partial\,e}\bigg]=&\bigg[\;\big(\mathcal{H}(e-b)-1\big)\times\Big(\mathcal{H}\big(\mathcal{R}(e-b)+b-\gamma_e\big)-1\Big)
  \\[0ex]
  &\;\;\;\;\;\;\;\;\times\mathcal{H}\Big(-\mathcal{R}\big(\mathcal{R}(e-b)+b-\gamma_e\big)+\mathcal{R}(e-b)+b-\gamma_b-\gamma_d\Big),
  \\[0ex]
  &\;\;\;\;e-\mathcal{H}(-\mathcal{R}(\mathcal{R}(e-b)+b-\gamma_e)+\mathcal{R}(e-b)+b-\gamma_b-\gamma_d)\;\bigg]\;\text{.}
\end{aligned}
\]

\hypertarget{expanding-phi}{%
\paragraph{\texorpdfstring{Expanding \(\phi\)}{Expanding \textbackslash phi}}\label{expanding-phi}}

Recall the definition of \(\phi=(\beta_u-\beta_l)\times\max{\Bigg(\gamma_d,\Bigg[\sum_{j=1}^{\max{(Q)}}\,\max{(\bm{\lambda}_j,l_j)}\Bigg]\Bigg)}\).

As for the final expansion, we want to expand the denominator and \(\phi\) together, i.e., \(\phi\times\max{(\bm{\lambda_q},l_q)^{-1}}\). All these expansions will lead to very large expressions, and we are starting here to add substitution letters. The expansion with parts \(j,k\) is:

\[
\begin{aligned}
  f\Big(s_b^{(q)}+\Big[\;\dots\;\times&\overbrace{\phi\times\max{(\bm{\lambda_q},l_q)}^{-1}}^{\text{(will become}\;[j-k]\text{ after expansion)}}\;\Big]\Big)
  \\[1ex]
  f\Big(s_b^{(q)}+\Big[\;\dots\;\times&\Big(\beta_u\times\max{(\bm{\lambda_q},l_q)}^{-1}-\beta_l\times\max{(\bm{\lambda_q},l_q)}^{-1}\Big)\times\max{(\gamma_d,\max{(\bm{\lambda}_1,l_1)}+\dots+\max{(\bm{\lambda}_4,l_4)})}
  \\[1ex]
  f\Big(s_b^{(q)}+\Big[\;\dots\;\times&\Bigg(\overbrace{\bigg(\beta_u\times\max{(\bm{\lambda_q},l_q)}^{-1}\times\overbrace{\max{(\gamma_d,\max{(\bm{\lambda}_1,l_1)}+\dots+\max{(\bm{\lambda}_4,l_4)})}\bigg)}^{\phi^{(s)}}}^{j}
  \\[0ex]
  &\;\;\;\;-\overbrace{\bigg(\beta_l\times\max{(\bm{\lambda_q},l_q)}^{-1}\times\overbrace{\max{(\gamma_d,\max{(\bm{\lambda}_1,l_1)}+\dots+\max{(\bm{\lambda}_4,l_4)})}\bigg)}^{\phi^{(s)}}}^{k}\Bigg)
\end{aligned}
\]

Previously, we split \(\phi\) into \(\phi^{(e)}\) and \(\phi^{(s)}\), which represent the portions \emph{extent} and \emph{scale} of it. The extent only consists of \((\beta_u-\beta_l)\), and we have already derived them. The scale expression is a nested \(\max\) with a summation inside, that affects \(\forall\,l\), regardless of what \(q\) equals.

\[
\begin{aligned}
  \phi^{(s)}=&\;\max{\Bigg(\gamma_d,\Bigg[\sum_{i=1}^{\max{(Q)}}\,\max{(\bm{\lambda}_i,l_i)}\Bigg]\Bigg)}
  \\[1ex]
  =&\;\max{\Big(\gamma_d,\max{(\bm{\lambda}_1,l_1)}+\dots+\max{(\bm{\lambda}_4,l_4)}\Big)}\;\text{, (for}\;q=4\text{),}
  \\[1ex]
  =&\;\overbrace{\mathcal{R}\Big(\Big[\;\overbrace{\mathcal{R}(l_1-\bm{\lambda}_1)+\bm{\lambda}_1}^{\text{(first }\max\text{ of summation)}}+\;\dots\;+\overbrace{\mathcal{R}(l_4-\bm{\lambda}_4)+\bm{\lambda}_4}^{\text{(}n\text{-th }\max\text{ of summation)}}\;\Big]-\gamma_d\Big)+\gamma_d}^{\text{(outer }\max\text{; recall that }\max{(x,y)=\mathcal{R}(y-x)+x}\text{)}}\;\text{.}
\end{aligned}
\]

In the above expansion, \(\gamma_d\) is a constant defined at initialization time, and the same is true for \(\bm{\lambda}\). However, \(\phi^{(s)}\) will appear in each part of the model as we see later, and its gradient is sensitive \(\forall\,l\), i.e., not just for, e.g., \(l_q\) or \(\forall\,l_r,r<q\). However, since it is the summation that will be affected, and since it does the same \(\forall\,l\), the gradient, conveniently, will be the same for each.

\[
\begin{aligned}
  \nabla\,\phi^{(s)}=&\;\frac{\partial\,\phi^{(s)}}{\partial\,l_q,\forall\,q\in Q}\;\text{,}
  \\[1ex]
  =&\;\mathcal{H}\Big(l_q-\bm{\lambda}_q\Big)\times \mathcal{H}\Big(\Big[\;\overbrace{\mathcal{R}(l_1-\bm{\lambda}_1)+\bm{\lambda}_1+\;\dots\;+R(l_4-\bm{\lambda}_4)+\bm{\lambda}_4}^{\sum_{i=1}^{\max{(Q)}}\,\max{(\bm{\lambda}_i,l_i)}}\;\Big]-\gamma_d\Big)\;\text{.}
\end{aligned}
\]

\hypertarget{expanding-m_q4c}{%
\paragraph{\texorpdfstring{Expanding \(m_{q=4}^c\)}{Expanding m\_\{q=4\}\^{}c}}\label{expanding-m_q4c}}

The goal is to fully expand the current sub-model into a big sum, where we can effortlessly derive each term separately. We also have now all of the sub-expressions and their partial gradient. Here we continue to assign substitution letters so that we can derive the model more easily.

\[
\begin{aligned}
  m_{q=4}^c=&\;f\bigg(s_b^{(q)}+\frac{(x-\beta_l-\phi_q)\times\delta_q\times\phi}{\max{(\bm{\lambda_q},l_q)}}\bigg)
  \\[1ex]
  =&\;f\bigg(s_b^{(q)}+\frac{(\delta_qx-\delta_q\beta_l-\delta_q\phi_q)\times\phi}{\max{(\bm{\lambda_q},l_q)}}\bigg)
  \\[1ex]
  =&\;f\Big(s_b^{(q)}+\Big[\;\overbrace{\delta_qx-\delta_q\beta_l-\overbrace{\delta_q\max{(\bm{\lambda_1},l_1)}-\delta_q\max{(\bm{\lambda_2},l_2)}-\delta_q\max{(\bm{\lambda_3},l_3)}}^{\phi_{q=4}}}^{(a-b-c-d-e)}\;\times\overbrace{\phi\times\max{(\bm{\lambda_q},l_q)}^{-1}}^{(j-k)}\;\Big]\Big)
  \\
  \vdots
  \\
  =&\;f\Big(s_b^{(q)}+aj-ak-bj+bk-cj+ck-dj+dk-ej+ek\Big)\;\text{.}
\end{aligned}
\]

The final expression for our model is now much more convenient to work with. It is worth noting that summands with \(c\) exist if \(q>1\), those with \(d\) exist if \(q>2\) and so on. Those are essentially what distinguishes each \(q\)-th sub-model from each other sub-model. For example, sub-model \(m_{q=1}^c\) has none of these summands, whereas sub-model \(m_{q=2}^c\) has those with \(c\), and sub-model \(m_{q=3}^c\) additionally has those with \(d\) and so on. That is why we chose an example with \(q\) up to \(4\), to better illustrate the differences.

\[
\begin{aligned}
  f\Bigg(&s_b^{(q)}
  \\[0ex]
  &+aj=\Big[\;\delta_q\times x\times \overbrace{\beta_u\times\max{(\bm{\lambda}_q,l_q)}^{-1}\times\max{\Big(\gamma_d,\max{(\bm{\lambda}_1,l_1)}+\dots+\max{(\bm{\lambda}_4,l_4)}\Big)}}^{j}\;\Big]
  \\[1ex]
  &-ak=\Big[\;\delta_q\times x\times\overbrace{\beta_l\times\max{(\bm{\lambda}_q,l_q)}^{-1}\times\max{\Big(\gamma_d,\;\dots\;\Big)}}^{k}\;\Big]
  \\[1ex]
  &-bj=\Big[\;\delta_q\times\beta_l\times \beta_u\times\max{(\dots)}^{-1}\times\max{(\dots)}\;\Big]
  \\[1ex]
  &+bk=\Big[\;\delta_q\times{\beta_l}^2\times{\dots}\times{\dots}\;\Big]
  \\[1ex]
  &-cj=\Big[\;\delta_q\times\max{(\bm{\lambda}_1,l_1)}\times\overbrace{\beta_u\times{\dots}\times{\dots}}^{j}\;\Big]
  \\[1ex]
  &+ck=\Big[\;\delta_q\times\max{(\bm{\lambda}_1,l_1)}\times\overbrace{\beta_l\times{\dots}\times{\dots}}^{k}\;\Big]
  \\[1em]
  &\overbrace{-dj=[\dots]+dk=[\dots]}^{\text{(if }q>1\text{)}}
  \\[1ex]
  &\overbrace{-ej=[\dots]+ek=[\dots]}^{\text{(if }q>2\text{)}}
  \\
  &\;\;\;\;\vdots
  \\
  &\;\text{(}\pm\;\text{additional terms for larger}\;q\text{)}\Bigg)\text{.}
  \\[1ex]
\end{aligned}
\]

The derivation of this model is now straightforward, as we can tackle each summand separately. Also, we have already created the partial derivatives of all sub-expressions earlier. Before we assemble the actual analytical gradient, we will test our model.

\hypertarget{testing-the-model}{%
\subsubsection{Testing the model}\label{testing-the-model}}

The main features of the proposed model are, that it is \textbf{self-regularizing} and optionally allows for half- or full-open window time warping. In code, we will call the model just \texttt{SRBTW}, and have the optional openness as setters and getters.

Before we go any further, we should test the model. When this notebook was started, it was called \emph{Closed-window Optimization}. However, we made some additions to the proposed model that allow partially- or fully-open window optimization. In this section, we will do some testing with the pattern and data we have, focusing on the \textbf{adaptive} (\textbf{A}) variable. We will be using the four intervals as defined earlier

\begin{verbatim}
##      q beta_l beta_u lambda_q   l_q l_prime_q l_q_c psi phi_q delta_t_q  sb_q
## [1,] 1      0      1        0 0.250     0.250 0.250   1 0.000     0.085 0.000
## [2,] 2      0      1        0 0.275     0.275 0.275   1 0.250     0.540 0.250
## [3,] 3      0      1        0 0.250     0.250 0.250   1 0.525     0.250 0.525
## [4,] 4      0      1        0 0.225     0.225 0.225   1 0.775     0.125 0.775
##       se_q  tb_q  te_q
## [1,] 0.250 0.000 0.085
## [2,] 0.525 0.085 0.625
## [3,] 0.775 0.625 0.875
## [4,] 1.000 0.875 1.000
\end{verbatim}

\includegraphics{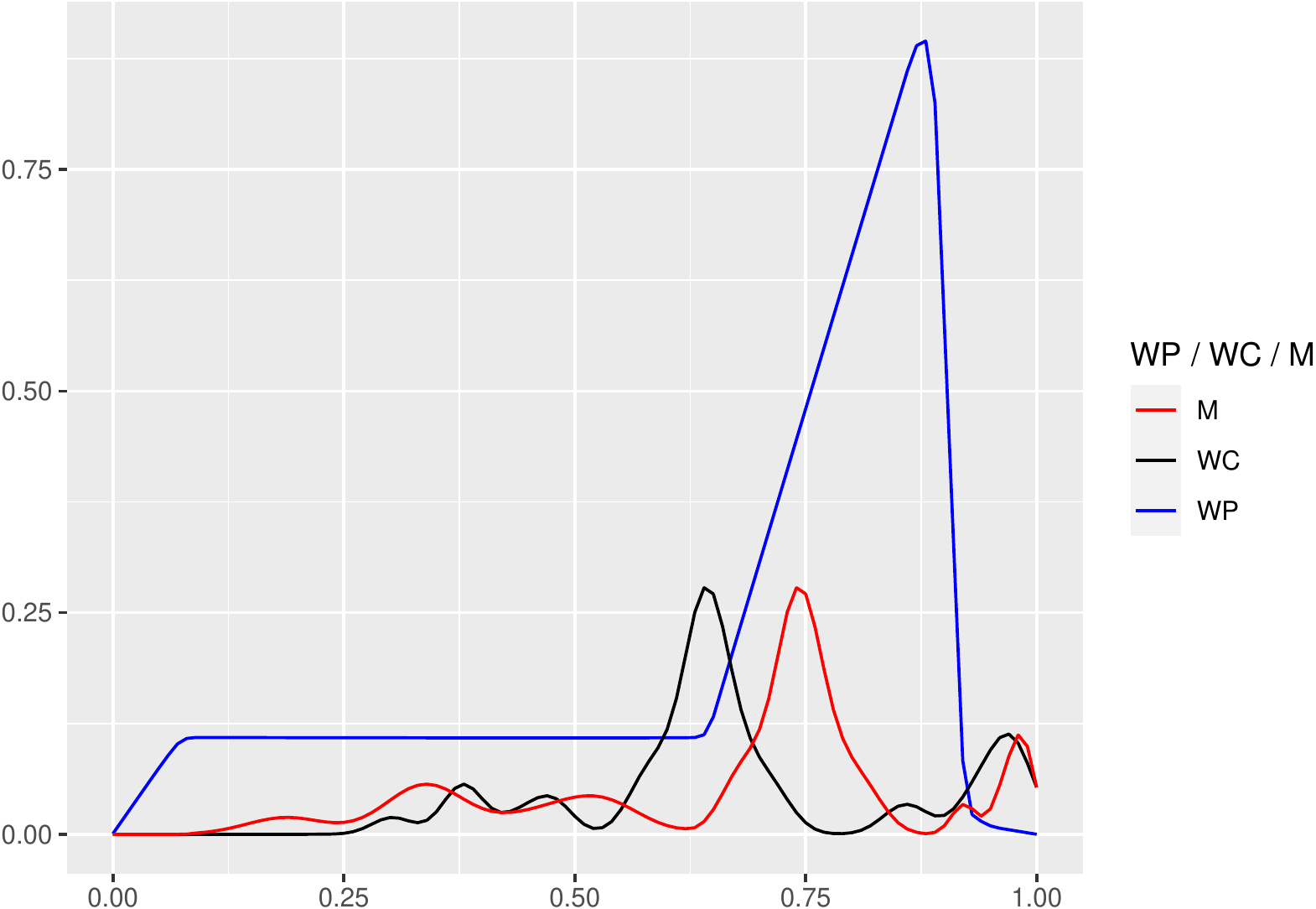}

Let's try some optimization using box-bounds and closed open and begin:

\begin{verbatim}
## $par
## [1] 0.4119631 0.1651927 0.0737804 0.3831419
## 
## $value
## [1] 1.272031
## 
## $counts
## function gradient 
##       45       45 
## 
## $convergence
## [1] 0
## 
## $message
## [1] "CONVERGENCE: REL_REDUCTION_OF_F <= FACTR*EPSMCH"
\end{verbatim}

\includegraphics{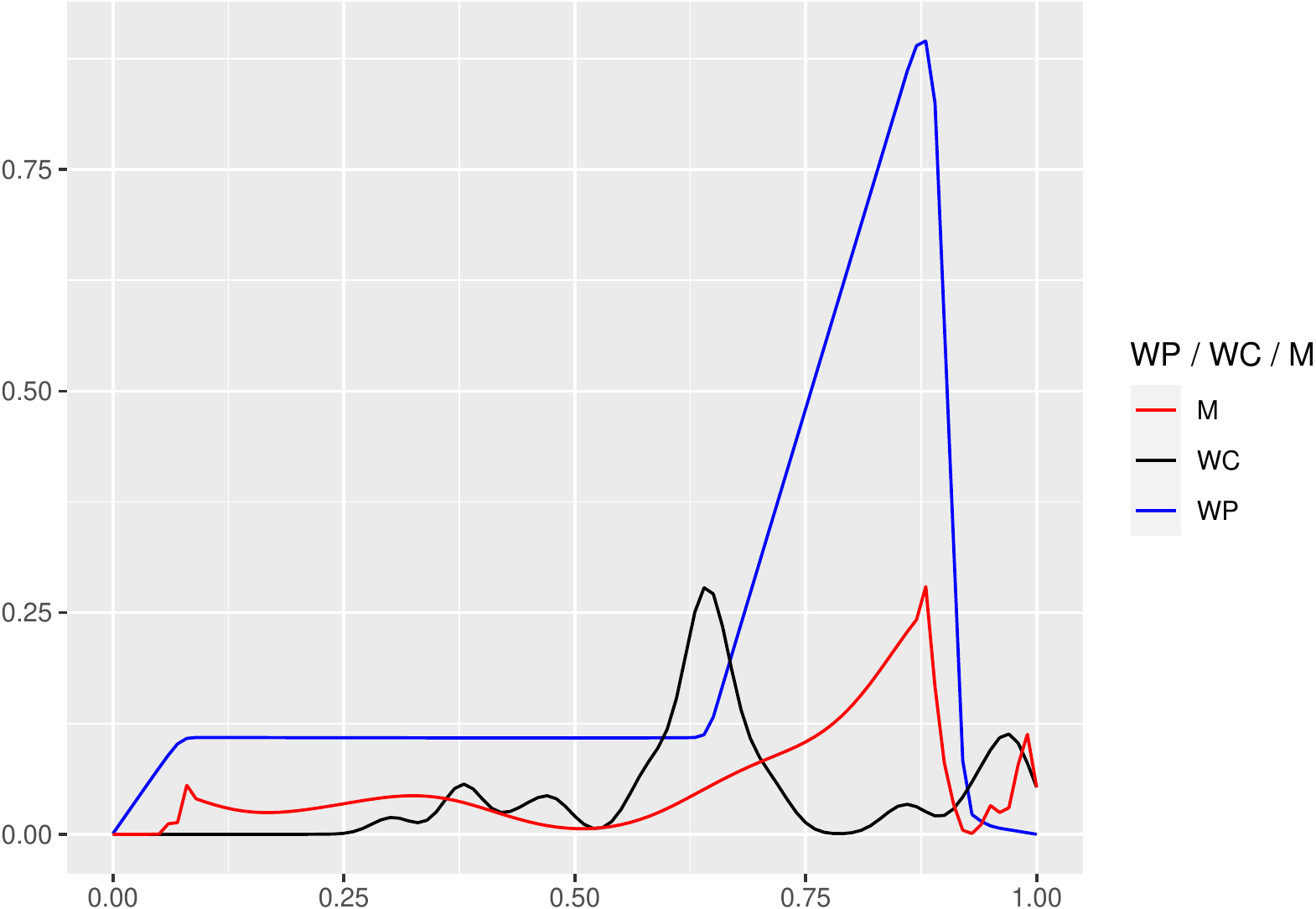}

In the meantime, I have implemented some new models as \textbf{R6-classes}. We should test that we can reach the same optimization result with these new structures. Some rationale for the architecture's design are the following:

Self-Regularizing Boundary Time Warping (\textbf{SRBTW}) is implicitly multi-variate and multi-dataseries capable. When defining a loss that concerns a variable, we will refer to it as a \textbf{Data-loss}. A loss that concerns the choice of model parameters will be called \textbf{Regularizer}. Since we have potentially many variables and intervals, a fully configured model may have many of either kind, and each of these are \textbf{objectives}; SRBTW, by design, hence is a \textbf{Multilevel-Model} and poses a \textbf{Multi-objective} optimization problem.

This is comparable to the output-layer of a neural network, so we can also view a fully-configured SRBTW as a vector-valued function, where each entry in the vector is a loss. The gradient then would be the \textbf{Jacobian}-matrix, i.e.,

\[
\begin{aligned}
  \mathcal{O}(\operatorname{SRBTW})=&\;f:\mathbb{R}^m\to\mathbb{R}^n\;\text{,}
  \\[1ex]
  &\;\text{where}\;m\;\text{is the number of parameters (begin, end, interval-lengths etc.),}
  \\[0ex]
  &\;\text{and}\;n\;\text{is the number of defined singular objectives,}
  \\[1ex]
  \mathbf{J}_f=&\;\begin{bmatrix}
    \frac{\partial\,f_1}{\partial\,x_1} & \cdots & \frac{\partial\,f_1}{\partial\,x_m} \\
    \vdots & \ddots & \vdots \\
    \frac{\partial\,f_n}{\partial\,x_1} & \cdots & \frac{\partial\,f_n}{\partial\,x_m}
  \end{bmatrix}
\end{aligned}
\]

Consider our Fire Drill example: The Warping Pattern has four variables and four intervals. If we want to fit a single project to it, we may want to define a single model and weight per variable and interval, which would result in already 16 objectives here. I wrote that SRBTW is implicitly capable of handling multiple variables and/or dataseries. Implicit because this complexity is reduced to having one SRBTW handle exactly two signals, the Warping Pattern and -Candidate, and then defining a \emph{singular} loss per each. It is perfectly fine to reuse an SRBTW model across intervals of the same variable, but in practice the overhead of doing so may be reduced by having each objective use its own instance. We provide a \textbf{linear scalarizer} to handle one or more objectives, each with their own weight. Since the linear scalarizer reduces any number of objectives to a single objective, computing it results in a scalar loss, and the Jacobian becomes a gradient again.

\[
\begin{aligned}
  \mathcal{O}^{(\text{LS})}(\operatorname{SRBTW})=&\;w_1\mathcal{L}_1(\cdot)+\;\dots\;+w_n\mathcal{L}_n(\cdot)\;\text{, (linear scalarization)}
  \\[1ex]
  =&\;\mathbf{w}^\top\bm{\mathcal{L}}\;\text{, with gradient (for any parameter in}\;\mathcal{L}_i\text{)}
  \\[1ex]
  \nabla\,\mathcal{O}=&\;w_1\mathcal{L}'_1(\cdot)+\;\dots\;+w_n\mathcal{L}'_n(\cdot)\;\text{.}
\end{aligned}
\]

In practice, each weight is some combination of multiple other weights, in our case it will most likely be the product of the weights for the current data series, variable and interval. The linear scalarizer also makes it apparent that each single objective (as well as its gradient) can be computed \textbf{independently} and simultaneously, meaning that we can arbitrarily parallelize the computation.

Back to our example that captures only the \textbf{A}-variable, let's set up this scenario with the new classes:

\begin{Shaded}
\begin{Highlighting}[]
\NormalTok{dataLoss }\OtherTok{\textless{}{-}}\NormalTok{ SRBTW\_DataLoss}\SpecialCharTok{$}\FunctionTok{new}\NormalTok{(}
  \AttributeTok{srbtw =}\NormalTok{ srbtw, }\AttributeTok{intervals =} \DecValTok{1}\SpecialCharTok{:}\DecValTok{4}\NormalTok{, }\CommentTok{\# note how we cover 4 intervals}
  \AttributeTok{weight =} \DecValTok{1}\NormalTok{, }\CommentTok{\# should be 4 but we need to check that our results}
  \CommentTok{\# of cow\_og1\_test1p are identical to cow\_og1\_test3p later.}
  \AttributeTok{continuous =} \ConstantTok{TRUE}\NormalTok{, }\AttributeTok{params =} \FunctionTok{rep}\NormalTok{(}\DecValTok{1}\SpecialCharTok{/}\DecValTok{4}\NormalTok{, }\FunctionTok{length}\NormalTok{(vartheta\_l)))}

\NormalTok{dataLoss\_RSS }\OtherTok{\textless{}{-}} \ControlFlowTok{function}\NormalTok{(loss, listOfSms) \{}
  \CommentTok{\# \textquotesingle{}loss\textquotesingle{} is a reference to the data{-}loss itself.}
\NormalTok{  continuous }\OtherTok{\textless{}{-}}\NormalTok{ loss}\SpecialCharTok{$}\FunctionTok{isContinuous}\NormalTok{()}
\NormalTok{  err }\OtherTok{\textless{}{-}} \DecValTok{0}
  
  \ControlFlowTok{for}\NormalTok{ (sm }\ControlFlowTok{in}\NormalTok{ listOfSms) \{}
\NormalTok{    t }\OtherTok{\textless{}{-}}\NormalTok{ sm}\SpecialCharTok{$}\FunctionTok{asTuple}\NormalTok{()}
    \ControlFlowTok{if}\NormalTok{ (continuous) \{}
\NormalTok{      tempf }\OtherTok{\textless{}{-}} \ControlFlowTok{function}\NormalTok{(x) (t}\SpecialCharTok{$}\FunctionTok{wp}\NormalTok{(x) }\SpecialCharTok{{-}}\NormalTok{ t}\SpecialCharTok{$}\FunctionTok{mqc}\NormalTok{(x))}\SpecialCharTok{\^{}}\DecValTok{2}
\NormalTok{      err }\OtherTok{\textless{}{-}}\NormalTok{ err }\SpecialCharTok{+}\NormalTok{ cubature}\SpecialCharTok{::}\FunctionTok{cubintegrate}\NormalTok{(}
        \AttributeTok{f =}\NormalTok{ tempf, }\AttributeTok{lower =}\NormalTok{ t}\SpecialCharTok{$}\NormalTok{tb\_q, }\AttributeTok{upper =}\NormalTok{ t}\SpecialCharTok{$}\NormalTok{te\_q)}\SpecialCharTok{$}\NormalTok{integral}
\NormalTok{    \} }\ControlFlowTok{else}\NormalTok{ \{}
\NormalTok{      X }\OtherTok{\textless{}{-}} \FunctionTok{seq}\NormalTok{(}\AttributeTok{from =}\NormalTok{ t}\SpecialCharTok{$}\NormalTok{tb\_q, }\AttributeTok{to =}\NormalTok{ t}\SpecialCharTok{$}\NormalTok{te\_q, }\AttributeTok{length.out =} \DecValTok{250}\NormalTok{)}
\NormalTok{      y }\OtherTok{\textless{}{-}} \FunctionTok{sapply}\NormalTok{(}\AttributeTok{X =}\NormalTok{ X, }\AttributeTok{FUN =}\NormalTok{ t}\SpecialCharTok{$}\NormalTok{wp)}
\NormalTok{      y\_hat }\OtherTok{\textless{}{-}} \FunctionTok{sapply}\NormalTok{(}\AttributeTok{X =}\NormalTok{ X, }\AttributeTok{FUN =}\NormalTok{ t}\SpecialCharTok{$}\NormalTok{mqc)}
\NormalTok{      err }\OtherTok{\textless{}{-}}\NormalTok{ err }\SpecialCharTok{+} \FunctionTok{sum}\NormalTok{((y }\SpecialCharTok{{-}}\NormalTok{ y\_hat)}\SpecialCharTok{\^{}}\DecValTok{2}\NormalTok{)}
\NormalTok{    \}}
\NormalTok{  \}}
  
  \FunctionTok{log}\NormalTok{(}\DecValTok{1} \SpecialCharTok{+}\NormalTok{ err)}
\NormalTok{\}}

\NormalTok{dataLoss}\SpecialCharTok{$}\FunctionTok{setLossFunc}\NormalTok{(}\AttributeTok{lossFunc =}\NormalTok{ dataLoss\_RSS)}

\NormalTok{soo }\OtherTok{\textless{}{-}}\NormalTok{ SRBTW\_SingleObjectiveOptimization}\SpecialCharTok{$}\FunctionTok{new}\NormalTok{(}\AttributeTok{srbtw =}\NormalTok{ srbtw)}
\NormalTok{soo}\SpecialCharTok{$}\FunctionTok{addObjective}\NormalTok{(}\AttributeTok{obj =}\NormalTok{ dataLoss)}
\NormalTok{soo}\SpecialCharTok{$}\FunctionTok{setParams}\NormalTok{(}\AttributeTok{params =} \FunctionTok{rep}\NormalTok{(}\DecValTok{1}\SpecialCharTok{/}\DecValTok{4}\NormalTok{, }\FunctionTok{length}\NormalTok{(vartheta\_l)))}
\NormalTok{soo}\SpecialCharTok{$}\FunctionTok{compute}\NormalTok{()}
\end{Highlighting}
\end{Shaded}

\begin{verbatim}
## [1] 0.07086299
\end{verbatim}

Let's attempt to optimize this:

\begin{verbatim}
## $par
## [1] 0.3548296 0.2611416 0.0157521 0.3507654
## 
## $value
## [1] 0.04087562
## 
## $counts
## function gradient 
##       25       25 
## 
## $convergence
## [1] 0
## 
## $message
## [1] "CONVERGENCE: REL_REDUCTION_OF_F <= FACTR*EPSMCH"
\end{verbatim}

\includegraphics{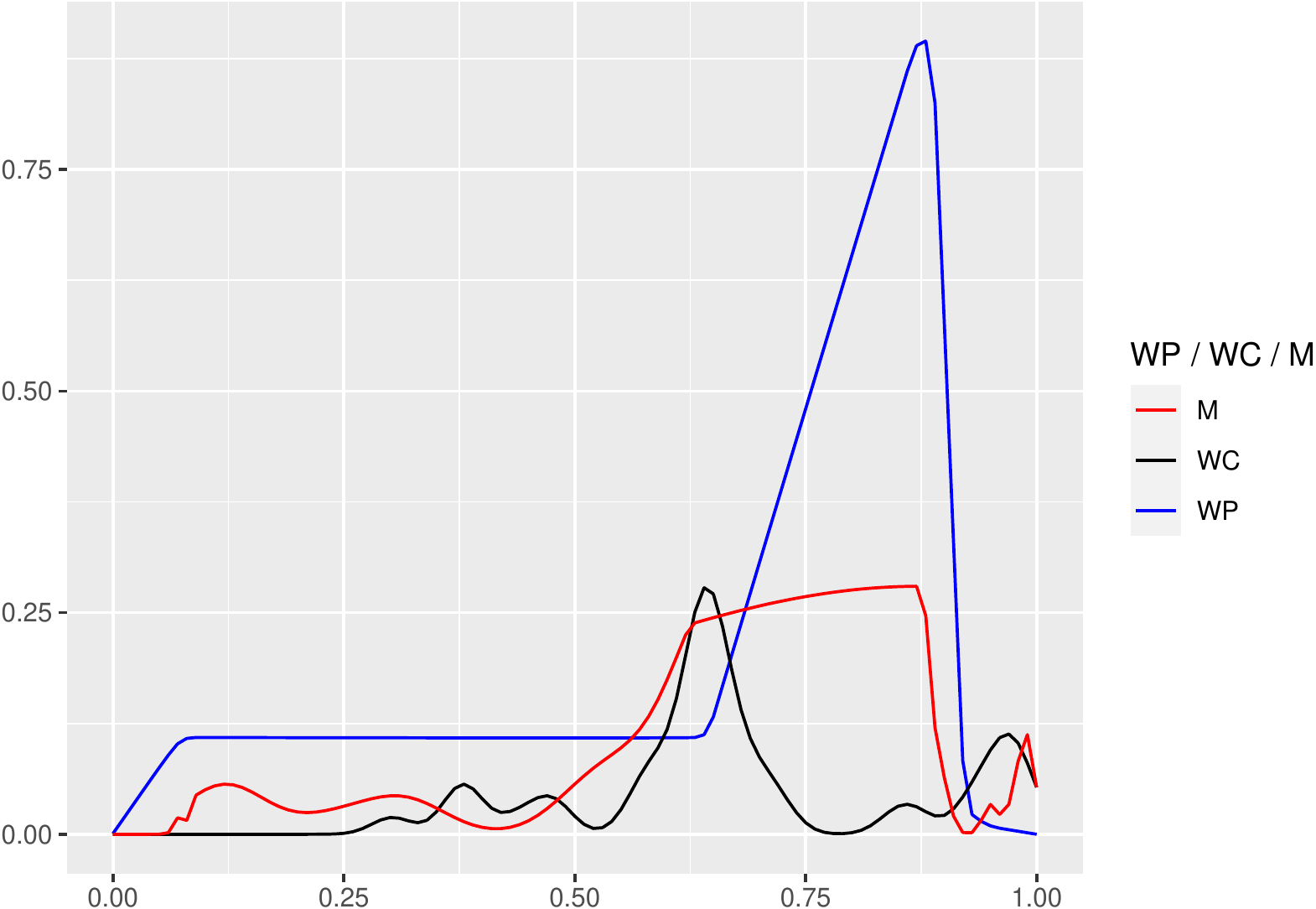}

While this fit looks subjectively worse compared to the previous fit using JSD, it is actually better w.r.t. the RSS-loss used here. The loss here is 0.040876. If we use the parameters from the JSD-fit and compute the loss for the RSS, it becomes:

\begin{Shaded}
\begin{Highlighting}[]
\NormalTok{srbtw}\SpecialCharTok{$}\FunctionTok{setAllParams}\NormalTok{(}\AttributeTok{params =}\NormalTok{ cow\_og1\_test1p}\SpecialCharTok{$}\NormalTok{par)}
\NormalTok{dataLoss}\SpecialCharTok{$}\FunctionTok{setLossFunc}\NormalTok{(}\AttributeTok{lossFunc =}\NormalTok{ dataLoss\_RSS)}
\NormalTok{soo}\SpecialCharTok{$}\FunctionTok{compute}\NormalTok{()}
\end{Highlighting}
\end{Shaded}

\begin{verbatim}
## [1] 0.05608667
\end{verbatim}

This is a clear demonstration of \textbf{SRBTW}'s strength, that it can use any arbitrary loss or (weighted) combinations thereof, to optimize for distinct goals simultaneously. Here is the test that should yield the exact same results as we got with \texttt{cow\_og1\_test1p}:

\begin{Shaded}
\begin{Highlighting}[]
\NormalTok{dataLoss\_JSD }\OtherTok{\textless{}{-}} \ControlFlowTok{function}\NormalTok{(loss, listOfSms) \{}

\NormalTok{  tempf }\OtherTok{\textless{}{-}} \FunctionTok{Vectorize}\NormalTok{(}\ControlFlowTok{function}\NormalTok{(x\_) \{}
    \FunctionTok{stopifnot}\NormalTok{(}\FunctionTok{length}\NormalTok{(x\_) }\SpecialCharTok{==} \DecValTok{1} \SpecialCharTok{\&\&} \SpecialCharTok{!}\FunctionTok{is.na}\NormalTok{(x\_))}
\NormalTok{    q }\OtherTok{\textless{}{-}}\NormalTok{ srbtw}\SpecialCharTok{$}\FunctionTok{getQForX}\NormalTok{(x\_)}
\NormalTok{    sm }\OtherTok{\textless{}{-}}\NormalTok{ listOfSms[[q]]}
\NormalTok{    t }\OtherTok{\textless{}{-}}\NormalTok{ sm}\SpecialCharTok{$}\FunctionTok{asTuple}\NormalTok{()}
\NormalTok{    t}\SpecialCharTok{$}\FunctionTok{mqc}\NormalTok{(x\_)}
\NormalTok{  \})}

  \SpecialCharTok{{-}}\FunctionTok{log}\NormalTok{(}\FunctionTok{stat\_diff\_2\_functions\_symmetric\_JSD\_score}\NormalTok{()(listOfSms[[}\DecValTok{1}\NormalTok{]]}\SpecialCharTok{$}\FunctionTok{asTuple}\NormalTok{()}\SpecialCharTok{$}\NormalTok{wp,}
\NormalTok{    tempf))}
\NormalTok{\}}

\NormalTok{dataLoss}\SpecialCharTok{$}\FunctionTok{setLossFunc}\NormalTok{(}\AttributeTok{lossFunc =}\NormalTok{ dataLoss\_JSD)}
\end{Highlighting}
\end{Shaded}

\begin{verbatim}
## $par
## [1] 0.4119631 0.1651927 0.0737804 0.3831419
## 
## $value
## [1] 1.272031
## 
## $counts
## function gradient 
##       45       45 
## 
## $convergence
## [1] 0
## 
## $message
## [1] "CONVERGENCE: REL_REDUCTION_OF_F <= FACTR*EPSMCH"
\end{verbatim}

\includegraphics{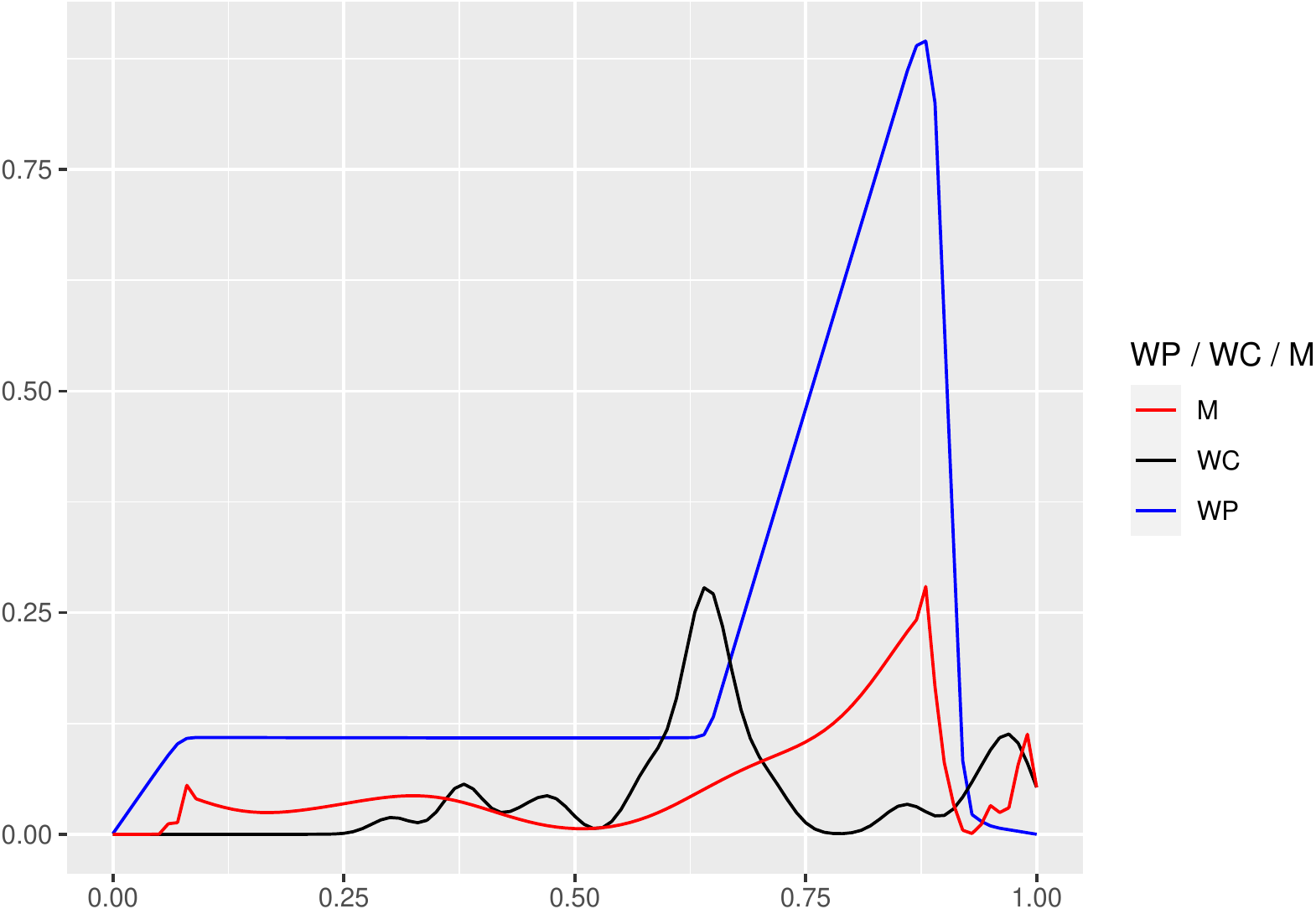}

These are the exact same results as in \texttt{cow\_og1\_test1p}!

\hypertarget{half--and-full-open-tests}{%
\paragraph{Half- and full-open tests}\label{half--and-full-open-tests}}

Let's do some tests where we leave either the begin or end open (or both) and see what happens.

\begin{Shaded}
\begin{Highlighting}[]
\NormalTok{srbtw}\SpecialCharTok{$}\FunctionTok{setOpenBegin}\NormalTok{(}\AttributeTok{ob =} \ConstantTok{TRUE}\NormalTok{)}
\NormalTok{srbtw}\SpecialCharTok{$}\FunctionTok{setOpenEnd}\NormalTok{(}\AttributeTok{oe =} \ConstantTok{TRUE}\NormalTok{)}
\end{Highlighting}
\end{Shaded}

\begin{verbatim}
## $par
## [1] 0.2826341 0.2983571 0.1597518 0.3405548 0.2325628 0.8355568
## 
## $value
## [1] 0.7642083
## 
## $counts
## function gradient 
##       51       51 
## 
## $convergence
## [1] 0
## 
## $message
## [1] "CONVERGENCE: REL_REDUCTION_OF_F <= FACTR*EPSMCH"
\end{verbatim}

\includegraphics{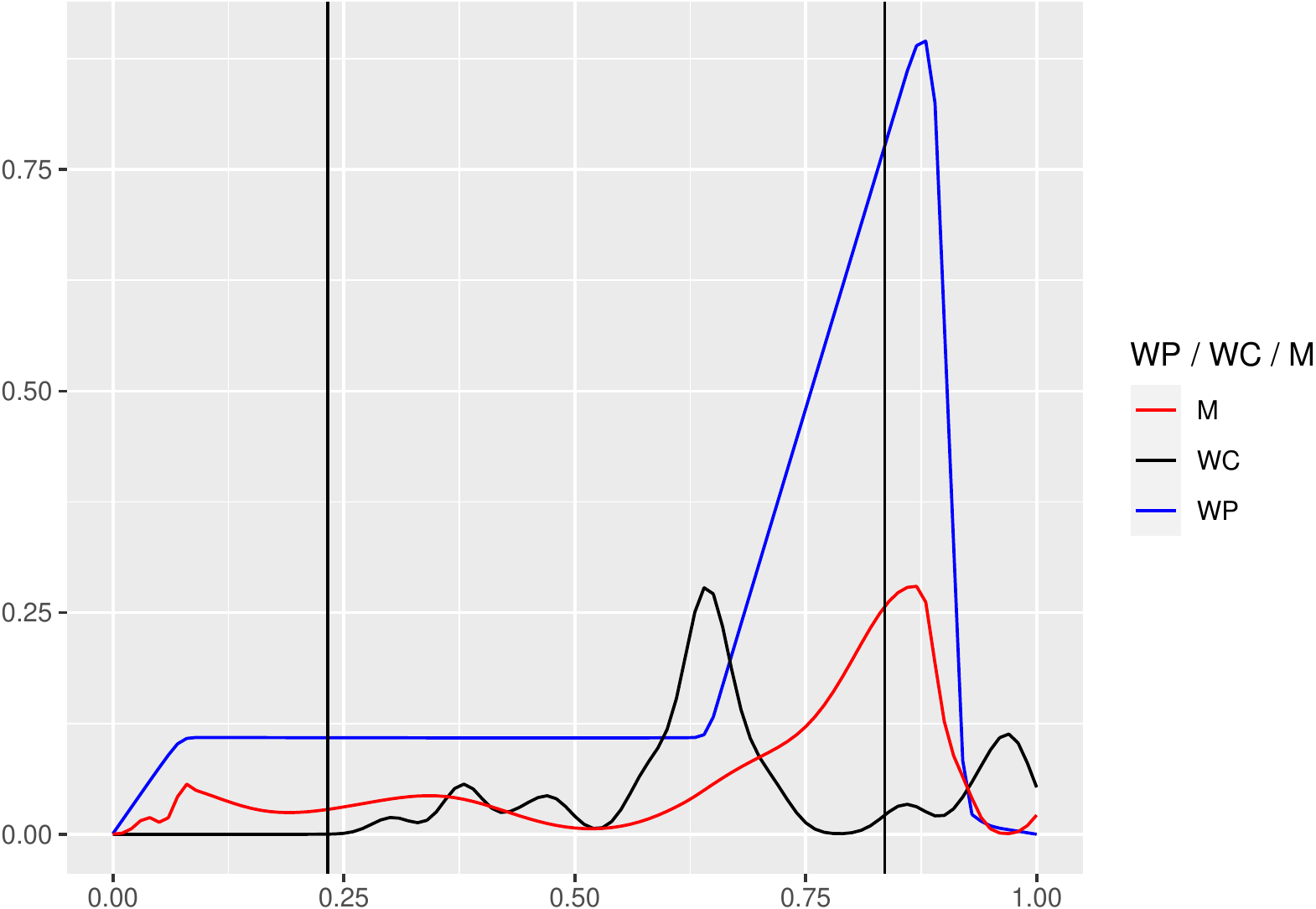}

This is still using the Jenson-Shannon divergence, and apparently it is a better fit. The black vertical lines indicate the used begin and end of the Warping Candidate, and \texttt{M} was sliced and laid over \texttt{WP}. The begin and end were 0.2325628, 0.8355568.

\begin{Shaded}
\begin{Highlighting}[]
\NormalTok{optextras}\SpecialCharTok{::}\FunctionTok{kktchk}\NormalTok{(cow\_og1\_test4p}\SpecialCharTok{$}\NormalTok{par, }\AttributeTok{fn =} \ControlFlowTok{function}\NormalTok{(x) \{}
\NormalTok{  soo}\SpecialCharTok{$}\FunctionTok{setParams}\NormalTok{(}\AttributeTok{params =}\NormalTok{ x)}
\NormalTok{  soo}\SpecialCharTok{$}\FunctionTok{compute}\NormalTok{()}
\NormalTok{\}, }\AttributeTok{gr =} \ControlFlowTok{function}\NormalTok{(x) \{}
\NormalTok{  soo}\SpecialCharTok{$}\FunctionTok{setParams}\NormalTok{(}\AttributeTok{params =}\NormalTok{ x)}
\NormalTok{  soo}\SpecialCharTok{$}\FunctionTok{computeGrad}\NormalTok{()}
\NormalTok{\})}
\end{Highlighting}
\end{Shaded}

Let's also try this with the RSS data-loss:

\begin{Shaded}
\begin{Highlighting}[]
\NormalTok{dataLoss}\SpecialCharTok{$}\FunctionTok{setLossFunc}\NormalTok{(}\AttributeTok{lossFunc =}\NormalTok{ dataLoss\_RSS)}
\end{Highlighting}
\end{Shaded}

\begin{verbatim}
## $par
## [1] 0.42615752 0.40798065 0.04777406 0.20491956 0.27012042 0.73161072
## 
## $value
## [1] 0.03670078
## 
## $counts
## function gradient 
##       77       77 
## 
## $convergence
## [1] 0
## 
## $message
## [1] "CONVERGENCE: REL_REDUCTION_OF_F <= FACTR*EPSMCH"
\end{verbatim}

\includegraphics{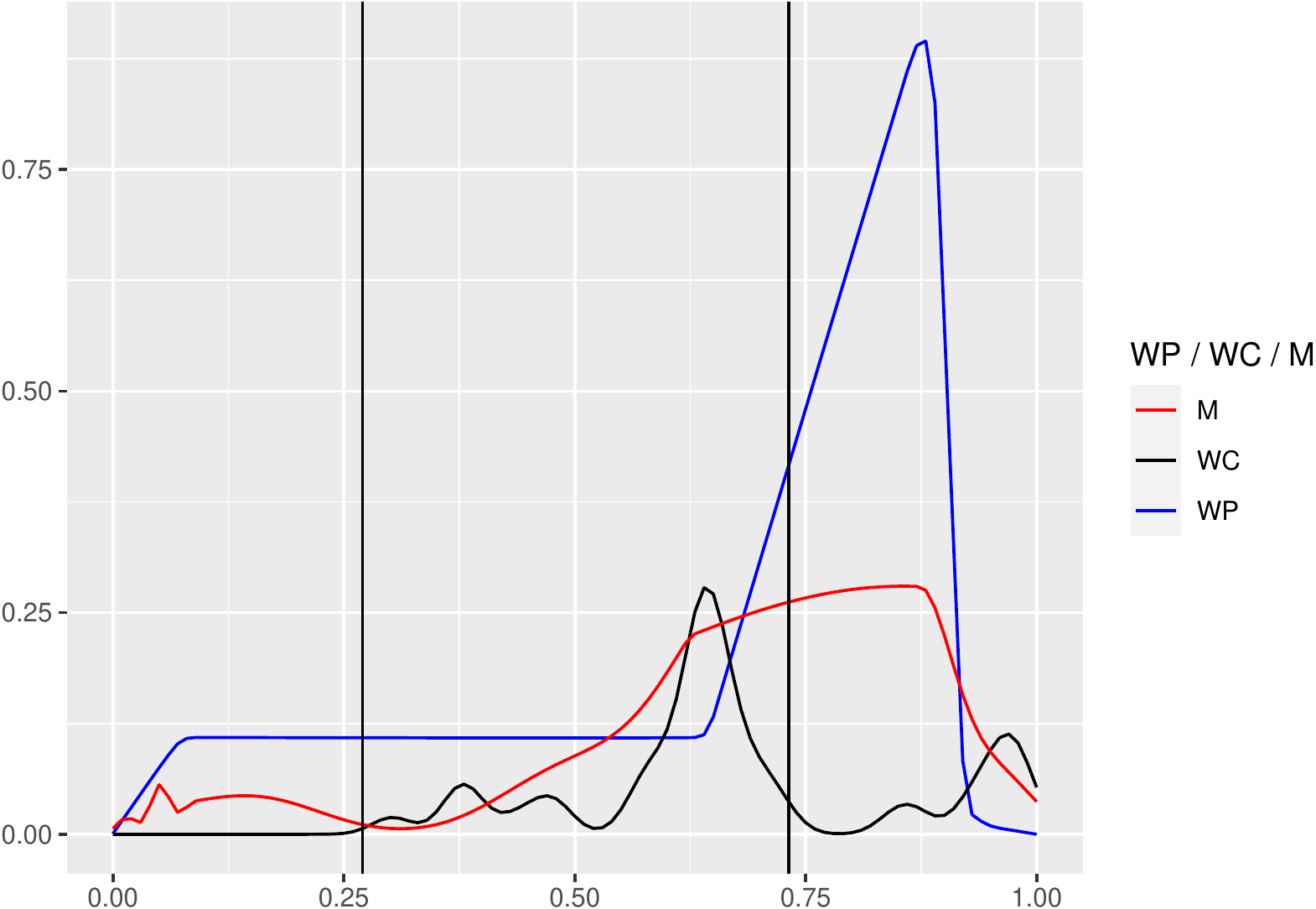}

That worked, too. I tested this with the discrete version, and it takes about twice as many iterations. However, we can also observe that the target-extent is smaller than in the previous example, which brings us to the next important matter: \textbf{Regularization}. But before that, let's make two half-open tests quickly, using JSD again:

\begin{Shaded}
\begin{Highlighting}[]
\NormalTok{srbtw}\SpecialCharTok{$}\FunctionTok{setOpenBegin}\NormalTok{(}\AttributeTok{ob =} \ConstantTok{FALSE}\NormalTok{)}
\NormalTok{srbtw}\SpecialCharTok{$}\FunctionTok{setOpenEnd}\NormalTok{(}\AttributeTok{oe =} \ConstantTok{TRUE}\NormalTok{)}

\NormalTok{dataLoss}\SpecialCharTok{$}\FunctionTok{setLossFunc}\NormalTok{(}\AttributeTok{lossFunc =}\NormalTok{ dataLoss\_JSD)}
\end{Highlighting}
\end{Shaded}

\begin{verbatim}
## $par
## [1] 0.2461881 0.3713805 0.1892241 0.4050136 0.8345426
## 
## $value
## [1] 0.7842093
## 
## $counts
## function gradient 
##       30       30 
## 
## $convergence
## [1] 0
## 
## $message
## [1] "CONVERGENCE: REL_REDUCTION_OF_F <= FACTR*EPSMCH"
\end{verbatim}

\includegraphics{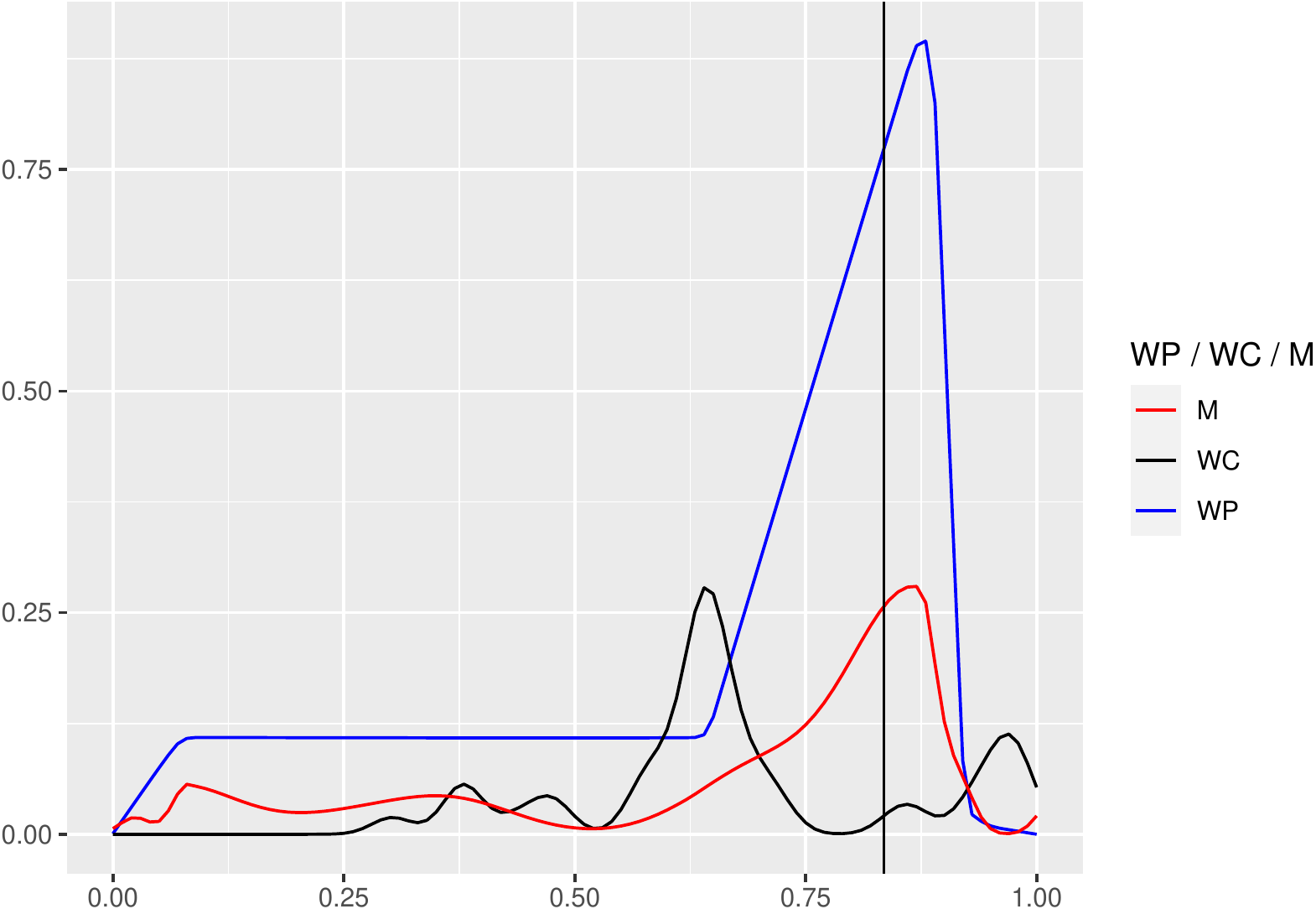}

Now we have a late cut-out, that basically ignores \texttt{1/6} of the signal at the end. This fit is better than the closed/closed one we made earlier, the loss here is 0.784209 and previously it was 1.272031. Let's do this once more, open begin, closed end this time:

\begin{Shaded}
\begin{Highlighting}[]
\NormalTok{srbtw}\SpecialCharTok{$}\FunctionTok{setOpenBegin}\NormalTok{(}\AttributeTok{ob =} \ConstantTok{TRUE}\NormalTok{)}
\NormalTok{srbtw}\SpecialCharTok{$}\FunctionTok{setOpenEnd}\NormalTok{(}\AttributeTok{oe =} \ConstantTok{FALSE}\NormalTok{)}
\end{Highlighting}
\end{Shaded}

\begin{verbatim}
## $par
## [1] 0.3219058 0.3399120 0.1826912 0.3834835 0.2325001
## 
## $value
## [1] 0.7643591
## 
## $counts
## function gradient 
##       35       35 
## 
## $convergence
## [1] 0
## 
## $message
## [1] "CONVERGENCE: REL_REDUCTION_OF_F <= FACTR*EPSMCH"
\end{verbatim}

\includegraphics{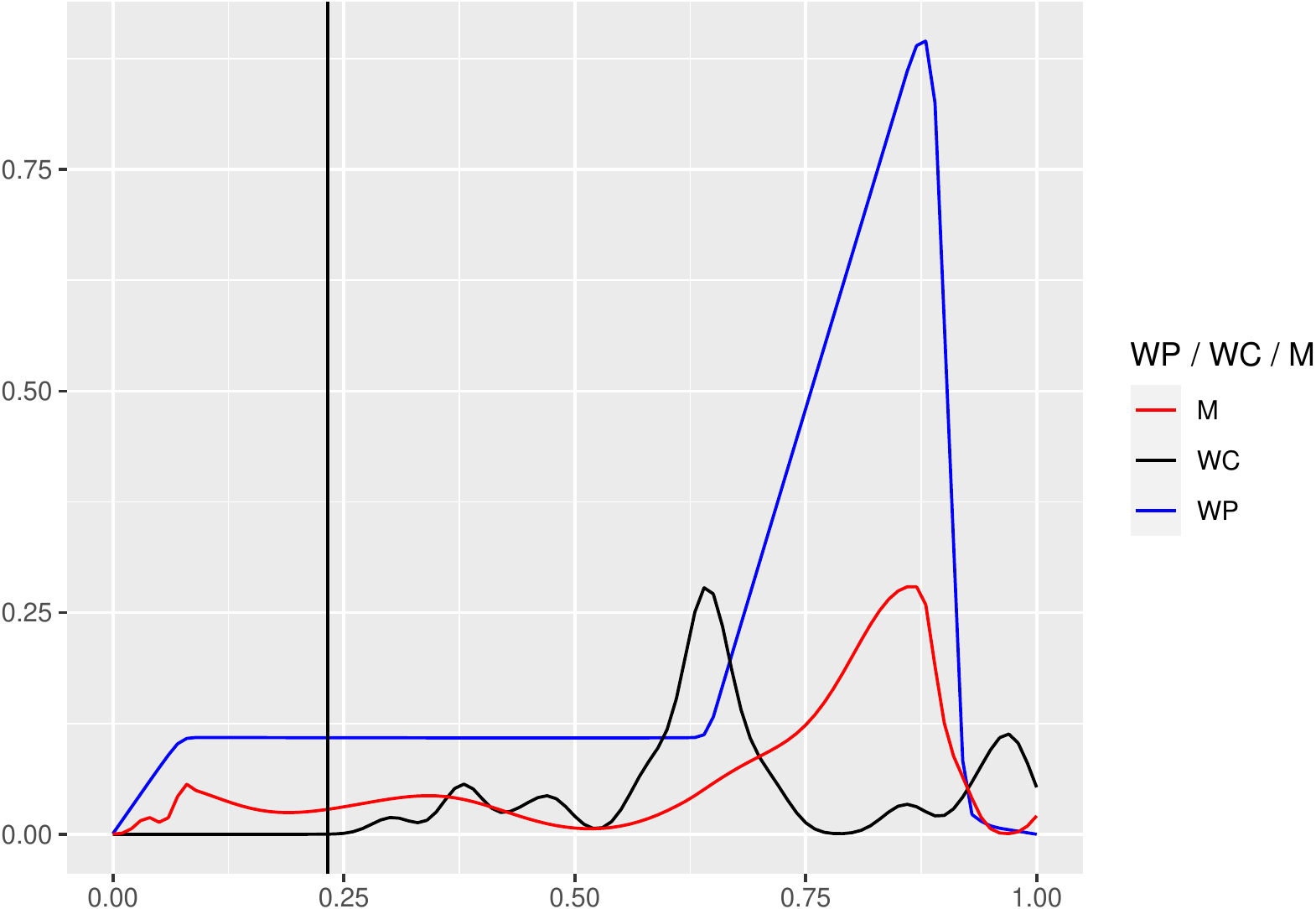}

Previously, I had wrongly set the begin-parameter to start with \(1\), and the final result did not move this parameter, which is the first obvious problem that warrants for regularization. However, after correction, I had let the parameter start from \(0\) and the result as can be seen is quite good again -- it finds a nice cut in while it has to keep the remainder of the signal. The loss here is the lowest thus far using JSD (0.764359).

\hypertarget{swapping-warping-pattern-and-candidate}{%
\paragraph{Swapping Warping Pattern and Candidate}\label{swapping-warping-pattern-and-candidate}}

This is something we should try as well. Remember that the difference between Warping Pattern and Warping Candidate is, that the former has to be matched wholly. Let's swap them and make an open/open test-fit. This test will tell us, which sub-support of the Warping Pattern actually matches the data we have.

\begin{Shaded}
\begin{Highlighting}[]
\NormalTok{vartheta\_l\_swap }\OtherTok{\textless{}{-}} \FunctionTok{rep}\NormalTok{(}\DecValTok{1}\SpecialCharTok{/}\DecValTok{15}\NormalTok{, }\DecValTok{15}\NormalTok{)}

\NormalTok{srbtw\_swap }\OtherTok{\textless{}{-}}\NormalTok{ SRBTW}\SpecialCharTok{$}\FunctionTok{new}\NormalTok{(}
  \AttributeTok{wp =}\NormalTok{ f,}
  \AttributeTok{wc =}\NormalTok{ r,}
  \AttributeTok{theta\_b =} \FunctionTok{seq}\NormalTok{(}\DecValTok{0}\NormalTok{, }\DecValTok{1}\NormalTok{, }\AttributeTok{length.out =} \DecValTok{16}\NormalTok{),}
  \AttributeTok{gamma\_bed =} \FunctionTok{c}\NormalTok{(}\DecValTok{0}\NormalTok{, }\DecValTok{1}\NormalTok{, }\DecValTok{0}\NormalTok{),}
  \AttributeTok{lambda =} \FunctionTok{rep}\NormalTok{(}\DecValTok{0}\NormalTok{, }\FunctionTok{length}\NormalTok{(vartheta\_l\_swap)),}
  \AttributeTok{begin =} \DecValTok{0}\NormalTok{,}
  \AttributeTok{end =} \DecValTok{1}\NormalTok{,}
  \AttributeTok{openBegin =} \ConstantTok{TRUE}\NormalTok{,}
  \AttributeTok{openEnd =} \ConstantTok{TRUE}
\NormalTok{)}

\NormalTok{srbtw\_swap}\SpecialCharTok{$}\FunctionTok{setParams}\NormalTok{(}\AttributeTok{vartheta\_l =}\NormalTok{ vartheta\_l\_swap)}

\NormalTok{dataLoss\_swap }\OtherTok{\textless{}{-}}\NormalTok{ SRBTW\_DataLoss}\SpecialCharTok{$}\FunctionTok{new}\NormalTok{(}
  \AttributeTok{srbtw =}\NormalTok{ srbtw\_swap, }\AttributeTok{intervals =} \DecValTok{1}\SpecialCharTok{:}\DecValTok{4}\NormalTok{, }\CommentTok{\# note how we cover 4 intervals}
  \AttributeTok{weight =} \DecValTok{1}\NormalTok{, }\CommentTok{\# should be 4 but we need to check that our results}
  \CommentTok{\# of cow\_og1\_test1p are identical to cow\_og1\_test3p later.}
  \AttributeTok{continuous =} \ConstantTok{TRUE}\NormalTok{, }\AttributeTok{params =} \FunctionTok{c}\NormalTok{(vartheta\_l\_swap, }\DecValTok{0}\NormalTok{, }\DecValTok{1}\NormalTok{))}

\NormalTok{soo\_swap }\OtherTok{\textless{}{-}}\NormalTok{ SRBTW\_SingleObjectiveOptimization}\SpecialCharTok{$}\FunctionTok{new}\NormalTok{(}\AttributeTok{srbtw =}\NormalTok{ srbtw\_swap)}
\NormalTok{soo\_swap}\SpecialCharTok{$}\FunctionTok{addObjective}\NormalTok{(}\AttributeTok{obj =}\NormalTok{ dataLoss\_swap)}
\NormalTok{dataLoss\_swap}\SpecialCharTok{$}\FunctionTok{setLossFunc}\NormalTok{(}\AttributeTok{lossFunc =}\NormalTok{ dataLoss\_RSS)}
\end{Highlighting}
\end{Shaded}

\begin{verbatim}
## $par
##  [1] 0.000000000 0.000000000 0.000000000 0.000292625 0.089836915 0.089836915
##  [7] 0.089836915 0.089836915 0.089836915 0.089836915 0.089836915 0.089836915
## [13] 0.089836915 0.089836915 0.089836915 0.000000000 0.976815702
## 
## $value
## [1] 1.790928e-07
## 
## $counts
## function gradient 
##       31       31 
## 
## $convergence
## [1] 0
## 
## $message
## [1] "CONVERGENCE: REL_REDUCTION_OF_F <= FACTR*EPSMCH"
\end{verbatim}

\includegraphics{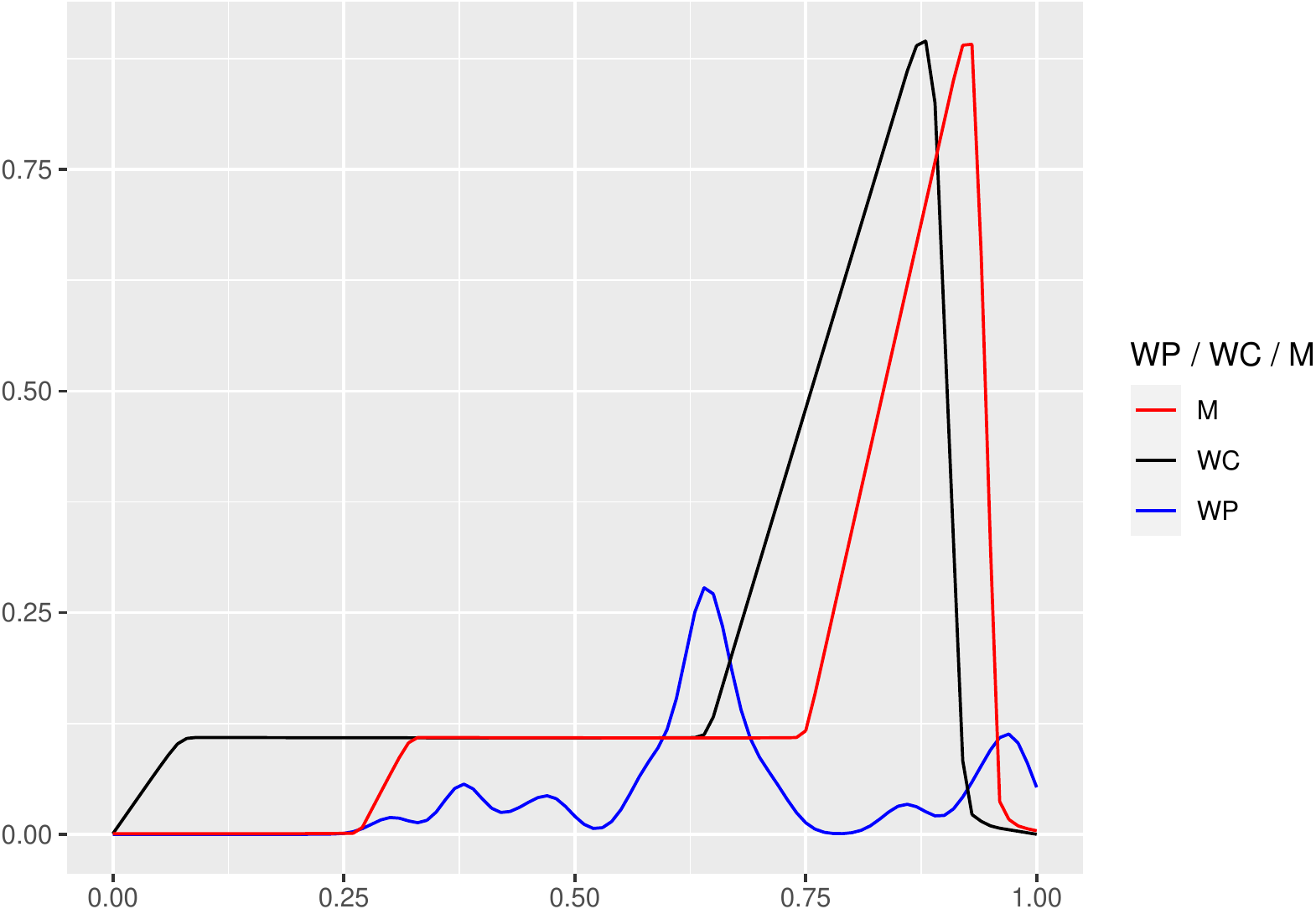}

This works great, too, but we have a somewhat unfavorable example here. The project data, which has now become the reference (WP, blue) is zero between \(\approx[0,0.26]\). We have attempted open/open optimization, but since the reference (now WC, black) does not have this at the beginning, the optimization finds a way around this by mapping the first three intervals of the blue WP to three times the length of zero, effectively moving the reference by that. There is some cut-off at the end, however ;) These zero-length intervals are still required, because only through those we learn the offset in the blue line, where the black/red data series start to match. For this example, the reference as WC matches (given the loss we used) the data as WP almost entirely (a small cut-off at the end). This means that the first three intervals are technically \emph{not} contained in the reference signal (now as WC). The result is a \emph{padding} in the beginning, and could have happened also at the end. Padding thus occurs whenever one or more consecutive lengths at the beginning or end are \(0\). In some cases, we would probably also get intermittent intervals with a zero length (zero-length intervals preceded or followed by non-zero-length intervals), which means that the match is \emph{partial}, and some interval of the WP has no counterpart in the WC, i.e., a section in the WP is missing in the WC.

\hypertarget{regularization}{%
\subsubsection{Regularization}\label{regularization}}

Regularization-losses are those that impose penalties on extreme parameters. SRBTW is self-regularizing w.r.t. interval-lengths and begin- and end-offsets. No interval can be of negative length (or smaller than a user-defined, yet positive, parameter). Begin and end are always ordered and always keep a minimum distance to each other (\(\gamma_d\)), while guaranteed not to extend beyond some user-defined lower and upper boundary. The tests up until now for the first optimization goal have shown these built-in regularizations alone are suitable for achieving very good fits. We also chose to build in these, so that no logarithmic barriers or (in-)equality constraints need to be used, as these require exponentially many additional iterations, especially when using a non-analytic gradient, leading to infeasibe large problems with already few (\(<\approx200\)) boundaries.

However, it may still be desirable to \emph{guide} the optimization process using some user-preferences. For example, one may want to avoid extreme interval-lengths. Or it may not be desirable to have the begin or end cut in too far, decimating the Warping Candidate too much. We formulate these two regularizers below but any other user-defined regularization is possible, too. Also, a regularizer is not strictly limited to compute a loss based on the parameters only, but it is more common, so we have made the distinction of \emph{data-loss} and \emph{regularizer} thus far, and we are going to continue to do so.

The nice thing about the architecture so far is, that SRBTW is already a \emph{multi-objective} optimization problem (using the linear scalarizer). We can just go ahead and add any regularizer as an additional objective to the overall loss. Each of these specifies its own weight, too.

\hypertarget{regularize-extreme-supports}{%
\paragraph{Regularize extreme supports}\label{regularize-extreme-supports}}

An extreme support is one that is much shorter than what would be allowed (given by \(\gamma_e-\gamma_b-\gamma_d\)). We can keep the regularizer and its gradient very simple.

\[
\begin{aligned}
  \mathcal{R}^{(\text{supp})}=&\;\frac{\beta_u-\beta_l}{\gamma_e-\gamma_b-\gamma_d}
  \\[1ex]
  =&\;\frac{\overbrace{\max{\Big(\gamma_b+\gamma_d,\min{\big(\gamma_e, \max{(b,e)}\big)}\Big)}}^{\beta_u}-\overbrace{\min{\Big(\gamma_e-\gamma_d,\max{\big(\gamma_b, \min{(b,e)}\big)}\Big)}}^{\beta_l}}{\gamma_e-\gamma_b-\gamma_d}\;\text{.}
\end{aligned}
\]

We had previously worked out the gradient for the sub-expression of \(\beta_u-\beta_l\), and can reuse it in this regularizer. Here, we showed the complete regularizer for both, open begin and -end BTW. If either of these is not open, \(\beta_l\) or \(\beta_u\) can be replaced with a constant. In that case, it is recommended to already account for the difference of that constant and its maximum allowed value (\(\gamma_b\) or \(\gamma_e\)). Note that its fraction is always between zero and one:

\[
\begin{aligned}
  \text{supp}\big(\mathcal{R}^{(\text{supp})}\big)=&\;\big\{\;x\in\mathbb{R}\;\rvert\;0\leq x\leq 1\;\big\}\;\text{, where values closer to zero}
  \\[0ex]
  &\;\text{mean more extreme intervals, so we propose a}\;\log\text{-loss:}
  \\[1ex]
  \mathcal{L}_{\mathcal{R}^{(\text{supp})}}=&\;-\log{\Big(\mathcal{R}^{(\text{supp})}\Big)}\mapsto[+0,\infty]\;\text{.}
\end{aligned}
\]

Until now I had not mentioned that I usually attempt to design data- and regularization-losses using logarithms, and, if possible, such that the result of a loss is \textbf{always positive}, usually with an upper bound of \(\infty\), that approaches \(+0\) as lower bound. Furthermore, and that depends on the objective, we most often deal with two kinds of losses. Either, the loss is some kind of ratio (or we can formulate it as such) or has a lower and upper bound. In this case, I usually scale the raw loss into the interval \([0,1]\) and use it in a negative logarithm (like the above regularizer). In the other case, the raw loss has only one bound (usually the lower bound, which is \(0\)). In that case, we can use the absolute raw loss as \(\log{(1+[\text{raw loss}])}\).

\hypertarget{regularize-extreme-intervals}{%
\paragraph{Regularize extreme intervals}\label{regularize-extreme-intervals}}

This is something that I had also done in some of the previous notebooks, but it is not compatible with SRBTW. Still, we can come up with something similar. First we need to answer the question, what is an extreme interval? This is subjective, and probably different from case to case. I want to suggest a quite general notion here that probably works in many cases.

We will assume that an extreme interval is one that deviates from its \emph{expected length}. A priori we do not know what amount of time time warping is required, so we assume none. That means, each interval's length is expected to be equal to or close to its corresponding Warping Pattern's counterpart (not the length but the ratio). Or, one could say that the expected length of each interval is the mean, which is the allowed extent divided by the number of intervals. Or, the user can define their own preference. Then for each interval, we could sum up the difference from the expected value.

\[
\begin{aligned}
  \bm{\kappa}\;\dots&\;\text{vector with expected lengths (ratios) for each interval,}
  \\[1ex]
  \mathcal{R}^{(\text{exint})}=&\;\sum_{q=1}^{\max{(Q)}}\,\Big(\bm{\vartheta}_q^{(l)}-\bm{\kappa}_q\Big)^2\;\text{, with gradient}
  \\[1ex]
  \frac{\partial\,\mathcal{R}^{(\text{exint})}}{\partial\,\bm{\vartheta}_q^{(l)}}=&\;2\sum_{q=1}^{\max{(Q)}}\,\Big(\bm{\vartheta}_q^{(l)}-\bm{\kappa}_q\Big)\;\text{and}\;\log\text{-loss}
  \\[1ex]
  \mathcal{L}_{\mathcal{R}^{(\text{exint})}}=&\;\log{\Big(1+\mathcal{R}^{(\text{exint})}\Big)}\mapsto[+0,\infty]\;\text{, with gradient}
  \\[1ex]
  \nabla\,\mathcal{L}_{\mathcal{R}^{(\text{exint})}}=\frac{\partial\,\mathcal{L}_{\mathcal{R}^{(\text{exint})}}}{\partial\,\bm{\vartheta}_q^{(l)}}=&\;\frac{\nabla\,\mathcal{R}^{(\text{exint})}}{1+\mathcal{R}^{(\text{exint})}}\;\text{.}
\end{aligned}
\]

In practice, this regularizer is never zero, and it should probably not have a large weight, as otherwise, it would actually suppress time warping. Also, one may ask why use \(\bm{\vartheta}_q^{(l)}\) and not \(l_q\), \(l'_q\) or \(l_q^{(c)}\) instead. Internally, SRBTW converts boundaries to interval-lengths, and then finally to ratios that are then scaled back using the actual extent given by \(\beta_u-\beta_l\). \(\bm{\vartheta}_q^{(l)}\). It is best if the regularizer directly concerns the parameters. Ideally, the lengths passed are close to ratios. If no box-bounds regularizer is used, those might be far off.

\hypertarget{regularize-box-bounds}{%
\paragraph{Regularize box-bounds}\label{regularize-box-bounds}}

This is something that is not explicitly required for SRBTW, as the box-bounds for all intervals' lengths are adhered to through the self-regularization process. The same goes for begin and end (if open). So far, we have however always used box-bounds, and the only reason for that was that this usually comes with a speedup, because of the vastly reduced feasible region. When box-constrained optimization cannot be used, it may thus be advisable to add some (mild) regularization. In the following definition, the loss for each parameter that is outside the bounds is strictly positive. Hint: This regularizer assumes there is a lower- and upper boundary for each parameter. Should that not be the case for some or all parameters, their specific \(i\)-th lower or upper bound can be replaced with \(-\infty,\infty\), respectively.

\[
\begin{aligned}
  \mathcal{R}^{(\text{bb})}=&\;\sum_{i=1}^{\left\lVert\,\bm{\theta}\,\right\rVert}\;\mathcal{H}\Big(b_i^{(l)}-\bm{\theta}_i\Big)*\big(b_i^{(l)}-\bm{\theta}_i\big) + \mathcal{H}\Big(\bm{\theta}_i-b_i^{(u)}\Big)*\big(\bm{\theta}_i-b_i^{(u)}\big)\;\text{, where}
  \\[1ex]
  b_i^{(l)},b_i^{(u)}\dots&\;\text{are the lower and upper bound for parameter}\;\bm{\theta}_i\text{, with gradient}
  \\[1ex]
  \frac{\partial\,\mathcal{R}^{(\text{bb})}}{\partial\,\bm{\theta}_i}=&\;\mathcal{D}(b_i^{(l)}-\bm{\theta}_i)\times(\bm{\theta}_i-b_i^{(l)})+\mathcal{D}(\bm{\theta}_i-b_i^{(u)})\times(\bm{\theta}_i-b_i^{(u)})
  \\[0ex]
  &\;+\mathcal{H}(\bm{\theta}_i-b_i^{(u)})-\mathcal{H}(b_i^{(l)}-\bm{\theta}_i)\;\text{, where}
  \\[1ex]
  \mathcal{D}(x)=&\;\begin{cases}
    1,&\text{if}\;x=0\text{,}
    \\
    0,&\text{otherwise,}
  \end{cases}\;\text{the Dirac-delta function, and suggested loss}
  \\[1ex]
  \mathcal{L}_{\mathcal{R}^{(\text{bb})}}=&\;\log{\Big(1+\mathcal{R}^{(\text{bb})}\Big)}\mapsto[+0,\infty]\;\text{, with gradient}
  \\[1ex]
  \nabla\,\mathcal{L}_{\mathcal{R}^{(\text{bb})}}=\frac{\partial\,\mathcal{L}_{\mathcal{R}^{(\text{bb})}}}{\partial\,\bm{\theta}_i}=&\;\frac{\nabla\,\mathcal{R}^{(\text{bb})}}{1+\mathcal{R}^{(\text{bb})}}\;\text{.}
\end{aligned}
\]

Also, instead of or additionally to this regularizer, we can sum up the number of violations. This can then be used, e.g., as base, exponent or multiplier. At this point I would recommend to use this regularizer for all parameters, that is all lengths (\(\bm{\vartheta}^{(l)}\)) as well as \(b,e\). Additionally, I recommend adding another instance of this regularizer for \(b,e\) that uses \(e-\gamma_d\) as upper boundary for \(b\), and \(b+\gamma_d\) as lower boundary for \(e\). This regularizer should be able to help the model figure out good values more quickly, and it should speed up the process if \(b>e\) (which is handled by the model but still not OK).

\hypertarget{extension-non-linear-time-warping}{%
\subsubsection{Extension: Non-linear time warping}\label{extension-non-linear-time-warping}}

We can easily make sr-BTW use non-linear time warping. Given the model \(m^c_q\), we replace the linear scaling function with an arbitrary function.

\[
\begin{aligned}
    \delta_q^{(t)}=&\;t_e^{(q)}-t_b^{(q)}\;\text{and}
    \\[1ex]
    \delta_q^{(s)}\equiv&\;l_q^{(c)}\;\text{,}
    \\[1ex]
    m^c_q(f,x,\dots)=&\;f\Bigg(\frac{\Big(x-t_b^{(q)}\Big)\times l_q^{(c)}}{\delta_q^{(t)}}+s_b^{(q)}\Bigg)\;\text{.}
\end{aligned}
\]

As we can see, the target-\(x\) is transformed in three (four) steps:

\begin{itemize}
\item
  \begin{enumerate}
  \def\labelenumi{\arabic{enumi}.}
  \tightlist
  \item
    -- Translation to \(0\) (\(x-t_b^{(q)}\)),
  \end{enumerate}
\item
  2.(a) -- Normalization according to the target-interval/-extent (\(\times{\delta_q^{(t)}}^{-1}\)),
\item
  2.(b) -- Scaling to the source-interval/-extent (\(\times l_q^{(c)}\)),
\item
  \begin{enumerate}
  \def\labelenumi{\arabic{enumi}.}
  \setcounter{enumi}{2}
  \tightlist
  \item
    -- Translation to offset in source-interval/-extent (\(+s_b^{(q)}\)).
  \end{enumerate}
\end{itemize}

So, really, the time warping happens in step \textbf{2}, when \(x\) is scaled \emph{within} the source-interval: Currently, this is a linear function of the form \(y=a*x+b\), where \(b=0\), as the function goes through the origin, thanks to steps \(1\)/\(3\). \(a\) results from putting the source- and target-extents into relation, so we have no additional parameters to learn.

The linear function hence maps a target-\(x\) to a source-\(x\), where the support is \(\Big[0\;,\;t_e^{(q)}-t_b^{(q)}\Big]\), and the co-domain is \(\Big[0\;,\;s_e^{(q)}-s_b^{(q)}\Big]\). The linear function is also strictly monotonically increasing, which is a requirement as we will show later. If the function had saddle points, i.e., a slope of \(0\), then this would mean no time warping is applied. A negative slope would result in negative time warping (i.e., while the target-\(x\) moves forward in time, the source-\(x\) would move backward), which is prohibited.

If we want to introduce arbitrary scaling functions, even non-linear functions, then it is helpful to introduce a few additional notions first:

\[
\begin{aligned}
    x'=&\;\Big(x-t_b^{(q)}\Big)\;\text{, pre-translation of}\;x\text{,}
    \\[1ex]
    s^{(\text{lin})}(x')=&\;x'\times\frac{\delta_q^{(s)}}{\delta_q^{(t)}}\;\text{, linear scaling used thus far,}
    \\[1ex]
    s^{(\text{nonlin})}(x')=&\;\dots\;\text{(some non-linear and strictly increasing scaling function),}
    \\[1ex]
    s'^{(\text{nonlin})}(x')=&\;\delta_q^{(s)}\times\frac{s^{(\text{nonlin})}(x')-s^{(\text{nonlin})}(0)}{s^{(\text{nonlin})}\Big(\underbrace{t_e^{(q)}-t_b^{(q)}}_{\delta_q^{(t)}}\Big)-s^{(\text{nonlin})}(0)}\;\text{, constrained version.}
\end{aligned}
\]

Any arbitrary scaling function \(s\) needs to be constrained, such that its constrained version \(s'\) satisfies the same properties as does the linear scaling function:

\begin{itemize}
\tightlist
\item
  The function is continuous,
\item
  same support and co-domain as the linear function,
\item
  goes through \([0,0]\) and \(\Big[t_e^{(q)}-t_b^{(q)},\;s_e^{(q)}-s_b^{(q)}\Big]\),
\item
  is strictly increasing.
\end{itemize}

Technically, it is not required that any scaling function is strictly increasing, only that \(s(0)\prec s(x),\,\forall\,x>0\), so the function may have saddle-points. Otherwise, its constrained version would potentially divide by \(0\).

If we now introduce arbitrary scaling functions (here using a constrained non-linear example) into our submodel formulation, we get:

\[
\begin{aligned}
    m_q^{c}(f,x',\dots)=&\;f\Big(s'^{(\text{nonlin})}\Big(x-t_b^{(q)}\Big)+s_b^{(q)}\Big)
\end{aligned}\;\text{.}
\]

Note that, while omitted here, any scaling function would be passed the pre-translated \(x\) as \(x'=x-t_b^{(q)}\), as well as all extent-parameters, namely \(s_b^{(q)},s_e^{(q)},t_b^{(q)},t_e^{(q)}\). If the scaling function introduces parameters that should be optimized for, e.g., if it is a polynomial, then its gradient must reflect that, and hence the submodel's gradient does change, too. It would be perfectly fine to use a different kind of scaling function in each interval.

Example with a non-linear polynomial, where we want to capture the portion of the function in the interval \([0,1]\), which is then scaled to the current target-extent of \(\Big[0,\delta_q^{(t)}\Big]\):

\[
\begin{aligned}
    s(x')=&\;\frac{mx'}{\delta_q^{(t)}}+\frac{nx'^2}{\delta_q^{(t)}}\;\text{,}
    \\[1ex]
    =&\;\frac{x'\times(m+nx')}{\delta_q^{(t)}}\;\text{, with constrained version}\;s'\text{:}
    \\[1ex]
    s'(x')=&\;\delta_q^{(s)}\times\frac{s\big(x'\big)-s(0)}{s\Big(\delta_q^{(t)}\Big)-s(0)}\;\text{,}
    \\[1ex]
    =&\;\frac{\delta_q^{(s)}\times s\big(x'\big)}{s\Big(\delta_q^{(t)}\Big)}\;\text{, as}\;s(0)=0\text{, fully expanded as}
    \\[1ex]
    =&\;\frac{\delta_q^{(s)}\times x'\times\big(m+nx'\big)}{\delta_q^{(t)}\times\Big(m+n\times\delta_q^{(t)}\Big)}\;\text{.}
\end{aligned}
\]

It is important that the scaling function covers some characteristic interval \([0,k]\), where \(k>0\). The value of \(k\) is not important, only that the interval starts with \(0\). In other words, any non-constrained scaling function \(s(x)\) needs to satisfy these criteria:

\begin{itemize}
\tightlist
\item
  The function is continuous,
\item
  has a support of \([0,k]\), where \(k>0\),
\item
  is monotonically increasing and satisifies \(s(0)\prec s(x),\,\forall\,x>0\).
\end{itemize}

The co-domain may arbitrary, as the constraining function re-maps it to \(\Big[0,\delta_q^{(s)}\Big]\).

In the previous example, we chose a scaling function that has two learnable parameters, \(m,n\). Also, the parameter \(\delta_q^{(s)}\equiv\,l_q^{(c)}\), which depends on other parameters of the model. So, the complete gradient requires deriving for these three kinds of parameters, and we will omit the set of partial derivatives for the last one here:

\[
\begin{aligned}
    \nabla\,s'(x')=&\;\Bigg[\frac{\partial\,s}{\partial\,m}\;,\;\frac{\partial\,s}{\partial\,n}\;,\;\Bigg(\frac{\partial\,s}{\partial\,\delta_q^{(s)}}\Bigg)\Bigg]\;\text{,}
    \\[1ex]
    =&\;\Bigg[\frac{\delta_q^{(s)}\times n\times x'\times\Big(\delta_q^{(t)}-x'\Big)}{\delta_q^{(t)}\times\Big(m+n\times\delta_q^{(t)}\Big)^2}\;,\;\frac{\delta_q^{(s)}\times m\times x'\times\Big(x'-\delta_q^{(t)}\Big)}{\delta_q^{(t)}\times\Big(m+n\times\delta_q^{(t)}\Big)^2}\;,\;\dots\Bigg]\;\text{.}
\end{aligned}
\]

\hypertarget{optimization-goal-ii}{%
\subsection{Optimization goal II}\label{optimization-goal-ii}}

This goal is actually specific to the current case, but in general the purpose is to a) verify, that the choice of data- and regularization-loss is apt for matching data that is expected to be somewhat similar to a given pattern, and b) to find upper- and/or lower-bounds (best/worst possible match) such that we can use the model for \textbf{scoring} when fitting to real-world data.

An important step is to validate whether the selected sub-models and scores are suitable to detect and score the reference pattern. This can be verified by setting the query signal equal to the reference signal, i.e., \(\text{WC} = \text{WP}\), and then having the multilevel model use starting intervals that are different from the reference intervals. Everything else is identical to the first optimization problem. If the selected data- and regularization-losses were chosen appropriately, the MLM converges to the reference intervals, i.e., each sub-model matches best what it should match. Ideally, this test is repeated a number of times, with randomly initialized query intervals.

\hypertarget{scoring-a-loss}{%
\subsubsection{Scoring a loss}\label{scoring-a-loss}}

If we have the upper and lower bound for some loss \(\mathcal{L}(\cdot)\), we can define a \textbf{score-operator}, \(\mathsf{S}^{\oplus}\), as:

\[
\begin{aligned}
  \mathcal{L}(\cdot)\;\dots&\;\text{data- or regularization-loss, and also its gradient}\;\nabla\,\mathcal{L}(\cdot)\text{,}
  \\[1ex]
  \omega,\beta^{(l)},\beta^{(u)}\;\dots&\;\text{weight, lower- and upper bound for loss}\;\mathcal{L}(\cdot)\text{,}
  \\[1ex]
  \mathsf{S}^{\oplus}(\cdot)=&\;\omega\Bigg(1-\frac{\mathcal{L}(\cdot)-\beta^{(l)}}{\beta^{(u)}-\beta^{(l)}}\Bigg)\;\mapsto[+0,1]\;\text{,}
  \\[0ex]
  &\;\text{score-operator, where larger scores are better}\;(\oplus)\text{,}
  \\[1ex]
  \nabla\,\mathsf{S}^{\oplus}\big(\mathcal{L}_q(\cdot)\big)=&\;\frac{\omega\nabla\,\mathcal{L}_q(\cdot)}{\beta^{(l)}-\beta^{(u)}}\;\text{, gradient of this operator.}
\end{aligned}
\]

The score of some loss really just is a scaling of it into the range \([0,1]\), where, depending on the definition, zero or one is the best (in the previous definition \(1\) would be the best, hence we wrote \(\mathsf{S}^{\oplus}\) -- otherwise, we would use the notion \(\mathsf{S}^{\ominus}\) for an operator where lower scores are better). If we choose not to negate the score by leaving out the \(1-\), the gradient is almost the same, except for that we subtract the lower bound from the upper bound, i.e., \(\beta^{(u)}-\beta^{(l)}\).

\textbf{Scoring?} --- The goal is to align (to time-warp) some data to a given pattern. If we attempt to align the pattern to itself, using some unfavorable starting values, we would expect it to perfectly match when the optimization is done. But a match depends on how the loss is defined, and what is its weight. If a single variable in a single interval is captured by a singular loss, and that loss is appropriate for matching the variable, then it should ideally be \(0\) (or whatever is the lowest possible value). Recall that \textbf{SRBTW} is just a model for aligning two data series, it is up to the user chose an appropriate loss that is \emph{adequate to pursue the user's objectives}. Such an adequate loss is one that reaches its global minimum in case of fitting the pattern to itself.

If two or more losses are combined by, e.g., a linear scalarizer, and if these losses pursue objectives that are somewhat conflicting with each other (not necessarily orthogonal), then the lower bound for that aggregation is larger than or equal to the sum of the singular minimal possible losses. It may not be trivial to determine the new lower bound for an aggregated loss. We can find this lower bound through optimization, and this is described in subsection \ref{ssec:match_wp}. In practice, the lowest possible aggregated loss lies on a \textbf{Pareto}-boundary and there may be many optimal losses on that boundary.

As for the upper bound of an aggregated loss, we too will have a Pareto-boundary. However, we cannot determine it computationally. The compromise is to sum up the worst possible losses of all singular losses. There are some losses that have a finite upper bound, and some that only have \(\infty\) as upper bound. However, our goal is to come up with losses that have a well-defined upper bound for our problem, and these are presented in the next sub-section.

\hypertarget{aggregation-of-scores}{%
\subsubsection{\texorpdfstring{Aggregation of scores\label{ssec:score_agg}}{Aggregation of scores}}\label{aggregation-of-scores}}

The mean aggregation operator for scores calculates a weighted sum. However, the average score is achieved simply by dividing by the number of scores. We suggest a slightly altered version which normalizes w.r.t. the \emph{magnitude} of the weights, and not just their amount.

\[
\begin{aligned}
  \mathsf{A}\Big(\bm{\mathsf{S}}^{\oplus},\bm{\omega}\Big)=&\;\Bigg(\bigg[\prod_{i=1}^{\left\lVert\,\bm{\mathsf{S}}^{\oplus}\,\right\rVert}\,1+\bm{\omega}_i\bm{\mathsf{S}}^{\oplus}_i(\cdot)\bigg]-1\Bigg)\times\underbrace{\Bigg(\bigg[\prod_{i=1}^{\left\lVert\,\bm{\omega}\,\right\rVert}\,1+\bm{\omega}_i\bigg]-1\Bigg)^{-1}}_{k}\;\text{,}
  \\[0ex]
  &\;\text{where}\;0<\bm{\omega}_i\leq1\;\land\;\left\lVert\,\bm{\mathsf{S}}^{\oplus}\,\right\rVert=\left\lVert\,\bm{\omega}\,\right\rVert\;\text{, i.e., one weight for each score,}
  \\[1ex]
  =&\;k\times\bigg[\prod_{i=1}^{\left\lVert\,\bm{\mathsf{S}}^{\oplus}\,\right\rVert}\,1+\bm{\omega}_i\bm{\mathsf{S}}^{\oplus}_i(\cdot)\bigg]-k\;\text{, with gradient}
  \\[1ex]
  \nabla\,\mathsf{A}\Big(\bm{\mathsf{S}}^{\oplus},\bm{\omega}\Big)=&\;k\times\Bigg[\sum_{i=1}^{\left\lVert\,\bm{\mathsf{S}}^{\oplus}\,\right\rVert}\,\bm{\omega}_i\nabla\,\bm{\mathsf{S}}^{\oplus}_i(\cdot)\times\bigg[\prod_{j=1}^{\left\lVert\,\bm{\mathsf{S}}^{\oplus}\,\right\rVert}\,1+\bm{\omega}_j\bm{\mathsf{S}}^{\oplus}_j(\cdot),\;\forall\,j\neq i\,\bigg]\Bigg]\;\text{.}
\end{aligned}
\]

\hypertarget{losses-with-finite-upper-bounds}{%
\subsubsection{Losses with finite upper bounds}\label{losses-with-finite-upper-bounds}}

Early on I had defined losses (back then called metrics) that all had one thing in common: their finite upper bound was either global, or we could pose our problem in a way such that a local finite upper bound could be specified. For many such metrics, this was done by scaling the data within an interval into the \textbf{unit-square}. In the following, we present some of these metrics, redefined as losses, together with their upper and lower local bounds. In general, we exploit the fact that each interval of the Warping Pattern is delimited by some rectangular area, and within that area, which not necessarily needs to be a (unit-)square, many definable losses have a clear upper bound.

Some of the scores will exploit the fact that in time-warping, the degrees of freedom are on the time- or x-axis only. When we define an optimization problem using SRBTW, we usually know the lower- and upper bounds of the y-axis. Even later for optimization goal III, when we allow the y-values to adjust, this will happen within \textbf{previously specified box-bounds}. Some of the losses described below will exploit these facts.

\hypertarget{area-between-curves}{%
\paragraph{Area between curves}\label{area-between-curves}}

This was one of the very first metrics I had implemented. The area between curves (ABC) within an interval has been proven to be quite useful. It is easy to compute, both for the continuous and discrete case. The discrete case is equivalent to the mean absolute error and approaches the continuous loss with sufficiently large number of samples. The ABC exploits the lower and upper bound for the y-axis (known a priori), and scales the rectangle of an interval using the current Warping Pattern's interval's extent (\(\delta_q\) below), such that the loss is always computed in a rectangle that always has the same size.

\[
\begin{aligned}
  y_q^{\min},y_q^{\max}\;\dots&\;\text{lower and upper bounds for the }y\text{-axis in the }q\text{-th WP-interval,}
  \\[1ex]
  \mathbf{x}_q^{(\text{WP})}\;\dots&\;\text{its support (going from}\;\bm{\theta}_{q+1}\;\text{to}\;\bm{\theta}_{q}\text{),}
  \\[1ex]
  \delta_q=&\;\bm{\theta}_{q+1}-\bm{\theta}_{q}\equiv\max{\Big(\mathbf{x}_q^{(\text{WP})}\Big)}-\min{\Big(\mathbf{x}_q^{(\text{WP})}\Big)}\;\text{, the extent of the support,}
  \\[1ex]
  \rho_q=&\;\delta_q\times\big(y_q^{\max}-y_q^{\min}\big)\;\text{, the total area of that interval,}
  \\[1ex]
  \rho_q^{(\text{WP})}=&\;\int\displaylimits_{\mathbf{x}_q^{(\text{WP})}}\,r(x)-y_q^{\min}\,dx\;\text{, note that}\;\forall\,x\to y_q^{\min}\leq r(x)\leq y_q^{\max}\text{, or}
  \\[1ex]
  \rho_q^{(\text{WP}_d)}=&\;\frac{1}{N}\sum_{i=1}^{N}\,r\Big(\mathbf{x}_{q,i}^{(\text{WP})}\Big)\;\text{,}
  \\[0ex]
  &\;\text{the (discrete) area between the reference signal (WP) and the x-/time-axis (at}\;y_q^{\min}\text{),}
  \\[1ex]
  \beta_l=&\;0\;\text{, the metric's lower bound, which is always}\;0\text{,}
  \\[1ex]
  \beta_u=&\;\max{\bigg(\rho_q^{\big(\text{WP}_{(d)}\big)},\rho_q-\rho_q^{\big(\text{WP}_{(d)}\big)}\bigg)}\;\text{, the metric's upper bound,}
  \\[1em]
  \mathcal{L}_q^{(\text{ABC}_c)}=&\;\beta_u^{-1}\times\int\displaylimits_{\mathbf{x}_q^{(\text{WP})}}\,\left\lVert\,r(x_q)-m_q^{(c)}\Big(x_q\Big)\,\right\rVert\;\text{, the continous ABC-loss,}
  \\[1ex]
  \mathcal{L}_q^{(\text{ABC}_d)}=&\;\frac{1}{N\times\beta_u}\sum_{i=1}^{N}\,\left\lVert\,r\Big(\mathbf{x}_{q,i}^{(\text{WP})}\Big)-m_q^{(c)}\Big(\mathbf{x}_{q,i}^{(\text{WC})}\Big)\,\right\rVert\;\text{, (discrete case), with gradient}
  \\[1ex]
  \nabla\,\mathcal{L}_q^{(\text{ABC}_d)}=&\;\frac{1}{N\times\beta_u}\sum_{i=1}^{N}\,\frac{\nabla\,m_q^{(c)}\Big(\mathbf{x}_{q,i}^{(\text{WC})}\Big)\times\bigg(m_q^{(c)}\Big(\mathbf{x}_{q,i}^{(\text{WC})}\Big)-r\Big(\mathbf{x}_{q,i}^{(\text{WP})}\Big)\bigg)}{\left\lVert\,m_q^{(c)}\Big(\mathbf{x}_{q,i}^{(\text{WC})}\Big)-r\Big(\mathbf{x}_{q,i}^{(\text{WP})}\Big)\,\right\rVert}\;\text{.}
\end{aligned}
\]

Note that no scaling back of the intervals is required, because the discrete version samples equidistantly with equally many samples (\(N\)) from both, the WP and the WC, within their current intervals. In the continuous case, the supports of WP and WC are identical, as that is what SRBTW does (move a candidate-interval to the pattern-interval). The discrete ABC loss is also known as Sum of absolute differences (SAD), \(L_1\)-norm, \emph{Manhattan}- or Taxicab-norm.

\hypertarget{residual-sum-of-squares}{%
\paragraph{Residual sum of squares}\label{residual-sum-of-squares}}

RSS, also known as sum of squared distance (SSD) and probably as many more synonyms, is usually always the first loss I use, as it is computationally cheap, easy to compute, fast and robust (delivers good results in many cases). Until now I had not been thinking about upper boundaries for it. However, with regard to that, it should be very similar to the ABC-loss, as the RSS' behavior is asymptotic. When, as in our case, we know the maximum y-extent, together with the (constant) support.

\[
\begin{aligned}
  \rho_q=&\;\delta_q\times\big(y_q^{\max}-y_q^{\min}\big)^2\;\text{, the maximum possible loss in the WP-interval's area,}
  \\[1ex]
  \rho_q^{(\text{WP})}=&\;\int\displaylimits_{\mathbf{x}_q^{(\text{WP})}}\,\Big(r(x)-y_q^{\min}\Big)^2\;\text{, note that}\;\forall\,x\to y_q^{\min}\leq r(x)\leq y_q^{\max}\text{, or}
  \\[1ex]
  \rho_q^{(\text{WP}_d)}=&\;\frac{1}{N}\sum_{i=1}^{N}\,\bigg(r\Big(\mathbf{x}_{q,i}^{(\text{WP})}\Big)-y_q^{\min}\bigg)^2\;\text{,}
  \\[0ex]
  &\;\text{the (discrete) squared loss between the reference signal (WP) and the x-/time-axis (at}\;y_q^{\min}\text{),}
  \\[1ex]
  \beta_l,\beta_u\;\dots&\;\text{(no change),}
  \\[1em]
  \mathcal{L}_q^{(\text{RSS}_c)}=&\;\beta_u^{-1}\times\int\displaylimits_{\mathbf{x}_q^{(\text{WP})}}\,\bigg(r(x_q)-m_q^{(c)}\Big(x_q\Big)\bigg)^2\;\text{, the continous RSS-loss,}
  \\[1ex]
  \mathcal{L}_q^{(\text{RSS}_d)}=&\;\frac{1}{N\times\beta_u}\sum_{i=1}^{N}\,\bigg(r\Big(\mathbf{x}_{q,i}^{(\text{WP})}\Big)-m_q^{(c)}\Big(\mathbf{x}_{q,i}^{(\text{WC})}\Big)\bigg)^2\;\text{, (discrete case), with gradient}
  \\[1ex]
  \nabla\,\mathcal{L}_q^{(\text{RSS}_d)}=&\;-\frac{2}{N\times\beta_u}\sum_{i=1}^{N}\,\nabla\,m_q^{(c)}\Big(\mathbf{x}_{q,i}^{(\text{WC})}\Big)\times\bigg(r\Big(\mathbf{x}_{q,i}^{(\text{WP})}\Big)-m_q^{(c)}\Big(\mathbf{x}_{q,i}^{(\text{WC})}\Big)\bigg)\;\text{.}
\end{aligned}
\]

The difference really just is to replace the norm with a square, and the then-simplified gradient.

\hypertarget{more-simple-metrics}{%
\paragraph{More simple metrics}\label{more-simple-metrics}}

For more simple metrics see \footnote{\url{https://web.archive.org/web/20200701202102/https://numerics.mathdotnet.com/distance.html}}. However, we will focus on metrics that we have upper bounds for.

\[
\begin{aligned}
  {\text{SSD/RSS}} &: (x, y) \mapsto \|x-y\|_2^2 = \langle x-y, x-y\rangle = \sum_{i=1}^{n} (x_i-y_i)^2
  \\[0ex]
  {\text{MAE}} &: (x, y) \mapsto \frac{d_{\mathbf{SAD}}}{n} = \frac{\|x-y\|_1}{n} = \frac{1}{n}\sum_{i=1}^{n} \left\lVert\,x_i-y_i\,\right\rVert
  \\[0ex]
  {\text{MSE}} &: (x, y) \mapsto \frac{d_{\mathbf{SSD}}}{n} = \frac{\|x-y\|_2^2}{n} = \frac{1}{n}\sum_{i=1}^{n} (x_i-y_i)^2
  \\[0ex]
  {\text{Euclidean}} &: (x, y) \mapsto \|x-y\|_2 = \sqrt{d_{\mathbf{SSD}}} = \sqrt{\sum_{i=1}^{n} (x_i-y_i)^2}
  \\[0ex]
  {\infty\text{/Chebyshev}} &: (x, y) \mapsto \|x-y\|_\infty = \lim_{p \rightarrow \infty}\bigg(\sum_{i=1}^{n} |x_i-y_i|^p\bigg)^\frac{1}{p} = \max_{i} \left\lVert\,x_i-y_i\,\right\rVert
  \\[0ex]
  {\text{p/Minkowski}} &: (x, y) \mapsto \|x-y\|_p = \bigg(\sum_{i=1}^{n} \left\lVert\,x_i-y_i\,\right\rVert^p\bigg)^\frac{1}{p}
  \\[0ex]
  {\text{Canberra}} &: (x, y) \mapsto \sum_{i=1}^{n} \frac{\left\lVert\,x_i-y_i\,\right\rVert}{\left\lVert\,x_i\,\right\rVert+\left\lVert\,y_i\,\right\rVert}
  \\[0ex]
  {\text{cosine}} &: (x, y) \mapsto 1-\frac{\langle x, y\rangle}{\|x\|_2\|y\|_2} = 1-\frac{\sum_{i=1}^{n} x_i y_i}{\sqrt{\sum_{i=1}^{n} x_i^2}\sqrt{\sum_{i=1}^{n} y_i^2}}
  \\[0ex]
  {\text{Pearson}} &: (x, y) \mapsto 1 - \text{Corr}(x, y)
\end{aligned}
\]

I have copy-pasted these here, and we should check later for which of these there may be an upper bound.

\hypertarget{correlation-between-curves-1}{%
\paragraph{Correlation between curves}\label{correlation-between-curves-1}}

This is also one I have used early on, and also as a score. The correlation is an intuitively elegant loss for estimating whether two curves have a similar shape. If one curve was to follow the shape of the other, the correlation would be high, a strongly negative correlation would mean one curve exhibits the opposite behavior. The correlation is between \([1,-1]\). While in some cases we would not be interested in negative correlation (and hence treat any correlation \(\leq 0\) as no correlation), I think that in many cases it is better for the \emph{training} of our model to not do that. The correlation is a good candidate for combination with other losses that, for example, measure the difference in magnitude (like ABC or RSS), as it itself does not care for that. The correlation is a \textbf{ratio-loss}, and does not depend on local finite upper bounds, it thus has \textbf{global} bounds.

When I say correlation, I mean \emph{Pearson}'s \textbf{sample}-correlation, just to be clear. In the following, we define the discrete version only, and we do it in a way that this loss is strictly positive, like the others presented so far.

\[
\begin{aligned}
  \operatorname{cor}(\mathbf{a},\mathbf{b})=&\;\frac{\sum_{i=1}^N\,(\mathbf{a}_i-\overline{\mathbf{a}})\times(\mathbf{b}_i-\overline{\mathbf{b}})}{\sqrt{\sum_{i=1}^{N}\,(\mathbf{a}_i-\overline{\mathbf{a}})^2}\times\sqrt{\sum_{i=1}^{N}\,(\mathbf{b}_i-\overline{\mathbf{b}})^2}}
  \\[1ex]
  =&\;\operatorname{cor}\bigg(r\Big(\mathbf{x}_{q}^{(\text{WP})}\Big)\;,\;m_q^{(c)}\Big(\mathbf{x}_{q}^{(\text{WC})}\Big)\bigg)
  \\[1ex]
  =&\;\operatorname{cor}\Bigg(\frac{\sum_{i=1}^N\,\bigg(r\Big(\mathbf{x}_{q,i}^{(\text{WP})}\Big)-\overline{r\Big(\mathbf{x}_{q}^{(\text{WP})}\Big)}\bigg)\times\bigg(m_q^{(c)}\Big(\mathbf{x}_{q,i}^{(\text{WC})}\Big)-\overline{m_q^{(c)}\Big(\mathbf{x}_{q}^{(\text{WC})}\Big)}\bigg)}{\sqrt{\sum_{i=1}^{N}\,\bigg(r\Big(\mathbf{x}_{q,i}^{(\text{WP})}\Big)-\overline{r\Big(\mathbf{x}_{q}^{(\text{WP})}\Big)}\bigg)^2}\times\sqrt{\sum_{i=1}^{N}\,\bigg(m_q^{(c)}\Big(\mathbf{x}_{q,i}^{(\text{WC})}\Big)-\overline{m_q^{(c)}\Big(\mathbf{x}_{q}^{(\text{WC})}\Big)}\bigg)^2}}\Bigg)\;\text{.}
\end{aligned}
\]

The gradient of this, using some substitutions for the \emph{constant} Warping Pattern, is:

\[
\begin{aligned}
  \bm{\tau}=&\;\sum_{i=1}^N\,\bigg(r\Big(\mathbf{x}_{q,i}^{(\text{WP})}\Big)-\overline{r\Big(\mathbf{x}_{q}^{(\text{WP})}\Big)}\bigg)\;\text{, and}
  \\[1ex]
  \bm{\gamma}_i=&\;m_q^{(c)}\Big(\mathbf{x}_{q,i}^{(\text{WC})}\Big)\;\text{,}
  \\[1ex]
  \bm{\hat{\gamma}}_i=&\;\nabla\,m_q^{(c)}\Big(\mathbf{x}_{q,i}^{(\text{WC})}\Big)\;\text{,}
  \\[1ex]
  \rho=&\;\overline{m_q^{(c)}\Big(\mathbf{x}_{q}^{(\text{WC})}\Big)}
  \\[1ex]
  \hat{\rho}=&\;\overline{\nabla\,m_q^{(c)}\Big(\mathbf{x}_{q}^{(\text{WC})}\Big)}
  \\[1ex]
  \nabla\,\operatorname{cor}(\cdot)=&\;\Bigg(\frac{\Big[\sum_{i=1}^N\,\bm{\tau}\times\big(\bm{\hat{\gamma}}_i-\hat{\rho}\big)\Big]\times\sqrt{\sum_{i=1}^N(\bm{\gamma}_i-\rho)^2}}{\sqrt{(\bm{\tau}^\top\bm{\tau})}\times\Big[\sum_{i=1}^N\,\big(\bm{\gamma}_i-\rho\big)\times\big(\bm{\hat{\gamma}}_i-\bm{\hat{\gamma}}_i\times\hat{\rho}\big)\Big]}\Bigg)\;\text{.}
\end{aligned}
\]

The actual loss shall map to \(\mathcal{L}^{(\text{Corr})}\to[0+,2]\), so we define it as:

\[
\begin{aligned}
  \mathcal{L}_q^{(\text{RSS})}=&\;1-\operatorname{cor}(\cdot)\;\text{, with gradient}
  \\[1ex]
  \nabla\,\mathcal{L}_q^{(\text{RSS})}=&\;-\nabla\operatorname{cor}(\cdot)\;\text{.}
\end{aligned}
\]

\hypertarget{ratio-between-curves-arc-lengths}{%
\paragraph{Ratio between curves' arc-lengths}\label{ratio-between-curves-arc-lengths}}

Comparing the arc-lengths of two curves is the next loss that we can define such that it has a global upper bound. This is true for any loss that is a \textbf{ratio}. First, we obtain two measurements, one from the WP and one from the WC (here: the arc-length). We then set both measurements into a relation that we subtract from \(1\). We make sure that each ratio ideally is \(1\), and goes to \(0\) the worse it gets. This is guaranteed by always dividing the minimum of both measurements by the maximum.

The arc-length can be obtained continuously and discretely.

\[
\begin{aligned}
  \operatorname{arclen}_{(c)} f(x)=&\;\int\displaylimits_{\mathbf{x}_q^{(\text{WP})}}\,\sqrt{1+\bigg[\frac{\partial}{\partial\,x}f(x)\bigg]^2}dx\;\text{,}
  \\[0ex]
  &\;\text{continuous arc-length of some function}\;f\;\text{with support}\;\mathbf{x}_q^{(\text{WP})}\text{,}
  \\[1ex]
  \operatorname{arclen}_{(d)} f(x)=&\;\lim_{N\to\infty}\,\sum_{i=1}^N\,\left\lVert\,f(\delta_i)-f(\delta_{i-1})\,\right\rVert\;\text{, where}
  \\[0ex]
  \delta_i=&\;\min{\big(\operatorname{supp}(f)\big)}+i\times\Big(\max{\big(\operatorname{supp}(f)\big)}-\min{\big(\operatorname{supp}(f)\big)}\Big)\times N^{-1}\;\text{, with gradient}
  \\[1ex]
  \nabla\,\operatorname{arclen}_{(d)}f(x)=&\;\lim_{N\to\infty}\,\sum_{i=1}^N\,\frac{\Big(f(\delta_i)-f(\delta_{i-1})\Big)\times\Big(f'(\delta_i)-f'(\delta_{i-1})\Big)}{\left\lVert\,f(\delta_i)-f(\delta_{i-1})\,\right\rVert}\;\text{.}
\end{aligned}
\]

The arc-length is the first \textbf{ratio}-loss, so we will define a generic ratio-loss where we can plug in any other loss that measures the same property of two signals and puts them into relation. The actual loss is then defined as:

\[
\begin{aligned}
  \mathcal{L}^{(\text{ratio})}=&\;1-\frac{\min{\Big(\mathcal{L}^{(\text{WP})}(\cdot)}\;,\;\mathcal{L}^{(\text{WC})}(\cdot)\Big)}{\max{\Big(\mathcal{L}^{(\text{WP})}(\cdot)}\;,\;\mathcal{L}^{(\text{WC})}(\cdot)\Big)}\;\mapsto[+0,1]\;\text{,}
  \\[0ex]
  &\;\text{where}\;\mathcal{L}^{(\text{WP})}(\cdot)>0\;,\;\mathcal{L}^{(\text{WC})}(\cdot)>0\;\text{measure some property/loss of the}
  \\[0ex]
  &\;\text{Warping Pattern and -Candidate, with gradient}
  \\[1ex]
  \nabla\,\mathcal{L}_q^{(\text{ratio})}=&\;\mathcal{H}\Big(\mathcal{L}^{(\text{WP})}(\cdot)-\mathcal{L}^{(\text{WC})}(\cdot)\Big)\times\frac{\mathcal{L}^{(\text{WC})}(\cdot)\times \nabla\,\mathcal{L}^{(\text{WP})}(\cdot)-\mathcal{L}^{(\text{WP})}(\cdot)\times \nabla\,\mathcal{L}^{(\text{WC})}(\cdot)}{\Big[\mathcal{L}^{(\text{WP})}(\cdot)\Big]^2}
  \\[1ex]
  &\;+\mathcal{H}\Big(\mathcal{L}^{(\text{WC})}(\cdot)-\mathcal{L}^{(\text{WP})}(\cdot)\Big)\times\frac{\mathcal{L}^{(\text{WP})}(\cdot)\times\nabla\,\mathcal{L}^{(\text{WC})}(\cdot)-\mathcal{L}^{(\text{WC})}(\cdot)\times \nabla\,\mathcal{L}^{(\text{WP})}(\cdot)}{\Big[\mathcal{L}^{(\text{WC})}(\cdot)\Big]^2}\;\text{.}
\end{aligned}
\]

We used the Heaviside step function to check the condition \(\mathcal{L}^{(\text{WP})}(\cdot)>\mathcal{L}^{(\text{WC})}(\cdot)\) (and vice versa). It is now straightforward to see that we can effortlessly plug in the arc-length operator defined previously. We have defined the ratio-loss such that a perfect ratio of \(\frac{1}{1}\) results in a loss of \(0\). As the denominator approaches \(\infty\), the loss approaches \(1\), i.e., \(\lim_{\text{denom}\to\infty}\mathcal{L}^{(\text{ratio})}=1\). The ratio-loss is \textbf{only defined for strictly positive} losses.

Here is a short list with more properties that may be useful for a ratio comparison:

\begin{itemize}
\tightlist
\item
  Standard deviation, Variance
\item
  Signal measures: RMS, Kurtosis, Impulse-factor, Peak (Peak for for strictly positive signals)
\item
  Any other unary aggregation that is positive (or an absolute value can be obtained), for example \(\min,\max\), mean, median etc.
\item
  discrete/differential Entropy (but we can define binary entropy-based losses, such as KL- and JSD-divergence, joint-/cross-entropy, mutual information etc.)
\end{itemize}

\hypertarget{jensenshannon-divergence}{%
\paragraph{Jensen--Shannon divergence}\label{jensenshannon-divergence}}

Most scores so far were \emph{low-level} or mid-level, meaning that they capture usually just a single simple property of a signal. Especially losses that discard many properties are low-level, such as \(\min,\max\) for example. The area between curves, correlation and arc-length ratio already are somewhat mid-level, as they aggregate more than one (or a few) properties into a loss. Early on I had tried to use entropy-based losses and -divergences. Coming across the Kullback-Leibler divergence I early saw the need for such losses that have an upper bound. The (symmetric) KL-divergence only has application specific bounds, and I cannot see how to develop one in our case. However, the Jenson--Shannon divergence will compare two probability distributions, and does so with a \textbf{global upper bound} of \(\ln{(2)}\) (or \(1\) when using the base-2 logarithm). Since it is a divergence, the global lower bound is \(0\) in any case, i.e., no divergence.

Using entropy-based measures technically requires the variables compared to be probability distributions (or functions thereof). This means that the sum/integral of these must be \(1\). As of our examples, we have some modeled some variables using probability densities, and some are just metrics (univariate events). Strictly speaking, such measurements are not applicable at all for the latter type. Also, for the former type, we would strictly speaking still need to consider any variable in question as a partial probability distribution, as we are looking at a specific interval of it. In practice so far, however, it turns out that for both kind of variables we can obtain excellent results by ignoring the nature of the variable and simply treating it as discrete or continuous probability distribution, as long as we ascertain that for any realization it is strictly positive and that it sums up/integrates to \(1\). That means that even if we look at a sub-support (interval) of a variable, we treat that current interval as the entire probability distribution of the variable in question. This approach is probably further justified by how SRBTW works: Given a Warping Pattern, we sub-divide it using come \textbf{constant} boundaries into \textbf{independent} intervals. For the scope of any single loss, the other intervals do not exist, so that we can view the captured variable as the entire probability distribution of it.

I had previously also used entropy-based measures with success, esp.~the mutual information (MI). It may be worth looking more into such available measures. However, those can be quite similar, and with the JSD we already have a well-working measure. The JSD has been used previously in Generative Adversarial Nets to minimize the divergence between probability distributions (Goodfellow et al. 2014).

In the following, we will first define how to normalize two continuous or discrete signals (note that mixed mode is not supported) such that they sum up/integrate to \(1\). Then we will define the Jensen--Shannon divergence for either type.

\[
\begin{aligned}
  \mathsf{N}^{(\text{c})}(f,\mathbf{x}^{(\text{supp})},a)=&\;f(a)\times\bigg[\int\displaylimits_{\mathbf{x}^{(\text{supp})}}\,f(x)\bigg]^{-1}\;\text{, where}\;\min{\big(\mathbf{x}^{(\text{supp})}\big)}\leq a\leq\max{\big(\mathbf{x}^{(\text{supp})}\big)}\;\land\;\forall\,f(a)\geq 0\;\text{,}
  \\[1ex]
  &\;\text{(continuous to-probability transform operator),}
  \\[1ex]
  \mathsf{N}^{(\text{d})}(\mathbf{y})=&\;\frac{\mathbf{y}-\min{(\mathbf{y})}+1}{\sum_{i=1}^{\left\lVert\,\mathbf{y}\,\right\rVert}\,\big[\mathbf{y}-\min{(\mathbf{y})}+1\big]_i}\;\text{,}
  \\[1ex]
  &\;\text{(discrete to-probability transform operator).}
\end{aligned}
\]

In practice, so far, we get more robust results with discrete vectors (typically using between \(1e3\) to \(1e4\) samples), even when those were sampled from the same functions we would use in the continuous case. Therefore, we will (for now at least) only show the discrete JSD.

\[
\begin{aligned}
  \mathbf{y}^{(\text{n})}=&\;\mathsf{N}^{(\text{d})}(\mathbf{y})\;\text{, and}
  \\[0ex]
  \mathbf{\hat{y}}^{(\text{n})}=&\;\mathsf{N}^{(\text{d})}(\mathbf{\hat{y}})\;\text{, to-probability normalized samples from two signals,}
  \\[1ex]
  \operatorname{KL}_{(\text{d,symm})}(\mathbf{y}\;\|\;\mathbf{\hat{y}})=&\;\sum_{i=1}^{\left\lVert\,\mathbf{y}\,\right\rVert}\,\mathbf{y}_i\times\log{\bigg(\frac{\mathbf{y}_i}{\mathbf{\hat{y}}_i}\bigg)}+\mathbf{\hat{y}}_i\times\log{\bigg(\frac{\mathbf{\hat{y}}_i}{\mathbf{y}_i}\bigg)}
  \\[1ex]
  \operatorname{JSD}_{(\text{d})}(\mathbf{y}\;\|\;\mathbf{\hat{y}})=&\;\frac{1}{2}\operatorname{KL}_{(\text{d,symm})}\Big(\mathbf{y}\;\|\;\frac{\mathbf{y}+\mathbf{\hat{y}}}{2}\Big)+\frac{1}{2}\operatorname{KL}_{(\text{d,symm})}\Big(\mathbf{\hat{y}}\;\|\;\frac{\mathbf{y}+\mathbf{\hat{y}}}{2}\Big)\;\text{.}
\end{aligned}
\]

\hypertarget{match-wp-against-itself}{%
\subsubsection{\texorpdfstring{Match WP against itself\label{ssec:match_wp}}{Match WP against itself}}\label{match-wp-against-itself}}

The WP can be matched best by the reference pattern itself, i.e., nothing matches the reference better than it itself. A side-effect of this optimization goal thus is to obtain the maximum possible score, given the selected sub-models and their losses. The maximum score may then later be used to normalize scores when fitting the then \emph{calibrated} MLM to actual project data, as now we have upper bounds for all scores.

As I wrote before, let's make some tests where we use random values as starting parameters and check whether the model can converge to the pattern. We make tests using equidistantly-spaced and randomized lengths.

\begin{Shaded}
\begin{Highlighting}[]
\NormalTok{cow\_og2\_test1p }\OtherTok{\textless{}{-}} \FunctionTok{loadResultsOrCompute}\NormalTok{(}\AttributeTok{file =} \StringTok{"../results/cow\_og2\_test1p.rds"}\NormalTok{, }\AttributeTok{computeExpr =}\NormalTok{ \{}
  \FunctionTok{set.seed}\NormalTok{(}\DecValTok{1337}\NormalTok{)}
\NormalTok{  useTests }\OtherTok{\textless{}{-}} \FunctionTok{list}\NormalTok{(}
    \FunctionTok{list}\NormalTok{(}\AttributeTok{p =} \FunctionTok{rep}\NormalTok{(}\DecValTok{1}\SpecialCharTok{/}\DecValTok{20}\NormalTok{, }\DecValTok{20}\NormalTok{), }\AttributeTok{n =} \StringTok{"eq{-}20"}\NormalTok{),}
    \FunctionTok{list}\NormalTok{(}\AttributeTok{p =} \FunctionTok{rep}\NormalTok{(}\DecValTok{1}\SpecialCharTok{/}\DecValTok{40}\NormalTok{, }\DecValTok{40}\NormalTok{), }\AttributeTok{n =} \StringTok{"eq{-}40"}\NormalTok{),}
    \FunctionTok{list}\NormalTok{(}\AttributeTok{p =} \FunctionTok{runif}\NormalTok{(}\DecValTok{20}\NormalTok{), }\AttributeTok{n =} \StringTok{"unif{-}20"}\NormalTok{),}
    \FunctionTok{list}\NormalTok{(}\AttributeTok{p =} \FunctionTok{runif}\NormalTok{(}\DecValTok{40}\NormalTok{), }\AttributeTok{n =} \StringTok{"unif{-}40"}\NormalTok{),}
    \FunctionTok{list}\NormalTok{(}\AttributeTok{p =} \FunctionTok{rbeta}\NormalTok{(}\DecValTok{20}\NormalTok{, }\DecValTok{1}\NormalTok{, }\DecValTok{2}\NormalTok{), }\AttributeTok{n =} \StringTok{"beta\_1\_2{-}20"}\NormalTok{),}
    \FunctionTok{list}\NormalTok{(}\AttributeTok{p =} \FunctionTok{rbeta}\NormalTok{(}\DecValTok{40}\NormalTok{, }\DecValTok{1}\NormalTok{, }\DecValTok{2}\NormalTok{), }\AttributeTok{n =} \StringTok{"beta\_1\_2{-}40"}\NormalTok{),}
    
    \FunctionTok{list}\NormalTok{(}\AttributeTok{p =} \FunctionTok{rep}\NormalTok{(}\DecValTok{1}\SpecialCharTok{/}\DecValTok{30}\NormalTok{, }\DecValTok{30}\NormalTok{), }\AttributeTok{b =} \FunctionTok{runif}\NormalTok{(}\DecValTok{1}\NormalTok{), }\AttributeTok{n =} \StringTok{"eq\_ob{-}30"}\NormalTok{),}
    \FunctionTok{list}\NormalTok{(}\AttributeTok{p =} \FunctionTok{rep}\NormalTok{(}\DecValTok{1}\SpecialCharTok{/}\DecValTok{20}\NormalTok{, }\DecValTok{20}\NormalTok{), }\AttributeTok{e =} \FunctionTok{runif}\NormalTok{(}\DecValTok{1}\NormalTok{), }\AttributeTok{n =} \StringTok{"eq\_oe{-}20"}\NormalTok{),}
    \FunctionTok{list}\NormalTok{(}\AttributeTok{p =} \FunctionTok{rep}\NormalTok{(}\DecValTok{1}\SpecialCharTok{/}\DecValTok{20}\NormalTok{, }\DecValTok{20}\NormalTok{), }\AttributeTok{b =} \FunctionTok{runif}\NormalTok{(}\DecValTok{1}\NormalTok{, }\AttributeTok{max =}\NormalTok{ .}\DecValTok{5}\NormalTok{), }\AttributeTok{e =} \FunctionTok{runif}\NormalTok{(}\DecValTok{1}\NormalTok{, }\AttributeTok{min =}\NormalTok{ .}\DecValTok{5}\NormalTok{), }\AttributeTok{n =} \StringTok{"eq\_ob\_oe{-}20"}\NormalTok{),}
    \FunctionTok{list}\NormalTok{(}\AttributeTok{p =} \FunctionTok{rbeta}\NormalTok{(}\DecValTok{40}\NormalTok{, }\DecValTok{1}\NormalTok{, }\DecValTok{2}\NormalTok{), }\AttributeTok{b =} \FunctionTok{runif}\NormalTok{(}\DecValTok{1}\NormalTok{, }\AttributeTok{max =}\NormalTok{ .}\DecValTok{5}\NormalTok{), }\AttributeTok{e =} \FunctionTok{runif}\NormalTok{(}\DecValTok{1}\NormalTok{, }\AttributeTok{min =}\NormalTok{ .}\DecValTok{5}\NormalTok{), }\AttributeTok{n =} \StringTok{"beta\_1\_2\_ob\_oe{-}40"}\NormalTok{)}
\NormalTok{  )}
  
  \ControlFlowTok{for}\NormalTok{ (i }\ControlFlowTok{in} \FunctionTok{seq\_len}\NormalTok{(}\AttributeTok{length.out =} \FunctionTok{length}\NormalTok{(useTests))) \{}
\NormalTok{    u }\OtherTok{\textless{}{-}}\NormalTok{ useTests[[i]]}
    \FunctionTok{print}\NormalTok{(}\FunctionTok{paste}\NormalTok{(}\StringTok{"Calculating"}\NormalTok{, u}\SpecialCharTok{$}\NormalTok{n, }\StringTok{"..."}\NormalTok{))}
\NormalTok{    params }\OtherTok{\textless{}{-}} \FunctionTok{list}\NormalTok{(}\AttributeTok{vtl =}\NormalTok{ u}\SpecialCharTok{$}\NormalTok{p)}
\NormalTok{    params[}\StringTok{"b"}\NormalTok{] }\OtherTok{\textless{}{-}} \ControlFlowTok{if}\NormalTok{ (}\StringTok{"b"} \SpecialCharTok{\%in\%} \FunctionTok{names}\NormalTok{(u)) u}\SpecialCharTok{$}\NormalTok{b }\ControlFlowTok{else} \DecValTok{0}
\NormalTok{    params[}\StringTok{"e"}\NormalTok{] }\OtherTok{\textless{}{-}} \ControlFlowTok{if}\NormalTok{ (}\StringTok{"e"} \SpecialCharTok{\%in\%} \FunctionTok{names}\NormalTok{(u)) u}\SpecialCharTok{$}\NormalTok{e }\ControlFlowTok{else} \DecValTok{1}
\NormalTok{    params[}\StringTok{"ob"}\NormalTok{] }\OtherTok{\textless{}{-}} \StringTok{"b"} \SpecialCharTok{\%in\%} \FunctionTok{names}\NormalTok{(u)}
\NormalTok{    params[}\StringTok{"oe"}\NormalTok{] }\OtherTok{\textless{}{-}} \StringTok{"e"} \SpecialCharTok{\%in\%} \FunctionTok{names}\NormalTok{(u)}
    
\NormalTok{    u}\SpecialCharTok{$}\NormalTok{r }\OtherTok{\textless{}{-}} \FunctionTok{do.call}\NormalTok{(}\AttributeTok{what =}\NormalTok{ cow\_og2, }\AttributeTok{args =}\NormalTok{ params)}
\NormalTok{    useTests[[i]] }\OtherTok{\textless{}{-}}\NormalTok{ u}
\NormalTok{  \}}
  
\NormalTok{  useTests}
\NormalTok{\})}
\end{Highlighting}
\end{Shaded}

\includegraphics{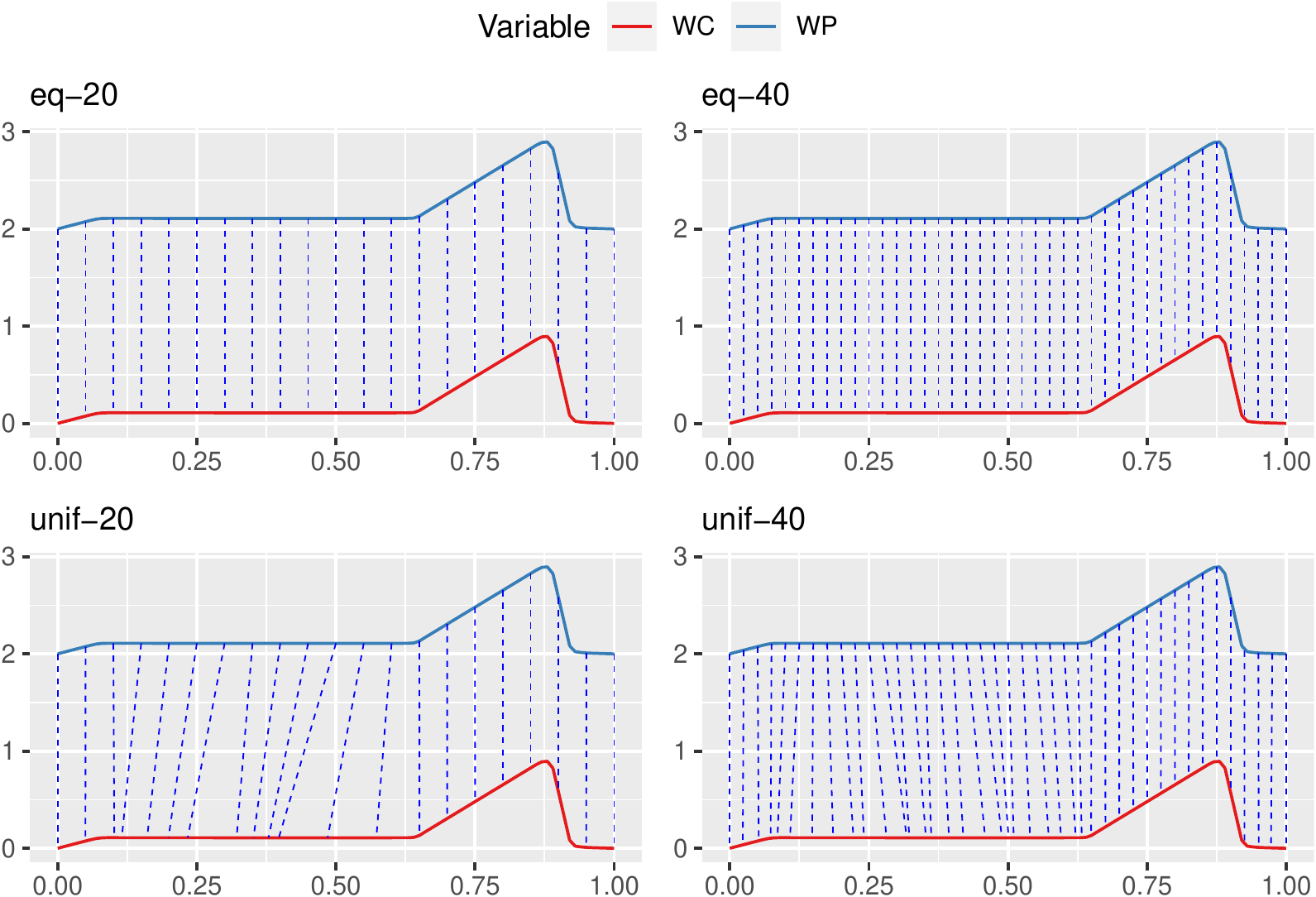} \includegraphics{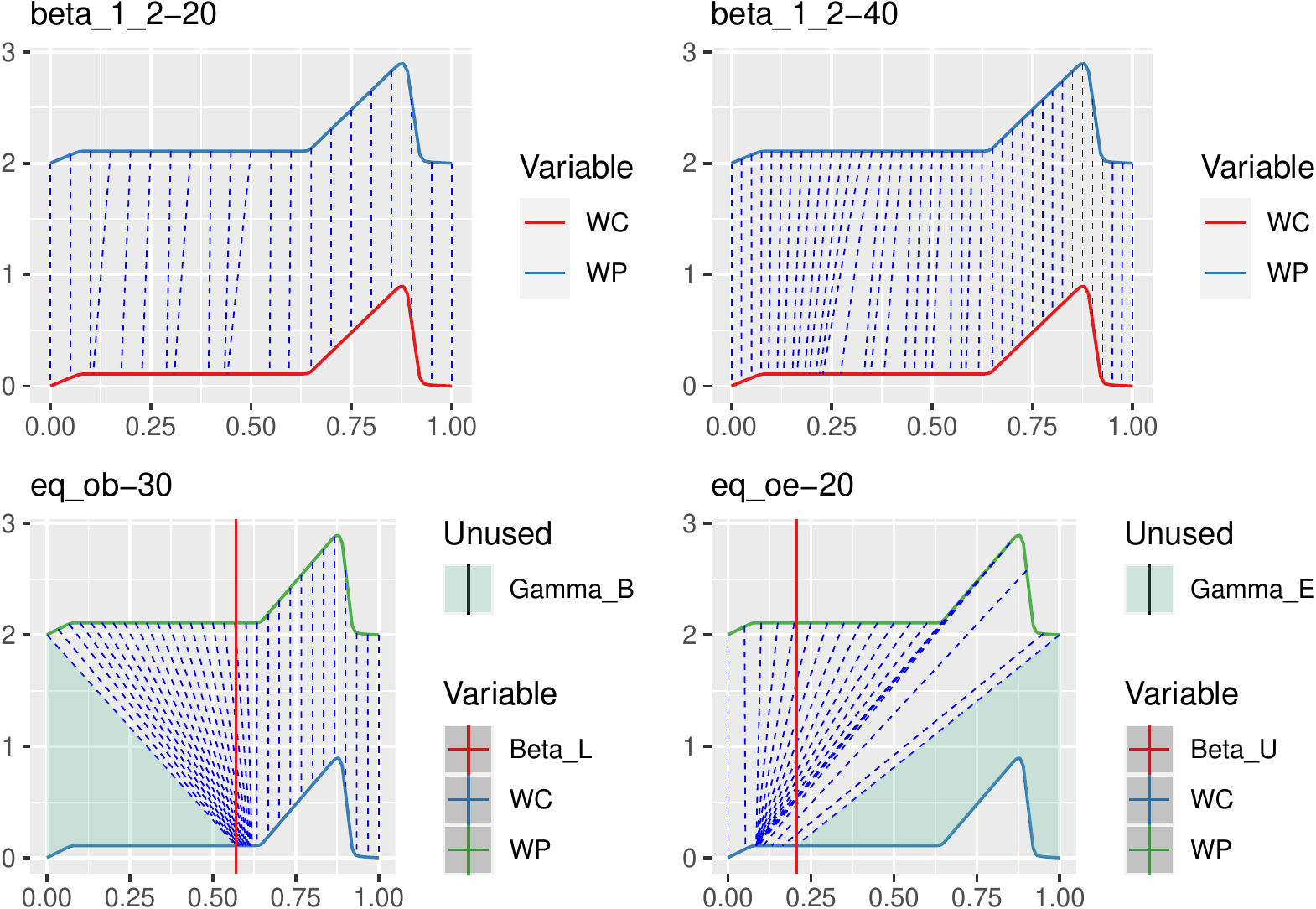} \includegraphics{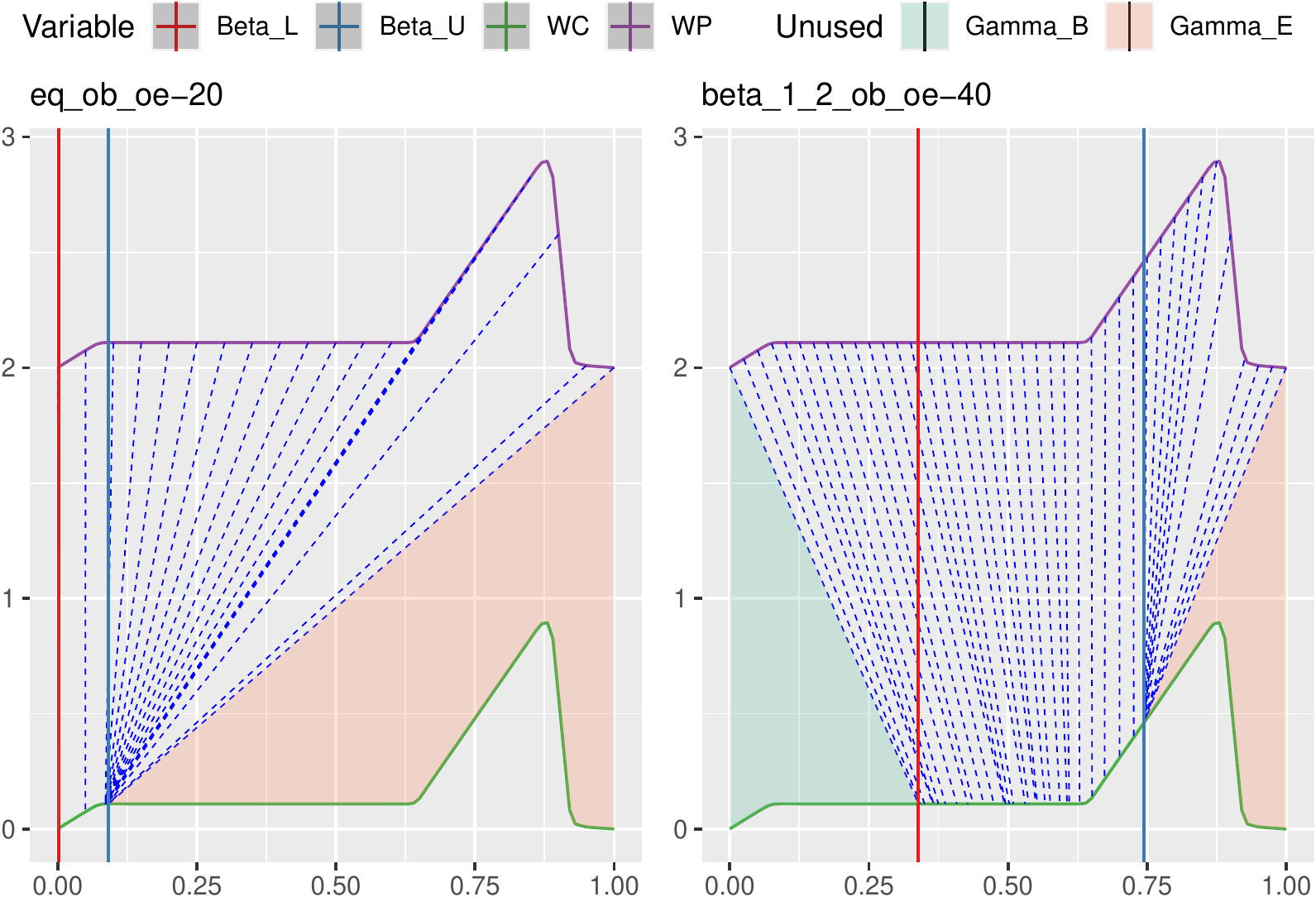}

In all of the above cases, I used the discrete RSS data-loss, which itself uses a constant number of samples of \(250\) per interval. Also I did \textbf{not} use any regularization. This means, for example, that even a very short interval will be sampled from \(250\) times, and if, for example, the WP interval was in the Long Stretch, and the WC interval is too, but it is much shorter, the loss will be equally low because of this setup. Especially when using open begin and/or end, regularization is strongly suggested

The first two equidistantly-spaced examples resulted in perfect fits, no real surprise there. Examples 3 and 4 are quite OK, but they have a couple of intervals that are very short. Within the Long Stretch, it is virtually impossible for any of the intervals there to find a good match, as the function looks the same to the left and to the right (i.e., the gradient). Where the WP is not flat, the intervals actually find a good match again (i.e., we observe vertical lines in the non-flat phases). Examples 5 and 6 show the same behavior (\texttt{beta\_1\_2-20} and \texttt{beta\_1\_2-40}). It is worth noting that these two achieved quite a low error, which again is a hint at how ill-posed our setup here was (see previous paragraph).

In examples 7 and 8 we use an open begin and end, respectively. In example 7, we see a perfect beginning with the \emph{Fire Drill} phase. Before that, almost all intervals have a loss of zero (except for the first \textasciitilde3-4). This is because of the peculiarity I mentioned earlier (\(250\) samples no matter the length). The begin was set initially at 0.5202 and did not move much -- I suppose it was just too far off for the optimization to find the real beginning \textbf{without} regularization. I suppose the same is true for example number 8, although that the starting-parameter for the end was 0.6569. However, that starting point is already slightly in the Fire Drill phase, and mapped initially to the last interval, which is flat. So the optimization probably moved it farther to the left to decrease that loss, as it would have gotten much stronger otherwise.

If we divide the loss by the number of intervals, we can actually make apples-to-apples comparisons.

\begin{table}

\caption{\label{tab:cow-og2-losses}Overview of incurred alignment losses over various setups.}
\centering
\begin{tabular}[t]{lllrrrrllrr}
\toprule
Type & ob & oe & b\_init & e\_init & b & e & loss & meanLoss & optFn & optGr\\
\midrule
eq-20 & - & - & 0.000 & 1.000 & 0.000 & 1.000 & 0.00e+00 & 0.00e+00 & 21 & 21\\
eq-40 & - & - & 0.000 & 1.000 & 0.000 & 1.000 & 0.00e+00 & 0.00e+00 & 21 & 21\\
unif-20 & - & - & 0.000 & 1.000 & 0.000 & 1.000 & 6.74e-04 & 3.37e-05 & 123 & 123\\
unif-40 & - & - & 0.000 & 1.000 & 0.000 & 1.000 & 3.49e-03 & 8.72e-05 & 117 & 117\\
beta\_1\_2-20 & - & - & 0.000 & 1.000 & 0.000 & 1.000 & 5.07e-05 & 2.53e-06 & 123 & 123\\
\addlinespace
beta\_1\_2-40 & - & - & 0.000 & 1.000 & 0.000 & 1.000 & 1.35e-03 & 3.37e-05 & 123 & 123\\
eq\_ob-30 & Y & - & 0.520 & 1.000 & 0.570 & 1.000 & 1.14e+00 & 3.80e-02 & 135 & 135\\
eq\_oe-20 & - & Y & 0.000 & 0.657 & 0.000 & 0.205 & 5.78e+00 & 2.89e-01 & 45 & 45\\
eq\_ob\_oe-20 & Y & Y & 0.225 & 0.700 & 0.002 & 0.091 & 5.78e+00 & 2.89e-01 & 99 & 99\\
beta\_1\_2\_ob\_oe-40 & Y & Y & 0.289 & 0.533 & 0.339 & 0.744 & 5.69e+00 & 1.42e-01 & 97 & 97\\
\bottomrule
\end{tabular}
\end{table}

The type in above table reveals some details: `eq' means starting lengths of equal length, `unif' mean they wer uniformly sampled from \([0,1]\) (with out scaling them back), `beta' means they were drawn from a Beta-distribution with \(\alpha=1\) and \(\beta=2\) (also, no scaling back). `ob' and `oe' in the name indicate that the begin and/or end was open. Finally, the number of intervals is at the end of the type.

\textbf{Conclusion}: Yes, if the loss is well defined and mechanism for taking interval-lengths, begin and end into account (extreme value regularization), then this will absolutely work well.

\hypertarget{optimization-goal-iii}{%
\subsection{Optimization goal III}\label{optimization-goal-iii}}

sr-BTW and sr-BAW allow to interchange Warping Pattern and Warping Candidate. Recall that only the former is treated as constant and needs to be used wholly (the WC only needs to be used wholly in closed/closed matching).

\begin{itemize}
\tightlist
\item
  \textbf{Pattern as WP, Data as WC}: This is the ``ordinary'' approach. It allows us to set boundaries on the WP that we think make sense for the pattern, it will thus preserve what was in the original intervals and apply warping to them. This approach is suitable for checking how well the data matches the pattern. If the goal were to adjust the WP, we could still use this approach and eventually inverse the warping. I.e., if we know how to warp the Data to the Pattern, we know how to warp the Pattern to the Data. That is not implemented yet but should be straightforward.
\item
  \textbf{Data as WP, Pattern AS WC}: This means that we still can choose boundaries, but they must be chosen on the data. This makes less sense as we do not know how to subdivide the data into intervals, so generally, you probably just want equidistantly-spaced boundaries. So, if we would choose our original Fire Drill AP as WC, we could not preserve the original segmentation into the 4 characteristic intervals, as the optimization would highly likely choose a different segmentation. It would be possible to restore the original segmentation after optimization, however. This would only work if we choose to have 4 intervals.
\end{itemize}

A note on highly multivariate optimization: In section \ref{ssec:srbtaw} we introduce the Multilevel-model (\textbf{MLM}) that supports arbitrary many variables for either, Warping Pattern or Warping Candidate. The algorithms introduced here, namely \texttt{sr-BTW} and \texttt{sr-BAW}, are only univariate: they support exactly one variable for the WP and one for the WC. In the suggested MLM, we will use one instance of an algorithm per \textbf{unique} pair of variables. All of the instances \textbf{share the same} set of parameters, so that the MLM appears as a single model. This is a prerequisite for multivariate fitting and illustrated with a practical example of ours: Suppose we want to adjust the \textbf{A}-variable of our Fire Drill AP, such that the distance between it and the A-variables of, say, 5 other projects is minimized. The MLM will have 5 instances of the algorithm that share the same parameters during optimization. The data from the 5 projects will be adjusted simultaneously. Once we are finished, we inverse the learned parameters so that we can modify the Fire Drill AP's A-variable. If we had used different parameters for each algorithm, there would be no clear path for the inversion of the parameters.

\hypertarget{ideas}{%
\subsubsection{Ideas}\label{ideas}}

\begin{itemize}
\tightlist
\item
  Use the AP as WP with the 4 intervals that we have defined and warp all data to it. If we know how to warp the data to the AP, we can inverse the results to apply warping to the pattern instead (see above). This approach also allows us to give low or no weight/loss to, e.g., the Long Stretch phase. If we were to do that, the other phases would extent into the Long Stretch, leaving is with a gap in each project that we can afterwards analyze. This will help us to learn about the Long Stretch phases!
\item
  Use the AP as query instead. This means that the project data is constant, and that we subdivide it into some intervals by some strategy or just equidistantly-spaced. Using, e.g., AIC/BIC, we find the right amount of boundaries. This approach cannot consider the original intervals of the AP as we defined them, as the optimization chooses new intervals (also, it would most likely end up using a different number of intervals). This approach could result in a vastly different pattern that does not preserve too many shapes from the original pattern, but it may still worth testing. In the case where we use exactly 4 intervals, it would be possible to restore the original segmentation.
\item
  A third approach: Start with an empty Warping Candidate: A straight horizontal line located at \(0.5\) for each variable and variably many equidistantly-spaced intervals (use information criterion to find best number). This is the approach where we throw away our definition of the Fire Drill AP and create one that is solely based on data. The resulting warped candidate is a new pattern then.
\end{itemize}

\hypertarget{approach-1}{%
\subsubsection{Approach}\label{approach-1}}

The approach that I suggest is to learn a \textbf{piece-wise linear} function, that will be overlaid with the signal. For each interval, we will learn a separate slope. However, instead of learning a slope directly, we will learn the terminal y-translation at the end of the interval (with the initial translation being \(0\)). A linear function is given by \(y=ax+b\), and we will only learn \(a\). \(b\), the intercept, is for each interval implicitly assumed to be \(0\), but in reality is the sum of all preceding terminal y-translations. This is required because the resulting piece-wise linear function needs to be continuous (otherwise, when adding it to the signal, the signal would also become discontinuous). The first interval has no such preceding y-translation, so we will learn one using the additional parameter, \(v\), that represents this. The advantage is, that we can simply \textbf{add} this piece-wise function to each interval, in our case, to each \(m_q^c(x,\dots)\). We can think of \(v\) and the ordered set of terminal translations as a set of perpendicular vectors, so that for each \(q\)-th interval, we simply sum up \(v\) and the set up to and including the preceding interval, to get the intercept.

I call this approach \textbf{self-regularizing boundary amplitude warping} (sr-BAW). In the style of the sub-model definition for sr-BTW, \(m_q^c\), we will call the sub-models for amplitude warping \(n_q^c\).

\hypertarget{sub-model-formulation-1}{%
\subsubsection{Sub-Model formulation}\label{sub-model-formulation-1}}

In each interval, we will call this function \(t_q\):

\[
\begin{aligned}
  \bm{\vartheta}^{(y)}\;\dots&\;\text{ordered set of terminal}\;y\text{-translations for each interval,}
  \\[1ex]
  a_q=&\;\frac{\bm{\vartheta}_q^{(y)}}{\iota_q}\;\text{, the slope for the}\;q\text{-th interval,}
  \\[0ex]
  &\;\text{where}\;\iota_q\;\text{is the length of the}\;q\text{-th \textbf{target}-interval,}
  \\[1ex]
  v\;\dots&\;\text{the global translation for the first interval,}
  \\[1ex]
  \phi_q^{(y)}=&\;\begin{cases}
    0,&\text{if}\;q=1,
    \\
    \sum_{i=1}^{q-1}\,\bm{\vartheta}^{(y)}_i,&\text{otherwise,}
  \end{cases}
  \\[0ex]
  &\;\text{(the intercept for the}\;q\text{-th interval), and}
  \\[1ex]
  \rho_q\;\dots&\;\text{starting}\;x\text{-offset of the}\;q\text{-th \textbf{target}-interval,}
  \\[1ex]
  t_q(x,\dots)=&\;a_q(x-\rho_q)+v+\phi_q^{(y)}\;\text{, linear amplitude warping in}\;q\text{-th interval.}
\end{aligned}
\]

The amount of parameters for the y-translations is \(1\) plus the amount of intervals (or the same as the amount of boundaries). For SRBTW, we had previously used the notion of \(\phi_q\) to calculate the offset for each \(q\)-th interval. We use a similar notion here, \(\phi_q^{(y)}\), to denote the intercept of each \(q\)-th interval.

\hypertarget{self-regularization}{%
\paragraph{Self-regularization}\label{self-regularization}}

Again, we would like this approach to be \textbf{self-regularizing}. That means, that any resulting final y-coordinate should never be less or more than the defined box-bounds. In DTW, the degrees of freedom are limited to warping time, i.e., the x-axis. Time can (linearly) compress and stretch, but never is the \textbf{amplitude} of the signal altered. This would not make much sense for DTW, either, as it is a purely discrete algorithm, that manipulates \textbf{each} discrete point of a query-signal. If DTW would alter the magnitude, the result would simply be the reference signal. Boundary Time Warping, on the other hand, instead of manipulating points, manipulates everything between the points (intervals) through linear affine transformation. This is the reason why the degrees of freedom are limited to time in DTW. That also means, for DTW, that time warping only works, when the query signal is within the bounds (y) of the reference signal. For that reason, time warping problems often have the application-specific y-limits that directly come from the reference signal, which is known a priori. It is also a possibility that the bounds are specified by the user.

In SRBAW it is possible to specify limits on a per-interval basis, so we will have two vectors, one with lower- and one with upper bounds for each interval. It would also be possible to specify the limits as a function (per interval). But at this point it is not required so we will not make that extra effort.

It is important to notice that in Boundary Time/Amplitude Warping, we cannot (must not) make any assumptions about the signal within an interval. For example, we cannot know its interval-global minimum or maximum. At the same time, we want to enforce box-bounds on the warped signal. The solution is to enforce these only after the warped signal has been computed. That means, that we need to wrap the warping function. We do this below by wrapping \(n_q(\dots)\) into the \(\min\)/\(\max\) expression \(n_q^c(\dots)\).

\[
\begin{aligned}
  f(x,\dots)\;\dots&\;\text{some signal function, could be, e.g.,}\;m_q^c(x,\dots)\text{,}
  \\[1ex]
  n_q(x,\dots)=&\;f(x,\dots)+\overbrace{a_q(x-\rho_q)+v+\phi_q^{(y)}}^{t_q(x,\dots)}\;\text{, sub-model that adds amplitude-warping to}\;f(x,\dots)\text{,}
  \\[1ex]
  \bm{\lambda}^{(\text{ymin})},\bm{\lambda}^{(\text{ymax})}\;\dots&\;\text{vectors with lower and upper}\;y\text{-bounds for each interval}
  \\[0ex]
  &\;\text{(which could also be functions of}\;q\;\text{and/or}\;x\text{),}
  \\[1ex]
  n_q^c(x,\dots)=&\;\max{\bigg(\bm{\lambda}_q^{(\text{ymin})}, \;\min{\Big(n_q(x,\dots),\bm{\lambda}_q^{(\text{ymax})}\Big)}\bigg)}\;\text{, the}\;\min\text{/}\max\text{-constrained sub-model.}
\end{aligned}
\]

\hypertarget{gradient-of-the-sub-model}{%
\paragraph{Gradient of the sub-model}\label{gradient-of-the-sub-model}}

We need to pay attention when defining the gradient, for cases when some parameters depend on parameters of the underlying model. For example, \(l_q^{(c)}\) depends on \(\forall\,l\in\bm{\vartheta}^{(l)}\) as well as \(b,e\) (if open begin and/or end), while \(\phi_q\) depends on \(\forall\,l_i^{(c)},i<q\). If we derive w.r.t. any of the sr-BAW parameters, i.e., \(v\) or any of the terminal translations in \(\bm{\vartheta}^{(y)}\), then this will eliminate any \(m_q^c\), but we need to be careful when deriving for parameters that affect any \(l_q^{(c)}\).

\hypertarget{testing}{%
\paragraph{Testing}\label{testing}}

We will define some piece-wise linear function and use the signal of the \textbf{A}-variable, then we warp the amplitude and see how that looks.

\begin{Shaded}
\begin{Highlighting}[]
\FunctionTok{print}\NormalTok{(baw\_v)}
\end{Highlighting}
\end{Shaded}

\begin{verbatim}
## [1] 0.05
\end{verbatim}

\begin{Shaded}
\begin{Highlighting}[]
\FunctionTok{print}\NormalTok{(theta\_baw)}
\end{Highlighting}
\end{Shaded}

\begin{verbatim}
## [1] 0.0 0.1 0.3 0.5 0.8 1.0
\end{verbatim}

\begin{Shaded}
\begin{Highlighting}[]
\FunctionTok{print}\NormalTok{(vartheta\_l\_baw)}
\end{Highlighting}
\end{Shaded}

\begin{verbatim}
## [1] 0.2 0.2 0.2 0.2 0.2
\end{verbatim}

\begin{Shaded}
\begin{Highlighting}[]
\FunctionTok{print}\NormalTok{(baw\_vartheta\_y)}
\end{Highlighting}
\end{Shaded}

\begin{verbatim}
## [1]  0.025  0.100 -0.100  0.050  0.025
\end{verbatim}

\begin{Shaded}
\begin{Highlighting}[]
\FunctionTok{ggplot}\NormalTok{() }\SpecialCharTok{+} \FunctionTok{xlim}\NormalTok{(}\DecValTok{0}\NormalTok{, }\DecValTok{1}\NormalTok{) }\SpecialCharTok{+} \FunctionTok{stat\_function}\NormalTok{(}\AttributeTok{fun =}\NormalTok{ f, }\AttributeTok{color =} \StringTok{"red"}\NormalTok{) }\SpecialCharTok{+} \FunctionTok{stat\_function}\NormalTok{(}\AttributeTok{fun =}\NormalTok{ baw,}
  \AttributeTok{color =} \StringTok{"black"}\NormalTok{) }\SpecialCharTok{+} \FunctionTok{stat\_function}\NormalTok{(}\AttributeTok{fun =}\NormalTok{ n, }\AttributeTok{color =} \StringTok{"blue"}\NormalTok{)}
\end{Highlighting}
\end{Shaded}

\includegraphics{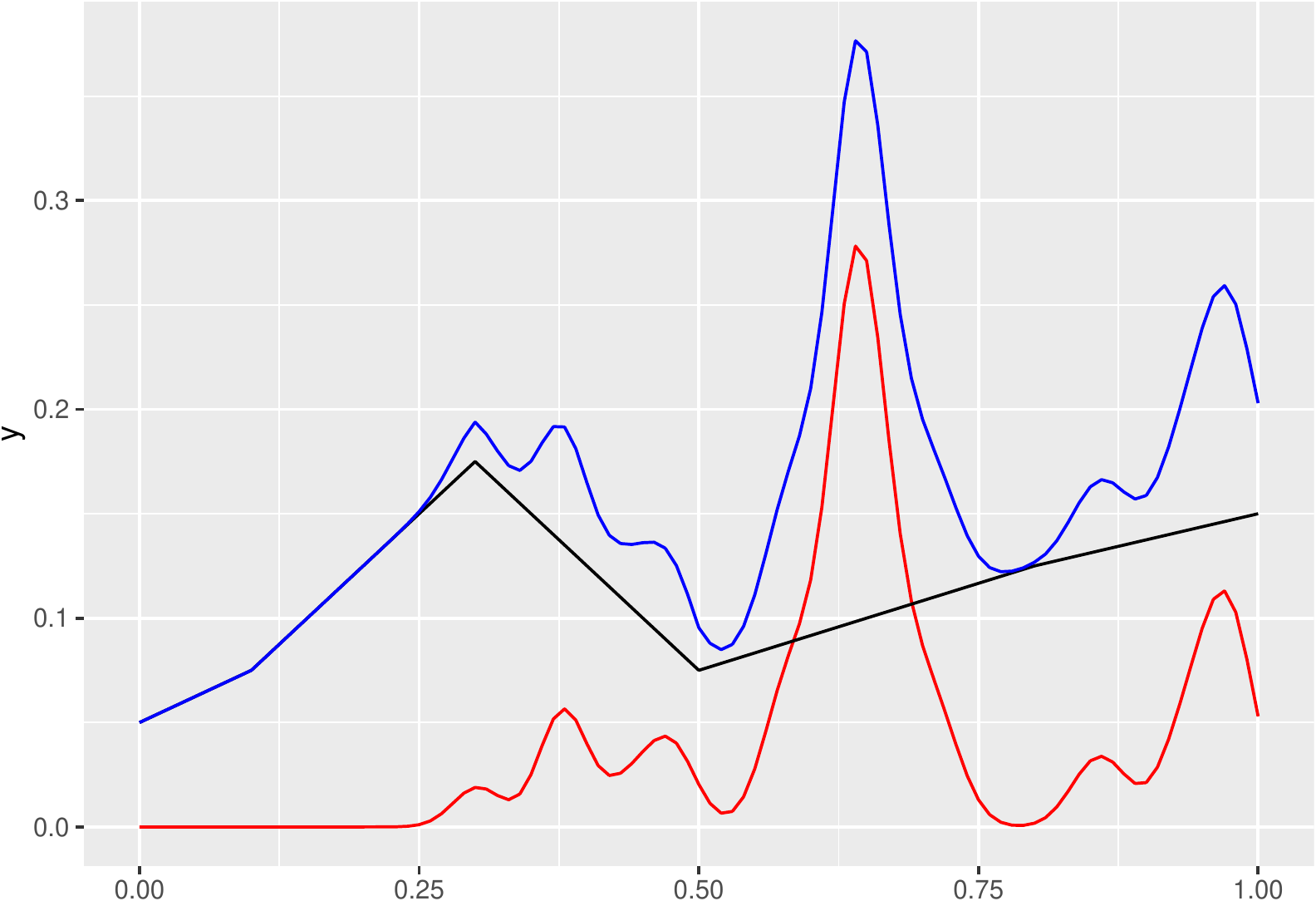}

What we see in the above plot is the following:

\begin{itemize}
\tightlist
\item
  \textbf{Red}: The \textbf{A}-variable, as it appeared in the dataset,
\item
  \textbf{Black}: the piece-wise linear function, with initial offset \texttt{baw\_v} (0.05), and per-interval terminal translations of 0.025, 0.1, -0.1, 0.05, 0.025. The interval-boundaries were equidistantly placed at 0, 0.1, 0.3, 0.5, 0.8, 1,
\item
  \textbf{Blue}: Red+Black, i.e., the piece-wise linear function applied to the signal from the dataset.
\end{itemize}

\hypertarget{architecture-of-sr-btw-sr-baw}{%
\subsubsection{Architecture of sr-BTW + sr-BAW}\label{architecture-of-sr-btw-sr-baw}}

So far, I had designed sr-BTW as an R6-class, and also a sub-model of it. This design was not made to be very performant, but to be very expressive and verbose, and I tried to be as close to the terms I use in this notebook as possible. BTW and BAW are stand-alone models, and they can be used independently from each other. If used together, the most straightforward way, however, is to \textbf{wrap} BTW in BAW, as I had shown it using the \(n_q^c\)-model, where the signal is actually \(m_q^c\). This is how we are going to do it here for optimization goal III.

\hypertarget{current-architecture}{%
\paragraph{Current architecture}\label{current-architecture}}

Our architecture here so far has a few drawbacks and does not exploit some commonalities. However, for the scope of the development of sr-BTW and sr-BAW, we will continue with this approach and just organically implement things as we need them. I plan to release everything as an R-package on CRAN, and for that purpose I will make a new repository where I implement it using the new architecture described below. I already took some minor parts of the new architecture and implemented it.

\hypertarget{future-architecture}{%
\paragraph{Future architecture}\label{future-architecture}}

In the future, we still want to use R6-classes. One important goal with the future architecture is extensibility. It must be possible for 3rd-party users to extend the code, especially with custom (data-)losses and regularizers. In the following, I show a class hierarchy of how it is planned to be eventually.

\begin{itemize}
\tightlist
\item
  \texttt{abstract\ class\ Differentiable} -- this will be the mother-class of everything that is differentiable, i.e., models, objectives etc. The Differentiable is just a functional piece that itself cannot be computed.

  \begin{itemize}
  \tightlist
  \item
    Main methods: \texttt{get0Function()} (returns the function that can be called), \texttt{get1stOrderPd(name)} and \texttt{get2ndOrderPd(name)} return functions for the the first- resp. second-order partial derivatives for the parameters specified by \texttt{name} (or all if \texttt{name=NULL}). The idea behind this class is to provide a standardized interface for any function that has zero or arbitrary many parameters that can be derived for once or twice. If we think of a specific loss, let's say the RSS, it needs to obtain the function for the warped candidate, and its gradient needs the partial derivatives of the WC. I.e., we need the actual functions and not just the results. This class will also allow us to provide some helper classes for, e.g., encapsulating a signal-generating function more robustly.
  \item
    More methods: \texttt{getParamNames()}, \texttt{getNumParams()}, \texttt{getNumOutputs()}, \texttt{getOutputNames()} -- note how there is no setter for the actual parameters, as the \texttt{Differentiable}'s purpose is to describe the encapsulated function. It is important to know what the parameters are one can derive by, especially later when we combine many objectives and not each and every has the same parameters.
  \end{itemize}
\item
  \texttt{abstract\ class\ Model\ extends\ Differentiable} -- A model is a mathematical formulation of how to, e.g., transform some data using some parameters. So that means we will have 1st- and 2nd-order partial derivatives next to the actual model function. The \texttt{Model} also encompasses methods for getting statistical properties, such as residual or the likelihood (after fitting).

  \begin{itemize}
  \tightlist
  \item
    Main methods: \texttt{setParams(p)}, \texttt{getParams()}/\texttt{coefficients()} for setting concrete values for some or all of the model's non-constant parameters (those that are altered during optimization).
  \item
    Statistical methods: \texttt{AIC()}/\texttt{BIC()}, \texttt{coefficients()}, \texttt{residuals()}, \texttt{likelihood()}
  \item
    More methods: \texttt{plot()} for example (this method is inherited from R6)
  \item
    Sub-classes: \textbf{\emph{\texttt{srBTAW}}} -- Our main \textbf{multilevel-model} (described below)
  \end{itemize}
\item
  \texttt{abstract\ class\ Objective\ extends\ Differentiable} -- An objective is something that can be computed; however, it does not yet have a notion of a model or concrete parameters. Instead, it introduces methods that allow (numeric) computation of the underlying \texttt{Differentiable}. An \texttt{Objective} has no notion of a weight, or whether it should be minimized or maximized, for example.

  \begin{itemize}
  \tightlist
  \item
    Main methods: \texttt{compute(x)}, \texttt{compute1stOrderPd(x,\ name)}, \texttt{compute2ndOrderPd(x,\ name)} -- given some \(\mathbf{x}\), calls the methods of the underlying \texttt{Differentiable}. The \texttt{name}-arguments allow to specify specific partial derivatives (return all if null). Those can then be used to construct, e.g., a Jacobian. These methods return lists of functions, and the idea is that the name of the index represents the index of the output-parameter, and then which variable it was derived for first, then second.
  \item
    Additionally: \texttt{compute1stOrderPd\_numeric(x,\ name)} and \texttt{compute2ndOrderPd\_numeric(x,\ name)} -- these methods compute the 1st- and 2nd-order partial derivatives numerically. We will use the same naming scheme as previously. Note that 1st-order PDs would result in a 2-dimensional matrix for vector-valued functions, and in a 3-dimensional tensor for the second order.
  \item
    Sub-classes (all extend \texttt{Objective}):

    \begin{itemize}
    \tightlist
    \item
      \texttt{Loss}: Adds a weight, and also it knows whether we minimize or maximize,
    \item
      \texttt{LogLoss}: Meta-class that takes another \texttt{Loss} and returns its result in a logarithmic way. For our project at hand, we usually return strictly positive logarithmic losses.
    \item
      \texttt{DataLoss}: Specialized objective that computes a loss solely based on the data. This is more of a semantic class.
    \item
      \texttt{Regularizer}: Like \texttt{DataLoss}, this is more of a semantic class that computes a loss solely based on the model's parameters.
    \item
      \texttt{Score}: Another meta-class that takes another \texttt{Objective} (or, e.g., \texttt{LogLoss}), together with some lower and upper bounds, and transforms it into a score.
    \item
      \texttt{abstract\ class\ MultiObjective}: Generic meta-class that can combine multiple other objectives, losses, regularizers etc. It is abstract and generic because A) it does not define an implementation on how to do that and B) because it provides some common facilities to do so, for example it will allow to compute objectives in parallel.

      \begin{itemize}
      \tightlist
      \item
        \texttt{LinearScalarizer\ extends\ MultiObjective}: A concrete example of a \texttt{MultiObjective} that adds together all defined objectives using their weight. This is mostly what we will be using for our purposes.
      \item
        \texttt{ScoreMagnitudeScalarizer\ extends\ MultiObjective}: Previously described in section \ref{ssec:score_agg}, this is a score-aggregator that can be used to train the MLM with one or more \texttt{Score}s.
      \end{itemize}
    \end{itemize}
  \end{itemize}
\end{itemize}

\hypertarget{srbtaw-extends-model-a-multilevel-model}{%
\subsubsection{\texorpdfstring{\texttt{srBTAW\ extends\ Model} -- a Multilevel-model\label{ssec:srbtaw}}{srBTAW extends Model -- a Multilevel-model}}\label{srbtaw-extends-model-a-multilevel-model}}

sr-BTW and sr-BAW are mere algorithms, and not quite models. Either can take a reference- and a query-signal (Warping Pattern and Warping Candidate) and compute a warped version of the candidate using the boundaries and terminal y-translations. Instead of inheriting from \texttt{Objective}, we chose to have the actual model called \textbf{\emph{\texttt{srBTAW}}} to inherit from \texttt{Differentiable}, and to encapsulate a single \texttt{Objective} that can be computed. The abstract methods of \texttt{Differentiable} are forwarded to the encapsulated \texttt{Objective}. This has the advantage of the encapsulated objective actually representing the entire model, also we can use a single objective or, e.g., a \texttt{MultiObjective}, making this de facto a \textbf{multilevel-model}, that supports arbitrary many variables (one or more) for either, Warping Pattern and Warping Candidate.

The main purpose of \texttt{srBTAW} is to encapsulate all available data (variables) and intervals, and to provide helper methods (factories) for creating losses. It will allow us to reuse instances of \texttt{SRBTW} and \texttt{SRBAW}/\texttt{SRBTWBAW}, as any of these create a pair of one WP and one WC.

\hypertarget{testing-1}{%
\subsubsection{Testing}\label{testing-1}}

As I wrote in the previous section, we will just wrap sr-BTW with sr-BAW. So I sub-classes it and its sub-model and made the required extensions. In the following, we will conduct very similar tests to the ones I did with sr-BTW, except for the fact that we allow the y-intercepts to adjust now, too.

\includegraphics{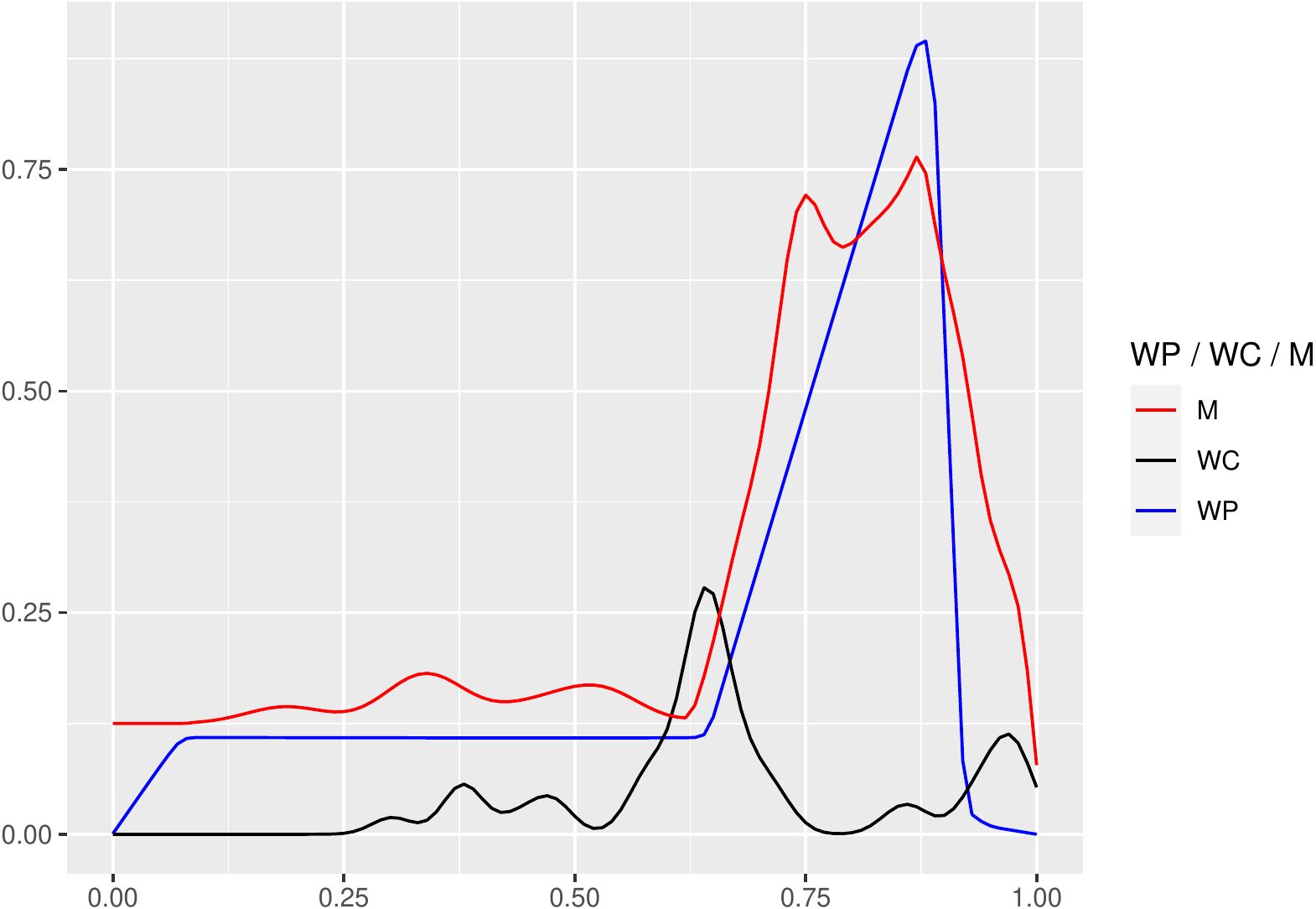}

I tried to put in values for the y-translation that would make the WC approximately close to the WP. It is still off quite a bit, but it should be already much closer (lower loss) than we had achieved in the first optimization where amplitude warping was not allowed. Now if we carefully check all of the values and the plot, we'll see that it works. Now what we are going to do is to repeat some of the optimizations, using the same losses as earlier, and then we are going to compare those fits.

\begin{verbatim}
## [1] 0.01653856
\end{verbatim}

Previously, the error at this point was \(\approx 0.0709\), it is already much lower. What we should attempt now, is to find even better parameters using optimization:

\begin{verbatim}
## $par
## [1]  0.000000000  0.728221204  0.120183423  0.575038549  0.004043294
## [6]  0.115330334 -0.048865578  0.737305823 -2.007297658
## 
## $value
## [1] 9.303243e-05
## 
## $counts
## function gradient 
##       90       90 
## 
## $convergence
## [1] 0
## 
## $message
## [1] "CONVERGENCE: REL_REDUCTION_OF_F <= FACTR*EPSMCH"
\end{verbatim}

\includegraphics{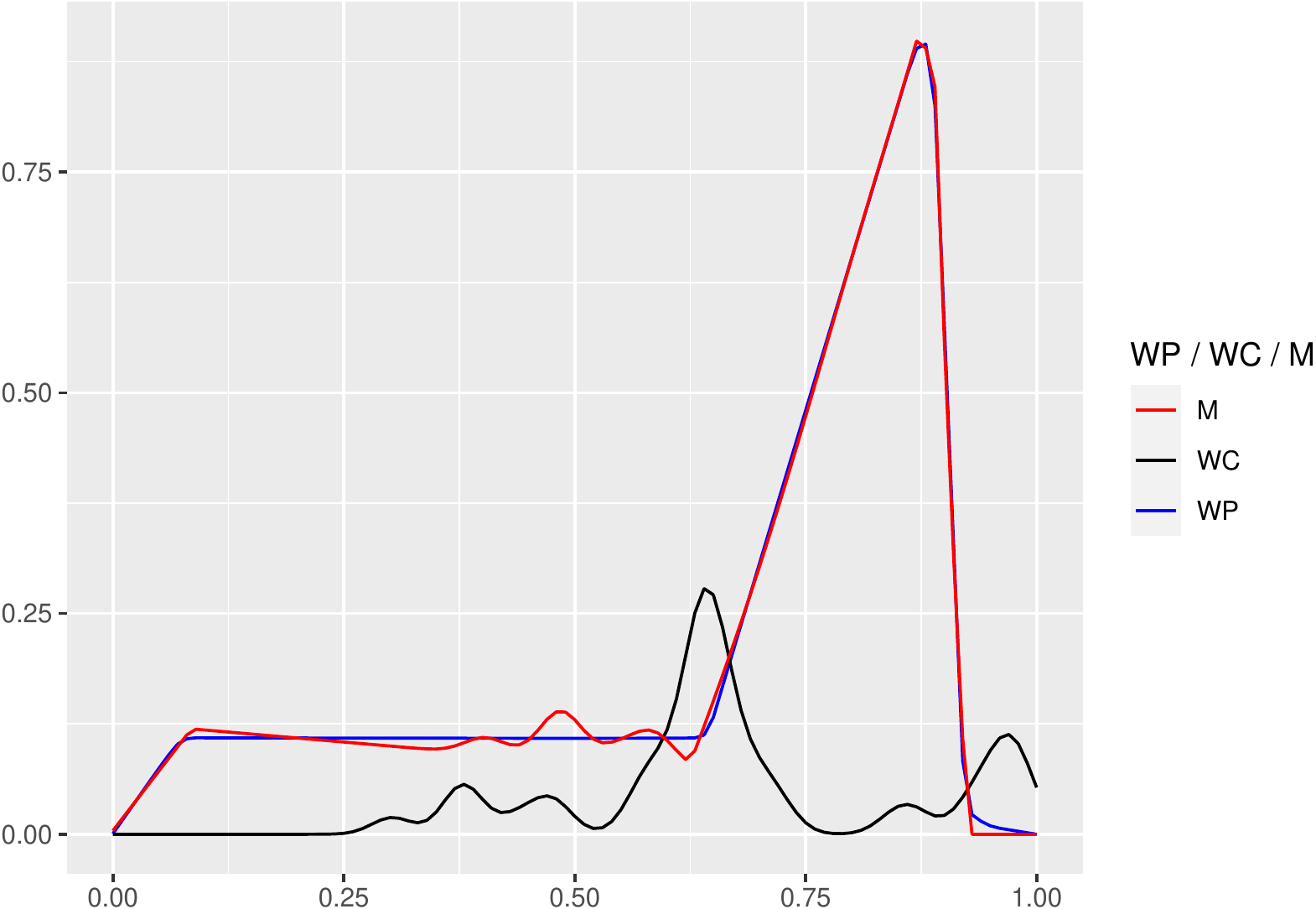} \includegraphics{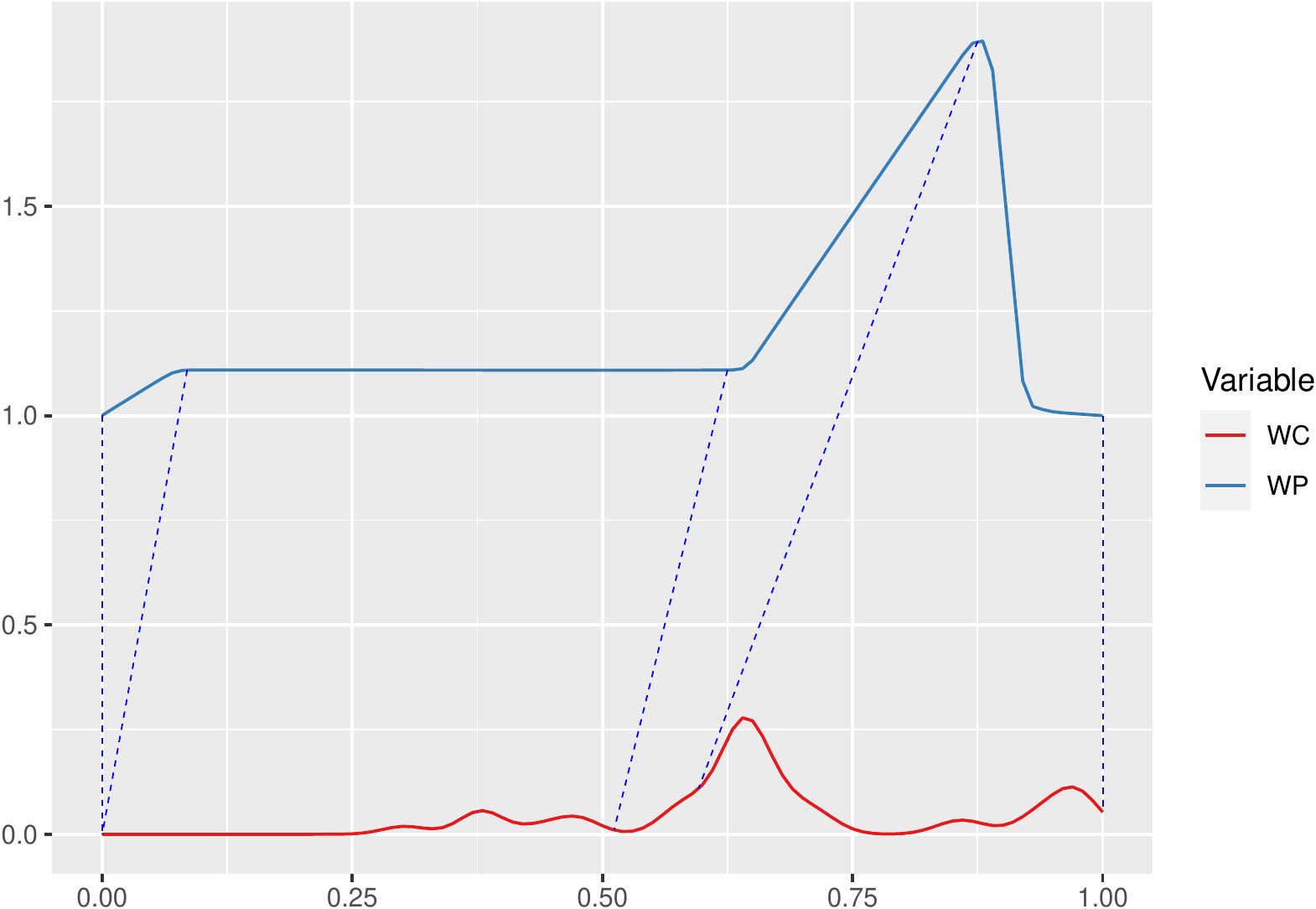}

Ok, that worked extraordinarily well. However, we also gave it quite good starting values to be fair. Using just zeros as starting values for \(v\) and all terminal y-translations worked also well. If we look at the learned parameters we notice a peculiarity of this model: The last terminal y-translation is almost minus three -- this allows the model in the last interval to decline steeply, and to actually make this ``flat foot'' at the end that the Warping Pattern has, because the signal now effectively becomes zero in the last section. That is why I had formulated the \(\min,\max\) self-regularization on the final signal, and not the terminal y-translations, as these might be rather extreme in order to achieve a good fit.

Let's also make an open/open test and see what happens:

\begin{Shaded}
\begin{Highlighting}[]
\NormalTok{srbtwbaw}\SpecialCharTok{$}\FunctionTok{setOpenBegin}\NormalTok{(}\AttributeTok{ob =} \ConstantTok{TRUE}\NormalTok{)}
\NormalTok{srbtwbaw}\SpecialCharTok{$}\FunctionTok{setOpenEnd}\NormalTok{(}\AttributeTok{oe =} \ConstantTok{TRUE}\NormalTok{)}
\end{Highlighting}
\end{Shaded}

\begin{verbatim}
## $par
##  [1]  0.000000000  0.628317037  0.111016404  0.416655367  0.000000000
##  [6]  0.940912174  0.006752451  0.112084465 -0.047787539  0.726855230
## [11] -2.055975386
## 
## $value
## [1] 9.045725e-05
## 
## $counts
## function gradient 
##      116      116 
## 
## $convergence
## [1] 1
## 
## $message
## [1] "NEW_X"
\end{verbatim}

\includegraphics{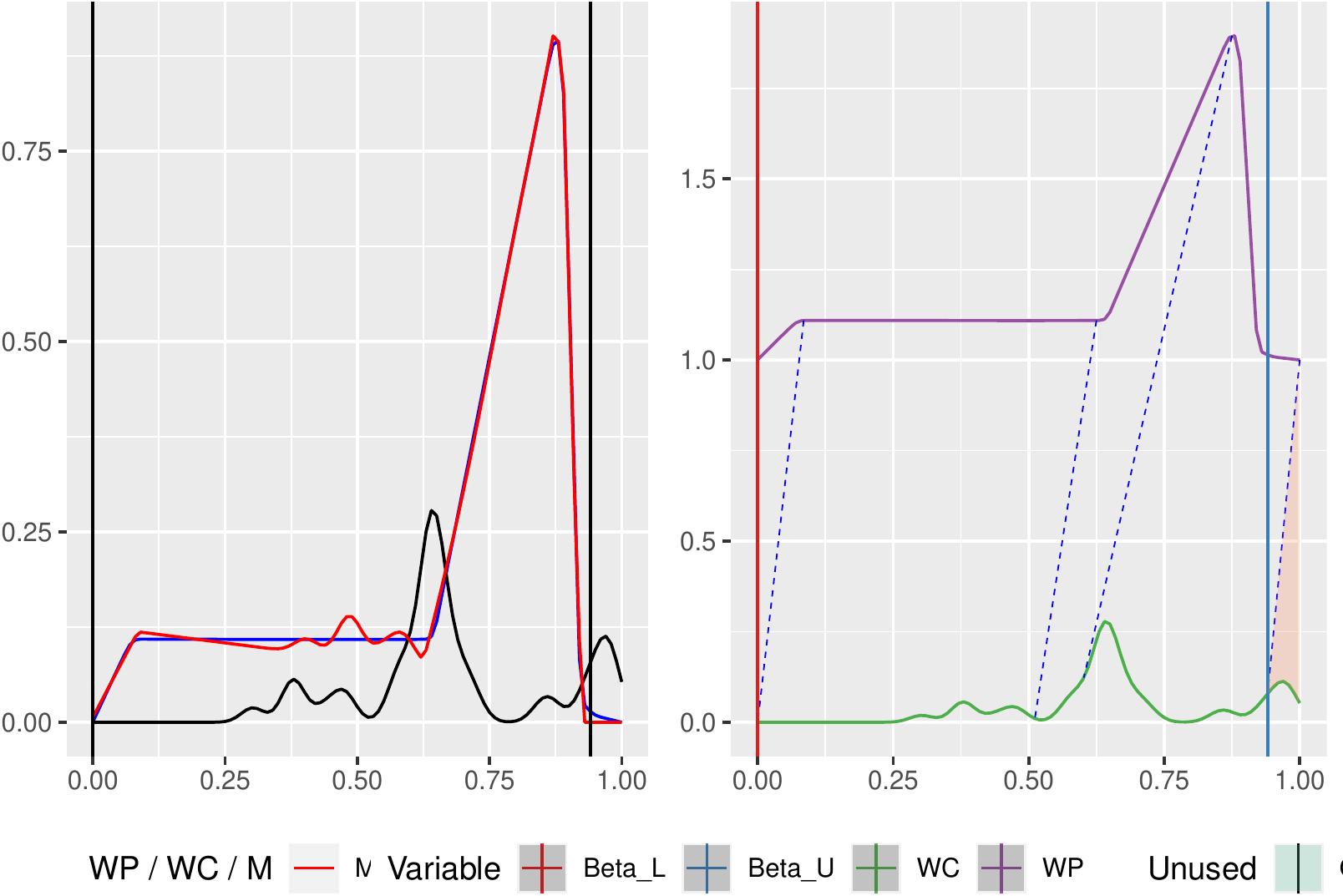}

Ok, cut-in is still at \(0\), but cut-out is at 0.9409. The error is as low as in the previous test.

\hypertarget{regularization-1}{%
\subsubsection{Regularization}\label{regularization-1}}

Regularization for sr-BAW is probably more difficult, as we cannot impose regularization directly on the parameters as I see it. Well, we could, but as I just wrote before this, it may be necessary for \(v\) and the terminal y-translations to assume extreme values (well beyond \(\bm{\lambda}^{(\text{ymin})},\bm{\lambda}^{(\text{ymax})}\)) to achieve a good fit. In other words, I would not interfere with these too much at the moment. In fact, we did above optimization with the minimum/maximum machine double, effectively alleviating box-bounds for these parameters. I suppose that it may be an idea to regularize all y-translations together, and to look at them as a distribution. Then, extreme values could be penalized. However, this is probably application specific.

What seems to be more promising is to impose regularization on the output of the model. Using \(\bm{\lambda}^{(\text{ymin})},\bm{\lambda}^{(\text{ymax})}\), we can limit the \textbf{output} of the model to be within box-bounds. However, if the model were to push too much of the signal onto either boundary, we would get a less optimal fit in many cases. An approach to alleviating this is described in the next subsection.

\hypertarget{regularize-extreme-outputs}{%
\paragraph{Regularize extreme outputs}\label{regularize-extreme-outputs}}

Another reason that makes regularization of the terminal y-translations difficult, is the fact that we cannot (and must not) make any assumptions about the signal in an interval -- we simply do not know how it behaves. Therefore, I suggest a regularizer that penalizes the output of the model instead. The more of its output is close to either box-bound, the higher the cost. This would require a somewhat \emph{hybrid} regularizer, as we need to consider the box-bounds, the y-translations and the data. A simple first version could take the \textbf{average} of the output signal and score that:

\[
\begin{aligned}
  \mathbf{\hat{y}},f(x)\;\dots&\;\text{complete discrete output of some model, or the model itself as function,}
  \\[1ex]
  \mu^{(c)}=&\;\frac{1}{\max{\big(\operatorname{supp}(f)\big)}-\min{\big(\operatorname{supp}(f)\big)}}\times\int\displaylimits_{\operatorname{supp}(f)}\,f(x)\;dx\;\text{, mean of function(model), or}
  \\[1ex]
  \mu^{(d)}=&\;\frac{1}{N}\sum_{i=1}^{N}\,\mathbf{\hat{y}}_i\;\text{, mean of some discrete output,}
  \\[1ex]
  \beta_l,\beta_u\;\dots&\;\text{lower- and upper bounds for the output,}
  \\[1ex]
  \mathcal{R}^{(\text{avgout})}=&\;1-\frac{\min{\big(\mu-\beta_l,\beta_u-\mu\big)}}{\frac{1}{2}\big(\beta_u-\beta_l\big)}\mapsto[0+,1]\;\text{, or, alternatively:}
  \\[1ex]
  \mathcal{R}^{(\text{avgout})}=&\;2\,\sqrt{\bigg(\frac{1}{2}\big(\beta_u-\beta_l\big)-\mu\bigg)^2}\mapsto[0+,1]\;\text{, and the log-loss:}
  \\[1ex]
  \mathcal{L}_{\mathcal{R}^{(\text{avgout})}}=&\;-\log{\Big(1-\mathcal{R}^{(\text{avgout})}\Big)}\mapsto[0+,\infty]\;\text{.}
\end{aligned}
\]

Note that both versions of the regularizer are defined as ratios, and that, at most, their loss can be \(1\) (and the least possible loss would be \(0\)). Also, \(\frac{1}{2}\big(\beta_u-\beta_l\big)\) can be replaced with some user-defined, arbitrary expectation, i.e., it does not necessarily need to be the mean of the upper and lower bound (but it needs to be within them).

\hypertarget{trapezoidal-overlap-regularization}{%
\paragraph{Trapezoidal-overlap regularization}\label{trapezoidal-overlap-regularization}}

Using the suggested y-translations results in a rectangular interval becoming a \textbf{trapezoid}. Even more specific, the trapezoid is a \textbf{parallelogram}, which has the \textbf{same} area as the original interval, as the base of corresponds to the interval's length. I suggest using another kind of regularizer that measures the overlap of these skewed and translated parallelograms with their original interval (as a ratio).

\[
\begin{aligned}
  \mathbf{x}_q\;\dots&\;\text{support for the}\;q\text{-th interval},
  \\[1ex]
  \beta_l,\beta_u\;\dots&\;\text{lower and upper boundary for}\;y\;\text{(coming from}\;\bm{\lambda}^{(\text{ymin})},\bm{\lambda}^{(\text{ymax})}\text{),}
  \\[1ex]
  b_{\beta_l}(x),b_{\beta_u}(x)\;\dots&\;\text{functions over the lower- and upper boundaries,}
  \\[1ex]
  o_q=&\;v+\phi_q^{(y)}\;\text{, initial translation in the interval,}
  \\[1ex]
  s,t\;\dots&\;\text{intersections of}\;t_q(x,\dots)\;\text{with lower- and upper boundary,}
  \\[0ex]
  &\;\text{(}s=\min{(\mathbf{x}_q)}\;\text{and}\;t=\max{(\mathbf{x}_q)}\;\text{if no intersetion),}
  \\[1ex]
  \delta_q^{(\max)}\;\dots&\;\text{maximum possible difference in interval}\;q\text{, i.e., no overlap,}
  \\[1ex]
  \delta_q=&\;\begin{cases}
    \text{if}\;o_q<\beta_l,&\begin{cases}
      \text{if}\;o_q+\bm{\vartheta}_q^{(y)}>\beta_l,&\int_s^{\max{(\mathbf{x}_q)}}\,\min{\big(b_{\beta_u}(x),t_q(x)\big)}-b_{\beta_l}(x)\,dx,
      \\
      \text{else}&\min{\Big(\delta_q^{(\max)},\int_s^t\,b_{\beta_l}(x)-t_q(x)\Big)},
    \end{cases}
    \\
    \text{else if}\;o_q>\beta_u,&\begin{cases}
      \text{if}\;o_q+\bm{\vartheta}_q^{(y)}<\beta_u,&\int_{\min{(\mathbf{x}_q)}}^{t}\,\min{\big(b_{\beta_u}(x),t_q(x)\big)-b_{\beta_l}(x)}\,dx,
      \\
      \text{else}&0,
    \end{cases}
    \\
    \text{else}&\begin{cases}
      \text{if}\;o_q+\bm{\vartheta}_q^{(y)}>\beta_u,&\int_s^t\,\min{\big(b_{\beta_u}(x),t_q(x)\big)}-b_{\beta_l}(x)\,dx,
      \\
      \text{else}&\int_s^t\,t_q(x)-b_{\beta_l}(x)\,dx,
    \end{cases}
  \end{cases}
  \\[1ex]
  \mathcal{R}^{(\text{trapez})}=&\;{\max{(Q)}}^{-1}\times\sum_{i=1}^{\max{(Q)}}\,\frac{\delta_q}{\delta_q^{(\max)}}\mapsto[0+,1]\;\text{, and the log-loss:}
  \\[1ex]
  \mathcal{L}_{\mathcal{R}^{(\text{trapez})}}=&\;-\log{\Big(1-\mathcal{R}^{(\text{trapez})}\Big)}\mapsto[0+,\infty]\;\text{.}
\end{aligned}
\]

\hypertarget{tikhonovl_2-regularization-of-translations}{%
\paragraph{\texorpdfstring{Tikhonov/\(L_2\) regularization of translations}{Tikhonov/L\_2 regularization of translations}}\label{tikhonovl_2-regularization-of-translations}}

A somewhat effortless way to regularize extreme y-translations is to use the (squared) \(L_2\)-norm for the parameters \(v\) and \(\bm{\vartheta}^{(y)}\), as we would prefer less extreme values for these translations. However, during optimization we also would prefer allowing these parameters to cover a wider range than others, to allow more adapted solutions. This regularization has to be used carefully therefore.

\[
\begin{aligned}
  \bm{\omega}=&\;\{v\,\frown\,\bm{\vartheta}^{(y)}\}\;\text{, vector containing all y-translations,}
  \\
  \lVert \bm{\omega}\rVert_2=&\;\sqrt{\sum_{i=1}^{\left\lVert\,\bm{\omega}\,\right\rVert}\,\bm{\omega}_i^2}\;\text{, the}\;L_2\text{-norm, alternatively}
  \\[1ex]
  \lVert \bm{\omega}\rVert_2^2=&\;\sum_{i=1}^{\left\lVert\,\bm{\omega}\,\right\rVert}\,\bm{\omega}_i^2\;\text{, the squared}\;L_2\text{-norm, with regularizer/loss}
  \\[1ex]
  \mathcal{L}_{\mathcal{R}^{(L_2)}}=&\;\log{\Big(1+\lVert \bm{\omega}\rVert_2^2\Big)}\;\text{.}
\end{aligned}
\]

\hypertarget{testing-new-srbtaw-model}{%
\subsubsection{\texorpdfstring{Testing new \texttt{srBTAW} model}{Testing new srBTAW model}}\label{testing-new-srbtaw-model}}

I did implement the new architecture and wrapped some of the older classes, and it's finally time to make a test.

\includegraphics{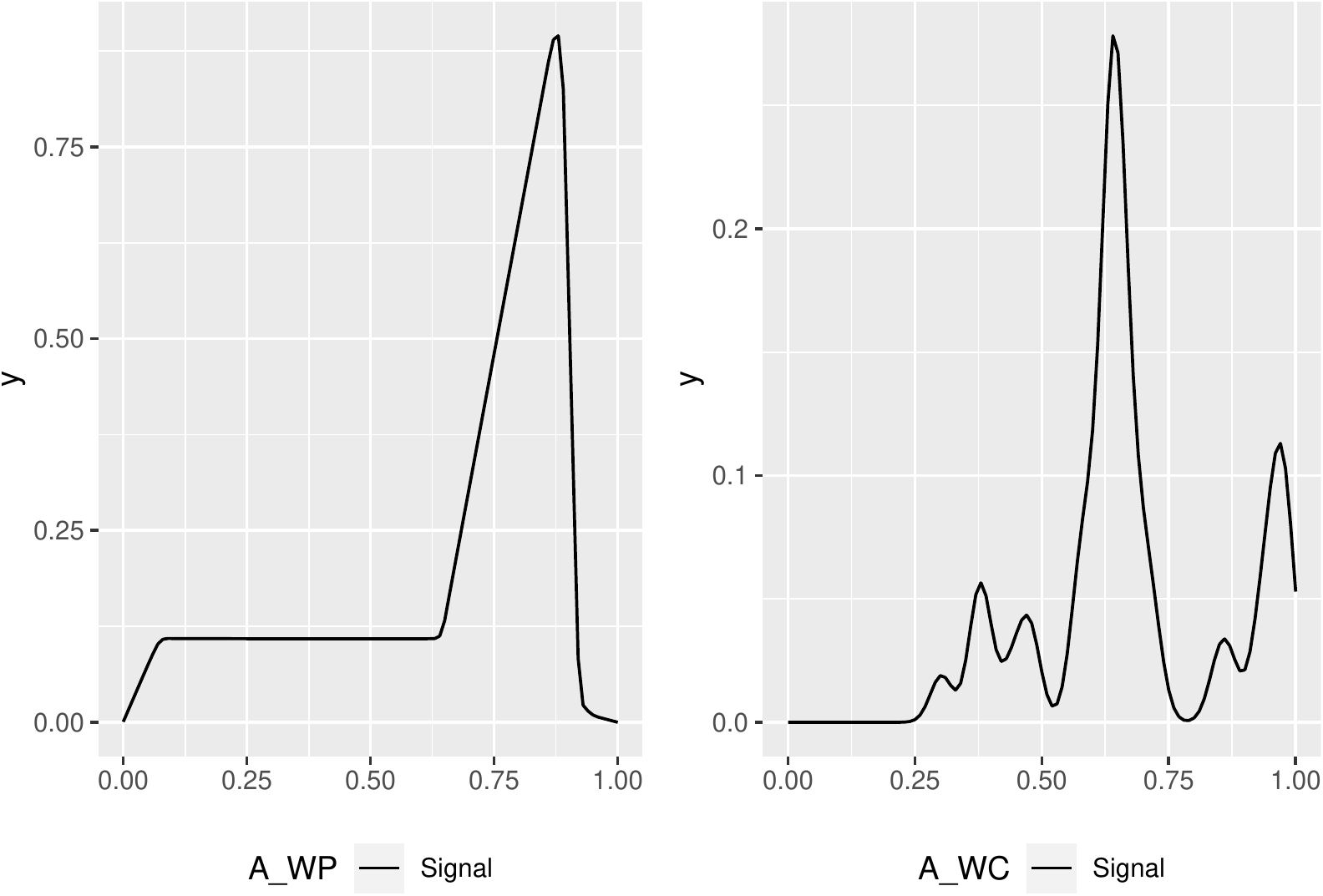}

Finally, we can set the parameters and start fitting!

We got our MLM and the two signals (one WP/WC each). The next steps are adding the signal to the model, as well as initializing an appropriate loss (objective).

The following optimization works equally well with a \textbf{continuous} version of the same loss, but computing is approx. ten times slower and needs about 50\% more iterations. However, it converges much faster to its final minimum, whereas we see a rather gradual improvement for the discrete RSS. In the following, we plot the parameters' course over all optimization steps. Note that values for \(\bm{\vartheta}^{(l)}\) (\texttt{vtl\_1} etc.) are not normalized as to sum up to \(1\).

\begin{verbatim}
## $par
##            v        vtl_1        vtl_2        vtl_3        vtl_4        vty_1 
##  0.004969021  0.001137423  0.808313993  0.131842724  0.637675363  0.114318356 
##        vty_2        vty_3        vty_4 
## -0.048637996  0.736088557 -1.998226471 
## 
## $value
## [1] 9.319825e-05
## 
## $counts
## function gradient 
##      114      114 
## 
## $convergence
## [1] 1
## 
## $message
## [1] "NEW_X"
\end{verbatim}

\includegraphics{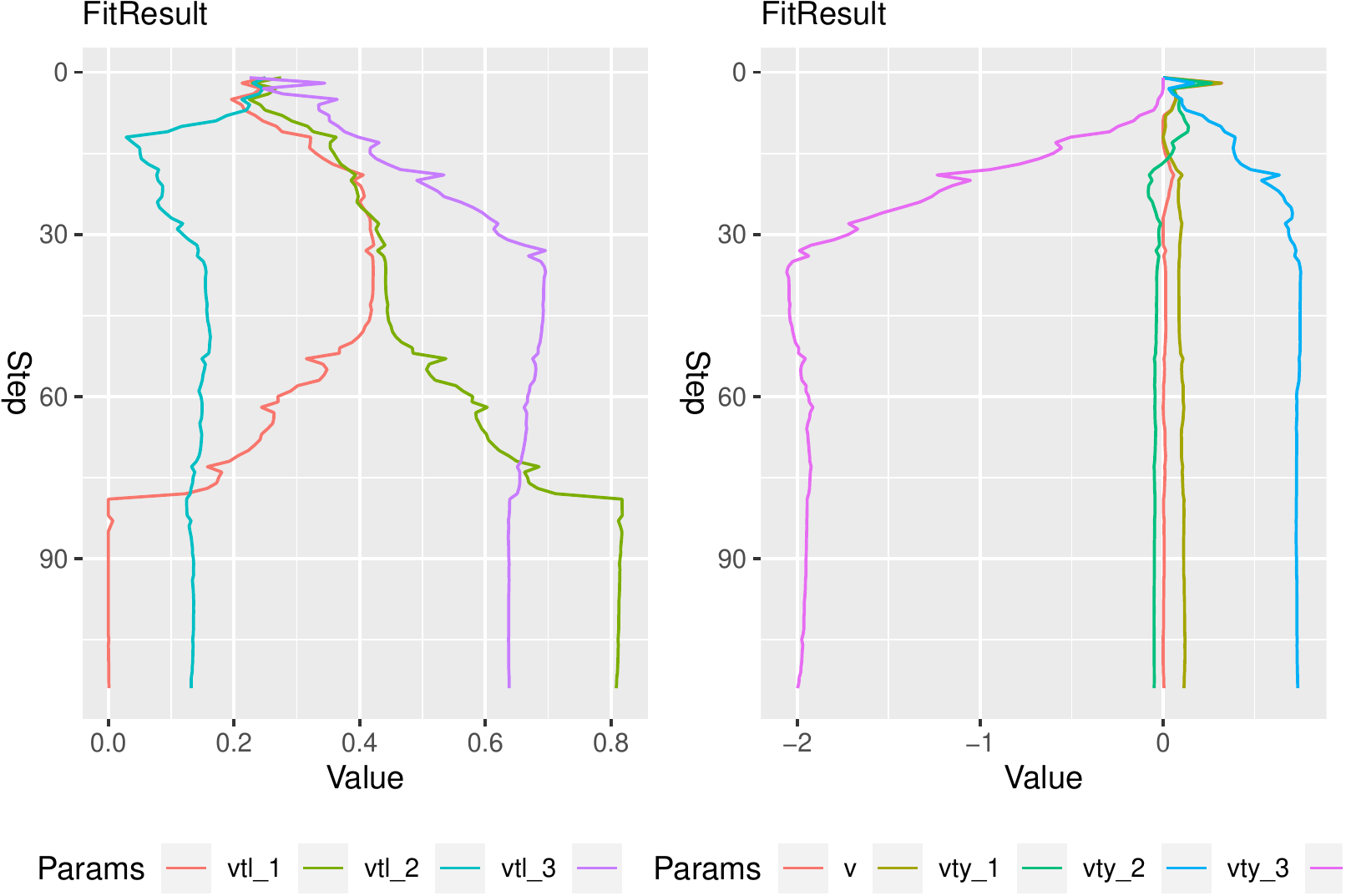}

The warping- and dual-original plots of this fit are in the top of the following figure. We can see that without regularization we obtain a very good fit. However, the first interval has a length of almost \(0\), which might not be desirable in practice.

Therefore, I have added very mild (weight of \(0.1\)) regularization for extreme intervals and ran it again, the results are in the second row of below plot. I suppose it is fair to say that this fit is the overall best so far: It selects the flat beginning to map to the first interval, then the slight fluctuations to the Long Stretch, before mapping the begin of the next interval to the signal's local minimum shortly before the peak, which is perfectly mapped to the reference's peak.

In the third row, I have added very mild trapezoidal-overlap regularization (weight of \(0.05\)). The fit is very good, too, but the first two intervals were selected even slightly better with extreme-interval regularization only. Also the peak appears to be selected slightly better there.

\includegraphics{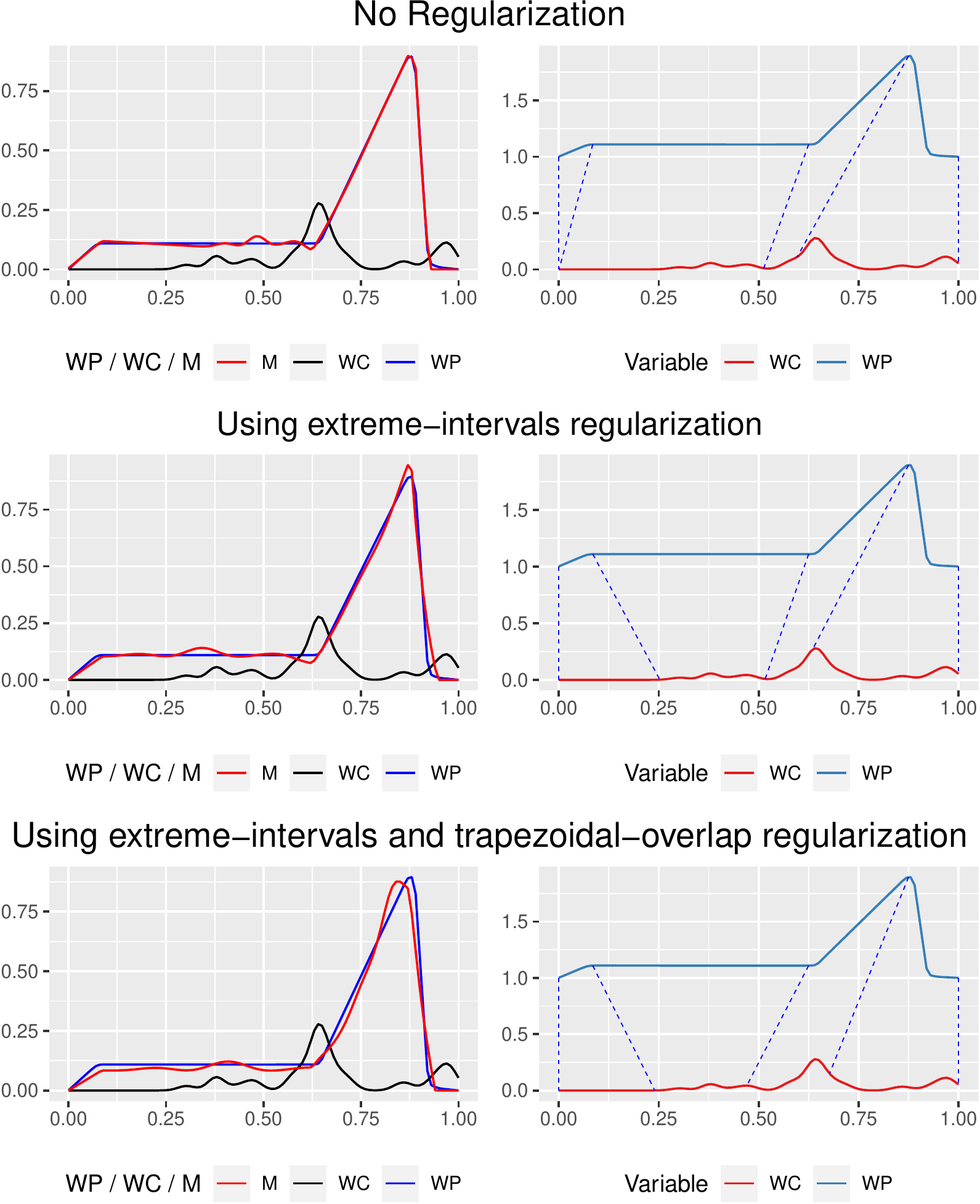}

This is (within margin of error) the same result as we got previously (first row), so this means that our new architecture indeed works! Like in the previous result, we have one interval with length \(\approx0\), and generally, the chosen lengths could align better with the reference. However, this is a human error, as the model did exactly what it was supposed to. In other words, what is missing, is \textbf{regularization}. The red \texttt{M}-function could probably not be closer to the blue warping pattern (\texttt{WP}). It is an extremely good solution that also requires extreme parameters, like the almost zero-length first interval and the y-translations for the last two intervals. If, for the use-case, the goal is to obtain \texttt{M} such that it is most close to \texttt{WP}, regardless of how extreme the parameters may be, then this is our winner. In practice, however, such extreme parameters are often less desirable, so I have tested two additional cases with added regularization.

My advice is the following: Since \texttt{srBTAW} is self-regularizing to some degree, always try first \textbf{without} any additional regularization. Regularizations should be added one-by-one to better observe their impact, and they should be balanced. A regularizer is always an additional optimization goal, do not forget that. If very short (almost or equal to zero length) intervals are acceptable, additional regularization is not necessary. To avoid these, extreme-interval regularization may be added. It should have a low weight compared to the data-losses to begin with (start with an order of magnitude smaller) and then may be adjusted from there.

For most cases it is probably advisable to start with regularizations for the interval-lengths, and not the y-translations. Looking at the first plot in the group above, the third and fourth intervals require quite extreme y-translations, esp.~since the absolute peak is mapped before it happens. Imposing just y-translations regularizations leads to a result similar as in the first plot. However, excessive y-translations are costly quickly, and can be avoided by aligning the intervals better in the first place. Y-translation regularization appears to have a stronger guiding-effect, and higher weights can lead to the signal not being able to properly translate. Also, and that depends on the regularizer used, penalizing y-translations should not be necessary in most cases, as we need/want the process to assume more extreme values in order to get a good fit. However, adding this kind of regularization is specific to the use-case and might still be beneficial. The three kinds of regularizers I had introduced earlier (Tikhonov, Average-output and Trapezoidal-overlap) are targeting three different properties and should cover any eventual needs.

\hypertarget{other-thoughts}{%
\subsection{Other thoughts}\label{other-thoughts}}

Short unsorted list of things to consider.

Check progress of pattern: SRBTW allows for open-end matching, so that it can be used to detect by how much (time wise into the project) Fire Drill matches. Or, we can inverse the process and match the AP to the project, to see which phases exist and score them.

In optimization goal III, we should also try to adapt the reference AP without actually defining a loss for the Long Stretch. With such an adapted pattern, we will find out what goes actually on in that phase in the projects we have, as the other intervals will match, leaving a gap for the LS.

Choosing a good number of boundaries \emph{a priori}: This depends on the problem, e.g., classification, and on whether a search can be performed for it. A search can be done using some information criterion (IC). Using some IC, then computing a best fit using \(1\dots n\) boundaries, we should get a somewhat convex function, and can choose the number of boundaries using the minimum of that function. Sometimes, however, we cannot do this and need to make a decision a priori. I have done some tests using the \textbf{entropy} of the signals, by treating them as probability density functions (which we showed works quite well for fitting a model). I have tried to compute it continuously (differential entropy) and discretely, by sampling from the signal. The latter case gives more usable values, yet it is much more volatile w.r.t. the number of samples. For example, a typical entropy (Shannon-bits) for \(1e3\) samples is \(\approx10\), and for \(1e5\) it is already \(\approx16\). Given the entropies of two signals, we could calculate the starting-number \(n\) of boundaries like \(n=(H^{(\text{WP})}\times H^{(\text{WC})})^{1-\frac{1}{e}}\). This gives OK-values for entropies in the range of approx. \(\approx[5,20]\) and is just one idea. Eventually, we probably would need to do an empirical evaluation with many examples, where we employ a full search using some IC. Then, we could attempt to come up with some rule that best approximates this value.

Choosing a priori: One idea could be to select some number of equidistant boundaries/intervals and then replace the signal in each interval with a linear function that goes from start to end. Then, we would compute a loss and AIC between this function and the original interval to find the best number of intervals.

Another idea: Calculate arc length over full support, then divide it by a constant number of intervals. Then, start from the begin of the support and go a partial arc length (calculate the end coordinate as we know the begin and the arc length already). Repeat to the end. This method will subdivide the function into parts of equal arc length.

Classification: One additional way for classification would be an open/open approach using only one interval. If the selected section is much shorter than the original signal, the match would be bad (for non-repetitive signals).

\clearpage

\hypertarget{technical-report-commit-classification-models-based-on-keywords-and-source-code-density}{%
\section{\texorpdfstring{Technical Report: Commit Classification models based on Keywords and Source Code Density\label{tr:comm-class-models}}{Technical Report: Commit Classification models based on Keywords and Source Code Density}}\label{technical-report-commit-classification-models-based-on-keywords-and-source-code-density}}

This is the self-contained technical report for training models suitable for commit classification using keywords and source code density.

\hypertarget{introduction-4}{%
\subsection{Introduction}\label{introduction-4}}

In this notebook, we will train and store some best best models for commit classification, as these will be detrimental to detecting maintenance activities in software projects. The models will be based on latest work from (Hönel et al. 2020).

Throughout this notebook, we will build a few models that are all similar. Our latest work indicated that including up to three previous generations of commits is beneficial. We will evaluate models that include 1, 2 or 3 previous generations.

Likewise, we want to go into the other direction, looking forward at children, using similar amounts of commits. We will call those models \emph{stateful} in this notebook, and only the model without any adjacent commits is called \emph{stateless}, but they all belong to the same \emph{class}. Finally, having a few best models, the overall classifier shall use the best model for the available data.

For finding the best models, the process is separated into two steps: First, do a k-fold cross-validation to find the best model and hyperparameters. Then, use the findings to train a model using the entire data. Also, we will store the associated scaler/pre-processor for that model.

Also note that each of the following models was already optimized w.r.t. to some aspects of the training, like using an already oversampled dataset. Also, we are using a very high split, as the resulting model will also be using all data. Using many folds and repeats, we make sure that overfitting is not a problem.

All complementary data and results can be found at Zenodo (Hönel, Pícha, et al. 2023). This notebook was written in a way that it can be run without any additional efforts to reproduce the outputs (using the pre-computed results). This notebook has a canonical URL\textsuperscript{\href{https://github.com/MrShoenel/anti-pattern-models/blob/master/notebooks/comm-class-models.Rmd}{{[}Link{]}}} and can be read online as a rendered markdown\textsuperscript{\href{https://github.com/MrShoenel/anti-pattern-models/blob/master/notebooks/comm-class-models.md}{{[}Link{]}}} version. All code can be found in this repository, too.

\hypertarget{stateless-model}{%
\subsection{Stateless model}\label{stateless-model}}

The stateless model shall be used whenever there is \textbf{no} data available from the parents or children.

\hypertarget{load-and-prepare-the-data}{%
\subsubsection{Load and prepare the data}\label{load-and-prepare-the-data}}

\begin{Shaded}
\begin{Highlighting}[]
\CommentTok{\# the stateless data:}
\NormalTok{data\_sl }\OtherTok{\textless{}{-}} \ControlFlowTok{if}\NormalTok{ (}\ConstantTok{FALSE}\NormalTok{) \{}
  \FunctionTok{getDataset}\NormalTok{(}\StringTok{"antipat\_gt\_all"}\NormalTok{)}
\NormalTok{\} }\ControlFlowTok{else}\NormalTok{ \{}
  \FunctionTok{readRDS}\NormalTok{(}\StringTok{"../data/antipat\_gt\_all.rds"}\NormalTok{)}
\NormalTok{\}}

\CommentTok{\# remove SHAs:}
\NormalTok{data\_sl }\OtherTok{\textless{}{-}}\NormalTok{ data\_sl[, }\SpecialCharTok{!}\NormalTok{(}\FunctionTok{names}\NormalTok{(data\_sl) }\SpecialCharTok{\%in\%} \FunctionTok{c}\NormalTok{(}\StringTok{"SHA1"}\NormalTok{, }\StringTok{"ParentCommitSHA1s"}\NormalTok{))]}
\CommentTok{\# factorize the labels:}
\NormalTok{data\_sl}\SpecialCharTok{$}\NormalTok{label }\OtherTok{\textless{}{-}} \FunctionTok{factor}\NormalTok{(}\AttributeTok{x =}\NormalTok{ data\_sl}\SpecialCharTok{$}\NormalTok{label, }\AttributeTok{levels =} \FunctionTok{sort}\NormalTok{(}\FunctionTok{unique}\NormalTok{(data\_sl}\SpecialCharTok{$}\NormalTok{label)))}
\end{Highlighting}
\end{Shaded}

The zero-variance predictors should be removed (if any).

\begin{Shaded}
\begin{Highlighting}[]
\NormalTok{nzv\_sl }\OtherTok{\textless{}{-}}\NormalTok{ caret}\SpecialCharTok{::}\FunctionTok{nearZeroVar}\NormalTok{(}\AttributeTok{x =}\NormalTok{ data\_sl, }\AttributeTok{saveMetrics =} \ConstantTok{TRUE}\NormalTok{, }\AttributeTok{names =} \ConstantTok{TRUE}\NormalTok{)}

\FunctionTok{print}\NormalTok{(}\FunctionTok{paste0}\NormalTok{(}\StringTok{"Zero{-}variance predictors to be removed are: "}\NormalTok{, }\FunctionTok{paste}\NormalTok{(}\FunctionTok{names}\NormalTok{(data\_sl)[nzv\_sl}\SpecialCharTok{$}\NormalTok{zeroVar],}
  \AttributeTok{collapse =} \StringTok{", "}\NormalTok{)))}
\end{Highlighting}
\end{Shaded}

\begin{verbatim}
## [1] "Zero-variance predictors to be removed are: IsInitialCommit, IsMergeCommit, NumberOfParentCommits"
\end{verbatim}

\begin{Shaded}
\begin{Highlighting}[]
\NormalTok{data\_sl }\OtherTok{\textless{}{-}}\NormalTok{ data\_sl[, }\SpecialCharTok{!}\NormalTok{nzv\_sl}\SpecialCharTok{$}\NormalTok{zeroVar]}
\end{Highlighting}
\end{Shaded}

\hypertarget{define-how-the-training-works}{%
\subsubsection{Define how the training works}\label{define-how-the-training-works}}

For each type of model, we will use a pre-defined train control.

Instead of sampling during training, we'll work with a resample of the entire dataset, using the \emph{synthetic minority over-sampling technique}
(Chawla et al. 2002).

\begin{Shaded}
\begin{Highlighting}[]
\NormalTok{numFolds }\OtherTok{\textless{}{-}} \DecValTok{5}
\NormalTok{numRepeats }\OtherTok{\textless{}{-}} \DecValTok{5}

\NormalTok{tc\_sl }\OtherTok{\textless{}{-}}\NormalTok{ caret}\SpecialCharTok{::}\FunctionTok{trainControl}\NormalTok{(}
  \AttributeTok{method =} \StringTok{"repeatedcv"}\NormalTok{, }\AttributeTok{p =} \FloatTok{0.9}\NormalTok{,}
  \AttributeTok{returnResamp =} \StringTok{"all"}\NormalTok{, }\AttributeTok{savePredictions =} \StringTok{"all"}\NormalTok{, }\AttributeTok{classProbs =} \ConstantTok{TRUE}
\NormalTok{  , }\AttributeTok{number =}\NormalTok{ numFolds, }\AttributeTok{repeats =}\NormalTok{ numRepeats}
\NormalTok{  , }\AttributeTok{seeds =} \FunctionTok{get\_seeds}\NormalTok{(}\AttributeTok{nh =} \DecValTok{200}\NormalTok{, }\AttributeTok{amount =} \DecValTok{2} \SpecialCharTok{*}\NormalTok{ numFolds }\SpecialCharTok{*}\NormalTok{ numRepeats)}
  \CommentTok{\#, sampling = "smote"}
\NormalTok{)}
\end{Highlighting}
\end{Shaded}

\hypertarget{tuning-of-several-models}{%
\subsubsection{Tuning of several models}\label{tuning-of-several-models}}

We do this step to find which models work well with our data. Later, we can try to combine the best models into a meta-model.

\begin{Shaded}
\begin{Highlighting}[]
\FunctionTok{set.seed}\NormalTok{(}\DecValTok{1337}\NormalTok{)}
\CommentTok{\# Let\textquotesingle{}s preserve 100 instances from the original data as validation data:}
\NormalTok{p }\OtherTok{\textless{}{-}}\NormalTok{ caret}\SpecialCharTok{::}\FunctionTok{createDataPartition}\NormalTok{(}\AttributeTok{y =}\NormalTok{ data\_sl}\SpecialCharTok{$}\NormalTok{label, }\AttributeTok{p =} \FloatTok{0.95}\NormalTok{, }\AttributeTok{list =} \ConstantTok{FALSE}\NormalTok{)}

\NormalTok{train\_sl }\OtherTok{\textless{}{-}}\NormalTok{ data\_sl[p, ]}
\NormalTok{valid\_sl }\OtherTok{\textless{}{-}}\NormalTok{ data\_sl[}\SpecialCharTok{{-}}\NormalTok{p, ]}


\CommentTok{\# As described above, we can use an oversampled dataset for this model.}
\CommentTok{\# However, most recent changes indicate this may or may not be beneficial.}
\NormalTok{train\_sl }\OtherTok{\textless{}{-}} \FunctionTok{balanceDatasetSmote}\NormalTok{(}\AttributeTok{data =}\NormalTok{ train\_sl, }\AttributeTok{stateColumn =} \StringTok{"label"}\NormalTok{)}
\end{Highlighting}
\end{Shaded}

\begin{Shaded}
\begin{Highlighting}[]
\CommentTok{\# Caret itself needs e1071}
\FunctionTok{library}\NormalTok{(e1071)}

\FunctionTok{library}\NormalTok{(gbm)}
\FunctionTok{library}\NormalTok{(plyr)}

\CommentTok{\# LogitBoost}
\FunctionTok{library}\NormalTok{(caTools)}

\CommentTok{\# C5.0}
\FunctionTok{library}\NormalTok{(C50)}

\CommentTok{\# ranger, rf}
\FunctionTok{library}\NormalTok{(ranger)}
\FunctionTok{library}\NormalTok{(dplyr)}
\FunctionTok{library}\NormalTok{(randomForest)}

\CommentTok{\# naive\_bayes}
\FunctionTok{library}\NormalTok{(naivebayes)}

\CommentTok{\# mlp, mlpMl etc.}
\FunctionTok{library}\NormalTok{(RSNNS)}

\CommentTok{\# nnet}
\FunctionTok{library}\NormalTok{(nnet)}

\CommentTok{\# svmPoly, svmRadial etc.}
\FunctionTok{library}\NormalTok{(kernlab)}

\CommentTok{\# xgbTree, xgbLinear, xgbDART}
\FunctionTok{library}\NormalTok{(xgboost)}


\NormalTok{results\_sl }\OtherTok{\textless{}{-}} \FunctionTok{loadResultsOrCompute}\NormalTok{(}\StringTok{"../results/sl.rds"}\NormalTok{, }\AttributeTok{computeExpr =}\NormalTok{ \{}
  \FunctionTok{doWithParallelCluster}\NormalTok{(}\AttributeTok{expr =}\NormalTok{ \{}
\NormalTok{    resList }\OtherTok{\textless{}{-}} \FunctionTok{list}\NormalTok{()}
\NormalTok{    methods }\OtherTok{\textless{}{-}} \FunctionTok{c}\NormalTok{(}\StringTok{"gbm"}\NormalTok{, }\StringTok{"LogitBoost"}\NormalTok{, }\StringTok{"C5.0"}\NormalTok{, }\StringTok{"rf"}\NormalTok{, }\StringTok{"ranger"}\NormalTok{, }\StringTok{"naive\_bayes"}\NormalTok{,}
      \StringTok{"mlp"}\NormalTok{, }\StringTok{"nnet"}\NormalTok{, }\StringTok{"svmPoly"}\NormalTok{, }\StringTok{"svmRadial"}\NormalTok{, }\StringTok{"xgbTree"}\NormalTok{, }\StringTok{"xgbDART"}\NormalTok{, }\StringTok{"xgbLinear"}\NormalTok{,}
      \StringTok{"null"}\NormalTok{)}

    \ControlFlowTok{for}\NormalTok{ (method }\ControlFlowTok{in}\NormalTok{ methods) \{}
\NormalTok{      resList[[method]] }\OtherTok{\textless{}{-}}\NormalTok{ base}\SpecialCharTok{::}\FunctionTok{tryCatch}\NormalTok{(\{}
\NormalTok{        caret}\SpecialCharTok{::}\FunctionTok{train}\NormalTok{(label }\SpecialCharTok{\textasciitilde{}}\NormalTok{ ., }\AttributeTok{data =}\NormalTok{ train\_sl, }\AttributeTok{trControl =}\NormalTok{ tc\_sl, }\AttributeTok{preProcess =} \FunctionTok{c}\NormalTok{(}\StringTok{"center"}\NormalTok{,}
          \StringTok{"scale"}\NormalTok{), }\AttributeTok{method =}\NormalTok{ method, }\AttributeTok{verbose =} \ConstantTok{FALSE}\NormalTok{)}
\NormalTok{      \}, }\AttributeTok{error =} \ControlFlowTok{function}\NormalTok{(cond) cond)}
\NormalTok{    \}}

\NormalTok{    resList}
\NormalTok{  \})}
\NormalTok{\})}
\end{Highlighting}
\end{Shaded}

\hypertarget{several-models-correlation-and-performance}{%
\paragraph{Several models: correlation and performance}\label{several-models-correlation-and-performance}}

The following will give us a correlation matrix of the models' predictions. The goal is to find models with high performance and unrelated predictions, so that they can be combined. The correlations are shown in tables \ref{tab:model-prediction-correlation1} and \ref{tab:model-prediction-correlation2}.

\begin{table}

\caption{\label{tab:model-prediction-correlation1}Correlations of models' predictions (part 1).}
\centering
\begin{tabular}[t]{lrrrrrrr}
\toprule
  & gbm & LogitBoost & C5.0 & rf & ranger & naive\_bayes & mlp\\
\midrule
gbm & 1.0000 & -0.1080 & -0.0689 & -0.0565 & 0.1566 & 0.0235 & 0.0652\\
LogitBoost & -0.1080 & 1.0000 & -0.0281 & -0.0979 & -0.0054 & -0.0587 & 0.1126\\
C5.0 & -0.0689 & -0.0281 & 1.0000 & -0.0565 & -0.2741 & -0.2880 & 0.4716\\
rf & -0.0565 & -0.0979 & -0.0565 & 1.0000 & -0.1473 & 0.3730 & -0.0736\\
ranger & 0.1566 & -0.0054 & -0.2741 & -0.1473 & 1.0000 & -0.0075 & -0.4465\\
\addlinespace
naive\_bayes & 0.0235 & -0.0587 & -0.2880 & 0.3730 & -0.0075 & 1.0000 & 0.0956\\
mlp & 0.0652 & 0.1126 & 0.4716 & -0.0736 & -0.4465 & 0.0956 & 1.0000\\
nnet & 0.2570 & 0.2848 & 0.0728 & 0.0485 & 0.1069 & -0.1905 & 0.1516\\
svmPoly & -0.1260 & -0.2447 & -0.1573 & -0.2841 & 0.2096 & 0.2341 & 0.0764\\
svmRadial & 0.3147 & -0.1108 & 0.1220 & 0.4284 & -0.2944 & 0.3539 & -0.0435\\
\addlinespace
xgbTree & -0.0134 & 0.2494 & -0.2393 & 0.2115 & -0.1651 & 0.0449 & 0.0007\\
xgbDART & -0.2682 & -0.0111 & 0.1322 & -0.3900 & 0.1307 & 0.2583 & 0.0966\\
xgbLinear & 0.0560 & -0.2253 & 0.0391 & 0.2076 & -0.2322 & -0.1731 & -0.0618\\
null & NA & NA & NA & NA & NA & NA & NA\\
\bottomrule
\end{tabular}
\end{table}

\begin{table}

\caption{\label{tab:model-prediction-correlation2}Correlations of models' predictions (part 2).}
\centering
\begin{tabular}[t]{lrrrrrrr}
\toprule
  & nnet & svmPoly & svmRadial & xgbTree & xgbDART & xgbLinear & null\\
\midrule
gbm & 0.2570 & -0.1260 & 0.3147 & -0.0134 & -0.2682 & 0.0560 & NA\\
LogitBoost & 0.2848 & -0.2447 & -0.1108 & 0.2494 & -0.0111 & -0.2253 & NA\\
C5.0 & 0.0728 & -0.1573 & 0.1220 & -0.2393 & 0.1322 & 0.0391 & NA\\
rf & 0.0485 & -0.2841 & 0.4284 & 0.2115 & -0.3900 & 0.2076 & NA\\
ranger & 0.1069 & 0.2096 & -0.2944 & -0.1651 & 0.1307 & -0.2322 & NA\\
\addlinespace
naive\_bayes & -0.1905 & 0.2341 & 0.3539 & 0.0449 & 0.2583 & -0.1731 & NA\\
mlp & 0.1516 & 0.0764 & -0.0435 & 0.0007 & 0.0966 & -0.0618 & NA\\
nnet & 1.0000 & -0.4363 & 0.0694 & 0.0475 & -0.2216 & -0.0691 & NA\\
svmPoly & -0.4363 & 1.0000 & -0.2885 & -0.1363 & 0.5035 & 0.0248 & NA\\
svmRadial & 0.0694 & -0.2885 & 1.0000 & 0.2385 & -0.1412 & 0.1629 & NA\\
\addlinespace
xgbTree & 0.0475 & -0.1363 & 0.2385 & 1.0000 & -0.1642 & -0.0954 & NA\\
xgbDART & -0.2216 & 0.5035 & -0.1412 & -0.1642 & 1.0000 & -0.3693 & NA\\
xgbLinear & -0.0691 & 0.0248 & 0.1629 & -0.0954 & -0.3693 & 1.0000 & NA\\
null & NA & NA & NA & NA & NA & NA & 1\\
\bottomrule
\end{tabular}
\end{table}

Show for each model the performance during training, and also predict on our validation data to get an idea of their goodness.

\hypertarget{several-models-train-candidates}{%
\paragraph{Several models: train candidates}\label{several-models-train-candidates}}

Using a selection of the best models, we will train a corresponding best model using the best-working hyperparameters. These models will then be evaluated below and used in the stacking attempts.

\begin{Shaded}
\begin{Highlighting}[]
\NormalTok{models\_sl }\OtherTok{\textless{}{-}} \FunctionTok{loadResultsOrCompute}\NormalTok{(}\AttributeTok{file =} \StringTok{"../results/models\_sl.rds"}\NormalTok{, }\AttributeTok{computeExpr =}\NormalTok{ \{}
\NormalTok{  res }\OtherTok{\textless{}{-}} \FunctionTok{list}\NormalTok{()}

  \ControlFlowTok{for}\NormalTok{ (modelName }\ControlFlowTok{in} \FunctionTok{names}\NormalTok{(results\_sl)) \{}
\NormalTok{    m }\OtherTok{\textless{}{-}}\NormalTok{ results\_sl[[modelName]]}

\NormalTok{    res[[modelName]] }\OtherTok{\textless{}{-}} \FunctionTok{caretFitOneModeltoAllData}\NormalTok{(}\AttributeTok{method =}\NormalTok{ modelName, }\AttributeTok{tuneGrid =}\NormalTok{ m}\SpecialCharTok{$}\NormalTok{bestTune,}
      \AttributeTok{data =}\NormalTok{ train\_sl)}
\NormalTok{  \}}

\NormalTok{  res}
\NormalTok{\})}
\end{Highlighting}
\end{Shaded}

As for predicting on validation data, we will use the models that were fit to the entire training data (see table \ref{tab:overview1}).

\begin{Shaded}
\begin{Highlighting}[]
\FunctionTok{generateModelOverview}\NormalTok{(results\_sl, models\_sl, }\AttributeTok{validationData =}\NormalTok{ valid\_sl, }\AttributeTok{label =} \StringTok{"overview1"}\NormalTok{,}
  \AttributeTok{caption =} \StringTok{"Overview of trained models and their performance using validation data."}\NormalTok{)}
\end{Highlighting}
\end{Shaded}

\begin{table}

\caption{\label{tab:overview1}Overview of trained models and their performance using validation data.}
\centering
\begin{tabular}[t]{lrrlrrrr}
\toprule
model & trAcc & trKap & predNA & valAcc\_withNA & valKap\_withNA & valAcc & valKap\\
\midrule
gbm & 0.8031 & 0.7046 & FALSE & 0.8060 & 0.6971 & 0.8060 & 0.6971\\
LogitBoost & 0.8219 & 0.7283 & TRUE & 0.7612 & 0.6200 & 0.8667 & 0.7907\\
C5.0 & 0.8036 & 0.7054 & FALSE & 0.7761 & 0.6480 & 0.7761 & 0.6480\\
rf & 0.8023 & 0.7035 & FALSE & 0.7463 & 0.6057 & 0.7463 & 0.6057\\
ranger & 0.8185 & 0.7277 & FALSE & 0.8209 & 0.7217 & 0.8209 & 0.7217\\
\addlinespace
naive\_bayes & 0.5038 & 0.2558 & FALSE & 0.5672 & 0.3076 & 0.5672 & 0.3076\\
mlp & 0.7646 & 0.6469 & FALSE & 0.7463 & 0.6063 & 0.7463 & 0.6063\\
nnet & 0.7585 & 0.6377 & FALSE & 0.7164 & 0.5541 & 0.7164 & 0.5541\\
svmPoly & 0.7619 & 0.6429 & FALSE & 0.7463 & 0.6098 & 0.7463 & 0.6098\\
svmRadial & 0.7496 & 0.6244 & FALSE & 0.6716 & 0.4898 & 0.6716 & 0.4898\\
\addlinespace
xgbTree & 0.8201 & 0.7302 & FALSE & 0.7164 & 0.5546 & 0.7164 & 0.5546\\
xgbDART & 0.8160 & 0.7240 & FALSE & 0.7164 & 0.5507 & 0.7164 & 0.5507\\
xgbLinear & 0.8169 & 0.7254 & FALSE & 0.7015 & 0.5273 & 0.7015 & 0.5273\\
null & 0.3333 & 0.0000 & FALSE & 0.2090 & 0.0000 & 0.2090 & 0.0000\\
\bottomrule
\end{tabular}
\end{table}

\hypertarget{manual-stacking-of-models}{%
\subsection{Manual stacking of models}\label{manual-stacking-of-models}}

While there are methods to train an ensemble classifier, we are attempting this first manually. Using some of the best and most uncorrelated models from the previous section, we will train a meta model based on these models' outputs. For that, we need a dataset. It will be generated by predicting class probabilities from each single model.

\begin{Shaded}
\begin{Highlighting}[]
\NormalTok{data\_stack\_train\_sl }\OtherTok{\textless{}{-}} \FunctionTok{data.frame}\NormalTok{(}\FunctionTok{matrix}\NormalTok{(}\AttributeTok{ncol =} \DecValTok{0}\NormalTok{, }\AttributeTok{nrow =} \FunctionTok{nrow}\NormalTok{(train\_sl)))}
\NormalTok{data\_stack\_valid\_sl }\OtherTok{\textless{}{-}} \FunctionTok{data.frame}\NormalTok{(}\FunctionTok{matrix}\NormalTok{(}\AttributeTok{ncol =} \DecValTok{0}\NormalTok{, }\AttributeTok{nrow =} \FunctionTok{nrow}\NormalTok{(valid\_sl)))}

\CommentTok{\# The name of the models to use from the previous section: stack\_manual\_models}
\CommentTok{\# \textless{}{-} names(results\_sl)[ !(names(results\_sl) \%in\% c(\textquotesingle{}naive\_bayes\textquotesingle{}, \textquotesingle{}mlp\textquotesingle{},}
\CommentTok{\# \textquotesingle{}nnet\textquotesingle{}, \textquotesingle{}svmPoly\textquotesingle{}, \textquotesingle{}svmRadial\textquotesingle{}, \textquotesingle{}xgbTree\textquotesingle{}, \textquotesingle{}xgbLinear\textquotesingle{}))] stack\_manual\_models}
\CommentTok{\# \textless{}{-} c(\textquotesingle{}LogitBoost\textquotesingle{}, \textquotesingle{}gbm\textquotesingle{}, \textquotesingle{}xgbDART\textquotesingle{}, \textquotesingle{}mlp\textquotesingle{}) \# \textless{}{-} This appears to work best}
\NormalTok{stack\_manual\_models }\OtherTok{\textless{}{-}} \FunctionTok{c}\NormalTok{(}\StringTok{"LogitBoost"}\NormalTok{, }\StringTok{"gbm"}\NormalTok{, }\StringTok{"ranger"}\NormalTok{)  }\CommentTok{\# \textless{}{-} This appears to work best}


\ControlFlowTok{for}\NormalTok{ (modelName }\ControlFlowTok{in}\NormalTok{ stack\_manual\_models) \{}
\NormalTok{  m }\OtherTok{\textless{}{-}}\NormalTok{ models\_sl[[modelName]]}

\NormalTok{  preds }\OtherTok{\textless{}{-}} \FunctionTok{tryCatch}\NormalTok{(\{}
    \FunctionTok{predict}\NormalTok{(m, train\_sl[, }\SpecialCharTok{!}\NormalTok{(}\FunctionTok{names}\NormalTok{(train\_sl) }\SpecialCharTok{\%in\%} \FunctionTok{c}\NormalTok{(}\StringTok{"label"}\NormalTok{))], }\AttributeTok{type =} \StringTok{"prob"}\NormalTok{)}
\NormalTok{  \}, }\AttributeTok{error =} \ControlFlowTok{function}\NormalTok{(cond) cond)}

\NormalTok{  preds\_valid }\OtherTok{\textless{}{-}} \FunctionTok{tryCatch}\NormalTok{(\{}
    \FunctionTok{predict}\NormalTok{(m, valid\_sl[, }\SpecialCharTok{!}\NormalTok{(}\FunctionTok{names}\NormalTok{(valid\_sl) }\SpecialCharTok{\%in\%} \FunctionTok{c}\NormalTok{(}\StringTok{"label"}\NormalTok{))], }\AttributeTok{type =} \StringTok{"prob"}\NormalTok{)}
\NormalTok{  \}, }\AttributeTok{error =} \ControlFlowTok{function}\NormalTok{(cond) cond)}

  \ControlFlowTok{if}\NormalTok{ (}\FunctionTok{any}\NormalTok{(}\FunctionTok{class}\NormalTok{(preds) }\SpecialCharTok{\%in\%} \FunctionTok{c}\NormalTok{(}\StringTok{"simpleError"}\NormalTok{, }\StringTok{"error"}\NormalTok{, }\StringTok{"condition"}\NormalTok{))) \{}
    \FunctionTok{print}\NormalTok{(}\FunctionTok{paste0}\NormalTok{(}\StringTok{"Cannot predict class probabilities for: "}\NormalTok{, modelName))}
\NormalTok{  \} }\ControlFlowTok{else}\NormalTok{ \{}
    \FunctionTok{colnames}\NormalTok{(preds) }\OtherTok{\textless{}{-}} \FunctionTok{paste0}\NormalTok{(}\FunctionTok{colnames}\NormalTok{(preds), }\StringTok{"\_"}\NormalTok{, modelName)}
    \FunctionTok{colnames}\NormalTok{(preds\_valid) }\OtherTok{\textless{}{-}} \FunctionTok{paste0}\NormalTok{(}\FunctionTok{colnames}\NormalTok{(preds\_valid), }\StringTok{"\_"}\NormalTok{, modelName)}

\NormalTok{    data\_stack\_train\_sl }\OtherTok{\textless{}{-}} \FunctionTok{cbind}\NormalTok{(data\_stack\_train\_sl, preds)}
\NormalTok{    data\_stack\_valid\_sl }\OtherTok{\textless{}{-}} \FunctionTok{cbind}\NormalTok{(data\_stack\_valid\_sl, preds\_valid)}
\NormalTok{  \}}
\NormalTok{\}}

\CommentTok{\# Let\textquotesingle{}s append the label{-}column:}
\NormalTok{data\_stack\_train\_sl}\SpecialCharTok{$}\NormalTok{label }\OtherTok{\textless{}{-}}\NormalTok{ train\_sl}\SpecialCharTok{$}\NormalTok{label}
\NormalTok{data\_stack\_valid\_sl}\SpecialCharTok{$}\NormalTok{label }\OtherTok{\textless{}{-}}\NormalTok{ valid\_sl}\SpecialCharTok{$}\NormalTok{label}
\end{Highlighting}
\end{Shaded}

Now that we have the data prepared for our manual ensemble, let's attempt to train some models.

\hypertarget{manual-neural-network}{%
\subsubsection{Manual neural network}\label{manual-neural-network}}

Before going back to caret, let's try a neural network the manual way.

\begin{Shaded}
\begin{Highlighting}[]
\FunctionTok{library}\NormalTok{(neuralnet)}
\FunctionTok{library}\NormalTok{(e1071)}

\NormalTok{nnet }\OtherTok{\textless{}{-}} \FunctionTok{loadResultsOrCompute}\NormalTok{(}\StringTok{"../results/nnet.rds"}\NormalTok{, }\AttributeTok{computeExpr =}\NormalTok{ \{}
  \FunctionTok{set.seed}\NormalTok{(}\DecValTok{49374}\NormalTok{)}
\NormalTok{  neuralnet}\SpecialCharTok{::}\FunctionTok{neuralnet}\NormalTok{(}\AttributeTok{formula =}\NormalTok{ label }\SpecialCharTok{\textasciitilde{}}\NormalTok{ ., }\AttributeTok{data =}\NormalTok{ data\_stack\_train\_sl, }\AttributeTok{act.fct =} \ControlFlowTok{function}\NormalTok{(x) }\FloatTok{1.5} \SpecialCharTok{*}
\NormalTok{    x }\SpecialCharTok{*} \FunctionTok{sigmoid}\NormalTok{(x), }\AttributeTok{hidden =} \FunctionTok{c}\NormalTok{(}\DecValTok{3}\NormalTok{), }\AttributeTok{threshold =} \FloatTok{0.005}\NormalTok{, }\AttributeTok{stepmax =} \FloatTok{2e+05}\NormalTok{, }\AttributeTok{lifesign =} \ControlFlowTok{if}\NormalTok{ (}\ConstantTok{FALSE}\NormalTok{)}
    \StringTok{"full"} \ControlFlowTok{else} \StringTok{"minimal"}\NormalTok{)}
\NormalTok{\})}
\end{Highlighting}
\end{Shaded}

The network has the following structure:

\begin{Shaded}
\begin{Highlighting}[]
\FunctionTok{plot}\NormalTok{(nnet, }\AttributeTok{rep =} \StringTok{"best"}\NormalTok{)}
\end{Highlighting}
\end{Shaded}

\includegraphics{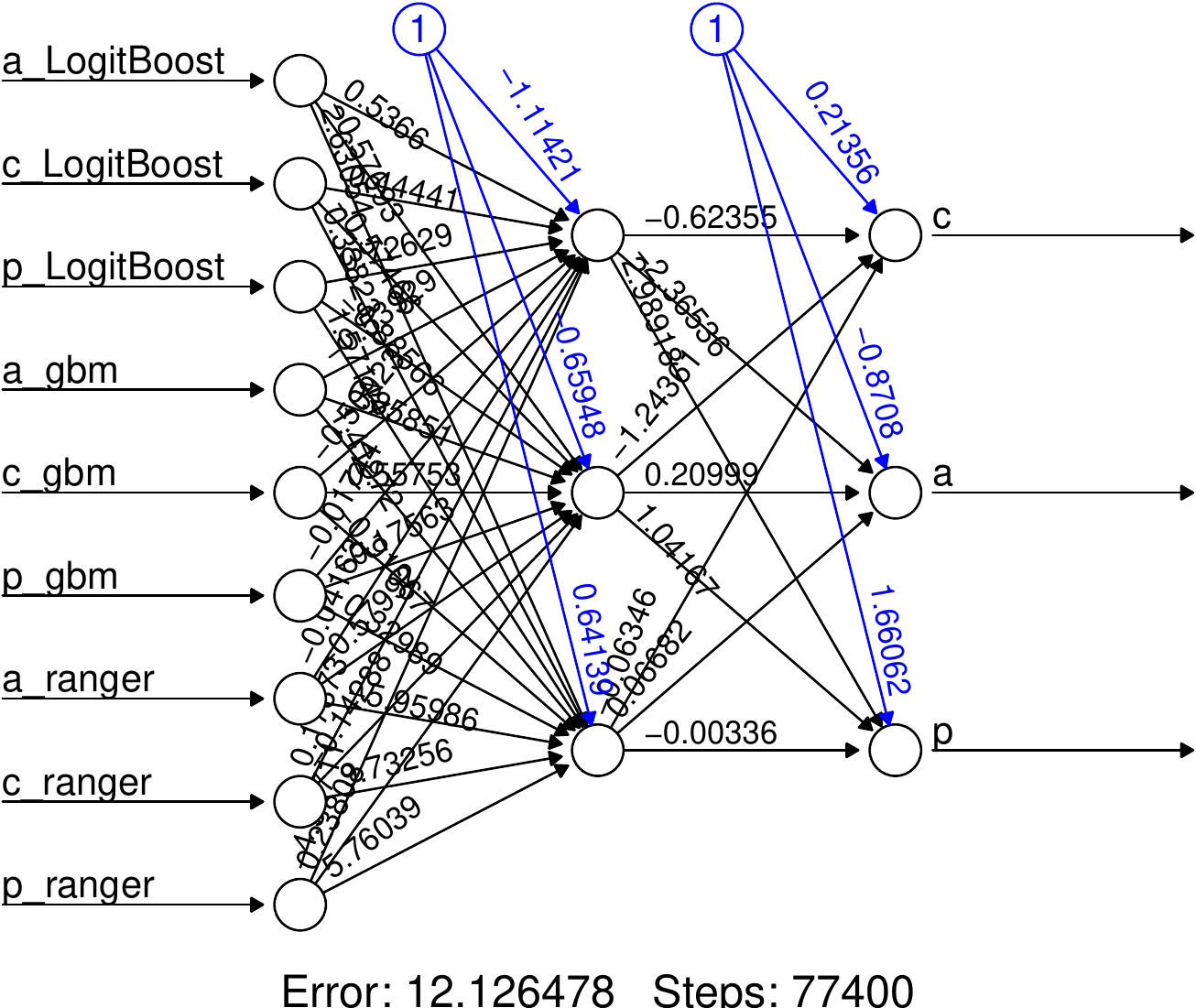}

\begin{Shaded}
\begin{Highlighting}[]
\NormalTok{nnet\_pred }\OtherTok{\textless{}{-}} \FunctionTok{predict}\NormalTok{(nnet, data\_stack\_valid\_sl)}
\FunctionTok{colnames}\NormalTok{(nnet\_pred) }\OtherTok{\textless{}{-}} \FunctionTok{levels}\NormalTok{(valid\_sl}\SpecialCharTok{$}\NormalTok{label)}

\NormalTok{nnet\_pred\_label }\OtherTok{\textless{}{-}} \FunctionTok{factor}\NormalTok{(}\AttributeTok{x =} \FunctionTok{levels}\NormalTok{(valid\_sl}\SpecialCharTok{$}\NormalTok{label)[}\FunctionTok{apply}\NormalTok{(nnet\_pred, }\DecValTok{1}\NormalTok{, which.max)],}
  \AttributeTok{levels =} \FunctionTok{levels}\NormalTok{(valid\_sl}\SpecialCharTok{$}\NormalTok{label))}

\NormalTok{caret}\SpecialCharTok{::}\FunctionTok{confusionMatrix}\NormalTok{(valid\_sl}\SpecialCharTok{$}\NormalTok{label, nnet\_pred\_label)}
\end{Highlighting}
\end{Shaded}

\begin{verbatim}
## Confusion Matrix and Statistics
## 
##           Reference
## Prediction  a  c  p
##          a 12  2  0
##          c  1 21  5
##          p  1  3 22
## 
## Overall Statistics
##                                          
##                Accuracy : 0.8209         
##                  95% CI : (0.708, 0.9039)
##     No Information Rate : 0.403          
##     P-Value [Acc > NIR] : 2.801e-12      
##                                          
##                   Kappa : 0.7217         
##                                          
##  Mcnemar's Test P-Value : 0.6077         
## 
## Statistics by Class:
## 
##                      Class: a Class: c Class: p
## Sensitivity            0.8571   0.8077   0.8148
## Specificity            0.9623   0.8537   0.9000
## Pos Pred Value         0.8571   0.7778   0.8462
## Neg Pred Value         0.9623   0.8750   0.8780
## Prevalence             0.2090   0.3881   0.4030
## Detection Rate         0.1791   0.3134   0.3284
## Detection Prevalence   0.2090   0.4030   0.3881
## Balanced Accuracy      0.9097   0.8307   0.8574
\end{verbatim}

While it works, the results are not better than those from the individual models.

\hypertarget{manual-stack-ms-using-caret}{%
\subsubsection{Manual stack (ms) using caret}\label{manual-stack-ms-using-caret}}

Let's attempt to learn a meta-model using caret.

For the next overview, again, fit the selected single models using their best tune and all available training data.

Now show the overview (table \ref{tab:overview2}):

\begin{Shaded}
\begin{Highlighting}[]
\FunctionTok{generateModelOverview}\NormalTok{(results\_ms, models\_ms, }\AttributeTok{validationData =}\NormalTok{ data\_stack\_valid\_sl,}
  \AttributeTok{label =} \StringTok{"overview2"}\NormalTok{, }\AttributeTok{caption =} \StringTok{"Results of using a manual stacking of models."}\NormalTok{)}
\end{Highlighting}
\end{Shaded}

\begin{table}

\caption{\label{tab:overview2}Results of using a manual stacking of models.}
\centering
\begin{tabular}[t]{lrrlrrrr}
\toprule
model & trAcc & trKap & predNA & valAcc\_withNA & valKap\_withNA & valAcc & valKap\\
\midrule
gbm & 0.9928 & 0.9892 & FALSE & 0.8209 & 0.7217 & 0.8209 & 0.7217\\
LogitBoost & 0.9931 & 0.9896 & FALSE & 0.7910 & 0.6768 & 0.7910 & 0.6768\\
ranger & 0.9929 & 0.9894 & FALSE & 0.8209 & 0.7217 & 0.8209 & 0.7217\\
mlp & 0.9932 & 0.9898 & FALSE & 0.8358 & 0.7449 & 0.8358 & 0.7449\\
nnet & 0.9931 & 0.9896 & FALSE & 0.8358 & 0.7449 & 0.8358 & 0.7449\\
\addlinespace
svmRadial & 0.9940 & 0.9910 & FALSE & 0.7612 & 0.6288 & 0.7612 & 0.6288\\
\bottomrule
\end{tabular}
\end{table}

The overview for all models, using oversampled training data, was this (table \ref{tab:overview3}):

\begin{Shaded}
\begin{Highlighting}[]
\NormalTok{results\_ms\_all }\OtherTok{\textless{}{-}} \FunctionTok{readRDS}\NormalTok{(}\StringTok{"../results/ms\_all.rds"}\NormalTok{)}
\NormalTok{models\_ms\_all }\OtherTok{\textless{}{-}} \FunctionTok{readRDS}\NormalTok{(}\StringTok{"../results/models\_ms\_all.rds"}\NormalTok{)}
\end{Highlighting}
\end{Shaded}

\begin{Shaded}
\begin{Highlighting}[]
\FunctionTok{generateModelOverview}\NormalTok{(results\_ms\_all, models\_ms\_all, }\AttributeTok{validationData =}\NormalTok{ data\_stack\_valid\_sl,}
  \AttributeTok{label =} \StringTok{"overview3"}\NormalTok{, }\AttributeTok{caption =} \StringTok{"Results of using all models and using oversampled data."}\NormalTok{)}
\end{Highlighting}
\end{Shaded}

\begin{table}

\caption{\label{tab:overview3}Results of using all models and using oversampled data.}
\centering
\begin{tabular}[t]{lrrlrrrr}
\toprule
model & trAcc & trKap & predNA & valAcc\_withNA & valKap\_withNA & valAcc & valKap\\
\midrule
gbm & 0.9932 & 0.9898 & FALSE & 0.8209 & 0.7217 & 0.8209 & 0.7217\\
LogitBoost & 0.9931 & 0.9896 & FALSE & 0.7910 & 0.6768 & 0.7910 & 0.6768\\
C5.0 & 0.9922 & 0.9883 & FALSE & 0.8060 & 0.6998 & 0.8060 & 0.6998\\
ranger & 0.9932 & 0.9898 & FALSE & 0.8209 & 0.7217 & 0.8209 & 0.7217\\
rf & 0.9932 & 0.9898 & FALSE & 0.8209 & 0.7217 & 0.8209 & 0.7217\\
\addlinespace
naive\_bayes & 0.9909 & 0.9863 & FALSE & 0.8060 & 0.6973 & 0.8060 & 0.6973\\
mlp & 0.9929 & 0.9894 & FALSE & 0.7910 & 0.6750 & 0.7910 & 0.6750\\
nnet & 0.9929 & 0.9894 & FALSE & 0.8358 & 0.7449 & 0.8358 & 0.7449\\
svmPoly & 0.9942 & 0.9913 & FALSE & 0.8060 & 0.6999 & 0.8060 & 0.6999\\
svmRadial & 0.9941 & 0.9912 & FALSE & 0.7761 & 0.6534 & 0.7761 & 0.6534\\
\addlinespace
xgbTree & 0.9931 & 0.9896 & FALSE & 0.8060 & 0.6973 & 0.8060 & 0.6973\\
xgbDART & 0.9937 & 0.9906 & FALSE & 0.8060 & 0.6973 & 0.8060 & 0.6973\\
xgbLinear & 0.9929 & 0.9894 & FALSE & 0.8209 & 0.7217 & 0.8209 & 0.7217\\
\bottomrule
\end{tabular}
\end{table}

It appears that the manual stacking was slightly useful, and we decide to use the \texttt{nnet} meta-model, that is based on the single models LogitBoost, gbm, ranger, as the final models. Remember that the single models produce predictions as to the class membership on the original data, and these are fed into the meta-model (the pipeline is: predict class memberships (once using each single model), combine all votes into a new dataset, predict final label based on these votes (using the meta model)).

\begin{Shaded}
\begin{Highlighting}[]
\NormalTok{create\_final\_model }\OtherTok{\textless{}{-}} \ControlFlowTok{function}\NormalTok{() \{}
  \CommentTok{\# The meta{-}model from the manual stacking:}
\NormalTok{  meta\_model }\OtherTok{\textless{}{-}}\NormalTok{ models\_ms}\SpecialCharTok{$}\NormalTok{nnet}
  \CommentTok{\# The single models from earlier training:}
\NormalTok{  single\_models }\OtherTok{\textless{}{-}}\NormalTok{ models\_sl[stack\_manual\_models]}

\NormalTok{  predict\_class\_membership }\OtherTok{\textless{}{-}} \ControlFlowTok{function}\NormalTok{(data, }\AttributeTok{modelList =}\NormalTok{ single\_models, }\AttributeTok{labelCol =} \StringTok{"label"}\NormalTok{) \{}
\NormalTok{    dataLabel }\OtherTok{\textless{}{-}} \ControlFlowTok{if}\NormalTok{ (labelCol }\SpecialCharTok{\%in\%} \FunctionTok{colnames}\NormalTok{(data)) \{}
\NormalTok{      data[[labelCol]]}
\NormalTok{    \} }\ControlFlowTok{else}\NormalTok{ \{}
      \FunctionTok{matrix}\NormalTok{(}\AttributeTok{ncol =} \DecValTok{0}\NormalTok{, }\AttributeTok{nrow =} \FunctionTok{nrow}\NormalTok{(data))}
\NormalTok{    \}}
\NormalTok{    data }\OtherTok{\textless{}{-}}\NormalTok{ data[, }\SpecialCharTok{!}\NormalTok{(}\FunctionTok{names}\NormalTok{(data) }\SpecialCharTok{\%in\%}\NormalTok{ labelCol)]}
\NormalTok{    dataCM }\OtherTok{\textless{}{-}} \FunctionTok{data.frame}\NormalTok{(}\FunctionTok{matrix}\NormalTok{(}\AttributeTok{ncol =} \DecValTok{0}\NormalTok{, }\AttributeTok{nrow =} \FunctionTok{nrow}\NormalTok{(data)))}

    \ControlFlowTok{for}\NormalTok{ (modelName }\ControlFlowTok{in} \FunctionTok{names}\NormalTok{(modelList)) \{}
\NormalTok{      m }\OtherTok{\textless{}{-}}\NormalTok{ modelList[[modelName]]}
\NormalTok{      temp }\OtherTok{\textless{}{-}}\NormalTok{ stats}\SpecialCharTok{::}\FunctionTok{predict}\NormalTok{(m, data, }\AttributeTok{type =} \StringTok{"prob"}\NormalTok{)}
      \FunctionTok{colnames}\NormalTok{(temp) }\OtherTok{\textless{}{-}} \FunctionTok{paste0}\NormalTok{(}\FunctionTok{colnames}\NormalTok{(temp), }\StringTok{"\_"}\NormalTok{, modelName)}
\NormalTok{      dataCM }\OtherTok{\textless{}{-}} \FunctionTok{cbind}\NormalTok{(dataCM, temp)}
\NormalTok{    \}}

    \FunctionTok{return}\NormalTok{(}\FunctionTok{cbind}\NormalTok{(dataCM, dataLabel))}
\NormalTok{  \}}

\NormalTok{  predict }\OtherTok{\textless{}{-}} \ControlFlowTok{function}\NormalTok{(data, }\AttributeTok{labelCol =} \StringTok{"label"}\NormalTok{, }\AttributeTok{type =} \FunctionTok{c}\NormalTok{(}\StringTok{"raw"}\NormalTok{, }\StringTok{"prob"}\NormalTok{, }\StringTok{"both"}\NormalTok{)) \{}
\NormalTok{    type }\OtherTok{\textless{}{-}} \ControlFlowTok{if}\NormalTok{ (}\FunctionTok{missing}\NormalTok{(type))}
\NormalTok{      type[}\DecValTok{1}\NormalTok{] }\ControlFlowTok{else}\NormalTok{ type}
\NormalTok{    dataCM }\OtherTok{\textless{}{-}} \FunctionTok{predict\_class\_membership}\NormalTok{(}\AttributeTok{data =}\NormalTok{ data, }\AttributeTok{labelCol =}\NormalTok{ labelCol)}
\NormalTok{    res }\OtherTok{\textless{}{-}} \FunctionTok{data.frame}\NormalTok{(}\FunctionTok{matrix}\NormalTok{(}\AttributeTok{ncol =} \DecValTok{0}\NormalTok{, }\AttributeTok{nrow =} \FunctionTok{nrow}\NormalTok{(data)))}

\NormalTok{    doRaw }\OtherTok{\textless{}{-}}\NormalTok{ type }\SpecialCharTok{==} \StringTok{"raw"}
\NormalTok{    doProb }\OtherTok{\textless{}{-}}\NormalTok{ type }\SpecialCharTok{==} \StringTok{"prob"}

\NormalTok{    asRaw }\OtherTok{\textless{}{-}}\NormalTok{ stats}\SpecialCharTok{::}\FunctionTok{predict}\NormalTok{(meta\_model, dataCM, }\AttributeTok{type =} \StringTok{"raw"}\NormalTok{)}
\NormalTok{    asProb }\OtherTok{\textless{}{-}}\NormalTok{ stats}\SpecialCharTok{::}\FunctionTok{predict}\NormalTok{(meta\_model, dataCM, }\AttributeTok{type =} \StringTok{"prob"}\NormalTok{)}
    \ControlFlowTok{if}\NormalTok{ (}\FunctionTok{is.factor}\NormalTok{(data[[labelCol]])) \{}
      \FunctionTok{colnames}\NormalTok{(asProb) }\OtherTok{\textless{}{-}} \FunctionTok{levels}\NormalTok{(data[[labelCol]])}
\NormalTok{    \}}

    \ControlFlowTok{if}\NormalTok{ (doRaw) \{}
      \FunctionTok{return}\NormalTok{(asRaw)}
\NormalTok{    \} }\ControlFlowTok{else} \ControlFlowTok{if}\NormalTok{ (doProb) \{}
      \FunctionTok{return}\NormalTok{(asProb)}
\NormalTok{    \}}

    \CommentTok{\# Both:}
\NormalTok{    res }\OtherTok{\textless{}{-}} \FunctionTok{cbind}\NormalTok{(res, asRaw)}
    \FunctionTok{colnames}\NormalTok{(res) }\OtherTok{\textless{}{-}}\NormalTok{ labelCol}
\NormalTok{    res }\OtherTok{\textless{}{-}} \FunctionTok{cbind}\NormalTok{(res, asProb)}
    \FunctionTok{return}\NormalTok{(res)}
\NormalTok{  \}}

  \FunctionTok{return}\NormalTok{(}\FunctionTok{list}\NormalTok{(}\AttributeTok{meta\_model =}\NormalTok{ meta\_model, }\AttributeTok{single\_models =}\NormalTok{ single\_models, }\AttributeTok{predict\_class\_membership =}\NormalTok{ predict\_class\_membership,}
    \AttributeTok{predict =}\NormalTok{ predict))}
\NormalTok{\}}


\NormalTok{final\_model }\OtherTok{\textless{}{-}} \FunctionTok{create\_final\_model}\NormalTok{()}

\FunctionTok{saveRDS}\NormalTok{(final\_model, }\AttributeTok{file =} \StringTok{"../results/final\_model.rds"}\NormalTok{)}
\end{Highlighting}
\end{Shaded}

A quick test of the final model:

\begin{Shaded}
\begin{Highlighting}[]
\NormalTok{caret}\SpecialCharTok{::}\FunctionTok{confusionMatrix}\NormalTok{(final\_model}\SpecialCharTok{$}\FunctionTok{predict}\NormalTok{(train\_sl), train\_sl}\SpecialCharTok{$}\NormalTok{label)}
\end{Highlighting}
\end{Shaded}

\begin{verbatim}
## Confusion Matrix and Statistics
## 
##           Reference
## Prediction   a   c   p
##          a 519   1   0
##          c   0 514   2
##          p   1   5 518
## 
## Overall Statistics
##                                           
##                Accuracy : 0.9942          
##                  95% CI : (0.9891, 0.9974)
##     No Information Rate : 0.3333          
##     P-Value [Acc > NIR] : <2e-16          
##                                           
##                   Kappa : 0.9913          
##                                           
##  Mcnemar's Test P-Value : 0.3496          
## 
## Statistics by Class:
## 
##                      Class: a Class: c Class: p
## Sensitivity            0.9981   0.9885   0.9962
## Specificity            0.9990   0.9981   0.9942
## Pos Pred Value         0.9981   0.9961   0.9885
## Neg Pred Value         0.9990   0.9943   0.9981
## Prevalence             0.3333   0.3333   0.3333
## Detection Rate         0.3327   0.3295   0.3321
## Detection Prevalence   0.3333   0.3308   0.3359
## Balanced Accuracy      0.9986   0.9933   0.9952
\end{verbatim}

\begin{Shaded}
\begin{Highlighting}[]
\FunctionTok{head}\NormalTok{(final\_model}\SpecialCharTok{$}\FunctionTok{predict}\NormalTok{(valid\_sl, }\AttributeTok{type =} \StringTok{"both"}\NormalTok{))}
\end{Highlighting}
\end{Shaded}

\begin{tabular}{l|r|r|r}
\hline
label & a & c & p\\
\hline
c & 0.0012107 & 0.9972771 & 0.0015122\\
\hline
c & 0.0008457 & 0.9979259 & 0.0012284\\
\hline
c & 0.0137335 & 0.9410197 & 0.0452468\\
\hline
a & 0.9983508 & 0.0006940 & 0.0009553\\
\hline
p & 0.0275879 & 0.3715442 & 0.6008679\\
\hline
c & 0.0011113 & 0.9973306 & 0.0015581\\
\hline
\end{tabular}

\begin{Shaded}
\begin{Highlighting}[]
\NormalTok{caret}\SpecialCharTok{::}\FunctionTok{confusionMatrix}\NormalTok{(final\_model}\SpecialCharTok{$}\FunctionTok{predict}\NormalTok{(valid\_sl), valid\_sl}\SpecialCharTok{$}\NormalTok{label)}
\end{Highlighting}
\end{Shaded}

\begin{verbatim}
## Confusion Matrix and Statistics
## 
##           Reference
## Prediction  a  c  p
##          a 12  1  1
##          c  1 22  3
##          p  1  4 22
## 
## Overall Statistics
##                                           
##                Accuracy : 0.8358          
##                  95% CI : (0.7252, 0.9151)
##     No Information Rate : 0.403           
##     P-Value [Acc > NIR] : 3.986e-13       
##                                           
##                   Kappa : 0.7449          
##                                           
##  Mcnemar's Test P-Value : 0.9862          
## 
## Statistics by Class:
## 
##                      Class: a Class: c Class: p
## Sensitivity            0.8571   0.8148   0.8462
## Specificity            0.9623   0.9000   0.8780
## Pos Pred Value         0.8571   0.8462   0.8148
## Neg Pred Value         0.9623   0.8780   0.9000
## Prevalence             0.2090   0.4030   0.3881
## Detection Rate         0.1791   0.3284   0.3284
## Detection Prevalence   0.2090   0.3881   0.4030
## Balanced Accuracy      0.9097   0.8574   0.8621
\end{verbatim}

\hypertarget{creating-an-ensemble-es-using-caretensemble}{%
\subsubsection{Creating an ensemble (es) using caretEnsemble}\label{creating-an-ensemble-es-using-caretensemble}}

The last thing to do is creating an ensemble using \texttt{caretEnsemble}. \textbf{NOTE}: Unfortunately, that package does not supports multi-class problems, and we will not attempt to modify the problem to fit the model at this time. The following tests below do not work.

\begin{Shaded}
\begin{Highlighting}[]
\FunctionTok{library}\NormalTok{(caretEnsemble)}

\NormalTok{tc\_sl\_es }\OtherTok{\textless{}{-}}\NormalTok{ caret}\SpecialCharTok{::}\FunctionTok{trainControl}\NormalTok{(}\AttributeTok{method =} \StringTok{"cv"}\NormalTok{, }\AttributeTok{savePredictions =} \StringTok{"final"}\NormalTok{, }\AttributeTok{classProbs =} \ConstantTok{TRUE}\NormalTok{)}
\end{Highlighting}
\end{Shaded}

Now let's create a list of models we would like to use.

\hypertarget{create-a-linear-ensemble}{%
\paragraph{Create a linear ensemble}\label{create-a-linear-ensemble}}

Using the list of trained models from the previous section, we create an ensemble that is a linear combination of all models.

\begin{Shaded}
\begin{Highlighting}[]
\CommentTok{\# model\_es\_linear \textless{}{-} caretEnsemble::caretStack( all.models = es\_list, method =}
\CommentTok{\# \textquotesingle{}glm\textquotesingle{}, \#metric = \textquotesingle{}Accuracy\textquotesingle{}, trControl = caret::trainControl( classProbs =}
\CommentTok{\# TRUE) ) summary(model\_es\_linear)}
\end{Highlighting}
\end{Shaded}

\hypertarget{some-tests-using-angular}{%
\subsection{Some tests using Angular}\label{some-tests-using-angular}}

Let's load the data we extracted using \emph{Git-Tools} (Hönel 2022) from the Angular repository (begin 2020 - now). After loading, we will predict the maintenance activity and save the file.

\begin{Shaded}
\begin{Highlighting}[]
\CommentTok{\# GitTools.exe {-}r C:\textbackslash{}temp\textbackslash{}angular\textbackslash{} {-}o \textquotesingle{}C:\textbackslash{}temp\textbackslash{}angular.csv\textquotesingle{} {-}s \textquotesingle{}2019{-}01{-}01}
\CommentTok{\# 00:00\textquotesingle{}}
\NormalTok{dateFormat }\OtherTok{\textless{}{-}} \StringTok{"\%Y{-}\%m{-}\%d \%H:\%M:\%S"}
\NormalTok{angularFile }\OtherTok{\textless{}{-}} \StringTok{"../data/angular.csv"}
\NormalTok{angular }\OtherTok{\textless{}{-}} \FunctionTok{read.csv}\NormalTok{(angularFile)}
\NormalTok{temp }\OtherTok{\textless{}{-}}\NormalTok{ final\_model}\SpecialCharTok{$}\FunctionTok{predict}\NormalTok{(}\AttributeTok{data =}\NormalTok{ angular, }\AttributeTok{type =} \StringTok{"both"}\NormalTok{)}
\NormalTok{angular}\SpecialCharTok{$}\NormalTok{label }\OtherTok{\textless{}{-}}\NormalTok{ temp}\SpecialCharTok{$}\NormalTok{label}
\NormalTok{angular}\SpecialCharTok{$}\NormalTok{prob\_a }\OtherTok{\textless{}{-}}\NormalTok{ temp}\SpecialCharTok{$}\NormalTok{a}
\NormalTok{angular}\SpecialCharTok{$}\NormalTok{prob\_c }\OtherTok{\textless{}{-}}\NormalTok{ temp}\SpecialCharTok{$}\NormalTok{c}
\NormalTok{angular}\SpecialCharTok{$}\NormalTok{prob\_p }\OtherTok{\textless{}{-}}\NormalTok{ temp}\SpecialCharTok{$}\NormalTok{p}

\NormalTok{angular}\SpecialCharTok{$}\NormalTok{CommitterTimeObj }\OtherTok{\textless{}{-}} \FunctionTok{strptime}\NormalTok{(angular}\SpecialCharTok{$}\NormalTok{CommitterTime, }\AttributeTok{format =}\NormalTok{ dateFormat)}
\NormalTok{angular}\SpecialCharTok{$}\NormalTok{AuthorTimeObj }\OtherTok{\textless{}{-}} \FunctionTok{strptime}\NormalTok{(angular}\SpecialCharTok{$}\NormalTok{AuthorTime, }\AttributeTok{format =}\NormalTok{ dateFormat)}

\FunctionTok{write.csv}\NormalTok{(angular, }\AttributeTok{file =}\NormalTok{ angularFile, }\AttributeTok{row.names =} \ConstantTok{FALSE}\NormalTok{)}

\FunctionTok{table}\NormalTok{(angular}\SpecialCharTok{$}\NormalTok{label)}
\end{Highlighting}
\end{Shaded}

\begin{verbatim}
## 
##    a    c    p 
## 1533 5819 3862
\end{verbatim}

Let's attempt some straightforward density plots for each activity, using the relative timestamp of each commit.

\begin{Shaded}
\begin{Highlighting}[]
\NormalTok{ggplot2}\SpecialCharTok{::}\FunctionTok{ggplot}\NormalTok{(}
  \CommentTok{\#data = angular[angular$AuthorTimeUnixEpochMilliSecs \textless{}= 1.57e12, ],}
  \AttributeTok{data =}\NormalTok{ angular[angular}\SpecialCharTok{$}\NormalTok{AuthorTimeObj }\SpecialCharTok{\textless{}} \FunctionTok{as.POSIXct}\NormalTok{(}\StringTok{"2019{-}11{-}01"}\NormalTok{), ],}
\NormalTok{  ggplot2}\SpecialCharTok{::}\FunctionTok{aes}\NormalTok{(}
\NormalTok{    AuthorTimeUnixEpochMilliSecs, }\AttributeTok{color =}\NormalTok{ label, }\AttributeTok{fill =}\NormalTok{ label)) }\SpecialCharTok{+}
\NormalTok{  ggplot2}\SpecialCharTok{::}\FunctionTok{geom\_density}\NormalTok{(}\AttributeTok{size =} \DecValTok{1}\NormalTok{, }\AttributeTok{alpha =} \FloatTok{0.5}\NormalTok{)}\CommentTok{\#, position = "fill")}
\end{Highlighting}
\end{Shaded}

\includegraphics{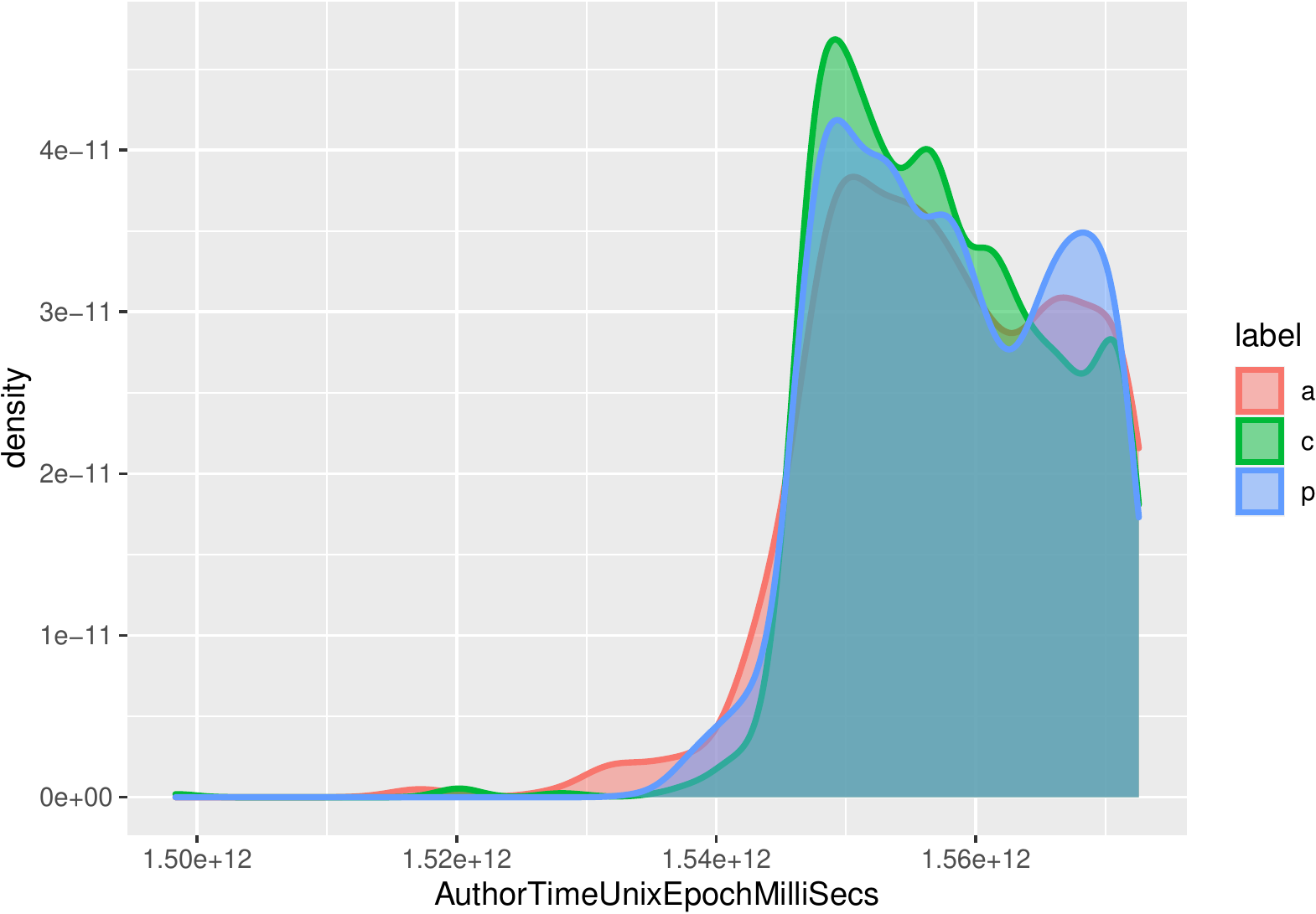}

\begin{Shaded}
\begin{Highlighting}[]
\NormalTok{ggplot2}\SpecialCharTok{::}\FunctionTok{ggplot}\NormalTok{(}
  \CommentTok{\#data = angular[angular$AuthorTimeUnixEpochMilliSecs \textless{}= 1.57e12, ],}
  \AttributeTok{data =}\NormalTok{ angular[angular}\SpecialCharTok{$}\NormalTok{AuthorTimeObj }\SpecialCharTok{\textgreater{}=} \FunctionTok{as.POSIXct}\NormalTok{(}\StringTok{"2019{-}11{-}01"}\NormalTok{), ],}
\NormalTok{  ggplot2}\SpecialCharTok{::}\FunctionTok{aes}\NormalTok{(}
\NormalTok{    AuthorTimeUnixEpochMilliSecs, }\AttributeTok{color =}\NormalTok{ label, }\AttributeTok{fill =}\NormalTok{ label)) }\SpecialCharTok{+}
\NormalTok{  ggplot2}\SpecialCharTok{::}\FunctionTok{geom\_density}\NormalTok{(}\AttributeTok{size =} \DecValTok{1}\NormalTok{, }\AttributeTok{alpha =} \FloatTok{0.5}\NormalTok{)}\CommentTok{\#, position = "fill")}
\end{Highlighting}
\end{Shaded}

\includegraphics{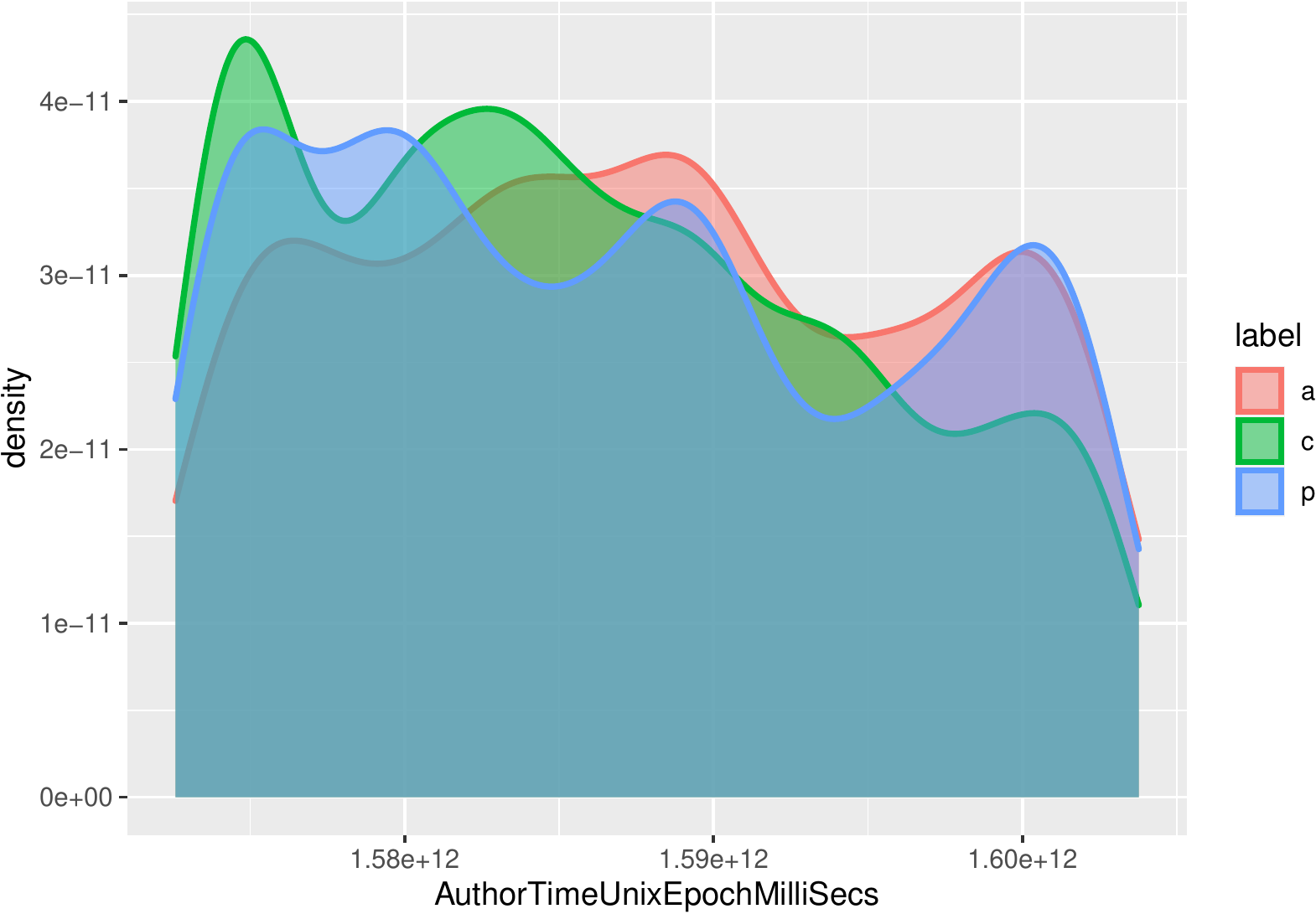}

It appears that the activities after 2019-11-01 are much more balanced. Let's look at a much smaller window:

\begin{Shaded}
\begin{Highlighting}[]
\NormalTok{temp }\OtherTok{\textless{}{-}}\NormalTok{ ggplot2}\SpecialCharTok{::}\FunctionTok{ggplot}\NormalTok{(}\AttributeTok{data =}\NormalTok{ angular[angular}\SpecialCharTok{$}\NormalTok{AuthorTimeObj }\SpecialCharTok{\textgreater{}} \FunctionTok{as.POSIXct}\NormalTok{(}\StringTok{"2020{-}02{-}03"}\NormalTok{) }\SpecialCharTok{\&}
\NormalTok{  angular}\SpecialCharTok{$}\NormalTok{AuthorTimeObj }\SpecialCharTok{\textless{}=} \FunctionTok{as.POSIXct}\NormalTok{(}\StringTok{"2020{-}02{-}23"}\NormalTok{), ], ggplot2}\SpecialCharTok{::}\FunctionTok{aes}\NormalTok{(AuthorTimeUnixEpochMilliSecs,}
  \AttributeTok{color =}\NormalTok{ label, }\AttributeTok{fill =}\NormalTok{ label))}

\NormalTok{temp }\SpecialCharTok{+}\NormalTok{ ggplot2}\SpecialCharTok{::}\FunctionTok{geom\_density}\NormalTok{(}\AttributeTok{size =} \DecValTok{1}\NormalTok{, }\AttributeTok{alpha =} \FloatTok{0.4}\NormalTok{)}
\end{Highlighting}
\end{Shaded}

\includegraphics{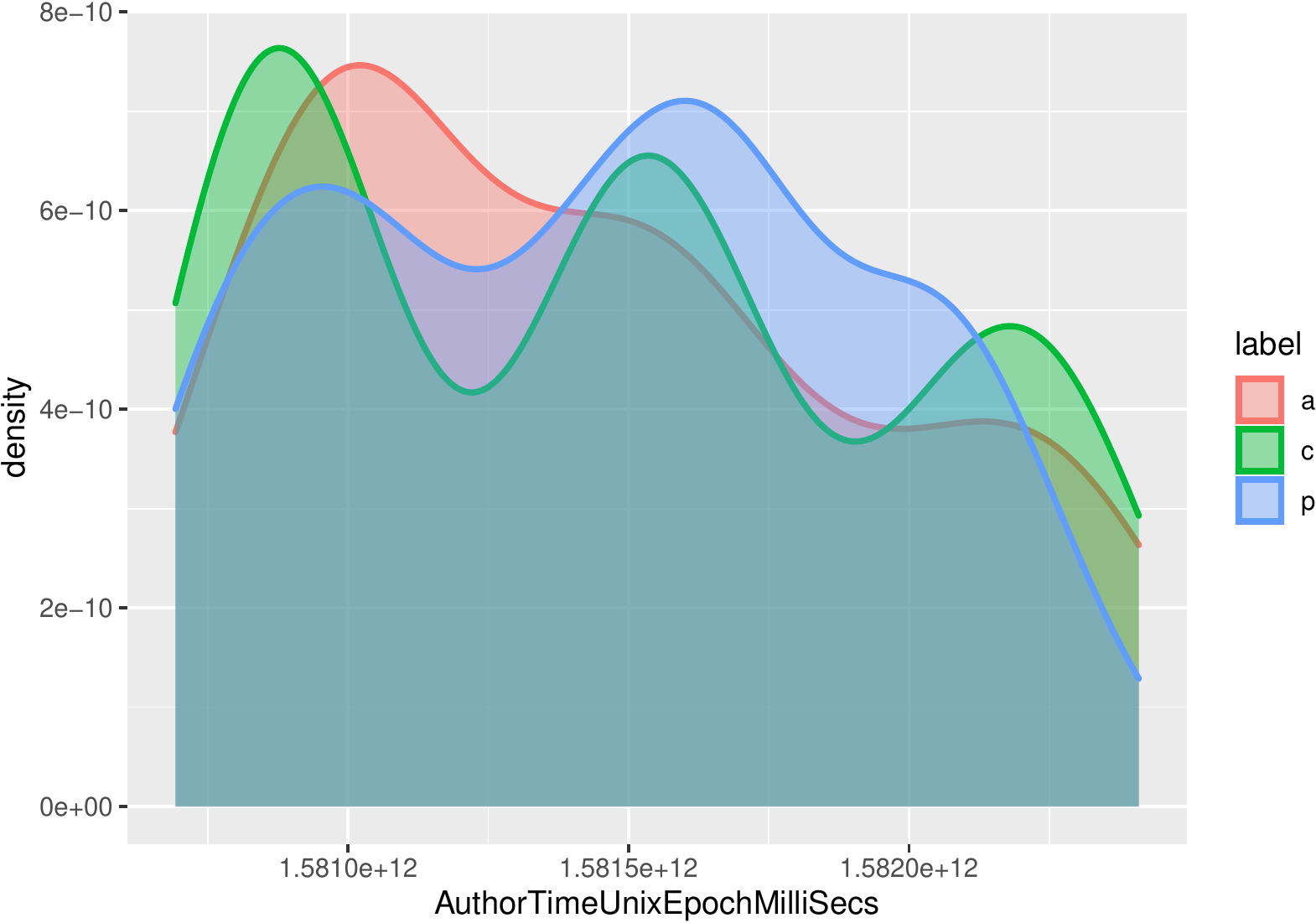}

\begin{Shaded}
\begin{Highlighting}[]
\NormalTok{temp }\SpecialCharTok{+}\NormalTok{ ggplot2}\SpecialCharTok{::}\FunctionTok{geom\_density}\NormalTok{(}\AttributeTok{size =} \DecValTok{1}\NormalTok{, }\AttributeTok{alpha =} \FloatTok{0.4}\NormalTok{, }\AttributeTok{position =} \StringTok{"fill"}\NormalTok{)}
\end{Highlighting}
\end{Shaded}

\includegraphics{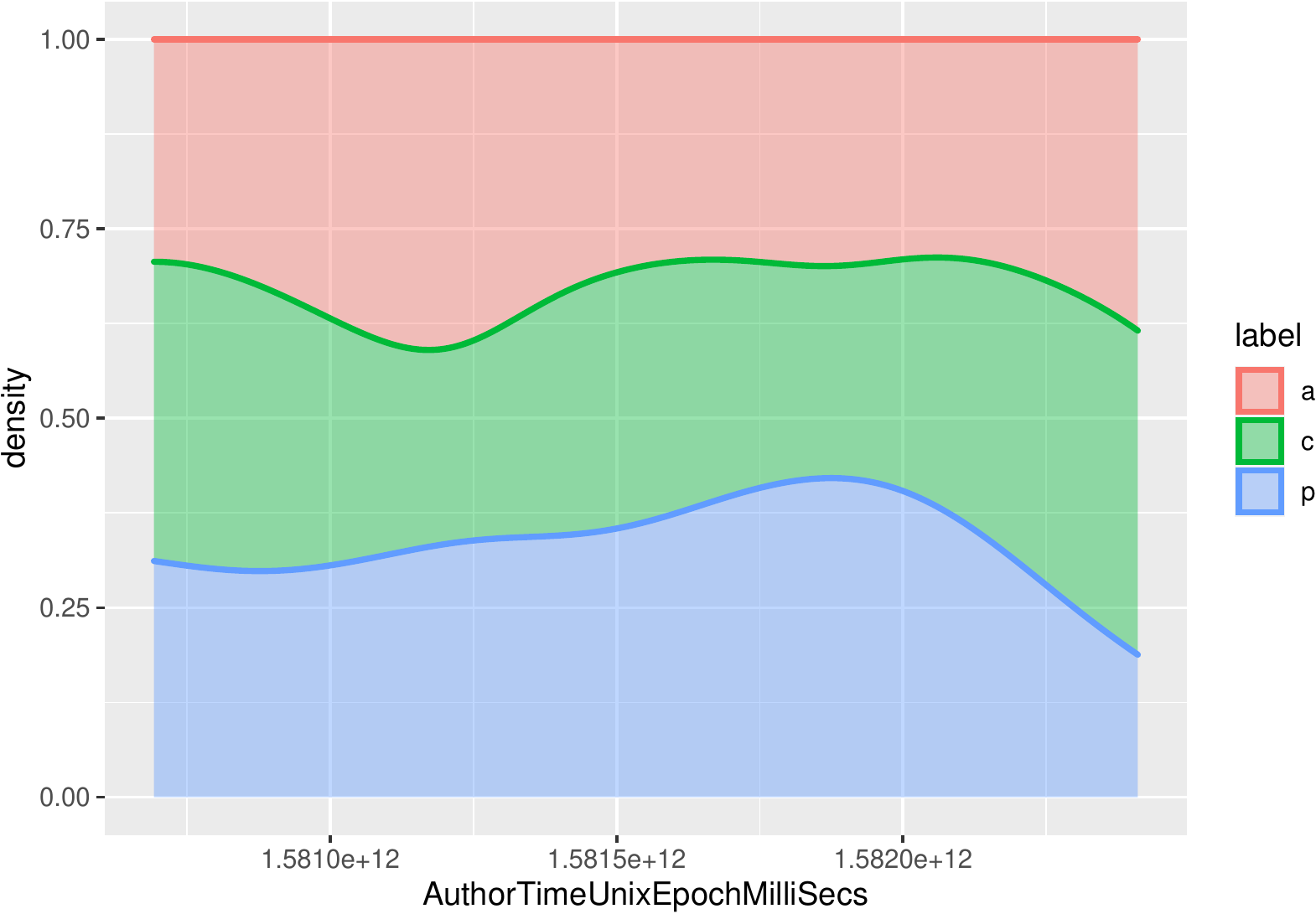}

\begin{Shaded}
\begin{Highlighting}[]
\NormalTok{temp }\SpecialCharTok{+}\NormalTok{ ggplot2}\SpecialCharTok{::}\FunctionTok{geom\_density}\NormalTok{(}\AttributeTok{size =} \DecValTok{1}\NormalTok{, }\AttributeTok{alpha =} \FloatTok{0.4}\NormalTok{, }\AttributeTok{kernel =} \StringTok{"rectangular"}\NormalTok{)}
\end{Highlighting}
\end{Shaded}

\includegraphics{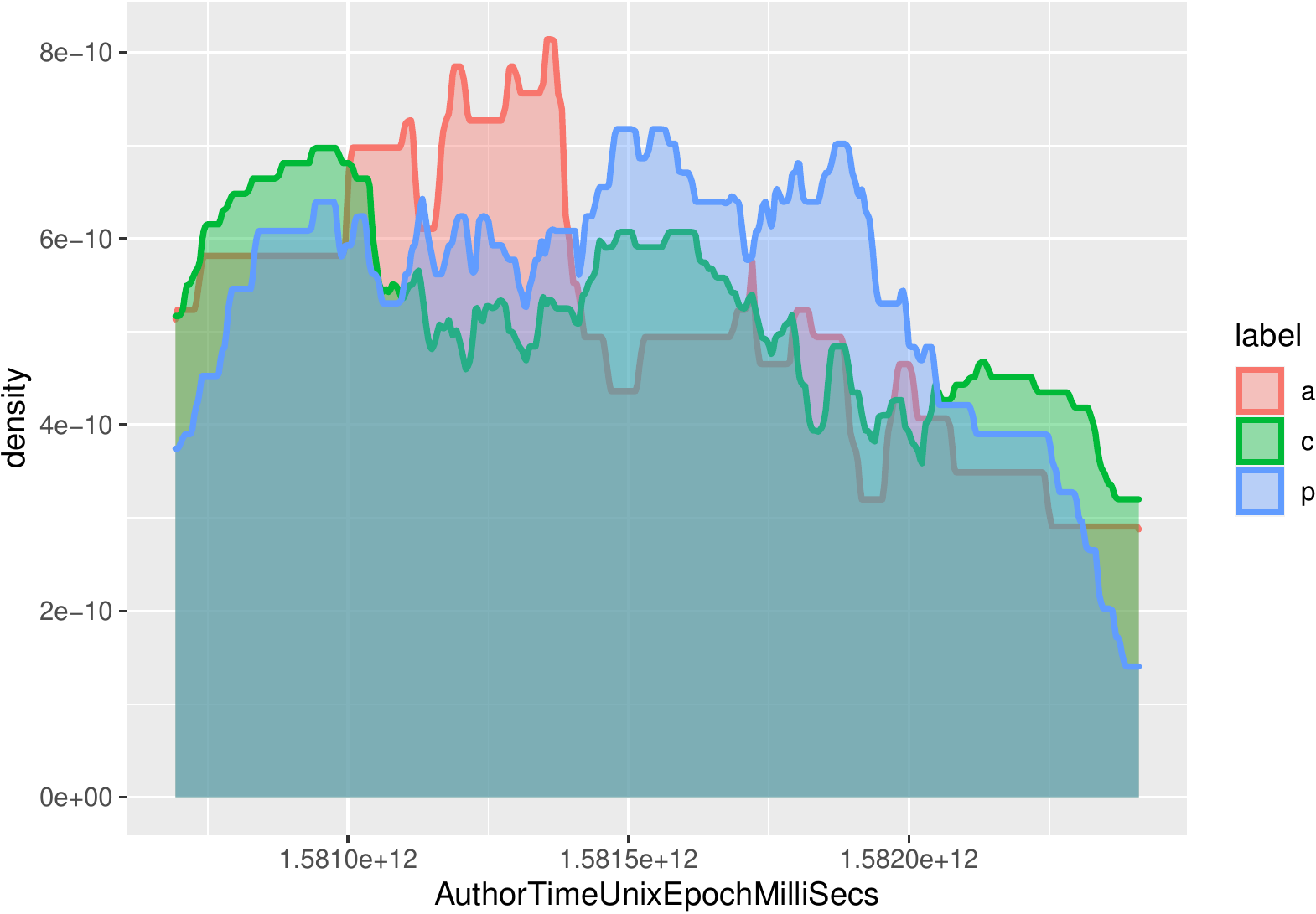}

\begin{Shaded}
\begin{Highlighting}[]
\NormalTok{temp }\SpecialCharTok{+}\NormalTok{ ggplot2}\SpecialCharTok{::}\FunctionTok{geom\_density}\NormalTok{(}\AttributeTok{size =} \DecValTok{1}\NormalTok{, }\AttributeTok{alpha =} \FloatTok{0.4}\NormalTok{, }\AttributeTok{kernel =} \StringTok{"rectangular"}\NormalTok{, }\AttributeTok{position =} \StringTok{"fill"}\NormalTok{)}
\end{Highlighting}
\end{Shaded}

\includegraphics{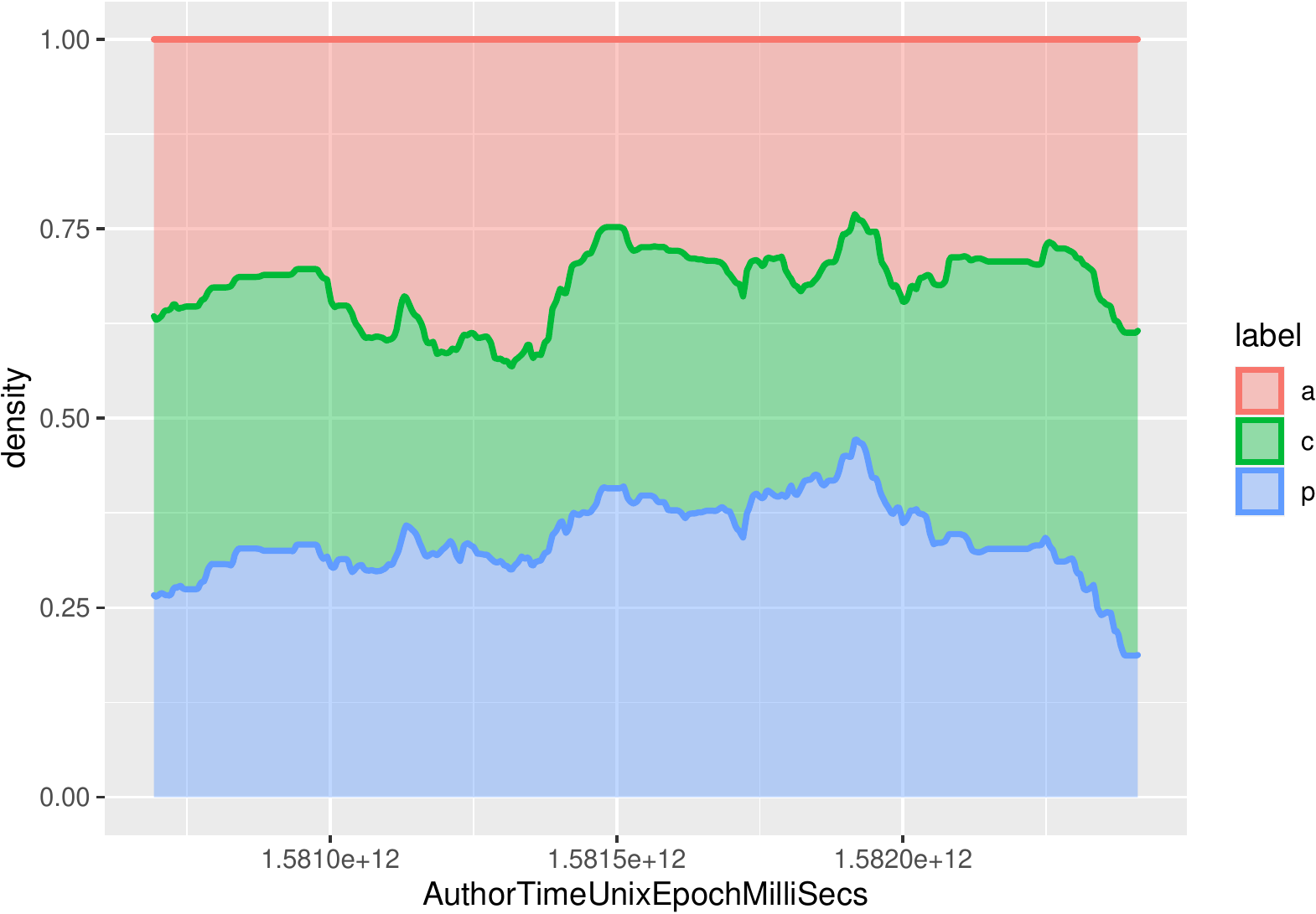}

The above plot is a 3-week snapshot, with weeks starting at Monday, 00:00, and ending at Sunday, 23:59. It appears that each week starts with a somewhat corrective phase. We can see this more easily when using a rectangular Kernel in a non-filled plot.

Here are some attempts with a rolling mean over the class-probabilities:

\begin{Shaded}
\begin{Highlighting}[]
\FunctionTok{library}\NormalTok{(zoo)}

\NormalTok{data }\OtherTok{\textless{}{-}}\NormalTok{ AirPassengers}
\FunctionTok{plot}\NormalTok{(data, }\AttributeTok{main =} \StringTok{"Simple Moving Average (SMA)"}\NormalTok{, }\AttributeTok{ylab =} \StringTok{"Passengers"}\NormalTok{)}
\FunctionTok{lines}\NormalTok{(}\FunctionTok{rollmean}\NormalTok{(data, }\DecValTok{5}\NormalTok{), }\AttributeTok{col =} \StringTok{"blue"}\NormalTok{)}
\FunctionTok{lines}\NormalTok{(}\FunctionTok{rollmean}\NormalTok{(data, }\DecValTok{40}\NormalTok{), }\AttributeTok{col =} \StringTok{"red"}\NormalTok{)}
\FunctionTok{legend}\NormalTok{(}\DecValTok{1950}\NormalTok{, }\DecValTok{600}\NormalTok{, }\AttributeTok{col =} \FunctionTok{c}\NormalTok{(}\StringTok{"black"}\NormalTok{, }\StringTok{"blue"}\NormalTok{, }\StringTok{"red"}\NormalTok{), }\AttributeTok{legend =} \FunctionTok{c}\NormalTok{(}\StringTok{"Raw"}\NormalTok{, }\StringTok{"SMA 5"}\NormalTok{, }\StringTok{"SMA 40"}\NormalTok{),}
  \AttributeTok{lty =} \DecValTok{1}\NormalTok{, }\AttributeTok{cex =} \FloatTok{0.8}\NormalTok{)}
\end{Highlighting}
\end{Shaded}

\includegraphics{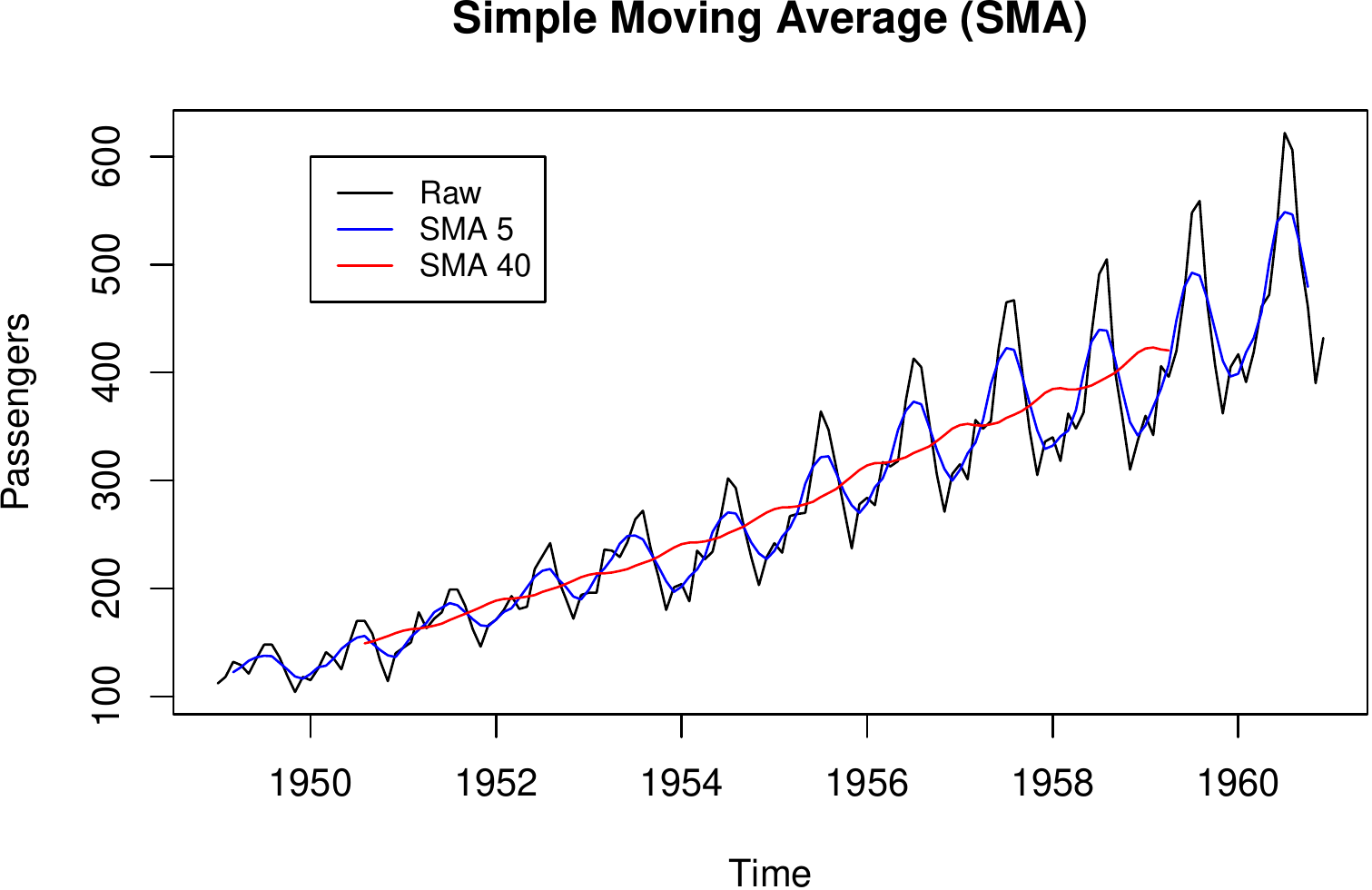}

\begin{Shaded}
\begin{Highlighting}[]
\NormalTok{data }\OtherTok{\textless{}{-}}\NormalTok{ angular[angular}\SpecialCharTok{$}\NormalTok{AuthorTimeObj }\SpecialCharTok{\textgreater{}} \FunctionTok{as.POSIXct}\NormalTok{(}\StringTok{"2020{-}02{-}03"}\NormalTok{) }\SpecialCharTok{\&}\NormalTok{ angular}\SpecialCharTok{$}\NormalTok{AuthorTimeObj }\SpecialCharTok{\textless{}=} \FunctionTok{as.POSIXct}\NormalTok{(}\StringTok{"2020{-}02{-}23"}\NormalTok{),]}

\FunctionTok{plot}\NormalTok{(}\FunctionTok{list}\NormalTok{(}
  \CommentTok{\#x = as.Date(data$AuthorTimeObj),}
  \AttributeTok{x =} \DecValTok{1}\SpecialCharTok{:}\FunctionTok{nrow}\NormalTok{(data),}
  \AttributeTok{y =}\NormalTok{ data}\SpecialCharTok{$}\NormalTok{prob\_a}
\NormalTok{))}
\FunctionTok{lines}\NormalTok{(}\FunctionTok{rollmean}\NormalTok{(data}\SpecialCharTok{$}\NormalTok{prob\_c, }\DecValTok{10}\NormalTok{), }\AttributeTok{col=}\StringTok{\textquotesingle{}blue\textquotesingle{}}\NormalTok{)}
\end{Highlighting}
\end{Shaded}

\includegraphics{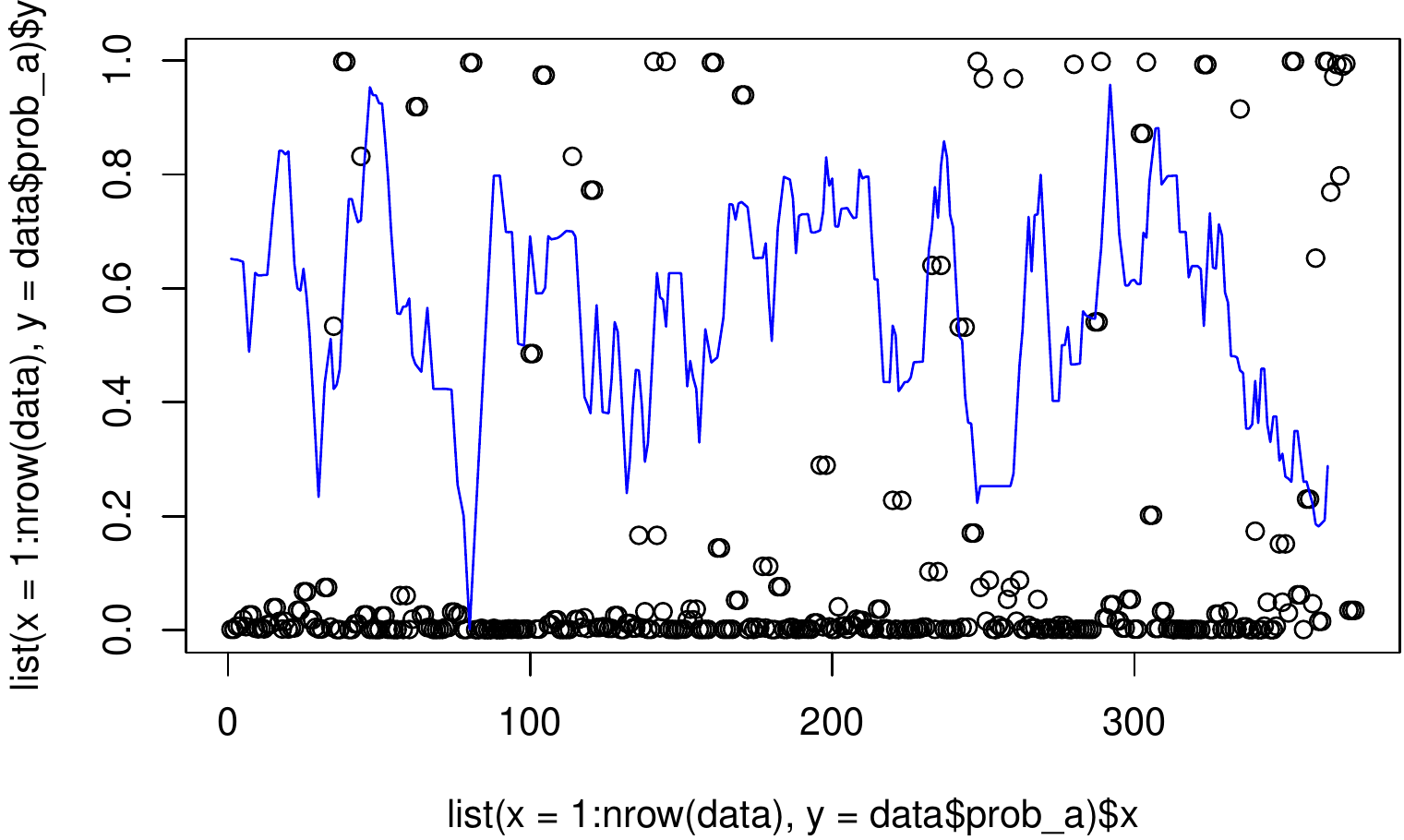}
\clearpage
\hypertarget{references}{%
\section*{References}\label{references}}
\addcontentsline{toc}{section}{References}

\hypertarget{refs}{}
\begin{CSLReferences}{1}{0}
\leavevmode\vadjust pre{\hypertarget{ref-akaike1981likelihood}{}}%
Akaike, Hirotugu. 1981. {``Likelihood of a Model and Information Criteria.''} \emph{Journal of Econometrics} 16 (1): 3--14.

\leavevmode\vadjust pre{\hypertarget{ref-rpkg_Matrix}{}}%
Bates, Douglas, and Martin Maechler. 2019. \emph{Matrix: Sparse and Dense Matrix Classes and Methods}. \url{https://CRAN.R-project.org/package=Matrix}.

\leavevmode\vadjust pre{\hypertarget{ref-rpkg_plsRglm}{}}%
Bertrand, F., and M. Maumy-Bertrand. 2014. \emph{Partial Least Squares Regression for Generalized Linear Models}. manual. \url{https://fbertran.github.io/homepage/}.

\leavevmode\vadjust pre{\hypertarget{ref-branco16}{}}%
Branco, Paula, Rita P. Ribeiro, and Luis Torgo. 2016. {``{UBL}: An r Package for Utility-Based Learning.''} \url{https://doi.org/10.48550/arXiv.1604.08079}.

\leavevmode\vadjust pre{\hypertarget{ref-brier1950verification}{}}%
Brier, Glenn W et al. 1950. {``Verification of Forecasts Expressed in Terms of Probability.''} \emph{Monthly Weather Review} 78 (1): 1--3.

\leavevmode\vadjust pre{\hypertarget{ref-brown1998refactoring}{}}%
Brown, William J, Raphael C Malveau, Hays W McCormick III, and Thomas J Mowbray. 1998. {``Refactoring Software, Architectures, and Projects in Crisis.''} John Wiley; Sons, Inc, Canada.

\leavevmode\vadjust pre{\hypertarget{ref-byrd1995lbfgsb}{}}%
Byrd, Richard H., Peihuang Lu, Jorge Nocedal, and Ciyou Zhu. 1995. {``A Limited Memory Algorithm for Bound Constrained Optimization.''} \emph{SIAM Journal on Scientific Computing} 16 (5): 1190--1208. \url{https://doi.org/10.1137/0916069}.

\leavevmode\vadjust pre{\hypertarget{ref-chawla2002smote}{}}%
Chawla, Nitesh V, Kevin W Bowyer, Lawrence O Hall, and W Philip Kegelmeyer. 2002. {``SMOTE: Synthetic Minority over-Sampling Technique.''} \emph{Journal of Artificial Intelligence Research} 16: 321--57.

\leavevmode\vadjust pre{\hypertarget{ref-cicchetti1981}{}}%
Cicchetti, D V, and S A Sparrow. 1981. {``{{D}eveloping criteria for establishing interrater reliability of specific items: applications to assessment of adaptive behavior}.''} \emph{Am. J. Ment. Defic.} 86 (2): 127--37.

\leavevmode\vadjust pre{\hypertarget{ref-cohen1968weighted}{}}%
Cohen, Jacob. 1968. {``Weighted Kappa: Nominal Scale Agreement Provision for Scaled Disagreement or Partial Credit.''} \emph{Psychological Bulletin} 70 (4): 213.

\leavevmode\vadjust pre{\hypertarget{ref-darst_using_2018}{}}%
Darst, Burcu F., Kristen C. Malecki, and Corinne D. Engelman. 2018. {``Using Recursive Feature Elimination in Random Forest to Account for Correlated Variables in High Dimensional Data.''} \emph{{BMC} Genetics} 19 (1): 65. \url{https://doi.org/10.1186/s12863-018-0633-8}.

\leavevmode\vadjust pre{\hypertarget{ref-delgado14}{}}%
Delgado, Manuel Fernández, Eva Cernadas, Senén Barro, and Dinani Gomes Amorim. 2014. {``Do We Need Hundreds of Classifiers to Solve Real World Classification Problems?''} \emph{J. Mach. Learn. Res.} 15 (1): 3133--81. \url{https://doi.org/10.5555/2627435.2697065}.

\leavevmode\vadjust pre{\hypertarget{ref-Fleiss1981}{}}%
Fleiss, Joseph L. 1981. \emph{Statistical Methods for Rates and Proportions}. 2nd ed. Probability \& Mathematical Statistics s. Nashville, TN: John Wiley \& Sons.

\leavevmode\vadjust pre{\hypertarget{ref-gerber2019}{}}%
Gerber, Florian, and Reinhard Furrer. 2019. {``optimParallel: An r Package Providing a Parallel Version of the l-BFGS-b Optimization Method.''} \emph{The R Journal} 11 (1): 352--58. \url{https://doi.org/10.32614/RJ-2019-030}.

\leavevmode\vadjust pre{\hypertarget{ref-giorgino2009}{}}%
Giorgino, Toni. 2009. {``Computing and Visualizing Dynamic Time Warping Alignments in {R}: The {dtw} Package.''} \emph{Journal of Statistical Software} 31 (7): 1--24. \url{https://doi.org/10.18637/jss.v031.i07}.

\leavevmode\vadjust pre{\hypertarget{ref-goodfellow2014generative}{}}%
Goodfellow, Ian J., Jean Pouget-Abadie, Mehdi Mirza, Bing Xu, David Warde-Farley, Sherjil Ozair, Aaron Courville, and Yoshua Bengio. 2014. {``Generative Adversarial Networks.''} \url{https://arxiv.org/abs/1406.2661}.

\leavevmode\vadjust pre{\hypertarget{ref-green1993nonparametric}{}}%
Green, Peter J, and Bernard W Silverman. 1993. \emph{Nonparametric Regression and Generalized Linear Models: A Roughness Penalty Approach}. Crc Press.

\leavevmode\vadjust pre{\hypertarget{ref-gwet2008ac1}{}}%
Gwet, Kilem Li. 2008. {``Computing Inter-Rater Reliability and Its Variance in the Presence of High Agreement.''} \emph{British Journal of Mathematical and Statistical Psychology} 61 (1): 29--48. \url{https://doi.org/10.1348/000711006X126600}.

\leavevmode\vadjust pre{\hypertarget{ref-gwet2014}{}}%
---------. 2014. \emph{Handbook of Inter-Rater Reliability: The Definitive Guide to Measuring the Extent of Agreement Among Raters}. 4th ed. Advanced Analytics.

\leavevmode\vadjust pre{\hypertarget{ref-rpkg_gamboost}{}}%
Hofner, Benjamin, Luigi Boccuto, and Markus Göker. 2015. {``Controlling False Discoveries in High-Dimensional Situations: Boosting with Stability Selection.''} \emph{{BMC} Bioinformatics} 16 (1): 144. \url{https://doi.org/10.1186/s12859-015-0575-3}.

\leavevmode\vadjust pre{\hypertarget{ref-honel2020gitdens}{}}%
Hönel, Sebastian. 2022. {``{Git Density 2022.10: Analyze git repositories to extract the Source Code Density and other Commit Properties},''} October. \url{https://doi.org/10.5281/zenodo.7148940}.

\leavevmode\vadjust pre{\hypertarget{ref-honel2021technical}{}}%
---------. 2023. {``Technical Reports Compilation: Detecting the Fire Drill Anti-Pattern Using Source Code and Issue-Tracking Data.''} \emph{CoRR}. arXiv. \url{https://doi.org/10.48550/arXiv.2104.15090}.

\leavevmode\vadjust pre{\hypertarget{ref-honel2020using}{}}%
Hönel, Sebastian, Morgan Ericsson, Welf Löwe, and Anna Wingkvist. 2020. {``Using Source Code Density to Improve the Accuracy of Automatic Commit Classification into Maintenance Activities.''} \emph{J. Syst. Softw.} 168: 110673. \url{https://doi.org/10.1016/j.jss.2020.110673}.

\leavevmode\vadjust pre{\hypertarget{ref-honel2022pareto}{}}%
Hönel, Sebastian, and Welf Löwe. 2022. {``An Approach to Ordering Objectives and Pareto Efficient Solutions.''} arXiv. \url{https://doi.org/10.48550/arXiv.2205.15291}.

\leavevmode\vadjust pre{\hypertarget{ref-honel2023embedded}{}}%
Hönel, Sebastian, Petr Picha, Morgan Ericsson, Premek Brada, Welf Löwe, and Anna Wingkvist. 2023. {``Activity-Based Detection of Pattern-Like Phenomena: An Embedded Case Study of the Fire Drill.''} \emph{E-Informatica Software Engineering Journal}, February.

\leavevmode\vadjust pre{\hypertarget{ref-honel_picha_2021}{}}%
Hönel, Sebastian, Petr Pícha, Premek Brada, Lenka Rychtarova, and Jakub Danek. 2023. {``{Detection of the Fire Drill anti-pattern: 15 real-world projects with ground truth, issue-tracking data, source code density, models and code}.''} Zenodo. \url{https://doi.org/10.5281/zenodo.4734053}.

\leavevmode\vadjust pre{\hypertarget{ref-James1984}{}}%
James, Lawrence R., Robert G. Demaree, and Gerrit Wolf. 1984. {``Estimating Within-Group Interrater Reliability with and Without Response Bias.''} \emph{Journal of Applied Psychology} 69 (1): 85--98. \url{https://doi.org/10.1037/0021-9010.69.1.85}.

\leavevmode\vadjust pre{\hypertarget{ref-klein2018agreement}{}}%
Klein, Daniel. 2018. {``Implementing a General Framework for Assessing Interrater Agreement in Stata.''} \emph{The Stata Journal} 18 (4): 871--901. \url{https://doi.org/10.1177/1536867X1801800408}.

\leavevmode\vadjust pre{\hypertarget{ref-landis1977application}{}}%
Landis, J Richard, and Gary G Koch. 1977. {``An Application of Hierarchical Kappa-Type Statistics in the Assessment of Majority Agreement Among Multiple Observers.''} \emph{Biometrics}, 363--74.

\leavevmode\vadjust pre{\hypertarget{ref-vandermaaten08a}{}}%
Maaten, Laurens van der, and Geoffrey Hinton. 2008. {``Visualizing Data Using t-SNE.''} \emph{Journal of Machine Learning Research} 9 (86): 2579--2605. \url{http://jmlr.org/papers/v9/vandermaaten08a.html}.

\leavevmode\vadjust pre{\hypertarget{ref-mockus2000change}{}}%
Mockus, Audris, and Lawrence G. Votta. 2000. {``Identifying Reasons for Software Changes Using Historic Databases.''} In \emph{2000 International Conference on Software Maintenance, {ICSM} 2000, San Jose, California, USA, October 11-14, 2000}, 120--30. {IEEE} Computer Society. \url{https://doi.org/10.1109/ICSM.2000.883028}.

\leavevmode\vadjust pre{\hypertarget{ref-ONeill2017}{}}%
ONeill, Thomas A. 2017. {``An Overview of Interrater Agreement on Likert Scales for Researchers and Practitioners.''} \emph{Frontiers in Psychology} 8 (May). \url{https://doi.org/10.3389/fpsyg.2017.00777}.

\leavevmode\vadjust pre{\hypertarget{ref-rpkg_ipred}{}}%
Peters, Andrea, and Torsten Hothorn. 2019. \emph{Ipred: Improved Predictors}. \url{https://CRAN.R-project.org/package=ipred}.

\leavevmode\vadjust pre{\hypertarget{ref-picha2022Firedrill}{}}%
Picha, Petr, Sebastian Hönel, Premek Brada, Morgan Ericsson, Welf Löwe, Anna Wingkvist, and Jakub Danek. 2022. {``Process Anti-Pattern Detection in Student Projects -- a Case Study.''} In \emph{EuroPLoP'22: European Conference on Pattern Languages of Programs 2022, Irsee, Deutschland, July 6 - 10, 2022}. {ACM}. \url{https://doi.org/10.1145/3551902.3551965}.

\leavevmode\vadjust pre{\hypertarget{ref-pukelsheim1994threesigma}{}}%
Pukelsheim, Friedrich. 1994. {``The Three Sigma Rule.''} \emph{The American Statistician} 48 (2): 88--91. \url{https://doi.org/10.2307/2684253}.

\leavevmode\vadjust pre{\hypertarget{ref-Regier2013}{}}%
Regier, Darrel A, William E Narrow, Diana E Clarke, Helena C Kraemer, S Janet Kuramoto, Emily A Kuhl, and David J Kupfer. 2013. {``{DSM-5} Field Trials in the United States and Canada, Part {II}: Test-Retest Reliability of Selected Categorical Diagnoses.''} \emph{Am. J. Psychiatry} 170 (1): 59--70. \url{https://doi.org/10.1176/appi.ajp.2012.12070999}.

\leavevmode\vadjust pre{\hypertarget{ref-royston1982}{}}%
Royston, J. P. 1982. {``Algorithm AS 181: The w Test for Normality.''} \emph{Journal of the Royal Statistical Society. Series C (Applied Statistics)} 31 (2): 176--80. \url{http://www.jstor.org/stable/2347986}.

\leavevmode\vadjust pre{\hypertarget{ref-selberg1940}{}}%
Selberg, Henrik L. 1940. {``Zwei Ungleichungen Sur Ergänzung Des Tchebycheffschen Lemmas.''} \emph{Scandinavian Actuarial Journal} 1940 (3-4): 121--25. \url{https://doi.org/10.1080/03461238.1940.10404804}.

\leavevmode\vadjust pre{\hypertarget{ref-shapiro1965}{}}%
Shapiro, S. S., and M. B. Wilk. 1965a. {``An Analysis of Variance Test for Normality (Complete Samples).''} \emph{Biometrika} 52 (3/4): 591--611. \url{http://www.jstor.org/stable/2333709}.

\leavevmode\vadjust pre{\hypertarget{ref-shapiro1965normal}{}}%
---------. 1965b. {``An Analysis of Variance Test for Normality (Complete Samples).''} \emph{Biometrika} 52 (3-4): 591--611. \url{https://doi.org/10.1093/biomet/52.3-4.591}.

\leavevmode\vadjust pre{\hypertarget{ref-silva2015software}{}}%
Silva, Pedro, Ana M Moreno, and Lawrence Peters. 2015. {``Software Project Management: Learning from Our Mistakes {[}Voice of Evidence{]}.''} \emph{IEEE Software} 32 (03): 40--43.

\leavevmode\vadjust pre{\hypertarget{ref-stephens1974}{}}%
Stephens, M. A. 1974. {``EDF Statistics for Goodness of Fit and Some Comparisons.''} \emph{Journal of the American Statistical Association} 69 (347): 730--37. \url{https://doi.org/10.1080/01621459.1974.10480196}.

\leavevmode\vadjust pre{\hypertarget{ref-swanson1976dimensions}{}}%
Swanson, E. Burton. 1976. {``The Dimensions of Maintenance.''} In \emph{Proceedings of the 2nd International Conference on Software Engineering, San Francisco, California, USA, October 13-15, 1976}, edited by Raymond T. Yeh and C. V. Ramamoorthy, 492--97. {IEEE} Computer Society. \url{https://doi.org/10.5555/800253.807723}.

\leavevmode\vadjust pre{\hypertarget{ref-chebyshev1867}{}}%
Tchébychef, Pafnuty. 1867. {``Des Valeurs Moyennes (Traduction Du Russe, n. De Khanikof.''} \emph{Journal de Mathématiques Pures Et Appliquées}, 177--84. \url{http://eudml.org/doc/234989}.

\leavevmode\vadjust pre{\hypertarget{ref-torgo13}{}}%
Torgo, Luís, Rita P. Ribeiro, Bernhard Pfahringer, and Paula Branco. 2013. {``{SMOTE} for Regression.''} In \emph{Progress in Artificial Intelligence - 16th Portuguese Conference on Artificial Intelligence, {EPIA} 2013, Angra Do Heroísmo, Azores, Portugal, September 9-12, 2013. Proceedings}, edited by Luís Correia, Luís Paulo Reis, and José Cascalho, 8154:378--89. Lecture Notes in Computer Science. Springer. \url{https://doi.org/10.1007/978-3-642-40669-0/_33}.

\leavevmode\vadjust pre{\hypertarget{ref-venables2002sstat}{}}%
Venables, W. N., and B. D. Ripley. 2002. \emph{Modern Applied Statistics with s}. Fourth. New York: Springer. \url{http://www.stats.ox.ac.uk/pub/MASS4}.

\leavevmode\vadjust pre{\hypertarget{ref-vysochanskij1980justification}{}}%
Vysochanskij, DF, and Yu I Petunin. 1980. {``Justification of the 3\(\sigma\) Rule for Unimodal Distributions.''} \emph{Theory of Probability and Mathematical Statistics} 21 (25-36).

\leavevmode\vadjust pre{\hypertarget{ref-xi2000bearing}{}}%
Xi, Fengfeng, Qiao Sun, and Govindappa Krishnappa. 2000. {``Bearing Diagnostics Based on Pattern Recognition of Statistical Parameters.''} \emph{Journal of Vibration and Control} 6 (3): 375--92. \url{http://www.acoustics.asn.au/conference_proceedings/ICSVS-1997/pdf/scan/sv970356.pdf}.

\end{CSLReferences}

\end{document}